\documentclass[%
twocolumn,
amsmath,amssymb,
aps,
longbibliography, superscriptaddress
]{revtex4-2}
\usepackage[utf8]{inputenc}
\usepackage{amsmath}
\usepackage{amsfonts}
\usepackage{amssymb}
\usepackage{braket}
\usepackage{graphicx}
\usepackage{float}
\usepackage[percent]{overpic}
\usepackage{tikz}
\usepackage{bm}
\usetikzlibrary{arrows,decorations.pathmorphing,backgrounds,positioning,fit,petri,decorations.pathreplacing,decorations.markings}
\usepackage{xcolor}
\definecolor{mygreen1}{rgb}{0, 0.4, 0}
\usepackage[margin=0.7in]{geometry}
\newcommand{\mbeq}{\overset{!}{=}}
\usepackage{pifont}
\usepackage{hyperref}
\hypersetup{
	colorlinks=true,
	linkcolor=blue,
	filecolor=magenta,
	urlcolor=cyan,
	citecolor=blue
}
\newcommand{\xmark}{\ding{55}}
\newcommand{\cmark}{\ding{51}}
\tikzset{
	on each segment/.style={
		decorate,
		decoration={
			show path construction,
			moveto code={},
			lineto code={
				\path [#1]
				(\tikzinputsegmentfirst) -- (\tikzinputsegmentlast);
			},
			curveto code={
				\path [#1] (\tikzinputsegmentfirst)
				.. controls
				(\tikzinputsegmentsupporta) and (\tikzinputsegmentsupportb)
				..
				(\tikzinputsegmentlast);
			},
			closepath code={
				\path [#1]
				(\tikzinputsegmentfirst) -- (\tikzinputsegmentlast);
			},
		},
	},
	mid arrow/.style={postaction={decorate,decoration={
				markings,
				mark=at position .5 with {\arrow[#1]{stealth}}
	}}},
}
\begin{document}
	
	\title{Excitations in the higher lattice gauge theory model for topological phases III: the (3+1)-dimensional case}
	\author{Joe Huxford}
	\affiliation{Rudolf Peierls Centre for Theoretical Physics, Clarendon Laboratory, Oxford OX1 3PU, UK}
	\affiliation{Department of Physics, University of Toronto, Ontario M5S 1A7, Canada}
	\author{Steven H. Simon}
\affiliation{Rudolf Peierls Centre for Theoretical Physics, Clarendon Laboratory, Oxford OX1 3PU, UK}

	\begin{abstract}
	
	In this, the third paper in our series describing the excitations of the higher lattice gauge theory model for topological phases, we will examine the 3+1d case in detail. We will explicitly construct the ribbon and membrane operators which create the topological excitations, and use these creation operators to find the pattern of condensation and confinement. We also use these operators to find the braiding relations of the excitations, and to construct charge measurement operators which project to states of definite topological charge.

	\end{abstract}
	
		\maketitle
	\tableofcontents

		\section{Introduction}

		Over the past decade, there has been significant consideration given to topological phases in 3+1d, in addition to the 1+1d and 2+1d cases that have been more heavily studied in the past. So far, results for 3+1d topological phases range from the construction of classes of commuting projector models \cite{Walker2012, Wan2015, Williamson2017, Bullivant2017} to a potential classification of the bosonic phases in the absence of symmetry \cite{Lan2018, Lan2019}. 3+1d phases are intriguing for several reasons. Firstly, we live in a 3+1d world, and so there is a natural interest in studying such phases. Secondly, the properties of 3+1d topological phases are quite different from their 2+1d cousins. Unlike their 2+1d counterparts, point-like particles in 3+1d are not expected to have non-trivial braiding with each-other (outside of the usual bosonic and fermionic cases) \cite{Doplicher1971, Doplicher1974}. However, in these higher-dimensional phases there exist loop-like excitations with non-trivial loop-braiding properties, so that exchange processes involving the loop-like excitations can result in a non-trivial transformation for the state of the system. Unfortunately, at present there are few examples of toy models where these loop-like excitations are constructed explicitly and the properties of the excitations (both point-like and loop-like) are studied in more depth. In this series of papers, we study one such toy model  \cite{Bullivant2017}, based on higher lattice gauge theory \cite{Pfeiffer2003, Baez2010, Gukov2013, Kapustin2014, Kapustin2017, Bullivant2017}, and aim to provide a detailed description of the excitations, and the conserved topological charge which they carry.

		As we have already seen in Refs. \cite{HuxfordPaper1, HuxfordPaper2}, the Hamiltonian model for topological phases based on higher lattice gauge theory \cite{Bullivant2017} hosts rich physics, including non-trivial loop braiding (in the 3+1d case), condensation and confinement. In Ref. \cite{HuxfordPaper1}, we gave a brief qualitative description of these features, while in Ref. \cite{HuxfordPaper2} we examined the 2+1d model. Now we will give a more explicit and mathematically detailed treatment of the 3+1d model, with full proofs presented in the Supplemental Material. Our approach focuses heavily on the so-called ribbon and membrane operators, which produce (as well as move and annihilate) the point-like and loop-like excitations of the model respectively. Because loops are extended objects, they sweep through a surface as they move, rather than just a line as point particles would. Therefore, to produce and move loop excitations we need an operator that is defined across a general membrane rather than a line or ribbon. The ribbon and membrane operators can be used to find the braiding statistics \cite{Kitaev2003, Levin2005} of the excitations, while closed ribbon and membrane operators can be used to measure topological charge \cite{Bombin2008}. In addition, certain properties of the excitations, such as whether they are confined or not, can be obtained directly from these operators. In this paper, we therefore aim to provide and justify the mathematical forms of the ribbon and membrane operators, and demonstrate how we can extract all of the previously mentioned information about the excitations from them.

		\subsection{Structure of this paper}

		In this paper, we will consider the 3+1d model in various cases (a summary of these cases is presented in Section \ref{Section_Recap_3d}, along with a brief reminder of the Hamiltonian model). For each case we define the membrane operators and find the effects of braiding in turn, before moving on to the next case. In Sections \ref{Section_3D_MO_Tri_Trivial} and \ref{Section_3D_Braiding_Tri_Trivial} we consider the case where one of the maps describing the model, $\rhd$, is trivial (Case 1 from Table \ref{Table_Cases}). In Section \ref{Section_3D_MO_Tri_Trivial} we construct the ribbon and membrane operators for the theory and discuss the pattern of condensation and confinement exhibited by the excitations that they produce. In Section \ref{Section_3D_Braiding_Tri_Trivial} we use these operators to work out the braiding properties of these excitations, which involves passing loop or point particles through loops. In Sections \ref{Section_3D_MO_Fake_Flat} and \ref{Section_3D_Braiding_Fake_Flat} we repeat the construction of ribbon operators and the braiding for another special case, called the fake-flat case (Case 3 from Table \ref{Table_Cases}), while in Sections \ref{Section_3D_MO_Central} and \ref{Section_3D_Braiding_Central} we repeat it for another special case (Case 2 from Table \ref{Table_Cases}), which generalizes the $\rhd$ trivial case.

		Having found the membrane operators and effects of braiding, in Section \ref{Section_3D_Topological_Sectors} we move on to consider the topological charges of the model. These topological charges are conserved quantities carried by the excitations of the model. In 2+1d, to measure the topological charge in a spatial region, we simply put an operator on the boundary of that region. This boundary will be topologically equivalent to a circle (or multiple circles). However, in 3+1d there are more topologically distinct surfaces which can enclose our regions of interest. We can use these different surfaces to measure the loop-like and point-like charge carried by the excitations. Using a sphere as our surface of measurement, we can determine the point-like charge contained within the sphere. We present the corresponding charge measurement operators in Section \ref{Section_Sphere_Charge_Reduced}, and also find the charge carried by some simple excitations. However, to measure loop-like charge we need some surface with non-contractible loops. An important example is the torus, which we look at in some detail in Section \ref{Section_Torus_Charge}. We compare the number of topological charges that we can measure with the torus to the ground state degeneracy of the 3-torus and find that they are equivalent, in the broad cases that we look at, as previously reported in Ref. \cite{Bullivant2020}. Finally, in Section \ref{Section_conclusion_3d}, we summarize our results and propose further avenues of research based on this work.

		In the Supplemental Material, we present the proofs of our results that were too lengthy to include in the main text. We demonstrate the commutation relation between the energy terms and the ribbon and membrane operators in Section \ref{Section_ribbon_membrane_energy_commutation} (using some results from the 2+1d case discussed in Ref. \cite{HuxfordPaper2}). Then in Section \ref{Section_topological_membrane_operators} we demonstrate that the non-confined ribbon and membrane operators are \textit{topological}, meaning that we can deform them through unexcited regions of space without affecting their action, provided that we keep the locations of any excitations they produce fixed. In Section \ref{Section_magnetic_condensation}, we show that some of the magnetic loop excitations are condensed, and can be produced by operators only acting near the excitations (meaning that they cannot carry loop-like topological charge). Next, in Section \ref{Section_braiding_supplement} we find the braiding relations of the various excitations by explicitly calculating the appropriate commutation relations between the membrane and ribbon operators. Finally, in Section \ref{Section_topological_sectors_supplement} we construct the measurement operators for topological charge and demonstrate that they are projectors.

		\section{Summary of the model}
		\label{Section_Recap_3d}

		In this section we will remind the reader of the Hamiltonian model we are studying, the higher lattice gauge theory model introduced in Ref. \cite{Bullivant2017}. We hope that this will provide a convenient place for the reader to refer back to for definitions of the various terms in the Hamiltonian, along with several useful identities.

		We are considering the model defined on a 3d lattice, representing the spatial degrees of freedom, with a Hamiltonian controlling the time evolution. The edges of the lattice are directed, while the plaquettes have a circulation and a base-point (a privileged vertex which we can think of as the start of the circulation). The edges are labeled by elements of a group $G$, and the plaquettes are labeled by elements of a second group, $E$. These groups are part of a \textit{crossed module}, which consists of the two groups and two maps, $\partial$ and $\rhd$. Here $\partial$ is a group homomorphism from $E$ to $G$, while $\rhd$ is a group homomorphism from $G$ to the automorphisms on $E$. That is, for each element $g \in G$, $g \rhd$ is a group isomorphism from $E$ to itself (so for $e \in E$, $g \rhd e$ is an element of $E$). These maps satisfy two additional constraints, called the Peiffer conditions \cite{Pfeiffer2003, Baez2002, Bullivant2017}:
		\begin{align}
		\partial(g \rhd e) &= g \partial(e)g^{-1} \ \forall g \in G, e \in E \label{Equation_Peiffer_1}\\
		\partial(e) \rhd f &= efe^{-1} \ \forall e,f \in E \label{Equation_Peiffer_2}.
		\end{align}
		
		The Hamiltonian is given by a sum of projectors, with terms for the vertices, edges, plaquettes and blobs (3-cells) of the lattice \cite{Bullivant2017}:
		\begin{equation}
		H = - \hspace{-0.1cm} \sum_{\text{vertices, }v} \hspace{-0.4cm} A_v \: - \sum_{\text{edges, } i} \hspace{-0.2cm} \mathcal{A}_i \: -\hspace{-0.3cm}\sum_{\text{plaquettes, }p} \hspace{-0.5cm} B_p \: - \sum_{\text{blobs, }b} \hspace{-0.2cm} \mathcal{B}_b. \label{Equation_Hamiltonian_3d}
		\end{equation}
	
		The vertex terms are a sum of vertex transforms, and can be thought of as projecting to states that are 1-gauge invariant. That is
		$$A_v = \frac{1}{|G|} \sum_{g \in G} A_v^g,$$
		where the vertex transforms have the algebra $A_v^g A_v^h =A_v^{gh}$, which implies that $A_v^g A_v =A_v$. This ensures that the ground states (which are eigenstates of $A_v$ with eigenvalue one) are invariant under the vertex transforms:
		$$A_v^g \ket{GS} = A_v^g A_v \ket{GS} =A_v \ket{GS} = \ket{GS}.$$
	
		As for the specific action of the vertex transforms, they act on the edges adjacent to the vertex, as well as any plaquette whose base-point is that vertex. For an edge $i$ (initially labeled by $g_i$) or plaquette $p$ (initially labeled by $e_p$), we have
		\begin{align}
		A_v^g: g_{i} &\rightarrow \begin{cases} gg_{i} &\text{ if $v$ is the start of $i$}\\
		g_ig^{-1} &\text{ if $v$ is the end of $i$}\\
		g_{i} &\text{ otherwise} \end{cases} \notag\\
		A_v^g : e_p &\rightarrow \begin{cases} g \rhd e_p &\text{if $v$ is the base-point of $p$}\\
		e_p &\text{otherwise.} \label{Equation_vertex_transform_definition}\end{cases}
		\end{align}
	
		Similarly, the edge term is a sum of edge transforms (2-gauge transforms)
		$$\mathcal{A}_i = \frac{1}{|E|} \sum_{e \in E} \mathcal{A}_i^e,$$
		which satisfy a similar algebra to the vertex transforms: $\mathcal{A}_i^e \mathcal{A}_i^f =\mathcal{A}_i^{ef}$. This ensures that individual edge transforms can be absorbed into the corresponding edge term, and into the ground state: $\mathcal{A}_i^e \mathcal{A}_i = \mathcal{A}_i$ and $\mathcal{A}_i^e \ket{GS} = \ket{GS}$. An edge transform $\mathcal{A}_i^e$ applied on an edge $i$ acts on the label of edge $i$ itself, as well as the labels of the adjacent plaquettes:
		\begin{align}
		\mathcal{A}_i^e: g_{j} &\rightarrow \begin{cases} \partial(e) g_{j} &\text{ if $i=j$}\\
		g_{j} &\text{ otherwise} \end{cases} \notag \\
		\mathcal{A}_i^e : e_p &\rightarrow \begin{cases} e_p [g(v_0(p) - s(i)) \rhd e^{-1}] &\text{if $i$ is on $p$ and}\\& \text{aligned with $p$}\\
		[g(\overline{v_0(p) - s(i)}) \rhd e] e_p &\text{if $i$ is on $p$ and}\\& \text{aligned against $p$}\\
		e_p &\text{otherwise.} \end{cases} \label{Equation_edge_transform_definition}
		\end{align}
		Here $s(i)$ is the source of edge $i$, which is the vertex attached to $i$ that $i$ points away from (with the vertex on the other end of $i$ being called the target). $g(v_0(p) - s(i))$ is the path element for the path from the base-point of plaquette $p$ to this source, running around the plaquette and aligned with the plaquette. On the other hand $g(\overline{v_0(p) - s(i)})$ is the path element for the path around the plaquette from $v_0(p)$ to $s(i)$, but this time anti-aligned with the plaquette.

		The next energy term is the plaquette term, $B_p$, which enforces so-called \textit{fake-flatness}. This is similar to the plaquette term from Kitaev's Quantum Double model, in that it restricts which labels the boundary of a plaquette can have. Unlike the term from the Quantum Double model however, the plaquette term in higher lattice gauge theory relates the label of the boundary to the surface label of the plaquette itself, rather than requiring the boundary label to be trivial as for the Quantum Double model. For a plaquette whose boundary (starting at the base-point and aligned with the circulation of the plaquette) has path label $\hat{g}_p$, and whose surface label is $\hat{e}_p$, the plaquette term $B_p$ acts as 
		\begin{equation}
		B_p =\delta(\partial(\hat{e}_p)\hat{g}_p, 1_G).
		\end{equation}
		 A plaquette which satisfies this Kronecker delta is called fake-flat, and a surface made from fake-flat plaquettes is also called fake-flat. Such a fake-flat surface will satisfy a similar condition on its surface and boundary labels. As we showed in Ref. \cite{HuxfordPaper2}, for a surface $m$ whose constituent plaquettes satisfy fake-flatness, the overall surface element will satisfy
		 $$\partial(\hat{e}(m))\hat{g}_{dm}=1_G,$$
		 where $\hat{g}_{dm}$ is the group element associated to the boundary of $m$ and the total surface element is constructed by combining individual surface elements, as explained in Ref. \cite{Bullivant2017} and as we will summarize shortly. Note that this fake-flatness condition enables the presence of closed paths with non-trivial label in the ground state (indeed such closed paths are created by the edge transforms), indicating that some magnetic fluxes proliferate in the ground state and so are condensed.

		The final energy term is the blob term, $\mathcal{B}_b$, which enforces that the surface element of the boundary of the blob (calculated from the plaquettes on that blob using the rules for combining surfaces explained in Ref. \cite{Bullivant2017}, which we will describe shortly) is equal to the identity element $1_E$. That is 
		\begin{equation}
		\mathcal{B}_b = \delta(\hat{e}(b),1_E), \label{Equation_blob_term_definition}
		\end{equation}
		 for blob $b$ with surface element $\hat{e}(b)$.

\begin{figure}[h]
		 	\begin{center}
		 		
		 		\includegraphics[width=\linewidth]{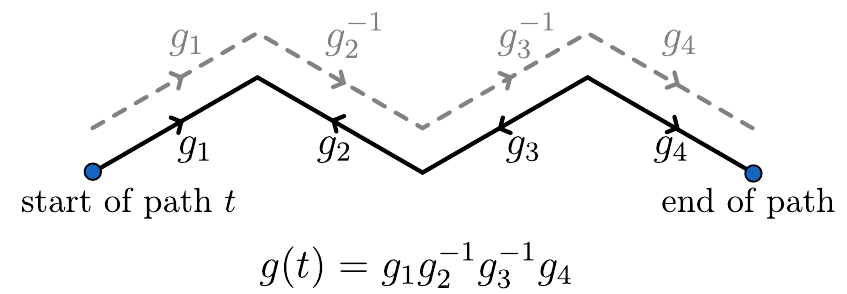}
		 		
		 		\caption{(Copy of Figure 37 from Ref. \cite{HuxfordPaper1}) An electric ribbon operator measures the value of a path and assigns a weight to each possibility, creating excitations at the two ends of the path. In order to find the group element associated to the path, we must first find the contribution of each edge to the path. In this example, the edges along the path are shown in black. Some of the edges are anti-aligned with the path and so we must invert the elements associated to these edges to find their contribution to the path. This is represented by the grey dashed lines, which are labeled with the contribution of each edge to the path.}
		 		\label{path_image_paper_3}
		 	\end{center}
		 \end{figure}

		 A key idea in the higher lattice gauge theory model is that we can compose edges into paths and plaquettes into surfaces, with composite objects appearing throughout the description of the model as well as in the ribbon and membrane operators. We will therefore briefly review the rules for this kind of combination. First, consider composing edges into paths. If two edges (or more general paths) lie end-to-end, then we can combine them into one path, with a group label given by the product of the elements for the two edges. Then to combine multiple edges into a path, we take a product of all of the constituent edge labels, with the first edge on the path appearing on the left of the product. If one or more of the edges points against the path (for example, the edges labeled by $g_2$ and $g_3$ in Figure \ref{path_image_paper_3}), then we include the edge label in the path element with an inverse.
		 
\begin{figure}[h]
		 	\begin{center}
		 		\includegraphics[width=\linewidth]{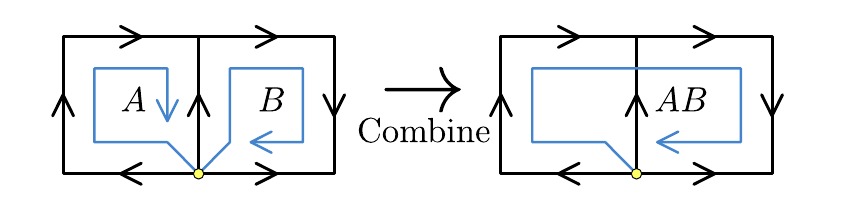}
		 		
		 		\caption{Two adjacent surfaces can be combined into one if their base-point (represented by the yellow dot) and circulation (represented by the blue arrow in the middle of each plaquette) match. The label of the combined surface is given by the product of the two individual elements in reverse order. That is, if the surfaces $A$ and $B$ have labels $e_A$ and $e_B$ respectively, then the combined surface has label $e_{AB}=e_B e_A$. If two adjacent surfaces do not have the same base-point and orientation then we can still combine them by using a set of rules that describe what happens when we change the orientation or move the base-point of a surface.}
		 		\label{combine_two_surfaces_elements}
		 	\end{center}
		 \end{figure}

		 Next consider composition of plaquettes, or more generally surfaces. Surfaces have both an orientation and a privileged vertex, called the base-point, which we can view as the start of the circulation. We represent this by drawing a circulating arrow in the plaquette which connects to the boundary at the base-point, as illustrated in Figure \ref{combine_two_surfaces_elements}. When we combine two adjacent plaquettes, we must ensure that the base-points and orientations of the plaquettes both agree. If they do, as in the example shown in Figure \ref{combine_two_surfaces_elements}, we can combine the plaquettes into a single surface whose label is a product of the two plaquettes. Contrary to the case of paths, the plaquette appearing first in the circulation is represented in the rightmost position of the product.
		 
\begin{figure}[h]
		 	\begin{center}	
		 		\includegraphics[width=\linewidth]{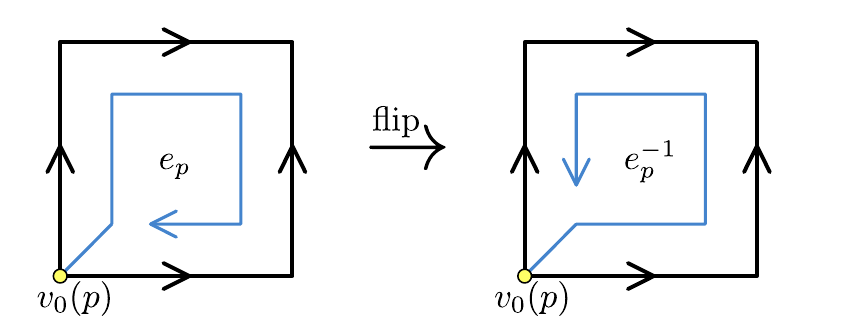}

		 		\caption{Given a plaquette with label $e_p$, the label of the corresponding plaquette with the opposite orientation is $e_p^{-1}$. Note that when we reverse the orientation of a plaquette, we leave its base-point, here $v_0(p)$, in the same position.}
		 		\label{flip_plaquette}
		 	\end{center}	
		 \end{figure}

\begin{figure}[h]
		 	\begin{center}	
		 		\includegraphics[width=\linewidth]{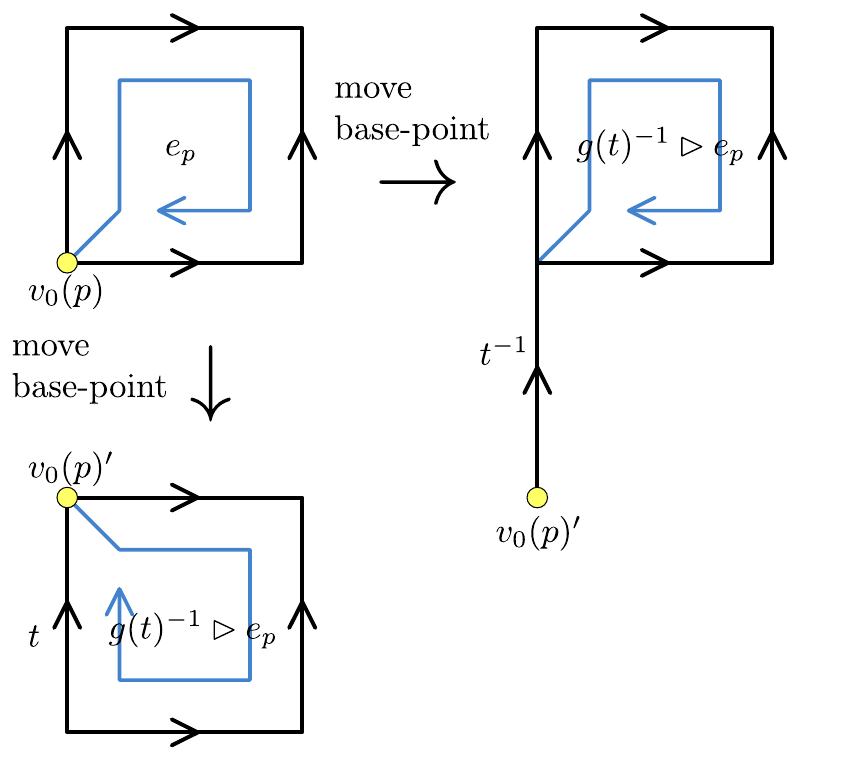}

		 		\caption{We can move the base-point of a surface, either along the boundary of the surface (resulting in the case shown in the bottom-left) or away from the surface (in which case we say that we whisker the surface, and obtain the situation shown in the top-right). When we move the base-point of plaquette $p$ along a path $t$ in this way, the surface label goes from $e_p$ to $g(t)^{-1} \rhd e_p$.}
		 		\label{move_basepoint}
		 	\end{center}	
		 \end{figure}

		 While this simple procedure works if the base-points and orientations of the two plaquettes agree, we will often want to combine adjacent plaquettes for which this is not the case (similar to how we want to combine edges into paths even if their orientations are not all aligned). In this case, we need a procedure for changing the base-point and orientation of a plaquette, and describing the label the plaquette would have with this new decoration. As described in Ref. \cite{HuxfordPaper1}, we can reverse the orientation of a plaquette (while keeping its base-point fixed) by inverting its group label, as shown in Figure \ref{flip_plaquette}. If we want to move the base-point along a path $t$, as shown in Figure \ref{move_basepoint}, then we must act on that plaquette element with $g(t)^{-1} \rhd$, so that the plaquette label goes from $e_p$ to $g(t)^{-1} \rhd e_p$. When moving the base-point in this way, we can either move it along the boundary of the plaquette, as shown in the bottom-left of Figure \ref{move_basepoint}, or we can move it away from the boundary, as shown in the top-right of Figure \ref{move_basepoint}. Combining these two procedures, we see that the general formula for the label of a composite surface $m$ is
		 \begin{equation}
		 \hat{e}(m)= \prod_{p \in m} g(v_0(m)-v_0(p)) \rhd e_p^{\sigma_p},
		 \end{equation}
		 where the $p \in m$ are the constituent plaquettes; $v_0(p)$ is the original base-point of plaquette $p$; $v_0(m)$ is the base-point of the combined surface; and $\sigma_p$ is 1 if the circulation of plaquette $p$ matches the surface and $-1$ otherwise. Note that this formula hides certain complexities, such as the order of the product and the precise definition of the paths $(v_0(m)-v_0(p))$, but often we care about situations where these details do not matter (for example if $E$ is Abelian the order does not matter, and if the surface is also fake-flat then the paths only need to be defined up to deformation).

\begin{figure}[h]
	\begin{center}	
		\includegraphics[width=\linewidth]{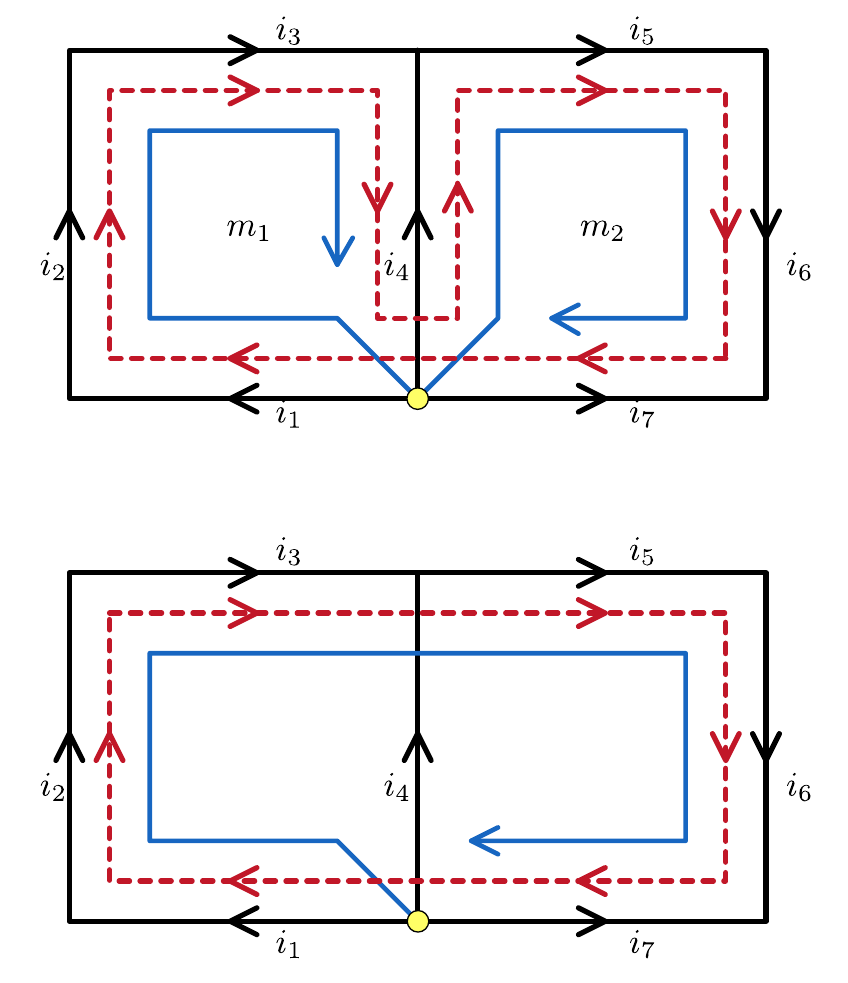}

		\caption{When we combine two surfaces (top image) into one (bottom image), the boundary of the combined surface is the product of the two individual boundaries (here the boundary path is represented as the dashed red line). This boundary can be simplified by removing edges that appear twice consecutively in the boundary with opposite orientation. In this case the combined boundary includes $i_4^{-1}i_4$ in the top image (this is the section that dips down in the image), which can be removed to give the boundary shown in the bottom image.}
		\label{combine_boundaries}
	\end{center}	
\end{figure}

		 One useful way of checking whether we have correctly composed two surfaces it to examine the boundary of the combined surface. The boundary of a surface made by composing two other surfaces is the product of the two individual boundaries, once we have ensured that the orientations and base-points of the two surfaces agree. This product of the boundaries follows from the rules for composing paths given previously. As an example, consider Figure \ref{combine_boundaries}. The boundary of the left surface ($m_1$) in the top image is $i_1 i_2 i_3 i_4^{-1}$ (note that the boundary starts at the base-point and follows the orientation of the plaquette), while the boundary of the right surface ($m_2$) is $i_4i_5i_6 i_7^{-1}$. Here $i_x$ represents an edge, rather than an edge label. The boundary of the combined surface is therefore $i_1 i_2 i_3 i_4^{-1} i_4i_5i_6 i_7^{-1}$. This is the path shown in red in the upper image of Figure \ref{combine_boundaries}. We can simplify this path by removing the section $i_4^{-1}i_4$, to give the boundary $i_1 i_2 i_3 i_4^{-1} i_4i_5i_6 i_7^{-1}$ shown in the lower image. This rule for combining boundaries ensures that the total surface satisfies fake-flatness, if the two constituent surfaces do. If the surface label of surface $m_x$ (for $x= 1$ or 2) is $e_x$, and the boundary bd$(m(x))$ has label $t_x$, then the surface $m_1$ satisfies fake-flatness when 
		 $$\partial(e_1)t_1=1_G$$
		 and the surface $m_2$ satisfies fake-flatness when
		 $$\partial(e_2)t_2=1_G.$$
		 
		 The combined surface has label $e_2e_1$ and boundary label $t_1t_2$ (note the opposite order of composition for the paths and surfaces). If the two constituent surfaces satisfy fake-flatness, then the combined surface label satisfies
		 \begin{align*}
		 \partial(e_2e_1)t_1t_2&= \partial(e_2)(\partial(e_1)t_1)t_2\\
		 &=\partial(e_2) 1_G t_2\\
		 &=1_G,
		 \end{align*}
		 so the total surface satisfies a fake-flatness condition, as we claimed earlier.

		Next, we want to remind the reader of the various special cases in which we consider the model. In the most general case of the model, the projectors in the Hamiltonian (specifically the edge and blob terms) no longer commute \cite{Bullivant2017}. Furthermore, there are inconsistencies with regards changing the branching structure of the lattice (reversing the orientation of edges or plaquettes, or moving the base-points of plaquettes around), as we showed in the Appendix of Ref. \cite{HuxfordPaper1}. This only occurs when $\rhd$ is non-trivial and there are fake-flatness violations (i.e., plaquette excitations), so the ground state space is always well-defined. This may be similar to how the plaquette terms in the string-net model become poorly defined when neighbouring vertex terms are not satisfied and are usually set to zero in such cases \cite{Levin2005}.  Regardless, these inconsistencies make it difficult to define the plaquette excitations, which form flux tubes. We therefore consider the higher lattice gauge theory model in various special cases that remove these inconsistencies, or at least make them more manageable. In the first such special case (Case 1 in Table \ref{Table_Cases}), we take the map $\rhd$ to be trivial, so that each map $g \rhd$ is the identity map ($g \rhd e =e $ for all $g \in G$ and $e \in E$). This leads to a model very similar in character to a 3+1d version of Kitaev's Quantum Double model, but with some additional excitations (and with condensation and confinement). Because of the Peiffer conditions, taking this form for $\rhd$ enforces that $E$ be an Abelian group and $\partial$ maps onto the centre of $G$. In the second special case (Case 2 in Table \ref{Table_Cases}), we instead take these properties as our starting point, allowing $\rhd$ to be general, subject to the constraint that $E$ is Abelian and $\partial$ maps to the centre of $G$. In the final case (Case 3 in Table \ref{Table_Cases}), we do not place any conditions on our crossed module, but instead restrict the Hilbert space to only include fake-flat states (states where all of the plaquette terms are satisfied). This prevents any inconsistencies, but means that some of the excitations are missing.

		\begin{table}[h]
			\begin{center}
				\begin{tabular}{ |c|c|c|c|c| } 
					\hline
					& & & & Full\\
					Case & $E$ & $\rhd$ & $\partial(E)$ & Hilbert \\ 
					& & & &Space\\
					\hline
					1 & Abelian & Trivial & $\subset$ centre($G$) & Yes\\ 
					2 & Abelian & General & $\subset$ centre($G$) & Yes\\ 
					3 & General & General & General & No \\
					\hline
				\end{tabular}
				
				\caption{A reminder of the special cases of the model}
				\label{Table_Cases}
			\end{center}
			
		\end{table}
	
		We mentioned that when $\rhd$ is trivial, the higher lattice gauge theory model becomes similar to a 3+1d version of Kitaev's quantum double model, i.e., to regular lattice gauge theory. Indeed, there are two subcases where the model is equivalent to lattice gauge theory. The first of these is when the group $E$ is the trivial group $\set{1_E}$ and so $\partial$ maps to the identity of $G$. In this case, the blob and edge energy terms become trivial, while the vertex and plaquette terms become the corresponding lattice gauge theory terms. This directly gives the lattice gauge theory Hilbert space and Hamiltonian. The other limit is where the group $G$ is trivial (which also implies that $\partial$ maps to the identity element, because $G$ only has one element). In this case, the vertex and plaquette terms become trivial. Under a change of basis from group elements of $E$ to irreps of $E$, the remaining energy terms become the lattice gauge theory terms on the dual lattice, where the group is the group of irreps of $E$. Specifically, the blob energy term of higher lattice gauge theory becomes the vertex term of lattice gauge theory, while the edge term becomes the plaquette term.  More generally, if both groups $G$ and $E$ are non-trivial, but $\rhd$ is trivial and $\partial$ maps to the identity of $G$, we can see that each of the energy terms given in Equations \ref{Equation_vertex_transform_definition}-\ref{Equation_blob_term_definition} only affects the variables corresponding to one group. This means that these variables decouple and we can think of higher lattice gauge theory as two decoupled lattice gauge theory models in this case. This also helps us to interpret more general cases of the model. If we change $\partial$ to map to a larger subgroup, the two lattice gauge theories interact and produce condensation and confinement, as we will discuss later.

		\section{Ribbon and membrane operators in the $\rhd$ trivial case}
		\label{Section_3D_MO_Tri_Trivial}
		Firstly, we consider Case 1 from Table \ref{Table_Cases}, the case where $\rhd$ is trivial ($g \rhd e=e \: \: \forall e \in E, \: g \in G$), which enforces that $E$ is Abelian.
		
		\subsection{Electric excitations}
		\label{Section_Electric_Tri_Trivial}
		The first type of excitation to consider is the electric excitations. The ribbon operators that produce the electric excitations in 3+1d have the same form as the ones for the 2+1d case that we considered in Ref. \cite{HuxfordPaper2}. That is, an electric ribbon operator measures the group element of a path and assigns a weight depending on the measured group element. As we claimed in Ref. \cite{HuxfordPaper1}, an electric ribbon operator applied on a path $t$ has the form
		\begin{equation}
		\hat{S}^{\vec{\alpha}}(t)= \sum_{g \in G} \alpha_{g} \delta( \hat{g}(t), g),
		\end{equation}
		where $\alpha$ is an arbitrary set of coefficients for each group element $g \in G$ and different choices for these coefficients describe different operators in a space of ribbon operators. A useful basis for this space has basis operators that are labeled by irreps of the group $G$ and the matrix indices for that irrep. These basis electric ribbon operators have the form
		\begin{equation}
		\hat{S}^{R,a,b}(t)= \sum_{g \in G} [D^{R}(g)]_{ab} \delta( \hat{g}(t), g),
		\end{equation}
		where $R$ is an irrep of $G$, $D^{R}(g)$ is the associated matrix representation of element $g$, and $a$ and $b$ are the matrix indices. As we proved in Ref. \cite{HuxfordPaper2} (in the 2+1d case, although the proof also holds for 3+1d), the operators labeled by non-trivial irreps excite the vertices at the ends of the path, whereas the operator labeled by the trivial irrep is the identity operator (which of course does not create any excitations). In addition, just as in the 2+1d case discussed in Ref. \cite{HuxfordPaper2}, the irreps that have non-trivial restriction to the image of $\partial$ label confined excitations. The electric ribbon operators labeled by such irreps cause the edges along the path to be excited, so the excitations produced at the ends of the ribbon are confined.

		\subsection{Magnetic excitations}
		\label{Section_3D_Tri_Trivial_Magnetic_Excitations}
\begin{figure}[h]
			\begin{center}	
				\includegraphics[width=\linewidth]{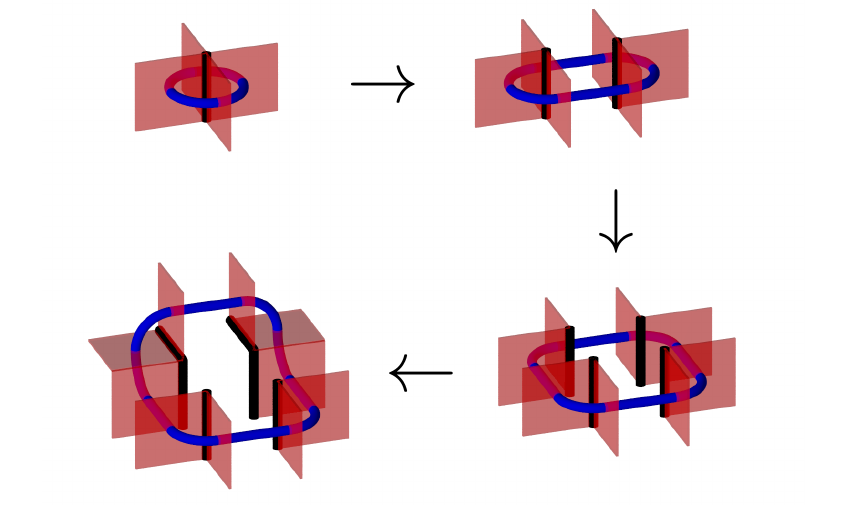}

				\caption{(Copy of Figure 40 from Ref. \cite{HuxfordPaper1}) In order to excite one of the plaquettes in the lattice and produce a magnetic excitation, we change the label one of the edges (black cylinders) on boundary of the plaquette. However, this excites all of the plaquettes (the excited plaquettes are the squares, shown in red) adjacent to that edge, as shown in the first image. Note that these plaquettes lie on a closed loop through their centres. If we change another edge label to try to prevent some of the plaquette excitations, we will excite the other plaquettes adjacent to that edge, as shown in the second image. Repeating the process, by changing the edges shown in black excites plaquettes along a closed loop (blue). Changing more edges simply changes the shape of this loop (unless we change the edge labels back and shrink the loop to nothing). This tells us that the magnetic excitations are loop-like.}
				\label{magnetic_step_by_step_2}
			\end{center}	
		\end{figure}

		Unlike the electric excitations, the magnetic excitations in 3+1d are significantly different from their counterparts in 2+1d. Whereas in 2+1d the magnetic excitations are point particles that are produced in pairs by a ribbon operator (as we described in Ref. \cite{HuxfordPaper2}), in 3+1d the elementary magnetic excitation is a ``flux tube" at the boundary of a membrane. That is, the magnetic excitations are loop-like. We can see that the magnetic excitations must be loop-like by trying to excite a single plaquette. We try changing the value of a single edge belonging to that plaquette. However, as shown in Figure \ref{magnetic_step_by_step_2}, in 3+1d each edge belongs to multiple plaquettes (as opposed to two plaquettes when there are only two spatial dimensions). Therefore, changing the label of an edge excites all of the plaquettes around that edge. We can then put one of these plaquettes back into a lower energy state by changing the label of another edge on that plaquette, but this in turn excites all of the other plaquettes attached to that edge (see the second image in Figure \ref{magnetic_step_by_step_2}). We see that these excited plaquettes lie on a closed loop that pierces their centres, as shown by the blue loops in Figure \ref{magnetic_step_by_step_2}. This is made more clear by considering changing more edges. Instead of changing edges along a line, we consider changing edges across some surface (such as the four edges shown in the third image of Figure \ref{magnetic_step_by_step_2}, which lie on a square). Changing these edges excites the plaquettes on the boundary of that surface, much as the ribbon operators in 2+1d excite particles at the ends of some path.

\begin{figure}[h]
			\begin{center}
				\includegraphics[width=\linewidth]{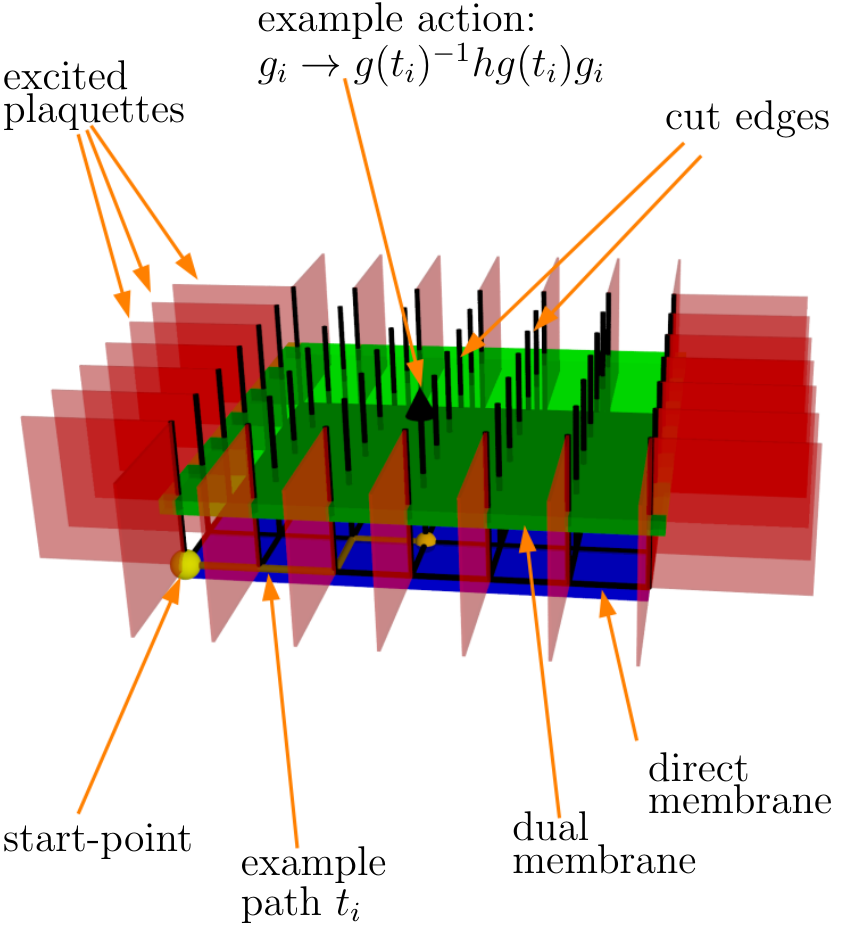}
					
				\caption{(Copy of Figure 41 from Ref. \cite{HuxfordPaper1}) Here we give an example of the membranes for the flux creation operator (magnetic membrane operator). The dual membrane (green) cuts through the edges changed by the operator. The direct membrane (blue) contains a vertex at the end of each of these cut edges (such as the orange sphere). A path from a privileged start-point to the end of the edge (such as the example path, $t_i$) determines the action on the edge. This action leads to the plaquettes around the boundary of the membrane being excited.}
				\label{fluxmembrane2}
			\end{center}
		\end{figure}

		The fact that we produce a loop excitation by changing edges across a surface rather than just along a path indicates that our creation operator is a membrane operator. While we have given a rough idea of the action of the membrane operator in the above discussion, we will now be more specific. In order to define the operator that produces a magnetic excitation, we must specify the region (the ``membrane") that this operator acts on. First we specify a membrane that passes through the centres of plaquettes and cuts through edges. The edges cut by the membrane (``cut edges") are acted on by the operator, as shown in Figure \ref{fluxmembrane2}. This membrane is called the dual membrane, and is analogous to the dual path for the magnetic ribbon operator in 2+1d. In addition to the dual membrane, we must specify a ``direct membrane". The cut edges terminate on this membrane. That is, the direct membrane contains one vertex at the end of each of the cut edges, as shown in Figure \ref{fluxmembrane2} (in special cases, with tightly folded membranes, both ends may be on the direct membrane and an edge may be cut twice by the dual membrane). We must also specify a set of paths to the vertices on the direct membrane. These paths go from a common start-point to the base of each cut edge (that is, to the vertex that lies on the direct membrane). We call this common start-point of the paths the start-point of the membrane or of the membrane operator. These features of the membrane operator are illustrated in Figure \ref{fluxmembrane2}.

		 The fact that we specify two membranes as part of the magnetic membrane operator indicates that our ``membrane operator" really acts on a ``thickened" membrane, much as the ribbon operators in 2+1d can be considered as acting on thickened strings, that is on ribbons (i.e., their support has some finite thickness). Regardless, we will continue to refer to these operators as membrane operators. This unfortunately means that our use of the term membrane is somewhat ambiguous. Sometimes we mean a ``thickened membrane" and sometimes just an unthickened membrane. Generally we try to use membrane to refer to the region on which our membrane operators act, whether those regions are thickened or otherwise. If we want to refer to a surface that may not be part of a membrane operator, we will call this a surface. If we want to refer to part of a thickened membrane, we will use the terms direct and dual membranes.

		Having specified these features of the membrane, we can now describe the action of the magnetic membrane operator. The membrane operator acts on the edges cut by the dual membrane, in a way that depends on the direct membrane and the paths we defined. This is analogous to how the action of the magnetic ribbon operator in the 2+1d case depends on a direct path and a dual path (see Ref. \cite{HuxfordPaper2}). The membrane operator is labeled by a group element $h \in G$, but the label of each cut edge $i$ is left-multiplied by $g(s.p-v_i)^{-1}hg(s.p-v_i)$ or right-multiplied by the inverse, where $g(s.p-v_i)$ is the group element associated to the path specified from the start-point $s.p$ to the vertex $v_i$ on the direct membrane that is attached to the cut edge $i$. Whether left-multiplication or right-multiplication by the inverse is used depends on the orientation of the edge. The edge label is left-multiplied if the edge points away from the direct membrane (as with the example edge in Figure \ref{fluxmembrane2}) and is right-multiplied by the inverse element if it points towards the membrane. That is, the action of the membrane operator on an edge $i$ with initial label $g_i$ is given by
		\begin{equation}
		C^h(m):g_i = \begin{cases} &g(s.p-v_i)^{-1}hg(s.p-v_i)g_i \\ & \hspace{0.4cm} \text{if $i$ points away from the} \\ & \hspace{0.5cm} \text{ direct membrane} \\ & g_ig(s.p-v_i)^{-1}h^{-1}g(s.p-v_i) \\& \hspace{0.4cm} \text{if $i$ points towards the} \\ & \hspace{0.5cm} \text{ direct membrane.} \end{cases} \label{Equation_magnetic_membrane_on_edges_main_text}
		\end{equation}

		The only difference compared to the action of the magnetic ribbon operator in 2+1d is that the operator acts on a general membrane, rather than just a ribbon. In particular, this means that instead of having a direct path along a ribbon, we have multiple paths across a membrane. For a given edge $i$, there are many potential choices for the path from the start-point to the edge. However, the action of the membrane operator is unaffected by deforming any of these paths over a region satisfying fake-flatness. This is because deforming the path in this way only changes the path element $g(s.p-v_i)$ by an element $\partial(e)$ in $\partial(E)$. This factor of $\partial(e)$ is in the centre of $G$ and so it does not affect the expression $g(s.p-v_i)^{-1}hg(s.p-v_i)$, which just gains a factor of $\partial(e)$ and a factor of $\partial(e)^{-1}$ which can be moved together and cancelled.

		Now that we have described the action of the membrane operator, we can discuss which of the energy terms are excited by the membrane operator. Firstly, note that if the membrane operator is labeled by $1_G$ then the membrane operator is just the identity operator and so will not excite any of the energy terms. From now on we will assume that we are talking about a membrane operator labeled by some non-trivial element of $G$. In this case the membrane operator excites the ``boundary plaquettes", which are plaquettes where only one edge on the plaquette is changed by the membrane operator (rather than two for the ``bulk" plaquettes). These boundary plaquettes lie around the perimeter of the membrane (they are the red plaquettes in Figure \ref{fluxmembrane2}) and we can construct a closed path around the boundary of the dual membrane which passes through these plaquettes.

		In addition to these plaquettes, the magnetic membrane operator may also excite the privileged start-point vertex that we defined previously, just as we saw in the 2+1d case for the magnetic ribbon operator in Ref. \cite{HuxfordPaper2}. In order for the magnetic membrane operator to produce an eigenstate of this vertex energy term when acting on an initially unexcited region, we need to construct a linear combination of the magnetic membrane operators labeled by elements of $G$. If the coefficients for this linear combination are a function of conjugacy class (that is we have an equal sum over all elements of the conjugacy class), the vertex is not excited. On the other hand, if the coefficients within each conjugacy class sum to zero, then the vertex is excited. In any other case (such as when we do not take a superposition of our operators), the start-point is neither definitely excited nor definitely unexcited, because we do not produce an energy eigenstate. While in Figure \ref{fluxmembrane2} the start-point is next to the loop-like excitation (at the edge of the membrane), the start-point can be displaced any distance from the excited loop (or even away from the membrane). The position of the start-point can be interpreted in terms of the picture of the ribbon and membrane operators creating and moving excitations. We can think of the membrane as corresponding to the process where we nucleate a loop at the start-point and then grow and move the loop along the membrane to its final position. The fact that the start-point may be excited far from the loop suggests that it can be treated as an additional particle. Therefore, when we produce a loop, we may also have to produce a point-particle. This is similar to how point-like charges must be produced in pairs in order to conserve topological charge. Indeed we will see in Section \ref{Section_Sphere_Charge_Reduced} that some loop-like excitations carry a non-trivial ``point-like" conserved charge, which must be balanced by the charge carried by the additional excitation at the start-point.

\begin{figure}[h]
			\begin{center}
					\includegraphics[width=\linewidth]{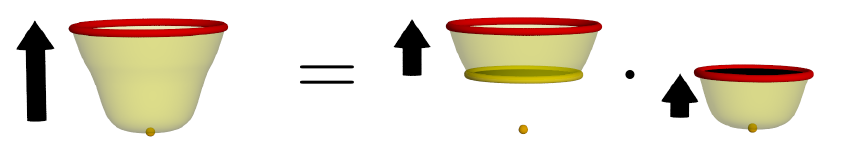}

				\caption{Given a flux creation operator (left), we can split it into two parts. One of these parts (the rightmost picture, comprised of the inner part of the original membrane) is another flux creation operator that nucleates the loop and moves it part way along the membrane. The second part (the outer part) takes that existing loop and moves it to the final position.}
				\label{membrane_splitting}
			\end{center}
		\end{figure}

		The magnetic membrane operator described above is a creation operator for our flux tube, which runs around the perimeter of the membrane. The membrane operator creates a flux tube (and its associated vertex excitation) from the vacuum. Another relevant operator is the one that moves an existing flux tube to a new position. The movement operator can be thought of as an ordinary membrane operator, but with an additional hole whose boundary fits the loop that we wish to move. To see that this is a movement operator, consider splitting a creation operator into two parts, an inner part, which is another creation operator, and an outer part applied on a membrane equivalent to a tube or annulus, as shown in Figure \ref{membrane_splitting}. We know that the overall membrane operator produces and moves an excitation to the boundary of the outer part, while the inner part produces and moves an excitation to the boundary of the inner part. Because the overall membrane operator is a combination of the inner membrane operator and outer membrane operator, this means that the outer membrane operator must take the excitation from the boundary of the inner part and move it to the boundary of the outer part, to match the action of the total membrane operator.

 		For instance, consider the example shown in Figure \ref{membrane_splitting}. In this figure, the yellow membranes indicate the membrane on which the operator is applied (so that if we zoomed in we would see the structure from Figure \ref{fluxmembrane2}). The yellow spheres are the start-points for the membranes (note that they are all in the same position). The opaque tori indicate the excitations they create. The two membranes on the right are displaced horizontally to indicate an order in which operators are applied, rather than spatial displacement. On the left-hand side of the Figure, we have an operator that simply creates and moves an excitation to the final position (indicated by the red torus). We can split this operator into the two operators shown on the right-hand side. The rightmost operator on the right-hand side is the inner part of the original membrane operator and so is another flux creation operator. Therefore, this operator also creates an excitation and moves it to its boundary, the lower red torus. In order for this decomposition of operators to agree with the total operator on the left-hand side, we see that the middle operator, which represents the outer part of the original membrane operator, must move an excitation from the lower position (the yellow torus) to the final position (the red torus). While in this context the outer membrane operator moves an existing excitation, we can also apply it when there are no existing excitations. In this case the operator instead creates two opposite fluxes at the two ends of the operator. We therefore do not need to distinguish between movement and creation operators, because the movement operators are also creation operators, although they create multiple loop-like excitations.
		
		More generally, we can put many holes in the membrane to produce many loop excitations. Indeed, we can think of the membrane operator that we originally defined as a closed, topologically spherical membrane with a single hole in it. Then the excited loops (or single loop in the ordinary case) are at the boundaries of these holes. A topologically spherical membrane operator would produce no excitations. Indeed, as we prove in Section \ref{Section_Topological_Magnetic_Tri_Trivial} in the Supplemental Material, such a spherical membrane operator will act trivially if it is contractible and encloses no other excitations.

		When producing an excitation in a given location, there are many choices for the position of the membrane. This is because the excitation is produced at the boundary of the membrane and many different membranes share the same boundary. However, the membrane operator is \textit{topological} in the following sense. We can freely deform the membrane on which the operator is applied through the lattice, while keeping the positions of any excitations produced by the operator fixed, as long as we do not deform the membrane over any existing excitations. When we do this, the action of the membrane operator is preserved. That is, given an initial state $\ket{\psi}$ and a magnetic membrane operators $C^h(m)$ applied on a membrane $m$, if we can deform the membrane $m$ into a new membrane $m'$ without crossing any excitations in $\ket{\psi}$, or moving the excitations produced by $C^h(m)$, then we have $C^h(m) \ket{\psi} =C^h(m') \ket{\psi}$. This means that, like the ribbons in 2+1d, the membrane is invisible when acting on the ground state; it does not matter precisely where we put the membrane. However, when we act on a state that already has excitations, the position may matter. Indeed this fact is vital when considering braiding and leads to the non-trivial braiding relations that we will see in Section \ref{Section_3D_Braiding_Tri_Trivial}.

		It is not just the magnetic membrane operators that have this property under deformation, but all of the non-confined membrane and ribbon operators, as we will prove in Section \ref{Section_topological_membrane_operators} in the Supplemental Material. We therefore call the non-confined ribbon and membrane operators topological. However, in reality this topological nature is a combined property of the ground state and the operators, because we can only freely deform the membranes over a space that does not contain any excitations.

		Having obtained the membrane operators, we can find their algebra, which can give us the fusion rules for the excitations (although to formally obtain the fusion rules we should organise our excitations according to their topological charge first). Just as in 2+1d, two magnetic operators applied on the same space combine by multiplication of their flux labels. The precise way in which this occurs depends on the position of the start-point of the common membrane $m$. We may have $C^g(m) C^h(m)=C^{gh}(m)$, but we could also have $C^g(m) C^h(m)=C^{hg}(m)$ if the paths from the start-point to the membrane $m$ themselves intersect with the dual membrane. This is because in this case the action of the membrane operator $C^h(m)$ affects the path labels $g(s.p(m)-v_i)$ that determine the action of $C^g(m)$ (see Equation \ref{Equation_magnetic_membrane_on_edges_main_text}), leading to the membrane operator $C^g(m)$ instead acting like $C^{hgh^{-1}}(m)$. We also note that, just as we described for the $E$-valued membrane operators in 2+1d in Ref. \cite{HuxfordPaper2} (see Section III D), we can have partial fusion of the excitations. In this case the two magnetic membrane operators share part of their membrane and boundary, but the membranes are not completely identical (i.e., we have some $C^g(m_2)$ instead of $C^g(m)$), which can lead to only sections of the excited strings merging.

\begin{figure}[h]
			\begin{center}
				\includegraphics[width=\linewidth]{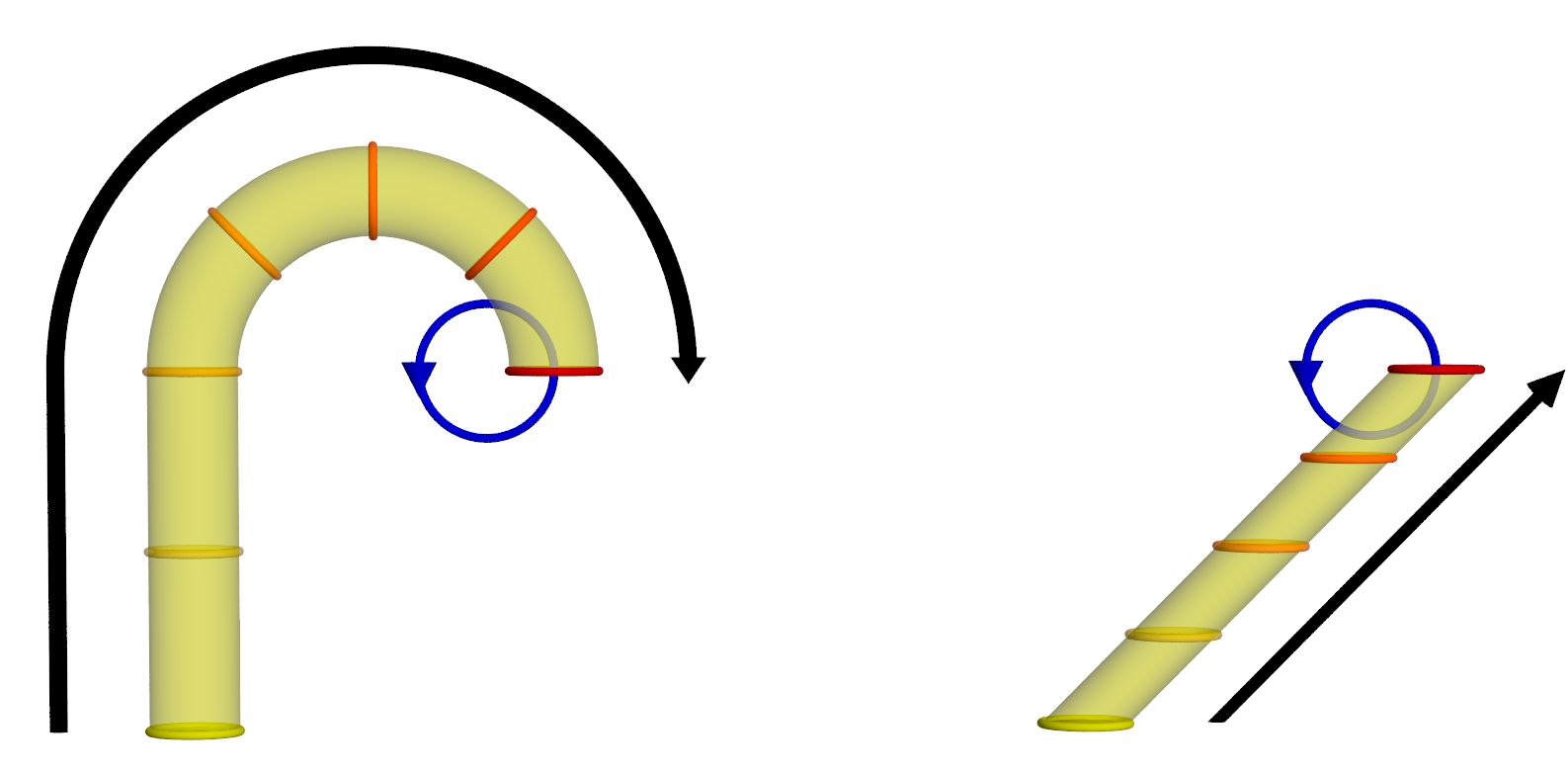}

				\caption{Two membranes that would move a loop excitation from the same initial location to the same final location. The membrane on the left flips the loop during its motion (intermediate positions of the loop are shown along the membrane), whereas the right membrane moves the loop without flipping it. For the same original loop, measuring the flux along the blue path in the left figure gives us the inverse of the flux along the blue path on the right.}
				\label{flipping_paths}
			\end{center}
		\end{figure}

		As well as fusion, loop excitations have another important relationship between the different excitations. We can flip the orientation of a loop excitation and ask what the resulting label should be in terms of an unflipped loop. By flip the loop, we mean that we turn the loop over during its motion using its membrane operator. Then we determine what membrane operator would produce an equivalent flux tube by producing a loop without flipping it over during its motion. An example of the relevant membranes is shown in Figure \ref{flipping_paths}. In the case of the magnetic excitations, flipping the loop over gives a loop labeled by the inverse of the original label. This indicates that to specify a flux, the orientation of the flux tube is important. This is a feature not seen in point-particles and highlights that measuring the topological charge of a loop excitation is not as simple as it is for point excitations.
		
		\subsection{$E$-valued loop excitations}
		\label{Section_E_Loop_Excitations_3D}
		Magnetic excitations are not the only loop-like excitations that we find in this model. In Ref. \cite{HuxfordPaper2}, we showed that in 2+1d we could have loop-like excitations that arise when our group $E$ is non-trivial, which we called $E$-valued loop excitations. These excitations persist in the 3+1d case, and the membrane operators that produce these excitations in 3+1d are very similar to the operators in 2+1d. Just as in 2+1d, the membrane operator measures the surface label $\hat{e}(m)$ of some membrane $m$ and assigns a weight depending on the value measured. It is convenient to consider a basis for this space of membrane operators where the weights are given by the irreducible representations of the group $E$. That is, we can define basis operators
		\begin{equation}
		L^{\mu}(m) = \sum_{e \in E} \mu(e) \delta(\hat{e}(m),e) \label{Equation_E_membrane_irrep_Abelian}
		\end{equation}
		where $\mu$ is an irrep of $E$ and $\mu(e)$ is the phase representing the element $e\in E$. Note that the irreps of $E$ are 1D because $E$ is Abelian when $\rhd$ is trivial. Then any of these operators that are labeled by non-trivial representations produce a loop of excited edges on the boundary of the membrane, whereas the operator labeled by the trivial irrep is the identity operator. We can fuse these excitations, with the resulting label being given by the product of the irreps under the multiplication $(\mu \cdot \nu) (e)= \mu(e) \cdot \nu(e)$ for irreps $\mu$ and $\nu$ of $E$.

\begin{figure}[h]
			\begin{center}
				\includegraphics[width=\linewidth]{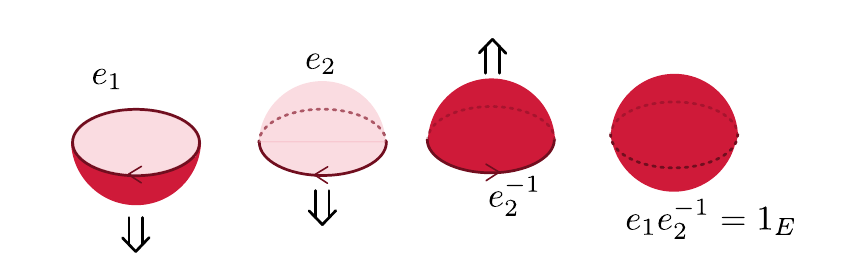}

				\caption{Given two different surfaces with the same boundary, such as the first two surfaces in the figure, their labels must be the same if we can deform one into another without crossing any excitations. This is because the volume over which we deform them must have trivial boundary surface label.}
				\label{gluing_surfaces_1}
			\end{center}
		\end{figure}

		Unlike in 2+1d, there are many different membranes that have the same boundary and so produce a loop excitation in the same location. However, much like the magnetic membrane operators, the $E$-valued membrane operators are topological, meaning that we can deform the membrane without changing the action of the membrane operator. For the $E$-valued membrane the topological nature is relatively intuitive and derives from the fact that closed, contractible surfaces are forced to have trivial label in the ground state by the blob condition in the Hamiltonian. The $E$-valued membrane measures the value of some surface. Given two such surfaces with the same boundary, we can consider the difference between their labels by inverting the orientation of one surface and combining it with the other surface by gluing the surfaces along their common boundary. If the two surfaces can be deformed into one-another, this gluing procedure produces a contractible closed surface that encloses no excitations. However, in the ground state such a surface must have trivial label, due to the blob energy terms. Therefore, the two original surfaces must have the same label and so the two original operators give the same result. This is shown in Figure \ref{gluing_surfaces_1}. In the leftmost diagram we have one surface, the boundary of which is our loop excitation. This surface is labeled by $e_1$. In the next diagram, we have another surface with the same boundary, labeled by $e_2$. We can invert this surface, reversing its orientation and changing the label from $e_2$ to $e_2^{-1}$ as shown in the third picture. Then we can glue these surfaces together to obtain a sphere labeled by $e_1 e_2^{-1}$. However, this resulting surface is contractible, so its label must be $1_E$ if it encloses no excitations. Therefore, $e_1 e_2^{-1}=1_E$. This indicates that the two different surfaces have the same label. From this, we see that deforming the surface, without crossing an excitation, does not affect the action of the membrane operator.
		
		\subsection{Blob excitations}
		
		\label{Section_3D_Blob_Excitation_Tri_Trivial}
\begin{figure}[h]
			\begin{center}
					\includegraphics[width=\linewidth]{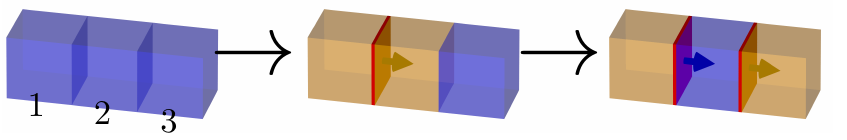}
				
				\caption{(Copy of Figure 38 from Ref. \cite{HuxfordPaper1}) We consider a series of blobs in the ground state (leftmost image). In the ground state, all of the blob terms are satisfied, which we represent here by colouring the blobs blue (dark gray in grayscale). Changing the label of the plaquette between blobs 1 and 2 excites both adjacent blobs, as can be seen in the middle image (we represent excited blobs by colouring them orange, or lighter gray in grayscale). Multiplying another plaquette label on blob 2 to try to correct it just moves the right-hand excitation from blob 2 to blob 3 (rightmost image). In each step, the plaquettes whose labels we changed are indicated by the (red) squares and their orientations are indicated by an arrow.}
				\label{blob_ribbon_operator_sequential}
			\end{center}
		\end{figure}

		In addition to the three types of excitation we have considered so far (and which we already saw in the 2+1d case discussed in Ref. \cite{HuxfordPaper2}), we have a fourth simple excitation in 3+1d, called the blob excitation (or 2-flux excitation). The blob excitations correspond to violations of the 2-flatness of blobs, also called 3-cells (i.e., to violations of the blob energy terms). As we described in Ref. \cite{HuxfordPaper1} (in Section III B), we can consider creating two blob excitations by changing a chain of plaquettes along a dual path in the lattice. The blob energy term forces the total surface label around the blob (which is a certain product of surface elements around the blob) to be $1_E$ when the blob is unexcited. Then to excite a blob the naive thing to do is to multiply a single plaquette by some group element, $e^{-1}$ for example. However, each plaquette belongs to two adjacent blobs, both of which will be excited by changing the plaquette, as is shown in Figure \ref{blob_ribbon_operator_sequential}. We can correct the 2-holonomy of one of these blobs by changing the label of another plaquette on that blob, but this excites yet another blob, as shown in Figure \ref{blob_ribbon_operator_sequential}. We can repeat this process with another plaquette, this only moves one of the blob excitations around. That is, by changing the labels of a series of plaquettes appropriately, we can produce a pair of blob excitations and move one of these excitations along a path. This is exactly the behaviour we expect of a \textit{ribbon operator}.
		
\begin{figure}[h]
			\begin{center}
					\includegraphics[width=\linewidth]{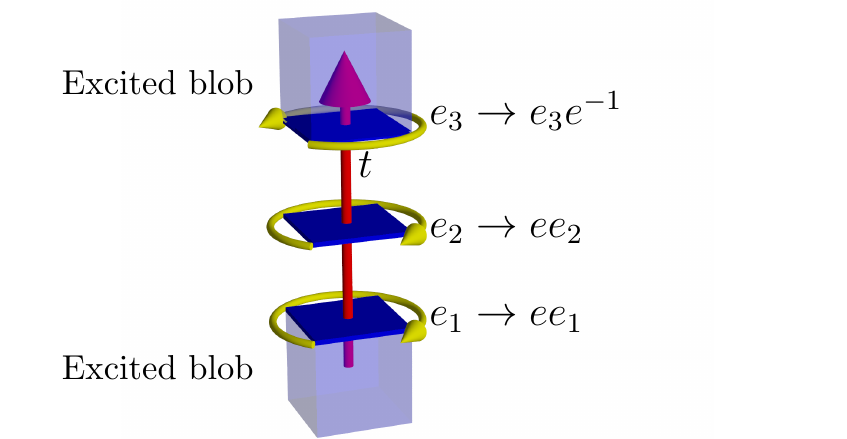}
				
				\caption{In the $\rhd$ trivial case, the blob ribbon operator multiplies the labels of the plaquettes pierced by the ribbon by $e$ or $e^{-1}$, depending on the orientation of the plaquette. Here the circulation of a plaquette is shown by the curved (yellow) arrows, and this can be converted into a direction using the right hand rule. The plaquette label is multiplied by $e^{-1}$ if the orientation of the plaquette matches that of the ribbon and by $e$ if the orientation is anti-aligned with the ribbon (note that the order of multiplication does not matter when $\rhd$ is trivial, because $E$ is Abelian, but the order is chosen in this figure to match the more general case).}
				\label{Effect_blob_ribbon_tri_trivial}
			\end{center}
		\end{figure}

		Having discussed the rough idea behind the blob ribbon operator, we will now be somewhat more precise about the action of the operator. Each blob ribbon operator is labeled by an element of $E$, for example $e$. We must also specify a (dual) path for the blob ribbon operator to act on. We denote the blob ribbon operator labeled by $e$ and acting on the path $t$ by $B^{e}(t)$. The path passes between the centres of blobs, much as a path on the lattice passes from one vertex to the next. Because the path travels between blobs, the path will pierce plaquettes and it is these pierced plaquettes that the operator will act on. The operator does so by multiplication of the plaquette label by $e^{-1}$ if the orientation of the plaquette is aligned with the direction of the ribbon and $e$ if it is aligned with the ribbon, where we use the right-hand rule to convert the clockwise or anticlockwise circulation of the plaquette into a direction in order to compare it with the orientation of the ribbon. An example of the action of the blob ribbon operator is illustrated in Figure \ref{Effect_blob_ribbon_tri_trivial}. This action excites the blob in which the ribbon originates and the blob in which it terminates. If the label of the operator $e$ is in $\text{ker}(\partial)$ then these two blob terms are the only excited energy terms. However, if $e$ is not in the kernel the plaquettes pierced by the ribbon are also excited, so the particles produced are confined (there is an energy cost that increases with the length of the ribbon). The plaquettes are excited because the plaquette operator checks that the image under $\partial$ of the plaquette element matches the path around the plaquette. Therefore, multiplying the plaquette label by an element $e$ with non-trivial $\partial(e)$ (i.e., an element outside the kernel of $\partial$) will cause this plaquette condition to be violated.

		Much like the other operators we have considered so far, blob ribbon operators can be combined by applying one after the other on the same path. This process leads to fusion of the simple excitations produced by the operators. The excitations fuse in a similar way to the magnetic ones: the ribbon algebra is given by $B^e(t) B^f(t)=B^{ef}(t)$.

		Before we move on to summarize the excitations, it is worth mentioning that the blob excitations in 3+1d replace the single-plaquette excitations from 2+1d. Recall from Ref. \cite{HuxfordPaper2} that we create the single-plaquette excitations by multiplying a plaquette label by an element of $E$. Because there are no blobs to excite in 2+1d (where the lattice is two-dimensional) this creates no excitations other than the plaquette. However, in 3+1d such an action produces blob excitations, as we saw from the action of the blob ribbon operator.
		
		\subsection{Condensation and confinement}
		\label{Section_3D_Condensation_Confinement_Tri_Trivial}
		
		In Refs. \cite{HuxfordPaper1} and \cite{HuxfordPaper2}, we described a type of transition between different higher lattice gauge theory models called condensation-confinement transitions. During this transition, some particle types become confined, so that it costs energy to separate a confined particle from its antiparticle, and others become ``condensed". A condensed excitation can be produced \textit{locally} and so carries trivial topological charge in the condensed phase. We can consider a model with no confinement where $\partial \rightarrow 1_G$. Note that in this uncondensed model, with both $\rhd$ and $\partial$ trivial, the two gauge groups ($G$ and $E$) decouple and the model can be treated as a tensor product of two independent lattice gauge theories, as we discussed in Section \ref{Section_Recap_3d}. Then changing this $\partial$ so that it maps onto a non-trivial subgroup of $G$, while keeping the groups $G$ and $E$ constant, results in certain topological charges condensing. In particular, the magnetic excitations labeled by $h \in \partial(E)$ and the $E$-valued loop excitations that are labeled by trivial irreps of the kernel become condensed. To see what we mean by this, consider the $E$-valued membrane operators, which have the form
	$$\sum_{e \in E} \alpha_e \delta( e, \hat{e}(m)).$$
	
	If the membrane $m$ satisfies fake-flatness, the surface label of the membrane is related to the label of its boundary $\text{bd}(m)$ through $\partial(\hat{e}(m))\hat{g}(\text{bd}(m))=1_G$. Then, if the coefficients $\alpha_e$ are only a function of $\partial(e)$, and so are not sensitive to the kernel of $\partial$, we can write the membrane operator (when acting on a fake-flat state) as
	\begin{align*}
	\sum_{e \in E} \alpha_e& \delta( e, \hat{e}(m))\\		
	&=\sum_{e_k \in \ker(\partial)} \sum_{q \in E/\ker(\partial)} \alpha_q \delta( qe_k, \hat{e}(m)),
	\end{align*}
	where the $q$ are representative elements from the cosets of $\ker(\partial)$ in $E$ and $\alpha_{qe_k}=\alpha_q$ because the coefficient is not sensitive to factors in the kernel. Then
		\begin{align*}
	\sum_{e \in E} \alpha_e& \delta( e, \hat{e}(m))\\	
		&= \sum_{q \in E/\ker(\partial)} \alpha_q \sum_{e_k \in \ker(\partial)}\delta( qe_k, \hat{e}(m))\\
		&=\sum_{q \in E/\ker(\partial)} \alpha_q \delta(\partial(q),\partial(\hat{e}(m))\\
			&=\sum_{q \in E/\ker(\partial)} \alpha_q \delta(\partial(q),\hat{g}(\text{bd}(m))^{-1})\\
		&= \sum_{g \in \partial(E)} \beta_g \delta(g, \hat{g}(\text{bd}(m))),
	\end{align*}	
where $\beta_{\partial(e)^{-1}}=\alpha_e$. This is just an electric ribbon operator applied on the boundary of $m$ and so is local to the excitation produced by the membrane operator. Rather than local in the usual sense of being restricted to a small spatial region, we mean that the operator only acts near the excitation. We see that the excitation produced by the membrane operator can be produced locally and so is condensed. That is, the $E$-valued membrane operators which are not sensitive to the kernel of $\partial$ produce condensed excitations. When we consider the irrep basis for the membrane operators, the operators labeled by irreps with trivial restriction to the kernel correspond to condensed excitations. For the magnetic excitations, it is the fluxes with label in $\partial(E)$ that are condensed. It is slightly more complicated to show directly that the membrane operators associated to these condensed excitations are equivalent to local operators, so we postpone this proof until Section \ref{Section_magnetic_condensation} of the Supplemental Material. However, as we discussed in Section \ref{Section_Recap_3d}, the plaquette terms allow for closed paths with label in $\partial(E)$  in the ground state rather than just closed paths with trivial label and these non-trivial paths are created by the edge transforms. Therefore, it is no surprise that fluxes with label in $\partial(E)$ should be condensed. We note that this condensation of fluxes is similar to that for a related field theory model used to discuss condensation and confinement in regular gauge theory \cite{Gukov2013}, where fluxes in the subgroup $\pi_1(H)$ (equivalent to $\partial(E)$) are condensed. This reinforces the connection between the higher lattice gauge theory model and (partially) condensed lattice gauge theory

As these magnetic and $E$-valued loops condense, some of the point-like particles in the model become confined. As we showed in Section S-I A of the Supplemental Material of Ref. \cite{HuxfordPaper2} (with the proof being the same for the 2+1d and 3+1d cases), the confined electric ribbon operators $\sum_g \alpha_g \delta( \hat{g}(t),g)$ are those for which $\sum_{e \in E} \alpha_{\partial(e)g}=0$ for all $g \in G$. This means that a basis ribbon operator, given by
$$S^{R,a,b}(t) = \sum_{g \in G} [D^R(g)]_{ab} \delta( \hat{g}(t),g),$$
for irrep $R$ of $G$ and matrix indices $a$ and $b$, is confined if $R$ has a non-trivial restriction to the subgroup $\partial(E)$ of $G$. Similar to the condensation of fluxes, this confinement is equivalent to that in Ref. \cite{Gukov2013}, where the electric operators are confined if they are sensitive to factors in a subgroup $\pi_1(H)$ (which is equivalent to $\partial(E)$ here). So far, this pattern of condensation and confinement is analogous to the 2+1d case described in Ref. \cite{HuxfordPaper2}, but in the 3+1d case we have an extra type of excitation, the blob excitation. As we discussed in the previous subsection, the blob excitations with label outside the kernel of $\partial$ are also confined, because the corresponding ribbon operators multiply the plaquette labels of the plaquettes pierced by the ribbon by a factor which breaks the fake-flatness condition. These confined blob excitations are important when discussing the condensation of the magnetic excitations. The magnetic condensation is slightly different when there are three spatial dimensions compared to the case where there are only two, because the magnetic excitations in 3+1d are loops. Rather than being equivalent to strictly local (i.e., unextended) operators when acting on the ground state, the magnetic membrane operator is instead equivalent to a (confined) blob excitation operator acting on a path that runs around the boundary of the magnetic membrane. This blob ribbon operator is not local in the usual sense, given that the operator is linearly extended, but it is instead local to the excitation. This is analogous to how the condensed $E$-valued membrane operators act equivalently to (confined) electric ribbon operators applied around the boundary of the membrane.
		
		\subsection{Summary of excitations}
		
\begin{figure}[h]
			\begin{center}
				\includegraphics[width=\linewidth]{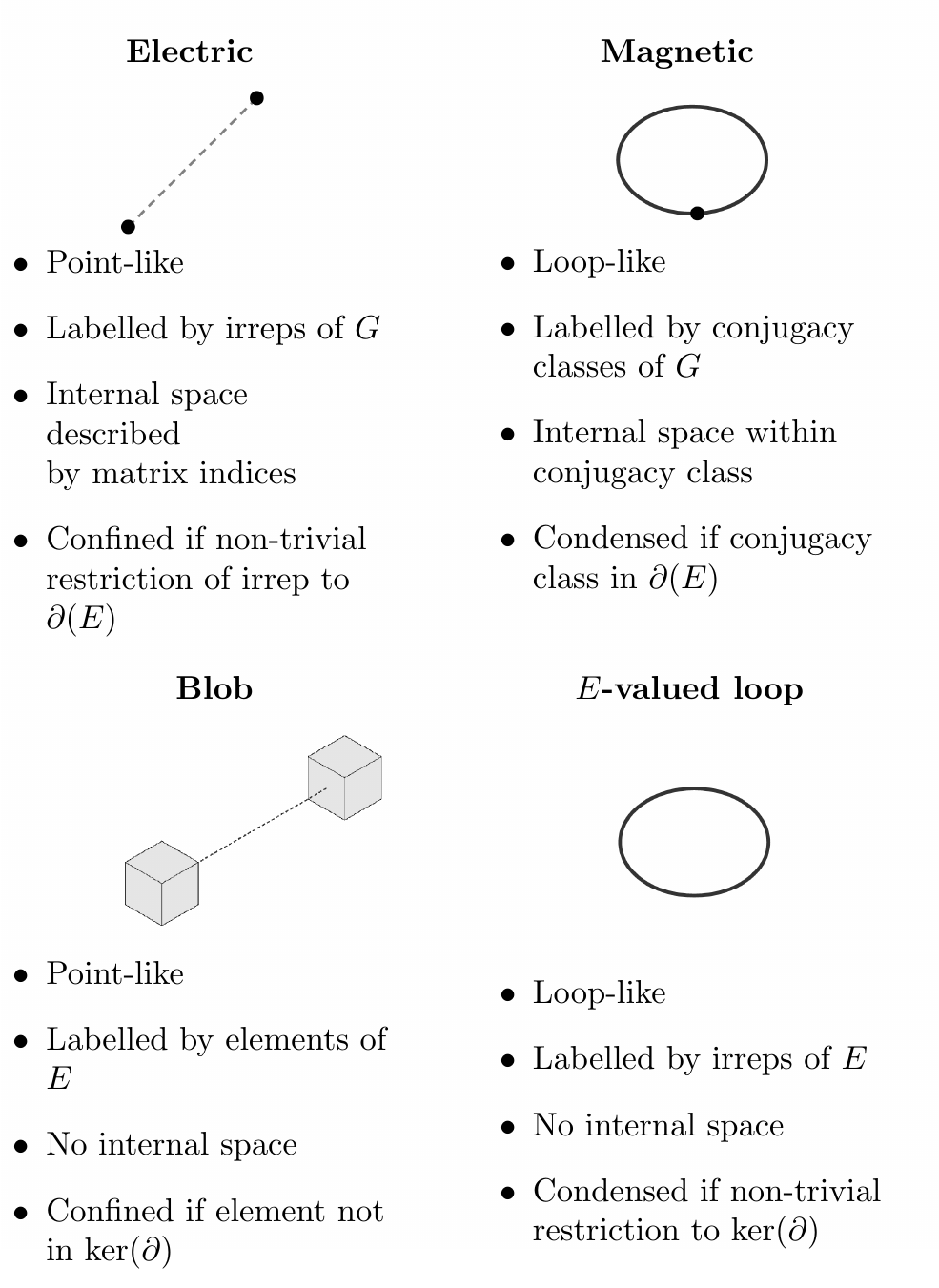}
				
				\caption{A summary of the excitations when $\rhd$ is trivial}
				\label{excitation_summary_tri_trivial}
			\end{center}
		\end{figure}

		Given the large number of excitations that we have seen so far, it may be useful to briefly summarize them. The simple excitations and their confinement and condensation properties for the case described above are summarized in Figure \ref{excitation_summary_tri_trivial}.
		
		\section{Braiding in the $\rhd$ trivial case}
		\label{Section_3D_Braiding_Tri_Trivial}
		
		Now that we have obtained the membrane and ribbon operators that produce the various excitations of our theory, we can use these operators to obtain the braiding relations of the excitations. We find that the non-trivial braiding is between the magnetic flux tubes and the electric charges; between the flux tubes and other flux tubes (though this is only non-trivial if $G$ is non-Abelian); and between the blob excitations and $E$-valued loops. We will describe all of these in more detail in the following sections. First, we will look at the relations involving the magnetic fluxes and electric charges. To describe this braiding, it is convenient to separately consider the cases where $G$ is Abelian and non-Abelian, starting with the Abelian case.
		
		\subsection{Abelian case}
		
		\subsubsection{Flux-charge braiding}
		\label{Section_Abelian_flux_charge_braiding}
\begin{figure}[h]
			\begin{center}
					\includegraphics[width=\linewidth]{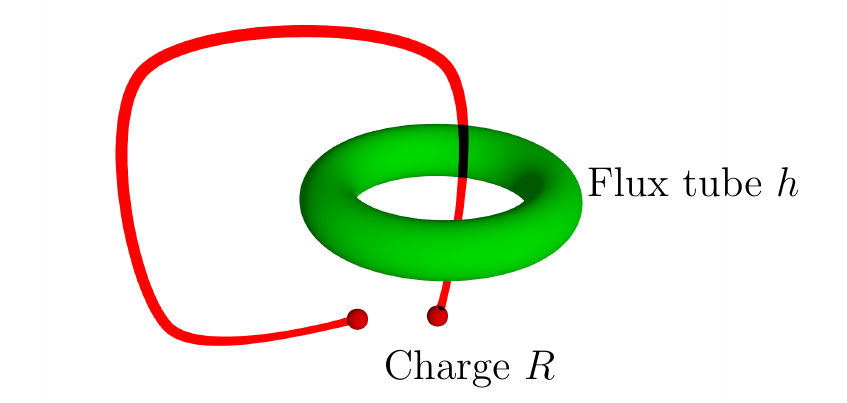}
				
				\caption{(Copy of Figure 35 from Ref. \cite{HuxfordPaper1}) Schematic view of braiding a charge through a loop. The red line tracks the motion of the charge.}
				\label{chargethroughloop2}
			\end{center}
		\end{figure}

		The first non-trivial braiding that we consider is the braiding involving our magnetic fluxes and electric charges. Recall that the magnetic fluxes are loop-like particles, whereas the electric charges are point-like. Because of this, the appropriate braiding between these two types of particle is to pass the electric charge through the loop and back around to its original position, as shown in Figure \ref{chargethroughloop2}. It is also possible to pass the electric charge around the loop (without passing through), just as if the loop were a point particle, but this process is found to be trivial in this model. Indeed, generally we find that in the higher lattice gauge theory model, any braiding where one particle (loop-like or otherwise) is moved around another particle (but not through a loop-particle) is trivial. This is because such an operation can always be performed by membrane (and ribbon) operators which never intersect and so commute. This means that each excitation is oblivious to the presence of the other and so the motion has the same result as moving through the vacuum.
		
		Now that we have discussed what the relevant braiding move is, we need to find how the excitations transform under such a move. The braiding relation is conveniently calculated by considering a commutation of operators as follows. Consider starting with a state that has no excitations and then applying a magnetic membrane operator that produces a flux tube. Then consider acting with an electric string operator to produce a pair of electric charges and move one along the path of the string, with this path passing through the loop excitation. In this case the electric excitation has undergone the braiding move we described earlier. We want to compare this situation to a similar one in which the electric excitation has not braided with the loop excitation. To do so, consider reversing the order of operators that we apply. Instead of first acting with the magnetic membrane and then with the electric string operator, we first apply the electric operator. This produces a pair of electric excitations and moves one of them along the ribbon. However, there is no magnetic excitation present at this stage, so no braiding occurs. Then we act with the magnetic membrane to produce our flux tube. In this situation, the excitations end up in the same location as when we applied the operators in the original order, but no braiding has occurred. Comparing these two situations therefore gives us the braiding relation. This means that to describe the braiding, we just need to find the relationship between the two possible orderings of the operators. That is, we need to calculate the commutation relations of the magnetic membrane operator and the electric ribbon operator.

		In the case where $G$ is Abelian, it is simple to calculate the commutation relation described above. Let the path of the electric ribbon operator be $t$ and consider a magnetic membrane operator $C^h(m)$ applied on a membrane which intersects with the path $t$. The path $t$ intersects with the membrane $m$ at some edge $i$ in $t$. The label of the path up to edge $i$ is not affected by the magnetic membrane, because it does not intersect with it. We denote this part of the path by $t_1$. The path after $i$, which we call $t_2$, is similarly unaffected. However, the label of the edge $i$ itself is multiplied by either $h$ or $h^{-1}$, depending on the relative orientation of the edge and the membrane. Then the total path $t$ is the composition of $t_1$, the edge $i$ and $t_2$, which we write as $t=t_1it_2$ (if the edge $i$ points along the path, otherwise $t=t_1i^{-1}t_2$). This means that the path label operator satisfies the following commutation relation with the magnetic membrane operator:
		\begin{align*}
		\hat{g}(t) C^h(m)&=\hat{g}(t_1)\hat{g}_i\hat{g}(t_2)C^h(m)\\
		&=C^h(m) \hat{g}(t_1)h^{\pm 1}\hat{g}_i\hat{g}(t_2),
		\end{align*}
		where the inverse depends on the orientation of the membrane. Because we are looking at the case where $G$ is Abelian, we can extract the factor $h^{\pm 1}$ to the front of the path operator and combine the sections of path to obtain
		$$\hat{g}(t)C^{h}(m)\ket{GS}=C^{h}(m) h^{\pm 1} \hat{g}(t) \ket{GS}.$$
		
		Now consider an electric ribbon operator, which has the form
		$$\sum_{g \in G} \alpha_g \delta(g, \hat{g}(t)),$$
		where $\alpha_g$ is an arbitrary set of coefficients. This ribbon operator then satisfies the commutation relation
		\begin{align}
		\sum_g \alpha_g& \delta(g,\hat{g}(t))C^{h}(m)\ket{GS} \notag \\
		&= C^{h}(m) \sum_g \alpha_g\delta( g,h^{\pm 1}\hat{g}(t))\ket{GS} \notag \\
		&=C^{h}(m) \sum_{g'=h^{\mp 1}g}\alpha_{h^{\pm 1}g'} \delta( g',\hat{g}(t))\ket{GS} \label{Equation_braiding_electric_magnetic_Abelian_1}
		\end{align}
		with the magnetic membrane operator. This relation is simplified when we consider an electric ribbon operator whose coefficients are described by an irrep $R$ of $G$. As discussed in Section \ref{Section_Electric_Tri_Trivial}, the electric ribbon operators labeled by irreps of $G$ form a basis for the space of electric ribbon operators, and so we can decompose any electric ribbon operator into a sum of such irrep-labeled ribbon operators. When $G$ is Abelian, all of the irreps are 1D, and so the basis ribbon operator labeled by irrep $R$ is given by
		$$S^R(t)= \sum_{g \in G} R(g) \delta(g, \hat{g}(t)),$$
		where $R(g)$ is the representation of element $g$ in the irrep, and is a phase because the irreps are 1D when $G$ is Abelian. Substituting this into Equation \ref{Equation_braiding_electric_magnetic_Abelian_1}, we see that the electric ribbon operator labeled by an irrep $R$ and the magnetic membrane operator satisfy the commutation relation
		\begin{align}
		S^R(t)&C^{h}(m)\ket{GS} \notag \\
		&=\sum_g R(g) \delta(g,\hat{g}(t))C^{h}(m)\ket{GS} \notag \\
		&=C^{h}(m) \sum_{g'=h^{\mp 1}g} R(h^{\pm 1}g') \delta( g',\hat{g}(t))\ket{GS} \notag \\
		&= C^{h}(m) \sum_{g'} R(h^{\pm 1})R(g') \delta( g',\hat{g}(t))\ket{GS} \notag \\
		&= R(h^{\pm 1}) C^{h}(m) \sum_{g'} R(g') \delta( g',\hat{g}(t))\ket{GS} \notag \\
		&= R(h^{\pm 1}) C^{h}(m)S^R(t)\ket{GS}.
		\end{align}
		
		This is the same as the unbraided version $C^{h}(m) S^R(t)$, except that we have gained a phase of $R(h)$ or $R(h^{-1})$. Therefore, under braiding the state obtains a simple phase $R(h)$ (or the inverse) and so the braiding is Abelian. The phase is $R(h)$ if the electric ribbon's path meets the direct membrane before the dual membrane, and the inverse otherwise, as we show in Section \ref{Section_electric_magnetic_braiding_3D_tri_trivial} of the Supplemental Material. Note that this is the same result that we would expect for conventional discrete gauge theory (see, e.g., Refs. \cite{Bais1980,Bucher1992} and our discussion of how this relates to higher gauge theory in Section II of Ref. \cite{HuxfordPaper1}).
		
		\subsubsection{Flux-flux braiding}
		\label{Section_Flux_Flux_Braiding_Tri_Trivial_Abelian}
\begin{figure}[h]
			\begin{center}
				\includegraphics[width=\linewidth]{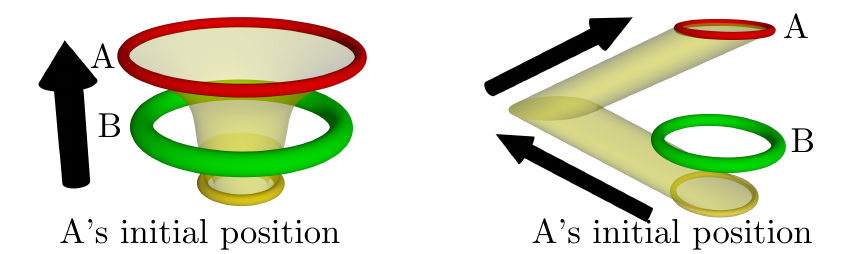}
		
				\caption{Schematic of a braid move (left) and a permutation move (right). The translucent membrane is the surface swept by the red loop (the loop at the top of each image), which is also the membrane on which we apply the corresponding membrane operator.}
				\label{LoopMoves}
				
			\end{center}
		\end{figure}

\begin{figure}[h]
			\begin{center}
				\includegraphics[width=\linewidth]{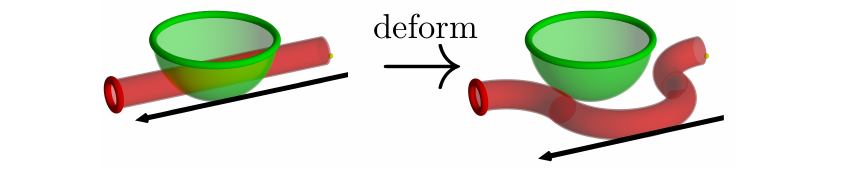}
			
				\caption{Exchange of two excitations is implemented by membranes which can be freely deformed so that they do not intersect. This means that the corresponding commutation relation of operators is trivial.}
				\label{Exchange_deform}
			\end{center}
		\end{figure}

		Just as a point particle can be braided with a loop-like excitation in two ways, so can two loops be braided in multiple ways. The allowed patterns of motion can be built from two types of movement. Firstly, we can move the loops around each-other (in the same way as we can move two point particles around each-other), which we call permutation. Secondly, we can pull a loop through another loop (or over it, which is equivalent to pulling the second loop through the first), just as we saw when braiding point particles with loops. These two moves are shown in Figure \ref{LoopMoves}. In this figure, in each case the red loop is moved, with the path of its motion being represented by the yellow membrane. The arrows indicate the direction of motion. In this model, the permutation move is trivial in that it is the same regardless of whether the green loop is present or not. This is because the motion can be performed by membrane operators acting on membranes that never intersect. Then because the membranes do not intersect, the membrane operators commute. Even if we choose to use membrane operators that do intersect, as in Figure \ref{Exchange_deform}, the membranes can be deformed so that they do not intersect, by using the topological property of membrane operators. In the example shown in Figure \ref{Exchange_deform}, a loop (shown as a small red torus in the figure) is moved along a surface (indicated by the red surface attached to the loop) that intersects twice with a green membrane. Although the red loop intersects the green membrane, the red loop does not pass through the larger green loop excitation created by this green membrane. This motion is performed by an operator placed on the red surface. The red and green membranes can be deformed so that one goes around the other, using the topological property of the membrane operators. Then because the membranes do not intersect (and indeed can be deformed so that they never come close), the corresponding membrane operators commute and so permutation is trivial. This is also true for permutation involving non-confined point particles or the $E$-valued loops: in 3+1d the permutation can be performed by ribbon or membrane operators which do not intersect.

		\begin{figure}[h]
			\begin{center}
					\includegraphics[width=\linewidth]{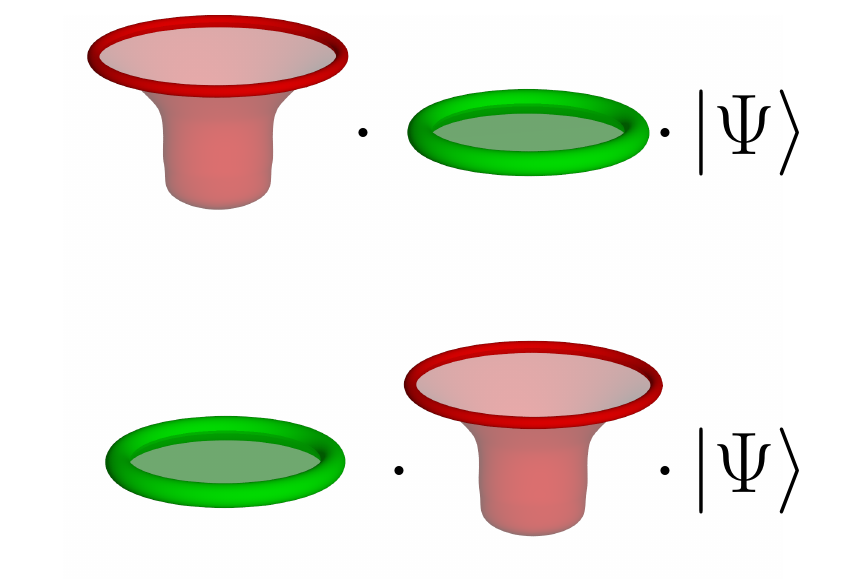}
			
				\caption{(Copy of Figure 43 from Ref. \cite{HuxfordPaper1}) The commutation of operators used to calculate the braiding. The partially transparent surfaces indicate the membranes for the operators, while the opaque loops indicate the excited regions, which are the boundaries of the membranes.}
				\label{LoopLoopCommutation}
			\end{center}
		\end{figure}

			As we did with the flux-charge braiding, we can express the braiding relation between two loops in terms of the commutation relation between creation operators. The flux tubes are created and moved by membrane operators, so the appropriate commutation relation is between two membrane operators, as indicated in Figure \ref{LoopLoopCommutation}. In the Abelian case, the two magnetic membrane operators commute, and so the loop braiding between two magnetic fluxes is trivial. This is because in this case, the magnetic membrane operator simply multiplies each cut edge by the label of the magnetic operator. This is in contrast to the non-Abelian case, where the action of the membrane operator on each edge depends on the value of the path from the start-point to that edge. This means that, in the Abelian case, the two membrane operators only share support when their membranes cut some of the same edges, so that the two membrane operators directly change the same edge label. Even then, the action on a shared edge is the same regardless of the order in which the operators act. Consider the action of two membrane operators $C^{h_1}(m_1)$ and $C^{h_2}(m_2)$ on such a shared edge $i$ (i.e., one cut by the dual membranes of both operators). If we first act with the membrane operator labeled by $h_2$ and then by the operator labeled by $h_1$, the edge label goes from $g_i$ to $h_1 h_2 g_i$ (possibly with inverses on $h_1$ or $h_2$, depending on the relative orientation of the membranes and the edge). On the other hand, when $C^{h_1}(m_1)$ acts first on the edge, followed by $C^{h_2}(m_2)$, the total effect on the edge label is given by $g_i \rightarrow h_2 h_1 g_i$. This is the same as $h_1h_2g_i$ (because $G$ is Abelian), so the two membrane operators commute and the braiding is trivial.

		\subsection{Non-Abelian case}
		\subsubsection{Flux-charge braiding}
		\label{Section_Flux_Charge_Braiding}
		In the case where $G$ is non-Abelian, the braiding relations between the magnetic fluxes and electric charges are a little more complex, although they still match our expectations from conventional gauge theory (see, e.g., Refs. \cite{Bais1980,Bucher1992} and our discussion in Section II of Ref. \cite{HuxfordPaper1}). Recall from the Abelian case that the electric string operator fails to commute with the magnetic membrane operator because the latter operator changes the label of one (or possibly more) of the edges along the path of the electric ribbon. In the Abelian case, the action on the path element was simple. The affected edge was multiplied by a fixed element $h$ or the inverse, and this factor could be brought to the front of the product of group elements that make up the path element, so that the entire path element was also multiplied by $h$ or the inverse. In the non-Abelian case, this is no longer true. Firstly, any changes to the edge cannot simply be extracted to the front of the path element by commutation. Secondly, the action of the membrane on the individual edge that is changed is more complex, depending on the path from the start-point of the magnetic membrane to the affected edge. This means that, rather than multiply the affected edge label by a fixed element $h$, we multiply the label by an element within the conjugacy class of $h$, with this element depending on the path element from the start-point of the membrane to the edge. However, this path element depends on the state that we act on, and even in the ground state the element is not generally fixed (the ground state is made of a superposition of states with different values of this path element) and so we must leave this path element as an operator. Therefore, the braiding relation is not generally well defined. To illustrate this idea, consider performing exactly the same braiding as in the Abelian case, by passing an electric ribbon operator through a magnetic membrane operator. Again we split the path $t$ of the electric ribbon into the path $t_1$ before the intersection; the edge $i$ along which the ribbon and membrane intersect; and the path $t_2$ after the intersection. Of these parts, only the group element $\hat{g}_i$ assigned to the edge $i$ is affected. For now, assume that the edge $i$ points along the path $t$, so that $t=t_1 i t_2$. If this path passes through the direct membrane of the magnetic membrane operator before the dual membrane (i.e., for a particular choice of relative orientation of ribbon and membrane), we have that
		 $$\hat{g}_iC^h(m)=C^h(m) \hat{g}(t_s)^{-1}h\hat{g}(t_s)\hat{g}_i,$$
		 where $t_s$ is the path from the start-point of the membrane to the crossing point and we have assumed that the ribbon is aligned so that the path reaches the direct membrane of the magnetic membrane operator before the dual membrane. Then for the entire path element, we have that
		 \begin{align*}
		\hat{g}(t)C^h(m)&= \hat{g}(t_1)\hat{g}_i \hat{g}(t_2)C^h(m)\\
		&=C^h(m)\hat{g}(t_1) \hat{g}(t_s)^{-1}h\hat{g}(t_s) \hat{g}_i \hat{g}(t_2).
		\end{align*}
		 
		 If we had taken edge $i$ to point against the path, we would have a similar result, because the edge element $\hat{g}_i$ would appear with an inverse in the path element, but the edge element would be right-multiplied by the inverse factor $\hat{g}(t_s)^{-1}h^{-1}\hat{g}(t_s)$ by the membrane operator, so we would obtain
		\begin{align*}
		 \hat{g}(t)C^h(m)&= \hat{g}(t_1)\hat{g}_i^{-1} \hat{g}(t_2)C^h(m)\\
		 &=C^h(m)\hat{g}(t_1) (\hat{g_i}\hat{g}(t_s)^{-1}h^{-1}\hat{g}(t_s))^{-1} \hat{g}(t_2)\\
		 &=C^h(m)\hat{g}(t_1)\hat{g}(t_s)^{-1}h\hat{g}(t_s) \hat{g}_i^{-1} \hat{g}(t_2).
		\end{align*}
		 
		 We can combine these cases by introducing $\hat{g}_i^{\sigma_i}$, where $\sigma_i$ is 1 if the edge and path align and $-1$ otherwise. Then we have
		 \begin{align*}
		 \hat{g}(t)C^h(m)&= \hat{g}(t_1)\hat{g}_i \hat{g}(t_2)C^h(m) \\
		 &=C^h(m)\hat{g}(t_1) \hat{g}(t_s)^{-1}h\hat{g}(t_s) \hat{g}_i^{\sigma_i} \hat{g}(t_2).
		 \end{align*}
		 
		 By inserting the identity in the form $\hat{g}(t_1)^{-1} \hat{g}(t_1)$, we can write the commutation relation as
		\begin{align}
		 \hat{g}(t)&C^h(m) \notag\\
		 &=C^h(m)\hat{g}(t_1) \hat{g}(t_s)^{-1}h\hat{g}(t_s)\hat{g}(t_1)^{-1} \hat{g}(t_1) \hat{g}_i^{\sigma_i} \hat{g}(t_2) \notag \\
		 &=C^h(m)(\hat{g}(t_s)\hat{g}(t_1)^{-1})^{-1} h(\hat{g}(t_s)\hat{g}(t_1)^{-1}) \hat{g}(t).
		\end{align}
		   
		 This is similar to the commutation relation from the Abelian case, except that the path element gains a factor of
		 $$(\hat{g}(t_s)\hat{g}(t_1)^{-1})^{-1} h(\hat{g}(t_s)\hat{g}(t_1)^{-1}),$$
		 instead of simply $h$. This factor is an operator and has no definite value in general, so the effect on the electric excitation depends on which configuration within the ground state we consider and we cannot extract a definite braiding relation. However, there is one special case where we can obtain a definite braiding relation. When the electric string starts at the start-point of the magnetic membrane, then the path sections $t_s$ and $t_1$ start and end at the same points as each-other. Provided that these path sections can be deformed into one-another without crossing over any excitations, the fake-flatness condition imposed by the plaquette energy terms ensures that $\hat{g}(t_s)=\hat{g}(t_1)$ up to a potential factor of $\partial(e)$ for some $e \in E$. Such factors of $\partial(e)$ do not affect 
		  $$(\hat{g}(t_s)\hat{g}(t_1)^{-1})^{-1} h(\hat{g}(t_s)\hat{g}(t_1)^{-1})$$
		because $\partial(e)$ is in the centre of $G$, so the factor of $\partial(e)$ and $\partial(e)^{-1}$ from $\hat{g}(t_s)$ and $\hat{g}(t_s)^{-1}$ cancel. Therefore,
		\begin{align}
	(\hat{g}(t_s)\hat{g}(t_1)^{-1})^{-1} h(\hat{g}(t_s)\hat{g}(t_1)^{-1}) \hat{g}(t)&= \partial(e)^{-1}h \partial(e)\hat{g}(t) \notag\\
		&=h\hat{g}(t). \label{Equation_path_magnetic_commutation_same_start_point}
		\end{align}
		
		This relation then gives us a simple braiding relation in the ``same-site" (or same start-point) case. Note that we gave this braiding relation for a particular choice of relative orientation for the ribbon and membrane operator, and if we reversed this relative orientation then the element $g(t)$ would instead be multiplied by $h^{-1}$ (as we show in Section \ref{Section_electric_magnetic_braiding_3D_tri_trivial} in the Supplemental Material). This braiding relation is a simple extension of the Abelian case. Having said that, the non-Abelian nature of the group does still have some relevance when we use our irrep basis for the electric operators. Recall that the electric excitations are labeled by irreps of the group $G$, combined with matrix indices for the irrep. When $G$ is Abelian, these representations are 1D and we need not worry about matrix indices. However, when $G$ is non-Abelian some of these irreps are not 1D. We can look at the effect of braiding on an electric ribbon labeled by a representation $R$ and its indices, for which we find that:
		\begin{align}
		&\sum_{g \in G} [D^{R}(g)]_{ab} \delta (g, \hat{g}(t)) C^h(m) \ket{GS} \notag \\
		& \ = \sum_{g \in G} C^h(m) [D^{R}(g)]_{ab} \delta (g, h\hat{g}(t))\ket{GS} \notag \\
		& \ =\sum_{g \in G} C^h(m) [D^{R}(g)]_{ab} \delta (h^{-1}g, \hat{g}(t))\ket{GS} \notag \\
		& \ =\sum_{g'=h^{-1}g} C^h(m) [D^{R}(hg')]_{ab} \delta (g', \hat{g}(t))\ket{GS} \notag \\
		& \ = C^h(m)  \sum_{c=1}^{|R|} [D^{R}(h)]_{ac}  \sum_{g' \in G} [D^{R}(g')]_{cb} \delta (g', \hat{g}(t)) \ket{GS}. \label{Equation_electric_magnetic_irrep_braiding}
		\end{align}
		
		We see from Equation \ref{Equation_electric_magnetic_irrep_braiding} that the braiding mixes electric ribbon operators labeled by different matrix indices but the same representation. The fact that the representation is left invariant suggests that the representations label the purely electric topological sectors, given that braiding cannot mix different sectors.

		The importance of the start-points of the membrane and ribbon operators when it comes to braiding can be interpreted in terms of gauge theory. Just as we discussed for the 2+1d case in Ref. \cite{HuxfordPaper2}, the start-point of the magnetic membrane can be seen as a unique point in that the flux tube produced by the magnetic membrane operator has a definite flux label with respect to this point even within the conjugacy class. When we give a flux tube a flux label, we must specify the path with respect to which we measure this flux. The path must link with the flux tube, but smoothly deforming the path should not change the flux label measured (where by smoothly deform, we mean pulling through the space represented by the lattice to another position on the lattice). An exception to this is the start-point of the path. Moving this start-point can change the flux label that we would assign to the flux tube (see, e.g., Ref. \cite{Alford1992}). Suppose that the flux label measured with respect to a particular start-point is $h$. Then if we measure the flux label of the flux tube starting from a different point, the result is related to $h$ by conjugation by a path element for a path between the two start-points. In our model this path element is not usually well-defined, because the energy eigenstates are usually linear combinations of states with different group elements assigned to the path. We say that the path element is generally operator valued. Therefore, the flux tube does not generally have definite flux with respect to points other than the start-point (with an exception if the flux label is in the centre of $G$). This non-definite flux is reflected in the flux-flux braiding. This idea is explained more clearly in the context of field theory in Ref. \cite{Alford1992}.

		A second interpretation of the start-point dependent fusion comes from anyon theory. We only have definite fusion in the case where the membrane operators share a start-point. Then, as in the 2+1d case, we may expect that we only have definite braiding when the fusion channel is definite: that is, when the operators involved have a common start-point.

		\subsubsection{Flux-flux braiding}
		\label{Section_Flux_Flux_Braiding_Tri_Trivial}
		As with the flux-charge braiding, the general result of braiding two magnetic fluxes is more complicated when $G$ is non-Abelian. However, as with the flux-charge braiding, there is a special case with simple braiding relations, when the two magnetic membrane operators have the same start-point. This makes sense if we think of the base-point as the point of definite flux, because we only expect a definite braiding result if the two flux tubes have definite flux when measured with respect to the same point.

		We consider the case where one flux tube (which we will call the inner loop) is passed through another loop (which we call the outer loop). The inner loop takes the role of the red loop in the left diagram of Figure \ref{LoopMoves}. Then in the same start-point case the label of the inner loop is conjugated by the label of the outer loop, while the label of the outer loop is unchanged. That is, if the label of the inner membrane operator is $k$ and the outer membrane operator is labeled by $h$, under braiding the label of the excitation from the inner membrane becomes $h^{-1}kh$ and the label of the outer membrane is unchanged. The simplest way to obtain this braiding relation is to use the topological nature of the operators. We can freely deform the membranes, as long as they do not cross an excitation and no excitations are moved in doing so. We can therefore pull the membrane that produces the inner loop fully through the other membrane (as there are no excitations within the membrane itself). However, when we do this we must keep the start-point fixed (because it may be excited), so the start-point is not moved through the outer membrane. However, recall that the action of the magnetic membrane operator depends on a set of paths from the start-point to the edges being changed by the membrane operator. As we deform the membrane, the start of these paths (the start-point) is fixed, while the ends are pulled through the outer membrane (if these ends were not already on the other side of the membrane). Therefore, all of these paths intersect the outer membrane. This means that the group elements of these paths will be changed by the action of the outer membrane operator, which will in turn affect the action of the inner membrane operator. To see how the inner membrane operator changes under the commutation relation we need to know how the labels of these paths are affected.

\begin{figure}[h]
			\begin{center}
				\includegraphics[width=\linewidth]{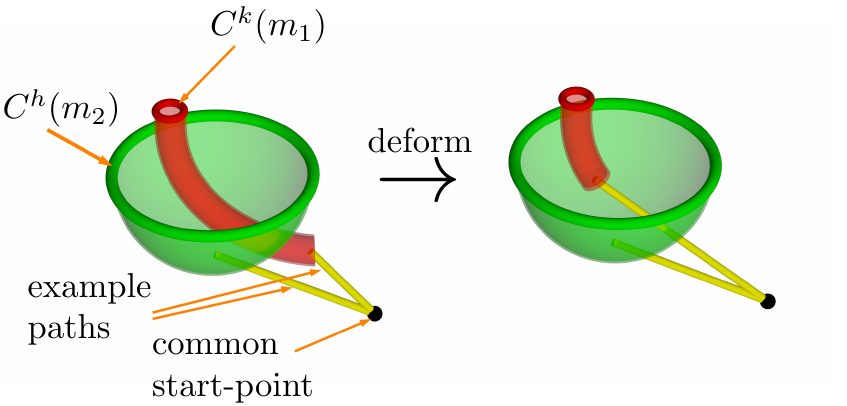}
				
				\caption{To calculate the braiding of two loops, we consider the situation shown on the left where we first apply a magnetic membrane operator $C^h(m_2)$, then apply a membrane operator $C^k(m_1)$ which intersects with the first membrane and pushes a loop excitation through that first membrane. We choose the start-points of these two membranes to be the same. Example paths from the common start-point to the two membranes are shown as (yellow) cylinders. We can use the topological property of the membrane operators to deform the inner (red) membrane, $m_1$, and pull it through the green membrane $m_2$ to obtain the image on the right-hand side. However, when we do so we must leave the start-point fixed. Therefore, the paths from the start-point to each point on the membrane $m_1$, such as the example path shown here, must pass through $m_2$. This leads to a non-trivial braiding relation in general.}
				\label{loop_loop_braiding_deform}
			\end{center}
		\end{figure}

		As an example, consider Figure \ref{loop_loop_braiding_deform}. In the left side of the figure we have the membranes in their original position, with the red membrane nucleating a loop at the common start-point and moving the red loop through the green one. To calculate the commutation relation, we deform the red membrane so that it is entirely pulled through the green membrane, as in the right side of Figure \ref{loop_loop_braiding_deform}. However, when we deform a membrane we must keep any excitations fixed, including the potentially excited start-point. Therefore, the start-point is fixed. This means that the paths from the start-point to the red membrane, like the example path to the red membrane in the right-hand side of the figure, pass through the green membrane and so can be affected by the green membrane operator. This is significant because these paths determine the action of the membrane operator, as explained in Section \ref{Section_3D_Tri_Trivial_Magnetic_Excitations}. An edge $i$ cut by the dual membrane of the magnetic membrane operator $C^k(m_1)$ has its edge label $g_i$ multiplied by a factor $g(s.p(m_1)-v_i)^{-1}kg(s.p(m_1)-v_i)$ (or the inverse), where $v_i$ is the vertex on the direct membrane that is attached to $i$. These factors have the form $g(t)^{-1}hg(t)$, where each path $t$ now intersects with the green membrane operator $C^h(m_2)$. We must therefore find the commutation relation of such a path label operator $g(t)$ with the membrane operator.

		We already saw how a magnetic membrane operator affects the label of paths that pierce the membrane when we looked at the charge flux-braiding. From the charge braiding calculation we know that the label of a path $t$ that starts at the common start-point and intersects the outer membrane changes from $g(t)$ to $hg(t)$, where $h$ is the label of the outer membrane operator being intersected (see Equation \ref{Equation_path_magnetic_commutation_same_start_point}). As discussed previously, this path element $g(t)$ appears in the action of the other (inner) membrane operator. If the inner membrane has label $k$, the membrane operator acts on an edge cut by the membrane by multiplying the edge label by $g(t)^{-1}kg(t)$, where $t$ is the path from the start-point to that edge. Under commutation with the outer membrane, when $g(t)$ changes to $hg(t)$, this action becomes multiplication by $g(t)^{-1} h^{-1}k h g(t)$, which is equivalent to the action of an unbraided membrane of label $h^{-1}k h$. Therefore, we see that the label of one of the flux tubes is conjugated by braiding. As we expect, the conjugacy class of the flux is invariant under braiding, but the flux element within the conjugacy class can be changed.

		It is worth noting that if we had not deformed one membrane to pull it entirely through the other, some of the paths from the start-point would not pierce the other membrane and so would be unaffected by the commutation relation. This means that the action of the membrane in the region from the start-point to the intersection of the membrane is unaltered. This reflects the fact that a membrane operator moves the loop excitation associated to it. Before the intersection, the membrane operator is moving the excitation before it has braided, so its label is the original label of the loop. After the intersection, braiding has occurred and so the label of the membrane (and the excitation) has changed. The precise point at which braiding has occurred is somewhat arbitrary in an anyon theory (although we can guarantee whether braiding has occurred if the excitations start and end in the same position). Similarly the choice of location for the membranes is somewhat arbitrary when the membrane operators act on the ground state because we can deform the membranes without affecting their action, and so we can change the location where the membrane operators intersect by deforming the membranes. This reflects the freedom in considering at which point during the motion the braiding transformation is applied (although if the excitations start and end in the same position, then the membrane operators will definitely intersect if braiding occurs, regardless of how we deform the membranes).

		Having obtained the braiding relation, it is useful to consider how braiding affects a linear combination of magnetic membrane operators with label within a certain conjugacy class, such as $\sum_h \alpha_h C^h(m)$. If the magnetic membrane operator is an equal superposition of operators labeled by each element of a conjugacy class, then the conjugation of the labels by the braiding only permutes the labels within the conjugacy class, which has no effect when the coefficient for each element is the same. Therefore, the overall membrane operator transforms trivially under braiding. For a general superposition, the conjugation (and so permutation of the labels) does affect the operator and so the braiding is non-trivial. These conditions match the conditions for the start-point of the membrane operator to be unexcited or excited. A magnetic operator with an unexcited start-point is an equal superposition of magnetic operators with labels within a conjugacy class, and so is unaffected by braiding through other magnetic excitations. On the other hand, a magnetic operator with an excited start-point will transform non-trivially when it braids through some other magnetic excitations. This is because, if the start-point is excited, the magnetic excitation carries a point-like charge, which enables it to braid non-trivially with other magnetic excitations when passed through them. Note that the same condition does not hold for the outer membrane operator, which can affect the inner membrane operator even if it does not have an excited start-point. This is because we can shrink the inner loop to a point before braiding it without affecting the braiding relation, whereas the loop-like character of the outer loop is essential for the braiding.

		It is important to note that the precise form of the braiding relation depends on the orientation of the loops involved. Flipping a magnetic excitation is equivalent to changing its label from $h$ to $h^{-1}$. Therefore, if we were to flip the orientation of the outer membrane from our earlier calculation, then the label of the inner membrane would change from $k$ to $hkh^{-1}$ under braiding rather than changing to $h^{-1}kh$. Flipping the orientation of the inner one does not change the expression, because the transformation is the same when we invert both sides: $k^{-1} \rightarrow h^{-1}k^{-1}h \implies k \rightarrow h^{-1}kh$. Therefore, if the orientation of both loops is flipped, the braiding transformation is
		\begin{equation}
		k \rightarrow hkh^{-1}.
		\label{braid_relation_magnetic_flipped}
		\end{equation}
		
		\subsubsection{Linking}
		\label{Section_linking}
		In addition to the non-trivial loop-loop braiding, there is another feature of loop excitations not present for point excitations. Two loop excitations may be \textit{linked}. In this case, depending on the labels of the two excitations, there may be an energetically costly ``linking string" that joins the two loops. This situation is indicated in Figure \ref{linking}. As we show in Section \ref{Section_linking_appendix} in the Supplemental Material, this linking string is present between two linked magnetic excitations when their labels do not commute. If the two loops are labeled by $g$ and $h$ and their membranes have the same start-point, then the two loops are linked by a string with a label similar to $ghg^{-1}h^{-1}$. The exact label depends on the relative orientations of the two loops and which path we choose to use to define the flux of the linking string (so $g$ or $h$ could appear with an inverse in the label of the linking string, or the label could be conjugated by $g$ or $h$).

		\begin{figure}[h]
			\begin{center}
				\includegraphics[width=\linewidth]{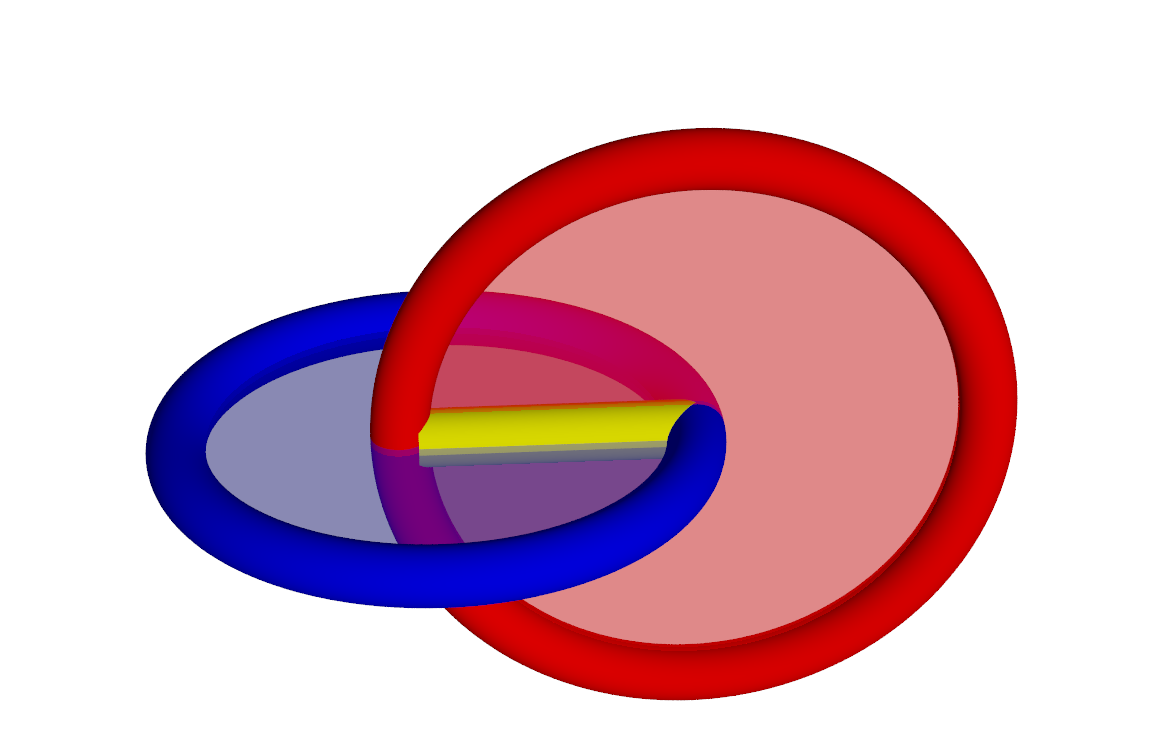}
					
				\caption{Two linked loops may have an energetically costly linking string between them, here indicated by the short (yellow) string.}
				\label{linking}
			\end{center}
		\end{figure}

\begin{figure}[h]
			\begin{center}
				\includegraphics[width=\linewidth]{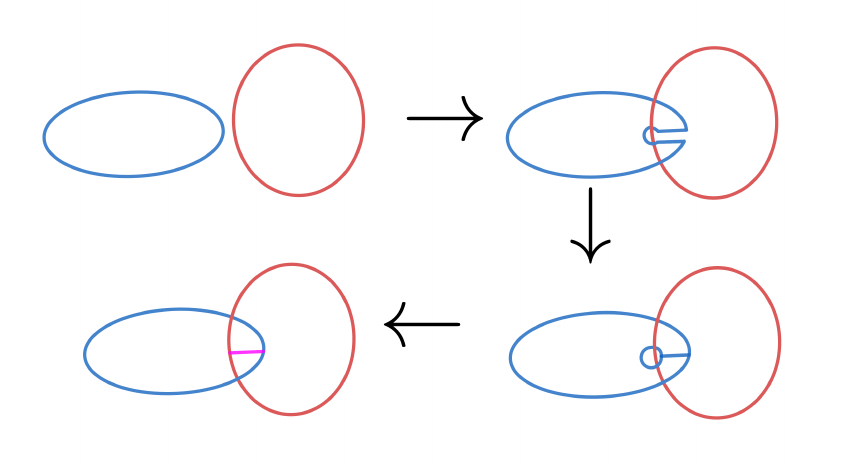}
			
				\caption{Given two loops (blue and red here) that cannot pass through each-other, we can push them together. To do this we must deform the boundary of one of the loops (second figure). The deformation then encloses the other loop (third figure). We can consider this deformed section of the loop as a new object, which is the linking string (coloured pink in the fourth figure), whose flux label will depend on the labels of the two linked loops.}
				\label{linking_process}
			\end{center}
		\end{figure}

		This linking string indicates that there is an obstruction to pushing the two loops through one-another and so pushing them through results in an energetically costly linking string. One way of viewing this is that the two strings are unable to pass through each-other. Therefore, instead of the loops being pushed through each-other, one is deformed to accommodate the other, as shown in Figure \ref{linking_process}. In Figure \ref{linking_process}, part of one of the loops envelops the other one and folds back on itself, as seen in the bottom-right of the figure (this becomes the pink string in the lower-left of the figure). We can consider this part of the two loops as the linking string. It is possible to work out the flux label of the linking string from this picture by writing a path that links with the thin section as a combination of the paths defining our two original fluxes (recall from Section \ref{Section_Flux_Charge_Braiding} that a flux tube is defined along with a path linking with that flux tube, which measures the flux value). A more complete argument for this (in terms of generic fluxes, rather than in terms of this specific model) is given in Ref. \cite{Alford1992}.
		
\begin{figure}[h]
			\begin{center}
				\includegraphics[width=\linewidth]{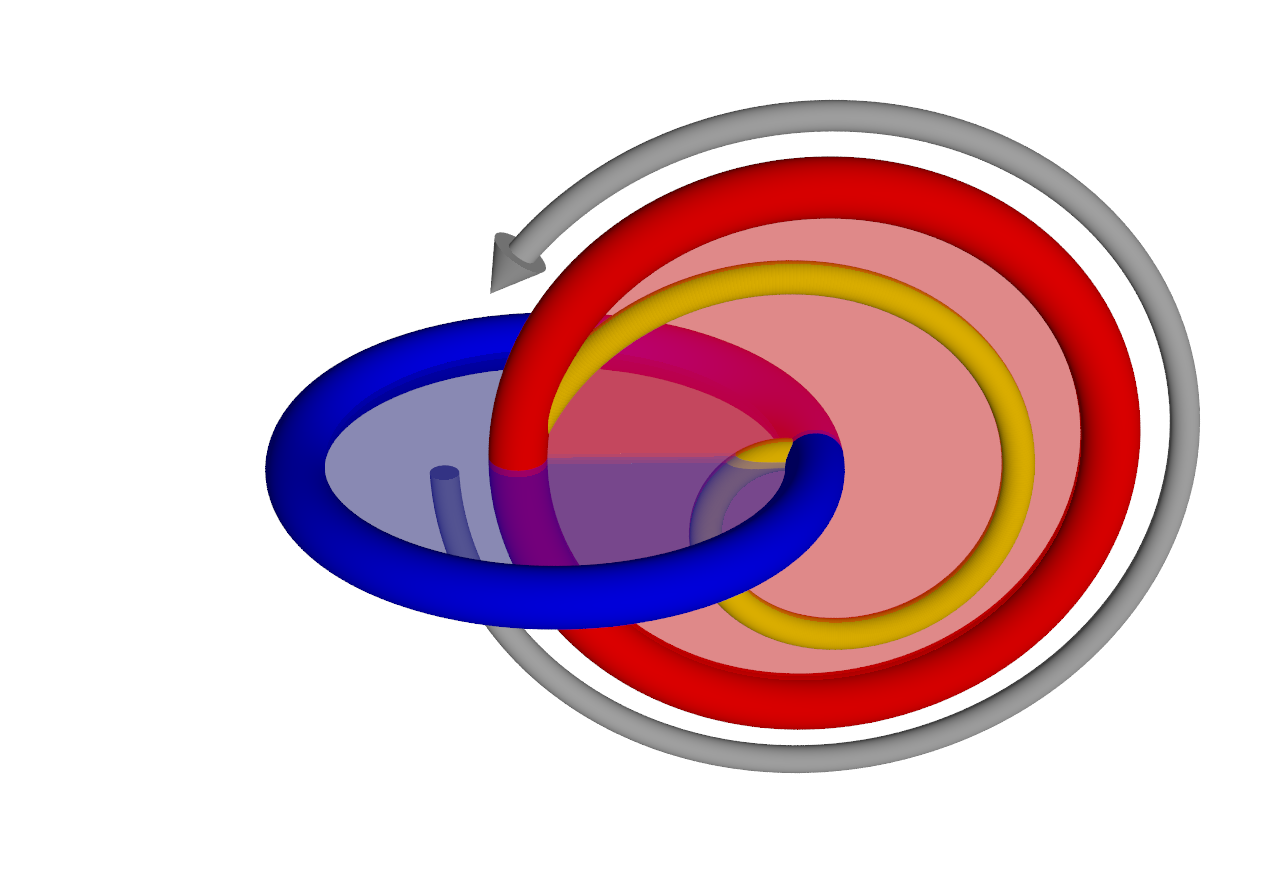}

				\caption{Rotating a linked loop drags the linking string and conjugates the label of part of the membrane of the rotating loop.}
				\label{linkrotation}
			\end{center}
		\end{figure}

		The presence of the linking string indicates that further relative motion of the two loops is non-trivial, because such motion will move the string (the position of which can be detected through the energy terms). For example, we can consider rotating one of the loops by a full rotation. This would be a trivial motion if the loops were unlinked. However, when the loops are linked the linking string will follow this rotation, as shown in Figure \ref{linkrotation}, indicating that the motion is non-trivial. In order to implement this rotation, we make the direct membrane paths (the paths that we defined when constructing the membrane operator) spiral outwards, wrapping around the linking (blue) loop. If we wanted to write the label of these paths in terms of the label of an unwrapped path, we would gain a factor that accounts for the flux of the blue loop. This in turn affects the action of the membrane operator, changing the effective label of parts of the membrane operator by conjugation by the flux of the blue loop.

		\subsubsection{Three-loop braiding}
\begin{figure}[h]
			\begin{center}
				\includegraphics[width=\linewidth]{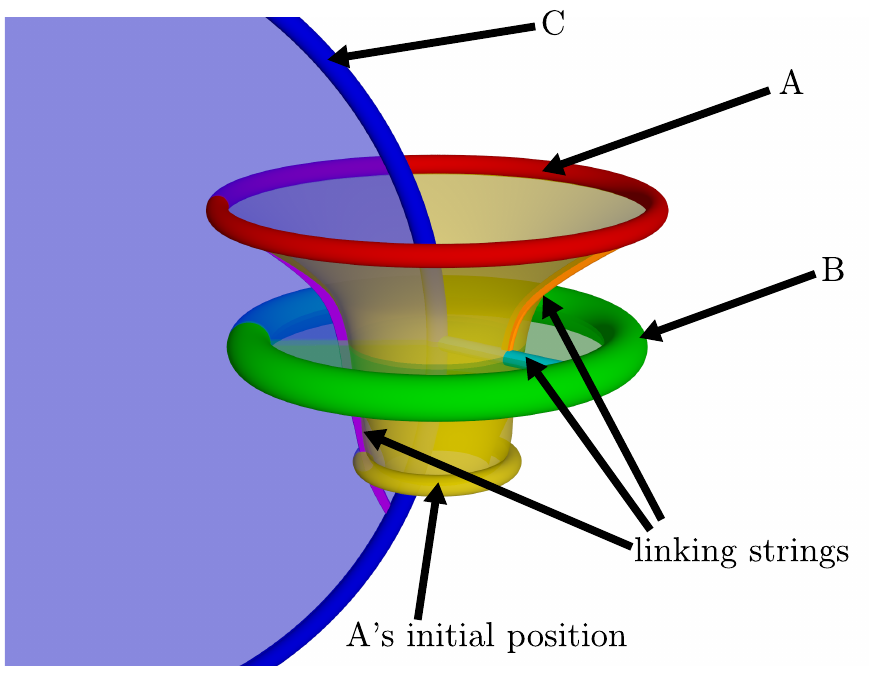}
				
				\caption{An example of three-loop braiding. The open strings (orange, purple and cyan) are possible linking strings }
				\label{threeloopbraiding}
			\end{center}
		\end{figure}

		It has become clear \cite{Wang2014, Jiang2014} that when considering loops, the simple case where two loops pass through each-other (two-loop braiding, shown in the left side of Figure \ref{LoopMoves}) that we have described so far does not fully describe the general topological properties of loops. A more general example of braiding is where two loops pass through each-other while both loops are linked to a third loop (see Figure \ref{threeloopbraiding}). This is known as three-loop \cite{Wang2014} or necklace braiding \cite{Bullivant2020a}.

		In this model however, the result of three-loop braiding is similar to that of the ordinary braiding. The only difference is that the two loops may also drag linking strings with the third loop. This is shown in Figure \ref{threeloopbraiding}. The transformation of the loop labels is otherwise the same as in the ordinary case. One exception is that the linking string of two magnetic excitations can cancel the confining string of a confined blob excitation (provided that the labels of the linking string and blob excitation agree, as described in Section \ref{Section_linking_appendix} in the Supplemental Material), which enables those blob excitations to move and braid freely while attached to a linked flux, which they would not normally be able to do.

		\subsection{Loop-blob braiding}
		\label{Section_Loop_Blob_Braiding_Tri_Trivial}
\begin{figure}[h]
			\begin{center}
				\includegraphics[width=\linewidth] {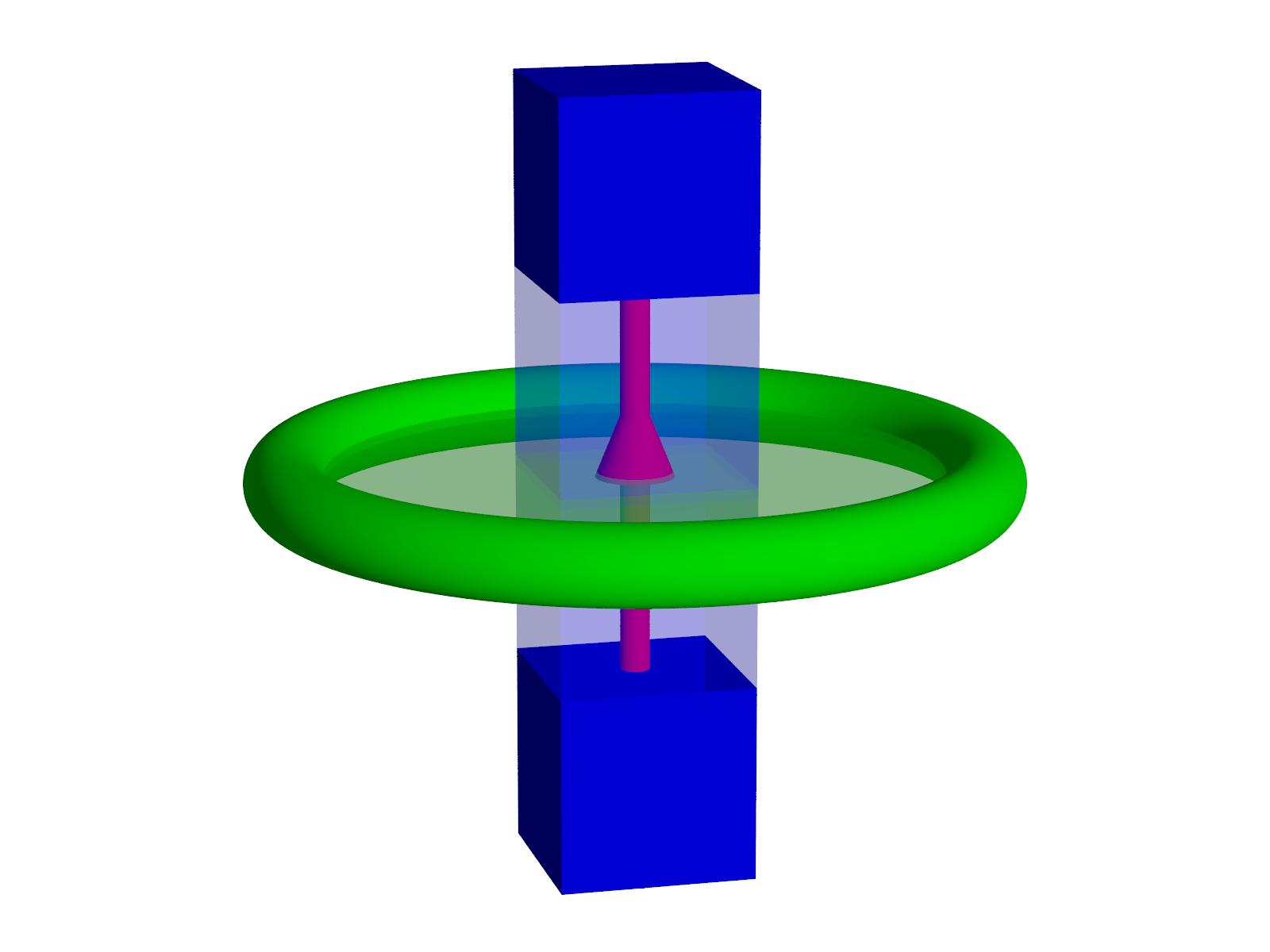}

				\caption{Schematic of blob-loop braiding. The blue cubes (dark gray in grayscale) represent the blob excitations at the ends of the ribbon operator (whose ribbon is represented by the translucent cuboid and the arrow). The ribbon operator moves one of the blob excitations through the loop-like excitation produced by an $E$-valued membrane operator applied on the translucent (green) membrane.}
				\label{blobloop}
			\end{center}
		\end{figure}

		So far we have considered the excitations that are described by the group $G$. The final non-trivial braiding is between the two types of excitation that are associated to the group $E$, the $E$-valued loops and blob excitations. In this case it is easy to find the braiding relations by looking at the effect of the blob ribbon operator on the surface measured by the operator that produces the loop. We consider a situation where the blob excitation passes through the loop excitation. To implement this situation on our lattice, we apply both an $E$-valued membrane operator and a blob ribbon operator whose path intersects with that membrane, as shown in Figure \ref{blobloop}.
		
\begin{figure}[h]
			\begin{center}
					\includegraphics[width=\linewidth]{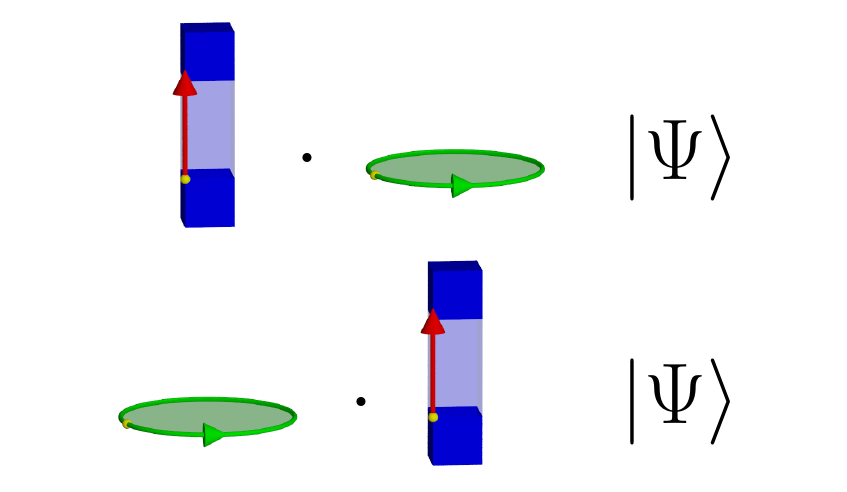}
		
				\caption{In order to determine the blob-loop braiding relations, we compare the situation shown in the top line, where we first apply an $E$-valued membrane operator and then a blob ribbon operator that intersects with that membrane, to the situation shown in the bottom line, where we apply the operators in the opposite order. Here $\ket{\psi}$ is any state with no other excitations near the support of the two operators (e.g., a ground state).}
				\label{blobloopdetail}
			\end{center}
		\end{figure}

		In order to compute the braiding relation, we compare the situation where we first create the loop and then push the blob through, thus performing the braiding move, with the one where we push the blob through empty space before producing the loop. The relevant commutation relation is shown in Figure \ref{blobloopdetail}.
		
		As we saw in Section \ref{Section_3D_Blob_Excitation_Tri_Trivial}, the blob ribbon operator with label $e$ multiplies the labels of the plaquettes pierced by its path by $e$ or $e^{-1}$, depending on the relative orientation of the plaquette and ribbon. This action is not sensitive to the presence of the $E$-valued loop excitation and so the blob ribbon operator is unaffected by the commutation. On the other hand, the membrane operator for the loop excitation is affected by the action of the blob ribbon operator. Recall that the $E$-valued membrane operator measures the surface element of the membrane on which it is applied. This surface element, $\hat{e}(m)$, is a product of the elements of individual plaquettes: $\hat{e}(m)=\prod_{\text{plaquettes in m}} \hat{e}_{\text{plaquette}}^{\pm 1}$, where the $\pm$ accounts for the relative orientation of the surface and plaquette. We defined the blob ribbon operator to intersect the membrane, and so it will affect the label of one of the plaquettes in this product. If the blob ribbon operator pierces the membrane $m$ through a plaquette $q$, then the label $e_q$ of that plaquette is multiplied by $e$ or $e^{-1}$, depending on the relative orientation of the plaquette and the ribbon. This in turn means that the contribution $e_q^{\pm 1}$ of the plaquette to the surface $m$ will be multiplied by $e$ or $e^{-1}$, depending on the relative orientation of the membrane and the ribbon. The orientation of the membrane matters rather than that of the plaquette, because the $\pm 1$ in the expression for the surface label accounts for the relative orientation of plaquette and membrane (if the plaquette is anti-aligned with the membrane, the inverse in $e_q^{- 1}$ converts a factor of $e$ from the ribbon operator into $e^{-1}$ if the orientation of the ribbon opposes the plaquette but matches the membrane, or vice-versa if the orientation of the ribbon matches the plaquette but opposes the membrane). If the orientation of the membrane matches the orientation of the blob ribbon operator, $e_q^{\pm 1}$ will be multiplied by $e^{-1}$. This indicates that $\hat{e}(m)B^e(t) =B^e(t) e^{- 1} \hat{e}(m)$ in this case. Then, considering the basis operator for our space of $E$-valued membrane operators labeled by an irrep $\gamma$ of $E$ (as defined in Equation \ref{Equation_E_membrane_irrep_Abelian}), the commutation relation with the blob ribbon operator $B^e(t)$ is given by
		\begin{align*}
		B^e(t) L^{\gamma}(m)&\ket{GS}=B^e(t) \sum_{e' \in E} \gamma({e'}) \delta(e',\hat{e}(m)) \ket{GS}\\
		&= \sum_{e' \in E} \gamma({e'}) \delta(e',e\hat{e}(m)) B^e(t) \ket{GS}\\
		&=\sum_{e' \in E} \gamma({e'}) \delta(e^{-1} e',\hat{e}(m)) B^e(t) \ket{GS}\\
		&=\sum_{e''=e^{-1}e'} \gamma(ee'') \delta(e'',\hat{e}(m)) B^e(t) \ket{GS}\\
		&=\gamma(e) \sum_{e'' \in E} \gamma(e'') \delta(e'',\hat{e}(m)) B^e(t) \ket{GS},
		\end{align*}
		where we used the fact that $E$ is Abelian to take $\gamma$ as a 1D irrep and separate $\gamma(ee'')$ into $\gamma(e)$ and $\gamma(e'')$. Therefore,
		\begin{equation}
				B^e(t) L^{\gamma}(m)\ket{GS}= \gamma(e) L^{\gamma}(m) B^e(t) \ket{GS}.
		\end{equation}
		Having the $E$-valued membrane operator on the left of the product (and the blob ribbon operator on the right) corresponds to the unbraided case (because in this case the blob excitation moved before the loop excitation is present), so the braiding of our two excitations results in accumulating a phase of $\gamma(e)$. A similar argument holds in the case where the membrane operator and blob ribbon operator are anti-aligned, except that we should replace $e$ by its inverse.

		It is worth noting that the blob excitations with label not in the kernel of $\partial$ (that is the confined blob excitations, as we saw in Section \ref{Section_3D_Blob_Excitation_Tri_Trivial}) braid non-trivially with the condensed $E$-valued loop excitations (those with trivial representation of the kernel), while those with label in the kernel braid trivially with them. This is because the condensed $E$-valued loop excitations have trivial representation of the kernel: $\gamma(e_K)=1$ for $e_K$ in the kernel. Therefore, the phase gained is 1 when a condensed loop braids with an unconfined blob (which carries a label in the kernel). This matches our expectation that only the confined excitations can braid non-trivially with the condensed excitations.

		\subsection{Summary of braiding when $\rhd$ is trivial}
		For convenience, we summarize the excitations that braid non-trivially with each-other in Table \ref{Table_Braiding_Tri_Trivial}. We can see that the excitations split into two sets. The electric and magnetic excitations (the excitations corresponding to the group $G$) braid non-trivially with each-other and the blob and $E$-valued loop excitations (the excitations corresponding to $E$) braid non-trivially with each-other, but there is no non-trivial braiding between the two sets.

		\begin{table}[h]
			\begin{center}
				\begin{tabular}{ |c|c|c|c|c| } 
					\hline
					Non-Trivial& &Magnetic & & $E$-valued \\
					Braiding?& Electric & flux & Blob & loop \\ 
					\hline
					Electric & \xmark & \cmark & \xmark & \xmark\\ 
					\hline
					Magnetic & && &\\ 
					
					flux & \cmark & \cmark & \xmark & \xmark \\
					\hline 
					Blob& \xmark & \xmark & \xmark & \cmark\\
					\hline
					$E$-valued & & & & \\
					loop  & \xmark & \xmark & \cmark & \xmark\\
					\hline
				\end{tabular}
				
				\caption{A summary of which excitations braid non-trivially in Case 1, where $\rhd$ is trivial. A tick indicates that at least some of the excitations of each type braid non-trivially with each-other, while a cross indicates that there is no non-trivial braiding between the two types. Notice that the table has a block-diagonal structure, with non-trivial braiding only in the blocks.}
				\label{Table_Braiding_Tri_Trivial}
			\end{center}
			
		\end{table}

		\section{Ribbon and membrane operators in the fake-flat case}
		
		Now that we have considered the first of our special cases, where $\rhd$ is trivial, we move on to another of our special cases (Case 3). We consider the case where our groups $G$ and $E$, as well as our maps $\rhd$ and $\partial$, are completely general, but we restrict our Hilbert space to only allow fake-flat configurations. Many of the features of the excitations are common between the two cases, so we will examine the differences between them rather than repeating our previous discussion entirely.
		
		\label{Section_3D_MO_Fake_Flat}
		\subsection{Electric excitations}
		The electric excitations are unchanged by taking $\rhd$ non-trivial. Just as in the $\rhd$ trivial case, we measure the value of a path and assign a weight according to the value of the path. This creates two point-like excitations at the ends of the path. The operators are best labeled by irreps of the group $G$, with non-trivial irreps giving the excitations and the trivial irrep giving the identity operator. The excitations labeled by irreps with a non-trivial restriction to the image of $\partial$ are confined.
		
		\subsection{$E$-valued loop excitations}
		\label{Section_3D_Loop_Tri_Non_Trivial}
		
		Next we consider the $E$-valued loop excitations, which are produced by membrane operators that measure the surface label of a membrane:
		$$L^{\vec{\alpha}}(m) = \sum_{e \in E} \alpha_e \delta(\hat{e}(m),e).$$
		Here $\hat{e}(m)$ is the surface label of the membrane $m$ and the $\alpha_e$ are a general set of coefficients. This operator has the same form as the corresponding operator for the $\rhd$ trivial case. However, there is a slight difference when we consider our irrep basis for the space of $E$-valued membrane operators (see Equation \ref{Equation_E_membrane_irrep_Abelian} for the $\rhd$ trivial case). Because our group $E$ may now be non-Abelian, the irreps are generally not 1D. This means that our basis for the $E$-valued membrane operators must include the matrix indices for those irreps, so that our general basis operator takes the form
		\begin{equation}
		L^{\mu,a,b}(m)= \sum_{e \in E} [D^{\mu}(e)]_{ab} \delta(\hat{e}(m),e), \label{Equation_E_membrane_irrep_basis_non_Abelian}
		\end{equation}
		where $\mu$ is an irrep of $E$, and $a$ and $b$ are the matrix indices.

		In addition to this slight difference in the presentation of the basis operators, we also have some physical differences compared to the $\rhd$ trivial case. When $\rhd$ is non-trivial, whenever we measure a surface we must specify a base-point with respect to which we measure that surface label \cite{Bullivant2017}. Because the membrane operator for the $E$-valued loop excitations involves measuring a surface, we must specify a base-point for the measurement. We call that base-point the start-point of our membrane operator. Similarly to the start-point of the magnetic membrane operator, this start-point may be excited by the action of the $E$-valued membrane operator. Recall from Section \ref{Section_Recap_3d} that a vertex transform at the base-point of a surface affects the value of that surface, with $A_v^g$ taking the surface label from $e$ to $g \rhd e$. This means that the vertex term (which is made of a sum of vertex transforms) at the start-point of our membrane operator may not commute with our membrane operator, which may result in the vertex being excited (while the other vertex terms are still left unexcited). Whether the start-point vertex is excited or not depends on whether the coefficients of $\delta(\hat{e}(m),e)$ are a function of the $\rhd$-classes of $E$, where two elements $e_1$ and $e_2$ are in the same $\rhd$-class if there exists a $g \in G$ such that $e_2 = g \rhd e_1$ (this is an equivalence relation). If the coefficient for each element $e \in E$ is equal to the coefficient for each element related by the $\rhd$ action, such as $g \rhd e$, then the start-point is not excited. On the other hand, if for each $\rhd$-class the coefficients for the elements within that $\rhd$-class sum to zero, then the start-point is excited. We note that the irrep basis given in Equation \ref{Equation_E_membrane_irrep_basis_non_Abelian} does not provide a good description for this phenomenon, because the coefficients $[D^{\mu}(e)]_{ab}$ given by the matrix elements of an irrep do not transform in a particular way under an $\rhd$ action, and this action can even cause mixing between irreps. Generally, to get membrane operators which either definitely excite the start-point or definitely leave it unexcited, we must consider linear combinations of the basis operators given in Equation \ref{Equation_E_membrane_irrep_basis_non_Abelian}.

\begin{figure}[h]
			\begin{center}
				\includegraphics[width=\linewidth]{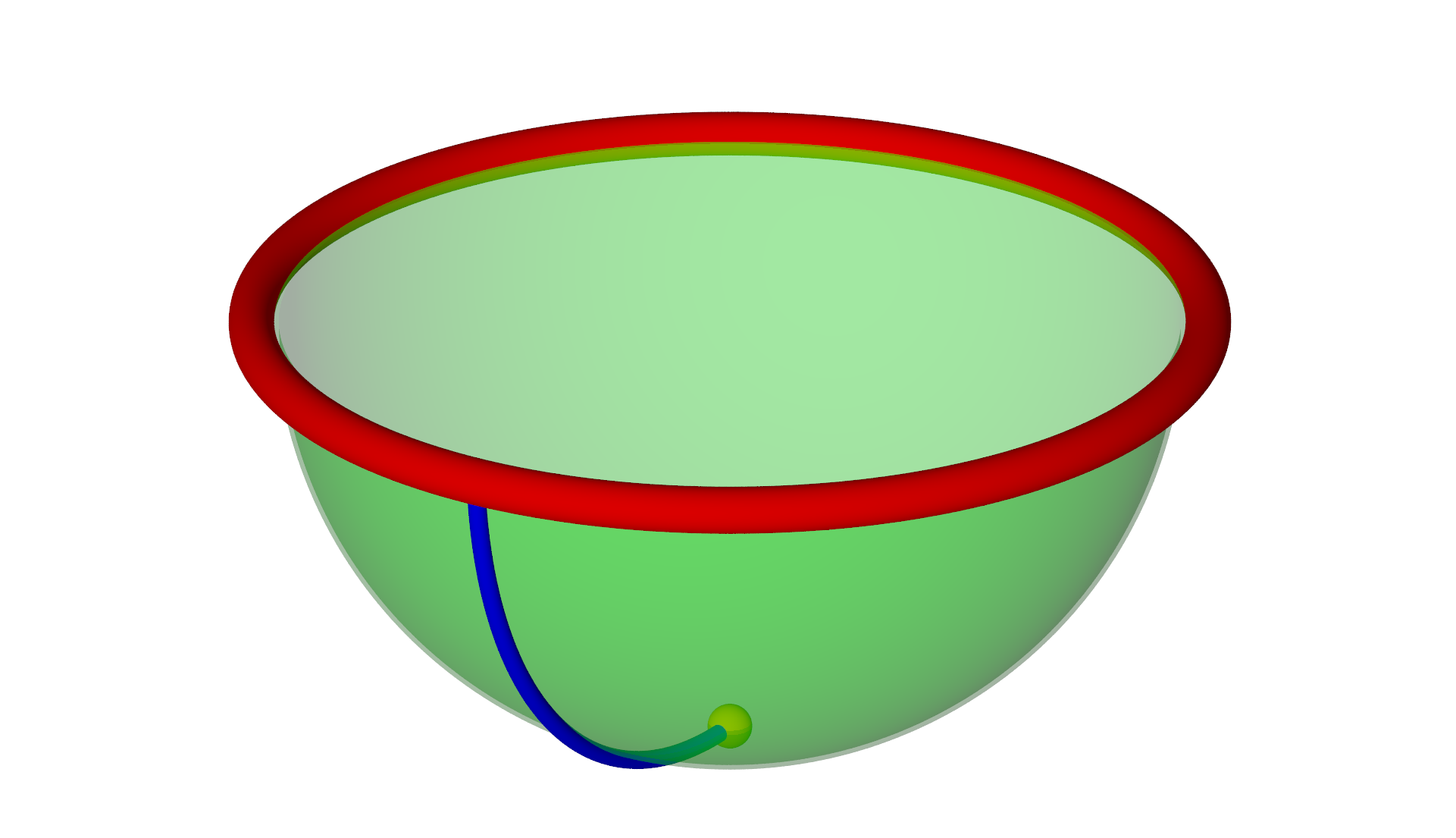}

				\caption{The boundary of the surface (green) measured by the membrane operator starts and ends at the start-point (yellow sphere). If this start-point is away from the naive boundary of the membrane (represented by the red torus) there may be a line of excited edges (blue path) connecting the start-point to the loop-like excitation, because the surface label measured by the membrane operator generally transforms non-trivially under such edge transforms when $E$ is non-Abelian. This indicates that for some loop-like excitations there is an energetic cost to moving the excitation away from the start-point.}
				\label{E_membrane_sp_away}
			\end{center}
		\end{figure}

		In addition to the start-point, we also need to be careful when determining the boundary of the surface, which supports the loop-like excitation. The boundary of the surface always starts and ends at the start-point of the membrane operator, because this start-point is the base-point of the surface. This means that if we nucleate a loop-like excitation at the start-point and then try to move it away from the start-point, part of the boundary still connects to the start-point, as shown in Figure \ref{E_membrane_sp_away}. This section of the boundary is attached to the start-point and the edges in this section appear twice in the boundary, with opposite orientation each time (see Section S-I C in the Supplemental Material of Ref. \cite{HuxfordPaper2} for more details about the boundary of surfaces), just like a whiskering path of the type considered in Section I D of Ref. \cite{HuxfordPaper1} (also shown in Figure \ref{move_basepoint}). For this reason, we will refer to such a section of boundary as a whiskered section. As we showed in Section S-I C of the Supplemental Material of Ref. \cite{HuxfordPaper2}, the edges along such sections of boundary may be excited if $E$ is non-Abelian (whereas if $E$ is Abelian these edges are never excited). Whether these edges are excited by a particular membrane operator or not depends on the coefficients associated to that membrane operator. For an $E$-valued membrane operator 
		$$L^{\vec{\alpha}}(m) = \sum_{e \in E} \alpha_e \delta(\hat{e}(m),e)$$
		with a general set of coefficients $\alpha_e$, the edges on the whiskered section of boundary are left unexcited if the coefficients are a function of conjugacy class, so that $\alpha_e = \alpha_{fef^{-1}}$ for all $e, f \in E$. For example, if the coefficients are given by the characters of an irrep $\mu$ of $E$ (which are invariant under conjugation), then the edges will be unexcited. On the other hand, the edges will be excited if the coefficients within each conjugacy class sum to zero. One important thing to note is that if the start-point is not excited, then the edges from the start-point to the loop are not excited either (as we may expect, because the start-point becomes unimportant when it is unexcited). To see this, we note that if the start-point is unexcited, then the coefficients of the membrane operator are invariant under the $\rhd$ action: $\alpha_e = \alpha_{g \rhd e}$ for all $g \in G$ and $e \in E$. In particular, this means that the coefficients are invariant under an action of the form $\partial(f) \rhd$ for any $f\in E$: $\alpha_e = \alpha_{\partial(f) \rhd e}$. However, because of the Peiffer condition Equation \ref{Equation_Peiffer_2}, $\partial(f) \rhd e = fef^{-1}$. That is, the $\rhd$ action from an element in $\partial(E)$ is equivalent to conjugation. Therefore, if the coefficient is invariant under any $\rhd$ action, it is also invariant under conjugation, and so if the start-point is unexcited then so are any edges on the whiskered section of the boundary.

		This idea that there may be a string of excited edges from the start-point to the loop-like excitation is significant for the behaviour of the excitation. If the edges are not excited then it indicates that we can move the loop-like excitation created by the membrane operator without dragging any excitations from the start-point. That is, the excitation is not confined. On the other hand, if the edges are excited then we cannot move the loop-like excitation away from the start-point without incurring an energy cost. This means that the excitation is confined. However, note that the additional energy cost is associated to a confining string which connects the loop to the start-point. This additional energy does not depend on the size of the loop itself. As we discuss in Section \ref{Section_Sphere_Charge_Reduced} (although we do not consider the fake-flat case in that section, we would expect a similar result), the loop-like excitations of this model can also carry point-like charge (which is balanced by the charge carried by the start-point). The fact that the confining energy cost depends on the separation of the loop and start-point (rather than the area of the membrane enclosed by the loop) suggests that it is the point-like charge that is confined, rather than the loop-like charge. This is further supported by the fact that the membrane operators that do not produce an excitation at the start-point never produce confined loop-like excitations.

		\subsection{Blob excitations}
		\label{Section_3D_Blob_Fake_Flat}
\begin{figure}[h]
			\begin{center}
			\includegraphics[width=\linewidth]{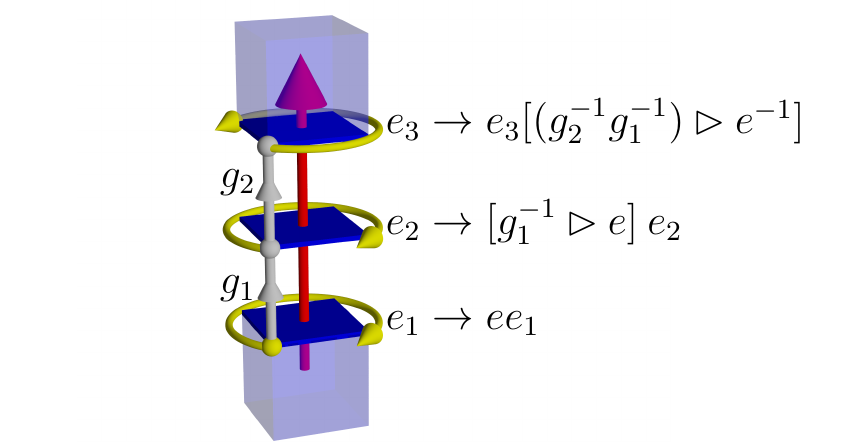}
			
				\caption{When $\rhd$ is non-trivial, the effect of the blob ribbon operator on a plaquette depends on the (inverse of the) path label for a path from a designated start-point of the operator (yellow sphere labeled $s.p$) to the base-point of the plaquette (the grey sphere attached to each plaquette). As in the $\rhd$ trivial case, the orientation of the plaquette (indicated by the curved yellow arrows) determines whether the plaquette label is left-multiplied by $e$ or right-multiplied by $e^{-1}$.}
				\label{effectbloboperator}
			\end{center}
		\end{figure}

		The blob excitations are changed significantly by taking $\rhd$ non-trivial and enforcing fake-flatness at the level of the Hilbert space. Firstly, as we saw in Section \ref{Section_3D_Blob_Excitation_Tri_Trivial}, in the $\rhd$ trivial case some of the blob ribbon operators excite the plaquettes along their length (and would still do so when we take $\rhd$ to be non-trivial). These confined ribbon operators must be thrown out in the fake-flat case because they violate fake-flatness. This means that the labels of the blob ribbon operators are restricted to lie in the kernel of $\partial$. Secondly, the action of each blob ribbon operator is more complicated. In addition to the path between the centres of blobs, we must specify a path on the lattice from a privileged vertex, called the start-point of the operator, to the base-points of each affected plaquette. We call this path the direct path, and call the original path (which pierces the affected plaquettes) the dual path. We can either have a single direct path that runs through each of these base-points (so that the path to each base-point is an extension of the path to the previous base-point), or instead have a set of paths, one for each pierced plaquette. Now instead of simply multiplying the plaquette labels by $e$ or $e^{-1}$, the blob ribbon operator left-multiplies the label of each plaquette $p$ pierced by the dual path by $g(s.p-v_0(p))^{-1} \rhd e$ or right-multiplies the plaquette elements by the inverse, where $(s.p-v_0(p))$ is the path from the start-point of the ribbon operator to the base-point of the affected plaquette and $g(s.p-v_0(p))$ is the corresponding group element. As in the $\rhd$ trivial case, whether the element or its inverse are used depends on the orientation of the plaquette with respect to the dual path of the blob ribbon operator, with the inverse used if the plaquette aligns with the dual path. A simple example of this action is shown in Figure \ref{effectbloboperator}.
		
		The way the direct path affects the action of the blob ribbon operator is similar to how the direct path affects the action of the magnetic ribbon in 2+1d (as we saw in Ref. \cite{HuxfordPaper2}), except that instead of conjugation by the path element we have this $\rhd$ action. If we were allowed blob ribbon operators labeled by a general element of $E$ (rather than just an element in the kernel of $\partial$), the precise path chosen for this direct path would be significant. This is because when we deform a path $t$ over a fake-flat surface, while keeping the start and end points fixed, the path label $g(t)$ is altered by a factor of the form $\partial(f)$ for some $f \in E$. When $E$ is general, this additional factor of $\partial(f)$ causes a non-trivial difference in expressions of the form $g(t)^{-1} \rhd e$, such as those which appear in the action of the blob ribbon operator. Specifically, from the Peiffer conditions (Equations \ref{Equation_Peiffer_1} and \ref{Equation_Peiffer_2} in Section \ref{Section_Recap_3d}) we have $(\partial(f)g(t)^{-1}) \rhd e = f [g(t)^{-1} \rhd e] f^{-1}$. However, when we restrict the element $e$ labeling the blob ribbon operator to be in the kernel of $\partial$, we also ensure that $e$ is in the centre of the group $E$. To see that elements in the kernel of $\partial$ are also in the centre of $E$, we again use the Peiffer conditions. Given that $e_k$ is an element of the kernel of $\partial$, the second Peiffer condition (Equation \ref{Equation_Peiffer_2}) tells us that
		\begin{align*}
		e_k f e_k^{-1}=\partial(e_k) \rhd f &= 1_G \rhd f =f\\
		& \implies e_k f = f e_k \: \: \forall f \in E.
		\end{align*}
		That is, elements in the kernel must commute with all elements of $E$. We also note that if $e_k$ is an element of the kernel, then so is $g \rhd e_k$ for any $g \in G$. This is because $\partial(g \rhd e_k)=g \partial(e_k) g^{-1}=g g^{-1}=1_G$, from the first Peiffer condition (Equation \ref{Equation_Peiffer_1}). Therefore, for any element $e_k$ in the kernel of $\partial$, $g \rhd e_k$ is also in the kernel of $\partial$ and so is in the centre of $E$. This means that in our earlier expression, $f [g(t)^{-1} \rhd e] f^{-1}$ is equal to $g(t)^{-1} \rhd e$ when $e$ is in the kernel, and so the additional factor of $\partial(f)$ from deforming the path $t$ is irrelevant, at least when we act on states that obey fake-flatness. Therefore, when we restrict our blob excitations to the non-confined ones, which are labeled by elements of the kernel of $\partial$, the precise choice of paths from the start-point to the base-points of the affected plaquettes does not matter. This insensitivity to the path is only for smooth deformation over fake-flat regions, so if the lattice supports non-contractible cycles then different choices of path may give different actions for the ribbon operator, with these different actions being equivalent to taking different labels for all or part of the ribbon operator.

		Much as we saw with the magnetic excitation in 2+1d, this dependence of the action of the blob ribbon operator on the value of (sections of) the direct path may lead to the blob ribbon operator exciting the start-point of the operator. This is because vertex transforms at the start-point can affect the path label of the direct path. As we show in Section \ref{Section_Blob_Ribbon_Fake_Flat} of the Supplemental Material, the start-point vertex is not excited if the ribbon operator is an equal superposition of all the ribbon operators labeled by elements in an $\rhd$-class (the sets of elements related by the $\rhd$ action), but is excited if the coefficients for the elements in each $\rhd$-class sum to zero.

		\subsection{Condensation and confinement}

		In Section \ref{Section_3D_Condensation_Confinement_Tri_Trivial}, we discussed the pattern of confinement and condensation exhibited by the excitations of the higher lattice gauge theory model when $\rhd$ is trivial. In addition, we explained that for any pair of groups $G$ and $E$ that form a valid crossed module with $\rhd$ trivial, there is a family of crossed modules (and so lattice models) differentiated from each-other by different maps $\partial$ (assuming that the two groups can support different homomorphisms $\partial$). In each family, the model described by the crossed module for which $\partial$ maps only to the identity of $G$ is an ``uncondensed model" where there is no condensation or confinement present. Then the transition to other models described by the same groups $G$ and $E$ (but different $\partial$) is a condensation-confinement transition. When $\rhd$ is non-trivial, however, the picture is less clear. This is because, given a model described by an arbitrary crossed module, we cannot necessarily construct a corresponding uncondensed model. To see this, consider a generic crossed module $(G,E, \partial, \rhd)$. Now suppose that $E$ is a non-Abelian group. This means that there are some pairs of elements $e,f \in E$ such that $efe^{-1} \neq f$. From the Peiffer condition Equation \ref{Equation_Peiffer_2} (see Section \ref{Section_Recap_3d}), this means that $\partial(e) \rhd f \neq f$ for this pair. However, there is condensation and confinement when $\partial(E)$ is not the trivial group (as we discussed in the $\rhd$ trivial case in Section \ref{Section_3D_Condensation_Confinement_Tri_Trivial} and will describe in the fake-flat case later in this section). If there were an ``uncondensed" model with $\partial \rightarrow 1_G$, then $\partial(e) \rhd f \neq f$ from the non-Abelian nature of the group $E$ implies that $1_G \rhd f \neq f$. However, this is incompatible with the definition of $\rhd$ as a group homomorphism from $G$ to endomorphisms on $E$ (see Section \ref{Section_Recap_3d}), because this definition means that the identity element $1_G$ should be mapped to the identity map on $E$. Therefore, there is no such crossed module and so no ``uncondensed" model corresponding to the pair of groups $G$ and $E$. Because the Hilbert space is fixed by the lattice and the groups $G$ and $E$, for a model described by a general crossed module there does not seem to be a corresponding uncondensed model with the same Hilbert space (at least not in the space of higher lattice gauge theory models, though there may well be another model giving the ``uncondensed" phase). This means that we are unable to describe the general model in terms of a condensation-confinement transition in this work. However, we can still describe the pattern of confinement (i.e., which excitations cost energy to separate from their antiparticle) and condensation (i.e., which operators act equivalently to ``local" operators on the ground state), which we aim to do briefly in this section.

		The first excitations to consider are the electric excitations, some of which are confined due to their ribbon operators exciting the edge terms along the ribbon. These excitations have the same pattern of confinement as in the $\rhd$ trivial case considered in Section \ref{Section_3D_Condensation_Confinement_Tri_Trivial}. Namely, the confined electric ribbon operators $\sum_g \alpha_g \delta( \hat{g}(t),g)$ have coefficients which satisfy $\sum_{e \in E} \alpha_{\partial(e)g}=0$ for all $g \in G$ and the unconfined ribbon operators have coefficients which satisfy $\alpha_{\partial(e)g}= \alpha_g$ for all $g \in G$ and $e \in E$ (while general ribbon operators can be split into contributions from the two cases and leave the edges along the ribbon in a superposition of excited and unexcited states).

		We next consider the blob excitations. Some of these, namely those created by blob ribbon operators with label outside the kernel of $\partial$, would be confined, but because the mechanism for this confinement is the violation of the plaquette terms that enforce fake-flatness, these ribbon operators must be excluded from the fake-flat model.

		So far, this pattern of confinement is the same as for the $\rhd$ trivial case described in Section \ref{Section_3D_Condensation_Confinement_Tri_Trivial}. However, unlike in that case, some of the $E$ -valued loops are also confined, as we described in Section \ref{Section_3D_Loop_Tri_Non_Trivial}. That is, the $E$-valued membrane operators 
		$$\sum_{e \in E} \alpha_e \delta( \hat{e}(m),e)$$
		whose coefficients $\alpha_e$ are sensitive to conjugation (i.e., $\alpha_e$ does not equal $\alpha_{fef^{-1}}$ for some pair $e,f \in E$) may produce an excited string (in addition to the loop excitation itself) as the loop is moved away from the start-point of the membrane. In particular, if the coefficients satisfy $\sum_{f \in E} \alpha_{fef^{-1}} =0$ for each $e \in E$, then the string is definitely excited (whereas the string is definitely not excited if $\alpha_e = \alpha_{fef^{-1}}$ for all $e, f \in E$). As we noted in Section \ref{Section_3D_Loop_Tri_Non_Trivial}, this confinement appears to correspond to confinement of the point-like charge carried by the loop excitation, rather than of the loop-like charge, because the confinement energy does not depend on the area of the loop, and the confinement can only occur when the start-point of the membrane is excited (an excited start-point seems to indicate the presence of a non-trivial point-like charge, as we are able to show in Section \ref{Section_3D_sphere_charge_examples} for the less general case where $\partial$ maps to the centre of $G$ and $E$ is Abelian).

		Next we consider the pattern of condensation evident in the model in this fake-flat case (Case 3 from Table \ref{Table_Cases}). By this, we mean that we want to look at which excitations can be produced by operators that are local to the excitation itself. That is, condensed point-like excitations can be produced by operators that act only on a few degrees of freedom near the excitations, and condensed loop-like excitations can be produced by operators near the loops (which have linear extent, so the operators need not be local in the traditional sense). Because we are unable to construct the magnetic excitations when we restrict to fake-flatness, the only condensed excitations remaining are $E$-valued loop excitations. Just as in the $\rhd$ trivial case, the condensed $E$-valued loop excitations are those that are produced by membrane operators which are not sensitive to surface elements in the kernel of $\partial$. This is because if the membrane operator is only sensitive to $\partial(\hat{e}(m))$ (i.e., is not sensitive to the kernel of $\partial$), then it has the same action as an electric ribbon operator measuring the boundary label (whereas a membrane operator sensitive to the kernel of $\partial$ can resolve information not obtainable from the boundary path element).

		One interesting fact about this pattern of condensation is that it can coexist with the confinement of the $E$-valued loop excitations. Recall that the condition for the $E$-valued loop excitation to be confined is that $ \sum_{f \in E} \alpha_{fef^{-1}} = 0$ for all $e \in E$. This is not mutually exclusive with the condition that the excitation is condensed ($\alpha_e = \alpha_{e e_k}$ for any $e \in E$ and $e_k \in \ker(\partial)$). For example, consider a crossed module of the form $(G,E=G, \partial = \text{id}, \rhd \rightarrow \text{ conj.})$, for which the two groups $G$ and $E$ are the same, with $\partial$ being the identity map, while $\rhd$ maps to conjugation (i.e., $g \rhd e =g e g^{-1}$). In this case, the only element of the kernel of $\partial$ is the identity element $1_G=1_E$. Therefore, any $E$-valued membrane operator will satisfy the condensation condition (i.e., any coefficients $\alpha_e$ satisfy $\alpha_e = \alpha_{ee_k}$ for all $e_k$ in the kernel of $\partial$, because $e_k$ can only be the identity element). This means that any coefficients satisfying the confinement condition $ \sum_{f \in E} \alpha_{fef^{-1}} = 0$ for this crossed module will also trivially satisfy the condensation condition. How can it be that an excitation is simultaneously confined and condensed? This is because, as we discussed previously in this section, it is the point-like charge carried by the excitation that is confined. On the other hand, the way we have defined condensation means that the condensation must be of the loop-like charge: we have shown that the loop excitation can be produced by an operator local to the excitation (which has linear extent), but this does not mean that the point-like charge can be produced locally in a point-like sense (i.e., with support only on a few degrees of freedom). This fact demonstrates that in future study of condensation and confinement in 3+1d, we must be careful to consider what exactly we mean by condensation or confinement, and which charges (not just excitations) undergo condensation.

		\section{Braiding in the fake-flat case}
		\label{Section_3D_Braiding_Fake_Flat}
		
		Next we discuss the braiding relations in this special case, Case 3 from Table \ref{Table_Cases}, in which we restrict to fake-flat configurations. The fact that we are unable to include the magnetic excitations (because they violate fake-flatness) means that the braiding relations are rather simple. Indeed, the only remaining non-trivial braiding is between the $E$-valued loop excitations and the blob excitations. Just as in the 2+1d case considered in Ref. \cite{HuxfordPaper2}, however, the signatures of the magnetic excitations are still present in the ground states of manifolds with non-contractible cycles, which can have labels outside of $\partial(E)$. Before we discuss the braiding proper, we will briefly describe how the excitations transform as they are moved around such non-contractible cycles.
		
		\subsection{Moving excitations around non-contractible cycles}
		
		The first type of excitation that we wish to consider moving around a non-contractible cycle is the electric excitation. The transformation obtained by moving an electric excitation around such a cycle is the same in the 3+1d case as in the 2+1d case considered in Ref. \cite{HuxfordPaper2}. Namely, if we compare an electric ribbon operator applied on a path $s$ to one that is applied on the path $s \cdot t$ obtained by concatenating the original path with a non-contractible closed path $t$ is
		\begin{equation}
		S^{R,a,b}(s \cdot t) = \sum_{c=1}^{|R|} [D^R(\hat{g}(t))]_{cb} S^{R,a,c}(s)
		\end{equation}
		where $S^{R,a,b}(s \cdot t)$ is the electric ribbon operator labeled by irrep $R$ of $G$ and matrix indices $a$ and $b$. We see that there is mixing between different electric ribbon operators labeled by the same irrep $R$, with this mixing controlled by the matrix $D^R(\hat{g}(t))$ representing the path element $\hat{g}(t)$ in irrep $R$. The path element $\hat{g}(t)$ is an operator, and the ground states are not typically eigenstates of this operator even for closed paths $t$.
		
		\begin{figure}[h]
			\begin{center}
				\includegraphics[width=\linewidth]{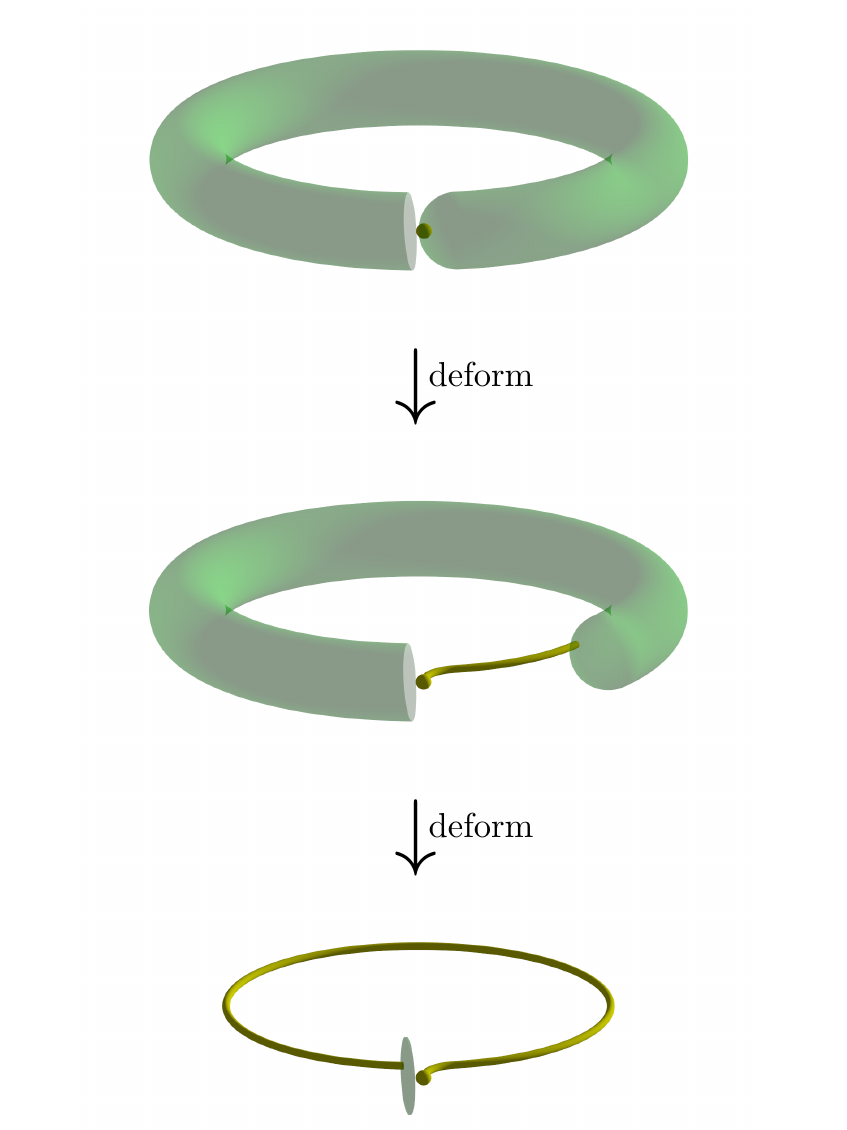}
		
				\caption{Given an $E$-valued membrane operator (green) wrapping around a closed cycle, we can deform the section wrapping around the cycle and shrink it down to nothing. This just leaves a whiskering string (yellow) connecting the start-point (yellow sphere) to the small part of the membrane remaining (the green disk in the final image).}
				\label{E_valued_membrane_whisker_deform}
			\end{center}
		\end{figure}

	In a similar way, we can find how the $E$-valued loop-like excitations transform as they are moved around a non-contractible cycle. In order to do so, we compare $E$-valued membrane operators applied on two membranes $m$ and $m'$ which are the same except that $m'$ is whiskered around the non-contractible cycle $t$. This is because the membrane operators are topological, and so a membrane travelling around the cycle can be deformed by shrinking the section of the membrane around the cycle down to nothing, so that only a whiskering path $t$ remains, as shown in Figure \ref{E_valued_membrane_whisker_deform}. Then the $E$-valued membrane operator $L^{\mu,a,b}(m')$ applied on this whiskered membrane is given by 
		$$L^{\mu,a,b}(m') = \sum_{e \in E} [D^{\mu}(e)]_{ab} \delta(e, \hat{e}(m')), $$
	where $\mu$ is the irrep of $E$ labeling the membrane operator (and $a$ and $b$ are the matrix indices labeling the operator). The surface element $\hat{e}(m')$ can be written in terms of the surface element of the unwhiskered membrane $m$ using the rules for whiskering surfaces given in Ref. \cite{Bullivant2017}. We have
		$$\hat{e}(m') =\hat{g}(t) \rhd \hat{e}(m),$$
	where $t$ is the closed cycle, which is also the path $(s.p(m')-s.p(m))$ between the start-points of the two membranes. Then we have
		\begin{align}
		L^{\mu,a,b}(m') &= \sum_{e \in E} [D^{\mu}(e)]_{ab} \delta(e, \hat{e}(m')) \notag\\
		&= \sum_{e \in E} [D^{\mu}(e)]_{ab} \delta(e, \hat{g}(t) \rhd \hat{e}(m))\notag \\
		&= \sum_{e \in E} [D^{\mu}(e)]_{ab} \delta( \hat{g}(t)^{-1} \rhd e, \hat{e}(m))\notag \\
		&= \sum_{e'= \hat{g}(t)^{-1} \rhd e \in E} [D^{\mu}(\hat{g}(t) \rhd e')]_{ab} \delta( e', \hat{e}(m)).
		\end{align}
	The matrix $D^{\mu}(\hat{g}(t) \rhd e')$ can be considered as the matrix representation for element $e'$ in a new irrep, $\hat{g}(t) \rhd \mu$. Therefore, we see that moving the $E$-valued loop excitation around the path $t$ mixes the irreps related by this $\rhd$ action. We say that irreps related by the action of $g \rhd$ for some $g \in G$ belong to the same $\rhd$-Rep class of irreps.

	The final type of excitation to consider passing around a non-contractible cycle is the blob excitation. Recall from Section \ref{Section_3D_Blob_Fake_Flat} that the action of a blob ribbon operator $B^e(r)$ on a plaquette $p$ pierced by the ribbon $r$ is (choosing the plaquette to be aligned with $r$ for simplicity)
	$$B^e(r) :e_p = e_p [g(s.p(r)-v_0(p))^{-1} \rhd e^{-1}].$$

	Taking the path $s.p(r)-v_0(p)$ to be $t \cdot s$, where $t$ is a closed non-contractible cycle, we can write this action as
		\begin{align*}
		B^e(r) :e_p &= e_p [g(t \cdot s)^{-1} \rhd e^{-1}]\\
		&= e_p [(g(s)^{-1}g(t)^{-1}) \rhd e^{-1}]\\
		&= e_p [g(s)^{-1} \rhd (g(t)^{-1} \rhd e^{-1})]
		\end{align*}
	which is the same as the action of a blob ribbon operator that does not wrap around the cycle (so that the path $(s.p(r)-v_0(p))$ is just $s$) except with $e$ replaced by $g(t)^{-1} \rhd e$. Note that here we have taken the direct path of the ribbon operator to wrap around the non-contractible cycle, but not the dual path. If we also let the dual path wrap around the cycle then the plaquettes pierced by $s$ obtain two factors acting on the plaquette label, one from the ribbon before it wraps the cycle (corresponding to the label $e$) and one from the ribbon after it wraps the cycle (corresponding to the label $g(t)^{-1} \rhd e$ as above).

	\subsection{Loop-blob braiding}
	\label{Section_3D_Loop_Blob_Braiding_Fake_Flat}
	Compared to the loop-blob braiding that we saw in Section \ref{Section_Loop_Blob_Braiding_Tri_Trivial}, the loop-blob braiding in this special case is slightly more complicated. This is because the action of the ribbon and membrane operators involved now depends on the values of various paths on the lattice. For example, as we saw in Section \ref{Section_3D_Blob_Fake_Flat}, the blob ribbon operator multiplies plaquette elements by a group element $\hat{g}(s.p-v_0(p))^{-1} \rhd e$ which depends on the label of a path. This label is really an operator, because the value of the path label depends on what state we are acting on. In particular, the ground state does not have a definite value of this label, instead being made up of a linear combination of states with different path labels. Because of this, we may expect that the braiding does not generally give us a definite result and that the braiding relation may depend on such operator-valued labels. However, as with the braiding of the flux tubes that we saw in Section \ref{Section_Flux_Flux_Braiding_Tri_Trivial}, the braiding relations are simple for particular cases where the start-points of the operators match. Therefore, we are most interested in these same start-point commutation relations. To understand these, it will be useful to first discuss the interpretation of the blob and loop excitations in 2-gauge theory.

\begin{figure}[h]
			\begin{center}
			\includegraphics[width=0.8\linewidth]{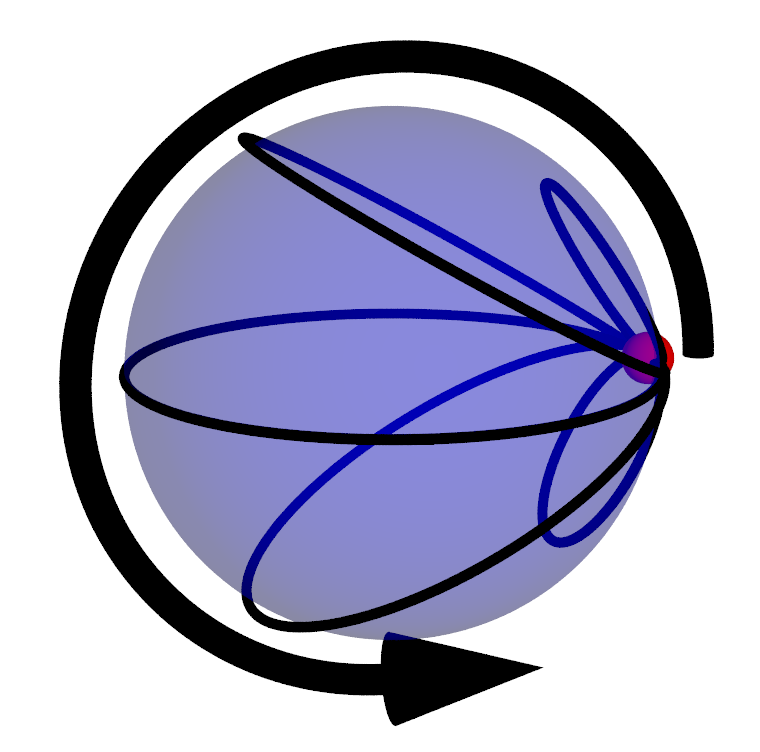}

				\caption{(Copy of Figure 25 from Ref. \cite{HuxfordPaper1}) The 2-holonomy of a surface (in this case a sphere) can be measured by a transport process. A small loop is created at the base-point (the small red sphere), then dragged over the surface (the larger blue sphere), as indicated by the arrow.}
				\label{surfaceholonomy1}
			\end{center}
		\end{figure}

	Similar to lattice gauge theory, it is useful to consider the gauge invariants of higher lattice gauge theory, which can be built from quantities associated to closed loops and surfaces \cite{Bullivant2017}. In addition to the ``1-flux" of loops, which is also present in ordinary gauge theory, we have the ``2-flux" or 2-holonomy of closed surfaces. The 2-flux itself, described by an element of $E$, is not a gauge-invariant quantity. However, the 2-flux of a closed surface is only changed within certain equivalence classes of elements by the gauge transforms \cite{Bullivant2017} (as we will see for tori and spheres in Sections \ref{Section_3D_Topological_Charge_Torus_Tri_nontrivial} and \ref{Section_sphere_topological_charge_appendix_full} in the Supplemental Material), and so these classes in $E$ are gauge-invariant, with the identity element in particular belonging to a class of its own. This means that the 2-flux is still a useful quantity. In this model, the blob excitations are associated to non-trivial 2-flux on a surface enclosing the blob excitation. The boundary of the excited blob is itself a surface with non-trivial 2-flux, because an excited blob by definition has a non-trivial surface label on its boundary. To measure this 2-flux, we must pass a loop over the closed surface whose 2-flux we wish to measure. When we measure the 2-flux, we must specify the base-point with respect to which we measure the 2-flux. The choice of base-point is equivalent to a gauge choice, so choosing a different base-point can give a different element for the 2-flux, within the same $\rhd$-class (i.e., the new element is related to the old one by a $\rhd$ action). Once we have chosen the base-point, our measurement loop must be nucleated at that base-point before being passed over the surface, as shown in Figure \ref{surfaceholonomy1}. In this model, the $E$-valued loop excitations measure 2-flux, as we can see from the fact that the corresponding measurement operators assign a weight to each possible surface label.
		
	This idea about measuring a 2-flux has important ramifications for our braiding. We have seen that the blob excitations are non-trivial 2-fluxes and the $E$-valued loop excitations measure surface elements, so these loop excitations can measure the 2-flux of the blob excitations. This is why there is non-trivial braiding between these two types of excitation. When we compare the situation where the loop excitation is passed over the blob to the situation where it is not (i.e., we compare the braided case to the unbraided case), we measure the 2-flux of the blob excitation. However, the blob ribbon operator produces excitations that have definite 2-flux only with respect to the start-point of the blob ribbon operator. Similarly, the $E$-valued loop measures 2-flux with respect to the start-point of the membrane operator that creates the loop excitation. Therefore, we expect a definite braiding relation when the start-point for our blob ribbon operator matches the start-point for our $E$-valued loop membrane operator. Note that when $\rhd$ is trivial, the start-points lose meaning (we do not need a direct path for the blob ribbon operators and the surface does not need a base-point for the $E$-valued loop) and the loop can be nucleated at any point before being passed over the blob excitation (rather than at the specified start-point of the blob ribbon operator). This is why the braiding is simple when $\rhd$ is trivial (as discussed in Section \ref{Section_Loop_Blob_Braiding_Tri_Trivial}) and it is not necessary to fix the positions of the start-points in that case.

		While we have discussed braiding of the blobs and the loops so far in terms of passing the loop over the blob, we can equally move the blob excitation through the loop instead. These are equivalent, but it is slightly easier to calculate the latter situation. The relevant commutation relation to calculate this braiding relation is shown in Figure \ref{blobloopdetail}. Again, we must ensure that the start-points of each operator are in the same location.

		The result of this same-site braiding, where we pass a ribbon operator $B^e(t)$ through an $E$-valued membrane operator $\delta(e_m, \hat{e}(m))$, is then
		 $$B^e(t) \delta(e_m, \hat{e}(m)) = \delta(e_m e^{-1}, \hat{e}(m)) B^e(t),$$
		as illustrated in Figure \ref{blobloopresult} (and proven in Section \ref{Section_braiding_fake-flat_appendix} in the Supplemental Material). The operators $\delta(e_m, \hat{e}(m))$ for each label $e_m \in E$ form a basis for our space of $E$-valued membrane operators, but we want to consider the commutation of one of the basis operators labeled by an irrep of $E$ instead. We have that 
		\begin{align}
		B^e(t)& \sum_{e_m \in E} [D^\alpha(e_m)]_{ab} \delta(\hat{e}(m),e_m) \notag \\
		&= \sum_{e_m \in E} [D^\alpha(e_m)]_{ab} \delta(\hat{e}(m),e_m e^{-1}) B^e(t)\notag \\
		&= \sum_{e'=e_m e^{-1} \in E} [D^\alpha(e' e)]_{ab} \delta(\hat{e}(m),e')B^e(t) \notag\\
		&= \sum_{e' \in E} \sum_{c=1}^{|\alpha|} [D^{\alpha}(e')]_{ac} [D^{\alpha}(e)]_{cb} \delta(\hat{e}(m),e')B^e(t) \notag\\
		&=\sum_{c=1}^{|\alpha|} [D^{\alpha}(e)]_{cb} \sum_{e' \in E} [D^{\alpha}(e')]_{ac} \delta(\hat{e}(m),e')B^e(t). \label{Equation_loop_blob_braiding_fake_flat}
		\end{align}
		
		If $\alpha$ is a 1D representation the braiding therefore results in the accumulation of a phase of $\alpha(e)$. If $\alpha$ is higher-dimensional, then it would seem that there is mixing between the different matrix indices. However, recall from Section \ref{Section_3D_Blob_Fake_Flat} that in order to ensure fake-flatness we restricted the label $e$ of the blob ribbon operators to be in the kernel of $\partial$ and therefore the centre of $E$. The matrix representation of an element of the centre is a scalar multiple of the identity from Schur's Lemma, so we can write $[D^{\alpha}(e)]_{cb} = \delta_{cb} [D^{\alpha}(e)]_{11}$ (where the index $1$ could be replaced with any index). This means that the braiding relation \ref{Equation_loop_blob_braiding_fake_flat} simplifies to
		\begin{align}
		B^e(t) \sum_{e_m \in E}& [D^\alpha(e_m)]_{ab} \delta(\hat{e}(m),e_m) \notag \\
		&= [D^{\alpha}(e)]_{11} \sum_{e' \in E} [D^{\alpha}(e')]_{ab} \delta(\hat{e}(m),e')B^e(t), \label{Equation_llop_brod_braiding_fake_flat_2}
		\end{align}
		so again we only accumulate a phase $[D^{\alpha}(e)]_{11}$ (this matrix element must be a phase because the matrix is diagonal and unitary, and in fact the matrix element can be used to define an irrep of the kernel of $\partial$).

		We see that the braiding relation between the $E$-valued loops and blob excitations is similar to the result we found for the braiding of magnetic fluxes and charges that we discussed in Section \ref{Section_Flux_Charge_Braiding}, except that in this case the irrep labels the loop-like, rather than point-like, excitation.
		
		\begin{figure}[h]
			\begin{center}
			\includegraphics[width=\linewidth]{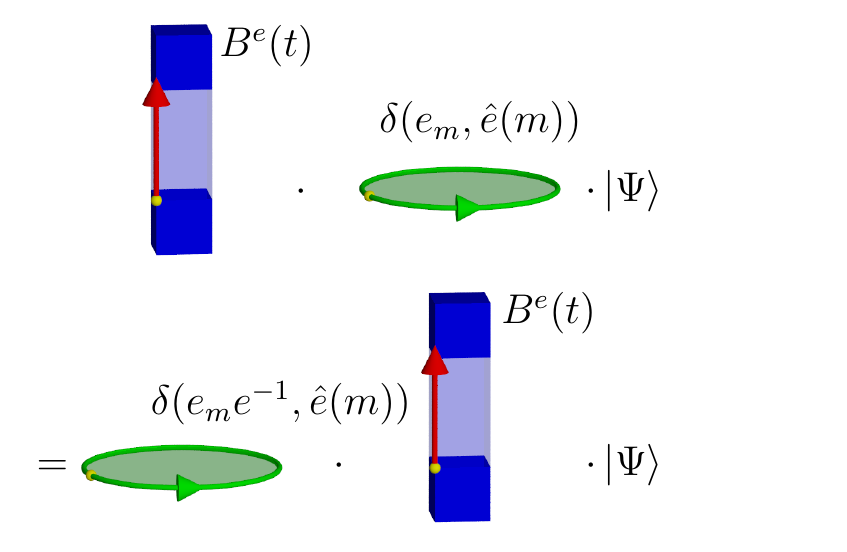}
				
				\caption{The result of loop-blob braiding ($\ket{\psi}$ is a state with no excitations near the two operators)}
				\label{blobloopresult}
			\end{center}
		\end{figure}

		\subsection{Summary of braiding in the fake-flat case}
		
		In Table \ref{Table_Braiding_Fake_Flat}, we summarize the braiding in the fake-flat case by indicating which excitations can braid non-trivially. Note that when we enforce fake-flatness on the level of the Hilbert space, there are no magnetic excitations, although we can still have non-trivial flux labels around non-contractible loops on manifolds that are not simply connected (e.g., the 3-torus).

	\begin{table}[h]
	\begin{center}
		\begin{tabular}{ |c|c|c|c|c| } 
			\hline
			Non-Trivial& & & $E$-valued  & Around\\
			Braiding?& Electric & Blob & loop & handle \\ 
			\hline
			Electric & \xmark & \xmark & \xmark& \cmark \\ 
			\hline

			Blob& \xmark  & \xmark & \cmark& \cmark \\
			\hline
			$E$-valued & & & & \\
			loop & \xmark  & \cmark & \xmark& \cmark\\
			\hline
		\end{tabular}
		
		\caption{A summary of the non-trivial braiding in the fake-flat case. In the fake-flat case, there are no magnetic excitations and the non-trivial braiding between excitations only involves the blob excitations and $E$-valued loops. However there are non-trivial results from moving excitations around handles (non-contractible cycles), which can support non-trivial 1-flux.}
		\label{Table_Braiding_Fake_Flat}
	\end{center}
	
\end{table}
		
		\section{Ribbon and membrane operators in the case where $\partial \rightarrow$ centre($G$) and $E$ is Abelian}
		
		\label{Section_3D_MO_Central}
		So far, we have examined the excitations of the higher lattice gauge theory model in two cases. Firstly, we looked at the case where $\rhd$ is trivial, where we can find all of the excitations of the model. However, in this case the membrane operators corresponding to the 2-gauge field (labeled by the group $E$) are simple, and produce no excitations at the start-point of the corresponding operators. Secondly, we looked at the case where $\rhd$ is general, but where we restrict our Hilbert space to the fake-flat subspace, thus having to exclude the magnetic excitations. This gave us more interesting excitations from our 2-gauge field, allowing the $E$-valued membrane operators and blob ribbon operators to produce additional excitations at the start-points of the operators. However, excluding the magnetic excitations removes the interesting features from the 1-gauge field excitations. In the following sections, we consider a generalization of the $\rhd$ trivial case, which will allow us to keep all of the excitations while also gaining many of the features from the $\rhd$ general case. This means that we will be able to see how the magnetic excitation interacts with the more general 2-gauge excitations. We consider the case where $E$ is Abelian and $\partial$ maps onto the centre of $G$ (Case 2 in Table \ref{Table_Cases}). Note that this case includes the $\rhd$ trivial case as a sub-case, so this is a strict generalization of the situation considered in Sections \ref{Section_3D_MO_Tri_Trivial} and \ref{Section_3D_Braiding_Tri_Trivial}.

\begin{figure*}[t!]
			\begin{center}
				\includegraphics[width=\linewidth]{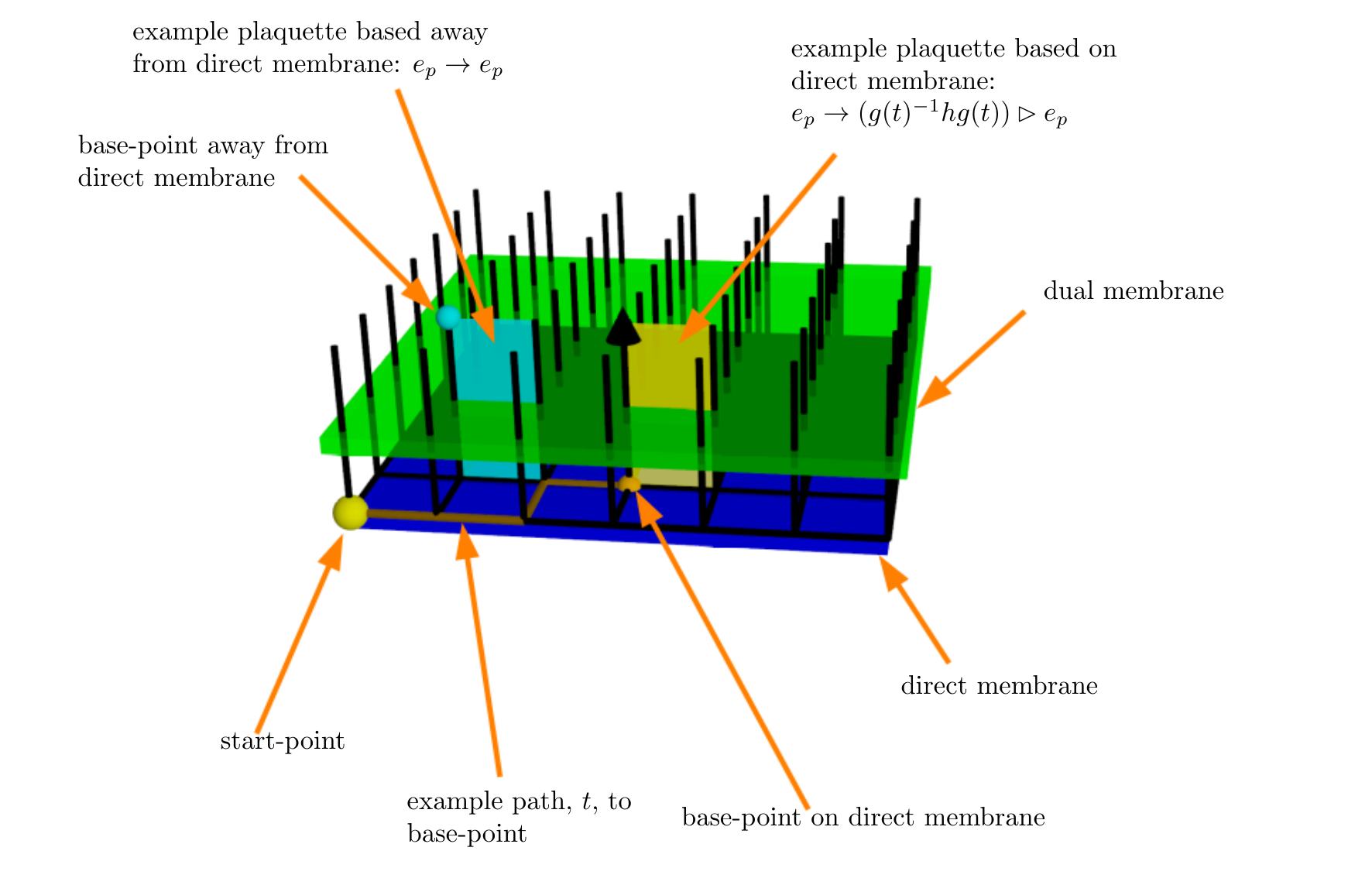}
				
				\caption{In addition to changing the edges cut by the dual membrane, when $\rhd$ is non-trivial the magnetic membrane operator affects the plaquettes cut by the dual membrane if their base-points lie on the direct membrane}
				\label{modmembranecutplaquettespart1}
			\end{center}
		\end{figure*}

		In this case, many of the features of the general crossed module (but fake-flat) case are preserved, despite our restrictions on the crossed module. The electric, $E$-valued loop and blob excitation creation operators are all the same as in the general crossed module (but fake-flat) case (see Section \ref{Section_3D_MO_Fake_Flat}), except that we also allow blob ribbon operators with labels outside the kernel of $\partial$ and the irreps that label the basis $E$-valued membrane operators are 1D. Because of this, we will not describe the operators that produce these excitations again. On the other hand, we can include excitations analogous to the magnetic excitations from the $\rhd$ trivial case. The membrane operators that produce these magnetic excitations are significantly altered from the $\rhd$ trivial case that we considered in Section \ref{Section_3D_Tri_Trivial_Magnetic_Excitations}, however. This alteration of the membrane operators is necessary to ensure that each membrane operator commutes with the various energy terms, apart from those near the boundary of the membrane. The magnetic membrane operators affect the edge labels in the same way as in the $\rhd$ trivial case, but they also affect the plaquette labels near the membrane. Recall that to specify our magnetic membrane we had to define a dual membrane, with the operator changing the labels of the edges cut by the dual membrane, and a direct membrane, with paths on the direct membrane controlling how the cut edges were changed (see Figure \ref{fluxmembrane2}). As well as edges, the dual membrane cuts through the plaquettes between these edges (such as the vertical plaquettes in Figure \ref{modmembranecutplaquettespart1}). In this new case, the magnetic membrane operator changes the label of the ``cut" plaquettes if their base-points lie on the direct membrane. We say that plaquettes whose base-points lie on the direct membrane are based on the direct membrane. The action of the operator on these plaquettes depends on paths on the direct lattice, in a similar way to the action on the edges. Given a cut plaquette based on the direct membrane, the action of the membrane operator on the plaquette depends on the label of a path from the start-point of the membrane to the base-point of the plaquette. An example of this type of path is shown in Figure \ref{modmembranecutplaquettespart1}. We denote the group element assigned to the path between the start-point and the base-point of plaquette $p$ by $g(s.p-v_0(p))$, where $s.p$ is the privileged start-point of the membrane operator and $v_0(p)$ is the base-point of the plaquette $p$. If $p$ is cut by the dual membrane and based on the direct membrane, then its label is changed from $e_p$ to $(g(s.p-v_0(p))^{-1}hg(s.p-v_0(p))) \rhd e_p$. As we mentioned previously, the magnetic membrane operator only acts on a cut plaquette in this way if the plaquette is based on the direct membrane. That is, if the plaquette has its base-point away from the direct membrane, then the label of the plaquette is not affected by this $\rhd$ action. It may seem arbitrary that the plaquette label is only changed in this way if its base-point is on the direct membrane. However, this is analogous to the action of vertex transforms, which only affect plaquettes that are based at the vertex on which we apply the transform (except instead of only affecting surfaces based at the vertex, the membrane operator affects surfaces based on that surface). Indeed, we demonstrate that closed magnetic membrane operators are closely related to the vertex transforms in Sections \ref{Section_Topological_Magnetic_Tri_Trivial} and \ref{Section_Topological_Magnetic_Tri_Nontrivial} of the Supplemental Material.

\begin{figure}[h]
			\begin{center}
				\includegraphics[width=\linewidth]{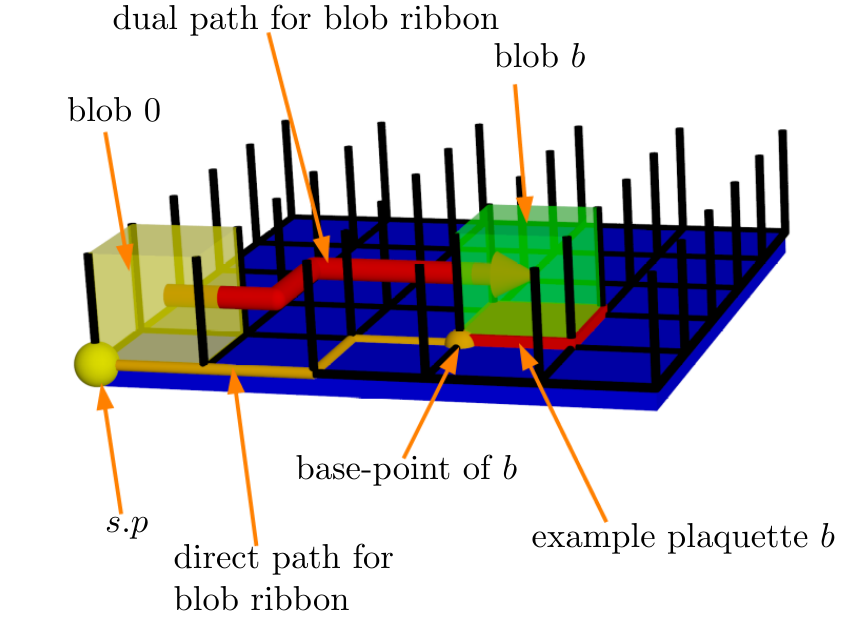}
					
				\caption{When we define the magnetic membrane operator $C^h_T(m)$, we must include blob ribbon operators. There is one blob ribbon operator per plaquette on the direct membrane (which is represented by the large blue surface here, while the dual membrane is omitted and would be above this surface, cutting through the vertical edges). Here we show an example, corresponding to the plaquette $b$ (red square). The dual path for the blob ribbon operator runs from the privileged blob 0 to the blob, blob $b$, which is attached to the plaquette $b$ and cut by the dual membrane (not shown for clarity, but it would be above the direct membrane and bisect the vertical edges). The direct path for the ribbon operator runs from the start-point of the membrane, $s.p$, to the base-point of plaquette $b$.}
				\label{blobsonmem}
			\end{center}
		\end{figure}

\begin{figure}[h]
		\begin{center}
		\includegraphics[width=\linewidth]{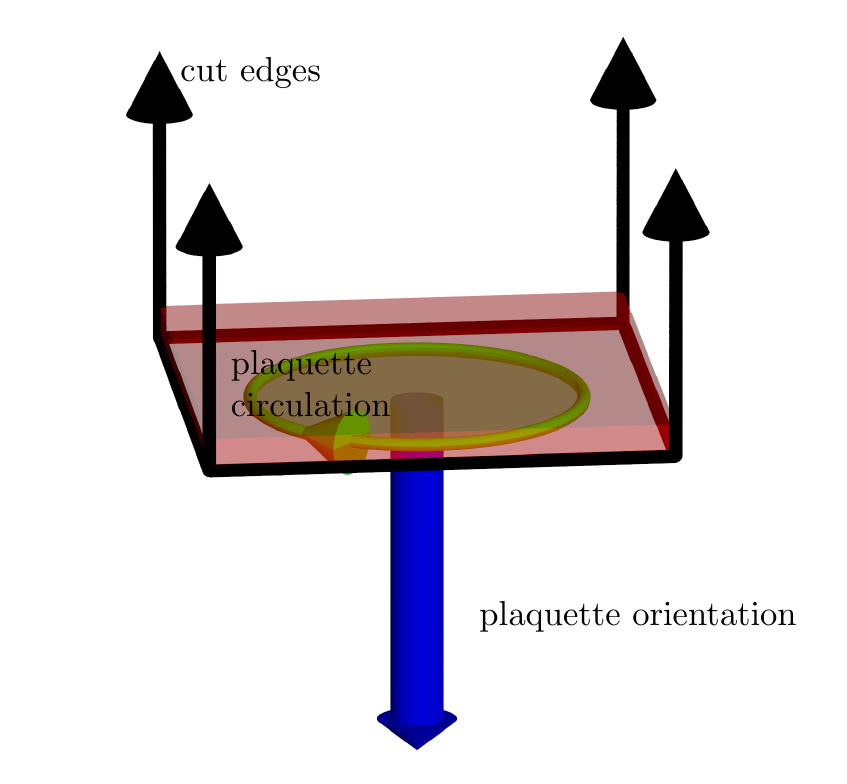}
			
			\caption{We consider the case where the plaquettes on the membrane point downwards, away from the cut edges. To obtain the case where some of the plaquettes point upwards, we must invert the labels of those plaquettes. Note that the orientation of the plaquette is related to the circulation by the right-hand rule.}
			\label{modmembraneorientation}
		\end{center}
	\end{figure}

		The $\rhd$ action on the plaquettes is not the only additional feature to the magnetic membrane operator in this new special case. Changing some of the edge labels by multiplication and plaquette labels by this $\rhd$ action leaves the blob conditions for blobs cut by the dual membrane unsatisfied (recall that the blob condition enforces that the total surface label of the blob is trivial). To correct this and ensure that the membrane operator commutes with the blob energy terms near the bulk of the membrane, blob ribbon operators (of the type considered in Section \ref{Section_3D_Blob_Fake_Flat}) are added to the membrane operator. For every plaquette that is entirely on the direct membrane (not cut by the dual membrane, but instead lying flat on the direct membrane), we have one such blob ribbon operator associated to that plaquette. We call the associated plaquette the base plaquette for that blob ribbon operator. The blob ribbon operators all start at the same privileged blob, which we call blob 0 and must define when specifying the magnetic membrane operator. The blob ribbon operators end at the blob that is connected to the base plaquette and cut by the dual membrane, as shown in Figure \ref{blobsonmem}. The label of this blob ribbon operator, for a base plaquette $b$ on the direct membrane and with orientation away from the dual membrane (downwards in Figure \ref{blobsonmem}, as shown in Figure \ref{modmembraneorientation}), is given by $f(b)=[g(s.p-v_{0}(b))\rhd e_b] [(h^{-1}g(s.p-v_{0}(b)))\rhd e_b^{-1}]$, where $e_b$ is the label of the base plaquette and $v_0(b)$ is the base-point of the plaquette. If the plaquette has the opposite orientation, we must invert the label $e_b$ in this expression. After incorporating this additional action for the magnetic membrane operator, we write the total magnetic membrane operator (denoted by $C^h_T(m)$, where $T$ indicates that it is the total operator) as 
		\begin{equation}
		C^h_T(m)=C^h_{\rhd}(m) \prod_{\text{plaquette }b \in m} B^{f(b)}(\text{blob }0 \rightarrow \text{blob }b),
		\label{total_magnetic_membrane_operator}
		\end{equation}
		where blob $b$ is the blob attached to base plaquette $b$ and cut by the dual membrane (note, however, that the same blob may be attached to multiple base plaquettes). In Equation \ref{total_magnetic_membrane_operator}, $C^h_{\rhd}(m)$ performs the action of the membrane operator on the edges and the $\rhd$ action on the plaquettes, while the $B^{f(b)}(\text{blob }0 \rightarrow \text{blob }b)$ operators are the added blob ribbon operators (see Section \ref{Section_3D_Blob_Fake_Flat} for a description of blob ribbon operators).

		Even with these modifications, the magnetic membrane operator still excites more energy terms than in the $\rhd$ trivial case (i.e., more than just the boundary plaquettes and potentially the start-point vertex). Firstly, the privileged blob, blob 0, is not generally left in an energy eigenstate. This is because the blob ribbon operators that we added to the magnetic membrane all originate in this blob and so change the surface label of the blob, from $1_E$ in the ground state to $[h \rhd \hat{e}(m)^{-1}] \hat{e}(m)$, where $\hat{e}(m)$ is the total surface label of the direct membrane. $\hat{e}(m)$ is an operator, which means that blob 0 is not generally left in an energy eigenstate. Secondly, the edges around the boundary of the direct membrane are potentially excited. This is because the labels of the added blob ribbon operators depend on the labels of the plaquettes on our direct membrane. Edges in the bulk of the membrane are attached to two plaquettes on the membrane and so the edge transform affects the blob ribbon operators associated with both plaquettes. These effects on the labels of the two ribbon operators (together with a contribution from the edge transform on the plaquettes cut by the dual membrane) cancel out, so that the edge transform commutes with the membrane operator, as we show in Section \ref{Section_Magnetic_Tri_Nontrivial_Commutation}. On the other hand, edges on the boundary of the membrane are only attached to one plaquette on the membrane, so there is no such cancellation. These boundary edges are therefore not generally left in an energy eigenstate.
		
		\subsection{Condensation and confinement}
		\label{Section_condensation_confinement_partial_central}
\begin{figure*}[ht]
			\begin{center}
				\includegraphics[width=\linewidth]{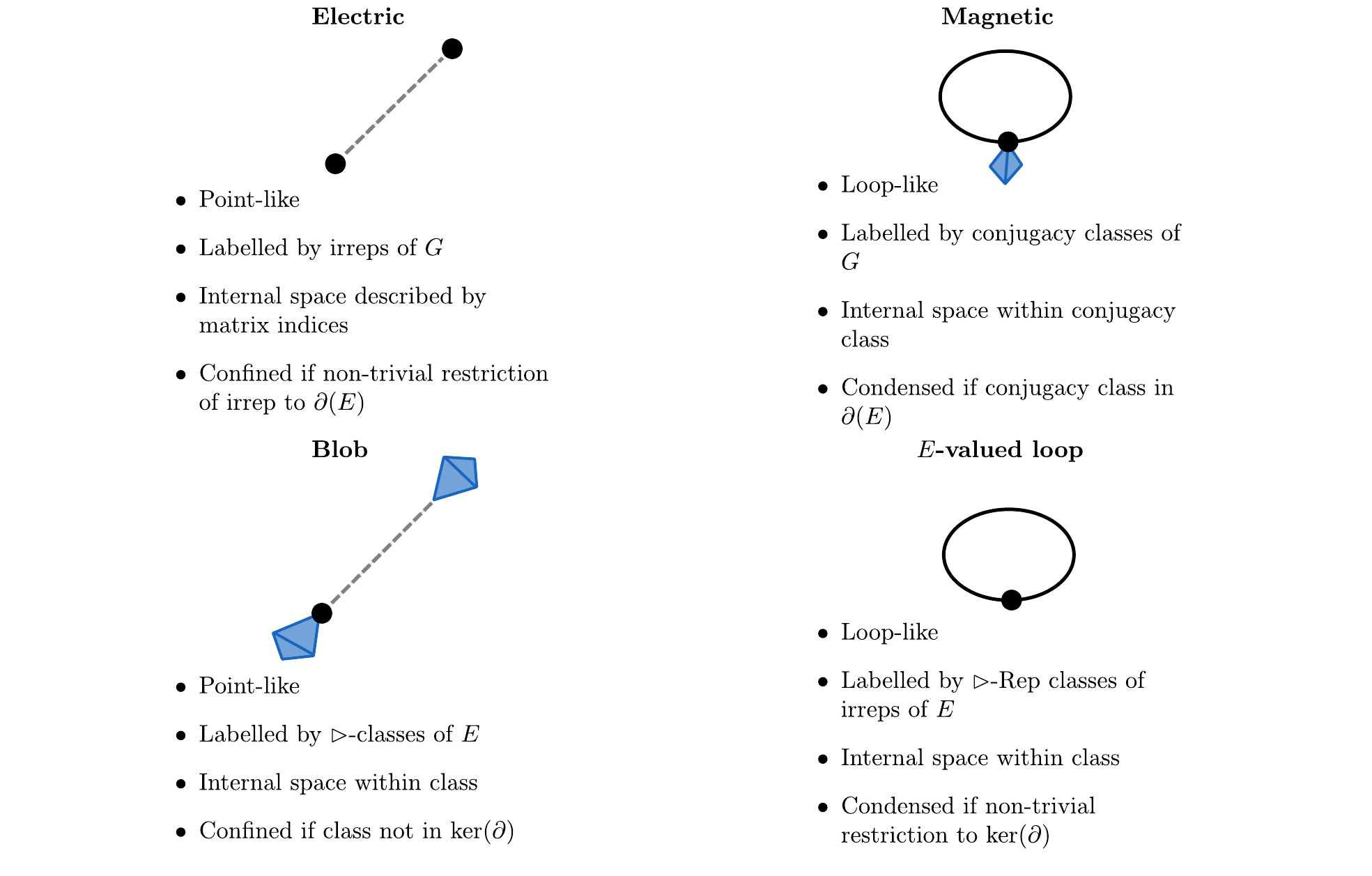}

				\caption{A summary of the excitations in the $E$ Abelian, $\partial \rightarrow \text{centre}(G)$ case}
				\label{Excitation_summary}
			\end{center}
		\end{figure*}

		In the previously considered $\rhd$ trivial (see Section \ref{Section_3D_Condensation_Confinement_Tri_Trivial}) and fake-flat cases (see Section \ref{Section_3D_MO_Fake_Flat}), we saw that many of the excitations are confined, meaning that it costs energy to separate a pair of excitations, in addition to the energy required to produce the pair. We also found that other particle types are condensed, meaning that they can be produced by local operators (local to the excitation, in the sense discussed in Section \ref{Section_3D_Condensation_Confinement_Tri_Trivial}) and so carry trivial topological charge. The pattern of condensation and confinement in the $\partial \rightarrow$ Centre($G$) case is the same as in the $\rhd$ trivial case discussed in Section \ref{Section_3D_Condensation_Confinement_Tri_Trivial}. The blob excitations with label not in the kernel of $\partial$ are confined, as are the electric excitations labeled by irreps of $G$ that have non-trivial restriction to the subgroup $\partial(E)$ of $G$. On the other hand, the condensed excitations are the magnetic excitations with label in the image of $\partial$ and the $E$-valued loops that are labeled by irreps of $E$ which are trivial on the kernel of $\partial$. These properties, along with the other properties of the excitations in this case, are summarised in Figure \ref{Excitation_summary}.
		
		\section{Braiding in the case where $\partial \rightarrow$ centre($G$) and $E$ is Abelian}
		\label{Section_3D_Braiding_Central}

		Now that we have described the membrane and ribbon operators that produce our excitations, we can consider the braiding relations of these excitations. Any braiding not involving the magnetic excitations is the same as in the fake-flat case described in Section \ref{Section_3D_Braiding_Fake_Flat}. Namely, there is non-trivial braiding between the blob excitations and the $E$-valued loops, with the same start-point braiding resulting in an accumulation of phase. The result is a phase, rather than the more general transformation given in Equation \ref{Equation_loop_blob_braiding_fake_flat} for the case considered in Section \ref{Section_3D_Loop_Blob_Braiding_Fake_Flat}, because the irreps of $E$ are 1D when $E$ is Abelian.

		Unlike for the fake-flat case, we can find the magnetic excitations and so describe their braiding relations. However, rather than using the magnetic membrane operator directly, it is convenient when considering braiding to combine the magnetic membrane operator with an $E$-valued membrane operator. We multiply the magnetic membrane operator by an $E$-valued membrane operator such as $\delta(e_m,\hat{e}(m))$, acting before the magnetic membrane operator. That is, we construct membrane operators of the form $C^h_T(m)\delta(e_m,\hat{e}(m))$, which we denote by $C^{h,e_m}_T(m)$. We note that combining the magnetic membrane operator with this $E$-valued membrane operator in this way does not excite regions of the lattice not already excited by the magnetic membrane operator, because both membrane operators only cause excitations near the boundary of the membrane, and possibly at blob 0 and the start-point of the membrane. We will shortly explain why we perform this combination of membrane operators (in essence, it gives the loop-like excitation a well-defined 2-flux), but before we discuss this we shall consider the combination of the membrane operators in more detail.

\begin{figure}[h]
			\begin{center}
				\includegraphics[width=\linewidth]{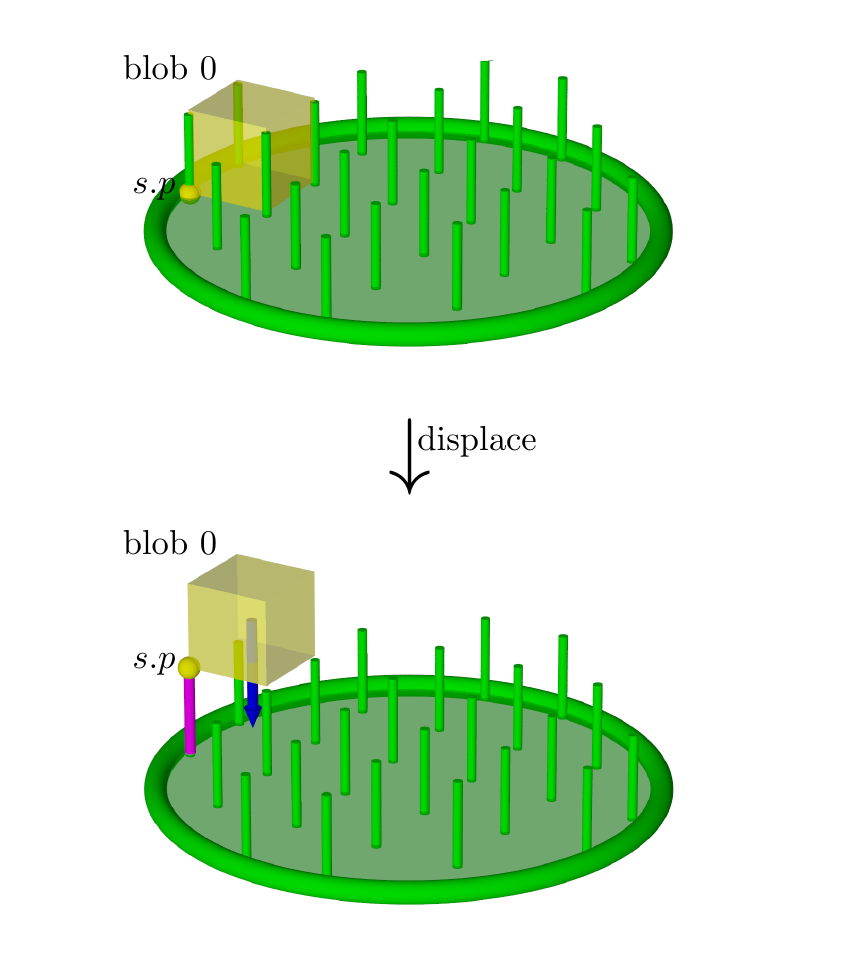}
				
				\caption{Rather than place blob 0 and the start-point (represented by the yellow cube and sphere respectively) of the membrane operator on the direct membrane (green) itself, as in the upper image, we displace them away from the membrane as shown in the lower image. Then blob 0 and the start-point are on the other side of the edges cut by the dual membrane (where the edges are represented by the green cylinders). This allows us to deform the membrane away from the start-point and blob 0 (downwards in the figure) using the topological property of the magnetic and $E$-valued membrane operators which make up the higher-flux membrane operator.}
				\label{higher_flux_displacement_main_text}
			\end{center}
		\end{figure}

		In order to combine the magnetic and $E$-valued membrane operators, there are some details that we must specify. The first of these is the relative orientation of the two membrane operators. We take the orientation of the $E$-valued membrane operator to point away from the dual membrane of the magnetic membrane operator. The second detail is more subtle. In addition to combining this $E$-valued membrane operator with the magnetic membrane operator, we move blob 0 and the start-point of the membrane operator so that they are displaced slightly away from the membrane itself, as shown in Figure \ref{higher_flux_displacement_main_text}. We do this because the $E$-valued membrane may cause an excitation at the start-point, which would prevent us from using the topological nature of the magnetic membrane operator to deform the membrane. It is not necessary to move blob 0 in this way, but it will be convenient when considering topological charge to have a clear separation between the point-like excitations (blob 0 and the start-point) and the loop-like excitation at the boundary of the membrane. A third detail is our convention for the overall orientation of the total membrane operator. Because we have displaced the start-point away from the membrane in a particular direction, it is sensible to define the orientation of the membrane to be consistent with this displacement. That is we imagine that the loop excitation is nucleated at the start-point and moves away from the start-point along the membrane. Therefore, the loop excitation would be oriented downwards in Figure \ref{higher_flux_displacement_main_text}, matching the orientation of the $E$-valued membrane operator.

		Having considered these details about the combined membrane operator, we now explain why this combination was useful. As we mentioned in Section \ref{Section_3D_MO_Central}, the action of the magnetic membrane operator generally causes the privileged blob, blob 0, of the membrane to acquire a non-trivial 2-flux (non-trivial surface label). However, this surface label is given in terms of an operator (the surface label $\hat{e}(m)$ of the membrane itself) and so is not well-defined. Including the $E$-valued membrane operator ensures that the 2-flux of the privileged blob 0 after the action of the magnetic membrane operator is well-defined. Giving blob 0 a definite 2-flux is significant because we expect the loop-like excitation to also carry a non-trivial 2-flux to balance the 2-flux of blob 0, and we expect the value of this 2-flux to be important in braiding relations. As we show in Section \ref{Section_braiding_higher_flux} of the Supplemental Material, the surface label of blob 0 after the action of the combined membrane operator (with displaced start-point, which does affect the label) is given by $e_m^{-1} [h^{-1} \rhd e_m]$. As we show in Section \ref{Section_braiding_higher_flux} (see Equation \ref{Equation_2_flux_higher_flux_excitation_appendix}), the 2-flux of the loop excitation labeled by $h$ and $e_m$ is given by
		\begin{equation}
		\tilde{e}_m = e_m [h^{-1} \rhd e_m^{-1}],
		\end{equation}
		which is the inverse of the 2-flux carried by blob 0. Due to the fact that the excitations produced by this combined membrane operator carry both ordinary magnetic flux and this 2-flux, we call the combined membrane operator a ``higher-flux membrane operator" and call the excitations higher-flux excitations. If we wish to instead consider the original magnetic membrane operator (without the attached $E$-valued membrane operator), we can simply sum over each value of $e_m$, because this gives us a complete sum of projectors $\delta(e_m,\hat{e}(m))$. That is $\sum_{e_m \in E} \delta(e_m,\hat{e}(m)) =1$ and so
		\begin{equation}
		\sum_{e_m \in E} C^{h,e_m}_T(m) = C^h_T(m).\label{Equation_higher_flux_to_magnetic}
		\end{equation}

		Having constructed this higher-flux membrane operator, we can now use it to find the braiding relations involving the higher-flux excitations. Because the action of the higher-flux membrane operator on the edges is the same as that of the magnetic membrane operator from the $\rhd$ trivial case, the braiding relation between the magnetic and electric excitations is the same as in that case (which is described in Section \ref{Section_3D_Braiding_Tri_Trivial}). However, as we will see shortly, the braiding between the higher-flux loop excitation and the other excitations is significantly altered. In particular, because the higher-flux excitations can carry a non-trivial 2-flux, we expect non-trivial braiding relations with the $E$-valued loops, which measure 2-flux.

		\subsection{Braiding of the higher-flux excitations with blob excitations}
		\label{Section_magnetic_blob_braiding}

\begin{figure}[h]
			\begin{center}
			\includegraphics[width=\linewidth]{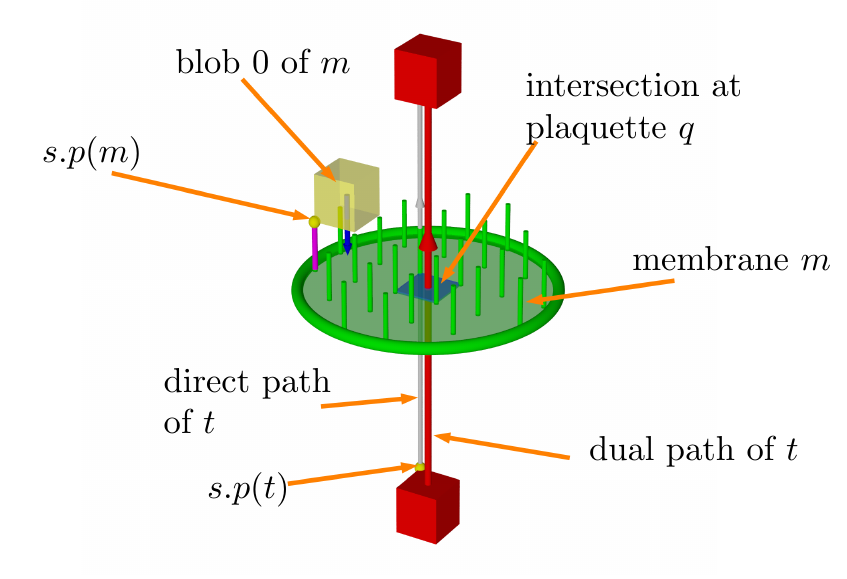}
		
				\caption{We consider a blob ribbon operator $B^e(t)$ (between the two red blobs) that passes through a higher-flux membrane operator, $C^{h,e_m}_T(m)$ (where $m$ is shown in green). The ribbon $t$ pierces the membrane $m$ through a plaquette $q$ (blue square).}
				\label{blob_operator_through_magnetic_1}
			\end{center}
		\end{figure}

		The first braiding relation we examine is between the higher-flux excitations and the blob excitations. We consider a blob ribbon operator $B^e(t)$, applied on a ribbon $t$, piercing the membrane of a higher-flux membrane operator $C^{h,e_m}_T(m)$, applied on a membrane $m$. The ribbon $t$ intersects the membrane $m$ at a plaquette $q$, as shown in Figure \ref{blob_operator_through_magnetic_1}. Note that the orientation of the operators is significant. We first look at the case where the blob ribbon operator pierces the direct membrane of the magnetic membrane operator before the dual membrane. If the membrane is oriented downwards, as in Figure \ref{blob_operator_through_magnetic_1} (note that the point-like excitation is above the loop) then the ribbon is oriented upwards (at the point of intersection at least).

\begin{figure}[h]
			\begin{center}
			\includegraphics[width=\linewidth]{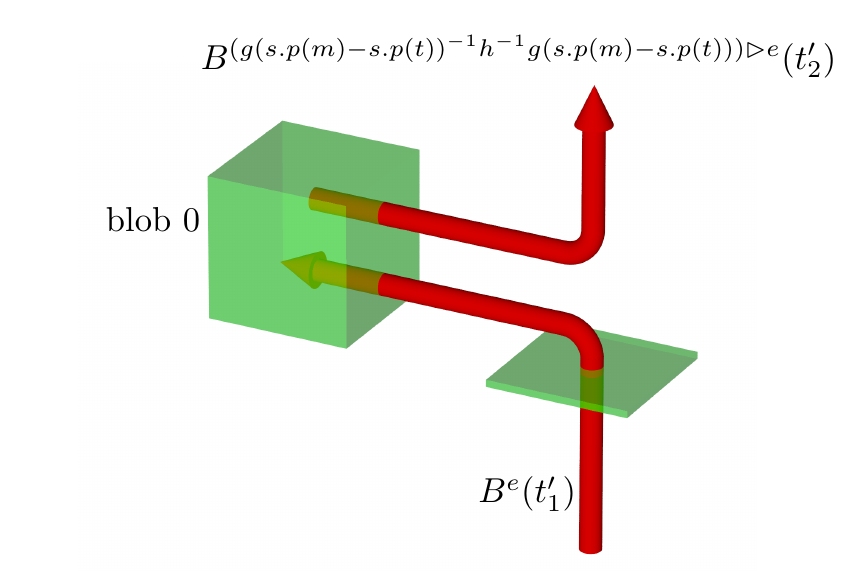}
				
				\caption{The blob ribbon operators after commutation}
				\label{blob_ops_after_commutation}
			\end{center}
		\end{figure}

		We have seen in previous cases that braiding is frequently well defined only when the start-points of the membrane and ribbon operators match. However, we shall first examine the general case where the start-points are arbitrary. As usual, we can relate the braiding relation to a commutation relation between the two operators. We compare the case where the magnetic membrane is produced first, and then the blob excitation moved through it, to the reverse case. We find that, as demonstrated in Section \ref{Section_braiding_higher_flux_blob} in the Supplemental Material,
		\begin{align}
		B^e&(t)C^{h,e_m}_T(m)\ket{GS} \notag\\
		&= C^{h,e_m[\hat{g}(s.p(m)-s.p(t)) \rhd e] }_T(m) B^e(t_1') \notag\\
		& \hspace{0.5cm} B^{(\hat{g}(s.p(t)-s.p(m)) h^{-1} \hat{g}(s.p(t)-s.p(m))^{-1}) \rhd e}(t_2') \ket{GS}. \label{higher_flux_blob_commutation_1}
		\end{align}
		In this expression, we note that the original ribbon operator is split into two parts, on ribbons $t_1'$ and $t_2'$, which transform differently under the braiding. Here $t_1'$ starts at the original origin of ribbon $t$ and ends at blob 0 of the membrane $m$ (corresponding to the part of the ribbon before the intersection with the membrane, except that it is diverted to end at blob 0 of $m$), while $t_2'$ starts at blob 0 and ends at the original end of ribbon $t$ (corresponding to the part of the ribbon after the intersection), as shown in Figure \ref{blob_ops_after_commutation}. We therefore see that, under commutation, the ribbon $t$ is diverted to pass through blob 0 of the membrane, and after passing through this blob it changes label from $e$ to $(\hat{g}(s.p(t)-s.p(m)) h^{-1}, \hat{g}(s.p(t)-s.p(m))^{-1}) \rhd e$.

	 The fact that the label of the blob ribbon operator before the intersection is unaffected by the commutation relation is perhaps unsurprising, because this part of the operator corresponds to the motion of the blob excitation before it braids with the magnetic excitation and so before it has undergone its transformation. This can be seen from the fact that the blob ribbon operator actually creates two excitations and the one which is not moved should not be affected by the other one moving through the loop excitation. Another thing to note is that, as long as the blob ribbon operator is not confined, we can deform the ribbons of the blob ribbon operators without changing their action, as long as we keep the end-points fixed. Because of this, it doesn't matter at which plaquette $q$ our magnetic membrane and blob ribbon operators intersect.

	 In addition to the transformation undergone by the blob ribbon operator, the $E$ label of the higher-flux excitation changes from $e_m$ to $e_m [\hat{g}(s.p(m)-s.p(t))\rhd e]$. This transformation of the $E$-label of the membrane operator is simply the standard braiding relation between a blob excitation and an $E$-valued membrane, as we saw in Section \ref{Section_3D_Braiding_Fake_Flat}. We note that this result and the other results given in this section, are proven fully in Section \ref{Section_braiding_higher_flux_blob} in the Supplemental Material.

		If we give the magnetic membrane operator and the blob ribbon operator the same start-point, the braiding relation that we explained above simplifies and we are able to remove the operator $\hat{g}(s.p(m)-s.p(t))$ from the relation. We move the start-points together, without moving them through the higher-flux membrane (which would alter the commutation relations found so far). In this case the blob label goes from $e$ before the braiding to $h^{-1} \rhd e$ afterwards (at least in the part after the intersection) and the $E$ label of the higher-flux membrane operator goes from $e_m$ beforehand to $e_m e$. If we had used the opposite orientation, obtained by reversing the direction of the blob ribbon operator, instead the blob label $e$ becomes $h \rhd e$ and the membrane label $e_m$ becomes $e_m [h \rhd e^{-1}]$. Again, this result is proven in Section \ref{Section_braiding_higher_flux_blob} of the Supplemental Material.

		If we want to consider the braiding of the original magnetic excitation, produced by the membrane operator $C^h(m) = \sum_{e_m \in E} C^{h,e_m}(m)$, we simply need to sum over the $E$-valued label $e$ of the higher-flux membrane operator. Then we have, from Equation \ref{higher_flux_blob_commutation_1}
		\begin{align}
		B^e&(t)C^{h}_T(m)\ket{GS} \notag\\
		&= \sum_{e_m \in E} B^e(t) C^{h, e_m}_T(m) \notag\\
		&= \sum_{e_m \in E} C^{h,e_m[\hat{g}(s.p(m)-s.p(t)) \rhd e] }_T(m) B^e(t_1') \notag \\ & \hspace{0.5cm} B^{(\hat{g}(s.p(t)-s.p(m)) h^{-1}\hat{g}(s.p(t)-s.p(m))^{-1}) \rhd e}(t_2') \ket{GS} \notag\\
		&= \sum_{e'_m =e_m [\hat{g}(s.p(m)-s.p(t))\rhd e]} C^{h,e'_m}_T(m)B^e(t_1) \notag \\
		& \hspace{0.5cm} B^{(\hat{g}(s.p(t)-s.p(m)) h^{-1}\hat{g}(s.p(t)-s.p(m))^{-1}) \rhd e}(t_2') \ket{GS} \notag\\
		&=C^h_T(m)B^e(t_1) \notag \\
		& \hspace{0.5cm} B^{(\hat{g}(s.p(t)-s.p(m)) h^{-1}\hat{g}(s.p(t)-s.p(m))^{-1}) \rhd e}(t_2') \ket{GS},
		 \label{higher_flux_blob_commutation_2}
		\end{align}
		from which we see that the magnetic excitation is unchanged by the braiding, whereas the blob excitation is affected in the same way as in the braiding with the higher-flux excitation.

		\subsection{Braiding with other higher-flux excitations}
		\label{Section_higher_flux_higher_flux_braiding}
\begin{figure}[h]
			\begin{center}
				\includegraphics[width=\linewidth]{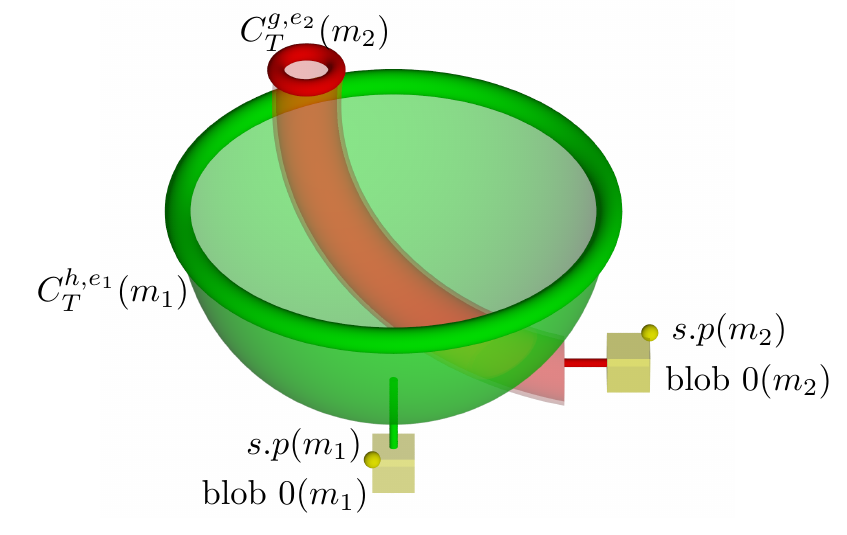}
				
				\caption{We consider the braiding move where we pull one higher-flux loop excitation (small red torus) through another (large green torus). This can be implemented using higher-flux membranes applied on the (green and red) membranes in the figure. If we first apply the membrane operator $C^{h,e_1}_T(m_1)$ on the larger (green) membrane, then $C_T^{g,e_2}(m_2)$ on the narrower (red) membrane, then we are considering the case where we first produce the larger (green) loop excitation then move the smaller (red) one through it. Comparing this to the opposite order of operators gives us the braiding relation. }
				\label{Braid_move_loops}
			\end{center}
		\end{figure}

		Next, we consider the braiding between two higher-flux excitations. As we described in Section \ref{Section_Flux_Flux_Braiding_Tri_Trivial_Abelian}, there are two kinds of braiding for loops. The first, which we call permutation, involves moving two loops around each-other without passing through one another. The other, which we term braiding, involves passing one though the other. As we discussed in Section \ref{Section_Flux_Flux_Braiding_Tri_Trivial_Abelian}, the permutation move is trivial in this model. Therefore, we just consider the braiding move. In this motion, shown in Figure \ref{Braid_move_loops}, one of the magnetic loop excitations (indicated by a small red ring) is moved along the red surface and through another loop (indicated by a large green loop attached to a large green surface). To calculate the braiding relation we apply membrane operators on these surfaces and examine the commutation relations between the membrane operators.

		We define the membrane operators as indicated in Figure \ref{Braid_move_loops}, but then we use the topological nature of the magnetic membrane operators to pull $m_2$ through $m_1$ (as we did in the $\rhd$ trivial case in Section \ref{Section_Flux_Flux_Braiding_Tri_Trivial}), while keeping the start-point and blob 0 fixed. In Section \ref{Section_braiding_higher_flux_higher_flux} in the Supplemental Material, we show that this leads to the commutation relation
		\begin{align}
		C&^{g,e_2}_T(m_2)C^{h,e_1}_T(m_1) \ket{GS} \notag \\
		&=C^{h,e_1 \big[g((1)-(2)) \rhd ([h_{[2-1]} \rhd e_2^{-1}] [(h_{[2-1]} g^{-1}) \rhd e_2])\big]}_T(m_1) \notag \\
		&\hspace{0.5cm} C_\rhd^{h_{[2-1]}gh_{[2-1]}^{-1}}(m_2) \notag \\
		&\hspace{0.5cm}  \bigg(\prod_{\substack{\text{plaquette }\\ p \in m_2}} B^{[h_{[2-1]}^{-1}\rhd e_{p|2} ] [(g^{-1}h_{[2-1]}^{-1})\rhd e_{p|2}^{-1}]}((2)-(1))\notag \\
		&\hspace{0.5cm}  B^{e_{p|2} [(h_{[2-1]} g^{-1} h_{[2-1]}^{-1}) \rhd e_{p|2}^{-1}]}((1)-p) \bigg) \notag \\
		&\hspace{0.5cm}  \delta(h_{[2-1]} \rhd e_2, \hat{e}(m_2))\ket{GS}, \label{Equation_higher_flux_braiding_main_text}
		\end{align}
		where $g((1)-(2))$ is the path element for the path between the two start-points of the membranes and 
		\begin{align}
		h_{[2-1]}&=g((1)-(2))^{-1}hg((1)-(2)) \notag \\
		& =g((2)-(1))hg((2)-(1))^{-1}.
		\end{align}

		To simplify the expression, we used $e_{p|2}$ to denote the label of the plaquette $p$ when we move its base-point to the start-point of $m_2$. This quantity is equivalent to $g(s.p(m_2)-v_0(p)) \rhd e_p$. Furthermore, we used $B^{...}((2)-(1))$ to denote a blob ribbon operator that runs from blob 0 of $m_2$ to blob 0 of $m_1$ and $B^{...}((1)-p)$ to denote a blob ribbon operator running from blob 0 of $m_1$ to the blob on $m_2$ that is attached to plaquette $p$. These blob ribbon operators may seem complicated, but the situation is analogous to the braiding of blob ribbon operators with the magnetic membranes. Each of the blob ribbon operators that we added to the magnetic membrane operator has a similar commutation relation with the magnetic membrane operator as an ordinary blob ribbon operator. Namely, the blob ribbon operator splits into two parts, one that runs from blob 0 of $m_2$ to blob 0 of $m_1$ and one which runs from blob 0 of $m_1$ to the final destination of the original blob ribbon operator. The only difference from the ordinary blob ribbon operator braiding is that the label of the blob ribbon operator is an operator (depending on the label of a plaquette on $m_2$), which causes an additional apparent change to the label even for the part of the blob ribbon operator before the intersection. However, this does not reflect a real change to the blob ribbon operators before the intersection. This is because we can combine the blob ribbon operators that pass from blob 0 of $m_2$ to blob 0 of $m_1$ into a single blob ribbon operator, and use $\delta(h_{[2-1]} \rhd e_2, \hat{e}(m_2))$ to fix its label in terms of $e_2$ instead of an operator. As shown in Section \ref{Section_braiding_higher_flux_higher_flux} in the Supplemental Material, this gives us
		\begin{align}
		C&^{g,e_2}_T(m_2)C^{h,e_1}_T(m_1) \ket{GS} \notag \\
		&=C^{h,e_1 \big[g((1)-(2)) \rhd ([h_{[2-1]} \rhd e_2^{-1}] [(h_{[2-1]} g^{-1}) \rhd e_2])\big]}_T(m_1) \notag \\
		&\hspace{0.5cm} C_\rhd^{h_{[2-1]}gh_{[2-1]}^{-1}}(m_2) \notag \\
		&\hspace{0.5cm} B^{e_2 [g^{-1}\rhd e_2^{-1}]}((2)-(1))\notag \\
		&\hspace{0.5cm} \bigg(\prod_{\substack{\text{plaquette }\\ p \in m_2}} B^{e_{p|2} [(h_{[2-1]} g^{-1} h_{[2-1]}^{-1}) \rhd e_{p|2}^{-1}]}((1)-p) \bigg) \notag \\
		&\hspace{0.5cm}\delta(h_{[2-1]} \rhd e_2, \hat{e}(m_2))\ket{GS}. \label{Equation_higher_flux_braiding_main_text_3}
		\end{align}

		Now the section of blob ribbon operator between blob 0 of each membrane, which is the part of the ribbon operator before the intersection of the membranes, is labeled by $e_2 [g^{-1}\rhd e_2^{-1}]$. This is the same label as it would have if we combined the blob ribbons on these sections in the absence of the second higher flux membrane $C^{h,e_1}_T(m_1)$. That is, this part of the blob ribbon operator is unaffected by the braiding, as we may expect given that this section of ribbon operator is the part before the intersection of the membranes. We can contrast this with the sections of the blob ribbon operators after the intersection (which should be affected by the braiding) which have their labels changed from 
		$$e_{p|2} [g^{-1} \rhd e_{p|2}^{-1}]$$
		to 
		$$e_{p|2} [(h_{[2-1]} g^{-1} h_{[2-1]}^{-1}) \rhd e_{p|2}^{-1}].$$
		
		We see that the only change is that we replace $g$ with $h_{[2-1]} g^{-1} h_{[2-1]}^{-1}$. This matches how the label $g$ of the higher-flux membrane operator transforms under braiding (as we see from the operator $C_\rhd^{h_{[2-1]}gh_{[2-1]}^{-1}}(m_2)$ in Equation \ref{Equation_higher_flux_braiding_main_text_3}). That is, the labels of the blob ribbon operators after the intersection (i.e., after braiding) are the labels we expect given the label of the magnetic part of the membrane operator after intersection.

		Apart from this splitting of the blob ribbon operators at the intersection of the membranes, we see that the labels of the two membrane operators change (as we mentioned previously for $g$). We have that
		\begin{align*}
		h &\rightarrow h,\\
		e_1 &\rightarrow e_1 \bigl[g((1)-(2)) \rhd ([h_{[2-1]} \rhd e_2^{-1}]\\
		& \hspace{1cm} [(h_{[2-1]}g^{-1}) \rhd e_2])\bigl],\\
		g &\rightarrow h_{[2-1]}gh_{[2-1]}^{-1},\\
		e_2 &\rightarrow h_{[2-1]} \rhd e_2.
		\end{align*}
	
		As usual for our braiding, when the start-points of the operators are not the same, we have operators in our braiding relations. When we take the start-points to be the same, these relations simplify to
		\begin{align}
		h &\rightarrow h, \notag\\
		e_1 &\rightarrow e_1 [h \rhd e_2^{-1}] [(hg^{-1}) \rhd e_2], \notag\\
		g &\rightarrow hgh^{-1}, \notag\\
		e_2 &\rightarrow h \rhd e_2. \label{Loop_Braiding_Central}
		\end{align}
		This removes any operators from the labels, so that we have definite braiding. Note that the transformation of the 1-flux labels ($h$ and $g$) is the same as for the braiding of two magnetic excitations in the $\rhd$-trivial case, as given in Equation \ref{braid_relation_magnetic_flipped} in Section \ref{Section_Flux_Flux_Braiding_Tri_Trivial}, replacing $k$ with $g$ (that equation in particular because we used the specific orientation of the loops also used to find that equation). However, the expression for the change of the $E$ labels is not so easy to interpret. It is easier to understand these results if we change variables, from the surface labels of the membranes to the 2-fluxes possessed by the loop-like excitations, as we discussed at the start of Section \ref{Section_3D_Braiding_Central}. The 2-flux of the loop excitation, $\tilde{e}_1$, is related to the $1$-flux label, $h$, and the surface label $e_1$ of the membrane operator by $\tilde{e}_1 = e_1 [h^{-1} \rhd e_1^{-1}]$. Therefore, we define
		$$\tilde{e}_1=e_1 [h^{-1} \rhd e_1^{-1}],$$
		$$\tilde{e}_2=e_2 [g^{-1} \rhd e_2^{-1}].$$
		
		Then, from our braiding relations in Eqs. \ref{Loop_Braiding_Central}, under braiding these 2-fluxes transform according to
		\begin{align}
		\tilde{e}_1 &\rightarrow e_1 [h \rhd \tilde{e}_2^{-1}] [h^{-1} \rhd e_1^{-1}]\tilde{e}_2 \notag\\
		& = \tilde{e}_1 [h \rhd \tilde{e}_2^{-1}] \tilde{e}_2, \notag\\
		\tilde{e}_2 &\rightarrow [h \rhd e_2] [(hg^{-1}h^{-1}) \rhd (h \rhd e_2^{-1})] \notag\\
		&= h \rhd (e_2 [g^{-1} \rhd e_2^{-1}]) \notag\\
		&= h \rhd \tilde{e}_2. \label{Loop_Braiding_Central_Flux}
		\end{align}
		
		This means that the product of the two fluxes transforms as
		$$\tilde{e}_1 \tilde{e}_2\rightarrow \tilde{e}_1 [h \rhd \tilde{e}_2^{-1}] \tilde{e}_2 [h \rhd \tilde{e}_2]= \tilde{e}_1 \tilde{e}_2$$
		under the braiding, which indicates that the product of these 2-fluxes is conserved.

		Putting this together, we can see that the 1-fluxes and 2-fluxes of the loop-like excitations transform as 
		$$((g,\tilde{e}_2),(h,\tilde{e}_1))\rightarrow ((h,\tilde{e}_1\tilde{e}_2 [h\rhd \tilde{e}_2^{-1}]), (hgh^{-1}, h \rhd \tilde{e}_2))$$
		under braiding, where the fact that one loop is moved past the other during the braiding is represented by swapping the order of their labels in the brackets.

		We also wish to work out the inverse transformation, which describes the reversed braiding process. Denoting the result of the forwards transformations as primed versions, we have from Eqs. \ref{Loop_Braiding_Central_Flux}:
		\begin{align*}
		&\tilde{e}_2'= h \rhd \tilde{e}_2\text{, with } h'=h \implies \tilde{e}_2={h'}^{-1} \rhd \tilde{e}_2'\\
		&g'=hgh^{-1} \implies g={h'}^{-1}g'h'\\
		&\tilde{e}_1'=\tilde{e}_1 [h \rhd \tilde{e}_2^{-1}] \tilde{e}_2 \implies \tilde{e}_1=\tilde{e}_1'\tilde{e}_2' [{h'}^{-1} \rhd {\tilde{e}_2}^{\prime -1}]
		\end{align*}
		The inverse transformation is therefore
		\begin{align}
		((h'&,\tilde{e}_1'),(g',\tilde{e}_2')) \rightarrow \notag\\
		& (({h'}^{-1}g'h',{h'}^{-1}\rhd \tilde{e}_2'),(h', \tilde{e}_1'\tilde{e}_2' {h'}^{-1}\rhd {\tilde{e}_2}^{ \prime -1})).
		\end{align}
		
		This matches the braiding proposed in Ref. \cite{Bullivant2018} for higher gauge theory based on discussions of the loop braid group. It is important to note that the braiding depends on the result of fusing the two excitations. Given two loops with 1-flux and 2-flux given by $(h,\tilde{e}_1)$ for the first loop and $(g, \tilde{e}_2)$ for the second loop, there are many possible fusion products. The fact that the products of 1-fluxes $hg \rightarrow hgh^{-1} h =hg$ (swapping the order after braiding to account for the swapping of loop positions) and of 2-fluxes $\tilde{e}_1 \tilde{e}_2$ are conserved indicates that these are the total 1-flux and 2-flux of the combined loops, and so we have obtained the braiding when they fuse to give the labels $(hg, \tilde{e}_1 \tilde{e}_2)$. We could equally have considered the braiding in a different situation, such as when the start-points of the two operators are in different positions, for which the loops $(h,\tilde{e}_1)$ and $(g, \tilde{e}_2)$ fuse to give different total quantum numbers than $(hg, \tilde{e}_1 \tilde{e}_2)$.

		As we did when considering the braiding of the higher-flux with the blob excitation, we can also consider the braiding of our original magnetic excitation, before we pinned an additional $E$-valued loop to it. As described by Equation \ref{Equation_higher_flux_to_magnetic}, we can obtain the original magnetic membrane operators from the higher-flux membrane operators by summing over all possible elements of $e$ for the $E$ label. That is, we consider
		\begin{align*}
		C^g_T&(m_2)C^h_T(m_1) \ket{GS}\\
		&= C^g_T(m_2)\sum_{e_2 \in E} \delta(e_2, \hat{e}(m_2))\\
		& \hspace{0.5cm} C^h_T(m_1) \sum_{e_1 \in E} \delta(e_1, \hat{e}(m_1))\ket{GS}\\
		&=\sum_{e_2 \in E}\sum_{e_1 \in E}C^{g,e_2}_T(m_2)C^{h,e_1}_T(m_1) \ket{GS}.
		\end{align*}
		Using Equation \ref{Equation_higher_flux_braiding_main_text}, we see that this gives us
		\begin{align*}
		&C^g_T(m_2)C^h_T(m_1) \ket{GS}\\
		&=\sum_{e_1 \in E} \sum_{e_2 \in E} \\
		& \hspace{0.5cm} C^{h,e_1 \big[g((1)-(2)) \rhd ([h_{[2-1]} \rhd e_2^{-1}] [(h_{[2-1]} g^{-1}) \rhd e_2])\big]}_T(m_1) \notag \\
		&\hspace{0.5cm} C_\rhd^{h_{[2-1]}gh_{[2-1]}^{-1}}(m_2) \notag \\
		&\hspace{0.5cm} \bigg(\prod_{\substack{\text{plaquette }\\ p \in m_2}} B^{[h_{[2-1]}^{-1}\rhd e_{p|2} ] [(g^{-1}h_{[2-1]}^{-1})\rhd e_{p|2}^{-1}]}((2)-(1))\notag \\
		&\hspace{0.5cm} B^{e_{p|2} [(h_{[2-1]} g^{-1} h_{[2-1]}^{-1}) \rhd e_{p|2}^{-1}]}((1)-p) \bigg) \notag \\
		&\hspace{0.5cm} \delta(h_{[2-1]} \rhd e_2, \hat{e}(m_2))\ket{GS}.
		\end{align*}
		
		Then we have 
		\begin{align*}
		\sum_{e_1 \in E}C&^{h,e_1 \big[g((1)-(2)) \rhd ([h_{[2-1]} \rhd e_2^{-1}] [(h_{[2-1]} g^{-1}) \rhd e_2])\big]}_T(m_1) \\
		&=C^h_T(m_1),
		\end{align*}
		because summing over $e_1$ gives us an equal sum over all Kronecker deltas $\delta(\hat{e}(m_1),e)$, regardless of the actual value of $e_2$. This gives us
		\begin{align*}
		C^g_T&(m_2)C^h_T(m_1) \ket{GS}\\
		&=\sum_{e_1 \in E} \sum_{e_2 \in E} C^{h}_T(m_1) C_\rhd^{h_{[2-1]}gh_{[2-1]}^{-1}}(m_2) \notag \\
		&\hspace{0.5cm} \bigg( \prod_{\substack{\text{plaquette }\\ p \in m_2}} B^{[h_{[2-1]}^{-1}\rhd e_{p|2} ] [(g^{-1}h_{[2-1]}^{-1})\rhd e_{p|2}^{-1}]}((2)-(1))\notag \\
		&\hspace{0.5cm} B^{e_{p|2} [(h_{[2-1]} g^{-1} h_{[2-1]}^{-1}) \rhd e_{p|2}^{-1}]}((1)-p) \bigg) \notag \\
		&\hspace{0.5cm} \delta(h_{[2-1]} \rhd e_2, \hat{e}(m_2))\ket{GS}.
		\end{align*}
		
		We can similarly use the sum over $e_2 \in E$ to remove the other Kronecker delta. We have 
		$$\sum_{e_2 \in E}\delta(h_{[2-1]} \rhd e_2, \hat{e}(m_2)) = 1$$
		because $h_{[2-1]} \rhd e_2$ runs over all $e \in E$. This gives us the final result
		\begin{align}
		C^g_T&(m_2)C^h_T(m_1) \ket{GS} \notag\\
		&=\sum_{e_1 \in E} \sum_{e_2 \in E} C^{h}_T(m_1) C_\rhd^{h_{[2-1]}gh_{[2-1]}^{-1}}(m_2) \notag \\
		&\hspace{0.5cm} \bigg( \prod_{\substack{\text{plaquette }\\ p \in m_2}} B^{[h_{[2-1]}^{-1}\rhd e_{p|2} ] [(g^{-1}h_{[2-1]}^{-1})\rhd e_{p|2}^{-1}]}((2)-(1))\notag \\
		&\hspace{0.5cm} B^{e_{p|2} [(h_{[2-1]} g^{-1} h_{[2-1]}^{-1}) \rhd e_{p|2}^{-1}]}((1)-p) \bigg) \ket{GS}.
		\end{align}
		
		Then looking at the effect of braiding on the $G$-valued label, we see that the result is simply conjugation of one of the magnetic flux labels by the other. The labels of the blob ribbons corresponding to $m_2$ are also changed before and after the intersection of the two membranes. Just as we discussed for the higher-flux membrane operators earlier in this section, the label after the intersection reflects the change to the label $g$ of the membrane operator applied on $m_2$, while the change to the label before the intersection is only an apparent change due to the operator label.

		\subsection{Braiding with $E$-valued loops}
		\label{Section_mod_mem_E_loop_braiding}
		The final braiding relation to consider is the braiding between these higher-flux excitations and the $E$-valued loops. We can obtain this relation from the calculation for two higher-flux membrane operators in the previous section, because the $E$-valued loops are simply higher-flux excitations with trivial $G$-label. We therefore simply need to take the special case of that calculation when one of the $G$ elements is $1_G$. Rather than repeating the full equations, we will only present the results in the same-start-point braiding case.

		First we consider the case where the red excitation shown in Figure \ref{Braid_move_loops} is a higher-flux excitation, produced by a membrane operator $C^{g, e_{\text{mag}}}_T(m_2)$, while the green excitation is a pure $E$-valued loop, labeled by $e_m$. In this case the label of the $E$-valued loop transforms as $e_m \rightarrow e_m e_{\text{mag}}^{-1} [g^{-1} \rhd e_{\text{mag}}]$ under the braiding, while the labels of the magnetic membrane operator are unaffected by the braiding. When we change to consider our irrep basis for the $E$-valued loops (given in Equation \ref{Equation_E_membrane_irrep_Abelian}) this transformation gives us a phase of
		\begin{equation}
		 \gamma( e_{\text{mag}}^{-1} [g^{-1} \rhd e_{\text{mag}}])^{-1},
		 \end{equation}
		  where $\gamma$ is the irrep of $E$ labeling the $E$-valued loop. Note that if we consider the ordinary magnetic excitation by averaging over $ e_{\text{mag}}$, the braiding relation is different for each value of $ e_{\text{mag}}$, so the different terms in the sum accumulate different transformations. This is part of the reason why it was necessary to consider the higher-flux membrane instead of the magnetic one in the first place. The result of this different transformation for the different $E$-labels is that, even if the excitation is initially an ordinary magnetic excitation, with an equal superposition of the different $E$-labels, it will not necessarily remain so after braiding, instead becoming an uneven superposition of the different higher-flux excitations, with labels coupled to the state of the $E$-valued loop.

		Now we consider the opposite case where the $E$-valued loop passes through the magnetic one. In this case, the red excitation from Figure \ref{Braid_move_loops} is a pure $E$-valued loop excitation produced by the membrane operator $\delta(\hat{e}(m_2),e_m)$, while the green excitation is a higher-flux loop excitation produced by the membrane operator $C^{h,e_1}_T(m_1)$. In this case the label $e_m$ of the $E$-valued loop transforms as $e_m \rightarrow h \rhd e_m$ under the braiding, while the labels of the higher-flux operator are again unaffected by the braiding move. In our irrep basis for the $E$-valued membrane operators, this transformation actually changes the irrep $\gamma$ labeling the membrane operator to a different irrep $h^{-1} \rhd \gamma$ in the same $\rhd$-Rep class of irreps (where two irreps of $E$, $\alpha$ and $\beta$, are in the same $\rhd$-Rep class if there exists a $g \in G$ such that $\alpha(g \rhd e)= \beta(e)$ for all $e \in E$). This suggests that the irreps of $E$ are not by themselves good labels for the topological charge, because the irreps are not invariant under braiding, and instead the $\rhd$-Rep classes should be important (although the condensation further affects the topological charge). However, as we see in Section \ref{Section_3D_sphere_charge_examples}, the point-like charge of the loop-like excitations actually depends on how the coefficients of the membrane operator transforms under the $\rhd$ action, and so this topological charge has some dependence on quantities within the $\rhd$ class as well. We will discuss the topological charge in more detail in Section \ref{Section_3D_Topological_Sectors}.
		
		\subsection{Summary of braiding in this case}
		
		Table \ref{Table_Braiding_Central} summarizes which types of excitation can have non-trivial braiding relations in the case where $E$ is Abelian and $\partial$ maps to the centre of $G$, where non-trivial braiding between the types of excitation is indicated by ticks. Note that the higher-flux excitations have potentially non-trivial braiding relations with every class of excitation.

		\begin{table}[h]
			\begin{center}
				\begin{tabular}{ |c|c|c|c|c| } 
					\hline
					Non-Trivial& &Higher- & & $E$-valued \\
					Braiding?& Electric & flux & Blob & loop \\ 
					\hline
					Electric & \xmark & \cmark & \xmark & \xmark\\ 
					\hline
					Higher- & && &\\ 
					
					flux & \cmark & \cmark & \cmark & \cmark \\
					\hline 
					Blob& \xmark & \cmark & \xmark & \cmark\\
					\hline
					$E$-valued &  & &  & \\
					loop & \xmark& \cmark & \cmark& \xmark \\
					\hline
				\end{tabular}
				
				\caption{A summary of which excitations braid non-trivially in Case 2, where the group $E$ is Abelian and $\partial$ maps onto the centre of $G$. A tick indicates that at least some of the excitations of each type braid non-trivially with each-other, while a cross indicates that there is no non-trivial braiding between the two types. }
				\label{Table_Braiding_Central}
			\end{center}
			
		\end{table}
		
		\section{Topological charge}
		\label{Section_3D_Topological_Sectors}
		
		In Ref. \cite{HuxfordPaper1} we explained the concept of topological charge and in Ref. \cite{HuxfordPaper2} we presented a detailed construction of the measurement operators for topological charge in the 2+1d case. To briefly restate our explanation from Ref. \cite{HuxfordPaper1}, topological charge is a quantity that is conserved, so that the only way to change the topological charge in a region is to apply an operator that connects this region to the rest of our lattice. Further conditions are imposed on the topological charge relevant to our model by requiring that the ground state is the topological vacuum. There is a significant difference between the 2+1d and 3+1d cases, however. Whereas we consider the topological charge in regions isomorphic to disks (or unions and differences of disks, like annuli) when there are only two spatial dimensions, there are more topologically distinct regions to consider when there are three spatial dimensions. For example, we have both topological balls and solid tori. The charge in these regions should be measured by operators on the surfaces of the regions, i.e., on spheres and tori. This variety of regions is related to our excitations. We have both point-like and loop-like excitations, both of which should carry topological charge. While we expect point-like charges to be fully measured by spheres, the sphere has no features that would allow it to distinguish between a loop and a point. On the other hand, a torus has handles which can link with a loop-like excitation and we expect this to allow the torus to distinguish between point particles and loops. We therefore expect that the loop excitations should carry a charge that is not measured by the sphere (in addition to some charge that can be measured by the sphere). Therefore, we need to include the toroidal measurement surfaces as well.

		In order to measure the topological charge held within or without a particular surface, we follow a similar procedure to the one used for the 2+1d case in Ref. \cite{HuxfordPaper2}, except that the boundary for our region is a surface rather than a path. We take our surface of interest and apply every closed membrane or ribbon operator that we can on this surface, before considering only the sums of these operators that will commute with the Hamiltonian. Combinations of ribbon and membrane operators can produce any linear operator on the Hilbert space. This is because we can consider ribbon and membrane operators that act only on a single edge or plaquette. An electric ribbon operator acting on a single edge can measure any value for that edge, while a magnetic membrane operator acting only on a single edge can multiply that edge by any group label. Together these allow us to freely control the label of any edge. Similarly an $E$-valued membrane operator can measure the label of a single plaquette and a blob ribbon operator can multiply its label by any value. Combinations of these operators can therefore control the label of every edge and plaquette in the lattice. However, when we restrict to operators that commute with the Hamiltonian, we are left only with closed ribbon and membrane operators. This restriction of commuting with the Hamiltonian is because our measurement operator should not by itself produce or move topological charge.

\begin{figure}[h]
			\begin{center}
				\includegraphics[width=\linewidth]{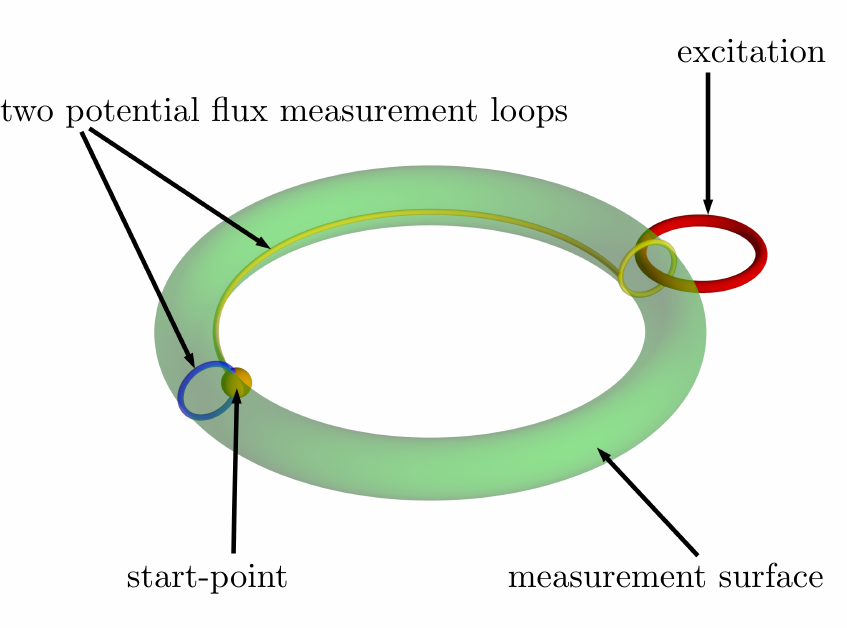}
				
				\caption{If an excitation pierces the measurement surface, then the charge within the surface is ill-defined. In the case shown in this figure, measuring the 1-flux along the two potential loops may give different results (to the point of not giving the same topological charge).}
				\label{measurement_intersected}
			\end{center}
		\end{figure}

		 We consider this process of measuring the charge for sphere-like and torus-like surfaces. Theoretically, we could do the same for an arbitrary surface. However, because the simple excitations of the model are either point-like or loop-like, it does not appear necessary to consider the charge measured by higher-genus surfaces. Nevertheless, this may not be the case and it would be interesting to construct the higher-genus measurement operators, but we leave this for future study. One subtlety with measuring a loop-like excitation is that, because the loops are extended, the excitations may pierce the measurement surface and not be wholly contained within or without the surface. As an example, consider the situation shown in Figure \ref{measurement_intersected}. As part of the measurement procedure, we must choose closed paths on which to measure any magnetic flux enclosed by the torus. However, in the presence of excitations on the surface itself, two choices of loops to measure on (for example, the blue or yellow paths in the figure) would give different results. This is because different loops may or may not link with the excitation (the thicker red torus). Because both choices are supposed to measure the charge within the torus (the partially transparent green torus), this leads to a contradiction. If we want to measure the charge held within a surface without knowing what excitations are present and where they are, this presents a difficulty. Therefore, we include in our charge measurement operators a projector to the space where the surface has no excitations. This sidesteps the above issue, but it does mean that we cannot measure the charge of confined excitations (which always cause excitations on a surface enclosing them) using this procedure.

		\subsection{Topological charge within a sphere in the case where $\partial \rightarrow$ centre($G$) and $E$ is Abelian}
		\label{Section_Sphere_Charge_Reduced}
		
		Before we look at the charge measured by a torus, which is sensitive to both loop-like and point-like charge, we will first examine the charge measured by a sphere. We will do this in the case where $E$ is Abelian and $\partial$ maps onto the centre of $G$ (Case 2 in Table \ref{Table_Cases}), which includes the $\rhd$ trivial case (Case 1 in Table \ref{Table_Cases}) as a sub-case. Because a sphere has no non-contractible cycles, the sphere should only be sensitive to point-like charge. Nonetheless, the sphere charge is interesting, not only because it lets us look at the properties of point particles, but because loop excitations also possess point-like charge. As we explained in the previous section, to measure the charge within a sphere we first project to the case where there are no excitations on the measurement surface. Then we consider which independent closed ribbon and membrane operators we can apply on this surface.
		
		While it may seem that we can independently apply ribbon operators around any closed loop on the surface of the sphere, this is not the case. Any ribbon operators are either topological or confined (or can be written as a linear combination of ribbon operators of the two types), as we show in Section \ref{Section_topological_membrane_operators} in the Supplemental Material. If a ribbon operator is confined, then applying it leads to excitations on the measurement surface, which we do not allow. On the other hand, if a ribbon operator is topological, then because all closed paths on the sphere are contractible on the spherical surface (and we do not allow excitations on the surface), the ribbon can be contracted to nothing without affecting the action of the ribbon operator. This means that applying a closed topological ribbon operator on the surface of the sphere is equivalent to applying the identity operator (at least in the subspace on which we apply measurement operators). Therefore, any ribbon operators that we are allowed to apply (the non-confined ones) act trivially. 
		 
		This leaves us only with the membrane operators $C^{h}_T(m)$ and $L^e(m)$, where $C^h_T(m)$ is the total magnetic membrane operator defined in Section \ref{Section_3D_MO_Central} (see Eq. \ref{total_magnetic_membrane_operator}) and $L^e(m)$ is the $E$-valued membrane operator $\delta(e, \hat{e}(m))$. We consider applying these two operators over the sphere. Although we apply both operators on the same sphere, when we define the membrane $m$ for each operator to act on we need to define a start-point for the membrane. It would seem that we could choose the start-points of the membranes to be different for the two membrane operators, giving us many potential measurement operators. However, this is not the case because of the requirement that the total measurement operator commute with the energy terms on the sphere. As we have discussed previously, and prove in Section \ref{Section_Magnetic_Tri_Nontrivial_Commutation} in the Supplemental Material for this work and Section S-I C in the Supplemental Material for Ref. \cite{HuxfordPaper2}, both the magnetic membrane operator and $E$-valued membrane operator commute with the vertex transforms except those at the start-points. If the two operators have different start-points, then each must individually commute with their specific start-point vertex transform (rather than their combination having to commute with a mutual start-point transform). However, when a membrane operator commutes with the start-point vertex transforms, the start-point of the operator becomes arbitrary. That is, if the start-point is not excited we can move the start-point without affecting the action of the membrane operator, because parallel transport of a vertex is equivalent to applying a vertex transform (see Section \ref{Section_magnetic_membrane_central_change_sp} in the Supplemental Material for a proof of this for the magnetic membrane operator and Section S-I C in the Supplemental Material of Ref. \cite{HuxfordPaper2} for a proof for the $E$-valued membrane operator). This means that we can move the start-points to be in the same location anyway, without affecting the action of the two membrane operators. Therefore, without loss of generality, we can consider the two start-points of the membrane operators to be in the same location.
		
		The most general operator we can apply is a linear combination of terms with the form $C^{h}_T(m) L^e(m)$ for different labels $h$ and $e$, where $m$ is the spherical membrane that we are measuring the charge within. We might also consider products that include multiples of one or more of the two types of operators, such as $C^h_T(m) L^e(m)L^f(m) C^g_T(m)$. However, because the two types of operator commute, we can always collect the separate instances of each type of operator, to give us terms like $C^h_T(m) C^g_T(m) L^e(m)L^f(m)$. Then we can use the algebra of the membrane operators to combine them, which just gives us $\delta(e,f)C^{hg}_T(m) L^e(m)$ for the above example. This is just an example of a linear combination of terms of the form $C^{h}_T(m) L^e(m)$, so we only need consider such terms. We take the membrane $m$ to be oriented inwards to match the orientation of the surface label of the direct membrane used in $C^{h}_T(m)$. Taking the opposite orientation would be equivalent to using $e^{-1}$ instead of $e$. We also do not displace blob 0 and the start-point of $C^h_T(m)$ and $L^e(m)$ from the membrane, contrary to the approach used in Section \ref{Section_3D_Braiding_Central} when considering the braiding of the higher-flux excitations. This choice does not matter, because the fact that we enforce the start-point and blob 0 to be unexcited by the combined action of the measurement operator means that we can freely move the start-point and blob 0 around without affecting the total action of the measurement operator. Moving the start-point is equivalent to applying a vertex transform at the start-point, which is trivial when the start-point is unexcited, and we show in Section \ref{Section_magnetic_change_blob_0} of the Supplemental Material that moving blob 0 is trivial when that blob is unexcited.
		
		Having found that the operator we apply must have the form $\sum_{h \in G} \sum_{e \in E} \alpha_{h,e} C^{h}_T(m) L^e(m)$, where $\alpha_{h,e}$ are a set of coefficients, we next have to find which coefficients lead to the operator commuting with the energy terms on the sphere. In Section \ref{Section_sphere_topological_charge_appendix_full} of the Supplemental Material, we show that requiring commutation with the energy terms leads to two types of restrictions for the coefficients. Some of these restrictions enforce that the coefficient $\alpha_{h,e}$ must be zero for certain labels (i.e., certain pairs of label $(h,e)$ for $C^{h}_T(m)L^{e}(m)$ are disallowed), while other conditions mean that the coefficients of two pairs $(h_1,e_1)$ and $(h_2,e_2)$ must be the same (i.e., $\alpha_{h_1,e_1}= \alpha_{h_2,e_2}$). As a shorthand, we write $(h_1,e_1) \overset{\mathrm{S}}{\sim} (h_2,e_2)$ for two pairs that are subject to this latter type of restriction (must have equal coefficients). Then the restrictions that we find are
		\begin{align}
		h \rhd e &=e, \\
		\partial(e)&=1_G, \\
		h &\overset{\mathrm{S1}}{\sim} \partial(f)h \ \forall f \in E, \intertext{and}
		(h,e) &\overset{\mathrm{S2}}{\sim} (ghg^{-1}, g \rhd e) \ \forall g \in G. 
	    \end{align}
		
		The two equivalence relations $\overset{\mathrm{S1}}{\sim}$ and $\overset{\mathrm{S2}}{\sim}$ together form the equivalence relation $\overset{\mathrm{S}}{\sim}$, where any pairs of labels $(h_1,e_1)$ and $(h_2,e_2)$ related by $\overset{\mathrm{S}}{\sim}$ must have equal coefficients. We can write $\overset{\mathrm{S}}{\sim}$ explicitly as
		\begin{equation}
		(h,e) \overset{\mathrm{S}}{\sim} (\partial(f) ghg^{-1}, g \rhd e),
		\end{equation}
		for each $g \in G$ and $f \in E$ (i.e., $(h,e)$ is in the same equivalence class as $(h',e')$ if there exists any $g \in G$ and $e \in E$ such that $(h',e') = (\partial(f) ghg^{-1}, g \rhd e)$). Given all of these conditions for our measurement operators, we can construct a basis for the space of allowed measurement operators. As we show in Section \ref{Section_sphere_topological_charge_appendix_full} of the Supplemental Material, one such basis is given by a set of operators labeled by two objects. The first object, $C$, is a $\rhd$-class of the kernel of $\partial$. A $\rhd$-class of the kernel is a subset of the kernel consisting of elements related by the equivalence relation $e \overset{\rhd}{\sim} f$ if there exists a $g \in G$ such that $g \rhd e=f$. It is convenient to pick a representative element $r_C$ for each such class $C$. Then we define the centralizer of the class $C$ as $Z_{\rhd, {r_C}}=\set{ h \in G | h \rhd r_C = r_C}$. The second object that labels our basis operators is a class within this centralizer, this time described by the equivalence relation
		\begin{equation}
		h \overset{Z_{\rhd, {r_C}}}{\sim} xhx^{-1} \partial(w)
		\label{union_coset_relation_main_text}
		\end{equation}
		for any elements $x \in Z_{\rhd, {r_C}}$ and $w \in E$. Note that this equivalence relation is similar to $\overset{\mathrm{S}}{\sim}$, in that it gives the same form of relation, but only for elements $x \in Z_{\rhd,r_C}$ (for which $x \rhd r_C=r_C$) rather than general elements $g \in G$. The basis operator corresponding to a particular $\rhd$-class $C$ of the kernel and equivalence class $D$ (defined by Equation \ref{union_coset_relation_main_text}) of the associated centralizer is
		$$T^{D,C}(m)= \sum_{q \in Q_C} \sum_{d \in D} C^{qdq^{-1}}_T(m) L^{q \rhd r_C}(m),$$
		where $Q_C$ is a set of elements of $G$ that move us between the elements of the $\rhd$-class $C$, so that each element $e_i \in C$ has a unique $q_i \in Q_C$ such that $e_i = q_i \rhd r_C$. As we show in Section \ref{Section_sphere_topological_charge_appendix_full} of the Supplemental Material, any element $g\in G$ can be uniquely decomposed as a product of an element of $Q_C$ and an element of $Z_{\rhd,r_C}$. We can use this basis of operators to construct the projectors to definite topological charge within the sphere. These projectors are given by
		\begin{equation}
		T^{R,C}(m)=\frac{|R|}{|Z_{\rhd,r_C}|} \sum_{D \in (Z_{\rhd,r_C})_{cl}} \chi_R(D) T^{D,C}(m), \label{Equation_sphere_projector_definition_main_text}
		\end{equation}
		where $R$ is an irrep of the quotient group $Z_{\rhd,r_C}/\partial(E)$ with dimension $|R|$ and $(Z_{\rhd,r_C})_{cl}$ is the set of classes in the centralizer defined by the equivalence relation Equation \ref{union_coset_relation_main_text}. Note that the character $\chi_R$ of irrep $R$ is independent of the element $d \in D$ (because characters are a function of conjugacy class, and $R$ being an irrep of the quotient group means that it is also insensitive to factors of $\partial(w)$ from Equation \ref{union_coset_relation_main_text}). In Section \ref{Section_sphere_topological_charge_appendix_full} of the Supplemental Material we prove that the operators defined by Equation \ref{Equation_sphere_projector_definition_main_text} are indeed projectors and are orthogonal and complete in our space.

		\subsubsection{The point-like charge of simple excitations}
		\label{Section_3D_sphere_charge_examples}
		Having worked out the projectors for the charges, it will be instructive to use them to check the topological charge of some of our simple excitations (those produced by single ribbon or membrane operators). To do this we try enclosing these charges with our measurement operators.

		We first consider measuring the charge of an electric excitation at the end-point of a ribbon operator. To do this, we first need to create our electric excitation, by applying an electric ribbon operator to our ground state. Considering a ribbon operator labeled by irrep $X$ of $G$ and matrix indices $a$ and $b$, we obtain the state
		$$\sum_{g \in G} [D^{X}(g)]_{ab} \delta(\hat{g}(t),g) \ket{GS}.$$

		\begin{figure}[h]
			\begin{center}
				\includegraphics[width=\linewidth]{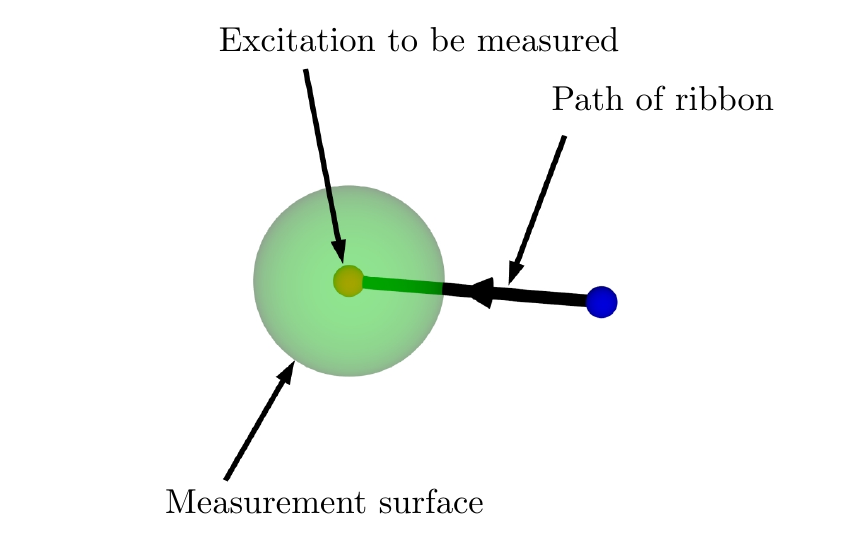}

				\caption{We measure the charge held at the end of an electric ribbon, using our spherical surface (large green sphere)}
				\label{3D_electric_measurement_2}
			\end{center}
		\end{figure}

		Next we want to measure this charge, by applying a measurement operator, as shown in Figure \ref{3D_electric_measurement_2}. Therefore, we want to calculate
		\begin{align*}
		T&^{R,C}(m) \sum_{g \in G} [D^{X}(g)]_{ab} \delta(\hat{g}(t),g) \ket{GS}\\
		& = \frac{|R|}{|Z_{\rhd,r_C}|} \sum_{D \in (Z_{\rhd,r_C})_{cl}} \chi_R(D) \sum_{d \in D} \sum_{q \in Q_C} C_T^{qdq^{-1}}(m)\\
		& \hspace{0.5cm} \delta(\hat{e}(m), q \rhd r_C) \sum_{g \in G} \delta(\hat{g}(t),g) [D^{X}(g)]_{ab} \ket{GS}.
		\end{align*}
		
		The operator $\delta(\hat{e}(m), q \rhd r_C)$ commutes with $\delta(\hat{g}(t),g)$, so we can commute $\delta(\hat{e}(m), q \rhd r_C)$ all the way to the right, so that it acts directly on the ground state. Then we have
		$$\delta(\hat{e}(m), q \rhd r_C)\ket{GS}=\delta(1_E, q\rhd r_C) \ket{GS},$$
		because $m$ is a sphere, and any contractible sphere in the ground state must have a surface label of $1_E$ due to the blob energy terms. We can write $\delta(q\rhd r_C,1_E)$ as $\delta(r_C,1_E)=\delta(C,\set{1_E})$ (using the fact that the identity is invariant under the $\rhd$ action and so is the only element of its $\rhd$-class). Therefore, we find that the result of measurement is zero unless the class that we are trying to measure is the trivial one. This is as we expect, because the electric excitations do not possess non-trivial 2-flux.

		Having found that the class $C$ must be trivial for a non-zero result, we can also simplify the other mathematical objects appearing in the projector. When $r_C=1_E$, we have that $h \rhd 1_E=1_E \ \forall h \in G$, which implies that $Z_{\rhd,r_C}=G$ and the quotient group $Z_{\rhd,r_C}/ \partial(E)$ is simply $G/\partial(E)$. In addition, the set $Q_C$ is the trivial group containing just the identity element, so we may drop the sum over $q \in Q_C$. Then the result of our measurement is
		\begin{align*}
		\delta&(C,\set{1_E}) \frac{|R|}{|G|} \sum_{D \in (Z_{\rhd,r_C})_{cl}} \chi_R(D) \sum_{d \in D} C_T^d(m)\\
		& \hspace{3cm}\sum_{g \in G} \delta(\hat{g}(t),g) [D^{X}(g)]_{ab} \ket{GS}\\
		&= \delta(C,\set{1_E}) \frac{|R|}{|G|} \sum_{d \in G} \sum_{g \in G} \chi_R(d) [D^{X}(g)]_{ab}\\
		& \hspace{3cm} C_T^d(m) \delta(\hat{g}(t),g) \ket{GS}.
		\end{align*}
		
		Then we just need to find the commutation relation between $C^d_T(m)$ and $\delta(\hat{g}(t),g)$. The calculation of this is analogous to the calculation performed to find the braiding relation between the electric and magnetic excitations (see Section \ref{Section_Flux_Charge_Braiding}), except that we have the opposite orientation of the magnetic membrane operator. We find that
		\begin{align*}
		C_T^d(m) &\delta(\hat{g}(t),g) \ket{GS}\\
		& = \delta(\hat{g}(t-m)d\hat{g}(t-m)^{-1}\hat{g}(t),g) C_T^d(m) \ket{GS},
		\end{align*}
		where $\hat{g}(t-m)$ is shorthand for $\hat{g}(s.p(t)-s.p(m))$, the path element for the path from the start-point of $t$ to the start-point of $m$. We also have that $C_T^d(m) \ket{GS}=\ket{GS}$ because the sphere is contractible and the operator is topological (so that we can deform the operator to nothing). Using these results in our previous expression gives
		\begin{align*}
		T^{R,C}(m) &\sum_{g \in G} [D^{X}(g)]_{ab} \delta(\hat{g}(t),g) \ket{GS}\\
		&= \delta(C, \set{1_E}) \frac{|R|}{|G|}\sum_{d \in G} \sum_{g \in G} \chi_R(d) [D^{X}(g)]_{ab}\\
		& \hspace{0.5cm} \delta(\hat{g}(t),\hat{g}(t-m)d^{-1}\hat{g}(t-m)^{-1}g) \ket{GS}.
		\end{align*}
		We then rewrite $\hat{g}(t-m)d\hat{g}(t-m)^{-1}$ as $d'$ and replace the sum over the dummy index $d$ with a sum of $d'$. Noting that the character $ \chi_R$ is a function of conjugacy class, so that $\chi_R(d') = \chi_R(d)$, we then see that
		\begin{align*}
		T^{R,C}(m) &\sum_{g \in G} [D^{X}(g)]_{ab} \delta(\hat{g}(t),g) \ket{GS}\\
		&=\delta(C, \set{1_E}) \frac{|R|}{|G|}\sum_{d' \in G} \sum_{g \in G} \chi_R(d') [D^{X}(g)]_{ab}\\
		& \hspace{0.5cm} \delta(\hat{g}(t),{d'}^{-1}g) \ket{GS}\\
		&= \delta(C, \set{1_E}) \frac{|R|}{|G|}\sum_{d' \in G} \sum_{g'={d'}^{-1}g \in G} \chi_R(d')\\
		& \hspace{0.5cm} [D^{X}(d'g')]_{ab} \delta(\hat{g}(t),g') \ket{GS}\\
		&= \delta(C, \set{1_E}) \frac{|R|}{|G|}\sum_{d',g' \in G} \sum_{c=1}^{|X|} \sum_{e=1}^{|R|} [D^R(d')]_{ee}\\
		& \hspace{0.5cm} [D^{X}(d')]_{ac} [D^{X}(g')]_{cb} \delta(\hat{g}(t),g')\ket{GS}.
		\end{align*}
		
		We now want to use the orthogonality relations for irreps of a group to simplify this. There is a slight complication in that $X$ is an irrep of $G$ whereas $R$ is an irrep of $G/\partial(E)$. However, $R$ induces a representation $R_G$ of $G$ defined by $R_G(g) =R(\tilde{g}\partial(E))$, where $\tilde{g}\partial(E)$ is the coset which $g$ belongs to. Therefore, each matrix from $R$ is copied $|\partial(E)|$ times in $R_G$. Given that $R$ is an irrep of $G/\partial(E)$, $R_G$ must also be irreducible (as a representation of $G$). This is because the same matrices appear in the two representations $R$ and $R_G$, so if $R$ cannot be reduced to a block diagonal form then neither can $R_G$. Then we can use $R_G$ instead of $R$ and apply the standard irrep orthogonality relations to obtain
		\begin{align}
		T^{R,C}(m) &\sum_{g \in G} [D^{X}(g)]_{ab} \delta(\hat{g}(t),g) \ket{GS} \notag\\
		&= \delta(C, \set{1_E}) \frac{|R|}{|G|} \sum_{g' \in G} \frac{|G|}{|R|} \delta_{e c}\delta_{e b} \delta(\overline{R}_G, X)\notag \\
		& \hspace{0.5cm} [D^{X}(g')]_{cb} \delta(\hat{g}(t),g')\ket{GS}\notag \\
		&= \delta(C,\set{1_E}) \delta(R_G, \overline{X}) \sum_{g' \in G}[D^{X}(g')]_{cb} \notag\\
		& \hspace{0.5cm} \delta(\hat{g}(t),g') \ket{GS}.
		\end{align}
		
		We see that the final result of applying the measurement operator is that we recover our original electric operator acting on the ground state, multiplied by $\delta(C,\set{1_E}) \delta(R_G, \overline{X})$. This indicates that the charge of the excitation is $(\set{1_E},\overline{X})$. If the irrep $X$ corresponds to a confined excitation, none of our measurement operators will give a non-zero result, because $R_G$ derives from an irrep of the quotient group, and so cannot have a non-trivial restriction to the image of $\partial$. This is a result of our requirement that we project out the states that have excitations on the measurement membrane itself, which naturally precludes the measurement of any confined excitations.

\begin{figure}[h]
			\begin{center}
				\includegraphics[width=\linewidth]{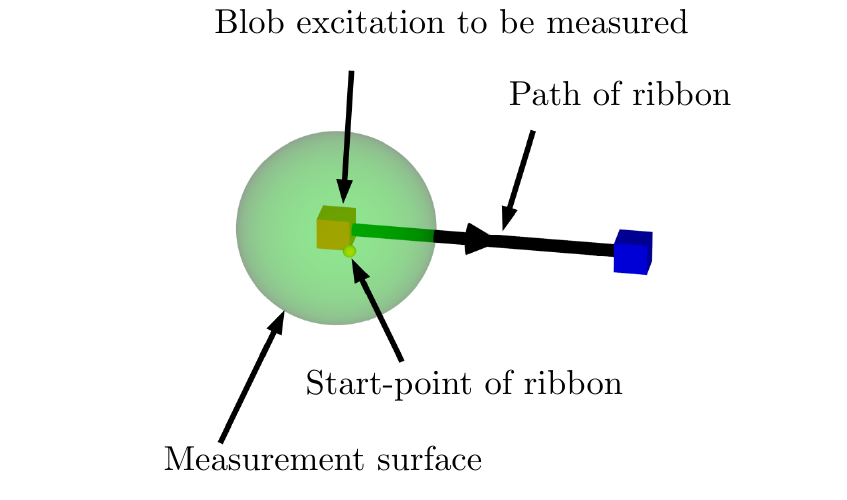}

				\caption{We measure the topological charge of the blob excitation at the start of the blob ribbon operator.}
				\label{3D_blob_measurement}
			\end{center}
		\end{figure}

		We can similarly check the charge of a blob excitation. We expect that the corresponding charge will be labeled by a non-trivial class $C$, because this class is associated to the 2-flux measured by the measurement operator. We may think that because the blob excitation can have an excited vertex, in addition to the excited blob, that this means that the representation labeling the charge should be non-trivial, just like with the electric excitation. However, we will see that the vertex excitation does not in this case result in a non-trivial representation. We measure the topological charge of the blob excitation at the start of the path of a blob ribbon operator, as indicated in Figure \ref{3D_blob_measurement}. We choose to measure the charge of the excitation at the start of the ribbon, rather than the end as we did with the electric excitation, because this will highlight the fact that the excited vertex enclosed by the measurement surface does not lead to a non-trivial representation $R$.
		
		We consider measuring the charge of a blob excitation produced by the blob ribbon operator 
		$$\sum_{e \in [\tilde{e}]_{\rhd}} \alpha_e B^e(t),$$
		where $[\tilde{e}]_{\rhd}$ is the $\rhd$-class containing $\tilde{e}$, which is defined by 
		\begin{equation}
		e \in [\tilde{e}]_{\rhd} \iff \exists g \in G \text{ s.t } e = g \rhd \tilde{e}.
		\label{rhd_class_def}
		\end{equation}
		
		In order to measure the charge, we apply a measurement operator $T^{R,C}$ on the state produced by acting with the blob ribbon operator on the ground state. That is, we examine a state
		\begin{align}
		T^{R,C}&(m) \sum_{e \in [\tilde{e}]_{\rhd}} \alpha_e B^e(t) \ket{GS} \notag\\
		&= \frac{|R|}{|Z_{\rhd,r_C}|} \sum_{D \in(Z_{\rhd,r_C})_{cl}} \chi_R(D) \sum_{d \in D} \sum_{q \in Q_C} C_T^{qdq^{-1}}(m)\notag\\
		& \hspace{1.5cm} \delta(\hat{e}(m), q \rhd r_C) \sum_{e \in [\tilde{e}]_{\rhd}} \alpha_e B^e(t) \ket{GS}, \label{blob_charge_measure_1}
		\end{align}
		where $\alpha_e$ is a set of coefficients that we keep general, so that we can show that the topological charge does not depend on the coefficients within the $\rhd$-class, only on the class itself.

		From a calculation analogous to the one for braiding between higher-flux excitations and blob excitations in Section \ref{Section_magnetic_blob_braiding}, we know that
		\begin{align}
		&C_T^{qdq^{-1}}(m) \delta(\hat{e}(m), q \rhd r_C) B^e(t) \ket{GS} \notag\\
		& = B^e(\text{start}(t)-\text{ blob }0) \notag \\
		& \hspace{0.5cm} B^{(g(m-t)^{-1}qdq^{-1}g(m-t)) \rhd e}(\text{blob }0- \text{end}(t)) C^{qdq^{-1}}_T(m) \notag\\
		& \hspace{0.5cm} \delta( \hat{e}(m),[q \rhd r_C] [\hat{g}(s.p(m)-s.p(t)) \rhd e^{-1}]) \ket{GS}, \label{blob_charge_measure_2}
		\end{align}
		where all of the blob ribbon operators have the same start-point as $t$ and $g(m-t)$ is shorthand for the path element $g(s.p(m)-s.p(t))$. Then, because the membrane $m$ is contractible, its surface element must be the identity in the ground state. Therefore, we have
		\begin{align}
		\delta( \hat{e}&(m),[q \rhd r_C] [\hat{g}(s.p(m)-s.p(t)) \rhd e^{-1}]) \ket{GS} \notag\\
		&= \delta([q \rhd r_C] [\hat{g}(s.p(m)-s.p(t)) \rhd e^{-1}], 1_E ) \ket{GS}. \label{higher_flux_on_ground_state}
		\end{align}
		
 Substituting the relations from Equations \ref{blob_charge_measure_2} and \ref{higher_flux_on_ground_state} into Equation \ref{blob_charge_measure_1}, we have
		\begin{align}
		&T^{R,C}(m) \sum_{e \in [\tilde{e}]_{\rhd}} \alpha_e B^e(t) \ket{GS} \notag \\
		&= \frac{|R|}{|Z_{\rhd,r_C}|} \sum_{D \in (Z_{\rhd,r_C})_{cl}} \chi_R(D) \sum_{d \in D} \sum_{q \in Q_C} \sum_{e \in [\tilde{e}]_{\rhd}} \alpha_e \notag \\
		& B^e(\text{start}(t) - \text{ blob }0) B^{(g(m-t)^{-1}qd) \rhd r_C}(\text{blob }0- \text{end}(t)) \notag\\
		& \delta(q \rhd r_C, \hat{g}(m-t) \rhd e) \ket{GS}, \label{blob_charge_measure_3}
		\end{align}
		where we used the Kronecker delta to rewrite the label of the second blob ribbon operator in terms of $r_C$. But then $d$ is an element of $Z_{\rhd,r_C}$, so $d \rhd r_C =r_C$. This means that $(\hat{g}(m-t)^{-1}qd) \rhd r_C=(\hat{g}(m-t)^{-1}q) \rhd r_C$. Then the Kronecker delta enforces that $q \rhd r_C= \hat{g}(m-t) \rhd e,$ so that
		\begin{align*}
		(\hat{g}(m-t)^{-1}q) \rhd r_C &= \hat{g}(m-t)^{-1} \rhd (q \rhd r_C)\\
		&=\hat{g}(m-t)^{-1} \rhd (\hat{g}(m-t) \rhd e)\\
		&=e.
		\end{align*}
		
		Substituting this into Equation \ref{blob_charge_measure_3}, we see that the result of our measurement is
		\begin{align*}
		T^{R,C}&(m) \sum_{e \in [\tilde{e}]_{\rhd}} \alpha_e B^e(t) \ket{GS}\\
		&=\frac{|R|}{|Z_{\rhd,r_C}|} \sum_{D \in (Z_{\rhd,r_C})_{cl}} \chi_R(D) \sum_{d \in D} \sum_{q \in Q_C} \sum_{e \in [\tilde{e}]_{\rhd}} \alpha_e\\
		& \hspace{0.5cm} B^e(\text{start}(t) - \text{ blob }0) B^{e}(\text{blob }0- \text{end}(t)) \\
		& \hspace{0.5cm} \delta(q \rhd r_C, \hat{g}(m-t) \rhd e) \ket{GS}.
		\end{align*}
		
	 We see that the labels of the two sections of the blob ribbon operator are the same, and so we can recombine them into a single ribbon operator applied on the original ribbon, $t$. We then have
		\begin{align*}
		T^{R,C}&(m) \sum_{e \in [\tilde{e}]_{\rhd}} \alpha_e B^e(t) \ket{GS}\\
		&=\frac{|R|}{|Z_{\rhd,r_C}|} \sum_{D \in (Z_{\rhd,r_C})_{cl}} \chi_R(D) \sum_{d \in D} \sum_{q \in Q_C} \sum_{e \in [\tilde{e}]_{\rhd}} \alpha_e\\
		& \hspace{0.5cm} B^e(t) \delta(q \rhd r_C, \hat{g}(m-t) \rhd e) \ket{GS}.
		\end{align*}
		
		Next, note that if the Kronecker delta 
		$$\delta(q \rhd r_C, \hat{g}(m-t) \rhd e)$$
		 is satisfied, then $e$ and $r_C$ must be in the same $\rhd$-class (they are related by the action of $q^{-1}\hat{g}(m-t)$), and so we can extract $\delta([\tilde{e}]_{\rhd},C)$ from the Kronecker delta to obtain
		 \begin{align*}
		 T^{R,C}&(m) \sum_{e \in [\tilde{e}]_{\rhd}} \alpha_e B^e(t) \ket{GS}\\
		 &=\frac{|R|}{|Z_{\rhd,r_C}|} \sum_{D \in (Z_{\rhd,r_C})_{cl}} \chi_R(D) \sum_{d \in D} \sum_{q \in Q_C} \sum_{e \in [\tilde{e}]_{\rhd}} \alpha_e\\
		 & \hspace{0.5cm} B^e(t) \delta([\tilde{e}]_{\rhd},C)\delta(q \rhd r_C, \hat{g}(m-t) \rhd e) \ket{GS}.
		 \end{align*}
		 In addition, the dummy variable $q$ now only appears in the expression 
		 $$\delta(q \rhd r_C, \hat{g}(m-t) \rhd e).$$ 
		 However, provided that $r_C$ and $e$ are in the same $\rhd$-class (as enforced by $\delta([\tilde{e}]_{\rhd},C)$), there is precisely one value of $q \in Q_C$ that satisfies $q \rhd r_C= \hat{g}(m-t) \rhd e$, and so we can remove the sum over $q$ along with the Kronecker delta, to obtain
		\begin{align*}
		T^{R,C}(m) \sum_{e \in [\tilde{e}]_{\rhd}}& \alpha_e B^e(t) \ket{GS}\\
		&=\delta([\tilde{e}]_{\rhd},C) \frac{|R|}{|Z_{\rhd,r_C}|} \sum_{d \in Z_{\rhd,r_C}} \chi_R(d)\\
		& \hspace{0.5cm} \sum_{e \in [\tilde{e}]_{\rhd}} \alpha_e B^e(t) \ket{GS}.
		\end{align*}
		
		Next, we wish to use orthogonality of characters to find the irrep $R$. To do so, we note that the character of the trivial irrep is one for all elements, and so 
		$$\sum_{d \in Z_{\rhd,r_C}} \chi_R(d)=\sum_{d \in Z_{\rhd,r_C}} \chi_R(d) \chi_{1_{\text{Rep}}}(d^{-1}).$$
		The index $d$ appears only in this expression, and so we can use orthogonality of characters to write
		\begin{align}
		T^{R,C}&(m) \sum_{e \in [\tilde{e}]_{\rhd}} \alpha_e B^e(t) \ket{GS} \notag\\
		&= \delta([\tilde{e}]_{\rhd},C) \frac{|R|}{|Z_{\rhd,r_C}|} \big(\sum_{d \in Z_{\rhd,r_C}} \chi_R(d) \chi_{1_{\text{Rep}}}(d^{-1})\big) \notag\\
		& \hspace{0.5cm} \sum_{e \in [\tilde{e}]_{\rhd}} \alpha_e B^e(t) \ket{GS} \notag\\
		&= \delta([\tilde{e}]_{\rhd},C) \frac{|R|}{|Z_{\rhd,r_C}|} \big(\delta(R,1_{\text{Rep}}) |Z_{\rhd,r_C}|\big) \notag \\
		& \hspace{0.5cm} \sum_{e \in [\tilde{e}]_{\rhd}} \alpha_e B^e(t) \ket{GS} \notag \\
		&= \delta([\tilde{e}]_{\rhd},C) \delta(R,1_{\text{Rep}}) \sum_{e \in [\tilde{e}]_{\rhd}} \alpha_e B^e(t) \ket{GS}.
		\end{align}
		This expression is just our original blob ribbon operator acting on the ground state, multiplied by $\delta([\tilde{e}]_{\rhd},C) \delta(R,1_{\text{Rep}})$. This indicates that our blob excitation has charge $([\tilde{e}]_{\rhd},1_{\text{Rep}})$. Note that because our measurement operator only runs over classes $C$ in the kernel of $\partial$, if the blob ribbon operator is confined we will always get zero when we act with our measurement operator. We also note that the representation $R$ is always the trivial representation, regardless of which set of coefficients $\alpha_e$ we have. This means that, as we stated earlier, even if the coefficients $\alpha_e$ are such that the blob ribbon operator excites the start-point vertex, this is not reflected in the charge of the excitation. The idea that the extra vertex excitation on an object may not correspond to an additional charge is something that is familiar from Kitaev's Quantum Double model in 2+1d \cite{Kitaev2003, Komar2017}.

\begin{figure}[h]
			\begin{center}
				\includegraphics[width=\linewidth]{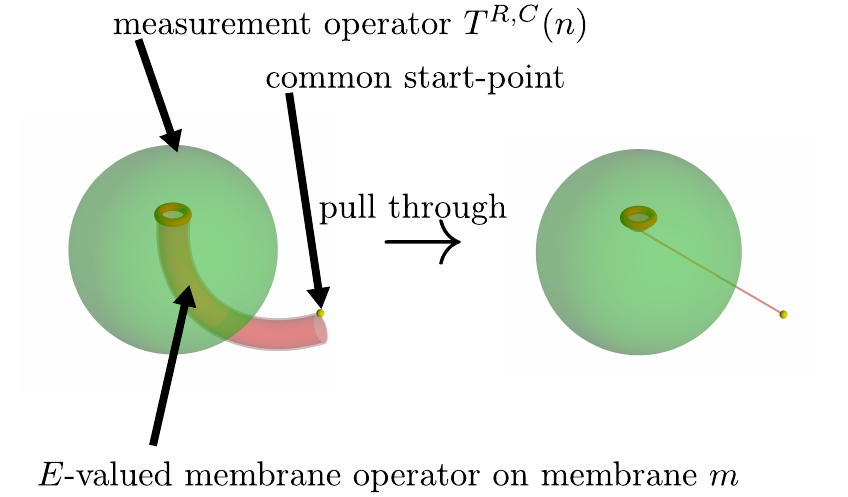}
		
				\caption{We measure the spherical charge of an $E$-valued loop. To simplify the calculation, we deform the $E$-valued membrane to pull it inside the measurement operator, while keeping the start-point outside.}
				\label{sphere_charge_of_loop}
			\end{center}
		\end{figure}

		We previously mentioned that the loop-like excitations of this model may also carry a point-like topological charge that can be measured by the spherical measurement operators. As an example, consider the $E$-valued loop excitations (we also examine the higher-flux excitations, but this is left to Section \ref{Section_point_like_charge_higher_flux} in the Supplemental Material due to the increased mathematical complexity of the calculation). We wish to measure the topological charge of such a loop excitation using our spherical measurement operator, as shown in Figure \ref{sphere_charge_of_loop}. Note that if the start-point were inside the measurement sphere, the entire membrane operator would be wholly within the sphere (or could be deformed to be within the sphere), so the membrane operator would commute with any measurement operator applied on that sphere. Therefore, the measurement operator would just measure the charge of the ground state, i.e., the vacuum charge. This means that the combined point-like charge of the start-point and the loop excitation is trivial, and so any point-like charge carried by the loop must be balanced by a charge belonging to the start-point.

		In order to find the charge of the loop-like excitation, we want to calculate
		$$T^{R,C}(n) \sum_{e \in E} a_e \delta(e, \hat{e}(m)) \ket{GS}.$$
		To do so, we must first evaluate
		$$C^{h}_T(n) L^{e_n}(n) \delta(\hat{e}(m),e) \ket{GS},$$
		where $h$ and $e_n$ are stand-ins for any label that can appear for the individual operators in our measurement operators. Firstly, we note that $L^{e_n}(n)= \delta(\hat{e}(n),e_n)$ commutes with $\delta(\hat{e}(m),e)$, because both are diagonal in the configuration basis (the basis where each edge is labeled by an element of $G$ and each plaquette is labeled by an element of $E$). On the other hand $\delta(\hat{e}(m),e)$ does not commute with the magnetic membrane operator $C^h_T(n)$. We can see this by writing the surface element $\hat{e}(m)$ in terms of the constituent plaquettes, as
		$$\hat{e}(m) = \prod_{p \in m} \hat{g}(s.p(m)-v_0(p)) \rhd \hat{e}_p,$$
		where $e_p$ is the label of plaquette $p$ and we have assumed each plaquette aligns with $m$ (otherwise we must replace the plaquette label with the inverse). We see that this depends on the group element associated to the path $(s.p(m)-v_0(p))$ from the start-point of the membrane to the base-point of the plaquette. This path passes through the membrane $n$ and so is affected by the magnetic membrane operator. As we prove in Section \ref{Section_electric_magnetic_braiding_3D_tri_trivial} of the Supplemental Material (see Equation \ref{Magnetic_electric_3D_braid_reverse}), such a path element satisfies the commutation relation
		\begin{align*}
		\hat{g}&(s.p(m)-v_0(p)) C^h_T(n)\\
		& = C^h_T(n)\hat{g}(s.p(m)-s.p(n))h^{-1}\hat{g}(s.p(m)-s.p(n))^{-1}\\
		& \hspace{0.5cm} \hat{g}(s.p(m)-v_0(p)).
		\end{align*}
		Defining 
		$$h_{[m-n]}^{\phantom {-1}}=\hat{g}(s.p(m)-s.p(n))h\hat{g}(s.p(m)-s.p(n))^{-1},$$
		this leads to the surface label satisfying the commutation relation
		\begin{align*}
		&C^h_T(n) \hat{e}(m) = h_{[m-n]}^{\phantom {-1}} \rhd \hat{e}(m) C^h_T(n),
		\end{align*}
		so that
		\begin{align*}
		C^h_T(n) \delta(\hat{e}(m),e)&= \delta(h_{[m-n]}^{\phantom {-1}} \rhd \hat{e}(m),e) C^h_T(n)\\
		&=\delta( \hat{e}(m),h_{[m-n]}^{-1} \rhd e) C^h_T(n).
		\end{align*}
		
		Then if we take the start-points of $m$ and $n$ to be the same (which has no effect on the result, because the start-point of the measurement operator can be changed without affecting the measurement operator), this becomes
		\begin{align*}
		C^{h}(n) L^{e_n}(n) &\delta(\hat{e}(m),e) \ket{GS}\\
		&= \delta(\hat{e}(m),h^{-1} \rhd e) C^{h} L^{e_n}(n) \ket{GS} \\
		&= \delta(\hat{e}(m),h^{-1} \rhd e) \delta(e_n,1_E) \ket{GS},
		\end{align*}
		where in the last line we used the fact that the contractible closed surface $n$ must have trivial label in the ground state, while the magnetic membrane operator applied on $n$ acts trivially on the ground state (again, because $n$ is closed and contractible). We can then use this result to evaluate the action of the measurement operator, to obtain
		\begin{align*}
		T^{R,C}&(n) \sum_{e \in E} a_e \delta(e, \hat{e}(m)) \ket{GS} \\
		&= \frac{|R|}{|Z_{\rhd,r_C}|} \sum_{D \in (Z_{\rhd,r_C})_{cl}} \chi_R(D) \sum_{d \in D} \sum_{q \in Q_C}\\
		& \hspace{1cm} C^{qdq^{-1}}(n)L^{q \rhd r_C}(n) \sum_{e \in E} a_e \delta(e, \hat{e}(m)) \ket{GS}\\
		&= \frac{|R|}{|Z_{\rhd,r_C}|} \sum_{D \in(Z_{\rhd,r_C})_{cl}} \chi_R(D) \sum_{d \in D} \sum_{q \in Q_C}\sum_{e \in E} a_e\\
		& \hspace{1cm} \delta(qd^{-1}q^{-1} \rhd e, \hat{e}(m)) \delta(e_n,1_E) \ket{GS}.
		\end{align*}
		
		Then $\delta(e_n,1_E)$ enforces that $C=\set{1_E}$ and so, just as with the calculation for the charge of the electric excitation, the groups involved in the projector simplify greatly. $Z_{\rhd,r_C}$, the group of elements in $G$ which stabilise $r_C$, becomes the whole group when $r_C=1_E$. This also means that the sum over representatives $q \in Q_C$ becomes trivial, with $q=1_G$ being the only element. This gives us
		\begin{align}
		T^{R,C}(n) &\sum_{e \in E} a_e \delta(e, \hat{e}(m)) \ket{GS} \notag\\
		&=\delta(C,\set{1_E}) \frac{|R|}{|G|} \sum_{d \in G} \chi_R (d) \sum_{e \in E} a_e \notag\\ 
		&\hspace{1cm}\delta(d^{-1} \rhd e, \hat{e}(m)) \ket{GS}\notag\\
		&=\delta(C,\set{1_E}) \frac{|R|}{|G|} \sum_{d \in G} \chi_R (d) \sum_{e'=d^{-1} \rhd e} a_{d \rhd e'}\notag\\
		&\hspace{1cm} \delta(e', \hat{e}(m))\ket{GS}\notag\\
		&= \delta(C,\set{1_E}) \frac{|R|}{|G|} \sum_{e' \in E} (\sum_{d \in G} \chi_R (d) a_{d \rhd e'}) \notag\\
		& \hspace{1cm}\delta(e', \hat{e}(m))\ket{GS}.
		\end{align}
		
		Examining the term in brackets, $(\sum_{d \in G} \chi_R (d) a_{d \rhd e'})$, we see that treating $a_{d \rhd e'}$ as a coefficient for $d$ will result in this coefficient being decomposed into irreps of $G$, describing the way in which the group $G$ acts on the coefficients by the $\rhd$ action.  Only the contribution from $\overline{R}$ survives, due to orthogonality of characters, meaning that the measurement only gives a non-zero result if the $a$ coefficients contain the irrep $\overline{R}$. Therefore, the $E$-valued loop can have a non-trivial spherical topological charge labeled by a representation of $G$, depending on the choice of $a$ coefficients. If the $a$ coefficients are invariant under the $\rhd$ action, so that $a_{d \rhd e'}=a_{e'}$ for all $d \in G$ and $e \in E$, then the term in brackets is zero unless $R$ is the trivial irrep, because $a$ must only contain the trivial irrep of $G$. We note that this is the same condition for the start-point to be unexcited, as discussed in Section \ref{Section_3D_Loop_Tri_Non_Trivial}, and so if the start-point is not excited then there is no point-like charge for the loop (this is to be expected, because if the start-point is not excited then it should not carry a charge to be balanced by the loop). However, if $a_{d \rhd e'} \neq a_{e'}$ in general, then there will be some contribution from a non-trivial irrep. In particular, if $\sum_{d \in G} a_{d \rhd e'} =0$ for all $e' \in E$ (which is the condition for the start-point to definitely be excited, as proven in Section S-I C of the Supplemental Material for Ref. \cite{HuxfordPaper2}) then the contribution from the trivial irrep is proportional to 
		$$\sum_{d \in G} \chi_{1_\text{Rep}} (d) a_{d \rhd e'}= \sum_{d \in G} a_{d \rhd e'} =0$$
		and so there is no contribution from the trivial charge measurement operator. This means that the point-like charge is definitely non-trivial if the start-point is excited.
		
		\subsection{Topological charge within a torus}
		\label{Section_Torus_Charge}

		Now we consider measuring the topological charge using a toroidal surface. To do this, we first choose such a surface to measure on. Then we project onto the space where the surface has no excitations on it, so that we only measure the charge if no objects intersect the surface itself. This is to avoid the case where a loop excitation is only partially inside the measurement surface, because then we cannot unambiguously define the charge within the torus, as explained earlier in Section \ref{Section_3D_Topological_Sectors}. A torus will allow us to measure both loop-like and point-like charge. One important thing to note is that the excitations that we measure need not lie inside the torus itself. Indeed we measure the loop-like charge of loop excitations that link with one of the cycles of the torus. For the meridian of the torus, those excitations will live inside the torus. However, the excitations that link with the longitude will be outside of the torus. This means that the torus surface can actually measure link-like excitations, made from loops outside the torus linking with those inside, rather than just loop-like excitations.

\begin{figure}[h]
			\begin{center}
				\includegraphics[width=\linewidth]{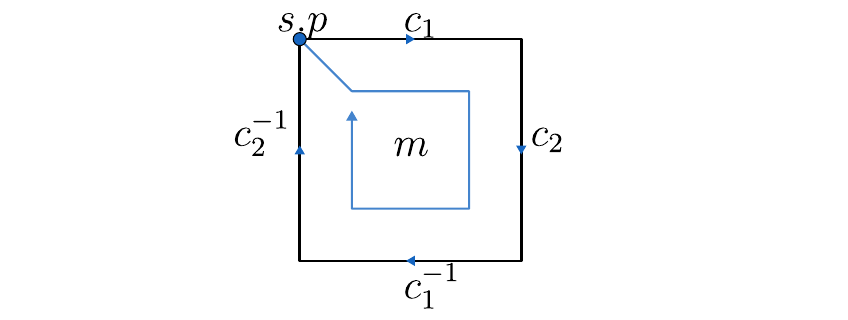}
				
				\caption{The surface of the torus is conveniently represented by a square with periodic boundary conditions. The edges of this square (which are glued due to the periodic boundary conditions) are referred to as the seams of the torus. We apply electric ribbon operators along these seams to measure the non-contractible cycles of the torus and apply an $E$-valued membrane operator on the surface. We will also apply blob ribbon operators around the two cycles and a magnetic membrane operator over the surface. The edges cut by the dual membrane of the magnetic membrane point outwards from the page.}
				\label{unfoldedtorus1}
			\end{center}
		\end{figure}

		The topological charges measured by the torus are more numerous and mathematically complicated to derive than those measured by the sphere. We therefore first consider the case where $\rhd$ is trivial (Case 1 from Table \ref{Table_Cases}) as an introduction. As illustrated in Figure \ref{unfoldedtorus1}, we represent the torus surface as a rectangle with opposite sides identified. These sides are then one particular choice for the two independent cycles of the torus. We will choose to apply any membrane operators on this rectangle with the boundary at the cycles, before closing the rectangle by gluing the opposite edges. This leaves ``seams" at the two cycles, which may have special properties because the action of the membrane operators on either side of the seam may not match. We can understand this by imagining taking a membrane operator applied on a rectangle and folding it up to glue the opposite edges together. There is no guarantee that a membrane operator acts the same on opposite sides of the rectangle, and this disparity may remain when we glue the sides together. We will see that this leads to additional joining conditions required to prevent additional excitations being present at these seams.

		To find the measurement operators, we have to first project onto the case where the surface itself is not excited, then we see what degrees of freedom are left over. After projecting onto all of the plaquettes on the surface being flat, these two cycles of the torus are still left undetermined. We therefore apply two closed electric operators $\delta(\hat{g}(c_1), g_{c_1})$ and $\delta(\hat{g}(c_2), g_{c_2})$, where $c_1$ and $c_2$ are the two cycles of the torus. We also apply a closed membrane operator $\delta(\hat{e}(m),e_m)$ on the torus, with the glued boundary of this torus being $c_1c_2c_1^{-1}c_2^{-1}$.

		 Requiring fake-flatness on the torus (this requirement follows from the plaquette terms) leads to the following constraint on the surface label $e_m$ of the torus and the labels $g_{c_1}$ and $g_{c_2}$ of the two cycles:
		$$\partial(e_m) g_{c_1} g_{c_2}g_{c_1}^{-1} g_{c_2}^{-1}=1_G.$$
		Together with the other conditions that we will discuss in this section, we prove this constraint in Section \ref{Section_3D_Topological_Charge_Torus_Tri_trivial} in the Supplemental Material. We can use the fact that $\rhd$ is trivial to rewrite this constraint in a simpler way. When $\rhd$ is trivial, conjugating an element $\partial(e) \in \partial(E)$ by any element $g \in G$ is trivial, because $g\partial(e)g^{-1}=\partial(g \rhd e)=\partial(e)$ for all $g \in G$ and $e \in E$. Then defining $[g,h]=ghg^{-1}h^{-1}$, we can write the above constraint in various ways. For example, we have
		\begin{align*}
		\partial(e_m)& g_{c_1} g_{c_2}g_{c_1}^{-1} g_{c_2}^{-1}=1_G\\
		&\implies \partial(e_m) =g_{c_2}g_{c_1}g_{c_2}^{-1}g_{c_1}^{-1}=[g_{c_2},g_{c_1}]\\
		&\implies (g_{c_2}g_{c_1})^{-1}\partial(e_m)g_{c_2}g_{c_1} =g_{c_2}^{-1}g_{c_1}^{-1}g_{c_2}g_{c_1}\\
		&\implies \partial((g_{c_2}g_{c_1})^{-1} \rhd e_m)=[g_{c_2}^{-1},g_{c_1}^{-1}]\\
		&\implies \partial(e_m)=[g_{c_2}^{-1},g_{c_1}^{-1}],
		\end{align*}
		where in the fourth line we used one of the Peiffer conditions (Equation \ref{Equation_Peiffer_1} in Section \ref{Section_Recap_3d}) and in the last line we used the fact that $\rhd$ is trivial. We can also write the condition as 
		\begin{equation}
	    \partial(e_m)^{-1}=[g_{c_1}^{-1},g_{c_2}^{-1}].
		\end{equation}
		
		Next we apply our magnetic membrane operator $C^h(m)$. Because we already projected to the subspace where the torus satisfies fake-flatness, some of the details of the operator are arbitrary; in particular the set of paths on the direct membrane, which affect the action on the edges, can be freely chosen, as long as these paths do not cross the seams of the membrane (we take this convention because two choices of path that differ by a non-contractible cycle may give different results and this gives us a consistent way of choosing the paths).

		Finally, we apply blob ribbon operators around the cycles, so that our measurement operator so far is given by
		\begin{align*}
		B^{e_{c_1}}(c_1)B^{e_{c_2}}(c_2)&C^h(m)
		\delta(\hat{e}(m),e_m)\\
		& \delta(\hat{g}(c_1),g_{c_1}) \delta(\hat{g}(c_2),g_{c_2}).
		\end{align*}
		In principle we could put closed blob ribbon operators anywhere on the membrane, rather than just on the cycles $c_1$ and $c_2$. However, any blob ribbon operators with label in the kernel of $\partial$ can be freely deformed on the surface without affecting the action of the ribbon operators, because these operators are topological and there are no edge excitations on the surface (edge excitations in particular are relevant, because we deform blob ribbon operators by applying edge transforms, which are trivial when the edges are unexcited). This means that any such blob ribbon operator that does not wrap around a non-contractible cycle may be contracted into nothing, while an operator that does wrap around a non-contractible cycle on the torus may be deformed to wrap around the chosen cycles $c_1$ and $c_2$ (if the ribbon operator wraps both cycles, or wraps one multiple times, we split it into multiple ribbon operators on the cycles). This is not true for the other blob ribbon operators (those with label outside the kernel), because they are confined and so are not topological. Instead we find that this confinement leads to their position being fixed and their labels being restricted (as described in Section \ref{Section_3D_Topological_Charge_Torus_Tri_trivial} of the Supplemental Material), in order not to create any excitations. This is because the magnetic membrane operator may create plaquette excitations on the seams of the torus, but these excitations can be removed if the confined blob ribbon operators lie along the seam and have appropriate labels to cancel the effect of the magnetic membrane operator. We cannot place confined ribbon operators elsewhere (away from these seams) without producing new excitations. The appropriate labels for blob ribbon operators $B^{e_{c_1}}(c_1)$ and $B^{e_{c_2}}(c_2)$, applied around the cycles $c_1$ and $c_2$ respectively, satisfy
		\begin{align}
		\partial(e_{c_2})&=[g_{c_1},h]  \\ 
		\partial(e_{c_1})&=[h,g_{c_2}].
		\end{align}

		So far we have restricted the labels by requiring that our operator does not violate the plaquette conditions. However, we also need the combined operator to commute with the vertex and edge transforms on the surface so that the operator does not create vertex and edge excitations. This forces us to use linear combinations of operators with different labels. In particular, we show in Section \ref{Section_3D_Topological_Charge_Torus_Tri_trivial} of the Supplemental Material that we need an equal sum of the operators with sets of labels in certain equivalence classes. If two sets of labels must appear with equal coefficients in the linear combination, we denote this with an equivalence relation. We find the relations
		\begin{align}
		(g_{c_1},g_{c_2},h)& \sim (x^{-1}g_{c_1}x,x^{-1}g_{c_2}x,x^{-1}hx) \: \: \forall x \in G \\
		g_{c_1}& \sim \partial(e) g_{c_1} \: \: \forall e \in E\\
		g_{c_2}& \sim \partial(e') g_{c_2} \: \: \forall e' \in E\\
		h &\sim \partial(e'')h \: \: \forall e'' \in E.
		\end{align}
		These conditions show a striking resemblance to the relations that appear in the calculation of the ground state degeneracy of the 3-torus in Ref. \cite{Bullivant2017} and indeed they map perfectly onto them in the $\rhd$ trivial case. This indicates that the number of topological charges we can measure within a 2-torus is the same as the ground state degeneracy of the 3-torus in the $\rhd$-trivial case, as found more generally in Ref. \cite{Bullivant2020}.

		We can repeat this calculation for the special case (Case 2 from Table \ref{Table_Cases}) where we only enforce that $E$ is Abelian and $\partial \rightarrow$ centre($G$). Following the same argument as for the previous case (with full proofs given in Section \ref{Section_3D_Topological_Charge_Torus_Tri_nontrivial} in the Supplemental Material), we obtain the restrictions
		\begin{align}
		\partial(e_m)&=[g_{c_2},g_{c_1}];\\
		\partial(e_{c_2})&=[g_{c_1},h];\\
		\partial(e_{c_1})&=[h,g_{c_2}];\\
		1_E &=[h \rhd e_m^{-1}] \: e_m e_{c_1}^{-1} [g_{c_1}^{-1} \rhd e_{c_1}] e_{c_2}^{-1} [g_{c_2}^{-1} \rhd e_{c_2}],
		\end{align}
		together with the equivalence relations
		\begin{align}
		&((g_{c_1},g_{c_2},h),(e_{c_1},e_{c_2},e_m)) \notag \\
		& \sim (g(g_{c_1},g_{c_2},h)g^{-1},g \rhd(e_{c_1},e_{c_2},e_m))\\
		&(g_{c_1},e_{c_2},e_m) \notag \\
		& \sim (\partial(e)^{-1}g_{c_1},e_{c_2} [h \rhd e] \: e^{-1}, e_m e^{-1} [g_{c_2}^{-1} \rhd e])\\
		&(g_{c_2},e_{c_1},e_m) \notag \\
		& \sim (\partial(r)g_{c_2},e_{c_1} [h \rhd r] \: r^{-1}, e_m r^{-1} [g_{c_1}^{-1} \rhd r])\intertext{and}
		&(h, e_{c_1},e_{c_2}) \notag \\
		& \sim (\partial(e)h, e_{c_1} [g_{c_2}^{-1} \rhd e] \: e^{-1}, e_{c_2} [g_{c_1}^{-1}\rhd e^{-1}] \: e).
		\end{align}
		
		These restrictions can again be mapped onto the ground state conditions given in Ref. \cite{Bullivant2017}, as we demonstrate in Section \ref{Section_3D_Topological_Charge_Torus_Tri_nontrivial} in the Supplemental Material. This indicates that again there are the same number of ground states on the 3-torus as there are charges that can be measured by the 2-torus. We note that this relationship between the ground state degeneracy on the manifold $M \times S^1$ and the charge sectors measured by the surface $M$ (with $M$ in this case being the 2-torus $T^2= S^1 \times S^1$ and $M \times S^1$ being the 3-torus) is something we may expect for a TQFT \cite{Atiyah1988}.

		We can use these conditions for the measurement operators to produce a set of projection operators that span the space of allowed measurement operators, just as we did for the spherical topological charge. Each such projector then corresponds to a particular value of topological charge. We find that these projectors are labeled by certain mathematical objects that were used by Bullivant et al. \cite{Bullivant2020} when examining the ground states of the HLGT model. Specifically, each projector is labeled by a class $C$ of a particular space (to be described shortly) and an irrep $R$ of a particular group. To define these objects, we must follow some of the workings from Ref. \cite{Bullivant2020}. We note that some of our notation is slightly different from that paper, in order to match notation that we have previously used (and that was used in Ref. \cite{Bullivant2017}). To understand $C$, we must first define boundary $\mathcal{G}$-colourings. These are sets of three elements $(g_y,g_z,e_x)$, where $g_y,g_z \in G$ and $e_x \in E$. If this set satisfies $g_z=g_y^{-1} \partial(e_x^{-1})g_zg_y$, it is called a boundary $\mathcal{G}$-colouring \cite{Bullivant2020}. These sets are then divided into classes, by the equivalence relation \cite{Bullivant2020}
		\begin{align}
		(g_y,g_z,e_x) \sim (a^{-1}&\partial(b_2^{-1})g_ya, \: a^{-1}\partial(b_1^{-1})g_za, \notag\\
		& a^{-1} \rhd (b_1^{-1}(g_z \rhd b_2^{-1})e_x(g_y \rhd b_1) b_2)), \label{Equation_sim_relation_Bullivant}
		\end{align}
		for each $a \in G$ and $b_1,b_2 \in E$. Then the label $C$ of the projector is one of these classes. The elements in $C$ are denoted by $(c_{y,i},c_{z,i},d_{x,i})$ and $(c_{y,1},c_{z,1},d_{x,1})$ is called the representative element of the class ($i$ is an index that runs from 1 to the size $|C|$ of the class $C$) \cite{Bullivant2020}.

		Expressions similar to the right-hand side of Equation \ref{Equation_sim_relation_Bullivant} will appear frequently in this section, so we introduce some shorthand from Ref. \cite{Bullivant2020}, defining
		\begin{equation}
		g^{k;f}= k^{-1}\partial(f)^{-1}g k, \label{Equation_superscript_notation_1_main_text}
		\end{equation}
		where $g$ and $k$ are elements of $G$ and $f$ is an element of $E$. We also introduce the notation from Ref. \cite{Bullivant2020} that
		\begin{equation}
		e^{k,h_1,h_2;f_1,f_2}= k^{-1} \rhd \big(f_{1}^{-1} (h_2 \rhd f_{2})^{-1} e [h_{1} \rhd f_{1}] f_{2} ), \label{Equation_superscript_notation_2_main_text}
		\end{equation}
		where $k$, $h_1$ and $h_2$ are elements of $G$ and $f_1$ and $f_2$ are elements of $E$. Then using this notation, Equation \ref{Equation_sim_relation_Bullivant} can be written as
		\begin{align}
		(g_y,g_z,e_x) \sim (g_y^{a;b_2}, \: g_z^{a;b_1},e_x^{a, g_y,g_z; b_1,b_2} ). \label{Equation_sim_relation_Bullivant_2}
		\end{align}

		In addition to the boundary $\mathcal{G}$-colourings, Bullivant et al. introduce sets of three elements $(g_x,e_y,e_z)$, where $g_x \in G$ and $e_y,e_z \in E$, with these sets of elements being called ``bulk $\mathcal{G}$-colourings" \cite{Bullivant2020}. These colourings are also divided into classes, this time using an equivalence relation that depends on the boundary colouring. The equivalence relation is \cite{Bullivant2020}
		$$(g_x,e_y,e_z) \underset{g_y,g_z}{\sim} (\partial( \lambda) g_x,[g_z \rhd \lambda] e_y \lambda^{-1}, [g_y \rhd \lambda] e_z \lambda^{-1})$$
		for each $\lambda \in E$. The corresponding set of equivalence classes is denoted by $\mathfrak{B}_{g_y,g_z}$. Then for a class $\mathcal{E}_{g_y,g_z}$ in the set $\mathfrak{B}_{g_y,g_z}$, the elements in $\mathcal{E}_{g_y,g_z}$ are denoted by 
		$$(s_{x,i},f_{y,i},f_{z,i}),$$
		for $i=1,2,...,|\mathcal{E}_{g_y,g_z}|.$ The element $(s_{x,1},f_{y,1},f_{z,1})$ is called the representative element for this class. A subset of these classes form a group called the stabiliser group of the class $C$, $Z_C$ \cite{Bullivant2020}:
		\begin{align*}
		Z_C :=& \{\mathcal{E}_C \in\mathfrak{B}_C |(c_{y,1},c_{z,1},d_{x,1}) =\\
		& (c_{y,1}^{s_{x,1};f_{z,1}},c_{z,1}^{s_{x,1};f_{y,1}} d_{x,1}^{s_{x,1},c_{y,1},c_{z,1};f_{y,1},f_{z,1}})\},
		\end{align*}
		where $\mathfrak{B}_C = \mathfrak{B}_{c_{y,1},c_{z,1}}$ and the subscript $C$ in $\mathcal{E}_C$ is to remind us that $\mathcal{E}_C$ belongs to $\mathfrak{B}_C$ (and is interchangeable with the subscript ${c_{y,1},c_{z,1}}$). The product for this group is defined so that the product of two classes $\mathcal{E}_C$ and $\mathcal{E}'_C \in Z_C$, $\mathcal{E}_C \cdot \mathcal{E}'_C$, is the equivalence class in $\mathfrak{B}_C$ whose representative element is
		$$(s_{x,1} s_{x,1}', f_{y,1} (s_{x,1} \rhd f_{y,1}'),f_{z,1}(s_{x,1} \rhd f_{z,1}')).$$
		The label $R$ of a projector to definite topological charge is an irrep of this stabilizer group.

		 Using the objects that we have discussed so far, we can finally define our projectors. First we define our product of individual membrane and ribbon operators:
		\begin{align}
		Y&^{(h, \: g_{c_1}, \: g_{c_2}, \: e_m, \: e_{c_1}, \: e_{c_2}^{-1})}(m) \notag\\
		&=B^{e_{c_1}}(c_1)B^{e_{c_2}^{-1}}(c_2)C^h_T(m)\notag\\
		& \hspace{0.4cm} \delta(\hat{e}(m),e_m) \delta(\hat{g}(c_1),g_{c_1}^{-1}) \delta(\hat{g}(c_2),g_{c_2}^{-1}).
		\end{align}
		Then we take appropriate linear combinations to construct the projector labeled by $R$ and $C$:
		\begin{align*}
		P&^{R,C}(m)\\
		&= \sum_{i=1}^{|C|} \sum_{\substack{\mathcal{E}_{c_{y,i},c_{z,i}}\\ \in \mathfrak{B}_{c_{y,i},c_{z,i}}}} \sum_{m=1}^{|R|} \sum_{\lambda \in E} \delta(c_{y,i},c_{y,i}^{s_{x,1};f_{z,1}})\\ 
		&\delta(c_{z,i},c_{z,i}^{s_{x,1};f_{y,1}})\delta(d_{x,i},d_{x,i}^{s_{x,1},c_{y,i},c_{z,i};f_{y,1},f_{z,1}})\\
		&D^R_{m,m}([\mathcal{E}_C^{\text{stab.}}]_{i,i}) \\
		&Y^{(\partial(\lambda) s_{x,1},\:c_{y,i}, \: c_{z,i}, \:d_{x,i},\: [c_{z,i} \rhd \lambda] f_{y,1} \lambda^{-1}, \: [c_{y,i} \rhd \lambda] f_{z,1} \lambda^{-1} )}(m).
		\end{align*}
		In this expression, $[\mathcal{E}_C^{\text{stab.}}]_{i,i}$ is the class in $\mathfrak{B}_C$ with representative element
		\begin{align*}
		\big(p_{x,i}^{-1}s_{x,1}p_{x,i}, \: \: p_{x,i}^{-1}& \rhd (q_{z,i}^{-1} f_{z,1}(s_{x,1} \rhd q_{z,i})),\\
		& p_{x,i}^{-1} \rhd (q_{y,i}^{-1} f_{y,1} [s_{x,1} \rhd q_{y,i}])\big),
		\end{align*}
		where the $p$ and $q$ elements are defined as representatives which satisfy
		$$(c_{y,1},c_{z,1},d_{x,1})=(c_{y,i}^{p_{x,i};q_{z,i}},c_{z,i}^{p_{x,i};q_{y,i}},d_{x,i}^{p_{x,i},c_{y,i},c_{z,i};q_{y,i},q_{z,i}})$$
		and $(p_{x,1},q_{y,1},q_{z,1})=(1_G,1_E,1_E)$ (that is the $p$ and $q$ move us around in the class $C$ to get from element $i$ to the representative element labeled by 1). In Section \ref{Section_3D_Topological_Charge_Torus_Projectors} of the Supplemental Material we perform the lengthy algebraic task of proving that the operators $P^{R,C}(m)$ labeled by the objects $R$ and $C$ form an orthogonal and complete set of projectors, indicating that the topological charges that we can measure with the torus are appropriately labeled by a class $C$ of boundary $\mathcal{G}$-colourings and an irrep $R$ of the corresponding stabiliser group.

		\section{Conclusion}
		\label{Section_conclusion_3d}
		In this work, we have discussed in detail the features of the higher lattice gauge theory model in 3+1d. We started by constructing the ribbon and membrane operators which create the simple excitations of the model. We found that there were two categories of excitation, those best labeled by objects related to the group $G$ and those best labeled by objects related to the other group, $E$. The former type of excitation, consisting of point-like electric excitations and loop-like magnetic flux tubes, are analogous to the excitations we expect from ordinary lattice gauge theory. The other type, consisting of point-like 2-gauge fluxes and loop-like 2-gauge charges, are related to properties of the surfaces of the lattice, instead of paths. We then considered the braiding properties of these excitations. When the map $\rhd$, defined as part of the crossed module, is trivial, these two types of excitations form separate sectors that do not have non-trivial braiding between them (only within each sector). However, when $\rhd$ is non-trivial (although we had to restrict to the case where $E$ is Abelian and $\partial$ maps to the centre of $G$) some magnetic excitations acquire a 2-flux and can braid non-trivially with all other types of excitation. This is reflected in the fact the membrane operator that produces the magnetic excitation must be modified to depend on the surface elements of the lattice.

		Another feature that we looked at was the condensation of certain excitations, and the accompanying confinement of others. We found that this was controlled by the map $\partial$, with no condensation or confinement when $\partial$ maps only to the identity element (at least in the case where $E$ is Abelian). By altering $\partial$ while keeping the groups fixed, we introduce condensation for some of the loop-like (magnetic and $E$-valued loop) excitations while causing some of the point-like (electric and blob) excitations to become confined. This can be thought of as a condensation-confinement transition, where the confined excitations are those that had non-trivial braiding with the condensed excitations. We also looked at the topological charge carried by the excitations, by constructing projectors to definite topological charge. The available charges depend on the surface of measurement, and we constructed the projectors for a spherical and toroidal surface. Similar to results found in Ref. \cite{Bullivant2020}, we found that the charges measured by the 2-torus surface matched the ground state degeneracy on the 3-torus. We saw that these 2-torus surfaces generally measured links, rather than simple loop-like excitations, suggesting that the number of inequivalent link-like excitations is equal to the ground state degeneracy.

		In Ref. \cite{HuxfordPaper1}, we already mentioned several potential avenues of interest for further study based on this work. Rather than repeat ourselves here, we would like to discuss one of these directions further. In this paper, we gave braiding relations in terms of the simple excitations produced by the membrane and ribbon operators. However, it would be useful to obtain the braiding relations and other properties in terms of the topological charge. To do so, it would first be necessary to find the fusion rules for the various topological charges. Because the topological charge depends on the measurement surface, we would need additional machinery to describe how different charges can fuse in a simple way. Once this has been done, we can consider a braiding process where the individual excitations are projected onto states of definite charge, and the overall system is similarly projected to definite total charge, in order to find the braiding relations satisfied by the charges (analogous to the approach used for 2+1d theories). In addition to better understanding this particular model, creating this machinery would give us a structure with which to study different models for topological phases in 3+1d and understand what properties we expect. We note that, as mentioned in Ref. \cite{HuxfordPaper1}, from some preliminary calculations we can see that different torus charges can only fuse if they have the same value of a certain ``threading flux", meaning that the quantum numbers passing through the two loops must be the same (for example if they are linked to the same excitation). When this threading flux is non-trivial, we are considering the fusion of two loop-like excitations while both are linked to another excitation. This means that the braiding of these charges would correspond to so-called ``necklace braiding" \cite{Bullivant2020a} (or three-loop braiding \cite{Wang2014}), meaning the braiding of two loops while linked to another. For general models this braiding can give different results compared to the usual two-loop braiding \cite{Wang2014, Jiang2014}. Therefore, having a structure in which to consider this process for generic 3+1d topological models would be most useful.
		
		Related to this idea, it would also be useful to understand better where the higher lattice gauge theory models fit into the wider landscape of 3+1d topological phases. It is conjectured that all 3+1d bosonic topological phases with bosonic point-like particles can be realised by Dijkgraaf-Witten theories \cite{Lan2018}. This implies that the higher lattice gauge theories should also be equivalent to lattice gauge theory, albeit with some additional non-topological content in the form of the condensed and confined excitations. It would be interesting to explicitly demonstrate this equivalence (or else disprove it), whether through a direct mapping or by considering the topological charges and other data as we described above and showing that they match between the two models. We have made some progress in this direction, but leave the results for future study.

\begin{acknowledgments}
	The authors would once again like to thank Jo\~{a}o Faria Martins and Alex Bullivant for helpful discussions about the higher lattice gauge theory model. We are also grateful to Paul Fendley for advice on the preparation of this series of papers. We acknowledge support from EPSRC Grant EP/S020527/1. Statement of compliance with EPSRC policy framework on research data: This publication is theoretical work that does not require supporting research data.
	
\end{acknowledgments}

\bibliography{referencesRevised}{}

	\onecolumngrid
	
	\newpage
	\begin{center}
	\textbf{\LARGE Supplemental Material}
	\end{center}
	
	\setcounter{equation}{0}
	\setcounter{figure}{0}
	\setcounter{table}{0}
	\makeatletter
	\renewcommand{\theequation}{S\arabic{equation}}
	\renewcommand{\thefigure}{S\arabic{figure}}
	\setcounter{section}{0}
	\renewcommand{\thesection}{S-\Roman{section}}
	
	\section{The energy of ribbon and membrane operators}
	\label{Section_ribbon_membrane_energy_commutation}
	In this section, we will demonstrate the commutation relations of the ribbon and membrane operators with the energy terms in the 3+1d case, building on our results for the 2+1d case in Ref. \cite{HuxfordPaper2} (see Section S-I in the Supplemental Material of that paper).
	
	\subsection{Electric ribbon operators and $E$-valued membrane operators}
	The commutation relations for the electric ribbon operators and $E$-valued membrane operators with the energy terms are the same in 3+1d as in 2+1d (in our proofs in Ref. \cite{HuxfordPaper2} the dimension did not affect any of our workings), except that in 3+1d we also have blob energy terms. These blob terms are diagonal in the configuration basis (where all edges and plaquettes are labelled by group elements of $G$ and $E$ respectively), as are the electric ribbon and $E$-valued membrane operators. Therefore, the electric ribbon operators and $E$-valued membrane operators commute with the blob energy terms.

	\subsection{Blob ribbon operators}
	\subsubsection{$\rhd$ trivial case}
	\label{Section_Blob_ribbon_proof_tri_trivial}
	Next we will demonstrate the commutation relations of the blob ribbon operators with the energy terms. We will first consider the case where $\rhd$ is trivial (Case 1 from Table \ref{Table_Cases} in Section \ref{Section_Recap_3d} of the main text).

	As explained in Section \ref{Section_3D_Blob_Excitation_Tri_Trivial} of the main text, the blob ribbon operator acts along a string that passes from blob to blob, cutting through plaquettes along the way. For a blob ribbon operator labelled by the group element $e \in E$, each plaquette passed through (pierced) by the ribbon is acted upon by pre-multiplication by $e$ or post-multiplication by $e^{-1}$, depending on the orientation of the plaquette. If the circulation of the plaquette is obtained by applying the right hand rule to the direction of the ribbon then we post-multiply by $e^{-1}$, but if it is obtained by using the anti-right hand rule we instead pre-multiply by $e$, as is illustrated by the example given in Figure \ref{Effect_blob_ribbon_tri_trivial} in Section \ref{Section_3D_Blob_Excitation_Tri_Trivial} of the main text. Because $E$ is Abelian when $\rhd$ is trivial, whether we pre-multiply or post-multiply is irrelevant, it only matters whether $e$ or $e^{-1}$ is used. Nonetheless, we use this language to match the other ribbon operators.

	We want to consider the commutation relations of the blob ribbon operator with the energy terms. First we consider the blob energy terms. Consider a blob of arbitrary shape in the middle of the ribbon, i.e., not one of the blobs that the ribbon originates or terminates in. Every time the ribbon enters the blob it must also exit the blob, since the ribbon does not originate or terminate in this blob. We therefore want to consider the combined effect of the blob ribbon entering and exiting the blob on the blob energy term. The blob energy term measures the total surface element of the boundary of the blob and checks that this element is the identity. The surface element of the blob, in the $\rhd$ trivial case, is a product of the individual plaquette elements on the surface, with an inverse if the orientation of the plaquette is opposite to the outwards normal of the blob (and no inverse if the orientation matches the outwards normal). That is, the 2-holonomy $H_2(B)$ of a blob $B$ is
	$$H_2(B)=\prod_{p \in \text{Bd}(B)} e_p^{\sigma_p},$$
	where Bd$(B)$ is the boundary surface of the blob and the $p$ in this surface are the constituent plaquettes that make up the boundary. $\sigma_p$ is a variable that is 1 or $-1$, depending on the orientation of the plaquette.

\begin{figure}[h]
		\begin{center}
			\begin{overpic}[width=0.7\linewidth]{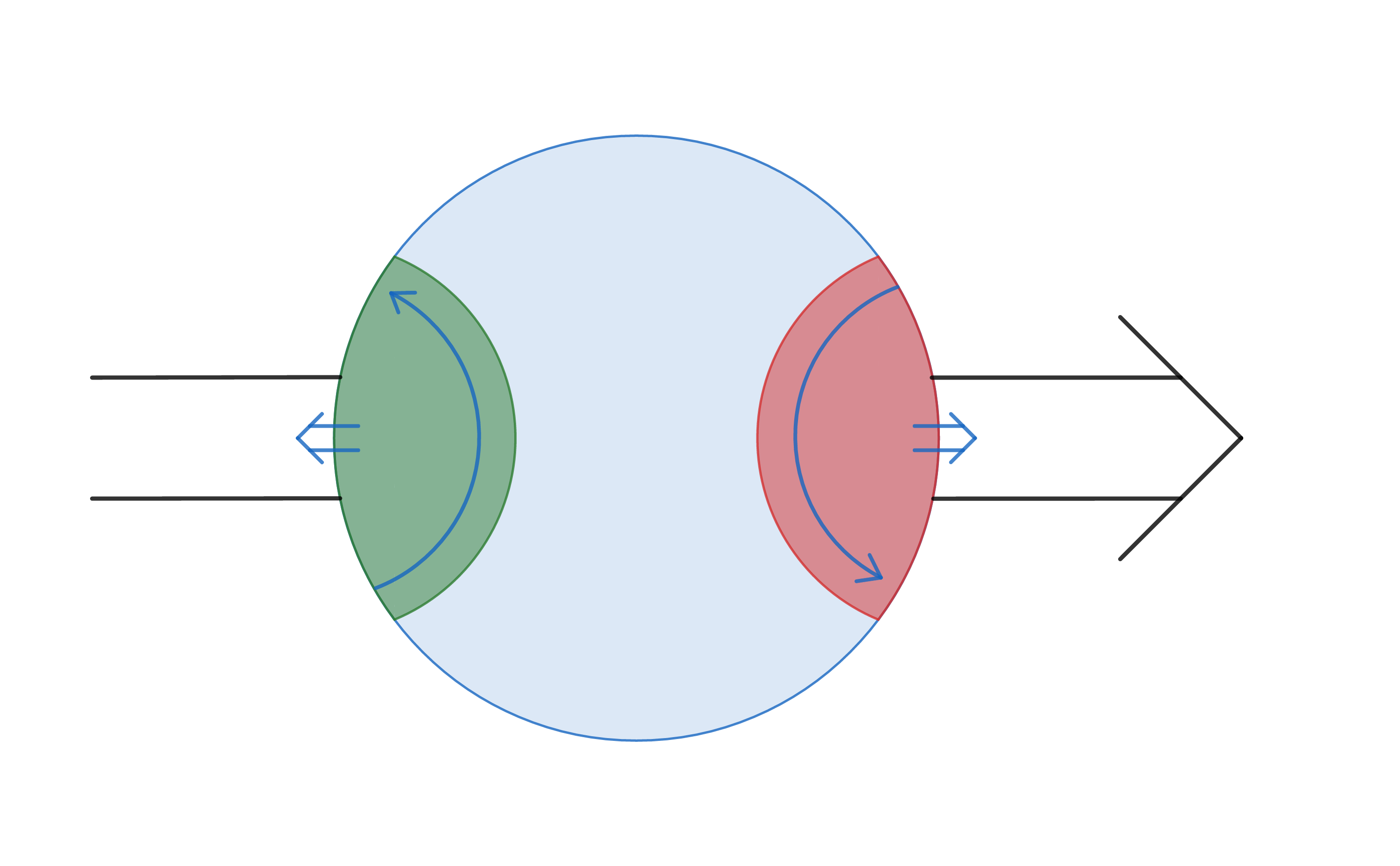}
				\put(40,4){blob $B$}
				\put(68,42){path of blob ribbon operator}
				\put(11,20){plaquette 1}
				\put(66,20){plaquette 2}
			\end{overpic}
			\caption{The blob ribbon operator enters a blob $B$ by piercing plaquette 1 (green) and exits it by piercing plaquette 2 (red). If the orientation of plaquette 1, determined by using the right-hand rule with its circulation, matches the outwards normal of the surface then its orientation is opposite to that of the ribbon. On the other hand if plaquette 2 has an orientation that matches the outwards normal to the blob then its orientation also matches the ribbon. This fact is responsible for the contributions to the blob energy term from the change to the two plaquettes cancelling.}
			\label{blob_plaquette_orientations}
		\end{center}
	\end{figure}

	The ribbon enters the blob $B$ by piercing a plaquette, which we will call plaquette 1, and exits by piercing another plaquette, plaquette 2, as shown in Figure \ref{blob_plaquette_orientations}. Suppose that these plaquettes have surface labels $e_1$ and $e_2$ respectively, before the action of the ribbon operator. If the orientation of plaquette 1 matches the outwards normal of the surface, then it does not match the orientation of the ribbon (whose direction is inwards to the surface here), as indicated in Figure \ref{blob_plaquette_orientations}. Therefore, in this case $e_1$ appears with no inverse in $H_2(B)$ and the plaquette label becomes $ee_1$ under the action of the blob ribbon operator. On the other hand, if the orientation of plaquette 1 does not match the outwards normal then its label appears as $e_1^{-1}$ in $H_2(B)$ but $e_1$ transforms to $e_1e^{-1}$ under the action of the blob ribbon operator. Therefore, the term $e_1^{-1}$ in $H_2(B)$ becomes $e e_1^{-1}$. We see that regardless of the orientation of the plaquette, the term corresponding to that plaquette in $H_2(B)$ gains a factor of $e$. Similarly we can consider plaquette 2. For this plaquette, the outwards normal of the blob matches the orientation of the ribbon. Therefore, if the orientation of plaquette 2 matches the outwards normal, it appears as $e_2$ in $H_2(B)$ and becomes $e_2 e^{-1}$ under the action of the ribbon operator. On the other hand if the orientation of the plaquette is in the opposite direction, it appears as $e_2^{-1}$ in the blob 2-holonomy and gains a factor of $e$ under the ribbon operator. Either way the term in $H_2(B)$ gains a factor of $e^{-1}$. This means that under the action of the blob ribbon operator $B^e(t)$ the blob 2-holonomy transforms as
	$$B^e(t):H_2(B) = B^e(t):\prod_{p \in \text{Bd}(B)} e_p^{\sigma_p} =...ee_1^{\pm 1}... e_2^{\pm 1} e^{-1}...= \prod_{p \in \text{Bd}(B)} e_p^{\sigma_p}=H_2(B),$$
	where we used the Abelian nature of $E$ to cancel the factors of $e$ and $e^{-1}$. From this we see that the blob 2-holonomy is left invariant by the action of the blob ribbon operator. So far we have only considered the case where each blob is passed through precisely once by the ribbon. However, if a blob is pierced more than twice, we can still pair up the times where the ribbon enters and exits the blob as above, and so $H_2(B)$ is still left invariant. We have therefore shown that the blob ribbon operator commutes with the blob energy terms, except perhaps for the blobs in which the ribbon originates and terminates.

	Next we will consider these final blob terms, corresponding to the blobs in which the ribbon originates or terminates. Whereas blobs in the middle of the ribbon are both entered and exited by the ribbon, the blob at the start of the ribbon is only exited by the ribbon (unless the ribbon doubles back and passes through this blob again in which case the pass-through does not affect the energy term, as discussed above). Therefore, the blob 2-holonomy only gains the transformation from the plaquette through which the ribbon exits the blob, which was called plaquette 2 in the previous considerations. This means that the 2-holonomy $H_2(B)$ gains a factor of $e^{-1}$ from the action of the blob ribbon operator. On the other hand, the blob in which the ribbon terminates is only entered by the ribbon and not exited by it, so we only gain the transformation corresponding to plaquette 1. Therefore, the two-holonomy transforms as $H_2(B) \rightarrow e H_2(B)$. This indicates that, when the blob ribbon operator acts on the ground state, it excites the two blobs at the origin and terminus of the ribbon, as long as $e$ is not the identity element of group $E$ (if $e$ is the identity element, then the blob ribbon operator is the identity operator).

	Next we consider the vertex energy terms, which are made of a sum of vertex transforms $A_v^g$. When $\rhd$ is trivial $A_v^g$ does not affect the plaquette labels or interact with them in any way; instead the vertex transform only acts on the edge labels. Conversely the blob ribbon operator does not interact with the edge labels in any way. Therefore, the vertex transforms commute with the blob ribbon operators.

	The next energy terms that we consider are the edge transforms. An edge transform $\mathcal{A}_i^f$ will act on plaquettes neighbouring the edge $i$ by pre-multiplying the plaquette label by $f$ or post-multiplying by $f^{-1}$ in the $\rhd$ trivial case. This means that the order in which we apply an edge transform and a blob ribbon operator could at most determine in which order a plaquette label is multiplied (e.g., $e_p \rightarrow e f e_p$ or $e_p \rightarrow fee_p$). However, because the group $E$ is Abelian the order of multiplication is irrelevant and so the edge transforms commute with the blob ribbon operators.

	The final energy terms to consider are the plaquette fake-flatness terms. Recall that the plaquette energy term measures the plaquette holonomy $H_1(p)=\partial(e_p)g(\text{boundary}(p))$ and checks if this is the identity element in $G$. The blob ribbon operator does not affect the path label $g(\text{boundary}(p))$ but it can affect the plaquette label $e_p$. If the ribbon pierces the plaquette then $e_p$ becomes $ee_p$ or $e_p e^{-1}$. Therefore, the action of the blob ribbon operator on the plaquette holonomy is
	$$B^e(t) H_1(p)= \partial(e)^{\pm 1} \partial(e_p)g(\text{boundary}(p)) = \partial(e)^{\pm 1} H_1(p).$$
	Therefore, if $\partial(e)=1_G$ the plaquette holonomy is invariant under the action of the blob ribbon operator, and so the plaquette is not excited by the ribbon operator. On the other hand, if $e$ is outside the kernel of $\partial$ then the plaquette holonomy is changed by the action of the blob ribbon operator and so the plaquette is excited by the action of the ribbon operator. In this latter case, every plaquette pierced by the ribbon is excited and so the blob excitations are confined.
	
	\subsubsection{Fake-flat case}
	\label{Section_Blob_Ribbon_Fake_Flat}
	Next we consider the case where our crossed module is general, but we restrict our Hilbert space to only include fake-flat configurations (Case 3 in Table \ref{Table_Cases} of the main text). In this case the blob ribbon operators must be modified, as we described in Section \ref{Section_3D_Blob_Fake_Flat} of the main text. Firstly, to avoid violating fake-flatness, we must restrict the label $e$ of the ribbon operator to satisfy $\partial(e)=1_G$; that is we only consider elements of $E$ in the kernel of $\partial$. Secondly, it is necessary to introduce a configuration dependent part to the action of the ribbon operator, so that the action depends on the states of the edges along the ribbon. Instead of simply multiplying the plaquette labels on the path by $e$ or the inverse, we must first specify a vertex as the start-point of the ribbon operator, which we label by $s.p$. Then the action on a plaquette $p$, with base-point $v_0(p)$, is to pre-multiply the plaquette label by $g(s.p-v_{0}(p))^{-1}\rhd e$ (or post-multiply by the inverse), where $s.p-v_0(p)$ is a path from the start-point of the ribbon to the base-point of $p$. An example of this action is shown in Figure \ref{effectbloboperator} in Section \ref{Section_3D_Blob_Fake_Flat} of the main text. Explicitly, for a blob ribbon operator $B^e(t)$ acting on a plaquette $p$ pierced by the ribbon, we have
	\begin{equation}
		B^e(t):e_p = \begin{cases} e_p [g(s.p-v_0(p))^{-1} \rhd e^{-1}] & \text{ if $p$ aligns with $t$} \\ [g(s.p-v_0(p))^{-1} \rhd e]e_p & \text{ if $p$ anti-aligns with $t$.} \end{cases}\label{Equation_blob_ribbon_fake_flat}
	\end{equation}

	The precise path chosen between the start-point and the base-point of each plaquette does not matter when we are considering fake-flat configurations. This is because, given two paths with the same start and end-points that can be smoothly deformed into one-another, the paths enclose a fake-flat surface and therefore have the same label up to some element $\partial(f)$ (although this is not the case if the paths together wrap some non-contractible cycle on the manifold). This means that given two paths $t_1$ and $t_2$ with the same start and end-points, which enclose a fake-flat surface, we have $g(t_1)= g(t_2)\partial(f)$ for some $f \in E$. If we chose $t_1$ as the path from the start-point of the ribbon operator to the base-point of the plaquette, then the plaquette label gains a factor of $g(t)^{-1} \rhd e$. On the other hand, if we chose $t_2$, then the plaquette label gains a factor of 
	$$g(t_2)^{-1} \rhd e = (\partial(f)g(t_1)^{-1}) \rhd e = f [g(t_1)^{-1} \rhd e] f^{-1},$$
	where the last equality follows from the second Peiffer condition (Equation \ref{Equation_Peiffer_2} in the main text). However, when the label $e$ of the blob ribbon operator is in the kernel of $\partial$ it is also in the centre of $E$ (and therefore so is $[g(t_1)^{-1} \rhd e]$). This also follows from the second Peiffer condition, which implies that, for any element $x \in E$, $x=1_G \rhd x= \partial(e)\rhd x = exe^{-1}$. Then the factor $g(t_2) \rhd e$ gained from choosing $t_2$ as the path from the start-point of the ribbon to the base-point of the plaquette is just
	$$g(t_2)^{-1} \rhd e =f [g(t_1)^{-1} \rhd e] f^{-1}= g(t_1)^{-1} \rhd e,$$
	which is the same result we would obtain from choosing $t_1$ instead. Therefore, it does not matter which of $g(t_1)$ and $g(t_2)$ is chosen for $g(s.p-v_{0}(p))^{-1}$, as long as the paths can be deformed into one-another. We note that this is only true because the element $e$ is taken to be in the kernel of $\partial$, and this restriction will be relevant throughout this section.

	Now we consider the commutation relations between the blob ribbon operator and the energy terms. We will first check the plaquette terms, which we enforce on the level of the Hilbert space rather than as an energetic constraint. Recall that the plaquette term for a plaquette $p$ is $\delta(1_G, H_1(p))$, where $H_1(p)= \partial(e_p)g(\text{boundary}(p))$ for plaquette label $e_p$. Under the action of the blob ribbon operator, $e_p$ becomes $(g(s.p-v_{0}(p))^{-1}\rhd e) \: e_p$ or $e_p \: (g(s.p-v_{0}(p))^{-1}\rhd e^{-1})$ (depending on the orientation of the plaquette). If $e$ is in the kernel of $\partial$, then so is $g \rhd e$ for all $g \in G$ and so $\partial\big([g(s.p-v_{0}(p))^{-1}\rhd e] \: e_p\big) = \partial\big(e_p \: [g(s.p-v_{0}(p))^{-1}\rhd e^{-1}]\big)=\partial(e_p)$. Therefore, the action of the blob ribbon operator preserves fake-flatness. On the other hand, if $e$ is not in the kernel of $\partial$ then $\partial(e_p)$ is not invariant under the action of the blob ribbon operator, and so fake-flatness is violated. This is why we must restrict $e$ to be in the kernel of $\partial$ when we consider the fake-flat case.

	Next we consider the other energy terms, which we do implement as ordinary energetic constraints. We first consider the vertex transforms. Unlike in the $\rhd$ trivial case, the vertex transforms can interact with the blob ribbon operator in a number of ways. Firstly, the vertex transforms can affect the path labels $g(s.p-v_{0}(p))$ which appear in the action of the blob ribbon operators. Secondly, the vertex transforms can directly affect the same plaquette labels on which the blob ribbon operator acts. Consider the action of the blob ribbon operator $B^e(t)$ on the plaquette (or plaquettes) pierced by the ribbon that has its base-point at a particular vertex $v_0(p)$. The action of the blob ribbon operator on this plaquette (or plaquettes) is to multiply the plaquette label by $g(s.p-v_{0}(p))^{-1}\rhd e$ or the inverse. The only vertex transforms which can affect the path label are the transforms at the start-point and at $v_0(p)$, because these are the two ends of the path. In addition, the only vertex transform that can affect the plaquette label directly is the transform at the base-point of the plaquette, $v_0(p)$. Therefore, when considering the action of the ribbon operator on the plaquette or plaquettes based at $v_0(p)$, the only vertex transforms which can fail to commute with this action are the vertex transforms at $s.p$ and at $v_0(p)$. We will first consider the transform at $v_0(p)$, $A_{v_0(p)}^x$, where we first assume that $v_0(p)$ is not also the start-point of the ribbon (we treat the case where these vertices are the same afterwards). This vertex transform changes the path label $g(s.p-v_{0}(p))$ to $g(s.p-v_{0}(p))x^{-1}$. In addition, the action of this transform on the plaquette is to take $e_p$ to $x \rhd e_p$. Therefore
	\begin{align*}
		B^e(t)A_{v_0(p)}^x: e_p &= \begin{cases} \big[ (g(s.p-v_{0}(p))x^{-1})^{-1}\rhd e \big] \: [x \rhd e_p] &\text{if $p$ is oriented against ribbon} \\
			[x \rhd e_p] \: \big[(g(s.p-v_{0}(p))x^{-1})^{-1}\rhd e^{-1}\big] &\text{if $p$ is oriented with ribbon} \end{cases}\\
		&=\begin{cases} \big[(xg(s.p-v_{0}(p))^{-1})\rhd e\big] \: [x \rhd e_p] &\text{if $p$ is oriented against ribbon} \\
			[x \rhd e_p] \: \big[(xg(s.p-v_{0}(p))^{-1})\rhd e^{-1}\big] &\text{if $p$ is oriented with ribbon} \end{cases}\\
		&=\begin{cases} x \rhd ([g(s.p-v_{0}(p))^{-1}\rhd e] e_p) &\text{if $p$ is oriented against ribbon} \\
			x \rhd (e_p \: [g(s.p-v_{0}(p))^{-1}\rhd e^{-1}]) &\text{if $p$ is oriented with ribbon} \end{cases}\\ 
		&= A_{v_0(p)}^x B^e(t) : e_p.
	\end{align*}
	This means that the vertex transform at $v_0(p)$ commutes with the action of the blob ribbon operator.

	Next we consider the vertex transform at the start-point of the ribbon, $s.p$, assuming that the base-point of the plaquette we are considering, $v_0(p)$, is not the start-point. In this case, the vertex transform at the start-point does not directly change the plaquette label, but does change the path element $g(s.p-v_{0}(p))$. In particular, the transform $A_{s.p}^x$ takes $g(s.p-v_{0}(p))$ to $xg(s.p-v_{0}(p))$. Therefore
	\begin{align*}
		B^e(t)A_{s.p}^x: e_p &= \begin{cases} [(xg(s.p-v_{0}(p)))^{-1} \rhd e] \: e_p &\text{if $p$ is oriented against the ribbon}\\ 
			e_p\: [(xg(s.p-v_{0}(p)))^{-1} \rhd e^{-1}] &\text{if $p$ is oriented along the ribbon}
		\end{cases}\\
		&= \begin{cases} \big[g(s.p-v_{0}(p))^{-1} \rhd (x^{-1} \rhd e)\big] \: e_p &\text{if $p$ is oriented against the ribbon}\\
			e_p \: \big[g(s.p-v_{0}(p))^{-1} \rhd (x^{-1} \rhd e)^{-1}\big] &\text{if $p$ is oriented along the ribbon}
		\end{cases}\\	 	
		&=A_{s.p}^x B^{x^{-1} \rhd e}(t):e_p	.											 
	\end{align*}

	This suggests that $B^e(t)A_{s.p}^g = A_{s.p}^g B^{g^{-1} \rhd e}(t)$, because this is true for the action on all plaquettes $p$ pierced by the ribbon and not based at the start-point. However, we have not yet considered plaquettes attached to the start-point itself and this relation must hold for all plaquettes for it to be an operator relation (it also must hold for any plaquettes not affected by the blob ribbon operator, but this holds trivially). That is, we now need to look at plaquettes for which the base-point $v_0(p)$ is also the start-point of the ribbon $s.p$. For these plaquettes, the path $(s.p-v_0(p))$ is the trivial path (or at least is a closed path, with label in the image of $\partial$ from fake-flatness) and so is not changed by the vertex transform. However, the vertex transform directly affects the plaquettes that have $s.p$ as their base-point. Therefore
	\begin{align*}
		B^e(t)A_{s.p}^g: e_p &= \begin{cases} e \: [g \rhd e_p] &\text{if $p$ is oriented against the ribbon}\\ 
			[g \rhd e_p] e^{-1} &\text{if $p$ is oriented along the ribbon}
		\end{cases}\\
		&= \begin{cases} g \rhd \big([g^{-1} \rhd e] \: e_p\big) &\text{if $p$ is oriented against the ribbon}\\
			g \rhd \big(e_p \: [g^{-1} \rhd e^{-1}]\big) &\text{if $p$ is oriented along the ribbon}
		\end{cases}\\	 	
		&=A_{s.p}^g B^{g^{-1} \rhd e}(t):e_p.											 
	\end{align*}
	We see that the action of the two operators on such plaquettes obeys the same commutation relation as the action on plaquettes not based at the start-point. Therefore, the relation $B^e(t)A_{s.p}^g: e_p =A_{s.p}^g B^{g^{-1} \rhd e}(t):e_p$ holds for all plaquettes, and so the operators satisfy the commutation relation 
	\begin{equation}
		B^e(t)A_{s.p}^g= A_{s.p}^g B^{g^{-1} \rhd e}(t). \label{Equation_commutation_blob_ribbon_start_point_transform}
	\end{equation}
	
	This means that the start-point vertex may be excited by the blob ribbon operator, depending on which linear combination of the blob ribbon operators $B^e(t)$ is taken. Consider a linear combination of the operators $B^{\vec{\alpha}}(t)= \sum_{e \in \text{ker}(\partial)} \alpha_e B^e(t)$. Then 
	\begin{align*}
		B^{\vec{\alpha}}(t) A_{s.p}^g&= \sum_{e \in \text{ker}(\partial)} \alpha_e B^e(t)A_{s.p}^g\\
		&= A_{s.p}^g \sum_{e \in \text{ker}(\partial)} \alpha_e B^{g^{-1} \rhd e}(t)\\
		&=A_{s.p}^g \sum_{\substack{e' =g^{-1} \rhd e\\ \in \text{ker}(\partial)}} \alpha_{g \rhd e'} B^{e'}(t).
	\end{align*}
	
	If $\alpha_{g \rhd e} = \alpha_e$ for all $e$ in the kernel of $\partial$, then we see that the vertex transform commutes with the ribbon operator. Because the vertex energy term is $A_{s.p} = \frac{1}{|G|} \sum_{g \in G} A_{s.p}^g$, this means that if $\alpha_e = \alpha_{g \rhd e}$ for all $g \in G$ and $e \in \text{ker}(\partial)$, then the ribbon operator commutes with the vertex transform at the start-point and so does not excite the vertex. On the other hand, consider applying the ribbon operator on the ground state, or some other state $\ket{\psi}$ for which the start-point is not excited. Then consider
	$$A_{s.p} B^{\vec{\alpha}}(t) \ket{\psi}.$$
	If this is zero it means that the ribbon operator $B^{\vec{\alpha}}(t)$ excites the start-point (recall that $A_{s.p}$ is a projector with the zero eigenvalues corresponding to excited states). If the start-point is not excited in $\ket{\psi}$ then $\ket{\psi}=A_{s.p} \ket{\psi}$. Therefore
	\begin{align*}
		A_{s.p} B^{\vec{\alpha}}(t) \ket{\psi}&= A_{s.p} B^{\vec{\alpha}}(t) A_{s.p}\ket{\psi}\\
		&= \frac{1}{|G|} \sum_g A^g_{s.p} \sum_{e \in \text{ker}(\partial)} \alpha_e B^e(t) A_{s.p}\ket{\psi}\\
		&= \frac{1}{|G|} \sum_g \sum_{e' \in \text{ker}(\partial)} \alpha_{g^{-1} \rhd e'} B^{e'}(t) A_{s.p}^g A_{s.p}\ket{\psi}\\
		&= \sum_{e' \in \text{ker}(\partial)} \frac{1}{|G|} \big( \sum_g \alpha_{g^{-1} \rhd e'}\big) B^{e'}(t) A_{s.p} \ket{\psi},
	\end{align*}
	where in the last line we used that $A_{s.p}^g A_{s.p} =A_{s.p}$. Then if $\sum_g \alpha_{g^{-1} \rhd e'}=0$, we have that $A_{s.p} B^{\vec{\alpha}}(t) \ket{\psi}=0$ and so the vertex is excited.

	Having examined the commutation relation between the blob ribbon operator and the vertex transforms, we next consider the edge transforms. The edge transform affects both path elements and surface elements, so we must check that neither of these effects fail to commute with the action of the blob ribbon operator. An edge transform $\mathcal{A}_i^f$ can affect the path label $g(s.p-v_0(p))$, by changing an edge label by $\partial(f)$. This means that if we act first with the edge transform and then the blob ribbon operator, instead of $g(s.p-v_{0}(p))^{-1}\rhd e$ in the action of the ribbon operator we will have $g_1 \partial(f)^{\pm 1} g_2 \rhd e$, where $g_1$ and $g_2$ represent the parts of the path element $g(s.p-v_{0}(p))^{-1}$ on either side of the edge. However, we will see that the edge transform does not affect the label $g(s.p-v_{0}(p))^{-1}\rhd e$. We have that $\partial(f) \rhd (g_2 \rhd e) = f (g_2\rhd e) f^{ -1}$ from the Peiffer conditions (see Equation \ref{Equation_Peiffer_2} in the main text). In addition the label $e$ of the blob ribbon operator is in the centre of $E$, which also means that $g_2 \rhd e$ is in the centre as well (as discussed in Section \ref{Section_3D_Blob_Fake_Flat} of the main text). This means that we can cancel the factors of $f$ and $f^{-1}$ and so the expression $g(s.p-v_{0}(p))^{-1}\rhd e$ is unchanged by the edge transform. Therefore, the action of the edge transforms on the edge labels commutes with the blob ribbon operators. We still need to consider the action of the edge transforms on the surface labels however, because both the blob ribbon operators and edge transforms act on the surface labels. The edge transform $\mathcal{A}_i^f$ multiplies the neighbouring surface labels by some $g \rhd f^{\pm 1}$, where $g$ is a path element as described in Section \ref{Section_Recap_3d} of the main text. Similarly, the blob ribbon operator $B^e(t)$ multiplies surface labels by some $k \rhd e^{\pm 1}$, where $k$ is again a path label as described at the beginning of this section. However, because $e$ is in the centre of $E$, each $k \rhd e^{\pm 1}$ is also in the centre of $E$. This means that the order in which we apply the multiplication from the blob ribbon operator and edge transform is irrelevant, because the factors commute. Therefore, the edge transforms (and so the edge energy terms, which are linear combinations of the edge transforms) commute with the blob ribbon operator.

	The final energy terms to consider are the blob energy terms. Just as in the $\rhd$ trivial case, we will consider the effect of the blob ribbon entering a blob on the blob 2-holonomy and show that this effect cancels with the effect of the ribbon leaving the blob. Therefore, only the blobs in which the ribbon originates or terminates (the blobs which have an imbalance between the ribbon entering and leaving the blob, in the simplest case only entering or leaving the blob) will be excited by the action of the blob ribbon operator. To see this, consider the blob 2-holonomy $H_2(B)$ of a blob $B$. This is given by
	$$H_2(B) = \prod_{p \in \text{Bd}(B)} g(v_0(B)-v_0(p)) \rhd e_p^{\sigma_p},$$
	where $v_0(B)$ is the base-point for the blob (which one is chosen does not matter). Note that, while the product over plaquettes should be taken in a specific order and the paths appearing must also be chosen appropriately (the order and precise paths must follow the rules for combining surfaces that we summarized in Section \ref{Section_Recap_3d}), we will not need to specify these details for this argument. The $\sigma_p$ are $1$ or $-1$ depending on the orientation of the plaquette $p$, as discussed in the $\rhd$ trivial case in Section \ref{Section_Blob_ribbon_proof_tri_trivial}. When the blob ribbon operator $B^e(t)$ enters the blob, it multiplies one plaquette by $g(s.p(t)-v_0(p))^{-1}\rhd e^{\pm 1}$. By the same argument as for the $\rhd$ trivial case in Section \ref{Section_Blob_ribbon_proof_tri_trivial}, while the plaquette can be multiplied by $g(s.p(t)-v_0(p))^{-1}\rhd e$ or the inverse, the term $e_p^{\pm 1}$ is always multiplied by $g(s.p(t)-v_0(p))^{-1}\rhd e$ (because the orientation of the plaquette relative to the blob normal is opposite to the orientation with respect to the blob ribbon). Therefore, the term $g(v_0(B)-v_0(p)) \rhd e_p^{\sigma_p}$ for the plaquette pierced by the blob ribbon when it enters the blob becomes
	$$g(v_0(B)-v_0(p)) \rhd \big( [g(s.p(t)-v_0(p))^{-1}\rhd e] \: e_p^{\sigma_p}\big)= \big[\big(g({v_0(B)-v_0(p)})g(v_0(p)-s.p(t))\big) \rhd e\big] \: \big[ g({v_0(B)-v_0(p)}) \rhd e_p^{\sigma_p}\big].$$

	Because $e$ is in the centre of $E$ (due to being in the kernel of $\partial$), the precise path $t$ appearing in an expression $g(t) \rhd e$ does not matter (or rather, any two paths $t_1$ and $t_2$ that can be deformed into one-another over a fake-flat surface will give the same result, as described in Section \ref{Section_3D_Blob_Fake_Flat} of the main text). Therefore, we can write $g(v_0(B)-v_0(p))g(v_0(p)-s.p(t))=g(v_0(B)-s.p(t))$ without needing to worry about the precise choice of path in each case. In addition, because $e$ is in the centre of $E$, we can extract the factor $\big(g(v_0(B)-v_0(p))g(v_0(p)-s.p(t))\big) \rhd e$ to the front of the product in $H_2(B)$. Putting these ideas together, the transformation of $H_2(B)$ from the ribbon entering the blob is
	\begin{equation}
		H_2(B) \rightarrow [g(v_0(B)-s.p(t)) \rhd e] \: H_2(B). \label{Equation_blob_ribbon_operator_enter_blob}
	\end{equation}
	
	Next we consider the effect of the ribbon leaving the blob on the blob 2-holonomy. Just as the ribbon entering the blob has the effect of multiplying one term $e_p^{\sigma_p}$ by $g(s.p(t)-v_0(p))^{-1}\rhd e$, the ribbon exiting the blob multiplies one $e_p^{\sigma_p}$ by $g(s.p(t)-v_0(p))^{-1}\rhd e^{-1}$ (because the orientation of the blob's normal matches the orientation of the ribbon in the place that the ribbon exits the blob). Just as before, we can extract this factor to the left and write $g(v_0(B)-v_0(p))g(s.p(t)-v_0(p))^{-1}= g(v_0(B)-s.p(t))$ to find that the effect of the ribbon leaving the blob is
	\begin{equation}
		H_2(B) \rightarrow [g(v_0(B)-s.p(t)) \rhd e^{-1}] H_2(B). \label{Equation_blob_ribbon_operator_exit_blob}
	\end{equation}
	
	We see that the effects of the blob ribbon entering and exiting a blob on the blob's 2-holonomy cancel, so the blob ribbon operator does not excite the blob energy terms for blobs in the middle of the ribbon. On the other hand, for the blobs at the origin and terminus of the ribbon there is no such cancellation, so the ribbon operator excites the blob energy terms.
	
	\subsubsection{The case where $\partial \rightarrow$ centre($G$) and $E$ is Abelian}
	\label{Section_Blob_Ribbon_Central}
	The final case that we consider is where the group $E$ is Abelian and the map $\partial$ from $E$ to $G$ maps on to the centre of $G$ (Case 2 in Table \ref{Table_Cases} of the main text). In this case, we can define the blob ribbon operators with any label from $E$, rather than just from the kernel of $\partial$. The blob ribbon operators are defined in the same way as in the fake-flat case considered in the previous section: when acting on a plaquette $p$ based at $v_0(p)$, the blob ribbon operator $B^e(t)$ acts on the plaquette label $e_p$ as
	\begin{equation}
		B^e(t) :e_p = \begin{cases} e_p \: [g(s.p-v_{0}(p))^{-1}\rhd e^{-1}] & \text{ if $p$ is oriented along the ribbon}\\
			[g(s.p-v_{0}(p))^{-1}\rhd e] \: e_p &\text{if $p$ is oriented against the ribbon.} \end{cases} \label{Equation_blob_ribbon_central}
	\end{equation}

	We want to consider the commutation relations of the ribbon operator with the energy terms. The commutation relations mostly follow from our arguments in the previous section. First we consider the plaquette term. If the label $e$ of the blob ribbon operator is not in the kernel of $\partial$, then the blob ribbon operator will change the $\partial(e_p)$ of all plaquettes $p$ through which it passes and so will excite the plaquettes, as described in the previous section. Unlike in the previous section, we do not need to throw out the ribbon operators which violate fake-flatness, because we allow fake-flatness violations in our Hilbert space. Instead, the ribbon operators which excite the plaquette terms create confined blob excitations, because there is an energy cost associated to the length of the ribbon used to create them.

	Note that, while some ribbon operators excite the plaquette terms, they only break fake-flatness by an element $\partial(e)$ for some $e \in E$ (because they change the plaquette surface label and not the edge labels around the plaquette). This is significant, because it means that two paths that enclose a surface where fake-flatness is violated in such a way only differ in label by some element $\partial(f)$ on top of the factor expected from deformation over the surface. In the case we are considering, an extra element of $\partial(E)$ in a path label does not affect certain expressions. In particular a factor of $\partial(e)$ in a path element $g(t)$ does not affect $g(t) \rhd e$ (from $\partial(f) \rhd e = fef^{-1} =e$, using $E$ Abelian) or $g(t)^{-1}hg(t)$ (from $\partial(f)^{-1}h\partial(f)=h$, using the fact that $\partial$ maps in to the centre of $G$). Because of this, such violations of fake-flatness do not affect the composition of surfaces (which relies on fake-flatness in order to be invariant under changing the branching structure) and so do not obstruct the sensible definition of blob energy terms (which rely on composing plaquettes into the surface of the blob) in the region of the ribbon.

	For the other energy terms, the commutation relations with the blob ribbon operator are exactly as we considered in Section \ref{Section_Blob_Ribbon_Fake_Flat}. We can follow that argument precisely, except without the requirement that the label of the blob ribbon operator $e$ be in the kernel of $\partial$, because we only used the fact that the label was in the centre of $E$ in the previous argument (and when $E$ is Abelian this is true for all $e \in E$). That is, the blob ribbon operator commutes with all edge transforms, all vertex transforms except potentially the one at the start-point of the ribbon and each blob energy term except the ones for the blobs in which the ribbon originates or terminates. The commutation relation of a blob ribbon operator $B^e(t)$ with a vertex transform $A_{s.p}^g$ as the start-point of $t$ is the same as Equation \ref{Equation_commutation_blob_ribbon_start_point_transform}:
	\begin{equation}
		B^e(t)A_{s.p}^g= A_{s.p}^g B^{g^{-1} \rhd e}(t). \label{Equation_commutation_blob_ribbon_start_point_transform_last_case}
	\end{equation}

	Similarly the transformation of the 2-holonomy of a blob entered by the ribbon is
	\begin{equation}
		H_2(B) \rightarrow [g(v_0(B)-s.p(t)) \rhd e] \: H_2(B), \label{Equation_blob_ribbon_operator_enter_blob_2}
	\end{equation}
	which is the same as Equation \ref{Equation_blob_ribbon_operator_enter_blob}, while the transformation of a blob exited by the ribbon is
	\begin{equation}
		H_2(B) \rightarrow [g(v_0(B)-s.p(t)) \rhd e^{-1}] H_2(B), \label{Equation_blob_ribbon_operator_exit_blob_2}
	\end{equation}
	which is the same as Equation \ref{Equation_blob_ribbon_operator_exit_blob}. These two transformations cancel for a blob which is both entered and exited by the ribbon (i.e., that is passed through by the ribbon).

	We have therefore shown that the blob ribbon operator $B^e(t)$ excites the blobs at the origin and terminus of the ribbon. In addition the blob ribbon operator excites the plaquettes pierced by the ribbon if $e$ is outside the kernel of $\partial$. Finally the start-point of the ribbon may be excited, depending on which linear combination $\sum_{e \in E} \alpha_e B^e(t)$ is considered. Note, however, that we have assumed that the state on which we act with the blob ribbon operator satisfies fake-flatness in the region on which we apply the ribbon (for example when considering the blob conditions along the ribbon). If it does not, the ribbon may cause some additional excitations.
	
	\subsubsection{Fusion of blob ribbon operators}

	We can combine two blob ribbon operators that are applied on the same ribbon (which means they must have the same start-point) into a single ribbon operator. Consider the action of two blob ribbon operators, $B^e(t)$ and $B^f(t)$, on a plaquette $p$ pierced by the ribbon $t$. We have
	\begin{align*}
		B^e(t)B^f(t):e_p &= B^e(t):\begin{cases} e_p [g(s.p-v_0(p))^{-1} \rhd f^{-1}] & \text{ if $p$ aligns with $t$} \\ [g(s.p-v_0(p))^{-1} \rhd f]e_p & \text{ if $p$ anti-aligns with $t$} \end{cases} \\
		&=\begin{cases} e_p [g(s.p-v_0(p))^{-1} \rhd f^{-1}][g(s.p-v_0(p))^{-1} \rhd e^{-1}] & \text{ if $p$ aligns with $t$} \\ [g(s.p-v_0(p))^{-1} \rhd e][g(s.p-v_0(p))^{-1} \rhd f]e_p & \text{ if $p$ anti-aligns with $t$} \end{cases} \\
		&=\begin{cases} e_p [g(s.p-v_0(p))^{-1} \rhd (f^{-1}e^{-1})] & \text{ if $p$ aligns with $t$} \\ [g(s.p-v_0(p))^{-1} \rhd (ef)]e_p & \text{ if $p$ anti-aligns with $t$} \end{cases} \notag\\
		&=B^{ef}(t):e_p.
	\end{align*}
	This holds for all plaquettes $p$, so it gives us the operator relation
	\begin{equation}
		B^e(t)B^f(t)=B^{ef}(t). \label{Equation_blob_ribbon_fake_flat_fusion}
	\end{equation}

	\subsubsection{Reversing the orientation of the dual path}
	\label{Section_blob_ribbon_invert_dual_path}

	When defining the blob ribbon operator in the preceding sections, we used the concept of a dual path. The plaquettes in the lattice which are pierced by the dual path are acted on by the blob ribbon operator. The relative orientation of the dual path and the plaquette determines the action of the blob ribbon operator on the plaquette. It is therefore useful to see what happens when we invert the orientation of this dual path. Given a plaquette $p$ pierced by a blob ribbon operator $B^e(t)$ the action of the blob ribbon operator on the plaquette is
	$$B^e(t) :e_p = \begin{cases} e_p \: [g(s.p-v_{0}(p))^{-1}\rhd e^{-1}] & \text{ if $p$ is oriented along the ribbon}\\
		[g(s.p-v_{0}(p))^{-1}\rhd e] \: e_p &\text{if $p$ is oriented against the ribbon}. \end{cases}$$
	
	Now consider what happens when we reverse the orientation of the dual path, to produce a new ribbon $\overline{t}$. A blob ribbon operator $B^x(\overline{t})$ applied on this ribbon acts on the plaquette $p$ as
	$$B^x(\overline{t}) :e_p = \begin{cases} e_p \: [g(s.p-v_{0}(p))^{-1}\rhd x^{-1}] & \text{ if $p$ is oriented along }\overline{t}\\
		[g(s.p-v_{0}(p))^{-1}\rhd x] \: e_p &\text{if $p$ is oriented against }\overline{t}. \end{cases}$$
	If we write this in terms of the orientation of the original ribbon, $t$, this becomes
	$$B^x(\overline{t}) :e_p = \begin{cases} e_p \: [g(s.p-v_{0}(p))^{-1}\rhd x^{-1}] & \text{ if $p$ is oriented against }t\\
		[g(s.p-v_{0}(p))^{-1}\rhd x] \: e_p &\text{if $p$ is oriented along }t. \end{cases}$$
	Then we can use the fact that we only consider blob ribbon operators whose labels lie in the centre of $E$ (either because $E$ is Abelian, or because we restrict to the fake-flat case when $E$ is non-Abelian), to rewrite this as
	$$B^x(\overline{t}) :e_p = \begin{cases} [g(s.p-v_{0}(p))^{-1}\rhd x^{-1}] e_p & \text{ if $p$ is oriented against }t\\
		e_p [g(s.p-v_{0}(p))^{-1}\rhd x] \: &\text{if $p$ is oriented along }t. \end{cases}$$
	Comparing this to the action of $B^e(t)$, we see that 
	\begin{align*}
		B^x(\overline{t}):e_p = B^{x^{-1}}(t):e_p.
	\end{align*}
	
	That is, the new ribbon operator acts in the same way on the plaquette as a blob ribbon operator on the original ribbon but with inverted label. This holds for all plaquettes $p$, so we can write this as an operator relation:
	\begin{equation}
		B^x(\overline{t}) = B^{x^{-1}}(t).
	\end{equation}
	If we invert this relation, we obtain 
	\begin{equation}
		B^x(t)= B^{x^{-1}}(\overline{t})
	\end{equation}
	
	That is, the ribbon operator $B^x(t)$ is unchanged if we simultaneously invert its label and the orientation of its dual path. In other words, we can invert the orientation of the dual path of a blob ribbon operator by inverting the label of the blob ribbon operator, without affecting the action of the ribbon operator. Note that we have not changed the direct path of the ribbon operator, so now the first plaquette affected by the reversed ribbon operator is at the end of the direct path.
	
	\subsubsection{Moving the start-point of a blob ribbon operator}
	\label{Section_blob_ribbon_move_sp}
\begin{figure}[h]
		\begin{center}
			\begin{overpic}[width=0.9\linewidth]{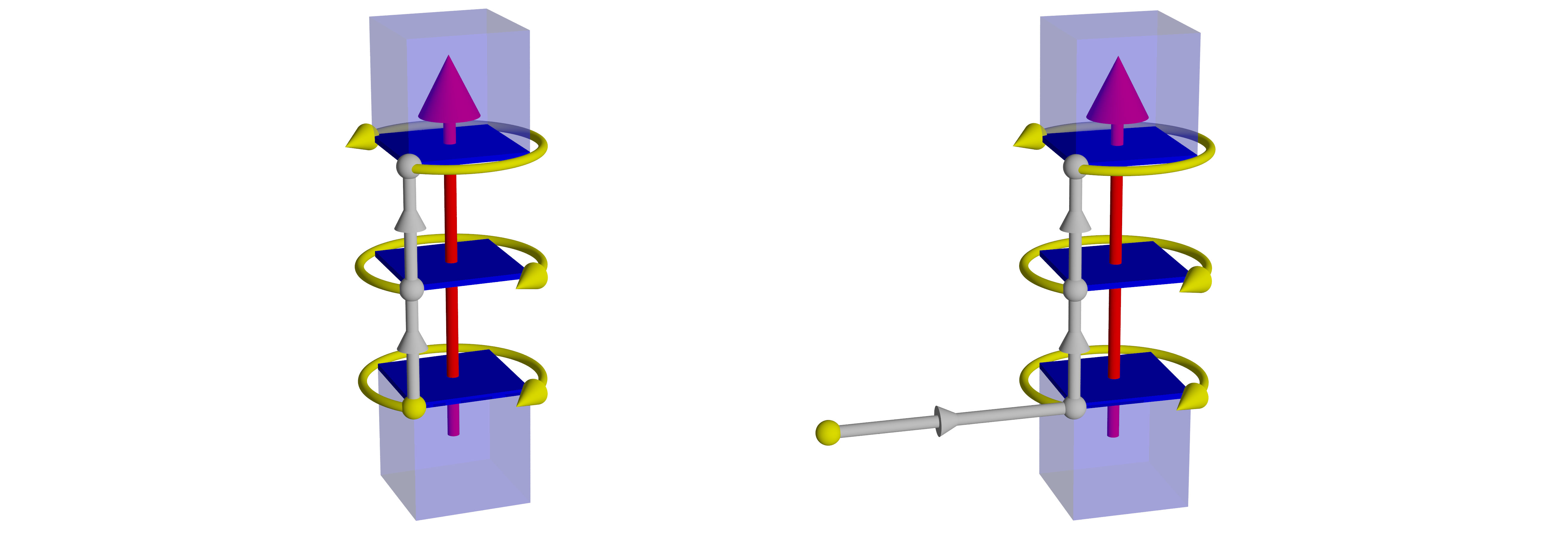}
				\put(45,15){\Huge $\rightarrow$}
				\put(44,20){move $s.p$}
				\put(20,7){$s.p(t)$}
				\put(45,5){$s.p(t')$}
				\put(18,20){$B^e(t)$}
				\put(60,20){$B^e(t')$}
			\end{overpic}
			\caption{We move the start-point of a blob ribbon operator from $s.p(t)$ to $s.p(t')$. This leaves the dual path (red arrow) of the ribbon operator fixed, but changes the direct path (grey edges). The dual path determines which plaquettes are affected by the ribbon operator, so the ribbon operators acts on the same plaquettes after we move the start-point, but the direct path partially determines the action on those plaquettes. We wish to see how changing the start-point affects the action on each plaquette.}
			\label{Blob_ribbon_move_sp}
		\end{center}
	\end{figure}

	When $\rhd$ is non-trivial, the action of the blob ribbon operator on a plaquette depends on a path from a special vertex, called the start-point of the ribbon operator, to the base-point of the plaquette being acted on. It is worth considering what happens to the ribbon operator when we change this start-point. Consider a blob ribbon operator acting on a ribbon $t$, whose start-point is $s.p(t)$. We then move the start-point to a new position, $s.p(t')$, while keeping the dual path of the ribbon fixed (i.e, the ribbon operator still acts on the same plaquettes), as shown in Figure \ref{Blob_ribbon_move_sp}. Let the initial blob ribbon operator be $B^e(t)$, while the final ribbon operator is $B^e(t')$. The action of the original ribbon operator on a plaquette $p$ pierced by the ribbon is 
	$$B^e(t): e_p = \begin{cases} e_p [g(s.p(t)-v_0(p))^{-1} \rhd e^{-1}] & \text{ if plaquette $p$ aligns with $t$} \\ [g(s.p(t)-v_0(p))^{-1} \rhd e] e_p & \text{ if plaquette $p$ anti-aligns with $t$,} \end{cases} $$
	where $v_0(p)$ is the base-point of the plaquette $p$.
	
	Similarly the action of the new ribbon operator $B^e(t')$ on the same plaquette is
	$$B^e(t'): e_p = \begin{cases} e_p [g(s.p(t')-v_0(p))^{-1} \rhd e^{-1}] & \text{ if plaquette $p$ aligns with $t'$} \\ [g(s.p(t')-v_0(p))^{-1} \rhd e] e_p & \text{ if plaquette $p$ anti-aligns with $t'$.} \end{cases} $$
	If we moved the start-point along the path $(s.p(t)-s.p(t'))$ when we changed the start-point, then we can write the path $(s.p(t')-v_0(p))$ as $(s.p(t')-s.p(t)) \cdot (s.p(t)-v_0(p)) = (s.p(t)-s.p(t'))^{-1} \cdot (s.p(t)-v_0(p))$. Therefore, denoting the label of path $(s.p(t)-s.p(t'))$ by $g(s.p(t)-s.p(t'))$, we have
	$$g(s.p(t')-v_0(p)) = g(s.p(t)-s.p(t'))^{-1} g(s.p(t)-v_0(p)).$$
	Substituting this into the action of $B^e(t')$, we see that
	\begin{align*}
		B^e(t'): e_p &= \begin{cases} e_p \big[\big( g(s.p(t)-s.p(t'))^{-1} g(s.p(t)-v_0(p)) \big)^{-1} \rhd e^{-1}\big] & \text{ if plaquette $p$ aligns with $t'$} \\ \big[\big( g(s.p(t)-s.p(t'))^{-1} g(s.p(t)-v_0(p)) \big)^{-1} \rhd e\big] e_p & \text{ if plaquette $p$ anti-aligns with $t'$.} \end{cases} \\
		&= \begin{cases} e_p \big[\big( g(s.p(t)-v_0(p))^{-1} g(s.p(t)-s.p(t'))\big) \rhd e^{-1}\big] & \text{ if plaquette $p$ aligns with $t'$} \\ \big[\big( g(s.p(t)-v_0(p))^{-1} g(s.p(t)-s.p(t'))\big) \rhd e\big] e_p & \text{ if plaquette $p$ anti-aligns with $t'$.} \end{cases}\\
		&= \begin{cases} e_p \big[ g(s.p(t)-v_0(p))^{-1} \rhd \big(g(s.p(t)-s.p(t')) \rhd e^{-1}\big)\big] & \text{ if plaquette $p$ aligns with $t'$} \\ \big[ g(s.p(t)-v_0(p))^{-1} \rhd \big( g(s.p(t)-s.p(t')) \rhd e\big)\big] e_p & \text{ if plaquette $p$ anti-aligns with $t'$.} \end{cases}\\
		&=B^{g(s.p(t)-s.p(t')) \rhd e}(t).
	\end{align*}
	
	We see that the ribbon operator $B^e(t')$ acts on plaquette $p$ in the same way as $B^{g(s.p(t)-s.p(t')) \rhd e}(t)$. This holds for all plaquettes (recall that $t$ and $t'$ have the same dual path and so the corresponding ribbon operators act on all the same plaquettes). Therefore, we can write that
	\begin{equation}
		B^e(t')=B^{g(s.p(t)-s.p(t')) \rhd e}(t). \label{Equation_blob_ribbon_change_sp_1}
	\end{equation}
	
	We have seen that moving the start-point of a blob ribbon operator, while keeping the label constant, changes how the operator acts. We can also change the start-point and simultaneously change the label of the operator, in order to keep the action of the operator fixed. That is, we can describe the same blob ribbon operator in different ways by choosing different start-points and taking appropriate labels for each (this is analogous to a choice of basis). For an initial blob ribbon operator $B^e(t)$, we wish to find the label $x$ of blob ribbon operator $B^x(t')$ such that $B^e(t)=B^x(t')$. To do this, we just need to invert Equation \ref{Equation_blob_ribbon_change_sp_1}. We therefore see that
	\begin{equation}
		B^e(t)=B^{g(s.p(t)-s.p(t'))^{-1} \rhd e}(t'). \label{Equation_blob_ribbon_change_sp_2}
	\end{equation}
	Note that the label $g(s.p(t)-s.p(t'))^{-1} \rhd e$ of the blob ribbon operator includes the path label $g(s.p(t)-s.p(t'))$, so the ribbon operator label is actually an operator. This reflects the idea that the 2-flux produced by the blob ribbon operator is well-defined with respect a particular start-point ($s.p(t)$ in this case), but not with respect other start-points (a fact that is important when we consider braiding).

	When considering Equation \ref{Equation_blob_ribbon_change_sp_1}, we can see a great similarity with Equation \ref{Equation_commutation_blob_ribbon_start_point_transform_last_case}, which describes the commutation relation between the blob ribbon operator and the vertex transform at the start-point. Indeed the transformation of the blob ribbon operator when we move the start-point along a path of label $g(s.p(t)-s.p(t'))$ is the same as the transformation when we apply the vertex transform $A_v^{g(s.p(t)-s.p(t'))^{-1}}$. This reflects the idea that we discussed in Ref. \cite{HuxfordPaper1}, that applying the vertex transform $A_v^g$ on a vertex $v$ is equivalent to parallel transporting that vertex along a path of label $g^{-1}$. For this reason, a blob ribbon operator that commutes with the vertex energy term at the start-point is also unaffected by moving its start-point. Mathematically, this occurs because a blob ribbon operator that commutes with the start-point vertex transform has the form $\sum_{g \in G} B^{g \rhd e}(t)$, and so any additional $\rhd$ action on the label can be absorbed into the sum over $g$.
	
	\subsubsection{Concatenation of blob ribbon operators}
	\label{Section_blob_ribbon_concatenate}
\begin{figure}[h]
		\begin{center}
			\begin{overpic}[width=0.9\linewidth]{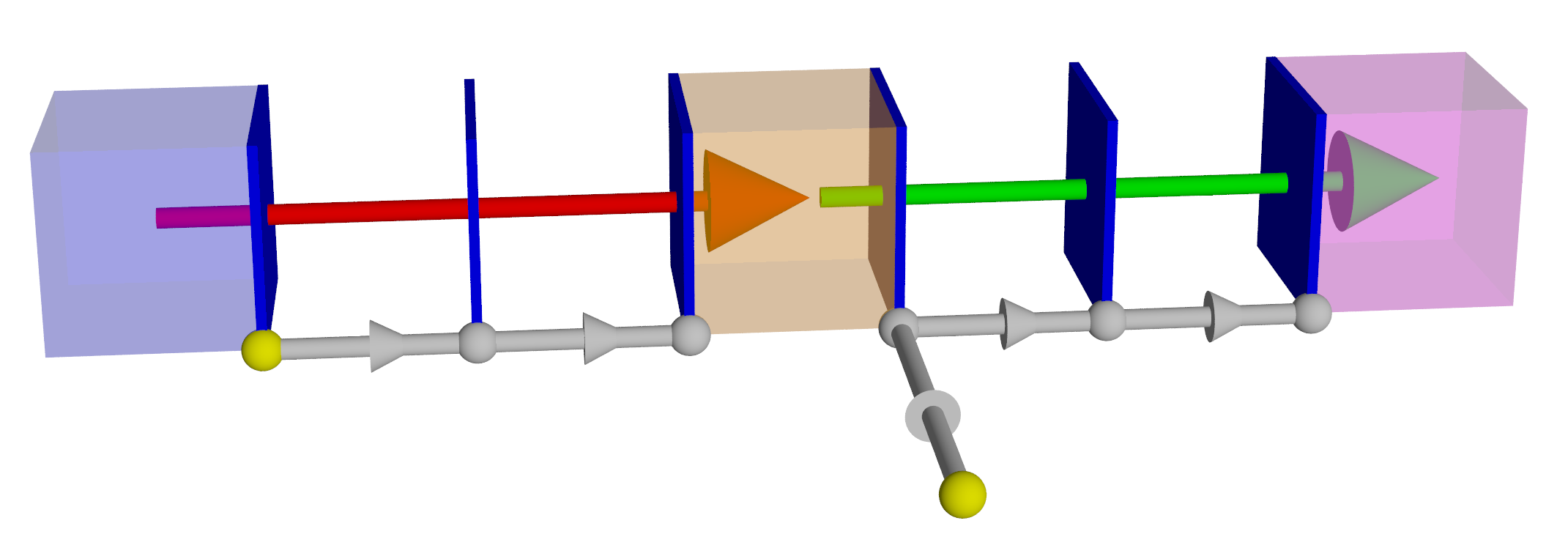}
				\put(15,9){$s.p(r)$}
				\put(64,2){$s.p(s)$}
				\put(25,30){$B^x(r)$}
				\put(62,30){$B^y(s)$}
			\end{overpic}
			\caption{We consider two blob ribbon operators, $B^x(r)$ and $B^y(s)$, whose dual paths (red and green arrows respectively) line up end to end. We wish to determine what conditions these ribbon operators must satisfy for us to be able to join them into a single blob ribbon operator.}
			\label{Blob_ribbon_concatenate_1}
		\end{center}
	\end{figure}

	In the previous section, we described the procedure by which we move the start-point of a blob ribbon operator. In this section, we will describe one reason we may wish to do this: the joining of two ribbons together into a longer one. Consider two blob ribbon operators $B^x(r)$ and $B^y(s)$, whose dual paths connect together so that they form an unbroken path through a series of plaquettes, as shown in Figure \ref{Blob_ribbon_concatenate_1}. We may wonder whether these two ribbon operators can be written as a single ribbon operator acting on the combined path. This may be true, but the two blob ribbon operators must satisfy certain conditions for this to be the case.
	
	Suppose that the product of the two blob ribbon operators can be written as a single blob ribbon operator $B^z(t)$, whose dual path is the concatenation of the dual paths of the other two ribbon operators. Then the action of $B^z(t)$ must agree with the action of $B^x(r)$ on the plaquettes pierced by $r$ and with the action of $B^y(s)$ on the plaquettes pierced by $s$. The action of $B^z(t)$ on a plaquette $p$ pierced by $t$ is
	$$B^z(t):e_p = \begin{cases} e_p [g(s.p(t)-v_0(p))^{-1} \rhd z^{-1}] & \text{if $p$ aligns with $t$}\\ [g(s.p(t)-v_0(p))^{-1} \rhd z] e_p & \text{if $p$ anti-aligns with $t$.} \end{cases}$$
	
	If the plaquette $p$ is pierced by $r$, then the action of $B^x(r)$ on $p$ is
	$$B^x(r):e_p = \begin{cases} e_p [g(s.p(r)-v_0(p))^{-1} \rhd x^{-1}] & \text{if $p$ aligns with $t$}\\ [g(s.p(r)-v_0(p))^{-1} \rhd x] e_p & \text{if $p$ anti-aligns with $t$,} \end{cases}$$
	where we used the fact that $r$ and $t$ have the same orientation to write the two cases in terms of the orientation of $t$ instead of $r$. We see that these actions are the same if the labels satisfy
	$$g(s.p(r)-v_0(p))^{-1} \rhd x= g(s.p(t)-v_0(p))^{-1} \rhd z.$$
	
	If we choose the start-point of $t$ to be the same as the start-point of $r$, then this condition becomes $x=z$, so the total ribbon operator must be $B^x(t)$. We now wish to see when the action of the other ribbon operator, $B^y(s)$, agrees with the action of $B^x(t)$ on the other section of path. The action of $B^y(s)$ on a plaquette pierced by $s$ is
	$$B^y(s):e_p = \begin{cases} e_p [g(s.p(s)-v_0(p))^{-1} \rhd y^{-1}] & \text{if $p$ aligns with $t$}\\ [g(s.p(s)-v_0(p))^{-1} \rhd y] e_p & \text{if $p$ anti-aligns with $t$.} \end{cases}$$
	
	This agrees with the action of $B^x(t)$ when the labels satisfy
	$$g(s.p(r)-v_0(p))^{-1} \rhd x= g(s.p(s)-v_0(p))^{-1} \rhd y,$$
	where $s.p(r)=s.p(t)$. If the paths $(s.p(r)-v_0(p))$ and $(s.p(s)-v_0(p))$ are the same, or can be deformed into one-another over a fake-flat surface, then the condition becomes $x=y$. A necessary condition for this is that the start-points of the two ribbons are the same (if the region is not fake-flat, or there are non-contractible cycles, then we must take further care that the direct paths of the two ribbons are compatible). On the other hand, if the start-points are different, then we can move the start-point of $s$ to match the start-point of $r$, and then the labels must match. To see this, we can write
	\begin{align*}
		g(s.p(s)-v_0(p))^{-1} \rhd y &= (g(s.p(s)-s.p(r))g(s.p(r)-v_0(p)))^{-1} \rhd y\\
		&= g(s.p(r)-v_0(p))^{-1} \rhd (g(s.p(s)-s.p(r))^{-1} \rhd y).
	\end{align*}
	
	We then note that $g(s.p(s)-s.p(r))^{-1} \rhd y$ is the label we must give the blob ribbon operator when we change its start-point from $s$ to $r$, if we wish to keep the action of the blob ribbon operator constant (see Equation \ref{Equation_blob_ribbon_change_sp_2}). Then the condition 
	$$g(s.p(r)-v_0(p))^{-1} \rhd x= g(s.p(s)-v_0(p))^{-1} \rhd y$$
	can be written as
	$$g(s.p(r)-v_0(p))^{-1} \rhd x = g(s.p(r)-v_0(p))^{-1} \rhd (g(s.p(s)-s.p(r))^{-1} \rhd y),$$
	which implies that $x= g(s.p(s)-s.p(r))^{-1} \rhd y$. That is, the label of the two blob ribbon operators that we wish to combine must match, once we move the start-points of the ribbon operators to the same position.

	\subsection{Magnetic membrane operators in the $\rhd$ trivial case}
	\label{Section_Magnetic_Membrane_Tri_trivial}
	In this section we derive the commutation relations of the magnetic membrane operator with the energy terms, in the case where $\rhd$ is trivial. Recall from Section \ref{Section_3D_Tri_Trivial_Magnetic_Excitations} of the main text that the magnetic membrane is characterised by two membranes: the dual membrane and the direct membrane. These are illustrated in Figure \ref{fluxmembrane2} in Section \ref{Section_3D_Tri_Trivial_Magnetic_Excitations}. The dual membrane intersects a set of edges and it is these edges that are acted on by the membrane operator. The direct membrane contains a set of vertices at one end of each of these edges. The action of the magnetic membrane operator on each edge depends on the path element for the path from a privileged start-point to the vertex in the direct membrane attached to the edge in question. The action of a magnetic membrane operator $C^h(m)$ on an edge element $g_i$ of an edge $i$ cut by the dual membrane is to either pre-multiply the edge element by $g(s.p-v_i)^{-1}hg(s.p-v_i)$ or post-multiply it by the inverse, where $g(s.p-v_i)$ is the relevant path element from the start-point of the membrane, $s.p$, to the vertex $v_i$ that lies on the direct membrane and is attached to edge $i$. If the edge points away from the vertex $v_i$ (i.e., away from the direct membrane) then it is pre-multiplied by $g(s.p-v_i)^{-1}hg(s.p-v_i)$, and if it instead points towards the vertex (towards the direct membrane) the edge element is instead post-multiplied by $g(s.p-v_i)^{-1}h^{-1}g(s.p-v_i)$. We can summarize this action on an edge $i$ cut by the dual membrane as
	\begin{equation}
		C^h(m):g_i = \begin{cases} g(s.p-v_i)^{-1}hg(s.p-v_i)g_i & \text{if $i$ points away from the direct membrane} \\ g_ig(s.p-v_i)^{-1}h^{-1}g(s.p-v_i) & \text{if $i$ points towards the direct membrane.} \end{cases} \label{Equation_magnetic_membrane_on_edges_appendix}
	\end{equation}

	Now we want to consider the commutation relations of this operator with the energy terms. We start by considering the vertex transforms. Just as in the 2+1d case considered in Ref. \cite{HuxfordPaper2} (see Section S-I B in the Supplemental Material for that paper), the vertex transforms may interact with the magnetic membrane operator in two ways. The first way is by directly affecting the edges whose labels are changed by the magnetic membrane operator: the edges that are cut by the dual membrane. An edge label is affected by vertex transforms at the two ends of the edge, so these transforms may fail to commute with the magnetic membrane operator. The second way in which a vertex transform may fail to commute with the membrane operator is if the vertex transform affects the path labels $g(s.p-v_i)$ that appear in the action of the magnetic membrane operator on an edge $i$ (see Equation \ref{Equation_magnetic_membrane_on_edges_appendix}). As we proved in Section S-I A of the Supplemental Material for Ref. \cite{HuxfordPaper2}, a vertex transform affects the labels of paths that start or end at that vertex, so we must consider the vertex transforms at the start-point of the membrane, $s.p$, and the vertex $v_i$ for each edge $i$ cut by the dual membrane. We note that because the vertex $v_i$ is attached to the edge $i$, a vertex transform at $v_i$ also directly affects the label of the edge $i$.

	Considering these two ways in which the vertex transforms can interact with the magnetic membrane operator, we see that there are three classes of vertex to consider. The first is the set of vertices which are attached to edges that are cut by the dual membrane, but which do not lie on the direct membrane. The transforms on these vertices directly affect the labels of the edges cut by the dual membrane, but do not affect the path labels which appear in the action of the membrane operator. The second class of vertices is the set of vertices which both lie on the direct membrane and are attached to edges cut by the dual membrane, except the start-point. The vertex transforms on these vertices directly affect the edges cut by the dual membrane and also affect the path labels which appear in the action of the membrane operator. The final type of vertex is the start-point vertex itself, which can also directly affect some edges, and also acts on the path labels for every edge cut by the membrane (because all of the paths $s.p-v_i$ start at the start-point).

	We start by considering the first class of vertices, those which do not lie on the direct membrane but are attached to edges cut by the dual membrane. Recall that each edge is attached to two vertices (the source and target of the edge), and for the edges intersected by the dual membrane one of these vertices lies on the direct membrane and one of them lies away from it (unless the membrane folds back on itself). However, the vertex transforms on the end of the edge not on the direct membrane will commute with the membrane operator. To see this, first note that the action of a vertex transform $A_v^x$ on an edge $i$ attached to the vertex is
	$$A_v^x : g_i = \begin{cases} xg_i & \text{if $i$ points away from the vertex}\\
		g_ix^{-1} & \text{if $i$ points towards the vertex.} \end{cases}$$
	
	That is, the vertex transform either left-multiplies the edge label by $x$ or right-multiplies it by $x^{-1}$, depending on the orientation of the edge. The membrane operator $C^h(m)$ also left-multiplies the label of an edge $i$ cut by the membrane by some element $g(t)^{-1}hg(t)$ or right-multiplies it by the inverse, depending on the orientation of the edge if the edge points towards the direct membrane. For the vertex attached to the edge but away from the direct membrane, if the edge points away from the direct membrane it points towards the vertex (and vice-versa). This means that, when the membrane operator right-multiplies the edge label by some element (because the edge points away from the membrane), the vertex transform left-multiplies it by another element (because the edge points towards the vertex), and vice-versa. Therefore, applying the magnetic membrane operator and then the vertex transform has the following action on the edge $i$:
	\begin{align*}
		C^h(m)A_v^x: g_i &= \begin{cases} g(s.p-v_i)^{-1}hg(s.p-v_i)g_i x^{-1} & \text{ if $i$ points away from the direct membrane (towards $v$)} \\ xg_ig(s.p-v_i)^{-1}h^{-1}g(s.p-v_i) & \text{ if $i$ points towards the direct membrane (away from $v$)} \end{cases}\\
		&= A_v^x C^h(m):g_i.
	\end{align*}
	
	Because one of the operators right-multiplies the edge label while the other left-multiplies it (and the group elements involved are independent), the contributions from the two operators commute. This means that the vertex transforms away from the direct membrane commute with the magnetic membrane operator.

\begin{figure}[h]
		\begin{center}
			\begin{overpic}[width=0.5\linewidth]{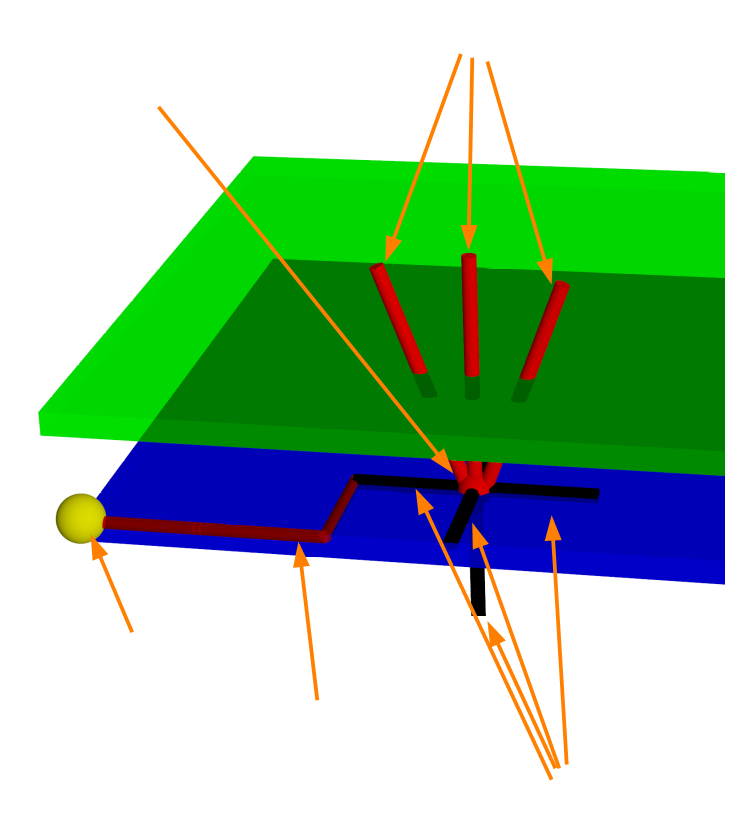}
				\put(48,94){\large shared edges}
				\put(55,2){\large other edges on vertex}
				\put(17,19){\large $s.p$}
				\put(30,10){\large $s.p-v_i$}
				\put(17,88){\large $v_i$}
				
			\end{overpic}
			\caption{We consider the action of the magnetic membrane operator and a vertex transform on the edges attached to a particular vertex $v_i$ in the direct membrane (blue). Some of the edges attached to $v_i$ are cut by the dual membrane (green) and so are affected by both the magnetic membrane operator and the vertex transform. These are called the shared edges.}
			\label{vertex_transform_mag_membrane}	
		\end{center}
	\end{figure}

	Next we consider vertex transforms on the vertices in the direct membrane, which can act on the same edges as the magnetic membrane operator and affect the path elements that appear in the action of the magnetic membrane operator. Just as we did for the ribbon in the 2+1d case in Ref. \cite{HuxfordPaper2} (see Section S-I B of the Supplemental Material), we group the edges that are cut by the dual membrane according to which vertex $v_i$ in the direct membrane they are attached to. We do this because edges attached to the same vertex are acted on by the membrane operator in a similar way. The action on these edges depends on a path element $g(s.p-v_i)$, so we want to consider which vertex transforms can change this path element. Recall that the only vertex transforms that affect a path label are the transforms at the start and end of the path. The start of this path is the start-point of the membrane, which is the same point for each path that we will consider (the start-point is a property of the membrane, independent of which vertex $v_i$ we consider). On the other hand, the end of this path is $v_i$, the vertex for which we are considering the attached edges, as shown in Figure \ref{vertex_transform_mag_membrane}. We therefore only need to consider the vertex transforms on these two vertices, starting with the ones at $v_i$.

	We will see that the vertex transforms at $v_i$ commute with the magnetic membrane operator, in exactly the same way that the analogous vertex transform commutes with the magnetic ribbon operator in the 2+1d case. The fact that we have three spatial dimensions rather than two makes no difference, because we do not need to worry about the geometry when considering the edges attached to a particular vertex. Nonetheless, we sketch the proof again here. We consider the situation where we apply the magnetic membrane operator followed by vertex transform on the vertex $v_i$ and compare this to the case where we apply the operators in the opposite order.

	We call the edges which are affected by both the magnetic membrane operator and the vertex transform (i.e., the edges attached to the vertex which are also cut by the dual membrane, which are the bright red edges in Figure \ref{vertex_transform_mag_membrane}) the shared edges. The other edges attached to the vertex are only affected by the vertex transform, so the action on these edges is the same regardless of the order in which we apply the operators. For a shared edge $i$, with label $g_i$, acting first with the magnetic membrane operator and then with the vertex transform gives
	\begin{align*}
		A_{v_i}^x C^h(m): g_i &=A_{v_i}^x: \begin{cases} g(s.p-v_i)^{-1}hg(s.p-v_i)g_i & \text{if } i \text{ points away from $v_i$}\\ g_i g(s.p-v_i)^{-1}h^{-1}g(s.p-v_i) &\text{if } i \text{ points towards $v_i$}\end{cases}\\
		&= \begin{cases} xg(s.p-v_i)^{-1}hg(s.p-v_i)g_i &\text{if } i \text{ points away from $v_i$}\\ g_i g(s.p-v_i)^{-1}h^{-1}g(s.p-v_i)x^{-1} &\text{if } i \text{ points towards $v_i$.} \end{cases}\\
	\end{align*}

	On the other hand, if we act with the operators in the opposite order $A_{v_i}^x$ takes the path element $g(s.p-v_i)$ to $g(s.p-v_i)x^{-1}$ before the membrane operator acts (because the vertex is at the end of the path), so that we have
	\begin{align*}
		C^h(m) A_{v_i}^x : g_i &=C^h: \begin{cases} xg_i&\text{if } i \text{ points away from $v_i$}\\ g_ix^{-1} & i \text{ points towards $v_i$}\end{cases}\\
		&=\begin{cases} (g(s.p-v_i)x^{-1})^{-1}h(g(s.p-v_i)x^{-1})xg_i &\text{if } i \text{ points away from $v_i$} \\ g_ix^{-1} (g(s.p-v_i)x^{-1})^{-1}h^{-1}(g(s.p-v_i)x^{-1}) &\text{if } i \text{ points towards $v_i$} \end{cases}\\
		&= \begin{cases} xg(s.p-v_i)^{-1}hg(s.p-v_i)g_i &\text{if } i \text{ points away from $v_i$}\\ g_i g(s.p-v_i)^{-1}h^{-1}g(s.p-v_i)x^{-1} &\text{if } i \text{ points towards $v_i$.} \end{cases}
	\end{align*}
	
	The combined action of the two operators is the same for either order, so the action of this vertex transform commutes with the action of the magnetic membrane operator on the edges attached to these vertices. Since the vertex transform does not affect any other edges that are cut by the membrane operator, or affect any path labels which appear in the magnetic membrane operator (other than the ones terminating at $v_i$), this means that the vertex transform commutes with the magnetic membrane operator.

	The final vertex transform to consider is the one at the privileged start-point of the membrane, $s.p$. Just like in the 2+1d case, the vertex transform $A_{s.p}^x$ takes all of the path labels $g(s.p-v_i)$ that appear in the action of the membrane operator to $xg(s.p-v_i)$ (as shown in Section S-I A of the Supplemental Material for Ref. \cite{HuxfordPaper2}) and so takes the factor $g(s.p-v_i)^{-1}hg(s.p-v_i)$ to $g(s.p-v_i)^{-1}x^{-1}hxg(s.p-v_i)$. This holds for all vertices $v_i$ except for the start-point itself. This means that when acting on an edge $i$ attached to a vertex $v_i$ that is not the start-point, we have
	\begin{align*}
		C^h(m) A_{s.p}^x:g_i &= \begin{cases} g(s.p-v_i)^{-1}x^{-1}hxg(s.p-v_i) g_i &\text{if } i \text{ points away from } v_i\\ g_i g(s.p-v_i)^{-1}x^{-1}h^{-1}xg(s.p-v_i) & \text{if } i \text{ points towards } v_i\end{cases}\\
		&= A_{s.p}^x C^{x^{-1}hx}(m):g_i
	\end{align*}
	which suggests that $C^h(m) A_{s.p}^x= A_{s.p}^x C^{x^{-1}hx}(m)$. However, for this commutation relation to hold it needs to be true for the action on every edge, including the edges attached to the start-point itself. For such an edge, initially labelled by $g_i$, we have
	\begin{align*}
		C^h(m)A_{s.p}^x:g_i &=\begin{cases} hx g_i & \text{if }i \text{ points away from $s.p$}\\ g_ix^{-1}h^{-1} & i \text{ points towards $s.p$}\end{cases}\\
		&= \begin{cases} x (x^{-1}hx) g_i &\text{if } i \text{ points away from $s.p$}\\ g_i(x^{-1}hx)^{-1}x^{-1} &\text{if } i \text{ points towards $s.p$}\end{cases}\\
		&= A_{s.p}^x C^{x^{-1}hx}(m):g_i.
	\end{align*}
	Therefore, the relation $C^h(m)A_{s.p}^x:g_i=A_{s.p}^x C^{x^{-1}hx}(m):g_i$ holds for every edge, meaning that we have the operator relation
	\begin{equation}
		C^h(m)A_{s.p}^x=A_{s.p}^x C^{x^{-1}hx}(m). \label{Equation_magnetic_membrane_start_point_transform}
	\end{equation} 
	
	Because the vertex transform at the start-point does not commute with the membrane operator, when we take linear combinations of the membrane operators with label in the conjugacy class of $h$, the start-point may be excited, just as in the 2+1d case examined in Ref. \cite{HuxfordPaper2} (specifically, it is not excited if the coefficients of the linear combination are a function of conjugacy class, and is excited if the coefficients within the conjugacy class sum to zero).

	Next we consider the edge transforms, which can seemingly interact with the magnetic membrane operator in two ways. Firstly, they change the value of a path element $g(s.p-v_i)$ appearing in the membrane operator if the edge on which we apply the transform lies on that path. However, just as we saw in the 2+1d case in Ref. \cite{HuxfordPaper2} (see Section S-I B of the Supplemental Material), while the edge transforms can affect the value of a path element, they always leave $g(t)^{-1}hg(t)$ invariant when $\rhd$ is trivial. This is because $g(t)$ gains a factor of $\partial(e)$ for some $e \in E$ from the edge transform, and $g(t)^{-1}$ gains a factor of $\partial(e)^{-1}$. However, when $\rhd$ is trivial factors in $\partial(E)$ are in the centre of $G$, so the two factors of $\partial(e)$ and $\partial(e)^{-1}$ can be commuted together and cancelled. The second way in which we may think that the edge transforms and magnetic membrane operator could interfere is by the transforms acting on the same edges that have their labels changed by the magnetic membrane operators (those edges cut by the dual membrane). However, the edge transforms on the edges cut by the dual membrane only change these edge labels by an element $\partial(e)$, which is again in the centre of $G$ and so commutes with the factors added by the magnetic membrane operator. Therefore, all edge transforms commute with the magnetic membrane operators.

\begin{figure}[h]
		\begin{center}
			\begin{overpic}[width=0.6\linewidth]{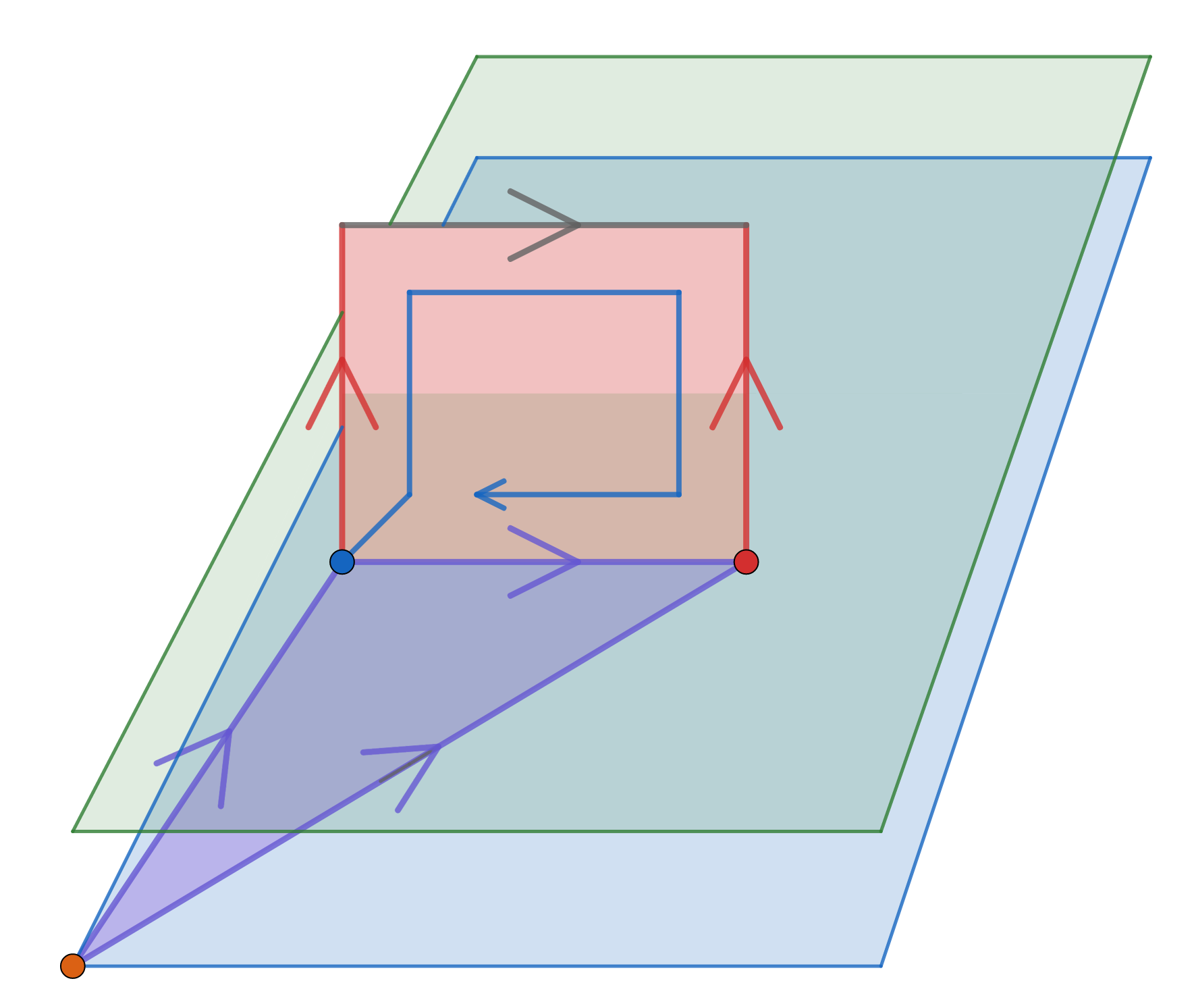}
				\put(26,57){$g_1$}
				\put(63,57){$g_2$}
				\put(41,66){$g_x$}
				\put(32,34.5){$g(v_1-v_2)$}
				\put(45,52){$e_p$}
				\put(25,37){$v_1$}
				\put(64,37){$v_2$}
				\put(2,1){$s.p$}
				\put(10,30){$g(s.p-v_1)$}
				\put(53,30){$g(s.p-v_2)$}
				
				\put(74.5,3){direct membrane}
				\put(74.5,14.5){dual membrane}
				
			\end{overpic}
			\caption{A generic plaquette $p$ (red) cut by the dual membrane (green). The red edges are the edges on the plaquette that are cut by the dual membrane. These edges could instead have the opposite orientation, but it is simple to check that this would not affect our conclusions. The vertices $v_1$ and $v_2$ are on the direct membrane and the paths from the start-point to these vertices have label $g(s.p-v_1)$ and $g(s.p-v_2)$. Together with the path $g(v_1-v_2)$ on the plaquette itself, these paths enclose a surface (purple) which satisfies fake-flatness in the ground state. Note that $g_x$ and $g(v_1-v_2)$ label paths and not individual edges (they could even contain no edges at all), so we have made no assumptions about the shape of the plaquette.}
			\label{internal_plaquette_mag_membrane}	
		\end{center}
	\end{figure}

\begin{figure}[h]
		\begin{center}
			\begin{overpic}[width=0.5\linewidth]{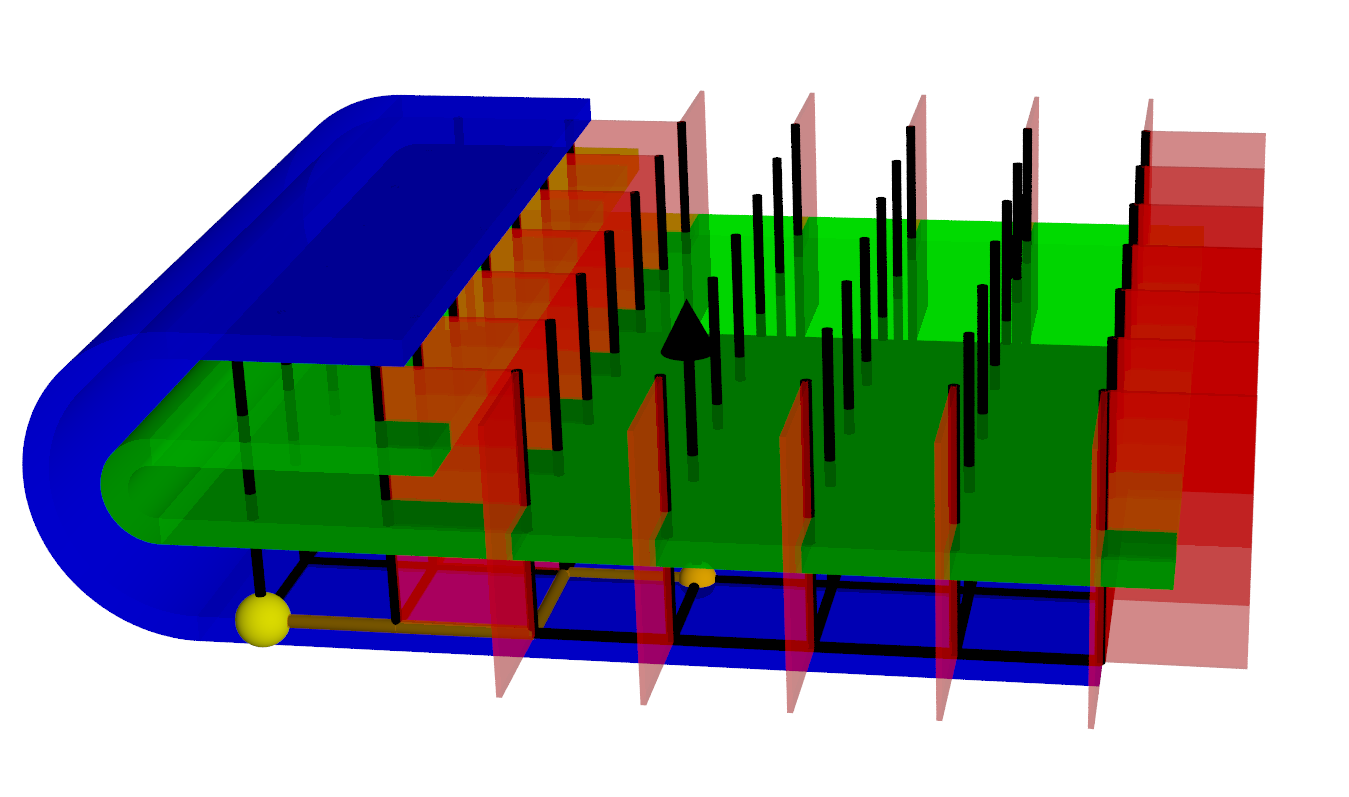}

			\end{overpic}
			\caption{If the dual membrane (green) folds over itself it can cut through the same edges or plaquettes more than once. In such cases the membrane operator can act on the same edges multiple times. In this case, each time the membrane cuts the plaquette causes a separate transformation of a type we have considered before.}
			\label{folded_membrane}	
		\end{center}
	\end{figure}

	Now consider the plaquette energy terms. The plaquettes that may be affected are those where some of the edges on the plaquette are cut by the dual membrane of the membrane operator. We divide such plaquettes into two classes. Firstly, we have the plaquettes which are pierced by the boundary of the dual membrane (the plaquettes which we claimed were excited by the membrane operator in Section \ref{Section_3D_Tri_Trivial_Magnetic_Excitations} of the main text). For these plaquettes, only one edge is cut by the dual membrane (assuming the dual membrane does not fold in on itself and cut the same plaquettes or edges multiple times, as in the case shown in Figure \ref{folded_membrane}). We refer to such plaquettes as boundary plaquettes. Secondly, we have the internal plaquettes, where the dual membrane cuts through two edges. We will see that the effect of the membrane operator on these two edges cancels in the expression for the plaquette holonomy and thus leaves the plaquette energy term unaffected. Consider such an internal plaquette cut by the membrane, with two edges affected by the membrane operator, as shown in Figure \ref{internal_plaquette_mag_membrane}. We are free to choose the base-point and orientation of the plaquette without affecting the energy term (as we showed in the Appendix of Ref. \cite{HuxfordPaper1}), so the plaquette in the figure is generic. The plaquette holonomy is then given by
	$$H_1(p)= \partial(e_p)g_1g_x g_2^{- 1} g(v_1-v_2)^{-1}.$$
	Under the action of the magnetic membrane operator $C^h(m)$, this becomes
	$$H_1(p) \rightarrow \partial(e_p) g(s.p-v_1)^{-1}hg(s.p-v_1) g_1 g_x g_2^{- 1} g(s.p-v_2)^{-1}h^{-1}g(s.p-v_2)g(v_1-v_2)^{-1}.$$

	Note that the path $(s.p-v_1) \cdot (v_1-v_2) \cdot (s.p-v_2)^{-1}$ is a closed loop. As long as this closed loop forms the boundary of a fake-flat surface (before we apply the transform), we know that $\partial(e)g(s.p-v_1)g(v_1-v_2)g(s.p-v_2)^{-1}=1_G$ for some $e \in E$ from the fake-flatness condition. This means that $g(s.p-v_2)^{-1}=g(v_1-v_2)^{-1}g(s.p-v_1)^{-1}\partial(e)^{-1}$. Making this substitution in the expression for the plaquette holonomy above gives
	\begin{align*}
		H_1(p)&\rightarrow \partial(e_p) g(s.p-v_1)^{-1}hg(s.p-v_1) g_1 g_x g_2^{- 1} \big(g(v_1-v_2)^{-1} g(s.p-v_1)^{-1}\partial(e)^{-1}\big) h^{-1}\\
		& \hspace{7cm} \big(\partial(e) g(s.p-v_1)g(v_1-v_2)\big) g(v_1-v_2)^{-1}
	\end{align*}
	Then using the fact that elements of $\partial(E)$ are in the centre of $G$, we can cancel the factors of $\partial(e)$ and its inverse, to obtain
	\begin{align*}
		H_1(p)&\rightarrow g(s.p-v_1)^{-1}hg(s.p-v_1) \partial(e_p) g_1 g_x g_2^{- 1} g(v_1-v_2)^{-1} g(s.p-v_1)^{-1}h^{-1} g(s.p-v_1)\\
		&=\big(g(s.p-v_1)^{-1}hg(s.p-v_1)\big) \big(\partial(e_p) g_1 g_x g_2^{-1} g(v_1-v_2)^{-1}\big) \big(g(s.p-v_1)^{-1}h^{-1} g(s.p-v_1)\big)\\
		&=\big(g(s.p-v_1)^{-1}hg(s.p-v_1)\big)H_1(p) \big(g(s.p-v_1)^{-1}hg(s.p-v_1)\big)^{-1},
	\end{align*}
	from which we see that the plaquette holonomy is simply conjugated by an element $g(s.p-v_1)^{-1}hg(s.p-v_1)$ by the action of the membrane operator. Conjugation preserves the identity element, so the magnetic membrane operator does not excite this internal plaquette. This means that the magnetic membrane operator $C^h(m)$ can only excite the boundary plaquettes.

	For these boundary plaquettes, only one edge on the plaquette is affected by the membrane operator, and so the plaquette holonomy only gains one factor of $g(s.p-v_1)^{-1}hg(s.p-v_1)$ or the inverse, meaning that there is no cancellation and the plaquette is excited unless $h=1_G$. In particular, the plaquette holonomy becomes
	\begin{align}
		H_1(p)&\rightarrow \big(g(s.p-v_1)^{-1}hg(s.p-v_1)\big)H_1(p) \label{Equation_action_magnetic_boundary_plaquette}
	\end{align}
	or 
	\begin{align*}
		H_1(p)&\rightarrow \big(g(s.p-v_1)^{-1}hg(s.p-v_1)\big)H_1(p),
	\end{align*}
	depending on which edge on the plaquette is cut by the membrane operator and the orientation of the plaquette.

	We should note that our argument that the internal plaquettes are not excited by the membrane operator relies on the path $(s.p-v_1) \cdot (v_1-v_2) \cdot (s.p-v_2)^{-1}$ enclosing a fake-flat surface. If this is not the case (for example, if these paths enclose an excited plaquette, created by another magnetic excitation) then the plaquette may be excited by the membrane operator. We already discussed some such cases in Section \ref{Section_linking} of the main text, where we consider the case where two magnetic flux tubes link together, and we will examine this linking phenomenon in more detail in Section \ref{Section_linking_appendix}. In using Figure \ref{internal_plaquette_mag_membrane}, we also implicitly made some assumptions about the shape of the membrane, in particular that it does not fold over and cut the same plaquette more than once. An example of a case where the membrane does cut the same plaquette more than once is shown in Figure \ref{folded_membrane}. However, we do not need to make such an assumption. These cases can be dealt with by separately considering each instance where the membrane cuts through the plaquette and then treating each instance in the way we have previously in this section (by pairing the edges cut each time the membrane passes through the plaquette and noting that the effect of such a pair on the plaquette holonomy cancels).


	The final energy terms to consider are the blob energy terms. When $\rhd$ is trivial the blob energy term only acts on the plaquette labels, which are not affected by the magnetic membrane operator. This means that the blob energy term commutes with the magnetic membrane operator. Combining this with our previous results for the commutation relations between the magnetic membrane operator and the energy terms, we see that the membrane operator only excites the boundary plaquettes around the membrane (the plaquettes pierced by the boundary of the dual membrane) and potentially the start-point vertex (depending on which linear combination of magnetic membrane operators we take).
	
	\subsubsection{Changing the start-point}
	\label{Section_magnetic_tri_trivial_move_sp}
	We have seen that the action of the magnetic membrane operator depends on our choice for a privileged vertex, which we called the start-point. The action of the magnetic membrane operator on an edge $i$ cut by the dual membrane is given by
	\begin{equation*}
		C^h(m):g_i = \begin{cases} g(s.p(m)-v_i)^{-1}hg(s.p(m)-v_i)g_i & \text{if $i$ points away from the direct membrane} \\ g_ig(s.p(m)-v_i)^{-1}h^{-1}g(s.p(m)-v_i) & \text{if $i$ points towards the direct membrane,} \end{cases}
	\end{equation*}
	which depends on the start-point through the path element $g(s.p(m)-v_i)$. It is therefore interesting to consider what happens when we change this start-point. Imagine moving the start-point from $s.p(m)$ to a new position, $s.p(m')$, along a path $(s.p(m)-s.p(m'))$. We denote the membrane with this new start-point by $m'$. The action of the membrane operator on an edge $i$ cut by the dual membrane is then
	\begin{equation*}
		C^h(m'):g_i = \begin{cases} g(s.p(m')-v_i)^{-1}hg(s.p(m')-v_i)g_i & \text{if $i$ points away from the direct membrane} \\ g_ig(s.p(m')-v_i)^{-1}h^{-1}g(s.p(m')-v_i) & \text{if $i$ points towards the direct membrane.} \end{cases}
	\end{equation*}
	
	When we move the start-point along $(s.p(m)-s.p(m'))$, we drag all of the paths $(s.p(m)-v_i)$ with the start-point, so that the new paths are $(s.p(m')-v_i)=(s.p(m')-s.p(m)) \cdot (s.p(m)-v_i) = (s.p(m)-s.p(m'))^{-1} \cdot (s.p(m)-v_i)$. This means that the path element $g(s.p(m')-v_i)$ can be written as $g(s.p(m')-v_i)=g(s.p(m)-s.p(m'))^{-1}g(s.p(m)-v_i)$. Substituting this into the action of $C^h(m')$, we see that
	\begin{align*}
		&C^h(m'):g_i = \begin{cases} g(s.p(m')-v_i)^{-1}hg(s.p(m')-v_i)g_i & \text{if $i$ points away from the direct membrane} \\ g_ig(s.p(m')-v_i)^{-1}h^{-1}g(s.p(m')-v_i) & \text{if $i$ points towards the direct membrane} \end{cases}\\
		&=\begin{cases} (g(s.p(m)-s.p(m'))^{-1}g(s.p(m)-v_i))^{-1}h(g(s.p(m)-s.p(m'))^{-1}g(s.p(m)-v_i))g_i & \text{if $i$ points away from} \\ & \text{the direct membrane} \\ g_i (g(s.p(m)-s.p(m'))^{-1}g(s.p(m)-v_i))^{-1}h^{-1}(g(s.p(m)-s.p(m'))^{-1}g(s.p(m)-v_i)) & \text{if $i$ points towards} \\ & \text{the direct membrane} \end{cases}\\
		&=\begin{cases} g(s.p(m)-v_i))^{-1}g(s.p(m)-s.p(m'))hg(s.p(m)-s.p(m'))^{-1}g(s.p(m)-v_i)g_i & \text{if $i$ points away from} \\ & \text{the direct membrane} \\ g_i g(s.p(m)-v_i)^{-1}g(s.p(m)-s.p(m'))h^{-1}g(s.p(m)-s.p(m'))^{-1}g(s.p(m)-v_i) & \text{if $i$ points towards}\\ & \text{the direct membrane} \end{cases}\\
		&=\begin{cases} g(s.p(m)-v_i)^{-1}[g(s.p(m)-s.p(m'))hg(s.p(m)-s.p(m'))^{-1}]g(s.p(m)-v_i)g_i & \text{if $i$ points away from} \\ & \text{the direct membrane} \\ g_i g(s.p(m)-v_i)^{-1}[g(s.p(m)-s.p(m'))h^{-1}g(s.p(m)-s.p(m'))^{-1}]g(s.p(m)-v_i) & \text{if $i$ points towards} \\ & \text{the direct membrane} \end{cases}\\
		&=C^{g(s.p(m)-s.p(m'))hg(s.p(m)-s.p(m'))^{-1}}(m):g_i,
	\end{align*}
	where in the last line we wrote the action in terms of a membrane operator on the original membrane $m$ (with original start-point $s.p(m)$). This holds for the action on all edges $i$, so we can write this in terms of the operators as
	\begin{equation}
		C^h(m')=C^{g(s.p(m)-s.p(m'))hg(s.p(m)-s.p(m'))^{-1}}(m)
		\label{Equation_magnetic_membrane_change_sp_1}
	\end{equation}
	
	We see that moving the start-point of the membrane operator is equivalent to changing the label of the operator, by conjugating the label by the path along which we move the start-point. Note the similarity with the transformation of the magnetic membrane operator under commutation with the vertex transform at the start-point, as described by Equation \ref{Equation_magnetic_membrane_start_point_transform}. Commutation with the vertex transform also conjugates the label of the magnetic membrane operator, in this case by the label of the vertex transform. As we discussed in Ref. \cite{HuxfordPaper1}, applying a vertex transform $A_v^g$ on a vertex is similar to parallel transporting that vertex along an edge labelled by $g^{-1}$, and so it is not surprising that moving the start-point of the membrane operator should have a similar effect to applying a vertex transform at the start-point. This also means that a magnetic membrane operator that is invariant under the vertex transforms (i.e., a magnetic membrane operator that produces no excitation at the start-point) is also invariant under a process by which we move the start-point.

	So far, we have considered moving the start-point of the membrane operator while keeping its label fixed, and have seen how this changes the action of the membrane operator. Just as we did with the blob ribbon operators in Section \ref{Section_blob_ribbon_move_sp}, it is also useful to consider how we can change the label as we move the start-point in order to keep the action of the membrane operator fixed. That is we want to find the label $x$ such that $C^x(m')=C^h(m)$. To find $x$, we use Equation \ref{Equation_magnetic_membrane_change_sp_1} and invert the transformation implied by that equation, to obtain
	\begin{equation}
		C^{g(s.p(m)-s.p(m'))^{-1}hg(s.p(m)-s.p(m'))}(m')=C^{h}(m). \label{Equation_magnetic_membrane_change_sp_2}
	\end{equation}

	\subsection{Magnetic membrane operators in the case where $\partial \rightarrow$ centre($G$) and $E$ is Abelian}
	\label{Section_Magnetic_Tri_Non_Trivial}

	We now consider the magnetic excitations in the case where $E$ is Abelian and $\partial(E)$ is in the centre of $G$ (Case 2 from Table \ref{Table_Cases} of the main text). Restricting the crossed module in this way allows us to simplify many expressions that we use throughout this section. Firstly, recall that the crossed module satisfies the Peiffer condition $\partial(e) \rhd f=efe^{-1}$ for all elements $e, \ f \in E$. When the group $E$ is Abelian, this relation simplifies to $\partial(e) \rhd f=f$. This has important consequences when we consider expressions such as $g(t) \rhd f$, where $g(t)$ is a path element. Two paths, $t_1$ and $t_2$, which can be deformed into one-another over a fake-flat surface differ in label only by some element $\partial(e) \in \partial(E)$. This is because the closed path $t_1t_2^{-1}$ is a closed path enclosing a fake-flat surface and so must have a label in $\partial(E)$. This means that, for any element $f \in E$, the two path elements satisfy $g(t_1)\rhd f = (\partial(e)g(t_2)) \rhd f = g(t_2) \rhd f$. Therefore, whenever we have a path element acting on a surface label in this way, we can freely deform the path that we are considering without changing the result, as long as we only deform the path across a fake-flat surface. One consequence of this and the Abelian nature of $E$ is that the calculation of the 2-holonomy of a blob becomes simple. For a blob $B$, based at a vertex $v_0(B)$, the blob 2-holonomy $H_2(B)$ is given by
	\begin{equation}
		H_2(B)=\prod_{\substack{\text{plaquettes } p\\ \in \text{Bd}(B)}} g(v_0(B)-v_{0}(p)) \rhd e_p^{\sigma_p},
		\label{Equation_blob_2_holonomy_appendix_1}
	\end{equation}
	where $\sigma_p$ is $+1$ if the plaquette's orientation (determined using the right-hand rule) points outwards from the blob and $\sigma_p$ is $-1$ if the orientation points inwards. In the case of the most general crossed module, the order of the product and the choice of paths for each $v_0(B)-v_0(p)$ are important (and should follow the rules for combining surfaces that we summarized in Section \ref{Section_Recap_3d}). However, because $E$ is Abelian, the order of multiplication is irrelevant. In addition, because we can deform the paths without affecting $g(v_0(B)-v_0(p)) \rhd e_p$, the precise choice of path does not matter provided that the region around the blob satisfies fake-flatness.

	Another useful result arises from the other Peiffer condition $\partial(g \rhd e) = g \partial(e) g^{-1}$ for all $g \in G$ and $e \in E$. Because we have restricted our crossed module so that $\partial(E)$ is within the centre of $G$, this Peiffer condition implies that $\partial(g\rhd e)=\partial(e)$. This means that expressions of the form $[h \rhd e] \: e^{-1}$ are in the kernel of $\partial$ for any $h \in G$ and $e \in E$. Another consequence of this restriction on $\partial$ is that, given a path $t$ and an expression of the form $g(t)^{-1}hg(t)$, we can freely deform the path $t$ over a fake-flat region without changing the expression. This is because deforming the path over such a region just changes the path label by an element $\partial(e)$. Then, because $\partial(e)$ is in the centre of $G$, the contributions of this factor to $g(t)$ cancels with its contribution to $g(t)^{-1}$:
	$$g(t)^{-1}hg(t) \rightarrow g(t)^{-1}\partial(e)^{-1}h\partial(e)g(t)=g(t)^{-1}hg(t).$$
	
	\subsubsection{Definition of the magnetic membrane operator}
	\label{Section_Magnetic_Tri_Non_Trivial_Definition}
	Having considered some of the consequences of restricting the crossed module in this way, we can discuss the magnetic membrane operator in this case. In order to define the magnetic membrane operator, we define two operators. Firstly we introduce the operator $C^h_{\rhd}(m)$. This operator acts on edge labels in the same way as the magnetic membrane operator $C^h(m)$ from the $\rhd$ trivial case (see Equation \ref{Equation_magnetic_membrane_on_edges_appendix} in Section \ref{Section_Magnetic_Membrane_Tri_trivial}), but also affects the surface labels of plaquettes that are cut by the dual membrane. For an unwhiskered plaquette that is cut by the dual membrane, the base-point of the plaquette either lies on the direct membrane or away from it, in which case it is separated from the direct membrane by the dual membrane. If a plaquette cut by the dual membrane of the operator has its base-point on the direct membrane of the membrane operator then the label of the plaquette goes from $e_p$ to $(g(s.p-v_0(p))^{-1}hg(s.p-v_0(p))) \rhd e_p$, where $(s.p-v_0(p))$ is a path from the start-point of the membrane operator to the base-point of the plaquette. If the base-point of the plaquette is away from the direct membrane, so that the dual membrane lies between the base-point and the direct membrane, then it is not affected. That is, the action of $C^h_{\rhd}(m)$ on an unwhiskered plaquette $p$ cut by the dual membrane of $m$ is
	\begin{equation}
		C^h_{\rhd}(m): e_p = \begin{cases} (g(s.p-v_0(p))^{-1}hg(s.p-v_0(p))) \rhd e_p & \text{ if $v_0(p)$ lies on the direct membrane} \\ e_p & \text{ otherwise,} \end{cases} \label{Equation_magnetic_membrane_rhd_action_appendix}
	\end{equation}
	where $v_0(p)$ is the base-point of plaquette $p$. This action is reminiscent of the vertex transforms, which only affect the surface labels of plaquettes which are based at the vertex on which we apply the transform. Note that we specified an unwhiskered plaquette, meaning plaquettes that do not have sections of boundary that enclose no surface and can be removed by moving the start-point. If we consider a whiskered plaquette, then the base-point of the plaquette could lie anywhere, and in particular the base-point could be whiskered through the direct membrane (on the other side to the dual membrane), and so neither lie on the direct membrane nor are separated from the direct membrane by the dual membrane. We will not consider such plaquettes directly, but we can define the action of the magnetic membrane operator on such plaquettes indirectly by demanding that the action of the membrane operator be consistent with the ``re-branching" procedure from Ref. \cite{HuxfordPaper1} where we move the base-point of a plaquette along path $t$ and simultaneously change its label from $e_p$ to $g(t)^{-1} \rhd e_p$ (we will show in Section \ref{Section_consistency_magnetic_membrane_operator} that the action of the magnetic membrane operator on unwhiskered plaquettes defined by Equation \ref{Equation_magnetic_membrane_rhd_action_appendix} is consistent with this procedure, but will not consider moving the base-point through the direct membrane).

\begin{figure}[h]
		\begin{center}
			\begin{overpic}[width=0.7\linewidth]{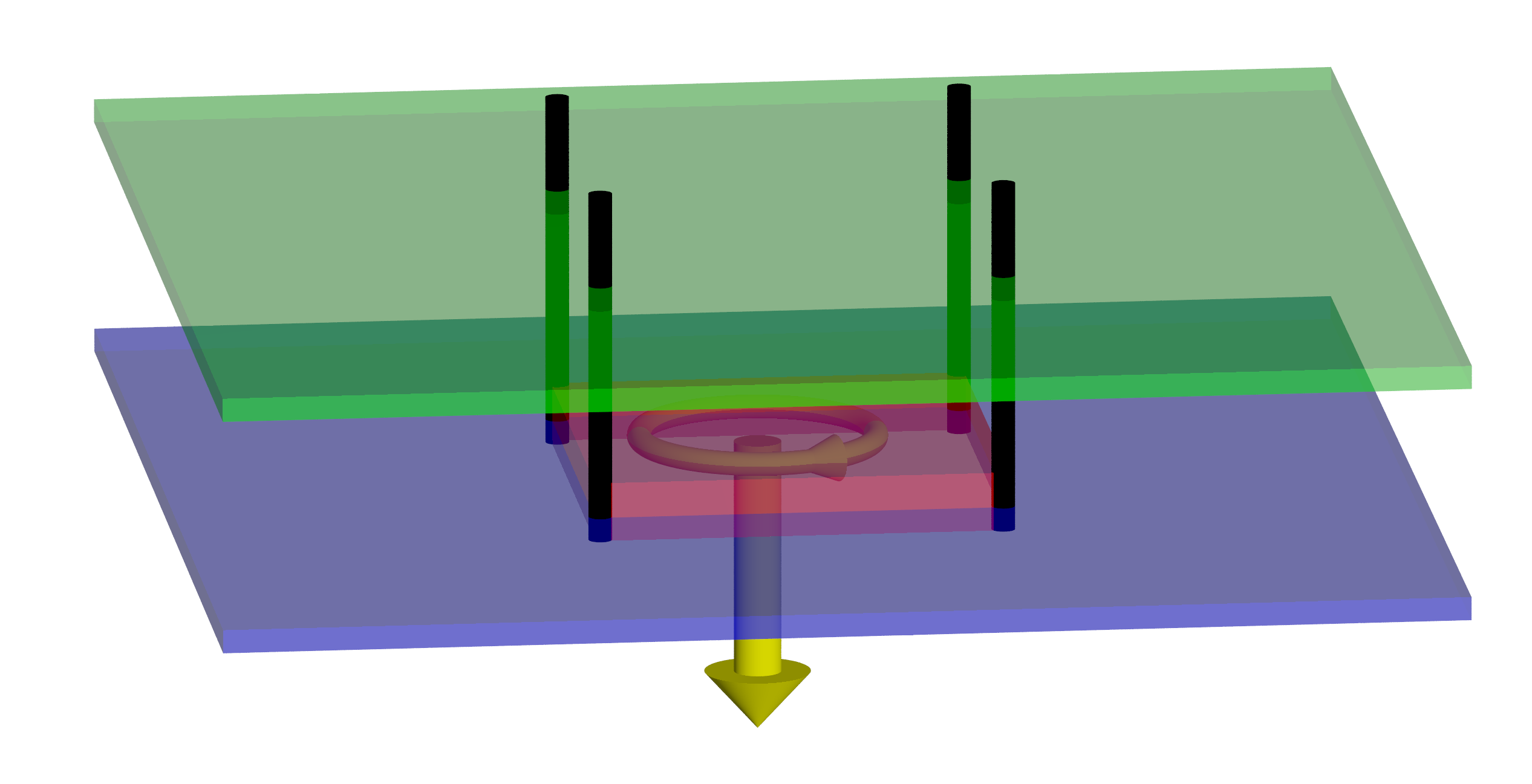}
				\put(55,5){orientation of plaquette}
				\put(98,10){direct membrane}
				\put(98,27){dual membrane}
			\end{overpic}
			\caption{The label of the blob ribbon operator depends on the label and orientation of the corresponding plaquette. The orientation of the plaquette is determined from its circulation using the right-hand rule. If the orientation of the plaquette is ``downwards'', away from the dual membrane as shown in the figure, then the label $f(p)$ of the blob ribbon operator is $[g(s.p-v_{0}(p))\rhd e_p] \: (h^{-1}g(s.p-v_{0}(p)))\rhd e_p^{-1}$. If the orientation is instead ``upwards'', then $e_p$ must be replaced with its inverse, which inverts $f(p)$.}
			\label{modmembraneorientationappendix}
		\end{center}
	\end{figure}

	Despite including this action on the plaquette labels, $C^h_{\rhd}(m)$ still does not have the desired commutation relations with the energy terms of our model. In particular, it would not commute with either the edge energy terms or the blob energy terms near the bulk of the membrane. Therefore, we must further modify the membrane operator to have the desired commutation relations. In order to construct the full operator, we must multiply $C^h_{\rhd}(m)$ by a series of blob ribbon operators, to produce the final magnetic membrane operator, which we denote by $C^h_T(m)$. Each of these blob ribbon operators acts along a ribbon that starts at the same blob, which we call blob 0. In the same way that we must specify the start-point of the membrane when we define $m$, we must define this blob 0. It is convenient to choose blob 0 to be attached to the start-point, but this is not necessary. We have one blob ribbon operator for each plaquette $p$ on the direct membrane of the membrane operator. Each such plaquette $p$ is attached to a blob, which we call blob $p$, which is cut through by the dual membrane. Then the ribbon for the ribbon operator associated to plaquette $p$ runs from blob 0 to blob $p$, and the start-point for the direct path of each such ribbon is the start-point of the membrane. An example of such a ribbon operator, with the various feature needed to define it, is illustrated in Figure \ref{blobsonmem} in Section \ref{Section_3D_MO_Central} of the main text. The label of the blob ribbon operator associated to a plaquette $p$, with base-point $v_0(p)$, is 
	\begin{equation}
		f(p)=[g(s.p-v_{0}(p))\rhd e_p^{\sigma_p}] \: [(h^{-1}g(s.p-v_0(p)))\rhd e_p^{-\sigma_p}],
		\label{Equation_magnetic_blob_ribbon_label}
	\end{equation}
	where $e_p$ is the label of the plaquette $p$, and $\sigma_p$ is $+1$ if the orientation of the plaquette, determined using the right hand rule, points away from the dual membrane as shown in Figure \ref{modmembraneorientationappendix} and is $-1$ if the orientation is the opposite to that shown in Figure \ref{modmembraneorientationappendix}. We note that the label $f(p)$ of these blob ribbon operators is in the kernel of $\partial$ because, as we discussed at the start of Section \ref{Section_Magnetic_Tri_Non_Trivial}, $\partial(e)=\partial(g \rhd e)$ for any $e \in E$ and $g \in G$ when $\partial$ maps to the centre of $G$. Therefore, $\partial(f_p) =\partial(e_p^{\sigma_p})\partial(e_p^{-\sigma_p})=1_G$. Including these blob ribbon operators, the total magnetic membrane operator is given by
	\begin{equation}
		C^h_T(m) = C^h_{\rhd}(m) \prod_{\substack{\text{plaquette }p \\ \text{on membrane}}} B^{f(p)}(\text{blob }0 \rightarrow \text{blob }p), \label{Equation_total_magnetic_membrane_definition_1}
	\end{equation}
	where $v_0(p)$ is the base-point of plaquette $p$ and $\text{blob }0 \rightarrow \text{blob }p$ is shorthand for the ribbon that starts at blob 0 and terminates in blob $p$, with start-point at the start-point of the membrane.

	\subsubsection{Consistency of the magnetic membrane operator with the re-branching procedures}
	\label{Section_consistency_magnetic_membrane_operator}
	Having defined our magnetic membrane operator we shall now consider commutation of the magnetic membrane operator with the energy terms. Due to the complexity of the magnetic membrane operator in the case where $\rhd$ is non-trivial but $\partial$ maps to the centre of $G$, it is inconvenient to consider every possible branching structure when examining the commutation relations between the energy terms and the edge transforms. However, we know from the Appendix of Ref. \cite{HuxfordPaper1} that the energy terms are invariant under various re-branching procedures, where we change certain features of the lattice, such as the orientations of edges and plaquettes, while simultaneously changing the labels on the lattice. In Ref. \cite{HuxfordPaper1} we explained that, if a membrane operator is also invariant under such procedures, we can check the commutation relations between the membrane operator and the energy terms with a particular choice of branching structure. The commutation relation would then hold for any choice of branching structure. Therefore, we wish to show that the magnetic membrane operator is indeed invariant under these procedures. That is, for each re-branching procedure $X$ from Ref. \cite{HuxfordPaper1} and magnetic membrane operator $C^h_T(m)$, we wish to show that $X^{-1}C^h_T(m)X=C^h_T(m)$.

	The first procedure to consider is denoted by $P_i$, where we flip the direction of an edge $i$ and simultaneously invert the label of that edge. Edge labels are relevant for the magnetic membrane operator in two ways. Firstly, the action of the membrane operator on various degrees of freedom depends on path labels (for example, through expressions such as $g(t)^{-1}hg(t)$). However, we showed in the Appendix of Ref. \cite{HuxfordPaper1} that path labels are invariant under the action of $P_i$, so $P_i$ will not affect the action of the magnetic membrane in this way. The second way in which $P_i$ could affect $C^h_T(m)$ is by flipping one of the edges that are cut by the dual membrane and so are acted on directly by the magnetic membrane operator. Recall that the action of the membrane operator $C^h_T(m)$ on such an edge $i$, with label $g_i$ is
	$$C^h_T(m): g_i = \begin{cases} g(s.p-v_i)^{-1}hg(s.p-v_i)g_i &\text{if } i \text{ points away from the direct membrane} \\ g_i g(s.p-v_i)^{-1}h^{-1}g(s.p-v_i) &\text{if } i \text{ points towards the direct membrane.} \end{cases}$$
	Therefore,
	\begin{align*}
		P_i^{-1}C^h_T(m)P_i :g_i & = P_i^{-1}C^h_T(m): g_i^{-1}\\
		&= P_i^{-1} \begin{cases} g_i^{-1} g(s.p-v_i)^{-1}h^{-1}g(s.p-v_i) &\text{if } i \text{ originally points away from the direct membrane} \\ g(s.p-v_i)^{-1}hg(s.p-v_i)g_i^{-1} &\text{if } i \text{ originally points towards the direct membrane.} \end{cases}\\
		&=\begin{cases} g(s.p-v_i)^{-1}hg(s.p-v_i)g_i &\text{if } i \text{ points away from the direct membrane} \\ g_i g(s.p-v_i)^{-1}h^{-1}g(s.p-v_i) &\text{if } i \text{ points towards the direct membrae.} \end{cases}\\
		&=C^h_T(m):g_i.
	\end{align*}
	
	We therefore see that the action of the magnetic membrane operator is consistent with flipping edges. The next procedure to consider is the one where we flip the orientation of a plaquette, while simultaneously inverting its label. We denote this operation, acting on a plaquette $p$, by $Q_p$. The magnetic membrane operator acts on the plaquette labels in three ways. Firstly, the labels of plaquettes in the direct membrane determines the label of the blob ribbon operators that we combine with $C^h_\rhd(m)$ to produce the full membrane operator, $C^h_T(m)$. Recall that the label of the blob ribbon operator corresponding to the plaquette $p$ in the direct membrane is 
	$$f(p)= [g(s.p-v_0(p))\rhd e_p^{\sigma_p}] \: (h^{-1}g(s.p-v_{0}(p)))\rhd e_p^{-\sigma_p},$$
	
	where $\sigma_p$ is $1$ if the plaquette $p$ points ``downwards", away from the dual membrane. If the plaquette $p$ points upwards, $\sigma_p$ is $-1$, which means that we must replace $e_p$ with its inverse in the expression for $f(p)$. Therefore, we can see that the label $f(p)$ is constructed in such a way that it is invariant under the operation $Q_p$. Reversing the orientation of the plaquette means that we must swap $e_p$ with $e_p^{-1}$ in the expression for $f(p)$. However, the procedure $Q_p$ also directly inverts the label $e_p$ of the plaquette. These two effects cancel out, leaving $f(p)$ unchanged when we apply $Q_p$.

	The second way in which $C^h_T(m)$ acts on plaquettes is through the action of the blob ribbon operators on the plaquettes pierced by the ribbon. Recall that the action of a blob ribbon operator $B^e(t)$ on a plaquette $p$ pierced by the ribbon depends on the orientation of the plaquette (determined from the circulation of the plaquette using the right-hand rule) compared to the direction of the ribbon:
	$$B^e(t):e_p = \begin{cases} e_p \: [g(s.p-v_0(p))^{-1} \rhd e^{-1}] &\text{if } p \text{ matches the direction of the ribbon}\\ [g(s.p-v_0(p))^{-1} \rhd e] \: e_p &\text{if } p \text{ is anti-aligned with the direction of the ribbon.} \end{cases}$$
	Therefore
	\begin{align*}
		Q_p^{-1}B^e(t)Q_p: e_p &= Q_p^{-1}B^e(t): e_p^{-1}\\
		&= \begin{cases}[g(s.p-v_0(p))^{-1} \rhd e] \: e_p^{-1} &\text{if } p \text{ originally matches the direction of the ribbon}\\ e_p^{-1} \: [g(s.p-v_0(p))^{-1} \rhd e^{-1}] &\text{if } p \text{ is originally anti-aligned with the direction of the ribbon} \end{cases}\\
		&= \begin{cases} e_p \: [g(s.p-v_0(p))^{-1} \rhd e^{-1}] &\text{if } p \text{ matches the direction of the ribbon}\\ [g(s.p-v_0(p))^{-1} \rhd e] \:e_p &\text{if } p \text{ is anti-aligned with the direction of the ribbon} \end{cases}\\
		&=B^e(t):e_p,
	\end{align*}
	from which we see that blob ribbon operators are invariant under $Q_p$. The third way in which the magnetic membrane operator interacts with the plaquettes is through the $\rhd$ action of $C^h_{\rhd}(m)$ on plaquettes. Recall that the action of $C^h_{\rhd}(m)$ on a plaquette $p$ that is cut by the dual membrane is
	$$C^h_{\rhd}(m): e_p = \begin{cases} \big(g(s.p-v_0(p))^{-1}hg(s.p-v_0(p))\big) \rhd e_p & \text{ if $v_0(p)$ lies on the direct membrane} \\ e_p & \text{ otherwise.} \end{cases}$$
	
	This means that
	\begin{align*}
		Q_p^{-1}C^h_{\rhd}(m)Q_p: e_p &= Q_p^{-1}C^h_{\rhd}(m): e_p^{-1}\\
		&= Q_p^{-1}: \begin{cases} \big(g(s.p-v_0(p))^{-1}hg(s.p-v_0(p))\big) \rhd e_p^{-1} & \text{ if $v_0(p)$ lies on the direct membrane} \\ e_p^{-1} & \text{ otherwise} \end{cases}\\
		&=\begin{cases} \big(g(s.p-v_0(p))^{-1}hg(s.p-v_0(p))\big) \rhd e_p & \text{ if $v_0(p)$ lies on the direct membrane} \\ e_p & \text{ otherwise} \end{cases}\\
		&= C^h_{\rhd}(m): e_p,
	\end{align*}
	from which we see that the action of $C^h_{\rhd}(m)$ is invariant under the plaquette flipping procedure. We have therefore shown that every part of $C^h_T(m)$ is invariant under this procedure.

\begin{figure}[h]
		\begin{center}
			\begin{overpic}[width=0.7\linewidth]{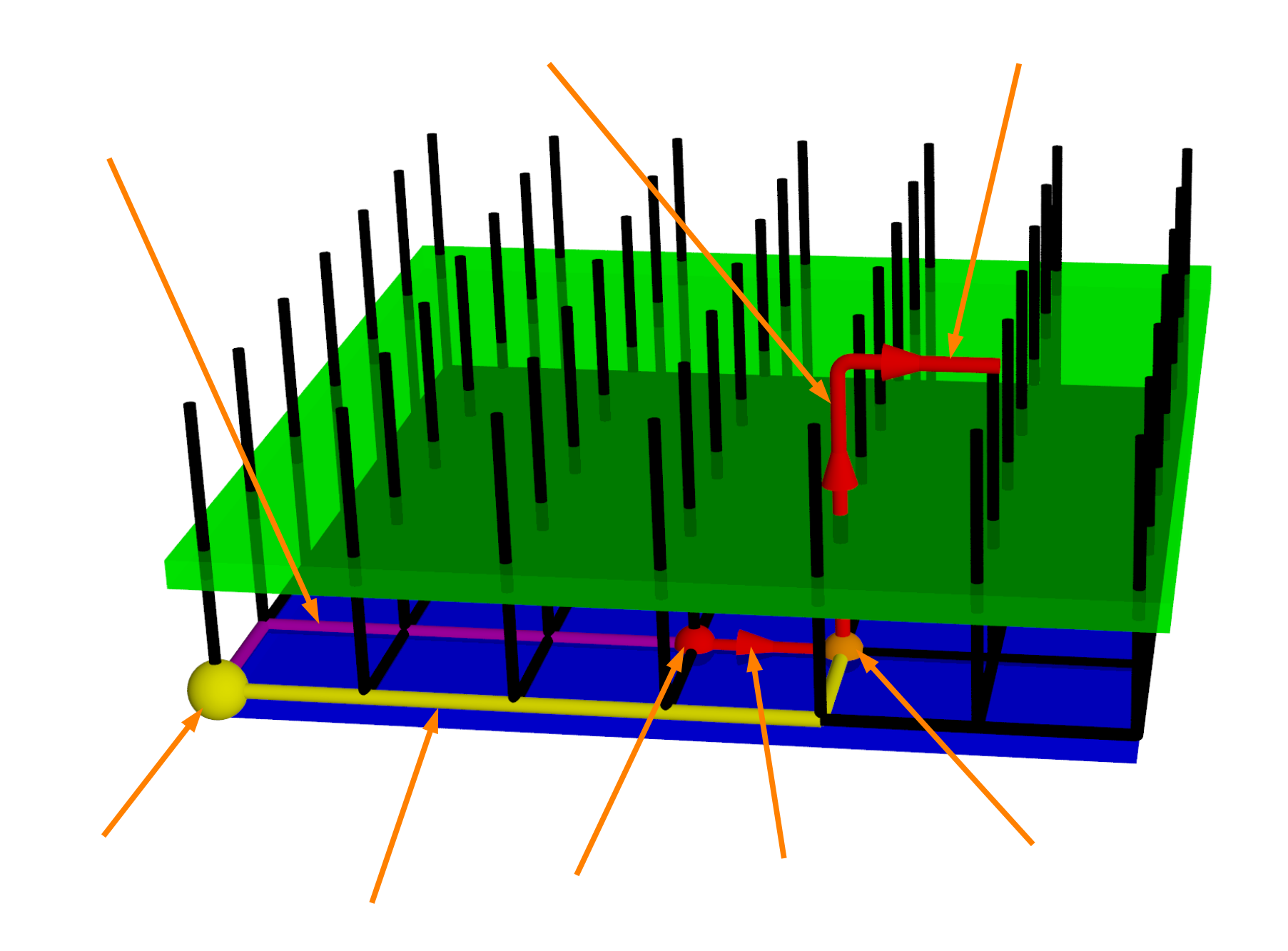}
				\put(4,8){$s.p$}
				\put(23,1){$(s.p-v_j)$}
				\put(42,4){$s.p(t)$}
				\put(61,6){$t_1$}
				\put(81,8){$v_j$}
				\put(3,62){$(s.p-s.p(t))$}
				\put(39,70){edge $j$}
				\put(79,70){$t_2$}	
			\end{overpic}
			\caption{We wish to consider how the magnetic membrane operator affects a path $t$ (red) that starts on the direct membrane at $s.p(t)$ and leaves the direct membrane by cutting through the dual membrane. To do this, we split the path $t$ into three parts. Edge $j$ is the edge which passes through the dual membrane and so is affected by the magnetic membrane operator (note that we have chosen an orientation for this edge, which we can do because the membrane operator is invariant under the edge-flipping procedure). $t_1$ is the path up to this edge and $t_2$ is the path after the edge. The yellow path $(s.p-v_j)$ determines the action of the membrane operator on the edge $j$. We find that the action of the membrane operator on the path $t$ depends on the path $(s.p-v_j)t_1^{-1}$, which can be smoothly deformed without affecting the action of the membrane operator (e.g., into the purple path $(s.p-s.p(t))$).}
			\label{magnetic_membrane_path_start}
		\end{center}
	\end{figure}

	The final procedure to consider is the one where we move the base-point of a plaquette $p$ along a path $t$, while simultaneously changing its label from $e_p$ to $g(t)^{-1} \rhd e_p$. We denote this operation by $E_p(t)$. We need to consider how this affects each of the ways that $C^h_T(m)$ interacts with the plaquettes, just as we did when considering the operation $Q_p$. To do this, we will need to understand how the magnetic membrane operator affects path elements, because when we apply $E_p(t)^{-1}$ to move the base-point of the plaquette back after acting with the membrane operator, we need to know how the element $g(t)$ is affected by the membrane operator. The membrane operator can affect any path $t$ which includes some of the edges cut by the dual membrane of the membrane operator. For example, consider a path which starts on the direct membrane and then passes through the dual membrane, as shown in Figure \ref{magnetic_membrane_path_start}. The path element is $g(t)=g(t_1)g_jg(t_2)$, where edge $j$ is the edge on the path $t$ which is cut by the dual membrane, $t_1$ is the section of $t$ up to this edge, and $t_2$ is the section after the edge. $C^h_T(m)$ affects this path by changing the edge element $g_j$ to $g(s.p-v_j)^{-1}hg(s.p-v_j)g_j$, where $v_j$ is the vertex attached to edge $j$ that lies on the direct membrane (the orange vertex in Figure \ref{magnetic_membrane_path_start}). Therefore, under the action of $C^h_T(m)$, the label $g(t)$ becomes
	\begin{align}
		C^h_T(m):g(t)&= g(t_1) g(s.p-v_j)^{-1}hg(s.p-v_j)g_j g(t_2) \notag\\
		&=\big(g(s.p-v_j)g(t_1)^{-1}\big)^{-1}h\big(g(s.p-v_j)g(t_1)^{-1}\big) g(t_1)g_j g(t_2)\notag\\
		&= \big(g(s.p-v_j)g(t_1)^{-1}\big)^{-1}h\big(g(s.p-v_j)g(t_1)^{-1}\big) g(t) \notag\\
		&=g(s.p-s.p(t))^{-1}hg(s.p-s.p(t)) g(t), \label{mag_membrane_path_start}
	\end{align}
	where $s.p(t)$ is the start of path $t$ and the path $s.p-s.p(t)$ is $(s.p-v_j)\cdot t_1^{-1}$. However, the exact path $s.p-s.p(t)$ does not matter if our initial state is fake-flat. This is because any path into which we can smoothly deform the path $s.p-s.p(t)$ over a fake-flat region will have a label which differs from $g(s.p-s.p(t))$ only by an element in $\partial(E)$. As discussed at the beginning of this section (Section \ref{Section_Magnetic_Tri_Non_Trivial}), such elements do not affect the expression $g(s.p-s.p(t))^{-1}hg(s.p-s.p(t))$ when $\partial(E)$ is in the centre of $G$.
	
\begin{figure}[h]
		\begin{center}
			\begin{overpic}[width=0.7\linewidth]{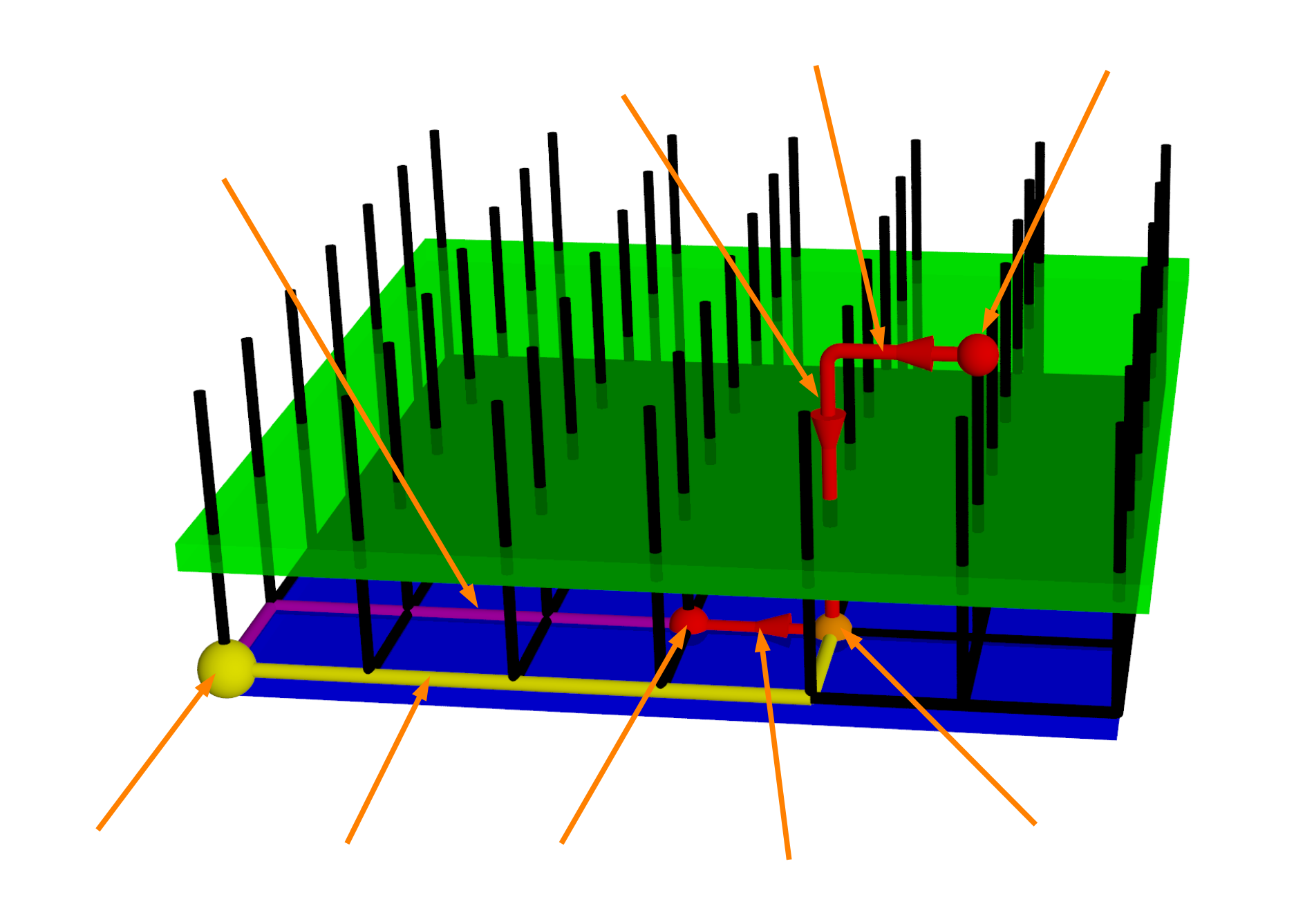}
				\put(6,6){$s.p$}
				\put(22,4){$(s.p-v_k)$}
				\put(40,4){$e.p(s)$}
				\put(59,3){$s_2$}
				\put(79,6){$v_k$}
				\put(10,58){$(s.p-e.p(s))$}
				\put(46,64){$k^{-1}$}
				\put(61,67){$s_1$}
				\put(82,66){$s.p(s)$}	
			\end{overpic}
			\caption{Just as we considered the effect of the magnetic membrane on a path that started on the direct membrane in Figure \ref{magnetic_membrane_path_start}, we now examine a path $s$ (red) that starts at $s.p(s)$ away from the direct membrane and cuts through the dual membrane to end on the direct membrane at $e.p(s)$. We split the path $s$ into three parts. Edge $k$ is the edge which passes through the dual membrane and so is affected by the magnetic membrane operator. We take the orientation of this edge to be opposite to the path $s$, so that the path $s$ includes $k^{-1}$. Then $s_1$ is the path up to this edge and $s_2$ is the path after the edge. That is, $s=s_1 \cdot k^{-1} \cdot s_2$. The action of the magnetic membrane on the edge $j$ depends on the value of a path from the start-point of the membrane, $s.p$, to the vertex $v_k$ attached to the edge and lying on the direct membrane. The result of this change to the edge is that $C^h_T(m) :g(s) =g(s)g(s.p-e.p(s))^{-1}h^{-1}g(s.p-e.p(s))$. }
			\label{magnetic_membrane_path_end}
		\end{center}
	\end{figure}

\begin{figure}[h]
		\begin{center}
			\begin{overpic}[width=0.6\linewidth]{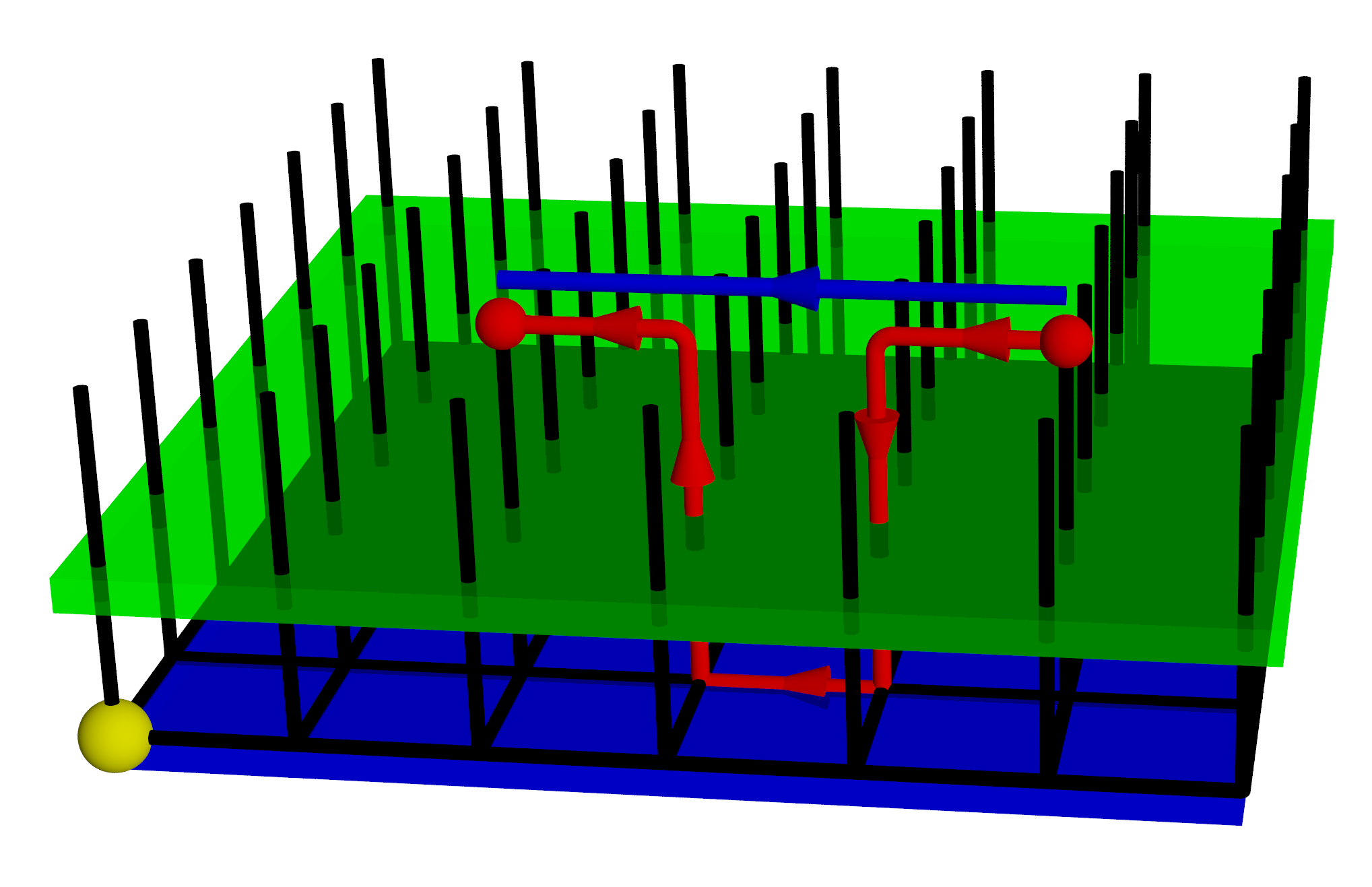}
				
			\end{overpic}
			\caption{Given a section of path, such as the (red) U-shaped one in this diagram, that passes through the dual membrane in both directions, the transformations from the two intersections with the dual membrane cancel. This means that the path transforms in the same way as a path which does not intersect with the dual membrane at all, such as the straight (blue) one in the diagram (we might also expect this from the fact that these two paths can be smoothly deformed into one-another over a fake-flat surface). }
			\label{magnetic_membrane_path_through}
		\end{center}
	\end{figure}

	Next consider a path $s$ which starts away from the direct membrane and then ends on the direct membrane, as shown in Figure \ref{magnetic_membrane_path_end}. In this case the path $g(s)=g(s_1)g_k^{-1}g(s_2)$. This transforms under the membrane operator as
	\begin{align}
		C^h_T(m):g(s) &= g(s_1)g_k^{-1} g(s.p-v_k)^{-1}h^{-1}g(s.p-v_k)g(s_2) \notag \\
		&=g(s_1)g_k^{-1} g(s_2) g(s_2)^{-1} g(s.p-v_k)^{-1}h^{-1}g(s.p-v_k)g(s_2) \notag\\
		&=g(s) \big(g(s.p-v_k)g(s_2)\big)^{-1}h^{-1}\big(g(s.p-v_k)g(s_2)\big).
	\end{align}
	Then note that $g(s_2)=g(v_k-e.p(s))$, where $e.p(s)$ is the end-point of $s$. Therefore
	\begin{align}
		C^h_T(m):g(s) &=g(s) \big(g(s.p-v_k)g(v_k-s.p(s))\big)^{-1}h^{-1}\big(g(s.p-v_k)g(v_k-e.p(s))\big) \notag\\
		&=g(s)g(s.p-e.p(s))^{-1}h^{-1}g(s.p-e.p(s)), \label{mag_membrane_path_end}
	\end{align}
	where $s.p-e.p(s)$ only needs to be defined up to smooth deformations. Note that we can also re-write this result as
	\begin{align}
		C^h_T(m):g(s) &= g(s)g(s.p-e.p(s))^{-1}h^{-1}g(s.p-e.p(s)) \notag \\
		&=\big(g(s.p-e.p(s))g(s)^{-1}\big)^{-1}h^{-1} \big(g(s.p-e.p(s))g(s)^{-1}\big)g(s) \notag \\
		&=\big(g(s.p-e.p(s))g(s.p(s)-e.p(s))^{-1}\big)^{-1}h^{-1} \big(g(s.p-e.p(s))g(s.p(s)-e.p(s))^{-1}\big)g(s) \notag \\
		&=g(s.p-s.p(s))^{-1}h^{-1}g(s.p-s.p(s))g(s), \label{mag_membrane_path_end_alternate}
	\end{align}
	which is similar to the transformation of the path $t$ that ends on the direct membrane, except that $h$ is replaced with $h^{-1}$. We can put these two transformations together to describe an arbitrary path $r$ which passes through the region cut by the dual membrane. Each time the path moves from the direct membrane out through the dual membrane, it gains a factor $g(s.p-s.p(r))^{-1}hg(s.p-s.p(r))$ to the left and each time it passes through the dual membrane from the outside to reach the direct membrane it gains a factor $g(s.p-s.p(r))^{-1}h^{-1}g(s.p-s.p(r))$ to the left. Note that if a path passes through the dual membrane twice in opposite directions, these transformations cancel. When considering the transformation of a path, this allows us to ignore sections which can be deformed so that they do not cross the membrane at all, as shown in Figure \ref{magnetic_membrane_path_through}. Instead we need to consider which side of the dual membrane the path starts on and which side it ends on (assuming the path does not go around the boundary of the dual membrane, but instead stays in the region near the bulk of the dual membrane like the paths we have considered so far).

	Having considered how the magnetic membrane operator interacts with path elements, we can now consider how the base-point changing procedure $E_p(t)$ interacts with the magnetic membrane operator. First consider how $E_p(t)$ interacts with the label $f(p)$ of the blob ribbon operator corresponding to plaquette $p$. This label is given by
	$$f(p)=[g(s.p-v_{0}(p))\rhd e_p] \: [\big(h^{-1}g({s.p}-v_{0}(p))\big)\rhd e_p^{-1}].$$
	Moving the base-point of plaquette $p$ means that we must change $v_0(p)$ (which is the original base-point of $p$) in the label $f(p)$ for the vertex $v_2$ which is at the end of the path $t$. When we move the base-point we also change the label $e_p$ to $g(v_0(p)-v_2)^{-1} \rhd e_p$. When we apply these two effects, the label $f(p)$ becomes
	$$[g(s.p-v_2)\rhd(g(v_2-v_0(p)) \rhd e_p)] \: [(h^{-1}g({s.p}-v_2)g(v_2-v_0(p)))\rhd e_p^{-1}].$$
	The path element $g({s.p}-v_2)g(v_2-v_0(p))$ is equal to $g(s.p-v_0(p))$, at least up to an element $\partial(k)$ for some $k \in E$. This is because, depending on the precise choice of paths, $(s.p-v_2) \cdot (v_2-v_0(p))$ may not be $(s.p-v_0(p))$ but can be deformed into it over a surface which will satisfy fake-flatness in the original state. Any additional factors of $\partial(k)$ do not affect the expression $g(s.p-v_0(p)) \rhd e_p$, as we discussed at the beginning of this section (Section \ref{Section_Magnetic_Tri_Non_Trivial}). From this we see that $E_p(t)$ does not affect the label $f(p)$ of the blob ribbon operators in $C^h_T(m)$.

	The next way in which $E_p(t)$ can interact with the membrane operators is directly through the blob ribbon operators. Recall that the action of a blob ribbon operator $B^e(t)$ on a plaquette $p$ pierced by the ribbon is
	$$B^e(t):e_p = \begin{cases} e_p \: [g(s.p-v_0(p))^{-1} \rhd e^{-1}] &\text{if } p \text{ matches the direction of the ribbon}\\ [g(s.p-v_0(p))^{-1} \rhd e] e_p &\text{if } p \text{ is anti-aligned with the ribbon,} \end{cases}$$
	which depends on the base-point of the plaquette $p$. Consider how this interacts with $E_p(v_0(p)-v_2)$. We have
	\begin{align*}
		E_p(t)^{-1}& B^e(t)E_p(t) :e_p\\
		&= E_p(t)^{-1} B^e(t): g(v_0(p)-v_2)^{-1} \rhd e_p\\
		&=E_p(t)^{-1}: \begin{cases} [g(v_0(p)-v_2)^{-1} \rhd e_p] \: [g(s.p-v_2)^{-1} \rhd e^{-1}] &\text{if } p \text{ is aligned with the ribbon}\\ [g(s.p-v_2)^{-1} \rhd e] \: [g(v_0(p)-v_2)^{-1} \rhd e_p] &\text{if } p \text{ is anti-aligned with the ribbon} \end{cases}\\
		&= \begin{cases}g(v_0(p)-v_2) \rhd\big( [g(v_0(p)-v_2)^{-1} \rhd e_p] \: [g(s.p-v_2)^{-1} \rhd e^{-1}]\big) &\text{if } p \text{ is aligned with the ribbon}\\g(v_0(p)-v_2) \rhd \big( [g(s.p-v_2)^{-1} \rhd e] \ [g(v_0(p)-v_2)^{-1} \rhd e_p]\big) &\text{if } p \text{ is anti-aligned with the ribbon} \end{cases}\\
		&= \begin{cases} [g(v_0(p)-v_2) \rhd \big(g(v_0(p)-v_2)^{-1} \rhd e_p\big)] \: [g(v_0(p)-v_2) \rhd \big(g(v_2-s.p) \rhd e^{-1}\big)] &\text{if } p \text{ is aligned with} \\ & \text{ the ribbon}\\ [g(v_0(p)-v_2) \rhd \big(g(v_2-s.p) \rhd e\big)] [g(v_0(p)-v_2) \rhd \big(g(v_0(p)-v_2)^{-1} \rhd e_p\big)] &\text{if } p \text{ is anti-aligned } \\ & \text{ with the ribbon} \end{cases}\\
		&=\begin{cases} e_p [\big(g(v_0(p)-v_2)g(v_2-s.p)\big) \rhd e^{-1}] &\text{if } p \text{ is aligned with the ribbon}\\ \big[\big(g(v_0(p)-v_2)g(v_2-s.p)\big) \rhd e\big] \: e_p &\text{if } p \text{ is anti-aligned with the ribbon} \end{cases}\\
		&= \begin{cases} e_p \: [g(s.p-v_0(p))^{-1} \rhd e^{-1}] &\text{if } p \text{ is aligned with the ribbon}\\ [g(s.p-v_0(p))^{-1} \rhd e] \: e_p &\text{if } p \text{ is anti-aligned with the ribbon,} \end{cases} 
	\end{align*}
	where $g(v_0(p)-v_2)g(v_2-s.p)=g(v_0(p)-s.p)$ (at least up to irrelevant factors in $\partial(E)$), so that
	$$E_p(t)^{-1} B^e(t)E_p(t) :e_p =B^e(t):e_p.$$
	That is, blob ribbon operators are invariant under the procedure of changing the base-point, provided that they act on a region which satisfies fake-flatness so that we do not need to worry about precisely defining the paths involved in each operator.

	Finally, we need to consider how changing the base-point of a plaquette affects the $\rhd$ action of $C^h_{\rhd}(m)$ on plaquettes cut by the dual membrane. This is interesting because the $\rhd$ action only affects plaquettes whose base-point lie on the direct membrane. This may have seemed somewhat arbitrary, but we will now show that this action is invariant under the operation of changing the base-point. From Equation \ref{Equation_magnetic_membrane_rhd_action_appendix}, the action on a plaquette $p$ cut by the dual membrane is
	\begin{equation*}
		C^h_{\rhd}(m): e_p = \begin{cases} (g(s.p-v_0(p))^{-1}hg(s.p-v_0(p))) \rhd e_p & \text{ if $v_0(p)$ lies on the direct membrane} \\ e_p & \text{ otherwise.} \end{cases}
	\end{equation*}
	
	Due to the prevalence of expressions such as $g(t)^{-1}hg(t)$ in the following argument, we define $h_{[g(t)]}=g(t)^{-1}hg(t)$. If we change the base-point of plaquette $p$ with $E_p(v_0(p)-v_2)$, then
	\begin{align*}
		E_p(v_0(p)-v_2)^{-1}&C^h_{\rhd}(m)E_p(v_0(p)-v_2): e_p\\
		&= E_p(v_0(p)-v_2)^{-1}C^h_{\rhd}(m): g(v_0(p)-v_2)^{-1} \rhd e_p\\
		&= E_p(v_0(p)-v_2)^{-1}: \begin{cases} h_{[g(s.p-v_2)]} \rhd (g(v_0(p)-v_2)^{-1} \rhd e_p) & \text{ if $v_2$ lies on the direct membrane} \\ g(v_0(p)-v_2)^{-1} \rhd e_p & \text{ otherwise.} \end{cases}
	\end{align*}
	When $E_p(v_0(p)-v_2)^{-1}$ moves the base-point back to $v_0(p)$ it is then important to keep track of how $C^h_{\rhd}(m)$ affected the path label for $v_0(p)-v_2$. If $v_0(p)$ is on the membrane, then because it is the start of the path $v_0(p)-v_2$, the path label $g(v_0(p)-v_2)$ gains a factor $g(s.p-v_0(p))^{-1}hg(s.p-v_0(p))=h_{[g(s.p-v_0(p))]}$ to the left, as demonstrated in Equation \ref{mag_membrane_path_start}. If $v_2$ is on the membrane, then because it is the end of the path $v_0(p)-v_2$, the path element $g(v_0(p)-v_2)$ gains a factor $g(s.p-v_2))^{-1}h^{-1}g(s.p-v_2)=h^{-1}_{[g(s.p-v_2)]}$ to the right, as demonstrated in Equation \ref{mag_membrane_path_end}. Therefore
	\begin{align*}
		&E_p(v_0(p)-v_2)^{-1}C^h_{\rhd}(m)E_p(v_0(p)-v_2): e_p \\
		&= E_p(v_0(p)-v_2)^{-1}: \begin{cases} h_{[g(s.p-v_2)]} \rhd (g(v_0(p)-v_2)^{-1} \rhd e_p) & \text{ if $v_2$ lies on the direct membrane} \\ g(v_0(p)-v_2)^{-1} \rhd e_p & \text{ otherwise} \end{cases}\\
		&= \begin{cases} (h_{[g(s.p-v_0(p))]}g(v_0(p)-v_2)h^{-1}_{[g(s.p-v_2)]}) \rhd ( h_{[g(s.p-v_2)]} \rhd (g(v_0(p)-v_2)^{-1} \rhd e_p)) & \text{if $v_2$ and $v_0(p)$ lie on the} \\ & \text{ direct membrane}\\
			(h_{[g(s.p-v_0(p))]}g(v_0(p)-v_2)) \rhd (g(v_0(p)-v_2)^{-1} \rhd e_p) & \text{if $v_0(p)$ lies on the direct} \\ & \text{ membrane and $v_2$ does not}\\
			(g(v_0(p)-v_2)h^{-1}_{[g(s.p-v_2)]}) \rhd ( h_{[g(s.p-v_2)]} \rhd (g(v_0(p)-v_2)^{-1} \rhd e_p)) & \text{if $v_2$ lies on the direct}\\ & \text{ membrane and $v_0(p)$ does not}\\
			g(v_0(p)-v_2) \rhd (g(v_0(p)-v_2)^{-1} \rhd e_p) & \text{if neither $v_0(p)$ nor $v_2$ lie}\\ & \text{ on the direct membrane}\end{cases}\\
		&=\begin{cases} (h_{[g(s.p-v_0(p))]}g(v_0(p)-v_2)) \rhd (g(v_0(p)-v_2)^{-1} \rhd e_p) & \text{if $v_2$ and $v_0(p)$ lie on the direct membrane}\\
			(h_{[g(s.p-v_0(p))]}) \rhd e_p & \text{if $v_0(p)$ lies on the direct membrane and $v_2$ does not}\\
			g(v_0(p)-v_2) \rhd (g(v_0(p)-v_2)^{-1} \rhd e_p) & \text{if $v_2$ lies on the direct membrane and $v_0(p)$ does not}\\
			e_p & \text{if neither $v_0(p)$ nor $v_2$ lie on the direct membrane}\end{cases}\\
		&= \begin{cases} h_{[g(s.p-v_0(p))]} \rhd e_p & \text{if $v_0(p)$ lies on the direct membrane} \\ e_p & \text{ otherwise} \end{cases}\\
		&=C^h_{\rhd}(m):e_p, 
	\end{align*}
	so the action of $C^h_{\rhd}(m)$ is invariant under $E_p(t)$. Note that we did not consider moving the base-point of the plaquette through the direct membrane (so that $v_2$ lies beneath the direct membrane in Figure \ref{magnetic_membrane_path_through}), because we will not consider plaquettes that are whiskered through the direct membrane (but we could define the action of the magnetic membrane operator on such plaquettes to be consistent with the re-branching procedure, from which we would see that the action is similar to that for plaquettes based on the direct membrane). Combined with our previous results, this means that $C^h_T(m)$ is invariant under $E_p(t)$, provided that the operator acts in a region that initially satisfies fake-flatness. We have already shown the invariance of this operator under the other re-branching procedures, so it is invariant under all such procedures. This means that, when considering the commutation of the magnetic membrane operator with the energy terms, we can freely choose a branching structure, and if the operators commute this for this branching structure then they will do so for all choices of the branching structure.

	\subsubsection{Commutation relations with energy terms}
	\label{Section_Magnetic_Tri_Nontrivial_Commutation}
	
	Having demonstrated that the magnetic membrane operator is invariant under changes to the branching structure of the lattice, we can now consider the commutation relations between the magnetic membrane operator and the energy terms. First we examine the plaquette energy terms. We separately consider the impact of $C^h_{\rhd}(m)$ and the added blob ribbon operators, starting with the ribbon operators. Note that the labels $f(p)$ of the blob ribbon operators, given in Equation \ref{Equation_magnetic_blob_ribbon_label}, satisfy $\partial(f(p))= \partial(e' \: [h^{-1} \rhd {e'}^{-1}])$ for $e'= g(s.p-v_{0}(p))\rhd e_p$. Therefore, from our discussion at the beginning of Section \ref{Section_Magnetic_Tri_Non_Trivial}, $\partial(f(p))=1_G$. That is, the labels of the added blob ribbon operators are in the kernel of $\partial$. The plaquette energy term of a plaquette $q$ only depends on the plaquette label $e_q$ through $\partial(e_q)$, which is unchanged by blob ribbon operators with label in the kernel. This means that the added blob ribbon operators commute with the plaquette term. Because of this, we need only consider the effect of $C^h_{\rhd}(m)$ on the plaquette term. This operator can affect the plaquette holonomy by changing the edges on the plaquette or by changing the plaquette label directly. The action of $C^h_{\rhd}(m)$ on the plaquette label at most takes it from label $e_q$ to a label of the form $(g(t)^{-1}hg(t)) \rhd e_q$. However, from our discussion at the start of this section, $\partial( x \rhd e_q)=\partial(e_q)$ for all $x \in G$. Therefore, the magnetic membrane operator does not affect $\partial(e_q)$ at all. This means that the only way in which the magnetic membrane operator interacts with the plaquette energy terms is by changing the edge elements. However, the action of $C^h_T(m)$ on the edge terms is the same as the operator $C^h(m)$ from the $\rhd$ trivial case. We can therefore follow the arguments for the $\rhd$ trivial case that we considered in Section \ref{Section_Magnetic_Membrane_Tri_trivial}. The more general $\rhd$ that we use here does not affect any of the arguments from the $\rhd$ trivial case and so we will not repeat the proof of the commutation relations here. The result is that the magnetic membrane operator only excites the plaquettes that are cut by the boundary of the dual membrane, as shown in Figure \ref{fluxmembrane2} in Section \ref{Section_3D_Tri_Trivial_Magnetic_Excitations} of the main text.

\begin{figure}[h]
		\begin{center}
			\begin{overpic}[width=0.5\linewidth]{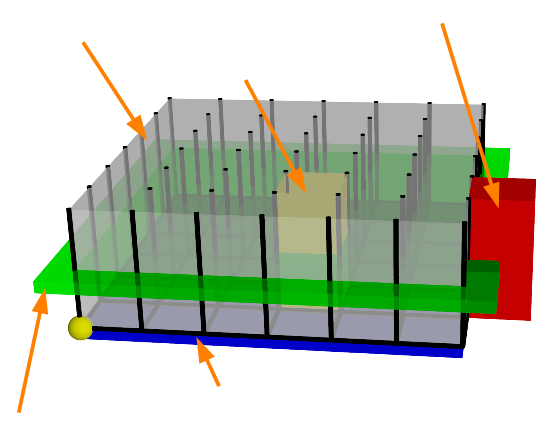}
				\put(0,-1){dual membrane}
				\put(30,4){direct membrane}
				\put(5,70){thickened membrane}
				\put(40,63){``bulk" blob}
				\put(70,72){``boundary" blob}
			\end{overpic}
			\caption{The blob energy terms that may be affected by the magnetic membrane operator are those corresponding to bulk blobs in the thickened membrane (grey region), such as the yellow blob indicated here, or those around the boundary which are cut by the edge of the dual membrane, such as the red blob.}
			\label{blobs_in_around_thickened_membrane}
		\end{center}
	\end{figure}

	Next, we examine the blob energy terms. The blob terms that can be affected by the membrane operator are those for which edges or plaquettes on the boundary of the blob are affected by the magnetic membrane operator. That means that we must consider blobs that are cut by the bulk of the dual membrane (we refer to these as blobs in the thickened membrane) and blobs around the boundary of the membrane (those that are intersected by the boundary of the dual membrane), as shown in Figure \ref{blobs_in_around_thickened_membrane}, as well as any additional blobs affected by the added blob ribbon operators.

	We will address the boundary blobs later. First we will consider blobs in the thickened membrane, other than the privileged 3-cell, blob 0. The impact of the magnetic membrane operator on such a blob can be split into contributions from the operator $C^h_{\rhd}(m)$ and the added blob ribbon operators. We'll start by examining the effect of the blob ribbon operators. We know from Section \ref{Section_Blob_Ribbon_Fake_Flat} that only blob ribbon operators where the ribbon starts or ends inside a blob change the 2-holonomy of the blob, $H_2 (\text{blob})$. Any ribbon operators that pass through the blob but do not terminate inside the blob do not change the blob 2-holonomy. For the magnetic membrane operator, all of the added blob ribbon operators start at blob 0, which means that for a blob other than blob 0, we only need to consider ribbons that end inside the blob. From Equation \ref{Equation_blob_ribbon_operator_enter_blob} in Section \ref{Section_Blob_Ribbon_Fake_Flat}, we know that a blob ribbon operator with label $f$ terminating in a blob $B$ affects the blob 2-holonomy as follows:
	$$H_2(B) \rightarrow [g(v_0(B)-s.p) \rhd f] \: H_2(B),$$
	where $v_0(B)$ is the base-point from which we calculate the blob 2-holonomy and $s.p$ is the start-point of the ribbon operator. In the case of the blob ribbon operators added to the magnetic membrane operator, the label $f$ of the blob ribbon operator associated to plaquette $p$ is
	$$f(p)=[g(s.p-v_{0}(p))\rhd e_p] [ (h^{-1}g(s.p-v_0(p)))\rhd e_p^{-1}\big],$$ 
	assuming that $p$ is oriented away from the dual membrane, as shown in Figure \ref{modmembraneorientationappendix}. This ribbon terminates in the blob $B$ that is attached to plaquette $p$ and lies within the thickened membrane, as indicated in Figure \ref{blobsonmem} in Section \ref{Section_3D_MO_Central} of the main text. The effect of the ribbon operator on the blob 2-holonomy is therefore
	\begin{align*}
		H_2(B) &\rightarrow g(v_0(B)-s.p) \rhd \big([g(s.p-v_{0}(p))\rhd e_p] \: [h^{-1}g(s.p-v_{0}(p))\rhd e_p^{-1}]\big) H_2(B)\\
		& = [g(v_0(B)-v_0(p)) \rhd e_p] \: [\big(g(s.p-v_0(B))^{-1}h^{-1} g(s.p-v_0(B))g(v_0(B)-v_0(p))\big) \rhd e_p^{-1}] \: H_2(B),
	\end{align*}
	where in the last step we used the fact that $g(v_0(B)-s.p)g(s.p-v_{0}(p))=g(v_0(B)-v_0(p))$, at least up to irrelevant factors in $\partial(E)$ that do not affect the $\rhd$ action. Recalling our definition $h_{[g(t)]}=g(t)^{-1}hg(t)$, we can write the change to the 2-holonomy more concisely as
	\begin{align}
		H_2(B) &\rightarrow [g(v_0(B)-v_0(p)) \rhd e_p] \: [h_{[g(s.p-v_0(B))]}^{-1} \rhd(g(v_0(B)-v_0(p)) \rhd e_p^{-1})] H_2(B) \notag \\
		&=\big[g(v_0(B)-v_0(p)) \rhd (e_p \: [h_{[g(s.p-v_0(p))]}^{-1} \rhd e_p^{-1}])\big]H_2(B). 
		\label{mod_mem_ribbon_operator_blob_holonomy}
	\end{align}

	Note that a blob $B$ may have multiple plaquettes on its surface that lie on the direct membrane, which we call ``base plaquettes". Each of these base plaquettes has a blob ribbon operator associated to it and so the 2-holonomy of blob $B$ is affected as in Equation \ref{mod_mem_ribbon_operator_blob_holonomy} above for each such plaquette. That is, the total action of the blob ribbon operators from $C^h_T(m)$ on a blob $B$ is
	\begin{align}
		\bigg( \prod_{\substack{\text{plaquette }p \\ \text{on membrane}}} &B^{f(p)}(\text{blob }0 \rightarrow \text{blob }p)\bigg): H_2(B) \notag \\
		&= \bigg( \prod_{b \text{ in the base of }B} g(v_0(B)-v_0(b)) \rhd (e_b \: [h_{[g(s.p-v_0(b))]}^{-1} \rhd e_b^{-1}]) \bigg) H_2(B) \label{mod_mem_ribbon_operator_blob_holonomy_2}
	\end{align}

\begin{figure}[h]
		\begin{center}
			\begin{overpic}[width=0.6\linewidth]{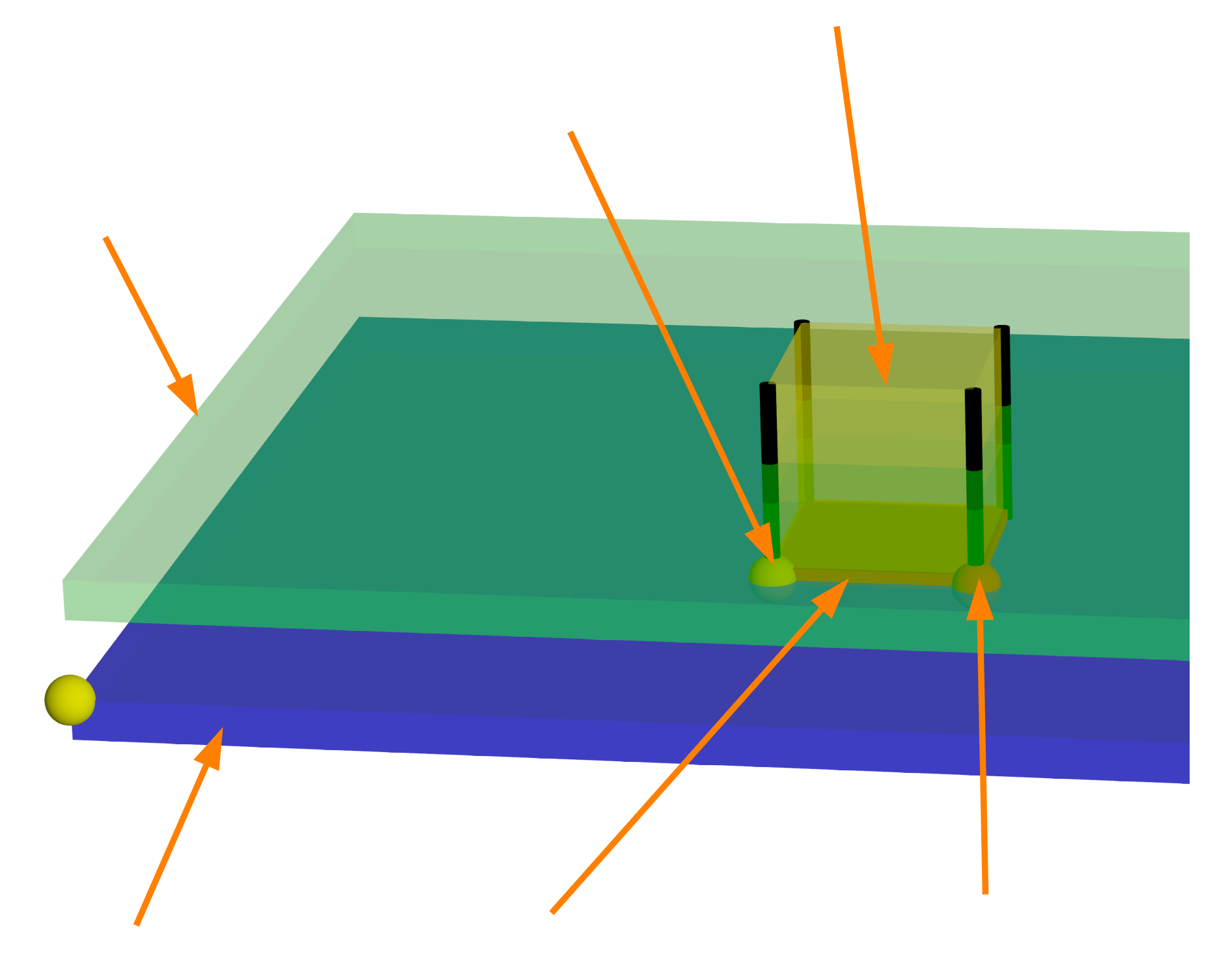}
				\put(30,3){plaquette on base}
				\put(67,79){blob}
				\put(0,1){direct membrane}
				\put(0,62){dual membrane}
				\put(38,71){base-point of blob}
				\put(70,3){base-point of base plaquette}
			\end{overpic}
			\caption{The ``base" plaquettes of a blob are those that lie on the direct membrane of the magnetic membrane operator. This means that the path from the base-point of the plaquette to the base-point of the blob (which we take to be on the direct membrane) can be deformed into one than never passes through the dual membrane.}
			\label{mod_mem_base_plaquette}
		\end{center}
	\end{figure}

	Now we consider the impact of the operator $C^h_{\rhd}(m)$ on the blob 2-holonomy. We can choose to calculate the 2-holonomy of the blob with any base-point without affecting the energy term, because changing the base-point of the blob only affects its 2-holonomy by an $\rhd$ action (which leaves the identity element invariant). We therefore choose the base-point of the blob ($v_0(B)$) to lie on the direct membrane. We now need to consider how $C^h_{\rhd}(m)$ affects the contribution of each plaquette to the blob 2-holonomy. To do so, we split the plaquettes on the blob $B$ into three types. Firstly we have the ``base" plaquettes, which lie on the direct membrane, as illustrated in Figure \ref{mod_mem_base_plaquette}. These are the same plaquettes that are associated to the blob ribbon operators we considered earlier. The labels of the base plaquettes are not affected by the $\rhd$ action of the membrane operator, because they are not cut by the dual membrane. However, the contribution to the blob holonomy from a plaquette $b$ is $g(v_0(B)-v_0(b))\rhd e_b^{\pm 1}$, which depends on the group element associated to a path from the base-point of the blob to the base-point of the plaquette. Both the base-point of the blob and the base-point of the plaquette lie on the direct membrane, which suggests that the path between them is unchanged by the action of the magnetic membrane operator. The path between these two points must pass through the dual membrane an even number of times and so can be deformed into a path which does not pass through the dual membrane. As we explained earlier in this section, this means that the path is not affected by the magnetic membrane operator.

\begin{figure}[h]
		\begin{center}
			\begin{overpic}[width=0.6\linewidth]{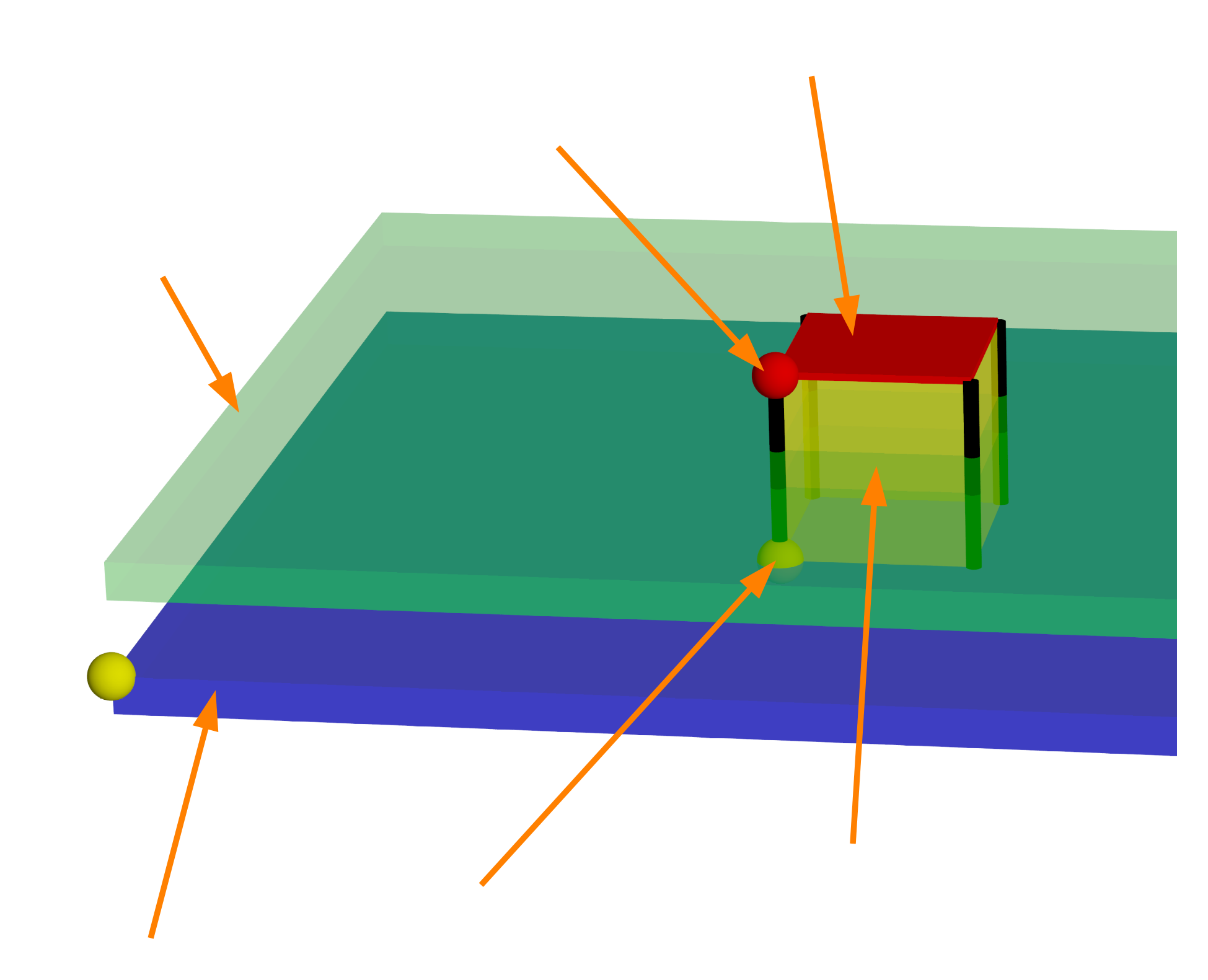}
				\put(30,5){base-point of blob}
				\put(63,77){top plaquette}
				\put(0,1){direct membrane}
				\put(0,59){dual membrane}
				\put(25,71){base-point of top plaquette}
				\put(68,8){blob}
			\end{overpic}
			\caption{The ``top" plaquettes of a blob cut by the dual membrane are those that are separated from the direct membrane by the edges (black cylinders) cut by the dual membrane. This means that the path from the base-point of the blob (which we choose to be on the direct membrane) to the base-point of the plaquette passes through the dual membrane.}
			\label{mod_mem_top_plaquette}
		\end{center}
	\end{figure}

	The next type of plaquette to consider are the ``top" plaquettes, which are not on the direct membrane and are also not cut by the dual membrane, but are attached to blobs which are cut by the dual membrane. We call these the top plaquettes, because if the direct membrane is at the base of the blob then these plaquettes are at the top of the blob. Such plaquettes are separated from the direct membrane by the edges that are cut by the dual membrane, as shown in Figure \ref{mod_mem_top_plaquette}. The $\rhd$ action does not affect the plaquette labels for these plaquettes because they are not cut by the dual membrane. However, this does not mean that the contribution to the blob 2-holonomy from these plaquettes is unchanged by the action of the membrane operator. These plaquettes have their base-point $v_0(p)$ separated from the base-point of the blob by the cut edges, so the path from $v_0(B)$ to $v_0(p)$ passes along the cut edges an odd number of times. Then we can determine how the path element is affected by the membrane operator from Equation \ref{mag_membrane_path_start}. We have
	$$g(v_0(B)-v_0(p)) \rhd e_p \rightarrow \big(g(s.p-v_0(B))^{-1}hg(s.p-v_0(B))g(v_0(B)-v_0(p))\big) \rhd e_p.$$
	Recalling our definition $h_{[g(t)]}=g(t)^{-1}hg(t)$, we can write this action more concisely as
	\begin{equation}
		g(v_0(B)-v_0(p)) \rhd e_p \rightarrow h_{[g(s.p-v_0(B))]} \rhd( g(v_0(B)-v_0(p)) \rhd e_p ).
		\label{mod_mem_top_plaquette_equation}
	\end{equation}

	The final type of plaquette to consider are the plaquettes that are cut by the dual membrane. The base-point $v_0(p)$ of such a plaquette, $p$, can either be on the direct membrane or be above it. If $v_0(p)$ is above the membrane then the plaquette label is not directly affected by the membrane operator. Instead, just as for the top plaquettes, the change to the contribution to the blob 2-holonomy is from the changes to the path elements. The calculation for this is the same as the one above for the top plaquettes, so just as for the top plaquettes we have
	\begin{equation}
		g(v_0(B)-v_0(p)) \rhd e_p \rightarrow \big(g(s.p-v_0(B))^{-1}hg(s.p-v_0(B))g(v_0(B)-v_0(p))\big) \rhd e_p
		\label{Equation_magnetic_side_plaquette_transform}
	\end{equation}
	for the side plaquettes that have base-point above the direct membrane.

	On the other hand, if $v_0(p)$ is on the membrane then the path element $g(v_0(B)-v_0(p))$ is unaffected by the membrane operator (just as in the discussion of the base plaquettes), but instead the plaquette label itself is affected. We have that $e_p \rightarrow g(s.p-v_0(p))^{-1}h g(s.p-v_0(p)) \rhd e_p$. Then
	\begin{align*}
		g(v_0(B)-v_0(p)) \rhd e_p &\rightarrow g(v_0(B)-v_0(p)) \rhd \big(g(s.p-v_0(p))^{-1}h g(s.p-v_0(p)) \rhd e_p\big)\\
		&=\big(g(v_0(B)-s.p)h g(s.p-v_0(B)) g(v_0(B) - v_0(p))\big) \rhd e_p\\
		&=\big(g(s.p-v_0(B))^{-1}h g(s.p-v_0(B)) g(v_0(B) - v_0(p))\big) \rhd e_p\\
		&= h_{[g(s.p-v_0(B))]} \rhd(g(v_0(B) - v_0(p)) \rhd e_p).
	\end{align*}
	Comparing this to Equation \ref{mod_mem_top_plaquette_equation} we see that the contribution from the side plaquettes transforms just like the contribution from the top plaquettes, even if the base-point of the side plaquette lies on the direct membrane. Only the base plaquettes transform differently (this is the reason why we have to add the blob ribbon operators whose labels depend on the base plaquettes). Therefore, under the action of $C^h_{\rhd}(m)$ the 2-holonomy of a blob, defined in Equation \ref{Equation_blob_2_holonomy_appendix_1}, transforms as
	\begin{align}
		C^h_{\rhd}(m): H_2(\text{blob } B) &= \bigg(\prod_{p \notin \text{base}(B)} h_{[g(s.p-v_0(B))]} \rhd(g(v_0(B) - v_0(p)) \rhd e_p) \bigg) \bigg(\prod_{b \in \text{base}(B)} g(v_0(B) - v_0(b)) \rhd e_b\bigg) \notag \\
		&= \biggl(h_{[g(s.p-v_0(B))]} \rhd \bigg(\prod_{p \notin \text{base}(B)} g(v_0(B) - v_0(p)) \rhd e_p \bigg)\biggl) \bigg(\prod_{b \in \text{base}(B)} g(v_0(B) - v_0(b)) \rhd e_b\bigg). \label{Equation_magnetic_blob_rhd_action}
	\end{align}
	
	Now we want to consider the total action of the membrane operator $C^h_T(m)$ on the blob 2-holonomy. As previously defined in Equation \ref{Equation_total_magnetic_membrane_definition_1}, the total magnetic membrane operator $C^h_T(m)$ is given by
	$$C^h_T(m) = C^h_{\rhd}(m) \prod_{\substack{\text{plaquette }p \\ \text{on membrane}}} B^{f(p)}(\text{blob }0 \rightarrow \text{blob }p),$$
	from which we see that we must first act with a series of blob ribbon operators on the blob 2-holonomy, before acting with $C^h_{\rhd}(m)$. Recalling that only the blob ribbon operators corresponding to the base of the blob contribute to the blob 2-holonomy, the total action on the blob 2-holonomy 
	$$H_2(\text{blob }B)=\prod_{\substack{\text{plaquettes } p\\ \in \text{Bd}(B)}} g(v_0(\text{blob }B)-v_{0}(p)) \rhd e_p$$
	is given by (using Equation \ref{mod_mem_ribbon_operator_blob_holonomy_2})
	\begin{align}
		C^h_T(m) :H_2(\text{blob }B) &= C^h_\rhd(m) \bigg( \prod_{b \in \text{base}(B)} B^{f(b)}(\text{blob }0 \rightarrow \text{blob }B)\bigg):H_2(\text{blob }B) \notag \\
		&= C^h_{\rhd}(m): \bigg( \prod_{b \in \text{base}(B)} g(v_0(B)-v_0(b)) \rhd (e_b \: [h_{[g(s.p-v_0(b))]}^{-1} \rhd e_b^{-1}]) \bigg) H_2(\text{blob }B). \label{Equation_magnetic_membrane_on_blob_1}
	\end{align}
	
	To calculate this, we must split up the contributions to the blob 2-holonomy from the base plaquettes and the side or top plaquettes, because these transform differently under the action of $C^h_{\rhd}(m)$. Note that, although the additional contributions from the blob ribbon operators have labels that depend on the base plaquettes, the blob ribbon operators actually affect the labels of the side plaquettes through which they enter blob $B$. This means that the contribution from the blob ribbon operators transform under $C^h_{\rhd}(m)$ like the side plaquettes (i.e., according to Equation \ref{Equation_magnetic_side_plaquette_transform}). That is, $C^h_{\rhd}(m)$ acts on the contribution
	$$\bigg(\prod_{b \in \text{base}(B)} g(v_0(B)-v_0(b)) \rhd (e_b \: [h_{[g(s.p-v_0(b))]}^{-1} \rhd e_b^{-1}]) \bigg)$$
	from the blob ribbon operators as
	\begin{align}
		C^h_{\rhd}(m) : \bigg(\prod_{b \in \text{base}}& g(v_0(B)-v_0(b)) \rhd (e_b \: [h_{[g(s.p-v_0(b))]}^{-1} \rhd e_b^{-1}]) \bigg) \notag \\
		&= h_{[g(s.p-v_0(B))]} \rhd \bigg(\prod_{b \in \text{base}(B)} g(v_0(B)-v_0(b)) \rhd (e_b \: [h_{[g(s.p-v_0(b))]}^{-1} \rhd e_b^{-1}])\bigg) \notag \\
		&= h_{[g(s.p-v_0(B))]} \rhd \bigg(\prod_{b \in \text{base}(B)} g(v_0(B)-v_0(b)) \rhd e_b\bigg) \notag \\
		& \hspace{1cm} h_{[g(s.p-v_0(B))]} \rhd \bigg(\prod_{b \in \text{base}} g(v_0(B)-v_0(b)) \rhd ( h_{[g(s.p-v_0(b))]}^{-1} \rhd e_b^{-1})\bigg) \notag \\
		&= h_{[g(s.p-v_0(B))]} \rhd \bigg(\prod_{b \in \text{base}(B)} g(v_0(B)-v_0(b)) \rhd e_b\bigg) \notag \\
		& \hspace{1cm} \bigg(\prod_{b \in \text{base}(B)} \big(h_{[g(s.p-v_0(B))]} g(v_0(B)-v_0(b)) h_{[g(s.p-v_0(b))]}^{-1}\big) \rhd e_b^{-1} \bigg). 
		\label{Equation_magnetic_membrane_on_blob_2}
	\end{align}
	
	We can see that the last term in Equation \ref{Equation_magnetic_membrane_on_blob_2} can be written as a product of elements of the form $$(h_{[g(s.p-v_0(B))]} g(v_0(B)-v_0(b)) h_{[g(s.p-v_0(b))]}^{-1}) \rhd e_b^{-1}.$$ Now consider the element of $G$, $h_{[g(s.p-v_0(B))]} g(v_0(B)-v_0(b)) h_{[g(s.p-v_0(b))]}^{-1}$. This can be written as
	\begin{align*}
		h_{[g(s.p-v_0(B))]} &g(v_0(B)-v_0(b)) h_{[g(s.p-v_0(b))]}^{-1}\\
		&= g(s.p-v_0(B))^{-1} h g(s.p-v_0(B)) g(v_0(B)-v_0(b)) g(s.p-v_0(b))^{-1}h^{-1} g(s.p-v_0(b))\\
		&= g(s.p-v_0(B))^{-1} h g(s.p-v_0(b))g(s.p-v_0(b))^{-1}h^{-1} g(s.p-v_0(b))\\
		&= g(v_0(B)-s.p) g(s.p-v_0(b))\\
		&=g(v_0(B)-v_0(b)).
	\end{align*}
	Substituting this into Equation \ref{Equation_magnetic_membrane_on_blob_2}, we obtain
	\begin{align}
		C^h_{\rhd}(m) : \bigg(\prod_{b \in \text{base}(B)}& g(v_0(B)-v_0(b)) \rhd (e_b \: h_{[g(s.p-v_0(b))]}^{-1} \rhd e_b^{-1}) \bigg) \notag \\
		&= \bigg[ h_{[g(s.p-v_0(B))]} \rhd \bigg( \prod_{b \in \text{base}(B)}g(v_0(B)-v_0(b)) \rhd e_b \bigg)\bigg] \bigg(\prod_{b \in \text{base}(B)} g(v_0(B)-v_0(b)) \rhd e_b^{-1} \bigg). \label{Equation_magnetic_membrane_on_blob_3}
	\end{align}
	This tells us how $C^h_{\rhd}(m)$ acts on the extra contribution to the blob 2-holonomy from the blob ribbon operators in Equation \ref{Equation_magnetic_membrane_on_blob_1}. We also need to know the action of $C^h_{\rhd}(m)$ on the $H_2(\text{blob }B)$ part of the last line of Equation \ref{Equation_magnetic_membrane_on_blob_1}. As discussed previously, this is given by Equation \ref{Equation_magnetic_blob_rhd_action}, which we can rewrite as:
	\begin{equation}
		C^h_{\rhd}(m): H_2(\text{blob }B) = \bigg[h_{[g(s.p-v_0(B))]} \rhd \bigg(\prod_{p \notin \text{base}(B)} g(v_0(B) - v_0(p)) \rhd e_p \bigg)\bigg] \bigg(\prod_{b \in \text{base}(B)} g(v_0(B) - v_0(b)) \rhd e_b\bigg). \label{Equation_magnetic_membrane_on_blob_4}
	\end{equation}
	
	Substituting Equations \ref{Equation_magnetic_membrane_on_blob_3} and \ref{Equation_magnetic_membrane_on_blob_4} into Equation \ref{Equation_magnetic_membrane_on_blob_1}, we see that the total action of the magnetic membrane operator on the blob 2-holonomy is
	\begin{align}
		C^h_T(m) &:H_2(\text{blob }B) = \bigg[ h_{[g(s.p-v_0(B))]} \rhd \bigg( \prod_{b \in \text{base}(B)}g(v_0(B)-v_0(b)) \rhd e_b \bigg)\bigg] \bigg(\prod_{b \in \text{base}(B)} g(v_0(B)-v_0(b)) \rhd e_b^{-1} \bigg)\notag\\
		& \hspace{3cm} \bigg[h_{[g(s.p-v_0(B))]} \rhd \bigg(\prod_{p \notin \text{base}(B)} g(v_0(B) - v_0(p)) \rhd e_p \bigg)\bigg] \bigg(\prod_{b \in \text{base}(B)} g(v_0(B) - v_0(b)) \rhd e_b\bigg)\notag \\
		&= \bigg[h_{[g(s.p-v_0(B))]} \rhd \bigg(\prod_{b \in \text{base}(B)} g(v_0(B)-v_0(b)) \rhd e_b \bigg)\bigg] \bigg[h_{[g(s.p-v_0(B))]} \rhd \bigg(\prod_{p \notin \text{base}(B)} g(v_0(B) - v_0(p)) \rhd e_p \bigg)\bigg] \notag\\
		& \hspace{1cm} \bigg( \prod_{b \in \text{base}(B)}g(v_0(B) - v_0(b)) \rhd e_b\bigg) \bigg(\prod_{b \in \text{base}(B)} g(v_0(B)-v_0(b)) \rhd e_b^{-1} \bigg) \notag \\
		&= h_{[g(s.p-v_0(B))]} \rhd \bigg(\prod_{\text{all } p \in \text{Bd}(B)} g(v_0(B) - v_0(p)) \rhd e_p \bigg) \notag\\
		&= h_{[g(s.p-v_0(B))]} \rhd H_2(\text{blob }B) \label{Equation_magnetic_membrane_on_blob_5}
	\end{align}

	This action preserves the identity element $1_E$ (because $g\rhd e =1_E \iff e =1_E$, for all $g \in G$), so the action of the membrane operator commutes with the blob energy terms (which check if the blob 2-holonomy is the identity) for these generic bulk blobs. We still need to check the blobs just outside the membrane and the special blob, blob 0. Consider blob 0 first, and assume for now that it is in the thickened membrane (we consider a case where it is displaced away from the thickened membrane in Section \ref{Section_braiding_higher_flux}). The blob ribbon operators added to the magnetic membrane operator only leave this blob, they do not enter it. From Equation \ref{Equation_blob_ribbon_operator_exit_blob} in Section \ref{Section_Blob_Ribbon_Fake_Flat}, the effect of one of the blob ribbon operators on the 2-holonomy of blob 0 is (setting the base-point of blob 0, $v_0(\text{blob }0)$, to the start-point of the membrane )
	\begin{align}
		B^f(t) : H_2(\text{blob }0) &= H_2(\text{blob }0) \:g(v_0(\text{blob }0)-s.p)\rhd f^{-1} \notag\\
		&=H_2(\text{blob }0) f^{-1}. \label{Equation_magnetic_blob_ribbon_blob_0}
	\end{align}
	
	Now we want to consider the contribution from all of the blob ribbon operators. Any blob ribbon operator which leaves blob 0 will contribute to the change in 2-holonomy. All of the blob ribbon operators, apart from those corresponding to plaquettes in the base of blob 0 itself, will leave blob 0. Each plaquette $p$ is associated to a blob ribbon operator of label 
	$$f(p)=g(s.p-v_0(p)) \rhd (e_p [h_{g(s.p-v_0(p))}^{-1} \rhd e_p^{-1}])$$ 
	(see Equation \ref{Equation_magnetic_blob_ribbon_label}, where we have taken $\sigma_p=1$ by taking all of the plaquettes to point away from the dual membrane). Therefore, under the action of the blob ribbon operators
	\begin{align*}
		H_2(\text{blob }0) &\rightarrow H_2(\text{blob }0) \prod_{\substack{p \in \text{ membrane,}\\ \notin \text{ base(blob 0)}}}f(p)^{-1}\\
		&=H_2(\text{blob }0) \bigg( \prod_{\substack{ p \in \text{ membrane,} \\\notin \text{ base(blob 0)}}} g(s.p-v_0(p)) \rhd (e_p^{-1} [h_{[g(s.p-v_0(p))]}^{-1} \rhd e_p])\bigg)\\
		&=H_2(\text{blob }0) \bigg(\prod_{\text{all }p \in \text{membrane}} [g(s.p-v_0(p)) \rhd e_p^{-1}] [h^{-1} \rhd \big(g(s.p-v_0(p)) \rhd e_p\big)]\bigg)\\
		& \quad \quad \bigg(\prod_{b \in \text{base(blob 0)}} [g(s.p-v_0(b)) \rhd e_b] \: [h^{-1} \rhd (g(s.p-v_0(b)) \rhd e_b^{-1})]\bigg),
	\end{align*}	
	where we inserted identity to complete the product to all plaquettes on the membrane, compensating with the inverse terms for the plaquettes on the base. Then noting that $\prod_p g(s.p-v_0(p)) \rhd e_p$ gives the total label of the membrane based at the $s.p$, which we denote by $e_m$, we have
	\begin{align*}
		H_2(\text{blob }0)&\rightarrow H_2(\text{blob }0) \: (e_m^{-1} \: [h^{-1}\rhd e_m]) \bigg(\prod_{b \in \text{base(blob 0)}} [g(s.p-v_0(b)) \rhd e_b] \: [h^{-1} \rhd \big(g(s.p-v_0(b)) \rhd e_b^{-1}\big)] \bigg).
	\end{align*}

	Note that the term 
	$$H_2(\text{blob }0) \prod_{b \in \text{base(blob 0)}}\big( (g(s.p-v_0(b)) \rhd e_b) \: (h^{-1} \rhd[ g(s.p-v_0(b)) \rhd e_b^{-1}])\big)$$
	is the same as the action of the ribbon operators on a generic blob given in Equation \ref{mod_mem_ribbon_operator_blob_holonomy_2} (replacing $h_{[g(t)]}$ with $h$ because the start-point of the blob is also the start-point of the membrane). Apart from the $e_m$ part, the action of the blob ribbon operators on blob 0 looks much like the action of the blob ribbon operators on any other blob in the thickened membrane. To complete the action of $C^h_T(m)$ we must now act with $C^h_{\rhd}(m)$ and consider the effect on the 2-holonomy of the blob. Firstly, the term
	$$H_2(\text{blob }0) \prod_{b \in \text{base}}\big( (g(s.p-v_0(b)) \rhd e_b) \: (h^{-1} \rhd[ g(s.p-v_0(b)) \rhd e_b^{-1}])\big)$$ 
	transforms in the same way as for a generic blob:
	\begin{align*}
		H_2(\text{blob }0) &\bigg(\prod_{b \in \text{base}} (g(s.p-v_0(b)) \rhd e_b) \: (h^{-1} \rhd[ g(s.p-v_0(b)) \rhd e_b^{-1}])\bigg) \\
		&\rightarrow (g(s.p-v_0(\text{blob }0))^{-1}hg(s.p-v_0(\text{blob }0))) \rhd H_2(\text{blob }0),
	\end{align*}
	and because $s.p=v_0(\text{blob } 0)$ this gives $h \rhd H_2(\text{blob }0)$. Compared to a generic blob, however, we have the extra factor $e_m \: [h^{-1} \rhd e_m]$, which we need to consider. This extra factor comes from changes to the labels of the side plaquettes of blob 0 from the blob ribbon operators. Therefore, this part transforms under $C^h_{\rhd}(m)$ in the same way as the side plaquette labels for the other blobs that we have considered:
	\begin{align*}
		e_m [h^{-1} \rhd e_m]&\rightarrow (g(s.p-v_0(\text{blob }0))^{-1}hg(s.p-v_0(\text{blob }0))) \rhd (e_m^{-1} [h^{-1} \rhd e_m])\\
		&= h \rhd (e_m^{-1} [h^{-1} \rhd e_m])\\
		&=[h \rhd e_m^{-1}] \: e_m.
	\end{align*}
	
	This means that the total transformation of the blob 2-holonomy for blob 0 is
	\begin{equation}
		H_2(\text{blob }0) \rightarrow [h \rhd H_2(\text{blob }0)]\: [h \rhd e_m^{-1}]\: e_m,
		\label{Equation_magnetic_blob_0}
	\end{equation}
	which gives $(h \rhd e_m^{-1}) \: e_m$ if $H_2(\text{blob }0)=1_E$ initially (i.e., if the blob is originally unexcited). From this we see that the action of the membrane operator may excite blob 0. However, the expression $(h \rhd e_m^{-1}) e_m$ depends on the total surface label of the direct membrane, which is measured by an operator. Therefore, for a generic label $h \in G$ for which there exists some $e_m \in E$ such that $h \rhd e_m \neq e_m$, the state resulting from acting with the magnetic membrane operator $C^h_T(m)$ on the ground state is not an eigenstate of the energy term at blob 0.

	The final type of blob to consider are those around the boundary of the thickened membrane. However, some of the plaquettes on the surface of these blobs are excited by the action of the magnetic membrane operator and so do not satisfy fake-flatness. Because of this, these blob energy terms are not well-defined (as described in Ref. \cite{Bullivant2017}, the blob terms are not gauge invariant when the plaquettes on the blob do not satisfy fake-flatness) and so we do not consider these blob terms further. Whether these blob terms are excited or not does not change the nature of the loop-like particle, because the blob terms are along the boundary of the membrane, where we already know that other excitations (the plaquette excitations) are produced by the membrane operator.

	The next energy terms to consider are the vertex transforms. As usual, we will separately examine the commutation relations between the vertex transform and the additional blob ribbon operators and the commutation relation between the transform and $C^h_{\rhd}(m)$. We have previously considered the commutation relations of vertex transforms with blob ribbon operators in Section \ref{Section_Blob_Ribbon_Fake_Flat}. However, there is an additional factor to consider for the blob ribbon operators that we added to the magnetic membrane operator. These blob ribbon operators have an operator as their label. That is, the blob ribbon operators are $B^{\hat{f}(b)}(\text{blob }0 \rightarrow \text{blob }b)$, where $\hat{f}(b)$ is an operator (its value depends on the label of a certain plaquette and path). Because of this, we need to consider how the vertex transform affects the label in addition to the usual commutation relations between vertex transforms and blob ribbon operators. First we want to consider when a vertex transform can affect the label. The label of the blob ribbon operator corresponding to plaquette $b$, with base-point $v_0(b)$, is
	$$f(b) = \big[g(s.p-v_0(b)) \rhd e_b\big] \big[h^{-1} g(s.p-v_0(b)) \rhd e_b^{-1}\big]\\
	=e_{b|s.p} \: [h^{-1} \rhd e_{b|s.p}^{-1}],$$
	where $e_{b|s.p}=g(s.p-v_0(p)) \rhd e_b$ is the label the plaquette would have if it was based at $s.p$ (i.e., if we whiskered the plaquette so that it was based at the start-point). The label of a plaquette based at a vertex $v$ is only affected by vertex transforms at $v$ \cite{Bullivant2017}. Therefore, the only vertex transform that can affect the label is the transform at the start-point. In addition, we know from Section \ref{Section_Blob_Ribbon_Fake_Flat} that only the vertex transform at the start-point of a constant-labelled blob ribbon operator may fail to commute with the ribbon operator. Putting these facts together, we see that only the vertex transforms at the start-point may fail to commute with the blob ribbon operators that we included in the magnetic membrane operator. The effect of such a vertex transform on the label of the blob ribbon operator is
	$$A_{s.p}^x:f(b)=\big[\big(xg(s.p-v_0(b))\big) \rhd e_b\big] \: \big[ \big(h^{-1}xg(s.p-v_0(b))\big) \rhd e_b^{-1}\big].$$
	
	On the other hand, the commutation between the vertex transform and a constant-labelled blob ribbon operator $B^e(\text{blob }0 \rightarrow \text{blob }b)$ is given by
	$$B^e(\text{blob }0 \rightarrow \text{blob }b)A_{s.p}^x =A_{s.p}^x B^{x^{-1} \rhd e},$$
	as we saw in Equation \ref{Equation_commutation_blob_ribbon_start_point_transform} in Section \ref{Section_Blob_Ribbon_Fake_Flat}. To put these two factors together, consider acting with the blob ribbon operator and the vertex transform on a basis state $\ket{\psi}$ where the plaquette $b$ has label $e_b$ and the path has the value $g(s.p-v_0(b))$. Then, denoting the operator-valued label of the blob ribbon operator by $\hat{f}(b)$ and the value this operator takes when acting on the state $\ket{\psi}$ by $f(b)$, we see that
	\begin{align*}
		B^{\hat{f}(b)}(\text{blob }0 \rightarrow \text{blob }b) A_{s.p}^x \ket{\psi}&=B^{A_{s.p}^x :f(b)} A_{s.p}^x \ket{\psi}\\
		&=B^{[x\rhd e_{b|s.p}] [(h^{-1}x) \rhd e_{b|s.p}^{-1}]}(\text{blob }0 \rightarrow \text{blob }b)A_{s.p}^x\ket{\psi}\\
		& =A_{s.p}^x B^{e_{b|s.p} [(x^{-1}h^{-1}x) \rhd e_{b|s.p}^{-1}]}(\text{blob }0 \rightarrow \text{blob }b) \ket{\psi},
	\end{align*}
	where the last line follows from the standard commutation relations for blob ribbon operators with a vertex transform at the start-point of the ribbon. We see that under commutation with the vertex transform, the label $h$ that appears in the blob ribbon operators becomes $x^{-1}hx$.

	This conjugation of the label $h$ is the same transformation that the label of a magnetic operator $C^h(m)$ undergoes under commutation with a vertex transform at the start-point in the $\rhd$ trivial case, suggesting that the same transformation will hold for $C^h_T(m)$. To check this, we need to verify that the label of $C^h_{\rhd}(m)$ also transforms in the same way. We will first show that the vertex transform at the start-point is the only one that may fail to commute with $C^h_{\rhd}(m)$. For the action of $C^h_{\rhd}(m)$ on the edge labels, this follows from the same reasoning that we gave in Section \ref{Section_Magnetic_Membrane_Tri_trivial} for the $\rhd$ trivial case, which we will not repeat here. For the $\rhd$ action on the plaquettes, which is given by
	\begin{equation*}
		C^h_{\rhd}(m): e_p = \begin{cases} (g(s.p-v_0(p))^{-1}hg(s.p-v_0(p))) \rhd e_p & \text{ if $v_0(p)$ lies on the direct membrane} \\ e_p & \text{ otherwise,} \end{cases} 
	\end{equation*}
	note that the only vertex transforms that may fail to commute are those applied at $v_0(p)$ (which can directly affect $e_p$ and also affect the path $g(s.p-v_0(p)$), and those applied at the start-point (which just affect the path element $g(s.p-v_0(p))$), assuming for now that $s.p$ and $v_0(p)$ are distinct. The vertex transform $A_{v_0(p)}^x$ at the base-point of plaquette $p$ acts on $e_p$ as $A_{v_0(p)}^x:e_p = x \rhd e_p$ and on the path element $g(s.p-v_0(p))$ as $A_{v_0(p)}^x:g(s.p-v_0(p))=g(s.p-v_0(p))x^{-1}$. Therefore,
	\begin{align*}
		C^h_{\rhd}(m)A_{v_0(p)}^x: e_p &= \begin{cases} (xg(s.p-v_0(p))^{-1}hg(s.p-v_0(p))x^{-1}) \rhd (x \rhd e_p) & \text{ if $v_0(p)$ lies on the direct membrane} \\ x \rhd e_p & \text{ otherwise} \end{cases} \\
		&=\begin{cases} (xg(s.p-v_0(p))^{-1}hg(s.p-v_0(p))) \rhd e_p & \text{ if $v_0(p)$ lies on the direct membrane} \\ x \rhd e_p & \text{ otherwise} \end{cases} \\
		&=\begin{cases} x \rhd \big((g(s.p-v_0(p))^{-1}hg(s.p-v_0(p))) \rhd e_p\big) & \text{ if $v_0(p)$ lies on the direct membrane} \\ x \rhd e_p & \text{ otherwise} \end{cases} \\
		&= A_{v_0(p)}C^h_{\rhd}(m):e_p,
	\end{align*}
	so the vertex transform commutes with the action of the membrane operator on this plaquette, as we claimed.

	Now consider the action of the vertex transform at the start-point. From the $\rhd$ trivial case discussed in Section \ref{Section_Magnetic_Membrane_Tri_trivial}, we know that for the action on the edges
	$$C^h_{\rhd}(m)A_{s.p}^x:g_i = A_{s.p}^x C^{x^{-1}hx}_{\rhd}(m):g_i,$$
	as indicated by Equation \ref{Equation_magnetic_membrane_start_point_transform}. A similar result holds for the action on the plaquettes, as we will now see. First consider the case of a plaquette $p$ which is cut by the dual membrane and for which $v_0(p)$ is not the start-point. Then the vertex transform $A_{s.p}^x$ takes the path element $g(s.p-v_0(p))$ to $xg(s.p-v_0(p))$, meaning that
	\begin{align*}
		C^h_{\rhd}(m)A_{s.p}^x: e_p &=\begin{cases} (g(s.p-v_0(p))^{-1}x^{-1}hxg(s.p-v_0(p))) \rhd e_p & \text{ if $v_0(p)$ lies on the direct membrane} \\ e_p & \text{ otherwise} \end{cases}\\
		&= A_{s.p}^x C^{x^{-1}hx}_{\rhd}(m):e_p.
	\end{align*} 
	If $s.p$ is also the base-point $v_0(p)$ for the plaquette, then the action of the membrane operator on the plaquette is simply
	$$C^h_{\rhd}(m): e_p =h \rhd e_p,$$
	because the path element $g(s.p-v_0(p))$ is trivial, and so
	\begin{align*}
		C^h_{\rhd}(m)A_{s.p}^x:e_p &= h \rhd (x \rhd e_p)\\
		&= (hx) \rhd e_p\\
		&= (xx^{-1}hx) \rhd e_p\\
		&= x \rhd \big((x^{-1}hx)\rhd e_p\big)\\
		&= A_{s.p}^x C^{x^{-1}hx}_{\rhd}(m):e_p.
	\end{align*}
	
	Putting these results together with those for the action of $C^h_{\rhd}(m)$ on the edges, we see that
	$$C^h_{\rhd}(m)A_{s.p}^x= A_{s.p}^x C^{x^{-1}hx}_{\rhd}(m).$$
	Combining this with the transformation of the blob ribbon operators under commutation with the vertex transform, we see that 
	\begin{equation}
		C^h_T(m) A_{s.p}^x =A_{s.p}^x C_{T}^{x^{-1}hx}(m), \label{Equation_magnetic_membrane_tri_nontrivial_start_point_transform}
	\end{equation}
	which is the same commutation relation as in the $\rhd$ trivial case. Therefore, the magnetic membrane operator commutes with all vertex transforms except for those applied at the start-point of the membrane.

\begin{figure}[h]
		\begin{center}
			\begin{overpic}[width=0.5\linewidth]{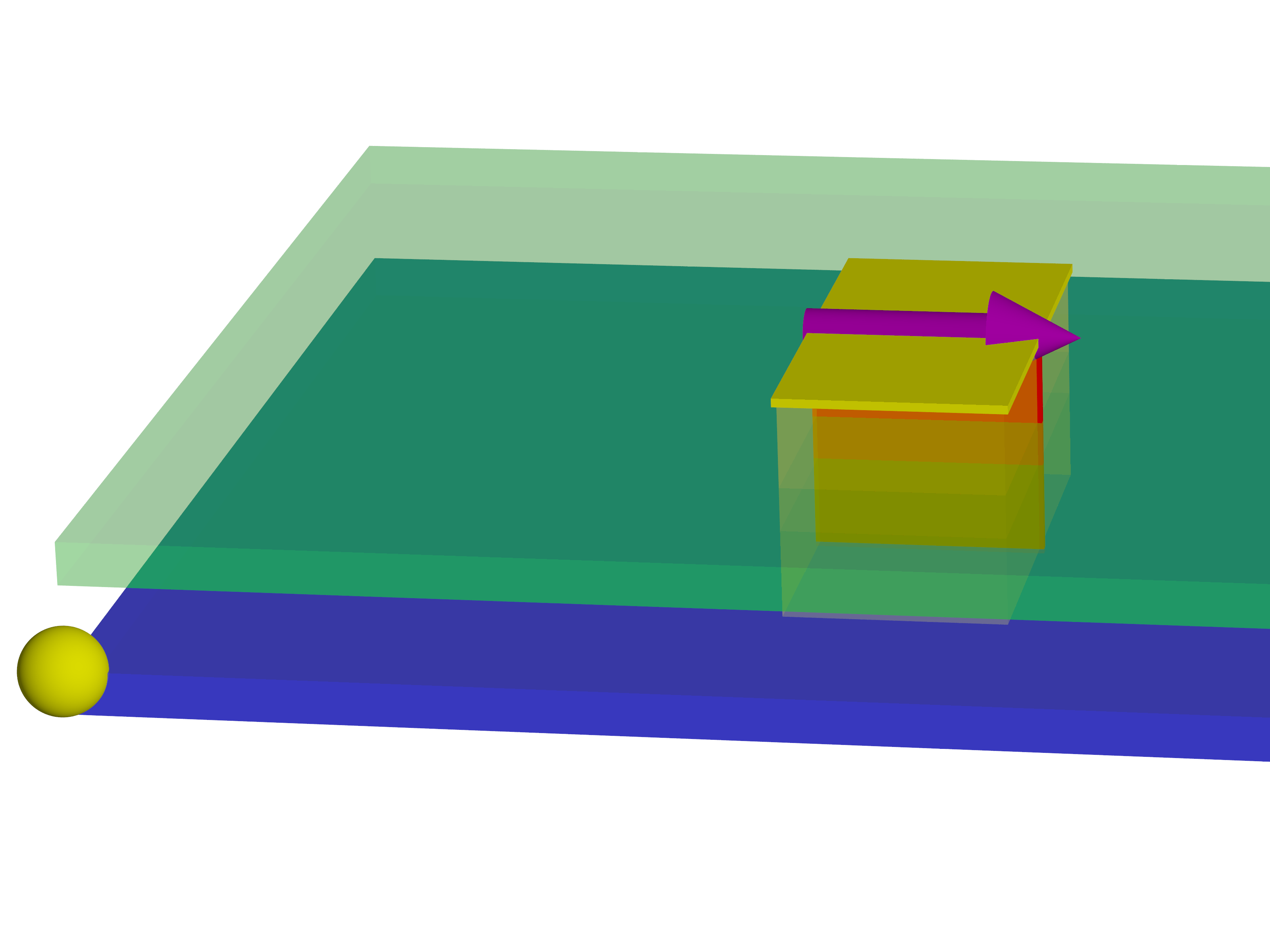}
				
			\end{overpic}
			\caption{Consider an edge just above the dual membrane, such as the edge in the figure (the purple arrow). In addition to affecting plaquettes such as the (yellow) ones above the dual membrane, an edge transform on the edge can affect plaquettes cut by the dual membrane, such as the (red) one under the edge.}
			\label{mod_mem_top_edge_example}
		\end{center}
	\end{figure}

\begin{figure}[h]
		\begin{center}
			\begin{overpic}[width=0.6\linewidth]{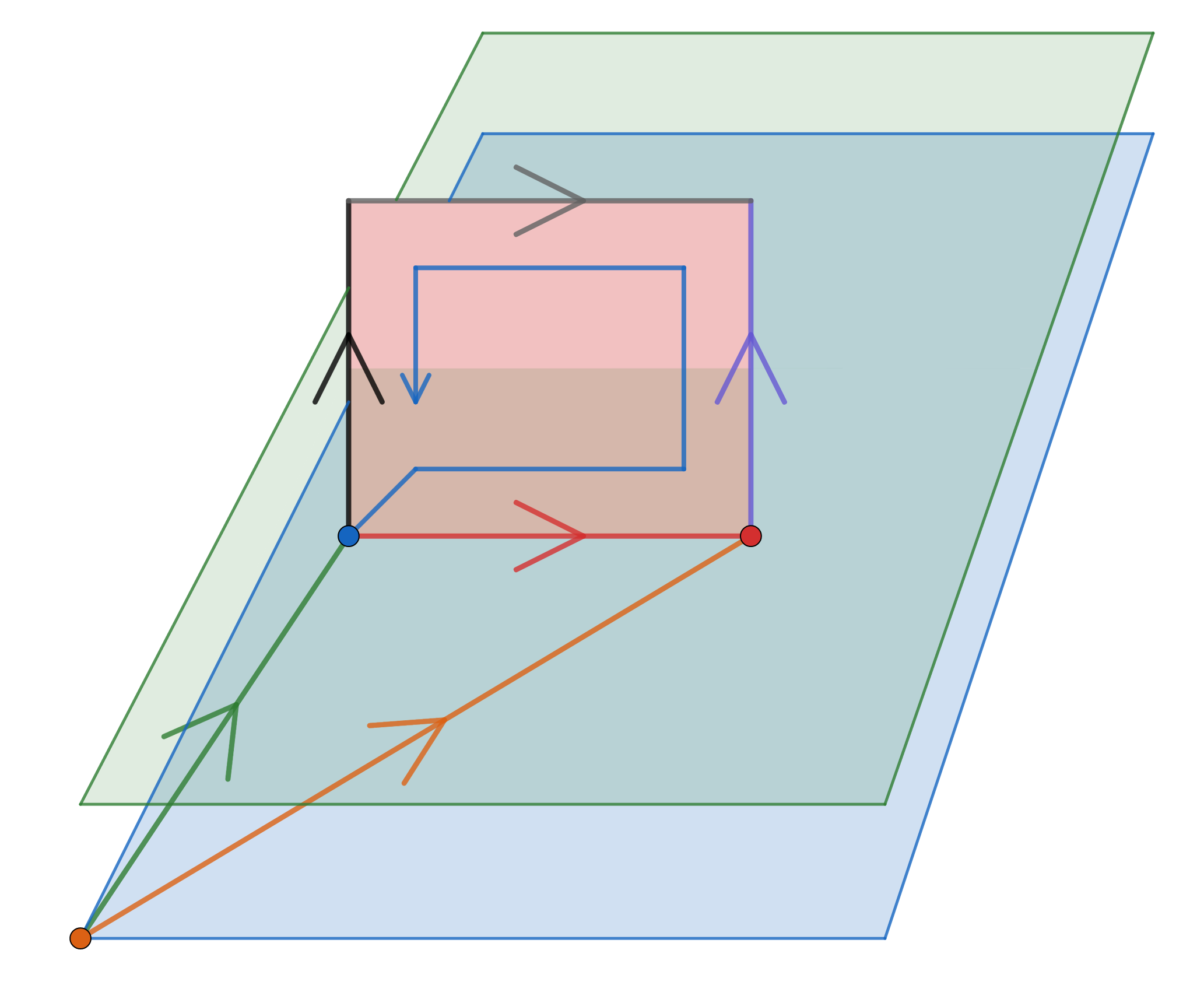}
				\put(2,4){$s.p$}
				\put(20,38){ $v_0(p)$}
				\put(63,38){ $s(i)$}
				\put(33,35){ $g(v_0(p)-s(i))$}
				\put(43,54){ $e_p$}
				\put(64,55){$i$}
				\put(6,30){$g(s.p-v_0(p))$}
				
			\end{overpic}
			\caption{Consider applying the magnetic membrane operator and an edge transform $\mathcal{A}_i^e$ on an edge $i$ cut by the dual membrane. The edge transform will affect adjacent plaquettes (such as the red shaded one), which are also cut by the dual membrane and so are also affected by the membrane operator. The simplest case to consider is the case where the source of edge $i$ (the red vertex, $s(i)$) and the base-point of the plaquette $p$ (the blue vertex, $v_0(p)$) are both on the direct membrane (the lower blue membrane). In this case the path $g(v_0(p)-v_i)$ (or $g(\overline{v_0(p)-v_{i+1}})$ if the plaquette has the opposite orientation to that shown in this figure) that appears in the edge transform is not affected by the magnetic membrane. Instead the non-commutativity of the edge transform with the magnetic membrane operator comes from the action of the magnetic membrane on the plaquette label itself. If we wish to consider other cases, where the vertices may not lie on the direct membrane, then we can use the invariance of the energy terms and membrane operator under the re-branching procedures discussed in the Appendix of Ref. \cite{HuxfordPaper1}.}
			\label{magnetic_membrane_edge_transform}	
		\end{center}
	\end{figure}

\begin{figure}[h]
		\begin{center}
			\begin{overpic}[width=0.75\linewidth]{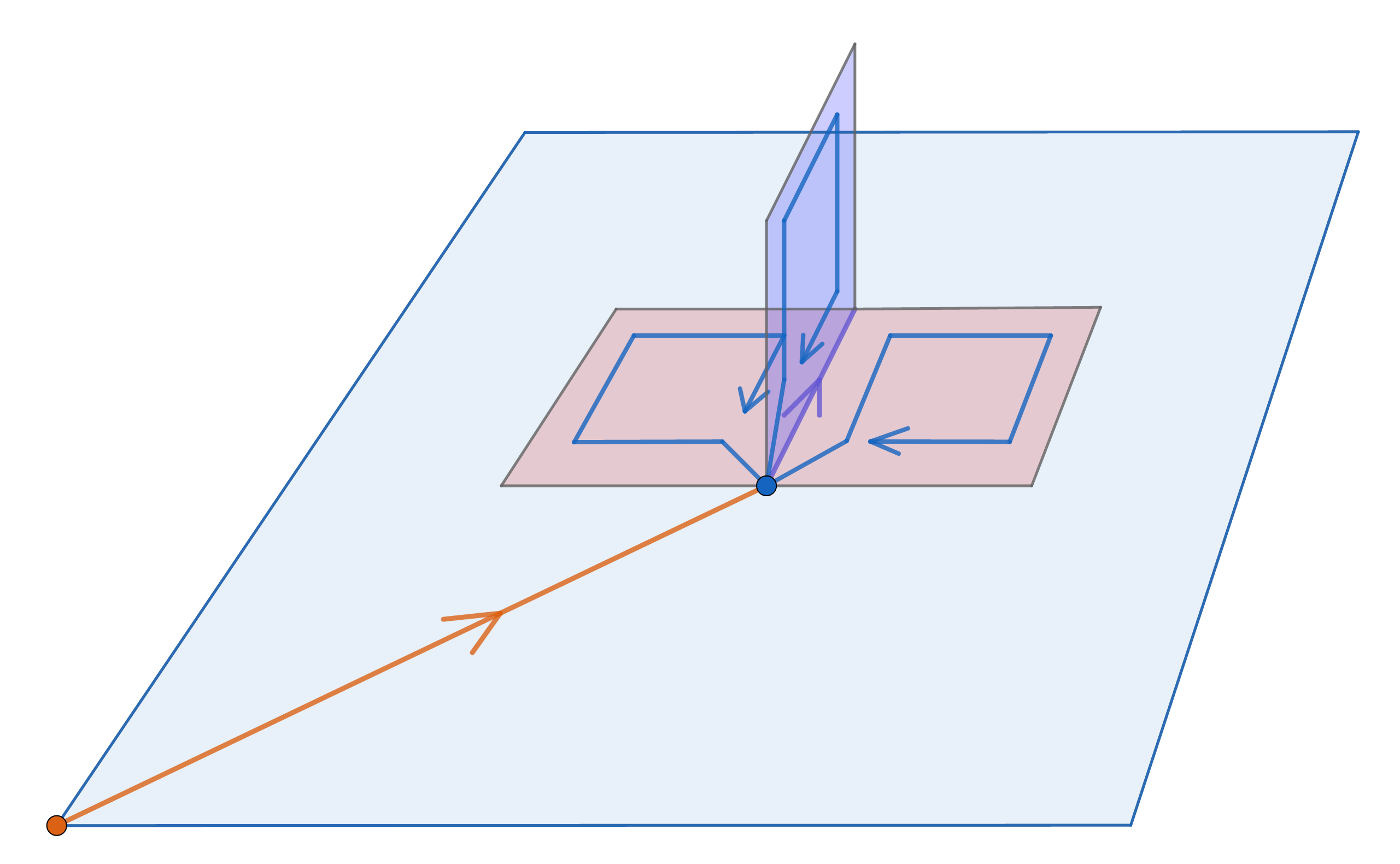}
				\put(55,24){\large $v_0$}
				\put(48,33){\large $e_1$}
				\put(65,33){\large $e_2$}
				\put(59,33){\large $i$}
				\put(-0.5,1){$s.p$}
				\put(83,4){direct membrane}
				\put(35,15){$g(s.p-v_0)$}
			\end{overpic}
			\caption{We consider an edge transform applied on an edge $i$ (purple) lying in the direct membrane (blue). Such an edge transform can affect plaquettes on the direct membrane, such as the red horizontal plaquettes; plaquettes that are cut by the dual membrane, such as the vertical plaquette; and plaquettes beneath the membrane (not shown, because these are not acted on by the magnetic membrane operator). We will consider the vertical plaquettes later, but first we examine the plaquettes lying in the direct membrane. For a given edge, $i$, there are two plaquettes that lie in the direct membrane and are attached to the edge (the plaquettes shown in red for edge $i$). Each of these plaquettes lying in the direct membrane has a blob ribbon operator associated with it, whose label depends on the label of the plaquette. Because the edge transform affects the labels of the adjacent plaquettes, the edge transform affects the labels of two blob ribbon operators associated the plaquettes. To simplify the calculation, we consider the case where the two plaquettes share a base-point, at $v_0$, and have the same orientation (this can be ensured using the re-branching procedures from the Appendix of Ref. \cite{HuxfordPaper1}).}
			\label{magnetic_membrane_edge_transform_direct_membrane}
		\end{center}
	\end{figure}

	Finally we must consider the edge transforms, which make up the edge energy terms. As usual, we first consider which edge transforms share support with the membrane operator. The action of the edge transform $\mathcal{A}_i^e$ on an edge $i'$ and plaquette $p$ is given by \cite{Bullivant2017}
	\begin{align}
		\mathcal{A}_i^e: g_{i'} &\rightarrow \begin{cases} \partial(e) g_{i'} &\text{ if $i=i'$}\\
			g_{i'} &\text{ otherwise} \end{cases} \label{Equation_edge_transform_edge_appendix}\\
		\mathcal{A}_i^e : e_p &\rightarrow \begin{cases} e_p (g(v_0(p) - s(i)) \rhd e^{-1}) &\text{if $i$ is on the plaquette $p$}\\& \text{and aligned with $p$}\\
			(g(\overline{v_0(p) - s(i)}) \rhd e) e_p &\text{if $i$ is on the plaquette $p$}\\& \text{and aligned against $p$}\\
			e_p &\text{ otherwise,} \end{cases} \label{Equation_edge_transform_plaquette_appendix}
	\end{align}
	where $g(v_0(p) - s(i))$ and $g(\overline{v_0(p) - s(i)})$ correspond to paths along the boundary of the plaquette from the base-point of the plaquette, $v_0(p)$, to the source of edge $i$ (with the overline indicating that the path is oriented against the plaquette). From this, we see that the edge transform $\mathcal{A}_i^e$ has support on the edge $i$ on which we apply the transform (the labels of which are changed by a factor $\partial(e)$), on plaquettes which are adjacent to the edge (which are changed by a factor of the form $g(t) \rhd e^{\pm 1}$), and on edges on the boundaries of those plaquettes (which determine the expression $g(t)$). Meanwhile, the magnetic membrane has support on the direct membrane (both the plaquettes, which determine the labels of the added blob ribbon operators, and the edges, which determine various path elements) and on the edges and plaquettes cut by the dual membrane. This means that we need to consider all of the edge transforms in the thickened membrane. That is, we need to consider edge transforms on the edges in the direct membrane, edges that are cut by the dual membrane, and edges which are above the dual membrane but which are attached to plaquettes cut by the dual membrane. Examples of these are shown in Figures \ref{magnetic_membrane_edge_transform_direct_membrane}, \ref{magnetic_membrane_edge_transform} and \ref{mod_mem_top_edge_example} respectively.

	First, we consider edges above the membrane, such as the one shown in Figure \ref{mod_mem_top_edge_example}. Such edge transforms do not affect the label $f(p)$ of the blob ribbon operators because they do not affect the base plaquettes or the path $s.p-v_0(b)$ that determines the label of a blob ribbon operator. In addition, edge transforms commute with blob ribbon operators that have non-operator labels, as shown in Section \ref{Section_Blob_Ribbon_Fake_Flat}. Putting these factors together, the edge transforms on edges above the membrane commute with the added blob ribbon operators.

	Next we must check that such an edge transform commutes with the action of $C^h_{\rhd}(m)$. The edge transform does not change the path labels that appear in $C^h_T(m)$ because these paths lie on the direct membrane, and even if it did change the paths, it would only be by an element in $\partial(E)$, which has no effect on terms such as $g(t)^{-1}hg(t)$ or $g(t) \rhd e$. However, this edge operator may affect one or more side plaquettes that are also potentially changed by $C^h_{\rhd}(m)$. That is, the edge lies on one or more plaquettes which are cut by the dual membrane (such as the red one in Figure \ref{mod_mem_top_edge_example}). In addition, $C^h_{\rhd}(m)$ may affect the path labels that appear in the action of the edge transform on one of these side plaquettes. Recall that $C^h_{\rhd}(m)$ changes the labels of plaquettes that are cut by the dual membrane and whose base-points are on the direct membrane. If the base-point $v_0(p)$ of a cut plaquette $p$ is above the dual membrane then the plaquette label is not changed by $C^h_{\rhd}(m)$. This means that we just need to worry about $C^h_{\rhd}(m)$ affecting the label of the path $v_0(p)-s(i)$ or $\overline{v_0(p)-s(i)}$ that appears in the edge transform. However, because the edge $i$ is above the dual membrane, the vertex $s(i)$ (which is the source of the edge) is also above the dual membrane. Given that both $v_0(p)$ and $v_i$ or $v_{i+1}$ are above the dual membrane, then the path between the two vertices can always be deformed into one that lies entirely above the dual membrane (and so is not cut by it), meaning that the expression $g(v_0(p)-s(i))\rhd e^{-1}$ or $g(\overline{v_0(p)-s(i)}) \rhd e$ that appears in the edge transform is not affected by the action of the membrane operator and so the edge transform commutes with the membrane operator. While we showed this for plaquettes whose base-points are above the dual membrane, the consistency of the membrane operator and edge transforms under the procedure where we move the base-point of plaquettes means that this must also hold for plaquettes whose base-points lie on the direct membrane (this can also be shown explicitly, but is left out here for the sake of brevity). Therefore, edge transforms applied on edges above the dual membrane commute with the magnetic membrane operator.

	Next we want to consider the edge transforms on the edges cut by the dual membrane, excluding those on the boundary of the thickened membrane (due to the fact these act on non-fake-flat plaquettes). The same argument that we gave for the top edges shows that an edge transform on the cut edges commutes with the blob ribbon operators (the transforms do not affect the labels of the ribbon operators and also commute with constant-labelled blob ribbon operators), so we only need to consider the commutation relation with $C^h_{\rhd}(m)$. Because we are considering an edge that is not on the boundary, every plaquette attached to the edge is cut by the dual membrane and so is affected by $C^h_{\rhd}(m)$ (either directly or through its edges).

	Consider one of the plaquettes affected by the edge transform, such as the red one shown in Figure \ref{magnetic_membrane_edge_transform}. We consider the case where, as in the figure, the edge on which we apply the edge transform points away from the direct membrane, so that its source lies on the direct membrane. In addition we assume that the base-point of the plaquette lies on the direct membrane. The fact that the edge energy term commutes even when the plaquette has a different branching structure will then be guaranteed by the invariance of the edge energy term (although not necessarily the individual edge transform, as we showed in the Appendix of Ref. \cite{HuxfordPaper1}) and the membrane operator under changes to the branching structure. Using the assumed branching structure, shown in Figure \ref{magnetic_membrane_edge_transform}, with the definition of the edge transform from Equation \ref{Equation_edge_transform_plaquette_appendix} and the action of $C^h_{\rhd}(m)$ from Equation \ref{Equation_magnetic_membrane_rhd_action_appendix}, we see that the combined action of the membrane operator and edge transform on the plaquette is 
	$$C^h_{\rhd}(m) \mathcal{A}_i^e :e_p = \big(g(s.p-v_0(p))^{-1}hg(s.p-v_0(p))\big) \rhd \big([g(v_0(p)-s(i)) \rhd e^{-1}] e_p\big).$$
	
	By comparison, if we apply the operators in the opposite order, the action of the two operators on the plaquette is
	\begin{align*}
		\mathcal{A}_i^e C^h_{\rhd}(m) :e_p &= \big[(g(s.p-v_0(p))^{-1}hg(s.p-v_0(p))) \rhd e_p\big] [g(v_0(p)-s(i)) \rhd e^{-1}]\\
		&= (g(s.p-v_0(p))^{-1}hg(s.p-v_0(p))) \rhd \big(e_p \big[\big(g(s.p-v_0(p))^{-1}hg(s.p-v_0(p))\big)^{-1} g(v_0(p)-s(i)) \rhd e^{-1}\big] \big)\\
		&=(g(s.p-v_0(p))^{-1}hg(s.p-v_0(p))) \rhd\\
		& \hspace{1cm} \big(e_p \big[\big(g(s.p-v_0(p))^{-1}h^{-1}g(s.p-v_0(p))g(v_0(p)-s(i))\big) \rhd e^{-1} \big]\big)\\
		&=(g(s.p-v_0(p))^{-1}hg(s.p-v_0(p))) \rhd\\
		& \hspace{1cm} \big(e_p \big[\big(g(s.p-v_0(p))^{-1}h^{-1}g(s.p-s(i))\big) \rhd e^{-1} \big]\big)\\
		&=(g(s.p-v_0(p))^{-1}hg(s.p-v_0(p))) \rhd\\
		& \hspace{1cm} \big(e_p \big[g(v_0(p)-s(i)) \rhd \big[\big(g(v_0(p)-s(i))^{-1}g(s.p-v_0(p))^{-1}h^{-1}g(s.p-s(i))\big ) \rhd e^{-1}\big] \big]\big)\\
		&=(g(s.p-v_0(p))^{-1}hg(s.p-v_0(p))) \rhd \big(e_p \big[g(v_0(p)-s(i)) \rhd [(g(s.p-s(i))^{-1}h^{-1}g(s.p-s(i))) \rhd e^{-1}]\big] \big)\\
		&=C^h_{\rhd}(m) \mathcal{A}_i^{\big(g(s.p-s(i))^{-1}h^{-1}g(s.p-s(i))\big)\rhd e}:e_p\\
		&=C^h_{\rhd}(m) \mathcal{A}_i^{h^{-1}_{[g(s.p-s(i))]}\rhd e}:e_p.
	\end{align*}
	
	In order for this relation to hold for the operators, rather than just the action on one plaquette, it must be hold for each plaquette and also for the action on the edge. Because the relation does not include any quantities particular to the plaquette (such as $v_0(p)$) it holds for all plaquettes adjacent to $i$ (with the assumed branching structure). It also holds for the action on the edge $i$ itself:
	\begin{align*}
		\mathcal{A}_i^e& C^h_{\rhd}(m) :g_i\\
		&= \partial(e)g(s.p-v_0(p))^{-1}hg(s.p-v_0(p))g_i\\
		&=g(s.p-v_0(p))^{-1}hg(s.p-v_0(p)) [g(s.p-v_0(p))^{-1}h^{-1}g(s.p-v_0(p))\partial(e)g(s.p-v_0(p))^{-1}hg(s.p-v_0(p))]g_i\\
		&= g(s.p-v_0(p))^{-1}hg(s.p-v_0(p))\partial( g(s.p-v_0(p))^{-1}h^{-1}g(s.p-v_0(p)) \rhd e)g_i\\
		&=C^h_{\rhd}(m) \mathcal{A}_i^{h^{-1}_{[g(s.p-s(i))]}\rhd e}:g_i,
	\end{align*}
	where the penultimate line follows from the Peiffer condition, Equation \ref{Equation_Peiffer_1} from the main text. Therefore, the relation
	$$C^h_{\rhd}(m) \mathcal{A}_i^{h^{-1}_{[g(s.p-s(i))]}\rhd e} \ket{\psi}=\mathcal{A}_i^e C^h_{\rhd}(m) \ket{\psi}$$
	holds for all states $\ket{\psi}$ that satisfy fake-flatness in the region of the membrane. Then, considering the edge energy term $\mathcal{A}_i$ we see that
	\begin{align*}
		\mathcal{A}_i C^h_{\rhd}(m) \ket{\psi}&= \frac{1}{|E|}\sum_{e \in E}\mathcal{A}_i^e C^h_{\rhd}(m) \ket{\psi}\\
		&=\frac{1}{|E|}\sum_{e \in E} C^h_{\rhd}(m) \mathcal{A}_i^{h^{-1}_{[g(s.p-s(i))]}\rhd e} \ket{\psi}\\
		&=C^h_{\rhd}(m) \frac{1}{|E|}\sum_{e'= h^{-1}_{[g(s.p-s(i))]}\rhd e\in E} \mathcal{A}_i^{e'} \ket{\psi}\\
		&=C^h_{\rhd}(m) \mathcal{A}_i \ket{\psi}.
	\end{align*}
	This implies that $\mathcal{A}_i$, the total edge operator, commutes with the membrane operator when acting on a fake-flat state. The condition of requiring fake-flatness is necessary to ensure that the choice of paths in the various operators is always consistent.

	Having considered edge transforms for edges above the membrane and edge transforms on the edges cut by the dual membrane, we now need to look at the edge operators for edges that lie on the direct membrane itself. Unlike the other two types of edge transform, these edge transforms can affect the labels $f(b)$ of the blob ribbon operators, in addition to affecting the plaquettes cut by the dual membrane. We will first consider the effect on the labels of the blob ribbon operators, before considering the action on the plaquettes cut by the dual membrane. Recall that each blob ribbon operator in $C^h_T(m)$ corresponds to a ``base" plaquette on the direct membrane, and the label of the blob ribbon operator corresponding to the plaquette $b$ is (as stated in Equation \ref{Equation_magnetic_blob_ribbon_label})
	$$f(b)=[g(s.p-v_{0}(b))\rhd e_b] [\big( h^{-1}g(s.p-v_0(b))\big)\rhd e_b^{-1}].$$
	
	An edge transform on the interior of the direct membrane affects the value of two such labels: those corresponding to the two plaquettes that lie on the direct membrane and are attached directly to the edge, as shown in Figure \ref{magnetic_membrane_edge_transform_direct_membrane}. That is, an edge transform on the direct membrane affects the labels $f(b)$ of blob ribbon operators corresponding to plaquettes $b$ which are adjacent to the edge and therefore have labels which are affected by the edge transform. We may imagine that the edge transform could also change a label $f(b)$ by changing the path variable that appears in the label $f(b)$. However, this is not the case because the edge transform can only change the path $g(s.p-v_0(b))$ by an element of $\partial(E)$, which does not affect $g(s.p-v_0(b)) \rhd e_b$, as we discussed at the start of Section \ref{Section_Magnetic_Tri_Non_Trivial}. Therefore, we need to consider the direct effect of the edge transform on the labels of the base plaquettes, which determine the labels $f(b)$ of the blob ribbon operators. For simplicity, we consider the case where the two affected plaquettes have the same base-point and where this base-point is the source of the edge on which we apply the edge transform (applying the re-branching procedures from the Appendix of Ref. \cite{HuxfordPaper1} if necessary to obtain this case).

	Using the notation from Figure \ref{magnetic_membrane_edge_transform_direct_membrane}, the plaquette labels transform under the edge transform as $e_1 \rightarrow e e_1$ and $e_2 \rightarrow e_2 e^{-1}$. Therefore, the labels $f_1$ and $f_2$ of the corresponding blob ribbon operators transform as 
	\begin{align*}
		f_1 \rightarrow& [g(s.p-v_0)\rhd(ee_1)] \: [(h^{-1}g(s.p-v_0)) \rhd (e_1^{-1}e^{-1})]\\
		=&f_1 [g(s.p-v_0)\rhd e]\: [(h^{-1}g(s.p-v_0)) \rhd e^{-1}],\\
		f_2 \rightarrow& f_2 [g(s.p-v_0)\rhd e^{-1}]\:[ (h^{-1}g(s.p-v_0)) \rhd e].
	\end{align*}

\begin{figure}[h]
		\begin{center}
			\begin{overpic}[width=0.75\linewidth]{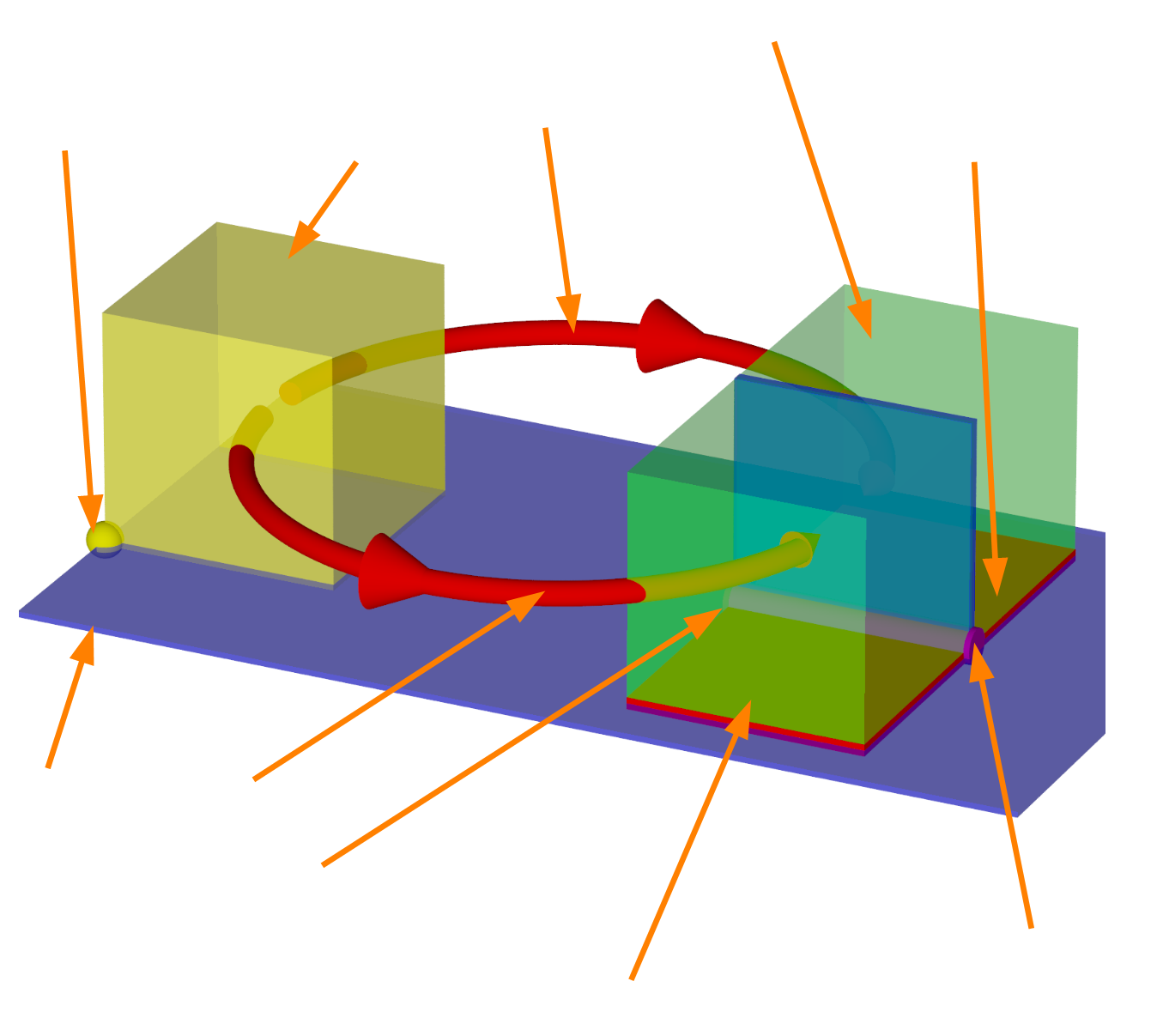}
				\put(3,77){ $s.p$}
				\put(18,20){(2)}
				\put(45,79){(1)}
				\put(31,75){blob 0}
				\put(20,14){blob 2}
				\put(64,86){blob 1}
				\put(52,2){plaquette 2}
				\put(82,76){plaquette 1}
				\put(88,7){edge $i$}
				\put(-1,20){\parbox{2cm}{\raggedright direct membrane}}
			\end{overpic}
			\caption{Under commutation with the edge transform on edge $i$ (purple), the blob ribbon operators produce two additional blob ribbon operators, on the paths (1) and (2) (shown in red). These paths travel from blob 0 (yellow) to the blobs 1 and 2 (green) on either side of the edge $i$ on which we apply the edge transform. We see that the paths (1) and (2) almost meet, but are separated by the plaquette or plaquettes attached to the edge $i$ itself (represented by the vertical blue square).}
			\label{edgeblobpathsmodmem}
		\end{center}
	\end{figure}

	In Sections \ref{Section_Blob_Ribbon_Fake_Flat} and \ref{Section_Blob_Ribbon_Central} we showed that edge transforms commute with ordinary blob ribbon operators, that is blob ribbon operators with constant (rather than operator) labels. Therefore, the non-commutativity of the blob ribbon operators and the edge transform comes only from the transformation of the blob ribbon labels. From this, we can deduce that the commutation relation between the set of blob ribbon operators in the magnetic membrane operator and the edge transform is
	\begin{align}
		\bigg(\prod_{b \in \text{membrane}}& B^{\hat{f}(b)}(\text{blob }0 \rightarrow \text{blob } b) \bigg) \mathcal{A}_i^e \notag\\
		&= \mathcal{A}_i^e \bigg(\prod_{b \in \text{membrane}} B^{f(b)}(\text{blob }0 \rightarrow \text{blob } b)\bigg) B^{[g(s.p-v_0)\rhd e] \: [(h^{-1}g(s.p-v_0)) \rhd e^{-1}]}(\text{blob }0 \rightarrow \text{blob }1) \notag\\
		& \quad \quad \quad B^{[g(s.p-v_0)\rhd e^{-1}] [(h^{-1}g(s.p-v_0)) \rhd e]}(\text{blob }0 \rightarrow \text{blob }2)
		\label{mod_mem_blobs_direct_edge_commutation}
	\end{align}	
	where blob 1 and blob 2 are the blobs attached to the plaquettes labelled $e_1$ and $e_2$ respectively and we split off the extra part of the blob ribbon operators using the relation $B^{k_1k_2}(t)=B^{k_1}(t)B^{k_2}(t)$. From now on, the ribbons $(\text{blob }0 \rightarrow \text{blob }1)$ and $(\text{blob }0 \rightarrow \text{blob }2)$ of the extra blob ribbon operators resulting from the edge transform will be denoted by the shorthand (1) and (2) respectively. These ribbons are illustrated in Figure \ref{edgeblobpathsmodmem}.
	
	From the fact that the ribbons of the two additional blob ribbon operators both start at blob 0, and end on adjacent blobs, we see that the two ribbons almost connect with each-other. If the two blob ribbon operators did terminate in the same blob, then it would be possible to deform them so that they act on the same ribbon. In Section \ref{Section_Topological_Blob_Ribbons}, we show that the non-confined blob ribbon operators are topological, by which we mean that smoothly deforming the ribbon on which we apply the blob ribbon operator (while keeping the end-points fixed) does not affect its action on the ground state (or equivalently, a non-confined blob ribbon acting on a closed ribbon is trivial when acting on the ground state). This means that if the two blob ribbon operators terminated in the same blob, we could smoothly deform one ribbon into the other without affecting the action of the operators. Then, because the labels of the two blob ribbon operators are the inverse of each-other, they would cancel if they acted on the same ribbon. Because the ribbons do not terminate on the same blob this does not seem possible. However, note that the plaquettes that separate the two blobs at the ends of the two blob ribbon operators (the blue vertical plaquette in Figure \ref{edgeblobpathsmodmem}) are attached directly to the edge on which we apply the transform (edge $i$ in Figure \ref{edgeblobpathsmodmem}). This suggests that these plaquettes may be further affected by the edge transform. We will soon see that the commutation of $\mathcal{A}_i^e$ with $C^h_{\rhd}$ provides the last piece of ribbon to connect the two ribbon operators and thus allows us to join the ribbon operators into a single trivial ribbon operator (when acting on the ground state).

	\begin{figure}[h]
		\begin{center}
			\begin{overpic}[width=0.6\linewidth]{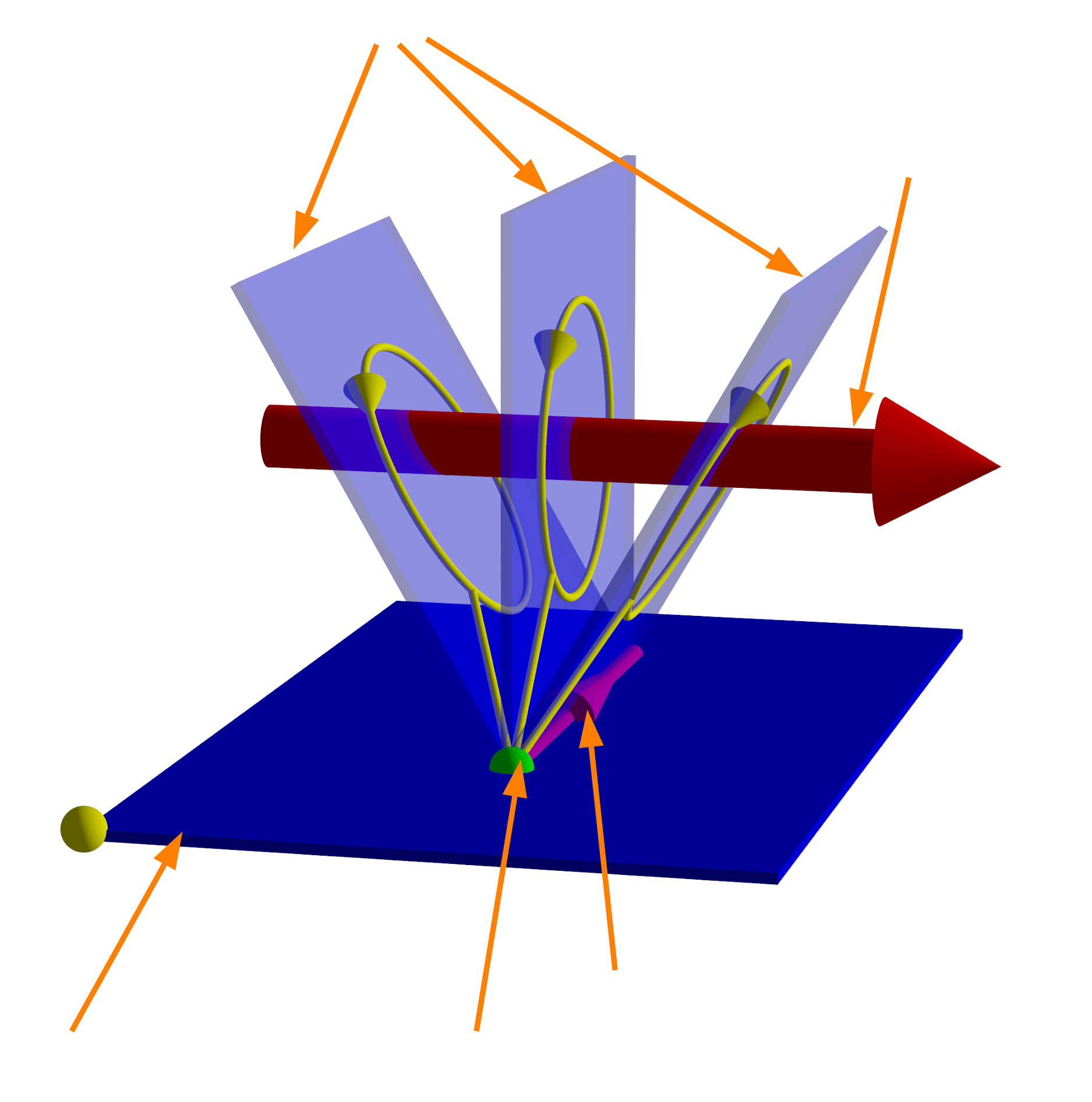}
				\put(0,4){direct membrane}
				\put(42,4.5){$v_i$}
				\put(51,10){edge $i$}
				\put(70,85){orientation of plaquettes}
				\put(20,98.5){\parbox{5cm}{\raggedright plaquettes $\set{q}$ attached to $i$ and cut by dual membrane}}
				
			\end{overpic}
			\caption{We consider the commutation relation between an edge transform on edge $i$ (purple) and $C^h_{\rhd}(m)$. These operators both act on the plaquettes $\set{q}$ (pale blue squares pierced by the red arrow) which are cut by the dual membrane and attached to edge $i$. We use the re-branching procedures discussed in the Appendix of Ref. \cite{HuxfordPaper1} to put all of the base-points of the plaquettes at the source of edge $i$, $v_i$ (green sphere). We also ensure that all of the plaquettes have the same orientation, with the circulation of each plaquette shown by the curved (yellow) arrows and the resulting orientation (found using the right-hand rule) shown by the large (red) arrow. Then we will see that the commutation of the membrane operator with the edge transform produces an additional action on these plaquettes, which is equivalent to a blob ribbon operator acting along the large (red) arrow.}
			\label{plaquettesonedge}
		\end{center}
	\end{figure}
	
	To that end, we now consider the commutation relation of the edge transform with $C^h_{\rhd}(m)$. For simplicity, we again put the base-point of each affected plaquette in the same position, at the source of $i$, and orient the plaquettes so that the edge has the same orientation as each plaquette, as shown in Figure \ref{plaquettesonedge}. Then for the cut plaquettes $\set{q}$ in the figure, with labels $\set{e_q}$, the combined action of the edge transform and $C^h_{\rhd}(m)$ on a plaquette $q$, if we first apply the membrane operators, is:
	$$\mathcal{A}_i^e C^h_{\rhd}(m):e_q =[(g(s.p-v_i)^{-1}hg(s.p-v_i))\rhd e_q] e^{-1},$$
	whereas for the other order of operators we have
	\begin{align}
		C^h_{\rhd}(m) \mathcal{A}_i^e:e_j &= (g(s.p-v_i)^{-1}hg(s.p-v_i))\rhd (e_j e^{-1}) \notag\\
		&=(\mathcal{A}_i^e C^h_{\rhd}(m):e_j) \cdot (e \: [g(s.p-v_i)^{-1}hg(s.p-v_i)\rhd e^{-1}])
		\label{mag_mem_direct_edge_commutation}
	\end{align}
	
	On the other hand, $C^h_{\rhd}(m)$ doesn't affect the plaquettes attached to edge $i$ that are not cut by the membrane (those plaquettes on or beneath the membrane), so the actions of the operators on these plaquettes commute. Because the commutation relation is different for the action on the different plaquettes, the commutation does not simply produce an edge transform of a different label as it has for some other commutation relations. This means that the membrane operator will not commute with the edge term even when we sum over the edge transforms to get the energy term $\mathcal{A}_i$. Instead, under commutation with $\mathcal{A}_i^e$ we produce an extra action just on the cut plaquettes, as we claimed earlier. Now we must check if that action does indeed allow us to complete the trivial ribbon operator that we mentioned earlier.

	The extra action of multiplication by $e [(g(s.p-v_i)^{-1}hg(s.p-v_i))\rhd e^{-1}]$ is equivalent to the action of a blob ribbon operator passing through these plaquettes, with start-point at $v_i$, with label $[g(s.p-v_i)^{-1}hg(s.p-v_i)\rhd e] e^{-1}$, as illustrated in Figure \ref{plaquettesonedge}. We wish to write this in terms of the action of a blob ribbon operator which has $s.p$ as its start-point, to match the other ribbon operators produced by the edge transform. This is because matching the start-points is a prerequisite for combining the blob ribbon operators, as explained in Section \ref{Section_blob_ribbon_concatenate}. We saw in Section \ref{Section_blob_ribbon_move_sp} that we can move the start-point of the direct path of a blob ribbon operator $B^x(t)$ without changing its dual path (the plaquettes it pierces) and without changing the action of the operator, if we change its label from $x$ to $g(s.p(t)-s.p(t'))^{-1} \rhd e$, where $s.p(t')$ is the new start-point. In the case of the blob ribbon operator shown in Figure \ref{plaquettesonedge}, we can change the start-point of the blob ribbon operator of label from $v_i$ to $s.p$ by changing its label from $e [g(s.p-v_i)^{-1}hg(s.p-v_i)\rhd e^{-1}]$ to
	\begin{align*}
		g(v_i-s.p)^{-1} \rhd([(g(s.p-v_i)^{-1}hg(s.p-v_i))\rhd e] \: e^{-1})&= g(s.p-v_i) \rhd([(g(s.p-v_i)^{-1}hg(s.p-v_i))\rhd e] \: e^{-1})\\
		&=[(hg(s.p-v_i)) \rhd e] \: [g(s.p-v_i) \rhd e^{-1}]\\
		&= h \rhd \big([g(s.p-v_i) \rhd e] \: [(h^{-1}g(s.p-v_i)) \rhd e^{-1}]\big).
	\end{align*}
	
\begin{figure}[h]
		\begin{center}
			\begin{overpic}[width=0.6\linewidth]{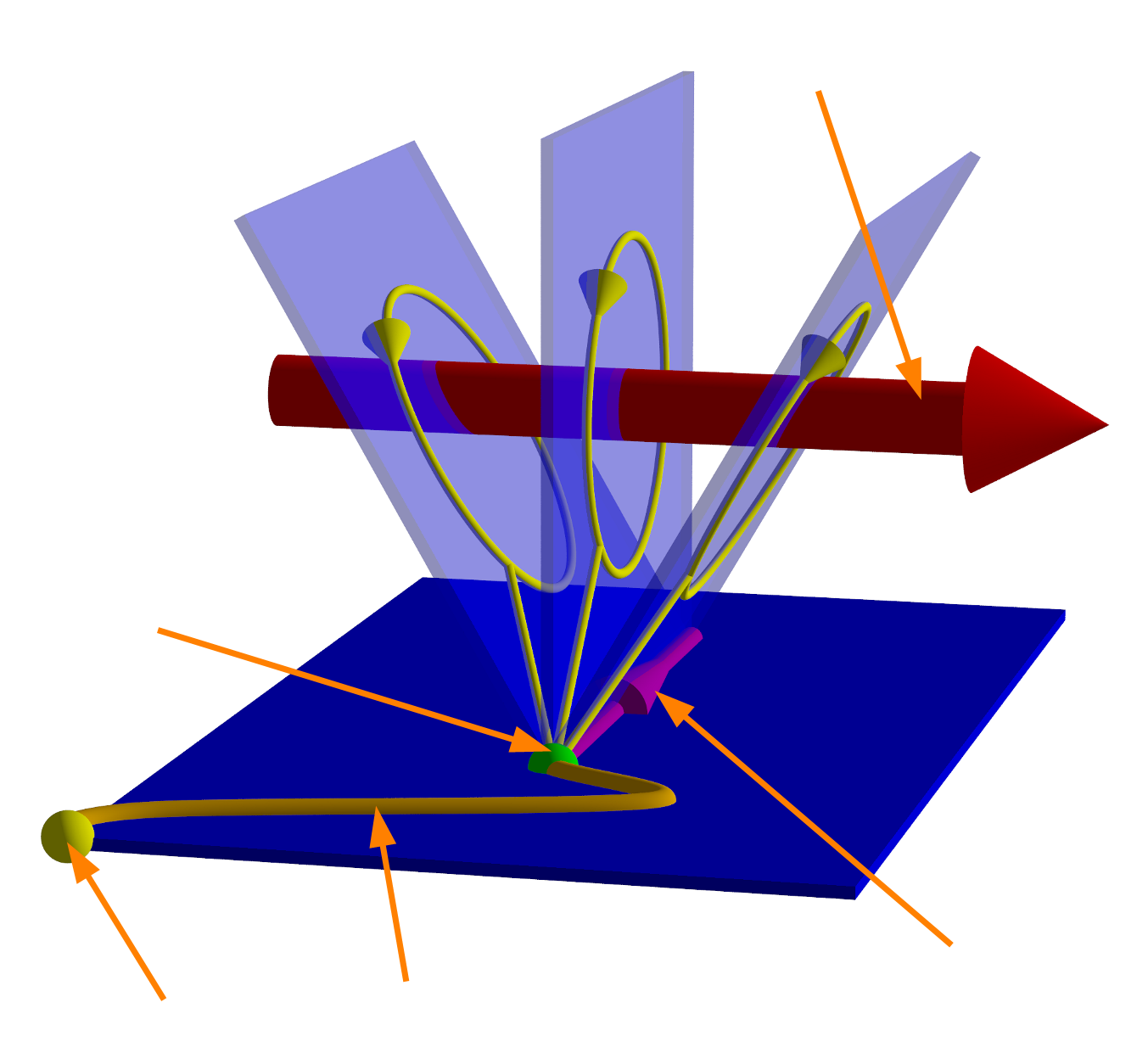}
				\put(14,2){$s.p$}
				\put(32,4.5){direct path of ribbon $(\delta)$}
				\put(10,37){$v_i$}
				\put(62,84){dual path of ribbon $(\delta)$}
				\put(84,9){edge $i$}
			\end{overpic}
			\caption{The extra action on the plaquettes cut by the dual membrane and attached to edge $i$ from the commutation of the edge transform $\mathcal{A}_i^e$ and the operator $C^h_{\rhd}(m)$ is equivalent to the action of a blob ribbon operator passing through these plaquettes. It is convenient to choose the start-point of this ribbon operator the same as the start-point of the membrane operator (we can change the start-point of a blob ribbon operator by also changing the label of the ribbon operator). The direct path for this ribbon is indicated by the curved path.}
			\label{ribbon_delta}
		\end{center}
	\end{figure}

	We denote the resulting ribbon, with start-point at $s.p$ and dual path passing through the plaquettes attached to edge $i$ by $(\delta)$ (see Figure \ref{ribbon_delta}). Notice that the label of the ribbon operator is $h \rhd \big([g(s.p-v_i) \rhd e] \: [(h^{-1}g(s.p-v_i)) \rhd e^{-1}]\big)$, which is very similar to the label of the blob ribbon operators produced by the commutation of the edge transform with the blob ribbon operators (see Equation \ref{mod_mem_blobs_direct_edge_commutation}).

	Having considered the commutation relations of the edge transforms with the added blob ribbon operators and with $C^h_{\rhd}(m)$, we now combine these to consider the commutation of the edge transform with the whole membrane operator $C^h_T(m)$. Putting the commutation relation we just obtained for $C^h_{\rhd}(m)$ with the commutation relation for the blob ribbon operators (Equation \ref{mod_mem_blobs_direct_edge_commutation}), we have
	\begin{align}
		C^h_T(m) \mathcal{A}_i^e =&C^h_{\rhd}(m) \bigg(\prod_{\substack{\text{plaquette } b \notag\\ \in \text{membrane}}}B^{\hat{f}(b)}(\text{blob }0 - \text{blob attached to }b) \bigg)\mathcal{A}_i^e \notag\\
		=&C^h_{\rhd}(m) \mathcal{A}_i^e \bigg(\prod_{b \in \text{membrane}} B^{f(b)}(\text{blob }0 - \text{blob }b)\bigg) \: B^{[g(s.p-v_i)\rhd e] \: [(h^{-1}g(s.p-v_i))\rhd e^{-1}]}(1)\notag \\
		& \hspace{1cm}B^{[g(s.p-v_i)\rhd e^{-1}] \: [(h^{-1}g(s.p-v_i))\rhd e]}(2)\notag\\
		=&\mathcal{A}_i^e B^{[hg(s.p-v_i)\rhd e] [g(s.p-v_i) \rhd e^{-1}]}(\delta) C^h_T(m) B^{[g(s.p-v_i)\rhd e] \: [(h^{-1}g(s.p-v_i))\rhd e^{-1}]}(1) \notag \\
		& \hspace{1cm} B^{[g(s.p-v_i)\rhd e^{-1}] \: [(h^{-1}g(s.p-v_i))\rhd e]}(2).
		\label{magnetic_membrane_base_edge_commutation_1}
	\end{align}
	
	We notice that one of the extra blob ribbon operators is to the left of $C^h_T(m)$ and the other two are to the right. We wish to group these operators together, so we need to commute the blob ribbon operator applied on $(\delta)$ past $C^h_T(m)$. We therefore need to know the commutation relation for this blob ribbon operators with $C^h_T(m)$. The blob ribbon operators in $C^h_T(m)$ will commute with the blob ribbon operator on $(\delta)$, because the blob ribbon operators do not affect each-others' labels and blob ribbon operators otherwise commute when $E$ is Abelian (since two blob ribbon operators simply multiply plaquette labels by factors which will commute). On the other hand, $C^h_{\rhd}(m)$ will not commute with the blob ribbon operator on $(\delta)$. This is because $C^h_{\rhd}(m)$ acts on each of the plaquettes cut by the dual membrane, which includes all of the plaquettes pierced by the dual path of the blob ribbon operator. The operator $C^h_{\rhd}(m)$ applies the map $g(s.p-v_0)^{-1}hg(s.p-v_0)\rhd$ on the label of such a plaquette if the base-point $v_0$ of the plaquette is on the direct membrane. Even if the base-point of the plaquette is not on the direct membrane, the magnetic membrane operator will change the path label for the path from the start-point of the ribbon $(\delta)$ to the base-point of that plaquette, which will then affect the action of the blob ribbon operator. For simplicity we can use the re-branching procedures to put the base-point of every plaquette pierced by the blob ribbon operators onto the direct membrane, so that we do not need to consider the two cases separately. Then we have:
	\begin{align}
		C^h_T(m) B^e(\delta) : e_p &= (g(s.p-v_0)^{-1}hg(s.p-v_0))\rhd (g(s.p-v_0)^{-1} \rhd e e_p) \notag\\
		&=[(g(s.p-v_0)h) \rhd e] [g(s.p-v_0)^{-1}hg(s.p-v_0)\rhd e_p] \notag\\
		B^e(\delta) C^h_T(m) :e_p &=[g^{-1}(s.p-v_0) \rhd e] [g(s.p-v_0)^{-1}hg(s.p-v_0)\rhd e_p] \notag\\
		&= C^h_T(m) B^{h^{-1}\rhd e}(\delta):e_p \notag\\
		& \implies C^h_T(m) B^e(\delta) = B^{h \rhd e}(\delta) C^h_T(m).
		\label{blob_ribbon_magnetic_commutation}
	\end{align}

	While we have shown this relation while considering the ribbon $(\delta)$, this commutation relation holds for any blob ribbon operator which shares a start-point with $m$ and acts entirely within the thickened membrane, including the blob ribbon operators that we included in $C^h_T(m)$ itself. Because $C^h_T$ does not affect the operator label of the blob ribbon operator that we wish to commute past it, we can use Equation \ref{blob_ribbon_magnetic_commutation} to rewrite Equation \ref{magnetic_membrane_base_edge_commutation_1} as
	\begin{align}
		C^h_T(m) \mathcal{A}_i^e&=\mathcal{A}_i^e C^h_T(m) B^{[g(s.p-v_0)\rhd e] [(h^{-1}g(s.p-v_0)) \rhd e^{-1}]}(\delta) \notag\\
		& \hspace{2cm} B^{[g(s.p-v_0)\rhd e] [(h^{-1}g(s.p-v_0)) \rhd e^{-1}]}(1) B^{[g(s.p-v_0)\rhd e^{-1}][(h^{-1}g(s.p-v_0)) \rhd e]}(2). \label{Equation_edge_transform_mag_membrane_three_ribbons}
	\end{align}
	
\begin{figure}[h]
		\begin{center}
			\begin{overpic}[width=0.6\linewidth]{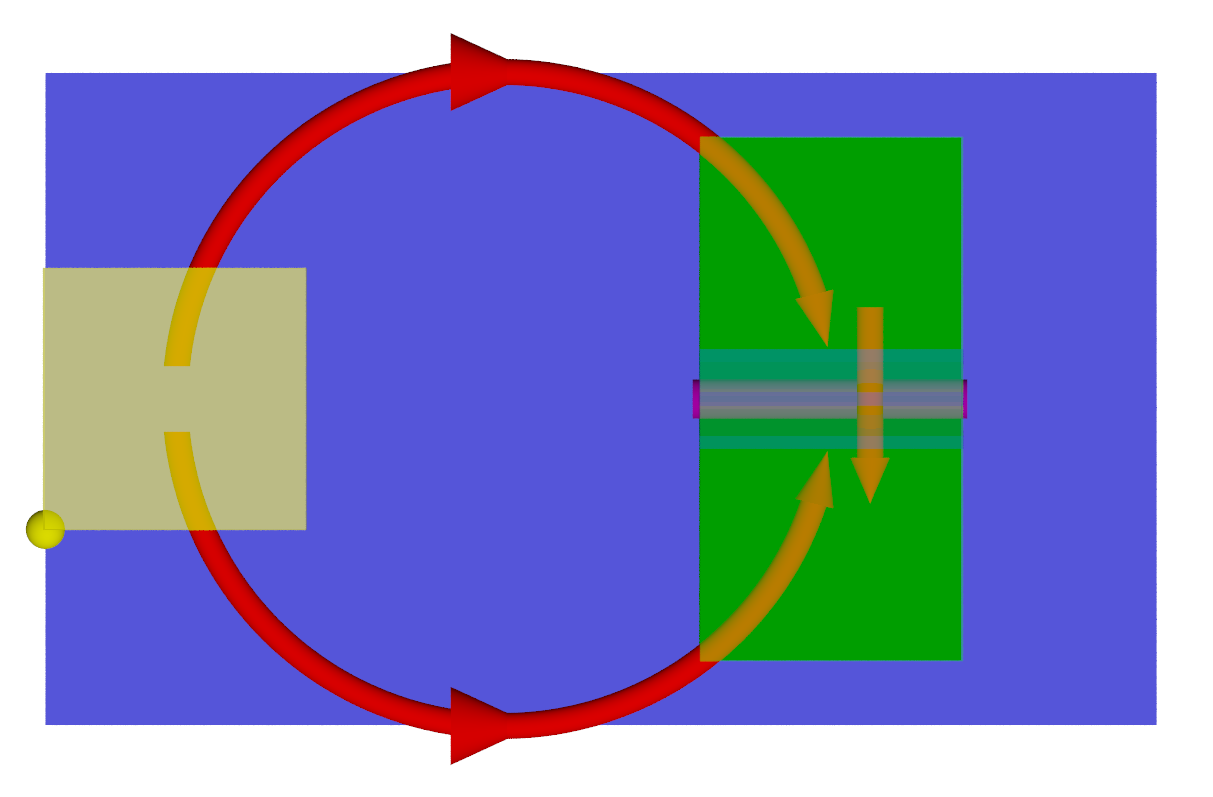}
				\put(43,62){\large (1)}
				\put(43,2){\large (2)}
				\put(73,28){\large ($\delta$)}
				
				\put(0,18){$s.p$}
				\put(70,48){blob 1}
				\put(70,18){blob 2}
				\put(5,32){blob 0}
			\end{overpic}
			\caption{The commutation of the edge transform on edge $i$ (purple) with $C^h_{\rhd}(m)$ produces an extra blob ribbon operator acting on the ribbon ($\delta$). Adding this path to Figure \ref{edgeblobpathsmodmem}, and taking a vertical view for clarity, we see that this path connects blobs 1 and 2 (green) and so bridges the gap between ribbons (1) and (2). ($\delta$) has the same orientation and start-point (yellow sphere) as (1) so we can combine the ribbon operators on (1) and $(\delta)$ into a single ribbon operator.}
			\label{path_delta_on_mod_mem}	
		\end{center}
	\end{figure}

	We now want to combine the three blob ribbon operators on ribbons (1), (2) and ($\delta$). The ribbons (1) and ($\delta$) have the same start-point (s.p) for the direct path, the same label and have dual paths that connect (i.e., that can be concatenated), as shown in Figure \ref{path_delta_on_mod_mem}. Therefore, we can combine the two blob ribbon operators on these ribbons into one blob ribbon operator with the same label and applied on the ribbon ($1 \cdot \delta$) produced by concatenating the dual paths. 
	
\begin{figure}[h]
		\begin{center}
			\begin{overpic}[width=0.6\linewidth]{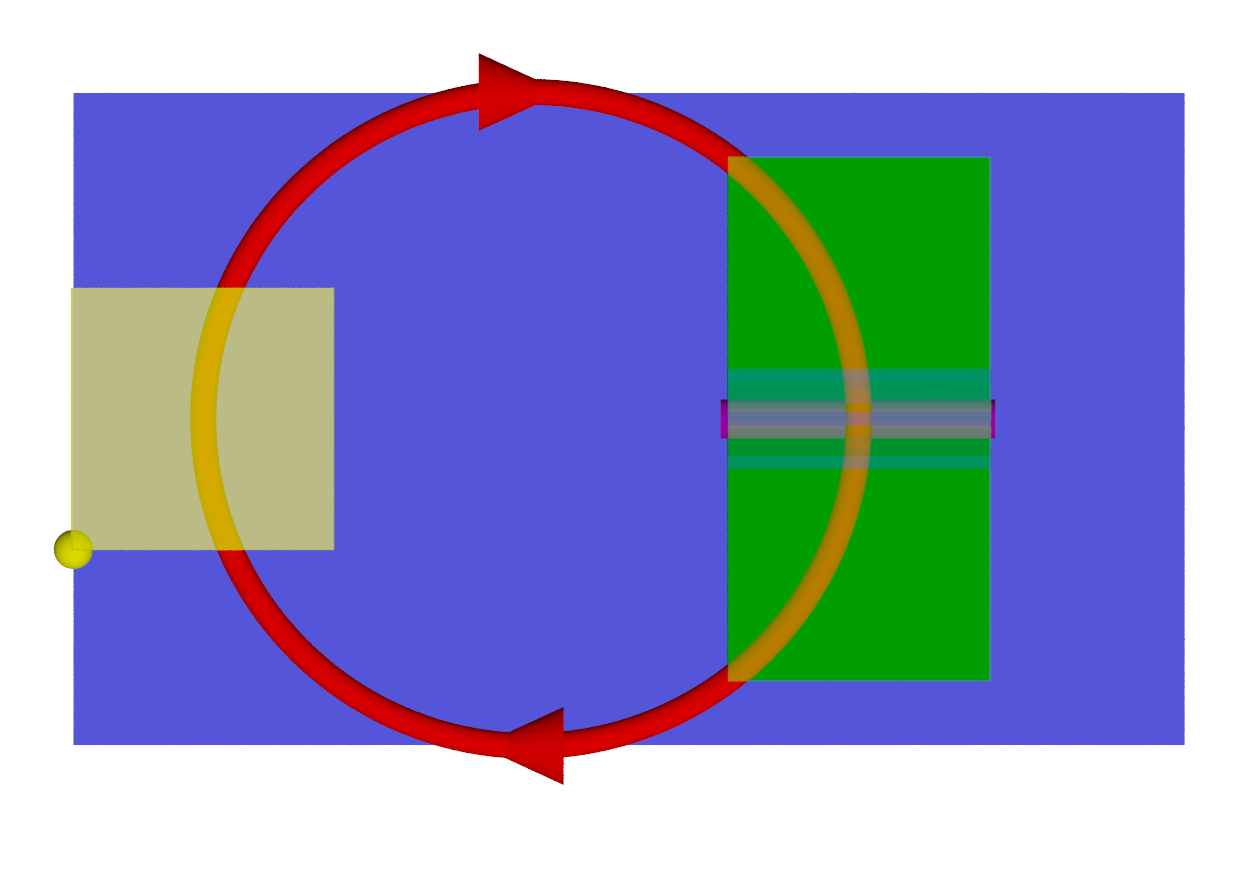}
				\put(7,35){blob 0}
				\put(1,24){$s.p$}
				\put(80,37){edge $i$}
				\put(70,45){blob 1}
				\put(70,27){blob 2}
				\put(20,5){closed ribbon $(1)\cdot(\delta)\cdot(2)^{-1}$}
			\end{overpic}
			\caption{The ribbon operators produced by the commutation of the edge transform and the magnetic membrane operator, on the ribbons (1), (2) and $(\delta)$ shown in Figure \ref{path_delta_on_mod_mem}, can be combined into a ribbon operator on a closed ribbon. As we show in Section \ref{Section_Topological_Blob_Ribbons}, such a ribbon operator (with label in the kernel of $\partial$) acts trivially on states for which we can contract the ribbon to nothing without crossing any excitations.}
			\label{closed_blob_ribbon_on_mod_mem}	
		\end{center}
	\end{figure}

	The blob ribbon operator on (2) has the inverse of the label of the other two ribbon operators. However, if we want to combine it with the other two, we also have to reverse the orientation of its dual path. We can swap the orientation of the dual path, while simultaneously inverting the label of the blob ribbon operator, without affecting its action, as discussed in Section \ref{Section_blob_ribbon_invert_dual_path}. Doing this swaps its label from $[g(s.p-v_0)\rhd e^{-1}](h^{-1}g(s.p-v_0) \rhd e)$ to
	\begin{align*}
		\big([g(s.p-v_0)\rhd e^{-1}][(h^{-1}g(s.p-v_0)) \rhd e]\big)^{-1} &= [(h^{-1}g(s.p-v_0)) \rhd e^{-1}] [g(s.p-v_0)\rhd e]\\
		&= [g(s.p-v_0)\rhd e] [(h^{-1}g(s.p-v_0)) \rhd e^{-1}],
	\end{align*}
	where in the last line we used the fact that $E$ is Abelian. We see that this label matches the labels of the blob ribbon operators on $(1)$ and $(\delta)$ from Equation \ref{Equation_edge_transform_mag_membrane_three_ribbons}. Together with the fact that these three ribbons share a start-point and now have dual paths that lie end-to-end, this means that we can combine the blob ribbon operator on (2) with the other blob ribbon operators. The resulting ribbon is closed (as shown in Figure \ref{closed_blob_ribbon_on_mod_mem}), and the blob ribbon operator has a label in the kernel of $\partial$ (and is therefore unconfined). This means that we can contract the closed ribbon into a trivial ribbon, as we show in Section \ref{Section_Topological_Blob_Ribbons}, as long as there are no excitations in the region enclosed by the ribbon operator. Therefore, $C^h_T(m)$ commutes with the edge transforms for edges that lie in the direct membrane. Combining this with our previous results, we see that $C^h_T(m)$ commutes with all of the edge transforms in the bulk of the membrane. 
	
\begin{figure}[h]
		\begin{center}
			\begin{overpic}[width=0.9\linewidth]{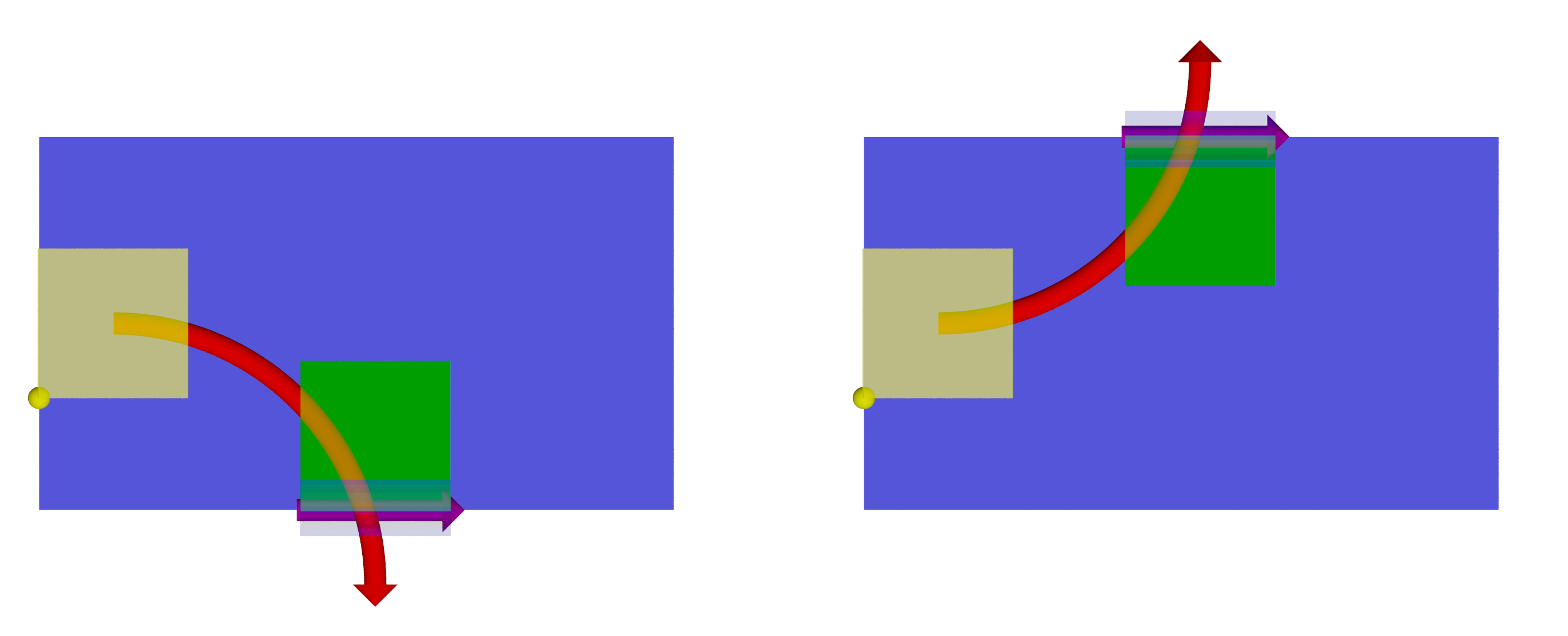}
				
				\put(28,4){\large $i$}
				\put(15,19){\large $(1 \cdot \delta)$}
				
				\put(70,20){\large $(2\cdot \delta^{-1})$}
				\put(80,34){\large $i$}
			\end{overpic}
			\caption{If we perform an edge transform on an edge $i$ on the boundary of the dual membrane (the purple edge in each side of the figure), then the commutation with the magnetic membrane operator produces a non-trivial blob ribbon operator on a ribbon that starts at blob 0 and exits the membrane near edge $i$. If the edge is anti-clockwise on the direct membrane, as in the left side, we produce blob ribbon operators on the ribbon $(1 \cdot \delta)$ shown, but no blob ribbon operator on ribbon $(2)$ to close the ribbon. If the edge is clockwise, we instead produce blob ribbon operators on the ribbon $(2 \cdot \delta^{-1})$, as shown in the right side of the figure.}
			\label{mod_mem_edge_transform_boundary}	
		\end{center}
	\end{figure}

	While we have considered all of the edges in the bulk of the membrane, we still need to consider edges on the boundary of the membrane. For edges that lie flat along the boundary of the direct membrane, we can use the reasoning from the case of edges in the bulk of the direct membrane. Because an edge on the boundary of the direct membrane only affects one plaquette in the direct membrane, it only affects the label of one of the blob ribbon operators in $C^h_T(m)$ and so only produces one of the extra blob ribbon operators from the commutation relation in Equation \ref{mod_mem_blobs_direct_edge_commutation}. The commutation relation between the edge transform and $C^h_{\rhd}(m)$ still produces the ribbon operator along the short ribbon $(\delta)$. This means that the commutation relation produces ribbon operators along $(1)$ and $(\delta)$ or $(2)$ and $(\delta)$, but not along all three ribbons. Therefore, we cannot combine the ribbon operators into a closed ribbon. This means that the commutation of the edge transform $\mathcal{A}_i^e$ with the membrane operator produces a non-trivial blob ribbon operator that travels from blob 0 past the edge on which we apply the edge transform, as shown in Figure \ref{mod_mem_edge_transform_boundary}. If the edge $i$ is anticlockwise on the direct membrane we have that
	$$C^h_T \mathcal{A}_i^e = \mathcal{A}_i^e B^{[hg(s.p-v_0)\rhd e] [g(s.p-v_0) \rhd e^{-1}]}(1 \cdot \delta) C^h_T,$$
	where the ribbon $(1 \cdot \delta)$ is shown in the left of Figure \ref{mod_mem_edge_transform_boundary}. On the other hand if the edge is clockwise on the direct membrane we have
	$$C^h_T \mathcal{A}_i^e = \mathcal{A}_i^e B^{[hg(s.p-v_0)\rhd e^{-1}] [g(s.p-v_0) \rhd e]}(2 \cdot \delta^{-1}) C^h_T,$$
	where the ribbon $(2 \cdot \delta^{-1})$ is shown in the right of Figure \ref{mod_mem_edge_transform_boundary}. This result from the commutation relation means that the boundary edge terms neither commute with the membrane operator, nor are definitely excited by it.
	
	In addition to the edge transforms on edges along the boundary of the direct membrane, we must consider the other edges on the boundary of the thickened membrane. Namely, we must consider the ``upwards" edges on the boundary (those cut by the dual membrane and on the boundary of the thickened membrane). These edges are adjacent to plaquettes that violate fake-flatness by a general element of $G$, so (as shown in the Appendix of Ref. \cite{HuxfordPaper1}) the edge transforms on such edges are not invariant under re-branching. We therefore cannot sensibly consider these edge transforms, except in special cases where the plaquettes are not excited. Such a special case can occur if we ``close" the membrane so that the boundary meets itself, depending on whether the membrane encloses any excitations (for example we can close a rectangular membrane into a torus by folding it so that the opposite edges touch). We will see an example of this in Section \ref{Section_3D_Topological_Charge_Torus_Tri_nontrivial}, where we consider using closed membrane operators to measure topological charge (we will see that there is a condition required for the boundary to join smoothly to itself with no resulting excitations).

	Similarly, blobs on the boundary of the membrane are adjacent to excited plaquettes and so the corresponding blob energy terms will not be well-defined, in that whether they are excited may depend on the branching structure. We therefore ignore these energy terms in the same way. It would be possible to formalize these ideas by defining the energy terms for blobs and edges to always be zero if any adjacent plaquettes do not satisfy fake-flatness, in which case these energy terms for the boundary would be excited by the magnetic membrane operator. However, we did not feel this was necessary because it does not affect the regions that are excited by the operator, only which energy terms within that region are excited. Regardless of how we treat these ill-defined energy terms, it is still true that plaquette energy terms around the boundary of the membrane are excited by the membrane operator, which gives rise to a loop-like excitation. Making the other energy terms in this region definitely excited would not change this or other important properties of the excitation.

	We have therefore shown that the magnetic membrane operator $C^h_T(m)$ commutes with the energy terms in the bulk of the membrane in the case where $\partial(E)$ is in the centre of $G$ and $E$ is Abelian. We also saw that, unless $h=1_G$ (in which case the membrane operator is the identity operator), the magnetic membrane operator excites the plaquettes around the boundary of the membrane. The violation of fake-flatness for these plaquettes causes other energy terms near the boundary to be ill-defined, though this does not affect the interpretation of the magnetic excitation as a loop-like one. Finally, the magnetic membrane operator may lead to excitations of the start-point of the membrane, as in the $\rhd$ trivial case, and a blob, blob 0, that we specify when defining the membrane. The start-point and blob 0 can be chosen to be far from the boundary of the membrane and therefore far from the loop-like excitation produced by the membrane operator. We therefore see that the membrane operator may produce a point-like excitation in addition to the loop-like one, which suggests that the loop-like excitation itself must sometimes carry a point-like charge, in order to balance the charge of the point-like excitation. Indeed, we will see exactly this in Section \ref{Section_point_like_charge_higher_flux}.
	
	\subsubsection{Changing the start-point}
	\label{Section_magnetic_membrane_central_change_sp}
	Just as in the $\rhd$ trivial case, the action of the magnetic membrane operator depends on a privileged vertex, called the start-point of the membrane. In this section, we will consider what happens when we change that start-point, just as we did for the $\rhd$ trivial case in Section \ref{Section_magnetic_tri_trivial_move_sp}. In order to do this, we will individually consider how the different parts of the magnetic membrane operator transform under changes to the start-point. As we described in Section \ref{Section_Magnetic_Tri_Non_Trivial_Definition}, the total magnetic membrane operator $C^h_T(m)$ is given by
	\begin{equation}
		C^h_T(m) = C^h_{\rhd}(m) \prod_{\substack{\text{plaquette }p \\ \text{on membrane}}} B^{f(p)}(\text{blob }0 \rightarrow \text{blob }p). \label{Equation_tri_non_trivial_magnetic_definition_restate}
	\end{equation}
	
	We will now consider the different parts of this operator in more detail, starting with $C^h_{\rhd}(m)$. This operator acts on an edge $i$ that is cut by the dual membrane of the membrane operator according to
	\begin{equation*}
		C^h_{\rhd}(m):g_i = \begin{cases} g(s.p(m)-v_i)^{-1}hg(s.p(m)-v_i)g_i & \text{if $i$ points away from the direct membrane} \\ g_ig(s.p(m)-v_i)^{-1}h^{-1}g(s.p(m)-v_i) & \text{if $i$ points towards the direct membrane,} \end{cases}
	\end{equation*}
	which is the same as the action of the magnetic membrane operator when $\rhd$ is trivial. In Section \ref{Section_magnetic_tri_trivial_move_sp} we considered moving the start-point of a magnetic membrane operator in the $\rhd$ trivial case. We found that moving the start-point from $s.p(m)$ to a new position $s.p(m')$, with the resulting membrane denoted by $m'$, results in the following action on the edges
	\begin{align*}
		C^h(m'):g_i &= \begin{cases} g(s.p(m')-v_i)^{-1}hg(s.p(m')-v_i)g_i & \text{if $i$ points away from the direct membrane} \\ g_ig(s.p(m')-v_i)^{-1}h^{-1}g(s.p(m')-v_i) & \text{if $i$ points towards the direct membrane.} \end{cases}\\
		&=C^{g(s.p(m)-s.p(m'))hg(s.p(m)-s.p(m'))^{-1}}(m):g_i.
	\end{align*}
	This will therefore also hold for the action of $C^h_{\rhd}(m)$ on the edges in this case (where $\rhd$ is not necessarily trivial). We therefore have
	\begin{equation*}
		C^h_{\rhd}(m'):g_i = C^{g(s.p(m)-s.p(m'))hg(s.p(m)-s.p(m'))^{-1}}_{\rhd}(m):g_i.
	\end{equation*}
	
	In addition to acting on the edges cut by the dual membrane, $C^h_{\rhd}(m)$ also acts on the plaquettes that are cut by the dual membrane. The action of the operator on such a plaquette $p$, with base-point $v_0(p)$, is
	\begin{equation*}
		C^h_{\rhd}(m): e_p = \begin{cases} (g(s.p(m)-v_0(p))^{-1}hg(s.p(m)-v_0(p))) \rhd e_p & \text{ if $v_0(p)$ lies on the direct membrane} \\ e_p & \text{ otherwise.} \end{cases} 
	\end{equation*}
	When we change the start-point from $s.p(m)$ to $s.p(m')$, this becomes
	\begin{equation}
		C^h_{\rhd}(m'): e_p = \begin{cases} (g(s.p(m')-v_0(p))^{-1}hg(s.p(m')-v_0(p))) \rhd e_p & \text{ if $v_0(p)$ lies on the direct membrane} \\ e_p & \text{ otherwise.} \end{cases} \label{Equation_magnetic_tri_non_trivial_move_sp_plaquette}
	\end{equation}
	
	We then split the path $(s.p(m')-v_0(p))$ into two sections, $(s.p(m')-s.p(m))$ and $(s.p(m)-v_0(p))$ in order to write $g(s.p(m')-v_0(p))=g(s.p(m')-s.p(m))g(s.p(m)-v_0(p))=g(s.p(m)-s.p(m'))^{-1}g(s.p(m)-v_0(p))$. Substituting this into Equation \ref{Equation_magnetic_tri_non_trivial_move_sp_plaquette} gives
	\begin{align*}
		C^h_{\rhd}(m'): e_p &= \begin{cases} \big((g(s.p(m)-s.p(m'))^{-1}g(s.p(m)-v_0(p)))^{-1}h & \text{if $v_0(p)$ lies on the direct membrane}\\
			\hspace{1cm} (g(s.p(m)-s.p(m'))^{-1}g(s.p(m)-v_0(p)))\big) \rhd e_p \\ e_p & \text{otherwise} \end{cases}\\
		&= \begin{cases} \big(g(s.p(m)-v_0(p))^{-1} [g(s.p(m)-s.p(m'))hg(s.p(m)-s.p(m'))^{-1}] & \text{if $v_0(p)$ lies on the direct}\\
			\hspace{1cm}g(s.p(m)-v_0(p))\big) \rhd e_p & \text{ membrane}\\ e_p & \text{otherwise} \end{cases}\\
		&= C^{g(s.p(m)-s.p(m'))hg(s.p(m)-s.p(m'))^{-1}}_{\rhd}(m):e_p.
	\end{align*}
	
	We see that for the action on the plaquettes, just as for the action on the edges, the action of the $C_{\rhd}$ operator with the new start-point is equivalent to the action of a $C_{\rhd}$ operator of label $g(s.p(m)-s.p(m'))hg(s.p(m)-s.p(m'))^{-1}$ with the old start-point. Because this holds for the action on both edges and plaquettes, we can write this as an operator relation:
	\begin{equation}
		C^h_{\rhd}(m')= C^{g(s.p(m)-s.p(m'))hg(s.p(m)-s.p(m'))^{-1}}_{\rhd}(m). \label{Equation_tri_non_trivial_magnetic_rhd_change_sp}
	\end{equation}
	
	Next we consider the effect of moving the start-point on the blob ribbon operators that are part of $C^h_T(m)$. Each plaquette $p$ in the direct membrane is associated to a blob ribbon operator
	$$B^{[g(s.p(m)-v_{0}(p))\rhd e_p^{\sigma_p}] \: [(h^{-1}g(s.p(m)-v_0(p)))\rhd e_p^{-\sigma_p}]}(t_p),$$
	where $\sigma_p$ is $+1$ if the plaquette is oriented away from the dual membrane and $-1$ otherwise. $t_p$ is the ribbon whose direct path starts at the start-point of the membrane and whose dual path runs from blob 0 of the membrane to the blob that is attached to plaquette $p$ and cut by the dual membrane. When we move the start-point of the membrane, we keep blob 0 fixed and so do not change the dual path of the ribbon operators, but we do change the direct path. We denote the new ribbon produced by moving the start-point from $s.p(m)$ to $s.p(m')$ by $t_p'$. Then the new ribbon operator is
	$$B^{[g(s.p(m')-v_{0}(p))\rhd e_p^{\sigma_p}] \: [(h^{-1}g(s.p(m')-v_0(p)))\rhd e_p^{-\sigma_p}]}(t_p').$$
	Now we write $g(s.p(m')-v_0(p))$ as $g(s.p(m)-s.p(m'))^{-1} g(s.p(m)-v_0(p))$ to rewrite this blob ribbon operator as
	$$B^{[(g(s.p(m)-s.p(m'))^{-1}g(s.p(m)-v_{0}(p)))\rhd e_p^{\sigma_p}] \: [(h^{-1}g(s.p(m)-s.p(m'))^{-1} g(s.p(m)-v_0(p)))\rhd e_p^{-\sigma_p}]}(t_p').$$
	
	In addition to the label changing due to the new start-point, the ribbon $t_p'$ is different from $t_p$. In Section \ref{Section_blob_ribbon_move_sp}, we described how changing the start-point of a blob ribbon operator from $s.p(t)$ to $s.p(t')$, and so changing the ribbon from $t$ to $t'$, affected the blob ribbon operator. We found that
	$$B^e(t')=B^{g(s.p(t)-s.p(t')) \rhd e}(t).$$
	In this case, we have $s.p(t)=s.p(m)$, $s.p(t')=s.p(m')$, $t=t_p$ and $t'=t_p'$. Therefore, we can write the new blob ribbon operator associated to plaquette $p$ as 
	\begin{align}
		&B^{[(g(s.p(m)-s.p(m'))^{-1}g(s.p(m)-v_{0}(p)))\rhd e_p^{\sigma_p}] \: [(h^{-1}g(s.p(m)-s.p(m'))^{-1} g(s.p(m)-v_0(p)))\rhd e_p^{-\sigma_p}]}(t_p') \notag \\
		& \hspace{1cm} = B^{g(s.p(m)-s.p(m')) \rhd \big([(g(s.p(m)-s.p(m'))^{-1}g(s.p(m)-v_{0}(p)))\rhd e_p^{\sigma_p}] \: [(h^{-1}g(s.p(m)-s.p(m'))^{-1} g(s.p(m)-v_0(p)))\rhd e_p^{-\sigma_p}]\big)}(t_p) \notag \\
		& \hspace{1cm} =B^{[g(s.p(m)-v_{0}(p))\rhd e_p^{\sigma_p}] [(g(s.p(m)-s.p(m'))h^{-1}g(s.p(m)-s.p(m'))^{-1} g(s.p(m)-v_0(p)))\rhd e_p^{-\sigma_p}]}, \label{Equation_tri_non_trivial_magnetic_blob_change_sp}
	\end{align}
	which has the same label as the original blob ribbon operator, but with $h$ replaced by $$g(s.p(m)-s.p(m'))hg(s.p(m)-s.p(m'))^{-1}.$$ 
	Putting this together with Equations \ref{Equation_tri_non_trivial_magnetic_definition_restate} and \ref{Equation_tri_non_trivial_magnetic_rhd_change_sp}, we see that
	\begin{align}
		C^h_T(m')=& \: C^{g(s.p(m)-s.p(m'))hg(s.p(m)-s.p(m'))^{-1}}_{\rhd}(m) \notag \\ & \hspace{-0.4cm} \bigg(\prod_{\substack{\text{plaquette }p \\ \text{on membrane}}} \hspace{-0.4cm} B^{[g(s.p(m)-v_{0}(p))\rhd e_p^{\sigma_p}] [(g(s.p(m)-s.p(m'))h^{-1}g(s.p(m)-s.p(m'))^{-1} g(s.p(m)-v_0(p)))\rhd e_p^{-\sigma_p}]}(\text{blob }0 \rightarrow \text{blob }p) \bigg) \notag\\
		=& \: C^{g(s.p(m)-s.p(m'))hg(s.p(m)-s.p(m'))^{-1}}_T(m). \label{Equation_tri_non_trivial_magnetic_change_sp_1}
	\end{align}
	That is, moving the start-point of the magnetic membrane operator from $s.p(m)$ to $s.p(m')$ is equivalent to conjugating the label of the membrane operator by $g(s.p(m)-s.p(m'))$, just as we saw in the $\rhd$ trivial case in Section \ref{Section_magnetic_tri_trivial_move_sp}. We note that this transformation is equivalent to the result from commuting the membrane operator past a vertex transform of label $g(s.p(m)-s.p(m'))^{-1}$ at the start-point of the original magnetic membrane operator (see Equation \ref{Equation_magnetic_membrane_tri_nontrivial_start_point_transform}). As we described in that section, this transformation is to be expected from the transformation of the membrane operator under vertex transforms at the start-point, because a vertex transform at the start-point is equivalent to parallel transport of the start-point. This means that if we take a linear combination of magnetic membrane operators that commutes with the vertex transforms (i.e., that doesn't excite the start-point), then this linear combination will also be unaffected by changes to the start-point.

	As we did for the blob ribbon operators and the magnetic membrane operator when $\rhd$ is trivial, we wish to consider what transformation of the label of the membrane operator should accompany a change in start-point in order to keep the action of the membrane operator fixed. That is, we wish to find the label $x$ such that $C^h_T(m)=C^x_T(m')$. From Equation \ref{Equation_tri_non_trivial_magnetic_change_sp_1}, we find that
	\begin{equation}
		C^h_T(m)=C^{g(s.p(m)-s.p(m'))^{-1}hg(s.p(m)-s.p(m'))}_T(m'). \label{Equation_tri_non_trivial_magnetic_change_sp_2}
	\end{equation}
	
	\subsubsection{Changing blob 0}
	\label{Section_magnetic_change_blob_0}
\begin{figure}[h]
		\begin{center}
			\begin{overpic}[width=0.9\linewidth]{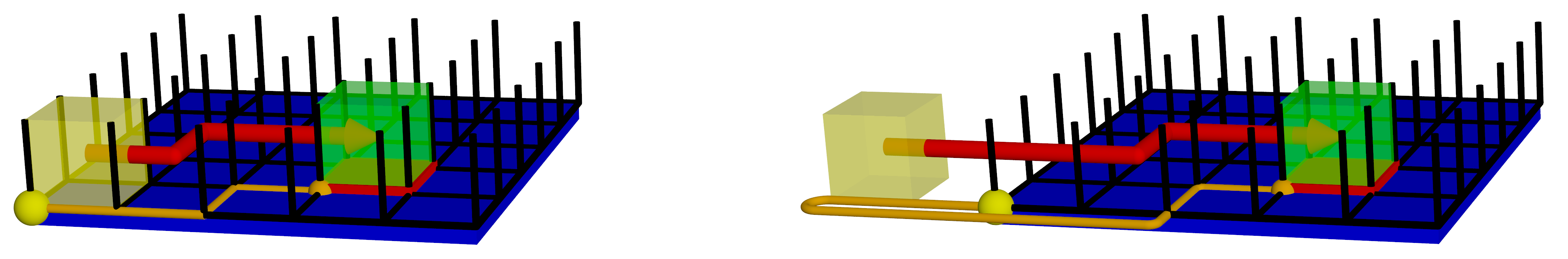}
				\put(41,8){\Huge $\rightarrow$}
				\put(38,12){move blob 0}
				\put(-4,12){blob 0($A$)}
				\put(53,12){blob 0($B$)}
			\end{overpic}
			\caption{We consider moving the position of blob 0 of a magnetic membrane operator. This affects the blob ribbon operators that are part of the magnetic membrane operator, by moving the start of the dual path of such a ribbon operator. In this figure, we show one particular blob ribbon operator, associated to the red plaquette. When we move blob 0 from position A to position B, we extend the dual path of the blob ribbon operator (the thicker red arrow). On the other hand, the start-point (yellow sphere) of the direct path (thinner orange path) stays fixed at the start-point of the membrane. This means we have to deform the direct path to pass by the new position of blob 0, so that the direct path passes through the base-points of all of the plaquettes pierced by the new dual path. However, the direct path to the base-points of the plaquettes pierced by the original dual path is equivalent to the original direct path (can be deformed into the original path).}
			\label{modified_membrane_move_blob_0_1}
		\end{center}
	\end{figure}

\begin{figure}[h]
		\begin{center}
			\begin{overpic}[width=0.7\linewidth]{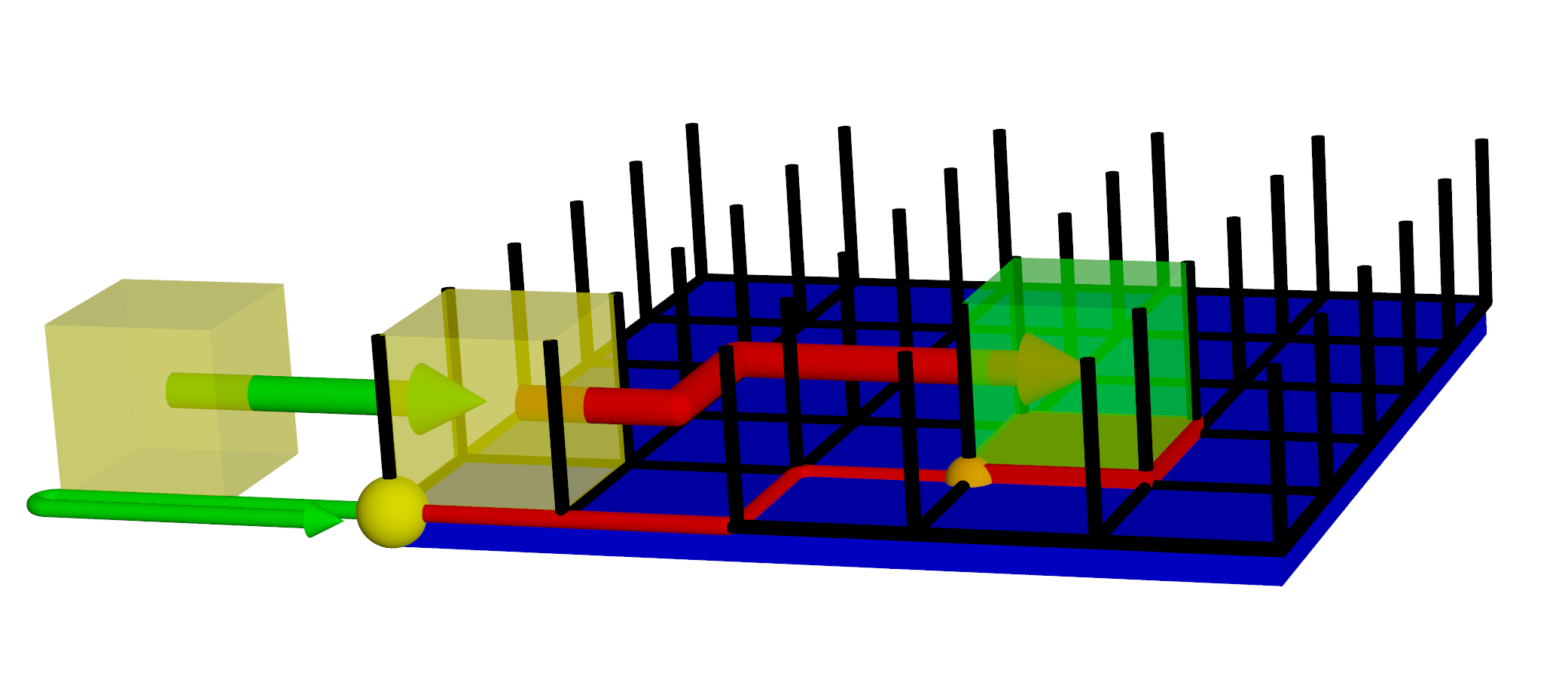}
				\put(7,22){blob 0($B$)}
				\put(27,22){blob 0($A$)}
				
			\end{overpic}
			\caption{We split the blob ribbon operator from the right of Figure \ref{modified_membrane_move_blob_0_1} into two ribbon operators, one corresponding to the original blob ribbon operator (from the left of Figure \ref{modified_membrane_move_blob_0_1}), shown in red (the darker arrow in grayscale), and one that acts on a ribbon that passes between the new and original positions of blob 0, shown in green (the lighter arrow in grayscale). In this figure the dual paths of each ribbon are the larger arrows, while the corresponding direct paths are the thinner paths of the same colour.}
			\label{modified_membrane_move_blob_0_2}
		\end{center}
	\end{figure}

	In addition to the privileged vertex, the start-point, the magnetic membrane operator depends on a privileged blob, called blob 0. We will now consider how changing this blob affects the magnetic membrane operator. Unlike when we move the start-point of the membrane operator, the $C^h_{\rhd}(m)$ part of the membrane operator is unaffected when we change blob 0. The choice of blob 0 only affects the blob ribbon operators that are part of the membrane operator. As we described in Section \ref{Section_Magnetic_Tri_Non_Trivial_Definition}, the dual path of each blob ribbon operator in $C^h_T(m)$ starts at blob 0. If we move blob 0 to a new position, we therefore add a section to the dual path that runs from the new position to the old one, as shown in Figure \ref{modified_membrane_move_blob_0_1}. For a plaquette $p$, which is associated to a blob ribbon operator $B^{f(p)}(t_p)$, moving blob 0 from position A to position B changes the blob ribbon operator to $B^{f(p)}(t_p')$. Then we can split this new blob ribbon operator into two sections, one corresponding to the original dual path and one including the extra section between $B$ and $A$, as shown in Figure \ref{modified_membrane_move_blob_0_2}. We can write this as
	$$B^{f(p)}(t_p')=B^{f(p)}(t_p)B^{f(p)}(B \rightarrow A).$$
	
	We now wish to repeat this process with each blob ribbon operator. The total magnetic membrane operator is given by
	\begin{equation}
		C^h_T(m) = C^h_{\rhd}(m) \prod_{\substack{\text{plaquette }p \\ \text{on membrane}}} B^{f(p)}(t_p).
	\end{equation}
	Therefore, changing blob 0 from A to B changes the magnetic membrane operator to
	\begin{align}
		C^h_T(m') &= C^h_{\rhd}(m) \prod_{\substack{\text{plaquette }p \\ \text{on membrane}}} B^{f(p)}(t_p') \notag \\
		&=C^h_{\rhd}(m) \bigg( \prod_{\substack{\text{plaquette }p \\ \text{on membrane}}} B^{f(p)}(t_p) B^{f(p)}(B \rightarrow A) \bigg)\notag \\
		&= C^h_{\rhd}(m) \bigg( \prod_{\substack{\text{plaquette }p \\ \text{on membrane}}} B^{f(p)}(t_p) \bigg) \bigg( \prod_{\substack{\text{plaquette }p \\ \text{on membrane}}}B^{f(p)}(B \rightarrow A) \bigg). \label{Equation_move_blob_0_total_1}
	\end{align}
	
	Because the ribbon $(B\rightarrow A)$ is the same for each plaquette $p$, we can combine the associated ribbon operators:
	\begin{align}
		\prod_{\substack{\text{plaquette }p \\ \text{on membrane}}}&B^{f(p)}(B \rightarrow A) =B^{(\prod_{p}f(p))}(B \rightarrow A). \label{Equation_move_blob_0_product_blob_ribbons}
	\end{align}
	Then we use the fact that $f(p)=(g(s.p-v_{0}(p))\rhd e_p^{\sigma_p}) \: (h^{-1}g(s.p-v_0(p)))\rhd e_p^{-\sigma_p}$, where $\sigma_p$ is $+1$ if the plaquette is oriented away from the dual membrane and $-1$ otherwise, to write
	\begin{align*}
		\prod_{\substack{\text{plaquette }p \\ \text{on membrane}}}f(p)&=\bigg(\prod_{\substack{\text{plaquette }p \\ \text{on membrane}}} (g(s.p-v_{0}(p))\rhd e_p^{\sigma_p}) \: (h^{-1}g(s.p-v_0(p)))\rhd e_p^{-\sigma_p} \bigg)\\
		&= \bigg(\prod_{\substack{\text{plaquette }p \\ \text{on membrane}}} g(s.p-v_{0}(p))\rhd e_p^{\sigma_p} \bigg) \bigg( h^{-1} \rhd \bigg[\prod_{\substack{\text{plaquette }p \\ \text{on membrane}}}\big( g(s.p-v_{0}(p))\rhd e_p^{\sigma_p} \big) \bigg]^{-1} \bigg).
	\end{align*}
	
	However, $\prod_{\substack{\text{plaquette }p \\ \text{on membrane}}} (g(s.p-v_{0}(p))\rhd e_p^{\sigma_p})$ is the total surface label of the direct membrane, $\hat{e}(m)$. Therefore
	\begin{align*}
		\prod_{\substack{\text{plaquette }p \\ \text{on membrane}}}f(p)&= \hat{e}(m) [h^{-1} \rhd \hat{e}(m)^{-1}].
	\end{align*}
	This means that the product of blob ribbon operators from Equation \ref{Equation_move_blob_0_product_blob_ribbons} can be written as
	\begin{align*}
		\prod_{\substack{\text{plaquette }p \\ \text{on membrane}}}B^{f(p)}(B \rightarrow A) 
		&=B^{\hat{e}(m) [h^{-1} \rhd \hat{e}(m)^{-1}]}(B \rightarrow A).
	\end{align*}

	Substituting this into Equation \ref{Equation_move_blob_0_total_1}, we see that the total magnetic membrane operator is
	\begin{align}
		C^h_T(m') &= C^h_{\rhd}(m) \bigg(\prod_{\substack{\text{plaquette }p \\ \text{on membrane}}} B^{f(p)}(t_p) \bigg) B^{\hat{e}(m) [h^{-1} \rhd \hat{e}(m)^{-1}]} (B \rightarrow A) \notag\\
		&= C^h_T(m)B^{\hat{e}(m) [h^{-1} \rhd \hat{e}(m)^{-1}]} (B \rightarrow A). \label{Equation_move_blob_0_result_1}
	\end{align}
	That is, changing blob 0 of the membrane is the same as introducing a new blob ribbon operator running between the two positions of blob 0. As we did when we considered changing the start-point, we also wish to consider how we can write the magnetic membrane operator $C^h_T(m)$ in terms of an operator with a different blob 0, without changing the action of the operator. Inverting Equation \ref{Equation_move_blob_0_result_1}, we see that
	\begin{align}
		C^h_T(m)&= C^h_T(m')B^{\hat{e}(m)^{-1} [h^{-1} \rhd \hat{e}(m)]} (B \rightarrow A) \notag\\
		&= C^h_T(m') B^{\hat{e}(m) [h^{-1} \rhd \hat{e}(m)^{-1}]} (A \rightarrow B). \label{Equation_move_blob_0_result_2}
	\end{align}
	
	We note that blob 0 is not excited by the membrane operator precisely when $\hat{e}(m) [h^{-1} \rhd \hat{e}(m)^{-1}]$ is trivial (see Equation \ref{Equation_magnetic_blob_0}), which is the same condition for the label of the extra blob ribbon operator in Equations \ref{Equation_move_blob_0_result_1} and \ref{Equation_move_blob_0_result_2} to be trivial. This makes sense, because if blob 0 is not excited then we may expect moving it to have no effect on the membrane operator. One particular case where this independence from the position of blob 0 occurs is when the membrane $m$ is closed, contractible and encloses no excitations, because then $\hat{e}(m)$ is the identity element $1_E$.
	
	\section{The topological nature of membrane and ribbon operators}
	\label{Section_topological_membrane_operators}
	
	An important property of the unconfined ribbon and membrane operators is that they are topological. By topological, we mean that we can deform the ribbon or membrane on which we apply these operators without changing their action on the ground state, provided that we keep the location of any excitations produced by the ribbon or membrane operator fixed. More generally, this holds for states other than the ground state provided that the region over which we deform the ribbon or membrane has no excitations. In this section we will prove that the ribbon and membrane operators that we have considered so far are topological in this sense, except for the ribbon operators corresponding to confined excitations. The ribbon operators corresponding to these confined excitations are not topological, as can be seen from the fact that they produce excitations along the length of the ribbon, the location of which can be detected through the energy terms. 
	
	\subsection{Electric ribbon operators}
	
	The non-confined electric ribbon operators are topological due to the fake-flatness property of the ground state. When we deform the path $t$ on which we apply the ribbon operator through an unexcited region, the path element only picks up a factor in $\partial(E)$ due to fake-flatness. The non-confined electric ribbon operators are precisely those which are insensitive to these factors, and so are unaffected by the deformation. We give a slightly more detailed explanation of this in Ref. \cite{HuxfordPaper2}, in Section S-II A of the Supplemental Material (while this is done in the context of the 2+1d model, the explanation holds in both 2+1d and 3+1d).

	\subsection{$E$-valued membrane operators}
	\label{Section_Topological_E_Membranes}

\begin{figure}[h]
		\begin{center}
			\includegraphics[width=0.7\linewidth]{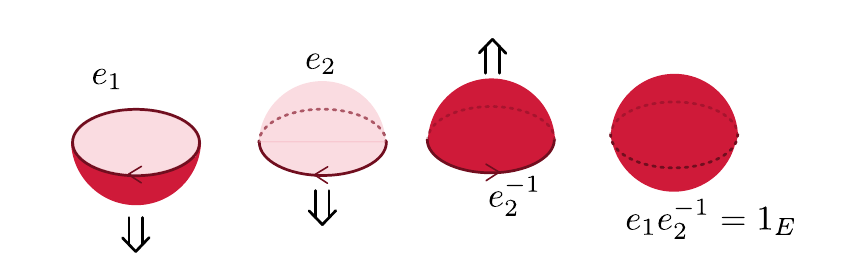}
			\caption{(Copy of Figure \ref{gluing_surfaces_1} from the main text.)  Given two different surfaces with the same boundary, if we can deform one into another without crossing any excitations then their labels must be the same.}
			\label{gluing_surfaces_1_appendix}
		\end{center}
	\end{figure}

	We can show that the $E$-valued membrane operators are topological using similar ideas to those we used for the electric ribbon operators. In fact, we described this in the main text in Section \ref{Section_E_Loop_Excitations_3D}. The $E$-valued membrane operators measure the total 2-holonomy of a membrane, with a weight for each possible group element assigned to that surface. That is, an $E$-valued membrane operator applied on a membrane $m$ has the form
	$$\sum_{e \in E} \alpha_e \delta( \hat{e}(m),e)$$
	where $\alpha_e$ is a set of coefficients and $\hat{e}(m)$ is the surface label of membrane $m$. Consider a membrane operator initially applied on a membrane $m_1$, and imagine smoothly deforming the membrane into a second membrane $m_2$. We do this by pulling the initial membrane $m_1$ across a solid, while keeping the boundary fixed. The surface label for the boundary of the solid swept by the membrane is $1_E$, provided this solid contains no excitations, because the blob terms enforce that the surface label of the blobs in the region is the identity. The boundary of the solid is formed by gluing the initial membrane $m_1$ with the final membrane $m_2$, but where the orientation of the final membrane is inverted. This indicates that the initial label of the membrane, $e(m_1)$ and the final label, $e(m_2)$, satisfy $e(m_1) e(m_2)^{-1}=1_E$, as shown in Figure \ref{gluing_surfaces_1_appendix}. Therefore, $e(m_1) =e(m_2)$. This means that the $e$-valued membrane operator acts the same regardless whether it acts on the initial membrane or the final one. One subtlety about this is that we must keep the entire boundary fixed when deforming the membrane. As we discussed in Section S-I C of the Supplemental Material of Ref. \cite{HuxfordPaper2}, the boundary of the membrane may not match the naive boundary, especially if the start-point of the membrane is away from this naive boundary (in which case there are a series of edges connecting the start-point to the naive boundary which appear twice in the actual boundary with opposite direction). This can cause an extra line of excitations that must be kept fixed when deforming the membrane, at least if $E$ is non-Abelian (as we showed in Section S-I C of the Supplemental Material of Ref. \cite{HuxfordPaper2} however, if the edges are not excited then we can freely deform this section of boundary). 
	
	\subsection{Blob ribbon operators}
	\label{Section_Topological_Blob_Ribbons}
\begin{figure}
		\begin{center}
			\begin{overpic}[width=0.75\linewidth]{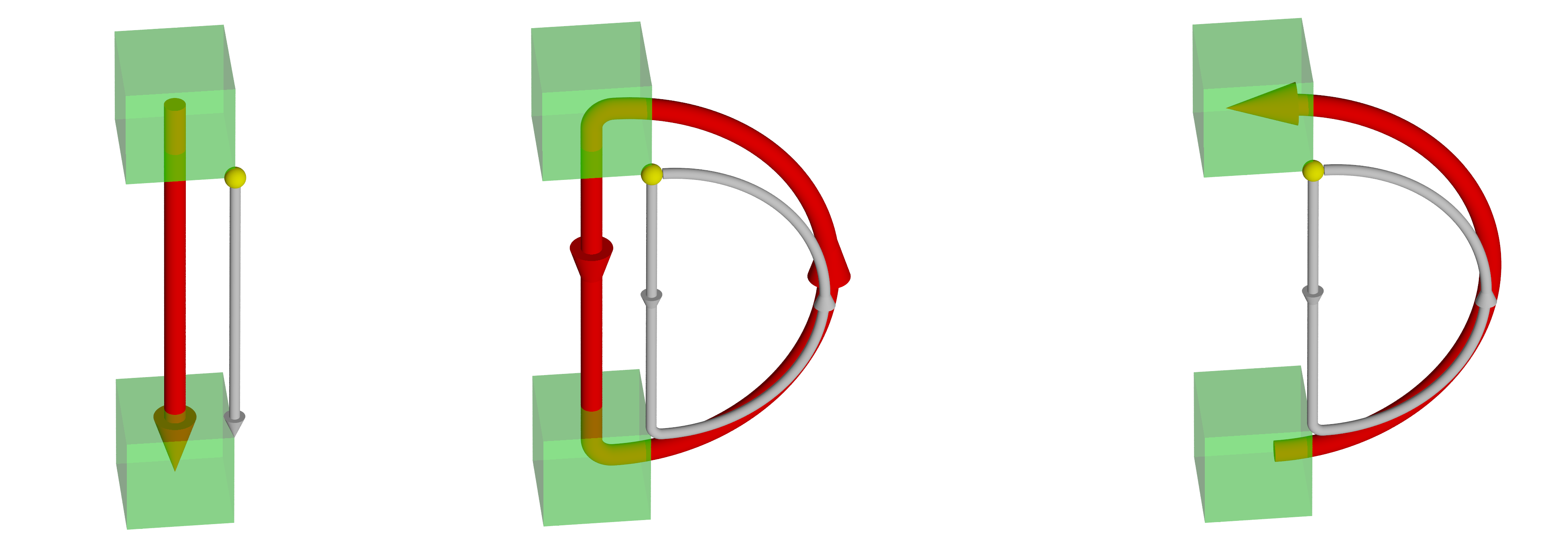}
				\put(0,22){$B^{e_k}(t_1)$}
				\put(23,22){$B^{e_k^{-1}}(t_1 \cdot t_2)$}
				\put(25,15){\Huge $\cdot$}
				\put(65,15){\Huge $=$}
				\put(97,22){$B^{e_k^{-1}}(t_2)$}
			\end{overpic}
			\caption{Given a blob ribbon operator of label $e_k$ on a ribbon $t_1$, with start-point $s.p$ (yellow sphere), we can add a blob ribbon operator on a closed path $t_1t_2$ with label $e_k$. This cancels the action of the original ribbon operator on the path $t_1$ but leaves the action of the closed ribbon operator on $t_2$. The remnant action is equivalent to a blob ribbon operator of label $e_k^{-1}$ applied on $t_2$, where the start-point of $t_2$ is still the original start-point, $s.p$. Here the dual paths of the ribbons are indicated by the thicker red arrows, while the direct paths are shown as thinner grey arrows. Note that because the start-point of $t_2$ is still $s.p$, the direct path for $t_2$ includes the direct path for $t_1$ (although it can be deformed freely).}
			\label{deform_ribbon_operator_step_1}
		\end{center}
	\end{figure}

\begin{figure}
		\begin{center}
			\begin{overpic}[width=0.6\linewidth]{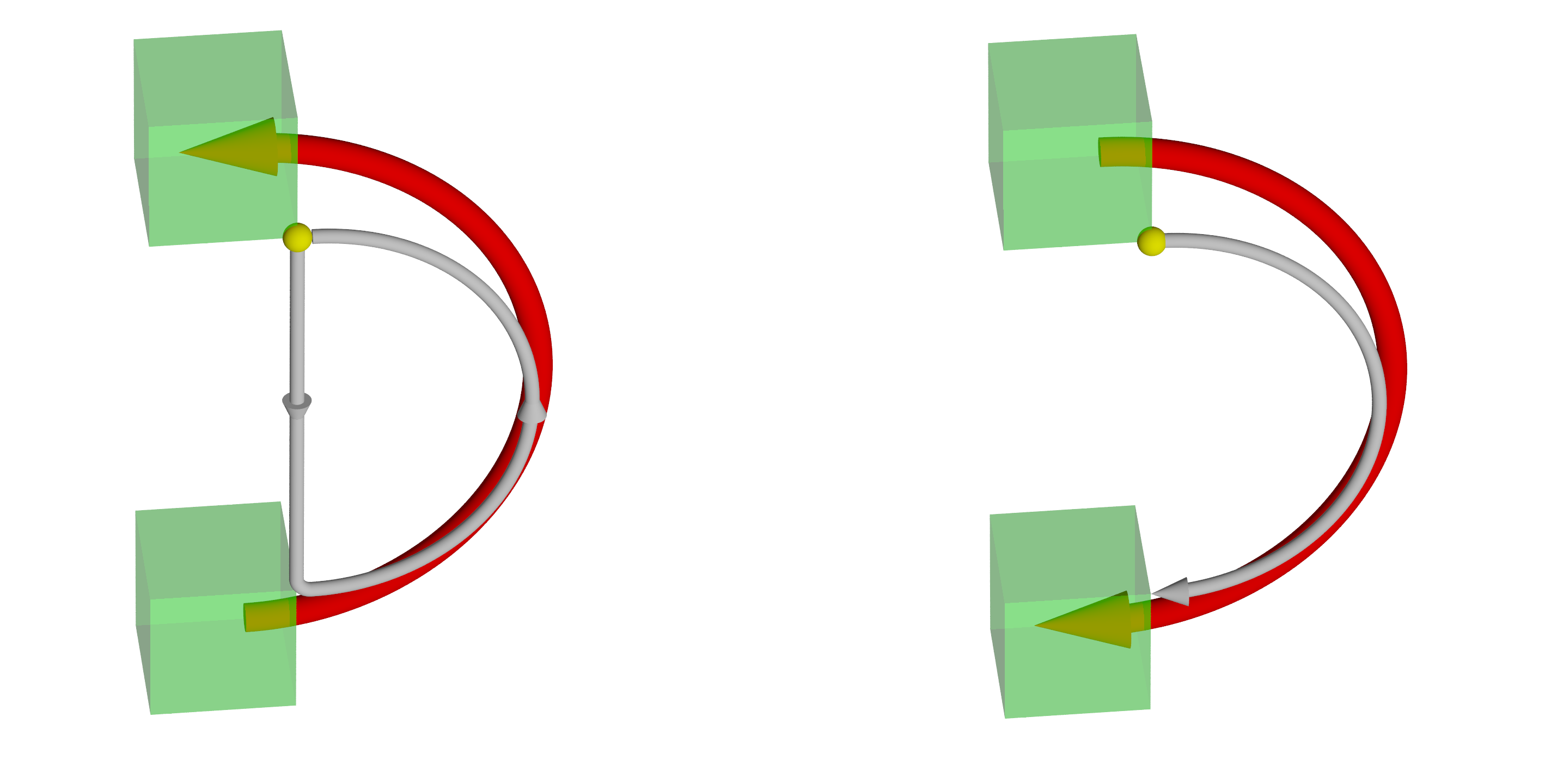}
				\put(30,40){$B^{e_k^{-1}}(t_2)$}
				\put(87,40){$B^{e_k}(t_2^{-1})$}
				\put(50,25){\Huge $=$}
			\end{overpic}
			\caption{We can reverse the orientation of the ribbon operator on $t_2$ by simultaneously inverting its label. The start-point of the ribbon operator is still $s.p$ (the yellow sphere), leading to the direct path represented by the thinner grey line. The resulting ribbon can be obtained from the original ribbon $t_1$ by deformation, and the corresponding blob ribbon operator also has the same label, $e_k$, as the original ribbon operator.}
			\label{deform_ribbon_operator_step_2}
		\end{center}
	\end{figure}

	The next operators to consider are the blob ribbon operators. However, as with the electric ribbon operators, not all of the blob ribbon operators are topological, because some of them are instead confined (as we discussed at the beginning of Section \ref{Section_topological_membrane_operators}, confined operators cannot be topological). We therefore wish to show that the non-confined ribbon operators are topological. In order to prove this, we will proceed in two steps. First we will show that a (non-confined) blob ribbon operator applied on a closed ribbon is trivial, provided that we can contract that ribbon to nothing without the ribbon crossing any excitations. Then we will show that combining an open ribbon operator with a closed ribbon operator produces an open ribbon operator on a new ribbon obtained from the original by deformation, as shown in Figures \ref{deform_ribbon_operator_step_1} and \ref{deform_ribbon_operator_step_2}. Because the actions of these closed ribbon operators are trivial on a state where they enclose no excitations, this means that the action of the open ribbon operator is the same before and after we deform the open ribbon. Therefore, the blob ribbon operator is topological.

\begin{figure}[h]
		\begin{center}
			\begin{overpic}[width=0.5\linewidth]{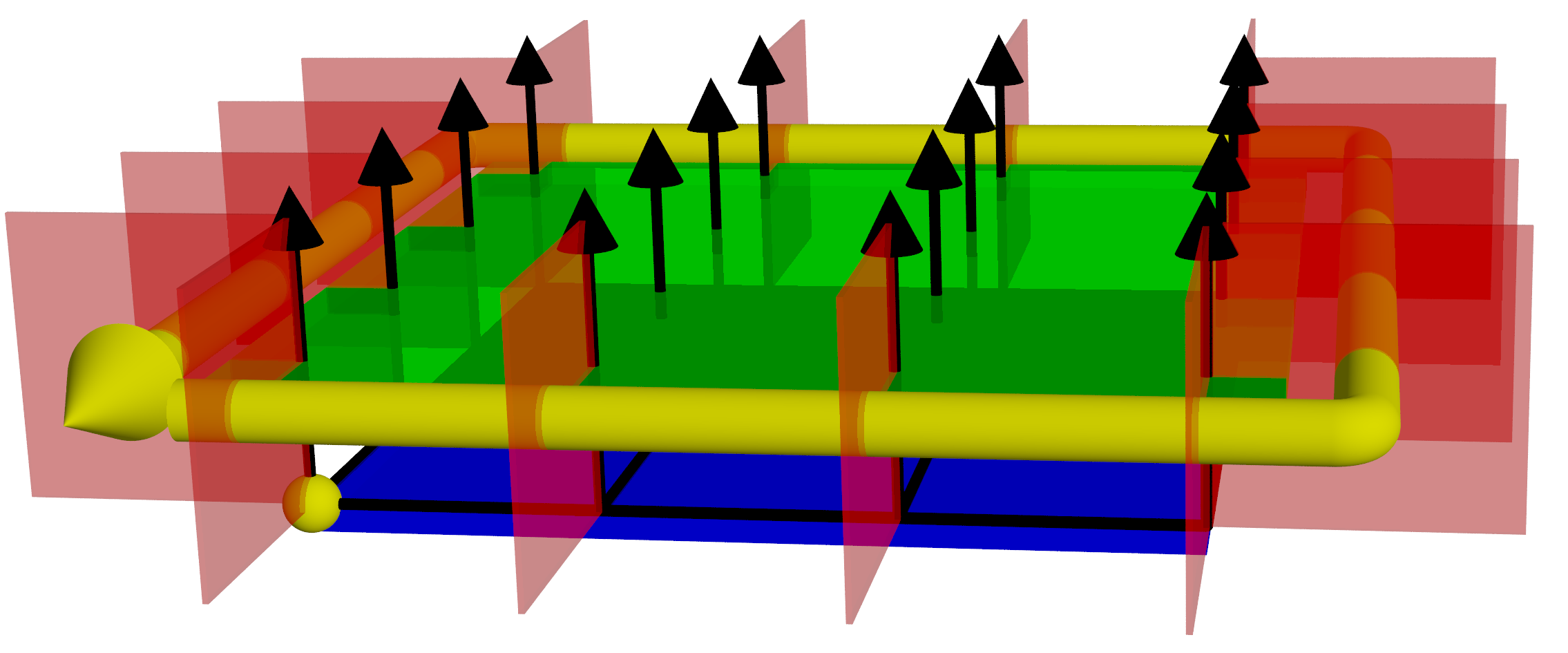}
				
			\end{overpic}
			\caption{In order to reproduce the action of a blob ribbon operator on a closed ribbon (represented by the large yellow arrow), we apply edge transforms on the edges cut by a membrane (the upper green surface). Having shown that the series of edge transforms that we use is invariant under flipping the orientations of the edges, it is convenient to choose the edges to have the same orientation with respect to the membrane. We therefore choose the edges to have their sources lying on another membrane (the lower blue membrane ) which does not intersect the original (upper) membrane. This means that the orientation of the edges cut by the upper membrane is upwards in this figure, as indicated by the black arrows.}
			\label{closed_ribbon_topological}	
		\end{center}
	\end{figure}

	The first step in this proof is to show that the non-confined closed blob ribbon operators act trivially on certain states. To do so, we will show that the closed ribbon operator acts in the same way as a series of edge transforms, which act trivially on states for which the edges are unexcited. In particular, we will show that the ribbon operator is equivalent to the action of edge transforms applied on edges cut by a surface whose boundary is the closed ribbon of the blob ribbon operator we wish to produce, as illustrated in Figure \ref{closed_ribbon_topological}. There is an entire family of surfaces with the same boundary, but we can use any such surface as long as there are no excitations on the surface in the initial state. We choose one such surface and define a normal for that surface. Now consider acting on each edge $i$ cut by the surface with the edge transform labelled by $g(s.p-s(i))^{-1} \rhd e_k^{\pm 1}$, where $s(i)$ is the source of the edge $i$ and $s.p$ is a privileged point which will become the start-point of the closed ribbon operator. $e_k$ is used if the direction of the edge matches the normal of the surface and $e_k^{-1}$ is used if it is anti-aligned. We claim that this series of edge transforms will reproduce the action of a closed ribbon operator of label $e_k$ that runs anticlockwise around the boundary of the surface. In order to do so, we first wish to show that the series of edge transforms is invariant under the re-branching procedures described in the Appendix of Ref. \cite{HuxfordPaper1}. This will allow us to fix the orientations of all of the edges and the base-points of all of the plaquettes affected by the transforms, while guaranteeing that our results will hold for any such choice.

	We showed in the Appendix of Ref. \cite{HuxfordPaper1} that the individual edge transforms are invariant under the procedures where we move the base-point of a plaquette or flip the orientation of a plaquette, but not under the procedure for which we flip the orientation of an edge (to make the edge transform invariant under the edge flip, we have to average over each edge transform to produce the edge energy term). In addition, none of the procedures for changing the branching structure affect the path element $g(s.p-s(i))$ that appears in the label of the edge transforms, except for the edge flipping procedure which changes the position of the source. This means that the operator that we apply is invariant under our procedures for changing the branching structure, except potentially for the edge-flipping procedure. We showed in the Appendix of Ref. \cite{HuxfordPaper1} that the edge-flipping procedure $P_i$ on edge $i$ took an edge transform $\mathcal{A}_i^e$ to $\mathcal{A}_i^{g_i \rhd e^{-1}}$. However, flipping the edge also swaps the position of the source $s(i)$ with that of the target of the edge, $t(i)$, and changes whether we apply the edge transform labelled by $g(s.p-s(i))^{-1} \rhd e_k$ or labelled by the inverse $g(s.p-s(i))^{-1} \rhd e_k^{-1}$ in our series of edge transforms. The inverse from flipping the edge will cancel with this inverse, so we just need to worry about the changing source. We have
	\begin{align*}
		P_i^{-1} \mathcal{A}_i^{g(s.p-s(i))^{-1} \rhd e_k^{\pm 1}} P_i&= \mathcal{A}_i^{g_i \rhd(g(s.p-t(i))^{-1} \rhd e_k^{\pm 1})}\\
		&=\mathcal{A}_i^{g(s(i)-t(i)) \rhd(g(s.p-t(i))^{-1} \rhd e_k^{\pm 1})}\\
		&=\mathcal{A}_i^{g(s.p-s(i))^{-1} \rhd e_k^{ \pm 1}}.
	\end{align*}
	This means that the configuration dependent edge transform that we apply is invariant under the edge flipping procedure, in addition to the other procedures for changing the branching structure. Therefore, we can freely pick a branching structure, and if our series of edge transforms reproduces the action of the blob ribbon operator on that branching structure then it will also do so for any choice of branching structure. For simplicity, we choose the sources of the edges cut by the membrane to lie on a second membrane which does not intersect the other membrane (analogous to the direct membrane of the magnetic membrane operators), as indicated in Figure \ref{closed_ribbon_topological}.

\begin{figure}[h]
		\begin{center}
			\begin{overpic}[width=0.6\linewidth]{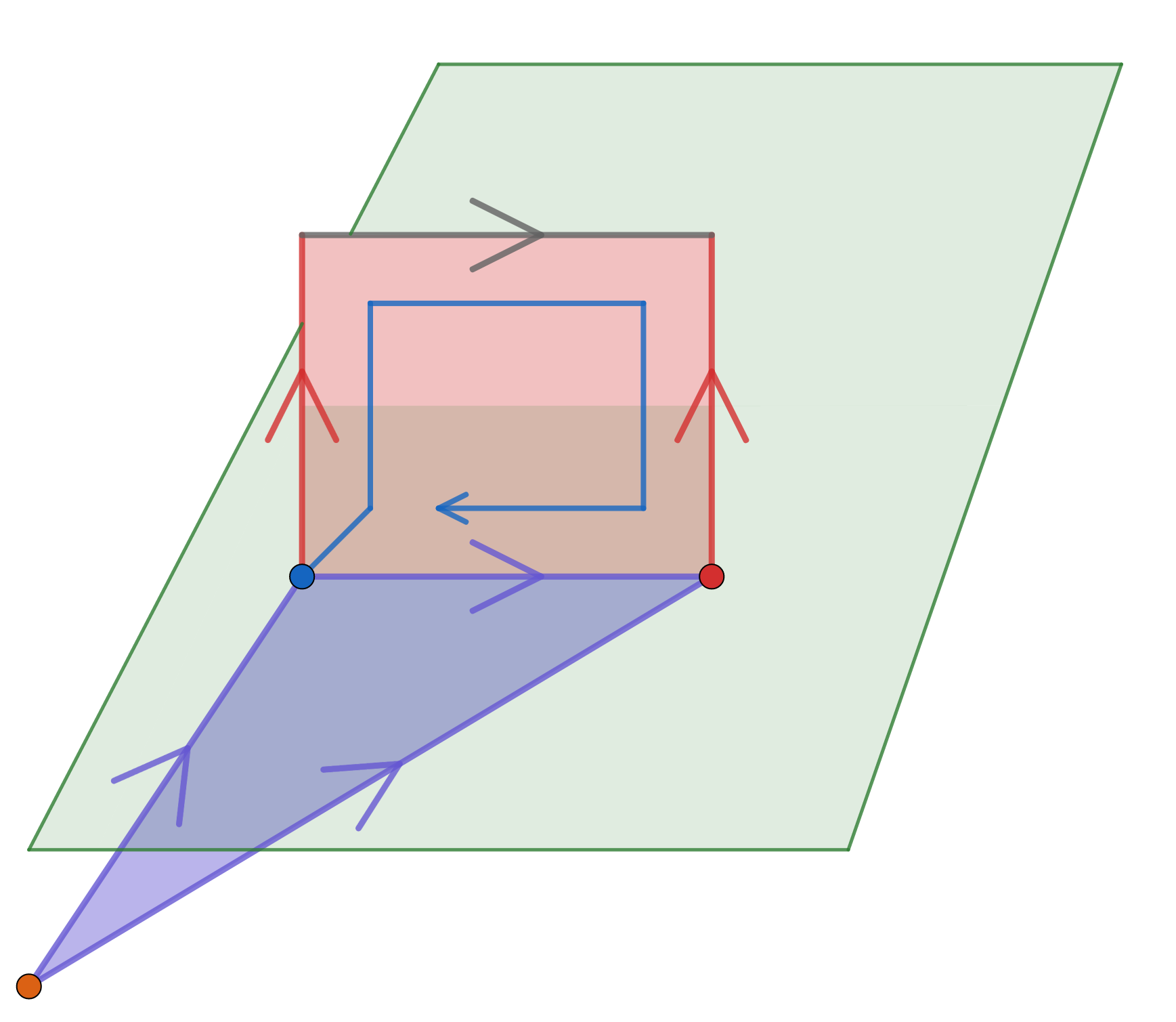}
				\put(-2,0){$s.p$}
				\put(9,36){$v_0(p)=s(i)$}
				\put(23,58){$i$}
				\put(62,36){$s(j)$}
				\put(62,58){$j$}
				\put(41,52){$p$}
				\put(6,26){$s.p-s(i)$}
				\put(43,26){$s.p-s(j)$}
				\put(32,34){$\overline{v_0(p)-s(j)}$}
				\put(74,20){membrane}
			\end{overpic}
			\caption{We consider the action of the series of edge transforms on a plaquette $p$ (red) cut by the bulk of the membrane (green). There are two edges, $i$ and $j$, for which the edge transforms affect the plaquette label. Provided that the region around the membrane satisfies fake-flatness (in this case the purple triangular region), the effects of these two edge transforms on the plaquette label cancel.}
			\label{blob_ribbon_topological_closed_internal_plaquette}
		\end{center}
	\end{figure}

	Because we are concerned with the unconfined excitations, we are interested in the case where $e_k$ is in the kernel of $\partial$ (considering the case of a general element of $E$ leads to another interesting result, as we show in Section \ref{Section_condensation_magnetic_centre_case}). Expressions of the form $g \rhd e_k$ are also in the kernel of $\partial$. Any edge transform with a label in the kernel does not affect the label of the edge on which we apply it, only changing the labels of the plaquettes attached to the edge. This means that the series of edge transforms that we apply does not affect any edge labels, so we can just consider the effect of this series of edge transforms on the plaquettes. There are two types of plaquettes to consider. First we have the ``internal" plaquettes, which are those plaquettes cut by the bulk of the membrane, for which two of the edges on the plaquette are cut by the membrane. Consider Figure \ref{blob_ribbon_topological_closed_internal_plaquette}, which shows such an internal plaquette, $p$. Then the two edges on which we apply the edge transforms are $i$ and $j$, where we have chosen the base-point of $p$, $v_0(p)$ to be the same as the source of $i$, $s(i)$. The effect of the two edge transforms applied on edges $i$ and $j$ on the plaquette $p$, with label $e_p$ is (using Equation \ref{Equation_edge_transform_definition} from the main text, which describes the action of the edge transform)
	\begin{align*}
		\mathcal{A}_j^{g(s.p-s(j))^{-1} \rhd e_k} \mathcal{A}_i^{g(s.p-s(i))^{-1} \rhd e_k}:e_p &=(\mathcal{A}_j^{g(s.p-s(j))^{-1} \rhd e_k}: e_p) g(s.p-s(i))^{-1} \rhd e_k^{-1}\\
		&=[g(\overline{v_0(p)-s(j)}) \rhd (g(s.p-s(j))^{-1} \rhd e_k)] e_p g(s.p-s(i))^{-1} \rhd e_k^{-1}\\
		&= [(g(s.p-s(j))g(\overline{v_0(p)-s(j)})^{-1})^{-1} \rhd e_k] e_p g(s.p-s(i))^{-1} \rhd e_k^{-1},
	\end{align*}
	where the factor $g(\overline{v_0(p)-s(j)})$ comes from the fact that the source of $j$ is distinct from $v_0(p)$, whereas there is no similar factor for $i$ because we chose the source of $i$ to coincide with $v_0(p)$. Then, we note that the two paths $(s.p-s(j)) \cdot (\overline{v_0(p)-s(j)})^{-1}$ and $s.p-s(i)$ enclose a surface between them (the purple surface in Figure \ref{blob_ribbon_topological_closed_internal_plaquette}). Therefore, these two paths differ only by a factor of $\partial(e)$, which does not affect an expression like $g(t) \rhd e_k$. This implies that $(g(s.p-s(j))g(\overline{v_0(p)-s(j)})^{-1})^{-1} \rhd e_k= g(s.p-s(i))^{-1} \rhd e_k$. This fact, together with the fact that $e_k$ is in the centre of $G$, means that
	\begin{align*}
		\mathcal{A}_j^{g(s.p-s(j))^{-1} \rhd e_k} \mathcal{A}_i^{g(s.p-s(i))^{-1} \rhd e_k}:e_p &= [(g(s.p-s(j))g(\overline{v_0(p)-s(j)})^{-1})^{-1} \rhd e_k] e_p [g(s.p-s(i))^{-1} \rhd e_k^{-1}]\\
		&= [g(s.p-s(i))^{-1} \rhd e_k] e_p [g(s.p-s(i))^{-1} \rhd e_k^{-1}]\\
		&= e_p,
	\end{align*}
	from which we see that the actions of the two edge transforms on the plaquette cancel out.
	
\begin{figure}[h]
		\begin{center}
			\begin{overpic}[width=0.6\linewidth]{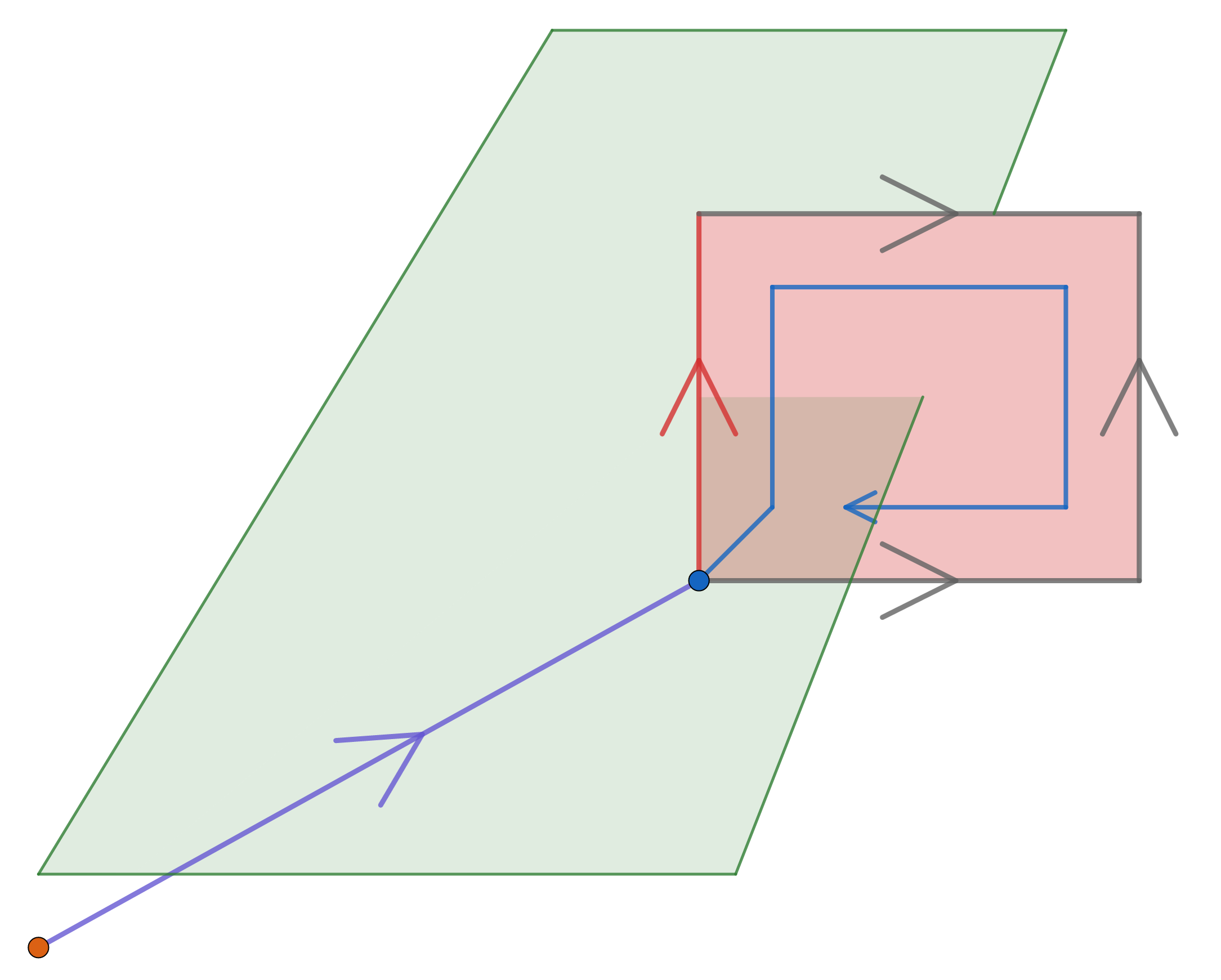}
				\put(-2,1){$s.p$}
				\put(40,32){$v_0(p)=s(i)$}
				\put(55,52){$i$}
				\put(77,48){$p$}
				\put(28,22){$s.p-s(i)$}
				\put(66,20){membrane}
			\end{overpic}
			\caption{We consider the action of the series of edge transforms on a plaquette (red) cut by the boundary of the membrane (green). Unlike for the plaquette considered in Figure \ref{blob_ribbon_topological_closed_internal_plaquette}, only one of the edges on the plaquette is cut by the membrane. This means that the plaquette label is only affected by one edge transform and so there is no cancellation. The effect of the edge transform on the plaquette is the same as a blob ribbon operator travelling around the boundary of the membrane.}
			\label{blob_ribbon_topological_closed_boundary_plaquette}
		\end{center}
	\end{figure}

	Next consider a plaquette on the boundary of the membrane, such as the one shown in Figure \ref{blob_ribbon_topological_closed_boundary_plaquette}. For such a plaquette, only one edge is cut by the membrane and so only one edge transform affects the plaquette. This means that we do not get cancellation of factors as we did for the internal plaquettes. Instead, using the notation from Figure \ref{blob_ribbon_topological_closed_boundary_plaquette}, we see that
	$$\mathcal{A}_i^{g(s.p-s(i))^{-1} \rhd e_k}:e_p = e_p [g(s.p-s(i))^{-1} \rhd e_k^{-1}].$$
	This action on a plaquette is the same as that of a blob ribbon operator of label $e_k$ and start-point $s.p$ piercing the plaquette by travelling anticlockwise around the membrane in Figure \ref{closed_ribbon_topological}. This will hold for all boundary plaquettes, regardless of their location on the boundary, because we can always choose the plaquette to be oriented such that the edge cut by the membrane is aligned with the boundary of the plaquette (any such plaquette can be obtained by dragging the plaquette shown in Figure \ref{blob_ribbon_topological_closed_boundary_plaquette} around the boundary of the membrane to the desired position). This means that the net result of the series of edge transforms that we apply is the same as applying a blob ribbon operator of label $e_k$ on a ribbon that passes around the boundary of the membrane.

	This shows that we can reproduce the effect of a closed blob ribbon operator using a series of configuration dependent edge transforms. Now we need to show that these edge transforms have no effect on a state with no excitations within the membrane. We already required that the region near the membrane satisfies fake-flatness, but now we will also assume that none of the edges we want to transform are excited. For such a state $\ket{\psi}$ we can write the action of the edge transform $\mathcal{A}_i^{g(s.p-s(i))^{-1} \rhd e_k}$ on the state as
	\begin{align*}
		\mathcal{A}_i^{g(s.p-s(i))^{-1}\rhd e_k} \ket{\psi} &=\mathcal{A}_i^{g(s.p-s(i))^{-1}\rhd e_k} \mathcal{A}_i \ket{\psi},
	\end{align*}
	using the fact that $\ket{\psi}$ is an eigenstate of $\mathcal{A}_i$ with eigenvalue $1$. Then, expanding the edge term gives us
	\begin{align*}
		\mathcal{A}_i^{g(s.p-s(i))^{-1}\rhd e_k} \ket{\psi} &=\mathcal{A}_i^{g(s.p-s(i))^{-1}\rhd e_k} \sum_{e \in E} \frac{1}{|E|} \mathcal{A}_i^e \ket{\psi}\\
		&=\sum_{e \in E} \frac{1}{|E|}\mathcal{A}_i^{[g(s.p-s(i))^{-1}\rhd e_k]e} \ket{\psi}.
	\end{align*}
	We note that in the last line we used the fact that $\mathcal{A}_i^e$ does not change the operator label ${g(s.p-s(i))^{-1}\rhd e_k}$ in order to combine the two edge transforms. This fact is necessary for the algebraic manipulation because $\mathcal{A}_i^{g(s.p-s(i))^{-1}\rhd e_k}$ originally acted after $\mathcal{A}_i^e$, whereas when we combine them they act simultaneously, so we need the order of application not to affect the label $g(s.p-s(i))^{-1}\rhd e_k$. Then we have
	\begin{align*}
		\mathcal{A}_i^{g(s.p-s(i))^{-1}\rhd e_k} \ket{\psi}&=\sum_{e'=g(s.p-s(i))^{-1}\rhd e_k}\frac{1}{|E|} \mathcal{A}_i^{e'} \ket{\psi}\\
		&=\mathcal{A}_i \ket{\psi}\\
		&=\ket{\psi}
	\end{align*}
	This indicates that we can absorb such an edge transform into the state $\ket{\psi}$, despite the fact that its label is itself an operator (i.e., the label is configuration-dependent). Using this process, we can absorb each configuration-dependent gauge transform from our series of transforms into the state in turn. This shows that the series of edge transforms does not affect the state, and so neither do the closed blob ribbon operators. We have therefore shown that the closed blob ribbon operators with label in the kernel of $\partial$ act trivially on a state for which we can contract the ribbon operator without crossing any excitations. Next we will use this to show that we can deform open blob ribbon operators without affecting their action on certain states, by demonstrating that combining an open ribbon operator with a closed ribbon operator allows us to deform the open ribbon.

	Consider applying a blob ribbon operator of label $e_k$ on an open ribbon $t_1$ (with start-point $s.p$). We then combine this with a closed ribbon operator with label $e_k^{-1}$ whose ribbon is $t_1$ followed by a new section of ribbon $t_2$ which closes the ribbon, as shown in Figure \ref{deform_ribbon_operator_step_1}. Both of the ribbon operators act on the section $t_1$. Any plaquette $p$ that is pierced by both of the ribbon operators transforms as
	$$e_p \rightarrow e_p [g(s.p-v_0(p))^{-1} \rhd e_k^{\sigma_p}] [g(s.p-v_0(p))^{-1} \rhd e_k^{- \sigma_p}]=e_p,$$
	where $\sigma_p$ is $-1$ if the plaquette's orientation matches the orientation of the ribbon and is $+1$ otherwise. Therefore, the combined action of the two ribbon operators is trivial for any plaquettes pierced by $t_1$. Now consider the other section of ribbon, $t_2$. Only the closed ribbon operator acts on plaquettes pierced by this section, so such a plaquette $p$ transforms as $e_p \rightarrow e_p [g(s.p-v_0(p))^{-1} \rhd e_k^{\sigma_p}]$. This is the same as the action of a ribbon operator with label $e_k^{-1}$ applied only on $t_2$, although note that the start-point for the direct path of the ribbon is still the original start-point $s.p$. This means that we can move the blob ribbon operator from $t_1$ to $t_2$ by applying a closed ribbon operator, as shown in Figure \ref{deform_ribbon_operator_step_1}. Note, however, that $t_2$ is not simply a deformed version of $t_1$, because it has the opposite orientation. In addition, the ribbon operator on $t_2$ has the inverse label $e_k^{-1}$. However, we showed in Section \ref{Section_blob_ribbon_invert_dual_path} that we can reverse the orientation of a ribbon operator by inverting its label. Therefore, as shown in Figure \ref{deform_ribbon_operator_step_2}, we have
	$$B^{e_k^{-1}}(t_2)=B^{e_k}(t_2^{-1}),$$
	where $t_2$ is the ribbon obtained from $t_2$ by reversing the orientation of its dual path (while keeping the start-point fixed). Note that the position of the direct path (or the individual paths $(s.p-v_0(p))$ which make up the direct path) only need to be defined up to deformation when the region on which we apply the operator satisfies fake-flatness.

	We have therefore shown that $B^{e_k}(t_1) B^{e_k^{-1}}(t_1 \cdot t_2) = B^{e_k}(t_2^{-1})$, where $t_1 \cdot t_2$ is a closed ribbon and $t_2^{-1}$ can be obtained from $t_1$ by deformation. This means that we can deform the original ribbon operator $B^{e_k}(t_1)$ by applying a closed ribbon operator. Because the closed ribbon operators act trivially on states with no excitations in the region, this means that deforming the ribbon does not change the action of the ribbon operator and so the blob ribbon operators with label in the kernel of $\partial$ are topological.

	\subsection{Magnetic membrane operators in the $\rhd$ trivial case}
	\label{Section_Topological_Magnetic_Tri_Trivial}
	Next we consider the magnetic membrane operators, starting with the case where $\rhd$ is trivial. We want to show that we can freely deform the membrane (by which we mean both the direct and dual membrane) of the magnetic membrane operator, without affecting its action under the conditions laid out at the start of Section \ref{Section_topological_membrane_operators}. Specifically, we want to show that we can deform the membrane as long as the deformation does not cause the membrane to cross any excitations and we keep the position of the excitations produced by the membrane operator fixed. Because the magnetic membrane operator produces excitations along the boundary of the membrane and may also produce an excitation at the start-point of the operator, it is necessary to keep both the boundary of the membrane and the start-point fixed as we deform the membrane.
	
	\begin{figure}[h]
		\begin{center}
			\begin{overpic}[width=0.75\linewidth]{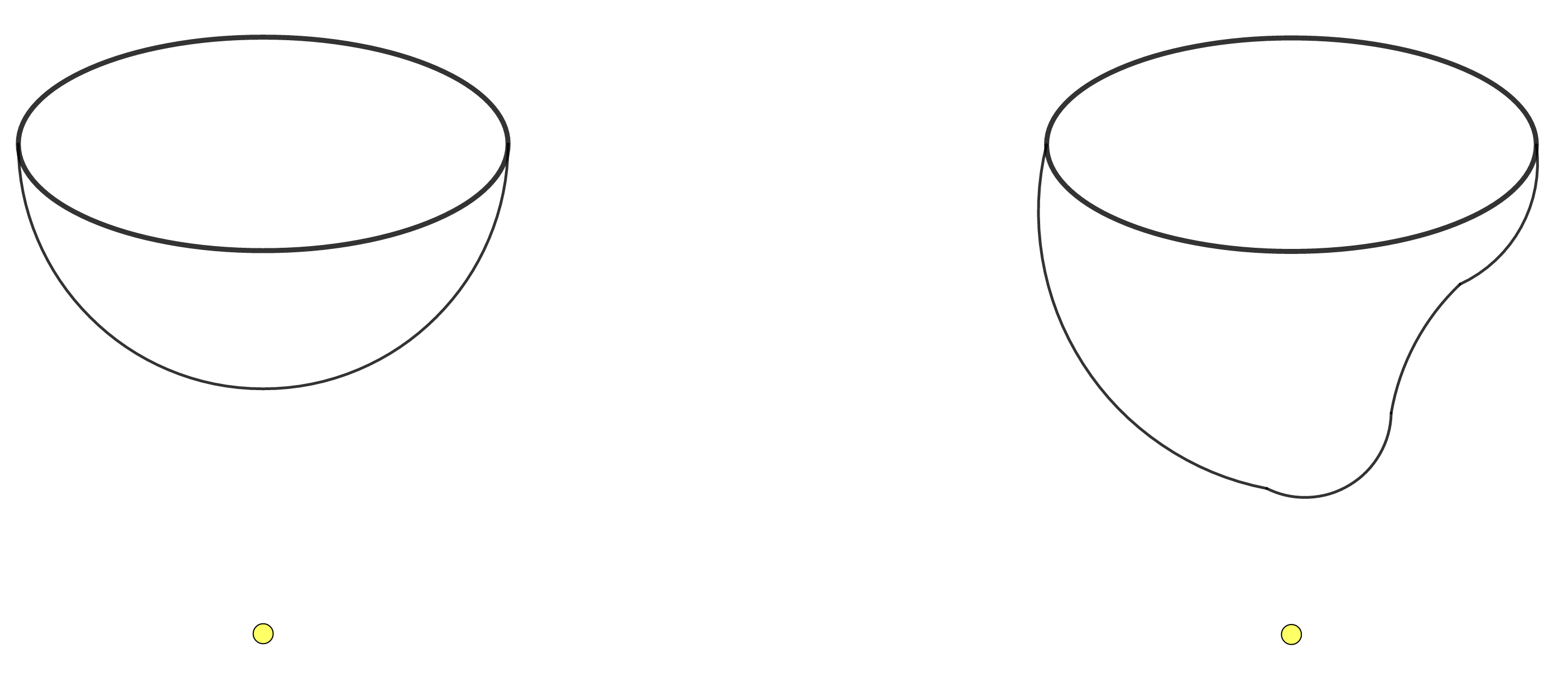}

				\put(34,35){$m$}
				\put(62,35){$m'$}
				\put(45,25){\Huge $\rightarrow$}
				\put(45,30){deform}
				\put(35,10){$C^h(m) \ket{\psi}=C^h(m') \ket{\psi}$}
				
			\end{overpic}
			\caption{We wish to show that magnetic membrane operators applied on two membranes (here $m$ and $m'$) which can be deformed into one-another have the same action on a state $\ket{\psi}$, provided that the region through which we deform the membrane has no excitations. Note that for this to be true, the start-point (shown as a yellow circle here) must be in the same place for each membrane.}
			\label{magnetic_membrane_two_membranes}
		\end{center}
	\end{figure}

	We will prove that the magnetic membrane operators are topological in two steps, taking a similar approach to our proof for the blob ribbon operators in Section \ref{Section_Topological_Blob_Ribbons}. First we will show that magnetic membrane operators applied on closed and contractible membranes act trivially on states for which the membranes enclose no excitations. Then we will show that combining magnetic membrane operators on open membranes with a series of contractible closed membrane operators allows us to deform the open membrane operators. Because these contractible closed membrane operators have trivial action, the deformation therefore has no effect on the action of the open membrane operator and so the magnetic membrane operators are topological.

	We begin by showing that the closed membrane operators act trivially when they enclose no excitations. Consider such a contractible closed membrane $c$ and a state $\ket{\psi}$ which has no excitations enclosed by $c$. We wish to show that 
	$$\hat{C}^{h}(c) \ket{\psi}=\ket{\psi}.$$
	To do this we will show that the closed membrane operator can be written as a series of vertex transforms, just as we did for closed magnetic ribbon operators in 2+1d in Ref. \cite{HuxfordPaper2} (see Section S-II B in the Supplemental Material). We claim that the operator $C^h(c)$ is equivalent to a product of vertex transforms on each of the vertices in the region $S$ enclosed by the dual membrane of $c$:
	$$C^h(c) = \prod_{ v \in S} A_v^{g(s.p-v)^{-1}hg(s.p-v)},$$
	where $s.p$ is the start-point of the membrane $c$. The reason that this is useful is because vertex transforms have no effect on a state $\ket{\psi}$ for which the vertex is unexcited. This is because, for an unexcited vertex $v$, we can pull out a factor of $A_v$ from the state and then absorb the vertex transform $A_v^g$ into $A_v$ (see Section \ref{Section_Recap_3d}), before reabsorbing $A_v$ into the state. By showing that the closed magnetic membrane operator is equivalent to a series of vertex transforms in the unexcited region enclosed by the membrane, we therefore show that the membrane operator acts trivially. We may be concerned that the vertex transforms involved have operator-valued labels (from the expressions $g(s.p-v)$) and so it is not immediately obvious that these transforms can be absorbed into the vertex terms. However, in Section S-II B in the Supplemental Material of Ref. \cite{HuxfordPaper2} we showed that these types of vertex transforms (with the particular form of the label $g(s.p-v)^{-1}hg(s.p-v)$) can still be absorbed into the vertex terms (and this proof did not rely on the lattice being two-dimensional).

\begin{figure}[h]
		\begin{center}
			\begin{overpic}[width=0.75\linewidth]{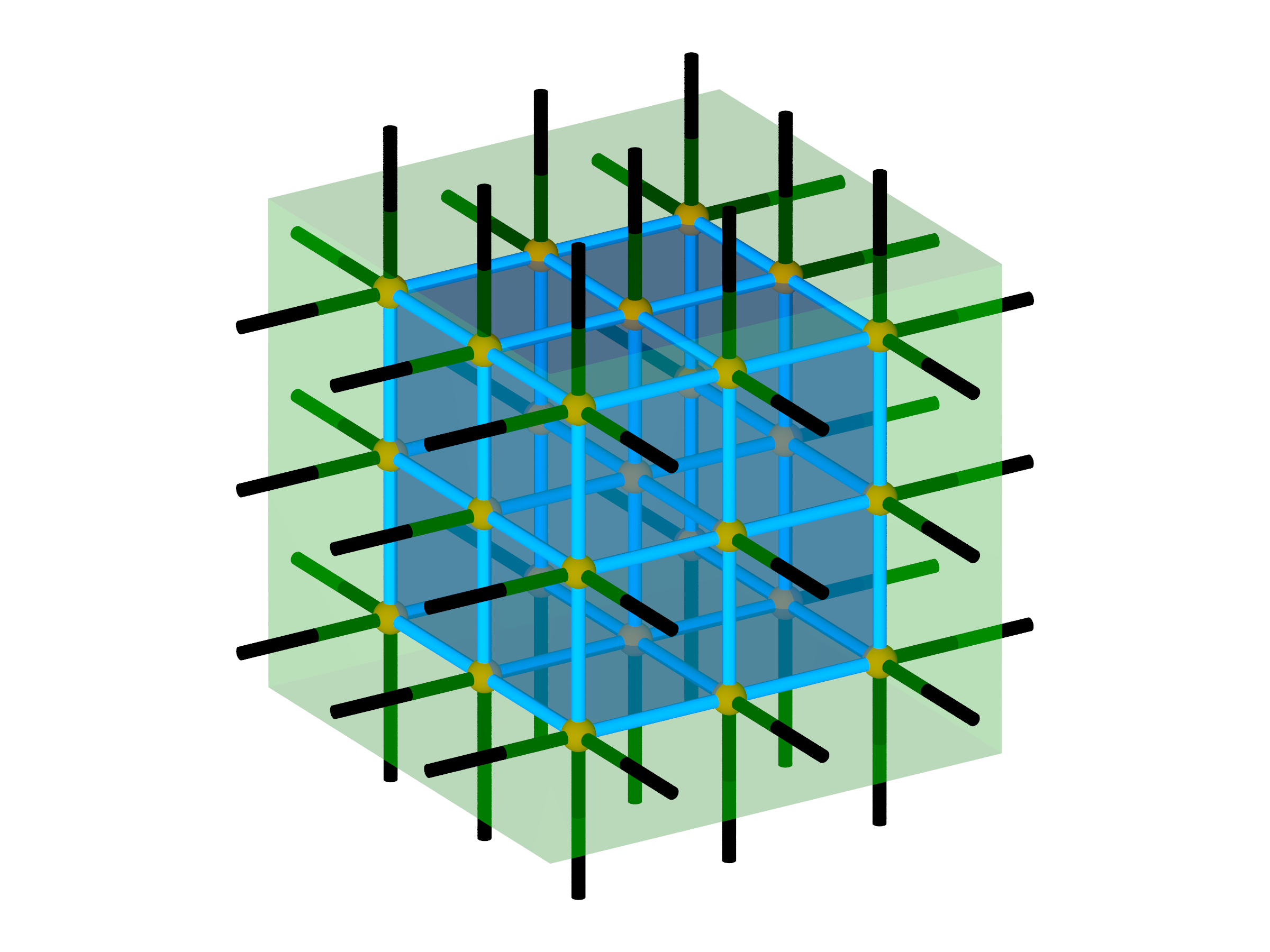}

			\end{overpic}
			\caption{In order to reproduce a closed magnetic membrane operator whose dual membrane is the outer (green) surface and whose direct membrane is the inner (blue) surface, we apply vertex transforms on the vertices (shown as spheres) enclosed by the dual membrane. If the vertex transforms do reproduce the membrane operator, they should act non-trivially on the edges which are cut by the dual membrane (these are the black edges), but not change the labels of the edges which lie on or within the blue surface and so do not pass the dual membrane (the lighter blue edges). As we will see, the vertex transforms do indeed act in this way and this occurs because the lighter edges are connected to two of the vertices on which we apply transforms (one on each end of the edge), while the black edges are only connected to one such vertex. }
			\label{closed_membrane_vertex_transforms_1}
		\end{center}
	\end{figure}

	We now need to prove that the closed membrane operator can indeed be written as a product of such vertex transforms. We claim that
	\begin{equation}
		C^h(c) = \prod_{v\text{ in }S} A_v^{\hat{g}(t)^{-1}h \hat{g}(t)}, \label{Equation_closed_magnetic_membrane_vertex_transforms_1}
	\end{equation}
	where $S$ is the region enclosed by the membrane $c$ (including vertices on the direct membrane itself) and $t$ is a path from the start-point of the membrane operator to the vertex in question (the precise path is not important because we are looking at the case where the region is initially fake-flat). An example of the membrane and the corresponding vertices on which we apply the transforms is shown in Figure \ref{closed_membrane_vertex_transforms_1}. In Equation \ref{Equation_closed_magnetic_membrane_vertex_transforms_1} we did not specify the order of the product, but we showed in Section S-II B of the Supplemental Material for Ref. \cite{HuxfordPaper2} (in the 2+1d case, but the same argument holds in this 3+1d case) that these vertex transforms commute, so we do not need to be more specific.

	There are two types of edge to consider in order to show that the action of the vertex transforms matches that of the membrane operator. Firstly, we have edges that are internal to the solid $S$. These are the edges for which both of the vertices that are attached to the edge are within the solid (the blue edges in Figure \ref{closed_membrane_vertex_transforms_1}). These edges are not cut by the dual membrane and so are not affected by the membrane operator. In order for the vertex transforms to match the action of the membrane operator, the vertex transforms must similarly leave the edge labels of these internal edges unchanged. Each edge $i$ is only affected by the two vertex transforms on the two ends of the edge, the source $s(i)$ and the target $t(i)$. Because both of these vertices are within the region $S$ for an internal edge, we apply vertex transforms on each of these vertices. Then under the action of the two vertex transforms, the label $g_i$ becomes
	\begin{equation}
		g_i \rightarrow g(s.p-s(i))^{-1} h g(s.p-s(i))g_i g(s.p-t(i))^{-1} h^{-1} g(s.p-t(i)). \label{Equation_mag_membrane_topological_vertex_transforms_1}
	\end{equation}

	We can use the same reasoning from the 2+1d case from Ref. \cite{HuxfordPaper2} to show that this expression is equal to $g_i$. The path corresponding to the label $g(s.p-s(i)) g_i g(s.p-t(i))^{-1}$ is a closed path from the start-point to the edge, along the edge and back to the start-point. Because we require that the region $S$ contains no excitations, this closed path will enclose a fake-flat surface (from the plaquette energy terms). This means that the path element must belong to the image of $\partial$. That is, for some element $e \in E$, we have $g(s.p-s(i))g_ig(s.p-t(i))^{-1} = \partial(e)$. We can then use the fact that $\partial(E)$ is in the centre of $G$ when $\rhd$ is trivial to write $\partial(e)=h^{-1}\partial(e)h$, so that $g(s.p-s(i))g_ig(s.p-t(i))^{-1} =h^{-1}\partial(e)h$. Substituting this into Equation \ref{Equation_mag_membrane_topological_vertex_transforms_1}, we see that the edge label transforms according to
	\begin{align*}
		g_i \rightarrow& g(s.p-s(i))^{-1} h [g(s.p-s(i))g_i g(s.p-t(i))^{-1}] h^{-1} g(s.p-t(i))\\
		=& g(s.p-s(i))^{-1} h [h^{-1}\partial(e)h] h^{-1} g(s.p-t(i))\\
		&= g(s.p-s(i))^{-1} \partial(e)g(s.p-t(i))\\
		&= g(s.p-s(i))^{-1} [g(s.p-s(i))g_ig(s.p-t(i))^{-1}] g(s.p-t(i))\\
		&=g_i
	\end{align*}
	under the action of the vertex transforms. That is, the label of an internal edge is left invariant by the vertex transforms, just as it is left invariant by the action of the magnetic membrane operator.

	The other edges to consider are those that are cut by the dual membrane of the magnetic membrane operator (the red edges in Figure \ref{closed_membrane_vertex_transforms_1}) and so are acted on non-trivially by the membrane operator. These edges are the ones for which only one of the vertices attached to the edge lies within the region $S$. This means that only one of the vertex transforms in our product of transforms affects the edge label (rather than two for the internal edges). Consider such an edge, edge $i$, which initially has label $g_i$. We denote the two ends of the edge by $s(i)$ for the source of $i$ and $t(i)$ for the target of $i$. If $s(i)$ is the vertex within $S$, then the edge label is only affected by a vertex transform on $s(i)$ (we do not apply one on $t(i)$). The action of the vertex transform $A_{s(i)}^{g(s.p-s(i))^{-1}hg(s.p-s(i))}$ on the label $g_i$ of the edge is 
	$$g_i \rightarrow g(s.p-s(i))^{-1} h g(s.p-s(i)) g_i.$$
	Similarly if the vertex inside $S$ is instead the target of $i$, then the edge label transforms as 
	$$g_i \rightarrow g_i g(s.p-t(i))^{-1}h^{-1}g(s.p-t(i))$$
	under $A_{t(i)}^{g(s.p-t(i))^{-1}hg(s.p-t(i))}$. This matches the action of the magnetic membrane operator (note that whichever vertex $s(i)$ or $t(i)$ is in the region $S$ is the vertex $v_i$ on the direct membrane that appears in the action of the membrane operator given in Equation \ref{Equation_magnetic_membrane_on_edges_main_text} of the main text), so the series of vertex transforms does indeed reproduce the action of the magnetic membrane operator on all edges.

	So far we have proven that the contractible closed magnetic membrane operator is equivalent to a product of configuration-dependent gauge transforms, which can be absorbed into the state, provided that the region enclosed by the membrane is unexcited. This in turn means that such membrane operators have no effect on states for which the membrane encloses no excitations. As an aside, this fact has interesting consequences for certain membrane operators. Recall that the open magnetic membrane operator could excite the start-point of the membrane and this would occur if we took certain sums of the operators. Consider a magnetic membrane operator that is a sum of membrane operators with label in a certain conjugacy class $[h]$ of $G$: $\sum_{x \in [h]} \alpha_x C^x(m)$. If the coefficients $\alpha$ sum to zero, then the start-point of the membrane operator is excited. In the case of the closed membrane operators however, we know that they act trivially on a state for which they enclose no excitations. Therefore, $\sum_{x \in [h]} \alpha_x C^x(c) = \sum_{x \in [h]} \alpha_x=0$. This indicates that given a magnetic membrane operator with an excitation at the start-point, it is not possible to close the membrane without annihilating the state. This means that it is not possible to collapse the loop-like excitation. This is because the loop-like excitations produced by such a membrane operator carry a point-like charge, which must be balanced by the charge at the start-point, and so collapsing the loop would violate topological charge conservation. We consider the point-like charge of the magnetic excitations in more detail (for a more general case than $\rhd$ trivial) in Section \ref{Section_point_like_charge_higher_flux}.

	Returning to our proof for the topological nature of the membrane operators, we have shown that the magnetic membranes acting on contractible closed membranes are trivial if they enclose no excitations. We now wish to use this result to show that the open membrane operators are topological. To do so we will show that applying contractible closed membrane operators allows us to deform the open ones. Suppose that we have a magnetic membrane operator of label $h$ acting on some (open) membrane $m$. That is, we consider $C^h(m) \ket{\psi}$ for some state $\ket{\psi}$. We want to relate this to the action of a membrane operator $C^h(m')$ on some other membrane $m'$ that shares its boundary and start-point with $m$. Consider applying to $\ket{\psi}$ another magnetic membrane operator, with label $h^{-1}$, on a closed membrane $c$ that includes the original membrane $m$ and which has the same start-point as $m$, before we act with the operator $C^h(m)$. That is, we consider the state $C^h(m) C^{h^{-1}}(c)\ket{\psi}$. Because the membrane $c$ includes $m$, any edge which is cut by the dual membrane of $m$ is also cut by the dual membrane of $m$. We consider the action of the two membrane operators on such an edge $i$. Recall that the action of a magnetic membrane operator $C^x(n)$ on an edge $i$ cut by the dual membrane of $n$ is
	$$C^x(n):g_i = \begin{cases} g(s.p-v_i)^{-1}xg(s.p-v_i)g_i &\text{if } i \text{ points away from the direct membrane} \\ g_i g(s.p-v_i)^{-1}x^{-1}g(s.p-v_i) &\text{if } i \text{ points towards the direct membrane.} \end{cases}$$

\begin{figure}[h]
		\begin{center}
			\begin{overpic}[width=0.75\linewidth]{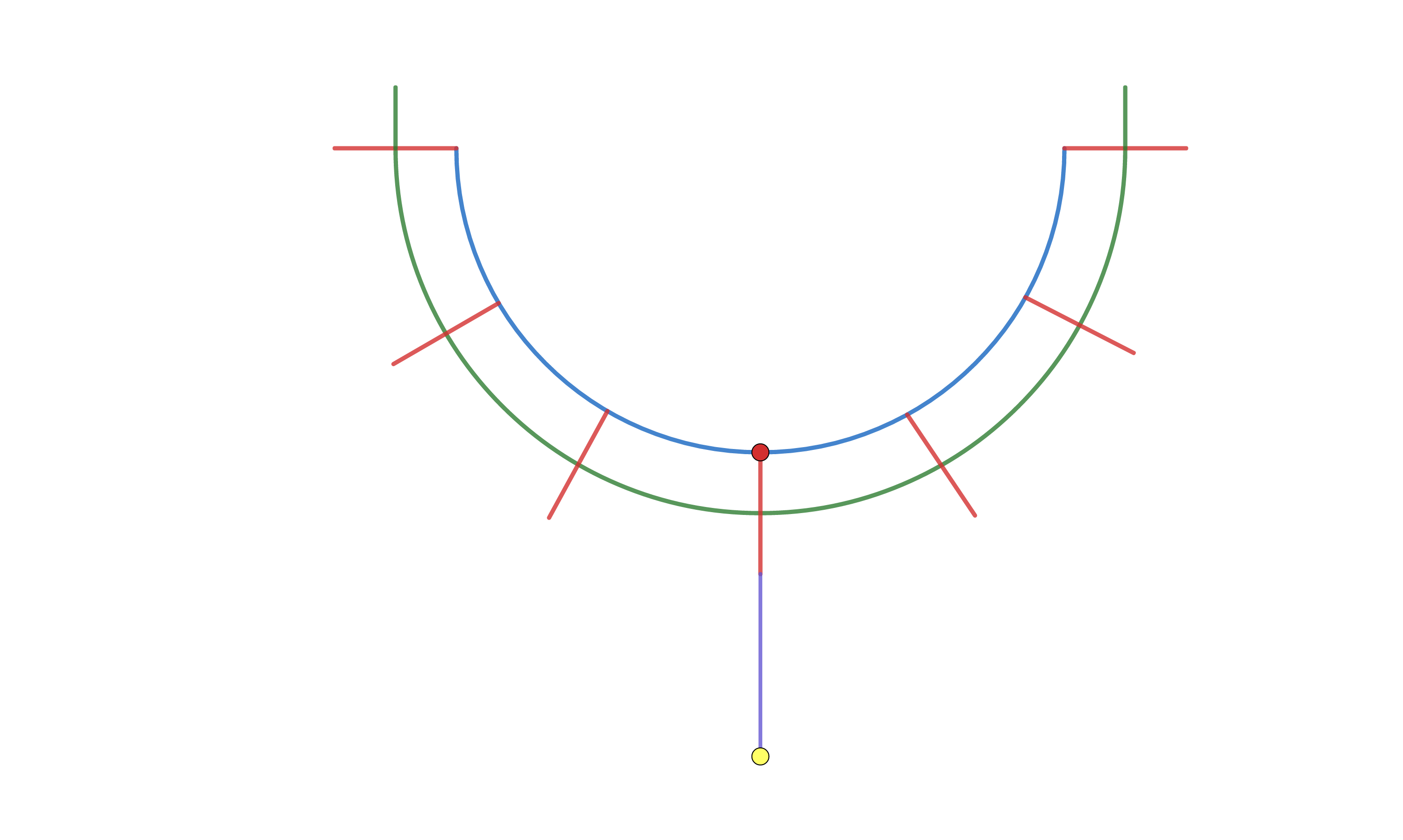}
				\put(55,5){$s.p$}
				\put(55,29){$v_i$}
				\put(55,20){edge $i$}
				\put(55,12){$s.p-v_i$}
				\put(34,43){direct membrane}
				\put(10,43){dual membrane}
				
			\end{overpic}
			\caption{Consider the case where the start-point (yellow) of a magnetic membrane operator is positioned away from the direct membrane of that operator. In this figure, we show a cross-section through such a magnetic membrane operator, with the direct membrane in blue and the dual membrane in green. The red edges are those cut by the dual membrane. We see that for an edge $i$, the path from the start-point to the vertex $v_i$, which is on the direct membrane and attached to edge $i$, may pass through one of the cut edges (in this example, the relevant edge is $i$ itself). Because the path element determines the action of the membrane operator, this means that if we apply two membrane operators in sequence on this region, the action of the first membrane operator may change the path label before the second operator acts. However, this does not change the action of the second membrane operator on the edge.}
			\label{magnetic_membrane_start_point_away}
		\end{center}
	\end{figure}

	This suggests that when we apply the two membrane operators in sequence we have
	\begin{align*}
		C^h(m)& C^{h^{-1}}(c):g_i
		\\&= \begin{cases} g(s.p-v_i)^{-1}hg(s.p-v_i) g(s.p-v_i)^{-1}h^{-1} g(s.p-v_i) g_i &\text{if } i \text{ points away from the direct membrane}\\
			g_i g(s.p-v_i)^{-1}h^{-1}g(s.p-v_i) g(s.p-v_i)^{-1}hg(s.p-v_i) &\text{if } i \text{ points towards the direct membrane} \end{cases}\\
		&=g_i, 
	\end{align*}
	so that the edge label is unchanged. There is something we must consider before we accept this result, however. When we act first with $C^{h^{-1}}(c)$, it is possible that the path element $g(s.p-v_i)$ may be affected by the membrane operator $C^{h^{-1}}(c)$ so that when we act with $C^h(m)$ the path element that appears in the action of $C^h(m)$ is different. This will occur if an edge affected by $C^{h^{-1}}(c)$ lies on the path $s.p-v_i$, as shown in Figure \ref{magnetic_membrane_start_point_away} (it may occur if the start-point is not located on the direct membrane). However, even if this is the case, we can use Equation \ref{mag_membrane_path_end} from Section \ref{Section_Magnetic_Tri_Non_Trivial}, which shows how a path element for a path that cuts through the dual membrane and ends on the direct membrane is affected by the membrane operator, to see that the path element is changed according to $g(s.p-v_i) \rightarrow g(s.p-v_i)g(s.p-v_i)^{-1}hg(s.p-v_i)=h g(s.p-v_i)$ (recall that the label of the membrane operator is $h^{-1}$, not $h$). This means that the term $g(s.p-v_i)^{-1} h g(s.p-v_i)$ which appears in the action of $C^h(m)$ will transform as 
	$$g(s.p-v_i)^{-1} h g(s.p-v_i) \rightarrow g(s.p-v_i)^{-1} h^{-1} h h g(s.p-v_i)= g(s.p-v_i)^{-1} h g(s.p-v_i)$$
	and so is unaffected by the action of $C^{h^{-1}}(c)$. This means that, even if the path element is changed by the closed membrane operator, the effect of the magnetic membrane operator on the edge is not, and our original naive answer for the combined action of the membrane operators is correct. As such, the edge $i$ is indeed left invariant by the two membrane operators. 
	
\begin{figure}[h]
		\begin{center}
			\begin{overpic}[width=0.75\linewidth]{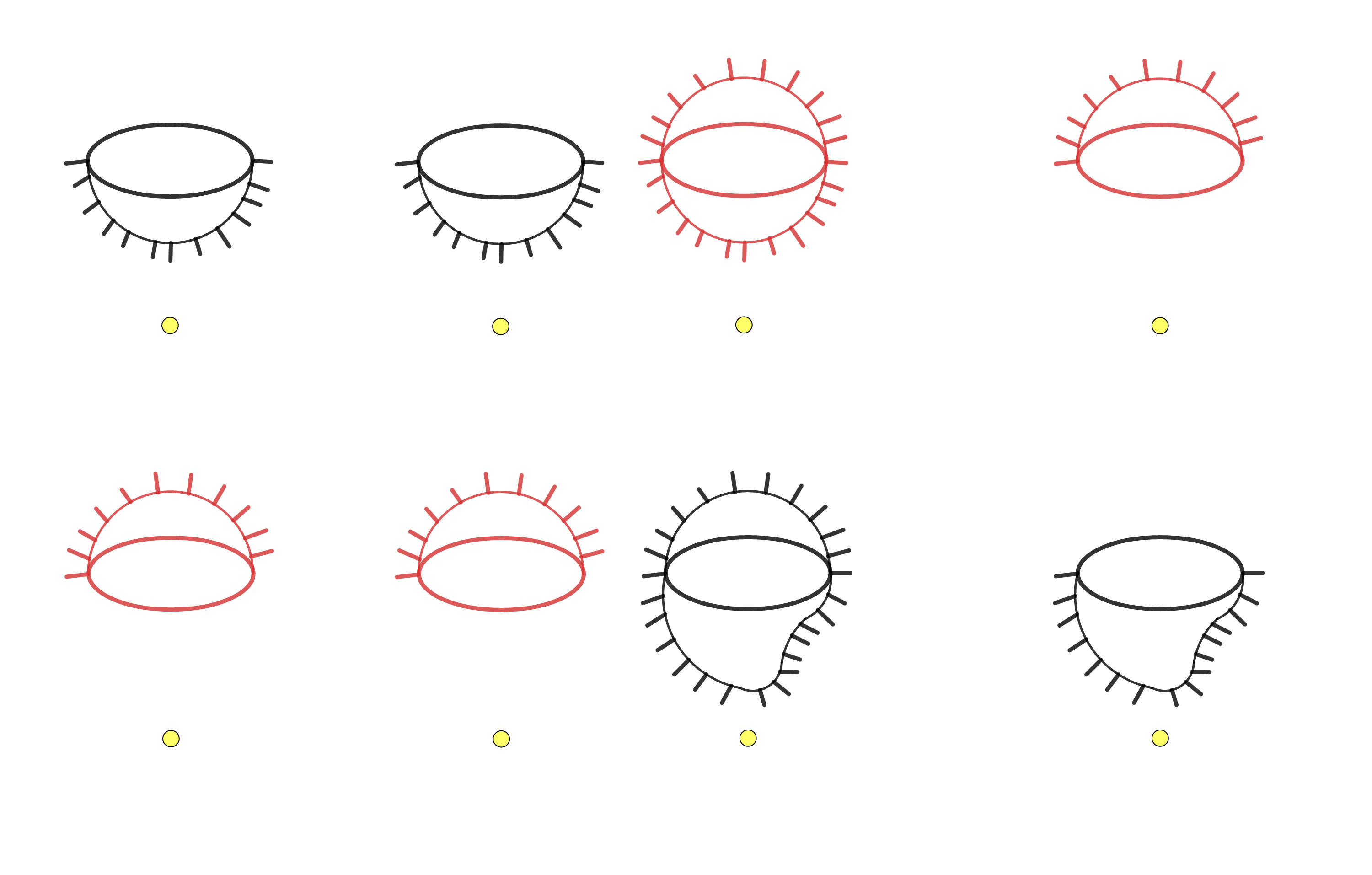}
				
				\put(19,50){\large $\cdot$ $\ket{\psi}=$}
				\put(65,50){\large $\cdot \ket{\psi}=$}
				\put(95,50){\large $\cdot \ket{\psi}$}
				\put(19,20){\large $\cdot$ $\ket{\psi}=$}
				\put(65,20){\large $\cdot \ket{\psi}=$}
				\put(95,20){\large $\cdot \ket{\psi}$}
				
				\put(12,56){\large $h$}
				\put(34,56){\large $h$}
				\put(55,60){\large $h^{-1}$}
				\put(85,60){\large $h^{-1}$}
				
				\put(12,30){\large $h^{-1}$}
				\put(34,30){\large $h^{-1}$}
				\put(55,30){\large $h$}
				\put(85,26){\large $h$}
				
				\put(45,50){\large $\cdot$}
				\put(45,20){\large $\cdot$}
			\end{overpic}
			\caption{We wish to show that an open magnetic membrane operator applied on a state $\ket{\psi}$ can be freely deformed, provided that it crosses no excitations in $\ket{\psi}$ during this deformation. Because the state $\ket{\psi}$ is invariant under the action of a closed magnetic membrane operator that encloses no excitations in the state, we can freely extract such an operator from the state, with inverse label to our open membrane operator. Then, by combining this closed membrane operator with the original open magnetic membrane operator of the inverse label, we can move the open membrane, while simultaneously inverting its label and orientation. This is indicated in the first line, where the membrane operators are represented by the spiked surfaces and the labels of these operators are indicated above the membranes. The spikes represent the edges cut by the dual membrane and so indicate the position of the dual membrane. The small (yellow) circles represent the start-points of the membranes, which are the same in each case. The black membrane is the original one that we wish to deform, by adding the red (gray in grayscale) closed membrane. We see that the relative orientation of the dual membrane and direct membrane is swapped by this (the spikes point upwards rather than downwards) and the label of the operator is also inverted. Repeating the process, as shown in the second line, gives us back an open membrane of the original label and orientation, but in a new location. That is, we can deform an open membrane by applying closed membrane operators.}
			\label{deform_open_membrane_closed_membrane}
		\end{center}
	\end{figure}

	We have therefore shown that the edges cut by both dual membranes are left unchanged, as if no membrane acted on these edges. Because the membrane $c$ includes the entirety of $m$, the only remaining part of the combined membrane operators is the action of $C^{h^{-1}}(c)$ on the edges that are cut by the dual membrane of $c$ but not by the dual membrane of $m$. This is equivalent to the action of a single membrane operator on a new membrane, consisting of the parts of $c$ not in $m$, with label $h^{-1}$, as shown in Figure \ref{deform_open_membrane_closed_membrane}. The resulting membrane operator is not a deformed version of the original however, because the label of the resulting membrane operator is $h^{-1}$ rather than $h$ and the orientation of the membrane has swapped, as indicated in Figure \ref{deform_open_membrane_closed_membrane}. Because the action of the closed membrane operator is trivial this indicates that we can reverse the orientation of a magnetic membrane operator by inverting its label, without affecting its action. This is a useful result in its own right, but we can also use this to show that we can deform the membrane (without reversing its orientation). To do so we simply repeat the previous process on the inverted membrane operator, this time adding a closed membrane of label $h$, whose membrane is the combination of $m'$ (the desired final membrane) and the leftover membrane from the previous closed membrane operator. The result of this is (following the previous argument) a membrane operator of label $h$ acting on membrane $m'$. That is, by applying the two closed membranes we deform our original membrane as we required.
	
	Therefore, we have shown that combining an open membrane operator with contractible closed membrane operators allows us to deform the open membrane. Combined with our argument that the closed membrane operators act trivially on states for which the membranes enclose no excitations, we see that we can deform the open membranes without affecting the action of the membrane operator. This means that the magnetic membrane operators are indeed topological.

	\subsection{Magnetic membrane operators in the case where $\partial \rightarrow$ centre($G$) and $E$ is Abelian}
	\label{Section_Topological_Magnetic_Tri_Nontrivial}
	In Section \ref{Section_Topological_Magnetic_Tri_Trivial}, we proved that the magnetic membrane operator is topological in the case where $\rhd$ is trivial. Now we wish to show that this is also true for the case where $\rhd$ is non-trivial, but we still restrict the crossed module so that $E$ is Abelian and $\partial$ maps to the centre of $G$. We will follow a similar approach to the one we used for the magnetic membrane operator when $\rhd$ is trivial. First, we will prove that a magnetic membrane operator $C^h_T(m)$ applied on a contractible closed membrane can be produced using only vertex and edge transforms, provided that the membrane encloses no excitations. These transforms can be absorbed into a state where these vertices and edges are not excited, so this will demonstrate that the magnetic membrane operator on a closed membrane acts trivially on such a state. Consider a closed membrane $m$, with start-point $s.p$. To reproduce the action of the membrane operator $C^h_T(m)$ we first perform a vertex transform $A_v^{g(s.p-v)^{-1}hg(s.p-v)}$ on each vertex $v$ enclosed by (or lying on) the direct membrane of the membrane operator that we wish to reproduce, as illustrated in Figure \ref{closed_membrane_vertex_transforms_1} in Section \ref{Section_Topological_Magnetic_Tri_Trivial}. This reproduces the action of the magnetic membrane operator on each edge, just as in the $\rhd$ trivial case we considered in Section \ref{Section_Topological_Magnetic_Tri_Trivial}. For internal edges (the blue edges in Figure \ref{closed_membrane_vertex_transforms_1}), the edge label is affected by two of the vertex transforms that we apply, one for each end of the edge. As we saw in Section \ref{Section_Topological_Magnetic_Tri_Trivial} the effects of the two vertex transforms on the edge will cancel, so the edge is unaffected by the series of vertex transforms. That is, for an edge $i$ for which we apply vertex transforms on each end of the edge, the label $g_i$ of the edge transforms as
	$$g_i \rightarrow g(s.p-s(i))^{-1}hg(s.p-s(i))g_ig(s.p-t(i))^{-1}h^{-1}g(s.p-t(i)),$$
	where $s(i)$ is the source of edge $i$ and $t(i)$ is the target. Looking at this expression, we see that $h$ and $h^{-1}$ are separated by a term $g(s.p-s(i))g_ig(s.p-t(i))^{-1}$. Both $g(s.p-t_i)$ and $g(s.p-s(i))g_i$ correspond to paths from the start-point to the target of $i$, where the path corresponding to $g(s.p-s(i))g_i$ is $(s.p-s(i)) \cdot i$ (with $\cdot$ denoting concatenation of paths). Because the paths $(s.p-t(i))$ and $(s.p-s(i)) \cdot i$ enclose a fake-flat region between them, the path elements $g(s.p-t(i))$ and $g(s.p-s(i))g_i$ differ only by an element of $\partial(E)$, which is in the centre of $G$ and so does not affect $g(s.p-t(i))^{-1}h^{-1}g(s.p-t(i))$. This means that we can replace $g(s.p-t(i))^{-1}h^{-1}g(s.p-t(i))$ in the transformation of the edge label with $g_i^{-1}g(s.p-s(i))^{-1}h^{-1}g(s.p-s(i))g_i$, to obtain
	$$g_i \rightarrow g(s.p-s(i))^{-1}hg(s.p-s(i))g_ig_i^{-1}g(s.p-s(i))^{-1}h^{-1}g(s.p-s(i))g_i=g_i.$$
	We see that the internal edges are left invariant by the vertex transforms, just as they are left unchanged by the action of $C^h_T(m)$.

	Next we consider the edges that are cut by the dual membrane. Such edges are adjacent to one vertex on the direct membrane and one vertex that is not enclosed by the membrane. This means that these edges will be affected by exactly one vertex transform, so the label $g_i$ of such an edge $i$ transforms as
	$$g_i \rightarrow g(s.p-s(i))^{-1}hg(s.p-s(i))g_i$$
	or
	$$g_i \rightarrow g_ig(s.p-t(i))^{-1}h^{-1}g(s.p-t(i))$$
	depending on whether the source or the target of the edge lies on the direct membrane. This also agrees with the action of $C^h_T(m)$, just as in the $\rhd$ trivial case. Therefore, the action on the edges can be reproduced using only vertex transforms.

	Now we need to consider whether we can also reproduce the action of the magnetic membrane on the plaquettes. First, recall that the magnetic membrane operator may produce a blob excitation at the privileged blob, blob 0, defined when we specify the membrane. The magnetic membrane operator $C^h_T(m)$ changes the 2-holomomy of this blob from $1_E$ in the ground state to $(h \rhd e_m^{-1}) \: e_m$, where $e_m$ is the total surface label of the direct membrane. However, if we are considering a closed contractible membrane, then the surface label $e_m$ must be the identity $1_E$ if the membrane does not enclose any excitations, due to the blob energy terms. This means that $[h \rhd e_m^{-1}] \: e_m =1_E$ and so the magnetic membrane operator does not excite blob 0. This means that the magnetic membrane operator produces no blob excitations at all. This is significant because it means that we can write the action of the membrane operator on the plaquettes as a series of closed blob ribbon operators, with operator-valued labels. In order to understand this, we consider changing the plaquette labels in a region in a way that preserves the blob 2-holonomies, but is otherwise arbitrary.

\begin{figure}[h]
		\begin{center}
			\begin{overpic}[width=0.75\linewidth]{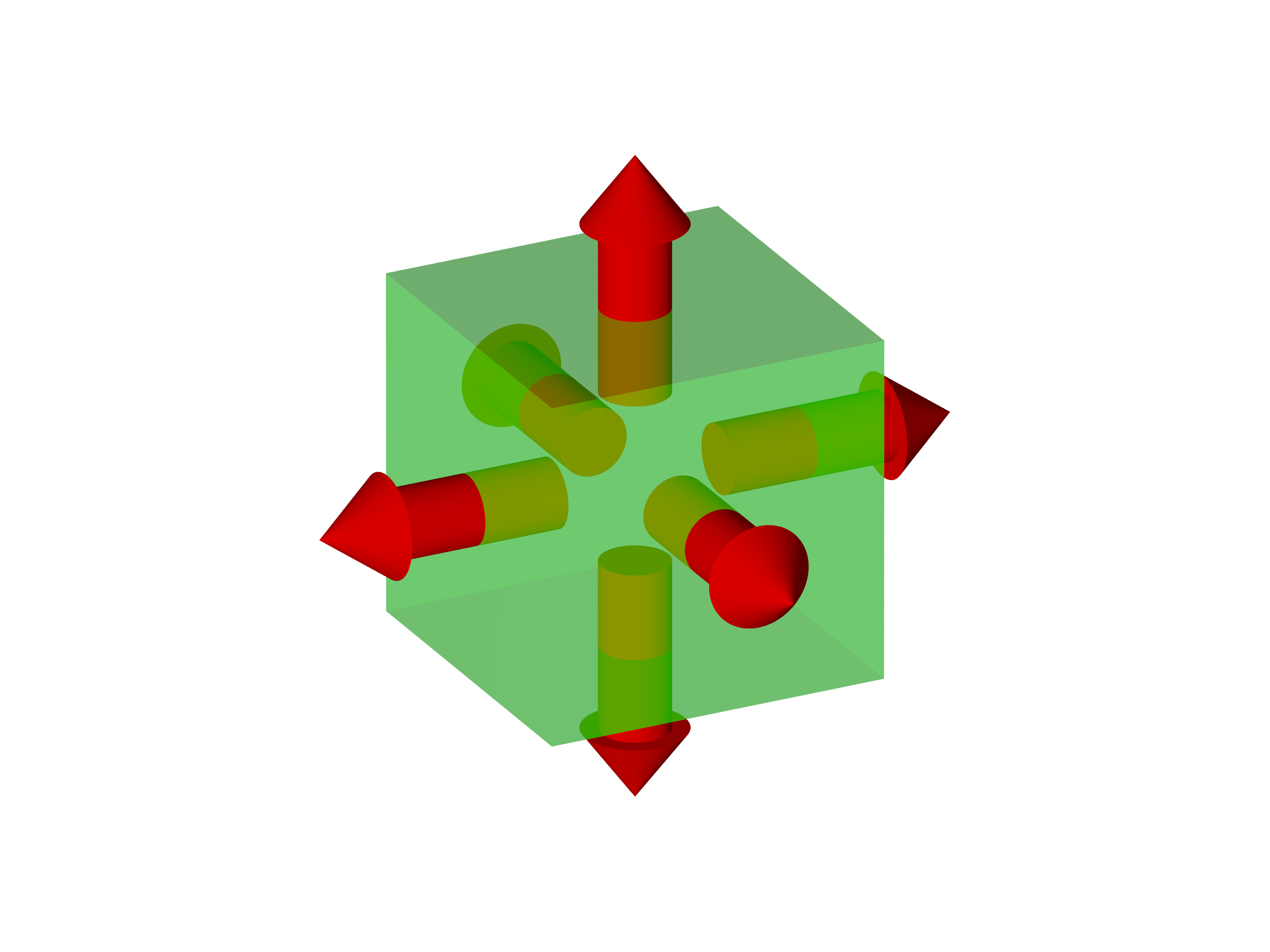}
				
			\end{overpic}
			\caption{The changes to the plaquettes on a blob can be expressed as the action of blob ribbon operators (represented here by red arrows) that only pass through those plaquettes.}
			\label{unfused_blob_ribbon_in_blob}
		\end{center}
	\end{figure}

\begin{figure}[h]
		\begin{center}
			\begin{overpic}[width=0.75\linewidth]{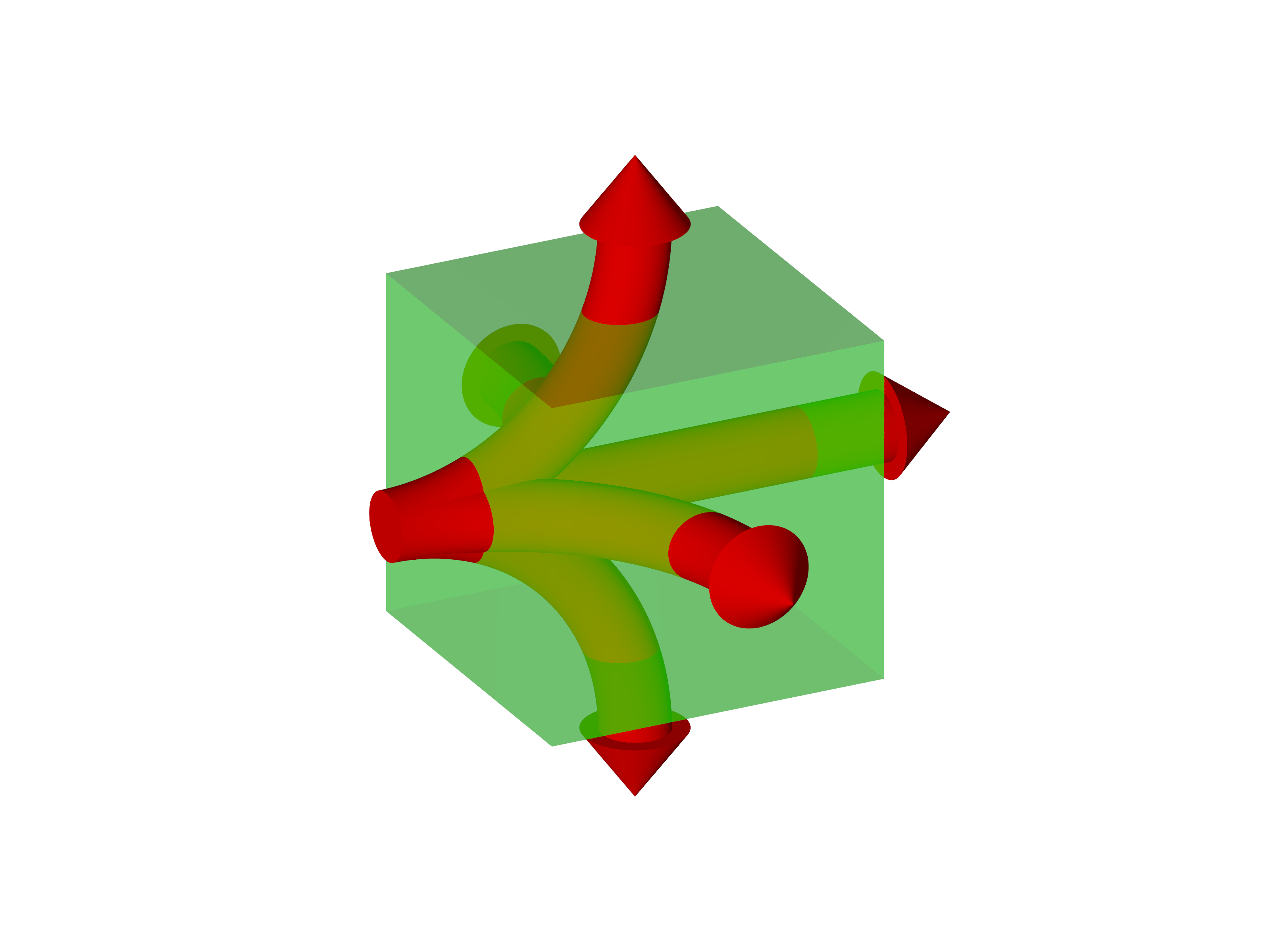}
				
			\end{overpic}
			\caption{If the blob 2-holonomy is preserved by the action on the plaquettes, then the blob ribbon operators representing that change must carry a net trivial 2-flux into the blob. If we treat one ribbon operator as incoming, then it must carry the same 2-flux as the outgoing ribbon operators. We can split this incoming operator into one for each outgoing one and connect the incoming and outgoing ribbons. Therefore, these blob ribbon operators can be fused together, with no blob ribbon operator terminating inside the blob. Repeating this for each blob, we see that the changes to the plaquettes can be written in terms of closed blob ribbon operators.}
			\label{fused_blob_ribbon_in_blob}
		\end{center}
	\end{figure}

	Consider a particular blob in such a region and consider changing the plaquettes on the boundary of the blob in an arbitrary way. We can write the action on each plaquette as the action of an operator-labelled blob ribbon operator which only passes through that plaquette. For example, if plaquette $p_1$ is initially labelled by $e_1$ and it is changed to $f_1$, then this is the same as the action of a blob ribbon operator of label $f_1^{-1}e_1$ (and start-point at that plaquette's base-point) passing through that plaquette. A similar thing holds for each plaquette, so we can think of an arbitrary change to the plaquette label as the action of a set of ribbon operators taking 2-flux into or out of the blob, as shown in Figure \ref{unfused_blob_ribbon_in_blob}. However, if the blob is not excited then the net change of 2-flux must be trivial, so that the 2-flux entering the blob is the same as the 2-flux leaving. Recalling from Section \ref{Section_blob_ribbon_invert_dual_path} that we can invert the orientation of the dual path of a blob ribbon operator by inverting its label, we can choose one blob ribbon operator to pass into the blob and the rest to pass out of the blob. Then, if we move the start-point of each ribbon operator to be the same, the label of the blob ribbon operator entering the blob is the same as the product of the labels of the ribbon operators leaving the blob. We can then split the ribbon operator that enters into one operator for each of the blob ribbon operators leaving the blob and connect each part entering the blob to the corresponding operator leaving the blob, as shown in Figure \ref{fused_blob_ribbon_in_blob}, so that none of the blob ribbon operators terminate in the blob. Repeating the idea for every blob in the affected region, we can write the action on the plaquettes as a series of closed blob ribbon operators, because these are the operators that do not terminate in any blob. Therefore, any action on the plaquettes which does not excite any blob energy terms can be expressed in terms of closed blob ribbon operators.

	This idea is particularly valuable when we consider the closed magnetic membrane operator, which does not produce any blob excitations. Recall that we reproduced the action of the magnetic operator on the edges with a series of vertex transforms. These vertex transforms also affect the plaquettes, but do not excite the blob energy terms. Then the action from the magnetic membrane operator that is left-over after we apply the vertex transforms is purely on the plaquette labels (though it can depend on the state of the edges). Note that this includes not only the action of the membrane operator on the plaquettes, but also the action needed to reverse the effects of the vertex transforms on the plaquettes. Denoting the combined action of all of the vertex transforms on the plaquettes by $\hat{A}(\text{plaquette})$ and the action of the membrane operator on the plaquettes by $\hat{C}(\text{plaquette})$, we have
	$$C^h_T(m)\ket{GS}= \hat{C}(\text{plaquette})\hat{A}(\text{plaquette})^{-1} \big(\prod_{v \in S} A_v^{g(s.p-v)^{-1}hg(s.p-v)}\big)\ket{GS},$$
	where $S$ is the region enclosed by the membrane. Then the remnant of the operator $C^h_T(m)$ after we extract the vertex transforms is $\hat{C}(\text{plaquette})\hat{A}(\text{plaquette})^{-1}$. Because neither the vertex transforms nor the magnetic membrane operators excite the blob energy terms, this left-over part of the action, which only affects the plaquette labels, also does not excite these terms and so can be expressed in terms of blob ribbon operators.

	We showed in Section \ref{Section_Topological_Blob_Ribbons} that non-confined closed blob ribbon operators were trivial when acting on a region without excitations. However, in that section we considered constant-labelled ribbon operators. The blob ribbon operators that reproduce the action of the magnetic membrane operator are operator-labelled, so we would need to use operator-labelled edge transforms to reproduce these blob ribbon operators (with a different operator label to the ones considered in Section \ref{Section_Topological_Blob_Ribbons}). It is therefore not obvious that these edge transforms can be absorbed into the ground state. However, we can use the fact that the membrane operator does not produce any edge excitations to solve this problem. Recall that the reason an operator-valued edge transform could fail to be absorbed into the associated energy term is the non-commutativity of the operator label with the transform. An operator valued transform $\mathcal{A}_i^{\hat{f}}$ can be written as $\sum_{e \in E} \mathcal{A}_i^e \delta(e, \hat{f})$ and so when considering $\mathcal{A}_i^{\hat{f}} \mathcal{A}_i$, representing the transform acting on an unexcited edge, we must commute $\mathcal{A}_i^e$ past $\delta(e, \hat{f})$ to reach $\mathcal{A}_i$. However, if we know that the edge is not excited after we act with the magnetic membrane operator, then we can instead apply the edge energy term from the left without affecting the state. Then the transform $\mathcal{A}_i^e$ does not need to be commuted past the operator label in order to be absorbed onto the edge energy term. That is, because the membrane operator does not excite the edge $i$, we can write
	$$\mathcal{A}_i C^h_T(m) \ket{\psi}=C^h_T(m) \ket{\psi}.$$
	Then, we know that the membrane operator can be written as a series of vertex and edge transforms with operator labels on some set of vertices and edges in the region (when acting on $\ket{\psi}$):
	\begin{equation}
		C^h_T(m)\ket{\psi}= \prod_i \mathcal{A}_i^{\hat{e}_i} \prod_v A_v^{\hat{g}_v}\ket{\psi}. \label{Equation_closed_magnetic_membrane_as_transforms_1}
	\end{equation}
	
	We now consider applying an edge energy term $\mathcal{A}_j$ to the left of both sides of Equation \ref{Equation_closed_magnetic_membrane_as_transforms_1}, where $j$ is one of the edges cut by the dual membrane so that $j$ is in the product over $i$ in Equation \ref{Equation_closed_magnetic_membrane_as_transforms_1}. In particular, we choose $j$ to be the edge that appears leftmost in the product over edges. We then have
	\begin{align}
		C^h_T(m) \ket{\psi}&= \prod_i \mathcal{A}_i^{\hat{e}_i} \prod_v A_v^{\hat{g}_v} \ket{\psi}\notag \\
		&\implies \mathcal{A}_jC^h_T(m) \ket{\psi} = \mathcal{A}_j \mathcal{A}_j^{\hat{e}_j}\prod_{i \neq j} \mathcal{A}_i^{\hat{e}_i}\prod_v A_v^{\hat{g}_v} \ket{\psi}, \label{Equation_closed_magnetic_membrane_as_transforms_2}
	\end{align}
	where we also extracted the factors corresponding to edge $j$ from the product over $i$. We know that the magnetic membrane operator does not produce any edge excitations, and in particular does not produce an edge excitation on $j$. This means that the state produced by acting with $C^h_T(m)$ on $\ket{\psi}$ is an eigenstate of $\mathcal{A}_j$ with eigenvalue 1: $\mathcal{A}_j C_h^T(m)\ket{\psi}=C^h_T(m) \ket{\psi}$. Inserting this into Equation \ref{Equation_closed_magnetic_membrane_as_transforms_2} gives us
	\begin{align*}
		C^h_T(m)\ket{\psi} &=\mathcal{A}_j \sum_{e \in E} \mathcal{A}_j^e \delta(e, \hat{e}_j) \prod_{i \neq j} \mathcal{A}_i^{\hat{e}_i}\prod_v A_v^{\hat{g}_v} \ket{\psi},
	\end{align*}
	
	We can then use the relation $\mathcal{A}_j \mathcal{A}_j^e =\mathcal{A}_j$ (this follows from the algebra of the edge transforms described in Section \ref{Section_Recap_3d} of the main text) to absorb the transform $\mathcal{A}_j^e$ on the right side of the equation into $\mathcal{A}_j$. This gives us
	\begin{align*}
		C^h_T(m) \ket{\psi} &=\mathcal{A}_j \sum_{e \in E} \delta(e, \hat{e}_j) \prod_{i \neq j} \mathcal{A}_i^{\hat{e}_i}\prod_v A_v^{\hat{g}_v} \ket{\psi}\\
		&=\mathcal{A}_j \prod_{i \neq j} \mathcal{A}_i^{\hat{e}_i}\prod_v A_v^{\hat{g}_v} \ket{\psi},
	\end{align*}
	which implies that we can absorb the leftmost operator-valued edge transform into the edge energy term. We can then repeat this for each edge to absorb all of the edge transforms into edge energy terms. We can then absorb the vertex transforms into the state $\ket{\psi}$, just as we did in the $\rhd$ trivial case (see Section \ref{Section_Topological_Magnetic_Tri_Trivial}), so Equation \ref{Equation_closed_magnetic_membrane_as_transforms_1} becomes
	$$C^h(m) \ket{\psi} = \prod_i \mathcal{A}_i \ket{\psi} = \ket{\psi}.$$
	That is, a closed magnetic membrane operator acting on a contractible surface is trivial, provided that the surface encloses no excitations.

\begin{figure}[h]
		\begin{center}
			\begin{overpic}[width=0.75\linewidth]{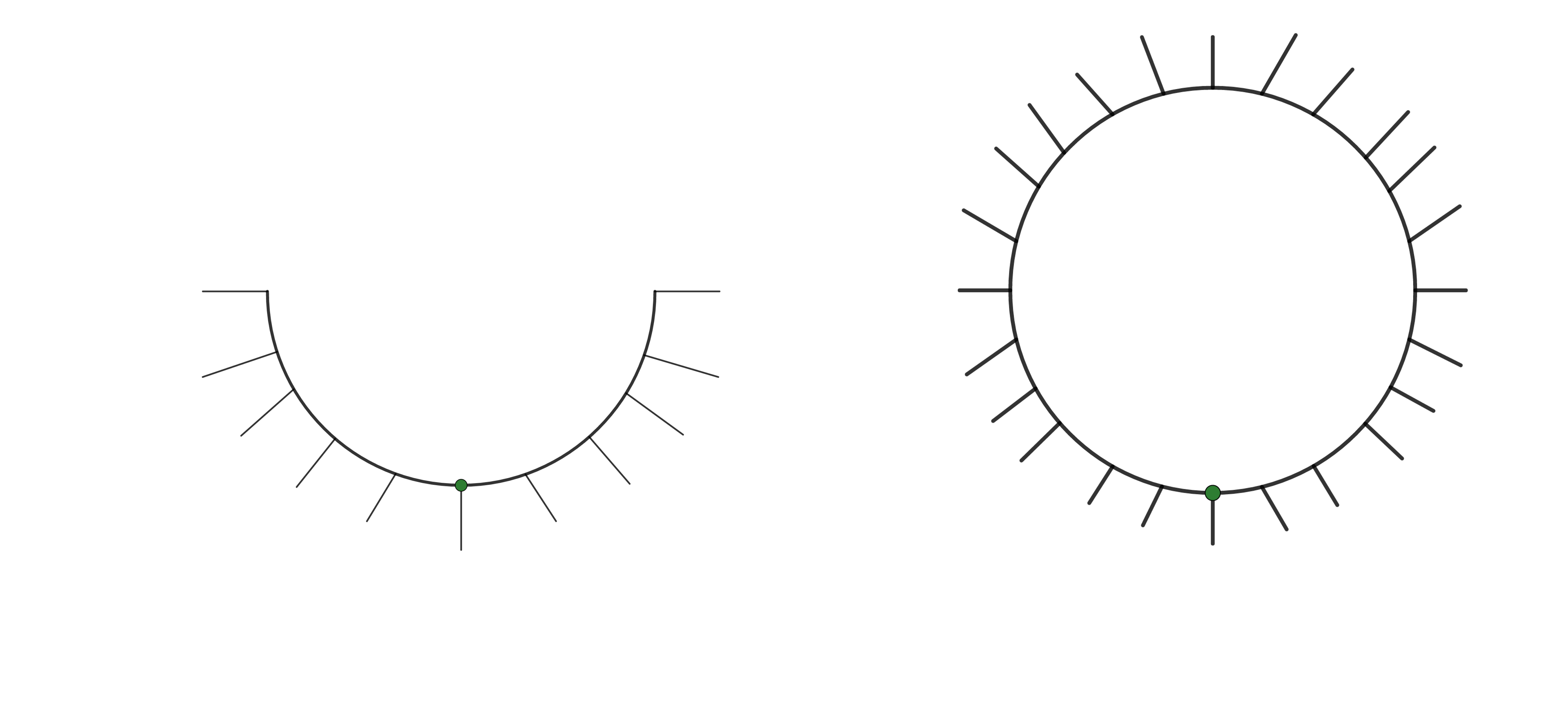}
				\put(15,28){\large $h$}
				\put(55,28){\large $h^{-1}$}
				\put(50,25){\Huge $\cdot$}	
			\end{overpic}
			
			\caption{We explicitly consider the product of a closed magnetic membrane operator $C^{h^{-1}}_T(c)$ (right) of label $h^{-1}$ and an open magnetic membrane operator $C^h_T(m)$ of label $h$, the cross-sections of which are illustrated here.}
			
			\label{Magnetic_membrane_closed_and_open}
		\end{center}
	\end{figure}

	Having shown that a contractible closed magnetic membrane operator acts trivially on states for which the membrane encloses no excitations, we next wish to show that we can deform open magnetic membrane operators by applying such closed ones. We consider applying a membrane operator $C^h_T(m)$ on an open membrane $m$. We wish to deform this membrane by applying a membrane operator $C^{h^{-1}}_T(c)$, acting on a contractible closed membrane $c$. Such an operator will act trivially on any state $\ket{\psi}$ for which the closed membrane encloses no excitations. That is, we can write $C^h_T(m) \ket{\psi}= C^h_T(m) C^{h^{-1}}_T(c) \ket{\psi}$. We choose the closed membrane $c$ so that it includes $m$. Now let us explicitly consider the product of the open and closed membrane operators, as indicated in Figure \ref{Magnetic_membrane_closed_and_open}.
	
	From the discussion in Section \ref{Section_Topological_Magnetic_Tri_Trivial}, we know that the action of $C^{h^{-1}}_T(c)$ cancels with the action of $C^h_T(m)$ on the edges cut by both dual membranes. In that Section we discussed the case where $\rhd$ is trivial, but the action on the edges is the same regardless of whether $\rhd$ is trivial. This just leaves a non-trivial action on the edges that are acted on by only one of the membrane operators, but not the other. Because $c$ includes the entirety of $m$, this means that the action of the two membrane operators on the edges is just the action of the closed membrane operator $C^h_T(c)$ on the edges cut by $c$ but not $m$. Now we want to consider whether this is also true for the action on the plaquettes. As described in Equation \ref{Equation_total_magnetic_membrane_definition_1}, a magnetic membrane operator $C^h_T(m)$ is given by a product of an operator $C^h_{\rhd}(m)$, which acts on the edges of the lattice and in a limited way on plaquettes cut by the dual membrane (with the $\rhd$ action defined in Equation \ref{Equation_magnetic_membrane_rhd_action_appendix}), and a series of blob ribbon operators:
	\begin{equation}
		C^h_T(m) = C^h_{\rhd}(m) \prod_{\substack{\text{plaquette }p \\ \text{on membrane}}}\big( B^{f(p)}(\text{blob }0 \rightarrow \text{blob }p) \big), 
	\end{equation}
	where
	\begin{align*}
		f(p)&=[g(s.p-v_{0}(p))\rhd e_p^{\sigma_p}] \: [(h^{-1}g(s.p-v_0(p)))\rhd e_p^{-\sigma_p}].
	\end{align*}
	In this expression, $\sigma_p$ is $+1$ if the plaquette points away from the dual membrane of the magnetic membrane operator and $-1$ if it points towards it. If we assume that the plaquettes point away from the dual membrane (or use the re-branching procedures from the Appendix of Ref. \cite{HuxfordPaper1} to force this) and introduce $e_p|_{s.p}= g(s.p-v_0(p)))\rhd e_p$, we can write the label as
	$$f(p)= e_p|_{s.p} [h^{-1} \rhd e_p^{-1}|_{s.p}].$$
	Then we can write the product of the open and closed magnetic membrane operators as
	\begin{align}
		C^h_T(m) C^{h^{-1}}_T(c)=& C^h_{\rhd}(m) \bigg( \prod_{\substack{\text{plaquette }p \\ \text{on membrane $m$}}} B^{e_p|_{s.p} [h^{-1} \rhd e_p^{-1}|_{s.p}]}(\text{blob }0 \rightarrow \text{blob }p) \bigg) \notag\\
		&\hspace{0.5cm} C^{h^{-1}}_{\rhd}(c) \bigg( \prod_{\substack{\text{plaquette }p \\ \text{on membrane $c$}}} B^{e_p|_{s.p} [h \rhd e_p^{-1}|_{s.p}]}(\text{blob }0 \rightarrow \text{blob }p) \bigg).
		\label{Equation_product_closed_open_magnetic_membranes}
	\end{align}

	Each blob ribbon operator in Equation \ref{Equation_product_closed_open_magnetic_membranes} is associated to a plaquette that lies in the direct membrane of either $c$ or $m$. Because the direct membrane of $c$ includes the direct membrane of $m$, the plaquettes in the direct membrane of $m$ are also in the direct membrane of $c$. This means that each plaquette in $m$ is associated to two blob ribbon operators, one from $C^h_T(m)$ and one from $C^{h^{-1}}_T(c)$. We can choose the closed membrane operator such that the blob ribbon operator in $C^{h^{-1}}_T(c)$ associated to such a plaquette $p$ is applied on precisely the same ribbon $(\text{blob }0 \rightarrow \text{blob }p)$ as the corresponding ribbon operator from $C^h_T(m)$. Then the two ribbon operators associated to plaquette $p$ are $B^{e_p|_{s.p} [h^{-1} \rhd e_p^{-1}|_{s.p}]}(t_p)$ from $C_T^h(m)$ and $B^{e_p|_{s.p} [h \rhd e_p^{-1}|_{s.p}]}(t_p)$ from $C_T^{h^{-1}}(m)$ (where $t_p$ is shorthand for $(\text{blob }0 \rightarrow \text{blob }p)$). We wish to show that these two blob ribbon operators cancel out.

	In order to cancel the two ribbon operators associated to a particular plaquette $p$, it is first necessary to put these ribbon operators next to each-other in Equation \ref{Equation_product_closed_open_magnetic_membranes}. In Equation \ref{Equation_product_closed_open_magnetic_membranes} we see that the blob ribbon operators from the two magnetic membrane operators are separated by the operator $C^h_{\rhd}(m)$ (and other blob ribbon operators, though the different blob ribbon operators commute). Therefore, we need to commute one of the blob ribbon operators past $C^h_{\rhd}(m)$ before we can consider cancelling them. We established in Section \ref{Section_Magnetic_Tri_Non_Trivial} that the blob ribbon operators do not commute with $C^h_{\rhd}(m)$, so we must first find the relevant commutation relation. If we just consider the blob ribbon operators corresponding to a particular shared plaquette $q$ in Equation \ref{Equation_product_closed_open_magnetic_membranes}, we wish to find
	$$C^h_{\rhd}(m) B^{e_q|_{s.p} [h^{-1} \rhd e_q^{-1}|_{s.p}]}(t_q) C^{h^{-1}}_{\rhd}(c)B^{e_q|_{s.p} [h \rhd e_q^{-1}|_{s.p}]}(t_q).$$
	Using the commutation relation in Equation \ref{blob_ribbon_magnetic_commutation} from Section \ref{Section_Magnetic_Tri_Non_Trivial}, we see that
	\begin{align*}
		C^h_{\rhd}(m) B^{e_q|_{s.p} [h^{-1} \rhd e_q^{-1}|_{s.p}]}(t_q) C^{h^{-1}}_{\rhd}(c)B^{e_q|_{s.p} [h \rhd e_q^{-1}|_{s.p}]}(t_q)&= C^h_{\rhd}(m)B^{e_q|_{s.p} [h^{-1} \rhd e_q^{-1}|_{s.p}]}(t_q)B^{(h^{-1}\rhd e_q|_{s.p}) e_q^{-1}|_{s.p}}(t_q)C^{h^{-1}}_{\rhd}(c)\\
		&=C^h_{\rhd}(m)C^{h^{-1}}_{\rhd}(c),
	\end{align*}
	so the two blob ribbon operators do indeed cancel out. We can perform this cancellation of blob ribbon operators for each plaquette that is shared by the direct membrane of both membrane operators. Because all of the plaquettes in the direct membrane of $m$ are also in the direct membrane of $c$, this just leaves the blob ribbon operators for the plaquettes that lie on the closed membrane $c$ but not the open one $m$. This means that the product of the two operators from Equation \ref{Equation_product_closed_open_magnetic_membranes} can be written as
	\begin{align}
		C^h_T(m) C^{h^{-1}}_T(c)&= C^h_{\rhd}(m) C^{h^{-1}}_{\rhd}(c) \bigg(\prod_{\substack{\text{plaquette }p \in c\\ \notin m}} B^{e_p|_{s.p} [h \rhd e_p^{-1}|_{s.p}]}(\text{blob }0 \rightarrow \text{blob }p) \bigg).
		\label{Equation_product_closed_open_magnetic_membranes_2}
	\end{align}

	Next we wish to combine the two $C^{h/h^{-1}}_{\rhd}$ operators. We already established that the action of the two membrane operators on the edges cut by both dual membranes cancels, leaving only the action of $C^{h^{-1}}_{\rhd}(c)$ on the edges cut by the dual membrane of $c$ but not $m$. Next we consider the $\rhd$ action of these operators on plaquettes cut by the dual membrane. As described in Section \ref{Section_Magnetic_Tri_Non_Trivial}, the magnetic membrane operator $C^h_{\rhd}(m)$ on a plaquette $q$ cut by the dual membrane is
	$$C^h_{\rhd}(m):e_p = \begin{cases} g(s.p(m)-v_0(p))^{-1}hg(s.p(m)-v_0(p)) \rhd e_p & \text{if $v_0(p)$ is on the direct membrane} \\ e_p & \text{otherwise.} \end{cases} $$
	The action of $C^{h^{-1}}_{\rhd}(c)$ on the same plaquette is therefore
	$$C^{h^{-1}}_{\rhd}(c):e_p =\begin{cases} g(s.p(m)-v_0(p))^{-1}h^{-1}g(s.p(m)-v_0(p)) \rhd e_p & \text{if $v_0(p)$ is on the direct membrane} \\ e_p & \text{otherwise,} \end{cases}$$
	where we used the fact that the start-points of $c$ and $m$ are the same. We can see that the actions of the two membrane operators are inverses of each-other, so they cancel. This means that the action of the two $C_{\rhd}$ operators cancels on both the edges and plaquettes in the region shared by the two membranes. Therefore, Equation \ref{Equation_product_closed_open_magnetic_membranes_2} becomes
	\begin{align*}
		C^h_T(m) C^{h^{-1}}_T(c)&=C^{h^{-1}}_{\rhd}(\text{unshared membrane}) \bigg( \prod_{\text{unshared plaquettes}} B^{e_p (h \rhd e_p^{-1})}(t_p) \bigg)\\
		&=C^{h^{-1}}_T(\text{unshared membrane}).
	\end{align*}

\begin{figure}[h]
		\begin{center}
			\begin{overpic}[width=0.75\linewidth]{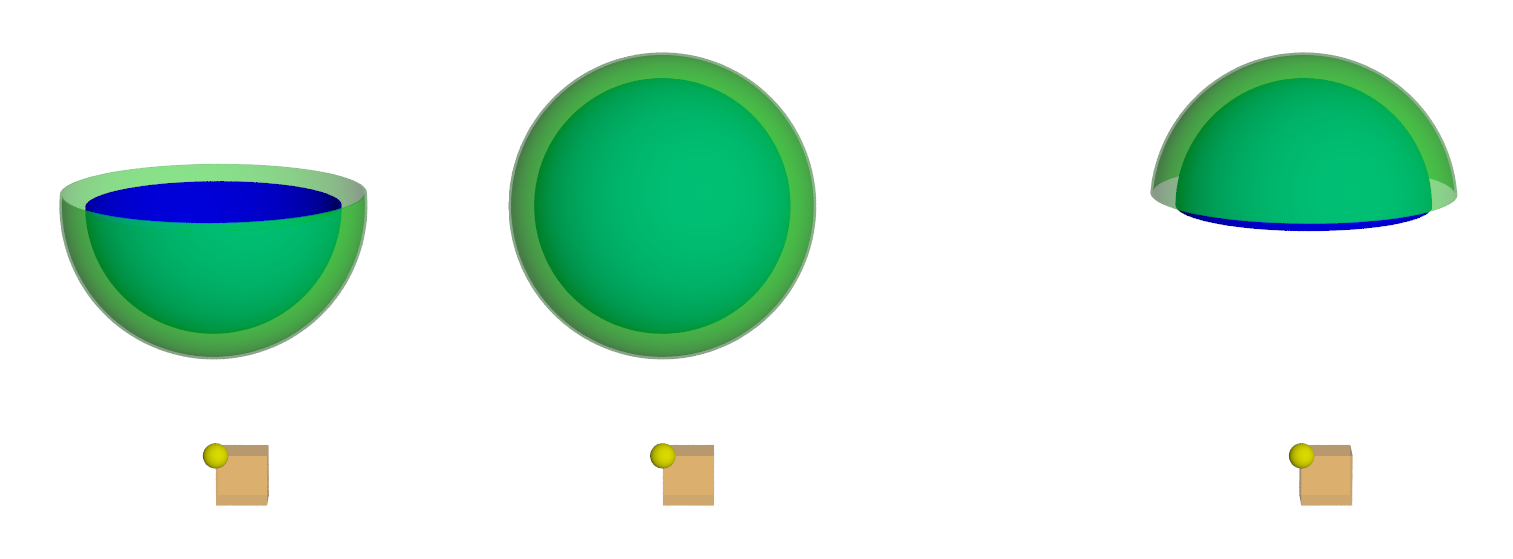}
				\put(28,22){\Huge $\cdot$}
				\put(63,22){\Huge $=$}
				\put(2,27){$C^h_T(m)$}
				\put(29,33){$C^{h^{-1}}_T(c)$}
				\put(70,33){$C^{h^{-1}}_T(r)$}
			\end{overpic}
			\caption{By combining an open membrane operator of label $h$ with a closed one of label $h^{-1}$, we can produce an open membrane operator of label $h^{-1}$ with the opposite orientation in a new position. The start-point (yellow sphere) and blob 0 (orange cube) of each membrane are in the same location. If we repeat the process, we can get back to an open membrane operator of label $h$ in a new position: we can deform our original membrane operator. In this figure, the dual membranes of the magnetic membrane operators are shown in green while the direct membranes are shown in blue. }
			\label{Chtopologicalfinal}
		\end{center}
	\end{figure}

	The result is a membrane operator of label $h^{-1}$ acting on the parts of $c$ not in $m$, as shown in Figure \ref{Chtopologicalfinal}. We denote this membrane by $r$. Note that the start-point and blob 0 do not need to be near the direct membrane of the membrane operator and even if they are initially we can deform the membrane away from these points. The membrane operator resulting from combining the open and closed membrane operators has the opposite orientation to the operator applied on $m$ and also the inverse label, indicating that we can reverse the orientation of a magnetic membrane operator by inverting its label. We can repeat this process, adding another membrane operator on a closed membrane combining $r$ and the desired final membrane, to again invert the orientation and label of the membrane operator and put the membrane into its final position, just as we did in the $\rhd$ trivial case in Section \ref{Section_Topological_Magnetic_Tri_Trivial}.
	
	\section{Condensation of magnetic excitations from the membrane operator picture}
	\label{Section_magnetic_condensation}
	In this section we will prove the claims we made in Sections \ref{Section_3D_Condensation_Confinement_Tri_Trivial} and \ref{Section_condensation_confinement_partial_central} of the main text, that some of the magnetic excitations are condensed. To do this, we will show that the corresponding membrane operators act like local operators on the ground state. Note that some of the $E$-valued loop excitations are also condensed, but we gave an argument to show this in Section \ref{Section_3D_Condensation_Confinement_Tri_Trivial}, so we will not repeat the argument here.

	\subsection{The case where $\rhd$ is trivial}
	\label{Section_magnetic_condensed_tri_trivial}
	We start with the case where $\rhd$ is trivial and wish to show that the magnetic excitations with label in $\partial(E)$ are condensed, as we claimed in Section \ref{Section_3D_Condensation_Confinement_Tri_Trivial}. We will follow a similar approach to the one we used for the 2+1d case in Ref. \cite{HuxfordPaper2} (see Section S-III). That is, we will demonstrate that any magnetic membrane operator with label in $\partial(E)$ has the same action as a series of edge transforms near the membrane, combined with a blob ribbon operator around the boundary of the membrane, at least when the region near the membrane is initially unexcited. These edge transforms can be absorbed into the state if the edges are initially unexcited, so the magnetic membrane operator is then equivalent to a blob ribbon operator applied around the boundary of the membrane. This ribbon operator is local to the excitation produced by the membrane operator (which again lies around the boundary of the membrane), so the magnetic excitation is condensed.

	The first step in showing this is to construct the relevant series of edge transforms. Whereas in the 2+1d case we applied an edge transform on each edge cut by the dual path of the ribbon operator, in 3+1d we perform an edge transform on each edge cut by the dual membrane of the magnetic membrane operator. Specifically, for each edge cut by the dual membrane, we will apply an edge transform with group label $e$ if the edge points away from the direct membrane and label $e^{-1}$ if the edge points towards the direct membrane. Let us compare the action of this series of edge transforms with that of a magnetic membrane operator with label $\partial(e)$, starting with the action on the edge labels. The effect of the series of edge transforms on an edge $i$ cut by the dual membrane is to take the edge label from $g_i$ to $\partial(e)g_i$ if the edge points away from the direct membrane and $\partial(e)^{-1}g_i$ if the edge points towards the direct membrane. By comparison, the action of the magnetic membrane on the edge is (using Equation \ref{Equation_magnetic_membrane_on_edges_appendix})
	\begin{align*}
		C^{\partial(e)}(m):g_i &= \begin{cases} g(s.p-v_i)^{-1}\partial(e)g(s.p-v_i) g_i & \text{ if $i$ points away from the direct membrane}\\
			g_i g(s.p-v_i)^{-1}\partial(e)^{-1}g(s.p-v_i) & \text{ if $i$ points towards from the direct membrane}\end{cases}\\
		&= \begin{cases} \partial(e) g_i & \text{ if $i$ points away from the direct membrane}\\
			\partial(e)^{-1}g_i & \text{ if $i$ points towards from the direct membrane,}\end{cases}\\
	\end{align*}
	where in the last line we used the fact that $\partial(e)$ belongs to the centre of $G$ when $\rhd$ is trivial (for any $e$). We therefore see that the action of the series of edge transforms on the edges matches the action of the magnetic membrane operator on those edges.

\begin{figure}[h]
		\begin{center}
			\begin{overpic}[width=0.6\linewidth]{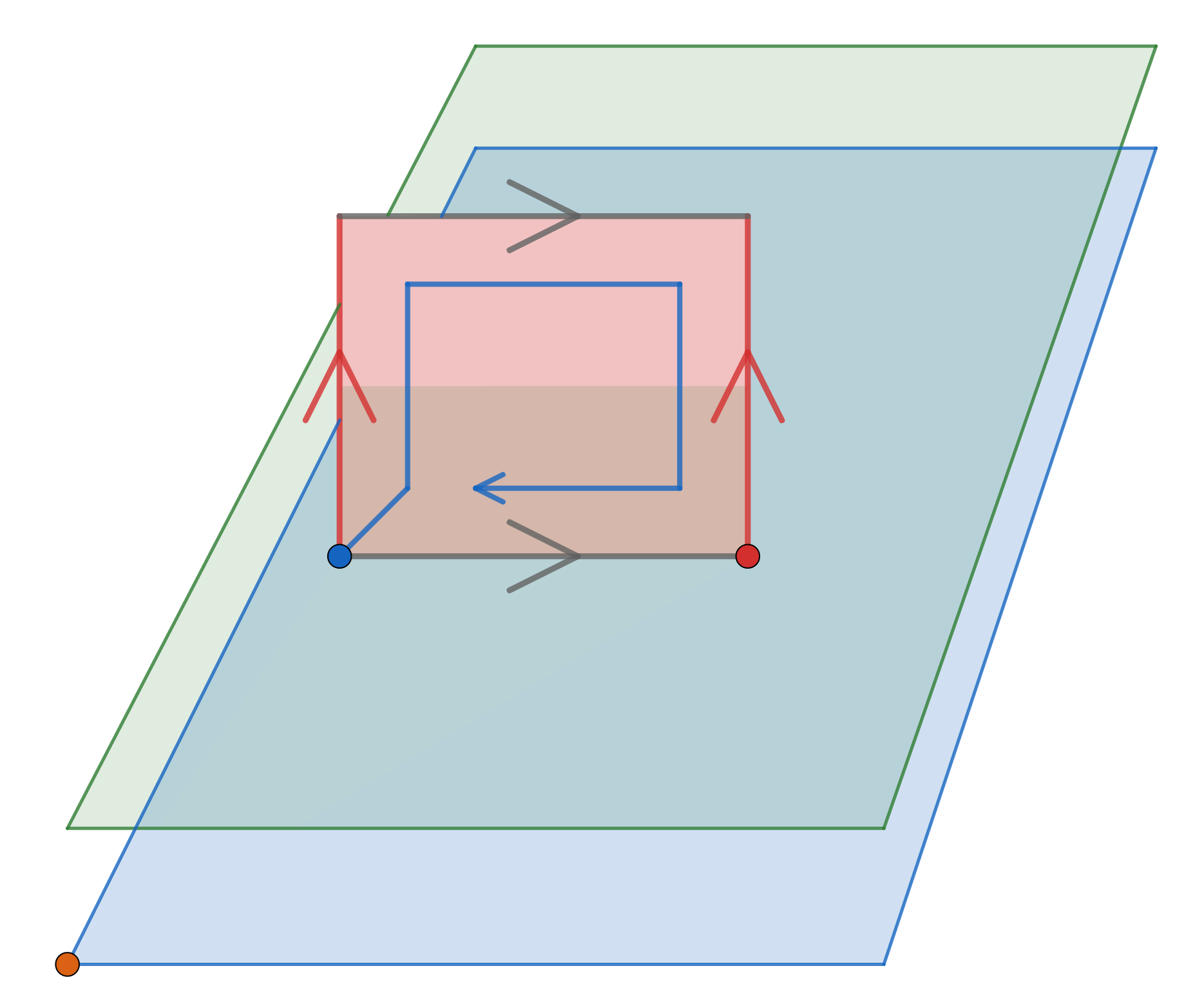}
				\put(0,1){$s.p$}
				\put(22,36){$s(i)$}
				\put(26,55){$i$}
				\put(64,36){$s(j)$}
				\put(63,55){$j$}
				\put(45,50){$p$}
				
				\put(74,14){dual membrane}
				\put(74,2){direct membrane}
			\end{overpic}
			\caption{Consider a plaquette $p$, which is cut through by the bulk of the dual membrane (green). This plaquette is adjacent to two edges, $i$ and $j$, which are cut by the dual membrane. In order to reproduce the magnetic membrane operator, we apply edge transforms on each of the edges cut by the dual membrane. Therefore, the label of $p$ is affected by two edge transforms and the contribution from these two edge transforms cancel.}
			\label{magnetic_membrane_condensed_tri_trivial_internal_plaquette}
		\end{center}
	\end{figure}

	Next we consider the action of these edge transforms on the plaquette labels. We distinguish between two types of plaquette. There are ``internal" plaquettes, which are cut by the bulk of the dual membrane, and boundary plaquettes, which are cut by the boundary of the dual membrane. Because the internal plaquettes are cut by the bulk of the membrane, the membrane bisects the entire plaquette and so two of the edges on such a plaquette are cut by the dual membrane. This means that the plaquette label is affected by two of the edge transforms from our series of transforms. The effects of these two edge transforms on the plaquette label cancel. To see this, consider the plaquette shown in Figure \ref{magnetic_membrane_condensed_tri_trivial_internal_plaquette}. This plaquette is adjacent to two edges, $i$ and $j$, which are cut by the dual membrane. The action of an edge transform on an adjacent plaquette is (from Equation \ref{Equation_edge_transform_definition} in the main text)
	\begin{align}
		\mathcal{A}_i^e : e_p &\rightarrow \begin{cases} e_p e^{-1} &\text{if $i$ is aligned with $p$}\\
			e e_p &\text{if $i$ is aligned against $p$.} \label{Equation_edge_transform_tri_trivial} \end{cases} 
	\end{align}

	Therefore, the edge transform $\mathcal{A}_i^e$ on edge $i$ contributes a factor $e^{-1}$ to the plaquette label while the transform $\mathcal{A}_j^e$ contributes a factor of $e$, which cancels. While we have chosen a particular orientation for these two edges, this result will also hold if we reversed either edge (and so holds for a generic internal plaquette). This is because reversing an edge inverts the contribution of the edge transform to the plaquette (see Equation \ref{Equation_edge_transform_tri_trivial}), but we would also apply an edge transform with the inverse label on that edge, by our definition for the series of edge transforms. These two inverses cancel, and so the action of the edge transform does not depend on the orientation of edge $i$ or $j$. Similarly, if we reversed the orientation of the plaquette, then the contribution from each of the two edges would be inverted (so that the transform on edge $i$ would contribute $e$ to the plaquette label and the transform on edge $j$ would contribute $e^{-1}$), but the two factors from the different edges would still cancel, indicating that the action of the series of edge transforms on the internal plaquette is also independent of the orientation of the plaquette.

\begin{figure}[h]
		\begin{center}
			\begin{overpic}[width=0.6\linewidth]{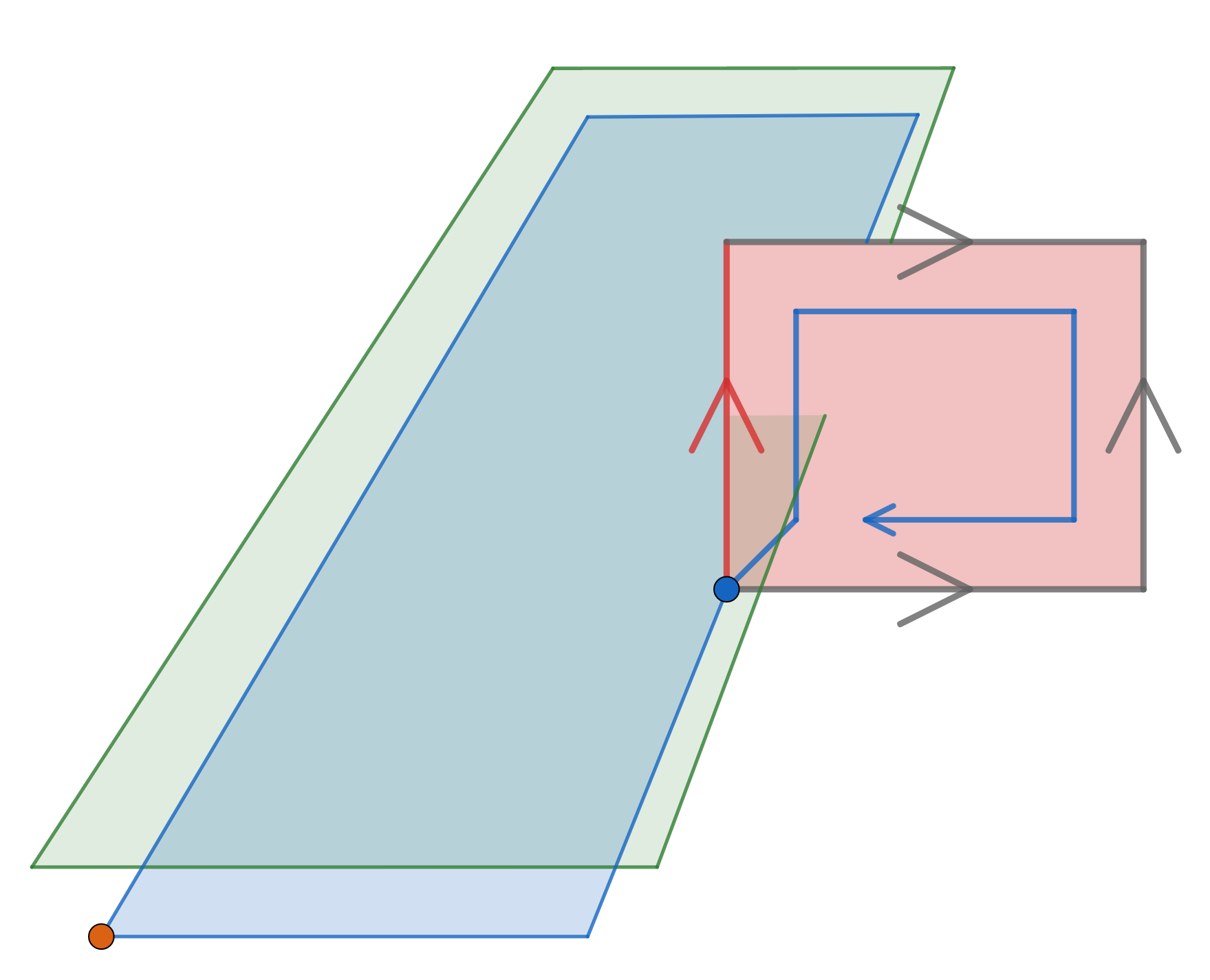}
				\put(3,1){$s.p$}
				\put(54,31){$s(i)$}
				\put(58,50){$i$}
				
				\put(75,46){$p$}
				
				\put(56,10){dual membrane}
				\put(50,3){direct membrane}
			\end{overpic}
			\caption{Consider a plaquette cut by the boundary of the dual membrane. Such a plaquette is only affected by one of the edge transforms that we apply in order to reproduce the action of the magnetic membrane operator. Therefore, there is no cancellation of different transforms. In this example, if the plaquette is initially labelled by $e_p$ then we apply the edge transform $\mathcal{A}_i^e$ which takes the label to $e_p e^{-1}$. If the orientation of the edge were flipped, we would apply the transform $\mathcal{A}_i^{e^{-1}}$, which would again take the plaquette label to $e^{-1}e_p=e_pe^{-1}$. On the other hand, if we flipped the plaquette instead, we would still apply $\mathcal{A}_i^e$, but this would result in the plaquette label becoming $ee_p$. We see therefore that the action on the plaquette only depends on the orientation of the plaquette, not edge $i$.}
			\label{magnetic_membrane_condensed_tri_trivial_boundary_plaquette}
		\end{center}
	\end{figure}

\begin{figure}[h]
		\begin{center}
			\begin{overpic}[width=0.5\linewidth]{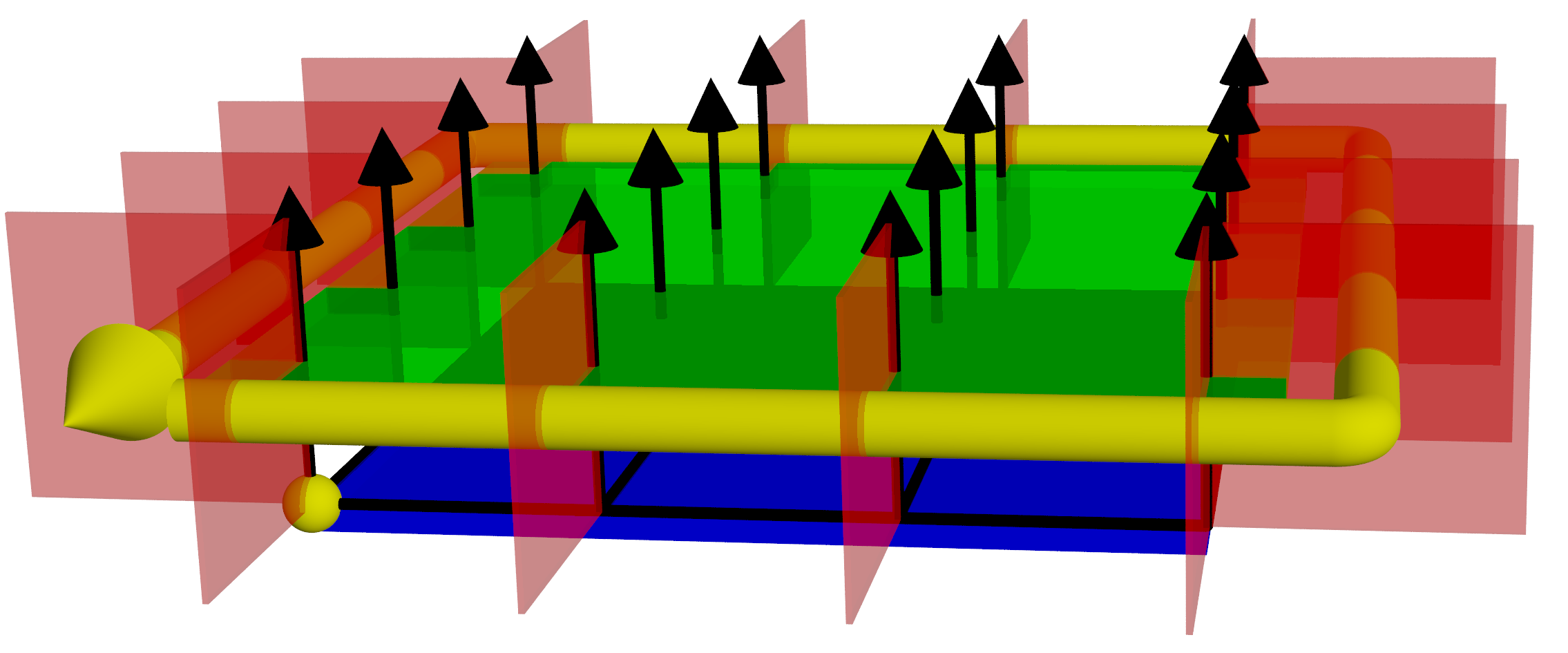}
				
			\end{overpic}
			\caption{We can reproduce the action of a magnetic membrane operator of label $\partial(e)$ in the bulk of the membrane with edge transforms on the edges (black arrows) cut by the dual membrane (the upper green surface). We find that the edge transforms reproduce the membrane operator up to the action of a (confined) blob ribbon operator of label $e$ (shown as a large yellow arrow) travelling anticlockwise around the membrane and passing through the boundary plaquettes (the red squares).}
			\label{magnetic_membrane_condensed_blob_ribbon}	
		\end{center}
	\end{figure}

	Now consider the boundary plaquettes, such as the one shown in Figure \ref{magnetic_membrane_condensed_tri_trivial_boundary_plaquette}. These plaquettes are only affected by one edge transform, because only one edge on each of these plaquettes is cut by the dual membrane. In the case shown in Figure \ref{magnetic_membrane_condensed_tri_trivial_boundary_plaquette}, we apply an edge transform $\mathcal{A}_i^e$ on the edge $i$ cut by the dual membrane, which will multiply the plaquette label by $e^{-1}$. If we were to reverse the orientation of the edge, this would not affect the contribution from the edge transform, just as we discussed for the internal plaquettes. This is because the inverted action of an edge transform under reversing the edge is countered by the fact that we would apply an edge transform of inverted label. However, unlike for the internal edges, reversing the orientation of the plaquette does matter. This is because flipping the orientation of the plaquette changes the contribution of the edge transform from $e^{-1}$ to $e$, according to Equation \ref{Equation_edge_transform_tri_trivial}. In the case of the internal plaquettes, there were two edge transforms and the contribution from each was inverted, which still resulted in cancellation. Here, however, there is only one edge transform. Putting this together, we see that whether the plaquette is multiplied by $e$ or $e^{-1}$ depends on the orientation of the plaquette and not the orientation of the edge. If the plaquette circulates up through the dual membrane, as in the example in Figure \ref{magnetic_membrane_condensed_tri_trivial_boundary_plaquette}, then the plaquette label is multiplied by $e^{-1}$ and if it circulates down through the dual membrane it is multiplied by $e$. Another way to say this is that, if we take the anticlockwise boundary of the membrane and place our right hand with thumb pointing along a section of the the boundary and fingers curled, then any boundary plaquette pierced by this section of boundary whose circulation matches our fingers is multiplied by $e^{-1}$ and any such plaquette with opposite orientation is multiplied by $e$. This is equivalent to the action of a blob ribbon operator of label $e$ travelling anticlockwise around the membrane, as shown in Figure \ref{magnetic_membrane_condensed_blob_ribbon}. We saw something very similar when we considered the topological nature of blob ribbon operators in Section \ref{Section_Topological_Blob_Ribbons}.

	Recall that we applied this series of edge transforms to reproduce the action of the magnetic membrane operator labelled by $\partial(e)$. The magnetic membrane operator does not act on any plaquette labels when $\rhd$ is trivial, so this action on the boundary plaquettes is not part of the magnetic membrane operator. To reproduce the action of the magnetic membrane operator, we must cancel this additional blob ribbon operator by applying a blob ribbon of label $e^{-1}$ around the boundary of the membrane in addition to the series of edge transforms. Then the edge transforms reproduce the action of the magnetic membrane operator on the edges and the (trivial) action of the membrane operator on the internal plaquettes, while the blob ribbon operator removes the unwanted action of the edge transforms on the boundary plaquettes. Applying this blob ribbon operator along with the edge transforms therefore recovers the effect of the magnetic membrane operator. Then because the edge transforms act trivially on a state for which the edges are unexcited, we find that the action of the magnetic membrane operator with label $\partial(e)$ on the ground state is equivalent to the action of a closed blob ribbon operator with label $e^{-1}$ around the outside of the membrane. This blob ribbon operator is local in the sense that it only has support near the boundary of the membrane (i.e., near the excitation), rather than across the bulk, so the magnetic membrane operator is condensed.

	\subsection{The case where $\partial \rightarrow$ centre$(G)$ and $E$ is Abelian}
	\label{Section_condensation_magnetic_centre_case}
	Next we consider the case where $\partial$ maps onto the centre of $G$ and $E$ is Abelian, but $\rhd$ may be non-trivial. We find that, just as in the $\rhd$ trivial case, the magnetic excitations labelled by elements in $\partial(E)$ are condensed. To see this, we will follow a similar procedure to the $\rhd$ trivial case considered in Section \ref{Section_magnetic_condensed_tri_trivial}. We will show that a series of edge transforms applied on the edges cut by the dual membrane of a membrane operator reproduce the action of the membrane operator in the bulk, with the difference in action along the boundary being equivalent to a blob ribbon operator. Consider a magnetic membrane operator $C^{\partial(e)}_T(m)$ applied on a membrane $m$. Because the label is in the image of $\partial$, the action of this operator is much simpler than that of other magnetic membrane operators. As stated in Equation \ref{Equation_total_magnetic_membrane_definition_1}, a magnetic membrane operator $C^h_T(m)$ is a product of an operator $C^h_{\rhd}(m)$ and a series of blob ribbon operators with labels of the form $ d [h^{-1} \rhd d^{-1}]$, for some (configuration dependent) $d \in E$. In the case of $C^{\partial(e)}_T(m)$, the blob ribbon operators have the form $d[\partial(e)^{-1} \rhd d^{-1}]$. From the second Peiffer condition, Equation \ref{Equation_Peiffer_2} in the main text, we know that $\partial(e)^{-1} \rhd d^{-1}= e^{-1}d^{-1}e$, which is equal to $d^{-1}$ because $E$ is Abelian. Therefore, $d[\partial(e)^{-1} \rhd d^{-1}]=dd^{-1}=1_E$, so the labels of the blob ribbon operators are trivial (meaning that the series of ribbon operators is just the identity operator).

	Next consider $C^{\partial(e)}_{\rhd}(m)$, which acts on both the edges and the plaquettes that are cut by the dual membrane. The action on the plaquettes is given by
	\begin{equation*}
		C^{\partial(e)}_{\rhd}(m): e_p = \begin{cases} (g(s.p-v_0(p))^{-1}\partial(e)g(s.p-v_0(p))) \rhd e_p & \text{ if $v_0(p)$ lies on the direct membrane} \\ e_p & \text{ otherwise,} \end{cases} 
	\end{equation*}
	from Equation \ref{Equation_magnetic_membrane_rhd_action_appendix}. Because $\partial(e)$ is in the centre of $G$, $g(s.p-v_0(p))^{-1}\partial(e)g(s.p-v_0(p))= \partial(e)$. Furthermore $\partial(e) \rhd e_p = ee_pe^{-1}$ from the second Peiffer condition (Equation \ref{Equation_Peiffer_2} in the main text), which means that $\partial(e) \rhd e_p = e_p$ when $E$ is Abelian. This means that the action of $C^{\partial(e)}_{\rhd}(m)$ on the plaquettes is given by
	\begin{align*}
		C^{\partial(e)}_{\rhd}(m): e_p &= \begin{cases} (g(s.p-v_0(p))^{-1}\partial(e)g(s.p-v_0(p))) \rhd e_p & \text{ if $v_0(p)$ lies on the direct membrane} \\ e_p & \text{ otherwise} \end{cases} \\
		&=\begin{cases} \partial(e) \rhd e_p & \text{ if $v_0(p)$ lies on the direct membrane} \\ e_p & \text{ otherwise,} \end{cases} \\
		&=e_p,
	\end{align*}
	which is trivial. Together with the fact that the blob ribbon operators in $C^{\partial(e)}_T(m)$ are trivial, this means that the membrane operator does not affect the plaquette labels at all. Furthermore, because $\partial$ maps onto the centre of $G$, the action of the operator on the edges is simplified. The action of the membrane operator $C^{\partial(e)}_T(m)$ on an edge $i$ cut by the dual membrane is (from \ref{Equation_magnetic_membrane_on_edges_appendix})
	\begin{equation*}
		C^{\partial(e)}_T(m):g_i = \begin{cases} g(s.p-v_i)^{-1}\partial(e)g(s.p-v_i)g_i & \text{if $i$ points away from the direct membrane} \\ g_ig(s.p-v_i)^{-1}\partial(e)^{-1}g(s.p-v_i) & \text{if $i$ points towards the direct membrane.} \end{cases} 
	\end{equation*}
	Because $\partial(e)$ is in the centre of $G$, this simply becomes
	\begin{equation*}
		C^{\partial(e)}_T(m):g_i = \begin{cases} \partial(e)g_i & \text{if $i$ points away from the direct membrane} \\ g_i\partial(e)^{-1} & \text{if $i$ points towards the direct membrane,} \end{cases} 
	\end{equation*}
	where even the order of multiplication does not matter.

	Having discussed the action of the magnetic membrane operator, we now wish to show that it can be reproduced in the bulk of the membrane by a series of edge transforms. We consider applying, on each edge $i$ cut by the dual membrane, an edge transform of label $g(s.p-s(i))^{-1} \rhd e$ if the edge points away from the direct membrane and $g(s.p-s(i))^{-1} \rhd e^{-1}$ if the edge points towards the direct membrane, where $s(i)$ is the source of edge $i$ and $s.p$ is the start-point of the membrane operator we wish to reproduce. Note that this series of edge transforms is exactly the same as the series we applied in Section \ref{Section_Topological_Blob_Ribbons} when showing that non-confined blob ribbon operators are topological, the only difference being that in that case we restricted $e$ to be in the kernel of $\partial$. We can therefore think of this calculation as a generalization of that one, where if $e$ is in the kernel we produce a magnetic membrane operator of label $1_G$ (i.e., a trivial operator) and a closed non-confined ribbon operator, thus demonstrating the trivial nature of the closed non-confined ribbon operator. On the other hand, we will see that if $e$ is not in the kernel the series of edge transforms produce a non-trivial membrane operator and a closed confined blob ribbon operator.

	The effect of the edge transforms on an edge $i$ that is cut by the dual membrane and points away from the direct membrane is to take its label from $g_i$ to $\partial(g(s.p-s(i))^{-1} \rhd e)g_i$, because such an edge is affected by an edge transform of label $g(s.p-s(i))^{-1} \rhd e$. Then, using the first Peiffer condition (Equation \ref{Equation_Peiffer_1} in the main text), we see that
	$$\partial(g(s.p-s(i))^{-1} \rhd e)g_i= g(s.p-s(i))^{-1}\partial(e)g(s.p-s(i))g_i.$$
	Because $\partial(e)$ is in the centre of $G$, $g(s.p-s(i))^{-1}\partial(e)g(s.p-s(i))g_i$ is equal to $\partial(e)g_i$. Therefore, the action of the series of edge transforms on the edge $i$ is to take the edge label from $g_i$ to $\partial(e)g_i$, which matches the action of the magnetic membrane operator. Similarly, if the edge points towards the direct membrane then it takes the label to $\partial(e)^{-1}g_i =g_i \partial(e)^{-1}$, which again matches the action of the magnetic membrane operator. Next we will show that the action of the edge transforms on the plaquettes that are away from the boundary of the membrane also matches that of the magnetic membrane (namely, they must act trivially on these plaquettes).

\begin{figure}[h]
		\begin{center}
			\begin{overpic}[width=0.75\linewidth]{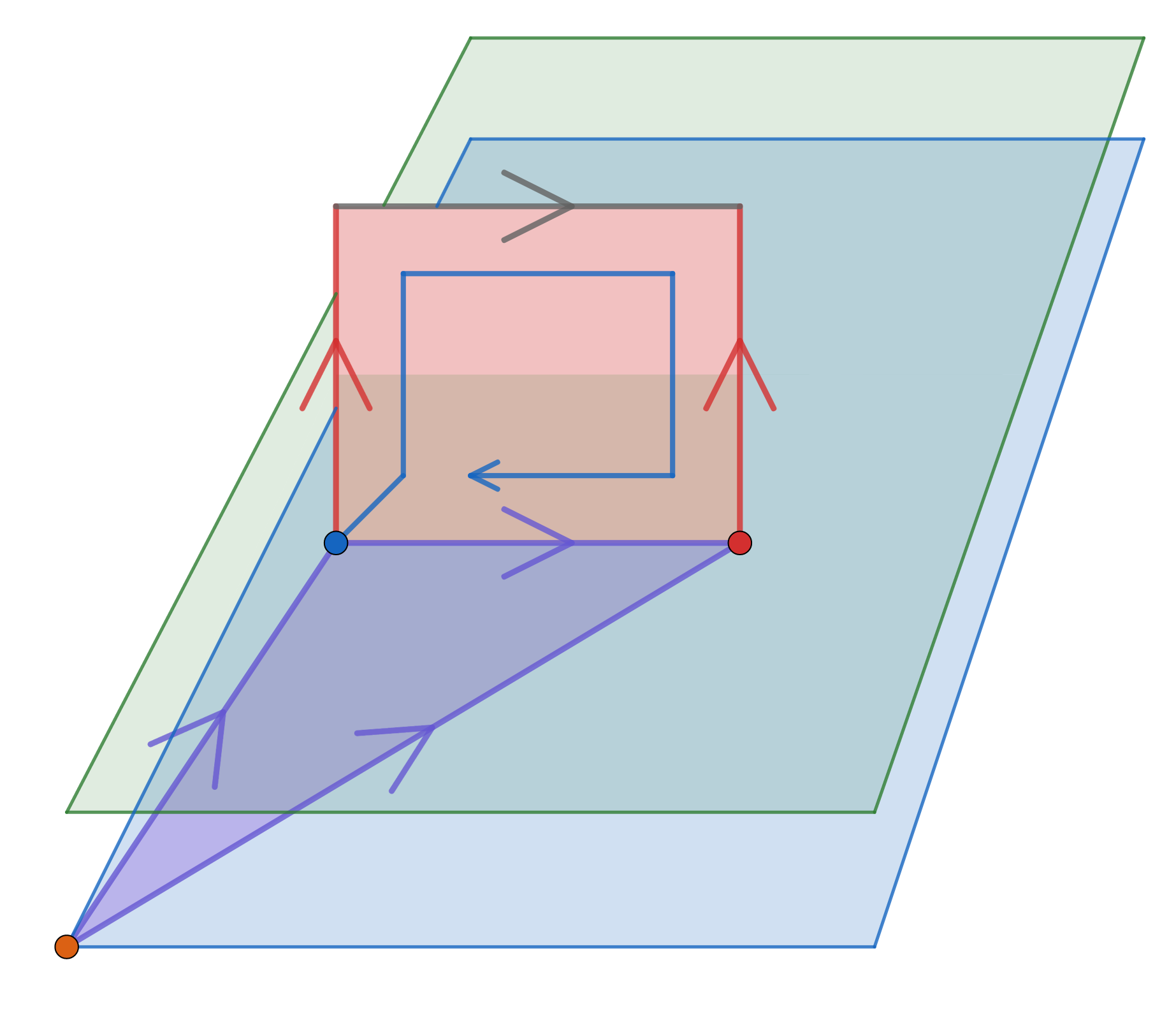}
				\put(0,5){$s.p$}
				\put(13,40){$v_0(p)=s(i)$}
				\put(27.2,58){$i$}
				\put(64,40){$s(j)$}
				\put(64,58){$j$}
				\put(45,54){$p$}
				\put(9,28){$s.p-s(i)$}
				\put(45,28){$s.p-s(j)$}
				\put(32,34){$\overline{v_0(p)-s(j)}$}
				\put(76,18){dual membrane}
				\put(76,6){direct membrane}
			\end{overpic}
			\caption{We consider the effect of the series of edge transforms on a plaquette cut by the bulk of the dual membrane. The plaquette $p$ is affected by two edge transforms, those on edges $i$ and $j$. We find that the effects of these two edge transforms cancel, provided that the (purple) shaded triangular region satisfies fake-flatness. Note that in this example we chose the base-point of the plaquette to be the source of $i$ and chose particular orientations for the edges and the plaquettes. These details do not affect our conclusions however.}
			\label{magnetic_membrane_condensed_tri_nontrivial_internal_plaquette}
		\end{center}
	\end{figure}

	Consider a plaquette that is cut by the bulk of the dual membrane of the magnetic membrane operator that we wish to reproduce. Such a plaquette $p$, like the example shown in Figure \ref{magnetic_membrane_condensed_tri_nontrivial_internal_plaquette}, is affected by two edge transforms in the series of operators that we use to reproduce the membrane operator. Let these two edges be $i$ and $j$, with sources $s(i)$ and $s(j)$, where we initially assume that the two edges point away from the direct membrane. Then the label $e_p$ of plaquette $p$ transforms as
	\begin{align*}
		e_p &\rightarrow [g(\overline{v_0(p)-s(j)}) \rhd (g(s.p-s(j))^{-1} \rhd e)] \: e_p \: [g(v_0(p)-s(i)) \rhd (g(s.p-s(i))^{-1} \rhd e^{-1})]\\
		&=[(g(\overline{v_0(p)-s(j)}) g(s.p-s(j))^{-1}) \rhd e] \: e_p \: [(g(v_0(p)-s(i))g(s.p-s(i))^{-1}) \rhd e^{-1}].
	\end{align*}
	
	The two path elements $g(\overline{v_0(p)-s(j)}) g(s.p-s(j))^{-1}$ and $g(v_0(p)-s(i))g(s.p-s(i))^{-1}$ both correspond to paths from $v_0(p)$ to the start-point $s.p$. Provided that these two paths enclose a fake-flat region between them (such as the purple region in Figure \ref{magnetic_membrane_condensed_tri_nontrivial_internal_plaquette}), the two path elements will differ only by an element in $\partial(E)$. This element will not affect expressions such as $(g(\overline{v_0(p)-s(j)}) g(s.p-s(j))^{-1}) \rhd e$, because the second Peiffer condition implies that $\partial(f) \rhd e = fef^{-1}$ for any pair $f,$ $e\in E$, and $E$ is Abelian so $\partial(f) \rhd e=e$. That is, factors in $\partial(E)$ act trivially via $\rhd$. Then we can replace both $g(\overline{v_0(p)-s(j)}) g(s.p-s(j))^{-1}$ and $g(v_0(p)-s(i))g(s.p-s(i))^{-1}$ with a path element $g(v_0(p)-s.p)$, without needing to define the precise position of this path. This gives us
	\begin{align*}
		e_p &\rightarrow[g(v_0(p)-s.p) \rhd e] \: [g(v_0(p)-s.p) \rhd e^{-1}] e_p\\
		&=e_p.
	\end{align*}
	
	We see that the contribution from the transform on edge $i$ to the plaquette label is $[g(v_0(p)-s.p) \rhd e^{-1}]$, while the contribution from edge $j$ is $[g(v_0(p)-s.p) \rhd e]$, which cancel. In doing this calculation, we assumed that the two edges point away from the direct membrane, but flipping an edge introduces one inverse from the action of the edge transform and one from our choice of edge transform in the series of transforms, which we defined to depend on the orientation of the edge. These two inverses cancel, so our conclusion that the transforms cancel holds regardless of the orientations of the edges. Similarly, if we were to reverse the orientation of the plaquette, we would swap the roles of edge $i$ and $j$, so $i$ contributes $[g(v_0(p)-s.p) \rhd e]$ and $j$ contributes the inverse, but this would not affect the result. This indicates that, much like the membrane operator that we wish to reproduce, the series of edge transforms has no effect on the plaquettes cut by the bulk of the dual membrane, regardless of the orientations of the edges and plaquettes.

	We showed earlier that the edge transforms reproduce the action of the magnetic membrane operator on the edges. Now that we have demonstrated that the edge transforms also reproduce the (trivial) action of the magnetic membrane operator on plaquettes in the bulk of the membrane. This means that the edge transforms can only differ in their action on the plaquettes cut by the boundary of the dual membrane. Each boundary plaquette is affected by only one edge transform, so the label of a plaquette $p$ gains only one of the factors of $g(v_0(p)-s.p) \rhd e$ or the inverse that we saw for the internal plaquettes. Just as in the $\rhd$ trivial case we considered in Section \ref{Section_magnetic_condensed_tri_trivial}, this factor depends on the orientation of the plaquette, but not on the orientation of the edge on which we apply the transform. For example, if we consider the plaquette shown in Figure \ref{magnetic_membrane_condensed_tri_trivial_boundary_plaquette}, the edge transform $\mathcal{A}_i^{g(s.p-s(i))^{-1} \rhd e}$ on edge $i$ would cause the plaquette label to gain a factor of $g(v_0(p)-s.p) \rhd e^{-1}$. If we reversed the edge label, we would apply the inverse edge transform $\mathcal{A}_i^{g(s.p-s(i))^{-1} \rhd e^{-1}}$, but the action of this edge transform would also be inverted due to the relative orientation of edge and plaquette. As we discussed previously for the internal plaquettes, the two inverses would cancel. On the other hand, if we invert the orientation of the plaquette, we invert the contribution of the edge transform to the plaquette without switching which edge transform we use, so the transform would instead cause the plaquette label to gain a factor of $g(v_0(p)-s.p) \rhd e$. Just as in the $\rhd$ trivial case, we can determine which factor the plaquette should gain by using the right-hand rule along the anticlockwise-directed boundary of the membrane. If the orientation of the plaquette matches the circulation from the right-hand rule, then the plaquette label gains a factor of $g(v_0(p)-s.p) \rhd e^{-1}= g(s.p-v_0(p))^{-1} \rhd e^{-1}$, and otherwise it gains the inverse factor. This action on the boundary plaquettes is the same as the action of a blob ribbon operator of label $e$, with start-point at the start-point of the membrane, travelling anticlockwise around the membrane (as shown in Figure \ref{magnetic_membrane_condensed_blob_ribbon}). Because the magnetic membrane operator has the same action as the edge transforms on the edge labels and the plaquette labels in the bulk, but the edge transforms have this additional action of a blob ribbon operator around the boundary of the membrane, we see that the magnetic membrane operator differs from the edge transforms only by an operator around the boundary of the membrane. Furthermore, these edge transforms act trivially on a state where the region of the membrane is unexcited. This is because we can write one of the edge transforms acting on such a state $\ket{\psi}$ as
	$$\mathcal{A}_i^{g(s.p-s(i))^{-1} \rhd e}\ket{\psi} = \mathcal{A}_i^{g(s.p-s(i))^{-1} \rhd e} \mathcal{A}_i \ket{\psi},$$
	using the fact that the state is an unexcited eigenstate of the edge term, and can then absorb the edge transform into the edge term. We might worry that this is not true because the edge transforms have operator labels, but as discussed in Section \ref{Section_Topological_Blob_Ribbons}, this is fine because the labels $g(s.p-s(i))^{-1} \rhd e$ of the edge transforms are not affected by the edge terms (recall that adding factors in $\partial(E)$ to a path element $g(t)$ does not affect expressions such as $g(t) \rhd e$).

	This means that the action of the membrane operator is equivalent to an operator (specifically a blob ribbon operator) that is local to the excitation produced by the membrane operator. This indicates that magnetic membrane operators with label in $\partial(E)$ are condensed.

	\section{Braiding statistics in 3+1d}
	\label{Section_braiding_supplement}
	In this section we will calculate the braiding relations of the excitations in the 3+1d model, in more detail than we considered them in the main text (Sections \ref{Section_3D_Braiding_Tri_Trivial}, \ref{Section_3D_Braiding_Fake_Flat} and \ref{Section_3D_Braiding_Central}). 
	
	\subsection{$\rhd$ trivial case}
	
	We start by considering the case where $\rhd$ is trivial, so that $g \rhd e=e$ for all elements $g \in G$ and $e \in E$. As described in Section \ref{Section_3D_Braiding_Tri_Trivial} of the main text, the non-trivial braiding relations in this case are between electric and magnetic excitations; between two magnetic excitations; and between the $E$-valued membrane and blob excitations. This is because the $E$-valued membrane and blob ribbon operators only act on the surface labels, whereas the electric ribbon and magnetic membrane operators only act on the edge labels. This means that the commutation relations between the ribbon or membrane operators acting on the surfaces and those acting on the edges are trivial and so the braiding relations are also trivial. The braiding between two electric excitations or two $E$-valued loop excitations is also trivial, because the corresponding ribbon or membrane operators are diagonal in the configuration basis (the basis where each edge and plaquette is labelled by an appropriate group element) and so the operators commute.
	
	\subsubsection{Braiding between electric and magnetic excitations}
	\label{Section_electric_magnetic_braiding_3D_tri_trivial}
	
	We look at the case where we pull an electric excitation up through a magnetic excitation and around to its initial position, as described in Section \ref{Section_Abelian_flux_charge_braiding} and illustrated in Figure \ref{chargethroughloop2} in the main text. In order to determine the result of this braiding, we compare the case where we first produce the magnetic excitation using a membrane operator and then produce and move the electric excitation with a ribbon operator to the case where we first move the electric excitation through the vacuum before producing the magnetic excitation. This means that we have to consider the commutation relation between an electric ribbon operator and magnetic membrane operator.

\begin{figure}[h]
		\begin{center}
			\begin{overpic}[width=0.95\linewidth]{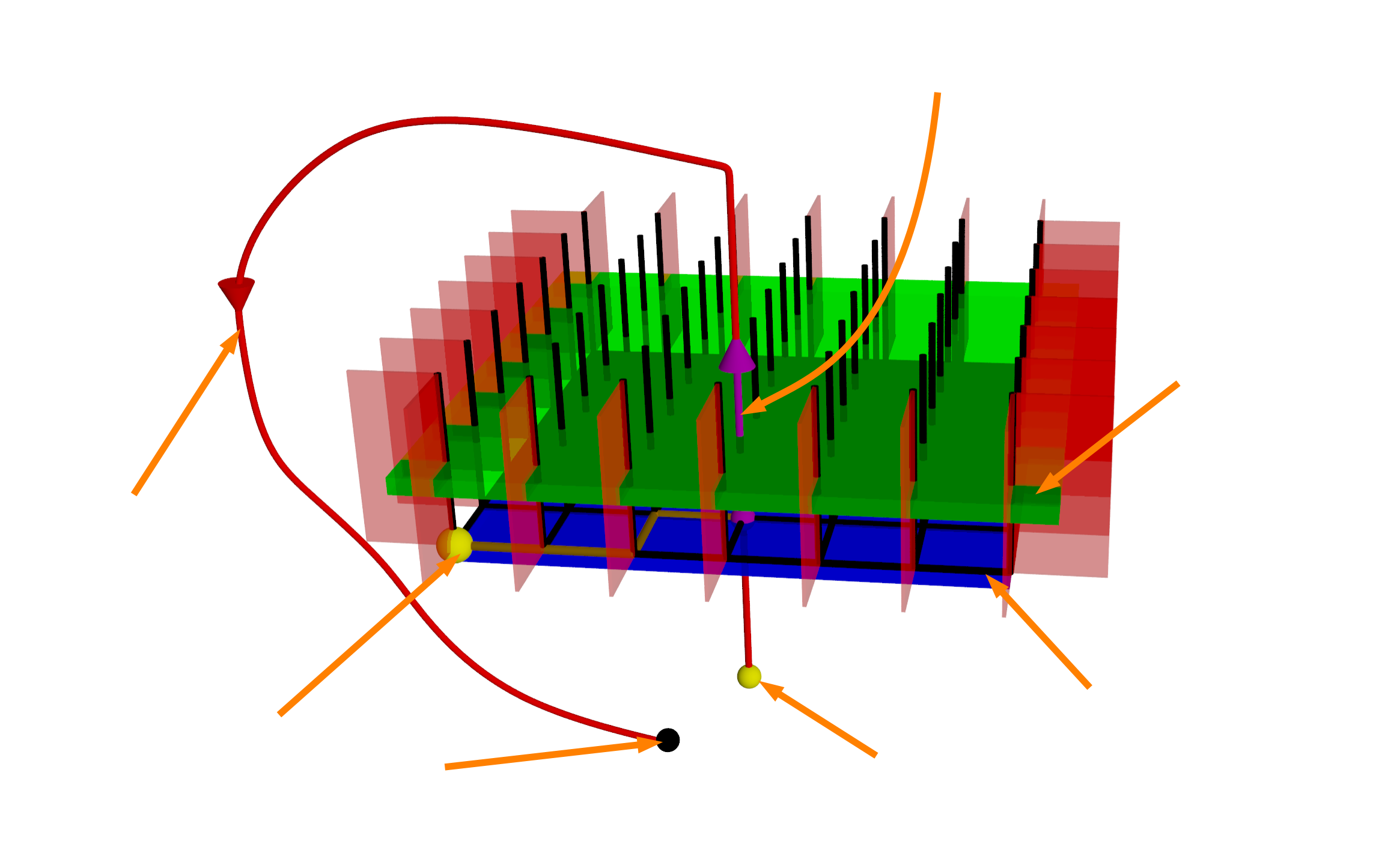}
				\put(8,24.5){\large $t$}
				\put(64,7){\large $s.p(t)$}
				\put(25.4,6){\large $e.p(t)$}
				\put(67.4,56.4){\large $i$}
				\put(86,35){\large dual membrane}
				\put(80,11){\large direct membrane}
				\put(13,9){\large $s.p(m)$}
			\end{overpic}
			\caption{We consider the case where an electric excitation (black sphere) is braided up through a magnetic excitation. The magnetic excitation is produced by the membrane operator $C^h(m)$ applied on the square membrane (the blue membrane is the direct membrane and the green membrane is the dual membrane, with the red squares being the plaquettes excited by the membrane operator). The electric excitation is moved by an electric ribbon operator applied on the path $t$. This path $t$ intersects the membrane on the purple edge, edge $i$. To determine how the electric ribbon operator interacts with the magnetic membrane operator, we must therefore consider the effect of the membrane operator on the edge $i$.}
			\label{electric_magnetic_braid_3D_appendix}
		\end{center}
	\end{figure}

	We consider the situation shown in Figure \ref{electric_magnetic_braid_3D_appendix}, where we have a magnetic membrane operator $C^h(m)$ applied on membrane $m$, and an electric ribbon operator on the path $t$. In addition to deciding the orientation of the braiding move (that is, whether we pull the electric excitation up through the loop or down through it), we must consider the orientation of the loop excitation itself (more precisely, it is the relative orientation of the flux tube and path of braiding that matters). Just as the magnetic flux particles in 2+1d have an orientation which determines whether their flux label corresponds to a clockwise or anticlockwise path around the particle, in 3+1d the flux tubes have an orientation. The orientation of the flux tube is determined by the magnetic membrane that produces it. For the example shown in Figure \ref{electric_magnetic_braid_3D_appendix}, we say that the orientation of the magnetic excitation is upwards. Therefore, we say that the orientation of the flux matches the orientation of the path. We will later consider the case where the path is reversed, which describes the situation where the particle braids against the orientation of the magnetic flux tube.

	We now consider the state $\delta(g, \hat{g}(t))C^h(m) \ket{GS}$, corresponding to the state where we first produce the magnetic flux tube and then braid the particle through it. We wish to commute the electric ribbon operator to the right to compare this to the case where we first move the electric excitation through the vacuum before producing the loop, thereby obtaining the braiding relation. In order to commute the electric operator $\delta(g, \hat{g}(t))$ to the right, we must know how the magnetic membrane operator affects the path element associated to $t$. The path $t$ intersects with the membrane along the edge $i$, shown in purple in Figure \ref{electric_magnetic_braid_3D_appendix}, and it is this edge that is affected by the magnetic membrane operator. The action of the magnetic membrane operator on an edge $i$ cut by the dual membrane is
	
	$$C^h(m):g_i = \begin{cases} g(s.p-v_i)^{-1}hg(s.p-v_i)g_i & \text{ if $i$ points away from the direct membrane} \\ g_i g(s.p-v_i)^{-1}h^{-1}g(s.p-v_i) & \text{ if $i$ points towards the direct membrane,} \end{cases}$$
	
	where $v_i$ is the vertex attached to edge $i$ and on the direct membrane of $m$. We split the path $t$ into three parts, corresponding to the sections of path before edge $i$ ($t_1$), after edge $i$ ($t_2$) and edge $i$ itself. Then $g(t)=g(t_1)g_i^{\sigma_i}g(t_2)$, where $\sigma_i$ is $+1$ if the edge $i$ points along the path (as in Figure \ref{electric_magnetic_braid_3D_appendix}) and $-1$ if it points against the path. The action of the membrane operator on the path element is then
	\begin{align}
		C^h(m):g(t)&=C^h(m): g(t_1)g_i^{\sigma_i}g(t_2) \notag \\
		&= g(t_1)g(s.p(m)-v_i)^{-1}hg(s.p(m)-v_i)g_i^{\sigma_i}g(t_2) \notag \\
		&=g(s.p(t)-v_i )g(s.p(m)-v_i)^{-1}hg(s.p(m)-v_i)g_i^{\sigma_i}g(t_2) \notag \\
		&=g(s.p(t)-s.p(m)) hg(s.p(m)-v_i) [g(v_i-s.p(t)) g(s.p(t)-v_i )]g_i^{\sigma_i}g(t_2), \notag \\
	\end{align}
	where we inserted the identity in the form of $g(v_i-s.p(t)) g(v_i-s.p(t))^{-1}=g(v_i-s.p(t)) g(s.p(t)-v_i )$. Noting that $g(s.p(t)-v_i)=g(t_1)$ (at least up to factors in $\partial(E)$ from deformation), this means that 	
	\begin{align}
		C^h(m):g(t)&=C^h(m): g(t_1)g_i^{\sigma_i}g(t_2) \notag \\
		&= g(s.p(t)-s.p(m)) hg(s.p(m)-v_i) g(v_i-s.p(t)) g(s.p(t)-v_i )g_i^{\sigma_i}g(t_2) \notag \\
		&= g(s.p(t)-s.p(m)) hg(s.p(m)-s.p(t)) g(t_1)g_i^{\sigma_i}g(t_2) \notag \\
		&=g(s.p(t)-s.p(m)) h g(s.p(t)-s.p(m))^{-1} g(t). \label{Equation_magnetic_membrane_electric_appendix_1}
	\end{align}

	We note that this is the same result as in the 2+1d case considered in Ref. \cite{HuxfordPaper2}, because the magnetic ribbon in the 2+1d model acts on individual edges in the same way as the magnetic membrane operator in 3+1d. We can use this result to determine the commutation relation between the electric ribbon and magnetic membrane operators. We have
	\begin{align}
		\delta(g, \hat{g}(t))C^h(m)\ket{GS} &= C^h(m)\delta(g, g(s.p(t)-s.p(m)) h g(s.p(t)-s.p(m))^{-1} \hat{g}(t) ) \ket{GS} \notag \\
		&=C^h(m) \delta(g(s.p(t)-s.p(m))h^{-1}g(s.p(t)-s.p(m))^{-1}g, \hat{g}(t))\ket{GS} \label{Equation_magnetic_membrane_electric_appendix_2}
	\end{align}

	We now wish to consider the basis for the electric ribbon operators described by irreps of the group $G$. We therefore consider an electric ribbon operator of the form $S^{R,a,b}(t)=\sum_{g \in G} [D^R(g)]_{ab} \delta(g, \hat{g}(t))$, where $R$ is an irrep of $G$ and $D^R(g)$ is the representation of element $g$ in $R$. The commutation relation between this and the magnetic membrane operator is then
	\begin{align}
		S^{R,a,b}(t)C^h(m)\ket{GS}&= \sum_{g \in G} [D^R(g)]_{ab} \delta(g,\hat{g}(t))C^h(m)\ket{GS} \notag \\
		&=C^h(m) \sum_{g \in G} [D^R(g)]_{ab} \delta(g(s.p(t)-s.p(m))h^{-1}g(s.p(t)-s.p(m))^{-1}g, \hat{g}(t)\ket{GS}\notag \\
		&= C^h(m) \sum_{\substack{g'= g(s.p(t)-s.p(m))h^{-1}\\ \hspace{0.2cm} \times g(s.p(t)-s.p(m))^{-1}g}} \hspace{-1cm} [D^R(g(s.p(t)-s.p(m))hg(s.p(t)-s.p(m))^{-1}g')]_{ab} \delta(g', \hat{g}(t))\ket{GS} \notag \\
		&= C^h(m) \sum_{g' \in G} \sum_{c=1}^{|R|} [D^R(g(s.p(t)-s.p(m))hg(s.p(t)-s.p(m))^{-1})]_{ac} [D^R(g')]_{cb} \delta(g', \hat{g}(t))\ket{GS} \notag \\
		&= C^h(m) \sum_{c=1}^{|R|} [D^R(g(s.p(t)-s.p(m))hg(s.p(t)-s.p(m))^{-1})]_{ac} S^{R,c,b}(t) \ket{GS}. \label{Equation_magnetic_membrane_electric_appendix_3}
	\end{align}
	
	This expressions indicates that the different electric ribbon operators labelled by the same irrep are mixed by the braiding (just as in the 2+1d case considered in Ref. \cite{HuxfordPaper2}). However, the coefficients for this mixing include the operator $g(s.p(t)-s.p(m))$, which means that we do not have well-defined braiding results. There are special cases for which we do get well defined results however (again, just as in the 2+1d case). In the case where $G$ is Abelian, so that $R$ is a 1D irrep of $G$, the braiding result simplifies to
	\begin{equation}
		S^{R}(t)C^h(m)\ket{GS}= R(h) C^h(m)S^R(t) \ket{GS}, \label{Equation_magnetic_membrane_electric_appendix_4}
	\end{equation}
	from which we see that we gain a simple phase $R(h)$ under braiding. When the electric ribbon operator has the same start-point as the membrane operator, the path $(s.p(t)-s.p(m))$ is a closed path, which obeys a fake-flatness restriction in the ground state. This means that $g(s.p(t)-s.p(m))$ must be in $\partial(E)$ and so is in the centre of $G$. Therefore, $g(s.p(t)-s.p(m))hg(s.p(t)-s.p(m))^{-1}=h$. This means that the braiding result becomes
	\begin{equation}
		S^{R,a,b}(t)C^h(m)\ket{GS}= C^h(m) \sum_{c=1}^{|R|} [D^R(h)]_{ac} S^{R,c,b}(t) \ket{GS}. \label{Equation_magnetic_membrane_electric_appendix_5}
	\end{equation}

	We can also consider the case where the electric ribbon passes through the magnetic membrane in the opposite direction (i.e., down through the membrane rather than up through it), by considering the path $t^{-1}=s$. Denoting the end-point of $s$, which is the start of $t$, by $e.p(s)$, we have
	\begin{align}
		C^h(m):g(s)&=C^h(m):g(t)^{-1} \notag \\
		&= [g(s.p(t)-s.p(m))hg(s.p(t)-s.p(m))^{-1} g(t)]^{-1}\notag \\
		&=g(t)^{-1}g(s.p(t)-s.p(m))h^{-1}g(s.p(t)-s.p(m))^{-1}\notag \\
		&=g(s)g(e.p(s)-s.p(m))h^{-1} g(e.p(s)-s.p(m))^{-1} \notag\\
		&=g(s.p(s)-e.p(s))g(e.p(s)-s.p(m))h^{-1} g(e.p(s)-s.p(m))^{-1}, \notag
	\end{align}
	where we wrote $g(s)$ as $g(s.p(s)-e.p(s))$. We then insert the identity in the form of $g(s.p(s)-s.p(m))^{-1} g(s.p(s)-s.p(m))$, to obtain
	\begin{align}
		C^h(m):g(s)&= g(s.p(s)-s.p(m))h^{-1}g(s.p(s)-s.p(m))^{-1} g(s.p(s)-s.p(m))g(e.p(s)-s.p(m))^{-1}\notag \\
		&=g(s.p(s)-s.p(m))h^{-1}g(s.p(s)-s.p(m))^{-1} g(s). \label{Magnetic_electric_3D_braid_reverse}
	\end{align}
	
	This is the same result as for an electric excitation braiding up through the magnetic membrane, except that we must replace $h$ with $h^{-1}$ (meaning that in the same start-point case, the path element just gains a factor of $h^{-1}$). This is what we expect from the fact that the transformations for the two directions of braiding should be inverses of each-other. In addition, this result also describes the case where we flip the orientation of the flux tube instead of inverting the path. The inverted group element in the braiding relation then reflects the fact that measuring the magnetic flux (by braiding a charge around it) against its orientation should result in an inverted element, when compared to measuring the flux along its orientation.
	
	\subsubsection{Braiding between two magnetic flux tubes}
	\label{Section_magnetic_magnetic_braiding_tri_trivial}
	
\begin{figure}[h]
		\begin{center}
			\begin{overpic}[width=0.3\linewidth]{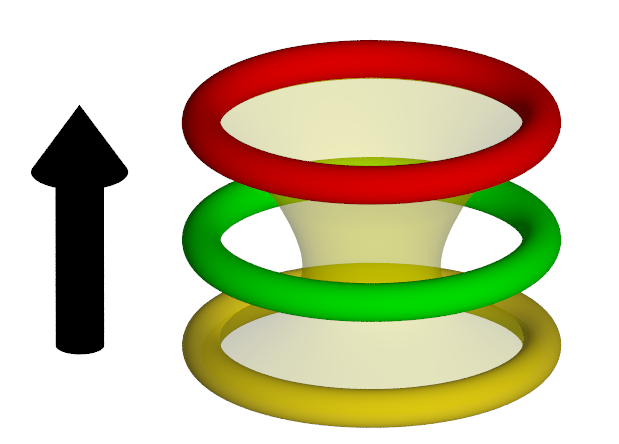}
				
			\end{overpic}
			\caption{We wish to consider the case where we pull one flux tube, initially in the position of the (yellow) lowest ring, up through another flux tube (the green tube in the middle position). The flux tube moves along the yellow surface to its final position at the upper (red) ring.}	
			\label{loop_braid_move_appendix}	
		\end{center}
	\end{figure}

\begin{figure}[h]
		\begin{center}
			\begin{overpic}[width=0.75\linewidth]{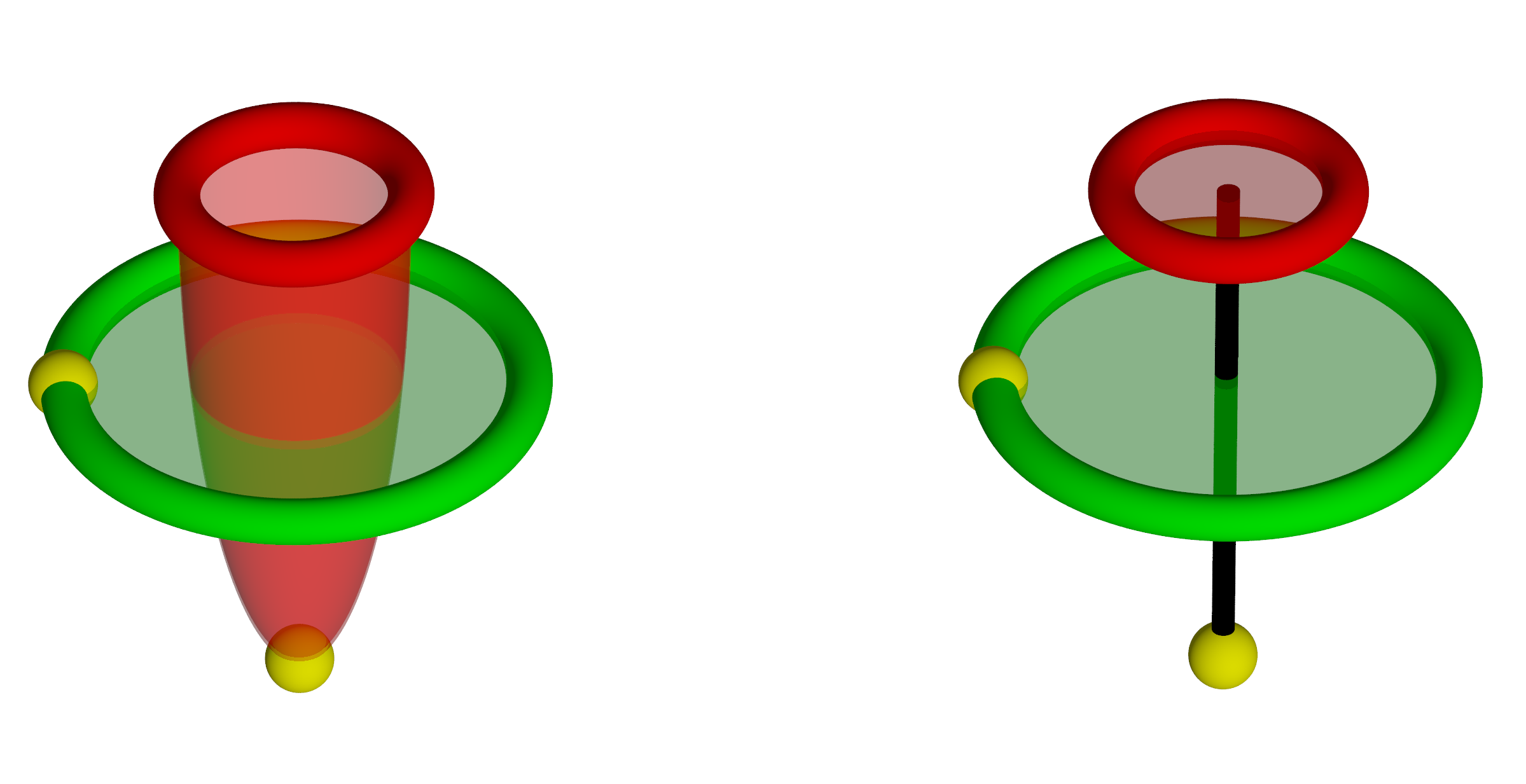}
				\put(0,17){$C^h(m)$}
				\put(3,40){$C^k(n)$}
				\put(-6,25){$s.p(m)$}
				\put(16,3){$s.p(n)$}
				
				\put(45,25){\Huge $\rightarrow$}
				\put(44,29){deform}
				
				\put(62,17){$C^h(m)$}
				\put(64,40){$C^k(n')$}
				\put(55,25){$s.p(m)$}
				\put(70,3){$s.p(n')=s.p(n)$}
				
			\end{overpic}
			\caption{In order to implement the braiding move shown in Figure \ref{loop_braid_move_appendix} using membrane operators, we consider applying two membrane operators, as shown in the left image. The first, on the green membrane (the lighter membrane in grayscale), nucleates and grows a flux tube. The second, on the red membrane (the darker membrane in grayscale), nucleates a second flux tube which is then moved through the first one. If we compare the case where we first produce the (lighter) green tube, then move the (darker) red tube through it, to the case where we first move the (darker) red tube through the vacuum before producing the (lighter) green loop, then we will obtain our braiding relation. Therefore, we need to consider the commutation relation between the two operators. In order to do so, it is convenient to use the topological property of the membrane operators to fully pull the (darker) red membrane through the (lighter) green one, as indicated in the right image. When we pull the (darker) red membrane through the (lighter) green one, we drag the paths (represented by the black cylinder) from the start-point to the edges through the (lighter) green membrane. Therefore, all of these paths are acted on by the membrane operator on the (lighter) green membrane. A more detailed version of the right image is shown in Figure \ref{Magnetic_membrane_braiding_orientation}, which indicates the orientation of the membranes.}
			\label{membrane_through_membrane_braid_alternate_appendix}
		\end{center}
	\end{figure}

	Having considered the braiding relation between an electric and a magnetic excitation, we next consider the braiding between two magnetic flux tubes. We consider the case shown in Figure \ref{loop_braid_move_appendix}, where we pull one flux tube through another. This situation can be implemented by applying two membrane operators, as indicated in the left side of Figure \ref{membrane_through_membrane_braid_alternate_appendix}. By applying the membrane operator $C^h(m)$ followed by $C^k(n)$, we first produce the outer excitation and move it into position, then produce and move the other flux tube through it. If we compare this to the case where we first produce the tube labelled by $k$ and move it through vacuum, before producing the flux tube labelled by $h$, then we will obtain the braiding relation. The simplest way to obtain the commutation relation of the two membrane operators is to use the topological nature of the membrane operators. As we proved in Section \ref{Section_Topological_Magnetic_Tri_Trivial}, we can deform the membrane on which we apply a magnetic membrane operator without changing the action of that operator, provided that we keep the position of any excitations fixed and do not deform the membrane across any excitations in the initial state. We can even use this to move the direct and dual membranes away from the start-point of the membrane (when we deform the membrane operator we must keep the start-point fixed, because it may be excited). We therefore consider deforming the membrane $n$ as shown in Figure \ref{membrane_through_membrane_braid_alternate_appendix}, to produce the new membrane $n'$, which is entirely on one side of $m$. Having done so, there is now no intersection between the direct and dual membranes of the two membrane operators. However, the paths from the start-point of $n'$, $s.p(n)$, to the edges affected by $C^k(n')$ all pass through the membrane $m$. Recall that the action of the magnetic membrane operator $C^k(n')$ on an edge $i$ cut by the dual membrane of the membrane operator, initially labelled by $g_i$, is
	$$C^k(n'): g_i= \begin{cases} g(s.p(n)-v_i)^{-1}kg(s.p(n)-v_i)g_i & \text{if $i$ points away from the direct membrane}\\ g_ig(s.p(n)-v_i)^{-1}k^{-1}g(s.p(n)-v_i) & \text{if $i$ points towards the direct membrane,} \end{cases}$$
	where $v_i$ is the end of $i$ which lies on the direct membrane. We see that the action of the operator on the edge depends on the path $(s.p(n)-v_i)$ from the start-point of the membrane to the edge. When we deformed the membrane $n$ through $m$, we ensured that all of these paths must pass through $m$, and so we must consider how these paths are affected by the presence of $C^h(m)$.

	\begin{figure}[h]
		\begin{center}
			\begin{overpic}[width=0.75\linewidth]{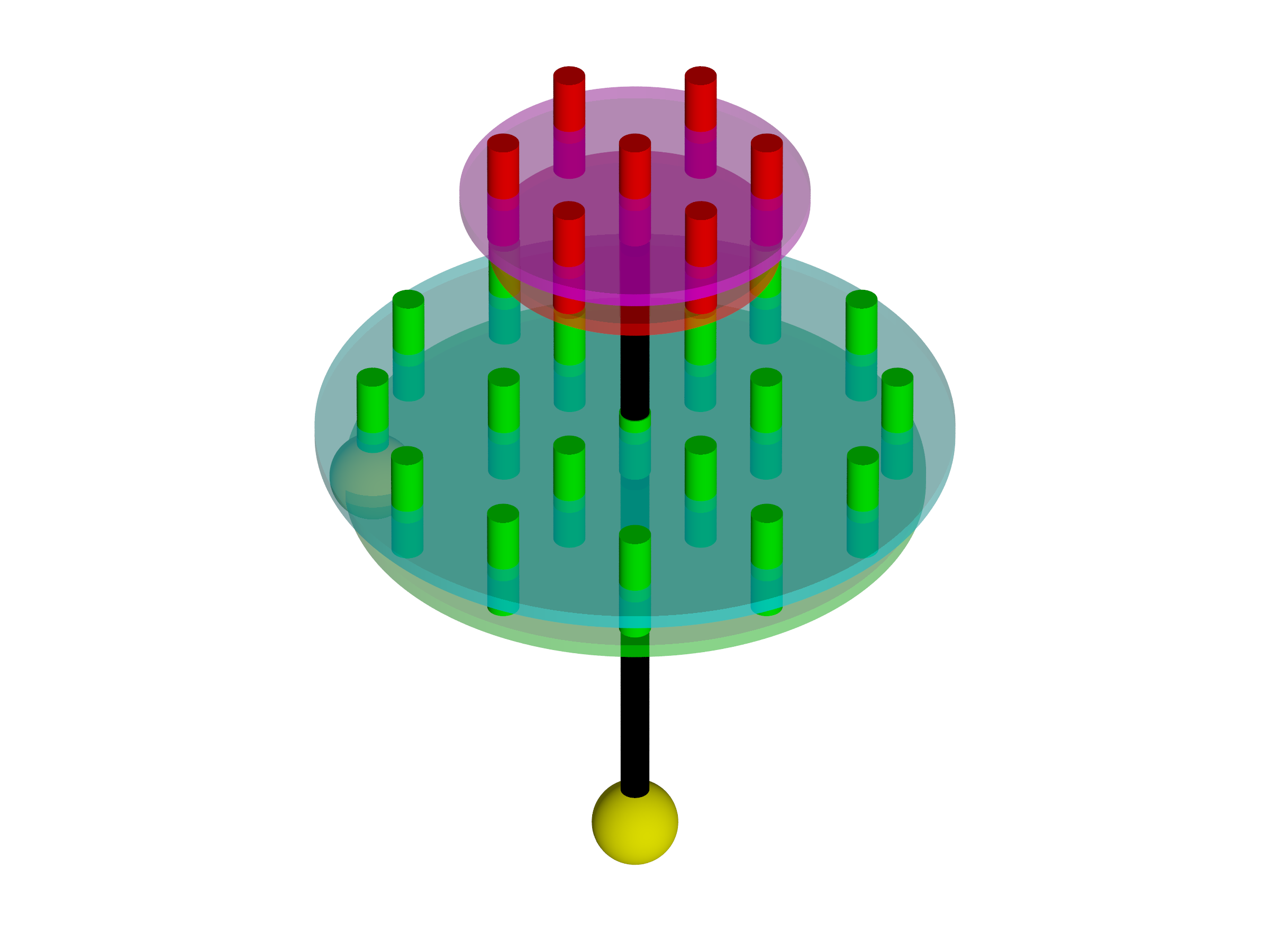}
				\put(77,40){$C^h(m)$}
				\put(64,60){$C^k(n')$}
			\end{overpic}
			\caption{When considering the braiding relations of fluxes, it is necessary to fix an orientation for the fluxes. We consider the case where both of the flux tubes are oriented ``upwards". That is, their dual membrane is above their direct membrane, so that the edges cut by the dual membrane stick upwards from the direct membrane. This means that paths passing up through the membrane $C^h(m)$ are modified by factors of the form $g(t)^{-1}hg(t)$ rather than of the form $g(t)^{-1}h^{-1}g(t)$.}
			\label{Magnetic_membrane_braiding_orientation}
		\end{center}
	\end{figure}
	
	In order to determine the action of $C^h(m)$ on the paths, it is necessary to fix an orientation for the membrane operators, by deciding the relative positions of the direct and dual membranes. We consider the case shown in Figure \ref{Magnetic_membrane_braiding_orientation}, where both membranes are oriented upwards (along the braiding path). This means that the paths from the start-point to the membrane $n'$ pass up through the membrane $m$, along the orientation of membrane $m$. We established in Section \ref{Section_electric_magnetic_braiding_3D_tri_trivial}, when we considered the braiding between an electric and magnetic excitation, that for this relative orientation of membrane $m$ and path $g(t)$, the action of $C^h(m)$ on the path is
	\begin{equation}
		C^h(m): g(t)= g(s.p(t)-s.p(m))hg(s.p(t)-s.p(m))^{-1}g(t) \label{Equation_magnetic_membrane_on_path_appendix_1}
	\end{equation}
	(see Equation \ref{Equation_magnetic_membrane_electric_appendix_1}). Therefore, taking $t$ to be the path $(s.p(n)-v_i)$ from the start-point of $n'$ (which is also the start-point of $n$) to one of the edges affected by the membrane operator on $n'$, and assuming for simplicity that $i$ points away from the direct membrane, we see that
	\begin{align}
		C^k(n')C^h(m):g_i &= \big(C^h(m):g(s.p(n)-v_i)\big)^{-1}k\big(C^h(m):g(s.p(n)-v_i)\big) g_i \notag \\
		&=\big(g(s.p(n)-s.p(m))hg(s.p(n)-s.p(m))^{-1}g(s.p(n)-v_i)\big)^{-1}k\notag \\
		&\hspace{1cm}\big(g(s.p(n)-s.p(m))hg(s.p(n)-s.p(m))^{-1}g(s.p(n)-v_i)\big) g_i\notag\\
		&= g(s.p(n)-v_i)^{-1}\big[g(s.p(n)-s.p(m))h^{-1}g(s.p(n)-s.p(m))^{-1}k\notag \\
		& \hspace{1cm} g(s.p(n)-s.p(m))hg(s.p(n)-s.p(m))^{-1}\big]g(s.p(n)-v_i) g_i\notag \\
		&=C^h(m)C^{g(s.p(n)-s.p(m))h^{-1}g(s.p(n)-s.p(m))^{-1}kg(s.p(n)-s.p(m))hg(s.p(n)-s.p(m))^{-1}}(n'):g_i. \label{Equation_magnetic_membranes_on_edge}
	\end{align}
	
	A similar result would hold if the edge $i$ pointed towards the direct membrane of $n'$, except we would have
	\begin{align}
		C^k(n')C^h(m):g_i &= g_i\big(C^h(m):g(s.p(n)-v_i)\big)^{-1}k^{-1}\big(C^h(m):g(s.p(n)-v_i)\big)
	\end{align}
	instead, which still gives
	\begin{align*}
		C^k(n')C^h(m):g_i &=C^h(m)C^{g(s.p(n)-s.p(m))h^{-1}g(s.p(n)-s.p(m))^{-1}kg(s.p(n)-s.p(m))hg(s.p(n)-s.p(m))^{-1}}(n'):g_i.
	\end{align*}
	Because Equation \ref{Equation_magnetic_membranes_on_edge} holds for the action on all edges $i$ that are affected by $C^k(n')$, and $C^k(n')$ does not interfere with the action of $C^h(m)$ (i.e., there are no other effects that cause the operators to fail to commute), Equation \ref{Equation_magnetic_membranes_on_edge} implies that 
	\begin{equation}
		C^k(n')C^h(m)\ket{GS}= C^h(m)C^{g(s.p(n)-s.p(m))h^{-1}g(s.p(n)-s.p(m))^{-1}kg(s.p(n)-s.p(m))hg(s.p(n)-s.p(m))^{-1}}(n')\ket{GS}. \label{Equation_magnetic_membranes_commute_1}
	\end{equation}
	
	As a final step, we can deform the membrane $n'$ back into its original position $n$. We can also simplify the notation by defining $h_{[n-m]}= g(s.p(n)-s.p(m))hg(s.p(n)-s.p(m))^{-1}$, allowing us to write Equation \ref{Equation_magnetic_membranes_commute_1} as
	\begin{equation}
		C^k(n)C^h(m)\ket{GS}= C^h(m)C^{h^{-1}_{[n-m]}kh^{\phantom {-1}}_{[n-m]}}(n)\ket{GS}. \label{Equation_magnetic_membranes_commute_2}
	\end{equation}
	
	We therefore see that the label $k$ of the membrane operator applied on $n$ is conjugated by an element $h_{[n-m]}$, while the label $h$ of the membrane operator applied on $m$ is left invariant. The presence of the operator $g(s.p(n)-s.p(m))$ in $h_{[n-m]}= g(s.p(n)-s.p(m))hg(s.p(n)-s.p(m))^{-1}$ reflects the fact that the two fluxes are defined with different start-points and so we should not expect a definite braiding result between the two flux tubes. If we consider the case where the start-points of the two flux tubes are the same, then the path $(s.p(n)-s.p(m))$ is closed. Then as long as this path is contractible, its label lies in $\partial(E)$ due to fake-flatness conditions. Because elements of $\partial(E)$ are in the centre of $G$, this means that $g(s.p(n)-s.p(m))hg(s.p(n)-s.p(m))^{-1}=h$. Therefore, the commutation relation \ref{Equation_magnetic_membranes_commute_1} becomes
	\begin{equation}
		C^k(n)C^h(m)\ket{GS}= C^h(m)C^{h^{-1}kh}(n)\ket{GS}. \label{Equation_magnetic_membranes_same_sp}
	\end{equation}
	We see that in this case, the label $k$ of the inner loop becomes $h^{-1}kh$ when it is pulled through the outer loop (while the label $h$ of the outer loop is again left invariant).

	We could also consider the case where the orientations of one or more of the magnetic flux tubes are flipped compared to the example in Figure \ref{Magnetic_membrane_braiding_orientation} (or where the inner flux tube is braided around and then down through the outer tube rather than up through it). The orientation of the membranes appeared in the calculation when we considered the effect of the outer magnetic membrane $C^h(m)$ on the path elements $g(s.p(n)-v_i)$ that appeared in the action of the inner magnetic membrane $C^k(n)$. If we have the opposite orientation for membrane $m$, then the effect of the magnetic membrane operator $C^h(m)$ on the path is (using Equation \ref{Magnetic_electric_3D_braid_reverse})
	$$C^h(m): g(s.p(n)-v_i)= g(s.p(n)-s.p(m))h^{-1}g(s.p(n)-s.p(m))^{-1}g(s.p(n)-v_i). $$
	Compared to Equation \ref{Equation_magnetic_membrane_on_path_appendix_1}, we simply replace $h$ with $h^{-1}$. Applying this change to the rest of the mathematical argument, we find that 
	\begin{equation}
		C^k(n)C^h(m)\ket{GS}= C^h(m)C^{h^{\phantom{-1}}_{[n-m]}kh^{-1}_{[n-m]}}(n)\ket{GS}, \label{Equation_magnetic_membranes_commute_reverse_1}
	\end{equation}
	so that the label $k$ of the inner loop is again conjugated, this time by the inverse element (compared to Equation \ref{Equation_magnetic_membranes_commute_2}). We would obtain a similar result if the inner loop was pulled down through the outer one (this is the same as the situation where we reverse both membranes, but viewed from a different angle). On the other hand reversing the orientation of $n$ does not affect the result.

	\subsubsection{Braiding of blob excitations and $E$-valued loops}
	\label{Section_braiding_blobs_E_loops_tri_trivial}
\begin{figure}[h]
		\begin{center}
			\begin{overpic}[width=0.75\linewidth]{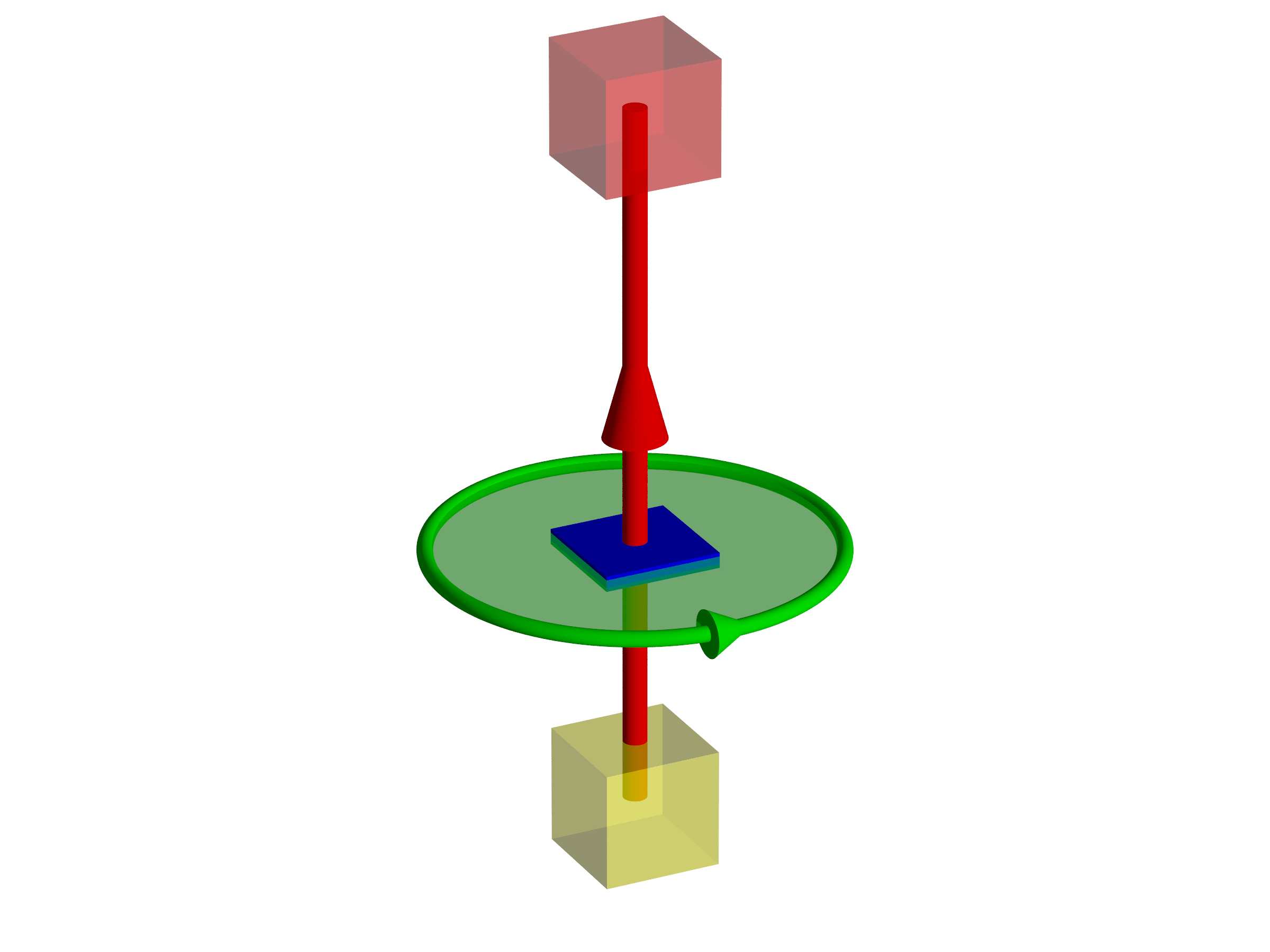}
				\put(55,33){$q$}
				\put(52,50){$t$}
				\put(68,30){$m$}	
			\end{overpic}
			\caption{We consider braiding a blob excitation (red) through an $E$-valued loop (green). To do so, we consider applying an $E$-valued membrane operator (on the green membrane $m$) to produce the loop excitation before applying a blob ribbon operator on the red path $t$ in the dual lattice to produce and move the blob excitation. Note that the orientation of the membrane operator, deduced from its boundary using the right-hand rule, is upwards, so that the orientation of the membrane matches that of the path. In order to work out the braiding relation between the blob and loop excitations, we must calculate the commutation relation of these operators. The operators interact at the point of their intersection, which is the (blue) square plaquette $q$.}
			\label{Blob_through_E_membrane_tri_trivial_appendix}
		\end{center}
	\end{figure}

	Having considered the braiding relations between the excitations that are associated to the group $G$ and the edges of the lattice, we now look at the excitations related to the group $E$ and the plaquette labels. In the case where $\rhd$ is trivial, the ribbon and membrane operators that produce these excitations only act on the surface labels of our lattice, with no dependence at all on the edge degrees of freedom. This means that the $E$-valued excitations braid trivially with the $G$-valued ones. However, they may still braid non-trivially with each-other. Just as we considered braiding an electric excitation up through a magnetic flux tube, we wish to find the result of pulling a blob excitation up through an $E$-valued loop excitation. In order to do so, we first apply an $E$-valued membrane operator $L^{e}(m)$ on a membrane $m$ to produce the $E$-valued loop excitation, before applying a blob ribbon operator $B^f(t)$ that passes through this membrane, as shown in Figure \ref{Blob_through_E_membrane_tri_trivial_appendix}. We then wish to compare this to the case where we instead move the blob excitation before producing the loop excitation. That is, we wish to take the state $B^f(t)L^e(m)\ket{GS}$ and commute $B^f(t)$ to the right of $L^e(m)$.
	
	The $E$-valued membrane operator $L^e(m)$ is $\delta(e, \hat{e}(m))$, where $e(m)$ is the total surface label of the membrane $m$ and $\hat{e}(m)$ is the associated operator. Then, in the case where $\rhd$ is trivial, this surface label is given by
	\begin{equation}
		e(m)=\prod_{\text{plaquette }p \in m}e_p^{\sigma_p},
		\label{Equation_surface_label_tri_trivial_1}
	\end{equation}
	where $\sigma_p$ is $+1$ if the orientation of the plaquette $p$ (determined from the orientation of its boundary) matches the orientation of the membrane at that point, and is $-1$ if the plaquette is anti-aligned with the membrane. We wish to determine how this surface label is affected by the action of the blob ribbon operator. If we denote the plaquette where the ribbon intersects with the membrane by $q$, then the non-commutativity of the blob ribbon operator and $E$-valued membrane operator comes from the action of the ribbon operator on the plaquette $q$. If the initial label of plaquette $q$ is $e_q$, then the action of the blob ribbon operator $B^f(t)$ on $q$ pierced by the ribbon is described by
	\begin{equation}
		B^f(t):e_q = \begin{cases} e_q f^{-1} & \text{if the orientation of $q$ matches the ribbon's orientation} \\ f e_q & \text{if the orientation of $q$ is opposite to that of the ribbon.} \end{cases}
		\label{Equation_blob_ribbon_on_intersection_plaquette_1}
	\end{equation}
	
	Because we are considering the case where the ribbon and the membrane have the same orientation, we can replace the orientation of the ribbon in Equation \ref{Equation_blob_ribbon_on_intersection_plaquette_1} with the orientation of the membrane. That is
	\begin{equation}
		B^f(t):e_q = \begin{cases} e_q f^{-1} & \text{if the orientation of $q$ is aligned with the orientation of the membrane} \\ f e_q & \text{if the orientation of $q$ is anti-aligned with the orientation of the membrane.} \end{cases}
		\label{Equation_blob_ribbon_on_intersection_plaquette_2}
	\end{equation}
	
	This manipulation is relevant because the contribution of plaquette $q$ to the surface label $e(m)$ is $e_q^{\sigma_q}$, which depends on the relative orientation of the plaquette and the membrane through the variable $\sigma_q$ (which is $1$ if the plaquette is aligned with the membrane and $-1$ if it is anti-aligned). From Equation \ref{Equation_blob_ribbon_on_intersection_plaquette_2}, the effect of the ribbon operator on the contribution $e_q^{\sigma_q}$ of plaquette $q$ to the surface label is
	$$B^f(t): e_q^{\sigma_q}= e_q^{\sigma_q}f^{-1},$$
	from which we see that the factor $f^{-1}$ gained by this contribution to the surface label is independent of the orientation of $q$. Therefore, the action of the ribbon operator on $e(m)$ is 
	\begin{align*}
		B^f(t):e(m) &= B^f(t): \prod_{\text{plaquette }p \in m}e_p^{\sigma_p}\\
		&= \big( \prod_{\text{plaquette }x \in m; \ x  \neq q} e_x^{\sigma_x} \big) e_q^{\sigma_q}f^{-1}\\
		&= f^{-1} \prod_{\text{plaquette }p \in m}e_p^{\sigma_p}\\
		&=f^{-1} e(m),
	\end{align*}
	where we used the fact that $E$ is Abelian when $\rhd$ is trivial to freely rearrange the product. This means that
	\begin{align*}
		\delta(e, \hat{e}(m))B^f(t) \ket{GS} &=B^f(t)\delta(e, f^{-1}\hat{e}(m)),
	\end{align*}
	and so
	\begin{align*}
		B^f(t)\delta(e, \hat{e}(m))\ket{GS} &= \delta(e, f\hat{e}(m)) B^f(t) \ket{GS}\\
		&=\delta(f^{-1}e, \hat{e}(m))B^f(t) \ket{GS}.
	\end{align*}

	It is interesting to examine the commutation relation for one of the irrep basis operators for the space of $E$-valued membrane operators. That is, we consider $L^R(m)= \sum_{e \in E} R(e) \delta(e, \hat{e}(m))$, where $R$ is an irrep of $E$. Because $E$ is Abelian, the irreps are 1D and so $R(e)$ needs no matrix indices. Then
	\begin{align*}
		B^f(t)L^R(m)\ket{GS} &= \sum_{e \in E} R(e) B^f(t)\delta(e, \hat{e}(m))\ket{GS}\\
		&=\sum_{e \in E} R(e) \delta(f^{-1}e, \hat{e}(m))B^f(t) \ket{GS}\\
		&= \sum_{e'=f^{-1}e \in E} R(fe') \delta(e', \hat{e}(m)) B^f(t) \ket{GS}\\
		&= R(f) \sum_{e' \in E} R(e') \delta(e', \hat{e}(m)) B^f(t) \ket{GS}\\
		&= R(f) B^f(t)L^R(m)\ket{GS},
	\end{align*}
	from which we see that the result of braiding a blob excitation with such an $E$-valued loop excitation is a simple phase.

	We have so far considered the case where the orientation of the membrane is aligned with that of the ribbon. If the orientation of the membrane is instead opposite to the ribbon, then the action of the blob ribbon operator on the surface is inverted compared to this case. That is, the action of $B^f(t)$ on the plaquette $q$ pierced by the ribbon is
	$$B^f(t): e_q^{\sigma_q}= fe_q^{\sigma_q}.$$
	Because $E$ is Abelian, this means that we simply need to replace $f^{-1}$ with $f$ in our previous braiding results. That is
	$$B^f(t)\delta(e, \hat{e}(m))\ket{GS} = \delta(fe, \hat{e}(m))B^f(t) \ket{GS}$$
	and so
	$$ B^f(t)L^R(m)\ket{GS} = R(f^{-1}) B^f(t)L^R(m)\ket{GS}. $$
	
	We note that these braiding relations are similar to the braiding between the magnetic and electric excitations. However, whereas in that case the loop-like magnetic flux tubes were labelled by an element of the group (in that case $G$) and the point-like electric excitations were labelled by irreps, in this case it is more natural to label the $E$-valued loops with irreps (of $E$) and the point-like blob excitations with elements of $E$. This reflects the idea that the blob excitations are associated with non-trivial 2-flux (just as the magnetic excitations are associated with non-trivial 1-flux), while the $E$-valued loop excitations are charges that measure this 2-flux. This idea is further strengthened by the fact that the $E$-valued loop excitations have trivial mutual braiding, because the $E$-valued membrane operators are all diagonal in the configuration basis (the basis where each edge is labelled by an element of $G$ and each plaquette by an element of $E$), whereas the magnetic excitations can have non-trivial mutual braiding.

	\subsubsection{Linking}
	\label{Section_linking_appendix}

\begin{figure}[h]
		\begin{center}
			\begin{overpic}[width=0.4\linewidth]{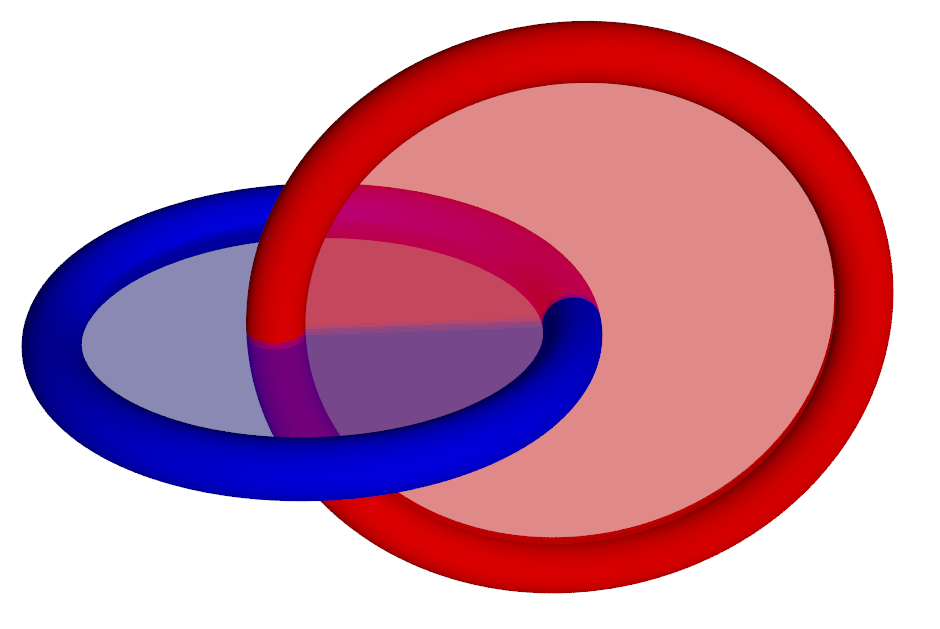}
				\put(-7,20){\large $C^k(n)$}
				\put(82,60){\large $C^h(m)$}
			\end{overpic}
			\caption{We consider the situation where the loop-like excitations (opaque tori) produced by two magnetic membrane operators (applied on the translucent membranes) link together. We consider first applying the membrane operator $C^h(m)$ on the red membrane, before applying the $C^k(n)$ on the blue membrane to produce an excitation that links with the one produced by $C^h(m)$.}
			\label{linking_picture_appendix}
			
		\end{center}
	\end{figure}

	In addition to their non-trivial braiding properties, the magnetic flux tubes may have non-trivial linking properties, as we discussed in Section \ref{Section_linking} of the main text. That is, when we try to push one loop through another (to link them), we may produce an energetic linking string between the two loops, as illustrated in Figures \ref{linking} and \ref{linking_process} in Section \ref{Section_linking}. In this section, we will explain how such a linking string occurs and derive the conditions for the fluxes to support such a linking string when $\rhd$ is trivial. In order to do this, we consider acting on the ground state with two magnetic membrane operators, with membranes chosen in such a way that the loop-like excitations produced by the two membrane operators link, as indicated in Figure \ref{linking_picture_appendix}. In order to see why this may produce a linking string, we must take a closer look at the action of the membrane operators. Suppose that we first act with the membrane operator $C^h(m)$ on the ground state, where $m$ is the red membrane in Figure \ref{linking_picture_appendix}, before acting with $C^k(n)$. Then, because the boundary of $m$ intersects with the membrane $n$, there is an excited plaquette (produced by $C^h(m)$) in the direct membrane of $n$, before we act with $C^k(n)$. This is relevant because we assumed that the direct membrane satisfied fake-flatness when we considered which energy terms are excited by the magnetic membrane in Section \ref{Section_Magnetic_Membrane_Tri_trivial}. The presence of an excited plaquette in the region of $C^k(n)$ may mean that $C^k(n)$ produces additional excitations.

\begin{figure}[h]
		\begin{center}
			\begin{overpic}[width=0.9\linewidth]{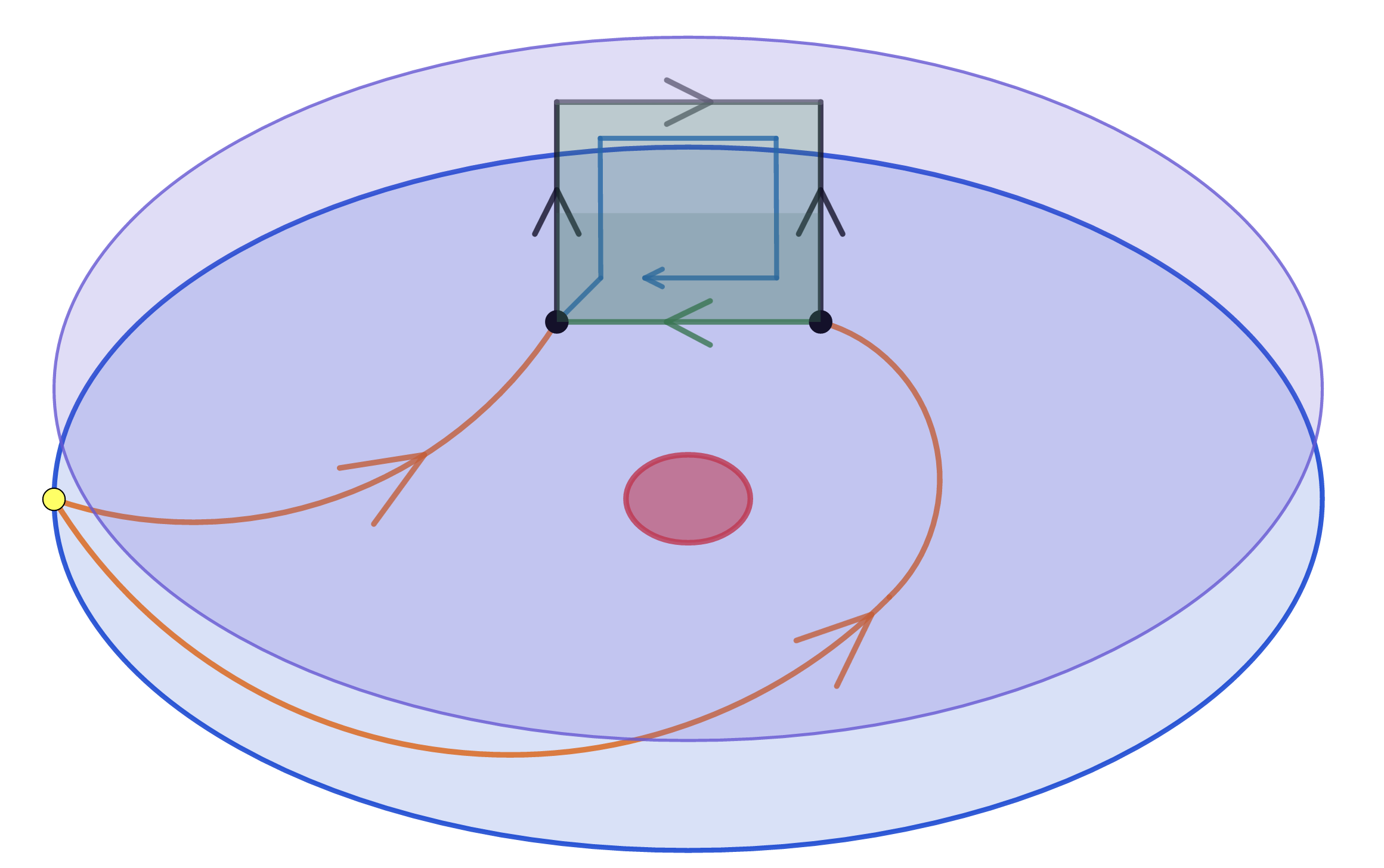}
				\put(23,27){$t_1$}
				\put(69,27){$t_2$}
				\put(48,36){$b$}
				\put(38,44){$i_1$}
				\put(61,44){$i_2$}
				\put(50,46){$p$}
				\put(53,56){$u$}
				\put(-2,24){$s.p(n)$}
				\put(37,39){$v_1$}
				\put(61,39){$v_2$}
				\put(86,9.5){direct membrane}
				\put(86,51){dual membrane}
			\end{overpic}
			\caption{We consider a plaquette $p$ that is cut by the bulk of the dual membrane (the purple upper membrane) of the magnetic membrane operator $C^k(n)$. Such a plaquette has two edges affected by the membrane operator, here $i_1$ and $i_2$, which are attached to the direct membrane (the lower blue membrane) at vertices $v_1$ and $v_2$ respectively. The action of the membrane operator on these edges is determined by the paths $t_1$ and $t_2$ from the start-point of the membrane to the vertices $v_1$ and $v_2$. The paths $t_1$ and $t_2$ form a closed path $t_2bt_1^{-1}$ when combined with the base $b$ of the plaquette $p$. If the surface enclosed by this closed path intersects another magnetic excitation (here the intersection is represented by the red oval), then the surface does not satisfy fake-flatness. This may lead to the plaquette $p$ being excited, depending on the labels of the two magnetic membrane operators, whereas in the absence of the second magnetic excitation, such a plaquette is not excited.}
			\label{linking_potentially_excited_plaquette}
			
		\end{center}
	\end{figure}

	In order to examine this possibility, we consider the action of the magnetic membrane operator $C^k(n)$ on the plaquette holonomy of a plaquette $p$ cut by the bulk of the dual membrane of $n$, in the presence of the excitations produced by $C^h(m)$. This is shown in Figure \ref{linking_potentially_excited_plaquette}. Two of the edges on the plaquette are cut by the dual membrane of $n$ and so are affected by the magnetic membrane operator $C^k(n)$. The action on each edge depends on a path from the start-point of the membrane to the edge. Denoting the affected edges by $i_1$ and $i_2$, and the paths to these edges by $t_1$ and $t_2$, the membrane operator acts on the edges according to
	$$C^k(n):g_{i_x} = \begin{cases} g(t_x)^{-1}kg(t_x)g_{i_x} & \text{ if $i_x$ points away from the direct membrane} \\ g_{i_x}g(t_x)^{-1}k^{-1}g(t_x) & \text{ if $i_x$ points towards the direct membrane,} \end{cases}$$
	where $x=1$ or $2$ for the two edges. Consider the case where, as shown in Figure \ref{linking_potentially_excited_plaquette}, the paths to the two edges, combined with the base $b$ of the plaquette $p$, encircle the excitation produced by $C^h(m)$. We call such a plaquette a ``linking plaquette" for the rest of this section. The total effect on the plaquette holonomy $H_1(p)=\partial(e_p)g_{i_1}g(u)g_{i_2}^{-1}g(b)$ of this linking plaquette is
	\begin{align}
		C^k(n):H_1(p) &= \partial(e_p) g(t_1)^{-1}kg(t_1)g_{i_1} g(u) g_{i_2}^{-1} g(t_2)^{-1}k^{-1}g(t_2)g(b) \notag \\
		&= g(t_1)^{-1}kg(t_1) \partial(e_p) g_{i_1} g(u) g_{i_2}^{-1} g(t_2)^{-1}k^{-1}g(t_2)g(b) \notag \\
		&= g(t_1)^{-1}kg(t_1) [ \partial(e_p) g_{i_1} g(u) g_{i_2}^{-1}g(b)] g(b)^{-1}g(t_2)^{-1}k^{-1}g(t_2)g(b). \label{Equation_linking_plaquette_holonomy_1}
	\end{align}
	
	Given that the plaquette holonomy initially satisfies fake-flatness (i.e., $H_1(p)=1_G$), Equation \ref{Equation_linking_plaquette_holonomy_1} becomes
	\begin{align}
		C^k(n):H_1(p) &= g(t_1)^{-1}kg(t_1) [1_G]g(b)^{-1}g(t_2)^{-1}k^{-1}g(t_2)g(b)\notag \\
		&= g(t_1)^{-1}kg(t_1) (g(t_2)g(b))^{-1}k^{-1}(g(t_2)g(b)). \label{Equation_linking_plaquette_holonomy_2}
	\end{align}

\begin{figure}[h]
		\begin{center}
			\begin{overpic}[width=0.6\linewidth]{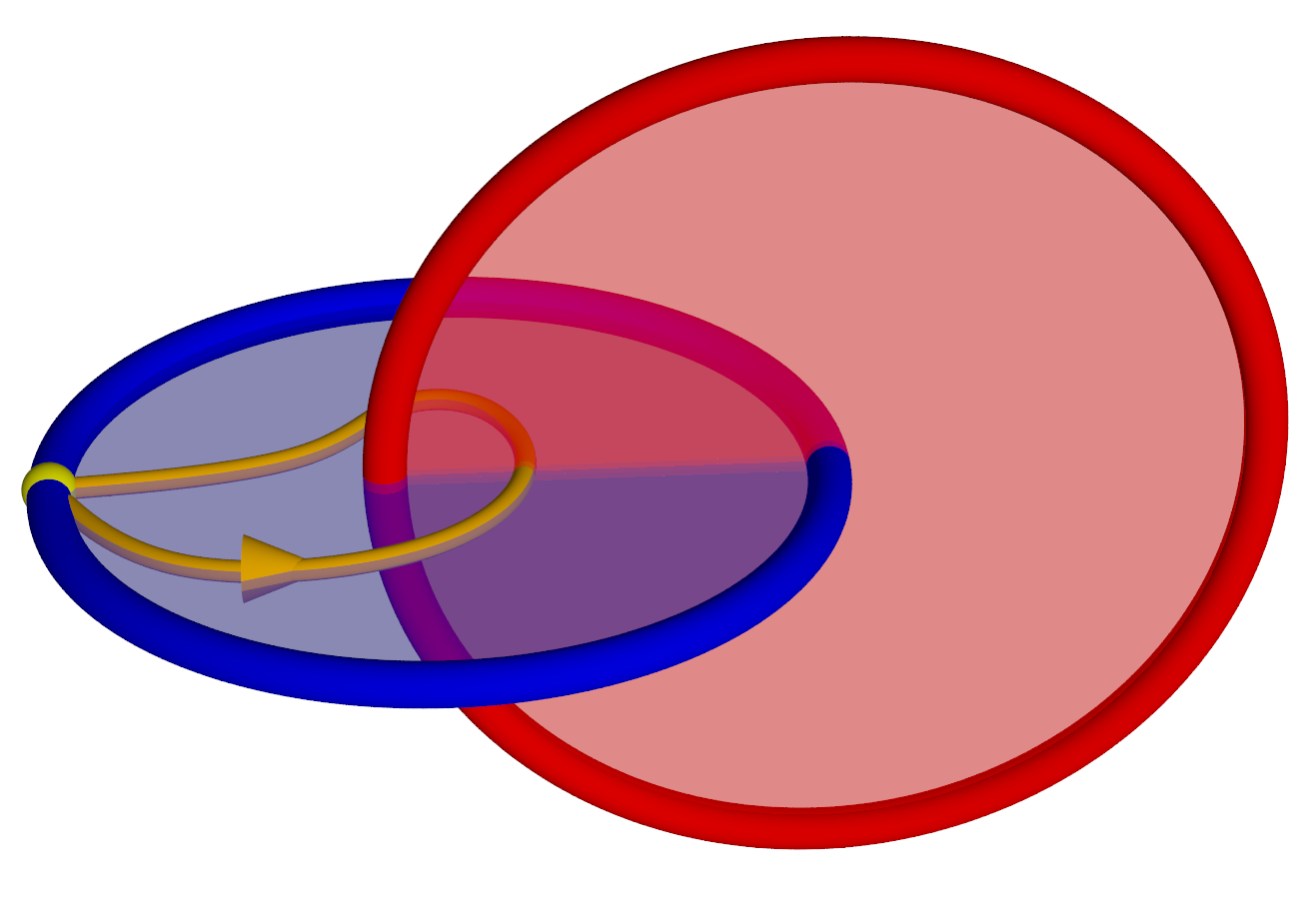}
				\put(0,18){\large $C^k(n)$}
				\put(88,60){\large $C^h(m)$}
				\put(12,29){\large $s$}
			\end{overpic}
			
			\caption{The closed path $s= t_2 \cdot b \cdot t_1^{-1}$ (here represented in orange) from Figure \ref{linking_potentially_excited_plaquette} passes through the membrane $m$ and is therefore acted on by the membrane operator $C^h(m)$. We can use our results from Section \ref{Section_electric_magnetic_braiding_3D_tri_trivial}, where we examined the braiding relations of electric and magnetic excitations, to determine how the path is affected. This in turn tells us how the plaquette holonomy of plaquette $p$ from Figure \ref{linking_potentially_excited_plaquette} is affected by the membrane operator $C^k(n)$.}
			\label{linking_closed_path}
		\end{center}
	\end{figure}

	Note that the paths $t_1$ and $t_2 \cdot b$ have the same start and end-points. This means that the path $t_2 \cdot b \cdot t_1^{-1}$ is a closed path. In the case where the direct membrane of $n$ satisfies fake-flatness, the path element $g(t_2)g(b)g(t_1)^{-1}$ associated to this path therefore satisfies the fake-flatness condition. This means that this group element belongs to $\partial(E)$ and so is in the centre of $G$. This means that Equation \ref{Equation_linking_plaquette_holonomy_2} becomes
	\begin{align*}
		C^k(n):H_1(p) &=g(t_1)^{-1}k [g(t_1) (g(t_2)g(b))^{-1}]k^{-1}(g(t_2)g(b))\\
		&= g(t_1)^{-1} [g(t_1) (g(t_2)g(b))^{-1}] k k^{-1}(g(t_2)g(b))\\
		&= g(t_1)^{-1} g(t_1) (g(t_2)g(b))^{-1}(g(t_2)g(b))\\
		&=1_G,
	\end{align*}
	so that the plaquette holonomy is left invariant and so the plaquette is not excited by the action of the magnetic membrane operator. However, we are interested in the case where the direct membrane does not satisfy fake-flatness, and instead the closed path encloses a plaquette excitation. Then $[g(t_1) (g(t_2)g(b))^{-1}]$ does not need to be in $\partial(E)$. This in turn means that plaquette holonomy may be changed by the membrane operator $C^k(n)$, and so the plaquette $p$ may be excited. In order to more definitively show whether the plaquette is excited, we need to consider the closed path $t_2 \cdot b \cdot t_1^{-1}$ in more detail. We call this path $s$ and refer to it as the closed base path associated to plaquette $p$. As indicated in Figure \ref{linking_closed_path}, $s$ passes through the membrane $m$, either with or against the orientation of the membrane. Therefore, as shown in Section \ref{Section_electric_magnetic_braiding_3D_tri_trivial}, the action of the membrane operator $C^h(m)$ on this path is (see Equations \ref{Equation_magnetic_membrane_electric_appendix_1} and \ref{Magnetic_electric_3D_braid_reverse})
	$$C^h(m):g(s) = \begin{cases} g(s.p(m)-s.p(s))^{-1} h g(s.p(m)-s.p(s)) g(s) & \text{ if $s$ aligns with the orientation of $m$} \\ g(s.p(m)-s.p(s))^{-1} h^{-1} g(s.p(m)-s.p(s)) g(s) & \text{ if $s$ aligns against the orientation of $m$,} \end{cases}$$
	so that
	\begin{equation}
		\hat{g}(s)C^h(m) \ket{GS} = \begin{cases} C^h(m) g(s.p(m)-s.p(s))^{-1} h g(s.p(m)-s.p(s)) \hat{g}(s) \ket{GS} & \text{ if $s$ aligns with $m$} \\ C^h(m) g(s.p(m)-s.p(s))^{-1} h^{-1} g(s.p(m)-s.p(s)) \hat{g}(s) \ket{GS} & \text{ if $s$ aligns against $m$}. \end{cases}
		\label{Equation_linking_closed_path_label_2}
	\end{equation}

	However, $s$ is a contractible closed path, so in the ground state its label is in $\partial(E)$ due to fake-flatness. It then gains a factor of $g(s.p(m)-s.p(s))^{-1} h g(s.p(m)-s.p(s))$ (or the inverse) from the action of $C^h(m)$, as described in Equation \ref{Equation_linking_closed_path_label_2}, and so then has the form $g(s.p(m)-s.p(s))^{-1} h^{\pm 1} g(s.p(m)-s.p(s))\partial(f)$ for some element $f \in E$. This means that the factor $g(s)^{-1}=g(t_1)(g(t_2)g(b))^{-1}$ appearing in the action of the membrane operator $C^k(n)$ on the plaquette holonomy in Equation \ref{Equation_linking_plaquette_holonomy_2} can be written as
	\begin{align}
		g(t_1)(g(t_2)g(b))^{-1} &=g(s.p(m)-s.p(s))^{-1} h^{\pm 1} g(s.p(m)-s.p(s))\partial(e) \notag \\
		&=g(s.p(m)-s.p(n))^{-1}h^{\pm 1}g(s.p(m)-s.p(n)) \partial(e) \label{Equation_linking_closed_path_label_3},
	\end{align}
	where $e=f^{-1}$ is some element of $E$ and $g(s.p(m)-s.p(n))^{-1}h^{\pm 1}g(s.p(m)-s.p(n))$ is the inverse of the factor from Equation \ref{Equation_linking_closed_path_label_2} (so the $\pm 1$ depends on the relative orientation of $m$ and $s$). By substituting this form into Equation \ref{Equation_linking_plaquette_holonomy_2}, we can write the action of $C^k(n)$ on the plaquette holonomy $H_1(p)$ as
	\begin{align}
		C^k&(n):H_1(p)= g(t_1)^{-1}kg(t_1) (g(t_2)g(b))^{-1}k^{-1}(g(t_2)g(b)) \notag\\
		&=g(t_1)^{-1}k [g(t_1) (g(t_2)g(b))^{-1}]k^{-1}(g(t_2)g(b))g(t_1)^{-1} g(t_1) \notag\\
		&=g(t_1)^{-1}k [g(s.p(m)-s.p(n))^{-1}h^{\pm 1}g(s.p(m)-s.p(n)) \partial(e)] k^{-1} \notag \\ & \hspace{2cm}[g(s.p(m)-s.p(n))^{-1}h^{\pm 1}g(s.p(m)-s.p(n)) \partial(e)]^{-1} g(t_1)\notag \\
		&= g(t_1)^{-1} [k g(s.p(m)-s.p(n))^{-1}h^{\pm 1}g(s.p(m)-s.p(n)) k^{-1} g(s.p(m)-s.p(n))^{-1}h^{ \mp 1}g(s.p(m)-s.p(n))] g(t_1). \label{Equation_linking_plaquette_holonomy}
	\end{align}

	Note that the term
	$$[k g(s.p(m)-s.p(n))^{-1}h^{\pm 1}g(s.p(m)-s.p(n)) k^{-1} g(s.p(m)-s.p(n))^{-1}h^{ \mp 1}g(s.p(m)-s.p(n))]$$
	is just the commutator of $k$ and $g(s.p(m)-s.p(n))^{-1}h^{\pm 1}g(s.p(m)-s.p(n))$. Therefore, the resulting plaquette holonomy is the identity element precisely when $k$ and $g(s.p(m)-s.p(n))^{-1}hg(s.p(m)-s.p(n))$ commute (which is the same condition for the braiding between the magnetic excitations to be trivial, as found in Section \ref{Section_magnetic_magnetic_braiding_tri_trivial}). In the case where the start-points are equal, the condition simplifies to requiring that $k$ and $h$ commute. This indicates that when $k$ and $h$ (or $g(s.p(m)-s.p(n))^{-1}hg(s.p(m)-s.p(n))$ if the start-points are not the same) do not commute, we may produce extra excitations when we act with $C^k(n)$. If the labels do not commute, we say that the two fluxes support a linking sting between them. However, all we have shown so far is that we have extra plaquette excitations; we have not yet shown that these extra excitations form a linking string between the two loop excitations. We have shown that, given that the labels of the two magnetic excitations do not commute, a plaquette cut by the dual membrane of $n$ is excited by the action of the membrane operator provided that the two paths to that plaquette, combined with the base of the plaquette itself, link with the excitation produced by $m$. That is, the linking plaquettes (which we defined just above \ref{Equation_linking_plaquette_holonomy_1}), such as plaquette $p$ from Figure \ref{linking_potentially_excited_plaquette}, are excited if the labels $k$ and $g(s.p(m)-s.p(n))^{-1}hg(s.p(m)-s.p(n))$ do not commute. However, we have not shown that such linking plaquettes always exist. When we define $C^k(n)$, we have many possible choices for the locations of the paths from the start-point to the affected edges and it is not immediately clear that all possible choices lead to linking plaquettes, and so to extra excited plaquettes. In order to prove this, we must show that, no matter how we choose the set of paths, some of the closed paths associated to the plaquettes cut by the dual membrane must inevitably link with the excitation.

\begin{figure}[h]
		\begin{center}
			\begin{overpic}[width=0.8\linewidth]{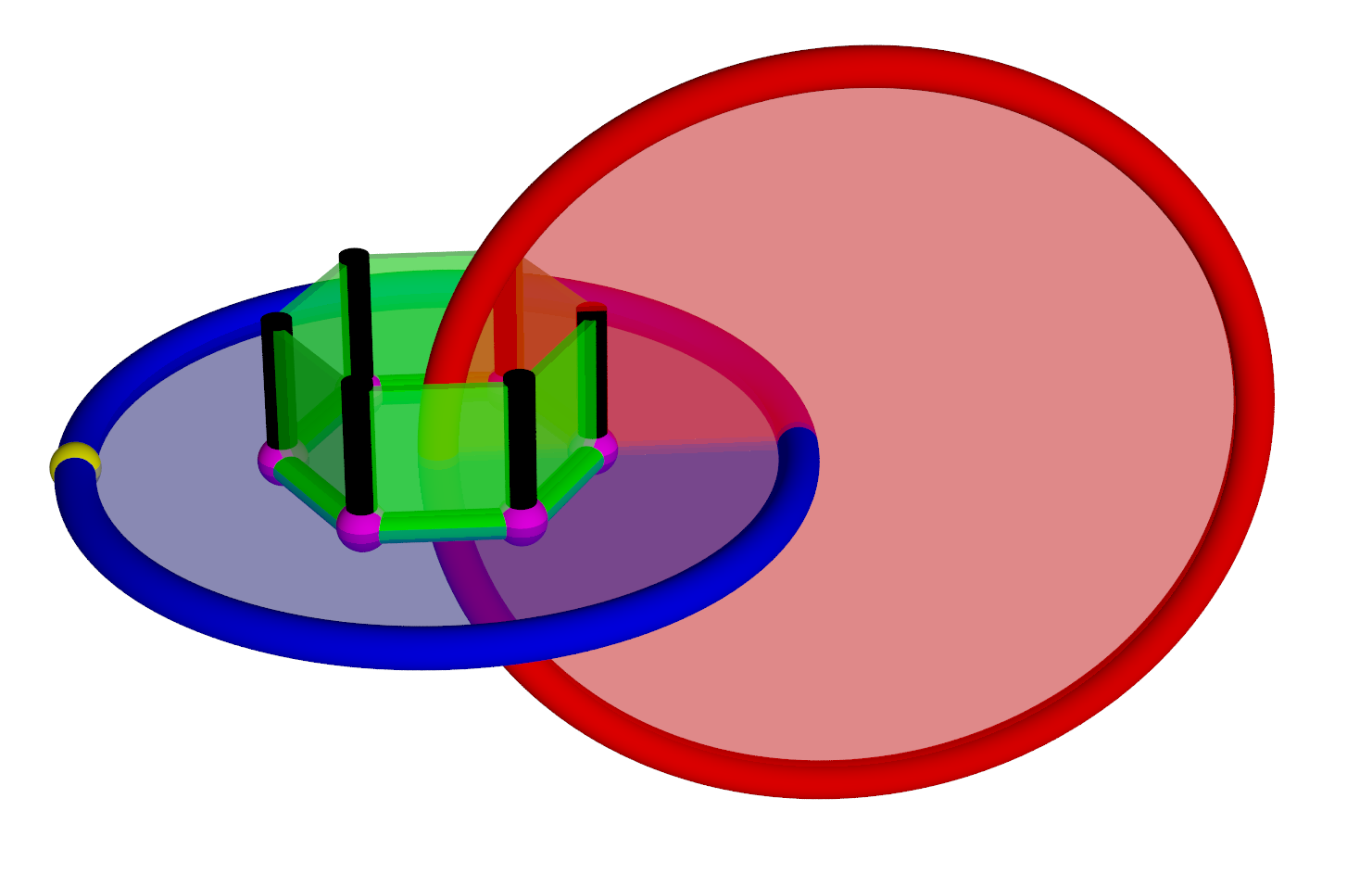}
				
			\end{overpic}
			\caption{Consider a ring of the plaquettes (green) affected by $C^k(n)$ that encircles the excitation produced by $C^h(m)$ (larger red torus). We will show that at least one of the plaquettes in this ring is a linking plaquette like plaquette $p$ from Figure \ref{linking_potentially_excited_plaquette}, in that the two paths from the start-point of $n$ to the plaquette join to the base of the plaquette to produce a closed path that links with the (red) flux tube.}
			\label{linking_chain_plaquettes}
		\end{center}
	\end{figure}

\begin{figure}[h]
		\begin{center}
			\begin{overpic}[width=\linewidth]{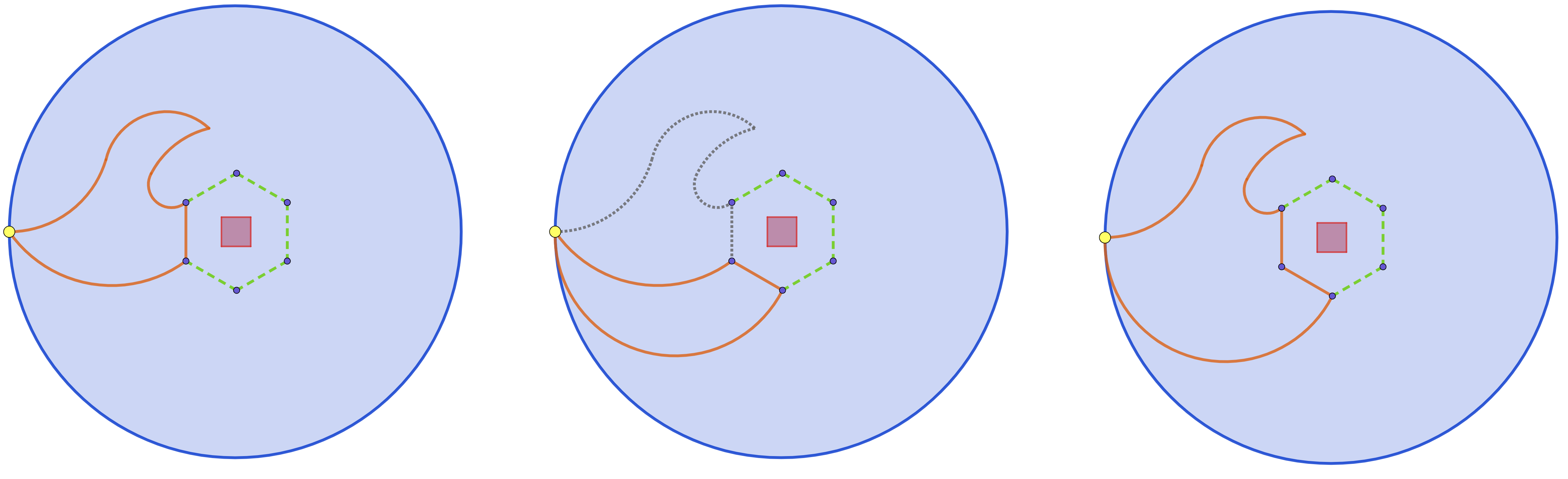}
				\put(32,15){\Huge $\cdot$}
				\put(65,15){\Huge $\rightarrow$}
				\put(64.2,19){combine}
			\end{overpic}
			
			\caption{We consider the ring of plaquettes from Figure \ref{linking_chain_plaquettes} (represented here by the dashed green hexagon) in a plan view. The (red) square represents the intersection of the membrane $n$ with the flux tube produced by $C^h(m)$, and is therefore an excited plaquette. The (yellow) circle on the left of each image is the start-point of $n$. Each plaquette (each side of the hexagon) is associated to a closed path (the closed base path for that plaquette). For example, the leftmost side of the hexagon is associated to the (orange) path in the leftmost image. We can combine the closed paths associated with neighbouring plaquettes. For instance, we can combine the closed path from the leftmost image (represented by the dotted path in the second image) with the closed path in the second image. This produces the closed path in the rightmost image. If the two constituent paths do not link with the flux tube, then neither must the combination of the two paths.}
			\label{linking_combining_paths_1}
		\end{center}
	\end{figure}

	In order to prove this result, we consider a set of plaquettes cut by the dual membrane of $n$ which form a closed chain around the flux tube produced by $m$, as shown in Figure \ref{linking_chain_plaquettes}. We will show that for any set of paths from the start-point of $m$ to the affected edges on these plaquettes, at least one of the plaquettes must be a linking plaquette (i.e., a plaquette for which the closed base path, which is the path from the start-point of the membrane to one of the affected edges on the plaquette, across the base of the plaquette and back along the path from the other affected edge on the plaquette to the start-point, encloses the flux tube, like the plaquette in Figure \ref{linking_potentially_excited_plaquette}). We denote the vertices attached to these edges and lying on the direct membrane by $v_x$, where $x$ runs from 1 to $N$, where $N$ is the number of these vertices on the cycle, as we progress anticlockwise around the cycle. Each vertex $v_x$ is adjacent to the vertices $v_{x-1}$ and $v_{x+1}$, where adjacency implies that the two vertices share a plaquette, and we define $v_{N+1}=v_1$. We then denote the paths from the start-point of the membrane to a vertex $v_x$ by $(s.p(n)-v_x)$. We also denote the path between neighbouring vertices $v_x$ and $v_{x+1}$, along the base of the connecting plaquette, by $(v_x-v_{x+1})$. Our claim is that at least one closed path of the form $(s.p(n)-v_x) \cdot (v_x-v_{x+1}) \cdot (s.p(n)-v_{x+1})^{-1}$ must encircle the flux tube produced by $C^h(m)$. These paths are the closed base paths associated to the plaquettes of the chain of plaquettes, analogous to the path $s$ associated to plaquette $p$ from Figure \ref{linking_potentially_excited_plaquette}, so if this is true then at least one plaquette is a linking plaquette. Suppose to the contrary that none of these closed base paths enclosed the flux, even though the path $(v_1-v_2) \cdot (v_2-v_3) \cdot... \cdot (v_N-v_1)$ does by construction. Then consider the product of two of these paths, $(s.p(n)-v_1)\cdot (v_1-v_{2}) \cdot (s.p(n)-v_{2})^{-1}$ and $(s.p(n)-v_2) \cdot (v_2-v_3) \cdot (s.p(n)-v_{3})^{-1}$, which is equivalent to 
	$$(s.p(n)-v_1)\cdot (v_1-v_{2}) \cdot (v_2-v_3) \cdot (s.p(n)-v_{3})^{-1},$$
	as shown in Figure \ref{linking_combining_paths_1}. From our assumption that neither of the closed paths $(s.p(n)-v_1)\cdot (v_1-v_{2}) \cdot (s.p(n)-v_{2})^{-1}$ and $(s.p(n)-v_2) \cdot (v_2-v_3) \cdot (s.p(n)-v_{3})^{-1}$ enclose the flux tube, the product cannot wrap around the flux tube either.
	
\begin{figure}[h]
		\begin{center}
			\begin{overpic}[width=0.9\linewidth]{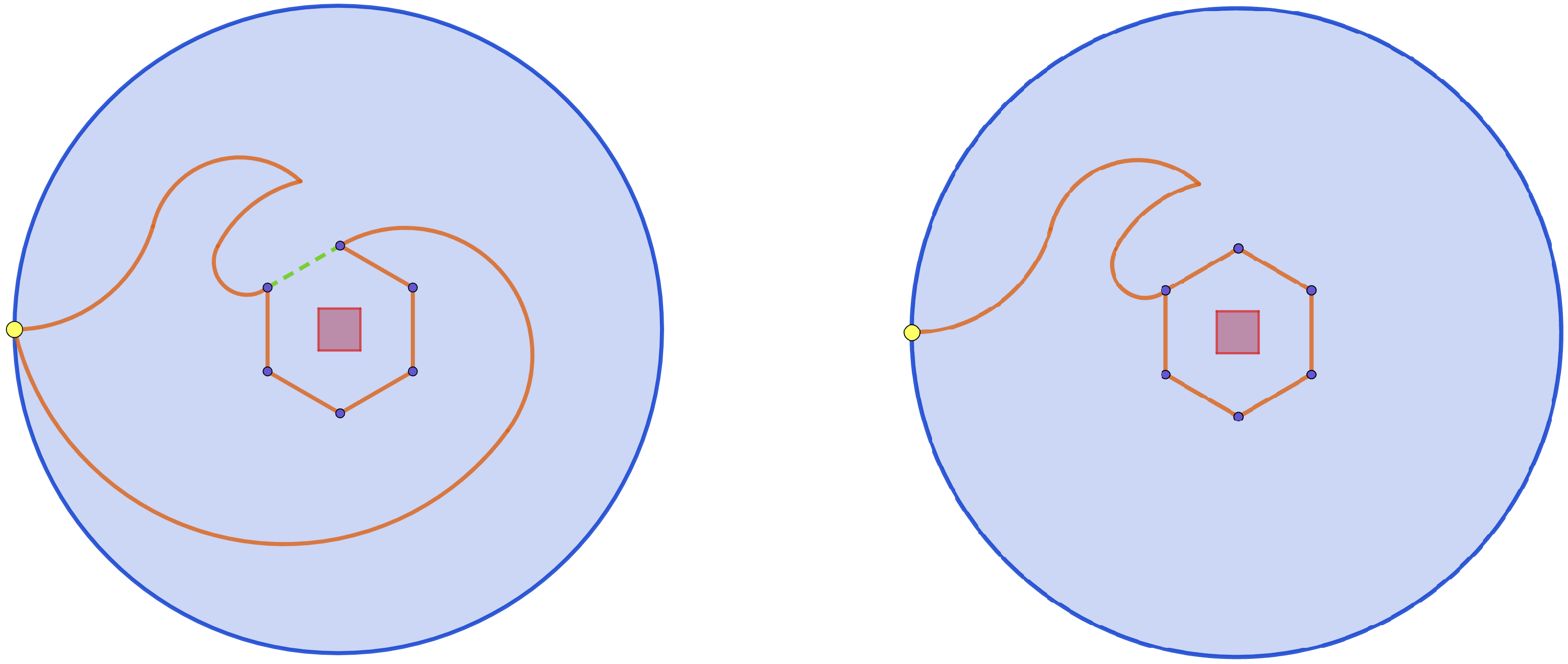}
				\put(48,20){\Huge $\rightarrow$}
			\end{overpic}
			
			\caption{We consider combining all of the closed paths associated with the plaquettes in the ring from Figures \ref{linking_chain_plaquettes} and \ref{linking_combining_paths_1}. Connecting the last path takes us from the (orange) solid path in the left image to the (orange) solid path in the right image. When we combine all of the closed paths associated to the plaquettes in the ring of plaquettes, we obtain a closed path which passes from the start-point of $n$ to one of the vertices, $v_1$, then all the way around the circle of plaquettes and back along the path from $v_1$ to the start-point. This closed path therefore links with the flux tube. This implies that the sum of the linking numbers of the closed paths associated with each plaquette is the linking number of the ring of plaquettes (i.e., $1$ or $-1$). This in turn means that one or more of the plaquettes in the ring of plaquettes is similar to the one from Figure \ref{linking_potentially_excited_plaquette}, with its associated closed path linking with the flux tube (in this case, it was the last plaquette, represented by the dashed green line in the left image), and so may be excited.}
			\label{linking_combining_paths_2}
		\end{center}
	\end{figure}

	We can repeat this argument, connecting all of the closed base paths to obtain 
	$$(s.p(n)-v_1)\cdot (v_1-v_{2}) \cdot (v_2-v_3)\cdot... \cdot (v_N-v_1) \cdot (s.p(n)-v_{1})^{-1},$$
	as shown in Figure \ref{linking_combining_paths_2}. This total path must not link with the flux tube by our assumption that none of the individual closed paths do. However, we defined the path $(v_1-v_{2}) \cdot (v_2-v_3)\cdot... \cdot (v_N-v_1)$ to link with the flux tube. This leads to a contradiction. Therefore, our assumption that none of the closed paths $(s.p(n)-v_x) \cdot (v_x-v_{x+1}) \cdot (s.p(n)-v_{x+1})^{-1}$ link with the flux is incorrect, and so at least one such path does link with the flux. This means that at least one plaquette in our circle of plaquettes is a linking plaquette, and so is excited (provided that the labels of the fluxes support additional excited plaquettes). This is true for any circle of plaquettes that links with the flux tube. This implies that, provided the labels of the two magnetic membrane operators are chosen appropriately, we will have multiple excited plaquettes. This does not, however, prove that these plaquettes lie along a string, since we have determined nothing about the location of the various excited plaquettes.

\begin{figure}[h]
		\begin{center}
			\begin{overpic}[width=0.4\linewidth]{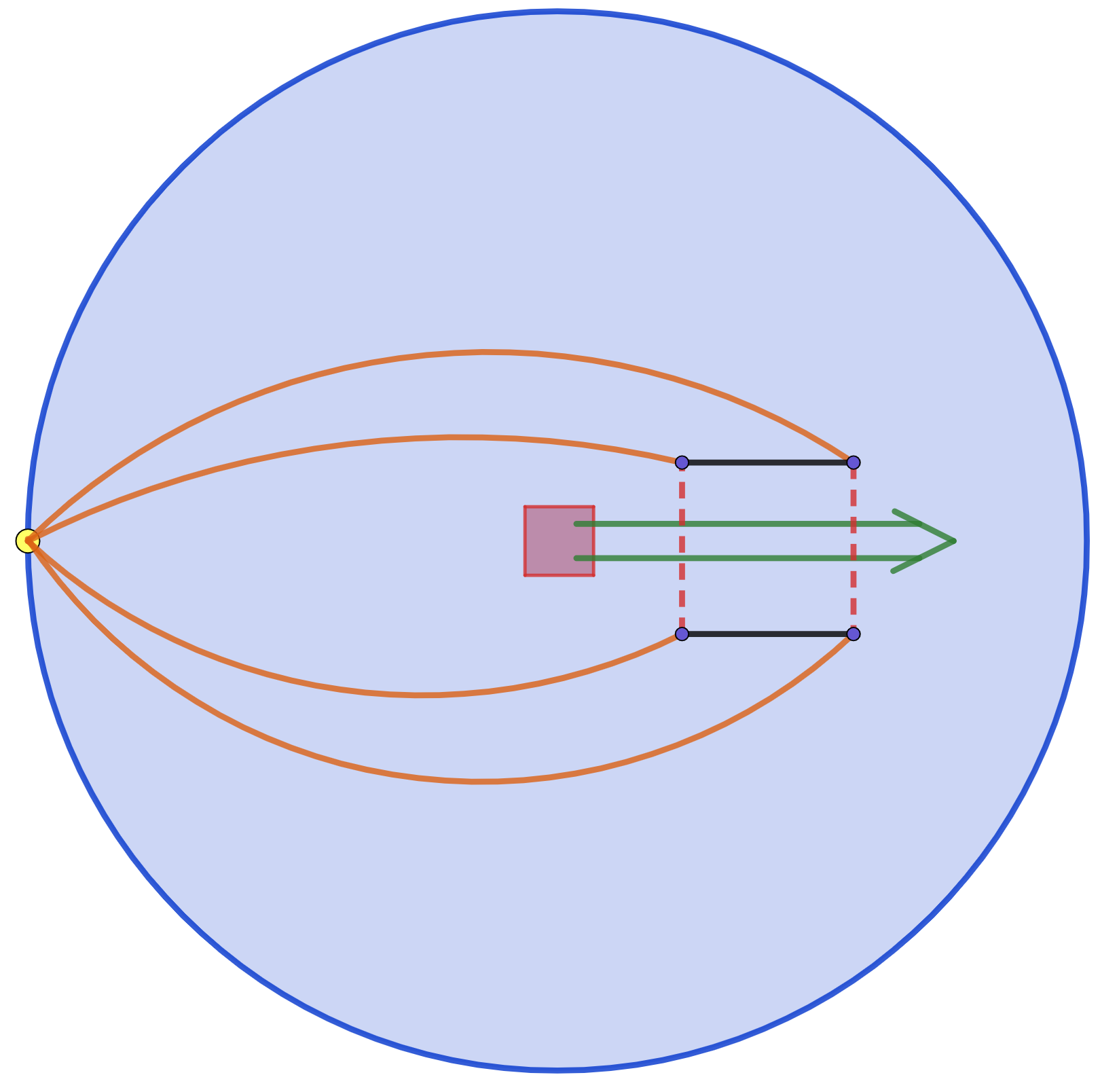}
				
			\end{overpic}
			\caption{Instead of considering a ring of plaquettes that links with the flux tube, consider a ring of plaquettes that does not enclose the tube. This ring is represented by the larger square in the figure. The linking numbers of the closed paths associated to each plaquette on the ring with the flux tube must sum to zero. This means that if one plaquette in the ring is associated with a closed base loop that links with the flux tube, so that the plaquette may be excited, another plaquette on the ring must also be associated with such a closed path. Indeed, the non-zero linking number is like a flux that must be conserved, with any linking-flux that enters the ring of plaquettes also leaving it. This means that the potentially excited plaquettes (represented by the dashed edges) lie along a string (or possibly a network of strings). In this figure, the string of flux entering and exiting the ring is represented by the (green) arrow.}
			\label{linking_unlinked_chain}
		\end{center}
	\end{figure}

\begin{figure}[h]
		\begin{center}
			\begin{overpic}[width=0.4\linewidth]{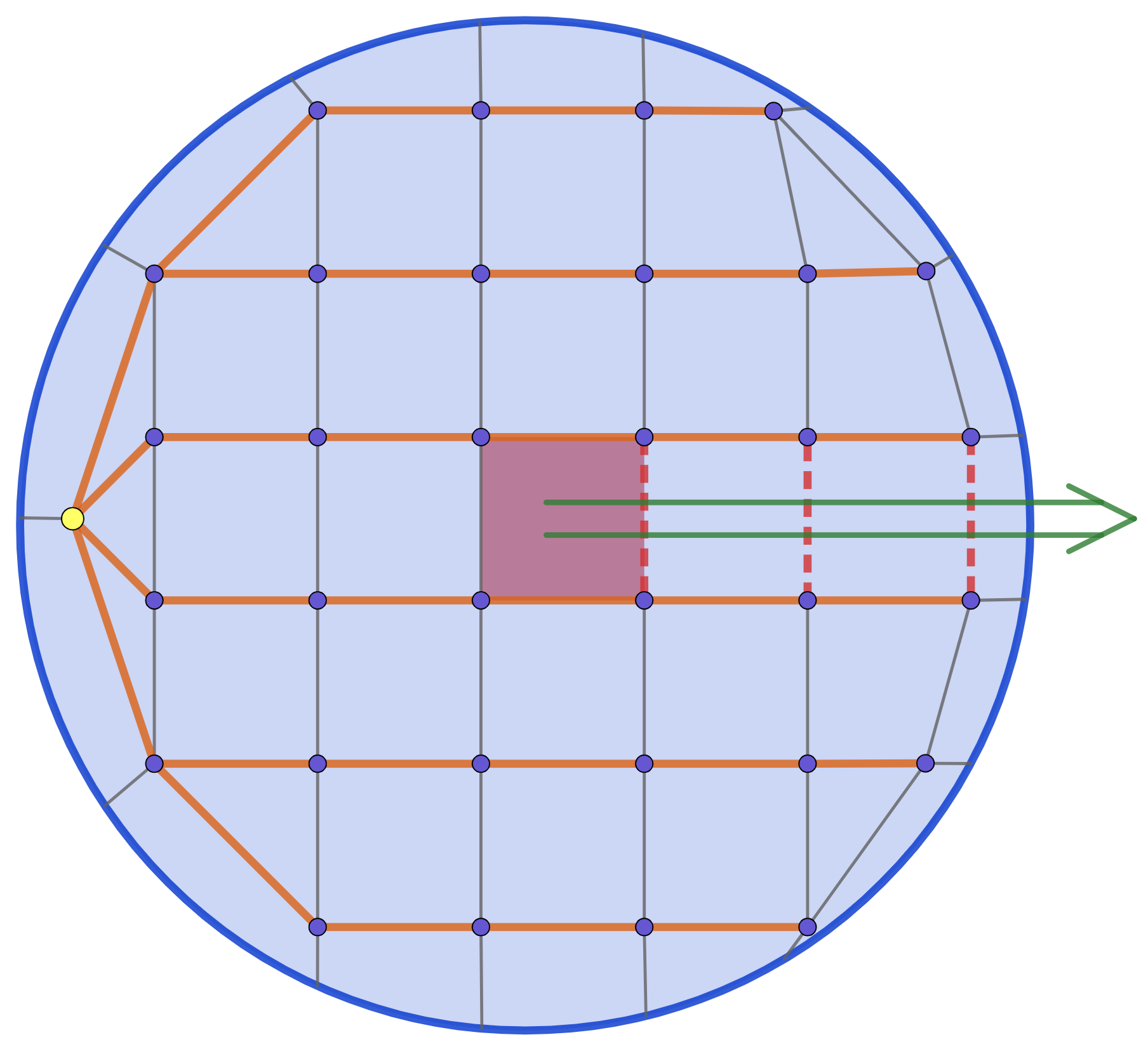}
				
			\end{overpic}
			\caption{We consider a particular example of the membrane $n$, where we have explicitly shown the vertices (the dots) and edges of this slice of the lattice. In this simple case, we choose the paths from the start-point (the left-most dot, indicated in yellow) to the vertices to lie along the orange paths. In this case, only the plaquettes whose bases are given by the red dashed edges are linking plaquettes (because the closed paths associated to these plaquettes enclose the flux) and so the linking causes a linking string that lies along the (green) arrow.}
			\label{linking_lattice_example}	
		\end{center}
	\end{figure}

	In order to confirm the existence of a linking string, we need to extend our argument about the chain of plaquettes. In addition to the fact that at least one of the plaquettes in the chain of plaquettes from Figure \ref{linking_chain_plaquettes} must be a linking plaquette, we can say that the linking numbers of the individual closed paths must sum to give the linking number of the circle of plaquettes (in this case 1 or $-1$ depending on our definition for the orientation of the flux tube). Furthermore, we can make a similar argument for any circle of plaquettes, including ones that don't link with the flux tube, such as the one shown in Figure \ref{linking_unlinked_chain}. In this case, the total linking number of the circle is zero, which means that the linking numbers of the closed base paths associated to the individual plaquettes in the circle must sum to zero (once we take into account the orientations of the closed paths). This means that, if one of the plaquettes in the circle is a linking plaquette, another must also be. The linking number acts like a conserved flux, where the excited plaquette caused by the intersection of the flux tube (red tube in Figure \ref{linking_picture_appendix}) with the membrane is the only source of linking-flux. This linking-flux must then exit the membrane through the plaquettes and causes excitations to the plaquettes it passes through. We represent this flow of flux with a string or strings (the linking string). Because this string starts at the source of linking-flux (the flux tube produced by $C^h(m)$) and ends at the boundary of the membrane $n$, this string connects the flux tubes produced by $C^h(m)$ and $C^k(n)$. In addition to a string (or strings) connecting the two flux tubes, it is possible to have closed loops of linking-flux. However, if we define the paths on $n$ from the start-point to the edges in the natural way, so that these paths represent the history of the loop as it grows outwards, then the linking-flux will have a simple form. For example, in the situation shown in Figure \ref{linking_lattice_example}, the linking-flux passes along a single line, with no closed loops or branching of the linking-flux. This gives us a single linking string connecting the two flux tubes. From Equation \ref{Equation_linking_plaquette_holonomy}, we know that a plaquette $p$ along this string has its plaquette holonomy modified to
	$$g(t_1)^{-1} [k g(s.p(m)-s.p(n))^{-1}h^{\pm 1}g(s.p(m)-s.p(n)) k^{-1} g(s.p(m)-s.p(n))^{-1}h^{ \mp 1}g(s.p(m)-s.p(n))] g(t_1)$$
	where the plus or minus depends on the orientation of the membrane $m$ and $g(t_1)$ is the path from the start-point of membrane $n$ to the base-point of the plaquette. Comparing this to Equation \ref{Equation_action_magnetic_boundary_plaquette}, which describes the action of a magnetic membrane operator on a boundary plaquette (one that is excited by the membrane operator), we see that this is equivalent to the action of a membrane operator of label
	$$k g(s.p(m)-s.p(n))^{-1}h^{\pm 1}g(s.p(m)-s.p(n)) k^{-1} g(s.p(m)-s.p(n))^{-1}h^{ \mp 1}g(s.p(m)-s.p(n))$$
	with start-point at $s.p(n)$ (note however that the label may be something else within the same conjugacy class, depending on how the paths from the start-point to the plaquettes are defined, or if a different start-point is used).

	One interesting case occurs when the label of the linking string is in the image of $\partial$. In this case, the linking string corresponds to a condensed magnetic excitation (see Section \ref{Section_magnetic_condensed_tri_trivial}), which means that it is equivalent to the string produced by a confined blob excitation. This raises the possibility that the linking string can be removed by applying a suitable blob ribbon operator. Suppose that the labels of the two magnetic membrane operators satisfy 
	$$k g(s.p(m)-s.p(n))^{-1}h^{\pm 1}g(s.p(m)-s.p(n)) k^{-1} g(s.p(m)-s.p(n))^{-1}h^{ \mp 1}g(s.p(m)-s.p(n)) = \partial(f)$$
	for some $f \in E$. Then from Equation \ref{Equation_linking_plaquette_holonomy}, the effect of the two magnetic membrane operators on a plaquette $p$ along the linking string is to take the (initially trivial) plaquette holonomy to
	$$H_1(p) \rightarrow g(t_1)^{-1} \partial(f) g(t_1).$$
	Because $\partial(E)$ is in the centre of $G$ when $\rhd$ is trivial (see Section \ref{Section_Recap_3d} of the main text), this simplifies to
	$$H_1(p) \rightarrow \partial(f).$$
	
\begin{figure}[h]
		\begin{center}
			\begin{overpic}[width=0.6\linewidth]{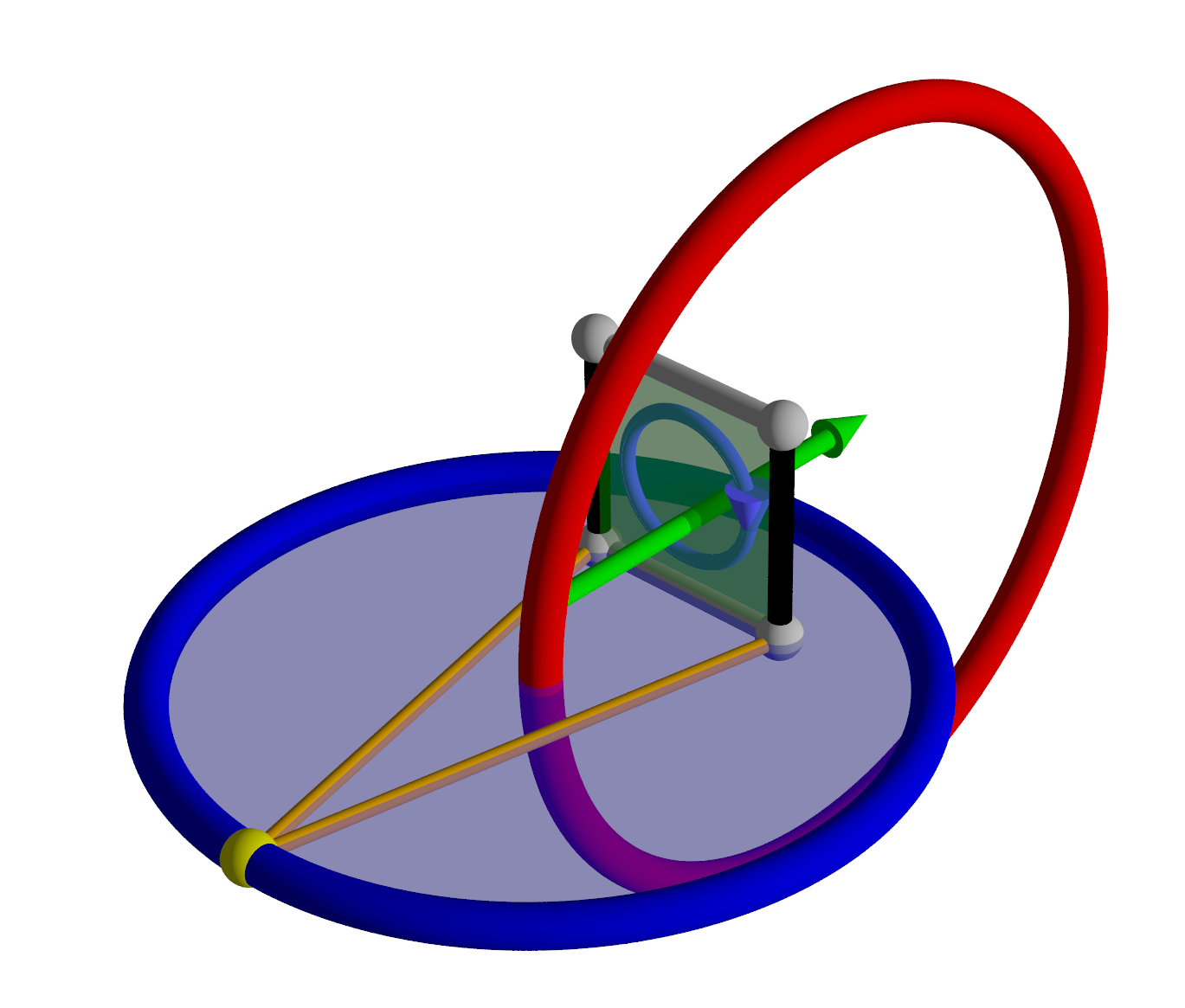}
				
			\end{overpic}
			\caption{When the flux label associated to the linking string lies in the image of $\partial$, it is possible to remove the linking string by applying a blob ribbon operator (represented here by the green arrow), which also produces blob excitations at the ends of the ribbon (not shown here).}
			\label{linking_remove_string}
		\end{center}
	\end{figure}

	Now consider applying a blob ribbon operator of label $f$ which passes along the linking string, and so through the plaquette $p$. We take the orientation of the ribbon operator to point from the loop produced by $C^h(m)$ to the boundary of membrane $n$, along the green arrow in Figure \ref{linking_lattice_example}, as shown in Figure \ref{linking_remove_string}. Then the effect of this blob ribbon operator on the plaquette shown in Figure \ref{linking_remove_string} is to take the plaquette label $e_p$ to $e_p f^{-1}$, because the plaquette's orientation matches that of the ribbon operator (we will discuss the opposite case shortly). Then the effect of this on the plaquette holonomy $H_1(p) = \partial(e_p)g_p$ (where $g_p$ is the path label of the boundary of $p$) is to gain a factor of $\partial(f^{-1})$. This factor cancels with the $\partial(f)$ from the linking string, and so leaves the plaquette unexcited. In order to remove the linking string entirely however, this must be true for each plaquette along the string, including for plaquettes with the opposite orientation. The result given in Equation \ref{Equation_linking_plaquette_holonomy} assumed that the orientation of the plaquette matched the linking string. If we reverse the orientation of the plaquette then the group label associated to the boundary of the plaquette is inverted. This means that the factor gained by the plaquette holonomy due to the linking string is also inverted, and so the appropriate transformation is
	$$H_1(p) \rightarrow \partial(f)^{-1}.$$
	
	The action of the blob ribbon on the plaquette label is also inverted when we switch the orientation of the plaquette. Because the orientation of the plaquette is now anti-aligned with the blob ribbon operator, the action of the blob ribbon operator on the plaquette label $e_p$ is to take $e_p$ to $fe_p$. This contributes a factor of $\partial(f)$ to the plaquette holonomy, which cancels with the factor from the linking string. We therefore see that regardless of the orientation of the plaquette, the blob ribbon operator removes the linking string. The ribbon operator will however produce blob excitations at the two ends of the ribbon, i.e., at the ends of the old linking string (these can be regarded as attached to the two flux tubes).
	
\begin{figure}[h]
		\begin{center}
			\begin{overpic}[width=0.5\linewidth]{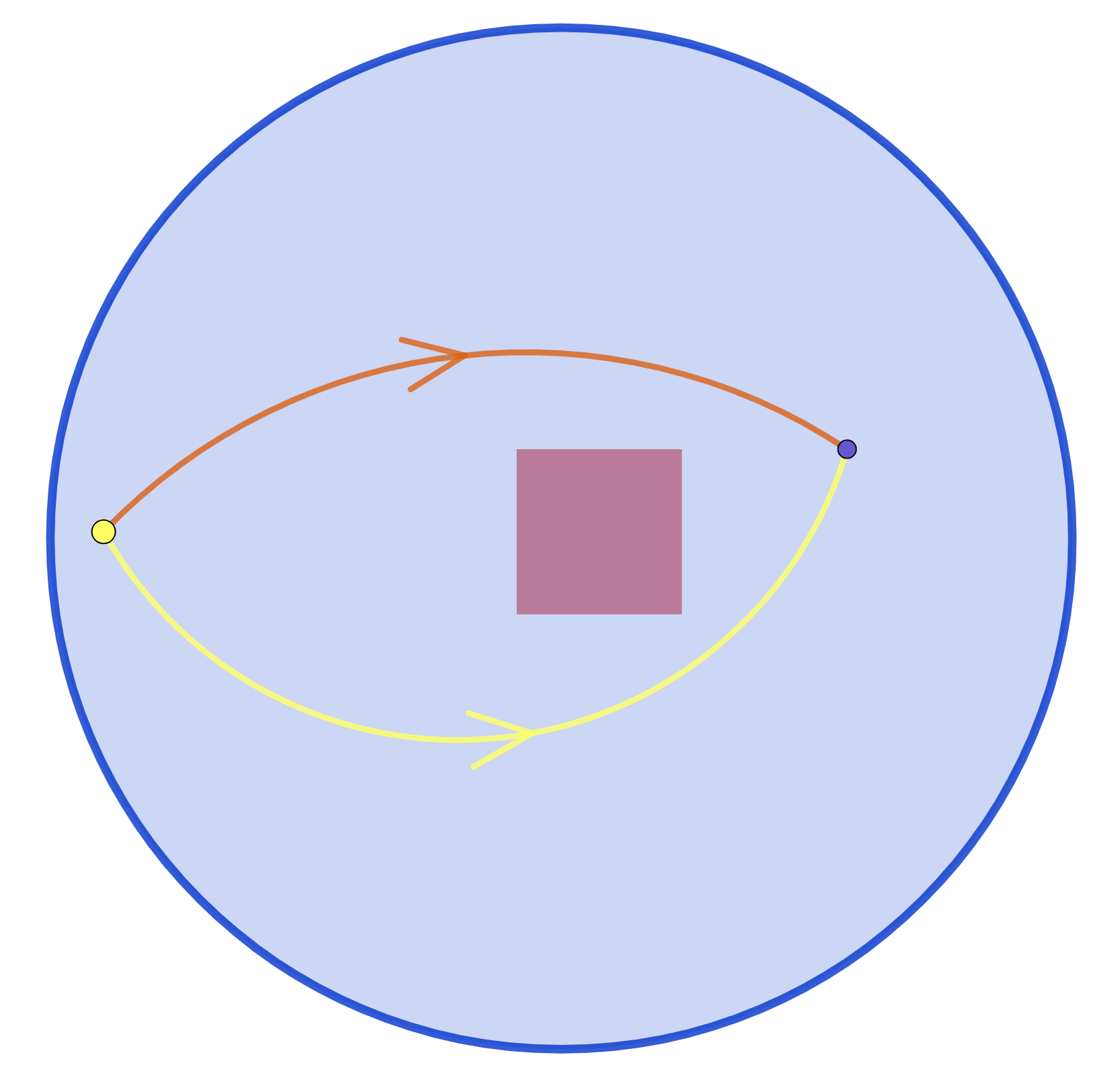}
				\put(12,48){$s.p$}
				\put(77,54){$v_i$}
				\put(50,67){$t_i$}
				\put(50,27){$t_i'$}
			\end{overpic}
			\caption{We wish to see what happens when we move the linking string, but first we need a way to do this. Starting from the situation shown in Figure \ref{linking_lattice_example}, we consider taking the path from the start-point (yellow) to a vertex $v_i$, and changing it from the (orange) path $t_i$ to the (yellow) one $t_i'$, which passes the other way around the flux (represented by the square). These two paths cannot be smoothly deformed into one another, without crossing the flux. As we will see, changing the path in this way has the effect of moving the linking string. }
			\label{linking_change_path}
		\end{center}
	\end{figure}

\begin{figure}[h]
		\begin{center}
			\begin{overpic}[width=0.9\linewidth]{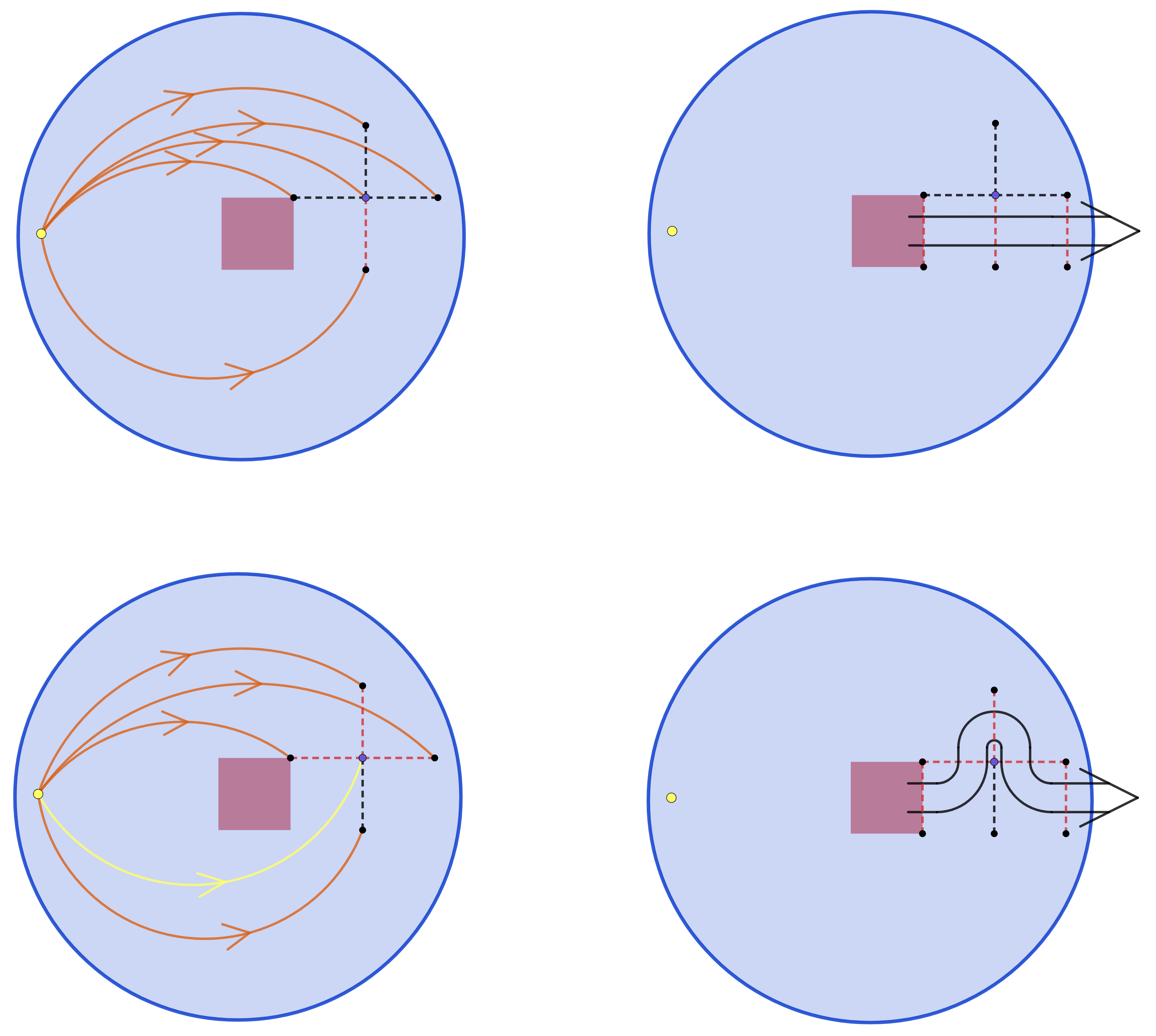}
				\put(20,43){\Huge $\downarrow$}
				\put(22,44){Change path}
				
				\put(74,43){\Huge $\downarrow$}
				\put(76,44){Change path}
				
				\put(32,71){$v_i$}
				\put(29,74.5){$t_i$}
				\put(32,22){$v_i$}
				\put(20,14){$t_i'$}
				
			\end{overpic}
			\caption{Changing the path to $v_i$ affects which of the adjacent plaquettes (represented in this top-down view by the dashed lines) are excited. Here excited plaquettes are red (or gray in grayscale) and unexcited plaquettes are black. Before we change the path (the situation shown in the top row), the path to vertex $v_i$ travels the same way around the flux as the paths to the other vertices above the flux, but the opposite way compared to the vertices below it. This leads to the linking string passing in a straight line rightwards from the flux, as shown in the top-right image (the linking string is represented by the double arrow). On the other hand, when we move the path to the vertex to pass the other way around the flux, the path instead agrees with the paths to the vertices below the flux. This leads to the plaquette below the vertex, which was previously excited by the linking string, being unexcited (represented by it being black). Instead the linking string passes around the vertex $v_i$, as shown in the bottom right image. If we wish to apply a blob ribbon operator to remove the linking string, it must follow this new path.}
			\label{linking_change_path_2}
		\end{center}
	\end{figure}

	In order to show that the linking string truly vanishes however, we should also show that its position becomes defined only up to deformations. That is, smoothly moving the position of the blob ribbon operator and the linking string together has no effect on the action of the combined membrane and ribbon operator. To do so, we must first understand how to deform the linking string. Recall that the linking string arises from a discontinuity in the paths from the start-point vertex to the other vertices on the membrane, such that the paths to adjacent vertices can pass the opposite way around the intruding flux tube (represented in red in Figure \ref{linking_remove_string} and previous figures). In the example shown in Figure \ref{linking_lattice_example}, some of the paths to the vertices pass above the flux tube (the intersection of the red flux tube with the membrane is represented by the red square) and some pass below it, and the linking string (represented by the green arrow) separates these types of vertices (at least, the ones to the right of the flux tube). However, we can consider disconnecting the path to one of these vertices, $v_i$, and passing it the other way around the flux tube instead, as shown in Figure \ref{linking_change_path}. We wish to consider what effect this has on the action on the edges attached to this vertex, and on the holonomies of the plaquettes surrounding it. Recall from earlier in this section that the action of the magnetic membrane operator excites a plaquette if the paths to the two vertices of the plaquette on the direct membrane travel opposite ways around the flux and so enclose the linked flux tube (when the paths are joined together by the base of the plaquette), as shown in Figure \ref{linking_potentially_excited_plaquette}. Considering Figure \ref{linking_change_path_2}, we can then see that changing the path to the vertex $v_i$ will swap which of the adjacent vertices have paths which travel the other way around the flux compared to the path to $v_i$. This in turn swaps which of the adjacent plaquettes are excited, therefore shifting the position of the linking string as shown in Figure \ref{linking_change_path_2}, by diverting it around the vertex. Then in order to remove the linking string, the blob ribbon operator that we apply must follow this new path. We are interested in how this change to the path to the vertex, and the associated change to the ribbon operator, changes the action of the combined membrane and ribbon operator. The original and new operators only differ in their action on the edges attached to the affected vertex, and on the adjacent plaquettes. We will therefore examine the action of the membrane and ribbon operators on these objects more closely.

	First, we consider the action of the membrane operator on the edge (or more generally edges) that are attached to the vertex $v_i$ and affected by the membrane operator. Denoting the original path to the vertex $v_i$ by $t_i$, and the new path by $t_i'$, the original action of the membrane operator on the edge (or edges) $i$ attached to this vertex is (assuming that the edge points away from the direct membrane)
	$$C^k(n):g_i = g(t_i)^{-1}kg(t_i) g_i,$$
	whereas the action of the new membrane operator is
	\begin{equation}
		C^k(n'):g_i = g(t_i')^{-1}kg(t_i')g_i, \label{Equation_linking_change_path_edge_action_1}
	\end{equation}
	where $n'$ refers to the membrane obtained from $n$ by changing the path to $v_i$ in this way. Now the two paths $t_i$ and $t_i'$ together enclose the flux (i.e., the closed path $t_i \cdot t_i^{\prime -1}$ encloses the linked flux tube), and so they satisfy
	\begin{equation}
		g(t_i)g(t_i')^{-1} = g(s.p(m)-s.p(n))^{-1}h^{\pm 1}g(s.p(m)-s.p(n)) \partial(e) \label{Equation_linking_change_path_closed_path_label}
	\end{equation}
	for some element $e \in E$, as stated in Equation \ref{Equation_linking_closed_path_label_3} (note that the orientation of the path $t_1 \cdot b^{-1}\cdot t_2^{-1} =s^{-1}$ in Equation \ref{Equation_linking_closed_path_label_3} is clockwise around the flux tube, just like the path $t_i \cdot t_i^{-1}$ in Equation \ref{Equation_linking_change_path_closed_path_label}). We can therefore write
	$$g(t_i')^{-1}=g(t_i)^{-1}g(s.p(m)-s.p(n))^{-1}h^{\pm 1}g(s.p(m)-s.p(n)) \partial(e)$$
	and use this in Equation \ref{Equation_linking_change_path_edge_action_1} to write the new action of the membrane operator as
	\begin{align*}
		C^k(n'):g_i &= g(t_i')^{-1}kg(t_i')g_i\\
		&= g(t_i)^{-1}g(s.p(m)-s.p(n))^{-1}h^{\pm 1}g(s.p(m)-s.p(n)) \partial(e) k \\
		& \hspace{0.5cm} \partial(e)^{-1} g(s.p(m)-s.p(n))^{-1}h^{\mp 1}g(s.p(m)-s.p(n)) g(t_i)g_i.
	\end{align*}
	We note that when $\rhd$ is trivial, $\partial(e)$ is always in the centre of $G$ for any $e \in E$ and so $\partial(e) k \partial(e)^{-1}=k$. This gives us
	\begin{align}
		C^k(n'):g_i &= g(t_i)^{-1}g(s.p(m)-s.p(n))^{-1}h^{\pm 1}g(s.p(m)-s.p(n)) k g(s.p(m)-s.p(n))^{-1}h^{\mp 1}g(s.p(m)-s.p(n)) g(t_i)g_i. \label{Equation_linking_change_path_edge_action_2}
	\end{align}
	
	Next, we recall that we have defined $f \in E$ such that 
	$$k g(s.p(m)-s.p(n))^{-1}h^{\pm 1}g(s.p(m)-s.p(n)) k^{-1} g(s.p(m)-s.p(n))^{-1}h^{ \mp 1}g(s.p(m)-s.p(n)) = \partial(f),$$
	with the existence of such an $f \in E$ being a prerequisite for being able to remove the linking string (and note that the $\pm 1$ are the same as in Equation \ref{Equation_linking_change_path_closed_path_label}). Then 
	$$g(s.p(m)-s.p(n))^{-1}h^{\pm 1}g(s.p(m)-s.p(n)) k^{-1} g(s.p(m)-s.p(n))^{-1}h^{ \mp 1}g(s.p(m)-s.p(n)) = k^{-1} \partial(f)$$
	and so
	\begin{align*}
		\partial(f)^{-1} k&=\big(g(s.p(m)-s.p(n))^{-1}h^{\pm 1}g(s.p(m)-s.p(n)) k^{-1} g(s.p(m)-s.p(n))^{-1}h^{ \mp 1}g(s.p(m)-s.p(n))\big)^{-1}\\
		&= g(s.p(m)-s.p(n))^{-1}h^{\pm 1}g(s.p(m)-s.p(n)) k g(s.p(m)-s.p(n))^{-1}h^{ \mp 1}g(s.p(m)-s.p(n)).\\
	\end{align*}
	Substituting this into Equation \ref{Equation_linking_change_path_edge_action_2}, we see that the new action of the membrane operator on the edge can be written as
	\begin{align}
		C^k(n'):g_i &= g(t_i')^{-1}kg(t_i')g_i \notag \\
		&= g(t_i)^{-1} \partial(f)^{-1} k g(t_i) g_i\notag \\
		&= \partial(f)^{-1} g(t_i)^{-1}kg(t_i)g_i, \label{Equation_linking_change_path_edge_action_3}
	\end{align}
	where in the last line we used the fact that $\partial(f)^{-1}$ is in the centre of $G$ to bring it to the front. This action differs from the action of the original membrane operator $C^k(n)$ only by this factor of $\partial(f)^{-1}$, which we note is equivalent to the factor we would gain from the action of an edge transform of label $f^{-1}$ on that edge. This suggests that the original and new operators are related by the action of an edge transform, which would be trivial for a state where the edge is unexcited (note that if there were multiple edges attached to $v_i$ and cut by the dual membrane, we would need to apply an edge transform on each such edge). For this to be true, however, the action on the plaquettes around the edge must also differ from the original action by the same edge transform. These plaquettes are only affected by the blob ribbon operator, so let us now consider the action of the ribbon operator on these plaquettes.

\begin{figure}[h]
		\begin{center}
			\begin{overpic}[width=\linewidth]{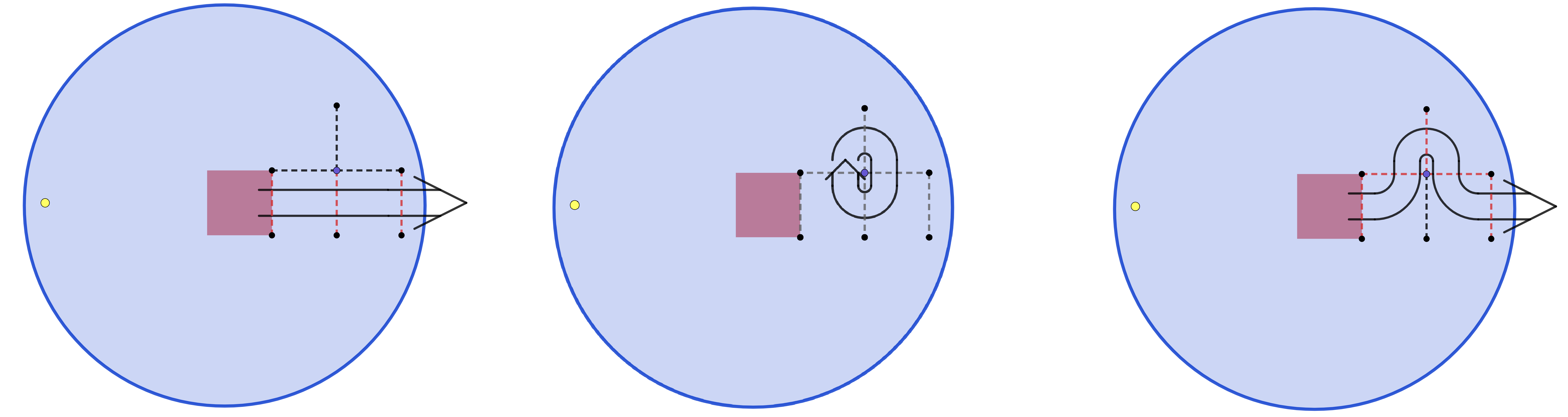}
				\put(32,12.5){\Huge $\cdot$}
				\put(63,12.5){\Huge $\rightarrow$}	
				
			\end{overpic}
			\caption{When we move the linking string, we must move the blob ribbon operator together with it. This can be done by applying a blob ribbon operator of the same label $f$ on a closed path around the edge $i$.}
			\label{linking_move_ribbon}
		\end{center}
	\end{figure}

	As shown in Figure \ref{linking_change_path_2}, before we move the linking string and blob ribbon operator, the ribbon operator passes through the plaquettes on one side of the edge, and after we move them, the ribbon operator passes through the plaquettes on the other side. The difference between these two actions is equivalent to a closed ribbon operator of label $f$ passing though all of the plaquettes around the edge in a clockwise direction, as shown in Figure \ref{linking_move_ribbon}. This is because two blob ribbon operators of the same label passing through the same plaquette with opposite directions will cancel. This means that the closed ribbon operator cancels with parts of the original ribbon operator. However, it also acts on plaquettes not affected by the original, therefore diverting the blob ribbon operator around the edge.

\begin{figure}[h]
		\begin{center}
			\begin{overpic}[width=0.5\linewidth]{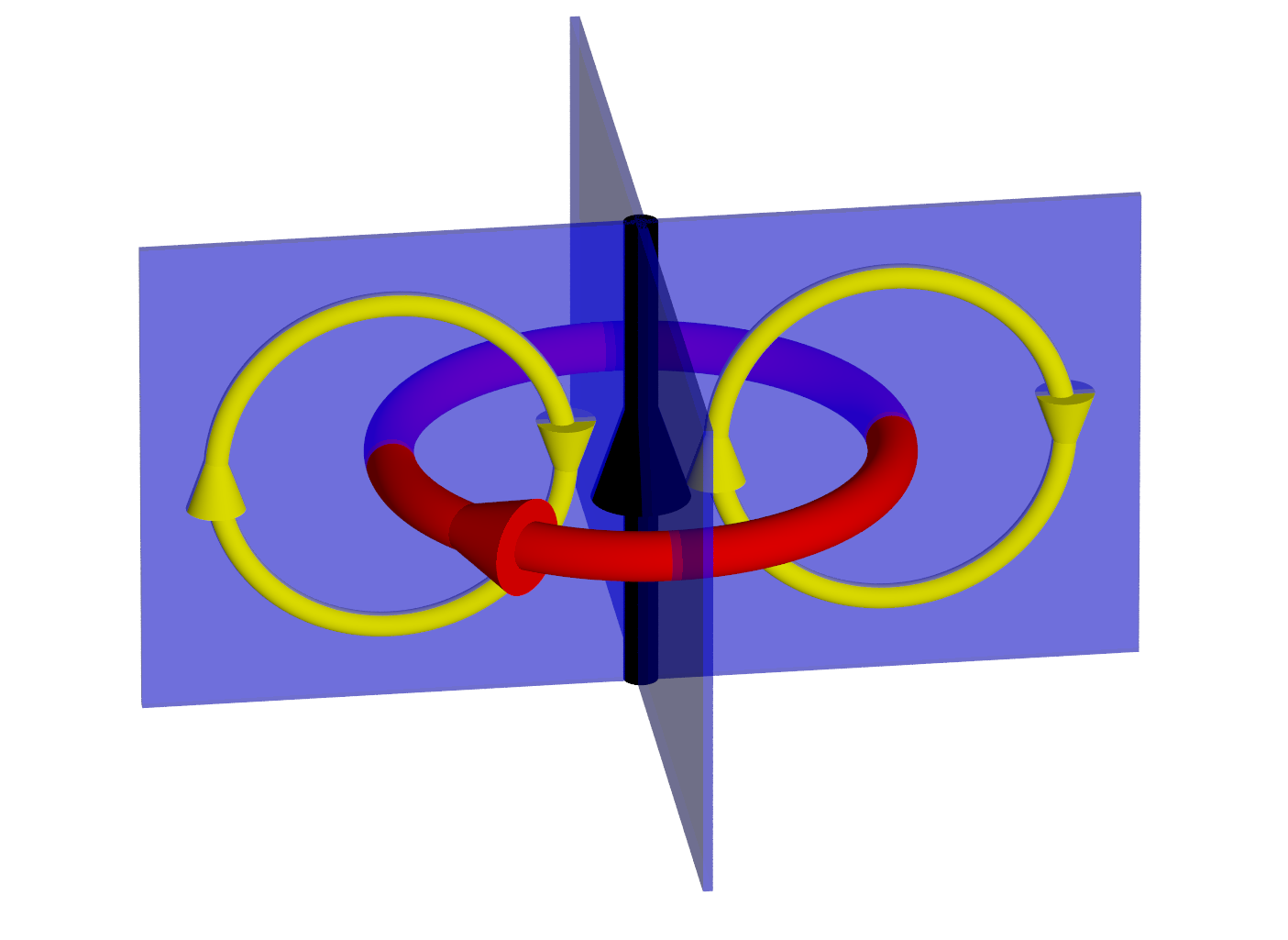}

			\end{overpic}
			\caption{The closed ribbon operator around the edge multiplies the labels of plaquettes oriented with the ribbon (such as the left one) by $f^{-1}$ and those oriented against it (such as the right one) by $f$, where the orientation of a plaquette is determined from its circulation (represented by the yellow arrows) using the right-hand rule. However, it can be seen that any plaquette oriented with the ribbon operator circulates against the upwards orientation of the edge (i.e., the part of the yellow arrow adjacent to the edge points against the edge) and any plaquette oriented against the ribbon circulates with the edge. This allows us to write the action of the closed ribbon operator on the plaquettes as the action of an edge transform $A_i^{f^{-1}}$.}
			\label{linking_closed_ribbon_around_edge}
		\end{center}
	\end{figure}

	Next we wish to show that the action of this closed blob ribbon operator is equivalent to that of an edge transform. The blob ribbon operator acts on a plaquette adjacent to the edge in a way that depends on the relative orientation of the plaquette and the ribbon. If the orientation of the plaquette (determined from the circulation by the right-hand rule) matches the orientation of the ribbon, then the plaquette label is multiplied by $f^{-1}$, and otherwise it is multiplied by $f$. From Figure \ref{linking_closed_ribbon_around_edge}, we see that the orientation of a plaquette around $i$ will match the orientation of the ribbon if the plaquette circulates downwards next to $i$. That is, the plaquette matches the orientation of the ribbon if the edge $i$ is aligned with the circulation of the plaquette at the edge. We can therefore see that the plaquette label is multiplied by $f$ if the plaquette is anti-aligned with the edge and multiplied by $f^{-1}$ if the plaquette is aligned with the edge. However, this is exactly the same action as an edge transform of label $f^{-1}$ on that edge (see Equation \ref{Equation_edge_transform_definition} in the main text, taking $\rhd$ to be trivial). Therefore, the additional closed blob ribbon operator needed to divert the blob ribbon operator acts in the same way as an edge transform $A_i^{f^{-1}}$ on the plaquettes. We also saw earlier that the additional action on the edge label $g_i$ associated to moving the linking string is equivalent to the action of the same edge transform on the edge label. Therefore, diverting the linking string and blob ribbon operator together is equivalent to the action of an edge transform, and so is trivial if the edge is unexcited. That is, the original membrane operator and the membrane operator with the deformed string act in the same way (at least in the absence of additional excitations other than the linking flux). We can then construct larger deformations of the linking string by repeating the process of changing the path to one of the vertices, each time leaving the action of the operators invariant. This tells us that the linking string really does disappear when we combine the string with the blob ribbon operator, in that its position becomes defined only up to smooth deformations.

	The fact that we can remove the linking string in this way is interesting for a couple of reasons. The first is that the linked magnetic loop excitations, which previously were confined by their mutual linking string, can move freely when we remove the string in this way. The second is that the blob excitations produced by the blob ribbon operator would usually be confined, but due to the cancellation of the confining string with the linking string, the blob excitations can move freely (provided that the fluxes move with them). Therefore, even though the individual excitations cannot move freely, the combined flux-blob system can, and in particular it can braid with other excitations normally, enabling new braiding relations that we cannot have without linking (because a confined excitation, like those that would be produced by the blob ribbon operator in the absence of the linking string, does not have topological braiding relations). In particular, we can move the flux-blob system through and around an $E$-valued membrane while remaining linked to the other flux (so this is an example of necklace or three-loop braiding), which results in the usual braiding relation between blob excitations and $E$-valued loops (i.e., for a blob ribbon operator labelled by element $f \in E$ and $E$-valued loop excitation labelled by irrep $\alpha$ of $E$, we gain a phase $\alpha(f)$), except that the blob excitations in question would usually be confined (and so would not have topological braiding).

	\subsection{Fake-flat case}
	\label{Section_braiding_fake-flat_appendix}
	Next we consider braiding in the case where we have a general crossed module, but we restrict our Hilbert space to the states satisfying fake-flatness on each plaquette. In this case, we do not have any magnetic excitations, because these correspond to violations of fake-flatness. In addition, the electric ribbon operators still commute with the $E$-valued membrane operators and blob ribbon operators, because the electric ribbon operator just measures the value of paths on the direct lattice, which are not affected by the action of these other operators. This means that we only have non-trivial braiding between the $E$-valued loop excitations and the blob excitations. This braiding is analogous to the braiding in the $\rhd$ trivial case, but it is complicated slightly due to the fact that the map $\rhd$ gives paths a way to act on plaquettes. Just as we did in Section \ref{Section_braiding_blobs_E_loops_tri_trivial} for the $\rhd$ trivial case, we consider acting on a ground state, first with an $E$-valued membrane operator $\delta(e, \hat{e}(m))$ that produces a loop excitation, then with a blob ribbon operator $B^f(t)$ that pushes a blob excitation through the loop. That is, we consider the state $B^f(t) \delta(e, \hat{e}(m))\ket{GS}$. In order to find the braiding relation between the blob excitation and $E$-valued loop excitation, we must compare this to the state $\delta(e, \hat{e}(m))B^f(t)\ket{GS}$ where we first move the blob excitation and then produce the loop excitation, as we described in Section \ref{Section_Loop_Blob_Braiding_Tri_Trivial} of the main text. To do so we must commute $\delta(e, \hat{e}(m))$ to the left of the expression $B^f(t) \delta(e, \hat{e}(m))\ket{GS}$.

	Unlike in the $\rhd$ trivial case, when we define the blob ribbon operator and $E$-valued membrane operator, we must define a start-point for each operator, and the choice of these start-points will affect the braiding relation. To see why this is the case, first consider the $E$-valued membrane operator. Whereas in the $\rhd$ trivial case, the surface label of $m$ is a simple product of the constituent plaquettes' elements (potentially with inverses, depending on orientation), when $\rhd$ is non-trivial we must take into account the base-points of each plaquette. These base-points must be moved to match the start-point of the surface in order to combine the plaquettes into a single surface. This means that the surface label is given by
	\begin{equation}
		e(m)= \prod_{\text{plaquettes } p \in m} g(s.p(m)-v_0(p)) \rhd e_p^{\sigma_p}, \label{Equation_total_surface}
	\end{equation}
	where $\sigma_p$ is $+1$ if the orientation of the plaquette matches that of the membrane and $-1$ otherwise. We see that this element depends on certain path elements $g(s.p(m)-v_0(p))$ from the start-point, $s.p(m)$, of the membrane to the base-point, $v_0(p)$, of the plaquette $p$. The action of the blob ribbon operator similarly depends on paths on the lattice, and its own start-point. For a plaquette $p$ pierced by the ribbon, the action of the blob ribbon operator on the plaquette label $e_p$ is
	\begin{equation}
		B^f(t): e_p= \begin{cases} e_q [g(s.p(t)-v_0(p))^{-1} \rhd f^{-1}] & \text{if the orientation of $p$ matches that of $t$ }\\ [g(s.p(t)-v_0(p))^{-1} \rhd f] e_p & \text{otherwise,}\end{cases}
		\label{Equation_blob_ribbon_on_plaquette_1}
	\end{equation}
	which depends on the path from the start-point of $t$, $s.p(t)$ to the base-point $v_0(q)$ of the plaquette $q$.

\begin{figure}[h]
		\begin{center}
			\begin{overpic}[width=0.6\linewidth]{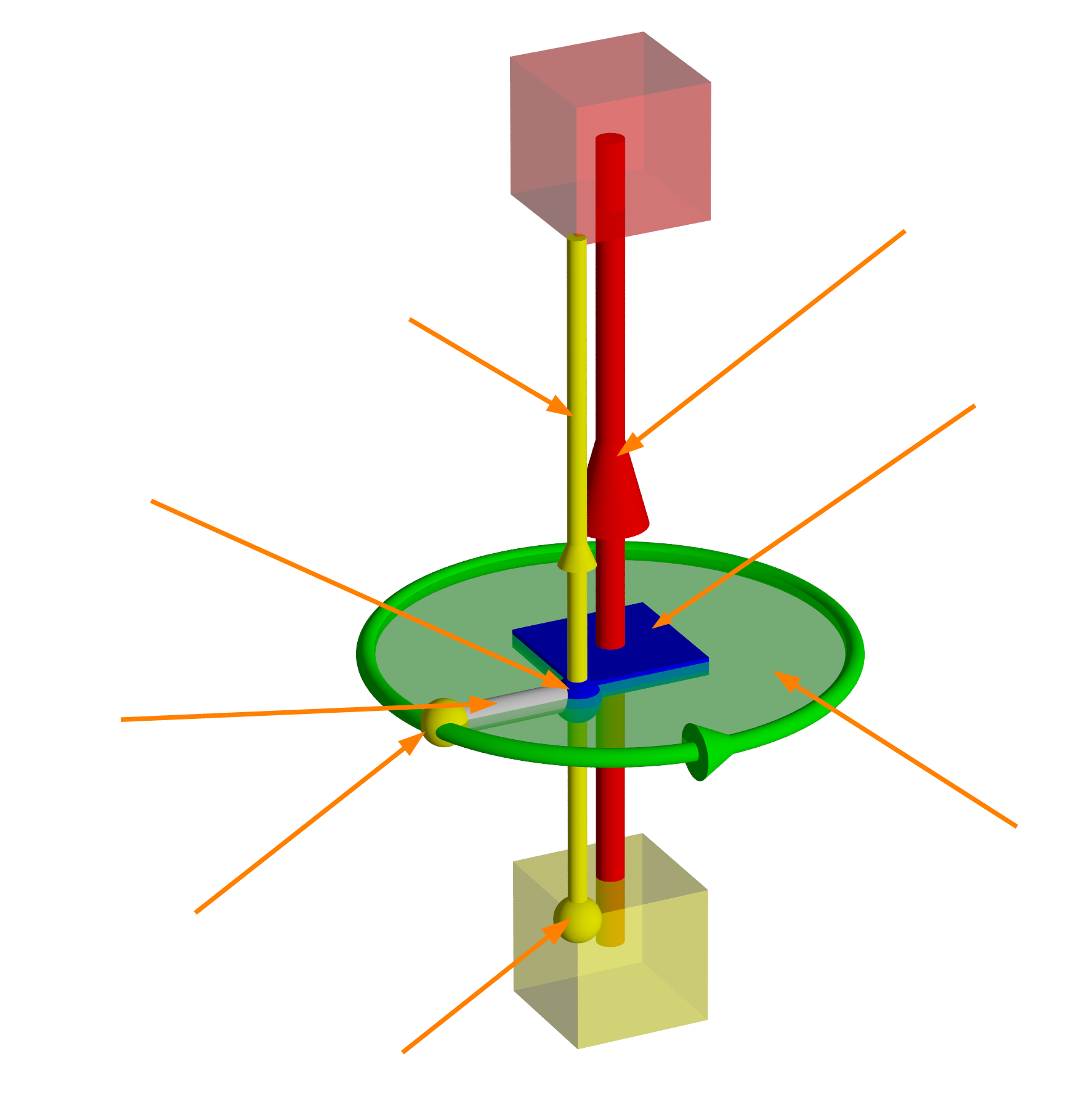}
				\put(83,80){dual path of $t$}
				\put(20,73){direct path of $t$}
				\put(6,55){$v_0(q)$}
				\put(90,64){plaquette $q$}
				\put(-11,34){$s.p(m)-v_0(q)$}
				\put(12,15){$s.p(m)$}
				\put(85,22){membrane $m$}
				\put(29,4){$s.p(t)$}
			\end{overpic}
			\caption{We consider braiding a blob excitation up through an $E$-valued loop excitation, where the $E$-valued excitation is oriented upwards. To do this, we apply an $E$-valued membrane operator $L^e(m)$, followed by a blob ribbon operator $B^f(t)$ piercing the membrane. The ribbon $t$ pierces the membrane at a plaquette $q$. When $\rhd$ is non-trivial, we must keep track of the start-points of the ribbon and membrane operators (yellow spheres). We find that the braiding result depends on a path element corresponding to a path between the two start-points. This path, $s.p(m)-s.p(t)$ passes from $s.p(m)$ to $v_0(q)$ along the grey path, before travelling to $s.p(t)$ backwards along the direct path of $t$ (yellow path).}
			\label{blob_through_E_mem_tri_non_trivial_appendix}
		\end{center}
	\end{figure}

	In order to determine the commutation relation between the ribbon and membrane operator, we must find how the ribbon operator affects the surface label measured by the membrane operator. Let the ribbon operator intersect the membrane at plaquette $q$. Then the ribbon operator affects the surface label through the contribution of plaquette $q$, $ g(s.p(m)-v_0(q)) \rhd e_q^{\sigma_q}$, to the surface label. If the path $t$ points along the orientation of the membrane $m$, as in the example in Figure \ref{blob_through_E_mem_tri_non_trivial_appendix}, then the orientation of $q$ matching the orientation of the ribbon implies that it also matches the orientation of the membrane, and so $\sigma_q=1$, whereas the orientation of the plaquette being opposed to the ribbon implies that it also opposes the orientation of the membrane, and so $\sigma_q=-1$. This inversion of the contribution of the plaquette to the surface label based on the relative orientation of the plaquette and membrane cancels the inverse in the effect of the blob ribbon operator based on the relative orientation of the plaquette and ribbon from Equation \ref{Equation_blob_ribbon_on_plaquette_1}. Therefore, we can write the effect of the blob ribbon operator on plaquette $q$ from Equation \ref{Equation_blob_ribbon_on_plaquette_1} as
	\begin{equation}
		B^f(t): e_q^{\sigma_q}= e_q^{\sigma_q} [g(s.p(t)-v_0(q))^{-1} \rhd f^{-1}].
		\label{Equation_blob_ribbon_on_plaquette_2}
	\end{equation}
	
	This means that the action of $B^f(t)$ on the contribution $g(s.p(m)-v_0(q)) \rhd e_q^{\sigma_q}$ of $q$ to the surface label (see Equation \ref{Equation_total_surface}) is
	\begin{align}
		B^f(t): g(s.p(m)-v_0(q)) \rhd e_q^{\sigma_q}&= g(s.p(m)-v_0(q)) \rhd (e_q^{\sigma_q}[g(s.p(t)-v_0(q))^{-1} \rhd f^{-1}] ) \notag \\
		&= [g(s.p(m)-v_0(q)) \rhd e_q^{\sigma_q}] [(g(s.p(m)-v_0(q))g(s.p(t)-v_0(q))^{-1}) \rhd f^{-1}] \notag \\
		&= [g(s.p(m)-v_0(q)) \rhd e_q^{\sigma_q}] [g(s.p(m)-s.p(t)) \rhd f^{-1}]. \label{Equation_blob_ribbon_on_plaquette_3}
	\end{align}

	We can then use Equation \ref{Equation_blob_ribbon_on_plaquette_3} to determine the action of $B^f(t)$ on the overall surface element $e(m)$. Because $E$ is not necessarily Abelian when $\rhd$ is non-trivial, we cannot simply separate the term corresponding to plaquette $q$ from the product over plaquettes in $m$. However, because we restrict to the fake-flat case, we are only allowed blob ribbon operators whose labels lie in the kernel of $\partial$, as described in Section \ref{Section_Blob_Ribbon_Fake_Flat}. Elements of $E$ that are in the kernel of $\partial$ are also in the centre of $E$, because for an element $e_k$ in the kernel and any element $x \in E$, the second Peiffer condition (Equation \ref{Equation_Peiffer_2} in the main text) becomes $x =e_k x e_k^{-1}$. This means that the label $f$ of the blob ribbon operator must be in the centre of $E$. Furthermore, given that the element $f$ is in the kernel of $\partial$, so is any expression of the form $g \rhd f$ (from the first Peiffer condition, Equation \ref{Equation_Peiffer_1} in the main text). This means that the factor $[g(s.p(m)-s.p(t)) \rhd f^{-1}]$ is in the centre of $E$ and so we can freely pull it to the front of $e(m)$. The action of the blob ribbon operator on the surface label is then given by
	\begin{align}
		B^f(t):e(m) &= B^f(t): \prod_{\text{plaquettes } p \in m} g(s.p(m)-v_0(p)) \rhd e_p^{\sigma_p} \notag \\
		&= \big[ \prod_{x \in m \text{ before }q} g(s.p(m)-v_0(x)) \rhd e_x^{\sigma_x}\big] [g(s.p(m)-v_0(q)) \rhd e_q^{\sigma_q}] [g(s.p(m)-s.p(t)) \rhd f^{-1}] \notag \\
		& \hspace{1cm} \big[\prod_{y \in m \text{ after }q}g(s.p(m)-v_0(y)) \rhd e_y^{\sigma_y}\big] \notag \\
		&=[g(s.p(m)-s.p(t)) \rhd f^{-1}]\big[\prod_{x \in m \text{ before }q}g(s.p(m)-v_0(x)) \rhd e_x^{\sigma_x}\big] [g(s.p(m)-v_0(q)) \rhd e_q^{\sigma_q}]\notag \\
		& \hspace{1cm} \big[\prod_{y \in m \text{ after }q}g(s.p(m)-v_0(y)) \rhd e_y^{\sigma_y}\big]\notag \\
		&=[g(s.p(m)-s.p(t)) \rhd f^{-1}] e(m). \label{Equation_blob_ribbon_on_surface_1}
	\end{align}
	Therefore, the commutation relation between the membrane and ribbon operator is
	$$\delta(e, \hat{e}(m))B^f(t) \ket{GS} =B^f(t) \delta(e, [g(s.p(m)-s.p(t)) \rhd f^{-1}] \hat{e}(m)) \ket{GS},$$
	and so
	\begin{align}
		B^f(t)\delta(e, \hat{e}(m))\ket{GS} &= \delta\big(e, [g(s.p(m)-s.p(t)) \rhd f^{-1}]^{-1} \hat{e}(m)\big) B^f(t) \ket{GS} \notag \\
		&= \delta\big([g(s.p(m)-s.p(t)) \rhd f^{-1}] e, \hat{e}(m)\big) B^f(t) \ket{GS}. \label{Equation_blob_ribbon_on_surface_2}
	\end{align}
	
	Just as we did in the $\rhd$ trivial case, it is instructive to consider an $E$-valued membrane operator of the form 
	$$L^{R,a,b}(m)= \sum_{e \in E} [D^R(e)]_{ab} \delta(e, \hat{e}(m)),$$
	where $R$ is an irrep of $E$ and $D^R(e)$ is the matrix representation for element $e$ (a matrix rather than a phase as in the $\rhd$ trivial case, because $E$ may be non-Abelian). Then
	\begin{align}
		B^f(t) L^{R,a,b}(m)\ket{GS} &= B^f(t) \sum_{e \in E} [D^R(e)]_{ab} \delta(e, \hat{e}(m))\ket{GS} \notag \\
		&=\sum_{e \in E} [D^R(e)]_{ab} \delta([g(s.p(m)-s.p(t)) \rhd f^{-1}] e, \hat{e}(m)) B^f(t) \ket{GS} \notag \\
		&=\sum_{e' =[g(s.p(m)-s.p(t)) \rhd f] e \in E} \hspace{-1cm} [D^R([g(s.p(m)-s.p(t)) \rhd f] e')]_{ab} \delta(e', \hat{e}(m)) B^f(t) \ket{GS} \notag \\
		&=\sum_{c=1}^{|R|} [D^R([g(s.p(m)-s.p(t)) \rhd f])]_{ac} \sum_{e' \in E} [D^R(e')]_{cb} \delta(e', \hat{e}(m)) B^f(t) \ket{GS} \notag \\
		&=\sum_{c=1}^{|R|} [D^R([g(s.p(m)-s.p(t)) \rhd f])]_{ac} L^{R,c,b}(m) B^f(t) \ket{GS}.
		\label{Equation_blob_ribbon_on_surface_3}
	\end{align}
	
\begin{figure}[h]
		\begin{center}
			\begin{overpic}[width=0.75\linewidth]{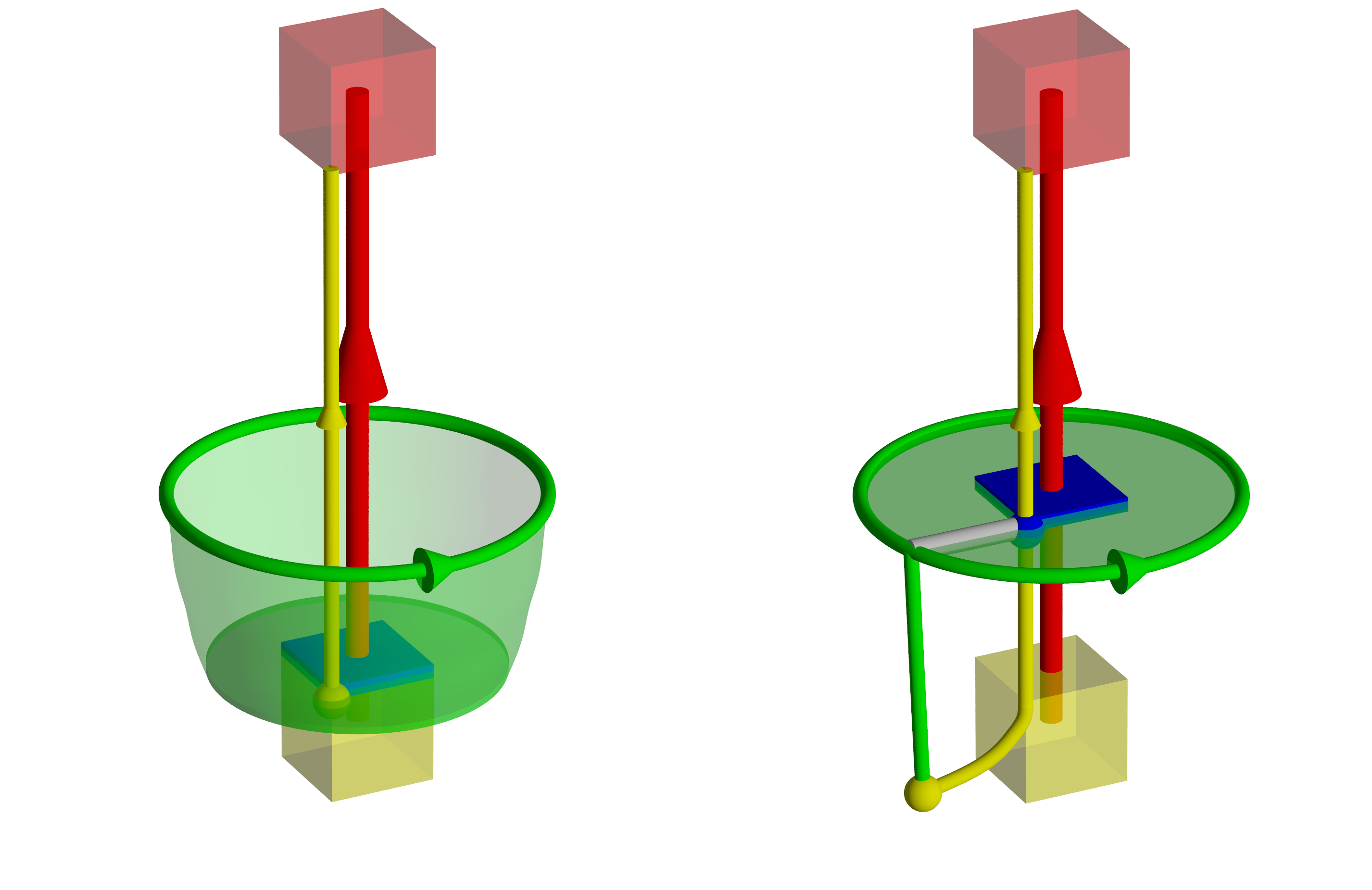}
				
			\end{overpic}
			\caption{The braiding relation between the blob excitation and $E$-valued loop excitation simplifies when the start-points of the ribbon and membrane operators are the same. Here we show two examples of such a situation. In the left image we see an example where the membrane touches the blob at the start of the ribbon. In the right figure we see an example where the start-point of the membrane is away from the membrane and the start-point of the ribbon is away from the starting blob. In either case, the path between the start-points that appears in the braiding relation is closed and contractible (the path could even be trivial, as in the left example), and so has a path element belonging to $\partial(E)$. The closed path element therefore acts trivially on elements of the kernel of $\partial$, such as the label of the blob ribbon operator, when acting via $\rhd$.}
			\label{blob_through_E_membrane_same_sp_two_examples}
		\end{center}
	\end{figure}

	It seems like this braiding results in mixing of the $E$-valued membrane operators labelled by the same irrep $R$ but different matrix indices $c,b$. However, we can simplify the braiding relation by noting that $f$ is in the kernel of $\partial$ and so is in the centre of $E$. This implies that $g(s.p(m)-s.p(t)) \rhd f$ is also in the kernel of $\partial$ (because $\partial(g(s.p(m)-s.p(t)) \rhd f) = g(s.p(m)-s.p(t)) \partial(f) g(s.p(m)-s.p(t))^{-1} =1_G$ from the first Peiffer condition, Equation \ref{Equation_Peiffer_1} in the main text), and so is in the centre of $E$. Schur's lemma implies that the (unitary) matrix representation of an element of the centre of the group is proportional to the identity matrix. This means that $[D^R([g(s.p(m)-s.p(t)) \rhd f])]_{ac}$ can be written as
	$$[D^R([g(s.p(m)-s.p(t)) \rhd f])]_{ac} = \delta_{ac} [D^R([g(s.p(m)-s.p(t)) \rhd f])]_{11}$$
	where the index 1 on the matrix is arbitrary. Then the braiding relation Equation \ref{Equation_blob_ribbon_on_surface_3} simplifies to
	\begin{align}
		B^f(t) L^{R,a,b}(m)\ket{GS} &= [D^R([g(s.p(m)-s.p(t)) \rhd f])]_{11} L^{R,a,b}(m) B^f(t) \ket{GS}.
	\end{align}
	which is simply the accumulation of a phase $[D^R([g(s.p(m)-s.p(t)) \rhd f])]_{11}$ (it must be a phase because the representation is unitary). However, the phase depends on the operator $g(s.p(m)-s.p(t))$, and so is not generally well-defined in a given state. If we consider the case where the start-points of $m$ and $t$ are the same, as in the examples shown in Figure \ref{blob_through_E_membrane_same_sp_two_examples}, this relation becomes simplified. Because $(s.p(m)-s.p(t))$ is then a closed path, $g(s.p(m)-s.p(t))$ must lie in the image of $\partial$, due to fake-flatness (provided that the path is contractible). In that case $g(s.p(m)-s.p(t)) \rhd f= \partial(x) \rhd f$ for some $x \in E$. However, one of our Peiffer conditions for the crossed module is that $\partial(x)\rhd f = xfx^{-1}$ for all $x$ and $f \in E$. Then, because $f$ is in the kernel of $\partial$ and so is in the centre of $E$, we have $xfx^{-1}=f$. This means that our braiding relation becomes
	\begin{equation}
		B^f(t) L^{R,a,b}(m)\ket{GS} = [D^R(f)]_{11} L^{R,a,b}(m) B^f(t) \ket{GS}, \label{Equation_blob_ribbon_on_surface_4}
	\end{equation}
	which no longer has an operator dependence in the braiding coefficients.
	
	The braiding relations given by Equations \ref{Equation_blob_ribbon_on_surface_2} through \ref{Equation_blob_ribbon_on_surface_4} hold when the path of the ribbon operator is aligned with the membrane orientation. In the case where the path is instead aligned against the orientation of the membrane, we must replace $f$ with $f^{-1}$ in the braiding relations. That is, the action of the ribbon operator on the plaquette $q$ through which it pierces the membrane $m$ is (c.f., Equation \ref{Equation_blob_ribbon_on_plaquette_2})
	\begin{equation}
		B^f(t): e_q^{\sigma_q}= [g(s.p(t)-v_0(q))^{-1} \rhd f] e_q^{\sigma_q}. \label{Equation_blob_ribbon_on_plaquette_reverse_1}
	\end{equation}
	
	Following the argument for the previous orientation, we then have the following commutation relation between the $E$-valued membrane labelled by the group element $e$ and the ribbon operator labelled by $f$:
	\begin{align}
		B^f(t)\delta(e, \hat{e}(m))\ket{GS} &= \delta(e, [g(s.p(m)-s.p(t)) \rhd f]^{-1} \hat{e}(m)) B^f(t) \ket{GS} \notag \\
		&= \delta([g(s.p(m)-s.p(t)) \rhd f] e, \hat{e}(m)) B^f(t) \ket{GS}. \label{Equation_blob_ribbon_on_surface_reverse_1}
	\end{align}
	For an $E$-valued membrane operator labelled by an irrep $R$ and matrix indices $a,b$ we see that
	\begin{equation}
		B^f(t) L^{R,a,b}(m)\ket{GS} =[D^R([g(s.p(m)-s.p(t)) \rhd f^{-1}])]_{11} L^{R,a,b}(m) B^f(t) \ket{GS}. \label{Equation_blob_ribbon_on_surface_reverse_2}
	\end{equation}
	Finally, if we take the start-points of the membrane and ribbon operators to be the same, we obtain
	\begin{equation}
		B^f(t) L^{R,a,b}(m)\ket{GS} = [D^R(f^{-1})]_{11} L^{R,a,b}(m) B^f(t) \ket{GS}. \label{Equation_blob_ribbon_on_surface_reverse_3}
	\end{equation}
	
	\subsection{The case where $\partial \rightarrow $ centre($G$) and $E$ is Abelian}
	\label{Section_braiding_higher_flux}
	
	In the case where $\rhd$ is non-trivial, the excitations produced by the magnetic membrane operators have rich braiding properties. Indeed, this class of excitations can braid non-trivially with every other class of excitation. In this section, we will derive the braiding relations of this class of excitation, in the case where $\partial$ maps to the centre of $G$ and $E$ is Abelian, but $\rhd$ is non-trivial. In order to consider the braiding properties of the flux tubes produced by the magnetic membrane operators, it is convenient to combine the magnetic membrane operator with an $E$-valued membrane operator, to obtain a ``higher-flux" membrane operator, as we mentioned in Section \ref{Section_3D_Braiding_Central} of the main text. We will start by further motivating this combination, before we consider the braiding relations of the resulting excitations.

\begin{figure}[h]
		\begin{center}
			\includegraphics[width=0.5\linewidth]{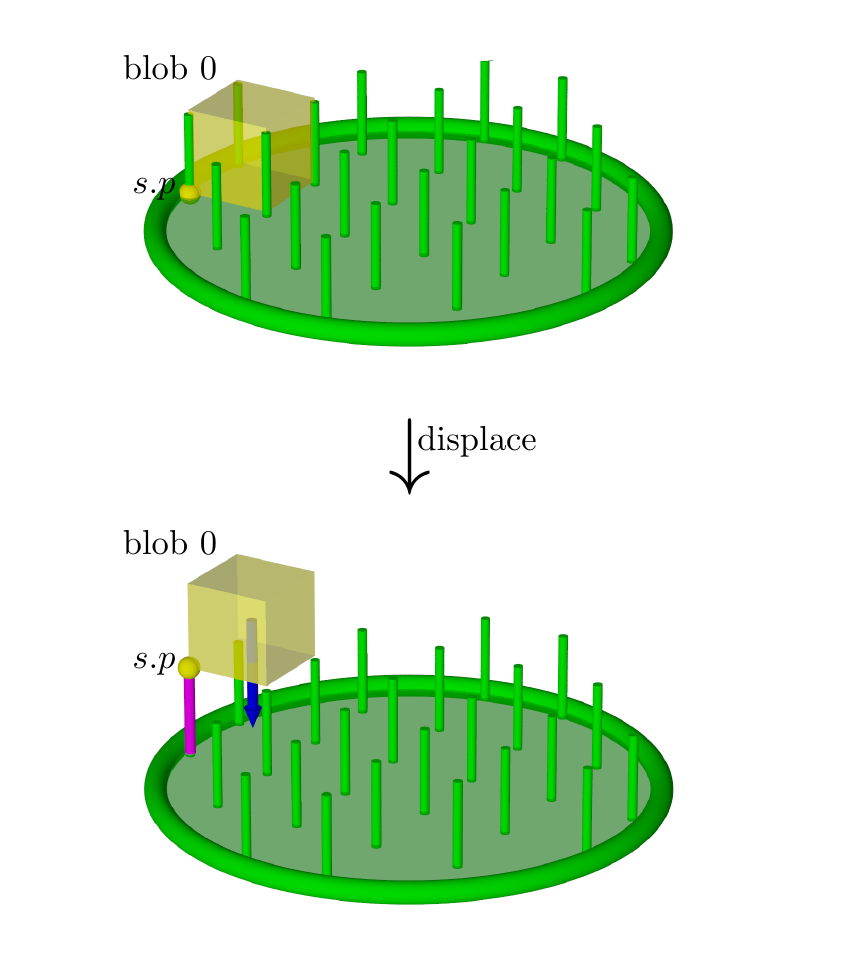}
			\caption{(Copy of Figure \ref{higher_flux_displacement_main_text} from the main text) Rather than place blob 0 and the start-point (represented by the yellow cube and sphere respectively) of the membrane operator on the direct membrane (green) itself, as in the upper image, we will always choose them to be slightly away from the membrane, as shown in the lower image. Then blob 0 and the start-point are on the other side of the edges cut by the dual membrane (where the edges are represented by the green cylinders). This allows us to deform the membrane away from the start-point and blob 0 (downwards in the figure) using the topological property of the magnetic and $E$-valued membrane operators which make up the higher-flux membrane operator.}
			\label{higher_flux_displacement_appendix}
		\end{center}
	\end{figure}

	We showed in Section \ref{Section_Magnetic_Tri_Non_Trivial} (see Equation \ref{Equation_magnetic_blob_0}) that the privileged blob, blob 0, of a magnetic membrane operator $C^h_T(m)$ obtains a 2-holonomy of $[h\rhd \hat{e}(m)^{-1}] \hat{e}(m)$, where $\hat{e}(m)$ is the total surface label of the direct membrane of the membrane operator. However, $\hat{e}(m)$ is an operator, so its value depends on the state on which we act. Furthermore, unless the membrane $m$ is closed, the ground state is not an eigenstate of the operator $\hat{e}(m)$ and so the 2-holonomy of blob 0 is not fixed after we act with the magnetic membrane operator, even when acting on the ground state. This is significant because the non-trivial 2-flux of blob 0 should be balanced by a non-trivial 2-flux associated to the loop excitation, as we will see in more detail shortly, which we expect to be important for the braiding relations. If the loop excitation carries an indefinite 2-flux, we might not expect to get well-defined braiding relations. We can give the loop excitation a definite 2-flux by acting first with an $E$-valued membrane $\delta(\hat{e}(m),e_m)$ to fix the surface value of the direct membrane, before we apply the magnetic membrane operator. We then consider the product of the two membrane operators as a single membrane operator, which we call the higher-flux membrane operator, given by $C^{h,e_m}_T(m)= C^h_T(m)\delta(\hat{e}(m),e_m)$. Combining the $E$-valued membrane operator with the magnetic membrane operator in this way does not violate any additional energy terms (compared to the ones that can be excited by a generic magnetic membrane operator), because the $E$-valued membrane may only excite the start-point and boundary edges of the direct membrane of $m$, which may already be excited by $C^h_T(m)$ (they are usually left in a state that is neither definitely excited nor definitely unexcited). We then ask if this combined operator $C^{h,e_m}_T(m) = C^h_T(m) \delta(\hat{e}(m),e_m)$ is topological. The two operators $C^h_T(m)$ and $\delta(\hat{e}(m),e_m)$ are topological, as we showed in Section \ref{Section_topological_membrane_operators}. However, recall that being topological means that we can deform the membranes of each operator, provided that the membrane does not cross any existing excitations. If we first act with $\delta(\hat{e}(m),e_m)$, which may produce an excitation at the start-point of $m$, then we cannot freely deform $C^h_T(m)$ through this point. When the start-point is on the membrane, as in the upper image in Figure \ref{higher_flux_displacement_appendix}, this prevents us from deforming $m$ away from the start-point. Mathematically, we know that in order to deform $C^h_T(m)$ we must apply vertex transforms on the vertices attached to edges cut by the dual membrane, but a vertex transform at the start-point does not commute with $\delta(\hat{e}(m),e_m)$ (it would induce an $\rhd$ action on the surface label), so applying the transforms would change the label $e_m$. In order to fix this, we choose to displace the start-point and blob 0 of the magnetic membrane operator slightly away from the membrane, as shown in the lower image in Figure \ref{higher_flux_displacement_appendix}. We note that we still cannot deform the membrane over its own start-point, but displacing the start-point makes it possible to deform the membrane away from the start-point. We also note that this change does not affect the pure magnetic membrane operator $C^h_T(m) = \sum_{e_m \in E} C^{h,e_m}_T(m)$, because if we start with the start-point on the membrane we can freely deform the membrane away from the start-point using the topological property of the magnetic membrane operator. We will use this adjustment of displacing the start-point and blob 0 away from the membrane throughout Section \ref{Section_braiding_higher_flux}.
	
	Having made this small change to the membrane, we wish to see what 2-flux blob 0 acquires due to the action of the higher-flux membrane operator, in terms of the labels $e_m$ and $h$ of the membrane operator. That is, we want to consider
	$$H_2(\text{blob 0}(m)) C^{h,e_m}_T(m) \ket{GS},$$
	where $H_2(\text{blob 0}(m))$ is the surface label (2-holonomy) of blob 0 of $m$. In order to find this, we split $C^{h,e_m}_T(m)$ into its constituent parts. We have
	$$C^{h,e_m}_T(m) = C^h_{\rhd}(m) \big[ \prod_{p \in m} B^{f(p)}(\text{blob 0}(m) \rightarrow \text{blob }p) \big] \delta( \hat{e}(m),e_m).$$
	
	Now that we have displaced blob 0 from the membrane itself, none of the plaquettes or edges on the boundary of blob 0 are cut by the dual membrane of $m$. Therefore, $H_2(\text{blob 0}(m))$ commutes with $C^h_{\rhd}(m)$. This tells us that
	\begin{align*}
		H_2( \text{blob 0}(m)) C^{h,e_m}_T(m) \ket{GS} &= H_2( \text{blob 0}(m)) C^h_{\rhd}(m) \big[\prod_{p \in m} B^{f(p)}(\text{blob 0}(m) \rightarrow \text{blob }p) \big]\delta( \hat{e}(m),e_m) \ket{GS}\\
		&= C^h_{\rhd}(m) H_2( \text{blob 0}(m)) \big[\prod_{p \in m} B^{f(p)}(\text{blob 0}(m) \rightarrow \text{blob }p) \big] \delta( \hat{e}(m),e_m) \ket{GS}.
	\end{align*}
	
	On the other hand, the 2-holonomy of blob 0 is affected by the action of the blob ribbon operators $$B^{f(p)}(\text{blob 0}(m) \rightarrow \text{blob }p).$$ Now that we have displaced blob 0 from $m$, all of these blob ribbon operators start in blob 0 and leave it (compared to the case where blob 0 is on $m$, where all of the ribbon operators leave blob 0 except those corresponding to the base of blob 0). The start-point of these ribbons is $s.p(m)$, which we also take as the base-point for blob 0. From Section \ref{Section_Magnetic_Tri_Nontrivial_Commutation} (see Equation \ref{Equation_magnetic_blob_ribbon_blob_0}), we know that this leads to the commutation relation
	$$H_2( \text{blob 0}(m))B^{f(p)}(\text{blob 0}(m) \rightarrow \text{blob }p) = B^{f(p)}(\text{blob 0}(m) \rightarrow \text{blob }p) H_2( \text{blob 0}(m)) f(p)^{-1},$$
	reflecting the fact that each blob ribbon operator removes a 2-flux of $f(p)$ from blob 0. Then we have
	\begin{align*}
		H_2( \text{blob 0}(m))& C^{h,e_m}_T(m) \ket{GS}\\
		&= C^h_{\rhd}(m) \big[\prod_{p \in m} B^{f(p)}(\text{blob 0}(m) \rightarrow \text{blob }p) \big] H_2( \text{blob 0}(m)) [\prod_{p \in m} f(p)^{-1}]\delta( \hat{e}(m),e_m) \ket{GS}.
	\end{align*}
	Here the label of each blob ribbon operator, $f(p)$, is given by
	$$f(p) = [g(s.p-v_0(p)) \rhd \hat{e}_p] [(h^{-1} g(s.p-v_0(p))) \rhd \hat{e}_p^{-1}],$$
	where $\hat{e}_p$ is the label of plaquette $p$, assuming that it is oriented away from the dual membrane of $m$ (otherwise we must replace $\hat{e}_p$ with its inverse). This means that
	\begin{align*}
		\prod_{p \in m} f(p)^{-1} &= \prod_{p \in m} \big([g(s.p-v_0(p)) \rhd \hat{e}_p] [(h^{-1} g(s.p-v_0(p))) \rhd \hat{e}_p^{-1}] \big)^{-1}\\
		&= \big( \prod_{p \in m} [g(s.p-v_0(p)) \rhd \hat{e}_p^{-1}] \big) \big[ h^{-1} \rhd \prod_{p \in m}\big( [g(s.p-v_0(p)) \rhd \hat{e}_p] \big) \big].
	\end{align*}
	
	We can then recognise the expression $\prod_{p \in m} [g(s.p-v_0(p)) \rhd \hat{e}_p]$ as the total surface label $\hat{e}(m)$ of the membrane $m$ (oriented away from the dual membrane). This gives us
	\begin{align*}
		H_2&( \text{blob 0}(m)) C^{h,e_m}_T(m) \ket{GS}\\
		&= C^h_{\rhd}(m) \big[\prod_{p \in m} B^{f(p)}(\text{blob 0}(m) \rightarrow \text{blob }p) \big] H_2( \text{blob 0}(m)) \hat{e}(m)^{-1} [h^{-1} \rhd \hat{e}(m)] \delta( \hat{e}(m),e_m) \ket{GS}\\
		&= C^h_{\rhd}(m) \big[\prod_{p \in m} B^{f(p)}(\text{blob 0}(m) \rightarrow \text{blob }p) \big] H_2( \text{blob 0}(m)) e_m^{-1} [h^{-1} \rhd e_m] \delta( \hat{e}(m),e_m) \ket{GS},
	\end{align*}
	where in the last line we used $\delta( \hat{e}(m),e_m)$ to replace the operator $\hat{e}(m)$ with the fixed group element $e_m$. We note that this expression is different from the expression we had when blob 0 and the start-point were attached directly to the membrane (in Section \ref{Section_Magnetic_Tri_Nontrivial_Commutation}), by an action of $h^{-1} \rhd$. In the absence of the $E$-valued membrane operator, this does not affect the action on the ground state, because the ground state contains an even superposition of states with different surface labels (eigenvalues of $\hat{e}(m)$). We know that this must be true because deforming the pure magnetic membrane operator does not affect the action of the membrane operator on the ground state, from the topological nature of the pure magnetic membrane operator. However, when we add the $E$-valued membrane operator, $\delta( \hat{e}(m),e_m)$ selects one of the values of the surface label and so this difference in label becomes significant. Then, because $H_2(\text{blob }0(m))$ commutes with $\delta( \hat{e}(m),e_m)$, we can move $H_2(\text{blob }0(m))$ so that it acts directly on the ground state. The surface label of blob 0 in the ground state must be the identity $1_E$ due to the blob energy terms. Therefore, we have
	\begin{align*}
		H_2( \text{blob 0}(m)) &C^{h,e_m}_T(m) \ket{GS}\\
		&=C^h_{\rhd}(m) \big[\prod_{p \in m} B^{f(p)}(\text{blob 0}(m) \rightarrow \text{blob }p) \big] \delta( \hat{e}(m),e_m) e_m^{-1} [h^{-1} \rhd e_m]H_2( \text{blob 0}(m))\ket{GS}\\
		&=C^h_{\rhd}(m) \big[\prod_{p \in m} B^{f(p)}(\text{blob 0}(m) \rightarrow \text{blob }p) \big] \delta( \hat{e}(m),e_m) e_m^{-1} [h^{-1} \rhd e_m]\ket{GS}\\
		&= C^{h,e_m}_T(m) e_m^{-1} [h^{-1} \rhd e_m]\ket{GS}.
	\end{align*}
	Then because $e_m^{-1} [h^{-1} \rhd e_m]$ is a constant, we see that the state $C^{h,e_m}_T(m) \ket{GS}$ is an eigenstate of $H_2( \text{blob 0}(m))$ and the eigenvalue, which is the definite surface label of blob 0, is $e_m^{-1} [h^{-1} \rhd e_m]$.

\begin{figure}[h]
		\begin{center}
			\begin{overpic}[width=0.9\linewidth]{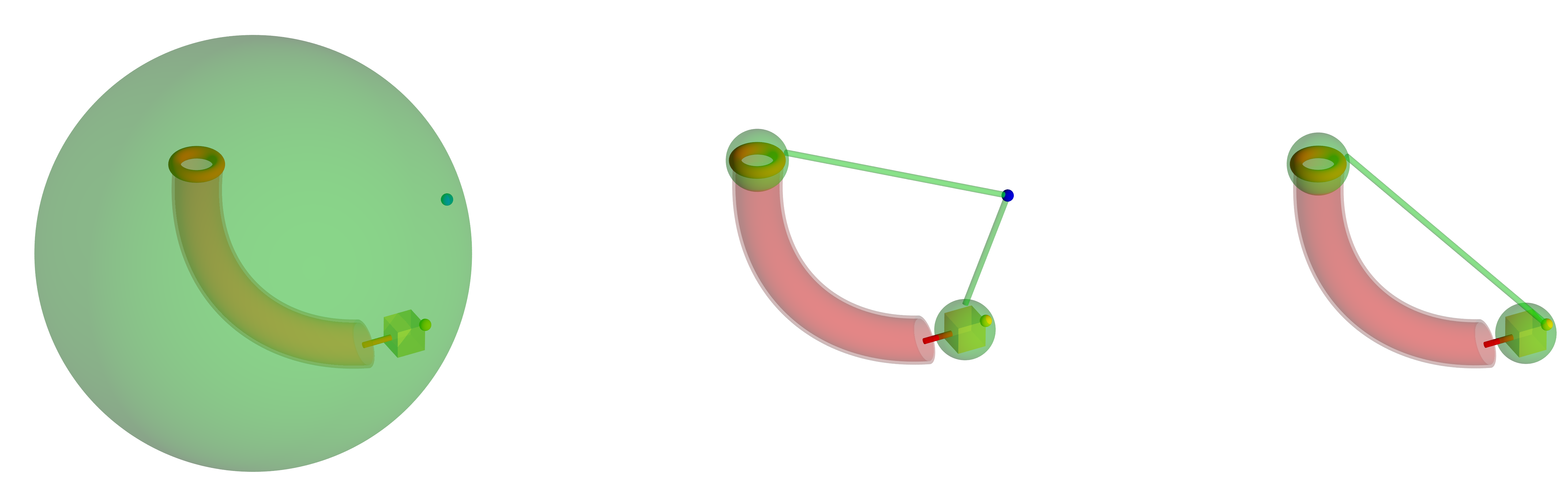}
				\put(35,15){\Huge $\rightarrow$}
				\put(70,15){\Huge $\rightarrow$}
				
				\put(35,18){deform}
				\put(69,18){move s.p}
			\end{overpic}
			\caption{In order to find the surface label of the higher-flux loop excitation, we consider a surface enclosing the entire membrane operator. Because it does not intersect with the membrane operator it must have the same label it has in the ground state, namely $1_E$. We then deform this surface into two parts, one enclosing blob 0 (the yellow cube) and the start-point (light yellow sphere) of the membrane operator, and the other enclosing the loop-like excitation. Then we move the base-point (the dark blue sphere in the leftmost and middle images) of the surface to match the start-point of the membrane operator. Moving the base-point in this way leaves the identity $1_E$ invariant. Then the product of the surface label of the loop-like excitation and blob 0 must be $1_E$.} 
			\label{higher_flux_surface}
		\end{center}
	\end{figure}

	Given that blob 0 carries such a non-trivial 2-holonomy, the loop excitation must also carry a non-trivial 2-holonomy to balance it. Now that we have displaced blob 0 from the membrane, it is easy to separate the contributions from the loop-like excitation and the point-like excitation of blob 0. We can measure the value of a surface which encloses both excitations and does not intersect with the membrane operator itself, as shown in the rightmost image in Figure \ref{higher_flux_surface}. The label of this surface must be $1_E$, because the surface does not intersect the membrane operator and so is unaffected by it, and the surface label is $1_E$ in the ground state (due to the blob energy terms). We can then deform the surface, as shown in Figure \ref{higher_flux_surface}, so that part of the surface encloses blob 0 and part of it encloses the loop excitation. We also move the base-point of the surface to match the start-point of the membrane. This changes the surface label from $e$ to some $g \rhd e$ (as discussed in Section S-I C in the Supplemental Material of Ref. \cite{HuxfordPaper2}), but the identity element is invariant under this $\rhd$ action and so the surface label is unchanged when we move the base-point. Then the total surface element $1_E$ is the product of the contributions from blob 0 and the loop-like excitation (a simple product because the two surfaces have no non-contractible loops which could carry a 1-flux, and we consider the contributions of each with respect to the start-point). Therefore, the contribution to the surface label from the loop-like excitation is the inverse of the contribution from blob 0, and so the 2-flux of the loop excitation is 
	\begin{equation}
		\tilde{e}_m= e_m [h^{-1} \rhd e_m^{-1}], \label{Equation_2_flux_higher_flux_excitation_appendix}
	\end{equation}
	when measured with respect to the start-point of the membrane operator.

	The reason this is significant is that it illustrates that the modified membrane operator is not simply associated with a non-trivial 1-holonomy about a closed path as in the $\rhd$ trivial case, but also with a non-trivial 2-holonomy around a closed surface. Unlike an $E$-valued loop-excitation, which only measures 2-flux, this loop excitation possesses 2-flux itself, like the blob excitations. By applying an additional $E$-valued membrane operator to the magnetic membrane operator, we can fix the value of this 2-flux.

	Having discussed the 2-flux carried by the higher-flux loop excitation, we will next demonstrate the braiding properties of the loop-like excitation, which we previously described in Section \ref{Section_3D_Braiding_Central} of the main text (except for the braiding with the electric excitations, which have the same relations as in the $\rhd$ trivial case from Section \ref{Section_electric_magnetic_braiding_3D_tri_trivial}, as discussed in Section \ref{Section_3D_Braiding_Central}).

	\subsubsection{Braiding with blob excitations}
	\label{Section_braiding_higher_flux_blob}

\begin{figure}[h]
		\begin{center}
			\begin{overpic}[width=0.5\linewidth]{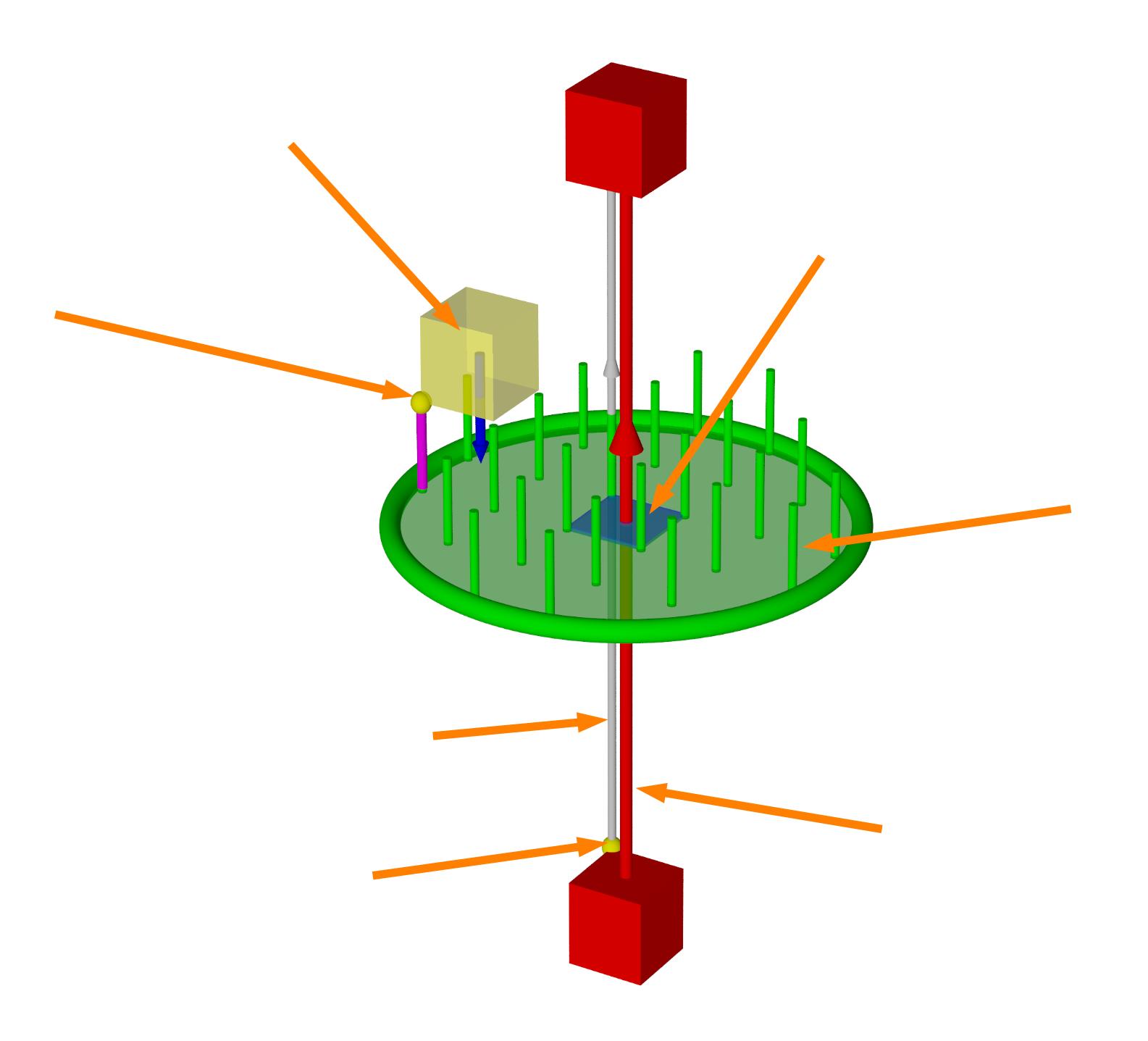}
				\put(-5,63){$s.p(m)$}
				\put(22,14){$s.p(t)$}
				\put(95,46){membrane $m$}
				\put(15,79){blob 0 of $m$}
				\put(13,26){direct path of $t$}
				\put(78,18){dual path of $t$}
				\put(70,70){plaquette $q$}
			\end{overpic}
			\caption{We consider a blob ribbon operator $B^e(t)$ that passes through a higher-flux membrane operator, $C^{h,e_m}_T(m)$. The ribbon $t$ pierces $m$ through a plaquette $q$ (blue square).}
			\label{blob_ribbon_through_higher_flux_overview}
		\end{center}
	\end{figure}

	We start by considering the braiding between the higher-flux excitations and the blob excitations. As usual for the braiding of a point-like excitation with a loop-like one, we wish to compare two situations to obtain the braiding relation. The first situation is where we first produce the loop-like excitation and then produce the point-like excitation and move it through the loop. The second case is where we first move the point-like excitation through empty space before producing the loop-like excitation. In order to set up these situations, we consider a blob ribbon operator $B^e(t)$ piercing the membrane of a higher-flux membrane operator $C^{h,e_m}_T(m)$, as shown in Figure \ref{blob_ribbon_through_higher_flux_overview}.

	As we have discussed before (e.g., in Section \ref{Section_Flux_Charge_Braiding} of the main text), we only expect well-defined braiding in the case were the start-points of the two operators coincide. However, it is instructive to consider first the case where the start-points do not have to be the same. We consider the state $B^e(t)C^{h,e_m}_T(m)\ket{GS}$, corresponding to the case where we first produce the loop excitation, then braid the blob excitation through it (we wish to compare this to the state where we apply the operators in the opposite order, corresponding to the case where no braiding occurs). Splitting the higher-flux membrane operator into its constituent parts, we can write this state as
	\begin{equation}
		B^e(t)C^{h,e_m}_T(m)\ket{GS}= B^e(t) C^h_{\rhd}(m) \big[ \prod_{p \in m} B^{[\hat{g}(s.p-v_0(p))\rhd \hat{e}_p] [(h^{-1} \hat{g}(s.p-v_0(p)))\rhd \hat{e}_p^{-1}]}(\text{blob }0 \rightarrow \text{blob }p) \big] \delta(\hat{e}(m),e_m) \ket{GS}, \label{Equation_blob_higher_flux_commutation_1}
	\end{equation}
	where each plaquette $p$ in the direct membrane of $m$ is associated to a blob ribbon operator
	$$B^{[\hat{g}(s.p-v_0(p))\rhd \hat{e}_p] [(h^{-1} \hat{g}(s.p-v_0(p)))\rhd \hat{e}_p^{-1}]}(\text{blob }0 \rightarrow \text{blob }p)$$
	and blob $p$ is the blob attached to plaquette $p$ and cut by the dual membrane of $m$ (as described in Section \ref{Section_3D_MO_Central} of the main text). We now wish to consider commuting the blob ribbon operator $B^e(t)$ to the right. The first step is to commute the blob ribbon operator past $C^h_{\rhd}(m)$. First, let us consider why these two operators do not commute. The action of the blob ribbon operator on a plaquette $A$, with label $e_A$, that is pierced by the ribbon is given by
	$$B^e(t): e_A = \begin{cases} e_A [\hat{g}(s.p(t)-v_0(A))^{-1} \rhd e^{-1}] & \text{ if the orientation of $A$ matches that of $t$} \\ [\hat{g}(s.p(t)-v_0(A))^{-1} \rhd e] e_A & \text{ if the orientation of $A$ is against that of $t$.} \end{cases}$$
	
\begin{figure}[h]
		\begin{center}
			\begin{overpic}[width=0.5\linewidth]{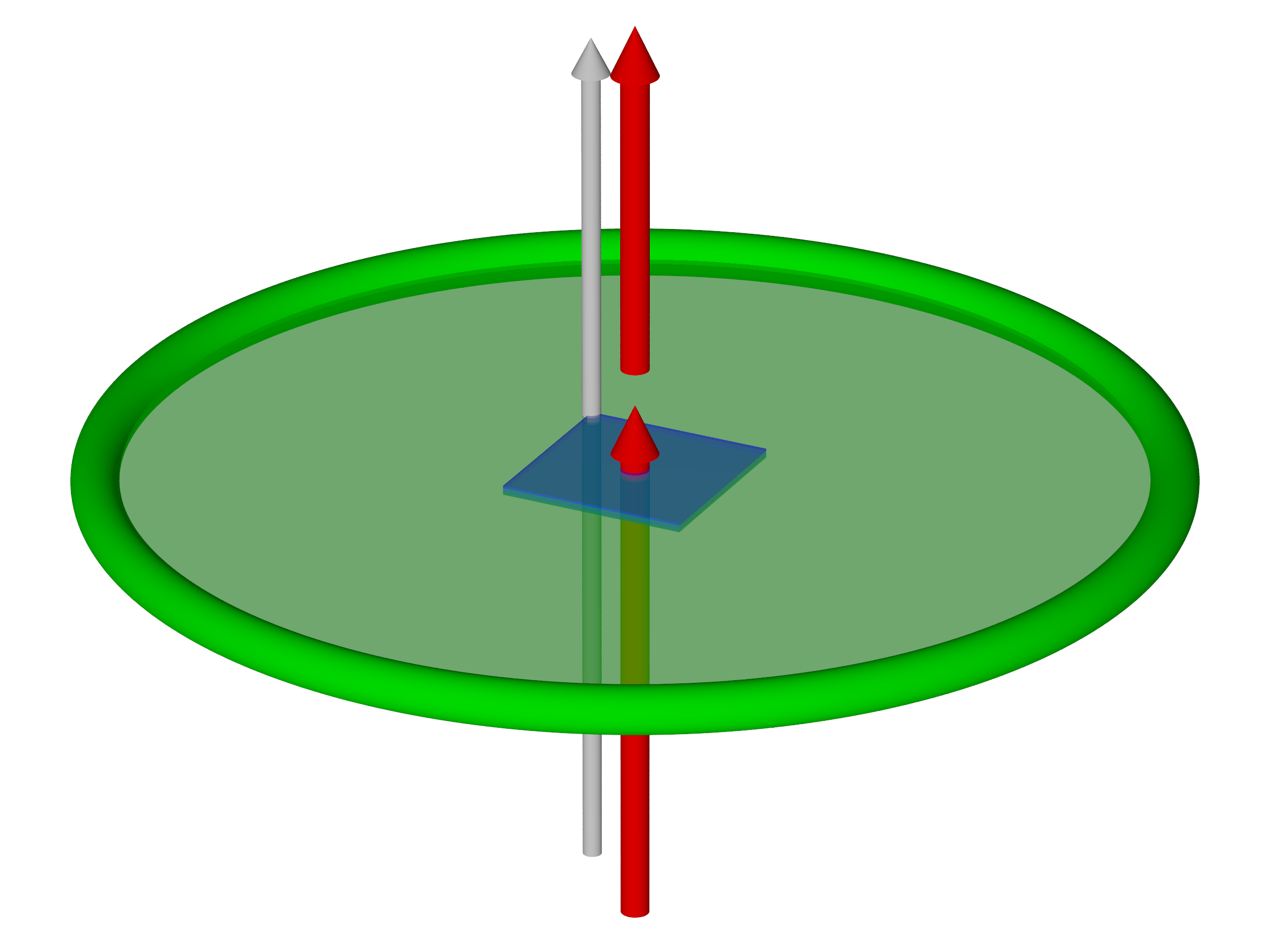}
				\put(52,10){$t_1$}
				\put(52,60){$t_2$}
				\put(58,35){plaquette $q$}
				\put(5,30){$m$}
				
			\end{overpic}
			\caption{We split the ribbon $t$ from Figure \ref{blob_ribbon_through_higher_flux_overview} into two parts. $t_1$ is the part up to (and including) the intersection of the ribbon with the direct membrane of $m$. This part of the ribbon is unaffected by the $C^h_{\rhd}(m)$ part of the magnetic membrane operator. The second part, $t_2$, is the rest of the ribbon. The direct path of $t_2$ (grey) passes through the membrane and so is affected by $C^h_{\rhd}(m)$ (note that the direct path of $t_1$ is the subset of the direct path of $t_2$ up to the intersection with the membrane). }
			\label{blob_through_higher_flux_split_ribbon}
		\end{center}
	\end{figure}

	We see that this action depends on a path element $\hat{g}(s.p(t)-v_0(A))$. The path $(s.p(t)-v_0(A))$ passes through the membrane $m$ if the plaquette $A$ is pierced by the ribbon after the intersection of the ribbon and the membrane (i.e., if the plaquette is above the membrane in Figure \ref{blob_ribbon_through_higher_flux_overview}). Therefore, the magnetic membrane operator affects this path element, as we discussed in Section \ref{Section_electric_magnetic_braiding_3D_tri_trivial} when considering the braiding of magnetic and electric excitations. In order to commute the blob ribbon operator $B^e(t)$ past $C^h_{\rhd}(m)$, we must split the ribbon operator into two parts. The first part, corresponding to the part of the ribbon before the intersection, is unaffected by the presence of the membrane operator. On the other hand, the second part of the ribbon operator corresponds to the part of the ribbon after the intersection and so is affected by the membrane operator. We denote the first part by $B^e(t_1)$ and the second by $B^e(t_2)$, where the ribbons $t_1$ and $t_2$ are illustrated in Figure \ref{blob_through_higher_flux_split_ribbon}.
	
	As already stated, the ribbon operator $B^e(t_1)$ is unaffected by the presence of $C^h_{\rhd}(m)$, so it commutes with this part of the membrane operator. In order to commute the other part, $B^e(t_2)$, past $C^h_{\rhd}(m)$, we must examine how $C^h_{\rhd}(m)$ affects the action of $B^e(t_2)$. This means understanding how $C^h_{\rhd}(m)$ affects the path element $\hat{g}(s.p(t)-v_0(A))$. In Section \ref{Section_electric_magnetic_braiding_3D_tri_trivial}, we considered the action of the magnetic membrane operator on a path element intersecting with the membrane, in the case where $\rhd$ is trivial. This also holds in the more general case we consider in this section, because the action of the membrane operator on the edges is the same in either case. Therefore, using Equation \ref{Equation_magnetic_membrane_electric_appendix_1}, we have
	\begin{equation}
		C^h_{\rhd}(m): \hat{g}(s.p(t)-v_0(A)) = \hat{g}(s.p(t)-s.p(m)) h \hat{g}(s.p(t)-s.p(m))^{-1} \hat{g}(s.p(t)-v_0(A)). \label{Equation_action_membrane_blob_ribbon_path}
	\end{equation}
	
	Defining $h_{[t-m]} = \hat{g}(s.p(t)-s.p(m)) h \hat{g}(s.p(t)-s.p(m))^{-1}$, we can write this as
	$$C^h_{\rhd}(m): \hat{g}(s.p(t)-v_0(A)) =h_{[t-m]} \hat{g}(s.p(t)-v_0(A)).$$
	This means that
	\begin{align*}
		B^e(t_2) C^h_{\rhd}(m): e_A &= \begin{cases} e_A [(C^h_{\rhd}(m):\hat{g}(s.p(t)-v_0(A)))^{-1} \rhd e^{-1}] & \text{ if the orientation of $A$ matches that of $t$} \\ [(C^h_{\rhd}(m):\hat{g}(s.p(t)-v_0(A)))^{-1} \rhd e] e_A & \text{ if the orientation of $A$ is against that of $t$} \end{cases}\\
		&= \begin{cases} e_A [(h_{[t-m]}^{\phantom{-1}}\hat{g}(s.p(t)-v_0(A)))^{-1} \rhd e^{-1}] & \text{ if the orientation of $A$ matches that of $t$} \\ [(h_{[t-m]}^{\phantom{-1}}\hat{g}(s.p(t)-v_0(A)))^{-1} \rhd e] e_A & \text{ if the orientation of $A$ is against that of $t$} \end{cases}\\
		&= \begin{cases} e_A [\hat{g}(s.p(t)-v_0(A))^{-1} \rhd (h_{[t-m]}^{-1} \rhd e^{-1})] & \text{ if the orientation of $A$ matches that of $t$} \\ [\hat{g}(s.p(t)-v_0(A))^{-1} \rhd (h_{[t-m]}^{-1} \rhd e)] e_A & \text{ if the orientation of $A$ is against that of $t$} \end{cases}\\
		&= C^h_{\rhd}(m) B^{h_{[t-m]}^{-1} \rhd e}(t_2):e_A.
	\end{align*}
	
\begin{figure}[h]
		\begin{center}
			\begin{overpic}[width=0.5\linewidth]{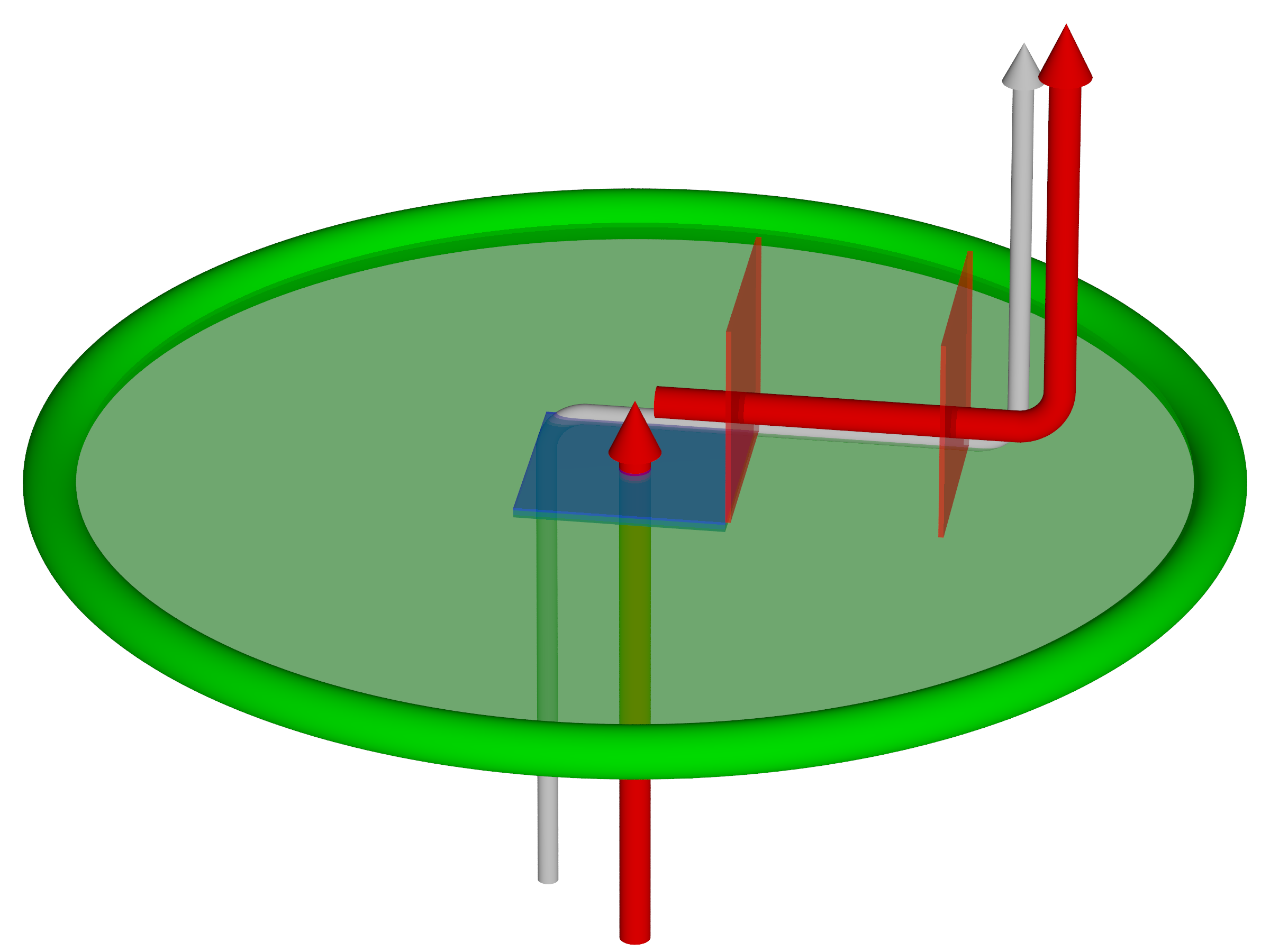}
				\put(52,6){$t_1$}
				\put(86,60){$t_2$}
			\end{overpic}
			\caption{It is possible for the ribbon $t_2$ to pass through plaquettes that are cut by the dual membrane of $m$, such as the red plaquettes in this figure. If the base-point of such a plaquette lies on the direct membrane, then the plaquette is subject to the $\rhd$ action of $C^h_{\rhd}(m)$, which may affect the commutation relation of $C^h_{\rhd}(m)$ and $B^e(t_2)$. However, if the base-point lies on the direct membrane, then the direct path to that plaquette defined in $B^e(t)$ is not affected by the action of $C^h_{\rhd}(m)$ on the edges (unlike other plaquettes pierced by $t_2$). It turns out that this leads to the same commutation relation for the action of the two operators on the plaquette as for other plaquettes pierced by $t_2$. We expect this from our freedom in changing the base-points of plaquettes, which enables us to move the base-point of a plaquette onto and off of the direct membrane.}
			\label{blob_through_higher_flux_turn}
		\end{center}
	\end{figure}

	Note that, depending on the ribbon $t_2$, it is possible that $t_2$ passes through some plaquettes whose base-points lie on the direct membrane of $m$, such as in the example shown in Figure \ref{blob_through_higher_flux_turn}. In this case, the path element $\hat{g}(s.p(t)-v_0(A))$ is not affected by the action of the magnetic membrane operator. However, instead the plaquette label itself is affected by the $\rhd$ action of $C^h_{\rhd}(m)$ (see Section \ref{Section_Magnetic_Tri_Non_Trivial}). In this case, we still obtain the same commutation relation for the action on the plaquette, as we require by consistency under changing the base-point of a plaquette. To be explicit, for such a plaquette, the action of $C^h_{\rhd}(m)$ on the plaquette is 
	\begin{align*}
		C^h_{\rhd}(m):e_A &= (\hat{g}(s.p(m)-v_0(A))^{-1} h\hat{g}(s.p(m)-v_0(A))) \rhd e_A\\
		&:= h_{[A-m]} \rhd e_A.
	\end{align*}
	
	Then acting with both $C^h_{\rhd}(m)$ and the blob ribbon operator on the plaquette $A$ gives us
	\begin{align*}
		B^e(t_2) C^h_{\rhd}(m)&: e_A = B^e(t_1): h_{[A-m]}^{\phantom{-1}} \rhd e_A\\
		&= \begin{cases}[h_{[A-m]}^{\phantom{-1}} \rhd e_A] [\hat{g}(s.p(t)-v_0(A))^{-1} \rhd e^{-1}] & \text{ if the orientation of $A$ matches that of $t$} \\ [\hat{g}(s.p(t)-v_0(A))^{-1} \rhd e] h_{[A-m]}^{\phantom{-1}} \rhd e_A & \text{ if the orientation of $A$ is against that of $t$} \end{cases}\\
		&= \begin{cases} h_{[A-m]}^{\phantom{-1}} \rhd \big(e_A [(h_{[A-m]}^{-1}\hat{g}(s.p(t)-v_0(A))^{-1}) \rhd e^{-1}]\big) & \text{ if the orientation of $A$ matches that of $t$} \\ h_{[A-m]}^{\phantom{-1}} \rhd \big([(h_{[A-m]}^{-1}\hat{g}(s.p(t)-v_0(A))^{-1}) \rhd e] e_A\big) & \text{ if the orientation of $A$ is against that of $t$} \end{cases}\\
		&= \begin{cases} h_{[A-m]}^{\phantom{-1}} \rhd \big(e_A [(\hat{g}(s.p(m)-v_0(A))^{-1} h^{-1}\hat{g}(s.p(m)-v_0(A))\hat{g}(s.p(t)-v_0(A))^{-1}) \rhd e^{-1}]\big) \\ \hspace{8cm} \text{ if the orientation of $A$ matches that of $t$} \\ h_{[A-m]}^{\phantom{-1}} \rhd \big([(\hat{g}(s.p(m)-v_0(A))^{-1} h^{-1}\hat{g}(s.p(m)-v_0(A))\hat{g}(s.p(t)-v_0(A))^{-1}) \rhd e] e_A\big) \\ \hspace{8cm} \text{ if the orientation of $A$ is against that of $t$.} \end{cases}\\
	\end{align*}
	
	We can then write $\hat{g}(s.p(m)-v_0(A))\hat{g}(s.p(t)-v_0(A))^{-1}=\hat{g}(s.p(m)-s.p(t))=\hat{g}(s.p(t)-s.p(m))^{-1}$, where we do not need to worry about the precise position of these paths when they act on fake-flat regions, to obtain
	\begin{align*}
		B^e(t_2) C^h_{\rhd}(m)&: e_A\\	
		&= \begin{cases} h_{[A-m]}^{\phantom{-1}} \rhd \big(e_A [( \hat{g}(s.p(t)-v_0(A))^{-1} \hat{g}(s.p(t)-s.p(m))h^{-1}\hat{g}(s.p(t)-s.p(m))^{-1}) \rhd e^{-1}]\big) \\ \hspace{8cm} \text{ if the orientation of $A$ matches that of $t$} \\ h_{[A-m]}^{\phantom{-1}} \rhd \big([( \hat{g}(s.p(t)-v_0(A))^{-1} \hat{g}(s.p(t)-s.p(m))h^{-1}\hat{g}(s.p(t)-s.p(m))^{-1}) \rhd e] e_A\big) \\ \hspace{8cm} \text{ if the orientation of $A$ is against that of $t$} \end{cases}\\
		&= C^h_{\rhd}(m) B^{h_{[t-m]}^{-1} \rhd e}(t_2):e_A.
	\end{align*}
	
	We see that, for either type of plaquette,
	$$B^e(t_2) C^h_{\rhd}(m): e_A = C^h_{\rhd}(m) B^{(\hat{g}(s.p(t)-s.p(m)) h^{-1} \hat{g}(s.p(t)-s.p(m))^{-1}) \rhd e}(t_2):e_A.$$
	This relation holds for all plaquettes pierced by $t_2$, and apart from the action on the plaquettes there is no interference between the two operators, so we can write this as a commutation relation between the operators: 
	\begin{equation}
		B^e(t_2) C^h_{\rhd}(m)= C^h_{\rhd}(m) B^{h_{[t-m]}^{-1} \rhd e}(t_2).
		\label{Equation_blob_ribbon_magnetic_commutation_appendix_1}
	\end{equation}
	Substituting this result into Equation \ref{Equation_blob_higher_flux_commutation_1}, we obtain
	\begin{align}
		B^e(t)&C^{h,e_m}_T(m)\ket{GS}= B^e(t) C^h_{\rhd}(m) \big[ \prod_{p \in m} B^{[\hat{g}(s.p-v_0(p))\rhd \hat{e}_p] [(h^{-1} \hat{g}(s.p-v_0(p)))\rhd \hat{e}_p^{-1}]}(\text{blob }0 \rightarrow \text{blob }p) \big] \delta(\hat{e}(m),e_m) \ket{GS} \notag \\
		&= B^e(t_1) B^e(t_2) C^h_{\rhd}(m) \big[ \prod_{p \in m} B^{[\hat{g}(s.p-v_0(p))\rhd \hat{e}_p] [(h^{-1} \hat{g}(s.p-v_0(p)))\rhd \hat{e}_p^{-1}]}(\text{blob }0 \rightarrow \text{blob }p) \big] \delta(\hat{e}(m),e_m) \ket{GS}\notag \\
		&= C^h_{\rhd}(m) B^e(t_1) B^{h_{[t-m]}^{-1} \rhd e}(t_2) \big[ \prod_{p \in m} B^{[\hat{g}(s.p-v_0(p))\rhd \hat{e}_p] [(h^{-1} \hat{g}(s.p-v_0(p)))\rhd \hat{e}_p^{-1}]}(\text{blob }0 \rightarrow \text{blob }p) \big] \delta(\hat{e}(m),e_m) \ket{GS}. \label{Equation_blob_higher_flux_commutation_2}
	\end{align}
	
	The next step is to commute the blob ribbon operators on ribbons $t_1$ and $t_2$ past the blob ribbon operators that are part of $C^h_T(m)$. Ordinarily, different blob ribbon operators would commute. However, the ribbon operators that are part of $C^h_T(m)$ have a label that depends on the labels of plaquettes on the direct membrane of $m$. Because the ribbon $t$ pierces this membrane, the labels of these plaquettes may be affected by the blob ribbon operators. More definitely, if the ribbon $t$ pierces the membrane $m$ once, at a plaquette $q$, then the label $e_q$ of plaquette $q$ is affected. The effect of the blob ribbon operator on this plaquette $q$ is
	\begin{equation}
		B^e(t_1):e_q =\begin{cases} e_q [\hat{g}(s.p(t)-v_0(q))^{-1} \rhd e^{-1}] &\text{if $q$ is oriented with the ribbon} \\ [\hat{g}(s.p(t)-v_0(q))^{-1} \rhd e] e_q &\text{if $q$ is oriented against the ribbon.} \end{cases} \label{Equation_blob_ribbon_action_on_higher_flux_plaquette}
	\end{equation}
	
	This change affects the label of the blob ribbon operator associated to the plaquette $p$. The blob ribbon operator associated to plaquette $q$ is 
	$$B^{[\hat{g}(s.p(m)-v_0(q))\rhd \hat{e}_q] [(h^{-1} \hat{g}(s.p(m)-v_0(q)))\rhd \hat{e}_q^{-1}]}(\text{blob }0 \rightarrow \text{blob }q),$$
	where $\hat{e}_q$ gives us the label of plaquette $q$ if it points downwards (which is against the orientation of the ribbon $t$ in this case), and if the plaquette pointed upwards we would use the inverse of the plaquette label. The effect of the blob ribbon operator $B^e(t_1)$ on the label of this blob ribbon operator is then
	\begin{align*}
		B^e&(t_1) : [\hat{g}(s.p(m)-v_0(q))\rhd \hat{e}_q] [(h^{-1} \hat{g}(s.p(m)-v_0(q)))\rhd \hat{e}_q^{-1}]\\
		&= [\hat{g}(s.p(m)-v_0(q))\rhd ([\hat{g}(s.p(t)-v_0(q))^{-1} \rhd e] \hat{e}_q)] [(h^{-1} \hat{g}(s.p(m)-v_0(q)))\rhd (\hat{e}_q^{-1} [\hat{g}(s.p(t)-v_0(q))^{-1} \rhd e]^{-1})].
	\end{align*}

	We can then use the fact that $E$ is Abelian (recall that we are considering Case 2 from Table \ref{Table_Cases} of the main text) to collect the terms corresponding to $e$ and those corresponding to the plaquette label $\hat{e}_q$, obtaining
	\begin{align*}
		B^e&(t_1) : [\hat{g}(s.p(m)-v_0(q))\rhd \hat{e}_q] [(h^{-1} \hat{g}(s.p(m)-v_0(q)))\rhd \hat{e}_q^{-1}]\\
		&= [\hat{g}(s.p(m)-v_0(q))\rhd \hat{e}_q] [(h^{-1} \hat{g}(s.p(m)-v_0(q)))\rhd \hat{e}_q^{-1}] \\
		& \hspace{0.5cm} [ (\hat{g}(s.p(m)-v_0(q))\hat{g}(s.p(t)-v_0(q))^{-1}) \rhd e] [(h^{-1} \hat{g}(s.p(m)-v_0(q))\hat{g}(s.p(t)-v_0(q))^{-1}) \rhd e^{-1}]\\
		&= [\hat{g}(s.p(m)-v_0(q))\rhd \hat{e}_q] [(h^{-1} \hat{g}(s.p(m)-v_0(q)))\rhd \hat{e}_q^{-1}] [\hat{g}(s.p(m)-s.p(t)) \rhd e] [(h^{-1}\hat{g}(s.p(m)-s.p(t))) \rhd e^{-1}],
	\end{align*}
	where we wrote $\hat{g}(s.p(m)-v_0(q))\hat{g}(s.p(t)-v_0(q))^{-1}$ as $\hat{g}(s.p(m)-s.p(t))$ (the exact position of the path does not matter, because deforming it over a fake-flat region only introduces a factor in $\partial(E)$ to the path element, which does not affect a term like $\hat{g}(s.p(m)-s.p(t)) \rhd e$ when $\partial(E)$ is in the centre of $G$). Therefore,
	\begin{align}
		B^e(t_1) &B^{[\hat{g}(s.p(m)-v_0(q))\rhd \hat{e}_q] [(h^{-1} \hat{g}(s.p(m)-v_0(q)))\rhd \hat{e}_q^{-1}]}(\text{blob }0 \rightarrow \text{blob }q) \notag \\
		&= B^{[B^e(t_1)]^{-1}:\big([\hat{g}(s.p(m)-v_0(q))\rhd \hat{e}_q] [(h^{-1} \hat{g}(s.p(m)-v_0(q)))\rhd \hat{e}_q^{-1}]\big)}(\text{blob }0 \rightarrow \text{blob }q) B^e(t_1) \notag \\
		&= B^{[\hat{g}(s.p(m)-v_0(q))\rhd \hat{e}_q] [(h^{-1} \hat{g}(s.p-v_0(q)))\rhd \hat{e}_q^{-1}] [\hat{g}(s.p(m)-s.p(t)) \rhd e^{-1}] [(h^{-1}\hat{g}(s.p(m)-s.p(t))) \rhd e]}(\text{blob }0 \rightarrow \text{blob }q) B^e(t_1) \notag \\
		&= B^{[\hat{g}(s.p(m)-v_0(q))\rhd \hat{e}_q] [(h^{-1} \hat{g}(s.p(m)-v_0(q)))\rhd \hat{e}_q^{-1}]}(\text{blob }0 \rightarrow \text{blob }q) \notag \\
		& \hspace {3cm} B^{[\hat{g}(s.p(m)-s.p(t)) \rhd e^{-1}][(h^{-1}\hat{g}(s.p(m)-s.p(t))) \rhd e]}(\text{blob }0 \rightarrow \text{blob }q) B^e(t_1). \label{Equation_blob_higher_flux_blob_commutation}
	\end{align}
	
	We see that the commutation generates an additional blob ribbon operator running from blob 0 to blob $q$ (which is the blob connected to plaquette $q$ and cut by the dual membrane). We will consider this blob ribbon operator in more detail later, after we have finished commuting the blob ribbon operator on $t$ to the right. Using the commutation relation Equation \ref{Equation_blob_higher_flux_blob_commutation} in Equation \ref{Equation_blob_higher_flux_commutation_2} gives us
	\begin{align}
		B^e(t)&C^{h,e_m}_T(m)\ket{GS} \notag \\
		=& C^h_{\rhd}(m) B^e(t_1) B^{h_{[t-m]}^{-1} \rhd e}(t_2)\big[ \prod_{p \in m} B^{[\hat{g}(s.p(m)-v_0(p))\rhd \hat{e}_p] [(h^{-1} \hat{g}(s.p(m)-v_0(p)))\rhd \hat{e}_p^{-1}]}(\text{blob }0 \rightarrow \text{blob }p) \big] \delta(\hat{e}(m),e_m) \ket{GS}\notag \\
		=& C^h_{\rhd}(m) \big[ \prod_{p \in m} B^{[\hat{g}(s.p(m)-v_0(p))\rhd \hat{e}_p] [(h^{-1} \hat{g}(s.p(m)-v_0(p)))\rhd \hat{e}_p^{-1}]}(\text{blob }0 \rightarrow \text{blob }p) \big]\notag \\
		& B^{[\hat{g}(s.p(m)-s.p(t)) \rhd e^{-1}] [(h^{-1}\hat{g}(s.p(m)-s.p(t))) \rhd e]}(\text{blob }0 \rightarrow \text{blob }q) B^e(t_1) B^{h_{[t-m]}^{-1} \rhd e}(t_2) \delta(\hat{e}(m),e_m) \ket{GS}\notag \\
		=& C^h_{T}(m)B^{[\hat{g}(s.p(m)-s.p(t)) \rhd e^{-1}] [(h^{-1}\hat{g}(s.p(m)-s.p(t))) \rhd e]}(\text{blob }0 \rightarrow \text{blob }q) B^e(t_1) B^{h_{[t-m]}^{-1} \rhd e}(t_2) \delta(\hat{e}(m),e_m) \ket{GS}. \label{Equation_blob_higher_flux_commutation_3}
	\end{align}

	The final step in the commutation relation is to commute the blob ribbon operators past the $E$-valued membrane operator $\delta(\hat{e}(m),e_m)$. Here $\hat{e}(m)$ is the total surface label, given by $\prod_{p \in m} \hat{g}(s.p(m)-v_0(p)) \rhd \hat{e}_p$, where the plaquette $p$ is taken to be oriented downwards in Figure \ref{blob_ribbon_through_higher_flux_overview} and the order of the product does not matter because $E$ is Abelian. We have already established that the ribbon only affects the plaquette $q$ on this membrane and we know how this label is affected. We can therefore use Equation \ref{Equation_blob_ribbon_action_on_higher_flux_plaquette} to determine how the blob ribbon operator acts on $\hat{e}(m)$. We have
	\begin{align*}
		B^e(t_1):& \hat{e}(m) = \big(\prod_{p \in m; \ p \neq q} \hat{g}(s.p(m)-v_0(p)) \rhd \hat{e}_p \big) [\hat{g}(s.p(m)-v_0(q)) \rhd (B^e(t): \hat{e}_q)]\\
		&= \big(\prod_{p \in m; \ p \neq q} \hat{g}(s.p(m)-v_0(p)) \rhd \hat{e}_p\big) [\hat{g}(s.p(m)-v_0(q)) \rhd ( [\hat{g}(s.p(t)-v_0(q))^{-1} \rhd e] \hat{e}_q)]\\
		&= \big(\prod_{p \in m; \ p \neq q} \hat{g}(s.p(m)-v_0(p)) \rhd \hat{e}_p\big) [\hat{g}(s.p(m)-v_0(q)) \rhd ( \hat{g}(s.p(t)-v_0(q))^{-1} \rhd e)] [\hat{g}(s.p(m)-v_0(q)) \rhd \hat{e}_q]\\
		&= \big(\prod_{p \in m; \ p \neq q} \hat{g}(s.p(m)-v_0(p)) \rhd \hat{e}_p\big)[\hat{g}(s.p(m)-v_0(q)) \rhd \hat{e}_q] [(\hat{g}(s.p(m)-v_0(q)) \hat{g}(v_0(q)-s.p(t))) \rhd e] \\
		&= \big(\prod_{p \in m} \hat{g}(s.p(m)-v_0(p)) \rhd \hat{e}_p\big) [\hat{g}(s.p(m)-s.p(t)) \rhd e]\\
		&= \hat{e}(m) [\hat{g}(s.p(m)-s.p(t)) \rhd e].
	\end{align*}
	Therefore, 
	\begin{align*}
		B^e(t_1) \delta(e_m, \hat{e}(m)) &= \delta(e_m, [B^e(t_1)]^{-1}:\hat{e}(m))B^e(t_1)\\
		&= \delta(e_m, \hat{e}(m)[\hat{g}(s.p(m)-s.p(t)) \rhd e]^{-1})B^e(t_1)\\
		&=\delta(e_m[\hat{g}(s.p(m)-s.p(t)) \rhd e], \hat{e}(m))B^e(t_1).
	\end{align*}
	
	The other blob ribbon operators (on $t_2$ and $(\text{blob }0 \rightarrow \text{blob }q)$) do not intersect the direct membrane of $m$. Therefore, Equation \ref{Equation_blob_higher_flux_commutation_3} becomes
	\begin{align}
		B^e(t)&C^{h,e_m}_T(m)\ket{GS} \notag \\
		&=C^h_{T}(m) \delta(\hat{e}(m),e_m[\hat{g}(s.p(m)-s.p(t)) \rhd e]) B^{[\hat{g}(s.p(m)-s.p(t)) \rhd e^{-1}] [(h^{-1}\hat{g}(s.p(m)-s.p(t))) \rhd e]}(\text{blob }0 \rightarrow \text{blob }q)\notag \\
		& \hspace{2cm} B^e(t_1) B^{(\hat{g}(s.p(t)-s.p(m)) h^{-1} \hat{g}(s.p(t)-s.p(m))^{-1}) \rhd e}(t_2) \ket{GS} \notag \\
		&=C^{h,e_m[\hat{g}(s.p(m)-s.p(t)) \rhd e] }_T(m) B^{[\hat{g}(s.p(m)-s.p(t)) \rhd e^{-1}] [(h^{-1}\hat{g}(s.p(m)-s.p(t))) \rhd e]}(\text{blob }0 \rightarrow \text{blob }q) \notag \\
		& \hspace{2cm} B^e(t_1) B^{(\hat{g}(s.p(t)-s.p(m)) h^{-1} \hat{g}(s.p(t)-s.p(m))^{-1}) \rhd e}(t_2) \ket{GS}. \label{Equation_blob_higher_flux_commutation_4}
	\end{align}
	
\begin{figure}[h]
		\begin{center}
			\begin{overpic}[width=0.75\linewidth]{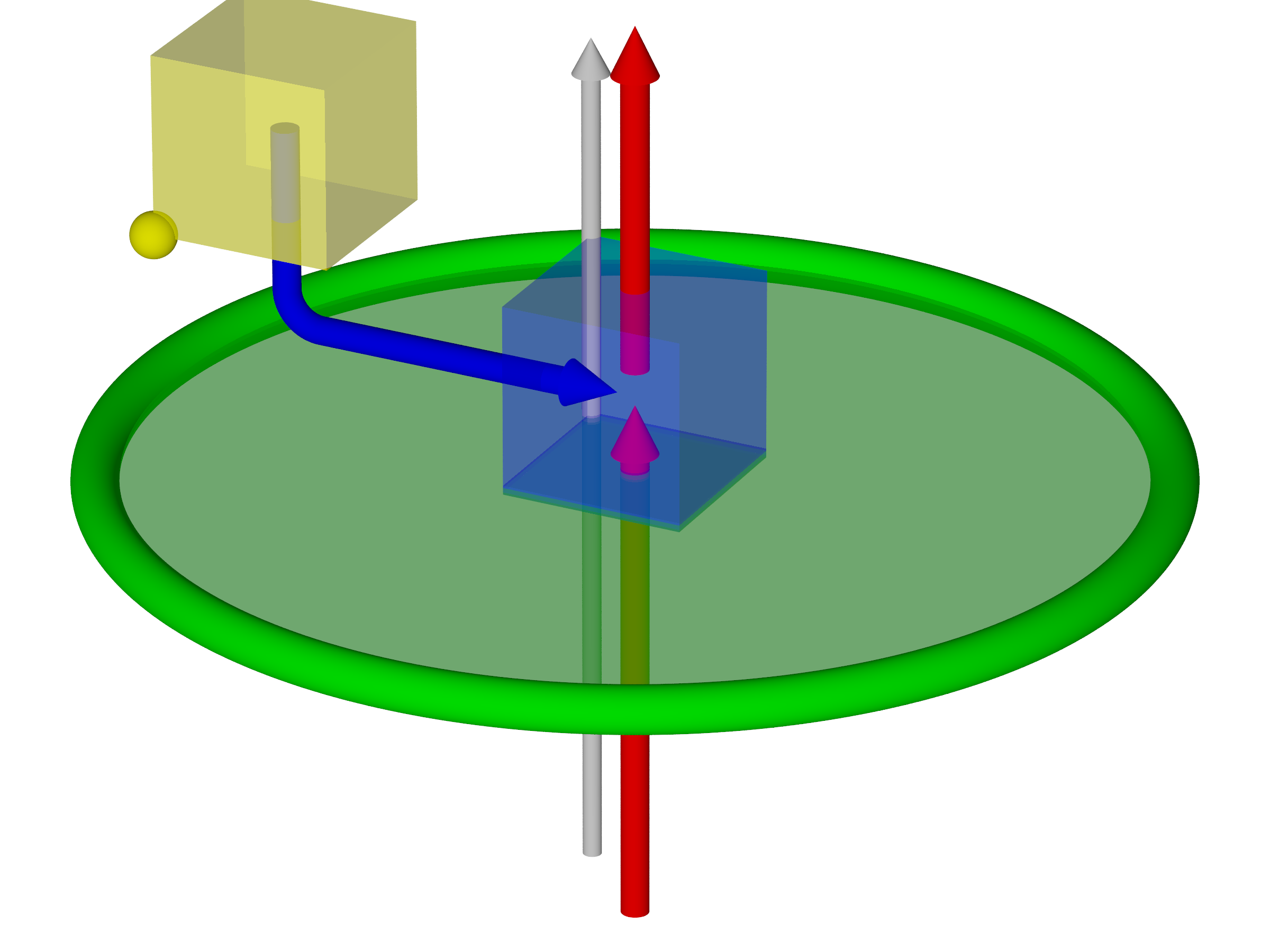}
				\put(52,10){$t_1$}
				\put(52,60){$t_2$}
				\put(23,42){(blob $0 -$blob $q$)}
			\end{overpic}
			\caption{Commuting the blob ribbon operator $B^e(t_1)$ past the blob ribbon operator from $C^h_T(m)$ associated to plaquette $q$ generates an additional blob ribbon operator. This blob ribbon operator acts along the blue path from blob 0 to the blob attached to plaquette $q$.}
			\label{blob_ribbon_through_higher_flux_additional_ribbon}
		\end{center}
	\end{figure}

	We see that the $G$ label of the higher-flux membrane is unchanged by the commutation, while the $E$ label goes from $e_m$ to $e_m[\hat{g}(s.p(m)-s.p(t)) \rhd e]$. We also still have this extra blob ribbon operator $$B^{[\hat{g}(s.p(m)-s.p(t)) \rhd e^{-1}] (h^{-1}\hat{g}(s.p(m)-s.p(t))) \rhd e}(\text{blob }0 \rightarrow \text{blob }q)$$ in addition to $B^e(t_1)$ and $B^{(\hat{g}(s.p(t)-s.p(m)) h^{-1} \hat{g}(s.p(t)-s.p(m))^{-1}) \rhd e}(t_2).$ These three ribbons are shown in Figure \ref{blob_ribbon_through_higher_flux_additional_ribbon}.

\begin{figure}[h]
		\begin{center}
			\begin{overpic}[width=0.75\linewidth]{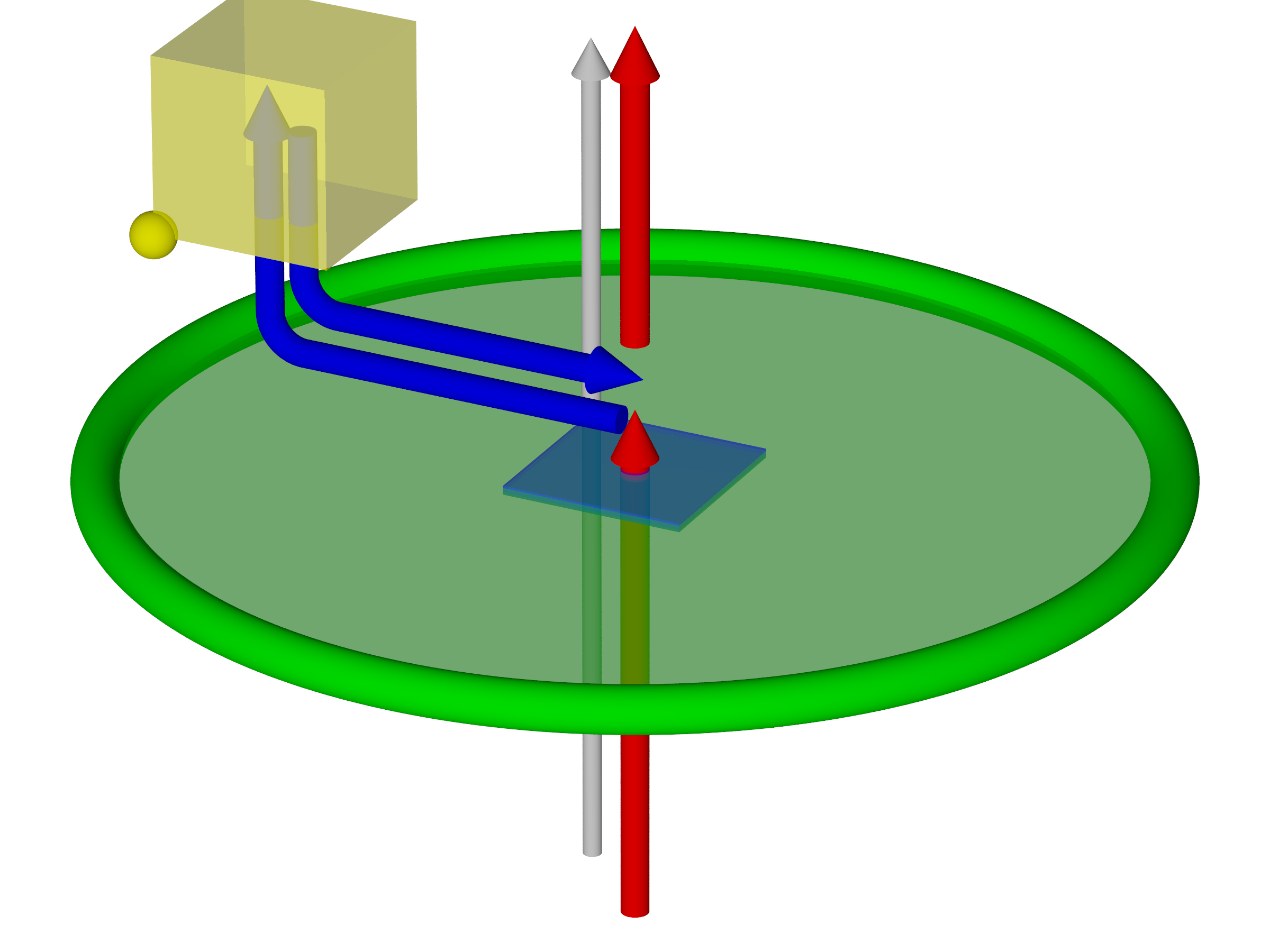}
				\put(52,10){$t_1$}
				\put(52,60){$t_2$}
				\put(23,40){(blob $q -$blob $0$)}
				\put(25,52){(blob $0 -$blob $q$)}
			\end{overpic}
			\caption{We split the extra blob ribbon operator from Figure \ref{blob_ribbon_through_higher_flux_additional_ribbon} into two parts and reverse the direction of one of these parts. Then it seems that these extra ribbons can be concatenated with the ribbons $t_1$ and $t_2$. This is indeed true, though we must first ensure that the start-points and labels of the ribbons that we connect are the same.}
			\label{blob_ribbon_through_higher_flux_four_ribbons}
		\end{center}
	\end{figure}

\begin{figure}[h]
		\begin{center}
			\begin{overpic}[width=0.98\linewidth]{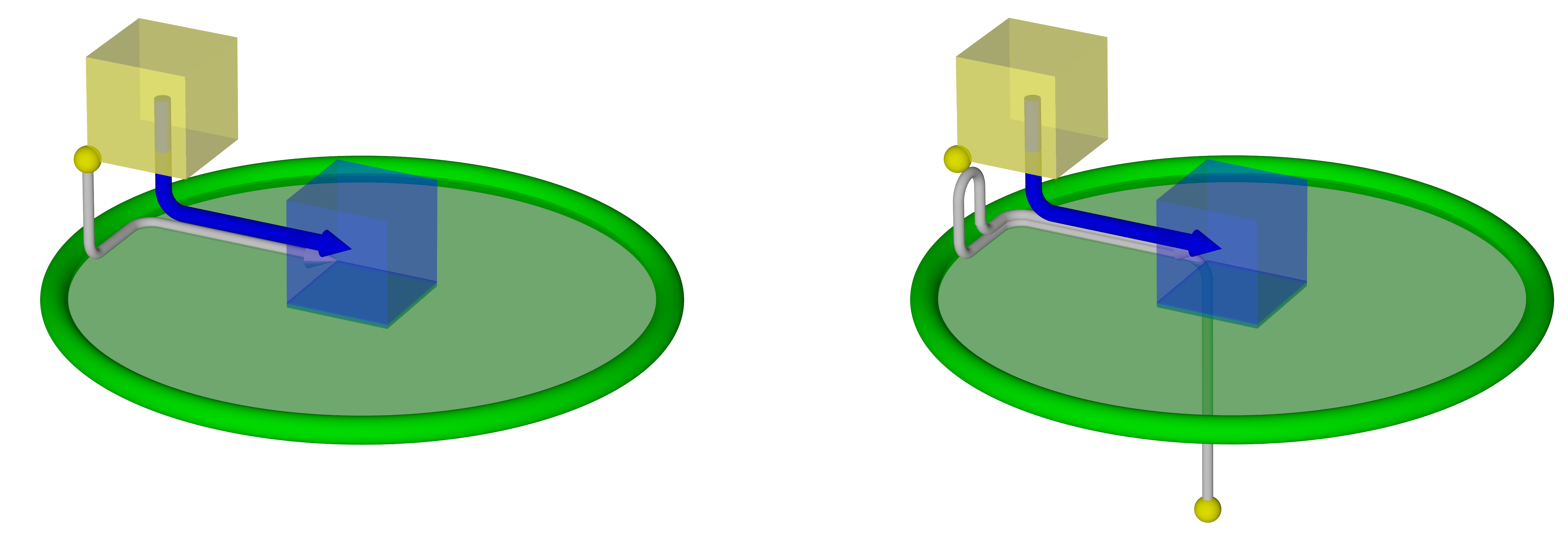}
				\put(49,18){\Huge $\rightarrow$}
				\put(78,3){$s.p(t)$}
				\put(47,21){change $s.p$}
				\put(0,26){$s.p(m)$}
			\end{overpic}
			\caption{In order to ensure that the start-points of the ribbons that we wish to combine are the same, we move the start-point of the two ribbons that run between blob 0 and blob $q$. Here we show this process for the ribbon that starts at blob 0 and terminates in blob $q$. On the left we have the original ribbon, with its direct path shown in light grey (the thinner tube) and its start-point at $s.p(m)$. Then we move the start-point to $s.p(t)$, so that the direct path is the light grey path in the right image. When we change the start-point in this manner, we must also change the label of the blob ribbon operator. It turns out that this change of label means that the blob ribbon operator has the same label as the ribbon operator on $t_2$, so that we can combine the ribbon operators. We can do a similar process for the ribbon that starts in blob $q$ and terminates in blob 0, allowing us to combine the associated ribbon operator with $B^e(t_1)$.}
			\label{blob_ribbon_through_higher_flux_change_sp}
		\end{center}
	\end{figure}

\begin{figure}[h]
		\begin{center}
			\begin{overpic}[width=0.75\linewidth]{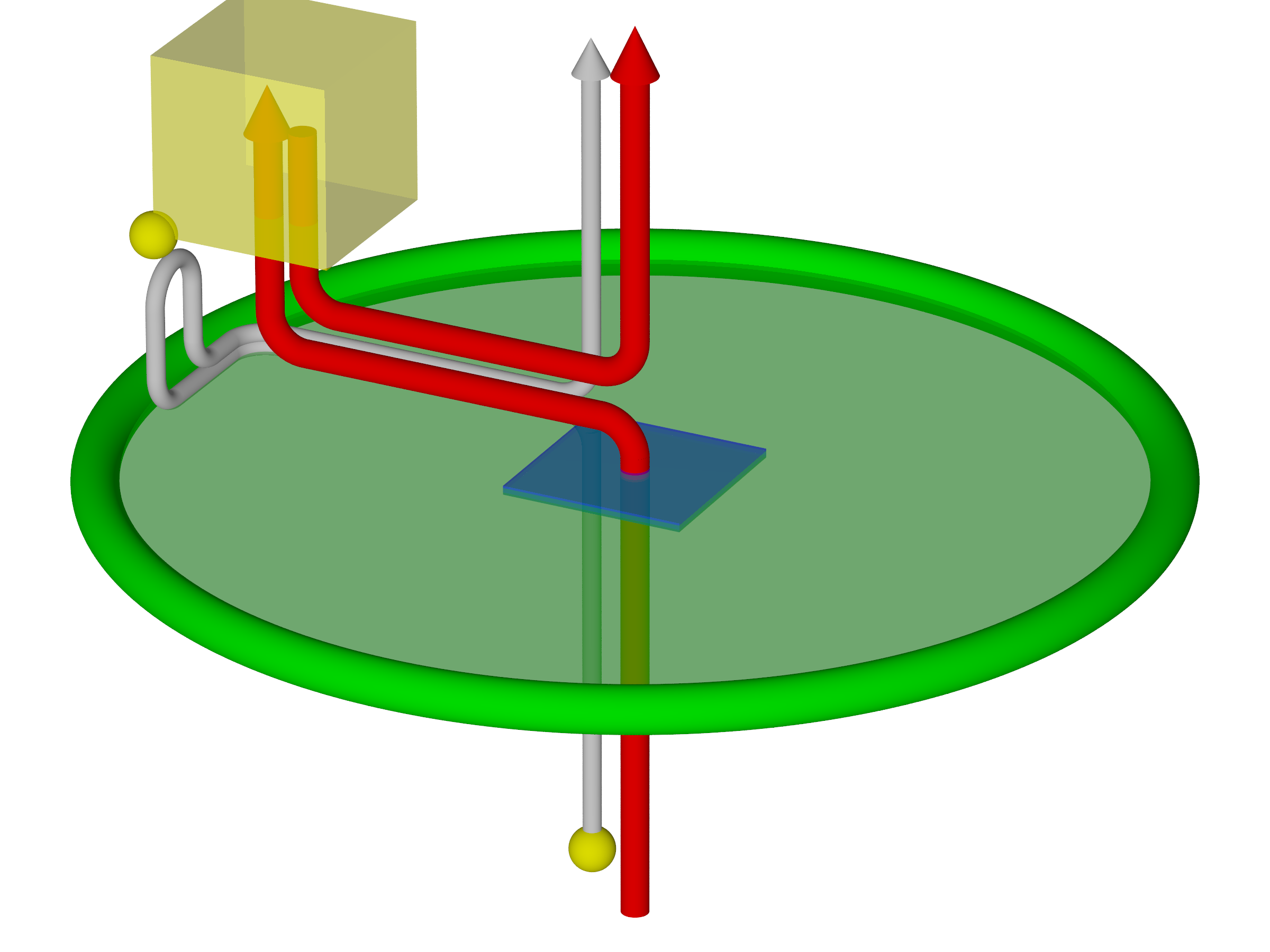}
				\put(52,10){$t_1'$}
				\put(52,60){$t_2'$}
				
			\end{overpic}
			\caption{After we change the start-point of the extra ribbon operators (on the blue ribbons from Figure \ref{blob_ribbon_through_higher_flux_four_ribbons}), we can connect them to the ribbon operators on $t_1$ and $t_2$. This leaves us with two ribbon operators, the first of which acts on a ribbon $t_1'$ whose dual path runs from the beginning of the original ribbon $t$ to blob 0 of $m$. The dual path of the second ribbon, $t_2'$, runs from blob 0 to the end of $t$ (the dual paths are represented by the thicker arrows). The thinner grey path represents the direct path of $t_2'$, while the direct path of $t_1'$ is just the first part of this grey path, up to the U-turn (i.e., the direct path of $t_2'$ includes the direct path of $t_1'$).}
			\label{blob_ribbon_through_higher_flux_fuse_ribbons}
		\end{center}
	\end{figure}

	Now consider splitting the blob ribbon operator $$B^{[\hat{g}(s.p(m)-s.p(t)) \rhd e^{-1}] [(h^{-1}\hat{g}(s.p(m)-s.p(t))) \rhd e]}(\text{blob }0 \rightarrow \text{blob }q)$$ into two pieces, $$B^{\hat{g}(s.p(m)-s.p(t)) \rhd e^{-1}} (\text{blob }0 \rightarrow \text{blob }q)$$ and $$B^{(h^{-1}\hat{g}(s.p(m)-s.p(t))) \rhd e}(\text{blob }0 \rightarrow \text{blob }q).$$ We can then reverse the orientation of the first of these, while simultaneously inverting its label, to give us $$B^{[\hat{g}(s.p(m)-s.p(t)) \rhd e]} (\text{blob }q \rightarrow \text{blob }0).$$ This gives us four blob ribbon operators in total, which are shown in Figure \ref{blob_ribbon_through_higher_flux_four_ribbons}. Note that it appears that these ribbons could be connected pairwise. Specifically, we can concatenate $t_1$ with $(\text{blob }q \rightarrow \text{blob }0)$ and $(\text{blob }0 \rightarrow \text{blob }q)$ with $t_2$. Furthermore, the labels of the associated blob ribbon operators seem to have a similar form. This suggests that we can fuse the four blob ribbon operators into two ribbon operators. In order to concatenate two ribbon operators in this manner, they must satisfy certain conditions, as described in Section \ref{Section_blob_ribbon_concatenate}. Firstly, the dual paths of the ribbons must connect together. This is satisfied by the ribbons we wish to connect, as illustrated in Figure \ref{blob_ribbon_through_higher_flux_four_ribbons}. Secondly the direct paths of the two ribbons must be compatible (in that the direct path of the second ribbon must contain the direct path of the first ribbon, at least up to deformation). Thirdly the labels of the two parts must match. These latter two conditions are not satisfied by our ribbons, but the two issues resolve each-other. We can extend the direct paths of the ribbons between blob 0 and blob $q$ to match the direct paths of $t_1$ and $t_2$. In order to do so, we must move the start-point of the two ribbons that pass between blob 0 to blob $q$ (the blue ribbons in Figure \ref{blob_ribbon_through_higher_flux_four_ribbons}) backwards along the direct path of $t$, as shown in Figure \ref{blob_ribbon_through_higher_flux_change_sp}. That is, we move the start-point of $(\text{blob }0 \rightarrow \text{blob }q)$ from $s.p(m)$ to $s.p(t)$. In order to preserve the action of a blob ribbon operator $B^x(\text{blob }0 \rightarrow \text{blob }q)$ when changing the start-point in this manner, we must simultaneously change the label of the blob ribbon operator from $x$ to 
	$\hat{g}(s.p(t)-s.p(m)) \rhd x$, as discussed in Section \ref{Section_blob_ribbon_move_sp}. Therefore, $B^{(h^{-1}\hat{g}(s.p(m)-s.p(t))) \rhd e}(\text{blob }0 \rightarrow \text{blob }q)$ becomes
	$$B^{(\hat{g}(s.p(t-s.p(m)))h^{-1}\hat{g}(s.p(m)-s.p(t))) \rhd e}(\text{blob }0 \rightarrow \text{blob }q| s.p(t)),$$
	where we use $|s.p(t)$ to indicate that the start-point of the ribbon is now $s.p(t)$. When we write the ribbon operator in this way, it has the same label as $B^{(\hat{g}(s.p(t)-s.p(m)) h^{-1} \hat{g}(s.p(t)-s.p(m))^{-1}) \rhd e}(t_2)$, as well as the same start-point. Therefore, as described in Section \ref{Section_blob_ribbon_concatenate}, we can connect these two ribbon operators. Similarly, the label of $B^{[\hat{g}(s.p(m)-s.p(t)) \rhd e]} (\text{blob }q \rightarrow \text{blob }0)$ becomes
	$$\hat{g}(s.p(t)-s.p(m)) \rhd (\hat{g}(s.p(m)-s.p(t)) \rhd e) = e,$$
	which is same as the label of $B^e(t_1)$. We can therefore connect these two blob ribbon operators together as well. This gives us the blob ribbon operators shown in Figure \ref{blob_ribbon_through_higher_flux_fuse_ribbons}, applied on ribbons that we denote by $t_1'$ and $t_2'$, where the dual path of $t_1'$ terminates in blob 0 and the dual path of $t_2'$ originates in blob 0. The labels of these ribbon operators do not depend on exactly which plaquette $q$ is at the intersection of the ribbon and membrane operator, as we require from the fact that we could deform the membrane (or the ribbon if the blob ribbon operator is not confined) to change the place where the ribbon and membrane intersect to a plaquette other than $q$.

	We can then use this process for combining the ribbon operators to write the overall commutation relation between the ribbon and membrane operators (see Equation \ref{Equation_blob_higher_flux_commutation_4}) as
	\begin{align}
		B^e(t)C^{h,e_m}_T(m)\ket{GS}&= C^{h,e_m[\hat{g}(s.p(m)-s.p(t)) \rhd e] }_T(m) B^e(t_1') B^{(\hat{g}(s.p(t)-s.p(m)) h^{-1} \hat{g}(s.p(t)-s.p(m))^{-1}) \rhd e}(t_2') \ket{GS}. \label{Equation_blob_higher_flux_commutation_5}
	\end{align}
	
	In addition to the changes to the $E$ label of the higher-flux membrane, which we discussed previously, the commutation affects the blob ribbon operators. The ribbon is diverted to pass into blob 0 of the membrane. Furthermore, after it enters blob 0, its label is changed from $e$ to $(\hat{g}(s.p(t)-s.p(m)) h^{-1} \hat{g}(s.p(t)-s.p(m))^{-1}) \rhd e$. The fact that the ribbon can be divided into two parts with two different labels reflects the fact that the ribbon operator produces two excitations, one at each end of the ribbon, and one of these excitations does not braid with the loop excitation. The label of the part of the ribbon before intersection is therefore unchanged by the commutation. On the other hand, the part after intersection is changed, to reflect the transformation of the braiding particle. This relation is simplified if we consider placing the start-points of the membrane and ribbon operators to be at the same position. We note that, when placing the start-points at the same position, we must be careful that we do not move the start-point of the blob ribbon operator through the magnetic membrane operator, which would alter the commutation relations we have considered up to now. The same start-point commutation relation is:
	\begin{align}
		B^e(t)C^{h,e_m}_T(m)\ket{GS}&= C^{h,e_m e }_T(m) B^e(t_1') B^{ h^{-1} \rhd e}(t_2')\ket{GS}. \label{Equation_blob_higher_flux_commutation_6}
	\end{align}
	
	It is worth noting that this transformation preserves the combined 2-flux of the higher-flux excitation and the blob excitation at the end of ribbon $t$. As we described at the start of Section \ref{Section_braiding_higher_flux}, the 2-flux of the higher-flux excitation with label $(h,e_m)$ is $e_m [h^{-1} \rhd e_m^{-1}]$. The 2-flux of the blob excitation produced by $B^e(t)$ is $e$. Therefore, the combined 2-flux of these excitations is $e_m [h^{-1} \rhd e_m^{-1}]e$ before braiding (this would be the result we would find if the ribbon and membrane did not intersect, so no braiding occurred). After the braiding, the label of the higher-flux membrane becomes $(h,e_m e)$, while the label of the blob ribbon becomes $h^{-1} \rhd e$. Therefore, the combined 2-flux is
	\begin{align*}
		e_m e [h^{-1} \rhd (e_m e)^{-1}] [h^{-1} \rhd e] &= e_m e [h^{-1} \rhd e^{-1}] [h^{-1} \rhd e_m^{-1}] [h^{-1} \rhd e]\\
		&= e_m e [h^{-1} \rhd e_m^{-1}],
	\end{align*}
	which is the same as the 2-flux before braiding.

\begin{figure}[h]
		\begin{center}
			\begin{overpic}[width=0.75\linewidth]{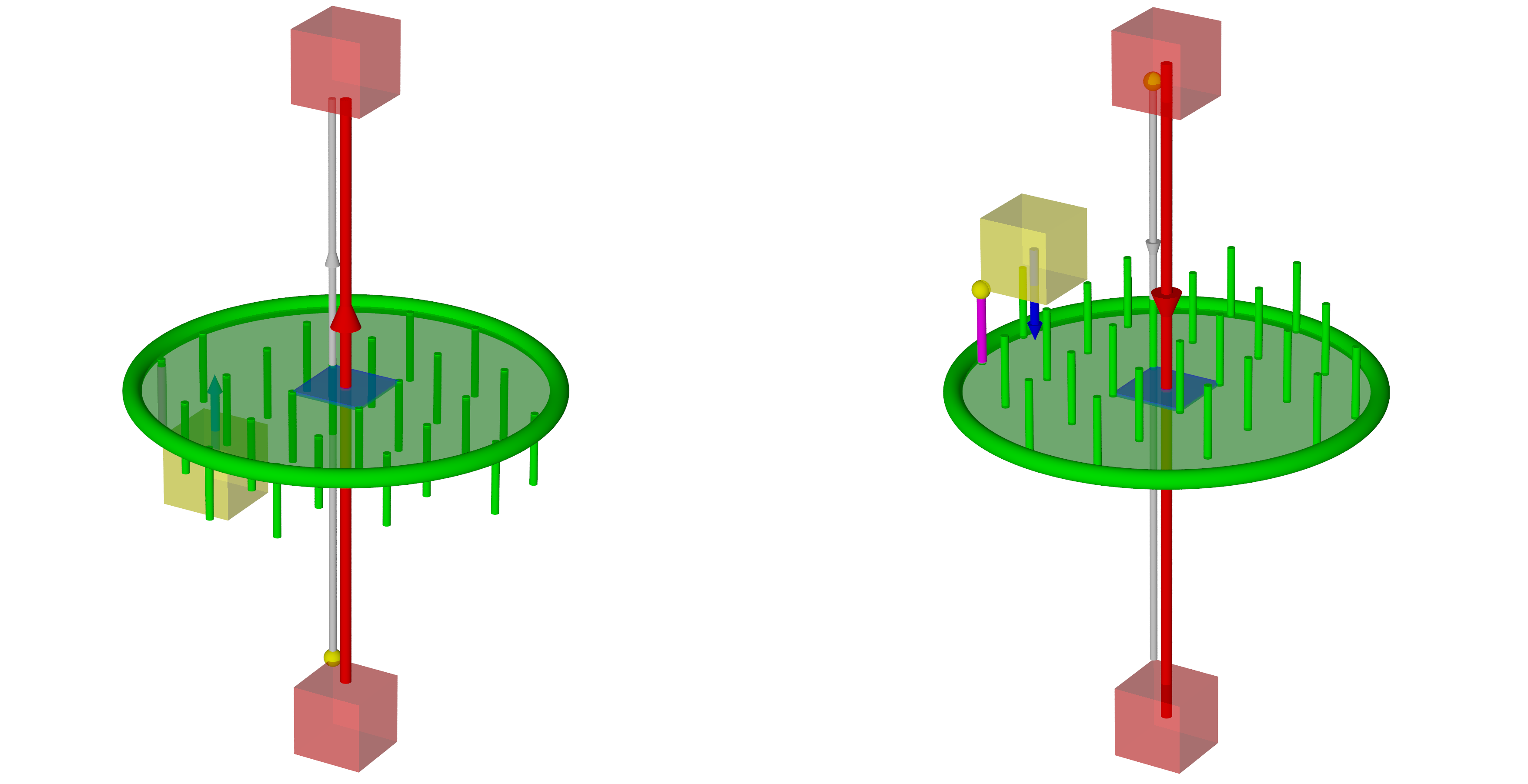}
				
			\end{overpic}
			\caption{We consider the case where the blob passes through the membrane with the opposite orientation compared to Figure \ref{blob_ribbon_through_higher_flux_overview}. This can either be because the membrane is reversed (as in the left image) or because the blob ribbon is reversed (as in the right image), though these are equivalent. We will consider the latter case, to make the figures clearer.}
			\label{blob_ribbon_through_higher_flux_reversed}
		\end{center}
	\end{figure}

	So far, we have considered the case where the blob ribbon operator passes through the membrane in one particular direction, as indicated in Figure \ref{blob_ribbon_through_higher_flux_overview}. In this orientation, the ribbon passes through the direct membrane of $m$ before the dual membrane. Next we consider the case where the blob ribbon operator passes through the membrane with the opposite orientation, as shown in Figure \ref{blob_ribbon_through_higher_flux_reversed}. Note that we can obtain this case from the previous one either by flipping the orientation of the membrane (as in the left of Figure \ref{blob_ribbon_through_higher_flux_reversed}) or reversing the orientation of the ribbon (as in the right of Figure \ref{blob_ribbon_through_higher_flux_reversed}). These are equivalent, but for the sake of keeping consistent figures we will consider the latter situation. We follow the same procedure as with the other orientation, starting by considering the state $B^e(t)C^{h,e_m}_T(m) \ket{GS}$. As before, we split the higher-flux membrane operator into its constituent parts:
	\begin{equation}
		B^e(t)C^{h,e_m}_T(m)\ket{GS}=B^e(t) C^h_{\rhd}(m) \big[ \prod_{p \in m} B^{f(p)}(\text{blob }0 \rightarrow \text{blob }p)\big] \delta(\hat{e}(m),e_m)\ket{GS}, \label{Equation_blob_higher_flux_commutation_reverse_1}
	\end{equation}
	where $f(p)=[\hat{g}(s.p(m)-v_0(p))\rhd \hat{e}_p] [(h^{-1} \hat{g}(s.p(m)-v_0(p)))\rhd \hat{e}_p^{-1}]$.
	
\begin{figure}[h]
		\begin{center}
			\begin{overpic}[width=0.7\linewidth]{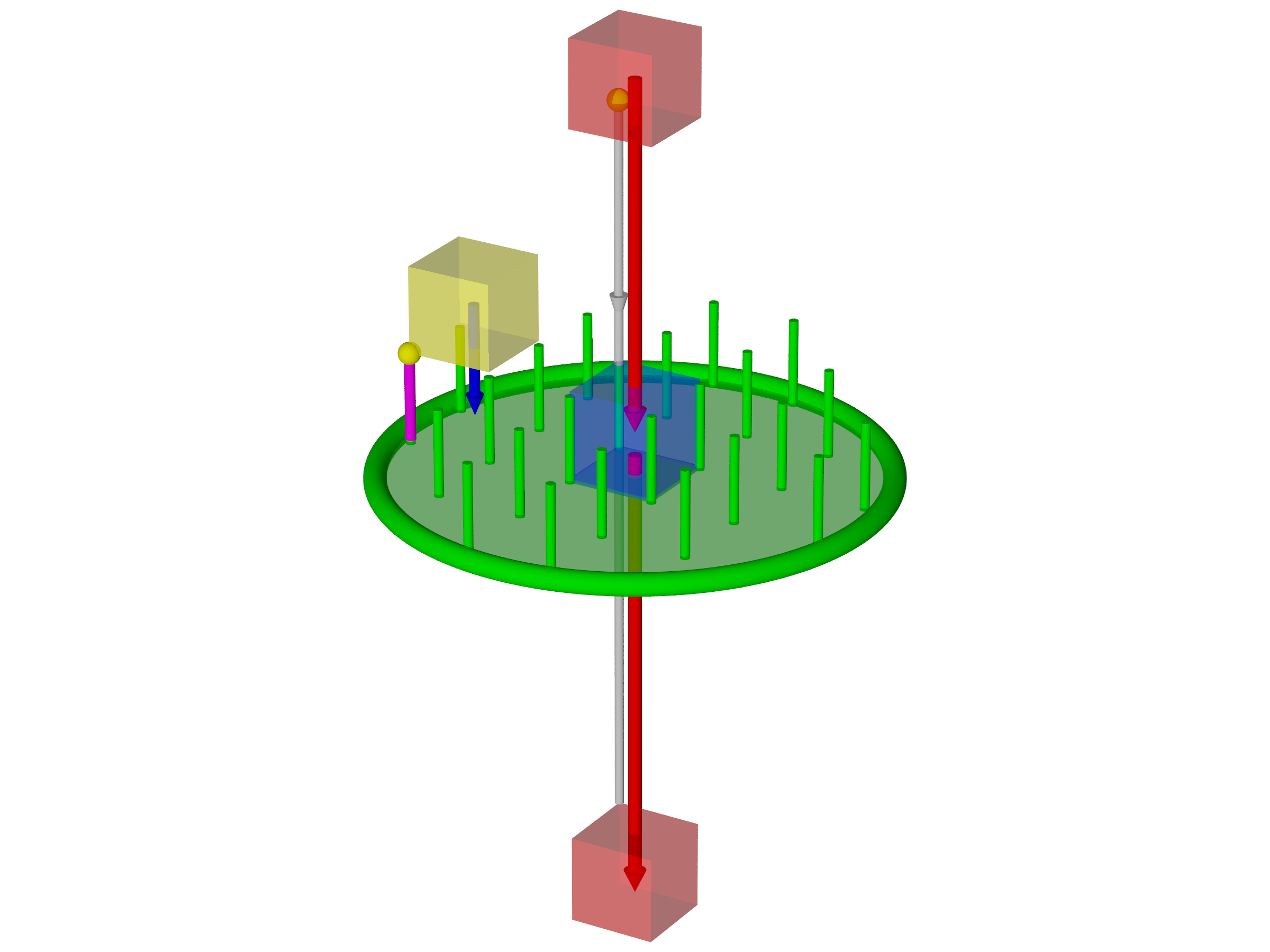}
				\put(51,55){$t_1$}
				\put(51,20){$t_2$}
			\end{overpic}
			\caption{We split the ribbon $t$ into two parts. $t_1$ is the part of the ribbon up to the intersection with the membrane, while $t_2$ is the part afterwards. Unlike with the previous orientation, this time $t_2$ pierces the direct membrane, through a plaquette $q$ (base of the blue cube).}
			\label{blob_ribbon_through_higher_flux_reversed_split_path}
		\end{center}
	\end{figure}

	Then, just as before, we split the blob ribbon operator $B^e(t)$ into two parts, applied on the ribbons $t_1$ and $t_2$ shown in Figure \ref{blob_ribbon_through_higher_flux_reversed_split_path}. $t_1$ corresponds to the part of the ribbon before the intersection whereas $t_2$ corresponds to the part after the intersection. Note that this time, it is the ribbon $t_2$ that pierces the direct membrane rather than $t_1$ (because the ribbon meets the dual membrane before the direct membrane with this orientation). These blob ribbon operators might not commute with the magnetic membrane operator, because the membrane operator affects the path element $\hat{g}(s.p(t)-v_0(A))$ that appears in the action of the blob ribbon operator on a plaquette $A$. If the path does not pass through the dual membrane of $m$, then the corresponding path element is unaffected. On the other hand, if the path passes through the membrane then the path element undergoes a transformation when acted on by the membrane operator. For this braiding orientation (see Figure \ref{blob_ribbon_through_higher_flux_reversed}), the path passes through the membrane in the opposite direction when compared to the case considered in Figure \ref{blob_ribbon_through_higher_flux_overview}, so we must use Equation \ref{Magnetic_electric_3D_braid_reverse} from Section \ref{Section_electric_magnetic_braiding_3D_tri_trivial} to determine the action of the magnetic membrane on this path. We have
	\begin{equation}
		C^h_{\rhd}(m): \hat{g}(s.p(t)-v_0(A)) = \hat{g}(s.p(t)-s.p(m)) h^{-1} \hat{g}(s.p(t)-s.p(m))^{-1} \hat{g}(s.p(t)-v_0(A)). \label{Equation_action_membrane_blob_ribbon_path_reversed}
	\end{equation}
	
	Now consider how this affects the two ribbon operators $B^e(t_1)$ and $B^e(t_2)$. The action of the ribbon operator $B^e(t_1)$ on most plaquettes pierced by the ribbon is unaffected by $C^h_{\rhd}(m)$ because $t_1$ is the section of ribbon before the intersection, and so the paths from the start-point of $t_1$ to most of the plaquettes pierced by $t_1$ do not intersect the membrane. The only exceptions to this are plaquettes whose base-points lie on the direct membrane of $m$ (we could also consider somewhat unnatural plaquettes that are whiskered through the direct membrane even though most of the plaquette is on the other side of the membrane, but we can always use consistency under changes to the base-point of plaquettes to ``unwhisker" these plaquettes, given that we only defined the action of the magnetic membrane operator on such plaquettes through this consistency requirement). On the other hand, we should consider plaquettes whose base-points lie on the direct membrane. It is possible for $t_1$ to pass through such plaquettes, just as $t_2$ did in the case of the opposite orientation of braiding (see Figure \ref{blob_through_higher_flux_turn}). For such a plaquette, $B$, the magnetic membrane operator acts on the path element $\hat{g}(s.p(t)-v_0(B))$ according to Equation \ref{Equation_action_membrane_blob_ribbon_path_reversed}. However, the magnetic membrane also acts directly on plaquettes whose base-points lie on the direct membrane (and which are cut by the dual membrane) with its $\rhd$ action. This $\rhd$ action on the plaquette will not commute with the action of the blob ribbon operator. It turns out that the combination of these two effects means that, while the blob ribbon operator may not commute individually with the action of $C^h_{\rhd}(m)$ on the edges or the action of $C^h_{\rhd}(m)$ on the plaquettes, it does commute with $C^h_{\rhd}(m)$ as a whole. For a plaquette $B$ whose base-point lies on the direct membrane and is cut by the dual membrane of $m$, the action of the membrane operator on the plaquette is
	$$C^h_{\rhd}(m):e_B =(\hat{g}(s.p(m)-v_0(B))^{-1}h\hat{g}(s.p(m)-v_0(B))) \rhd e_B.$$
	
	Defining $h_{[B-m]}=\hat{g}(s.p(m)-v_0(B))^{-1}h\hat{g}(s.p(m)-v_0(B))$, the combined action of the membrane operator and blob ribbon operator on the plaquette $B$ is
	\begin{align*}
		B^e(t_1) C^h_{\rhd}(m): e_B &= \begin{cases} [h_{[B-m]} \rhd e_B] [\big(C^h_{\rhd}(m):\hat{g}(s.p(t)-v_0(B))\big)^{-1} \rhd e^{-1}] & \text{if $B$ is aligned with $t$}\\
			[ \big(C^h_{\rhd}(m):\hat{g}(s.p(t)-v_0(B))\big)^{-1} \rhd e^{-1}] [h_{[B-m]} \rhd e_B] & \text{if $B$ is anti-aligned with $t$} \end{cases}\\
		&= \begin{cases} [h_{[B-m]} \rhd e_B] [\big(\hat{g}(s.p(t)-s.p(m)) h^{-1} \hat{g}(s.p(t)-s.p(m))^{-1} \hat{g}(s.p(t)-v_0(B))\big)^{-1} \rhd e^{-1}]\\ \hspace{9cm} \text{if $B$ is aligned with $t$}\\
			[ \big(\hat{g}(s.p(t)-s.p(m)) h^{-1} \hat{g}(s.p(t)-s.p(m))^{-1} \hat{g}(s.p(t)-v_0(B))\big)^{-1} \rhd e] [h_{[B-m]} \rhd e_B] \\ \hspace{9cm} \text{if $B$ is anti-aligned with $t$} \end{cases}\\
		&= \begin{cases} [h_{[B-m]} \rhd e_B] [\big( \hat{g}(s.p(t)-v_0(B))^{-1} \hat{g}(s.p(t)-s.p(m)) h \hat{g}(s.p(t)-s.p(m))^{-1} \big) \rhd e^{-1}]\\ \hspace{9cm} \text{if $B$ is aligned with $t$}\\
			[ \big( \hat{g}(s.p(t)-v_0(B))^{-1} \hat{g}(s.p(t)-s.p(m)) h \hat{g}(s.p(t)-s.p(m))^{-1} \big) \rhd e] [h_{[B-m]} \rhd e_B] \\\hspace{9cm} \text{if $B$ is anti-aligned with $t$} \end{cases}\\
		&= \begin{cases} [h_{[B-m]} \rhd e_B] [\big( \hat{g}(s.p(m)-v_0(B))^{-1} h \hat{g}(s.p(m)-v_0(B)) \hat{g}(s.p(t)-v_0(B))^{-1} \big) \rhd e^{-1}]\\ \hspace{9cm} \text{if $B$ is aligned with $t$}\\
			[ \big( \hat{g}(s.p(m)-v_0(B))^{-1} h \hat{g}(s.p(m)-v_0(B)) \hat{g}(s.p(t)-v_0(B))^{-1} \big) \rhd e] [h_{[B-m]} \rhd e_B]\\ \hspace{9cm} \text{if $B$ is anti-aligned with $t$} \end{cases}\\
		&=\begin{cases} h_{[B-m]} \rhd ( e_B [ \hat{g}(s.p(t)-v_0(B))^{-1} \rhd e^{-1}]) & \text{if $B$ is aligned with $t$}\\
			h_{[B-m]} \rhd ( [\hat{g}(s.p(t)-v_0(B))^{-1} \rhd e] e_B) & \text{if $B$ is anti-aligned with $t$} \end{cases}\\
		&= C^h_{\rhd}(m) B^e(t_1):e_B.
	\end{align*}

	We see that the actions of the ribbon operator $B^e(t_1)$ and membrane operator $C^h_{\rhd}(m)$ commute, even on these plaquettes with base-point on the direct membrane. We expect this result from the consistency of the ribbon and membrane operators under changes of the base-points of plaquettes (see Section \ref{Section_Magnetic_Tri_Non_Trivial}), which would allow us to move the base-point of a plaquette away from the direct membrane before acting with the ribbon and membrane operators, and then move the base-point back afterwards.

	We now consider the other ribbon operator, $B^e(t_2)$. The path $(s.p(t)-v_0(A))$ that appears in the action of this ribbon operator on a plaquette $A$ always passes through the membrane (unless we consider some unusual plaquette which is whiskered through the dual membrane of $m$). This means that the path transforms according to Equation \ref{Equation_action_membrane_blob_ribbon_path_reversed}. Comparing this equation to Equation \ref{Equation_action_membrane_blob_ribbon_path}, which describes the transformation of the path in the case of the opposite braiding orientation, we see that we just need to replace $h$ with $h^{-1}$. This means that (c.f. Equation \ref{Equation_blob_ribbon_magnetic_commutation_appendix_1})
	\begin{equation}
		B^{e}(t_2) C^h_{\rhd}(m)= C^h_{\rhd}(m) B^{(\hat{g}(s.p(t)-s.p(m)) h \hat{g}(s.p(t)-s.p(m))^{-1}) \rhd e}(t_2). \label{Equation_blob_ribbon_magnetic_commutation_appendix_1_reversed}
	\end{equation}
	Applying this to Equation \ref{Equation_blob_higher_flux_commutation_reverse_1}, we have
	\begin{align}
		B^e(t)&C^{h,e_m}_T(m)\ket{GS}=B^e(t) C^h_{\rhd}(m) \big[ \prod_{p \in m} B^{f(p)}(\text{blob }0 \rightarrow \text{blob }p)\big] \delta(\hat{e}(m),e_m)\ket{GS} \notag\\
		&= C^h_{\rhd}(m) B^e(t_1) B^{(\hat{g}(s.p(t)-s.p(m)) h \hat{g}(s.p(t)-s.p(m))^{-1}) \rhd e}(t_2)\big[ \prod_{p \in m} B^{f(p)}(\text{blob }0 \rightarrow \text{blob }p)\big] \delta(\hat{e}(m),e_m)\ket{GS}. \label{Equation_blob_higher_flux_commutation_reverse_2}
	\end{align}

	Next we must commute the blob ribbon operators on ribbon $t$ past the blob ribbon operators that are part of $C^h_T(m)$. The ribbon $t_2$ pierces the membrane at a plaquette $q$ and so affects the label $e_q$ of this plaquette. The action of the blob ribbon operator on this plaquette, taking the plaquette to be oriented to match $\hat{e}(m)$ (downwards in Figure \ref{blob_ribbon_through_higher_flux_reversed_split_path}), is
	\begin{equation}
		B^{(\hat{g}(s.p(t)-s.p(m)) h \hat{g}(s.p(t)-s.p(m))^{-1}) \rhd e}(t_2):e_q = e_q \big[\hat{g}(s.p(t)-v_0(q))^{-1} \rhd [\big(\hat{g}(s.p(t)-s.p(m)) h \hat{g}(s.p(t)-s.p(m))^{-1}\big) \rhd e]^{-1}\big] \label{Equation_blob_ribbon_action_on_higher_flux_plaquette_reverse}.
	\end{equation}
	Note that, compared to Equation \ref{Equation_blob_ribbon_action_on_higher_flux_plaquette}, which was the equivalent relation for the opposite orientation of ribbon, not only is the label $e$ inverted, but $e$ is also acted on by $(\hat{g}(s.p(t)-s.p(m)) h \hat{g}(s.p(t)-s.p(m))^{-1}) \rhd $.

	The action of the ribbon operator on the label of $q$ then affects the blob ribbon operator from $C^h_T(m)$ that is associated to plaquette $q$, whose label
	$$[\hat{g}(s.p(m)-v_0(q))\rhd \hat{e}_q] [(h^{-1} \hat{g}(s.p(m)-v_0(q)))\rhd \hat{e}_q^{-1}]$$
	depends on the surface element $e_q$ of plaquette $q$. Using Equation \ref{Equation_blob_ribbon_action_on_higher_flux_plaquette_reverse}, and denoting $\hat{g}(s.p(t)-s.p(m)) h \hat{g}(s.p(t)-s.p(m))^{-1}$ by $h_{[t-m]}$ for brevity, we see that
	\begin{align*}
		B&^{h_{[t-m]} \rhd e}(t_2): \big([\hat{g}(s.p(m)-v_0(q))\rhd \hat{e}_q] [(h^{-1} \hat{g}(s.p(m)-v_0(q)))\rhd \hat{e}_q^{-1}]\big)\\
		&= \big [\hat{g}(s.p(m)-v_0(q))\rhd \big(\hat{e}_q [\hat{g}(s.p(t)-v_0(q))^{-1} \rhd (h_{[t-m]} \rhd e)^{-1}]\big) \big]\\
		& \hspace{0.8cm} \big[(h^{-1}\hat{g}(s.p(m)-v_0(q)))\rhd \big (\hat{e}_q[\hat{g}(s.p(t)-v_0(q))^{-1} \rhd (h_{[t-m]} \rhd e)^{-1}]\big)^{-1} \big]\\
		&= [\hat{g}(s.p(m)-v_0(q))\rhd \hat{e}_q ] [(\hat{g}(s.p(m)-v_0(q))\hat{g}(s.p(t)-v_0(q))^{-1}) \rhd (h_{[t-m]} \rhd e)^{-1}] \\
		& \hspace{0.8cm} [(h^{-1}\hat{g}(s.p(m)-v_0(q)))\rhd \hat{e}_q^{-1}] [(h^{-1}\hat{g}(s.p(m)-v_0(q))\hat{g}(s.p(t)-v_0(q))^{-1}) \rhd (h_{[t-m]} \rhd e)] \\
		&= [\hat{g}(s.p(m)-v_0(q))\rhd \hat{e}_q ] [(h^{-1}\hat{g}(s.p(m)-v_0(q)))\rhd \hat{e}_q^{-1}]\\
		& \hspace{0.8cm} [(\hat{g}(s.p(m)-v_0(q))\hat{g}(s.p(t)-v_0(q))^{-1}) \rhd (h_{[t-m]} \rhd e)^{-1}]\\
		& \hspace{1.1cm} [(h^{-1}\hat{g}(s.p(m)-v_0(q))\hat{g}(s.p(t)-v_0(q))^{-1}) \rhd (h_{[t-m]} \rhd e)]\\
		&=[\hat{g}(s.p(m)-v_0(q))\rhd \hat{e}_q ] [(h^{-1}\hat{g}(s.p(m)-v_0(q)))\rhd \hat{e}_q^{-1}]\\ &\hspace{0.8cm}[\hat{g}(s.p(m)-s.p(t)) \rhd (h_{[t-m]} \rhd e)^{-1}] [(h^{-1}\hat{g}(s.p(m)-s.p(t))) \rhd (h_{[t-m]} \rhd e)]\\
		&=[\hat{g}(s.p(m)-v_0(q))\rhd \hat{e}_q ] [(h^{-1}\hat{g}(s.p(m)-v_0(q)))\rhd \hat{e}_q^{-1}]\\ &\hspace{0.8cm} [\big(\hat{g}(s.p(m)-s.p(t))h_{[t-m]}\big) \rhd e^{-1}] [\big(h^{-1}\hat{g}(s.p(m)-s.p(t))h_{[t-m]}\big) \rhd e].
	\end{align*}
	
	We can simplify this somewhat by noting that 
	$$\hat{g}(s.p(m)-s.p(t)))h_{[t-m]} = \hat{g}(s.p(m)-s.p(t)))\hat{g}(s.p(t)-s.p(m)) h \hat{g}(s.p(t)-s.p(m))^{-1}= h \hat{g}(s.p(t)-s.p(m))^{-1}.$$
	Then
	\begin{align*}
		B&^{h_{[t-m]} \rhd e}(t_2): \big([\hat{g}(s.p(m)-v_0(q))\rhd \hat{e}_q] [(h^{-1} \hat{g}(s.p(m)-v_0(q)))\rhd \hat{e}_q^{-1}]\big)\\
		&= [\hat{g}(s.p(m)-v_0(q))\rhd \hat{e}_q ] [(h^{-1}\hat{g}(s.p(m)-v_0(q)))\rhd \hat{e}_q^{-1}] \\
		& \hspace{0.8cm}[\big(h \hat{g}(s.p(t)-s.p(m))^{-1}\big) \rhd e^{-1}] [\big(h^{-1}h \hat{g}(s.p(t)-s.p(m))^{-1}\big) \rhd e]\\
		&=[\hat{g}(s.p(m)-v_0(q))\rhd \hat{e}_q ] [(h^{-1}\hat{g}(s.p(m)-v_0(q)))\rhd \hat{e}_q^{-1}] [\big(h \hat{g}(s.p(t)-s.p(m))^{-1}\big) \rhd e^{-1}] [ \hat{g}(s.p(t)-s.p(m))^{-1}\rhd e].
	\end{align*}
	
	We can then use this to determine the commutation relation between the blob ribbon operator on $t_2$ and the one associated to plaquette $q$. We have
	\begin{align*}
		&B^{h_{[t-m]} \rhd e}(t_2) B^{[\hat{g}(s.p(m)-v_0(q))\rhd \hat{e}_q] [(h^{-1} \hat{g}(s.p(m)-v_0(q)))\rhd \hat{e}_q^{-1}]}(\text{blob }0 \rightarrow \text{blob }q)\\
		&= B^{[B^{h_{[t-m]} \rhd e}(t_2)]^{-1}: \big([\hat{g}(s.p(m)-v_0(q))\rhd \hat{e}_q] [(h^{-1} \hat{g}(s.p(m)-v_0(q)))\rhd \hat{e}_q^{-1}]\big)}(\text{blob }0 \rightarrow \text{blob }q)B^{h_{[t-m]} \rhd e}(t_2),
	\end{align*}
	where the inverse in $[B^{h_{[t-m]} \rhd e}(t_2)]^{-1}$ is because the blob ribbon operator on $t_2$ is initially to the left of the other ribbon operator, and so acts last (meaning we must remove the action of $[B^{h_{[t-m]} \rhd e}(t_2)]^{-1}$ on the label of the other blob ribbon operator when we commute the operator on $t_2$ to the right). Then we have
	\begin{align*}
		&B^{h_{[t-m]} \rhd e}(t_2) B^{[\hat{g}(s.p(m)-v_0(q))\rhd \hat{e}_q] [(h^{-1} \hat{g}(s.p(m)-v_0(q)))\rhd \hat{e}_q^{-1}]}(\text{blob }0 \rightarrow \text{blob }q)\\
		&= B^{[\hat{g}(s.p(m)-v_0(q))\rhd \hat{e}_q] [(h^{-1} \hat{g}(s.p(m)-v_0(q)))\rhd \hat{e}_q^{-1}][(h \hat{g}(s.p(t)-s.p(m))^{-1}) \rhd e] [ \hat{g}(s.p(t)-s.p(m))^{-1}\rhd e^{-1}]}(\text{blob }0 \rightarrow \text{blob }q)\\
		& \hspace{0.5cm} B^{h_{[t-m]} \rhd e}(t_2)\\
		&=B^{[\hat{g}(s.p(m)-v_0(q))\rhd \hat{e}_q] [(h^{-1} \hat{g}(s.p(m)-v_0(q)))\rhd \hat{e}_q^{-1}]}(\text{blob }0 \rightarrow \text{blob }q)\\
		& \hspace{0.5cm} B^{[(h \hat{g}(s.p(t)-s.p(m))^{-1}) \rhd e] [ \hat{g}(s.p(t)-s.p(m))^{-1}\rhd e^{-1}]}(\text{blob }0 \rightarrow \text{blob }q)B^{h_{[t-m]} \rhd e}(t_2).
	\end{align*}
	
	Just as in the case of the opposite orientation, we see that we have produced an additional blob ribbon operator on the ribbon $(\text{blob }0 \rightarrow \text{blob }q)$, this time with label 
	$$[\big(h \hat{g}(s.p(t)-s.p(m))^{-1}\big) \rhd e] [ \hat{g}(s.p(t)-s.p(m))^{-1}\rhd e^{-1}].$$
	Using this in our overall commutation relation, Equation \ref{Equation_blob_higher_flux_commutation_reverse_2}, gives us
	\begin{align}
		B^e(t)&C^{h,e_m}_T(m)\ket{GS} \notag\\
		&=C^h_{\rhd}(m) \big[ \prod_{p \in m} B^{f(p)}(\text{blob }0 \rightarrow \text{blob }p)\big] B^e(t_1) B^{ [(h \hat{g}(s.p(t)-s.p(m))^{-1}) \rhd e] [\hat{g}(s.p(t)-s.p(m))^{-1} \rhd e^{-1}]}(\text{blob }0 \rightarrow \text{blob }q) \notag \\
		& \hspace{1cm} B^{ h_{[t-m]} \rhd e}(t_2)\delta(\hat{e}(m),e_m)\ket{GS}. \label{Equation_blob_higher_flux_commutation_reverse_3}
	\end{align}

	Next we want to commute the blob ribbon operators (except those in $ \big[ \prod_{p \in m} B^{f(p)}(\text{blob }0 \rightarrow \text{blob }p)\big]$) past $\delta(\hat{e}(m),e_m)$. The total surface label $\hat{e}(m)$ is given by
	$$\hat{e}(m)= \prod_{p \in m} \hat{g}(s.p(m)-v_0(p)) \rhd \hat{e}_p,$$
	where $\hat{e}_p$ is the label of $p$ assuming that it is oriented with $m$ (downwards in Figure \ref{blob_ribbon_through_higher_flux_reversed_split_path}). This time the ribbon on $t_2$ intersects with the direct membrane, and so will affect this surface label, while the ribbon on $t_1$ does not. $t_2$ intersects $m$ at a plaquette $q$. Therefore, using the action of the ribbon operator on plaquette $q$ (see Equation \ref{Equation_blob_ribbon_action_on_higher_flux_plaquette_reverse}), we have
	\begin{align*}
		B^{ h_{[t-m]} \rhd e}(t_2): \hat{e}(m) &= \hat{g}(s.p(m)-v_0(q)) \rhd [\hat{g}(s.p(t)-v_0(q))^{-1} \rhd (h_{[t-m]} \rhd e)^{-1}] \hat{e}(m)\\
		&= [(\hat{g}(s.p(t)-s.p(m)))^{-1}h_{[t-m]}) \rhd e^{-1}] \hat{e}(m)\\
		&= [(h\hat{g}(s.p(t)-s.p(m)^{-1})) \rhd e^{-1}] \hat{e}(m).
	\end{align*}
	Therefore
	\begin{align*}
		B^{h_{[t-m]} \rhd e}(t_2)\delta(e_m,\hat{e}(m))&=\delta(e_m,[B^{h_{[t-m]} \rhd e}(t_2)]^{-1}:\hat{e}(m)	B^{h_{[t-m]} \rhd e}(t_2)\\
		&= \delta(e_m, [(h\hat{g}(s.p(t)-s.p(m)^{-1})) \rhd e] \hat{e}(m)) B^{h_{[t-m]} \rhd e}(t_2)\\
		&=\delta([(h\hat{g}(s.p(t)-s.p(m)^{-1})) \rhd e^{-1}]e_m, \hat{e}(m) )B^{h_{[t-m]} \rhd e}(t_2).
	\end{align*}
	
	The overall commutation relation between the blob ribbon and higher-flux membrane operator (Equation \ref{Equation_blob_higher_flux_commutation_reverse_3}) then becomes
	\begin{align}
		B^e(t)&C^{h,e_m}_T(m)\ket{GS} \notag \\
		&= C^h_{\rhd}(m) \big[ \prod_{p \in m} B^{f(p)}(\text{blob }0 \rightarrow \text{blob }p)\big] \delta([(h\hat{g}(s.p(t)-s.p(m)^{-1})) \rhd e^{-1}]e_m, \hat{e}(m) ) \notag \\
		& \hspace{1cm} B^e(t_1) B^{ [(h \hat{g}(s.p(t)-s.p(m))^{-1}) \rhd e] [\hat{g}(s.p(t)-s.p(m))^{-1} \rhd e^{-1}]}(\text{blob }0 \rightarrow \text{blob }q) B^{ h_{[t-m]} \rhd e}(t_2) \ket{GS} \notag \\
		&= C^{h, (h\hat{g}(s.p(t)-s.p(m)^{-1})) \rhd e^{-1}]e_m }_T(m)B^e(t_1) B^{ [(h \hat{g}(s.p(t)-s.p(m))^{-1}) \rhd e] [\hat{g}(s.p(t)-s.p(m))^{-1} \rhd e^{-1}]}(\text{blob }0 \rightarrow \text{blob }q) \notag \\
		& \hspace{1cm} B^{ h_{[t-m]} \rhd e}(t_2) \ket{GS}. \label{Equation_blob_higher_flux_commutation_reverse_4}
	\end{align}
	
\begin{figure}[h]
		\begin{center}
			\begin{overpic}[width=0.7\linewidth]{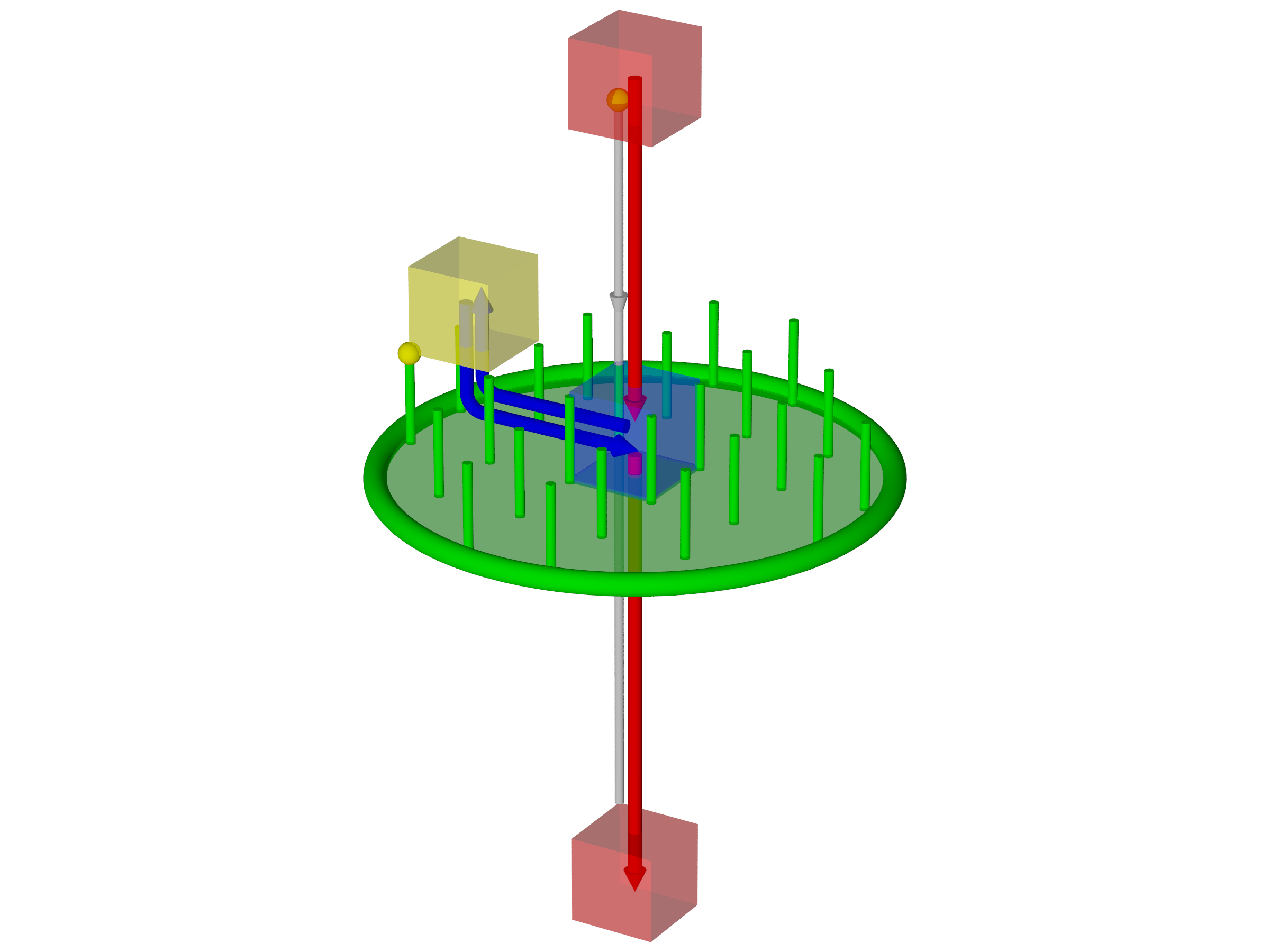}
				\put(51,55){$t_1$}
				\put(51,20){$t_2$}
				\put(26,50){blob 0}
				\put(53,40){blob $q$}
			\end{overpic}
			\caption{Just as was the case for the opposite braiding orientation, commuting the blob ribbon operator $B^e(t)$ past the magnetic membrane operator produced an additional blob ribbon operator. We split this additional operator into two parts and reverse the orientation of one of them, to give two blob ribbon operators acting on the blue ribbons shown in the figure. We can see that these extra ribbons can be connected to $t_1$ and $t_2$.}
			\label{blob_ribbon_through_higher_flux_reversed_four_ribbons}
		\end{center}
	\end{figure}

\begin{figure}[h]
		\begin{center}
			\begin{overpic}[width=0.75\linewidth]{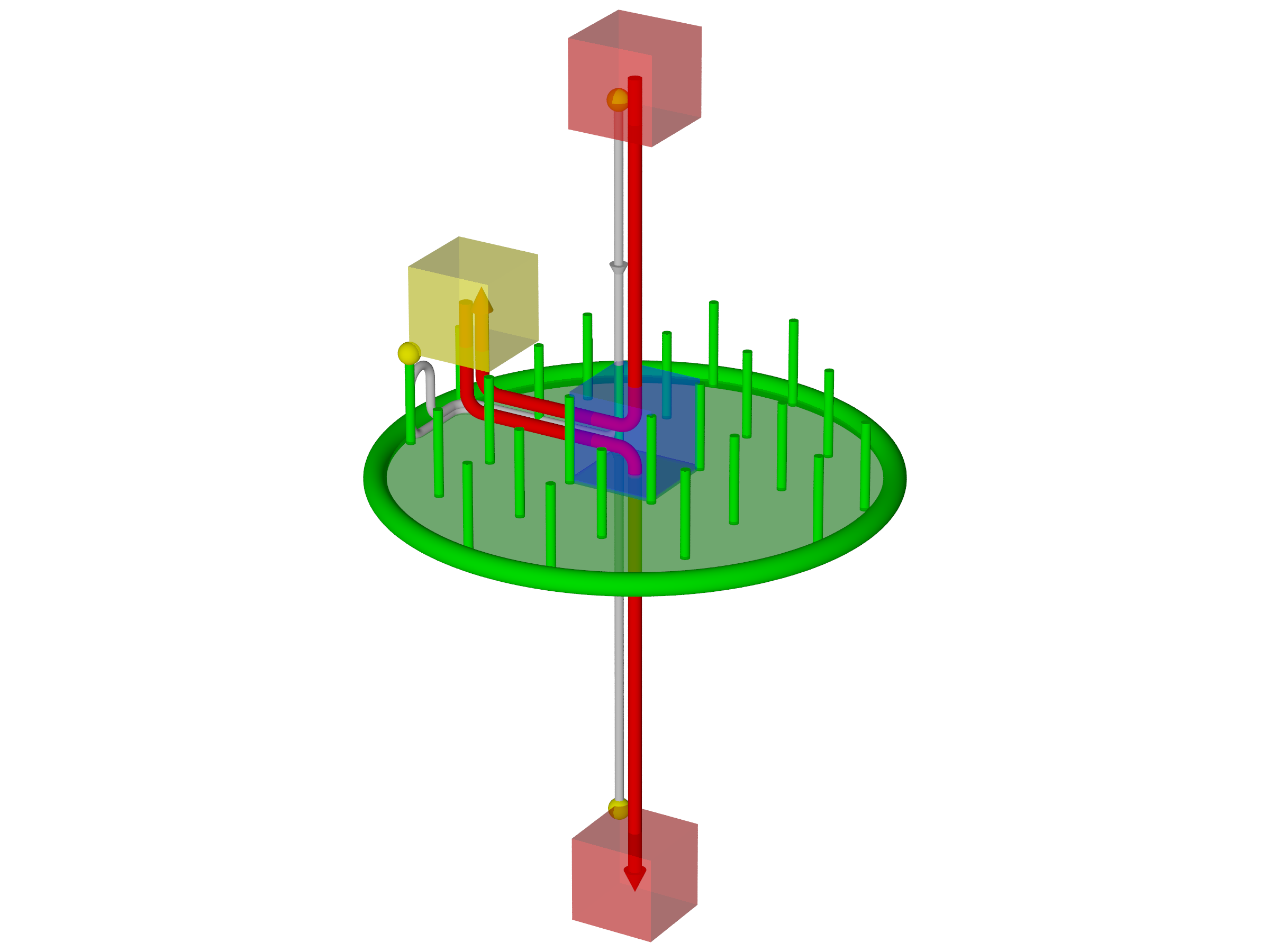}
				\put(51,55){$t_1'$}
				\put(51,20){$t_2'$}
			\end{overpic}
			\caption{By changing the start-points of the blue ribbons from Figure \ref{blob_ribbon_through_higher_flux_reversed_four_ribbons}, we can connect them to the ribbons $t_1$ and $t_2$. This leaves us with blob ribbon operators on the two ribbons shown in this figure.}
			\label{blob_ribbon_through_higher_flux_reversed_final_ribbons}
		\end{center}
	\end{figure}

	Just as in the case of the other braiding orientation, we have an additional blob ribbon operator, this time
	$$B^{ [(h \hat{g}(s.p(t)-s.p(m))^{-1}) \rhd e] [\hat{g}(s.p(t)-s.p(m))^{-1} \rhd e^{-1}]}(\text{blob }0 \rightarrow \text{blob }q).$$ 
	We can split this blob ribbon operator into two parts, 
	$$B^{ (h \hat{g}(s.p(t)-s.p(m))^{-1}) \rhd e} (\text{blob }0 \rightarrow \text{blob }q)$$ 
	and 
	$$B^{\hat{g}(s.p(t)-s.p(m))^{-1} \rhd e^{-1}}(\text{blob }0 \rightarrow \text{blob }q).$$
	We can then invert the direction of the latter part by inverting its label. This gives us $$B^{\hat{g}(s.p(t)-s.p(m))^{-1} \rhd e}(\text{blob }q \rightarrow \text{blob }0).$$ The ribbons associated to these blob ribbon operators are shown in Figure \ref{blob_ribbon_through_higher_flux_reversed_four_ribbons}. As we did previously when considering the other braiding orientation, we wish to connect the ribbons pairwise. We therefore move the start-points of the ribbons that pass between blobs 0 and $q$, from $s.p(m)$ to $s.p(t)$, so that all of the ribbon operators have the same start-point. This gives us the blob ribbon operators 
	$$B^{(\hat{g}(s.p(t)-s.p(m))h \hat{g}(s.p(t)-s.p(m))^{-1}) \rhd e} (\text{blob }0 \rightarrow \text{blob }q|s.p(t))$$ and $$B^{e}(\text{blob }q \rightarrow \text{blob }0|s.p(t)),$$
	where we used the notation $|s.p(t)$ to remind us that the start-points of the direct paths of the ribbons are now at $s.p(t)$. These ribbon operators have the same label (and start-point) as the ribbon operators on $t_2$ and $t_1$ respectively, and have dual paths that lie end-to-end with those ribbons. Therefore, we can concatenate the corresponding ribbons to produce two ribbon operators, $B^{h_{[t-m]} \rhd e}(t_2')$ and $B^e(t_1')$, where the ribbon $t_1'$ passes from the starting blob of ribbon $t$ to blob 0 and $t_2'$ passes from blob 0 to the ending blob of ribbon $t$, as shown in Figure \ref{blob_ribbon_through_higher_flux_reversed_final_ribbons}. Then our final commutation relation is
	\begin{align}
		B^e(t)C^{h,e_m}_T(m)\ket{GS}&= C^{h, [(h\hat{g}(s.p(t)-s.p(m)^{-1})) \rhd e^{-1}]e_m }_T(m)B^e(t_1') B^{ h_{[t-m]} \rhd e}(t_2') \ket{GS}. \label{Equation_blob_higher_flux_commutation_reverse_5}
	\end{align}
	
	This is very similar to the commutation relation for the other orientation, except that the label of the blob ribbon operator on $t_2'$ is $h_{[t-m]} \rhd e$ rather than $h^{-1}_{[t-m]} \rhd e$, reflecting the orientation of the 1-flux of the higher-flux excitation, and the surface label of the higher-flux excitation becomes $(h\hat{g}(s.p(t)-s.p(m)^{-1})) \rhd e^{-1}]e_m$ rather than $[\hat{g}(s.p(t)-s.p(m)^{-1}) \rhd e]e_m$. The change from $e$ to $e^{-1}$ simply reflects the relative orientation of the $E$-valued membrane part of the higher-flux membrane and the ribbon operator. On the other hand the additional $h \rhd$ action comes from the fact that the 1-flux of the higher-flux excitation acts on the blob ribbon operator, changing the 2-flux associated to it, and only afterwards does the blob ribbon operator act on the surface label of the higher-flux excitation. This changes the 2-flux of the higher-flux excitation. This is required to preserve the total 2-flux of the higher-flux and blob excitation. To see this, consider the case where the start-points of the membrane and ribbon are the same. In this case, Equation \ref{Equation_blob_higher_flux_commutation_reverse_5} becomes
	\begin{align}
		B^e(t)C^{h,e_m}_T(m)\ket{GS}&= C^{h, [h \rhd e^{-1}]e_m }_T(m)B^e(t_1') B^{ h \rhd e}(t_2') \ket{GS}. \label{Equation_blob_higher_flux_commutation_reverse_6}
	\end{align}
	
	Recall from the beginning of Section \ref{Section_braiding_higher_flux} that the 2-flux of the higher-flux excitation labelled by $(h,e_m)$ is given by $e_m [ h^{-1} \rhd e_m^{-1}]$, while the 2-flux of the blob excitation at the end of a blob ribbon operator with label $e$ is $e$. The total 2-flux of the blob ribbon and higher-flux excitation is therefore $e_m [h^{-1} \rhd e_m^{-1}] e$ before the braiding. After the braiding, the higher-flux excitation is labelled by $(h, [h \rhd e^{-1}]e_m)$, so its 2-flux is $[h \rhd e^{-1}]e_m [h^{-1} \rhd ([h \rhd e^{-1}]e_m)^{-1}]$. The label of the blob excitation is $h \rhd e$, so its 2-flux is also $h \rhd e$. This means that the total 2-flux is (using the Abelian nature of $E$, so that we do not need to worry about the order of products)
	\begin{align*}
		[h \rhd e^{-1}]e_m [h^{-1} \rhd ([h \rhd e^{-1}]e_m)^{-1}] [h \rhd e]&= [h \rhd e^{-1}] [h \rhd e] e_m [h^{-1} \rhd ([h \rhd e^{-1}]e_m)^{-1}]\\
		&= e_m [h^{-1} \rhd (e_m^{-1}[h \rhd e])]\\
		&=e_m [h^{-1} \rhd e_m^{-1}] e,
	\end{align*}
	which is the same as the total 2-flux before braiding.

	\subsubsection{Braiding with other higher-flux excitations}
	\label{Section_braiding_higher_flux_higher_flux}

\begin{figure}[h]
		\begin{center}
			\includegraphics[width=0.8\linewidth]{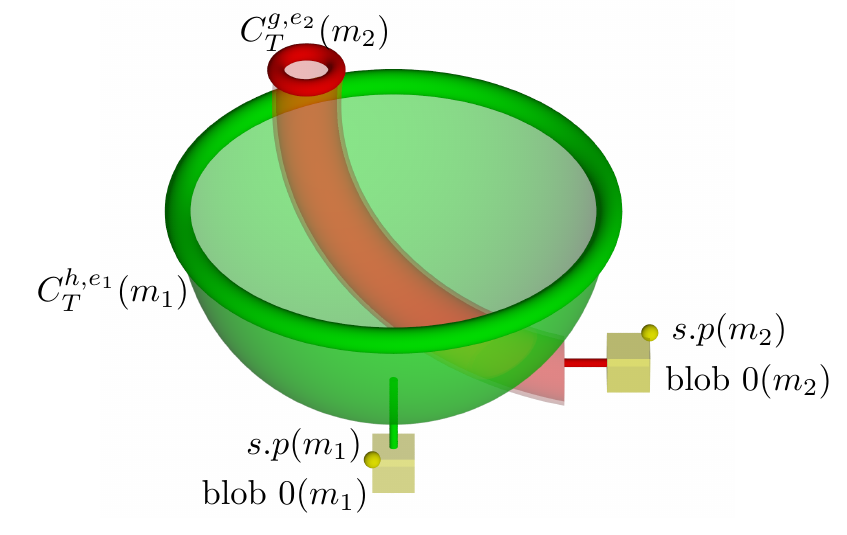}
			\caption{(Copy of Figure \ref{Braid_move_loops} from the main text.) We consider the braiding move where we pull one higher-flux loop excitation (small red torus) through another (large green torus). This can be implemented using higher-flux membranes applied on the (green and red) membranes in the figure. If we first apply the membrane operator $C^{h,e_1}_T(m_1)$ on the larger (green) membrane, then $C_T^{g,e_2}(m_2)$ on the narrower (red) membrane, then we are considering the case where we first produce the larger (green) loop excitation then move the smaller (red) one through it. Comparing this to the opposite order of operators gives us the braiding relation.}
			\label{higher_flux_membrane_through_membrane_appendix}
		\end{center}
	\end{figure}

	Having considered the braiding between the higher-flux and blob excitations, we now have the tools necessary to look at braiding between two higher-flux excitations. As we explained in Section \ref{Section_Flux_Flux_Braiding_Tri_Trivial_Abelian} of the main text, there are two kinds of braiding motion for loops. The first type, which we call permutation, involves moving two loops around each-other without passing through one another. The other involves passing one though the other. As explained in Section \ref{Section_Flux_Flux_Braiding_Tri_Trivial_Abelian}, the permutation move is trivial in this model (in the sense that it doesn't affect the label of the excitations, although the motion itself is non-trivial). Therefore, we consider only the second kind of motion. In order to implement this kind of braiding move, we consider the two membranes shown in Figure \ref{higher_flux_membrane_through_membrane_appendix}. As usual for a braiding relation, the order in which we apply membrane operators on these membranes determines whether braiding occurs or not. If we first apply a membrane operator on the green membrane in Figure \ref{higher_flux_membrane_through_membrane_appendix}, $m_1$, then apply a membrane operator on the red membrane, $m_2'$ (where we used the primed notation because we will shortly deform the membrane into a new one, $m_2$), then we are considering the case where we first produce the green loop before moving the red loop through it. Applying the operators in the other order produces and moves the red loop through empty space before producing the green loop. This means that we are interested in the commutation relation between two higher-flux membrane operators. We therefore consider the state
	\begin{equation}
		C_T^{k,e_2}(m_2') C_T^{h,e_1}(m_1) \ket{GS}.
	\end{equation}
	
	Note that when considering the braiding of the higher-flux loop excitations, we must be careful to fix their orientation. To do so, we choose the relative positions of the direct and dual membranes of the higher-flux membrane operators. We choose the dual membrane to be outside the direct membrane in Figure \ref{higher_flux_membrane_through_membrane_appendix}, so that the edges cut by the dual membrane stick outwards from the membranes.

\begin{figure}[h]
		\begin{center}
			\begin{overpic}[width=0.75\linewidth]{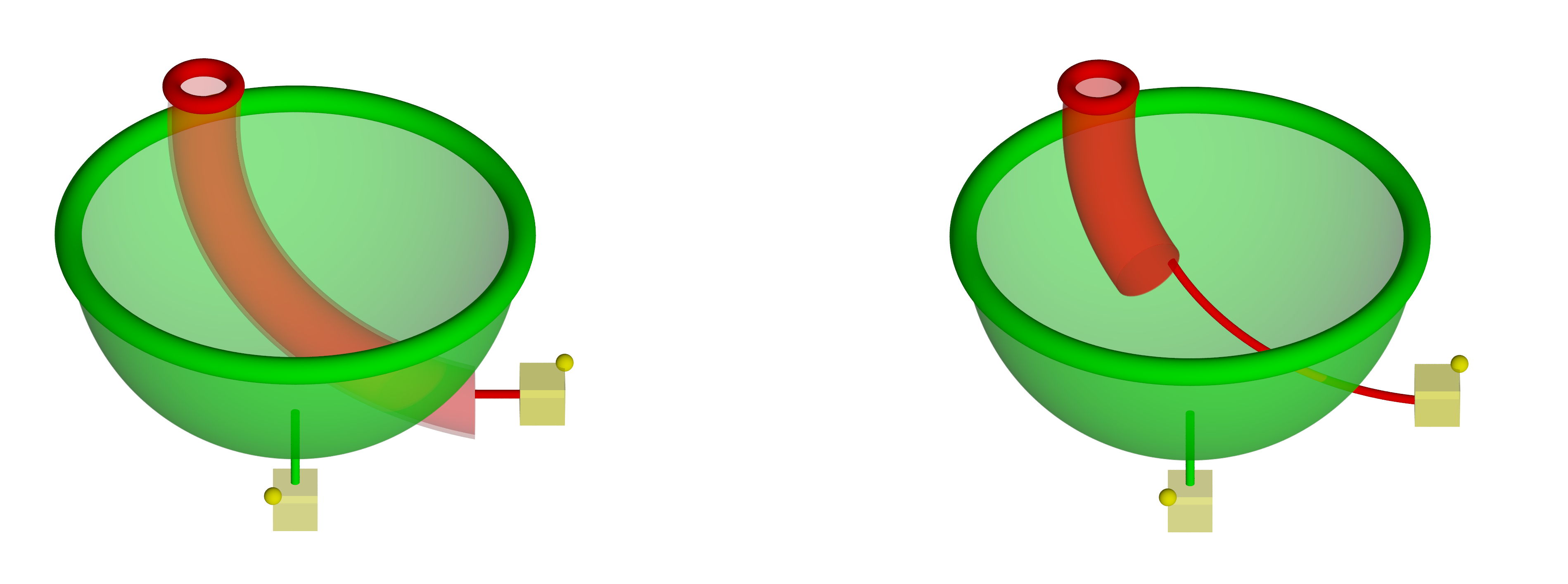}
				\put(45,12){\Huge $\rightarrow$}
				\put(42,17){deform $m_2'$}
				\put(-2,33){\large $C^{k,e_2}_T(m_2')$}
				\put(-5,8){\large $C^{h,e_1}_T(m_1)$}
				
				\put(55,33){\large $C^{k,e_2}_T(m_2)$}
				\put(53,8){\large $C^{h,e_1}_T(m_1)$}
			\end{overpic}
			\caption{In order to compute the commutation relation between the two membrane operators, it is convenient to use the topological property of the membrane operators in order to pull one completely through the other. Pulling $m_2'$ through $m_1$, so that there is no intersection between the membranes, results in the membrane $m_2$ shown in the image on the right. Note that when we perform this deformation, we must keep the positions of the blob 0 and the start-point of $m_2'$ fixed. This means that as we pull the membrane, we drag with it the paths and ribbons that run from the start-point and blob 0 to the membrane. These are represented by the (red) path connecting blob 0 to the membrane, and we see that this path still intersects with $m_1$. This is what provides a non-trivial commutation relation.}
			\label{higher_flux_braid_pull_through}
		\end{center}
	\end{figure}

	Similarly to our approach to the braiding relation for the $\rhd$ trivial case, we will use the topological nature of the membrane operators to calculate the commutation relation. First we deform the thinner membrane, $m_2'$, in order to pull it through the green membrane, as shown in Figure \ref{higher_flux_braid_pull_through}. This gives us a new membrane $m_2$, with an equivalent higher-flux membrane operator on this membrane. Note that when we perform this deformation we must leave the start-point and blob 0 of the membrane in the same position. This is important because the action of the membrane operator on $m_2$ (both the magnetic and the $E$-valued membrane parts of the higher-flux operator) depends on various paths from the start-point to the membrane $m_2$, which will now intersect with the other membrane $m_1$. The labels of these paths will be affected by the membrane operator on $m_1$, therefore changing the action of $C_T^{k,e_2}(m_2)$. Furthermore, $C^k_T(m_2)$ includes blob ribbon operators, which will pierce $m_1$ and so affect the labels of plaquettes on this membrane (the paths and ribbons intersecting with $m_1$ are illustrated by the red path in Figure \ref{higher_flux_braid_pull_through}). These facts are responsible for the commutation relation between the two higher-flux membrane operators being non-trivial, as we will see shortly.
	
	Now let us calculate the commutation relation between the two higher-flux membrane operators on $m_1$ and $m_2$. In order to do so, we first split $C^{k,e_2}_T(m_2)$ into its component parts:
	\begin{align}
		&C_T^{k,e_2}(m_2) C_T^{h,e_1}(m_1) \ket{GS} \notag\\
		&=C^k_{\rhd}(m_2) \big[\prod_{p \in m_2} B^{[\hat{g}(s.p(m_2)-v_0(p))\rhd e] [(k^{-1} \hat{g}(s.p(m_2)-v_0(p)))\rhd e^{-1}]}(\text{blob }0(m_2)-\text{blob }p) \big] \delta(e_2,\hat{e}(m_2)) C^{h,e_1}_T(m_1)\ket{GS}.
		\label{Equation_higher_flux_higher_flux_commutation_1}
	\end{align}

	We now start commuting $C^{h,e_1}_T(m_1)$ to the left. The first step is to consider
	$$\delta(e_2,\hat{e}(m_2))C^{h,e_1}_T(m_1).$$
	The operator $C^{h,e_1}_T(m_1)$ acts on the surface label $\hat{e}(m_2)$ by changing path labels that appear in $\hat{e}(m_2)$. Recall that the surface label $\hat{e}(m_2)$ is given by
	$$\hat{e}(m_2)=\prod_{p \in m_2} \hat{g}(s.p(m_2)-v_0(p))\rhd e_p.$$
	None of the plaquettes $p$ are affected by the membrane operator $C^{h,e_1}_T(m_1)$, because none of them intersect with the membrane $m_1$. However, the paths $(s.p(m_2)-v_0(p))$ do intersect with $m_1$, as we mentioned earlier. Because of this, the $C^h_{\rhd}(m_1)$ part of $C^{h,e_1}_T(m_1)$ acts on these paths as (using Equation \ref{Magnetic_electric_3D_braid_reverse})
	\begin{align}
		C^{h}_{\rhd}(m_1):\hat{g}(s.p(m_2)-v_0(p))&= \hat{g}(s.p(m_1)-s.p(m_2))^{-1}h^{-1}\hat{g}(s.p(m_1)-s.p(m_2))\hat{g}(s.p(m_2)-v_0(p)).
		\label{Equation_higher_flux_membrane_act_on_path_2}
	\end{align}
	
	This means that the action of $C^{h,e_1}_T(m_1)$ on the surface label $\hat{e}(m_2)$ is
	\begin{align*}
		C^{h,e_1}_T(m_1): \hat{e}(m_2) & = \big[ \prod_{p \in m_2} \big(C^{h,e_1}_T(m_1): \hat{g}(s.p(m_2)-v_0(p))\big)\rhd e_p \big] \\
		&= \big[\prod_{p \in m_2} \big(\hat{g}(s.p(m_1)-s.p(m_2))^{-1}h^{-1}\hat{g}(s.p(m_1)-s.p(m_2))\hat{g}(s.p(m_2)-v_0(p))\big)\rhd e_p \big]\\
		&= \big(\hat{g}(s.p(m_1)-s.p(m_2))^{-1}h^{-1}\hat{g}(s.p(m_1)-s.p(m_2))\big) \rhd \big[\prod_{p \in m_2} \hat{g}(s.p(m_2)-v_0(p))\rhd e_p \big]\\
		&= \big(\hat{g}(s.p(m_1)-s.p(m_2))^{-1}h^{-1}\hat{g}(s.p(m_1)-s.p(m_2))\big) \rhd \hat{e}(m_2).
	\end{align*}
	Therefore, the commutation relation between $\delta(e_2,\hat{e}(m_2))$ and $C^{h,e_1}_T(m_1)$ is
	\begin{align*}
		\delta(e_2,\hat{e}(m_2))C^{h,e_1}_T(m_1)&=C^{h,e_1}_T(m_1) \delta\big(e_2, (\hat{g}(s.p(m_1)-s.p(m_2))^{-1}h^{-1}\hat{g}(s.p(m_1)-(m_2)))\rhd \hat{e}(m_2)\big)\\
		&=C^{h,e_1}_T(m_1)\delta (\hat{g}(s.p(m_1)-s.p(m_2))^{-1}h\hat{g}(s.p(m_1)-s.p(m_2))\rhd e_2, \hat{e}(m_2)).
	\end{align*}
	
	We then define the short-hand
	\begin{equation}
		h^{\phantom{-1}}_{[2-1]}= \hat{g}(s.p(m_1)-s.p(m_2))^{-1}h \hat{g}(s.p(m_1)-s.p(m_2))= \hat{g}(s.p(m_2)-s.p(m_1))h \hat{g}(s.p(m_2)-s.p(m_1))^{-1}, 
	\end{equation}
	to write the commutation relation as
	\begin{align*}
		\delta(e_2,\hat{e}(m_2))C^{h,e_1}_T(m_1)&=C^{h,e_1}_T(m_1) \delta\big(e_2, h^{-1}_{[2-1]}\rhd \hat{e}(m_2)\big)\\
		&=C^{h,e_1}_T(m_1)\delta (h^{\phantom{-1}}_{[2-1]}\rhd e_2, \hat{e}(m_2)).
	\end{align*}
	Using this result in the overall commutation relation, Equation \ref{Equation_higher_flux_higher_flux_commutation_1}, gives 
	\begin{align}
		&C^{k,e_2}_T(m_2) C^{h,e_1}_T(m_1)\ket{GS} \notag\\
		&=C^k_{\rhd}(m_2) \big[\prod_{p \in m_2} B^{[\hat{g}(s.p(m_2)-v_0(p))\rhd e] [(k^{-1} \hat{g}(s.p(m_2)-v_0(p)))\rhd e^{-1}]}(\text{blob } 0(m_2)-\text{blob }p) \big] C^{h,e_1}_T(m_1) \delta(h^{\phantom{-1}}_{[2-1]} \rhd e_2, \hat{e}(m_2)),
		\label{Equation_higher_flux_higher_flux_commutation_2}
	\end{align}

	Next we consider
	$$B^{[\hat{g}(s.p(m_2)-v_0(p))\rhd e_p] [(k^{-1} \hat{g}(s.p(m_2)-v_0(p)))\rhd e_p^{-1}]}(\text{blob }0 (m_2) \rightarrow \text{blob }p) C^{h,e_1}_T(m_1),$$
	where we again wish to commute $C^{h,e_1}_T(m_1)$ to the left. In Section \ref{Section_braiding_higher_flux_blob} we considered the commutation relation between the blob ribbon operators and the higher-flux membrane operators. For this orientation of the ribbon operators, the relevant result from that section is Equation \ref{Equation_blob_higher_flux_commutation_reverse_5}. However, we must remember that these blob ribbon operators have an operator-valued label, and this label is itself affected by the commutation with $C^{h,e_1}_T(m_1)$. In order to deal with both sources of non-commutativity, we write each blob ribbon operator as
	\begin{align*}
		B&^{[\hat{g}(s.p(m_2)-v_0(p))\rhd e_p] [k^{-1} \hat{g}(s.p(m_2)-v_0(p))\rhd e_p^{-1}]}(\text{blob }0 (m_2) \rightarrow \text{blob }p)\\
		&= \sum_{e' \in E} B^{e'} (\text{blob }0 (m_2) \rightarrow \text{blob }p)\delta(e', [\hat{g}(s.p(m_2)-v_0(p))\rhd e_p] [k^{-1} \hat{g}(s.p(m_2)-v_0(p))\rhd e_p^{-1}]), 
	\end{align*}
	which will allow us to separately consider the commutation relation of the label of the blob ribbon operator and of the blob ribbon operator itself. Using Equation \ref{Equation_higher_flux_membrane_act_on_path_2}, which describes the transformation of the path element $\hat{g}(s.p(m_2)-v_0(p))$, we have
	\begin{align*}
		\sum_{e' \in E}& B^{e'}(\text{blob }0 (m_2) \rightarrow \text{blob }p) \delta(e', [\hat{g}(s.p(m_2)-v_0(p))\rhd e_p] [(k^{-1} \hat{g}(s.p(m_2)-v_0(p)))\rhd e_p^{-1}]) C^{h,e_1}_T(m_1)\\
		&= \sum_{e' \in E} B^{e'}(\text{blob }0 (m_2) \rightarrow \text{blob }p) C^{h,e_1}_T(m_1) \\
		& \hspace{0.5cm} \delta(e', [(h^{-1}_{[2-1]}\hat{g}(s.p(m_2)-v_0(p)))\rhd e_p] [(k^{-1} h^{-1}_{[2-1]}\hat{g}(s.p(m_2)-v_0(p)))\rhd e_p^{-1}]),
	\end{align*}
	where $h^{\phantom{-1}}_{[2-1]} = \hat{g}(s.p(m_1)-s.p(m_2))^{-1}h \hat{g}(s.p(m_1)-s.p(m_2))$. Then, using Equation \ref{Equation_blob_higher_flux_commutation_reverse_5}, which describes the commutation relation of a (constant-labelled) blob ribbon operator and higher-flux membrane operator, we have
	\begin{align}
		\sum_{e' \in E} &B^{e'}(\text{blob }0 (m_2) \rightarrow \text{blob }p) C^{h,e_1}_T(m_1) \delta(e', [(h^{-1}_{[2-1]}\hat{g}(s.p(m_2)-v_0(p)))\rhd e_p] [(k^{-1} h^{-1}_{[2-1]}\hat{g}(s.p(m_2)-v_0(p)))\rhd e_p^{-1}]) \notag\\
		=&\sum_{e' \in E} C^{h,e_1 [(h \hat{g}(s.p(m_1)-s.p(m_2)))\rhd {e'}^{-1}]}_T(m_1) B^{e'}(\text{blob }0 (m_2)- \text{blob }0(m_1)) \notag \\ &B^{h^{\phantom{-1}}_{[2-1]} \rhd e'}(\text{blob }0(m_1)-\text{blob }p) \delta(e', [(h^{-1}_{[2-1]}\hat{g}(s.p(m_2)-v_0(p)))\rhd e_p] [(k^{-1} h^{-1}_{[2-1]}\hat{g}(s.p(m_2)-v_0(p)))\rhd e_p^{-1}]). \label{Equation_blob_ribbon_operator_commute_higher_flux}
	\end{align}
	
	We then substitute $e' = [(h^{-1}_{[2-1]}\hat{g}(s.p(m_2)-v_0(p)))\rhd e_p] [(k^{-1} h^{-1}_{[2-1]}\hat{g}(s.p(m_2)-v_0(p)))\rhd e_p^{-1}]$ (from the Kronecker delta) into the labels of the blob ribbon operators and the higher-flux membrane operator on $m_1$. The label of the higher-flux membrane operator is then
	\begin{align*}
		e_1 & \: [(h \hat{g}(s.p(m_1)-s.p(m_2)))\rhd {e'}^{-1}] = e_1 \: [(\hat{g}(s.p(m_1)-s.p(m_2))h^{\phantom{-1}}_{[2-1]}) \rhd {e'}^{-1}]\\
		&= e_1 \: (\hat{g}(s.p(m_1)-s.p(m_2))h^{\phantom{-1}}_{[2-1]}) \rhd \big([(h^{-1}_{[2-1]}\hat{g}(s.p(m_2)-v_0(p)))\rhd e_p] [(k^{-1} h^{-1}_{[2-1]}\hat{g}(s.p(m_2)-v_0(p)))\rhd e_p^{-1}]\big)^{-1}\\
		&= e_1 \: \hat{g}(s.p(m_1)-s.p(m_2)) \rhd \big([(h^{\phantom{-1}}_{[2-1]}k^{-1} h^{-1}_{[2-1]}\hat{g}(s.p(m_2)-v_0(p)))\rhd e_p][ \hat{g}(s.p(m_2)-v_0(p))\rhd e_p^{-1}]\big).
	\end{align*}
	
	The label of the blob ribbon operator that runs from blob 0$(m_1)$ to blob $p$ is
	\begin{equation}
		h^{\phantom{-1}}_{[2-1]} \rhd e' = [\hat{g}(s.p(m_2)-v_0(p))\rhd e_p] [(h^{\phantom{-1}}_{[2-1]}k^{-1} h^{-1}_{[2-1]}\hat{g}(s.p(m_2)-v_0(p)))\rhd e_p^{-1}].
	\end{equation}
	Substituting these into Equation \ref{Equation_blob_ribbon_operator_commute_higher_flux} gives
	\begin{align}
		B^{[\hat{g}(s.p(m_2)-v_0(p))\rhd e_p] [k^{-1} \hat{g}(s.p(m_2)-v_0(p))\rhd e_p^{-1}]}&(\text{blob }0 (m_2) \rightarrow \text{blob }p)C^{h,e_1}_T(m_1) \notag\\
		&\hspace{-5cm} = C^{h,e_1 \hat{g}(s.p(m_1)-s.p(m_2)) \rhd ([(h^{\phantom{-1}}_{[2-1]}k^{-1} h^{-1}_{[2-1]}\hat{g}(s.p(m_2)-v_0(p)))\rhd e_p][ \hat{g}(s.p(m_2)-v_0(p))\rhd e_p^{-1}])}_T(m_1) \notag\\
		& \hspace{-4cm}B^{[(h^{-1}_{[2-1]}\hat{g}(s.p(m_2)-v_0(p)))\rhd e_p] [(k^{-1} h^{-1}_{[2-1]}\hat{g}(s.p(m_2)-v_0(p)))\rhd e_p^{-1}]}(\text{blob }0(m_2)- \text{blob }0 (m_1))\notag \\
		& \hspace{-4cm} B^{[\hat{g}(s.p(m_2)-v_0(p))\rhd e_p] [(h^{\phantom{-1}}_{[2-1]}k^{-1} h^{-1}_{[2-1]}\hat{g}(s.p(m_2)-v_0(p)))\rhd e_p^{-1}]}(\text{blob }0(m_1)- \text{blob }p). \label{Equation_blob_ribbon_operator_commute_higher_flux_2}
	\end{align}
	
	Note that the blob ribbon operator is diverted to pass through blob 0 of $m_1$, as described in Section \ref{Section_braiding_higher_flux_blob}. In addition it splits into two parts, one part that runs from blob 0 of $m_2$ to blob 0 of $m_1$ (which represents the part of the ribbon before braiding) and one that runs from blob 0 of $m_1$ to its original end-point, blob $p$. Equation \ref{Equation_blob_ribbon_operator_commute_higher_flux_2} holds for each blob ribbon operator in $C^h_T(m_2)$, so we can sequentially commute each of the blob ribbon operators past $C^{h,e_1}_T(m_1)$ in Equation \ref{Equation_higher_flux_higher_flux_commutation_2}. This then gives us the commutation relation 
	\begin{align}
		&C_T^{k,e_2}(m_2)C_T^{h,e_1}(m_1)\ket{GS} \notag\\
		&=C^k_{\rhd}(m_2) \big[ \prod_{p \in m_2} B^{[\hat{g}(s.p(m_2)-v_0(p))\rhd e_p] [k^{-1} \hat{g}(s.p(m_2)-v_0(p))\rhd e_p^{-1}]}(\text{blob }0 (m_2) \rightarrow \text{blob }p) \big] C^{h,e_1}_T(m_1) \notag\\
		& \hspace{0.5cm} \delta(h^{\phantom{-1}}_{[2-1]} \rhd e_2, \hat{e}(m_2))\ket{GS} \notag \\
		&= C^k_{\rhd}(m_2) C_T^{h,e_1 \big[\prod_{p \in m_2} \hat{g}(s.p(m_1)-s.p(m_2)) \rhd \big([(h^{\phantom{-1}}_{[2-1]}k^{-1} h^{-1}_{[2-1]}\hat{g}(s.p(m_2)-v_0(p)))\rhd e_p][ \hat{g}(s.p(m_2)-v_0(p))\rhd e_p^{-1}]\big)\big]}(m_1) \notag \\
		& \hspace{0.5cm} \bigg(\prod_{p \in m_2} B^{[(h^{-1}_{[2-1]}\hat{g}(s.p(m_2)-v_0(p)))\rhd e_p] [(k^{-1} h^{-1}_{[2-1]}\hat{g}(s.p(m_2)-v_0(p)))\rhd e_p^{-1}]}(\text{blob } 0(m_2)-\text{blob }0 (m_1))\notag \\
		& \hspace{0.5cm} B^{[\hat{g}(s.p(m_2)-v_0(p))\rhd e_p] [(h^{\phantom{-1}}_{[2-1]}k^{-1} h^{-1}_{[2-1]}\hat{g}(s.p(m_2)-v_0(p)))\rhd e_p^{-1}]}(\text{blob }0(m_1)-\text{blob }p) \bigg)\delta(h^{\phantom{-1}}_{[2-1]} \rhd e_2, \hat{e}(m_2))\ket{GS}. \label{Equation_higher_flux_higher_flux_commutation_3}
	\end{align}
	
	Now consider the $E$ label of the higher-flux membrane on $m_1$. This label is
	\begin{align*}
		e_1 \bigg(\prod_{p \in m_2} &\hat{g}(s.p(m_1)-s.p(m_2)) \rhd ([(h^{\phantom{-1}}_{[2-1]}k^{-1} h^{-1}_{[2-1]}\hat{g}(s.p(m_2)-v_0(p)))\rhd e_p][ \hat{g}(s.p(m_2)-v_0(p))\rhd e_p^{-1}])\bigg)\\
		&= e_1 \bigg(\prod_{p \in m_2} [( \hat{g}(s.p(m_1)-s.p(m_2)) h^{\phantom{-1}}_{[2-1]}k^{-1} h^{-1}_{[2-1]}) \rhd (\hat{g}(s.p(m_2)-v_0(p))\rhd e_p)] \\
		& \hspace{1cm} [\hat{g}(s.p(m_1)-s.p(m_2)) \rhd ( \hat{g}(s.p(m_2)-v_0(p))\rhd e_p^{-1})] \bigg)\\
		&= e_1 \big[( \hat{g}(s.p(m_1)-s.p(m_2)) h^{\phantom{-1}}_{[2-1]}k^{-1} h^{-1}_{[2-1]}) \rhd \big(\prod_{p \in m_2} \hat{g}(s.p(m_2)-v_0(p))\rhd e_p \big) \big]\\
		& \hspace{1cm} \big[\hat{g}(s.p(m_1)-s.p(m_2)) \rhd \big( \prod_{p \in m_2} \hat{g}(s.p(m_2)-v_0(p))\rhd e_p^{-1} \big) \big]\\
		&= e_1 [( \hat{g}(s.p(m_1)-s.p(m_2)) h^{\phantom{-1}}_{[2-1]}k^{-1} h^{-1}_{[2-1]}) \rhd \hat{e}(m_2)] [\hat{g}(s.p(m_1)-s.p(m_2)) \rhd \hat{e}(m_2)^{-1}],
	\end{align*}
	where $\hat{e}(m_2)= \prod_{p \in m_2} \hat{g}(s.p(m_2)-v_0(p))\rhd e_p$ is the total surface element of membrane $m_2$. Then using the constraint enforced by $\delta(h^{\phantom{-1}}_{[2-1]} \rhd e_2, \hat{e}(m_2))$, we can write the $E$ label of the membrane operator on $m_1$ as
	$$ e_1 [( \hat{g}(s.p(m_1)-s.p(m_2)) h^{\phantom{-1}}_{[2-1]}k^{-1}) \rhd e_2] [(\hat{g}(s.p(m_1)-s.p(m_2))h^{\phantom{-1}}_{[2-1]}) \rhd e_2^{-1}].$$
	
	Substituting this into Equation \ref{Equation_higher_flux_higher_flux_commutation_3} gives us
	\begin{align}
		C_T^{k,e_2}(m_2)&C_T^{h,e_1}(m_1)\ket{GS}\notag \\
		&= C^k_{\rhd}(m_2) C_T^{h, e_1 [( \hat{g}(s.p(m_1)-s.p(m_2)) h^{\phantom{-1}}_{[2-1]}k^{-1}) \rhd e_2] [(\hat{g}(s.p(m_1)-s.p(m_2))h^{\phantom{-1}}_{[2-1]}) \rhd e_2^{-1}]}(m_1) \notag \\
		& \bigg(\prod_{p \in m_2}B^{[(h^{-1}_{[2-1]}\hat{g}(s.p(m_2)-v_0(p)))\rhd e_p] [(k^{-1} h^{-1}_{[2-1]}\hat{g}(s.p(m_2)-v_0(p)))\rhd e_p^{-1}]}(\text{blob }0(m_2)- \text{blob }0 (m_1))\notag \\
		& B^{[\hat{g}(s.p(m_2)-v_0(p))\rhd e_p] [(h^{\phantom{-1}}_{[2-1]}k^{-1} h^{-1}_{[2-1]}\hat{g}(s.p(m_2)-v_0(p)))\rhd e_p^{-1}]}(\text{blob }0(m_1)- \text{blob }p) \bigg)\delta(h^{\phantom{-1}}_{[2-1]} \rhd e_2, \hat{e}(m_2))\ket{GS}. \label{Equation_higher_flux_higher_flux_commutation_4}
	\end{align}
	
	The final step to obtain the commutation relation is to commute $C^k_{\rhd}(m_2)$ past the higher-flux membrane operator on $m_1$. The only way that these two operators interact is through the higher-flux membrane operator on $m_1$ affecting the direct paths from the start-point of $m_2$ to the edges (and plaquettes) affected by $C^k_{\rhd}(m_2)$. The action of $C^k_{\rhd}(m_2)$ on an edge $i$, with label $g_i$, cut by the dual membrane of $m_2$ is
	$$C^k_{\rhd}(m_2): g_i = \begin{cases} \hat{g}(s.p(m_2)-v_i)^{-1}k\hat{g}(s.p(m_2)-v_i)g_i & \text{ if $i$ points away from the direct membrane}\\
		g_i \hat{g}(s.p(m_2)-v_i)^{-1}k^{-1}\hat{g}(s.p(m_2)-v_i) & \text{ if $i$ points towards the direct membrane}, \end{cases}$$
	where $v_i$ is the end of the edge on the direct membrane of $m_2$. Similarly, the action of $C^k_{\rhd}(m_2)$ on a plaquette $p$ that is cut by the dual membrane of $m_2$, and whose base-point $v_0(p)$ lies on the direct membrane of $m_2$ is
	$$C^k_{\rhd}(m_2): e_p = (\hat{g}(s.p(m_2)-v_0(p))^{-1}k\hat{g}(s.p(m_2)-v_0(p))) \rhd e_p,$$
	while the membrane operator acts trivially on plaquettes whose base-points do not lie on the direct membrane. In both the action on the edges and the plaquettes, the action depends on the quantity $\hat{g}(s.p(m_2)-v)^{-1}k\hat{g}(s.p(m_2)-v)$, where $v$ is a vertex on the direct membrane of $m_2$. From Section \ref{Section_electric_magnetic_braiding_3D_tri_trivial}, we know how the path element $\hat{g}(s.p(m_2)-v)$ is affected by the magnetic membrane operator $C^h_T(m_1)$. We have, using Equation \ref{Magnetic_electric_3D_braid_reverse}, that $C^{h,e}_T(m_1): \hat{g}(s.p(m_2)-v)=h_{[2-1]}^{-1}\hat{g}(s.p(m_2)-v)$, where
	$$h_{[2-1]}= \hat{g}(s.p(m_2)-s.p(m_1)) h \hat{g}(s.p(m_2)-s.p(m_1))^{-1}.$$
	This means that
	$$C^{h,e}_T(m_1): \hat{g}(s.p(m_2)-v)^{-1}k \hat{g}(s.p(m_2)-v)= \hat{g}(s.p(m_2)-v)^{-1}h^{\phantom{-1}}_{[2-1]}k h_{[2-1]}^{-1} \hat{g}(s.p(m_2)-v).$$
	We see that this leads to conjugation of the label $k$ by $h^{\phantom{-1}}_{[2-1]}$. Therefore
	\begin{equation}
		C^k_{\rhd}(m_2) C^h_T(m_1) = C^h_T(m_1) C^{h^{\phantom{-1}}_{[2-1]}kh_{[2-1]}^{-1}}_{\rhd}(m_2). \label{Equation_higher_flux_braiding_magnetic_commutation}
	\end{equation}
	In addition, the $E$ part of the higher-flux membrane operator on $m_1$ commutes freely past $C^k_{\rhd}(m_2)$ (its label is unaffected by $C^k_{\rhd}(m_2)$). Therefore
	$$C^k_{\rhd}(m_2)C^{h,x}_T(m_1)=C^{h,x}_T(m_1) C^{h^{\phantom{-1}}_{[2-1]}kh_{[2-1]}^{-1}}_{\rhd}(m_2),$$
	where $x=e_1 [( \hat{g}(s.p(m_1)-s.p(m_2)) h^{\phantom{-1}}_{[2-1]}k^{-1}) \rhd e_2] [(\hat{g}(s.p(m_1)-s.p(m_2))h^{\phantom{-1}}_{[2-1]}) \rhd e_2^{-1}]$. Substituting this result into the commutation relation Equation \ref{Equation_higher_flux_higher_flux_commutation_4} gives us the final result
	\begin{align}
		C_T^{k,e_2}(m_2)&C_T^{h,e_1}(m_1)\ket{GS}\notag \\
		&=C^{h,e_1 [\hat{g}(s.p(m_1)-s.p(m_2)) \rhd ([h^{\phantom{-1}}_{[2-1]} \rhd e_2^{-1}] [ (h^{\phantom{-1}}_{[2-1]} k^{-1}) \rhd e_2])]}_T(m_1) C_\rhd^{h^{\phantom{-1}}_{[2-1]}kh_{[2-1]}^{-1}}(m_2) \notag \\
		& \bigg( \prod_{p \in m_2}B^{[(h^{-1}_{[2-1]}\hat{g}(s.p(m_2)-v_0(p)))\rhd e_p] [(k^{-1} h^{-1}_{[2-1]}\hat{g}(s.p(m_2)-v_0(p)))\rhd e_p^{-1}]}(\text{blob }0(m_2)- \text{blob }0 (m_1))\notag \\
		& B^{[\hat{g}(s.p(m_2)-v_0(p))\rhd e_p] [(h^{\phantom{-1}}_{[2-1]}k^{-1} h^{-1}_{[2-1]}\hat{g}(s.p(m_2)-v_0(p)))\rhd e_p^{-1}]}(\text{blob }0(m_1)- \text{blob }p) \bigg)\delta(h^{\phantom{-1}}_{[2-1]} \rhd e_2, \hat{e}(m_2))\ket{GS}.
		\label{Equation_higher_flux_higher_flux_commutation_5}
	\end{align}
	
	Note that, as described in Section \ref{Section_higher_flux_higher_flux_braiding} of the main text, there is an apparent change to the label of the blob ribbon operators associated to membrane $m_2$, even in the section $(\text{blob }0(m_2)- \text{blob }0 (m_1))$ before the braiding. However, this apparent change is just an artefact caused by the operator label of the ribbon operator. To see that this change does not reflect a real change in label, consider the term corresponding to the blob ribbon operators before the intersection:
	$$B^{[(h^{-1}_{[2-1]}\hat{g}(s.p(m_2)-v_0(p)))\rhd e_p] [(k^{-1} h^{-1}_{[2-1]}\hat{g}(s.p(m_2)-v_0(p)))\rhd e_p^{-1}]}(\text{blob }0(m_2)- \text{blob }0 (m_1)).$$
	Because all of these blob ribbon operators are applied on the same ribbon (or can be deformed so that they lie on the same ribbon), they can be combined into a single blob ribbon operator with label
	\begin{align}
		\bigg(\prod_{p \in m_2}& [(h^{-1}_{[2-1]}\hat{g}(s.p(m_2)-v_0(p)))\rhd e_p] [(k^{-1} h^{-1}_{[2-1]}\hat{g}(s.p(m_2)-v_0(p)))\rhd e_p^{-1}] \bigg) \notag \\
		&= \big[h_{[2-1]}^{-1}\rhd \big( \prod_{p \in m_2} \hat{g}(s.p(m_2)-v_0(p)) \rhd e_{p} \big) \big] \big[(k^{-1}h_{[2-1]}^{-1})\rhd \big( \prod_{p \in m_2} \hat{g}(s.p(m_2)-v_0(p)) \rhd e_{p} \big)^{-1}\big]. \label{Equation_combined_blob_ribbon_label_1_appendix}
	\end{align}
	The expression 
	$$\prod_{p \in m_2} \hat{g}(s.p(m_2) -v_0(p)) \rhd e_p $$
	is just the surface element $\hat{e}(m_2)$ (note that we have defined all of the plaquettes on $m_2$ to share the same orientation). Therefore, we can write the label of the combined blob ribbon operator from Equation \ref{Equation_combined_blob_ribbon_label_1_appendix} as
	$$[h_{[2-1]}^{-1}\rhd \hat{e}(m_2) ] [(k^{-1}h_{[2-1]}^{-1})\rhd \hat{e}(m_2)^{-1}].$$
	Substituting this into Equation \ref{Equation_higher_flux_higher_flux_commutation_5} for the braiding relation, we obtain
	\begin{align}
		&C_T^{k,e_2}(m_2)C_T^{h,e_1}(m_1)\ket{GS}\notag \\
		&=C^{h,e_1 [\hat{g}(s.p(m_1)-s.p(m_2)) \rhd ([h^{\phantom{-1}}_{[2-1]} \rhd e_2^{-1}] [ (h^{\phantom{-1}}_{[2-1]} k^{-1}) \rhd e_2])]}_T(m_1) C_\rhd^{h^{\phantom{-1}}_{[2-1]}kh_{[2-1]}^{-1}}(m_2) \notag \\
		& \hspace{0.5cm} B^{[h^{-1}_{[2-1]}\rhd \hat{e}(m_2)] [(k^{-1} h^{-1}_{[2-1]}) \rhd \hat{e}(m_2)^{-1}]}(\text{blob }0(m_2)- \text{blob }0 (m_1))\notag \\
		& \hspace{0.5cm} \big[ \prod_{p \in m_2}B^{[\hat{g}(s.p(m_2)-v_0(p))\rhd e_p] [(h^{\phantom{-1}}_{[2-1]}k^{-1} h^{-1}_{[2-1]}\hat{g}(s.p(m_2)-v_0(p)))\rhd e_p^{-1}]}(\text{blob }0(m_1)- \text{blob }p) \big]\delta(h^{\phantom{-1}}_{[2-1]} \rhd e_2, \hat{e}(m_2))\ket{GS}.
		\label{Equation_higher_flux_braiding_appendix_2}
	\end{align}

	We can then use the $\delta(h^{\phantom{-1}}_{[2-1]} \rhd e_2, \hat{e}(m_2))$, which fixes the value of the surface element $\hat{e}(m_2)$, to write Equation \ref{Equation_higher_flux_braiding_appendix_2} as
	\begin{align}
		&C_T^{k,e_2}(m_2)C_T^{h,e_1}(m_1)\ket{GS}\notag \\
		&=C^{h,e_1 [\hat{g}(s.p(m_1)-s.p(m_2)) \rhd ([h^{\phantom{-1}}_{[2-1]} \rhd e_2^{-1}] [ (h^{\phantom{-1}}_{[2-1]} k^{-1}) \rhd e_2])]}_T(m_1) C_\rhd^{h^{\phantom{-1}}_{[2-1]}kh_{[2-1]}^{-1}}(m_2) \notag \\
		& \hspace{0.5cm} B^{e_2 [k^{-1} \rhd e_2^{-1}]}(\text{blob }0(m_2)- \text{blob }0 (m_1))\notag \\
		& \hspace{0.5cm} \big[ \prod_{p \in m_2}B^{[\hat{g}(s.p(m_2)-v_0(p))\rhd e_p] [(h^{\phantom{-1}}_{[2-1]}k^{-1} h^{-1}_{[2-1]}\hat{g}(s.p(m_2)-v_0(p)))\rhd e_p^{-1}]}(\text{blob }0(m_1)- \text{blob }p) \big]\delta(h^{\phantom{-1}}_{[2-1]} \rhd e_2, \hat{e}(m_2))\ket{GS}.
		\label{Equation_higher_flux_braiding_appendix_3}
	\end{align}

	We can then see that the blob ribbon operator before the intersection (representing the label before braiding) is labelled by $e_2 [k^{-1}\rhd e_2^{-1}]$, which is the same label as it would have without braiding (as we would expect). On the other hand the sections of the blob ribbon operators after braiding have their labels changed from 
	$$[\hat{g}(s.p(m_2)-v_0(p))\rhd e_p] [(k^{-1}\hat{g}(s.p(m_2)-v_0(p)))\rhd e_p^{-1}]$$ 
	to 
	$$[\hat{g}(s.p(m_2)-v_0(p))\rhd e_p] [(h^{\phantom{-1}}_{[2-1]}k^{-1} h^{-1}_{[2-1]}\hat{g}(s.p(m_2)-v_0(p)))\rhd e_p^{-1}].$$
	This reflects the fact that the label of the magnetic part of the higher-flux operator changes from $k$ to $h^{\phantom{-1}}_{[2-1]}k h^{-1}_{[2-1]}$ under braiding (i.e., the blob ribbon operator carries the label we expect for the label of the braided higher flux membrane).

	Apart from the splitting of the blobs at the intersection as we saw with the braiding between blob ribbon operators and higher-flux excitations, we see the following transformations for the labels of the two higher-flux operators
	\begin{align}
		h &\rightarrow h, \label{higher_flux_higher_flux_braid_result_1} \\
		e_1 &\rightarrow e_1 \ \hat{g}(s.p(m_1)-s.p(m_2)) \rhd ([h^{\phantom{-1}}_{[2-1]} \rhd e_2^{-1}][ (h^{\phantom{-1}}_{[2-1]}k^{-1}) \rhd e_2]),\\
		k &\rightarrow h^{\phantom{-1}}_{[2-1]}kh_{[2-1]}^{-1},\\
		e_2 &\rightarrow h^{\phantom{-1}}_{[2-1]} \rhd e_2. \label{higher_flux_higher_flux_braid_result_4}
	\end{align}
	We note that the $G$ labels transform in the same way as in Equation \ref{Equation_magnetic_membranes_commute_reverse_1}, which describes the same braiding of magnetic excitations in the case where $\rhd$ is trivial.

	Now we want to consider the case where the start-points of the two membrane operators are the same. In this case Equations \ref{higher_flux_higher_flux_braid_result_1} - \ref{higher_flux_higher_flux_braid_result_4} simplify to
	\begin{align}
		h &\rightarrow h, \label{higher_flux_higher_flux_braid_result_same_sp_1}\\
		e_1 &\rightarrow e_1 [h \rhd e_2^{-1}] [(hk^{-1}) \rhd e_2],\\
		k &\rightarrow hkh^{-1},\\
		e_2 &\rightarrow h \rhd e_2, \label{higher_flux_higher_flux_braid_result_same_sp_4}
	\end{align}
	which eliminates the operator dependence of the braiding results. It is interesting to consider these results in terms of the 2-flux of the two higher-flux excitations, rather than just the $E$ labels of the operators. Therefore, we define
	$$\tilde{e}_1=e_1 [h^{-1} \rhd e_1^{-1}],$$
	$$\tilde{e}_2=e_2 [k^{-1} \rhd e_2^{-1}],$$
	which are the 2-fluxes of the two loop excitations (as discussed at the start of Section \ref{Section_braiding_higher_flux}). Under the braiding described by Equations \ref{higher_flux_higher_flux_braid_result_same_sp_1} - \ref{higher_flux_higher_flux_braid_result_same_sp_4}, the flux $\tilde{e}_1$ transforms as
	\begin{align*}
		\tilde{e}_1&=e_1 [h^{-1} \rhd e_1^{-1}] \\
		&\rightarrow (e_1 [h \rhd e_2^{-1}] [ (hk^{-1}) \rhd e_2]) [h^{-1} \rhd ( e_1 [h \rhd e_2^{-1}] [ (hk^{-1}) \rhd e_2])^{-1}]\\
		&=e_1 [h \rhd e_2^{-1}] [ (hk^{-1}) \rhd e_2] [k^{-1} \rhd e_2^{-1}] e_2 [h^{-1} \rhd e_1^{-1}]\\
		&= e_1 [h^{-1} \rhd e_1^{-1}] e_2 [k^{-1} \rhd e_2^{-1}] [h \rhd(e_2^{-1} [k^{-1} \rhd e_2])]\\
		&= \tilde{e}_1 \tilde{e}_2 [h \rhd \tilde{e}_2^{-1}].
	\end{align*}
	
	Meanwhile, the flux $\tilde{e}_2$ transforms as
	\begin{align*}
		\tilde{e}_2&= e_2 [k^{-1} \rhd e_2^{-1}]\\
		& \rightarrow [h \rhd e_2] [(hkh^{-1})^{-1} \rhd (h \rhd e_2)^{-1}]\\
		&=[h \rhd e_2] [(hk^{-1}h^{-1}) \rhd (h \rhd e_2^{-1})]\\
		&= [h \rhd e_2] [(hk^{-1}) \rhd e_2^{-1}]\\
		&= h \rhd (e_2 [k^{-1} \rhd e_2^{-1}])\\
		&= h \rhd \tilde{e}_2.
	\end{align*}
	This means that the product of the 2-fluxes transforms as
	\begin{align*}
		\tilde{e}_1 \tilde{e}_2 \rightarrow& (\tilde{e}_1 \tilde{e}_2 [h \rhd \tilde{e}_2^{-1}]) [h \rhd \tilde{e}_2]\\
		&=\tilde{e}_1 \tilde{e}_2,
	\end{align*}
	so this product is conserved by the braiding. We will see shortly that this product represents the combined 2-flux of the loop excitations, indicating that the total 2-flux is conserved by the braiding process, as we may expect. If we write Equations \ref{higher_flux_higher_flux_braid_result_same_sp_1} - \ref{higher_flux_higher_flux_braid_result_same_sp_4} in terms of these 2-fluxes, the braiding relation is
	\begin{equation}
		((k,\tilde{e}_2),(h,\tilde{e}_1))\rightarrow ((h,\tilde{e}_1\tilde{e}_2 [h\rhd \tilde{e}_2^{-1}]), (hkh^{-1}, h \rhd \tilde{e}_2)), \label{higher_flux_higher_flux_braid_result_2_fluxes}
	\end{equation}
	where we swapped the position of the labels of the two loops in addition to applying the change to their labels from the braiding, in order to represent the fact that one loop moves past the other, so the order of the loops swaps. This is important when we consider the total 1-flux for example, where we need to be careful to take the product of the 1-fluxes in the correct order due to the non-Abelian nature of $G$. The 1-flux before the braiding is $hk$. After the braiding, the loops have swapped position and we must take the order of multiplication to be opposite, so we obtain $k'h'$, where $k'$ and $h'$ are the 1-flux labels of the two loops after braiding. Using $h'=h$ and $k' = hkh^{-1}$, this gives a 1-flux of $k'h'=hkh^{-1}h =hk$, indicating that the 1-flux is conserved.

	We are also interested in the inverse transformation, which corresponds to the opposite braiding orientation (either swapping the orientation of both membranes or performing the braiding in reverse). We will therefore invert the transformation in Equation \ref{higher_flux_higher_flux_braid_result_2_fluxes}. Denoting the labels resulting from the forwards transformations as primed versions of the labels, we have:
	\begin{align*}
		\tilde{e}_2'&= h \rhd \tilde{e}_2, \: h'=h\implies \tilde{e}_2={h'}^{-1} \rhd \tilde{e}_2'\\
		k'&=hkh^{-1} \implies k={h'}^{-1}k'h'\\
		\tilde{e}_1'&=\tilde{e}_1 [h \rhd \tilde{e}_2^{-1}] \tilde{e}_2 \implies \tilde{e}_1=\tilde{e}_1'\tilde{e}_2' [{h'}^{-1} \rhd {{\tilde{e}_2}}^{\prime -1}].
	\end{align*}
	Therefore, the inverse transformation is
	\begin{equation}
		((h',\tilde{e}_1'),(k',\tilde{e}_2')) \rightarrow (({h'}^{-1}k'h',[{h'}^{-1}\rhd \tilde{e}_2']),(h', \tilde{e}_1'\tilde{e}_2' [{h'}^{-1}\rhd {\tilde{e}_2}^{\prime -1}])),
	\end{equation}
	where we have again swapped the positions of the labels to represent the loops swapping positions. We note that this braiding relation matches the braiding found in Ref. \cite{Bullivant2018} based on discussions of the loop braid group.

\begin{figure}[h]
		\begin{center}
			\begin{overpic}[width=0.6\linewidth]{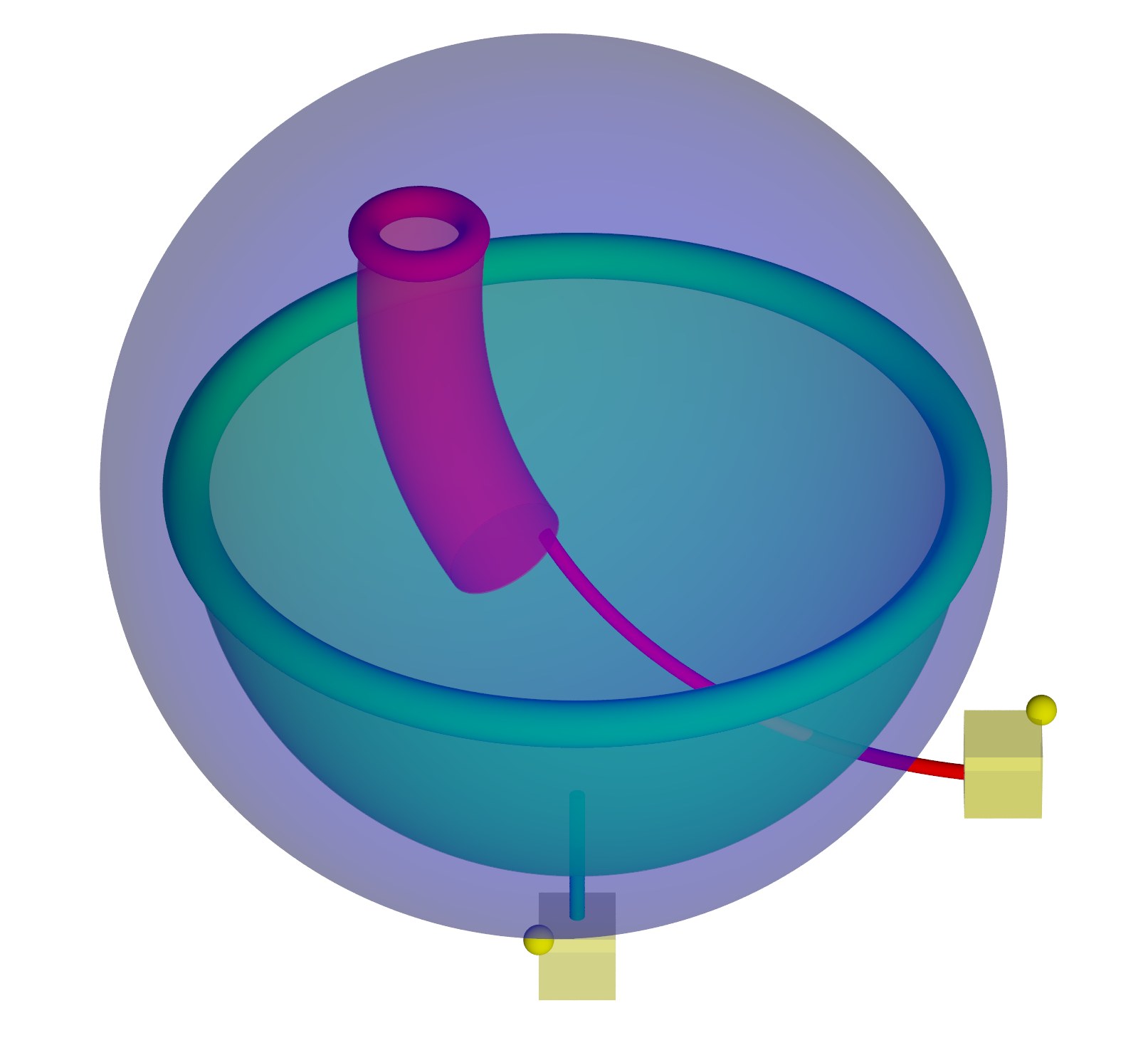}
				
			\end{overpic}
			\caption{We consider measuring the total surface label of a surface $c$ (the blue sphere) which encloses both of the loop-like excitations, but not the start-point or blob 0 of either of the higher-flux membrane operators. We will see that this has a surface label of $\tilde{e}_1 \tilde{e}_2$}
			\label{higher_flux_braid_combined_2_flux}
		\end{center}
	\end{figure}

	Next we consider the quantity $\tilde{e}_1 \tilde{e}_2$, which is preserved by the braiding. As we mentioned earlier, this is the combined 2-flux of the two loop-like excitations. In order to see this, consider a surface $c$ enclosing both loop-like excitations and oriented outwards, as shown in Figure \ref{higher_flux_braid_combined_2_flux}. We can find the label of this surface by applying an $E$-valued membrane operator $\delta(\hat{e}(c),x)$ on this surface, which will give 1 if the label of the surface $c$ is $x$ and 0 otherwise. Therefore, we consider the state
	\begin{align*}
		\delta(\hat{e}(c),x) C_T^{k,e_2}(m_2) C_T^{h,e_1}(m_1) \ket{GS}.
	\end{align*}
	In order to move $\delta(\hat{e}(c),x)$ to the right, so that it acts on the ground state (where $\hat{e}(c)$ gives $1_E$), we separate the higher-flux membrane operators into their constituent parts. First we write
	$$C_T^{k,e_2}(m_2) = C_{\rhd}^k(m_2) \big[ \prod_{p \in m_2} B^{f(p)}(\text{blob }0(m_2) \rightarrow \text{blob }p) \big] \delta(\hat{e}(m_2),e_2).$$

	We can choose the surface $c$ so that the entire membrane $m_2$ is within $c$ (except for the start-point and blob 0, which are outside of the membrane because we do not want to consider their 2-flux) as shown in Figure \ref{higher_flux_braid_combined_2_flux}. Therefore, the surface label $\hat{e}(c)$ is unaffected by $C^k_{\rhd}(m_2)$, which acts on the edges and plaquettes cut by the dual membrane of $m_2$. As a result, $\delta(\hat{e}(c),x)$ commutes with $C^k_{\rhd}(m_2)$. On the other hand, all of the blob ribbon operators $B^{f(p)}(\text{blob }0(m_2) \rightarrow \text{blob }p) $ pierce the membrane $c$. As we saw in Section \ref{Section_braiding_fake-flat_appendix}, when considering the braiding relation between blob excitations and $E$-valued loop excitations (in the case where the blob ribbon is anti-aligned with the surface normal), this leads to the commutation relation
	$$B^f(t)\delta(e, \hat{e}(c))\ket{GS} = \delta(e, [\hat{g}(s.p(c)-s.p(t)) \rhd f]^{-1} \hat{e}(c))$$
	and so
	$$ \delta(\hat{e}(c), x) B^{f(p)}(\text{blob }0(m_2) \rightarrow \text{blob }p) = B^{f(p)}(\text{blob }0(m_2) \rightarrow \text{blob }p) \delta( [\hat{g}(s.p(c)-s.p(m_2)) \rhd f(p)] \hat{e}(c), x).$$
	Therefore, 
	\begin{align*}
		\delta(\hat{e}(c), x)& \big[ \prod_{p \in m_2} B^{f(p)}(\text{blob }0(m_2) \rightarrow \text{blob }p) \big]\\
		&= \big[ \prod_{p \in m_2} B^{f(p)}(\text{blob }0(m_2) \rightarrow \text{blob }p) \big] \delta\bigl( \big[ \prod_{p \in m_2} \hat{g}(s.p(c)-s.p(m_2)) \rhd f(p) \big] \hat{e}(c), x\bigl)\\
		&= \big[ \prod_{p \in m_2} B^{f(p)}(\text{blob }0(m_2) \rightarrow \text{blob }p) \big] \delta\bigg( \hat{e}(c), \big(\hat{g}(s.p(c)-s.p(m_2)) \rhd \big[ \prod_{p \in m_2} f(p)^{-1} \big]\big) x\bigg).
	\end{align*}
	Now we wish to simplify this by considering $\prod_{p \in m_2 }f(p)^{-1}$. This is
	\begin{align*}
		\prod_{p \in m_2}f(p)^{-1} &= \big( \prod_{p \in m_2}[\hat{g}(s.p(m_2)-v_0(p)) \rhd \hat{e}_p^{-1}] [(k^{-1}\hat{g}(s.p(m_2)-v_0(p))) \rhd \hat{e}_p] \big)\\
		&= \big( \prod_{p \in m_2} [\hat{g}(s.p(m_2)-v_0(p)) \rhd \hat{e}_p] \big) ^{-1} \big[k^{-1} \rhd \big( \prod_{p \in m_2} [\hat{g}(s.p(m_2)-v_0(p)) \rhd \hat{e}_p] \big)\big].
	\end{align*}
	Using the definition of the surface operator $\hat{e}(m_2) = \prod_{p \in m_2} \hat{g}(s.p(m_2)-v_0(p)) \rhd \hat{e}_p$ (where we assume that all plaquettes $p$ are aligned with $m_2$) we can write this as
	\begin{equation}
		\prod_{p \in m_2}f(p)^{-1} = [k^{-1} \rhd \hat{e}(m_2)] \hat{e}(m_2)^{-1}.
		\label{Equation_product_f_1}
	\end{equation}
	
	Therefore, we have
	\begin{align*}
		\delta(\hat{e}(c),x)& C_T^{k,e_2}(m_2) C_T^{h,e_1}(m_1) \ket{GS}\\
		&= \delta(\hat{e}(c),x) C_{\rhd}^k(m_2) \big[ \prod_{p \in m_2} B^{f(p)}(\text{blob }0(m_2) \rightarrow \text{blob }p) \big] \delta(\hat{e}(m_2),e_2) C_T^{h,e_1}(m_1) \ket{GS}\\
		&= C_{\rhd}^k(m_2) \big[ \prod_{p \in m_2} B^{f(p)}(\text{blob }0(m_2) \rightarrow \text{blob }p) \big] \delta(\hat{e}(c), [\hat{g}(s.p(c)-s.p(m_2)) \rhd ([k^{-1} \rhd \hat{e}(m_2)] \hat{e}(m_2)^{-1}) ]x) \\
		& \hspace{0.5cm} \delta(\hat{e}(m_2),e_2) C_T^{h,e_1}(m_1) \ket{GS}.
	\end{align*}
	We can then use $\delta(\hat{e}(m_2),e_2)$ to replace the operator $m_2$ with the group label $e_2$, to obtain
	\begin{align*}
		\delta(\hat{e}(c),x)& C_T^{k,e_2}(m_2) C_T^{h,e_1}(m_1) \ket{GS}\\
		&= C_{\rhd}^k(m_2) \big[ \prod_{p \in m_2} B^{f(p)}(\text{blob }0(m_2) \rightarrow \text{blob }p) \big] \delta(\hat{e}(c), [\hat{g}(s.p(c)-s.p(m_2)) \rhd ([k^{-1} \rhd e_2] e_2^{-1}) ]x)\\
		& \hspace{0.5cm} \delta(\hat{e}(m_2),e_2) C_T^{h,e_1}(m_1) \ket{GS}.
	\end{align*}
	Then the operator $\delta(\hat{e}(c), [\hat{g}(s.p(c)-s.p(m_2)) \rhd ([k^{-1} \rhd e_2] e_2^{-1}) ]x)$ commutes with $\delta(\hat{e}(m_2),e_2) $(both are diagonal in the configuration basis), so 
	\begin{align*}
		\delta(\hat{e}(c),x) C_T^{k,e_2}(m_2)& C_T^{h,e_1}(m_1) \ket{GS}\\
		&= C_{\rhd}^k(m_2) \big[ \prod_{p \in m_2} B^{f(p)}(\text{blob }0(m_2) \rightarrow \text{blob }p) \big] \delta(\hat{e}(m_2),e_2) \\
		& \hspace{0.5cm} \delta(\hat{e}(c), [\hat{g}(s.p(c)-s.p(m_2)) \rhd ([k^{-1} \rhd e_2] e_2^{-1}) ]x) C_T^{h,e_1}(m_1) \ket{GS}\\
		&= C_T^{k,e_2}(m_2)\delta(\hat{e}(c), [\hat{g}(s.p(c)-s.p(m_2)) \rhd ([k^{-1} \rhd e_2] e_2^{-1}) ]x)C_T^{h,e_1}(m_1) \ket{GS}.
	\end{align*}
	
	We then just have to repeat the process, commuting the surface operator past $ C_T^{h,e_1}(m_1)$. The only difference is that now $ \delta(\hat{e}(c), [\hat{g}(s.p(c)-s.p(m_2)) \rhd ([k^{-1} \rhd e_2] e_2^{-1}) ]x)$ contains the path operator $\hat{g}(s.p(c)-s.p(m_2))$. However, this path does not intersect with $m_1$ (again, because $m_1$ is entirely enclosed within $c$ apart from its start-point and blob 0) and so is unaffected by $ C_T^{h,e_1}(m_1)$. Therefore, we can just repeat the same procedure to obtain
	\begin{align*}
		&\delta(\hat{e}(c),x) C_T^{k,e_2}(m_2) C_T^{h,e_1}(m_1) \ket{GS}\\
		&= C_T^{k,e_2}(m_2) \delta(\hat{e}(c), [\hat{g}(s.p(c)-s.p(m_2)) \rhd ([k^{-1} \rhd e_2] e_2^{-1}) ]x) C_T^{h,e_1}(m_1) \ket{GS}\\
		&= C_T^{k,e_2}(m_2) C_T^{h,e_1}(m_1) \delta(\hat{e}(c), [\hat{g}(s.p(c)-s.p(m_1)) \rhd ([h^{-1} \rhd e_1] e_1^{-1}) ] [\hat{g}(s.p(c)-s.p(m_2)) \rhd ([k^{-1} \rhd e_2] e_2^{-1}) ]x) \ket{GS}.
	\end{align*}
	
	Then because $c$ is a closed contractible membrane $\hat{e}(c)$ must be equal to $1_E$ in the ground state. Therefore, we only have a non-zero contribution when 
	$$x = [\hat{g}(s.p(c)-s.p(m_1)) \rhd ([h^{-1} \rhd e_1^{-1}] e_1) ] [\hat{g}(s.p(c)-s.p(m_2)) \rhd ([k^{-1} \rhd e_2^{-1}] e_2) ].$$
	
	If we take the start-points of $m_2$ and $m_1$ to be the same, and to be equal to $s.p(c)$, we find that the surface label $x$ must be
	$$[h^{-1} \rhd e_1^{-1}] e_1 [k^{-1} \rhd e_2^{-1}] e_2,$$
	which we can recognise as $\tilde{e}_1 \tilde{e}_2$. Therefore, $\tilde{e}_1 \tilde{e}_2$ is indeed the combined 2-flux of the two loop-like excitations and it makes sense that this should be preserved by the braiding.

	\subsubsection{Braiding with $E$-valued loops}
	
	As a subset of the results given for the braiding between two higher-flux loops, we can consider the case where one of the loops is an $E$-valued loop, by taking the magnetic part of that loop to be trivial (by replacing the $G$-valued label of the corresponding membrane operator by $1_G$). For the case where the higher-flux loop passes through the $E$-valued one, we must take $h=1_G$ in Equations \ref{higher_flux_higher_flux_braid_result_1} - \ref{higher_flux_higher_flux_braid_result_4}. This gives us, for the braiding between the excitations produced by $C^{k,e_2}_T(m_2)$ and $\delta(e_1, \hat{e}(m_1))$:
	\begin{align}
		e_1 &\rightarrow e_1 \hat{g}(s.p(m_1)-s.p(m_2)) \rhd ( e_2^{-1}[k^{-1} \rhd e_2]),\\
		k &\rightarrow k,\\
		e_2 &\rightarrow e_2. \label{higher_flux_E_loop_braid_result_4}
	\end{align}
	
	In the case where the start-points of the two membrane operators are the same, we see that the label $e_1$ of the $E$-valued membrane becomes
	$$e_1 \rightarrow e_1 e_{2}^{-1} [k^{-1} \rhd e_{2}],$$
	with the labels of the higher-flux membrane being unchanged. For the opposite case where the $E$-valued loop passes through the magnetic one we must take $k=1_G$ instead. Equations \ref{higher_flux_higher_flux_braid_result_1} - \ref{higher_flux_higher_flux_braid_result_4} then give us, for the braiding between the excitations produced by $\delta(e_2, \hat{e}(m_2)) $ and $C^{h,e_1}_T(m_2)$ :
	\begin{align}
		h &\rightarrow h, \label{E_loop_higher_flux_braid_result_1} \\
		e_1 &\rightarrow e_1,\\
		e_2 &\rightarrow h^{\phantom{-1}}_{[2-1]} \rhd e_2. \label{E_loop_higher_flux_braid_result_4}
	\end{align}
	We see that when the start-points are the same, the label $e_2$ of the $E$-valued loop excitation becomes $h \rhd e_2$, while the labels of the higher-flux excitation are again unchanged.
	
	\section{Topological sectors in 3+1d}
	\label{Section_topological_sectors_supplement}
	
	\subsection{Topological charge within a torus when $\rhd$ is trivial}
	\label{Section_3D_Topological_Charge_Torus_Tri_trivial}
	
\begin{figure}[h]
		\begin{center}
			\begin{overpic}[width=0.4\linewidth]{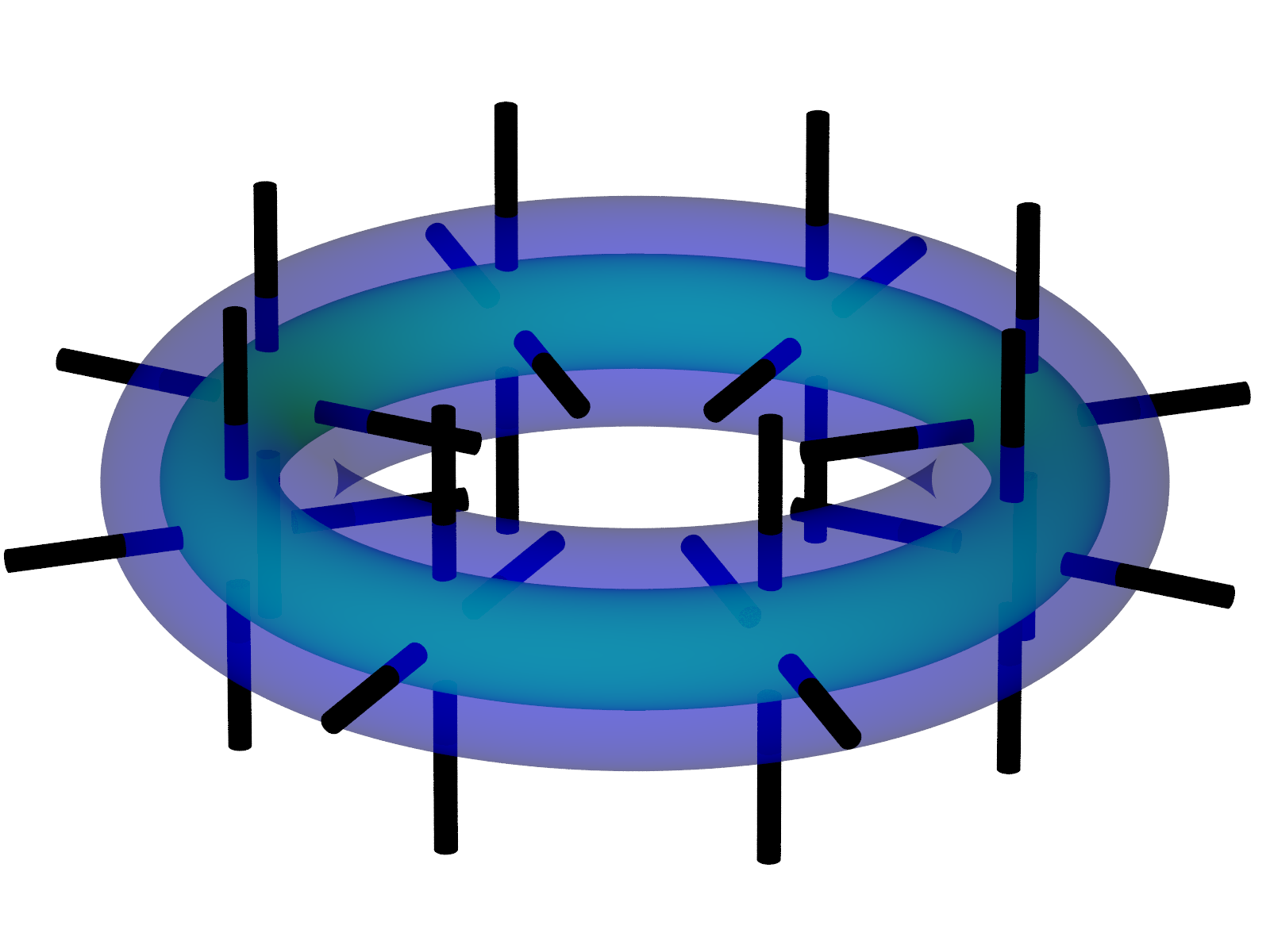}
				
			\end{overpic}
			\caption{We apply membrane and ribbon operators in the region between the outer (blue) and inner (green) tori. We will apply an $E$-valued membrane operator on the inner torus. In addition, we apply a magnetic membrane operator for which the inner torus is the direct membrane and the outer torus is the dual membrane (the black edges represent the edges cut by the dual membrane). We will also apply ribbon operators around the non-contractible cycles of the torus (not shown here).}
			\label{thickenedmembrane}
		\end{center}
	\end{figure}

	In Section \ref{Section_3D_Topological_Sectors} of the main text we explained how to measure the topological charge within a region. We then described the topological charges of the higher lattice gauge theory model that can be measured by a toroidal surface in Section \ref{Section_Torus_Charge} of the main text. In this section, we will explicitly construct the measurement operators and prove the results for the topological charge that we presented in Section \ref{Section_Torus_Charge} of the main text in the case where $\rhd$ is trivial. As we described in Section \ref{Section_3D_Topological_Sectors} of the main text, to measure the topological charge associated to a particular surface we first apply a projection operator that enforces the fact that measurement surface itself has no excitations on it. This means that we can only measure the charge if no objects intersect the surface itself (this is to avoid loop-like excitations being only partially inside the surface), and so we cannot measure the charge of confined excitations (which would drag an energetic string that crosses the surface). After doing that, we apply the most general operators possible to the surface, while enforcing that these operators do not produce any excitations, and see what different combinations are valid. This means applying closed membrane and ribbon operators, as explained in Section \ref{Section_3D_Topological_Sectors}. Considering a toroidal surface, such as the one shown in Figure \ref{thickenedmembrane}, we wish to determine which operators we can apply on its surface without producing any excitations. We will see that we can apply ribbon operators on the independent non-contractible paths on the torus and membrane operators whose membrane is the entire torus. In this section, unless otherwise stated, a path is considered contractible only if we can contract the path without it leaving the torus surface (for example, the longitude and meridian of the torus are not contractible, even though we could shrink them to nothing by moving them away from the torus). This distinction is important because we know that the surface contains no excitations (because of our projection), so topological operators can be deformed across this surface without affecting their action. On the other hand, the regions inside or away from the torus can contain excitations, so we cannot deform operators freely across such regions.

\begin{figure}[h]
		\begin{center}
			\includegraphics[width=0.7\linewidth]{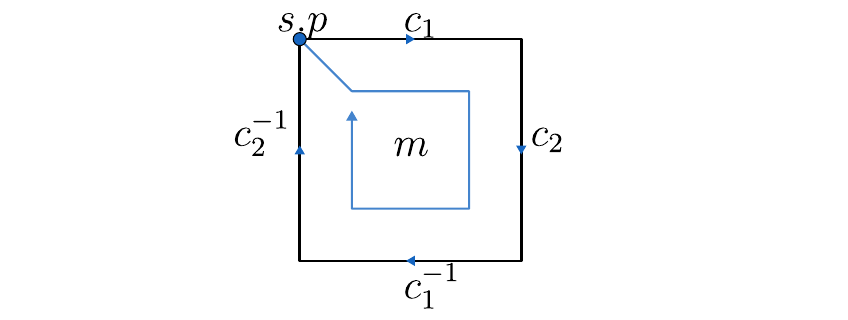}
			\caption{(Copy of Figure \ref{unfoldedtorus1} from the main text) The surface of the torus is conveniently represented by a square with periodic boundary conditions. The boundaries of this square (which are glued due to the periodic boundary conditions) are referred to as the seams of the torus. We apply electric ribbon operators along these seams to measure the non-contractible cycles of the torus and apply an $E$-valued membrane operator on the surface. We will also apply blob ribbon operators around the two cycles and a magnetic membrane operator over the surface. The edges cut by the dual membrane of the magnetic membrane point outwards from the page.}
			\label{unfoldedtorus1appendix}
		\end{center}
	\end{figure}

	First we consider how we can measure the group elements associated to closed paths on the torus. Enforcing that the plaquettes on the surface of the torus satisfy fake-flatness (i.e., that they are unexcited) restricts the value of contractible paths on the torus surface to lie within $\partial(E)$. However, the only electric ribbon operators which are sensitive to elements within $\partial(E)$ are confined, as we proved in Ref. \cite{HuxfordPaper2} (see Section S-I A of the Supplemental Material). Applying such electric ribbon operators on these cycles would cause the edges along the closed paths to be excited, while applying ribbon operators not sensitive to factors in $\partial(E)$ would have no effect. On the other hand, enforcing that the plaquettes satisfy fake-flatness does not restrict the values of the non-contractible cycles of the torus (which do not enclose a surface on the torus that would have to satisfy fake-flatness). We can therefore apply two closed electric operators $\delta(\hat{g}(c_1), g_{c_1})$ and $\delta(\hat{g}(c_2), g_{c_2})$ on the two independent cycles, $c_1$ and $c_2$, of the torus. In addition to measuring the paths on the torus, we can measure the group element associated to the surface of the torus itself by applying a closed $E$-valued membrane operator $\delta(\hat{e}(m),e_m)$. In order to do so, we imagine the torus as a square with opposite sides glued. We refer to the glued sides as the ``seams" of the torus. We will see that some operators behave differently near these seams compared to the bulk of the membrane. That is, there may be an obstruction to the square closing smoothly into a surface with no boundary features. It is convenient to choose these seams to lie in the same position as the cycles measured by the electric ribbon operators, so that the cycles form the boundary of the membrane. We give an illustration of this in Figure \ref{unfoldedtorus1appendix}.
	
\begin{figure}[h]
		\begin{center}
			\begin{overpic}[width=0.4\linewidth]{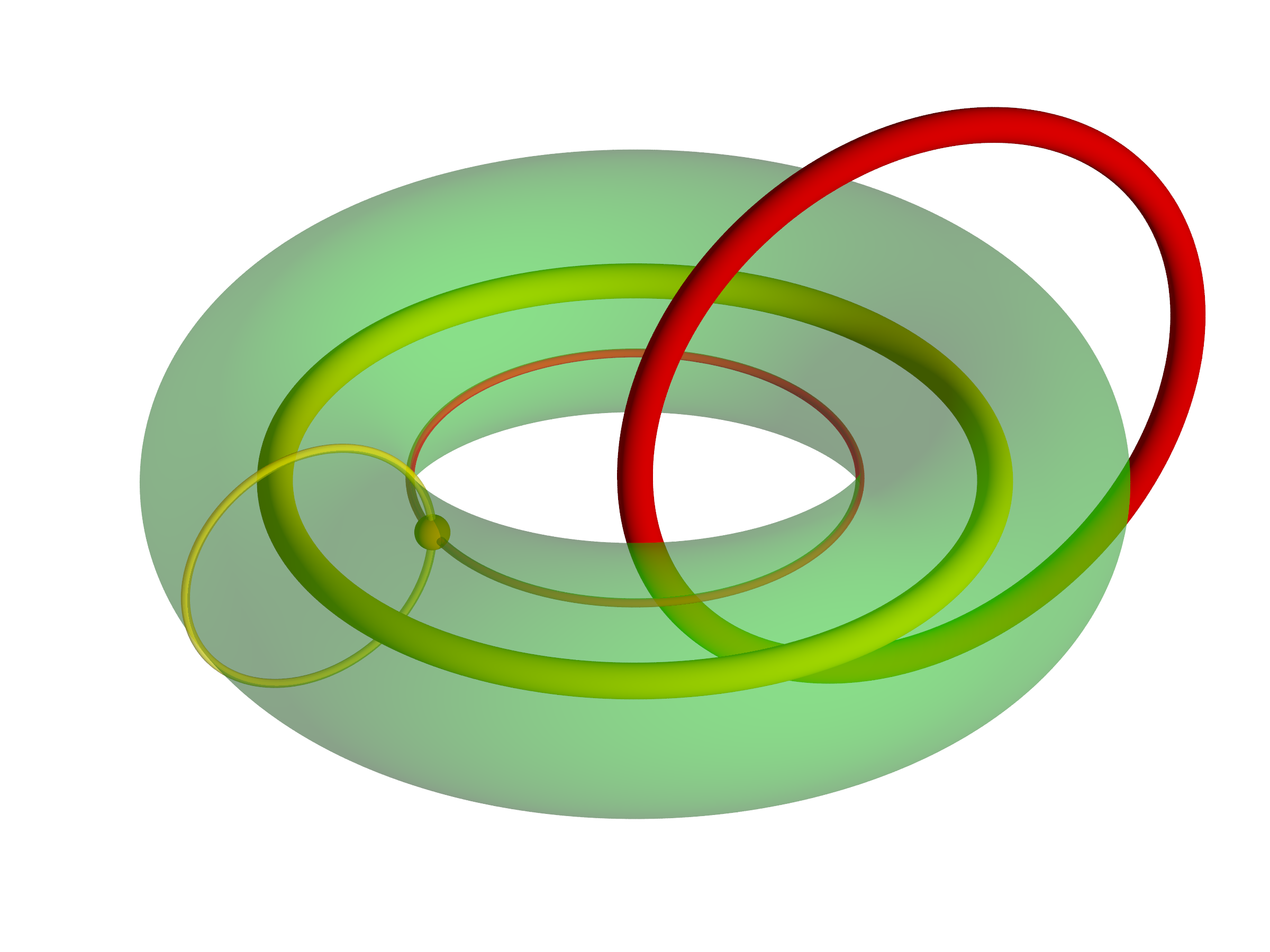}
				\put(12,22){\large $c_1$}
				\put(68,40){\large $c_2$}
			\end{overpic}
			\caption{The electric ribbon operators around the two cycles (the thin tori) measure the magnetic flux through the two cycles. A non-trivial flux around a cycle would be produced by a magnetic flux tube (or tubes) that links with that cycle, such as the thicker tori shown here (these are colour coded to match the colour of the cycle that would measure the flux). We can see that the tubes that produce a non-trivial flux around each cycle link together.}
			\label{torus_measure_link}
		\end{center}
	\end{figure}

\begin{figure}[h]
		\begin{center}
			\begin{overpic}[width=0.75\linewidth]{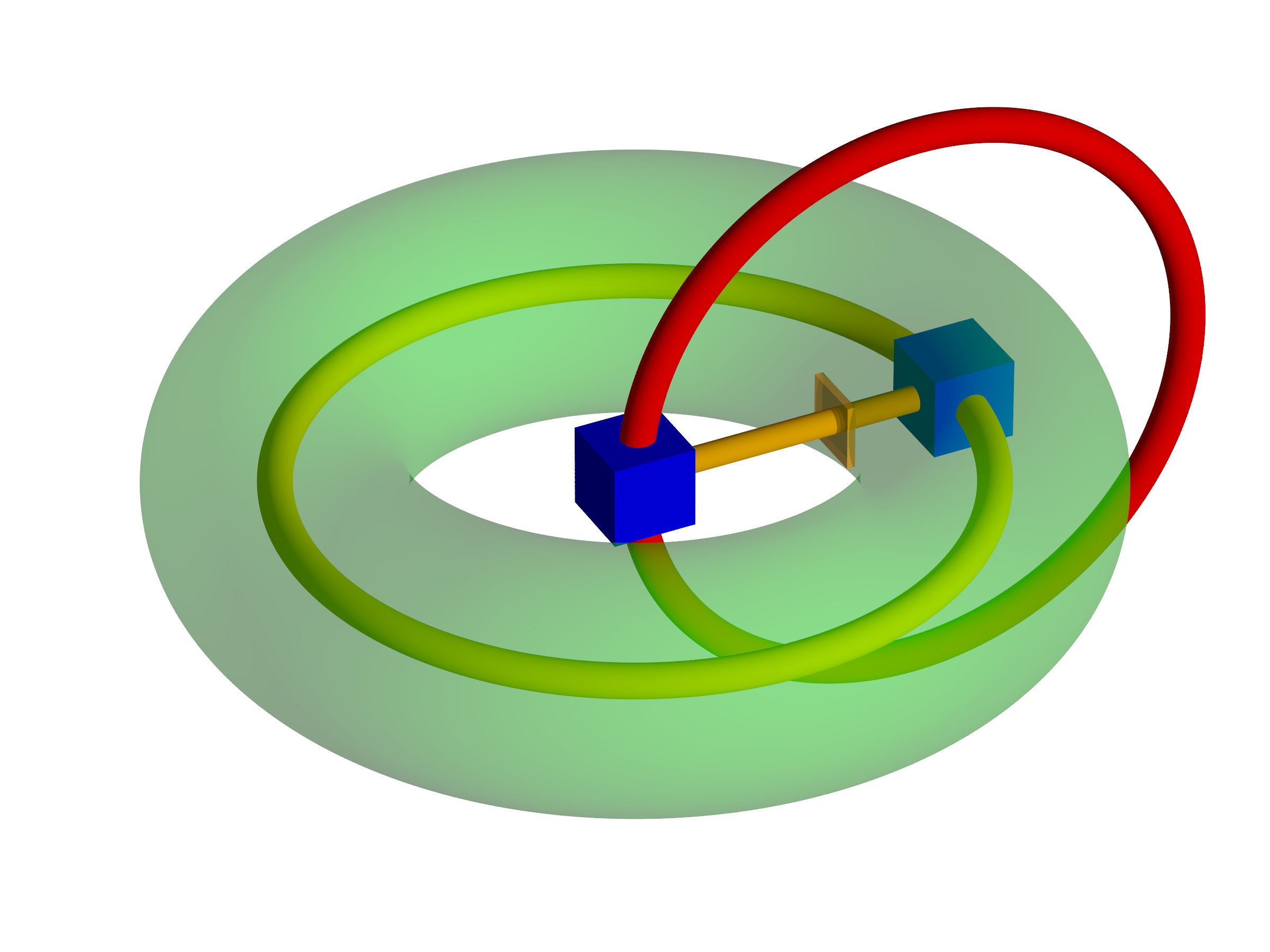}
			\end{overpic}
			\caption{Two linked flux tubes generally drag a linking string between them, represented here by the orange string connecting the two loops. This would leave an excitation on the surface of the measurement torus, as indicated by the orange patch. In some circumstances, this linking string can be removed by applying a suitable blob ribbon operator along the linking string. However, this moves 2-flux from inside the torus to outside, leaving two blob excitations (here indicated by blue cubes), one inside the torus and one outside the torus. This means that, in order to remove the linking string, there must be some net 2-flux inside the torus, as described by Equation \ref{Equation_Torus_Flatness_1}.}
			\label{torus_measure_link_string}
		\end{center}
	\end{figure}

	While we can apply the ribbon and membrane operators with any labels $g_{c_1}$, $g_{c_2}$ and $e_m$, not all choices of these labels are consistent with the projection onto an unexcited surface. If the torus surface is not excited, then the region bounded by the square in Figure \ref{unfoldedtorus1appendix} (which is the toroidal surface) must satisfy fake-flatness. This gives us the constraint
	$$\partial(e_m) g_{c_1} g_{c_2}g_{c_1}^{-1} g_{c_2}^{-1}=1_G.$$
	If the labels of the operators do not satisfy this constraint, then the electric ribbon and $E$-valued membrane operators will always gives zero when acting on a state where the torus surface is unexcited (because no such state simultaneously satisfies the Kronecker deltas in the three operators), and will therefore always give zero when acting after the projection operator that projects onto such unexcited states. This means that we should not consider any measurement operator whose labels do not satisfy this constraint as a valid measurement operator. We can rewrite the constraint as
	$$\partial(e_m)^{-1}=[g_{c_1},g_{c_2}],$$
	where we have defined the commutator $[g,h]=ghg^{-1}h^{-1}$. We can also write this condition in various other forms, using the restrictions we place upon the crossed module. One useful form is 
	\begin{equation}
		\partial(e_m)^{-1}=[g_{c_1}^{-1},g_{c_2}^{-1}],
		\label{Equation_Torus_Flatness_1}
	\end{equation}
	which we obtain from the following manipulations:
	\begin{align*}
		\partial(e_m)^{-1}= [g_{c_1},g_{c_2}]& \implies \partial(e_m)^{-1}=g_{c_1}g_{c_2}g_{c_1}^{-1}g_{c_2}^{-1}\\
		& \implies (g_{c_1}g_{c_2})^{-1}\partial(e_m)^{-1}g_{c_1}g_{c_2}=g_{c_1}^{-1}g_{c_2}^{-1}g_{c_1}g_{c_2}\\
		& \implies \partial( (g_{c_1}g_{c_2})^{-1} \rhd e_m)=[g_{c_1}^{-1},g_{c_2}^{-1}]\\
		&\implies \partial(e_m)=[g_{c_1}^{-1},g_{c_2}^{-1}],
	\end{align*}
	where in the penultimate line we used the Peiffer condition, Equation \ref{Equation_Peiffer_1} from the main text, and in the last line we used the fact that $\rhd$ is trivial. This condition has particular interest when we consider the meaning of the variables $g_{c_1}$, $g_{c_2}$ and $e_m$ in terms of the types of excitation that we wish to measure. We note that the electric ribbon operators around the two handles of the torus measure the 1-holonomy around these handles. A non-trivial 1-holonomy around the cycle $c_1$ would be produced by a flux tube (or tubes) inside the torus, whereas a non-trivial 1-holonomy around cycle $c_2$ would be produced by a flux tube or tubes that link with the torus, as shown in Figure \ref{torus_measure_link}. That means that the flux tubes measured by $c_1$ will link with the tubes measured by $c_2$. In the simplest case, there will be one flux tube of each type, as shown in Figure \ref{torus_measure_link} (more generally, there can be multiple loops of each type). Let us consider how the condition given in Equation \ref{Equation_Torus_Flatness_1} would arise when measuring the charge in this situation, where we have two linked flux tubes of the type considered in Section \ref{Section_linking_appendix}. In that section, we found that two magnetic flux tubes would drag a linking string if their labels do not commute (at least when the fluxes have the same start-point, otherwise the commutation relation involves the path between the start-points, as discussed beneath Equation \ref{Equation_linking_plaquette_holonomy}). If the labels do not commute, but the commutator is only an element of $\partial(E)$, then the linking string can be removed by applying a suitable blob ribbon operator, whose label is related to the commutator of the two flux labels via the map $\partial$. If we measured the charge of two linked tubes with the toroidal measurement operator, then any linking string would cause an excitation on the surface of the torus, which is forbidden for our measurement operator (hence the condition given in Equation \ref{Equation_Torus_Flatness_1}). Therefore, the measurement operator can only measure linked loops without a linking string, or linked loops for which we have removed the linking string by applying the appropriate blob ribbon operator. If we remove the linking string by applying a blob ribbon operator between the two linked tubes, then we leave a non-trivial 2-flux enclosed by the toroidal surface, due to the fact that the blob ribbon operator produces non-trivial 2-fluxes at its ends (one of which is enclosed by the measurement surface), as shown in Figure \ref{torus_measure_link_string}. This provides another way of looking at the condition Equation \ref{Equation_Torus_Flatness_1}; it puts a condition on the 2-flux that must be enclosed by the torus in order to remove the linking string between the two magnetic fluxes and so prevent any excitations on the torus surface. This condition agrees with the one we found when discussing the linking in Section \ref{Section_linking_appendix}: the image under $\partial$ of this 2-flux must match the commutator of the 1-fluxes. We can also have additional 2-flux from blob excitations inside the torus, but they must have a total 2-flux with label in the kernel, otherwise they are confined and so must drag a confining string which produces excitations of the surface of the torus (which is forbidden). This means that they do not affect the condition Equation \ref{Equation_Torus_Flatness_1}).
	
	So far, we have considered applying the operators which measure the group elements of paths and surfaces. However, we can also apply membrane and ribbon operators that change the group elements, namely magnetic membrane operators and blob ribbon operators. First, we apply the magnetic membrane operator $C^h(m)$ onto the torus surface. Recall from Section \ref{Section_3D_Tri_Trivial_Magnetic_Excitations} of the main text that when we define a magnetic membrane operator, we must define a start-point for the operator and a set of paths from that start-point to each of the edges affected by the operator. We will choose the start-point to lie at the beginning (which is also the end) of the cycles $c_1$ and $c_2$. The reason we can do this without missing out on possible measurement operators will be explained later, when we consider how to make the measurement operator commute with the vertex energy terms. Apart from this start-point, the operator is robust to the deformation of the paths from the start-point to the edges. This is because any factors within $\partial(E)$ obtained by deforming these paths over a fake-flat region do not affect the action of the operator (as described in Section \ref{Section_3D_Tri_Trivial_Magnetic_Excitations}) and the torus surface is guaranteed to be fake-flat due to our projection operator. However, the presence of non-contractible cycles (which are not guaranteed to satisfy fake-flatness) means that there are different choices for these paths which cannot be deformed into one-another over a fake-flat surface, and so which may give different operators. For now, we choose the paths so that they do not cross the seams of the cycle, and we will explain why we can do this later in this section.

	The final operators that we apply are blob ribbon operators, which we put around the two cycles of the torus. In principle, we could apply these operators on any closed cycle (closed because an open blob ribbon operator would produce blob excitations). However, this apparent freedom is somewhat illusory, due to the fact that blob ribbon operators are either topological (if their label lies in the kernel of $\partial$), or confined. Blob ribbon operators with label in the kernel of $\partial$ can be freely deformed (as shown in Section \ref{Section_Topological_Blob_Ribbons}), which means that it does not matter precisely where we place the operators (up to homotopy classes of paths). If the original blob ribbon operator did not wrap any of the non-contractible cycles on the torus surface, then we can deform the ribbon into a trivial one by contracting the ribbon across the torus surface (which we know to be unexcited, permitting the deformation). If the blob ribbon operator wrapped one of the cycles, then we can deform it into a standard position along one of the seams of the membrane. If the ribbon operator initially wrapped both cycles, or wrapped one cycle multiple times, we can split the ribbon operator into multiple ribbon operators that each wrap a cycle once.

\begin{figure}[h]
		\begin{center}
			\begin{overpic}[width=0.6\linewidth]{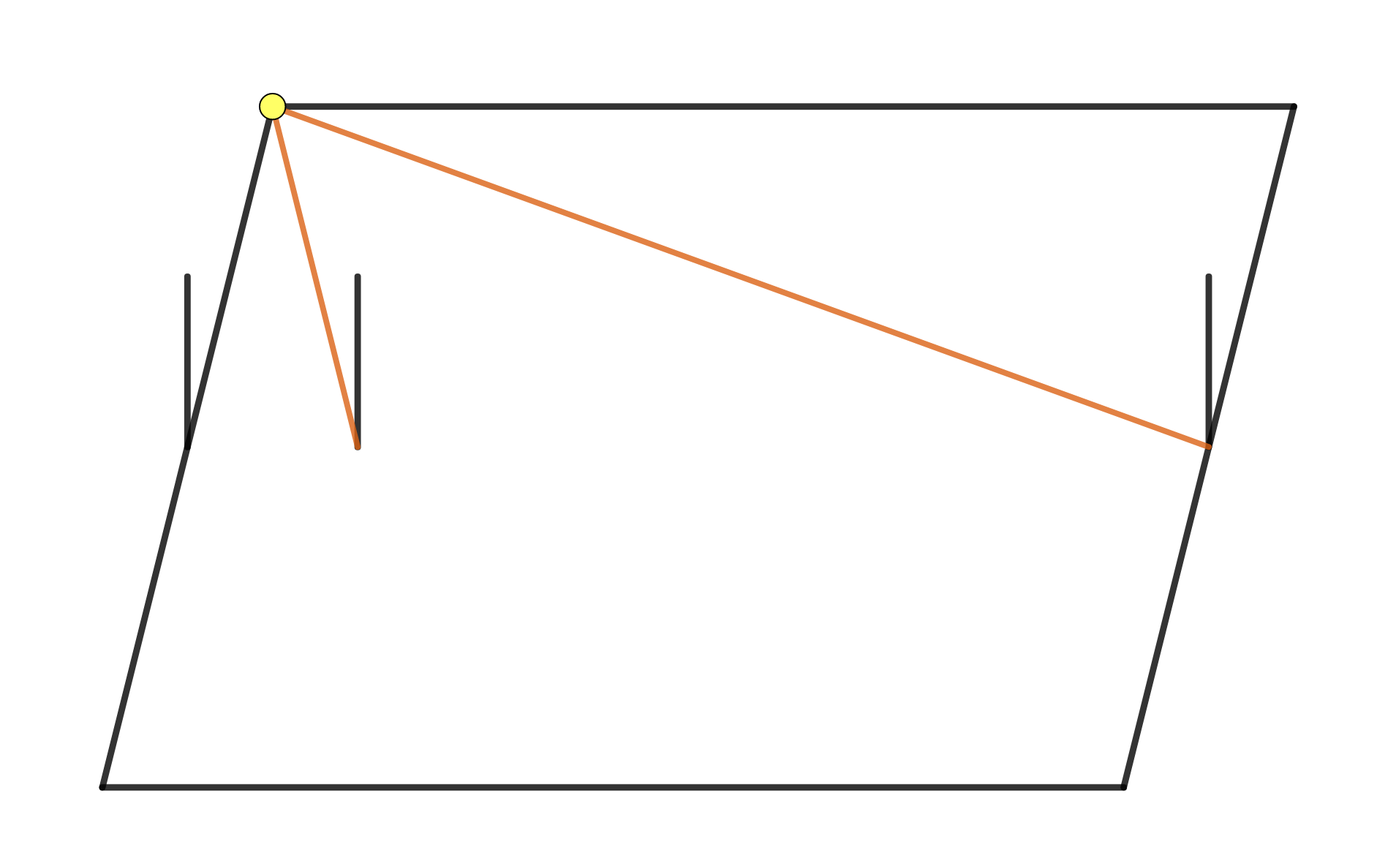}
				\put(13,55){\large $s.p$}
				\put(55,57){\large $c_1$}
				\put(9,31){\large $v_2$}
				\put(27,31){\large $v_1$}
				\put(50,46){\large $t_2$}
				\put(22.5,46){\large $t_1$}
				\put(90,31){\large $v_2$}
				\put(9,38){\large $i_2$}
				\put(27,38){\large $i_1$}
			\end{overpic}
			
			\caption{The action of the magnetic membrane operator on an edge depends on the path to the vertex on the direct membrane that is attached to that edge. We choose these paths so that they do not cross the seams of the torus. In the case of vertices lying on the seams of the membrane, such as $v_2$, the path can either wrap one of the cycles or not (i.e., the path can meet the left instance of $v_2$ or the right instance in this figure). We choose such paths to wrap the relevant cycle(s). Either way, there is a discontinuity in the paths near the seams. For example, the path to vertex $v_1$, which is near to $v_2$, cannot be deformed into a path that differs from $t_2$ only by the addition of a small path from $v_2$ to $v_1$.}
			\label{Torus_measurement_magnetic_discontinuity}
		\end{center}
	\end{figure}

	On the other hand, the blob ribbon operators with label outside the kernel cannot be deformed, because they are confined. However, the fact that such ribbon operators are confined leads to the position of their ribbons being fixed and their labels being restricted, in order to prevent the measurement operator from producing any excitations. To see why this is the case, consider the action of the magnetic membrane operator $C^h(m)$ on the membrane. As we described earlier, we choose the paths from the start-point of the membrane to each affected edge (which determine the action on the edges) so that the paths never cross the seams of the torus. We will later explain why we can do this, but for now we wish to examine the action of the magnetic membrane operator for this particular choice of paths. For the edges that lie on the seams, we must choose whether the paths wrap a cycle of the torus or not. We choose these paths so that they do, as illustrated in Figure \ref{Torus_measurement_magnetic_discontinuity}. This introduces a discontinuity near the seam, because the path to an edge near the seam may not cross much of the cycle, whereas the path to an edge on the seam wraps the entire cycle. This discontinuity would still occur if we had chosen the paths to the edges on the seam to not wrap the cycle, though it would occur on the right-side of the membrane in Figure \ref{Torus_measurement_magnetic_discontinuity} rather than the left. Indeed, such discontinuity is inevitable when considering the magnetic membrane operator on a torus. It is worth noting that such a discontinuity does not occur on a sphere, where there are no non-contractible cycles and so any paths with the same start and end-points are equivalent (i.e., have labels which differ only by elements in $\partial(E)$).

\begin{figure}[h]
		\begin{center}
			\begin{overpic}[width=0.7\linewidth]{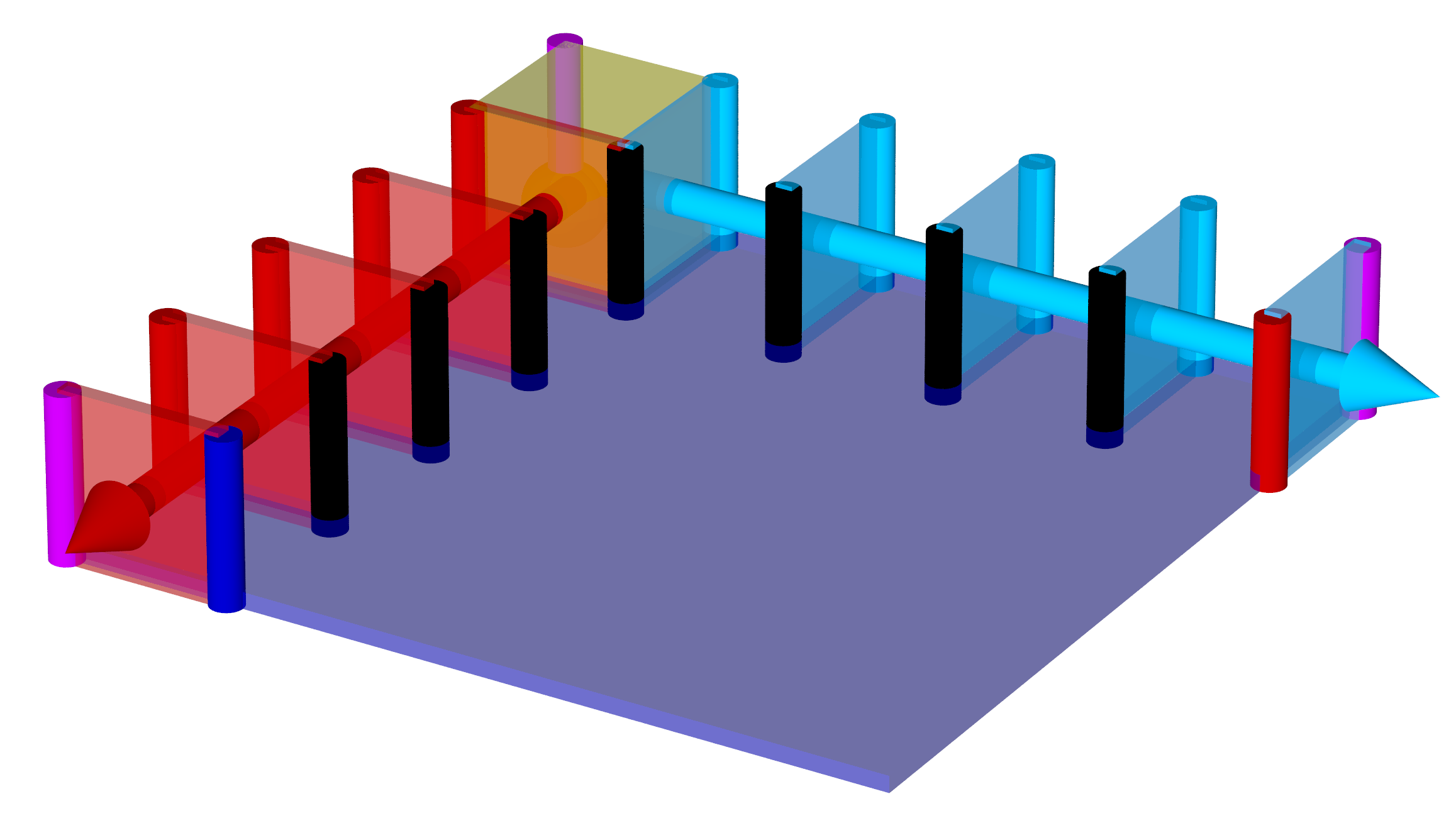}
				
			\end{overpic}
			\caption{Plaquettes near the seams of the torus, such as the red (darker gray in grayscale) and blue (lighter gray in grayscale) squares here, may be affected by the discontinuity of the paths in the magnetic membrane operators. This is because the edges on the seams of the torus differ from their neighbouring edges (represented in black) by a path that nearly wraps $c_1$ (or a parallel cycle) in the case of the red (darker gray) edges or nearly wraps $c_2$ in the case of the blue (lighter gray edges). The path to the (purple) edges on the corners of the torus wraps both cycles (note that all of these corner edges are actually the same edge, repeated due to the periodic boundary conditions). We can see that the plaquettes affected by each type of discontinuity are also the plaquettes that would be pierced by an arrow (blue or red arrow) passing from blob 0 (yellow cube in the top-left) along a path on the dual lattice close to the relevant seam. That is, they lie along the dual path of a blob ribbon operator. We will see that the plaquettes can be excited and this must be corrected by applying a blob ribbon operator along these paths.}
			\label{Torus_measurement_discontinuity_2}
		\end{center}
	\end{figure}

	The discontinuity in the paths defined in the magnetic membrane operator on the torus can lead to a discontinuity in the action of that operator on the edges themselves, which may result in additional plaquette excitations at the seams. The magnetic membrane operator only preserves the fake-flatness condition for plaquettes cut by the bulk of the membrane because two of the edges on the plaquette are affected by the membrane operator, and the effect of changing the two edge labels cancels out. This holds because the paths to the two edges differ from each-other by smooth deformation, plus composition with the base of the plaquette. However, near the seam the paths to the two edges (such as $i_1$ and $i_2$ in Figure \ref{Torus_measurement_magnetic_discontinuity}) may differ by a different, larger, path. This leads to potential excitations for plaquettes adjacent to the seams, as indicated in Figure \ref{Torus_measurement_discontinuity_2}.

	\begin{figure}[h]
		\begin{center}
			\begin{overpic}[width=0.75\linewidth]{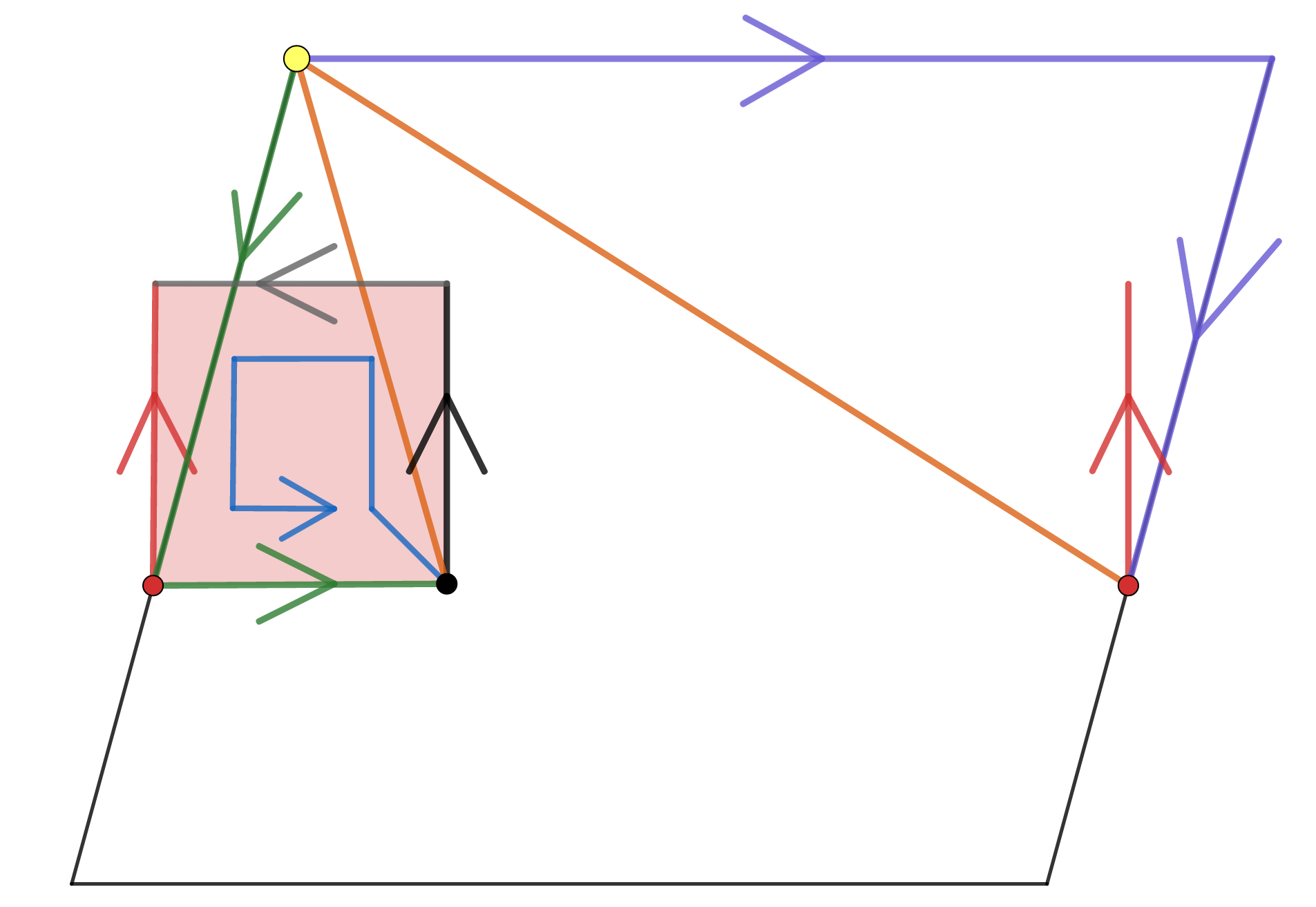}
				\put(17,65){\large $s.p$}
				\put(63,66){\large $c_1$}
				\put(7,24){\large $v_2$}
				\put(35,24){\large $v_1$}
				\put(60,42){\large $t_2$}
				\put(25,56){\large $t_1$}
				\put(87,24){\large $v_2$}
				\put(7,40){\large $i_2$}
				\put(35,40){\large $i_1$}
				\put(82,40){\large $i_2$}
				\put(24,21){\large $d$}
				\put(19,44){\large $u$}
				\put(15,52){\large $s$}
				\put(22,35){\large $p$}
				\put(95,52){\large $s$}
				
			\end{overpic}
			\caption{A plaquette near the seam $c_2$ may be affected by the discontinuity in the paths defined for the magnetic membrane operator. We consider a plaquette (shaded square) for which one of its affected edges, $i_2$ (the red arrow), lies on the seam $c_2$ and one, $i_1$ (the black arrow), is towards the bulk of the membrane. The path $t_2$ we define in the magnetic membrane operator to the vertex $v_2$ attached to $i_2$ differs greatly from the path $t_1$ to the vertex $v_1$ attached to the black edge. The path $t_1$ can be deformed into the path $sd$, while the path $t_2$ can be deformed into the path $c_1s$ without affecting the action of the magnetic membrane operator. We note that if the path to $v_2$ was instead the (green) path $s$, the paths to $v_1$ and $v_2$ would only differ by path $d$ which is local to the plaquette. Instead the paths differ by the cycle $c_1$ in addition to $d$.}
			\label{plaquette_near_c_2_tri_trivial}
		\end{center}
	\end{figure}

	We wish to examine the effect of this discontinuity on the plaquette holonomy in more detail. Consider a plaquette near the seam of the torus, such as the one shown in Figure \ref{plaquette_near_c_2_tri_trivial}. For such a plaquette, of the two edges on that plaquette cut by the dual membrane, one is attached to the $c_2$ boundary and one is not, as shown in Figure \ref{plaquette_near_c_2_tri_trivial}. Then under the action of the magnetic membrane operator (assuming that the edges point away from the direct membrane, which we can enforce with our procedures for changing the branching structure, as explained in the Appendix of Ref. \cite{HuxfordPaper1}) the edge labels on that plaquette transform as
	\begin{align*}
		g_1\rightarrow& g(t_1)^{-1}hg(t_1)g_1,\\
		g_2\rightarrow& g(t_2)^{-1}hg(t_2)g_2,
	\end{align*}
	where $g_1$ is the original label of $i_1$ and $g_2$ is the original label of $i_2$. Because $\partial(E)$ is in the centre of $G$, we are free to deform the paths $t_1$ and $t_2$ through the flat region near the torus surface without changing the expression $g(t_i)^{-1}hg(t_i)$ (any factors in $\partial(E)$ obtained by $g(t_i)$ as we deform $t_i$ can be cancelled with the inverse factor gained by $g(t_i)^{-1}$). We therefore deform the path $t_1$ into the green path $sd$ shown in Figure \ref{plaquette_near_c_2_tri_trivial}, and the path $t_2$ into the purple path $c_1s$ shown in the same figure (these deformed paths have the same group elements as the originals, up to factors in $\partial(E)$). Therefore, the transformation of the edge labels cut by the dual membrane can be written as
	\begin{align*}
		g_1\rightarrow& g(d)^{-1}g(s)^{-1}hg(s)g(d)g_1\\
		g_2\rightarrow& g(s)^{-1}g_{c_1}^{-1}hg_{c_1}g(s)g_2.
	\end{align*}
	
	We now examine how this change to the edge labels affects the 1-holonomy of the plaquette, which determines whether fake-flatness is satisfied. The 1-holonomy around the plaquette is originally given by
	$$H_{1}(p)=\partial(e_p)g_1g(u)g_2^{-1}g(d)=1_G,$$
	where the edges associated to each label are illustrated in Figure \ref{plaquette_near_c_2_tri_trivial} and the holonomy is originally the identity due to the plaquette condition. Under the transformation of the edges, the plaquette holonomy becomes
	\begin{align}
		H_1(p) & \rightarrow\partial(e_p) \big(g(d)^{-1}g(s)^{-1}hg(s)g(d)g_1\big)g(u)( g(s)^{-1}g_{c_1}^{-1}hg_{c_1}g(s)g_2)^{-1}g(d) \notag\\
		&=\partial(e_p) g(d)^{-1}g(s)^{-1} hg(s)g(d)g_1g(u)g_2^{-1} g(s)^{-1} g_{c_1}^{-1}h^{-1}g_{c_1}g(s) g(d). \notag
	\end{align}	
	
	We can use the fact that $\partial(e_p)$ is in the centre of $G$ to move it next to $g_1$, to obtain
	\begin{align}
		H_1(p) & \rightarrow g(d)^{-1}g(s)^{-1} hg(s)g(d)\partial(e_p) g_1g(u)g_2^{-1} g(s)^{-1} g_{c_1}^{-1}h^{-1}g_{c_1}g(s) g(d). \notag\\
		&=g(d)^{-1}g(s)^{-1} hg(s)g(d) \big[\partial(e_p) g_1g(u)g_2^{-1}g(d)\big]g(d)^{-1} g(s)^{-1} g_{c_1}^{-1}h^{-1}g_{c_1}g(s) g(d), \notag
	\end{align}	
	where we inserted the identity in the form of $g(d)g(d)^{-1}$. We then note that the term $\partial(e_p) g_1g(u)g_2^{-1}g(d)$ is the original 1-holonomy of the plaquette, which is the identity. We therefore have
	\begin{align}
		H_1(p) & \rightarrow g(d)^{-1}g(s)^{-1} hg(s)g(d) [1_G] g(d)^{-1} g(s)^{-1} g_{c_1}^{-1}h^{-1}g_{c_1}g(s) g(d) \notag\\
		&=g(d)^{-1}g(s)^{-1} h g_{c_1}^{-1}h^{-1}g_{c_1}g(s) g(d). \label{Equation_torus_seam_c_2_plaquette_transformation_1}
	\end{align}
	
	We see that this expression will not be the identity element if $h$ and $g_{c_1}$ do not commute, indicating that the action of the magnetic membrane operator can violate fake-flatness and therefore produces excitations on the surface of the torus (which we do not allow). However, now consider additionally acting with a blob ribbon operator that runs along the seam of the torus and pierces all plaquettes such as $p$ (i.e., a blob ribbon operator along the red arrow in Figure \ref{Torus_measurement_discontinuity_2}). Let the label of such a blob ribbon operator be $e_{c_2}$. Then the blob ribbon operator acts on $e_p$ as $e_p \rightarrow e_p e_{c_2}^{-1}$. Combining the action of the blob ribbon operator with the transformation on the edges caused by the magnetic membrane operator, we see that the plaquette holonomy becomes
	\begin{align*}
		\partial(e_p e_{c_2}^{-1} ) (g(d)^{-1}g(s)^{-1}hg(s)g(d)g_1)g(u)( g(s)^{-1}g_{c_1}^{-1}hc_1g(s)g_2)^{-1}g(d) &= \partial(e_{c_2}^{-1})g(d)^{-1}g(s)^{-1} h g_{c_1}^{-1}h^{-1}g_{c_1}g(s) g(d)\\
		&= g(d)^{-1}g(s)^{-1} (\partial(e_{c_2})^{-1} hg_{c_1}^{-1}h^{-1}g_{c_1})g(s) g(d).
	\end{align*}
	
	This expression gives the identity element (and so the plaquette satisfies fake-flatness) provided that $e_{c_2}$ satisfies $\partial(e_{c_2})^{-1} hg_{c_1}^{-1}h^{-1}g_{c_1}=1_G$, or equivalently $\partial(e_{c_2})=hg_{c_1}^{-1}h^{-1}g_{c_1}$. We can put this expression in various similar forms. For example
	\begin{align*} 
		\partial(e_{c_2})&= hg_{c_1}^{-1}h^{-1}g_{c_1}\\
		&\implies \partial(e_{c_2})=\partial(h^{-1} \rhd e_{c_2})=g_{c_1}^{-1}h^{-1}g_{c_1}h,
	\end{align*}
	where the first equality on the second line come from the fact that $\rhd$ is trivial and the second comes from the Peiffer condition $\partial(g \rhd e) =g \partial(e)g^{-1}$ for all $g \in G$ and $e \in E$. We can also write this condition more concisely using a commutator as
	\begin{equation}
		\partial(e_{c_2})=[g_{c_1}^{-1},h^{-1}]. 
		\label{Equation_c2_seam_condition_1}
	\end{equation}
	Therefore, we must apply a blob ribbon operator with this label in order to avoid creating excitations on the surface of the torus. This is not just a condition on $e_{c_2}$ however. If $g_{c_1}$ and $h$ commute up to an element of $\partial(E)$, then we can always find an $e_{c_2}$ that satisfies this condition. If their commutator is outside $\partial(E)$, however, then there is no label $e_{c_2}$ that satisfies the condition and we cannot avoid the plaquette being excited. This indicates that a pair $h$ and $g_{c_1}$ whose commutator is outside $\partial(E)$ is not compatible with a valid measurement operator.

	As stated by the condition in Equation \ref{Equation_c2_seam_condition_1}, we must put a confined blob ribbon operator around a particular strip of lattice adjacent to the seam $c_2$ of the torus to counter the plaquette excitations produced by the magnetic membrane operator (if $g_{c_1}$ and $h$ do not commute). We will see that a similar condition holds for the ribbon operator around the seam $c_1$. We cannot put a confined blob ribbon operator in a different position (for example, we cannot put such a ribbon operator parallel to $c_2$ but horizontally displaced from the strip shown in Figure \ref{Torus_measurement_discontinuity_2}) because the plaquettes outside the strip adjacent to $c_2$ already satisfy fake-flatness, meaning that a blob ribbon operator with $\partial(e) \neq 1_G$ would cause excitations. This means that we have restricted the position on which we are allowed to put confined blob ribbon operators and also restricted the label that they can have. Then, as already mentioned, any non-confined blob ribbon operators wrapping a cycle parallel to $c_2$ can be freely moved onto this strip (they are topological). This means we can just apply the blob ribbon operator on this strip of the lattice (and the corresponding strip near $c_1$) without losing potential measurement operators.

	At this point, it is worth briefly mentioning a similarity between this situation and one that we have seen previously, namely the linking of two magnetic excitations. In Section \ref{Section_linking_appendix}, we saw that when two magnetic flux tubes are linked, one of the tubes pierces the position of the magnetic membrane operator that produces the other flux tube and this leads to additional excited plaquettes. This occurs because the presence of the magnetic flux piercing the membrane leads to a discontinuity between paths on the membrane that pass around the flux in different ways. This is completely analogous to the current situation, where the paths on the membrane may pass either way around a non-contractible cycle. In this current case, the non-contractible cycle may have non-trivial flux if a magnetic flux tube is enclosed by this cycle, and in the case of linking the magnetic flux was explicitly present. Furthermore, we also saw that in certain cases (requiring the commutation of two flux labels up to an element in $\partial(E)$) the linking string (i.e., the additional excitations caused by the linking of two magnetic fluxes) could be removed by the application of a confined blob ribbon operator, just as we see here.

\begin{figure}[h]
		\begin{center}
			\begin{overpic}[width=0.7\linewidth]{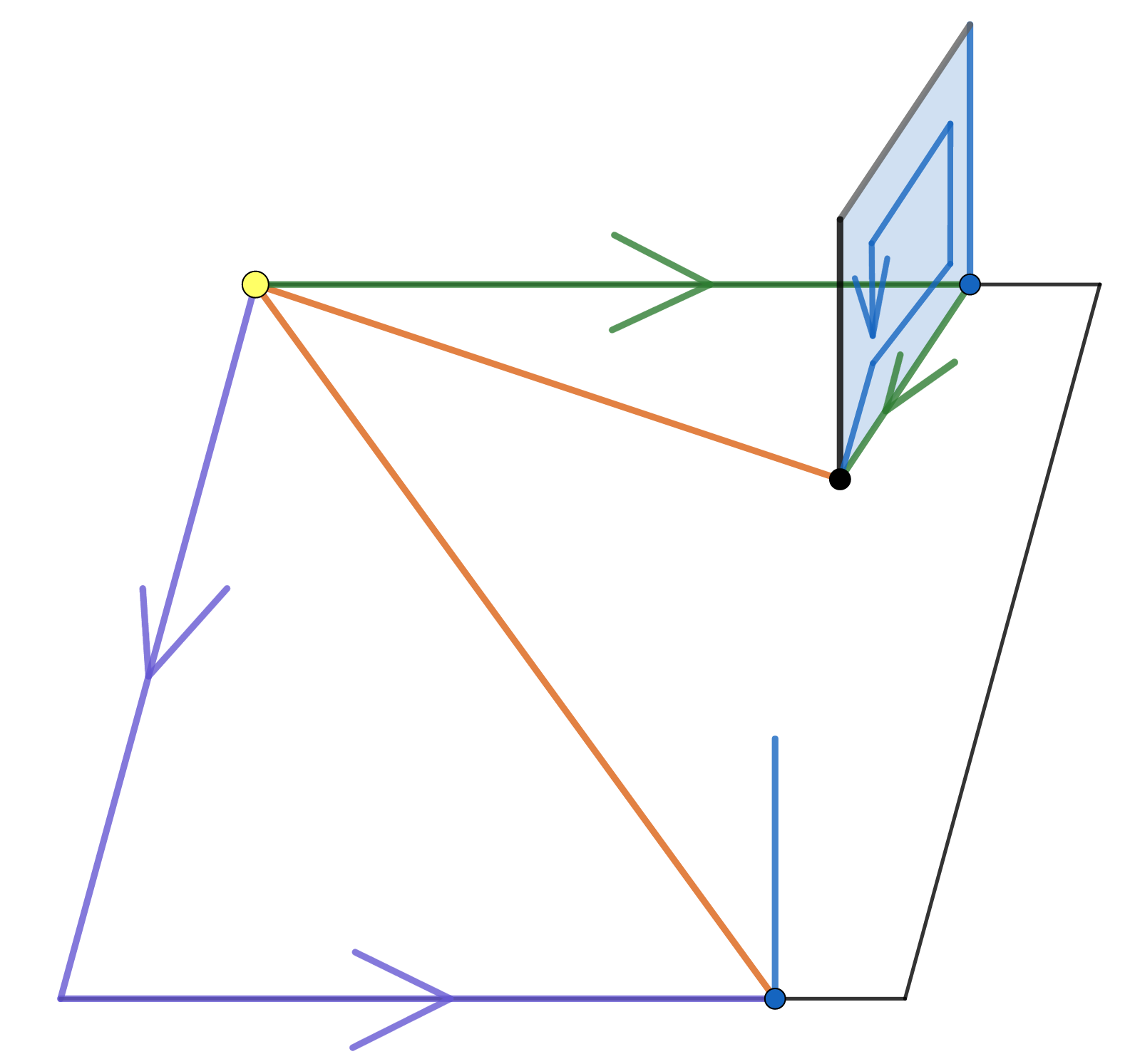}
				\put(17,67){\large $s.p$}
				\put(87,67){\large $v_2$}
				\put(76,49){\large $v_1$}
				\put(50,57){\large $t_1$}
				\put(41,38){\large $t_2$}
				\put(10,33){\large $c_2$}
				\put(41,3){\large $a$}
				\put(87,80){\large $i_2$}
				\put(71,63){\large $i_1$}
				\put(65,3){\large $v_2$}
				\put(70,16){\large $i_2$}
				\put(52,70){\large $a$}	
				\put(82,57){\large $b$}
				\put(52,70){\large $a$}	
				\put(80,72){\large $p$}
				\put(78,85){\large $u$}
				
			\end{overpic}
			\caption{Just as with plaquettes near $c_2$, a plaquette near $c_1$, such as the shaded plaquette, may be affected by a discontinuity in the action of the magnetic membrane operator. In the example plaquette, the edge $i_2$ (the blue edge) is attached to vertex $v_2$, which lies on the seam. The other edge that lies on the plaquette and is cut by the dual membrane of the magnetic membrane operator, $i_1$, lies towards the bulk of the membrane. This means that the paths (orange) to the two edges are very different. We can deform the path $t_2$ into the (purple) path $c_2 a$, while we can deform the path $t_1$ into the (green) path $ab$.}
			\label{plaquette_near_c_1_tri_trivial}	
		\end{center}
	\end{figure}

	Returning to the present circumstance of measuring the topological charge with a toroidal surface, we now consider the other cycle $c_1$. Just as the action of the magnetic membrane operator near the cycle $c_2$ restricts the position and label of blob ribbon operators around $c_2$, so too are the position and label of the blob ribbon operator around $c_1$ restricted. We will demonstrate this in the same way, by looking at plaquettes on the strip near $c_1$. Such a plaquette is shown in Figure \ref{plaquette_near_c_1_tri_trivial}.

	Just as we did in the previous case, we can deform the paths $t_1$ and $t_2$ from Figure \ref{plaquette_near_c_1_tri_trivial} in order to write their path elements in terms of common variables. We use such a deformation to write that $g(t_1)= g(a)g(b)$ and $g(t_2)=g(c_2)g(a)$, up to irrelevant factors in $\partial(E)$. Then the labels $g_1$ and $g_2$ of the edges $i_1$ and $i_2$ transform as
	\begin{align*}
		g_1 &\rightarrow (g(a)g(b))^{-1}h(g(a)g(b))g_1,\\
		g_2 & \rightarrow (g_{c_2}g(a))^{-1}h(g_{c_2}g(a))g_2
	\end{align*}
	under the action of the magnetic membrane operator. Therefore, the plaquette holonomy $H_1(p)$ of plaquette $p$ transforms under the magnetic membrane operator $C^h(m)$ according to
	\begin{align*}
		H_1(p)&= \partial(e_p) g(b)^{-1}g_2 g(u)g_1^{-1} =1_G\\
		&\rightarrow \partial(e_p) g(b)^{-1} [(g_{c_2}g(a))^{-1}h(g_{c_2}g(a))g_2] g(u) [(g(a)g(b))^{-1}h(g(a)g(b))g_1]^{-1}\\
		&=g(b)^{-1} g(a)^{-1}g_{c_2}^{-1}hg_{c_2}g(a) g(b) [\partial(e_p)g(b)^{-1}g_2 g(u)g_1^{-1}]g(b)^{-1}g(a)^{-1}h^{-1}g(a)g(b)\\
		&=g(b)^{-1} g(a)^{-1}g_{c_2}^{-1}hg_{c_2}g(a) g(b) [1_G]g(b)^{-1}g(a)^{-1}h^{-1}g(a)g(b)\\
		&=g(b)^{-1} g(a)^{-1}g_{c_2}^{-1}hg_{c_2}h^{-1}g(a)g(b).
	\end{align*}
	
	From this we see that the plaquette does not satisfy fake-flatness after the action of the membrane operator if $h$ and $g_{c_2}$ do not commute. As before, we can correct this by applying a blob ribbon operator along the seam, with label $e_{c_1}$. This affects the plaquette label $e_p$ as $e_p \rightarrow e_p e_{c_1}^{-1}$ so the total plaquette holonomy becomes
	$$\partial(e_{c_1})^{-1}g(b)^{-1} g(a)^{-1}g_{c_2}^{-1}hg_{c_2}h^{-1}g(a)g(b) =(g(a)g(b))^{-1}\partial(e_{c_1})^{-1}g_{c_2}^{-1}hg_{c_2}h^{-1}g(a)g(b).$$
	
	From this, we derive the condition that the blob ribbon operator must satisfy:
	\begin{align*}
		\partial(e_{c_1})&= (h g_{c_2}^{-1}h^{-1}g_{c_2})^{-1} = g_{c_2}^{-1}hg_{c_2} h^{-1}\\
		& \implies \partial(e_{c_1})= \partial(h^{-1} \rhd e_{c_1})= h^{-1}\partial(e_{c_1})h= h^{-1}g_{c_2}^{-1}hg_{c_2},
	\end{align*}
	where we use the fact that $\rhd$ is trivial along with the first Peiffer condition (Equation \ref{Equation_Peiffer_1} from the main text) to obtain the final line. We can write this condition with a commutator as
	\begin{equation}
		\partial(e_{c_1})=[h^{-1},g_{c_2}^{-1}].
		\label{Equation_c1_seam_condition_1}
	\end{equation}

\begin{figure}[h]
		\begin{center}
			\begin{overpic}[width=0.7\linewidth]{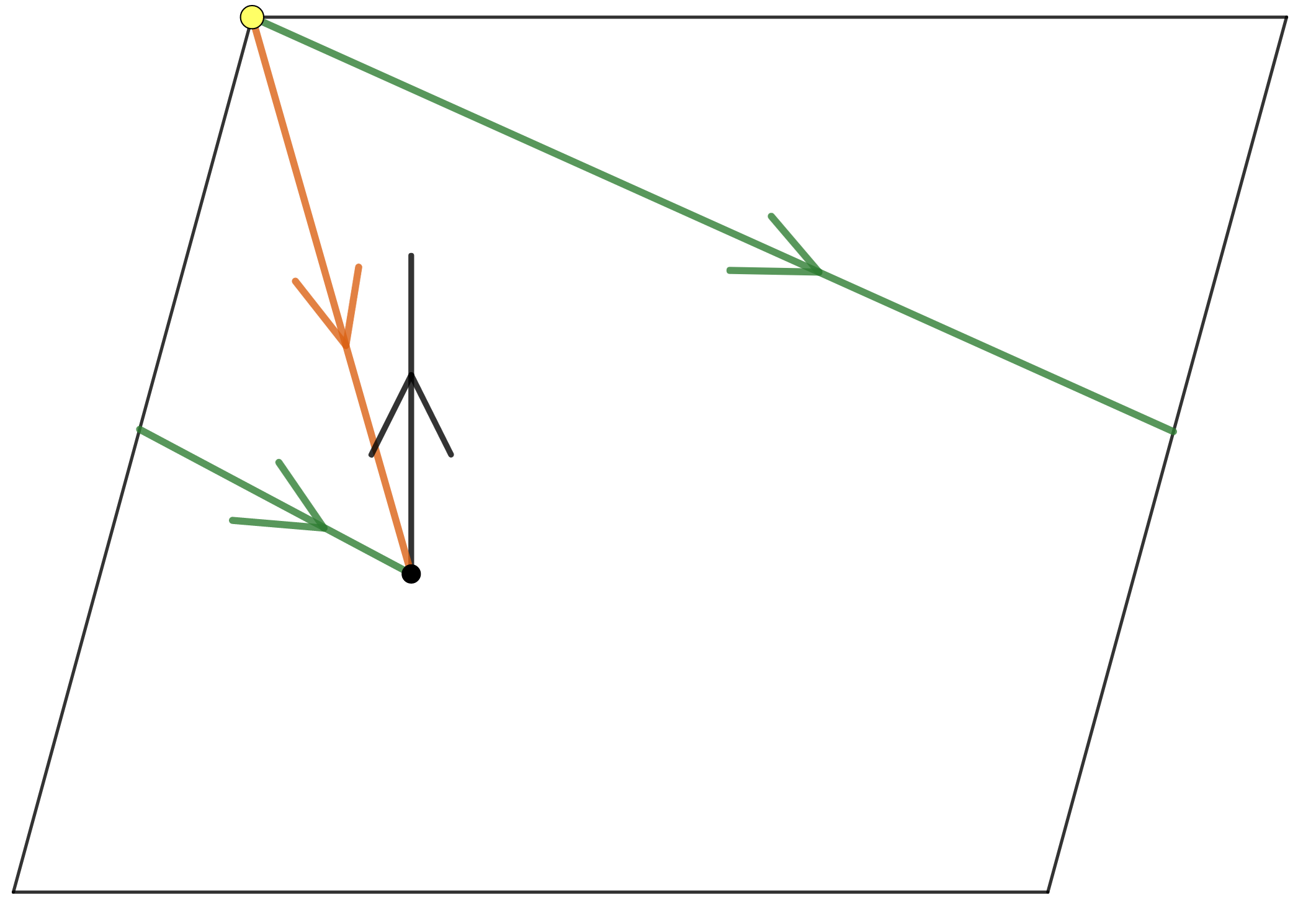}
				\put(32,43){$i$}
				\put(25,51){$t_i$}
				\put(50,56){$t_i'$}
				\put(13,67){$s.p$}
				\put(55,70){$c_1$}
				\put(33,25){$v_i$}

			\end{overpic}
			\caption{We consider one particular edge, edge $i$, that is cut by the dual membrane of the magnetic membrane operator. We originally chose the path from the start-point of the membrane to this edge so that the path does not cross the seam of the torus (the boundary of the square in the figure). An example of this type of path is indicated by the orange path, $t_i$. We now consider what happens if we change this path, choosing a new path which the original path cannot be smoothly deformed into. For example, we could consider the green path $t_i'$, which differs from the original path by a non-contractible path (if we travel along the green path and then the orange path in reverse, the resulting closed path could be deformed into the cycle $c_1$ of the torus for this particular choice of $t_i'$).}
			\label{torus_reconnect_edge_1}
		\end{center}
	\end{figure}

	Before we consider the other conditions that our measurement operator must satisfy, we wish to justify some of our earlier claims. The magnetic membrane operator is associated with a set of paths from the start-point of the membrane to the edges affected by the operator. We claimed that we could freely choose these paths without losing possible measurement operators, and chose the paths so that they do not cross the seams in the torus. We will now consider why we were allowed to choose these paths in this way. First, we note that deforming the paths smoothly in the region of the torus has no effect on the magnetic membrane operator, as we discussed in Section \ref{Section_Magnetic_Tri_Non_Trivial}. However, due to the non-contractible cycles of the torus, a given edge could be associated to different paths that cannot be deformed into each-other. We therefore wish to consider what happens to the action of the magnetic membrane operator when we change the path to a particular edge by more than a simple deformation, such as in the example shown in Figure \ref{torus_reconnect_edge_1}. 
	
\begin{figure}[h]
		\begin{center}
			\begin{overpic}[width=0.7\linewidth]{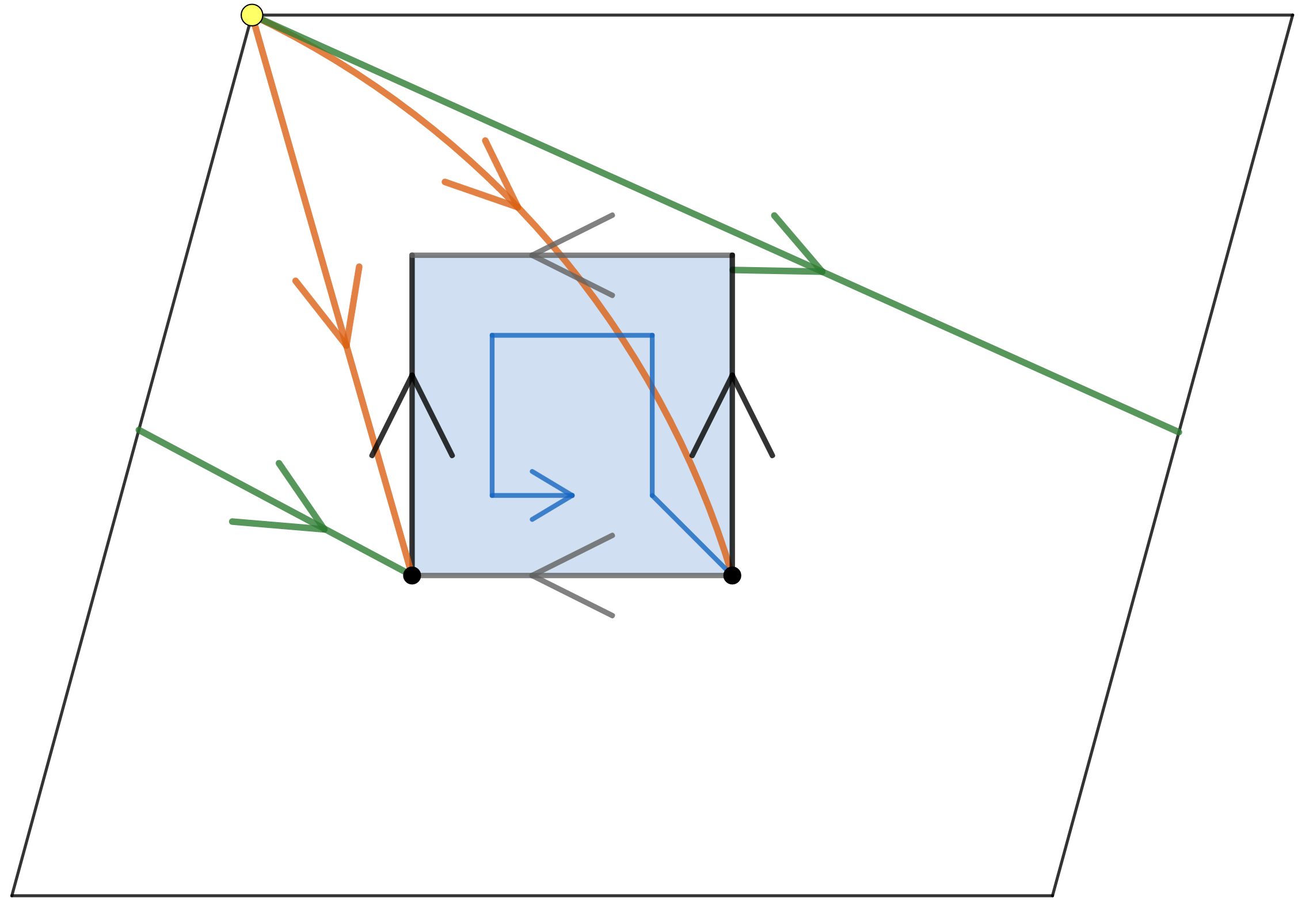}
				\put(29,43){$i$}
				\put(57,43){$j$}
				\put(43,38){$p$}
				\put(31,58){$t_j$}
				\put(21,51){$t_i$}
				\put(43,60){$t_i'$}
				\put(48,52){$u$}
				\put(48,24){$b$}

			\end{overpic}
			\caption{We consider a plaquette $p$ that is attached to the edge $i$ from Figure \ref{torus_reconnect_edge_1}. Changing the path from the start-point to the edge affects the action of the magnetic membrane operator on that edge and therefore changes the action on the plaquette holonomy. As a result of this, it is possible that the plaquette may not satisfy fake-flatness afterwards.}
			\label{torus_reconnect_edge_plaquette_off_seam}
			
		\end{center}
	\end{figure}

	We wish to see how changing the path $t_i$ to an edge $i$ by more than a simple deformation affects the action of the magnetic membrane operator on that edge (note that an analogous argument was made in the case of linking in Section \ref{Section_linking_appendix}, where we showed how changing the path to an edge by making it pass the other way around a flux tube affected the action of a magnetic membrane operator). We denote the original path to the edge by $t_i$ and the new path to the edge by $t_i'$ (examples of $t_i$ and $t_i'$ are shown in Figure \ref{torus_reconnect_edge_1}, but there are many distinct choices of $t_i'$ and the following argument holds regardless of which choice we take). These paths differ by a closed non-contractible path, which we call $s$. That is, $t_i' = s \cdot t_i$. Calling the labels of these paths $g(t_i')$, $g(t_i)$ and $g(s)$, we have $g(t_i') = g(s)g(t_i)$. We denote the new membrane, with the path $t_i$ changed to $t_i'$, by $m'$, so that the altered magnetic membrane operator is $C^h(m')$. Then the action of the altered magnetic membrane operator on the edge $i$, with initial label $g_i$, is (from the standard formula for the action of the magnetic membrane operator Equation \ref{Equation_magnetic_membrane_on_edges_appendix})
	\begin{align*}
		C^h(m'):g_i &= g(t_i')^{-1}hg(t_i')g_i\\
		&=g(t_i)^{-1}g(s)^{-1}hg(s)g(t_i)g_i,
	\end{align*} 
	where we have assumed that the edge $i$ points away from the direct membrane (but our reasoning will also hold for the opposite orientation due to invariance under the re-branching procedures discussed in the Appendix of Ref. \cite{HuxfordPaper1}). Now we must consider whether this change to the action on the edge generates a new and distinct measurement operator. In order for the resulting measurement operator to be valid, it must not excite any of the energy terms that we have considered so far. In particular, the measurement operator must not excite the plaquettes around the edge, whose plaquette holonomies depend on the edge label. We must therefore consider the action of the new magnetic membrane operator on the plaquette holonomy of the plaquettes around the edge $i$. We first consider a plaquette that is not on the seam and so is not pierced by the blob ribbon operators around each cycle (see the earlier discussion about the action of the magnetic membrane operator near the seams of the torus), such as the plaquette $p$ shown in Figure \ref{torus_reconnect_edge_plaquette_off_seam}. We take the circulation of this plaquette (and any other plaquettes adjacent to the edge) to be oriented against edge $i$. As with the orientation of $i$, this choice has no significance and we can invert the orientation of the plaquette using a procedure from the Appendix of Ref. \cite{HuxfordPaper1}. As well as acting on edge $i$, the magnetic membrane operator affects another edge on the plaquette, which we call edge $j$ and take to point away from the direct membrane. We denote the path from the start-point of the membrane operator to this edge by $t_j$, so that the action of the membrane operator on the edge $j$ is
	\begin{align*}
		C^h(m'): g_j = g(t_j)^{-1}hg(t_j)g_j.
	\end{align*}
	
	Now we consider how the action of the membrane operator affects the holonomy of the plaquette $p$, which is originally given by
	$$H_1(p)=\partial(e_p)g_j g(u)g_i^{-1}g(b)^{-1}=1_G,$$
	where $u$ and $b$ are the top and base of the plaquette respectively, as shown in Figure \ref{torus_reconnect_edge_plaquette_off_seam}. Using the action of the membrane operator on each edge, we have
	\begin{equation}
		C^h(m'):H_1(p)=\partial(e_p) g(t_j)^{-1}hg(t_j)g_j g(u)g_i^{-1} g(t_i)^{-1}g(s)^{-1}h^{-1}g(s)g(t_i) g(b)^{-1}. \label{Equation_torus_magnetic_change_paths_plaquette_off_seam_tri_trivial_1}
	\end{equation}
	
	Then, because we are considering a plaquette away from the seam, the original path $t_i$ can be deformed into the path $t_j \cdot b$ over a fake-flat surface. As we discussed in Section \ref{Section_Magnetic_Membrane_Tri_trivial}, when showing that the magnetic membrane operator commutes with the plaquette terms in the bulk of the membrane, this means that $g(t_i)$ is the same as $g(t_j)g(b)$, up to factors in $\partial(E)$ which do not affect $g(t_i)^{-1}hg(t_i)$. Using this relation in Equation \ref{Equation_torus_magnetic_change_paths_plaquette_off_seam_tri_trivial_1}, we see that
	\begin{align}
		C^h(m'):H_1(p)&=\partial(e_p) g(t_j)^{-1}hg(t_j)g_j g(u)g_i^{-1} (g(t_j)g(b))^{-1}g(s)^{-1}h^{-1}g(s)(g(t_j)g(b)) g(b)^{-1} \notag \\
		&= \partial(e_p) g(t_j)^{-1}hg(t_j)g_j g(u)g_i^{-1} g(b)^{-1}g(t_j)^{-1}g(s)^{-1}h^{-1}g(s)g(t_j) \notag \\
		&=g(t_j)^{-1}hg(t_j) [\partial(e_p)g_j g(u)g_i^{-1} g(b)^{-1}] g(t_j)^{-1}g(s)^{-1}h^{-1}g(s)g(t_j), \label{Equation_torus_magnetic_change_paths_plaquette_off_seam_tri_trivial_2}
	\end{align}
	where in the last step we used that $\partial(E)$ is in the centre of $G$. We then recognise the term $[\partial(e_p)g_j g(u)g_i^{-1} g(b)^{-1}]$ as the original $H_1(p)$, which is the identity due to the requirement that the region satisfies fake-flatness. Therefore, Equation \ref{Equation_torus_magnetic_change_paths_plaquette_off_seam_tri_trivial_2} becomes
	\begin{align}
		C^h(m'):H_1(p)&= g(t_j)^{-1}hg(t_j) g(t_j)^{-1}g(s)^{-1}h^{-1}g(s)g(t_j) \notag\\
		&= g(t_j)^{-1}hg(s)^{-1}h^{-1}g(s)g(t_j), \label{Equation_torus_magnetic_change_paths_plaquette_off_seam_tri_trivial_3}
	\end{align}
	which indicates that the plaquette may be excited due to the action of the magnetic membrane. This equation is very reminiscent of Equation \ref{Equation_torus_seam_c_2_plaquette_transformation_1}, where we considered a plaquette on the seam of the magnetic membrane operator. Indeed the origin of the potential excitation is the same, occurring due to a discontinuity in the paths on the magnetic membrane operator. Just as in the case of the seam, the plaquette $p$ is unavoidably excited if the commutator of $h$ and $g(s)$ is outside $\partial(E)$. In this case, our trial measurement operator is not valid and we must throw it out. However, this is not a separate condition to the ones that we found before we considered altering a path. The closed path $s$ starts and ends at the start-point of the membrane and so must be (or must be deformable into) a product of the two cycles of our torus, $c_1$ and $c_2$, each of which may appear any number of times in the product. For example, we could have $s=c_1 \cdot c_2 \cdot c_1 \cdot c_2^{-1}$. From Equations \ref{Equation_c2_seam_condition_1} and \ref{Equation_c1_seam_condition_1}, we know that the path elements of the two cycles must commute with $h$ up to elements in $\partial(E)$ in order for our measurement operator to be valid. Therefore, the same must be true of $g(s)$.

	Now suppose that our original measurement operator was valid, so that $g(s)$ commutes with $h$ up to an element of $\partial(E)$. Then we can write $hg(s)^{-1}h^{-1}g(s)=\partial(x)$, for some element $x \in E$. Then, because $\partial(E)$ is in the centre of $G$, $g(t_j)^{-1}hg(s)^{-1}h^{-1}g(s)g(t_j) = \partial(x)$, so Equation \ref{Equation_torus_magnetic_change_paths_plaquette_off_seam_tri_trivial_3} becomes
	\begin{align}
		C^h(m):H_1(p)&= \partial(x). \label{Equation_torus_magnetic_change_paths_plaquette_off_seam_tri_trivial_4}
	\end{align}
	
\begin{figure}[h]
		\begin{center}
			\begin{overpic}[width=0.7\linewidth]{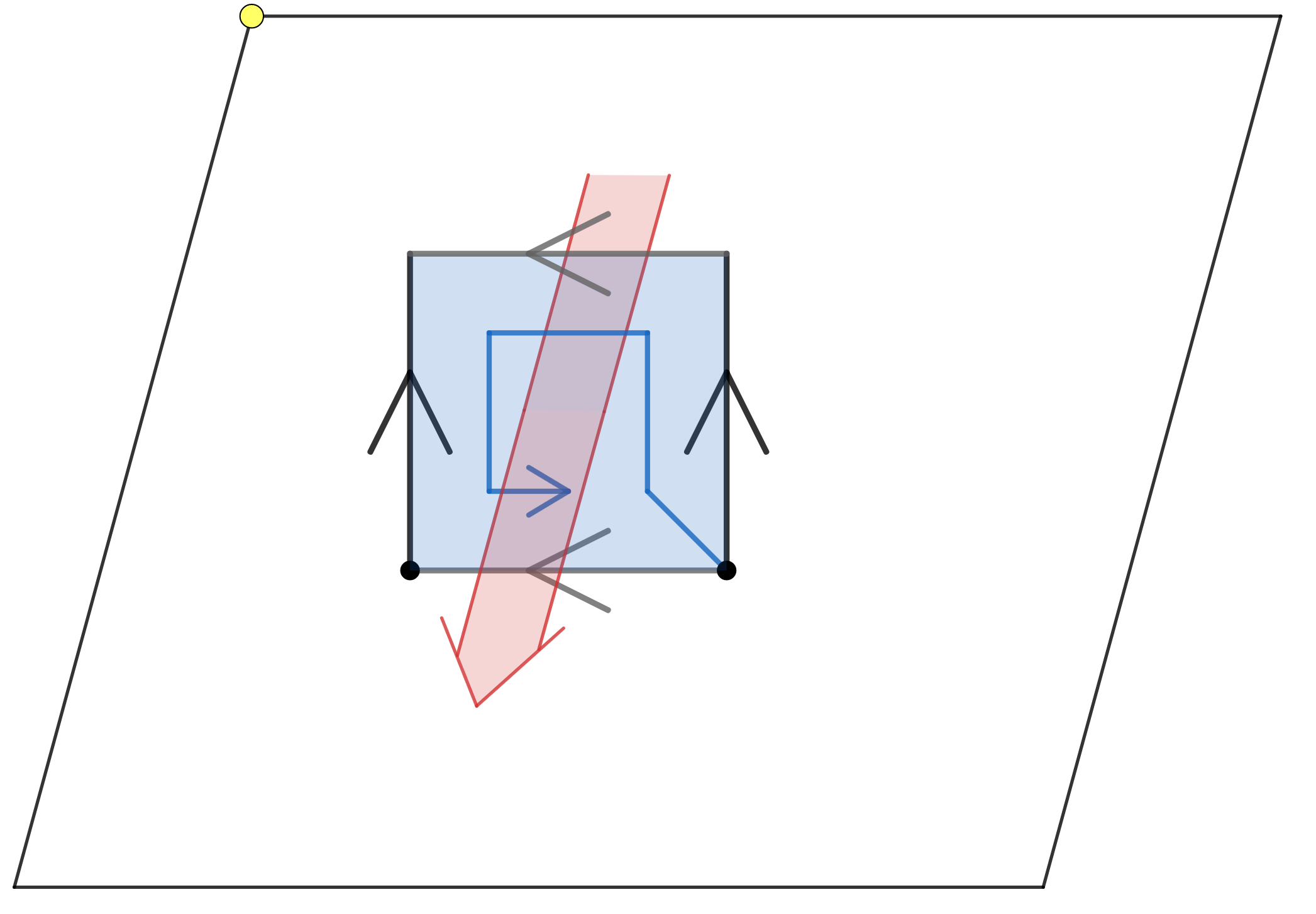}
				\put(28,42){$i$}
				\put(43,20){ribbon $r$}

			\end{overpic}
			\caption{As described previously, changing the path to edge $i$ may leave adjacent plaquettes violating fake-flatness. In order for the measurement operator to remain a valid measurement operator, it is necessary for the plaquette to remain fake-flat. This can be achieved by adding an additional blob ribbon operator that passes through the plaquette. In the example in this figure, corresponding to the plaquette from Figure \ref{torus_reconnect_edge_plaquette_off_seam}, the relevant ribbon $r$ is shown as a red arrow.}
			\label{torus_reconnect_edge_plaquette_ribbon}
			
		\end{center}
	\end{figure}

\begin{figure}[h]
		\begin{center}
			\begin{overpic}[width=0.5\linewidth]{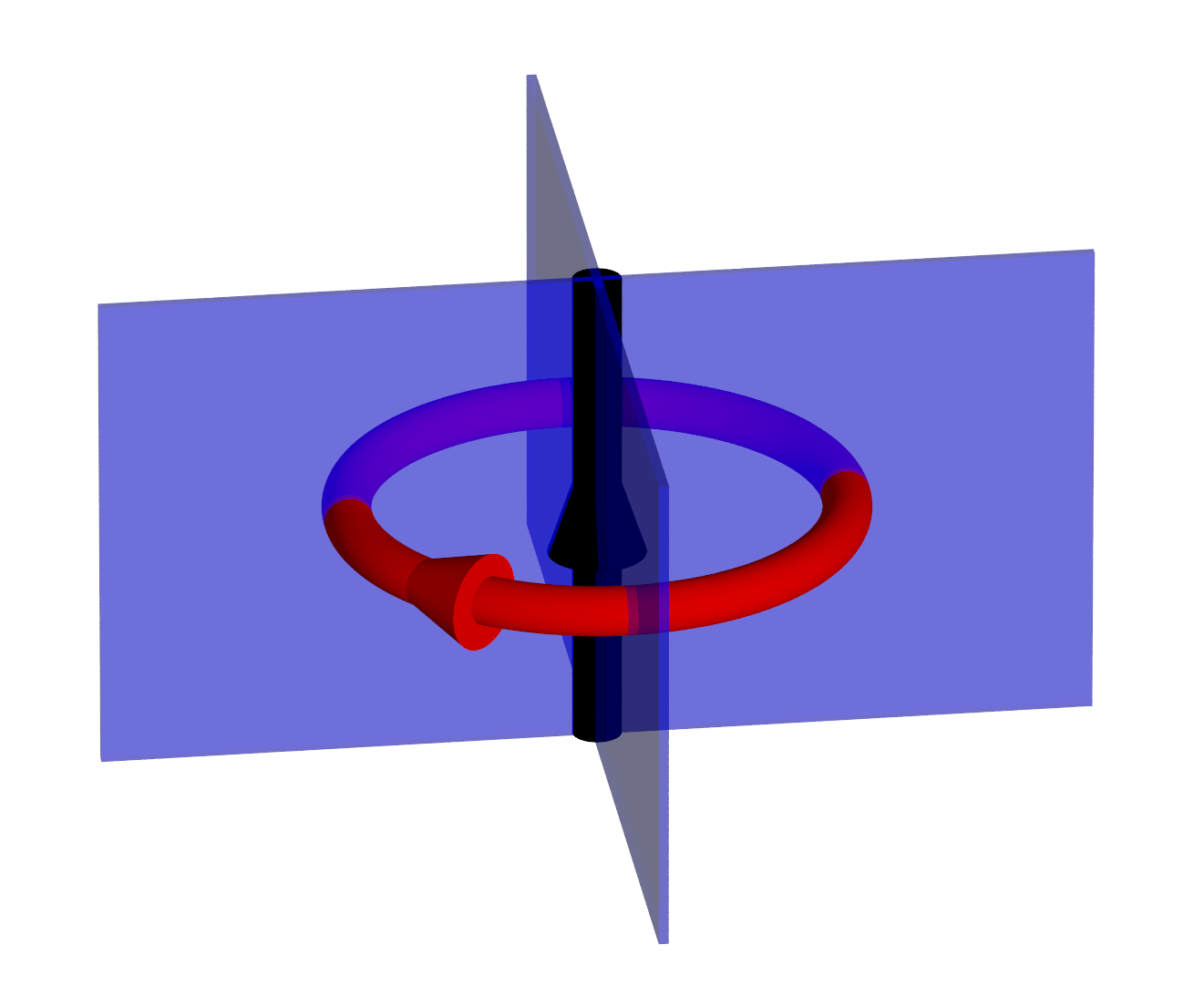}
				\put(53,55){\large edge $i$}
				\put(13,30){\large ribbon $r$}

			\end{overpic}
			\caption{We have seen that we must add blob ribbon operators that pass through each plaquette (blue squares) adjacent to edge $i$ in order to ensure that they satisfy fake-flatness. These blob ribbon operators have the same label and orientation and so can be combined into a single blob ribbon operator (represented by the red torus) encircling the edge.}
			\label{closed_blob_ribbon_around_edge}
		\end{center}
	\end{figure}

	In this case, just as we did for the seam, the plaquette holonomy can be corrected (i.e., made to be the identity) by applying a blob ribbon operator $B^x(r)$ that passes through the plaquette, as shown in Figure \ref{torus_reconnect_edge_plaquette_ribbon}. This blob ribbon operator acts on the plaquette label $e_p$ as $B^x(r):e_p= e_p x^{-1}$, and so $\partial(e_p)$ becomes $\partial(e_px^{-1})=\partial(e_p)\partial(x^{-1})$. Combining this with the action of the magnetic membrane operator on the plaquette, we have
	$$B^x(r)C^h(m'):H_1(p) = \partial(x^{-1})\partial(x)=1_G,$$
	so that, as we claimed, the blob ribbon operator corrects the plaquette holonomy. In Figure \ref{torus_reconnect_edge_plaquette_ribbon}, the ribbon travels clockwise around edge $i$ (if we imagine extending it into a closed ribbon around $i$). Note that if we had inverted the circulation of the plaquette, the plaquette holonomy produced by acting with $C^h(m')$ would be inverted (i.e., it would be $\partial(x)^{-1}$). However, the action of the blob ribbon operator of a given direction on the plaquette holonomy would also be inverted, so the blob ribbon operator that we need to apply to make the plaquette satisfy fake-flatness would be the same regardless of the circulation of the plaquette. In addition, this orientation does not depend on the position of the plaquette. We can imagine rotating the plaquette $p$ around $i$ and the ribbon must rotate with it, so the blob ribbon operator always passes through the plaquette in a clockwise manner around $i$ (clockwise with respect the upwards direction in Figure \ref{torus_reconnect_edge_plaquette_ribbon}). This means that we can correct the plaquette holonomy of all of the plaquettes around edge $i$ by applying a closed ribbon operator encircling $i$, as shown in Figure \ref{closed_blob_ribbon_around_edge}.

\begin{figure}[h]
		\begin{center}
			\begin{overpic}[width=0.7\linewidth]{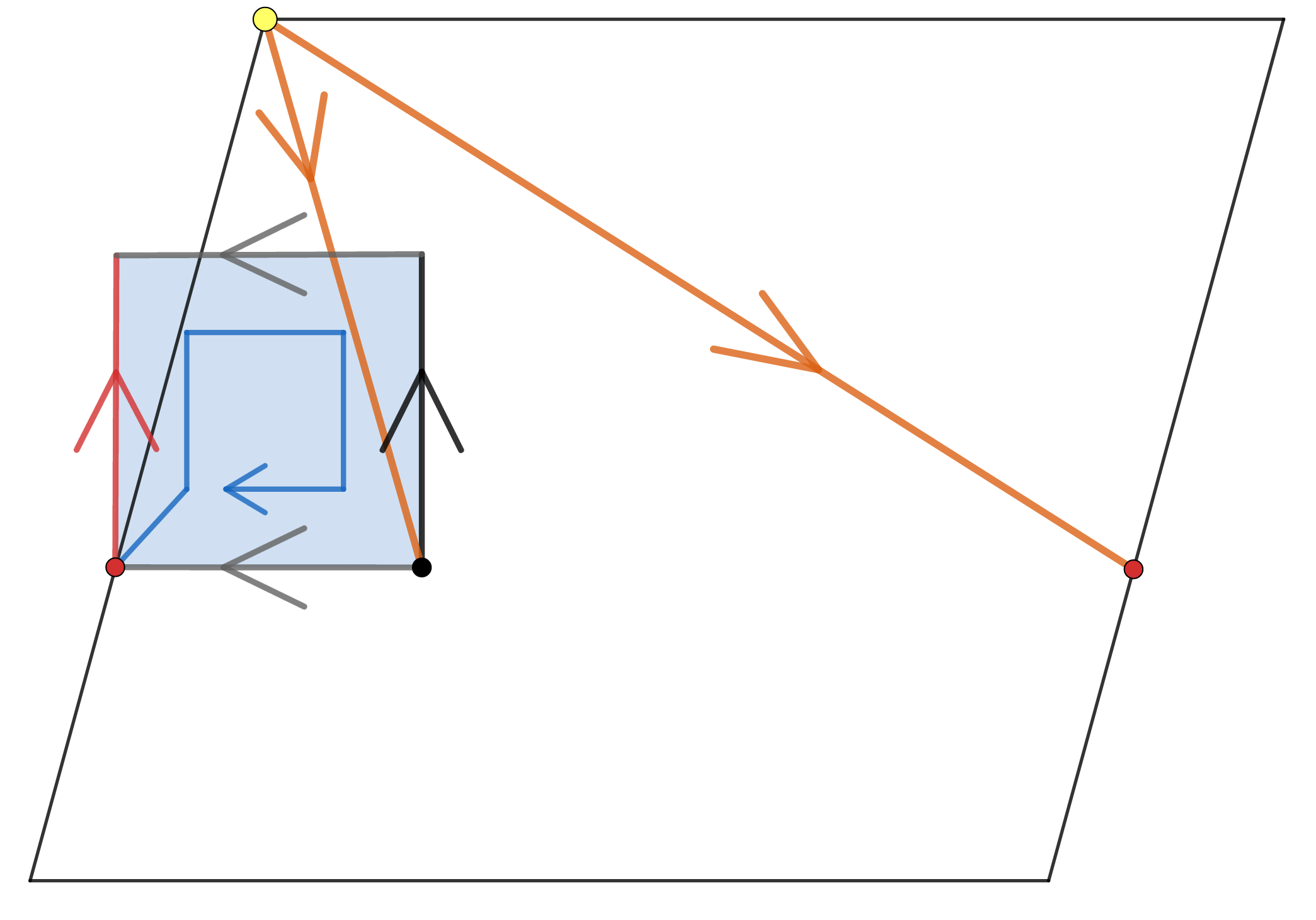}
				
				\put(33,42){$i$}
				\put(7,42){$j$}
				\put(26,52){$t_i$}
				\put(51,50){$t_j$}
				\put(20,38){$p$}
				\put(16,24){$b$}
				\put(16,51){$u$}

			\end{overpic}
			\caption{We can also consider a plaquette on one of the seams, such as the plaquette $p$ in this figure. When we change the path to edge $i$ from $t_i$ to $t_i'$ (as in Figure \ref{torus_reconnect_edge_1}), the plaquette holonomy is affected in the same way as for a plaquette away from the seam.}
			\label{torus_reconnect_edge_plaquette_on_seam}

		\end{center}
	\end{figure}

	In the previous workings, we assumed that the plaquette was not next to a seam of the torus when determining how the plaquette holonomy was affected by changing the path to edge $i$. However, we can apply the same reasoning to a plaquette next to the seam, such as the one shown in Figure \ref{torus_reconnect_edge_plaquette_on_seam}. A plaquette on the seam is pierced by one of the blob ribbon operators that wrap the cycles of the torus, as shown in Figure \ref{Torus_measurement_discontinuity_2}. As we discussed earlier, the action of the blob ribbon operator counters the discontinuity in the paths at the seam and ensures that the plaquette holonomy is preserved by the action of the original measurement operator. That is, under the action of our original measurement operator, the label $e_p$ of plaquette $p$ becomes $e_p'$, and the plaquette holonomy 
	$$H_1(p)= \partial(e_p)g_jg(u)g_i^{-1}g(b)^{-1}$$ becomes
	$$\partial(e_p') g(t_j)^{-1}hg(t_j)g_j g(u)g_i^{-1} g(t_i)^{-1}h^{-1}g(t_i) g(b)^{-1}=1_G,$$
	where $e_p'$ is chosen (by applying an appropriate blob ribbon operator on the seam) to make the plaquette holonomy the identity element. Then under the action of the new measurement operator, where we must replace $g(t_i)$ with $g(t_i')=g(s)g(t_i)$ in the expression above, the plaquette holonomy becomes 
	\begin{align*}
		C^h(m'): H_1(p)&= \partial(e_p') g(t_j)^{-1}hg(t_j)g_j g(u)g_i^{-1} g(t_i')^{-1}h^{-1}g(t_i') g(b)^{-1}\\
		&=\partial(e_p') g(t_j)^{-1}hg(t_j)g_j g(u)g_i^{-1} g(t_i)^{-1}g(s)^{-1}h^{-1}g(s)g(t_i) g(b)^{-1}.
	\end{align*}
	By inserting the identity in the form
	$$ (g(t_i)^{-1}h^{-1}g(t_i) g(b)^{-1} g(b) g(t_i)^{-1}hg(t_i))=1_G,$$
	we can write this as
	\begin{align}
		H_1(p) \rightarrow& \partial(e_p') g(t_j)^{-1}hg(t_j)g_j g(u)g_i^{-1} [g(t_i)^{-1}h^{-1}g(t_i) g(b)^{-1} g(b) g(t_i)^{-1}hg(t_i)] \notag \\
		& \hspace{1cm} g(t_i)^{-1}g(s)^{-1}h^{-1}g(s)g(t_i) g(b)^{-1} \notag \\
		=&\big(\partial(e_p') g(t_j)^{-1}hg(t_j)g_j g(u)g_i^{-1} g(t_i)^{-1}h^{-1}g(t_i) g(b)^{-1}\big) g(b) (g(t_j)g(b))^{-1}h(g(t_j)g(b))\notag\\ & \hspace{1cm} g(t_i)^{-1}
		g(s)^{-1}h^{-1}g(s)g(t_i) g(b)^{-1} \notag\\
		=&\big(1_G\big) g(b) (g(t_j)g(b))^{-1}h(g(t_j)g(b))(g(t_j)g(b))^{-1}g(s)^{-1}h^{-1}g(s)(g(t_j)g(b)) g(b)^{-1} \notag\\
		&= g(b)g(t_i)^{-1}hg(s)^{-1}h^{-1}g(s)g(t_i)b^{-1} \notag\\
		&=\partial(x), \label{Equation_torus_reconnect_edges_plaquette_on_seam_1}
	\end{align}
	where $\partial(x)=hg(s)^{-1}h^{-1}g(s)$ (which is equal to $ g(b)g(t_i)^{-1}hg(s)^{-1}h^{-1}g(s)g(t_i)b^{-1} $ using the fact that $\partial(x) \in \partial(E)$ is in the centre of $G$). This is the same result we saw for a plaquette away from the seam in Equations \ref{Equation_torus_magnetic_change_paths_plaquette_off_seam_tri_trivial_3} and \ref{Equation_torus_magnetic_change_paths_plaquette_off_seam_tri_trivial_4}. Therefore, this plaquette can also be made to be fake-flat by applying a blob ribbon operator of label $x$ running through the plaquette (in a clockwise direction around $i$). This is therefore true for each plaquette $p$ that is adjacent to $i$, so we can correct each plaquette by applying a blob ribbon operator on a closed path clockwise around $i$ that passes through each plaquette, as shown in Figure \ref{closed_blob_ribbon_around_edge}. Note that we could have chosen the blob ribbon operators through the individual plaquettes to have different labels $x'$ that satisfy $\partial(x')=\partial(x)$, rather than making one closed ribbon operator through all of them, if we only wanted to correct the plaquette holonomy (because the condition only depends on $\partial(x)$, not $x$ itself). However, if these labels were different, so that we cannot combine the individual blob ribbon operators into a single closed ribbon operator, then the blob holonomy of some of the blobs around the edge would be affected and so we would excite these blobs. Therefore, we must choose all of the blob ribbons to have the same label so that we can connect them into a closed ribbon (as long as the ribbon pieces have the same label, we could choose any label $x'$ satisfying $\partial(x')=\partial(x)$, but the following argument would not be affected by this change). Once we have added this additional blob ribbon operator, the measurement operator does not excite any plaquettes and so satisfies the conditions to be a measurement operator that we have considered so far.

	While adding the blob ribbon operator makes the new measurement operator satisfy the conditions for a valid measurement operator that we have considered so far, it does not mean that it is a distinct measurement operator. Indeed, we will now show that it is actually equivalent to the original measurement operator, from before we changed the path to the edge. This new measurement operator differs from the original only in its action on $i$ and the adjacent plaquettes (this difference is local, which already suggests that we have not generated a distinct measurement operator, because a topological measurement operator should not depend on local details). Now consider how the action on the degrees of freedom around $i$ differs between the two operators. The original magnetic membrane operator, which we denote $C^h(m)$, acted on the edge $i$ according to
	$$C^h(m):g_i = g(t_i)^{-1}hg(t_i)g_i$$
	and on the adjacent plaquettes trivially, i.e., according to
	$$C^h(m)e_p =e_p,$$
	where we have not included the action of the blob ribbon operators around the cycles on the plaquette if the plaquette is on the seam (because they are the same in either case). The new magnetic membrane operator acts on $g_i$ as
	\begin{align*}
		C^h(m'):g_i =& g(t'_i)^{-1}hg(t'_i)g_i\\
		=& g(t_i)^{-1}g(s)^{-1}hg(s)g(t_i)g_i.
	\end{align*}
	
	In order to compare this to the action of the original membrane operator, we insert the identity in the form
	$$1_G= g(t_i)^{-1}h^{-1}g(t_i) g(t_i)^{-1}hg(t_i),$$
	to obtain
	\begin{align*}
		C^h(m'):g_i =& g(t_i)^{-1}g(s)^{-1}hg(s)g(t_i) [g(t_i)^{-1}h^{-1}g(t_i) g(t_i)^{-1}hg(t_i)]g_i\\
		=& g(t_i)^{-1}g(s)^{-1}hg(s) h^{-1}g(t_i) (g(t_i)^{-1}hg(t_i) g_i)\\
		=& g(t_i)^{-1}g(s)^{-1}hg(s) h^{-1}g(t_i) \cdot (C^h(m):g_i).
	\end{align*}
	
	We then recognise that $g(t_i)^{-1}g(s)^{-1}hg(s) h^{-1}g(t_i)= \partial(x^{-1})$ from Equation \ref{Equation_torus_reconnect_edges_plaquette_on_seam_1}, so that
	\begin{align*}
		C^h(m'):g_i = \partial(x)^{-1} (C^h(m):g_i).
	\end{align*}
	This factor of $\partial(x)^{-1}$ is the same factor that we would gain by applying an edge transform $\mathcal{A}_i^{x^{-1}}$ on $i$, so we can write this as
	\begin{align*}
		C^h(m'):g_i = \mathcal{A}_i^{x^{-1}}C^h(m):g_i.
	\end{align*}
	
	Now consider the action of closed blob ribbon operator that we must apply around $i$ to ensure that fake-flatness is satisfied for the new membrane operator, on the plaquettes surrounding $i$. This is given by
	\begin{align*}
		B^x(r):e_p &=e_p x^{-1}\\
		&=x^{-1}e_p,
	\end{align*}
	due to the action of the closed blob ribbon operator around the edge $i$, where we again excluded the action of the blob ribbon operators around the cycles of the torus. However, this is the same as the action of the edge transform $\mathcal{A}_i^{x^{-1}}$ on the plaquettes (recall that we took the circulation of the plaquettes to be oriented against edge $i$). Because the additional action from $B^x(r)C^h(m')$ on the edge and the surrounding plaquettes when compared to the action of $C^h(m)$ is equivalent to an edge transform on edge $i$ (again note the equivalence to the argument given in Section \ref{Section_linking_appendix} for deforming a linking string removed by applying a blob ribbon operator), we can write
	$$B^x(r)C^h(m')=\mathcal{A}_i^{x^{-1}}C^h(m).$$
	However, the region on which we apply the measurement operator is originally unexcited and must remain unexcited by the action of the measurement operator (this is a requirement for our measurement operator, as described in Section \ref{Section_3D_Topological_Sectors} of the main text). Therefore, this edge transform must act trivially on our subspace. This means that, in our subspace, $B^x(r)C^h(m')=C^h(m)$ and so the two measurement operators are equivalent. Therefore, we can freely change the path from the start-point to edge $i$, but we will not generate distinct measurement operators. This is true for all edges $i$, and so it is true regardless of how we change each path in the magnetic membrane operator (in particular we could change the path to edges on the seams so that they do not cross the seam, in which case the additional blob ribbon operator $B^x(r)$ that we must apply to ensure fake-flatness will just divert the usual blob ribbon operator along the seam to pass around that edge).

	We have now applied all of the independent operators that we can. We denote the operator that applies the electric operators on the two cycles, the $E$-valued membrane operator on the torus, the flux membrane and the blob ribbon operators around the cycles by
	\begin{equation}
		T^{[e_{c_1},e_{c_2},e_m,g_{c_1},g_{c_2},h]}(m)=B^{e_{c_1}}(c_1)B^{e_{c_2}}(c_2)C^h(m) \delta(\hat{e}(m),e_m)\delta(\hat{g}(c_1), g_{c_1}) \delta(\hat{g}(c_2), g_{c_2}).
		\label{Equation_T_operator_definition_1}
	\end{equation}
	
	We must remember, however, that the $T^X$ operator (where $X$ stands in for the set of labels for the measurement operator) is not the entire story. Before applying the $T^X$ operator, we must first apply the projection operator that ensures that the membrane is unexcited. This will be relevant when we consider further restrictions on the allowed operators.

	So far, we have ensured that the $T^X$ operators satisfy the fake-flatness conditions and blob conditions (the blob conditions are satisfied because we use closed blob ribbon operators only) by restricting which sets of labels the measurement operators can have. Now we need to ensure that the vertex and edge energy terms on the membrane are also satisfied after applying our measurement operator $T^X$. Recall that an energy term $A_v$ or $\mathcal{A}_i$ is satisfied for a state $\ket{\alpha}$ if $A_v \ket{\alpha}= \ket{\alpha}$ (and similarly for $\mathcal{A}_i$). Requiring that this condition be satisfied for the state produced by the action of our measurement operator on a state $\ket{\psi}$ therefore means that $A_v T^X P \ket{\psi} = T^X P \ket{\psi}$, where $P$ is the projection operator that ensures the measurement surface is initially unexcited. This means that our operator must satisfy $A_vT^XA_v =T^X A_v$ for all vertices $v$ on our membrane, where the $A_v$ operator on the right comes from the projection operator (and similarly for the edge energy terms). We will see that this condition leads to us needing to take linear combinations of the $T^X$ operators with different labels. If the operator $T^X(m)$ satisfies a commutation relation of the form $A_v^g T^X(m) = T^{Y(g)}(m) A_v^g$ (we will find these relations slightly later in this section), then $A_v^g T^X(m) A_v = T^{Y(g)}(m) A_v^g A_v = T^{Y(g)}(m) A_v$, because $A_v^g A_v=A_v$ as described in Section \ref{Section_Recap_3d} of the main text. This means that 
	\begin{align*}
		A_v T^X(m) A_v &= \frac{1}{|G|} \sum_{g \in G} A_v^g T^X(m) A_v\\
		&= \frac{1}{|G|} \sum_{g \in G} T^{Y(g)}(m) A_v,
	\end{align*}
	from which we see that the commutation of the measurement operator with the vertex term generates a sum of $T$ operators with different labels. Therefore, in order to ensure that the vertex energy term is satisfied after we apply the operator, we must apply a linear combination of the $T$ operators. We can use the commutation relation above to generate these linear combinations. That is, the allowed combinations take the form $\frac{1}{|G|} \sum_{g \in G} T^{Y(g)}(m)$, where $Y(1_G)=X$ is the initial label set and the $Y(g)$ labels are generated by the vertex transform. To see that such a sum satisfies our constraint, we note that
	\begin{align*}
		A_v \frac{1}{|G|} \sum_{g \in G} T^{Y(g)}(m) A_v &= A_v A_v T^X(m) A_v,
	\end{align*}
	from our definition of $T^{Y(g)}(m)$. Then the vertex term $A_v$ is a projector, so it satisfies $A_v A_v=A_v$. Therefore,
	\begin{align*}
		A_v \frac{1}{|G|} \sum_{g \in G} T^{Y(g)}(m) A_v&= A_v T^X(m) A_v\\
		&= \frac{1}{|G|} \sum_{g \in G} T^{Y(g)}(m) A_v.
	\end{align*}
	
	In addition to performing this procedure with the vertex transforms, we must do the same for each edge transform for the edges on the membrane. Ensuring that the measurement operator commutes with these transforms therefore restricts which linear combinations of the $T$ operators can be included in a measurement operator. The group structure of the $A_v^g$ and $\mathcal{A}_i^e$ operators means that the linear combinations have equal coefficients for each label set generated by the vertex and edge transforms. That is, if two label sets are related by a commutation relation with an edge or vertex transform, such as $X$ and $Y(g)$ above, they must always appear together in any allowed superposition and appear with equal coefficient. This divides the different allowed label sets into classes that are related by the edge and vertex transforms, with each element of a particular class always appearing together in an allowed superposition. As a shorthand, we write that two label sets must always appear together (with equal coefficient) in any allowed linear combination by $ X \sim Y(g)$. This is an equivalence relation, where reflexivity follows from the fact that $X$ is related to itself by the trivial vertex transform, symmetry follows from the fact that the inverse of a vertex or edge transforms is another transform, and transitivity follows from the group structure of the transforms (in particular, from associativity).

	Now that we have established that the commutation relations between the candidate measurement operators and the vertex and edge transforms restrict the allowed linear combinations of the $T^X$ operators in a measurement operator, we will find these restrictions explicitly. We start by examining the conditions arising from the vertex transforms. The different ribbon and membrane operators in $T^X$ commute with all of the vertex transforms, except those at their respective start-points (as we showed in Section \ref{Section_ribbon_membrane_energy_commutation}). Because all of the start-points of the operators are chosen to be the same, we only have to consider the vertex transforms at this one vertex. Let us start by examining how this vertex transform affects the electric ribbon operators applied on the two cycles of the torus. For $c_1$, the effect of vertex transform $A_{s.p}^g$ is given by
	\begin{align*}
		\delta(g_{c_1},\hat{g}(c_1))A_{s.p}^g&=A_{s.p}^g \delta(g_{c_1},g\hat{g}(c_1)g^{-1})\\
		&=A_{s.p}^g \delta(g^{-1}g_{c_1} g,\hat{g}(c_1))\\
		& \implies g_{c_1} \rightarrow g^{-1}g_{c_1} g.
	\end{align*}
	The vertex transform also affects the label of the other cycle ($c_2$) in exactly the same way, so that $g_{c_2} \rightarrow g^{-1}g_{c_2} g$.

	The final operator in the measurement operator that fails to commute with the vertex transform at the start-point is the magnetic membrane operator. Using Equation \ref{Equation_magnetic_membrane_start_point_transform} from Section \ref{Section_Magnetic_Membrane_Tri_trivial}, we see that
	$$C^h(m) A_{s.p}^g = A_{s.p}^g C^{g^{-1}hg}(m).$$ 
	Therefore, the label of the magnetic membrane operator transforms as $h \rightarrow g^{-1}hg$. We therefore have
	$$T^{e_{c_1},e_{c_2},e_m,g_{c_1},g_{c_2},h}A_{s.p}^g =A_{s.p}^g T^{e_{c_1},e_{c_2},e_m,g^{-1}g_{c_1}g,g^{-1}g_{c_2}g,g^{-1}hg}.$$
	This means that our operator must be an equal sum of $T$ operators with label sets related by the equivalence relations
	\begin{equation}
		(g_{c_1},g_{c_2},h) \sim (g^{-1}g_{c_1}g, g^{-1}g_{c_2}g, g^{-1}hg), 
		\label{Equation_torus_vertex_transform_condition_1}
	\end{equation}
	for each $g \in G$.

	At this point, we can explain why we were allowed to place all of the start-points of the operators in the same position (the start of the two electric ribbon operators and the magnetic membrane operator). The reason for this is that, if the start-points were in different positions, then the operators would have to separately satisfy vertex transforms at each vertex. For example, consider the case where the start-point of the magnetic membrane is different from the start of the two ribbons. Then, in order to satisfy the vertex energy term we would need an equal sum of all of the magnetic membrane operators with label in a particular conjugacy class (as described in Section \ref{Section_Magnetic_Membrane_Tri_trivial}). However, if the magnetic membrane operator does satisfy this condition then its start-point is irrelevant and can be changed without affecting the action of the operator, as described in Section \ref{Section_magnetic_tri_trivial_move_sp}. Therefore, we could freely move this start-point to the same position as the start-points of the ribbons. In a similar way, if the start-point of one of the electric ribbons is away from the other start-points, then that ribbon operator must satisfy the vertex transforms at its start-point individually. However, because the vertex transform is equivalent to parallel transport of that vertex (and all paths attached to it, such as the electric ribbon), as described in Ref. \cite{HuxfordPaper1} (see Section I D 3), this means that the ribbon operator must also be insensitive to moving the start-point. This means that the allowed operators whose start-points are in different positions are counter-intuitively equivalent to a subset of the operators whose start-points are all in the same position. For a similar reason, the choice of this shared start-point does not matter: the overall measurement operator must satisfy the vertex transform at this start-point, which means that moving the start-point does not affect the operator.

\begin{figure}[h]
		\begin{center}
			\begin{overpic}[width=0.5\linewidth]{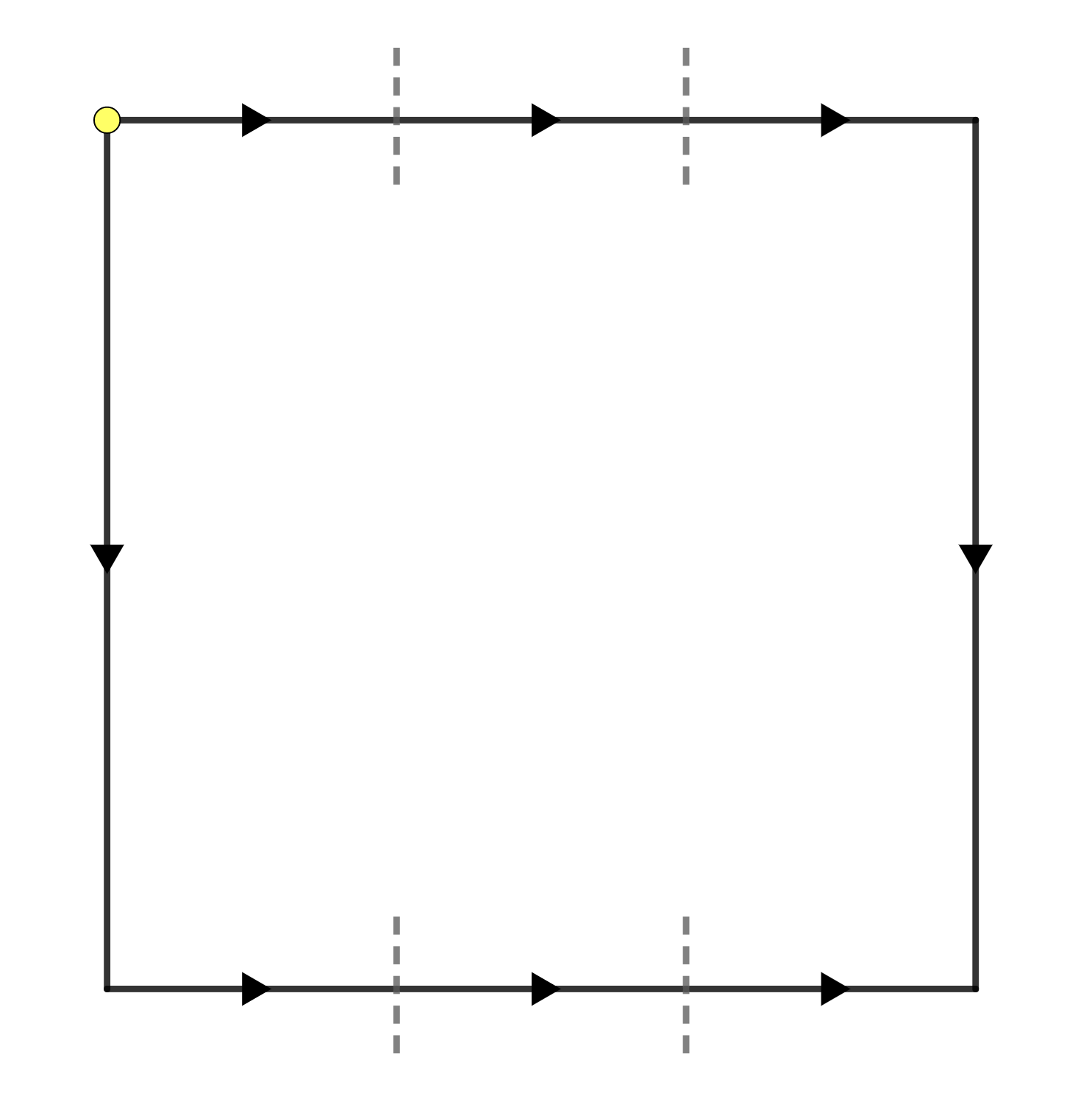}
				
				\put (43,92){\large edge $i$}
				
				\put(43,3){\large edge $i$}
				\put(2,50){\large $c_2$}
				\put(92,50){\large $c_2$}
			\end{overpic}
			\caption{Typically, an edge transform on the boundary of a membrane affects the surface label of that membrane. However, in the case of the torus (represented by a square with opposite sides glued) each edge appears twice in the boundary of the square, as indicated by the example edge $i$. Each of these appearances contributes to the change in the surface label, and when $\rhd$ is trivial these two contributions cancel.}
			\label{torus_charge_edge_transform_tri_trivial}
		\end{center}
	\end{figure}

	Having considered the vertex transforms, we next consider edge transforms that lie flat on the membrane. All edge transforms commute with the blob ribbon operators, as proven in Section \ref{Section_Blob_ribbon_proof_tri_trivial}. In addition, the magnetic membrane operator commutes with the edge transforms when $\rhd$ is trivial (as shown in Section \ref{Section_Magnetic_Membrane_Tri_trivial}). This leaves us to consider the $E$-valued membrane operators and the electric ribbon operators. First we consider the $E$-valued membrane operator. We examine the impact of an edge transform $\mathcal{A}_i^e$, applied on an edge $i$ on the boundary of $m$, on the surface label $\hat{e}(m)$. Considering Figure \ref{torus_charge_edge_transform_tri_trivial}, we see that the edge appears twice in the boundary of $m$, with opposite orientations with respect to that boundary. Given that $\rhd$ is trivial, we see from Equation \ref{Equation_edge_transform_definition} in the main text that the contribution of the edge transform $\mathcal{A}_i^e$ to the surface from the appearance of the edge that matches the orientation of the surface is $e^{-1}$ and the contribution from the appearance that is anti-aligned with the surface is $e$. These contributions cancel, so the edge transform leaves the surface label invariant and therefore commutes with the the $E$-valued membrane operator. Another way of thinking about this is that the edge is adjacent to two plaquettes in the surface and the contribution to each cancels once we have chosen those plaquettes to have the same orientation (if the two plaquettes are not aligned, the edge transform affects them both in the same way, but one of the plaquettes must appear in the surface label with an inverse, which leads to the contributions from the edge transform to the surface cancelling anyway).

	On the other hand, the edge transforms do change the path labels $g_{c_1}, g_{c_2}$ of the two cycles of the torus. An edge transform on $c_1$ changes $g_{c_1}$ by an element of $\partial(E)$ (while not affecting $g_{c_2}$) and similarly an edge transform on $c_2$ changes $g_{c_2}$ by such an element (while leaving $g_{c_1}$ unchanged). We therefore have 
	\begin{equation}
		g_{c_1} \sim \partial(e) g_{c_1}
		\label{Equation_torus_c1_edge_transform_1}
	\end{equation}
	for all $e \in E$ from the edges on $c_1$, and independently
	\begin{equation}
		g_{c_2} \sim \partial(e') g_{c_2}
		\label{Equation_torus_c2_edge_transform_1}
	\end{equation}
	from the edges on $c_2$, for each $e' \in E$.

	So far, we have seen restrictions to our measurement operators that arise from the incompatibility of the labels of the candidate measurement operators with the energy conditions. There is one more source of restriction, which we will see when we examine the last types of edges near the membrane. In addition to the edges lying in the direct membrane of the magnetic membrane operator, we must consider the edges that are cut by the dual membrane (point outwards from the direct membrane). For these particular edges, the edge transforms automatically commute with the operator $T^X$, because they commute with the different constituent operators of $T^X$ (unlike the edge transforms on the cycles, they do not change the path elements of the cycles). However, we saw in Section \ref{Section_magnetic_condensed_tri_trivial} that some of the magnetic membrane operators were condensed and could be produced using edge transforms (and an operator on the boundary). In a similar way we will see that combining the operator $T^X$ with a series of edge transforms on the outwards edges can change the labels of $T^X$ (specifically it changes the magnetic label, as we may expect from the fact that we are essentially producing a condensed magnetic membrane operator). Because the edge transforms act trivially when the edges are unexcited, we can produce these transforms without changing the state. That means that if multiplying $T^{X}(m)$ by a series of these edge transforms results in the operator $T^{Y}(m)$ then these two $T$ operators actually act the same in the space that we project to (where the membrane is unexcited), so they should not be treated as different measurement operators. We write this, as we did with previous conditions, as an equivalence relation between different label sets. While we write the conditions in the same way, the origin of the condition is different from those from the other transforms, because rather than requiring a summation over different labels to make the measurement operator commute with the transforms, this condition comes from the fact that the projection naturally generates all the equivalent labels.

	As we discussed in Section \ref{Section_magnetic_condensed_tri_trivial} when considering condensation, applying the edge transform $\mathcal{A}_i^e$ on every edge cut by the dual membrane of a magnetic membrane operator generates the operator $C^{\partial(e)}(m)$. In that case, we were considering open membrane operators, and the edge transforms also generated a closed ribbon operator around the boundary of the membrane operator. However, when we close the membrane into a torus this is no longer the case. If we consider the square shown in Figure \ref{unfoldedtorus1appendix}, we would have a blob ribbon operator $B^{e^{-1}}(t)$ around $c_2c_1c_2^{-1}c_1^{-1}$. When we close the square into a torus, the section on $c_2$ would cancel with the section on $c_2^{-1}$ (and similar for $c_1$). Therefore, the ribbon operator is trivial, and we just have $\prod_i \mathcal{A}_i^e =C^{\partial(e)}(m)$. Combining this condensed membrane operator with a membrane operator $C^h(m)$ from our measurement operator gives us
	$$C^h(m) \rightarrow C^h(m) C^{\partial(e)}(m)=C^{h\partial(e)}(m).$$
	This means that we have the following relation for the magnetic label of the measurement operator:
	\begin{equation}
		h \sim \partial(e'')h \ \forall e'' \in E.
		\label{Equation_torus_outwards_edge_transform_1}
	\end{equation}

	To summarize our results for this section, we have the following restrictions on the labels for a valid measurement operator:
	\begin{align}
		\partial(e_m)^{-1}&=[g_{c_1}^{-1},g_{c_2}^{-1}]\\
		\partial(e_{c_2})&=[g_{c_1}^{-1},h^{-1}] \\
		\partial(e_{c_1})&=[h^{-1},g_{c_2}^{-1}] \\
		(g_{c_1},g_{c_2},h) &\sim (g^{-1}g_{c_1}g, g^{-1}g_{c_2}g, g^{-1}hg) \ \forall g \in G \\
		g_{c_1} &\sim \partial(e) g_{c_1} \ \forall e \in E\\
		g_{c_2} &\sim \partial(e') g_{c_2} \ \forall e' \in E\\
		h &\sim \partial(e'')h \ \forall e'' \in E. 
	\end{align}

	These conditions show a striking resemblance to the relations and restrictions that appear for the calculation of the ground state degeneracy of the higher-lattice gauge theory model on the 3-torus, as given in Ref. \cite{Bullivant2017}. In fact, these conditions are exactly the same. For the calculation of the ground state degeneracy in the special case where $\rhd$ is trivial, we have \cite{Bullivant2017}:
	\begin{align}
		\partial(e)&=[x,z]\\ 
		\partial(f)&=[x,y]\\
		\partial(k)&=[y,z]\\
		(x,y,z) &\sim (axa^{-1},aya^{-1},aza^{-1})\\
		x &\sim \partial(e_x)x\\
		z &\sim \partial(e_z)z\\
		y &\sim \partial(e_y)y,
	\end{align}
	where $e$, $f$ and $k$ are elements of $E$ corresponding to surface elements on the 3-torus, $x$, $y$ and $z$ are elements of $E$ corresponding to cycles of the 3-torus, and $e_x,e_y,e_z \in E$ and $a \in G$ are dummy variables arising from energy terms. We can see that these ground state conditions are the same as the restrictions for the labels of the measurement operator if we make the identifications $e_m=e^{-1}$, $e_{c_2}=f$, $e_{c_1}=k$, $g_{c_1}=x^{-1}$, $g_{c_2}=z^{-1}$ and $h =y^{-1}$, and also make appropriate relations between the dummy parameters $(a,e_x,e_y,e_z)$ and $(g,e,e',e'')$. We can therefore say that the number of unconfined topological charges that can be measured by a toroidal surface in this way is equal to the ground state degeneracy of the model on the 3-torus, when $\rhd$ is trivial.

	\subsection{Topological charge within a torus when $E$ is Abelian and $\partial \rightarrow$ centre($G$)}
	\label{Section_3D_Topological_Charge_Torus_Tri_nontrivial}
	
	We next consider the toroidal measurement operators in the case where $E$ is Abelian, $G$ is general, $\partial$ maps to the centre of $G$ and $\rhd$ is general (Case 2 in Table \ref{Table_Cases} of the main text). We follow the same procedure as for the $\rhd$ trivial case from the previous section. The first step is to project to the space where there are no excitations in the region of the measurement operator. This means projecting to each closed loop on the torus satisfying fake-flatness and the blobs not being excited, as well as performing every gauge transform in the region.

	Just as in the $\rhd$ trivial case, after projecting to the space without excitations on the surface of the measurement operator, we then apply closed ribbon and membrane operators on that surface. We will apply an operator of the form,
	\begin{equation}
		T^{[e_{c_1},e_{c_2},e_m,g_{c_1},g_{c_2},h]}(m)=B^{e_{c_1}}(c_1)B^{e_{c_2}}(c_2)C^h(m) \delta(\hat{e}(m),e_m)\delta(\hat{g}(c_1), g_{c_1}) \delta(\hat{g}(c_2), g_{c_2}).
		\label{Equation_T_operator_definition_2}
	\end{equation}
	where we will explain the components of this operator as we go through this section. Firstly, we choose two independent non-contractible cycles on the torus ($c_1$ and $c_2$), on which we apply two closed electric operators $\delta(\hat{g}(c_1),g_{c_1})$ and $\delta(\hat{g}(c_1),g_{c_1})$. It is convenient to picture the torus as a square, with the edges glued together to form the ``seams" of the torus, as shown in Figure \ref{unfoldedtorus1appendix}. We choose these seams to match the cycles $c_1$ and $c_2$. Just as in the $\rhd$ trivial case, we could also apply electric ribbon operators on the contractible loops, because fake-flatness does not fully specify the labels of the contractible closed loops on the torus. However, fake-flatness does restrict the closed loops to have group element within $\partial(E)$ and any electric ribbon operator that is sensitive to elements within $\partial(E)$ is confined and so would excite the edge energy terms. Similarly, we could apply electric ribbon operators on the other non-contractible cycles, but these cycles are not independent of $c_1$ and $c_2$. This means that we can deform non-confined electric ribbon operators into ribbons which lie on $c_1$ or $c_2$ (possibly wrapping them multiple times), while the confined versions that we cannot deform would cause excitations. The position of confined ribbon operators must be carefully chosen so that these excitations are removed by the other membrane operators, as we shall see later, which leads to the electric ribbon operators only being applied on the $c_1$ and $c_2$. In addition to the electric ribbon operators, we apply a closed $E$-valued membrane operator $\delta(\hat{e}(m),e_m)$ on the direct membrane (the green surface in Figure \ref{thickenedmembrane}, or the square in Figure \ref{unfoldedtorus1appendix}), with orientation pointing inwards on the torus. It is convenient to treat the torus as a rectangle with opposite sides identified, as indicated in Figure \ref{unfoldedtorus1appendix}. When we close the rectangle into a torus by gluing the opposite edges, the membrane operators applied on the rectangle may produce excitations at the boundary of this rectangle, even though the resulting torus membrane is closed. This represents an obstruction to closing the square membrane into a torus, and so we must keep track of the location where we glue the edges, which we call the seam of the torus.

	Just as in the $\rhd$ trivial case, we have additional requirements on the labels of the operators that make up the measurement operator to make sure that the operator does not produce any excitations or annihilate all states for which the toroidal region is unexcited (the subspace we are interested in). The first of these conditions only involves the electric ribbon operators and $E$-valued membrane operator that we have applied so far. Requiring the overall torus surface to satisfy fake-flatness gives us the condition
	$$\partial(e_m)g_{c_1}g_{c_2}g_{c_1}^{-1}g_{c_2}^{-1}=1_G,$$ 
	which we can write as
	\begin{equation}
		\partial(e_m)^{-1}=[g_{c_1},g_{c_2}].
		\label{torus_flatness_condition_2}
	\end{equation}

	After applying the electric ribbon operators and $E$-valued membrane operator, we next apply a magnetic membrane operator $C^h_T(m)$, where the direct membrane of this operator is the torus surface (like for the $E$-valued membrane) and the dual membrane is a torus slightly outside of the direct membrane (recall from Section \ref{Section_3D_Tri_Trivial_Magnetic_Excitations} of the main text that the dual membrane lies on the dual lattice, and cuts through the edges affected by the magnetic membrane operator). Because the magnetic membrane operator is applied on an unexcited membrane and $\partial$ maps to the centre of $G$, the precise positions of the paths involved in the membrane operator do not matter, except that we require these paths not to cross the seams at the edge of the square. Just as in the $\rhd$ trivial case that we considered in Section \ref{Section_3D_Topological_Charge_Torus_Tri_trivial}, we could make a different choice for these paths, but it would not result in a new operator, as we will explain later. Similarly, we choose the ribbons for the blob ribbon operators that are part of $C^h_T(m)$ not to cross the seams.

	Finally, we apply blob ribbon operators around the two cycles of the torus. Just as we argued in Section \ref{Section_3D_Topological_Charge_Torus_Tri_trivial} for the $\rhd$ trivial case, any blob ribbon operators with labels in the kernel of $\partial$ can be deformed into blob ribbon operators around the two cycles, or can be contracted into trivial operators. In the $\rhd$ trivial case, we saw that energy constraints meant that we could only put the confined blob ribbon operators (those with label outside the kernel) along the seams of the torus, and limited the labels allowed to these blob ribbon operators. We will see that this also holds in the $\rhd$ non-trivial case. In order to show this, we must examine the action of the magnetic membrane operator $C^h_T(m)$ on the plaquettes near the seam of the membrane, where there is a discontinuity in its action (as explained in Section \ref{Section_3D_Topological_Charge_Torus_Tri_trivial}). These plaquettes might not satisfy fake-flatness because the path from the start-point of the membrane operator to one of the edges on the plaquette wraps around a non-contractible cycle whereas the path to the other edge does not, as shown in Figure \ref{plaquette_near_c_2_tri_trivial} in Section \ref{Section_3D_Topological_Charge_Torus_Tri_trivial}.

	We consider the plaquette holonomy $H_1(p)$ of such a plaquette. As described in Section \ref{Section_Magnetic_Tri_Non_Trivial}, the magnetic membrane operator can affect the plaquette labels themselves in addition to changing the labels of edges. However, while the magnetic membrane can affect the label $e_p$ of a plaquette (the $C^h_{\rhd}(m)$ part of the operator can induce a change through the map $\rhd$, and the additional blob ribbon operators in the membrane operator also affect plaquette labels), it only changes it by an element of the kernel of $\partial$ and so does not change the expression $\partial(e_p)$ which appears in the plaquette holonomy. Therefore, the change to the plaquette holonomy under the magnetic membrane operator comes solely from the changes to the edge labels. The magnetic membrane operator takes the label $g_x$ of the edge $i_x$ (with $x=1$ or $x=2$) to $g(t_x)^{-1}hg(t_x)g_x$, where $t_x$ is the path from the start-point of the membrane operator to the edge $i_x$. The plaquette holonomy $H_1(p)$ of the plaquette $p$ illustrated in Figure \ref{plaquette_near_c_2_tri_trivial} therefore transforms according to 
	\begin{align}
		H_1(p)&=\partial(e_p)g_1g(u)g_2^{-1}d=1_G \notag\\
		&\rightarrow \partial(e_p)g(t_1)^{-1}hg(t_1)g_1g(u)g_2^{-1}g(t_2)^{-1}h^{-1}g(t_2)d. \label{Equation_toroidal_charge_plaquette_near_c2}
	\end{align}
	
	When it comes to expressions such as $g(t)^{-1}hg(t)$, which involve conjugation of $h$ by a path element, we are allowed to deform the path $t$ over a flat region without changing this expression. This is because such a deformation only changes the path element $g(t)$ by an element in $\partial(E)$, which is also in the centre of $G$ and so does not affect the conjugation (see Section \ref{Section_Magnetic_Tri_Non_Trivial}). We deform the paths $t_1$ and $t_2$ into the green and purple paths $sd$ and $c_1s$ shown in Figure \ref{plaquette_near_c_2_tri_trivial}. Then, replacing the path elements $g(t_1)$ and $g(t_2)$ in Equation \ref{Equation_toroidal_charge_plaquette_near_c2} with these deformed path elements, we see that, just as in the $\rhd$ trivial case, the plaquette holonomy transforms as
	\begin{align*}
		H_1(p) &\rightarrow \partial(e_p) (g(d)^{-1}g(s)^{-1}hg(s)g(d)g_1)g(u)( g(s)^{-1}g_{c_1}^{-1}hg_{c_1}g(s)g_2)^{-1}g(d) \\
		&=g(d)^{-1}g(s)^{-1} hg(s)g(d)\partial(e_p) g_1g(u)g_2^{-1} g(s)^{-1} g_{c_1}^{-1}h^{-1}g_{c_1}g(s) g(d) \\
		&=g(d)^{-1}g(s)^{-1} hg(s)g(d) \big[\partial(e_p) g_1g(u)g_2^{-1}g(d)\big]g(d)^{-1} g(s)^{-1} g_{c_1}^{-1}h^{-1}g_{c_1}g(s) g(d). 
	\end{align*}
	
	We can then recognise the term $\partial(e_p) g_1g(u)g_2^{-1}g(d)$ as the original plaquette holonomy, which is $1_G$ from the fake-flatness condition. Then
	\begin{align}
		H_1(p) &\rightarrow g(d)^{-1}g(s)^{-1} hg(s)g(d) \big[1_G\big]g(d)^{-1} g(s)^{-1} g_{c_1}^{-1}h^{-1}g_{c_1}g(s) g(d) \notag\\
		&=g(d)^{-1}g(s)^{-1} h g_{c_1}^{-1}h^{-1}g_{c_1}g(s) g(d). \label{Equation_plaquette_holonomy_boundary_torus_1}
	\end{align}
	
	It is possible that the resulting plaquette holonomy is not equal to $1_G$, indicating that fake-flatness may be violated by the magnetic membrane operator in this boundary region (this represents an obstruction to the membrane operator closing smoothly). As we will see shortly, we can sometimes remove these additional excitations by applying a confined blob ribbon operator of appropriate label along the seam of the torus. In addition we know that the magnetic membrane operator does not produce plaquette excitations in the bulk region (from the properties of the membrane operator away from its boundary). This means that we cannot apply a confined blob ribbon operator on another path parallel to $c_2$, because it would produce excitations that cannot cancel with those produced by the membrane operator. The confined blob ribbon operators that wrap $c_2$ or equivalent cycles are therefore restricted to this boundary region, in order to prevent them from violating fake-flatness elsewhere (while non-confined ribbon operators are topological and can be deformed onto $c_2$ if they are applied on an equivalent cycle). Furthermore, the label of the action of the blob ribbon operators must be chosen so that the ribbon operators counter the effect of the magnetic membrane operator and restore fake-flatness in this region. A blob ribbon operator of label $e_{c_2}$ wrapping $c_2$ takes the plaquette label $e_p$ to $e_p ( g(t_1)^{-1} \rhd e_{c_2}^{-1})$, meaning that it contributes an extra factor of $\partial( g(t_1)^{-1} \rhd e_{c_2}^{-1})$ to the plaquette holonomy. This means that the total effect of the blob ribbon operator and magnetic membrane operator on the plaquette holonomy is (from Equation \ref{Equation_plaquette_holonomy_boundary_torus_1})
	\begin{align*}
		H_1(p)=1_G &\rightarrow \partial (g(t_1)^{-1} \rhd e_{c_2}^{-1}) g(d)^{-1}g(s)^{-1} h g_{c_1}^{-1}h^{-1}g_{c_1}g(s) g(d)\\
		&=\partial(e_{c_2}^{-1})g(d)^{-1}g(s)^{-1}h g_{c_1}^{-1}h^{-1}g_{c_1}g(s) g(d)\\
		&= g(d)^{-1}g(s)^{-1}\partial(e_{c_2}^{-1}) h g_{c_1}^{-1}h^{-1}g_{c_1}g(s) g(d),
	\end{align*}
	where we used the first Peiffer condition (Equation \ref{Equation_Peiffer_1} in the main text), together with the fact that $\partial(e)$ is in the centre of $G$, to write $\partial (g(t_1)^{-1} \rhd e_{c_2}^{-1})= \partial (e_{c_2}^{-1})$ and then commute it past $g(d)^{-1}g(s)^{-1}$. If we require the plaquette holonomy to be $1_G$ in order for that plaquette not to be excited, then we must have
	\begin{align*}
		\partial&(e_{c_2})^{-1}hg_{c_1}^{-1}h^{-1}g_{c_1}=1_G\\
		&\implies \partial(e_{c_2})=hg_{c_1}^{-1}h^{-1}g_{c_1}.
	\end{align*}
	Then we can conjugate both sides of this equation by multiplying by $g_{c_1}$ to the left and $g_{c_1}^{-1}$ to the right. Again using the fact that $\partial(E)$ is in the centre of $G$, we then have
	\begin{align*}
		g_{c_1}\partial(e_{c_2})g_{c_1}^{-1}=g_{c_1}[hg_{c_1}^{-1}h^{-1}g_{c_1}]g_{c_1}^{-1}
		&\implies \partial(e_{c_2})=g_{c_1}\partial(e_{c_2})g_{c_1}^{-1}\\
		& \implies\partial(e_{c_2}) =g_{c_1}hg_{c_1}^{-1}h^{-1} := [g_{c_1},h].
	\end{align*}
	That is, in order to ensure that the plaquette is not excited, the labels of the measurement operator must satisfy
	\begin{equation}
		\partial(e_{c_2})=[g_{c_1},h].
		\label{Equation_c2_seam_condition_tri_nontrivial}
	\end{equation}
	
	As we noted for the $\rhd$ trivial case in the previous section, this is not just a condition for the label of the blob ribbon operator. If the commutator of $g_{c_1}$ and $h$ is outside of the image of $\partial$, there is no label $e_{c_2}$ that can satisfy the condition in Equation \ref{Equation_c2_seam_condition_tri_nontrivial}, and so the labels $h$ and $g_{c_1}$ do not give us a valid measurement operator.

	Just as we considered plaquettes near $c_2$, we must consider boundary plaquettes near $c_1$. An example of such a plaquette is shown in Figure \ref{plaquette_near_c_1_tri_trivial} in Section \ref{Section_3D_Topological_Charge_Torus_Tri_trivial}. The plaquette holonomy for this plaquette is initially given by
	\begin{equation}
		H_1(p)=1_G=\partial(e_p)g(b)^{-1}g_2g(u)g_1^{-1}, \label{Equation_toroidal_charge_plaquette_near_c1}
	\end{equation}
	where $g_1$ is the initial label of edge $i_1$ and $g_2$ is the initial label of edge $i_2$. Under the action of the magnetic membrane operator, the edge labels $g_1$ and $g_2$ transform as
	$$g_x \rightarrow g(t_x)^{-1}hg(t_x)g_x,$$
	for $x=1$ and $x=2$. Just as in the $c_2$ case, we may deform the paths $t_1$ and $t_2$ into more convenient ones, without affecting how the edge labels transform. We deform the paths $t_1$ and $t_2$ into the green and purple paths in Figure \ref{plaquette_near_c_1_tri_trivial}, $ab$ and $c_2a$ respectively. Then the edge labels $g_1$ and $g_2$ transform according to
	\begin{align*}
		g_1 &\rightarrow (g(a)g(b))^{-1}h(g(a)g(b))g_1\\
		g_2 &\rightarrow (g_{c_2}g(a))^{-1}h(g_{c_2}g(a))g_2.
	\end{align*}
	
	Inserting these expressions into Equation \ref{Equation_toroidal_charge_plaquette_near_c1}, which gives the plaquette holonomy of plaquette $p$, we get
	\begin{align}
		H_1(p)&= \partial(e_p) g(b)^{-1}g_2 g(u)g_1^{-1} =1_G \notag \\
		&\rightarrow \partial(e_p) g(b)^{-1} [(g_{c_2}g(a))^{-1}h(g_{c_2}g(a))g_2] g(u) [(g(a)g(b))^{-1}h(g(a)g(b))g_1]^{-1} \notag \\
		&=g(b)^{-1} g(a)^{-1}g_{c_2}^{-1}hg_{c_2}g(a) g(b) \big[\partial(e_p)g(b)^{-1}g_2 g(u)g_1^{-1}\big]g(b)^{-1}g(a)^{-1}h^{-1}g(a)g(b) \notag \\
		&=g(b)^{-1} g(a)^{-1}g_{c_2}^{-1}hg_{c_2}g(a) g(b) \big[1_G\big]g(b)^{-1}g(a)^{-1}h^{-1}g(a)g(b) \notag \\
		&=g(b)^{-1} g(a)^{-1}g_{c_2}^{-1}hg_{c_2}h^{-1}g(a)g(b). \label{Equation_plaquette_holonomy_boundary_torus_2}
	\end{align}
	
	Again we see that the plaquette does not necessarily satisfy fake-flatness in the boundary region, unlike the plaquettes away from the boundary. This means that if we wish to apply confined blob ribbon operators around cycles equivalent to $c_1$, we must apply them on this boundary region, to avoid them exciting other plaquettes. Furthermore, the label of this blob ribbon operator must be chosen to force this boundary region to satisfy fake-flatness. The effect of the blob ribbon operator $B^{e_{c_1}}(c_1)$ wrapping $c_1$ on the plaquette from earlier is $B^{e_{c_1}}:e_p=e_p \: [g(t_1)^{-1} \rhd e_{c_1}^{-1}]$. Including this action with the effect of the magnetic membrane operator, the total effect on the plaquette holonomy is
	\begin{align*}
		H_1(p) &\rightarrow \partial(g(t_1)^{-1} \rhd e_{c_1})^{-1}g(b)^{-1} g(a)^{-1}g_{c_2}^{-1}hg_{c_2}h^{-1}g(a)g(b)\\ &=(g(a)g(b))^{-1}\partial(e_{c_1})^{-1}g_{c_2}^{-1}hg_{c_2}h^{-1}g(a)g(b).
	\end{align*}
	
	In order to make this be the identity element, so that the plaquette condition is satisfied, we require the label $e_{c_1}$ of the blob ribbon operator to satisfy the condition
	\begin{align*}
		\partial&(e_{c_1})^{-1}g_{c_2}^{-1}hg_{c_2}h^{-1}=1_G \\
		&\implies \partial(e_{c_1})=g_{c_2}^{-1}hg_{c_2}h^{-1}.
	\end{align*}
	Conjugating both sides of this equation by $g_{c_2}$ and using the fact that $\partial(E)$ is in the centre of $G$ gives us the condition
	\begin{align}
		g_{c_2}&\partial(e_{c_1})g_{c_2}^{-1}=g_{c_2} [g_{c_2}^{-1}hg_{c_2}h^{-1}]g_{c_2}^{-1} \notag \\
		&\implies \partial(e)=hg_{c_2}h^{-1}g_{c_2}^{-1} \notag\\
		&\implies \partial(e_{c_1})=[h,g_{c_2}]. \label{Equation_c1_seam_condition_tri_nontrivial}
	\end{align}
	
	So far, these conditions have matched the ones we derived in Section \ref{Section_3D_Topological_Charge_Torus_Tri_trivial} for the $\rhd$ trivial case. However, the magnetic membrane operator can produce additional excitations when $\rhd$ is non-trivial, and we must ensure that these excitations are prevented. Unlike in the $\rhd$ trivial case, the magnetic membrane operator can produce a blob excitation at the privileged blob 0 defined when we set up the magnetic membrane operator. We choose to put this blob at the start (and end) of the blob ribbon operators that we apply on each cycle, which we will shortly see allows us to prevent the excitation of blob 0. As we discuss later, this choice for the position does not result in a loss of generality, so we will not miss any distinct measurement operators.

	In order to prevent the excitation of blob 0, its 2-holonomy must be preserved (as $1_E$) by the action of the various membrane and ribbon operators which we apply. There are several contributions to the blob 2-holonomy of blob 0. First consider the effect of the magnetic operator $C^h_T(m)$ itself on $H_2(\text{blob }0)$. Using our results from Section \ref{Section_Magnetic_Tri_Non_Trivial} (see Equation \ref{Equation_magnetic_blob_0}), we see that 
	$$C^h_T(m): H_2(\text{blob }0)=[h \rhd H_2(\text{blob }0)] [h \rhd e_m^{-1}] e_m\\
	=[h \rhd e_m^{-1}] e_m.$$
	
	Now we consider the effect of the blob ribbon operators on the blob 2-holonomy. While each of the blob ribbons both starts in and terminates in blob 0, which normally results in the 2-holonomy of the blob being preserved, the ribbon wraps a non-contractible cycle in between starting in the blob and terminating in it. Because of this, and the fact that the action of the blob ribbon operator on a plaquette depends on the path that the ribbon operator travels up to that plaquette, the actions from exiting and entering the blob may not cancel. The blob 2-holonomy is given by
	$$H_2(\text{blob }0)=\prod_{\text{plaquettes }p \in \text{Bd(blob $0$)}} g(s.p-v_{0}(p))\rhd e_p^{\sigma_p},$$
	where $\sigma_p$ is $+1$ if the plaquette is oriented outwards on the blob (using the right-hand rule) and is $-1$ if the plaquette points inwards. Here the paths $(s.p-v_{0}(p))$ run from the start-point of the membrane, which is also the base-point of blob 0, to the base-point $v_0(p)$ of plaquette $p$, and are entirely local to blob 0. As we will see shortly, the reason that the blob ribbon operators may excite the blob is that the ribbon operators wrap around a non-contractible cycle, which means that the paths appearing in the blob ribbon operator may not match those in the 2-holonomy of blob 0.

	\begin{figure}[h]
		\begin{center}
			\begin{overpic}[width=0.75\linewidth]{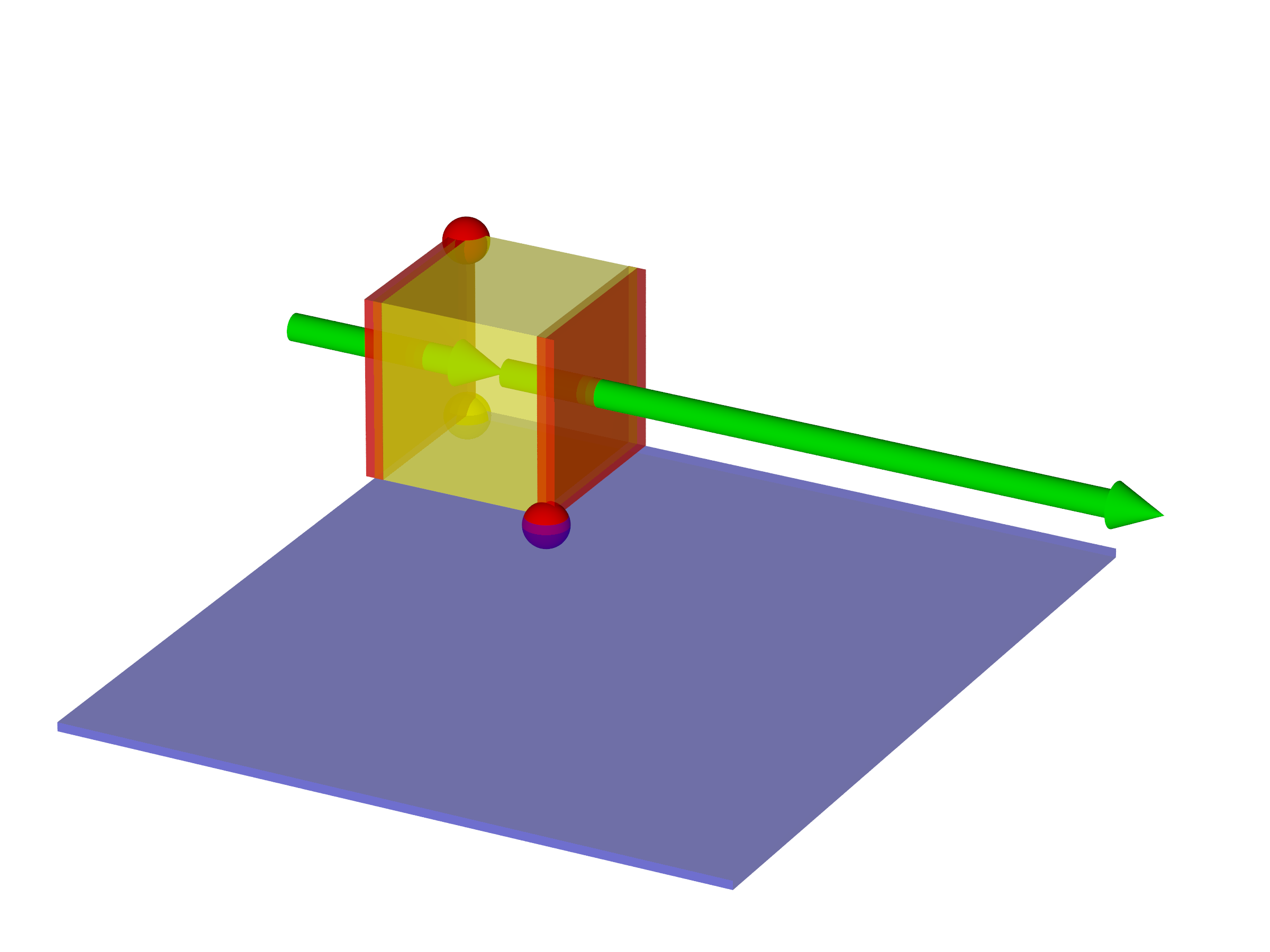}
				\put(52,46){\large $e_1$}
				\put(27,52){\large $e_2$}
				\put(60,45){\large $B^{e_{c_1}}(c_1)$ wraps $c_1$}
				\put(32,59){$v_0(2)$}
				\put(45,32){$v_0(1)$}
			\end{overpic}
			\caption{The blob ribbon operator $B^{e_{c_1}(c_1)}$, which wraps the cycle $c_1$, may excite blob 0, which is at the start of the cycle. This is because the blob ribbon operator passes along a non-contractible cycle before returning to the blob. The ribbon exits blob 0 (yellow cube) through the plaquette labelled by $e_1$ and returns through the plaquette labelled by $e_2$ after passing along most of the cycle. The red spheres $v_0(1)$ and $v_0(2)$ represent possible positions for the base-points of the two plaquettes, while the (yellow) sphere in the back-left corner of the membrane represents the start-point of the membrane. }
			\label{torus_charge_blob_wrapping_c1}
		\end{center}
	\end{figure}
	
	First consider the blob ribbon operator wrapping the cycle $c_1$, as shown in Figure \ref{torus_charge_blob_wrapping_c1}. The blob ribbon operator leaves blob 0 through a plaquette, which we call plaquette 1, with initial label $e_1$. We will assume that this plaquette points outwards from the blob, and so points along the ribbon (although this does not affect our result). From Equation \ref{Equation_blob_ribbon_central}, the effect of the blob ribbon operator on this plaquette label is
	$$B^{e_{c_1}}(c_1):e_1=[g({s.p-v_{0}(1)})^{-1} \rhd e_{c_1}^{-1}]e_1,$$
	where $g(s.p-v_0(1))$ is the path from the start-point to the base-point of the plaquette, $v_0(1)$, and is entirely local to blob 0. It is therefore equivalent (up to deformations, and so up to irrelevant factors in $\partial(E)$) to the path label that appears in the contribution of plaquette 1 to the blob 2-holonomy. Note that if we had taken the plaquette to point inwards on the blob, the plaquette label would have gained the factor $[g({s.p-v_{0}(1)})^{-1} \rhd e_{c_1}]$ instead of the inverse, but the contribution of the plaquette to the blob 2-holonomy would also be inverted, and these two effects would cancel.

	Now consider the plaquette through which the ribbon operator enters blob 0 (after wrapping the cycle), which we denote by plaquette 2, and which is initially labelled by $e_2$. We again take the plaquette to point outwards from the blob, which now means it points against the ribbon (because the ribbon enters the blob at plaquette 2, the ribbon points inwards here). Then the effect of the ribbon operator on plaquette 2 is
	$$B^{e_{c_1}}(c_1):e_2=[g({s.p-v_{0}(2) \text{(via $c_1$)}})^{-1} \rhd e_{c_1}] e_2,$$
	where the path $(s.p-v_0(2) \text{(via $c_1$)})$ nearly wraps the cycle $c_1$ and so is not local to the blob. Indeed the path $(s.p-v_0(2) \text{(via $c_1$)})$ differs from the path label $(s.p-v_0(2))$ that appears in the blob 2-holonomy (apart from by deformation) by the cycle $c_1$, so that the two paths cannot be deformed into one another. Instead 
	$$g(s.p-v_0(2) \text{(via $c_1$)})=g_{c_1} g(s.p-v_0(2) \text{(not via $c_1$)})$$ 
	up to factors in $\partial(E)$. This means that the total effect of the blob ribbon operators on the blob 2-holonomy is
	\begin{align*}
		H_2(\text{blob }0)&=\prod_{\text{plaquettes }p \in \text{Bd(blob $0$)}} g(s.p-v_{0}(p))\rhd e_p^{\sigma_p}\\
		&\rightarrow [g(s.p-v_{0}(1))\rhd (g(s.p-v_0(1))^{-1}\rhd e_{c_1}^{-1})] [g(s.p-v_0(2) (\text{not via $c_1$})) \rhd (g(s.p-v_0(2) \text{(via $c_1$)})^{-1}\rhd e_{c_1})]\\ & \hspace{1cm} H_2(\text{blob }0)\\
		&=e_{c_1}^{-1}\big[g({s.p-v_0(2) (\text{not via $c_1$})}) \rhd \big( (g({s.p-v_0(2) (\text{not via $c_1$})})^{-1} g_{c_1}^{-1})\rhd e_{c_1}\big)\big] H_2(\text{blob }0)\\
		&=e_{c_1}^{-1} [g_{c_1}^{-1}\rhd e_{c_1}] H_2(\text{blob }0).
	\end{align*}	
	
	The same is true (with appropriate relabelling of variables) for the blob ribbon operator wrapping the other cycle, $c_2$, so that the total effect of the magnetic membrane operator and blob ribbon operators on the blob 2-holonomy of blob 0 (which is originally $1_E$) is 
	\begin{align*}
		H_2(\text{blob }0) & \rightarrow [h \rhd e_m^{-1}] e_m e_{c_1}^{-1}[g_{c_1}^{-1}\rhd e_{c_1}] e_{c_2}^{-1} [g_{c_2}^{-1} \rhd e_{c_2}].
	\end{align*}
	Requiring that this expression be equal to $1_E$, in order to ensure that blob 0 is not excited, gives us the condition that
	\begin{equation}
		[h \rhd e_m^{-1}] e_m e_{c_1}^{-1} [g_{c_1}^{-1} \rhd e_{c_1}] e_{c_2}^{-1}[g_{c_2}^{-1} \rhd e_{c_2}]=1_E.
		\label{Equation_torus_blob_0_condition}
	\end{equation}
	
	At this point we should discuss our choice to place blob 0 at the point of origin of the two blob ribbon operators. It may seem that we have thrown out potential measurement operators by making this restriction. However, just as we discussed with the start-point in Section \ref{Section_3D_Topological_Charge_Torus_Tri_trivial}, choosing the special blobs of the operators to be in different locations does not lead to different operators. If we had chosen blob 0 of the magnetic membrane operator to lie in a different position from the origin of the ribbon operators, then the effect of the blob ribbon operators on their origin blob would have to cancel, and the effect of the magnetic membrane operator on blob 0 would separately have to leave the 2-holonomy invariant. This would give us two conditions. Furthermore, this would not give us a separate situation from the one where the origin blob for the ribbons is the same as blob 0 of the magnetic membrane operator. This is because blob 0 being unexcited by the magnetic membrane operator enables us to move blob 0 without affecting the action of the magnetic membrane operator (see Section \ref{Section_magnetic_change_blob_0} ), and so we could move blob 0 to the origin of the two ribbon operators, so that they coincide after all, without affecting the action of the measurement operator. Therefore, although it may seem counter-intuitive, the situation where the origin of the ribbon operators is different from blob 0 is in fact a subset of the cases where the origin is the same. Similarly, we could choose the origin of the two blob ribbon operators to be different, giving more independent conditions on the labels of these blob ribbon operators, but this would only give us a subset of the cases where they coincide. This is because we can move the origin blob of any blob ribbon operator by attaching another segment of blob ribbon operator that connects the original origin to the new origin to extend the ribbon (and similarly with the end of the ribbon). We considered such a concatenation of ribbon operators in Section \ref{Section_blob_ribbon_concatenate}. If the origin blob is not excited, then the label of the connecting segment required to move the origin blob is the identity element and so the additional segment has no effect on the state. This means that we could freely move the origin blob of the closed ribbon operators if they independently left their origin blobs unexcited, and so could move them to coincide.

	We also claimed that, just as in the $\rhd$ trivial case from Section \ref{Section_3D_Topological_Charge_Torus_Tri_trivial}, choosing different paths from the start-point of the magnetic membrane operator to the edges cut by the dual membrane does not lead to different measurement operators, even if those paths cannot be smoothly deformed into the original ones. We will now show that this is indeed the case, following the same argument as in the $\rhd$ trivial case. Consider an edge $i$ that is cut by the dual membrane of the magnetic membrane operator. This is associated with a path $t_i$ that starts at the start-point of the membrane and ends at the vertex that is attached to the edge and lies on the direct membrane of the membrane operator. We initially chose this path so that it does not cross the seams of the torus. Now we consider replacing this path with a different path, $t_i'$, which may cross the seams of the torus any number of times. An example is shown in Figure \ref{torus_reconnect_edge_1}.

	If the edge label is initially $g_i$, then the original action of the magnetic membrane operator on the edge is
	$$C^h_{\rhd}(m):g_i = g(t_i)^{-1}hg(t_i)g_i,$$
	assuming that the edge $i$ points away from the direct membrane (which we can ensure using the re-branching procedures from the Appendix of Ref. \cite{HuxfordPaper1}). Then if we change the path from $t_i$ to $t_i'$, this action is instead
	$$C^h_{\rhd}(m'):g_i = g(t_i')^{-1}hg(t_i')g_i,$$
	where we use $m'$ to denote the membrane once we have changed the path $t_i$ to $t_i'$. The membrane $m'$ is the same as $m$ in every other respect, and in particular the start-point $s.p(m')$ of the new membrane is the same as $s.p(m)$, the start-point of $m$. Defining the closed path $s= t_i' \cdot t_i^{-1}$, so that $g(s)=g(t_i')g(t_i)^{-1}$, we can write the action of the new membrane operator on edge $i$ as
	\begin{align*}
		C^h_{\rhd}(m'):g_i &= (g(s)g(t_i))^{-1}h(g(s)g(t_i))g_i\\
		&=g(t_i)^{-1}g(s)^{-1}hg(s)g(t_i)g_i.
	\end{align*}
	
	We now wish to consider how this change affects the holonomy of the plaquettes adjacent to edge $i$. We start by considering plaquettes away from the seams of the torus, such as the plaquette $p$ shown in Figure \ref{torus_reconnect_edge_plaquette_off_seam}. In addition to edge $i$, the boundary of this plaquette includes another edge, edge $j$, which is cut by the dual membrane of the magnetic membrane operator. The plaquette holonomy of $p$, $H_1(p)$, is initially given by 
	$$H_1(p)=\partial(e_p) g_j g(u) g_i^{-1}g(b)^{-1}.$$
	
	Under the action of the magnetic membrane operator $C^h_{\rhd}(m')$, the edge label $g_j$ becomes $g(t_j)^{-1}hg(t_j)g_j$, where $t_j$ is the path from the start-point of $m'$ to the edge. Note that we are not considering changing the path $t_j$, only $t_i$, so $t_j$ is the same in $m$ and $m'$. In Figure \ref{torus_reconnect_edge_plaquette_off_seam}, we took the base-point of the plaquette to lie on the direct membrane (though this choice and the choice for the orientation of the plaquette are not important and can be altered by using the re-branching procedures from the Appendix of Ref. \cite{HuxfordPaper1}), so the plaquette label $e_p$ will also be transformed as $e_p \rightarrow g(s.p(m)-v_0(p))^{-1}hg(s.p(m)-v_0(p)) \rhd e_p$. However, this does not change $\partial(e_p)$ (see the discussion at the start of Section \ref{Section_Magnetic_Tri_Non_Trivial}) and so does not affect the plaquette holonomy. Putting this together with the transformation of edge $i$, we see that the plaquette holonomy transforms as
	\begin{align*}
		C^h_{\rhd}(m'):H_1(p)&= \partial(e_p) g(t_j)^{-1}hg(t_j)g_j g(u) g_i^{-1}g(t_i')^{-1}h^{-1}g(t_i')g(b)^{-1}\\
		&= \partial(e_p) g(t_j)^{-1}hg(t_j)g_j g(u) g_i^{-1}g(t_i)^{-1}g(s)^{-1}h^{-1}g(s)g(t_i) g(b)^{-1}.
	\end{align*}
	
	On the other hand, under the original magnetic membrane operator, the plaquette holonomy becomes
	\begin{align}
		C^h_{\rhd}(m):H_1(p)&= \partial(e_p) g(t_j)^{-1}hg(t_j)g_j g(u) g_i^{-1}g(t_i)^{-1}h^{-1}g(t_i)g(b)^{-1}, \label{Equation_torus_charge_change_path_original}
	\end{align}
	which we know to be the identity. We wish to see how the transformation under $C^h_{\rhd}(m')$ differs from this. We can insert the identity in the form
	$$1_G= \big[g(t_i)^{-1}h^{-1}g(t_i) g(b)^{-1} g(b) g(t_i)^{-1}hg(t_i)\big],$$
	to write that transformation as
	\begin{align*}
		C^h_{\rhd}(m')&:H_1(p)= \partial(e_p) g(t_j)^{-1}hg(t_j)g_j g(u)g_i^{-1}g(t_i)^{-1}g(s)^{-1}h^{-1}g(s)g(t_i) g(b)^{-1}\\
		&= \partial(e_p) g(t_j)^{-1}hg(t_j)g_j g(u)g_i^{-1} \big[g(t_i)^{-1}h^{-1}g(t_i) g(b)^{-1} g(b) g(t_i)^{-1}hg(t_i)\big] g(t_i)^{-1} g(s)^{-1}h^{-1}g(s)g(t_i)g(b)^{-1}\\
		&= \big[ \partial(e_p) g(t_j)^{-1}hg(t_j)g_j g(u)g_i^{-1} g(t_i)^{-1}h^{-1}g(t_i) g(b)^{-1} \big] g(b) g(t_i)^{-1}hg(t_i) g(t_i^{-1}) g(s)^{-1}h^{-1}g(s)g(t_i)g(b)^{-1}.
	\end{align*}
	
	We can recognise the term
	$$ \big[ \partial(e_p) g(t_j)^{-1}hg(t_j)g_j g(u)g_i^{-1} g(t_i)^{-1}h^{-1}g(t_i) g(b)^{-1} \big]$$
	as the result of acting on the plaquette holonomy with the original membrane operator, so this term must be the identity element. Then we have
	\begin{align*}
		C^h_{\rhd}(m'):H_1(p)&= [C^h_{\rhd}(m):H_1(p)] g(b) g(t_i)^{-1}hg(s)^{-1}h^{-1}g(s)g(t_i)g(b)^{-1}\\
		&=[1_G] g(b) g(t_i)^{-1} [hg(s)^{-1}h^{-1}g(s)] g(t_i)g(b)^{-1}\\
		&= g(b) g(t_i)^{-1} [hg(s)^{-1}h^{-1}g(s)] g(t_i)g(b)^{-1}.
	\end{align*}
	
	This indicates that if $h$ and $g(s)$ do not commute, the plaquette is excited by the action of the new magnetic membrane operator. We now consider the path $s$ in more detail. It is a closed path lying on the torus that starts and ends at the start-point of the membrane. It must therefore be deformable into some product of the cycles of the torus, $c_1$ and $c_2$, where each cycle may appear any number of times in the product. Because the surface of the torus satisfies fake-flatness, this deformation only introduces factors in $\partial(E)$ (which are in the centre of $G$ and so do not affect commutators) to the path element. This in turn means that, up to factors in $\partial(E)$, the group element $g(s)$ must be a product of the group elements $g_{c_1}$ and $g_{c_2}$ that correspond to those cycles. We showed earlier that, in order for our measurement operator on the membrane $m$ to be valid, $g_{c_1}$ and $g_{c_2}$ must individually commute with $h$ up to elements in $\partial(E)$. The same is therefore true of $g(s)$. This means that $[hg(s)^{-1}h^{-1}g(s)]= \partial(x)$ for some $x \in E$. This implies that, just as in the $\rhd$ trivial case, we can correct the plaquette holonomy by applying an additional blob ribbon operator that passes through the plaquette. We therefore consider acting with a blob ribbon operator $B^x(r)$ on a ribbon that passes through the plaquette and whose start-point is the start-point of $m$, $s.p(m)$ (which is the same as $s.p(m')$). This acts on the plaquette as $B^{x}(r):e_p= e_p [g(s.p(m)-v_0(p))^{-1} \rhd x^{-1}]$. We also ensure that the direct path of the blob ribbon operator passes along $t_i$ before travelling to $v_0(p)$ on a path local to the plaquette, so that $g(s.p(m)-v_0(p))=g(t_i)g(v_i-v_0(p))$. We do this because each plaquette adjacent to $i$ will require a similar blob ribbon operator, and we wish to combine them into a single closed ribbon operator, which requires that the direct paths be compatible (see Section \ref{Section_blob_ribbon_concatenate}). The blob ribbon operator then introduces an additional factor of $\partial(g(s.p(m)-v_0(p))^{-1} \rhd x^{-1})$ to the plaquette holonomy (note that this factor is in the centre of $G$, so we can pull it to the front of the plaquette holonomy). Therefore, the plaquette holonomy becomes
	\begin{align*}
		B^x(r)&C^h_{\rhd}(m'):H_1(p)=\partial( g(s.p(m)-v_0(p))^{-1} \rhd x^{-1}) g(b) g(t_i)^{-1} [hg(s)^{-1}h^{-1}g(s)] g(t_i)g(b)^{-1},
	\end{align*}
	under the combined action of the blob ribbon operator and magnetic membrane operator. Then, substituting $[hg(s)^{-1}h^{-1}g(s)]=\partial(x)$ and using the fact that $\partial(x^{-1})= \partial( g(s.p(m)-v_0(p))^{-1} \rhd x^{-1})$ is in the centre of $G$, we have
	\begin{align*}
		B^x(r)C^h_{\rhd}(m'):H_1(p)&= \partial(x^{-1}) g(b) g(t_i)^{-1} \partial(x)g(t_i)g(b)^{-1}\\
		&= \partial(x)^{-1} \partial(x)\\
		&=1_G.
	\end{align*}
	
\begin{figure}[h]
		\begin{center}
			\begin{overpic}[width=0.75\linewidth]{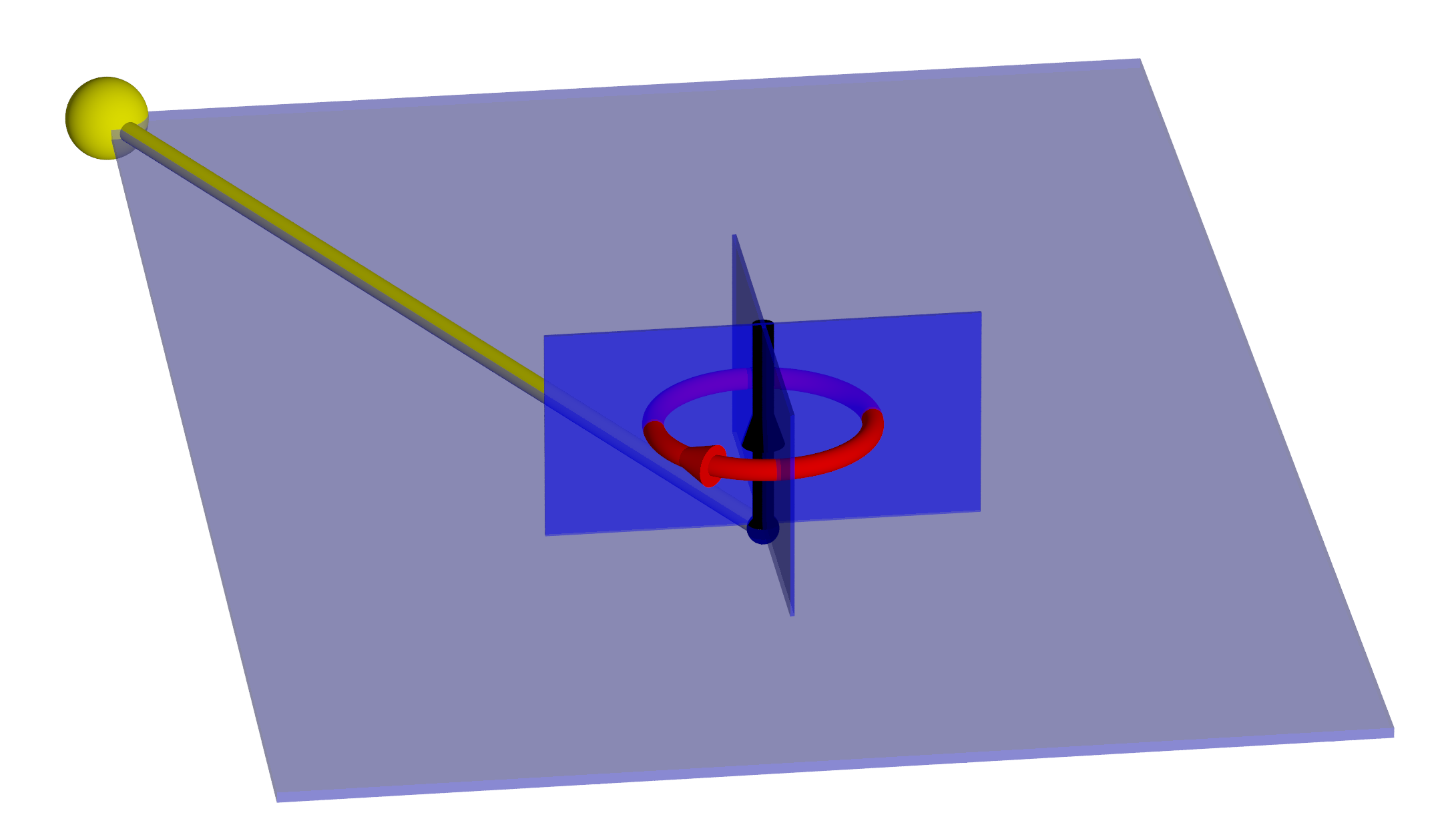}
				\put(5,53){$s.p(m)$}
				\put(54,35){edge $i$}
				\put(62,28){ribbon $r$}
			\end{overpic}
			\caption{When we change the path to edge $i$ that appears in the magnetic membrane operator, the magnetic membrane operator may excite the plaquettes surrounding the edge $i$. In order to prevent this, so that we may construct a valid measurement operator, we must introduce a blob ribbon operator whose dual path (red) encircles the edge $i$ and whose start-point is the start-point of the membrane (yellow sphere). The yellow path represents the first section of the direct path of the ribbon, which runs form the start-point of the membrane to the edge (depending on the base-points of the plaquettes around $i$, there may be additional sections running between the base-points of the plaquettes which are not shown).}
			\label{torus_reconnect_edge_tri_non_trivial_closed_ribbon}
		\end{center}
	\end{figure}

	Therefore, just as in the $\rhd$ trivial case, we can correct the plaquette holonomy by applying a blob ribbon operator through the plaquette, as shown in Figure \ref{torus_reconnect_edge_plaquette_ribbon}. We considered a plaquette away from the seam, but we can similarly consider a plaquette on the seam, which is pierced by an additional blob ribbon operator along the seam. In that case, the plaquette label $e_p$ is transformed to some other label $e_p'$ due to the action of the blob ribbon operator, but that blob ribbon operator was chosen to ensure that the plaquette holonomy is preserved under the action of the original magnetic membrane operator. This just means that, instead of
	\begin{align*}
		C^h_{\rhd}(m):H_1(p)&= \partial(e_p) g(t_j)^{-1}hg(t_j)g_j g(u) g_i^{-1}g(t_i)^{-1}h^{-1}g(t_i)g(b)^{-1}=1_G,
	\end{align*}
	we have
	\begin{align*}
		B^{e_{c_x}}(c_x)C^h_{\rhd}(m):H_1(p)&= \partial(e_p') g(t_j)^{-1}hg(t_j)g_j g(u) g_i^{-1}g(t_i)^{-1}h^{-1}g(t_i)g(b)^{-1}=1_G,
	\end{align*}
	for one of the cycles $c_x$, where $x=1$ or $2$. Then the action of the modified magnetic membrane operator, combined with the blob ribbon operator, gives us
	\begin{align*}
		B^{e_{c_x}}&(c_x)C^h_{\rhd}(m'):H_1(p)= \partial(e_p') g(t_j)^{-1}hg(t_j)g_j g(u)g_i^{-1}g(t_i)^{-1}g(s)^{-1}h^{-1}g(s)g(t_i) g(b)^{-1}\\
		&=[ \partial(e_p') g(t_j)^{-1}hg(t_j)g_j g(u)g_i^{-1} g(t_i)^{-1}h^{-1}g(t_i) g(b)^{-1} ] g(b) g(t_i)^{-1}hg(t_i) g(t_i^{-1}) g(s)^{-1}h^{-1}g(s)g(t_i)g(b)^{-1}\\
		&= [B^{e_{c_x}}(c_x)C^h_{\rhd}(m):H_1(p)] g(b) g(t_i)^{-1}hg(s)^{-1}h^{-1}g(s)g(t_i)g(b)^{-1}\\
		&=[1_G] g(b) g(t_i)^{-1} [hg(s)^{-1}h^{-1}g(s)] g(t_i)g(b)^{-1}\\
		&= g(b) g(t_i)^{-1} [hg(s)^{-1}h^{-1}g(s)] g(t_i)g(b)^{-1},
	\end{align*}
	and the argument otherwise proceeds exactly as in the case of a plaquette away from the seam. Therefore, all of the plaquettes adjacent to the edge can be made to satisfy fake-flatness by the addition of a blob ribbon operator of label $x$, where $\partial(x)=hg(s)^{-1}h^{-1}g(s)$, passing through the plaquette. This label and start-point of the ribbon operator is the same for each plaquette (and the orientations of the ribbons also agree, as we describe in the $\rhd$ trivial case in Section \ref{Section_3D_Topological_Charge_Torus_Tri_trivial}). We could have chosen different labels $x'$ satisfying $\partial(x)=\partial(x')$ for the ribbon operator associated to each plaquette, if all we required was to correct the plaquette holonomy of each plaquette, but this it would cause excitations on the blobs around the edge, so we had to have all of these labels be the same (though they could all be the same $x'$ for any $x' \in E$ satisfying $\partial(x)=\partial(x')$). This means that we can combine these ribbon operators into a single ribbon operator encircling the edge $i$, as shown in Figure \ref{torus_reconnect_edge_tri_non_trivial_closed_ribbon}. In order for our measurement operator on $m'$ to be valid, we must include this additional closed blob ribbon operator so that no plaquettes are excited by the action of the new measurement operator.

	Just as in the $\rhd$ trivial case, we wish to show that the difference in the action of the two measurement operators, the one on $m$ and the one on $m'$, is just equivalent to the action of an edge transform and so is trivial (because edge transforms act trivially on states where the edge is unexcited, and we demand that the measurement operator acts on an unexcited torus). The original operator acts on the edge $i$ as
	$$g_i \rightarrow g(t_i)^{-1}hg(t_i)g_i,$$
	while the action of the new operator on the edge is
	\begin{align*}
		g_i \rightarrow& g(t_i)^{-1}g(s)^{-1}hg(s)g(t_i)g_i.
	\end{align*}
	
	We can insert the identity in the form
	$$(g(t_i)^{-1}h^{-1}g(t_i)g(t_i)^{-1}hg(t_i))=1_G,$$
	to obtain
	\begin{align*}
		g_i \rightarrow& g(t_i)^{-1}g(s)^{-1}hg(s)g(t_i) (g(t_i)^{-1}h^{-1}g(t_i)g(t_i)^{-1}hg(t_i))g_i\\
		=& [ g(t_i)^{-1}g(s)^{-1}hg(s)g(t_i)g(t_i)^{-1}h^{-1}g(t_i) ] g(t_i)^{-1}hg(t_i)g_i\\
		=& [ g(t_i)^{-1}g(s)^{-1}hg(s)h^{-1}g(t_i) ] g(t_i)^{-1}hg(t_i)g_i.
	\end{align*}
	
	The term $g(s)^{-1}hg(s)h^{-1}$ is equal to $\partial(x)^{-1}$, where $\partial(x)= hg(s)^{-1}h^{-1}g(s)$. Therefore,
	\begin{align*}	
		g_i \rightarrow& g(t_i)^{-1}\partial(x)^{-1} g(t_i) (g(t_i)^{-1}hg(t_i)g_i)\\
		=& \partial(g(t_i)^{-1} \rhd x^{-1}) (g(t_i)^{-1}hg(t_i)g_i),
	\end{align*}
	where in the last step we used the first Peiffer condition, Equation \ref{Equation_Peiffer_1} in the main text. This is the same as the action of the old measurement operator, except with an additional factor of $\partial(g(t_i)^{-1} \rhd x^{-1})$. This is exactly the factor we would expect from an edge transform of label $g(t_i)^{-1} \rhd x^{-1}$ on the edge. Next we consider the plaquettes adjacent to the edge $i$, which are the other degrees of freedom affected differently by the two operators. These plaquettes gain an additional factor of 
	$$g(s.p(m)-v_0(p))^{-1} \rhd x^{-1}$$
	from the action of the closed blob ribbon operator that we must apply around edge $i$ (assuming that their orientation matches that of the blob ribbon operator, meaning that they circulate against edge $i$, otherwise we would replace $x$ with $x^{-1}$). We can then split the path element $g(s.p(m)-v_0(p))$ as $g(s.p(m)-v_0(p))=g(s.p(m)-v_i)g(v_i-v_0(p))$, where $g(s.p(m)-v_i)=g(t_i)$ and we do not include any factors in $\partial(E)$ that may appear from deformation of the paths. Then 
	\begin{align*}
		g(s.p(m)-v_0(p))^{-1} \rhd x^{-1}&= (g(s.p(m)-v_i)g(v_i-v_0(p)))^{-1} \rhd x^{-1}\\
		&= (g(t_i)g(v_i-v_0(p)))^{-1} \rhd x^{-1}\\
		&=(g(v_i-v_0(p))^{-1} g(t_i)^{-1}) \rhd x^{-1}\\
		&= g(v_0(p)-v_i) \rhd (g(t_i)^{-1} \rhd x^{-1}).
	\end{align*}
	
	However, this is just the factor that we would expect from an edge transform of label $g(t_i)^{-1} \rhd x^{-1}$ on the edge $i$ (see Equation \ref{Equation_edge_transform_definition} in the main text), where we note that plaquettes whose orientation matches that of the ribbon operator around $i$ anti-align with edge $i$. Therefore, the difference between the two measurement operators is just that of an edge transform, which acts trivially in the subspace where the membrane is initially unexcited, and so the measurement operators are not actually distinct.

	In addition to the paths from the start-point of the membrane to the various edges cut by the dual membrane, when $\rhd$ is non-trivial the action of the membrane operator depends on the paths from the start-point to the base-points of the plaquettes that are cut by the dual membrane and whose base-points lie on the direct membrane. Recall that the magnetic membrane operator $C^h_{\rhd}(m)$ acts on such a plaquette as $C^h_{\rhd}(m):e_p= \big(g(s.p(m)-v_0(p))^{-1}hg(s.p(m)-v_0(p))\big) \rhd e_p$, where $s.p(m)-v_0(p)$ is a path from the start-point of the membrane to the base-point of the plaquette. As discussed in Section \ref{Section_Magnetic_Tri_Non_Trivial}, we know that deforming this path has no effect on the action of the membrane operator. However, we now wish to show that changing this path by concatenating it with a non-contractible path leaves the action of the measurement operator on the torus invariant. We consider changing this path into the path $s \cdot (s.p(m)-v_0(p))$, where $s$ is a closed path starting and ending at $s.p(m)$. Then the new action of the operator is
	\begin{align}
		C^h_{\rhd}(m'):e_p&= \big(g(s.p(m)-v_0(p))^{-1}g(s)^{-1}hg(s)g(s.p(m)-v_0(p))\big) \rhd e_p. \label{Equation_torus_rhd_plaquettte_change_path}
	\end{align}
	
	We consider inserting the identity, in the form of
	$$1_G= (g(s.p(m)-v_0(p))^{-1}h^{-1}g(s.p(m)-v_0(p)) g(s.p(m)-v_0(p))^{-1}hg(s.p(m)-v_0(p))),$$
	into the expression $$\big(g(s.p(m)-v_0(p))^{-1}g(s)^{-1}hg(s)g(s.p(m)-v_0(p))\big).$$ Then we have
	\begin{align*}
		\big(g(s.p(m)&-v_0(p))^{-1}g(s)^{-1}hg(s)g(s.p(m)-v_0(p))\big)\\
		&= \big(g(s.p(m)-v_0(p))^{-1}g(s)^{-1}hg(s)g(s.p(m)-v_0(p))\big)\\
		& \hspace{1cm} \big(g(s.p(m)-v_0(p))^{-1}h^{-1}g(s.p(m)-v_0(p)) g(s.p(m)-v_0(p))^{-1}hg(s.p(m)-v_0(p))\big)\\
		&= \big(g(s.p(m)-v_0(p))^{-1}g(s)^{-1}hg(s)h^{-1}g(s.p(m)-v_0(p))\big) \big(g(s.p(m)-v_0(p))^{-1}hg(s.p(m)-v_0(p))\big).
	\end{align*}
	Inserting this into Equation \ref{Equation_torus_rhd_plaquettte_change_path} gives us
	\begin{align*}
		C^h_{\rhd}&(m'):e_p= \big[\big(g(s.p(m)-v_0(p))^{-1}g(s)^{-1}hg(s)h^{-1}g(s.p(m)-v_0(p))\big) \big(g(s.p(m)-v_0(p))^{-1}hg(s.p(m)-v_0(p))\big)\big] \rhd e_p\\
		&=\big( g(s.p(m)-v_0(p))^{-1}g(s)^{-1}hg(s)h^{-1}g(s.p(m)-v_0(p))\big) \rhd \big[(g(s.p(m)-v_0(p))^{-1}hg(s.p(m)-v_0(p))) \rhd e_p \big],
	\end{align*}
	which has an additional $\big( g(s.p(m)-v_0(p))^{-1}g(s)^{-1}hg(s)h^{-1}g(s.p(m)-v_0(p))\big) \rhd$ action compared to the action of the original operator $C^h_{\rhd}(m)$. Just as we discussed when we considered changing the paths to the edges, the closed path $s$ must be some combination of the cycles of the torus, and the group elements associated to these cycles commute with $h$ up to elements in $\partial(E)$. Therefore, $g(s)$ also commutes with $h$ up to an element in $\partial(E)$. This means that $g(s.p(m)-v_0(p))^{-1}g(s)^{-1}hg(s)h^{-1}g(s.p(m)-v_0(p)$ is an element of $\partial(E)$, which we denote by $\partial(y)$. Such a factor acts trivially via $\rhd$ when $E$ is Abelian, because the second Peiffer condition (Equation \ref{Equation_Peiffer_2} in the main text) tells us that $\partial(y) \rhd e =yey^{-1}$ for all $y$ and $e$ in $E$, and then the restriction that $E$ is Abelian ensures that this is trivial. Therefore, we have
	\begin{align*}
		C^h_{\rhd}(m'):e_p&= \partial(y) \rhd [(g(s.p(m)-v_0(p))^{-1}hg(s.p(m)-v_0(p))) \rhd e_p ]\\
		&=y [(g(s.p(m)-v_0(p))^{-1}hg(s.p(m)-v_0(p))) \rhd e_p]y^{-1}\\
		&=(g(s.p(m)-v_0(p))^{-1}hg(s.p(m)-v_0(p))) \rhd e_p\\
		&=C^h_{\rhd}(m):e_p.
	\end{align*}
	That is, changing the path from the start-point of the membrane to the base-point of the plaquette has no effect on the action of the magnetic membrane operator and so no effect on the measurement operator. This means that we cannot generate distinct measurement operators by changing the paths that appear in the magnetic membrane operator.

	The magnetic membrane operator also includes blob ribbon operators, one for each plaquette $p$ on the direct membrane, which we have chosen not to cross the seams. The blob ribbon operator associated to a plaquette $p$ passes from blob 0 of the membrane to the blob attached to plaquette $p$ which is cut through by the dual membrane. From Equation \ref{Equation_magnetic_blob_ribbon_label}, the label of this blob ribbon operator is 
	$$f(p)=[g(s.p-v_{0}(p))\rhd e_p^{\sigma_p}] \: [(h^{-1}g(s.p-v_0(p)))\rhd e_p^{-\sigma_p}],$$
	which depends on a path $s.p-v_{0}(p)$. However, this path should be chosen to match the section of direct path of the ribbon operator, up to deformation, in order to ensure that the 2-holonomy of the blob attached to plaquette $p$ is satisfied (we used this implicitly in Section \ref{Section_Magnetic_Tri_Nontrivial_Commutation}). Now assume that both this path and the ribbon itself are changed to cross the seam. For example, suppose that we replace $(s.p-v_0(p))$ with a new path $c_1^{-1} (s.p-v_0(p))$, which would cross the seam. Then the label $f(p)$ would become 
	$$f'(p)= [(g_{c_1}^{-1}g(s.p-v_0(p))) \rhd e_p^{\sigma_p}] [(h^{-1}g_{c_1}^{-1}g(s.p-v_0(p)))\rhd e_p^{-\sigma_p}].$$
	However, from Equation \ref{Equation_c2_seam_condition_tri_nontrivial}, we know that $g_{c_1}$ and $h$ must commute up to an element in $\partial(E)$, which does not affect the $\rhd$ action on $e_p$. Therefore
	\begin{align*}
		f'(p)&=[(g_{c_1}^{-1}g(s.p-v_0(p))) \rhd e_p^{\sigma_p}] [(g_{c_1}^{-1}h^{-1}g(s.p-v_0(p)))\rhd e_p^{-\sigma_p}]\\
		&=g_{c_1}^{-1} \rhd \big([g(s.p-v_{0}(p))\rhd e_p^{\sigma_p}] \: [(h^{-1}g(s.p-v_0(p)))\rhd e_p^{-\sigma_p}]\big)\\
		&= g_{c_1}^{-1} \rhd f(p).
	\end{align*}
	
	In addition to the label of the blob ribbon operator changing, the direct path is also changed. Consider the action of the original blob ribbon operator on a plaquette $q$. This is given by
	$$B^{f(p)}(\text{blob }0 \rightarrow \text{blob }p):e_q= e_q [g(s.p-v_0(q))^{-1} \rhd f(p)^{\pm 1}],$$
	where $\pm 1$ depends on the relative orientation of the plaquette $q$ and the ribbon. Now consider the action of the new blob ribbon operator, which passes the other way around the membrane and has label $f'(p)$, on a plaquette $k$. We have
	$$B^{f'(p)}(\text{blob }0 \rightarrow \text{blob }p):e_k= e_k [(g_{c_1}^{-1}g(s.p-v_0(k)))^{-1} \rhd f'(p)^{\pm 1}],$$
	where $g_{c_1}^{-1}$ is present because the direct path wraps the opposite away around the membrane compared to the original ribbon operator. Then using $f'(p)= g_{c_1}^{-1} \rhd f(p)$, we have
	\begin{align*}
		B^{f'(p)}(\text{blob }0 \rightarrow \text{blob }p):e_k &= e_k [(g_{c_1}^{-1}g(s.p-v_0(k)))^{-1} \rhd (g_{c_1}^{-1} \rhd f(p)^{\pm 1})]\\
		&=e_k [(g(s.p-v_0(k))^{-1} g_{c_1} g_{c_1}^{-1}) \rhd f(p)^{\pm 1})]\\
		&= e_k [g(s.p-v_0(k))^{-1} \rhd f(p)^{\pm 1}].
	\end{align*}
	
	This has the same form as the transformation of plaquette $q$ under the original blob ribbon operator. If the dual paths of the two ribbons were the same, this would mean that the two blob ribbon operators are the same as well. However, the dual paths of the two ribbons also pass the opposite away around the cycle $c_1$ (so the new ribbon crosses the seam), although they both start at blob 0 and end at the blob attached to plaquette $p$. If the cycle were contractible, we could deform the two ribbons into one-another, using the fact that the label of the ribbon operators in the magnetic membrane operators are in the kernel of $\partial$ and so the ribbon operators are topological. Instead, the two blob ribbon operators differ by a closed ribbon operator of label $f(p)$ that wraps around the cycle $c_1$ (this is analogous to the situation in Section \ref{Section_Topological_Blob_Ribbons}, where we showed that we could move a blob ribbon operator by applying a closed ribbon operator). This ribbon operator does not act trivially (unlike in the case from Section \ref{Section_Topological_Blob_Ribbons}) because it is applied on a non-contractible cycle, but can be combined with the existing ribbon operator on $c_1$ from our measurement operator. When we do this, we just change the label of the blob ribbon operator on $c_1$ by an element of the kernel of $\partial$, so the result of changing the blob ribbon operator associated to plaquette $p$ is just to obtain another measurement operator from our existing set of measurement operators, rather than to produce a wholly new operator. While in this case we chose to change the blob ribbon operator associated to $p$ by making it wrap the other way around cycle $c_1$, we could have done the same thing for cycle $c_2$ or any other closed path, with similar results.

	So far, we have ensured that the measurement operator does not excite any plaquette or blob energy terms. This resulted in a candidate measurement operator $T^{[e_{c_1},e_{c_2},e_m,g_{c_1},g_{c_2},h]}(m)$, where
	\begin{equation}
		T^{[e_{c_1},e_{c_2},e_m,g_{c_1},g_{c_2},h]}(m)=B^{e_{c_1}}(c_1)B^{e_{c_2}}(c_2)C^h_T(m) \delta(\hat{e}(m),e_m)\delta(\hat{g}(c_1), g_{c_1}) \delta(\hat{g}(c_2), g_{c_2}),
		\label{Equation_T_operator_definition_central}
	\end{equation}
	and the labels of this operator must satisfy the conditions
	\begin{align*}
		\partial(e_m)^{-1}&=[g_{c_1},g_{c_2}]\\
		\partial(e_{c_2})&=[g_{c_1},h]\\
		\partial(e_{c_1})&=[h,g_{c_2}]\\
		[h \rhd e_m^{-1}]& e_m e_{c_1}^{-1} [g_{c_1}^{-1} \rhd e_{c_1}] e_{c_2}^{-1}[g_{c_2}^{-1} \rhd e_{c_2}]=1_E.
	\end{align*}

	Next we need to consider the commutation of the measurement operator with the various vertex and edge transforms, to ensure that the measurement operator does not lead to the excitation of the corresponding energy terms. In Section \ref{Section_3D_Topological_Charge_Torus_Tri_trivial}, we described how this leads to the measurement operators being made up of a linear combination of $T^X$ operators with different labels. We saw that we needed to have equal sums of operators with labels that were related by certain equivalence relations, with these equivalence relations being generated by the commutation of a vertex or edge transform with the $T$ operator. Rather than repeat the entire argument from Section \ref{Section_3D_Topological_Charge_Torus_Tri_trivial}, we will just demonstrate which equivalence relations arise from the different transforms when $\rhd$ is non-trivial. First we consider the vertex transforms. The only one that fails to commute with any of the components of the measurement operator is the one at the start-point, $s.p$, of each of the membrane and ribbon operators. This vertex transform interacts with each of the individual membrane and ribbon operators that we applied (the two electric ribbon operators, the two blob ribbon operators, the $E$-valued membrane operator and the magnetic membrane operator). For the electric ribbon operators applied around the two cycles, we have the following commutations relation
	$$\delta(g_{c_i},\hat{g}(c_i))A_{s.p}^g =A^g_{s.p}\delta(g_{c_i},g\hat{g}(c_i)g^{-1}) = A^g_{s.p}\delta(g^{-1}g_{c_i}g,\hat{g}(c_i)),$$
	where $i$ takes the values 1 or 2 to represent the two different cycles. For the magnetic membrane operator $C^h_T(m)$, the vertex transform leads to conjugation of its label:
	$$C^h_T(m) A_{s.p}^g=A_{s.p}^g C_T^{g^{-1}hg}(m),$$
	as described by Equation \ref{Equation_magnetic_membrane_tri_nontrivial_start_point_transform} in Section \ref{Section_Magnetic_Tri_Non_Trivial}. For the blob ribbon operators
	$$B^{e_{c_i}}(c_i)A_v^g=A_v^g B^{g^{-1} \rhd e_{c_i}}(c_i),$$
	as we described in Section \ref{Section_Blob_Ribbon_Central} (see Equation \ref{Equation_commutation_blob_ribbon_start_point_transform_last_case}). Finally, for the $E$-valued membrane
	$$\delta(e_m,\hat{e}(m))A_v^g=A_v^g \delta(e_m,g\rhd \hat{e}(m))=A_v^g \delta(g^{-1} \rhd e_m, \hat{e}(m)),$$
	as demonstrated in Ref. \cite{HuxfordPaper2} (see Section S-I C of the Supplemental Material). These commutation relations therefore lead to the equivalence relation
	\begin{equation}
		(g_{c_1},g_{c_2},h,e_{c_1},e_{c_2},e_m) \sim (gg_{c_1}g^{-1},gg_{c_2}g^{-1},ghg^{-1},g\rhd e_{c_1}, g \rhd e_{c_2}, g \rhd e_m),
		\label{Equation_torus_vertex_condition_tri_nontrivial}
	\end{equation}
	(where we relabelled $g^{-1}$ to $g$ for simplicity). We then have one such relation for each group element $g \in G$ (from each of the vertex transforms $A_v^g$).

\begin{figure}[h]
		\begin{center}
			\begin{overpic}[width=0.75\linewidth]{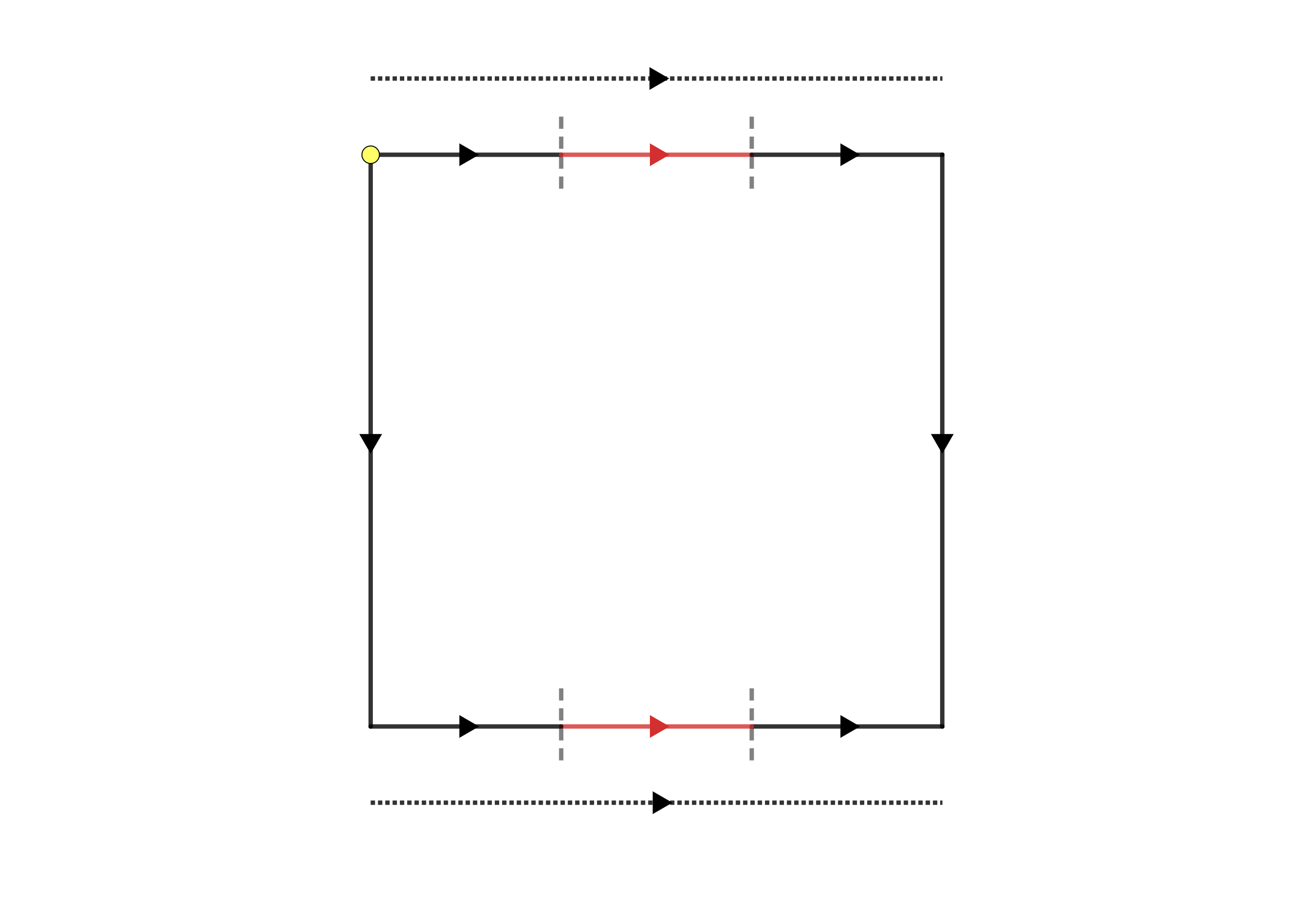}
				\put(46,53){\large edge $i$}
				\put(46,15.5){\large edge $i$}
				\put(24,35){\large $c_2$}
				\put(73,35){\large $c_2$}
				\put(49,65){\large $c_1$}
				\put(49,4){\large $c_1$}
			\end{overpic}
			\caption{We consider an edge transform on an edge $i$ along the cycle $c_1$ (where $c_1$ is the top and bottom sides of the square). This edge transform will affect the label of the electric ribbon operator applied on $c_1$, but will also affect the magnetic membrane operator and the $E$-valued membrane operator applied on the surface of the torus itself.}
			\label{torus_charge_edge_transform_c1}
		\end{center}
	\end{figure}

	Next we must consider the edge transforms in the region of the measurement operator. Of the edges on the direct membrane, only those on the ``seams" (the cycles) host edge transforms that may fail to commute with the measurement operator. The edge transforms in the bulk of the direct membrane commute with the membrane operators, due to the fact that they lie away from the boundary of the membrane, as discussed in Section \ref{Section_Magnetic_Tri_Non_Trivial} for the magnetic membrane operator and Section S-I C in Ref. \cite{HuxfordPaper2} for the $E$-valued membrane operator. These bulk transforms also commute with the blob ribbon operators (because those ribbon operators generally commute with edge transforms) and the electric ribbon operators (because the electric ribbon operators are only affected by transforms on the path of the operators). We therefore just need to consider edges along the boundaries of the membrane (the seam on the torus). We start by considering an arbitrary edge on the cycle $c_1$, as illustrated in Figure \ref{torus_charge_edge_transform_c1}.

	The commutation relation of the edge transform $\mathcal{A}_i^e$ on the edge $i$ with the electric operator along the cycle $c_1$ is
	\begin{align*}
		\delta(\hat{g}(c_1),g_{c_1})\mathcal{A}_i^e&=\mathcal{A}_i^e \delta((\mathcal{A}_i^e)^{-1}\hat{g}(c_1)\mathcal{A}_i^e,g_{c_1})\\
		&=\mathcal{A}_i^e \delta(\partial(e)\hat{g}(c_1),g_{c_1}),
	\end{align*}
	where we used the fact that $\partial(e)$ is in the centre of $G$ to pull it out to the front of the expression (from whichever edge in $c_1$ it is acting on). Then we have
	\begin{align*}
		\delta(\hat{g}(c_1),g_{c_1})\mathcal{A}_i^e&=\mathcal{A}_i^e \delta(\hat{g}(c_1),\partial(e)^{-1}g_{c_1}).
	\end{align*}

\begin{figure}[h]
		\begin{center}
			\begin{overpic}[width=0.7\linewidth]{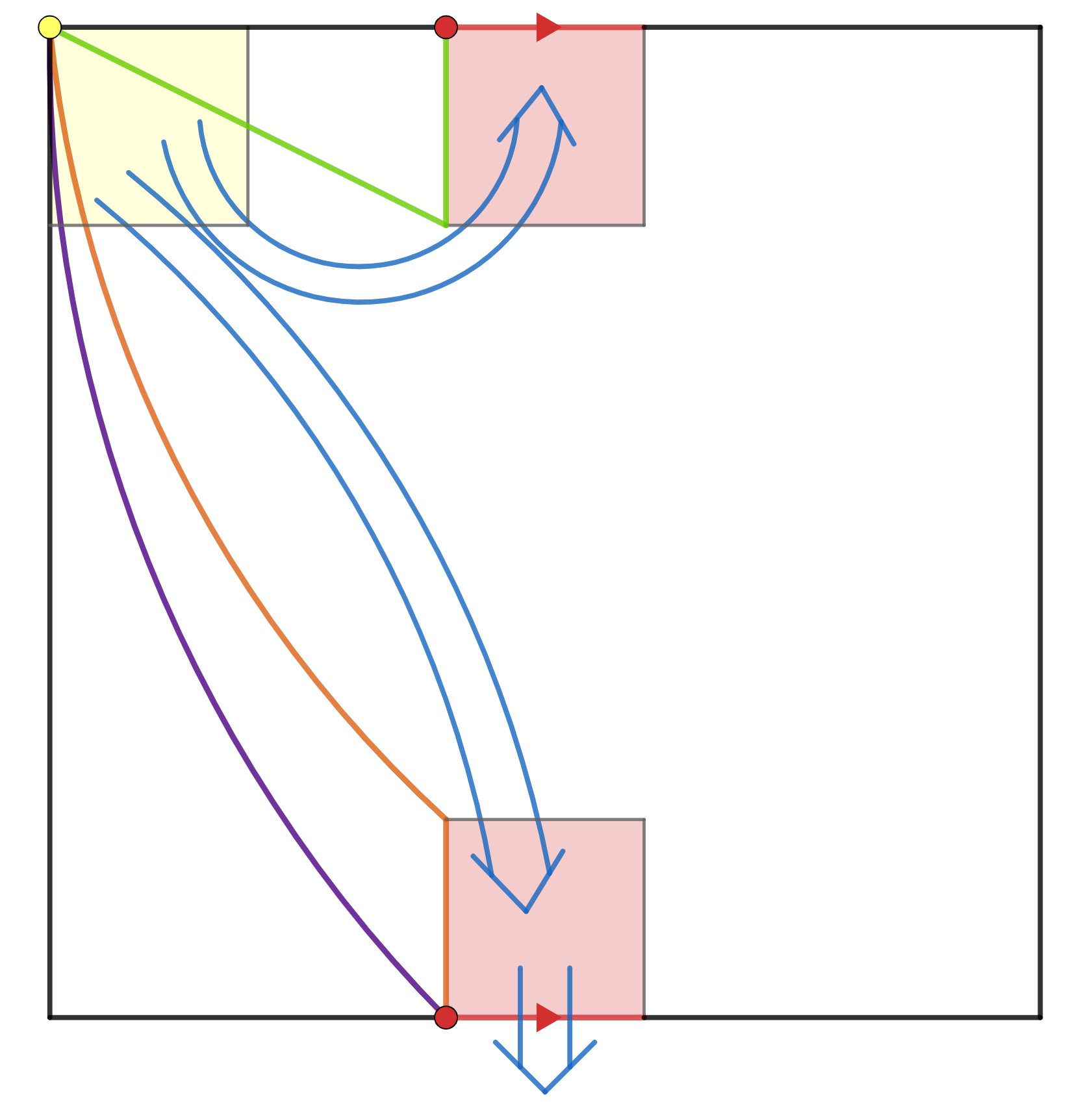}
				\put(0,97){$s.p$}
				\put(52,98){$i$}
				\put(37,99){$v_i$}
				\put(41,72){$(2)$}
				\put(28,88){$s.p-v_i(2)$}
				\put(41,48){$(1)$}
				\put(51,0){$(\delta)$}
				\put(52,9){$i$}
				\put(22,47){$s.p-v_i(1)$}
				\put(8.5,30){$s.p-v_i(\delta)$}	
			\end{overpic}
			\caption{The commutation of the edge transform on an edge $i$ along $c_1$ with the magnetic membrane operator produces blob ribbon operators along three ribbons: $(1)$, $(2)$ and $(\delta)$. The ribbons $(1)$ and $(2)$ pass from blob 0 (the base of which is represented by the yellow square in the top-left) to the plaquettes adjacent to $i$ (represented by the shaded red squares in the top-middle and bottom-middle sections), on either side of the seam. The ribbon $(\delta)$ is a short ribbon that passes across the seam, through the plaquettes connected to $i$ but above the shaded red plaquettes (see Figures \ref{ribbon_delta} and \ref{path_delta_on_mod_mem} in Section \ref{Section_Magnetic_Tri_Non_Trivial} for an illustration). This ribbon connects $(1)$ and $(2)$ and, with some manipulation of the ribbon operator applied on $(2)$, the three ribbon operators can be connected into a single ribbon operator applied on a closed ribbon. The labels of each of the ribbon operators depends on the corresponding path from the start-point to the source of the edge $i$, $v_i$. We can see that the paths corresponding to $(1)$ and $(\delta)$, which are $s.p-v_i(1)$ and $s.p-v_i(\delta)$ respectively, can be deformed into one-another, while the path $s.p-v_i(2)$ corresponding to $(2)$ differs from these paths by the cycle $c_2$ in addition to a smooth deformation.}
			\label{torus_charge_blob_ribbons_edge_c1_detail}
			
		\end{center}
	\end{figure}

\begin{figure}[h]
		\begin{center}
			\begin{overpic}[width=0.75\linewidth]{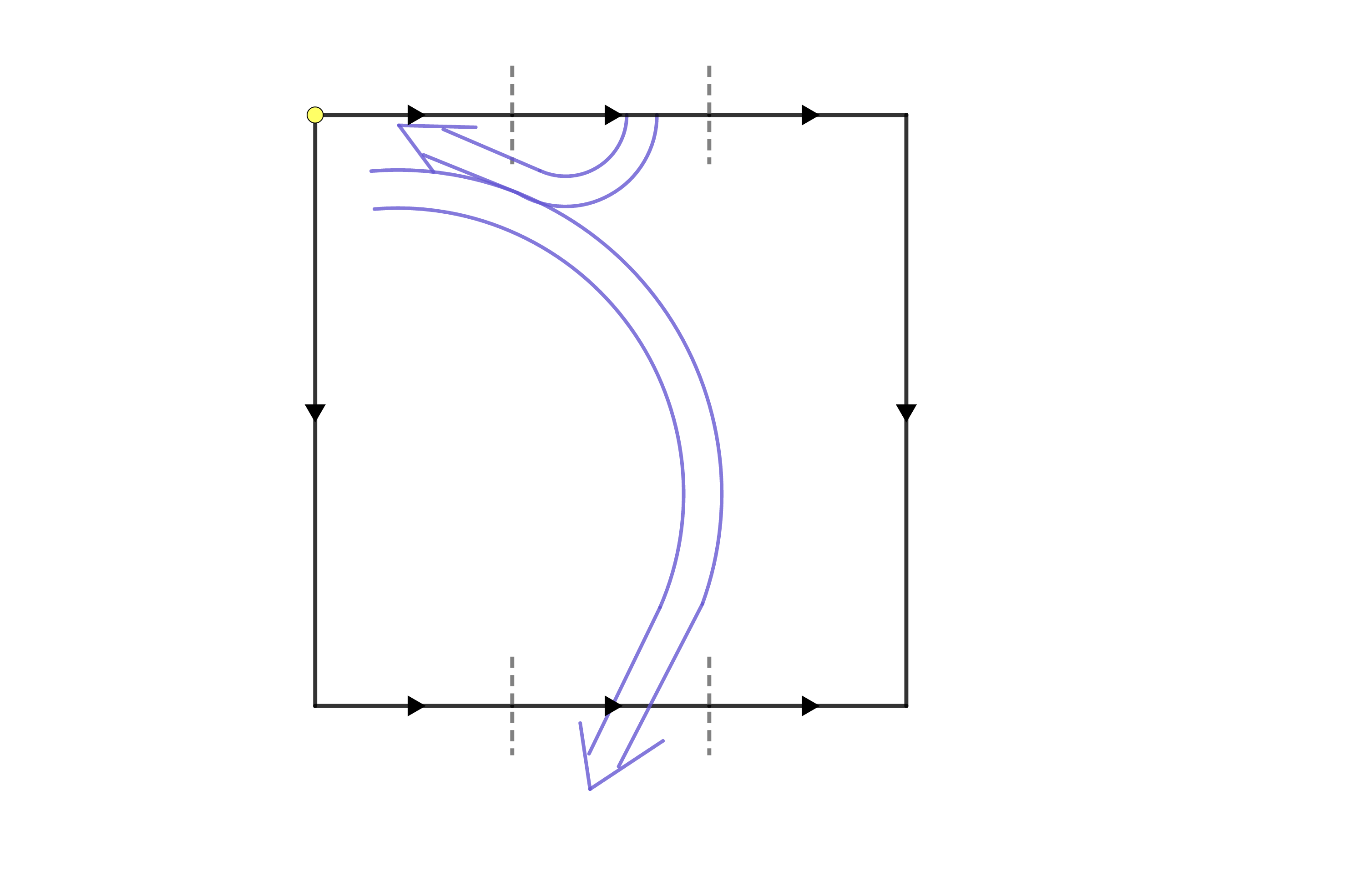}
				\put(47,50){\large $(2)^{-1}$}
				\put(54,25){\large $(1+\delta)$}
			\end{overpic}
			\caption{The commutation of the edge transform on edge $i$ with the magnetic membrane operator produces additional blob ribbon operators, as described in Section \ref{Section_Magnetic_Tri_Nontrivial_Commutation}. These blob ribbon operators lie on the ribbons $(1)$, $(2)$ and $(\delta)$. These ribbon operators fuse to give a closed ribbon operator. We can combine the blob ribbons on $(1)$ and $(\delta)$ simply (to give an operator on the path $(1+\delta)$), but to combine these operators with the one on $(2)$ we must perform some manipulations of that operator. The first step is to reverse the orientation of path (2) while simultaneously inverting the label of the blob ribbon operator.}
			\label{torus_charge_blob_ribbons_from_edge_c1}
		\end{center}
	\end{figure}

	We therefore see that the label of the electric ribbon operator transforms as 
	\begin{equation}
		g_{c_1}\rightarrow \partial(e)^{-1}g_{c_1}
		\label{Equation_torus_charge_edge_transform_cycle_1_electric}
	\end{equation}
	under the edge transform. In the $\rhd$ trivial case this was the only effect of the edge transform. When $\rhd$ is non-trivial however, the electric ribbon operator is not the only operator which fails to commute with the transform. Recall from Section \ref{Section_Magnetic_Tri_Non_Trivial} (see Figure \ref{mod_mem_edge_transform_boundary} and the discussion preceding it) that an edge transform applied on the boundary of a magnetic membrane operator generates a blob ribbon operator that passes from blob 0 past the boundary of the membrane. As explained in Section \ref{Section_Magnetic_Tri_Non_Trivial}, an edge transform generates blob ribbon operators when commuting with the magnetic membrane operator for two reasons. Firstly, the edge transform affects adjacent plaquettes lying on the direct membrane. Because the label of the plaquettes on the direct membrane determine the label of the blob ribbon operators that are part of the magnetic membrane operator, changing the label of these plaquettes changes the labels of the added blob ribbon operators. This change of label can be expressed as an additional blob ribbon operator. The ribbon for this ribbon operator passes from blob 0 to the blob attached to the affected plaquette. This picture becomes slightly different when we consider a torus made by gluing the sides of a square, because an edge on the seam of the torus is adjacent to two plaquettes on the torus, one either side of the seam. Considering the case of an edge $i$ along the cycle $c_1$, because $i$ is adjacent to two plaquettes, as shown in Figure \ref{torus_charge_blob_ribbons_edge_c1_detail}, the edge transform produces two blob ribbon operators by changing the plaquette labels. These blob ribbon operators act along the ribbons (1) and (2) shown in Figure \ref{torus_charge_blob_ribbons_edge_c1_detail}. From Equation \ref{magnetic_membrane_base_edge_commutation_1} in Section \ref{Section_Magnetic_Tri_Nontrivial_Commutation}, the labels of these blob ribbon operators are $[(h g(s.p-v_i(1))) \rhd e] \: [g(s.p-v_i(1)) \rhd e^{-1}]$ for the ribbon operator along (1) and $([(h g(s.p-v_i(2))) \rhd e] \: [g(s.p-v_i(2)) \rhd e^{-1}])^{-1}$ for the ribbon operator along (2), where the paths $(s.p-v_i(1))$ and $(s.p-v_i(2))$ are indicated in Figure \ref{torus_charge_blob_ribbons_edge_c1_detail}. Note that there is an additional $h\rhd$ action on these labels compared to Equation \ref{magnetic_membrane_base_edge_commutation_1}, because we take these ribbon operators to act after the magnetic membrane operator here (i.e., to the left of the magnetic membrane operator rather than the right as in Equation \ref{magnetic_membrane_base_edge_commutation_1} from Section \ref{Section_Magnetic_Tri_Nontrivial_Commutation}) and there is an $h \rhd$ action from commuting the blob ribbon operators past the magnetic membrane operator, as we discussed in Section \ref{Section_Magnetic_Tri_Nontrivial_Commutation} (note that the label of the ribbon operator on (1) in this section matches the label of the ribbon operator on $(\delta)$ in Equation \ref{magnetic_membrane_base_edge_commutation_1}, which is to the left of the magnetic membrane operator in that equation, for precisely this reason). Note also that the inverse in label between the two cases are due to one of the instances of edge $i$ being clockwise around the membrane and the other being anticlockwise. We can put the label of the ribbon operator applied on ribbon (2) in the same form as the label of ribbon (1) by reversing the orientation of ribbon (2) to give $(2)^{-1}$, as illustrated in Figure \ref{torus_charge_blob_ribbons_from_edge_c1}.

	The second type of blob ribbon operator produced by the commutation of the edge transform with the magnetic membrane operator comes from the action of the edge transform on the ``vertical" plaquettes that are cut by the dual membrane (see Figure \ref{ribbon_delta} in Section \ref{Section_Magnetic_Tri_Non_Trivial}). This produces a blob ribbon operator on a small ribbon, which we label $(\delta)$, which crosses the seam of the membrane, as illustrated in Figure \ref{torus_charge_blob_ribbons_edge_c1_detail}. The label of this ribbon operator is $[(h g(s.p-v_i(\delta))) \rhd e] \: [g(s.p-v_i(\delta)) \rhd e^{-1}]$ which is equal to $[(h g(s.p-v_i(1))) \rhd e] \: [g(s.p-v_i(1)) \rhd e^{-1}]$, because the path $(s.p-v_i(\delta))$ can be deformed into $(s.p-v_i(1))$. The fact that the ribbon operator applied on $(\delta)$ has the same label as the ribbon operator applied on $(1)$, together with the fact that these two ribbon operators have the same start-point ($s.p$) and have direct paths that agree, means that we can connect these two blob ribbon operators into a single one on the concatenated ribbon $(1+ \delta)$, as illustrated in Figure \ref{torus_charge_blob_ribbons_from_edge_c1}.

	We next need to demonstrate that the ribbon operator on $(1+ \delta)$ can also be connected to the ribbon operator on $(2)^{-1}$, in order to produce a closed ribbon operator. The ribbon operator on $(2)^{-1}$ has label $[(h g(s.p-v_i(2))) \rhd e] \: [g(s.p-v_i(2)) \rhd e^{-1}]$, which is different from the label of the other ribbon operator because the path $(s.p-v_i(2))$ cannot be deformed into the path $(s.p-v_i(1))$. This would imply that we cannot connect the two ribbon operators. However, the direct path for ribbon (2) also prevents it from being combined. This is because when we connect a ribbon operator to the end of another ribbon operator, the direct path of the ribbon at the end must include the direct path of the other ribbon (or be able to be deformed so that it includes this path). In this case the direct path of ribbon $(1+ \delta)$ wraps the cycle $c_2$, so to connect another piece of ribbon to the end of $(1 + \delta)$, we would need the direct path of the piece of ribbon to wrap around the cycle first, which the direct path of $(2)$ does not do. It turns out that these two issues cancel each-other out. We can change the direct path of the ribbon operator along $(2)^{-1}$ to wrap the cycle, by 
	using the procedure for changing the start-point of a blob ribbon operator described in Section \ref{Section_blob_ribbon_move_sp} (see Equation \ref{Equation_blob_ribbon_change_sp_2}) to move the start-point along $c_2^{-1}$. This changes the label of the ribbon operator to 
	\begin{align*}
		g_{c_2} \rhd [(h g(s.p-v_i(2))) \rhd e] \: [g(s.p-v_i(2)) \rhd e^{-1}]&=[(g_{c_2}h g(s.p-v_i(2)))\rhd e] \: [(g_{c_2}g(s.p-v_i(2))) \rhd e^{-1}]\\
		&=[(g_{c_2}hg_{c_2}^{-1}h^{-1} h g_{c_2}g(s.p-v_i(2)))\rhd e] \: [g_{c_2}g(s.p-v_i(2)) \rhd e^{-1}],
	\end{align*}
	where we inserted the identity in the form of $1_G=g_{c_2}^{-1}h^{-1} h g_{c_2}$ to obtain the second line. Then, note that $g_{c_2}g(s.p-v_i(2))=g(s.p-v_i(1))$ up to factors in $\partial(E)$ from deformation, and $g_{c_2}hg_{c_2}^{-1}h^{-1}$ belongs in $\partial(E)$ from the condition given in Equation \ref{Equation_c1_seam_condition_tri_nontrivial}. Elements in $\partial(E)$ do not affect expressions of the form $g(t) \rhd e$, so we see that the label of the blob ribbon operator is
	$$[(h g(s.p-v_i(1)))\rhd e] \: [g_{c_2}g(s.p-v_i(1)) \rhd e^{-1}],$$
	which matches the label of the ribbon operator applied on $(1+ \delta)$. This means that we can connect these two ribbon operators into a single closed ribbon operator with this label, wrapping around the cycle $c_2$. Therefore, the edge transform produces a closed blob ribbon operator that contributes to the label $e_{c_2}$ of the ribbon operator that wraps $c_2$ from our measurement operator. Denoting $(s.p-v_i(1))$ simply by $(s.p-v_i)$, we see that
	\begin{equation}
		e_{c_2} \rightarrow e_{c_2} [(hg(s.p-v_i))\rhd e] [g(s.p-v_i) \rhd e^{-1}]
		\label{Equation_torus_charge_edge_transform_cycle_1_blob_ribbon}
	\end{equation}
	under the action of $\mathcal{A}_i^e$.

\begin{figure}[h]
		\begin{center}
			\begin{overpic}[width=0.5\linewidth]{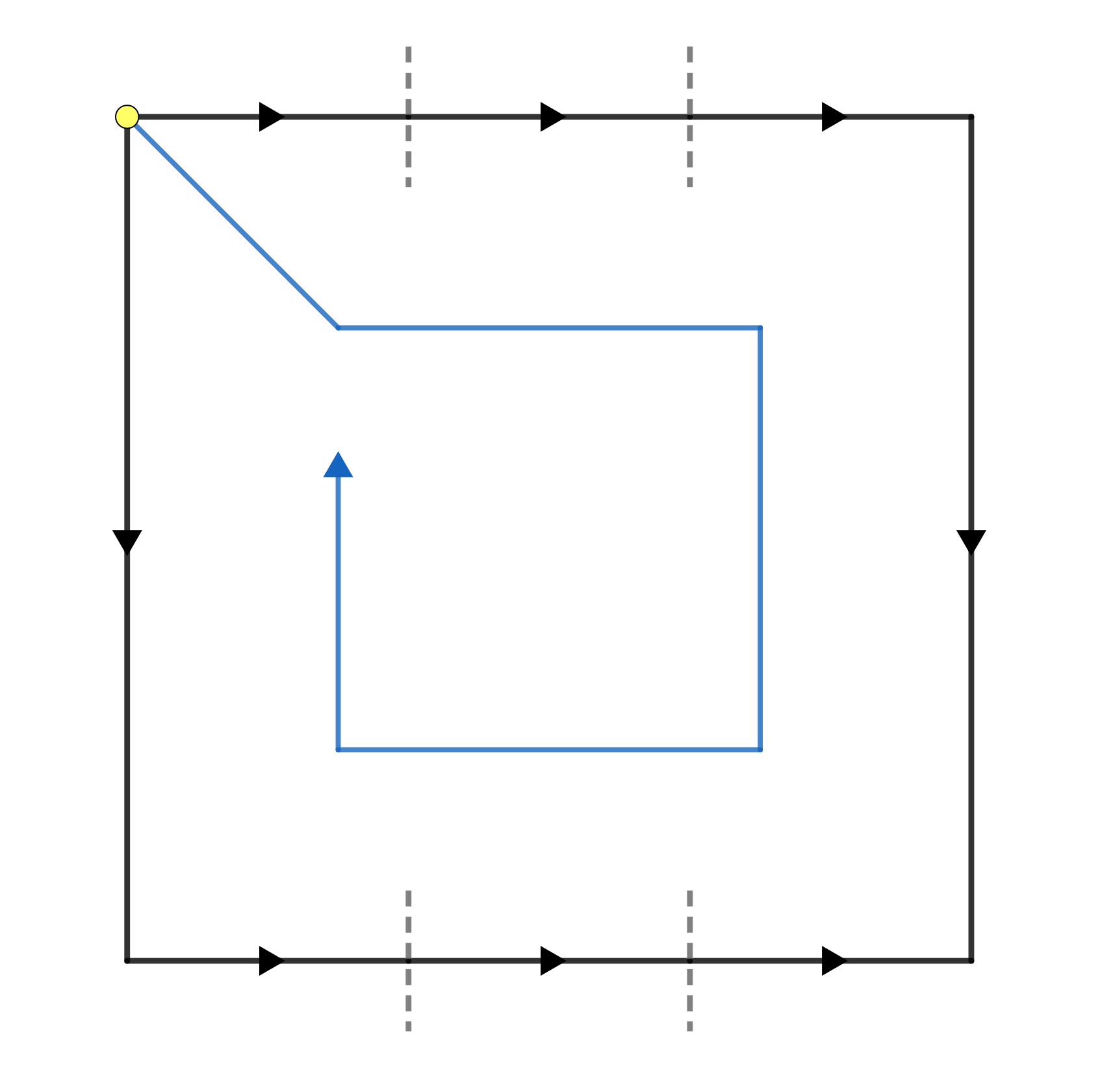}
				\put(22,92){\large $a$}
				\put (45,92){\large edge $i$}
				\put(22,6){\large $a$}
				\put(45,6){\large edge $i$}
				\put(5,50){\large $c_2$}
				\put(92,50){\large $c_2$}
				\put(47,50){\large $\hat{e}(m)$}
			\end{overpic}
			\caption{When $\rhd$ is non-trivial, the edge transform on edge $i$ affects the surface label of the torus. Because the edge $i$ appears more than once on the boundary $c_1 c_2c_1^{-1}c_2^{-1}$, we must use Equation S14 from the Supplemental Material of Ref. \cite{HuxfordPaper2} to determine how it affects the surface.}
			\label{torus_charge_edge_transform_c1_2}
			
		\end{center}
	\end{figure}

	The final impact of the edge transform is to change the surface label of the torus, due to the non-trivial cycles of the torus, which means that the edge transform does not commute with the $E$-valued membrane operator applied on the surface. We use Equation S14 from Ref. \cite{HuxfordPaper2} (see Section S-I C in the Supplemental Material), which gives the transformation properties of a surface under an edge transform when the edge in question appears more than once in the boundary of the surface. From this equation, we see that the transformation of the surface label operator $\hat{e}(m)$ is given by
	$$\hat{e}(m)A_i^e = A_i^e \hat{e}(m) [(g_{c_1}g_{c_2}g_{c_1}^{-1} g(a)) \rhd e] [g(a) \rhd e^{-1}],$$
	where $a$ is the path from the start-point to the first appearance of edge $i$ in a clockwise direction (with label $g(a)$) and $c_1c_2c_1^{-1} a$ is the path to the second appearance of the edge in the clockwise direction, as can be seen from Figure \ref{torus_charge_edge_transform_c1_2}. We can then use the condition $g_{c_1}g_{c_2}g_{c_1}^{-1}g_{c_2}^{-1} =\partial(e_m)^{-1}$ from Equation \ref{torus_flatness_condition_2} to rewrite $g_{c_1}g_{c_2}g_{c_1}^{-1} $ as $\partial(e_m)^{-1}g_{c_2}$. Then we have
	$$\hat{e}(m)A_i^e =A_i^e [(\partial(e_m)^{-1}g_{c_2} g(a)) \rhd e] \hat{e}(m) [g(a) \rhd e^{-1}],$$
	which becomes (using $\partial(e)\rhd f = f \ \forall e, \: f \in E$ when $E$ is Abelian)
	\begin{equation} 
		\hat{e}(m)A_i^e =A_i^e [(g_{c_2} g(a)) \rhd e] \hat{e}(m) [g(a) \rhd e^{-1}], \label{Equation_torus_edge_transform_c1_surface_label_1}
	\end{equation}
	where $c_2 a$ is the path from the start-point to the appearance of edge $i$ in the anticlockwise direction. This switch from the clockwise to anticlockwise path is something we can do as long as the surface satisfies fake-flatness, because this switch then just corresponds to a smooth deformation of the path over a fake-flat region. We see that the transformation of the surface label under the edge transform is trivial when $\rhd$ is trivial, which is why we did not need to consider this transformation when considering that case in Section \ref{Section_3D_Topological_Charge_Torus_Tri_trivial}.

	The path $c_2a$ can be deformed into the path $s.p-v_i(1)$ from Figure \ref{torus_charge_blob_ribbons_edge_c1_detail}. This means that the labels of these two paths differ only by an element of $\partial(E)$ and so $(g_{c_2}g(a)) \rhd e= g(s.p-v_i(1)) \rhd e$ (factors in $\partial(E)$ are irrelevant when $E$ is Abelian, as explained in Section \ref{Section_Magnetic_Tri_Non_Trivial}). Substituting this into Equation \ref{Equation_torus_edge_transform_c1_surface_label_1}, we see that the effect of the edge transform on the surface label can be written as
	$$\hat{e}(m) A_i^e =A_i^e [g(s.p-v_i(1)) \rhd e] [(g_{c_2}^{-1}g(s.p-v_i(1))) \rhd e^{-1}] \hat{e}(m).$$
	
	When considering the operator $\delta(\hat{e}(m),e_m)$, this means that
	\begin{align*}
		\delta(\hat{e}(m),e_m)A_i^e& = A_i^e \delta( [g(s.p-v_i(1)) \rhd e] [(g_{c_2}^{-1}g(s.p-v_i(1))) \rhd e^{-1}] \hat{e}(m),e_m)\\
		=& \delta( \hat{e}(m),[g(s.p-v_i(1)) \rhd e]^{-1} [(g_{c_2}^{-1}g(s.p-v_i(1))) \rhd e]e_m)
	\end{align*}
	and so the label $e_m$ transforms as
	$$ e_m \rightarrow [g(s.p-v_i(1)) \rhd e^{-1}] [(g_{c_2}^{-1} g(s.p-v_i(1))) \rhd e] e_m.$$

	We can simplify this expression by relabelling $g(s.p-v_i(1)) \rhd e =s$. Note that, because of the Peiffer conditions when $\partial$ maps to the centre of $G$, $\partial(e)=\partial(s)$. This gives us
	\begin{equation}
		e_m \rightarrow e_m s^{-1}[g_{c_2}^{-1} \rhd s].
		\label{Equation_torus_charge_edge_transform_cycle_1_surface_label}
	\end{equation}
	
	We then combine Equation \ref{Equation_torus_charge_edge_transform_cycle_1_surface_label} with Equations \ref{Equation_torus_charge_edge_transform_cycle_1_electric} and \ref{Equation_torus_charge_edge_transform_cycle_1_blob_ribbon}, which describe how the other labels of our measurement operator transforms under commutation with the edge transform, to find that the edge transform leads to the equivalence relation
	\begin{equation}
		(g_{c_1},e_{c_2},e_m) \sim (\partial(s)^{-1}g_{c_1},e_{c_2} [h\rhd s] s^{-1},e_m s^{-1}[g_{c_2} \rhd s]).
		\label{Equation_torus_edge_transform_c1_tri_non_trivial}
	\end{equation}

	\begin{figure}[h]
		\begin{center}
			\begin{overpic}[width=0.6\linewidth]{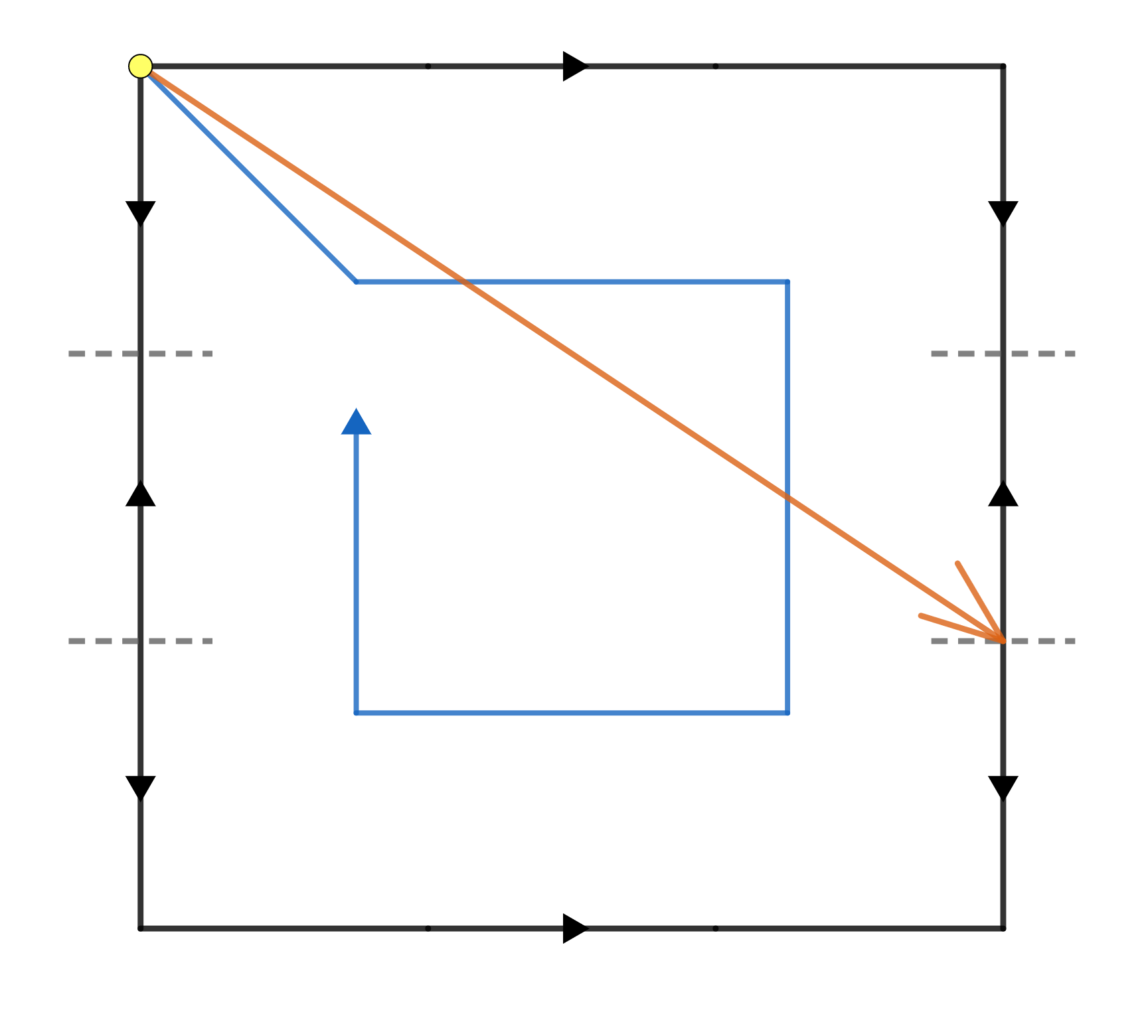}
				\put(8,70){\large $b$}
				\put (50,87){\large $c_1$}
				\put(90,70){\large $b$}
				\put(50,5){\large $c_1$}
				\put(0,46){\large edge $i$}
				\put(90,46){\large edge $i$}
				\put(4,20){\large $d^{-1}$}
				\put(90,20){\large $d^{-1}$}
				\put(89,32){\large $v_i$}
				\put(30,74){\large $s.p-v_i(1)$}
			\end{overpic}
			
			\caption{Having considered an edge transform on an edge along the cycle $c_1$, we next consider an edge along the cycle $c_2$. This has a very similar effect to the edge transform on $c_1$, as we may expect.}
			\label{torus_charge_edge_transform_c2}
		\end{center}
	\end{figure}
	
	Now we take a similar look at an edge transform applied on an edge $i$ on the other cycle, $c_2$, as illustrated in Figure \ref{torus_charge_edge_transform_c2}. First we examine how this edge transform affects the electric ribbon operator applied along the cycle $c_2$. We have that
	\begin{align}
		\delta(\hat{g}(c_2), g_{c_2})\mathcal{A}_i^e &=\mathcal{A}_i^e \delta(\partial(e)^{-1}\hat{g}(c_2),g_{c_2}) \notag\\
		&=\mathcal{A}_i^e \delta(\hat{g}(c_2),\partial(e)g_{c_2}) \notag\\
		&\implies g_{c_2} \rightarrow \partial(e)g_{c_2}, \label{Equation_torus_charge_edge_transform_cycle_2_electric}
	\end{align}
	where the inverse in the factor of $\partial(e)$ compared to the case of the other cycle, $c_1$, is simply because we chose the edge to be anti-aligned with $c_2$. Similar to the edge transform along $c_1$, the edge transform for an edge on $c_2$ generates a blob ribbon operator, this time around $c_1$, when it commutes past the magnetic membrane operator. The same argument as for the previous edge tells us that we will produce a closed blob ribbon operator. By considering the blob ribbon operator produced by the right instance of the edge in Figure \ref{torus_charge_edge_transform_c2}, using the results from Section \ref{Section_Magnetic_Tri_Non_Trivial} (see Equation \ref{magnetic_membrane_base_edge_commutation_1} and note that we must commute the blob ribbon operator to the left of the magnetic membrane operator,so it picks up an additional factor of $h \rhd$), we know that the label of the blob ribbon operator around $c_1$ transforms according to
	\begin{equation}
		e_{c_1} \rightarrow e_{c_1} [(h g(s.p-v_i(1))) \rhd e] [g(s.p-v_i(1)) \rhd e^{-1}].
		\label{Equation_torus_charge_edge_transform_cycle_2_blob_ribbon}
	\end{equation}

	Finally we consider the effect of the edge transform on the surface label $e_m$. Using Equation S14 from Section S-I C in the Supplemental Material of Ref. \cite{HuxfordPaper2}, we have that
	$$\hat{e}(m)A_i^e = A_i^e[(g_{c_2}g_{c_1} g(d)) \rhd e] \hat{e}(m) [(g_{c_1} g_{c_2} g_{c_1}^{-1} d) \rhd e^{-1}],$$
	where $g(d)$ is the group element assigned to the path $d$ from Figure \ref{torus_charge_edge_transform_c2}. The path $c_2c_1d$ can be deformed into $s.p-v_i(1)$, meaning that the labels of these paths differ at most by an element of $\partial(E)$. Therefore, the surface label operator transforms under the edge transform as
	\begin{equation}
		\hat{e}(m) A_i^e= A_i^e [g(s.p-v_i(1)) \rhd e] \hat{e}(m) [(g_{c_1} g_{c_2} g_{c_1}^{-1} g(d)) \rhd e^{-1}]. \label{Equation_torus_edge_transform_c2_surface_label_2}
	\end{equation}
	We can eliminate $g(d)$ from this expression by writing $g(d)= g_{c_1}^{-1}g_{c_2}^{-1}g(s.p-v_i(1))$ (up to irrelevant factors in $\partial(E)$), so that
	$$g_{c_1} g_{c_2} g_{c_1}^{-1} g(d) = g_{c_1} g_{c_2} g_{c_1}^{-1}g_{c_1}^{-1}g_{c_2}^{-1}g(s.p-v_i(1)).$$
	
	Because $g_{c_1}$ and $g_{c_2}$ commute up to factors in $\partial(E)$ (from Equation \ref{torus_flatness_condition_2}), we can collect the factors of $g_{c_1}$ and its inverse, to obtain
	$$g_{c_1} g_{c_2} g_{c_1}^{-1} g(d)= g_{c_1} g_{c_1}^{-1}g_{c_1}^{-1}g_{c_2} g_{c_2}^{-1}g(s.p-v_i(1))=g_{c_1}^{-1}g(s.p-v_i(1)),$$
	where this equality again holds up to irrelevant factors in $\partial(E)$. Inserting this relation into Equation \ref{Equation_torus_edge_transform_c2_surface_label_2}, we see that the surface transforms according to
	\begin{align*}
		\hat{e}(m)A_i^e &= A_i^e [g(s.p-v_i(1)) \rhd e] \hat{e}(m) [ (g_{c_1}^{-1} g(s.p-v_i(1))) \rhd e^{-1}]\\
		&\implies \delta(\hat{e}(m),e_m)A_i^e = A_i^e \delta([g(s.p-v_i(1)) \rhd e] \hat{e}(m) [ (g_{c_1}^{-1} g(s.p-v_i(1))) \rhd e^{-1}],e_m)\\
		& \hspace{3cm} = \delta(\hat{e}(m), [g(s.p-v_i(1)) \rhd e]^{-1} e_m[ (g_{c_1}^{-1} g(s.p-v_i(1))) \rhd e]),
	\end{align*}
	so the label $e_m$ transforms as
	$$e_m \rightarrow [g(s.p-v_i(1)) \rhd e^{-1}] [( g_{c_1}^{-1}g(s.p-v_i(1))) \rhd e] e_m$$
	under the action of the edge transform. Then, relabelling $g(s.p-v_i(1)) \rhd e=r$, this becomes
	\begin{equation}
		e_m \rightarrow r^{-1} [ g_{c_1}^{-1} \rhd r] e_m. \label{Equation_torus_charge_edge_transform_cycle_2_surface_label}
	\end{equation}
	
	Combining Equation \ref{Equation_torus_charge_edge_transform_cycle_2_surface_label} with Equations \ref{Equation_torus_charge_edge_transform_cycle_2_electric} and \ref{Equation_torus_charge_edge_transform_cycle_2_blob_ribbon}, we see that the edge transform generates the equivalence relation
	\begin{equation}
		(g_{c_2}, e_{c_1},e_m) \sim (\partial(r)g_{c_2},e_{c_1} [h \rhd r] \: r^{-1}, e_m r^{-1}[g_{c_1}^{-1} \rhd r]),
		\label{Equation_torus_edge_transform_c2_tri_non_trivial}
	\end{equation}
	where we used the fact that $\partial(r)= \partial(e)$ (which follows from the Peiffer conditions when $\partial(E)$ is in the centre of $G$) to write the transformation of $g_{c_2}$ in terms of $r$.

	Having considered the vertex transforms and the edge transforms for edges that lie on the direct membrane, we must finally consider the outwards edge transforms. That is, we consider the edge transforms on the edges that are cut by the dual membrane of the magnetic membrane operator. In Section \ref{Section_3D_Topological_Charge_Torus_Tri_trivial}, we saw that in the $\rhd$ trivial case, combining a series of these edge transforms with the measurement operator (as opposed to commuting them past the measurement operator as with the other transforms) changed the labels of the measurement operator. Because these edge transforms act trivially in the subspace where the membrane is initially unexcited, this means that the new measurement operator resulting from combining the old one with the edge transforms is equivalent to the old measurement operator. The same thing also occurs in this case, except that the edge transforms change more than just the label of the magnetic part of the measurement operator. We consider performing an edge transform on each outwards edge $i$, with label $g(s.p-v_i)^{-1} \rhd e$, where we use the procedures from the Appendix of Ref. \cite{HuxfordPaper1} to orient each edge so that its source lies on the direct membrane. Then for an edge $i$ with initial label $g_i$ affected by both the magnetic membrane operator and the edge transform, the edge label becomes $\partial(g(s.p-v_i)^{-1}\rhd e) g(s.p-v_i)^{-1}h g(s.p-v_i)g_i.$ We can then use the Peiffer condition (Equation \ref{Equation_Peiffer_1} from the main text) to write the resulting label as
	\begin{align*}
		\partial(g(s.p-v_i)^{-1}\rhd e) g(s.p-v_i)^{-1}h g(s.p-v_i)g_i&= (g(s.p-v_i)^{-1}\partial(e)g(s.p-v_i)) g(s.p-v_i)^{-1}h g(s.p-v_i)g_i\\
		&=g(s.p-v_i)^{-1} \partial(e)h g(s.p-v_i)g_i\\
		&=C^{\partial(e)h}_T(m):g_i.
	\end{align*}
	
	This tells us that the combined action of the magnetic membrane operator $C^h_T(m)$ and edge transforms on the edges is the same as the action of a magnetic membrane operator of label $C^{\partial(e)h}_T(m)$ (just as we saw in the $\rhd$ trivial case, where the magnetic membrane operator only acts on the edges). We saw something very similar in Section \ref{Section_condensation_magnetic_centre_case}, where we considered the condensation of the magnetic membrane operators. In that case, we used the same series of edge transforms to reproduce the magnetic membrane operator $C^{\partial(e)}_T(m)$, up to blob ribbon operators along the boundary. The same thing occurs here, as we will see by considering the plaquettes affected by the edge transforms. Note that the membrane operator $C^{\partial(e)h}_T(m)$ has the same effect on the plaquettes as $C^h_T(m)$, because the action on plaquettes is insensitive to elements in $\partial(E)$. This means that we can combine the action of the edge transforms on the edges with $C^h_T(m)$ to produce $C^{\partial(e)h}_T(m)$ and any action of the edge transforms on the plaquettes is treated as a separate operator.

	We now take a closer look at the action of the edge transforms on the plaquettes. In Section \ref{Section_condensation_magnetic_centre_case}, when considering the condensation of magnetic membrane operators, we applied exactly such a sequence of edge transforms on an open membrane and found that they left the plaquette labels invariant in the bulk of the membrane. This is because each plaquette cut through by the dual membrane is affected by two edge transforms. The path-dependent label of these edge transforms ensures that the two transforms cancel out. That is, for a plaquette $p$ with two edges $i$ and $j$ that are affected, the plaquette label $e_p$ transforms as
	\begin{align*}
		e_p \rightarrow& [g(\overline{v_0(p)-v_j}) \rhd (g(s.p-v_j)^{-1} \rhd e)] e_p [g(v_0(p)-v_i) \rhd (g(s.p-v_i)^{-1} \rhd e)]^{-1}\\
		=& [(g(\overline{v_0(p)-v_j})g(v_j-s.p)) \rhd e] e_p [(g(v_0(p)-v_i) g(v_i-s.p)) \rhd e)]^{-1}.
	\end{align*}
	
\begin{figure}[h]
		\begin{center}
			\begin{overpic}[width=0.6\linewidth]{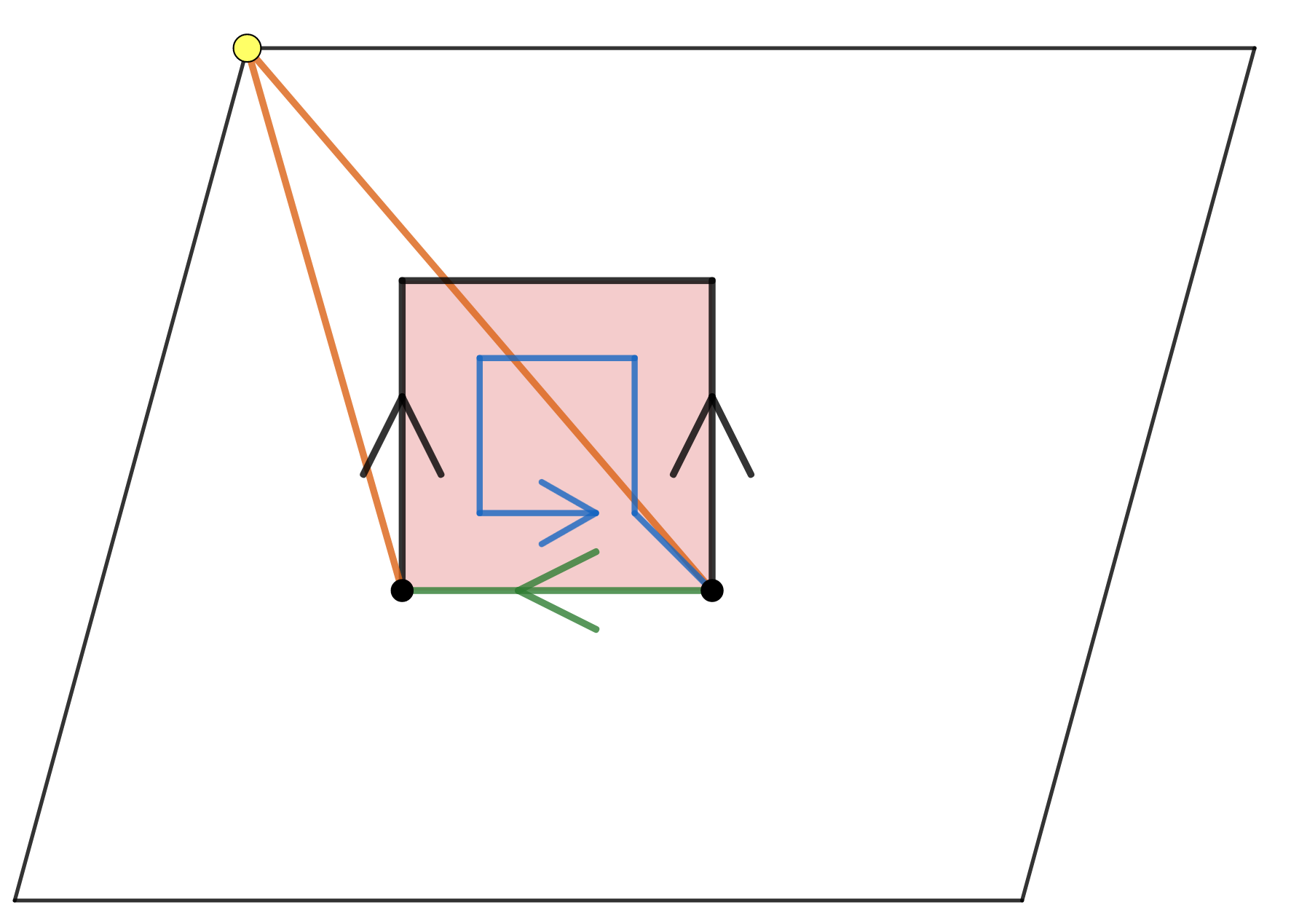}
				\put(14,67){$s.p$}
				\put(13,43){$(s.p-v_j)$}
				\put(32,54){$(s.p-v_i)$}
				\put(35,19){$(\overline{v_0(p)-v_j}) $}
				\put(42,37){$p$}
				\put(56,43){$i$}
				\put(29,43){$j$}
				
			\end{overpic}
			\caption{We claim that combining a series of edge transforms applied on all of the edges cut by the dual membrane, such as the two vertical black arrows in the figure, with the measurement operator changes the label of the magnetic component of the measurement operator as well as the labels of the blob ribbon operators, but does not have any additional effect. For this to be true, the edge transforms must have no effect on internal plaquettes, such as the red shaded one in the figure. Indeed, we know this to be true, because we already considered the same series of edge transforms in Section \ref{Section_condensation_magnetic_centre_case} when considering the condensation of the magnetic membrane operator. The shaded plaquette is affected by the two edge transforms on the vertical edges, the effects of which cancel due to the smooth variation in the paths (orange) from the start-point to the edges which determine the labels of the edge transforms. That is, the path $(s.p-v_i)$ from the start-point to edge $i$, when combined with the path segment $(\overline{v_0(p)-v_j})$ from the base of the plaquette, can be smoothly deformed into the path $(s.p-v_j)$ from the start-point to edge $j$.}
			\label{internal_plaquette_on_torus}	
		\end{center}
	\end{figure}

\begin{figure}[h]
		\begin{center}
			\begin{overpic}[width=0.6\linewidth]{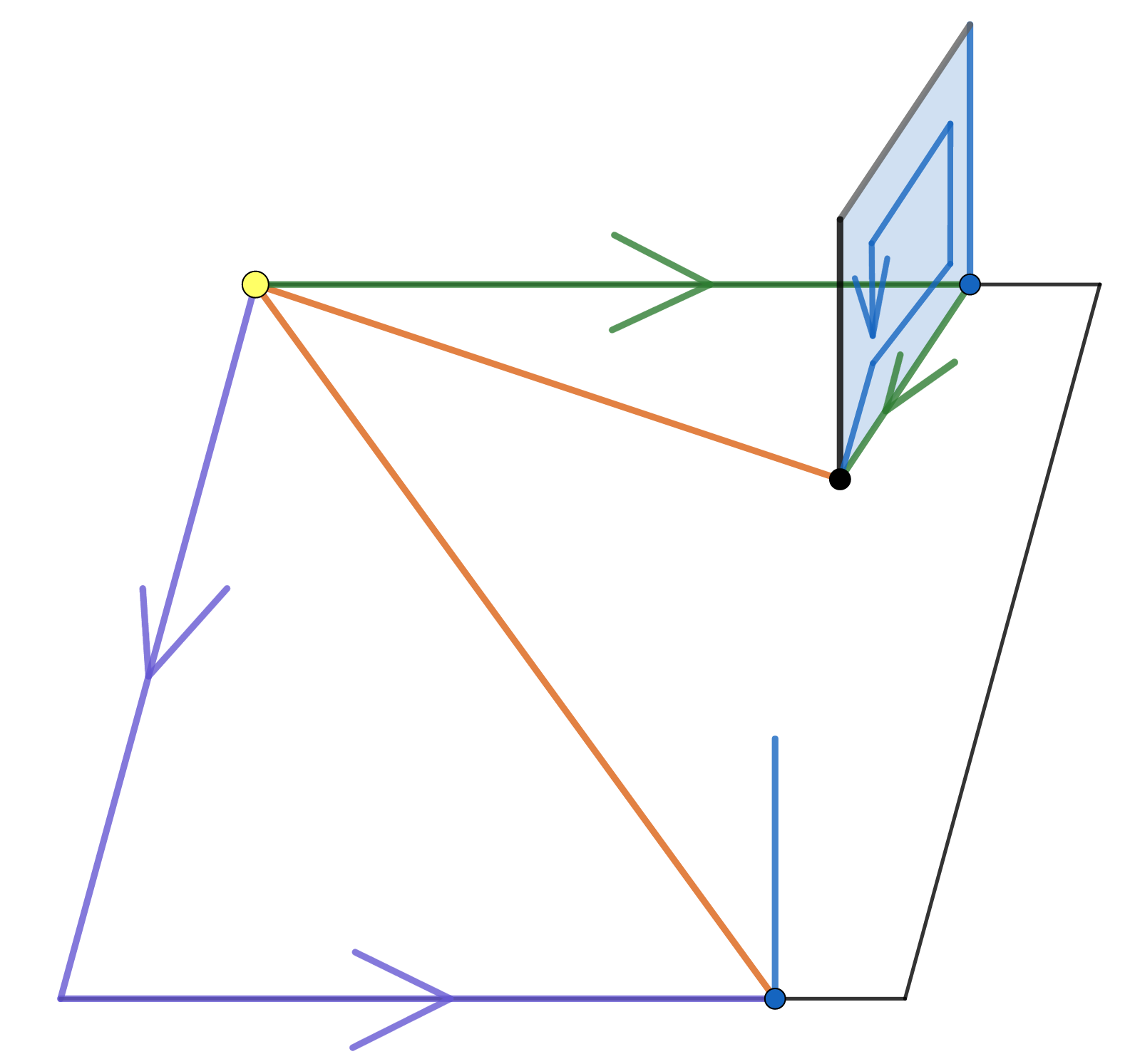}
				\put(17,67){\large $s.p$}
				\put(87,67){\large $v_2$}
				\put(76,49){\large $v_1$}
				\put(50,57){\large $t_1$}
				\put(41,38){\large $t_2$}
				\put(10,33){\large $c_2$}
				\put(41,3){\large $a$}
				\put(87,80){\large $i_2$}
				\put(71,63){\large $i_1$}
				\put(65,3){\large $v_2$}
				\put(70,16){\large $i_2$}
				\put(52,70){\large $a$}	
				\put(82,57){\large $b$}
				\put(52,70){\large $a$}	
				\put(80,72){\large $p$}
				\put(79,85){\large $u$}

			\end{overpic}
			\caption{While the paths from the start-point to the edges on which we apply the edge transforms vary smoothly in the bulk of the membrane, near the seams of the torus there is a discontinuity in these paths. We first consider plaquettes near the cycle $c_1$ (top). The blue shaded plaquette in the figure is affected by edge transforms on the vertical black and blue edges, $i_1$ and $i_2$ (which point upwards, with the arrows being omitted for space). The paths from the start-point to the edges which determine the label of the edge transforms vary largely between the two edges, because the path to the edge $i_2$ on the seam must cross the cycle $c_2$. This means that $t_1$ and $t_2$ differ by more than just a deformation and a local path. $t_1$ can be deformed into the green path $ab$, while $t_2$ can be deformed into the purple path $c_2\cdot a$. This difference in paths leads to the action of the edge transforms on the plaquettes being non-trivial, and this action can be expressed in terms of a blob ribbon operator passing through the plaquette.}	
			\label{boundary_plaquette_near_c1_tri_nontrivial}
		\end{center}
	\end{figure}

	Then, because the paths $(s.p-v_x)$ smoothly vary over the membrane, and the paths from $v_0(p)$ to $v_x$ are local to the plaquette, we can smoothly deform $(\overline{v_0(p)-v_j})\cdot(v_j-s.p) $ into $(v_0(p)-v_i)\cdot(v_i-s.p)$, as illustrated in Figure \ref{internal_plaquette_on_torus}. This means that the two paths differ only by elements of $\partial(E)$ and so give the same result when acting on $e$. Therefore, the action of the two edge transforms cancels. However, note that this relies on the smooth variation of the path $(s.p-v_i)$. Therefore, this does not necessarily hold for the plaquettes near the seam, where there is a discontinuity in this path (the same plaquettes for which the magnetic membrane operator has a discontinuity and which are affected by the blob ribbon operators). We now consider such plaquettes, starting with one adjacent to $c_1$, as shown in Figure \ref{boundary_plaquette_near_c1_tri_nontrivial}.

	The effect of the edge transforms on the edges $i_1$ and $i_2$ is to take the label of the plaquette from $e_p$ to
	$$[g(t_1)^{-1} \rhd e] e_p [g(b)^{-1} \rhd (g(t_2)^{-1} \rhd e^{-1})].$$
	We can see that $t_2$ may be deformed into the purple path $c_2 \cdot a$ from Figure \ref{boundary_plaquette_near_c1_tri_nontrivial}, with label $g_{c_2}g(a)$, while $t_1$ may be deformed into the green path $ab$, with label $g(a)g(b)$. Therefore, the final label of the plaquette is
	\begin{align*}
		[(g(a)g(b))^{-1} \rhd e] e_p [g(b)^{-1} \rhd ((g_{c_2}g(a))^{-1} \rhd e^{-1})] &= e_p [(g(a)g(b))^{-1} \rhd (e [g_{c_2}^{-1} \rhd e^{-1}])].
	\end{align*}
	This is the same as the action of a blob ribbon operator of label $e^{-1} [g_{c_2}^{-1} \rhd e]$ passing through the plaquette. This label is independent of which plaquette adjacent to $c_1$ that we consider. Therefore, we can write the additional action of the edge transforms on each such plaquette as a blob ribbon operator passing along the cycle. This can be combined with the blob ribbon operator that is part of the measurement operator, to take the label of that blob ribbon operator from $e_{c_1}$ to $e_{c_1} e^{-1} [g_{c_2}^{-1} \rhd e]$.

\begin{figure}[h]
		\begin{center}
			\begin{overpic}[width=0.6\linewidth]{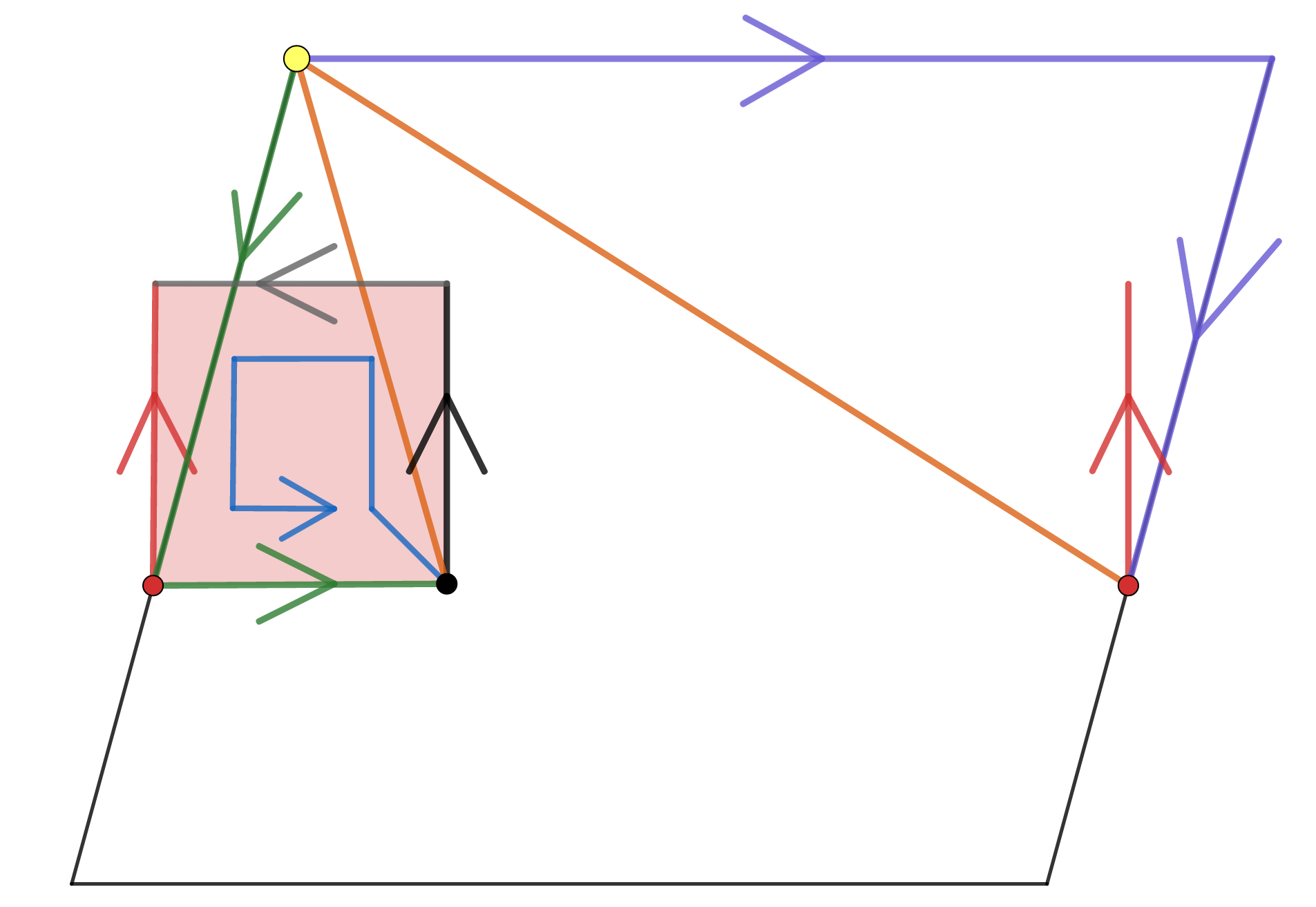}
				\put(17,65){\large $s.p$}
				\put(63,66){\large $c_1$}
				\put(7,24){\large $v_2$}
				\put(35,24){\large $v_1$}
				\put(60,42){\large $t_2$}
				\put(25,56){\large $t_1$}
				\put(87,24){\large $v_2$}
				\put(7,40){\large $i_2$}
				\put(35,40){\large $i_1$}
				\put(82,40){\large $i_2$}
				\put(24,21){\large $d$}
				\put(19,44){\large $u$}
				\put(15,52){\large $b$}
				\put(22,35){\large $p$}
				\put(95,52){\large $b$}
				
			\end{overpic}
			\caption{Having already considered the effect of the series of edge transforms on a plaquette near the seam at $c_1$, we now consider a plaquette near the seam at $c_2$. Just as before, the discontinuity in paths to the two vertical edges $i_1$ and $i_2$ (black and red respectively) on which we apply the edge transforms leads to a non-trivial action on the plaquette. For the purposes of calculation, it is convenient to deform the path $t_1$ to the green path $bd$ and $t_2$ to the purple path $c_1b$, which only changes the group elements associated to these paths by irrelevant factors in $\partial(E)$.}
			\label{boundary_plaquette_near_c2_tri_nontrivial}
		\end{center}
	\end{figure}

	In a similar way, we must consider plaquettes that are next to the cycle $c_2$, such as the plaquette shown in Figure \ref{boundary_plaquette_near_c2_tri_nontrivial}. The effect of the two edge transforms on this plaquette is to take the plaquette label from $e_p$ to 
	$$[g(d)^{-1} \rhd (g(t_2)^{-1} \rhd e)] e_p [g(t_1)^{-1} \rhd e^{-1}].$$
	Noting that we can deform the path $t_1$ into the green path $bd$ from Figure \ref{boundary_plaquette_near_c2_tri_nontrivial} and the path $t_2$ into the purple path $c_1b$, we can write the new plaquette label as
	$$[g(d)^{-1} \rhd ((g_{c_1}g(b))^{-1} \rhd e)] e_p [(g(b)g(d))^{-1} \rhd e^{-1}] = e_p [(g(b)g(d))^{-1} \rhd ([g_{c_1}^{-1} \rhd e] e^{-1})].$$
	This is the same as the action of a blob ribbon operator of label $e [g_{c_1}^{-1} \rhd e^{-1}]$ on the plaquette. This means that the additional action of the edge transforms on the plaquettes takes the label of the blob ribbon operator wrapping $c_2$ from $e_{c_2}$ to
	$$e_{c_2} e [g_{c_1}^{-1} \rhd e^{-1}].$$

	Putting this together with the transformation of the magnetic membrane operator and other blob ribbon operator under the edge transforms, we see that the edge transforms generate the equivalence relation
	\begin{equation}
		(h,e_{c_1}, e_{c_2}) \sim (\partial(e)h,e_{c_1} [g_{c_2}^{-1}\rhd e] e^{-1}, e_{c_2} [g_{c_1}^{-1}\rhd e^{-1}] e)
		\label{Equation_torus_edge_transform_outwards_tri_nontrivial}
	\end{equation}
	for our measurement operator. 
	
	We have now considered the restrictions on the measurement operator from all of the various energy terms. Putting all of our conditions together, we have
	\begin{align}
		\partial(e_m)^{-1}&=[g_{c_1},g_{c_2}] \label{C1} \tag{C1}\\
		\partial(e_{c_2})&=[g_{c_1},h]\label{C2} \tag{C2}\\
		\partial(e_{c_1})&=[h,g_{c_2}] \label{C3}\tag{C3}\\
		[h \rhd e_m^{-1}] &e_m e_{c_1}^{-1} [g_{c_1}^{-1} \rhd e_{c_1}] e_{c_2}^{-1} [g_{c_2}^{-1} \rhd e_{c_2}] =1_E \label{C4}\tag{C4}\\
		((g_{c_1},g_{c_2},h),(e_{c_1},e_{c_2},e_m)) &\sim (g(g_{c_1},g_{c_2},h)g^{-1},g \rhd(e_{c_1},e_{c_2},e_m)) \label{C5}\tag{C5}\\
		(g_{c_1},e_{c_2},e_m) &\sim (\partial(e)^{-1}g_{c_1},e_{c_2} [h \rhd e] e^{-1}, e_m e^{-1} [g_{c_2}^{-1} \rhd e]) \label{C6}\tag{C6}\\
		(g_{c_2},e_{c_1},e_m)& \sim (\partial(r)g_{c_2},e_{c_1}[h \rhd r] r^{-1}, e_m r^{-1} [(g_{c_1}g_{c_2}g_{c_1}^{-1}g_{c_1}^{-1}g_{c_2}^{-1}) \rhd r]) \label{C7}\tag{C7}\\
		(h, e_{c_1},e_{c_2}) &\sim (\partial(e)h, e_{c_1} [g_{c_2}^{-1} \rhd e] e^{-1}, e_{c_2} [ g_{c_1}^{-1}\rhd e^{-1}] e). \label{C8}\tag{C8}
	\end{align}
	
	Just as in the $\rhd$ trivial case, these conditions look similar to the ground state conditions for a 3-torus, as described in Ref. \cite{Bullivant2017}. We will now show that they are indeed equivalent, by finding an isomorphism between labels that maps our conditions onto the conditions that determine the ground state degeneracy. According to Ref. \cite{Bullivant2017}, the ground state degeneracy is calculated by counting the classes generated by the following conditions and equivalence relations on labels $(x,y,z,e,f,k)$:
	\begin{align}
		[x,z]=&\partial(e) \label{GS1} \tag{GS1}\\
		[x,y]=&\partial(f) \label{GS2}\tag{GS2}\\
		[y,z]=&\partial(k) \label{GS3}\tag{GS3}\\
		f [(yxy^{-1})\rhd k^{-1}]& [y \rhd e] [(yzy^{-1}) \rhd f^{-1}]k=e \label{GS4}\tag{GS4}\\
		(x,y,z,e,f,k) \sim & (axa^{-1}, aya^{-1}, aza^{-1}, a \rhd e, a \rhd f, a \rhd k) \label{GS5} \tag{GS5}\\
		\sim & (\partial(e_x)x,y,z, e_x e [z \rhd e_x^{-1}], e_x f [y \rhd e_x^{-1}], k) \label{GS6}\tag{GS6}\\
		\sim & (x,y, \partial(e_z)z, [x \rhd e_z] e e_z^{-1}, f, [y \rhd e_z] k e_z^{-1}) \label{GS7}\tag{GS7}\\
		\sim & (x, \partial(e_y)y,z, e, [x \rhd e_y] f e_y^{-1}, e_y k [z \rhd e_y^{-1}]).\label{GS8}\tag{GS8}
	\end{align}
	
	Note that as of the writing of this work, there appears to be a typo in the relation in Ref. \cite{Bullivant2017} corresponding to Equation \ref{GS7}, which we have corrected in the equations above after communication with the authors of Ref. \cite{Bullivant2017}. We try to match these to our conditions for the measurement operators by using the mapping
	$$e_m =e^{-1}, e_{c_1}=k^{-1}, e_{c_2}=f^{-1}, h=y, g_{c_1}=x^{-1}, g_{c_2}=z^{-1}.$$
	
	We will start by considering our conditions that are written in terms of equalities, Conditions \ref{C1} through \ref{C4}. These become
	\begin{align}
		\partial(e)&=[x^{-1},z^{-1}] \label{X1} \tag{X1}\\
		\partial(f^{-1})&=[x^{-1},y] \label{X2} \tag{X2}\\
		\partial(k^{-1})&=[y,z^{-1}] \label{X3} \tag{X3}\\
		[y \rhd e] e^{-1}& [x \rhd k^{-1}] k [z \rhd f^{-1}] f =1_E \label{X4} \tag{X4}
	\end{align}
	
	These relations may not immediately look the same as the relations that determine the ground state degeneracy, but there are some algebraic manipulations we can do to put these in the same form as the ground state relations. We start by considering the commutator relations. The relation $\partial(e)=[x^{-1},z^{-1}]$ is, when written explicitly
	$$\partial(e)=	x^{-1}z^{-1}xz.$$
	We can conjugate both sides of this expression by $zx$ and then use that $\partial(e)$ is in the centre of $G$ (due to our restrictions to the crossed module), to write this condition as
	\begin{align*}
		(zx)\partial(e)(zx)^{-1}&=zx[x^{-1}z^{-1}xz](zx)^{-1}\\
		&\implies \partial(e) = xz x^{-1}z^{-1} =[x,z].
	\end{align*}
	This means that our Condition \ref{X1} is the same as Condition \ref{GS1} from the ground state degeneracy calculation. For the other commutator relations, similar manipulations can be done. We can write Condition \ref{X2} as
	\begin{align*}
		\partial(f^{-1})&=[x^{-1},y]=x^{-1}yxy^{-1}\\
		&\implies\partial(f)=yx^{-1}y^{-1}x\\
		&\implies \partial(f)=x\partial(f)x^{-1}=xyx^{-1}y^{-1}\\
		& \implies \partial(f)=[x,y], 
	\end{align*}
	which agrees with Condition \ref{GS2} for the ground states. We can similarly write Condition \ref{X3} as
	\begin{align*}
		\partial(k^{-1})&=[y,z^{-1}]=yz^{-1}y^{-1}z\\
		&\implies \partial(k)=z^{-1}yzy^{-1}\\
		&\implies \partial(k)=z\partial(k)z^{-1}=yzy^{-1}z^{-1}\\
		& \implies \partial(k)=[y,z],
	\end{align*}
	which agrees with Condition \ref{GS3}.

	Now consider Condition \ref{X4}, which is the last restriction described by an equation. We can use the other conditions, together with the fact that $E$ is Abelian and $\partial$ maps to the centre of $G$, to simplify the corresponding ground state condition, Condition \ref{GS4}. Note that 
	$$yxy^{-1}x^{-1}=\partial(f)^{-1} \implies yxy^{-1}=\partial(f)^{-1}x,$$
	so that $$(yxy^{-1}) \rhd k^{-1} = (\partial(f)^{-1}x) \rhd k^{-1}=x \rhd k^{-1},$$
	where the last equality follows from the fact that elements in $\partial(E)$, like $\partial(f)^{-1}$, act trivially via $\rhd$ when $E$ is Abelian (this follows from the second Peiffer condition, Equation \ref{Equation_Peiffer_2} in the main text). Similarly $yzy^{-1}z^{-1}=\partial(k)$, so 
	$$(yzy^{-1}) \rhd f^{-1}=(\partial(k) z) \rhd f^{-1}=z \rhd f^{-1}.$$
	Therefore, the corresponding condition from the ground state calculation, Condition \ref{GS4}, which is originally
	$$f [(yxy^{-1}) \rhd k^{-1}] [y \rhd e] [(yzy^{-1}) \rhd f^{-1}]\:k =e,$$
	becomes
	$$f [x \rhd k^{-1}]\: [y \rhd e]\: [z \rhd f^{-1}]\: k\: e^{-1}=1_E.$$
	Because $E$ is Abelian, we are free to reorder the elements in this product, so Condition \ref{GS4} can be written as
	$$[y \rhd e] \: e^{-1} [x \rhd k^{-1}] k\: [z \rhd f^{-1}]\: f =1_E,$$
	which is the same as our condition, Condition \ref{X4}.

	Next we consider the conditions that are written as equivalence relations, Conditions \ref{C5} through \ref{C8}. After the mapping, these conditions become
	\begin{align}
		(x^{-1}, y, z^{-1}, e^{-1},f^{-1},k^{-1}) &\sim (gx^{-1}g^{-1},gyg^{-1},gz^{-1}g^{-1}, g \rhd e^{-1}, g \rhd f^{-1}, g \rhd k^{-1}) \label{X5} \tag{X5}\\
		(x^{-1},e^{-1}, f^{-1}) &\sim (\partial(b)^{-1}x^{-1}, e^{-1} b^{-1} [z \rhd b], f^{-1} [y \rhd b]b^{-1} ) \label{X6} \tag{X6}\\
		(z^{-1}, e^{-1}, k^{-1}) &\sim (\partial(r)z^{-1},e^{-1} r^{-1} [x \rhd r], k^{-1} [y \rhd r] r^{-1}) \label{X7} \tag{X7}\\
		(y,f^{-1},k^{-1}) &\sim (\partial(c)y, f^{-1} [x \rhd c^{-1}] c, k^{-1} [z \rhd c] c^{-1}). \label{X8} \tag{X8}
	\end{align}
	
	Looking at the first of these conditions, Condition \ref{X5}, we see that
	\begin{align*}
		(x^{-1},y,z^{-1},e^{-1},f^{-1},k^{-1}) &\sim (gx^{-1}g^{-1},gyg^{-1},gzg^{-1}, g \rhd k^{-1},g \rhd f^{-1},g \rhd e^{-1})\\
		&\implies (x,y,z,e,f,k) \sim (gxg^{-1},gyg^{-1}, gzg^{-1},g \rhd e, g \rhd f, g \rhd k),
	\end{align*}
	which agrees with the GSD calculation (Condition \ref{GS5}), when we replace the dummy index $g$ with $a$. Next we consider Condition \ref{X6}. This equivalence relation can be written as
	\begin{align*}
		(x^{-1},e^{-1}, f^{-1}) &\sim (\partial(b)^{-1}x^{-1}, e^{-1} b^{-1} [z \rhd b], f^{-1} [y \rhd b]b^{-1} )\\
		& \implies (x,e,f) \sim (x\partial(b), [z\rhd b^{-1}] b e, b [y \rhd b^{-1}] f)\\
		& \implies (x,e,f) \sim (\partial(b)x, e [z\rhd b^{-1}] b, f [y \rhd b^{-1}] b),
	\end{align*}
	where in the last step we used the fact that $E$ is Abelian and $\partial(E)$ is in the centre of $G$. This is then the same as Condition \ref{GS6} once we relabel the dummy index $b$ to $e_x$. Moving on to Condition \ref{X7}, we see that it can be rewritten as
	\begin{align*}
		(z^{-1}, e^{-1}, k^{-1}) &\sim (\partial(r)z^{-1},e^{-1} r^{-1} [x \rhd r], k^{-1} [y \rhd r] r^{-1})\\
		& \implies (z,e,k) \sim (z\partial(r^{-1}), [x\rhd r^{-1}] r e, r[y\rhd r^{-1}]k)\\
		&\implies (z,e,k) \sim (\partial(r^{-1})z, e r [x\rhd r^{-1}], k r[y\rhd r^{-1}])\\
		&\implies (z,e,k) \sim (\partial(e_z)z,e e_z^{-1} [x \rhd e_z], k e_z^{-1} [y \rhd e_z]),
	\end{align*}	
	where we relabelled the dummy index according to $r^{-1} = e_z$ in the last step. This agrees with the condition given in Condition \ref{GS7}. Finally we have Condition \ref{X8}, which can be rewritten as
	$$(y,f,k) \sim (\partial(c)y, [x \rhd c]f c^{-1}, [z\rhd c^{-1}]k c),$$
	which agrees with Condition \ref{GS8} of the ground state conditions when we relabel $c$ to $e_y$. Therefore, all of our conditions agree with the ground state conditions. This indicates that there are the same number of types of label for the unconfined excitations measured by the 2-torus as there are ground states on the 3-torus, as we may expect for a model describing a fully extended TQFT.

	\subsection{Construction of projectors for topological charge}
	\label{Section_3D_Topological_Charge_Torus_Projectors}
	In this section we follow our derivation of the space of topological measurement operators on a torus in Section \ref{Section_3D_Topological_Charge_Torus_Tri_nontrivial} by constructing a basis for this space consisting of projection operators that project to states with definite topological charge. At the moment, the space of measurement operators is defined in terms of operators of the form
	\begin{align*}
		T^{[e_{c_1}, e_{c_2},e_m,g_{c_1},g_{c_2},h]}(m) &= B^{e_{c_1}}(c_1)B^{e_{c_2}}(c_2)C^h_T(m)
		\delta(\hat{e}(m),e_m) \delta(\hat{g}(c_1),g_{c_1}) \delta(\hat{g}(c_2),g_{c_2}).
	\end{align*}

	Only certain combinations of the labels of these operators are allowed, as described by Conditions \ref{C1}-\ref{C4} in Section \ref{Section_3D_Topological_Charge_Torus_Tri_nontrivial}. Furthermore, we must take linear combinations of these, which must commute with the Hamiltonian, in order to obtain valid measurement operators. The allowed combinations are described by the equivalence relations \ref{C5}-\ref{C8} in Section \ref{Section_3D_Topological_Charge_Torus_Tri_nontrivial}. While following the equivalence relations allows us to construct well-defined measurement operators, the resulting operators may not be projectors. In order to find the projection operators, we have to change basis by taking appropriate combinations of these measurement operators. It is convenient to first introduce a set of operators which are equivalent to the $T$ operators, but with an isomorphism between the labels. We therefore define
	\begin{equation}
		Y^{(g_x,g_y,g_z,e_x,f_y,f_z)}(m)=T^{[f_y,f_z^{-1},e_x,g_y^{-1},g_z^{-1},g_x]}(m).
		\label{Equation_define_Y}
	\end{equation}
	Then we wish to construct the projectors using linear combinations of these $Y$ operators. As we described in Section \ref{Section_Torus_Charge} of the main text, these projectors are labelled by certain mathematical objects $R$ and $C$. These objects were introduced in Ref. \cite{Bullivant2020} for a tube algebra approach to studying the higher lattice gauge theory model, and to describe $R$ and $C$, we must first explain several concepts from Ref. \cite{Bullivant2020}.

	We will start by explaining how the space of $Y$ operators fits into the formalism of Ref. \cite{Bullivant2020}. The label of our $Y$ operator is given by a sextuple $(g_x,g_y,g_z,e_x,f_y,f_z)$. This sextuple can be split into two triples, $(g_y,g_z,e_x)$ and $(g_x,f_y,f_z)$. Given the space $(G,G,E)$ made of triples $(g_y,g_z,e_x)$, the boundary $\mathcal{G}$-colourings are defined as the triples satisfying \cite{Bullivant2020}
	\begin{equation}
		g_z=g_y^{-1} \partial(e_x^{-1})g_zg_y. \label{Equation_boundary_colouring_definition}
	\end{equation}
	
	This condition on the triple is similar to Condition \ref{C1} from Section \ref{Section_3D_Topological_Charge_Torus_Tri_nontrivial}, which describes a restriction to the labels of our $T$ operators. In fact, the conditions are equivalent, as we will verify later when we prove that our projectors satisfy all of the Conditions \ref{C1}-\ref{C8}. This indicates that these boundary $\mathcal{G}$-colourings provide an appropriate mathematical structure for the three labels $(g_y,g_z,e_x)$ of our measurement operators. The space of boundary $\mathcal{G}$-colourings can be divided into equivalence classes, using the equivalence relation \cite{Bullivant2020}
	\begin{align}
		(g_y,g_z,e_x) \overset{\mathrm{I}}{\sim} (a^{-1}\partial(b_2^{-1})g_ya, \: a^{-1}\partial(b_1^{-1})g_za, \:a^{-1} \rhd (b_1^{-1}(g_z \rhd b_2^{-1})e_x(g_y \rhd b_1) b_2)), \label{Equation_boundary_class_equivalence_relation}
	\end{align}
	for each $a \in G$ and $b_1,b_2 \in E$. We claim, and will show later, that a projector to definite topological charge is labelled by one of these classes of boundary $\mathcal{G}$-colourings (this is the object $C$ we mentioned earlier), together with another type of object that we will describe shortly. Given a class $C$, we denote an element of the class by
	$$(c_{y,i},c_{z,i},d_{x,i}),$$
	where $i$ is an index that runs from 1 to the size of the class, $|C|$. The element $(c_{y,1},c_{z,1},d_{x,1})$ is called the \textit{representative element} of class $C$ \cite{Bullivant2020}. Because the different elements are related to the representative element by the equivalence relation $\overset{\mathrm{I}}{\sim}$ from Equation \ref{Equation_boundary_class_equivalence_relation}, for each element of $C$ we can choose a particular set of $a$, $b_1$ and $b_2$ that connects that element to the representative one. That is, following Ref. \cite{Bullivant2020} we choose elements $p_{x,i} \in G$, $q_{y,i} \in E$ and $q_{z,i} \in E$ such that $(c_{y,i},c_{z,i},d_{x,i})$ satisfies
	\begin{equation}
		(c_{y,1},c_{z,1},d_{x,1})=(p_{x,i}^{-1}\partial(q_{z,i})^{-1}c_{y,1}p_{x,i}, \ p_{x,i}^{-1}\partial(q_{y,i}^{-1})c_{z,1}p_{x,i}, \ p_{x,i}^{-1} \rhd [q_{y,i}^{-1} (c_{z,i} \rhd q_{z,i}^{-1})d_{x,i} [c_{y,i} \rhd q_{y,i}] q_{z,i}]). \label{Equation_define_p_and_q}
	\end{equation}
	
	We can introduce further notation from Ref. \cite{Bullivant2020} in order to write this equation more compactly. We define
	\begin{equation}
		g^{k;f}= k^{-1}\partial(f)^{-1}g k, \label{Equation_superscript_notation_1}
	\end{equation}
	where $g$ and $k$ are elements of $G$, and $f$ is an element of $E$, and
	\begin{equation}
		e^{k,h_1,h_2;f_1,f_2}= k^{-1} \rhd \big(f_{1}^{-1} (h_2 \rhd f_{2})^{-1} e [h_{1} \rhd f_{1}] f_{2} ), \label{Equation_superscript_notation_2}
	\end{equation}
	where $k$, $h_1$ and $h_2$ are elements of $G$ and $f_1$, and $f_2$ are elements of $E$. Then we can write Equation \ref{Equation_define_p_and_q} as
	$$(c_{y,1},c_{z,1},d_{x,1})=(c_{y,i}^{p_{x,i};q_{z,i}}, \ c_{z,i}^{p_{x,i};q_{y,i}}, \ d_{x,i}^{p_{x,i},c_{y,i},c_{z,i};q_{y,i},q_{z,i}}).$$
	
	For each $i$, there may be many choices for $(p_{x,i},q_{y,i},q_{z,i})$ that would satisfy this relation. This means that the one that we choose is really a representative of a class of such elements. While we are generally free to pick this representative, we always take $p_{x,1}=1_G$, $q_{y,1}=1_E$ and $q_{z,1}=1_E$ (clearly the transformation that links $(c_{y,1},c_{z,1},d_{x,1})$ to itself is trivial).

	So far, we have introduced the boundary $\mathcal{G}$-colourings, described how the space of such colourings can be split into equivalence classes and introduced machinery to move between the elements of an individual class. However, this only describes the triple $(g_y,g_z,e_x)$ in our sextuple $(g_x,g_y,g_z,e_x,e_y,e_z)$ that represents the label of the $Y$ operator. In addition to the boundary $\mathcal{G}$-colourings, we have a space of ``bulk $\mathcal{G}$-colourings", which are triples $(g_x,e_y,e_z)$, where $g_x \in G$ and $e_y,e_z \in E$ \cite{Bullivant2020}. This space can be endowed with a structure that depends on the boundary $\mathcal{G}$-colourings. For a given boundary $\mathcal{G}$-colouring $(g_y,g_z,e_x)$, we split the space of bulk $\mathcal{G}$-colourings into classes defined by the equivalence relation \cite{Bullivant2020}
	\begin{equation}
		(g_x,e_y,e_z) \underset{{g_y,g_z}}{\overset{\mathrm{II}}{\sim}} (\partial( \lambda) g_x, [g_z \rhd \lambda] e_y \lambda^{-1}, [g_y \rhd \lambda] e_z \lambda^{-1}), \label{Equation_bulk_class_equivalence_relation}
	\end{equation}
	for all $\lambda \in E$. Note that this equivalence relation divides the space of bulk $\mathcal{G}$-colourings into classes, but the form of the relation depends on which boundary $\mathcal{G}$-colouring we are considering. This means that whenever we consider classes of bulk $\mathcal{G}$-colourings, we must specify the accompanying boundary $\mathcal{G}$-colouring. The set of equivalence classes corresponding to a particular boundary $\mathcal{G}$-colouring $(g_y,g_z,e_x)$ is denoted by $\mathfrak{B}_{g_y,g_z}$ \cite{Bullivant2020} (note that $e_x$ does not appear in the equivalence relation Equation \ref{Equation_bulk_class_equivalence_relation}). For a class $\mathcal{E}_{g_y,g_z}$ in $\mathfrak{B}_{g_y,g_z}$, the elements in $\mathcal{E}_{g_y,g_z}$ are denoted by 
	$$(s_{x,i},f_{y,i},f_{z,i}), $$
	where $i=1,2,...,|\mathcal{E}_{g_y,g_z}|$ and $|\mathcal{E}_{g_y,g_z}|$ is the size of the class. We call the element $(s_{x,1},f_{y,1},f_{z,1})$ the representative element for this class. So far, we have not put any restrictions on which triples we wish to consider, unlike for the boundary $\mathcal{G}$-colourings, which had to satisfy Equation \ref{Equation_boundary_colouring_definition}. However, just as we defined the equivalence relation among bulk $\mathcal{G}$-colourings to depend on a boundary $\mathcal{G}$-colouring, so we define a restricted set of bulk $\mathcal{G}$-colourings, which depends on the same boundary $\mathcal{G}$-colouring. Given a boundary $\mathcal{G}$-colouring $(g_y,g_z,e_x)$, the \textit{stabiliser group} $Z_{g_y,g_z}$ is made of the classes $\mathcal{E}_{g_y,g_z}$ in $\mathfrak{B}_{g_y,g_z}$ for which the representative element $(s_{x,1},f_{y,1},f_{z,1})$ satisfies \cite{Bullivant2020}
	\begin{align}
		(g_y,g_z,e_x) = (s_{x,1}^{-1} \partial(f_{z,1})^{-1}g_y s_{x,1}, \ s_{x,1}^{-1} \partial(f_{y,1})^{-1}g_z s_{x,1}, \ s_{x,1}^{-1} \rhd (f_{y,1}^{-1} [g_z \rhd f_{z,1}]^{-1}e_x [g_y \rhd f_{y,1}]f_{z,1})). \label{Equation_stabiliser_group_definition}
	\end{align}
	
	If the representative element satisfies this condition, so do the other elements, as we may expect from the freedom to choose a representative element for the class $\mathcal{E}_{g_y,g_z}$. To see this, we use Equation \ref{Equation_bulk_class_equivalence_relation} to write a general element of the class as 
	$$(s_{x,j},f_{y,j},f_{z,j})= (\partial(\lambda)s_{x,1}, \ [g_z \rhd \lambda]f_{y,1}\lambda^{-1}, \ [g_y \rhd \lambda]f_{z,1}\lambda^{-1})$$ for some $\lambda \in E$. Then we consider replacing the representative element $(s_{x,1},f_{y,1},f_{z,1})$ in the right-hand side of Equation \ref{Equation_stabiliser_group_definition}, to obtain
	\begin{equation}
		(s_{x,j}^{-1} \partial(f_{z,j})^{-1}g_y s_{x,j}, \ s_{x,j}^{-1} \partial(f_{y,j})^{-1}g_z s_{x,j}, \ s_{x,j}^{-1} \rhd (f_{y,j}^{-1} [g_z \rhd f_{z,j}]^{-1}e_x [g_y \rhd f_{y,j}]f_{z,j})). \label{Equation_stabiliser_group_non_representative_element}
	\end{equation}
	
	We want to write this in terms of the representative element $(s_{x,1},f_{y,1},f_{z,1})$, so that we can show that the other elements satisfy an equivalent condition. Looking at the first term, we have
	\begin{align*}
		s_{x,j}^{-1} \partial(f_{z,j})^{-1}g_y s_{x,j}&= (\partial(\lambda)s_{x,1})^{-1} \partial([g_y \rhd \lambda]f_{z,1}\lambda^{-1})^{-1} g_y (\partial(\lambda)s_{x,1})\\
		&=s_{x,1}^{-1}\partial(\lambda)^{-1} \partial(\lambda) \partial(f_{z,1})^{-1} \partial(g_y \rhd \lambda)^{-1} g_y (\partial(\lambda)s_{x,1})\\
		&=s_{x,1}^{-1} \partial(f_{z,1})^{-1} \partial(g_y \rhd \lambda^{-1}) g_y \partial(\lambda)s_{x,1}.
	\end{align*}
	Using the first Peiffer condition (Equation \ref{Equation_Peiffer_1} from the main text), we have $\partial(g_y \rhd \lambda^{-1})=g_y \partial(\lambda^{-1})g_y^{-1}$. Therefore
	\begin{align}
		s_{x,j}^{-1} \partial(f_{z,j})^{-1}g_y s_{x,j}&=s_{x,1}^{-1} \partial(f_{z,1})^{-1} \partial(g_y \rhd \lambda^{-1}) g_y \partial(\lambda)s_{x,1} \notag\\
		&=s_{x,1}^{-1} \partial(f_{z,1})^{-1} g_y \partial(\lambda^{-1})g_y^{-1}g_y \partial(\lambda)s_{x,1}\notag \\
		&=s_{x,1}^{-1} \partial(f_{z,1})^{-1} g_y s_{x,1}, \label{Equation_stabiliser_group_non_representative_element_first_term}
	\end{align}
	which is the same as the equivalent term in Equation \ref{Equation_stabiliser_group_definition}, meaning that this part of the condition is satisfied by $(s_{x,j},f_{y,j},f_{z,j})$ if it is satisfied by the representative element $(s_{x,1},f_{y,1},f_{z,1})$. Then the second term in Equation \ref{Equation_stabiliser_group_non_representative_element} is $s_{x,j}^{-1} \partial(f_{y,j})^{-1}g_z s_{x,j}$, which similarly gives us
	\begin{align}
		s_{x,j}^{-1} \partial(f_{y,j})^{-1}g_z s_{x,j}&= (\partial(\lambda)s_{x,1})^{-1} \partial([g_z \rhd \lambda]f_{y,1}\lambda^{-1})^{-1} g_z (\partial(\lambda)s_{x,1})\notag\\
		&=s_{x,1}^{-1}\partial(\lambda)^{-1} \partial(\lambda) \partial(f_{y,1})^{-1} \partial([g_z \rhd \lambda])^{-1} g_z (\partial(\lambda)s_{x,1})\notag\\
		&=s_{x,1}^{-1}\partial(f_{y,1})^{-1} g_z\partial(\lambda)^{-1}g_z^{-1} g_z \partial(\lambda)s_{x,1} \notag\\
		&=s_{x,1}^{-1}\partial(f_{y,1})^{-1} g_zs_{x,1}, \label{Equation_stabiliser_group_non_representative_element_second_term}
	\end{align}
	which matches the corresponding term in \ref{Equation_stabiliser_group_definition}. Finally the third term in Equation \ref{Equation_stabiliser_group_non_representative_element} is
	\begin{align}
		&s_{x,j}^{-1} \rhd (f_{y,j}^{-1} [g_z \rhd f_{z,j}]^{-1}e_x [g_y \rhd f_{y,j}]f_{z,j}) \notag\\
		&=(\partial(\lambda)s_{x,1})^{-1} \rhd \big( ([g_z \rhd \lambda]f_{y,1}\lambda^{-1})^{-1} [g_z \rhd ([g_y \rhd \lambda]f_{z,1}\lambda^{-1})]^{-1} e_x [g_y \rhd ([g_z \rhd \lambda]f_{y,1}\lambda^{-1})] ([g_y \rhd \lambda]f_{z,1}\lambda^{-1}) \big) \notag\\
		&= (s_{x,1}^{-1}\partial(\lambda)^{-1}) \rhd \big( \lambda f_{y,1}^{-1}[g_z \rhd \lambda]^{-1}[g_z \rhd \lambda] [g_z \rhd f_{z,1}^{-1}][(g_zg_y) \rhd \lambda^{-1} ] e_x [(g_yg_z) \rhd \lambda] [g_y \rhd f_{y,1}] [g_y \rhd \lambda^{-1}] [g_y \rhd \lambda]f_{z,1}\lambda^{-1} \big) \notag\\
		&=s_{x,1}^{-1} \rhd \big(\partial(\lambda)^{-1} \rhd \big( \lambda f_{y,1}^{-1} [g_z \rhd f_{z,1}^{-1}][(g_zg_y) \rhd \lambda^{-1} ] e_x [(g_yg_z) \rhd \lambda] [g_y \rhd f_{y,1}]f_{z,1}\lambda^{-1} \big) \big). \label{Equation_stabiliser_group_non_representative_element_third_term}
	\end{align}
	
	Because $(g_y,g_z,e_x)$ is a boundary $\mathcal{G}$-colouring, it satisfies Equation \ref{Equation_boundary_colouring_definition}, which means that $g_yg_z= \partial(e_x^{-1})g_zg_y$. Therefore, 
	\begin{align*}
		[(g_zg_y) \rhd \lambda^{-1} ] e_x [(g_yg_z) \rhd \lambda] &= [(g_zg_y) \rhd \lambda^{-1} ] e_x [(\partial(e_x^{-1})g_zg_y) \rhd \lambda].
	\end{align*}
	Using the second Peiffer condition (Equation \ref{Equation_Peiffer_2} in the main text), we can then write $$(\partial(e_x^{-1})g_zg_y) \rhd \lambda =\partial(e_x)^{-1} \rhd ((g_zg_y) \rhd \lambda)= e_x^{-1} [(g_zg_y) \rhd \lambda] e_x,$$
	so that
	\begin{align*}
		[(g_zg_y) \rhd \lambda^{-1} ] e_x [(g_yg_z) \rhd \lambda] &= [(g_zg_y) \rhd \lambda^{-1} ] e_x [(\partial(e_x^{-1})g_zg_y) \rhd \lambda]\\
		&= [(g_zg_y) \rhd \lambda^{-1} ] e_x e_x^{-1} [(g_zg_y) \rhd \lambda] e_x\\
		&=e_x.
	\end{align*}
	
	Substituting this into Equation \ref{Equation_stabiliser_group_non_representative_element_third_term}, we obtain
	\begin{align*}
		s_{x,j}^{-1} \rhd &(f_{y,j}^{-1} [g_z \rhd f_{z,j}]^{-1}e_x [g_y \rhd f_{y,j}]f_{z,j})\\
		&=s_{x,1}^{-1} \rhd \big(\partial(\lambda)^{-1} \rhd \big( \lambda f_{y,1}^{-1} [g_z \rhd f_{z,1}^{-1}]e_x [g_y \rhd f_{y,1}]f_{z,1}\lambda^{-1} \big)\big)\\
		&=s_{x,1}^{-1} \rhd \big(\lambda^{-1} \lambda f_{y,1}^{-1} [g_z \rhd f_{z,1}^{-1}]e_x [g_y \rhd f_{y,1}]f_{z,1}\lambda^{-1} \lambda \big)\\
		&=s_{x,1}^{-1} \rhd \big( f_{y,1}^{-1} [g_z \rhd f_{z,1}^{-1}]e_x [g_y \rhd f_{y,1}]f_{z,1} \big),
	\end{align*}
	where we again used the second Peiffer condition, Equation \ref{Equation_Peiffer_2} in the main text, to replace $\partial(\lambda)^{-1} \rhd$ by conjugation by $\lambda^{-1}$. The resulting expression matches the corresponding term in Equation \ref{Equation_stabiliser_group_definition}. Combining this with Equations \ref{Equation_stabiliser_group_non_representative_element_first_term} and \ref{Equation_stabiliser_group_non_representative_element_second_term}, we see that
	\begin{align*}
		(s_{x,j}^{-1} \partial(f_{z,j})^{-1}g_y s_{x,j},& s_{x,j}^{-1} \partial(f_{y,j})^{-1}g_z s_{x,j}, s_{x,j}^{-1} \rhd (f_{y,j}^{-1} [g_z \rhd f_{z,j}]^{-1}e_x [g_y \rhd f_{y,j}]f_{z,j}))\\
		&=(s_{x,1}^{-1} \partial(f_{z,1})^{-1}g_y s_{x,1}, s_{x,1}^{-1} \partial(f_{y,1})^{-1}g_z s_{x,1}, s_{x,1}^{-1} \rhd (f_{y,1}^{-1} [g_z \rhd f_{z,1}]^{-1}e_x [g_y \rhd f_{y,1}]f_{z,1})).
	\end{align*}
	
	This means that if the representative element $(s_{x,1},f_{y,1},f_{z,1})$ of a class $\mathcal{E}_{g_y,g_z}$ satisfies the condition 
	$$(g_y,g_z,e_x) = (s_{x,1}^{-1} \partial(f_{z,1})^{-1}g_y s_{x,1}, s_{x,1}^{-1} \partial(f_{y,1})^{-1}g_z s_{x,1}, s_{x,1}^{-1} \rhd (f_{y,1}^{-1} [g_z \rhd f_{z,1}]^{-1}e_x [g_y \rhd f_{y,1}]f_{z,1})),$$
	then so will an arbitrary element $(s_{x,j},f_{y,j},f_{z,j})$ of that class (and vice-versa). For that reason, we will sometimes refer to an element of a class belonging to a stabiliser group, rather than a class belonging to the stabiliser group as we have previously, by which we mean that the class containing that element belongs to the stabiliser group.

	We have been calling the stabiliser group a group, but we have not yet considered the product that is necessary to define a group. Given two classes $\mathcal{E}_{g_y,g_z}$ and $\mathcal{E}'_{g_y,g_z}$ in $Z_{g_y,g_z}$, with representative elements $(s_{x,1}, f_{y,1},f_{z,1})$ and $(s'_{x,1}, f'_{y,1},f'_{z,1})$ respectively, the product class $\mathcal{E}^{\text{prod.}}_{g_y,g_z}= \mathcal{E}_{g_y,g_z} \cdot \mathcal{E}'_{g_y,g_z}$ is the class in $Z_{g_y,g_z}$ with representative element \cite{Bullivant2020}
	\begin{equation}
		(s^{\text{prod.}}_{x,1}, f^{\text{prod.}}_{y,1},f^{\text{prod.}}_{z,1})=(s_{x,1} s_{x,1}', \ f_{y,1} [s_{x,1} \rhd f_{y,1}'],\ f_{z,1}[s_{x,1} \rhd f_{z,1}']). \label{Equation_stabiliser_group_product_definition}
	\end{equation}
	From this expression, we see that the inverse of the class $\mathcal{E}_{g_y,g_z}$ is the class with representative element
	$$(s_{x,1}^{-1}, \ s_{x,1}^{-1} \rhd f_{y,1}^{-1}, \ s_{x,1}^{-1} \rhd f_{z,1}^{-1}).$$
	In addition to defining this product on the classes, we will define it on the level of elements. Given two triples $(g,e_1,e_2)$ and $(g',e_1',e_2')$ in the group $(G,E,E)$, we define the product of those elements by 
	\begin{equation}
		(g,\ e_1,\ e_2)\cdot (g',\ e_1',\ e_2')=(gg',\ e_1 [g \rhd e_1'], \ e_2 [g \rhd e_2']). \label{Equation_triple_product_definition}
	\end{equation}

	The product on the level of elements respects the product on the level of classes in a stabiliser group. That is, given two classes $A$ and $B$ in a stabiliser group $Z_{g_y,g_z}$, whose product is $A\cdot B =C$, then the product of any element in $A$ with any element in $B$ is an element in $C$. To see this, consider the representative elements $(s_{x,1},f_{y,1},f_{z,1})$ and $(s'_{x,1},f'_{y,1},f'_{z,1})$ of classes $\mathcal{E}_{g_y,g_z}$ and $\mathcal{E'}_{g_y,g_z}$ respectively. From Equation \ref{Equation_bulk_class_equivalence_relation}, an arbitrary element of class $\mathcal{E}_{g_y,g_z}$ has the form
	$$(\partial(\lambda)s_{x,1}, [g_z \rhd \lambda] f_{y,1} \lambda^{-1}, [g_y \rhd \lambda]f_{z,1}\lambda^{-1}),$$
	and an arbitrary element of $\mathcal{E'}_{g_y,g_z}$ can be written as
	$$(\partial(\mu)s'_{x,1}, [g_z \rhd \mu] f'_{y,1} \mu^{-1}, [g_y \rhd \mu]f'_{z,1}\mu^{-1}),$$
	where $\lambda$ and $\mu$ are elements of $E$. The product of the elements of $\mathcal{E}_{g_y,g_z}$ and $\mathcal{E'}_{g_y,g_z}$ is given by
	\begin{align}
		(&\partial(\lambda)s_{x,1}, \ [g_z \rhd \lambda] f_{y,1} \lambda^{-1}, \ [g_y \rhd \lambda]f_{z,1}\lambda^{-1}) \cdot (\partial(\mu)s'_{x,1}, \ [g_z \rhd \mu] f'_{y,1} \mu^{-1}, \ [g_y \rhd \mu]f'_{z,1}\mu^{-1})\notag \\
		&= (\partial(\lambda)s_{x,1}\partial(\mu)s'_{x,1}, \ [g_z \rhd \lambda] f_{y,1} \lambda^{-1} (\partial(\lambda)s_{x,1}) \rhd ([g_z \rhd \mu] f'_{y,1} \mu^{-1}), \ [g_y \rhd \lambda]f_{z,1}\lambda^{-1} (\partial(\lambda)s_{x,1})\rhd([g_y \rhd \mu]f'_{z,1}\mu^{-1}) ). \label{Equation_product_bulk_in_product_class_1}
	\end{align}
	
	We wish to show that this product element lies in the product class $\mathcal{E}_{g_y,g_z} \cdot \mathcal{E'}_{g_y,g_z}$. From Equation \ref{Equation_stabiliser_group_product_definition}, the representative element of the product class $\mathcal{E}^{\text{prod.}}_{g_y,g_z}= \mathcal{E}_{g_y,g_z}\cdot \mathcal{E'}_{g_y,g_z}$ is $(s_{x,1}s'_{x,1}, f_{y,1} [s_{x,1} \rhd f'_{y,1}], f_{z,1} [s_{x,1} \rhd f'_{z,1}])$. This means that a general element of the product class has the form
	\begin{equation}
		(\partial(\nu) s_{x,1}s'_{x,1}, \ [g_z \rhd \nu] f_{y,1} [s_{x,1} \rhd f'_{y,1}] \nu^{-1}, \ [g_y \rhd \nu] f_{z,1} [s_{x,1} \rhd f'_{z,1}] \nu^{-1}), \label{Equation_bulk_product_class_element}
	\end{equation}
	for some $\nu \in E$. Now we will show that our product element
	$$(\partial(\lambda)s_{x,1}\partial(\mu)s'_{x,1}, \: [g_z \rhd \lambda] f_{y,1} \lambda^{-1} (\partial(\lambda)s_{x,1}) \rhd ([g_z \rhd \mu] f'_{y,1} \mu^{-1}), \:[g_y \rhd \lambda]f_{z,1}\lambda^{-1} (\partial(\lambda)s_{x,1})\rhd([g_y \rhd \mu]f'_{z,1}\mu^{-1}) )$$
	can be written in this form. We start by examining the first term, $\partial(\lambda)s_{x,1}\partial(\mu)s'_{x,1}$. The first Peiffer condition (Equation \ref{Equation_Peiffer_1} in the main text) tells us that $s_{x,1}\partial(\mu)=\partial(s_{x,1} \rhd\mu)s_{x,1}$, so that
	\begin{align}
		\partial(\lambda)s_{x,1}\partial(\mu)s'_{x,1} &= \partial(\lambda)\partial(s_{x,1}\rhd \mu)s_{x,1}s'_{x,1} \notag\\
		&= \partial(\lambda [s_{x,1}\rhd \mu])s_{x,1}s'_{x,1}. \label{Equation_product_bulk_in_product_class_workings_0}
	\end{align}
	
	This matches the corresponding term for an element of the product class (from Equation \ref{Equation_bulk_product_class_element}) if we take $\nu = \lambda [s_{x,1}\rhd \mu]$. Next, consider the second term of our product element, 
	$$[g_z \rhd \lambda] f_{y,1} \lambda^{-1} (\partial(\lambda)s_{x,1}) \rhd ([g_z \rhd \mu] f'_{y,1} \mu^{-1}).$$
	We can start by using the second Peiffer condition (Equation \ref{Equation_Peiffer_2} in the main text) to write that
	\begin{align*}
		[g_z \rhd \lambda] f_{y,1} \lambda^{-1} (\partial(\lambda)s_{x,1}) \rhd ([g_z \rhd \mu] f'_{y,1} \mu^{-1}) &= [g_z \rhd \lambda] f_{y,1} \lambda^{-1} [\partial(\lambda) \rhd (s_{x,1} \rhd ([g_z \rhd \mu] f'_{y,1} \mu^{-1}))]\\
		&= [g_z \rhd \lambda] f_{y,1} \lambda^{-1} [\lambda (s_{x,1} \rhd ([g_z \rhd \mu] f'_{y,1} \mu^{-1}))\lambda^{-1}]\\
		&= [g_z \rhd \lambda] f_{y,1} [s_{x,1} \rhd ([g_z \rhd \mu] f'_{y,1} \mu^{-1})]\lambda^{-1}.
	\end{align*}
	Next we expand the expression $[s_{x,1} \rhd ([g_z \rhd \mu] f'_{y,1} \mu^{-1})]$ to obtain
	\begin{align}
		[g_z \rhd \lambda] f_{y,1} [s_{x,1} \rhd ([g_z \rhd \mu] f'_{y,1} \mu^{-1})]\lambda^{-1} &= [g_z \rhd \lambda] f_{y,1} [(s_{x,1}g_z) \rhd \mu] [s_{x,1} \rhd f'_{y,1}] [s_{x,1} \rhd \mu^{-1}] \lambda^{-1}. \label{Equation_product_bulk_in_product_class_workings_1}
	\end{align}
	
	We wish to write this in terms of $\lambda [s_{x,1} \rhd \mu]$, which we identified as $\nu$, so the term $[(s_{x,1}g_z) \rhd \mu]$ appears problematic. However, we can use the condition for the class $\mathcal{E}_{g_y,g_z}$ to be in the stabiliser group $Z_{g_y,g_z}$, which tells us that $s_{x,1}^{-1}\partial(f_{y,1})^{-1}g_z s_{x,1}=g_z$ (see Equation \ref{Equation_stabiliser_group_definition}). Therefore, 
	\begin{align*}
		[(s_{x,1}g_z) \rhd \mu] &= [(\partial(f_{y,1})^{-1}g_z s_{x,1}) \rhd \mu]\\
		&= f_{y,1}^{-1} [(g_z s_{x,1}) \rhd \mu] f_{y,1},
	\end{align*}
	where in the last line we used the second Peiffer condition (Equation \ref{Equation_Peiffer_2} in the main text). Substituting this into Equation \ref{Equation_product_bulk_in_product_class_workings_1} gives us
	\begin{align}
		[g_z \rhd \lambda] f_{y,1} [(s_{x,1}g_z) \rhd \mu] [s_{x,1} \rhd f'_{y,1}] [s_{x,1} \rhd \mu^{-1}] \lambda^{-1} &= [g_z \rhd \lambda] f_{y,1} f_{y,1}^{-1} [(g_z s_{x,1}) \rhd \mu] f_{y,1} [s_{x,1} \rhd f'_{y,1}] [s_{x,1} \rhd \mu^{-1}] \lambda^{-1} \notag\\
		&= [g_z \rhd \lambda] [(g_z s_{x,1}) \rhd \mu] f_{y,1} [s_{x,1} \rhd f'_{y,1}] [s_{x,1} \rhd \mu^{-1}] \lambda^{-1} \notag \\
		&= [g_z \rhd ( \lambda [s_{x,1} \rhd \mu])] f_{y,1} [s_{x,1} \rhd f'_{y,1}] (\lambda [s_{x,1} \rhd \mu])^{-1}, \label{Equation_product_bulk_in_product_class_workings_2}
	\end{align}
	which we recognise from Equation \ref{Equation_bulk_product_class_element} as the corresponding term in an element of the product class, with $\nu = \lambda [s_{x,1} \rhd \mu]$. Following analogous steps with the third term of the product element
	$$[g_y \rhd \lambda]f_{z,1}\lambda^{-1} (\partial(\lambda)s_{x,1})\rhd([g_y \rhd \mu]f'_{z,1}\mu^{-1}),$$
	we obtain
	\begin{align}
		[g_y \rhd \lambda]f_{z,1}\lambda^{-1} (\partial(\lambda)s_{x,1})\rhd([g_y \rhd \mu]f'_{z,1}\mu^{-1})&= [g_y \rhd (\lambda [s_{x,1} \rhd \mu])] f_{z,1} [s_{x,1} \rhd f'_{z,1}] (\lambda [s_{x,1} \rhd \mu])^{-1} \label{Equation_product_bulk_in_product_class_workings_3},
	\end{align}
	which we recognise as the corresponding term from the element of the product class from Equation \ref{Equation_bulk_product_class_element}, with $\nu= \lambda [s_{x,1} \rhd \mu]$. Combining this with the previous two terms (from Equations \ref{Equation_product_bulk_in_product_class_workings_0} and \ref{Equation_product_bulk_in_product_class_workings_2}), we see that the product of our arbitrary elements from classes $\mathcal{E}_{g_y,g_z}$ and $\mathcal{E'}_{g_y,g_z}$ is
	\begin{align}
		(\partial(\lambda)s_{x,1}, \ [g_z \rhd \lambda] f_{y,1} \lambda^{-1}&, \ [g_y \rhd \lambda]f_{z,1}\lambda^{-1}) \cdot (\partial(\mu)s'_{x,1}, \ [g_z \rhd \mu] f'_{y,1} \mu^{-1}, \ [g_y \rhd \mu]f'_{z,1}\mu^{-1}) \notag \\
		&=(\partial(\lambda [s_{x,1}\rhd \mu])s_{x,1}s'_{x,1}, \ [g_z \rhd ( \lambda [s_{x,1} \rhd \mu])] f_{y,1} [s_{x,1} \rhd f'_{y,1}] (\lambda [s_{x,1} \rhd \mu])^{-1}, \notag\\ & \hspace{1cm} [g_y \rhd (\lambda [s_{x,1} \rhd \mu])] f_{z,1} [s_{x,1} \rhd f'_{z,1}] (\lambda [s_{x,1} \rhd \mu])^{-1}) \label{Equation_product_bulk_in_product_class_result}
	\end{align}
	and is indeed an element of the product class $\mathcal{E}_{g_y,g_z} \cdot \mathcal{E'}_{g_y,g_z}$.

	So far, we have discussed the bulk $\mathcal{G}$-colourings and stabiliser group associated to an individual $\mathcal{G}$-colouring. However, the boundary $\mathcal{G}$-colourings themselves fit into equivalence classes, so we can also associate a stabiliser group with a class of $\mathcal{G}$-colourings. This is done in the natural way, by using the quantities related to the representative element of the class of boundary $\mathcal{G}$-colourings \cite{Bullivant2020}. Given a class $C$ of boundary $\mathcal{G}$-colourings, with representative element $(c_{y,1},c_{z,1},d_{x,1})$, the bulk $\mathcal{G}$-colourings associated to the class are the same as the ones associated to the representative element. That is, the classes of bulk $\mathcal{G}$-colourings associated to class $C$ are defined by the equivalence relation Equation \ref{Equation_bulk_class_equivalence_relation} for $g_y=c_{y,1}$ and $g_z=c_{z,1}$. The set of these classes is denoted by $\mathfrak{B}_C$ (so that $\mathfrak{B}_C=\mathfrak{B}_{c_{y,1},c_{z,1}}$). Then the stabiliser group of $C$, which is denoted by $Z_C$, is made of the classes in $\mathfrak{B}_C$ for which the representative element $(s_{x,1},f_{y,1},f_{z,1})$ satisfies \cite{Bullivant2020}
	\begin{equation}
		(c_{y,1},c_{z,1},d_{x,1})=(s_{x,1}^{-1} \partial(f_{z,1})^{-1} c_{y,1}s_{x,1}, \: s_{x,1}^{-1} \partial(f_{y,1})^{-1} c_{z,1}s_{x,1}, \: s_{x,1}^{-1} \rhd (f_{y,1}^{-1} (c_{z,1} \rhd f_{z,1})^{-1} d_{x,1} [c_{y,1} \rhd f_{y,1}] f_{z,1} )). \label{Equation_stabiliser_group_Z_C_appendix}
	\end{equation}
	That is, the stabiliser group $Z_C$ associated to the class $C$ is given by $Z_C=Z_{c_{y,1},c_{z,1}}$. The precise elements of the stabiliser group $Z_C$ depend on the choice of representative element $(c_{y,1},c_{z,1},d_{x,1})$ of $C$, as can be seen from Equation \ref{Equation_stabiliser_group_Z_C_appendix}. However, this choice of representative is arbitrary. This means that the different stabiliser groups that we would get by choosing different representatives of $C$ should be somehow equivalent. This manifests in the form of an isomorphism between the stabiliser groups that would arise from the different choices of representative. Consider the stabiliser group $Z_{C,i}= Z_{c_{y,i},c_{z,i}}$, i.e., the stabiliser group associated to the $i$th element of $C$ rather than the representative element. Given an element $\mathcal{E}_{c_{y,i},c_{z,i}}$ of $Z_{C,i}$, we define the class of bulk $\mathcal{G}$-colourings $[\mathcal{E}_{C}^{\text{stab.}}]_{i,i}$ with representative element \cite{Bullivant2020} 
	\begin{equation}
		\big(p_{x,i}^{-1}s_{x,1}p_{x,i}, \: \: p_{x,i}^{-1} \rhd (q_{y,i}^{-1} f_{y,1} s_{x,1} \rhd q_{y,i}), \: \: p_{x,i}^{-1} \rhd (q_{z,i}^{-1} f_{z,1}(s_{x,1} \rhd q_{z,i}))\big), \label{Equation_definition_stabiliser_group_isomorphism}
	\end{equation}
	whose other elements are obtained from the equivalence relation Equation \ref{Equation_bulk_class_equivalence_relation} with $g_y=c_{y,1}$ and $g_z=c_{z,1}$. The double subscript ``$i,i$" in $[\mathcal{E}_{C}^{\text{stab.}}]_{i,i}$ is used because this object is an example of a more general construct with two independent subscripts defined in Ref. \cite{Bullivant2020}, but we will only need the case where the subscripts are the same. The class $[\mathcal{E}_{C}^{\text{stab.}}]_{i,i}$ defined in this way belongs to the stabiliser group $Z_C$, as shown in Ref. \cite{Bullivant2020}, so that Equation \ref{Equation_definition_stabiliser_group_isomorphism} gives us a map between the two stabiliser groups $Z_{C,i}$ and $Z_C$. To see that this mapping is indeed an isomorphism, consider two classes $\mathcal{E}_{c_{y,i},c_{z,i}}$ and $\mathcal{E'}_{c_{y,i},c_{z,i}}$ in $Z_{C,i}$. The representative elements of these two classes are denoted by $(s_{x,1}, f_{y,1},f_{z,1})$ and $(s'_{x,1}, f'_{y,1},f'_{z,1})$ respectively. Then under the mapping, these classes become $[\mathcal{E}_{C}^{\text{stab.}}]_{i,i}$ and $[\mathcal{E'}_{C}^{\text{stab.}}]_{i,i}$, with representative elements
	\begin{equation}
		\big(p_{x,i}^{-1}s_{x,1}p_{x,i}, \: p_{x,i}^{-1} \rhd (q_{y,i}^{-1} f_{y,1} [s_{x,1} \rhd q_{y,i}]), \: p_{x,i}^{-1} \rhd (q_{z,i}^{-1} f_{z,1}[s_{x,1} \rhd q_{z,i}])\big)
	\end{equation}
	and
	\begin{equation}
		\big(p_{x,i}^{-1}s'_{x,1}p_{x,i}, \: p_{x,i}^{-1} \rhd (q_{y,i}^{-1} f'_{y,1} [s'_{x,1} \rhd q_{y,i}]), \: p_{x,i}^{-1} \rhd (q_{z,i}^{-1} f'_{z,1}[s'_{x,1} \rhd q_{z,i}])\big).
	\end{equation}
	
	We wish to show that the product of the two classes is preserved by the mapping from $Z_C$ to $Z_{C,i}$, so that 
	$$[(\mathcal{E}_{c_{y,i},c_{z,i}} \cdot \mathcal{E'}_{c_{y,i},c_{z,i}})_C]_{i,i}=[\mathcal{E}_{C}^{\text{stab.}}]_{i,i} \cdot [\mathcal{E'}_{C}^{\text{stab.}}]_{i,i}.$$
	From Equation \ref{Equation_stabiliser_group_product_definition}, the representative element of $[\mathcal{E}^{\text{stab.}}_C]_{i,i} \cdot [\mathcal{E'}^{\text{stab.}}_C]_{i,i}$ is
	\begin{align}
		\big(p_{x,i}^{-1} s_{x,1}& p_{x,i} p_{x,i}^{-1} s'_{x,1} p_{x,i}, \notag \\ p_{x,i}^{-1}& \rhd (q_{y,i}^{-1} f_{y,1} [s_{x,1} \rhd q_{y,i}])[(p_{x,i}^{-1}s_{x,1}p_{x,i} p_{x,i}^{-1}) \rhd (q_{y,i}^{-1}f'_{y,1}[s_{x,1}' \rhd q_{y,i}])], \notag \\ & p_{x,i}^{-1} \rhd (q_{z,i}^{-1} f_{z,1} [s_{x,1} \rhd q_{z,i}])(p_{x,i}^{-1} s_{x,1} p_{x,i}) \rhd (p_{x,i}^{-1} \rhd (q_{z,i}^{-1} f'_{z,1}[s_{x,1}' \rhd q_{z,i}]))\big) \notag \\ &\hspace{1cm}= \big(p_{x,i}^{-1}s_{x,1}s'_{x,1}p_{x,i}, \notag \\ & \hspace{2cm}p_{x,i}^{-1} \rhd (q_{y,i}^{-1}f_{y,1} [s_{x,1} \rhd q_{y,i}] [s_{x,1} \rhd q_{y,i}^{-1}] [s_{x,1} \rhd f_{y,1}'] [(s_{x,1} s'_{x,1}) \rhd q_{y,i}]), \notag\\& \hspace{2.5cm} p_{x,i}^{-1} \rhd (q_{z,i}^{-1}f_{z,1} [s_{x,1} \rhd q_{z,i}] [s_{x,1} \rhd q_{z,i}^{-1}] [s_{x,1} \rhd f_{z,1}'] [(s_{x,1} s'_{x,1}) \rhd q_{z,i}])\big). \label{Equation_stabiliser_isomorphism_product_1}
	\end{align}
	
	From Equation \ref{Equation_stabiliser_group_product_definition}, the representative element of $\mathcal{E}_{c_{y,i},c_{z,i}} \cdot \mathcal{E'}_{c_{y,i},c_{z,i}}$ is given by 
	\begin{equation}
		(s^{\text{prod.}}_{x,1}, \ f^{\text{prod.}}_{y,1},\ f^{\text{prod.}}_{z,1})=(s_{x,1} s_{x,1}', \ f_{y,1} (s_{x,1} \rhd f_{y,1}'), \ f_{z,1}(s_{x,1} \rhd f_{z,1}')).
	\end{equation}
	Inserting this into Equation \ref{Equation_stabiliser_isomorphism_product_1} tells us that the representative element of $[\mathcal{E}^{\text{stab.}}_C]_{i,i} \cdot [\mathcal{E'}^{\text{stab.}}_C]_{i,i}$ is
	\begin{align*}
		&\big(p_{x,i}^{-1}s_{x,1}s'_{x,1}p_{x,i}, \\ & \quad p_{x,i}^{-1} \rhd (q_{y,i}^{-1}f_{y,1} [s_{x,1} \rhd q_{y,i}] [s_{x,1} \rhd q_{y,i}^{-1}] [s_{x,1} \rhd f_{y,1}'] [(s_{x,1} s'_{x,1}) \rhd q_{y,i}]),\\ & \quad p_{x,i}^{-1} \rhd (q_{z,i}^{-1}f_{z,1} [s_{x,1} \rhd q_{z,i}] [s_{x,1} \rhd q_{z,i}^{-1}] [s_{x,1} \rhd f_{z,1}'] [(s_{x,1} s'_{x,1}) \rhd q_{z,i}])\big)\\
		&=\big(p_{x,i}^{-1} s_{x,1}^{\text{prod.}}p_{x,i}, \ p_{x,i}^{-1} \rhd (q_{y,i}^{-1} f_{y,1} [s_{x,1} \rhd f'_{y,1}] [(s_{x,1}s'_{x,1}) \rhd q_{y,i}]), \ p_{x,i}^{-1} \rhd (q_{z,i}^{-1} f_{z,1} [s_{x,1} \rhd f'_{z,1}] [(s_{x,1}s'_{x,1}) \rhd q_{z,i}])\big)\\
		&=\big(p_{x,i}^{-1}s_{x,1}^{\text{prod.}} p_{x,i}, \ p_{x,i}^{-1} \rhd (q_{y,i}^{-1} f_{y,1}^{\text{prod.}} [s_{x,1}^{\text{prod.}} \rhd q_{y,i}]), \ p_{x,i}^{-1} \rhd (q_{z,i}^{-1} f_{z,1}^{\text{prod.}} [s_{x,1}^{\text{prod.}} \rhd q_{z,i}])\big),
	\end{align*}
	which is the representative of $[(\mathcal{E}^{\text{prod}})^{\text{ stab.}}_C]_{i,i}$. This therefore establishes a homomorphism between $Z_C$ and $Z_{C,i}$. The mapping is invertible, so it is also an isomorphism. This means that, as we expect, the structure of the stabiliser group does not depend on the arbitrary choice of representatives for $C$ and so the stabiliser group is a sensible object to consider.

	The stabiliser groups play an important role in our projectors to definite topological charge. Given a stabiliser group $Z_C$, we can construct the irreducible representations of that group. A projector to definite topological charge is labelled by a class $C$ of boundary $\mathcal{G}$-colourings and an irrep $R$ of the associated stabiliser group $Z_C$. We propose that the projector labelled by class $C$ of the boundary $\mathcal{G}$-colourings and irrep $R$ of the corresponding stabiliser group is given by
	\begin{align}
		P^{R,C}(m)=& N_{R,C} \sum_{i=1}^{|C|} \sum_{\substack{\mathcal{E}_{c_{y,i},c_{z,i}}\\ \in \mathfrak{B}_{c_{y,i},c_{z,i}}}} \sum_{m=1}^{|R|} \sum_{\lambda \in E} \delta(c_{y,i},c_{y,i}^{s_{x,1};f_{z,1}}) \delta(c_{z,i},c_{z,i}^{s_{x,1};f_{y,1}})\delta(d_{x,i},d_{x,i}^{s_{x,1},c_{y,i},c_{z,i};f_{y,1},f_{z,1}}) \notag\\
		&D^R_{m,m}([\mathcal{E}_C^{\text{stab.}}]_{i,i}) Y^{(\partial(\lambda) s_{x,1},\: c_{y,i}, \: c_{z,i}, \: d_{x,i}, \: [c_{z,i} \rhd \lambda] f_{y,1} \lambda^{-1}, \: [c_{y,i} \rhd \lambda] f_{z,1} \lambda^{-1} )}(m), \label{Equation_torus_projector_definition_appendix_1}
	\end{align}
	where $N_{R,C}= \frac{|R|}{|E||Z_C|}$ is a normalization factor. In the rest of this section we will demonstrate that the operators defined in this way are indeed orthogonal projectors and that they satisfy the conditions laid out in Section \ref{Section_3D_Topological_Charge_Torus_Tri_nontrivial} for a valid measurement operator (i.e., that the coefficients of the different $T$ operators satisfy the conditions laid out in Section \ref{Section_3D_Topological_Charge_Torus_Tri_nontrivial}, where the $T$ operators are related to the $Y$ operators by Equation \ref{Equation_define_Y}). First, we wish to check that the operators are a set of orthogonal projectors. That is, we wish to show that they satisfy $P^{R,C}(m) P^{R',C'}(m)=\delta(C,C') \delta(R,R') P^{R,C}(m)$. Before we do this, we should note that, as explained in Section \ref{Section_3D_Topological_Charge_Torus_Tri_trivial}, we must apply a projector onto the space where the torus is unexcited before we apply any measurement operators. This projector is not included in the $T$ operators, or indeed in the projectors $P^{R,C}(m)$, so we must first show that this preliminary projection does not affect our algebra when we consider applying two measurement operators in sequence. The total measurement operator that we apply, including the preliminary projection, takes the form
	$$\hat{O}=P^{R,C}(m) \hat{P}=\sum_{X} \alpha_{X} T^{X}(m) \hat{P},$$
	where $\hat{P}$ is the projector onto an unexcited surface, $X$ represents the labels of the $T$ operators and $\alpha_{X}$ is the set of coefficients for the different $T^X(m)$ operators in the measurement operator. Then considering the product of this with another measurement operator
	$$\hat{Q}= P^{R',C'}(m) \hat{P}= \sum_{X} \beta_{X} T^{X}(m) \hat{P},$$ 
	we have
	$$\hat{O} \hat{Q} = \sum_{X} \alpha_{X} T^{X}(m) \hat{P} \sum_{X'} \beta_{X'} T^{X'}(m) \hat{P}.$$
	This means that when we apply two measurement operators in sequence, there is a copy of the projector $\hat{P}$ sandwiched between the two sets of $T$ operators from the measurement operators. However, because our sum of $T$ operators will commute with the energy terms (including the projector, which is made up of energy terms) due to the conditions we set on the measurement operator, we have that 
	\begin{align*}
		\sum_{X} \alpha_{X} T^{X}(m) \hat{P} \sum_{X'} \beta_{X'} T^{X'}(m) \hat{P} &= \sum_{X} \alpha_{X} T^{X}(m) \sum_{X'} \beta_{X'} T^{X'}(m) \hat{P} \hat{P}\\
		&=\sum_{X} \alpha_{X} T^{X}(m) \sum_{X'} \beta_{X'} T^{X'}(m) \hat{P}\\
		&= P^{R,C}(m) P^{R',C'}(m) \hat{P},
	\end{align*}
	where $\hat{P}^2= \hat{P}$ because $\hat{P}$ is a projector. This means that when we apply two measurement operators in sequence, we can directly combine the $T$ operators, without worrying about the preliminary projection operator $\hat{P}$. Therefore, when demonstrating that our measurement operators are projectors, we can work with $P^{R,C}(m)$, rather than $P^{R,C}(m)\hat{P}$.

	We now consider a product of two of our measurement operators, $P^{R,C}(m)$ and $P^{R',C'}(m)$. This product can be written as a linear combination of products of two $T$ operators. In order to write this in terms of a single projection operator, we must decompose a product of two $T$ operators into a linear combination of $T$ operators. That is, we must find the algebra satisfied by the $T$ operators. We have
	\begin{align}
		T^{[e_1,e_2,e_3,g_1,g_2,g_3]}(m) T^{[f_1,f_2,f_3,k_1,k_2,k_3]}(m) =& B^{e_1}(c_1) B^{e_2}(c_2) C_T^{g_3}(m) \delta(\hat{e}(m),e_3) \delta(\hat{g}(c_1),g_1) \delta(\hat{g}(c_2),g_2) \notag\\
		& \: B^{f_1}(c_1) B^{f_2}(c_2) C_T^{k_3}(m) \delta(\hat{e}(m),f_3) \delta(\hat{g}(c_1),k_1) \delta(\hat{g}(c_2),k_2) \notag\\
		=&B^{e_1}(c_1) B^{e_2}(c_2) C_T^{g_3}(m) B^{f_1}(c_1) B^{f_2}(c_2) C_T^{k_3}(m) \delta(\hat{e}(m),e_3) \delta(\hat{e}(m),f_3) \notag\\
		& \: \delta(\hat{g}(c_1),g_1) 
		\delta(\hat{g}(c_1),k_1) \delta(\hat{g}(c_2),g_2) \delta(\hat{g}(c_2),k_2), \notag 
	\end{align}
	because all of the Kronecker deltas commute with all of the other operators, due to the fact that these other operators do not change the surface label or closed loop elements which appear in the Kronecker deltas. We can then use the pairs of similar Kronecker deltas to remove the operator from one member of each pair:
	\begin{align}
		T^{[e_1,e_2,e_3,g_1,g_2,g_3]}(m) T^{[f_1,f_2,f_3,k_1,k_2,k_3]}(m) = &B^{e_1}(c_1) B^{e_2}(c_2) C_T^{g_3}(m) B^{f_1}(c_1) B^{f_2}(c_2) C_T^{k_3}(m) \delta(\hat{e}(m),e_3) \delta(e_3,f_3) \notag\\
		& \: \delta(\hat{g}(c_1),g_1) 
		\delta(g_1,k_1) \delta(\hat{g}(c_2),g_2) \delta(g_2,k_2). \label{Equation_T_product_1}
	\end{align}

	The next step is to use the commutation relation for the magnetic membrane operator and the blob ribbon operators applied on the cycles $c_1$ and $c_2$. From Equation \ref{blob_ribbon_magnetic_commutation} in Section \ref{Section_Magnetic_Tri_Non_Trivial}, we have that
	\begin{equation}
		C^h_T(m) B^e(c_i) = B^{h \rhd e}(c_i) C^h_T(m). \label{Equation_torus_commute_magnetic_blob}
	\end{equation}
	Applying this relation to our product of $T$ operators in Equation \ref{Equation_T_product_1} gives us
	\begin{align*}
		T^{[e_1,e_2,e_3,g_1,g_2,g_3]}(m) T^{[f_1,f_2,f_3,k_1,k_2,k_3]}(m) =& B^{e_1}(c_1) B^{g_3 \rhd f_1}(c_1) B^{e_2}(c_2) B^{g_3 \rhd f_2}(c_2) C_T^{g_3}(m) C_T^{k_3}(m) \delta(\hat{e}(m),e_3) \delta(e_3,f_3)\\
		& \quad \delta(\hat{g}(c_1),g_1) 
		\delta(g_1,k_1) \delta(\hat{g}(c_2),g_2) \delta(g_2,k_2). 
	\end{align*}
	By bringing the blob ribbon operators together in this way, we can combine them using Equation \ref{Equation_blob_ribbon_fake_flat_fusion} to obtain
	\begin{align}
		T^{[e_1,e_2,e_3,g_1,g_2,g_3]}(m) T^{[f_1,f_2,f_3,k_1,k_2,k_3]}(m) =& B^{e_1[g_3 \rhd f_1]}(c_1) B^{e_2 [g_3 \rhd f_2]}(c_2) C_T^{g_3}(m) C_T^{k_3}(m) \delta(\hat{e}(m),e_3) \delta(e_3,f_3)\notag \\
		& \quad \delta(\hat{g}(c_1),g_1) 
		\delta(g_1,k_1) \delta(\hat{g}(c_2),g_2) \delta(g_2,k_2). 
		\label{Equation_T_product_2}
	\end{align}

	Next we will combine the magnetic membrane operators. To do so, we must separate $C^x_T(m)$ (where $x=g_3$ or $k_3$) into the operator $C^x_{\rhd}(m)$ and the attached blob ribbon operators, as described in Section \ref{Section_Magnetic_Tri_Non_Trivial}. Then we have
	\begin{align*}
		C_T^{g_3}(m) C_T^{k_3}(m)&= \big[ \prod_{b \in m} B^{[g_3 \rhd e_b|_{s.p}]e^{-1}_b|_{s.p}}(t_b) \big] C_{\rhd}^{g_3}(m) \big[ \prod_{b \in m} B^{[k_3 \rhd e_b|_{s.p}]e^{-1}_b|_{s.p}}(t_b) \big] C_{\rhd}^{k_3}(m),
	\end{align*}
	where $b$ runs over the plaquettes on the direct membrane of $m$ and $e_b|_{s.p}$ is the label plaquette $b$ has with respect to the start-point of the membrane operator, assuming that the plaquette is oriented away from the dual membrane. Then, using Equation \ref{Equation_torus_commute_magnetic_blob} to commute the blob ribbon operators associated with $C^{k_3}_T(m)$ past $C^{g_3}_{\rhd}(m)$, we have
	\begin{align*}
		C_T^{g_3}(m) C_T^{k_3}(m)&=\big( \prod_{b \in m} B^{[g_3 \rhd e_b|_{s.p}]e^{-1}_b|_{s.p}}(t_b) B^{g_3 \rhd ([k_3 \rhd e_b|_{s.p}]e^{-1}_b|_{s.p})}(t_b) \big) C_{\rhd}^{g_3}(m) C_{\rhd}^{k_3}(m).
	\end{align*}
	
	We can then combine the blob ribbon operators associated to the same plaquette $b$, using Equation \ref{Equation_blob_ribbon_fake_flat_fusion}, to obtain
	\begin{align*}
		C_T^{g_3}(m) C_T^{k_3}(m)&= \big(\prod_{b \in m} B^{[g_3 \rhd e_b|_{s.p}]e^{-1}_b|_{s.p}}(t_b) B^{[(g_3 k_3) \rhd e_b|_{s.p}][g_3 \rhd e^{-1}_b|_{s.p}]}(t_b) \big) C_{\rhd}^{g_3}(m) C_{\rhd}^{k_3}(m)\\
		&= \big(\prod_{b \in m} B^{[g_3 \rhd e_b|_{s.p}]e^{-1}_b|_{s.p}[(g_3 k_3) \rhd e_b|_{s.p}][g_3 \rhd e^{-1}_b|_{s.p}]}(t_b) \big) C_{\rhd}^{g_3}(m) C_{\rhd}^{k_3}(m)\\
		&= \big(\prod_{b \in m} B^{e^{-1}_b|_{s.p}[(g_3 k_3) \rhd e_b|_{s.p}]}(t_b) \big) C_{\rhd}^{g_3}(m) C_{\rhd}^{k_3}(m),
	\end{align*}
	where we used the fact that $E$ is Abelian to cancel the factors of $[g_3 \rhd e_b|_{s.p}]$ and $[g_3 \rhd e^{-1}_b|_{s.p}]$. We see that the blob ribbon operators combine into a single blob ribbon operator of the same form, but with $g_3$ and $k_3$ replaced by the product $g_3k_3$.

	Finally we must combine the two $C_{\rhd}$ operators. Given two magnetic membrane operators $C_{\rhd}^{g_3}(m)$ and $C_{\rhd}^{k_3}(m)$, applied on the same membrane, their combined action on an edge cut by the mutual dual membrane and with initial label $g_i$ is (choosing the edge to point away from the direct membrane for definiteness, although a similar calculation holds if the edge points in the opposite direction)
	\begin{align*}
		C_{\rhd}^{g_3}(m) C_{\rhd}^{k_3}(m) :g_i &= C_{\rhd}^{g_3}(m): g(s.p-v_i)^{-1}k_3g(s.p-v_i)g_i\\
		&= g(s.p-v_i)^{-1}g_3g(s.p-v_i)g(s.p-v_i)^{-1}k_3g(s.p-v_i)g_i\\
		&=g(s.p-v_i)^{-1}g_3k_3g(s.p-v_i)g_i\\
		&= C_{\rhd}^{g_3k_3}(m): g_i,
	\end{align*}
	provided that the path element $g(s.p-v_i)$ is not itself affected by the magnetic membrane operator (this holds in the case where the start-point lies on the direct membrane), which suggests that $C_T^{g_3}(m) C_T^{k_3}(m)=C_T^{g_3 k_3}(m)$. However, we must also check that this holds for the action of the operators on the plaquettes. Their $\rhd$ action on a plaquette with base-point on the direct membrane satisfies
	\begin{align*}
		C^{g_3}_{\rhd}(m) C^{k_3}_{\rhd}(m):e_p &= C^{g_3}_{\rhd}(m): (g(s.p-v_0(p))^{-1}k_3g(s.p-v_0(p))) \rhd e_p\\
		&= (g(s.p-v_0(p))^{-1}g_3g(s.p-v_0(p))) [(g(s.p-v_0(p))^{-1}k_3g(s.p-v_0(p))) \rhd e_p]\\
		&=((g(s.p-v_0(p))^{-1}g_3k_3g(s.p-v_0(p)))) \rhd e_p\\
		&=C^{g_3k_3}_{\rhd}(m) :e_p,
	\end{align*}
	so that
	\begin{equation}
		C^{g_3}_{\rhd}(m) C^{k_3}_{\rhd}(m)=C^{g_3k_3}_{\rhd}(m),
	\end{equation}
	and so
	\begin{equation}
		C^{g_3}_{T}(m) C^{k_3}_{\rhd}(m)=C^{g_3k_3}_{T}(m).
	\end{equation}
	We can then substitute this result into Equation \ref{Equation_T_product_2}, to obtain
	\begin{align}
		T^{[e_1,e_2,e_3,g_1,g_2,g_3]}(m) T^{[f_1,f_2,f_3,k_1,k_2,k_3]}(m) =& B^{e_1 [g_3 \rhd f_1]}(c_1) B^{e_2 [g_3 \rhd f_2]}(c_2) C_T^{g_3 k_3}(m) \delta(\hat{e}(m),e_3) \delta(e_3,f_3)\notag \\
		& \quad \delta(\hat{g}(c_1),g_1) \delta(g_1,k_1) \delta(\hat{g}(c_2),g_2) \delta(g_2,k_2) \notag \\
		=& \delta(g_1,k_1) \delta(g_2,k_2) \delta(e_3,f_3) T^{[g_1, \: g_2, \: g_3k_3, \: e_1 [g_3 \rhd f_1], \: e_2 [g_3 \rhd f_2], \: e_3]}(m). \label{Equation_T_product_3}
	\end{align}

	This completes our decomposition of the product, which will enable us to check if a given sum of $T$ operators is a projector. In order to do this, it is convenient to write Equation \ref{Equation_T_product_3} in terms of the $Y$ operators. Applying the isomorphism between the $T$ and $Y$ labels described by Equation \ref{Equation_define_Y}, we see that
	\begin{align}
		Y^{(g_x,g_y,g_z,f_x,f_y,f_z)}(m)& Y^{(k_x,k_y,k_z,w_x,w_y,w_z)}(m) = \delta(g_y,k_y) \delta(g_z,k_z) \delta(f_x,w_x) Y^{(g_x k_x, \: g_y, \: g_z, \:f_y [g_x \rhd w_y], \: f_z [g_x \rhd w_z])}(m). \label{Equation_Y_product_1}
	\end{align}

	Next we want to apply this decomposition of the product of $Y$ operators to a product of the projectors, to demonstrate that they are indeed orthogonal projectors. Taking a product of two of the projector operators defined in Equation \ref{Equation_torus_projector_definition_appendix_1}, labelled by the pairs $(R, C)$ and $(R', C')$, gives
	\begin{align}
		P^{R,C}(m)P^{R',C'}(m)&= N_{R,C} N_{R',C'} \sum_{i=1}^{|C|} \sum_{\substack{\mathcal{E}_{c_{y,i},c_{z,i}}\\ \in \mathfrak{B}_{c_{y,i},c_{z,i}}}} \sum_{m=1}^{|R|} \sum_{\lambda \in E} \sum_{j=1}^{|C'|} \sum_{\substack{\mathcal{E}'_{c_{y,j}',c_{z,j}'}\\ \in \mathfrak{B}_{c_{y,i}',c_{z,i}'}}} \sum_{n=1}^{|R'|} \sum_{\mu \in E} \delta(c_{y,i},c_{y,i}^{s_{x,1};f_{z,1}}) \delta(c_{z,i},c_{z,i}^{s_{x,1};f_{y,1}}) \notag \\ & \hspace{1cm}\delta(d_{x,i},d_{x,i}^{s_{x,1},c_{y,i},c_{z,i};f_{y,1},f_{z,1}}) 
		\delta(c'_{y,j},{c'}_{y,j}^{s'_{x,1};f'_{z,1}}) \delta(c'_{z,j},{c'}_{z,j}^{s'_{x,1};f'_{y,1}}) \delta(d'_{x,j},{d'}_{x,j}^{s'_{x,1},c'_{y,j},c'_{z,j};f'_{y,1},f'_{z,1}}) \notag \\
		& \hspace{1cm}D^R_{m,m}([\mathcal{E}_C^{\text{stab.}}]_{i,i}) D^{R'}_{n,n}([\mathcal{E'}_{C'}^{\text{stab.}}]_{j,j})Y^{(\partial(\lambda) s_{x,1},\: c_{y,i},\: c_{z,i}, \: d_{x,i}, \: [c_{z,i} \rhd \lambda] f_{y,1} \lambda^{-1},\: [c_{y,i} \rhd \lambda] f_{z,1} \lambda^{-1} )}(m) \notag \\
		& \hspace{1cm}Y^{(\partial(\mu) s'_{x,1}, \ c'_{y,j},\ c'_{z,j},\ d'_{x,j}, \ [c'_{z,l} \rhd \mu] f'_{y,1} \mu^{-1}, \ [c'_{y,j} \rhd \mu] f'_{z,1} \mu^{-1} )}(m). \label{Equation_torus_projector_product_1}
	\end{align}
	
	Our decomposition for the product of the two $Y$ operators (Equation \ref{Equation_Y_product_1}) means that we can replace the product of the two $Y$ operators in Equation \ref{Equation_torus_projector_product_1} with
	\begin{align*}
		\delta(&c_{y,i},{c'}_{y,j}) \delta(c_{z,i},c'_{z,j}) \delta(d_{x,i},d'_{x,j}) \\
		&Y^{\big(\partial(\lambda) s_{x,1} \partial(\mu) s'_{x,1},\ c_{y,i}, \ c_{z,i}, \ d_{x,i}, \ [c_{z,i} \rhd \lambda] f_{y,1} \lambda^{-1} [(\partial(\lambda) s_{x,1}) \rhd ([c'_{z,j} \rhd \mu] f'_{y,1} \mu^{-1})],\ [c_{y,i} \rhd \lambda] f_{z,1} \lambda^{-1} [(\partial(\lambda) s_{x,1}) \rhd ([c'_{y,j} \rhd \mu] f'_{z,1} \mu^{-1})] \big)}(m).
	\end{align*}
	
	As part of this expression we have the product 
	$$\delta (c_{y,i},c_{y,j}') \delta(c_{z,i},c_{z,j}') \delta(d_{x,i},d_{x,j}'),$$
	which is equivalent to
	$$\delta((c_{y,i},c_{z,i},d_{x,i}),(c'_{y,j},c'_{z,j},d'_{x,j})).$$
	
	We note that $(c_{y,i},c_{z,i},d_{x,i})$ is an element in the class $C$ of boundary $\mathcal{G}$-colourings, while $(c'_{y,j},c'_{z,j},d'_{x,j})$ is an element in class $C'$. The Kronecker delta therefore enforces both that $C=C'$ (because the delta only gives 1 when the two classes share an element, in which case the classes are equal by the partitioning of space into classes) and that $i=j$ (from equality of the particular elements in question). Because of this, we can simplify our product of projectors. We have
	\begin{align}
		&P^{R,C}(m)P^{R',C'}(m) \notag\\
		&= \delta(C,C') N_{R,C} N_{R',C}\sum_{i=1}^{|C|} \sum_{\substack{\mathcal{E}_{c_{y,i},c_{z,i}},\\ \mathcal{E'}_{c_{y,i},c_{z,i}} \\ \in \mathfrak{B}_{c_{y,i},c_{z,i}}}} \sum_{m=1}^{|R|} \sum_{n=1}^{|R'|} \sum_{ \lambda, \mu \in E} \delta(c_{y,i},c_{y,i}^{s_{x,1};f_{z,1}}) \delta(c_{y,i},c_{y,i}^{s'_{x,1};f'_{z,1}}) 
		\delta(c_{z,i},c_{z,i}^{s_{x,1};f_{y,1}}) \notag \\& \hspace{0.5cm}\delta(c_{z,i},c_{z,i}^{s'_{x,1};f'_{y,1}}) \delta(d_{x,i},d_{x,i}^{s_{x,1},c_{y,i},c_{z,i};f_{y,1},f_{z,1}}) \delta(d_{x,i},d_{x,i}^{s'_{x,1},c_{y,i},c_{z,i};f'_{y,1},f'_{z,1}}) 
		D^{R}_{m,m}([\mathcal{E}^{\text{stab.}}_C]_{i,i}) D^{R'}_{n,n}([\mathcal{E'}^{\text{stab.}}_C]_{i,i}) \notag \\
		& \hspace{0.5cm} Y^{(\partial(\lambda) s_{x,1} \partial(\mu) s'_{x,1}, \: c_{y,i}, \: c_{z,i}, \: d_{x,i},\: [c_{z,i} \rhd \lambda] f_{y,1} \lambda^{-1} [(\partial(\lambda) s_{x,1}) \rhd ([c_{z,i} \rhd \mu] f'_{y,1} \mu^{-1})],\: [c_{y,i} \rhd \lambda] f_{z,1} \lambda^{-1} [(\partial(\lambda) s_{x,1}) \rhd ([c_{y,i} \rhd \mu] f'_{z,1} \mu^{-1})] )}(m). \label{Equation_torus_projector_product_2}
	\end{align}
	
	Note that the dummy indices $\mathcal{E'}_{c_{y,i},c_{z,i}}$ and $\mathcal{E}_{c_{y,i},c_{z,i}}$ now belong to the same space $\mathfrak{B}_{c_{y,i},c_{z,i}}$ and so the use of the subscript ``$c_{y,i},c_{z,i}$" to highlight the different spaces for the two indices is unnecessary. Therefore, we may drop this subscript from now on (i.e., we use $\mathcal{E}$ and $\mathcal{E'}$ instead of $\mathcal{E}_{c_{y,i},c_{z,i}}$ and $\mathcal{E'}_{c_{y,i},c_{z,i}}$).

	Having established that the product of the two projectors is non-zero only when they are labelled by the same class $C$ of boundary $\mathcal{G}$-colourings, we next want to do the same for the representation labels $R$ and $R'$. We note that our expression contains the term $D^{R}_{m,m}([\mathcal{E}^{\text{stab.}}_C]_{i,i}) D^{R'}_{n,n}([\mathcal{E'}^{\text{stab.}}_C]_{i,i})$, which suggests that we can use an orthogonality relation for the irreps. The normal procedure for doing this is to eliminate one of the dummy indices $\mathcal{E}$ and $\mathcal{E'}$ for the product of these classes. Then the sum over the remaining index delivers the orthogonality relation. We will indeed use such an approach, but in order to do so we must first look at the label of the $Y$ operator. The label includes quantities that belong to the classes $\mathcal{E}$ and $\mathcal{E'}$. Therefore, in order to use an orthogonality relation for the matrix representations, we must replace the quantities pertaining to $\mathcal{E}$ and $\mathcal{E'}$ with the corresponding expressions belonging to the product class.

	Let us now consider the labels of $Y$ in more detail. The three labels which include quantities relating to the classes $\mathcal{E}$ and $\mathcal{E'}$ are 
	$$\partial(\lambda)s_{x,1} \partial(\mu)s'_{x,1},$$
	$$[c_{z,i} \rhd \lambda] f_{y,1} \lambda^{-1} [(\partial(\lambda) s_{x,1}) \rhd ([c_{z,i} \rhd \mu] f'_{y,1} \mu^{-1})],$$
	and
	$$[c_{y,i} \rhd \lambda] f_{z,1} \lambda^{-1} [(\partial(\lambda) s_{x,1}) \rhd ([c_{y,i} \rhd \mu] f'_{z,1} \mu^{-1})].$$
	The form of these labels is tied to the structure of the class $\mathcal{E}$. The representative element of class $\mathcal{E}$ is $(s_{x,1},f_{y,1},f_{z,1})$ and other elements of the class take the form $(\partial(\lambda)s_{x,1},[c_{z,i} \rhd \lambda] f_{y,1} \lambda^{-1}, [c_{y,i} \rhd \lambda] f_{z,1} \lambda^{-1} )$ for some $\lambda$ in $E$. Therefore, the expression involving $\lambda$ takes us from the representative element of the class to the other elements in the class. In addition, note that the classes that appear in our measurement operator are not generic. The Kronecker deltas at the start of the expression for the projector (see Equation \ref{Equation_torus_projector_definition_appendix_1}) enforce the condition that
	$$(c_{y,i},c_{z,i},d_{x,i})= (c_{y,i}^{s_{x,1};f_{z,1}}, \: c_{z,i}^{s_{x,1};f_{y,1}}, \: d_{x,i}^{s_{x,1},c_{y,i},c_{z,i};f_{y,1},f_{z,1}} ).$$
	
	Comparing this to Equation \ref{Equation_stabiliser_group_definition}, we see that $\mathcal{E}$ (and similarly $\mathcal{E'}$) lie in the stabiliser group $Z_{c_{y,i},c_{z,i}}=Z_{C,i}$. The stabiliser group has a product, as we explained earlier in this section. The product of the two classes $\mathcal{E}$ and $\mathcal{E'}$ has representative element \cite{Bullivant2020}
	\begin{equation}
		(s_{x_1}^{\text{prod.}}, f_{y,1}^{\text{prod.}},f_{z,1}^{\text{prod.}})=(s_{x,1}s'_{x,1}, \: f_{y,1} [s_{x,1} \rhd f_{y,1}'], \: f_{z,1} [s_{x,1} \rhd f'_{z,1}]).
		\label{Equation_product_class_representative_element}
	\end{equation}
	This means that the elements of the product class have the form
	\begin{equation}
		(\partial(\nu)s_{x,1}s'_{x,1}, \: [c_{z,i} \rhd \nu] f_{y,1} [s_{x,1} \rhd f_{y,1}'] \nu^{-1}, \: [c_{y,i} \rhd \nu] f_{z,1} [s_{x,1} \rhd f'_{z,1}] \nu^{-1}),
		\label{Equation_product_class_elements}
	\end{equation}
	which looks similar to the label of the $Y$ operator from the product of our projectors in Equation \ref{Equation_torus_projector_product_2}. We now wish to make this connection more explicit, so that we can exploit this group structure. Using Equation \ref{Equation_product_bulk_in_product_class_1} (taking $g_y=c_{y,i}$ and $g_z=c_{z,i}$), we can recognise the triple 
	$$(\partial(\lambda)s_{x,1} \partial(\mu)s'_{x,1}, \: [c_{z,i} \rhd \lambda] f_{y,1} \lambda^{-1} [(\partial(\lambda) s_{x,1}) \rhd ([c_{z,i} \rhd \mu] f'_{y,1} \mu^{-1})], \: [c_{y,i} \rhd \lambda] f_{z,1} \lambda^{-1} [(\partial(\lambda) s_{x,1}) \rhd ([c_{y,i} \rhd \mu] f'_{z,1} \mu^{-1} )])$$
	as the product
	$$(\partial(\lambda)s_{x,1}, [c_{z,i} \rhd \lambda] f_{y,1} \lambda^{-1}, [c_{y,i} \rhd \lambda] f_{z,1} \lambda^{-1}) \cdot (\partial(\mu)s'_{x,1}, [c_{z,i} \rhd \mu] f'_{y,1} \mu^{-1}, [c_{y,i} \rhd \mu] f'_{z,1} \mu^{-1})$$
	between an element of class $\mathcal{E} \in Z_{c_{y,i},c_{z,i}}$ and an element of class $\mathcal{E'} \in Z_{c_{y,i},c_{z,i}}$. Then using Equation \ref{Equation_product_bulk_in_product_class_result}, we can write this product as
	\begin{align*}
		(\partial(\lambda)s_{x,1}& \partial(\mu)s'_{x,1}, \: [c_{z,i} \rhd \lambda] f_{y,1} \lambda^{-1} [(\partial(\lambda) s_{x,1}) \rhd ([c_{z,i} \rhd \mu] f'_{y,1} \mu^{-1})], \: [c_{y,i} \rhd \lambda] f_{z,1} \lambda^{-1} [(\partial(\lambda) s_{x,1}) \rhd ([c_{y,i} \rhd \mu] f'_{z,1} \mu^{-1} )])\\
		&= (\partial(\lambda [s_{x,1}\rhd \mu])s_{x,1}s'_{x,1}, \: [c_{z,i} \rhd ( \lambda [s_{x,1} \rhd \mu])] f_{y,1} [s_{x,1} \rhd f'_{y,1}] (\lambda [s_{x,1} \rhd \mu])^{-1}, \\ & \hspace{1cm} [c_{y,i} \rhd (\lambda [s_{x,1} \rhd \mu])] f_{z,1} [s_{x,1} \rhd f'_{z,1}] (\lambda [s_{x,1} \rhd \mu])^{-1})\\
		&=(\partial(\lambda [s_{x,1}\rhd \mu])s^{\text{prod.}}_{x,1}, \: [c_{z,i} \rhd ( \lambda [s_{x,1} \rhd \mu])] f^{\text{prod.}}_{y,1} (\lambda [s_{x,1} \rhd \mu])^{-1}, \\ & \hspace{1cm} [c_{y,i} \rhd (\lambda [s_{x,1} \rhd \mu])] f^{\text{prod.}}_{z,1} (\lambda [s_{x,1} \rhd \mu])^{-1}),
	\end{align*}
	where the superscript ``prod." indicates the representative element of the product class. We can then recognise the total expression as an element of the product class (substituting $\nu=\lambda [s_{x,1}\rhd \mu]$ in Equation \ref{Equation_product_class_elements}). This means that we can write the relevant labels of our $Y$ operator from Equation \ref{Equation_torus_projector_product_2} in terms of the product class and $\lambda [s_{x,1} \rhd \mu]$, which we denote by $\nu$. Then Equation \ref{Equation_torus_projector_product_2} becomes
	\begin{align}
		P^{R,C}(m)&P^{R',C'}(m) \notag\\
		=& \delta(C,C') N_{R,C} N_{R',C} \sum_{i=1}^{|C|} \sum_{\substack{\mathcal{E}_{c_{y,i},c_{z,i}},\\ \mathcal{E'}_{c_{y,i},c_{z,i}} \\ \in \mathfrak{B}_{c_{y,i},c_{z,i}}}} \sum_{m=1}^{|R|} \sum_{n=1}^{|R'|} \sum_{ \lambda, \mu \in E} \delta(c_{y,i},c_{y,i}^{s_{x,1};f_{z,1}}) \delta(c_{y,i},c_{y,i}^{s'_{x,1};f'_{z,1}}) 
		\delta(c_{z,i},c_{z,i}^{s_{x,1};f_{y,1}}) \delta(c_{z,i},c_{z,i}^{s'_{x,1};f'_{y,1}}) \notag \\
		&\delta(d_{x,i},d_{x,i}^{s_{x,1},c_{y,i},c_{z,i};f_{y,1},f_{z,1}}) \delta(d_{x,i},d_{x,i}^{s'_{x,1},c_{y,i},c_{z,i};f'_{y,1},f'_{z,1}}) 
		D^{R}_{m,m}([\mathcal{E}^{\text{stab.}}_C]_{i,i}) D^{R'}_{n,n}([\mathcal{E'}^{\text{stab.}}_C]_{i,i})\notag \\
		&Y^{(\partial(\lambda [s_{x,1} \rhd \mu]) s^{\text{prod.}}_{x,1}, \: c_{y,i}, \: c_{z,i}, \: d_{x,i},\: [c_{z,i} \rhd (\lambda [s_{x,1} \rhd \mu])] f^{\text{prod.}}_{y,1} (\lambda [s_{x,1} \rhd \mu])^{-1},\: [c_{y,i} \rhd (\lambda [s_{x,1} \rhd \mu])] f^{\text{prod.}}_{z,1} (\lambda [s_{x,1} \rhd \mu])^{-1}) }(m)\notag \\
		=& \delta(C,C') N_{R,C} N_{R',C} \sum_{i=1}^{|C|} \sum_{\substack{\mathcal{E}_{c_{y,i},c_{z,i}},\\ \mathcal{E'}_{c_{y,i},c_{z,i}} \\ \in \mathfrak{B}_{c_{y,i},c_{z,i}}}} \sum_{m=1}^{|R|} \sum_{n=1}^{|R'|} \sum_{ \nu= \lambda [s_{x,1} \rhd \mu] } \sum_{\mu \in E} \delta(c_{y,i},c_{y,i}^{s_{x,1};f_{z,1}}) \delta(c_{y,i},c_{y,i}^{s'_{x,1};f'_{z,1}}) 
		\delta(c_{z,i},c_{z,i}^{s_{x,1};f_{y,1}}) \notag\\ &\delta(c_{z,i},c_{z,i}^{s'_{x,1};f'_{y,1}})\delta(d_{x,i},d_{x,i}^{s_{x,1},c_{y,i},c_{z,i};f_{y,1},f_{z,1}}) \delta(d_{x,i},d_{x,i}^{s'_{x,1},c_{y,i},c_{z,i};f'_{y,1},f'_{z,1}}) 
		D^{R}_{m,m}([\mathcal{E}^{\text{stab.}}_C]_{i,i}) D^{R'}_{n,n}([\mathcal{E'}^{\text{stab.}}_C]_{i,i}) \notag \\
		&Y^{(\partial(\nu) s^{\text{prod.}}_{x,1}, \: c_{y,i}, \: c_{z,i}, \: d_{x,i},\: [c_{z,i} \rhd \nu ] f^{\text{prod.}}_{y,1} \nu^{-1},\: [c_{y,i} \rhd \nu ] f^{\text{prod.}}_{z,1} \nu^{-1}) }(m). \label{Equation_torus_projector_product_3}
	\end{align}
	
	The index $\mu$ now only appears in the sum, so we can replace the sum over $\mu$ by multiplication by $|E|$. However, at this stage we cannot eliminate the classes $\mathcal{E}$ and $\mathcal{E'}$ in favour of the product class, because both $\mathcal{E}$ and $\mathcal{E'}$ appear elsewhere in \ref{Equation_torus_projector_product_3}. Firstly, the classes appear in the Kronecker deltas at the front of our expression for the product of projectors. This latter problem can be resolved relatively easily. As discussed earlier in the section, these deltas enforce that the classes $\mathcal{E}$ and $\mathcal{E'}$ are in the stabilizer group $Z_{C,i}=Z_{c_{y,i},c_{z,i}}$. We can therefore absorb these deltas into the sum, by only summing over $\mathcal{E}$ and $\mathcal{E'}$ that belong to the stabilizer group:
	\begin{align}
		P^{R,C}(m) P^{R',C'}(m) =& |E| \delta(C,C') N_{R,C} N_{R',C} \sum_{i=1}^{|C|} \sum_{\substack{\mathcal{E}, \mathcal{E'} \in \\ Z_{c_{y,i},c_{z,i}}}} \sum_{m=1}^{|R|} \sum_{n=1}^{|R'|} \sum_{\nu \in E} D^R_{m,m}([\mathcal{E}_C^{\text{stab.}}]_{i,i}) D^{R'}_{n,n}([\mathcal{E'}_C^{\text{stab.}}]_{i,i}) \notag \\
		&\quad Y^{(\partial(\nu) s^{\text{prod.}}_{x,1}, \: c_{y,i}, \: c_{z,i}, \: d_{x,i}, \: [c_{z,i} \rhd \nu] f_{y,1}^{\text{prod.}}{\nu}^{-1}, [c_{y,i} \rhd \nu]f_{z,1}^{\text{prod.}} {\nu}^{-1})}(m). \label{Equation_torus_projector_product_4}
	\end{align}
	
	This not only eliminates the explicit appearance of the classes in the Kronecker deltas, but also ensures that our sum over the classes is over a group (the stabilizer group), which will be necessary in order to use orthogonality of irreps. Summing only over elements of a group is also important because it means that we can use the group properties to eliminate the dummy index $\mathcal{E'}$ in favour of the product of $\mathcal{E}$ and $\mathcal{E'}$, while ensuring that we still sum over the same set of $Y$ operators (using the fact that group multiplication is one-to-one).

	The final place that the classes appear is in the matrix elements $D^{R}_{m,m}([\mathcal{E}^{\text{stab.}}_C]_{i,i})$ and $D^{R'}_{n,n}([\mathcal{E'}_C^{\text{stab.}}]_{i,i})$, but it is not that simple to extract this dependence on $\mathcal{E}$ and $\mathcal{E'}$ from these matrix elements. The arguments of the matrices are $[\mathcal{E}^{\text{stab.}}_C]_{i,i}$ and $[\mathcal{E'}^{\text{stab.}}_C]_{i,i}$, which are related to $\mathcal{E}$ and $\mathcal{E'}$ but are not the same. Recall that $\mathcal{E}$ and $\mathcal{E'}$ belong to the stabiliser group $Z_{c_{y,i},c_{z,i}}$, which we also denote by $Z_{C,i}$. On the other hand $[\mathcal{E}^{\text{stab.}}_C]_{i,i}$ and $[\mathcal{E'}^{\text{stab.}}_C]_{i,i}$ belong to $Z_C=Z_{c_{y,1},c_{z,1}}$. As we explained earlier in this section, the two groups $Z_C$ and $Z_{C,i}$ are isomorphic because they just correspond to different choices of representative for $C$ (where the isomorphism is directly described by Equation \ref{Equation_definition_stabiliser_group_isomorphism}). With this knowledge in hand, we now consider the term
	$$D^R_{m,m}([\mathcal{E}_C^{\text{stab.}}]_{i,i}) D^{R'}_{n,n}([\mathcal{E'}_C^{\text{stab.}}]_{i,i}) $$
	from our product of projectors (Equation \ref{Equation_torus_projector_product_4}). This term depends on the class $\mathcal{E}$ through the relationship between $\mathcal{E}$ and $[\mathcal{E}^{\text{stab.}}_C]_{i,i}$ (and similar for $\mathcal{E'}$). Because this relationship is an isomorphism, $\mathcal{E} \cdot \mathcal{E'} = \mathcal{E}^{\text{prod.}}$ implies that 
	$$[\mathcal{E}^{\text{stab.}}_C]_{i,i} \cdot [\mathcal{E'}^{\text{stab.}}_C]_{i,i} = [(\mathcal{E} \cdot \mathcal{E'})_C^{\text{stab.}}]_{i,i}= [(\mathcal{E}^{\text{prod.}})^{\text{stab.}}_C]_{i,i}.$$
	We use this to write
	$$D^{R'}_{n,n}([\mathcal{E'}_C^{\text{stab.}}]_{i,i}) = D^{R'}_{n,n}([(\mathcal{E}^{-1})_C^{\text{stab.}}]_{i,i}[(\mathcal{E}^{\text{prod.}})_C^{\text{\text{stab.}}}]_{i,i}).$$
	Substituting this into Equation \ref{Equation_torus_projector_product_4}, we have
	\begin{align*}
		P^{R,C}(m)P^{R',C'}(m) &= |E| \delta(C,C') N_{R,C} N_{R',C} \sum_{i=1}^{|C|} \sum_{m=1}^{|R|} \sum_{n=1}^{|R'|} \sum_{\mathcal{E}, \mathcal{E'} \in Z_{C,i}} \sum_{\nu \in E} D^{R}_{m,m}([\mathcal{E}_C^{\text{stab.}}]_{i,i}) D^{R'}_{nn}([(\mathcal{E}^{-1})_C^{\text{stab.}}(\mathcal{E}^{\text{prod.}})_C^{\text{stab.}}]_{i,i}) \\ &\hspace{0.5cm} Y^{(\partial(\nu)s^{\text{prod.}}_{x,1}, \: c_{y,i}, \: c_{z,i}, \: d_{x,i}, \: [c_{z,i} \rhd \nu] f^{\text{prod.}}_{y,1} {\nu}^{-1},\: [c_{y,i} \rhd \nu]f^{\text{prod.}}_{z,1} {\nu}^{-1})}(m).
	\end{align*}
	
	We can then replace the sum over $\mathcal{E'}$ with one over $\mathcal{E}^{\text{prod.}}$ and split the matrix representation $D^{R'}_{nn}([(\mathcal{E}^{-1})_C^{\text{stab.}}(\mathcal{E}^{\text{prod.}})_C^{\text{stab.}}]_{i,i})$ into contributions from $\mathcal{E}^{-1}$ and $\mathcal{E}^{\text{prod.}}$, to obtain
	\begin{align*}
		P^{R,C}(m)P^{R',C'}(m)&= \sum_{i=1}^{|C|} \sum_{m=1}^{|R|} \sum_{n,p=1}^{|R'|} \sum_{\mathcal{E}, \mathcal{E}^{\text{prod.}} \in Z_{C,i}} \sum_{\nu \in E} \delta(C,C')|E| N_{R,C} N_{R',C} D^R_{m,m}([\mathcal{E}_C^{\text{stab.}}]_{i,i}) D^{R'}_{np}([({\mathcal{E}^{-1}})_C^{\text{stab.}}]_{i,i}) \\ & \hspace{0.5cm} D^{R'}_{pn}([({\mathcal{E}^{\text{prod.}}})_C^{\text{stab.}}]_{i,i}) 
		Y^{(\partial(\nu)s^{\text{prod.}}_{x,1}, \: c_{y,i}, \: c_{z,i}, \: d_{x,i}, \: [c_{z,i} \rhd \nu] f^{\text{prod.}}_{y,1} {\nu}^{-1}, \: [c_{y,i} \rhd \nu]f^{\text{prod.}}_{z,1} {\nu}^{-1})}(m)\\
		&=\sum_{i=1}^{|C|} \sum_{\mathcal{E}^{\text{prod.}} \in Z_{C,i}} \sum_{\nu \in E} \delta(C,C')|E| N_{R,C} N_{R',C} \sum_{m=1}^{|R|} \sum_{n,p=1}^{|R'|} \big(\sum_{\mathcal{E} \in Z_{C,i}} D^R_{m,m}([\mathcal{E}_C^{\text{stab.}}]_{i,i}) D^{R'}_{np}([({\mathcal{E}^{-1}})_C^{\text{stab.}}]_{i,i}) \big) \\ & \hspace{0.5cm} D^{R'}_{pn}([({\mathcal{E}^{\text{prod.}}})_C^{\text{stab.}}]_{i,i}) 
		Y^{(\partial(\nu)s^{\text{prod.}}_{x,1}, \: c_{y,i}, \: c_{z,i}, \:d_{x,i}, \: [c_{z,i} \rhd \nu] f^{\text{prod.}}_{y,1} {\nu}^{-1}, \: [c_{y,i} \rhd \nu]f^{\text{prod.}}_{z,1} {\nu}^{-1})}(m).
	\end{align*}

	Then we may use the Grand Orthogonality Theorem for the matrix elements of irreps, giving us
	\begin{align*}
		P^{R,C}(m)P^{R',C'}(m)&= \sum_{i=1}^{|C|} \sum_{\mathcal{E}^{\text{prod.}} \in Z_{C,i}} \sum_{\nu \in E} \delta(C,C') N_{R,C} N_{R',C} |E|\sum_{m,n,p=1}^{|R|}\big(\frac{|Z_C|}{|R|} \delta_{mn} \delta_{mp} \delta(R,R')\big)D^R_{pn}([({\mathcal{E}^{\text{prod.}}})_C^{\text{stab.}}]_{i,i}) \\
		& \hspace{1cm} Y^{(\partial(\nu) s^{\text{prod.}}_{x,1}, \: c_{y,i}, \: c_{z,i}, \: d_{x,i},\: [c_{z,i} \rhd \nu] f^{\text{prod.}}_{y,1} {\nu}^{-1},\: [c_{y,i} \rhd \nu]f^{\text{prod.}}_{z,1} {\nu}^{-1})}(m) \\
		&= \frac{|E||Z_C|}{|R|} \delta(C,C') \delta(R,R') N_{R,C} N_{R,C} \sum_{i=1}^{|C|} \sum_{\mathcal{E}^{\text{prod.}} \in Z_{C,i}} \sum_{\nu \in E} \sum_{n=1}^{|R|} D^R_{nn} ([({\mathcal{E}^{\text{prod.}}})_C^{\text{stab.}}]_{i,i}) \\
		& \hspace{1cm} Y^{(\partial(\nu)s^{\text{prod.}}_{x,1}, \:c_{y,i}, \: c_{z,i}, \:d_{x,i}, \: [c_{z,i} \rhd \nu] f^{\text{prod.}}_{y,1} {\nu}^{-1}, \: [c_{y,i} \rhd \nu]f^{\text{prod.}}_{z,1} {\nu}^{-1})}(m).
	\end{align*}	
	
	From Equation \ref{Equation_torus_projector_definition_appendix_1}, we can recognise 
	\begin{align*}
		N_{R,C} \sum_{i=1}^{|C|} \sum_{\mathcal{E}^{\text{prod.}} \in Z_{C,i}} \sum_{\nu \in E} \sum_{n=1}^{|R|} D^R_{nn} ([({\mathcal{E}^{\text{prod.}}})_C^{\text{stab.}}]_{i,i}) Y^{(\partial(\nu)s^{\text{prod.}}_{x,1}, \: c_{y,i}, \: c_{z,i}, \: d_{x,i}, \: [c_{z,i} \rhd \nu] f^{\text{prod.}}_{y,1} {\nu}^{-1}, \: [c_{y,i} \rhd \nu]f^{\text{prod.}}_{z,1} {\nu}^{-1})}(m)
	\end{align*}
	as the projector $P^{R,C}$ (where we have absorbed the Kronecker deltas from Equation \ref{Equation_torus_projector_definition_appendix_1} by summing only over the stabiliser group $Z_{C,i}$) and so we have
	\begin{align}
		P^{R,C}(m)P^{R',C'}(m)&= \frac{|E||Z_C|}{|R|} P^{R,C}(m) \delta(R,R') \delta(C,C') N_{R,C}.
	\end{align}
	If we then substitute the normalization factor $N_{R,C}=\frac{|R|}{|E||Z_C|}$, we obtain
	\begin{align}
		P^{R,C}(m)P^{R',C'}(m)&= \delta(R,R') \delta(C,C') P^{R,C}(m).
	\end{align}

	Therefore, the set of operators
	\begin{align}
		P^{R,C}(m)&= \frac{|R|}{|E||Z_C|} \sum_{i=1}^{|C|} \sum_{\substack{\mathcal{E}_{c_{y,i},c_{z,i}}\\ \in \mathfrak{B}_{c_{y,i},c_{z,i}}}} \sum_{m=1}^{|R|} \sum_{\lambda \in E} \delta(c_{y,i},c_{y,i}^{s_{x,1};f_{z,1}}) \delta(c_{z,i},c_{z,i}^{s_{x,1};f_{y,1}})\delta(d_{x,i},d_{x,i}^{s_{x,1},c_{y,i},c_{z,i};f_{y,1},f_{z,1}}) \notag \\
		& \hspace{1cm}D^R_{m,m}([\mathcal{E}_C^{\text{stab.}}]_{i,i}) Y^{(\partial(\lambda) s_{x,1},\: c_{y,i}, \: c_{z,i}, \: d_{x,i}, \: [c_{z,i} \rhd \lambda] f_{y,1} \lambda^{-1}, \: [c_{y,i} \rhd \lambda] f_{z,1} \lambda^{-1} )}(m) \label{Equation_torus_projector_definition_appendix_normalized}
	\end{align}
	for the classes $C$ of boundary $\mathcal{G}$-colourings and irreps $R$ of the corresponding stabiliser group $Z_C$ do indeed form a set of orthogonal projectors.

	Next we need to demonstrate that we do not just have a set of projectors, but the correct set. The projectors we constructed are in one-to-one correspondence with the ground states on the 3-torus, because the pairs $(R,C)$ are in such a correspondence, as shown in Ref. \cite{Bullivant2020}). We also know from Section \ref{Section_3D_Topological_Charge_Torus_Tri_nontrivial} that the number of independent topological measurement operators is the same as the ground state degeneracy on the 3-torus. This indicates that we have the correct number of projectors, but we still need to check that these projectors are indeed valid measurement operators. We need to demonstrate that the projectors, which are linear combinations of the $T$ operators from Section \ref{Section_3D_Topological_Charge_Torus_Tri_nontrivial}, obey our restrictions on the allowed coefficients of the $T$ operators that we derived in Section \ref{Section_3D_Topological_Charge_Torus_Tri_nontrivial}.

	Recall from Section \ref{Section_3D_Topological_Charge_Torus_Tri_nontrivial} that we have two kinds of restriction. Firstly, the $T$ operators appearing in the projectors must have labels that obey Conditions \ref{C1}-\ref{C4} from Section \ref{Section_3D_Topological_Charge_Torus_Tri_nontrivial}. Secondly the coefficients of the $T$ operators in the projector must obey our equivalence relation rules, as described by Conditions \ref{C5}-\ref{C8} from Section \ref{Section_3D_Topological_Charge_Torus_Tri_nontrivial}. We begin by considering the first type of restriction. We wrote these conditions in terms of $T$ operators, but it is simple to convert these conditions into conditions for the $Y$ operators. If an operator $Y$ appearing in the projector has label $(v_x,v_y,v_z,w_x,w_y,w_z)$, then this label must satisfy
	\begin{align*}
		\partial(w_x)&=[v_z,v_y] \label{Y1} \tag{Y1}\\
		\partial(w_y)&=[v_z,v_x] \label{Y2} \tag{Y2}\\
		\partial(w_z)&=[v_y,v_x] \label{Y3} \tag{Y3}\\
		1_E &= [v_x \rhd w_x^{-1}] w_x w_y^{-1} [v_y \rhd w_y] w_z [v_z \rhd w_z^{-1}]. \label{Y4} \tag{Y4}
	\end{align*}

	From Equation \ref{Equation_torus_projector_definition_appendix_normalized}, we see that the labels of $Y$ that appear in the projector are $\partial(\lambda)s_{x,1}=v_x$, $c_{y,i}=v_y$, $c_{z,i}=v_z$, $d_{x,i}=w_x$, $[c_{z,i} \rhd \lambda] f_{y,1} \lambda^{-1}=w_y$ and $[c_{y,i} \rhd \lambda] f_{z,1} \lambda^{-1} =w_z$. In particular, $(c_{y,i},c_{z,i},d_{x,i})$ is the $i$th element of class $C$ of boundary $\mathcal{G}$-colourings, which immediately restricts the properties of these labels. Furthermore, the $Y$ operator with a particular set of labels only appears in the projector if the labels satisfy the Kronecker deltas from Equation \ref{Equation_torus_projector_definition_appendix_normalized}. From this, we can see which of our restrictions are enforced by the structure of the projector. The fact that the triple $(c_{y,i}, c_{z,i}, d_{x,i})$ is a boundary $\mathcal{G}$-colouring enforces that
	$$\partial(d_{x,i})=[c_{z,i},c_{y,i}].$$
	Then using the relations $d_{x,i} =w_x$, $c_{z,i}=v_z$ and $c_{y,i}=v_y$, we see that Condition \ref{Y1} is immediately satisfied by the projector.

	Now consider the constraints enforced by the Kronecker deltas at the front of our projector operator. $\delta(c_{y,i},c_{y,i}^{s_{x,1};f_{z,1}})$ implies that we only have non-zero coefficients if
	\begin{align*}
		c_{y,i}&= s_{x,1}^{-1} \partial(f_{z_1})^{-1} c_{y,i}s_{x,1}\\
		& \implies \partial(f_{z,1}) = c_{y,i}s_{x,1}c_{y,i}^{-1}s_{x,1}^{-1}\\
		& \implies \partial(c_{y,i} \rhd \lambda) \partial(f_{z,1}) \partial(\lambda^{-1}) = c_{y,i} \partial(\lambda) c_{y,i}^{-1} c_{y,i}s_{x,1}c_{y,i}^{-1}s_{x,1}^{-1} \partial(\lambda)^{-1}\\
		& \implies \partial([c_{y,i} \rhd \lambda] f_{z,1} \lambda^{-1}) = c_{y,i} \partial(\lambda)s_{x,1} c_{y,i}^{-1} s_{x,1}^{-1} \partial(\lambda)^{-1}.
	\end{align*}	
	Using $[c_{y,i} \rhd \lambda] f_{z,1} \lambda^{-1} =w_z$, $c_{y,i}=v_y$ and $\partial(\lambda)s_{x,1}=v_x$, we see that
	\begin{align*}
		\partial(w_z) &= v_y v_x v_y^{-1} v_x^{-1} = [v_y,v_x],
	\end{align*}
	so that Condition \ref{Y3} is satisfied. Next we use the delta $\delta(c_{z,i},c_{z,i}^{s_{x,1};f_{y,1}})$, which gives the restriction
	\begin{align*}
		c_{z,i}&=s_{x,1}^{-1}\partial(f_{y,1})^{-1}c_{z,i}s_{x,1}\\
		&\implies \partial(f_{y,1})=c_{z,i}s_{x,1}c_{z,i}^{-1}s_{x,1}^{-1}\\
		&\implies \partial(c_{z,i} \rhd \lambda) \partial(f_{y,1}) \partial(\lambda^{-1})= c_{z,i} \partial(\lambda) c_{z,i}^{-1} c_{z,i} s_{x,1} c_{z,i}^{-1} s_{x,1}^{-1} \partial(\lambda)^{-1}\\
		& \implies \partial([c_{z,i} \rhd \lambda] f_{y,1} \lambda^{-1})=c_{z,i} \partial(\lambda)s_{x,1} c_{z,i}^{-1} s_{x,1}^{-1} \partial(\lambda)^{-1}.
	\end{align*}
	
	Substituting in the labels of the $Y$ operator $w_y=[c_{z,i} \rhd \lambda] f_{y,1} \lambda^{-1}$, $v_x=\partial(\lambda)s_{x,1}$ and $v_z=c_{z,i}$, we see that the Kronecker delta ensures that
	\begin{align*}	
		\partial(w_y)=[v_z,v_x]
	\end{align*}
	for any contributing $Y$ operator, so Condition \ref{Y2} is also satisfied. Finally we look at
	$\delta(d_{x,i},d_{x,i}^{s_{x,1},c_{y,i},c_{z,i};f_{y,1},f_{z,1}}).$ This gives us the restriction that
	\begin{align*}
		d_{x,i}&=s_{x,1}^{-1} \rhd \big(f_{y,i}^{-1} (c_{z,i} \rhd f_{z,1})^{-1} d_{x,i} [c_{y,i} \rhd f_{y,1}] f_{z,1}\big).
	\end{align*}
	Applying the map $s_{x,1} \rhd$ to both sides of this equation gives us
	\begin{align*}
		s_{x,1} \rhd d_{x,i}&=f_{y,i}^{-1} (c_{z,i} \rhd f_{z,1})^{-1} d_{x,i} [c_{y,i} \rhd f_{y,1}] f_{z,1}
		& \implies [s_{x,1} \rhd d_{x,i}^{-1}] f_{y,i}^{-1} (c_{z,i} \rhd f_{z,1})^{-1} d_{x,i} [c_{y,i} \rhd f_{y,1}] f_{z,1}=1_E.
	\end{align*}
	
	Then, inserting the identity in the form of $\lambda^{-1} \lambda$, $[c_{z,i} \rhd \lambda^{-1}] [c_{z,i} \rhd \lambda]$, $[(c_{z,i} c_{y,i}) \rhd \lambda^{-1}] [(c_{z,i} c_{y,i}) \rhd \lambda]$ and $[c_{y,i} \rhd \lambda^{-1}] [c_{y,i} \rhd \lambda]$, we see that
	\begin{align*}
		[s_{x,1} \rhd d_{x,i}^{-1}]& (\lambda^{-1} \lambda) f_{y,i}^{-1} ([c_{z,i} \rhd \lambda^{-1}] [c_{z,i} \rhd \lambda]) (c_{z,i} \rhd f_{z,1})^{-1} ([(c_{z,i} c_{y,i}) \rhd \lambda^{-1}] [(c_{z,i} c_{y,i}) \rhd \lambda]) d_{x,i} [c_{y,i} \rhd f_{y,1}]\\
		& \hspace{3cm} ([c_{y,i} \rhd \lambda^{-1}] [c_{y,i} \rhd \lambda]) f_{z,1} =1_E,\\
		& \implies \lambda [s_{x,1} \rhd d_{x,i}^{-1}] \lambda^{-1} \lambda f_{y,i}^{-1} [c_{z,i} \rhd \lambda^{-1}] [c_{z,i} \rhd \lambda] (c_{z,i} \rhd f_{z,1})^{-1} [(c_{z,i} c_{y,i}) \rhd \lambda^{-1}] [(c_{z,i} c_{y,i}) \rhd \lambda] d_{x,i} [c_{y,i} \rhd f_{y,1}]\\
		& \hspace{3cm} [c_{y,i} \rhd \lambda^{-1}] [c_{y,i} \rhd \lambda] f_{z,1} \lambda^{-1}=1_E,
	\end{align*}	
	where we conjugated the overall expression by $\lambda$ in the last step. Then we can use the second Peiffer condition, Equation \ref{Equation_Peiffer_2} in the main text, to write $\lambda [s_{x,1} \rhd d_{x,i}^{-1}] \lambda^{-1}$ as $(\partial(\lambda) s_{x,1}) \rhd d_{x,i}^{-1}$. This, along with some algebraic manipulations, gives us
	\begin{align*}
		&[(\partial(\lambda)s_{x,1}) \rhd d_{x,i}^{-1}] \big([c_{z,i} \rhd \lambda] f_{y,1} \lambda^{-1}\big)^{-1} \big( [(c_{z,i} c_{y,i}) \rhd \lambda] [c_{z,i} \rhd f_{z,1}][c_{z,i} \rhd \lambda]^{-1}\big)^{-1} [(c_{z,i} c_{y,i}) \rhd \lambda] d_{x,i} [c_{y,i} \rhd f_{y,1}]\\
		& \hspace{3cm} [c_{y,i} \rhd \lambda^{-1}] ([c_{y,i} \rhd \lambda] f_{z,1} \lambda^{-1})=1_E.
	\end{align*}
	
	Next we want to extract the action of $c_{z,i} \rhd$ from $\big( [(c_{z,i} c_{y,i}) \rhd \lambda] [c_{z,i} \rhd f_{z,1}][c_{z,i} \rhd \lambda]^{-1}\big)^{-1}$ and $c_{y,i} \rhd$ from $[(c_{z,i} c_{y,i}) \rhd \lambda] d_{x,i} [c_{y,i} \rhd f_{y,1}]
	[c_{y,i} \rhd \lambda^{-1}] ([c_{y,i} \rhd \lambda]$, in order to write these in terms of $[c_{y,i} \rhd \lambda] f_{z,1} \lambda^{-1} =w_z$ and $[c_{z,i} \rhd \lambda] f_{y,1} \lambda^{-1}=w_y$ respectively. This is easy for the former term, but for the latter term there is an intrusive factor of $d_{x,i}$ and the product $c_{z,i} c_{y,i}$ is in the wrong order in $[(c_{z,i} c_{y,i}) \rhd \lambda]$ to extract $c_{y,i} \rhd$. To fix both of these problems, we write 
	$$c_{z,i}c_{y,i}= c_{z_i}c_{y,i} c_{z,i}^{-1}c_{y_i}^{-1} c_{y,i}c_{z,i}.$$
	Then we can use the relation $c_{z_i}c_{y,i} c_{z,i}^{-1}c_{y_i}^{-1}= \partial(d_{x,i})$ (from the fact that $(c_{y,i},c_{z,i},d_{x,i})$ is a boundary $\mathcal{G}$-colouring), to write

	\begin{align*}
		& [(\partial(\lambda) s_{x,1}) \rhd d_{x,i}^{-1}] ([c_{z,i} \rhd \lambda] f_{y,1} \lambda^{-1})^{-1} \big[c_{z,i} \rhd ([c_{y,i} \rhd \lambda] f_{z,1} \lambda^{-1})^{-1}\big] [(c_{z,i}c_{y,i}c_{z,i}^{-1}c_{y,i}^{-1}c_{y,i}c_{z,i}) \rhd \lambda] d_{x,i} [c_{y,i} \rhd f_{y,1}]\\
		& \hspace{3cm} [c_{y,i} \rhd \lambda^{-1}] ([c_{y,i} \rhd \lambda] f_{z,1} \lambda^{-1})=1_E\\
		&\implies [(\partial(\lambda) s_{x,1}) \rhd d_{x,i}^{-1}] ([c_{z,i} \rhd \lambda] f_{y,1} \lambda^{-1})^{-1} \big[c_{z,i} \rhd ([c_{y,i} \rhd \lambda] f_{z,1} \lambda^{-1})^{-1}\big] [(\partial(d_{x,i})c_{y,i}c_{z,i}) \rhd \lambda] d_{x,i} [c_{y,i} \rhd f_{y,1}]\\
		& \hspace{3cm} [c_{y,i} \rhd \lambda^{-1}] ([c_{y,i} \rhd \lambda] f_{z,1} \lambda^{-1})=1_E.
	\end{align*}
	
	Then we can use the second Peiffer condition (Equation \ref{Equation_Peiffer_2} in the main text) to replace $[(\partial(d_{x,i})c_{y,i}c_{z,i}) \rhd \lambda]$ with $d_{x,i}[(c_{y,i}c_{z,i}) \rhd \lambda]d_{x,i}^{-1}$ and write this as
	\begin{align*}
		&[(\partial(\lambda) s_{x,1}) \rhd d_{x,i}^{-1}] ([c_{z,i} \rhd \lambda] f_{y,1} \lambda^{-1})^{-1} \big[c_{z,i} \rhd ([c_{y,i} \rhd \lambda] f_{z,1} \lambda^{-1})^{-1}\big] d_{x,i} [(c_{y,i}c_{z,i}) \rhd \lambda] [c_{y,i} \rhd f_{y,1}]\\
		& \hspace{3cm} [c_{y,i} \rhd \lambda^{-1}] ([c_{y,i} \rhd \lambda] f_{z,1} \lambda^{-1})=1_E \\
		&\implies [(\partial(\lambda) s_{x,1}) \rhd d_{x,i}^{-1}] ([c_{z,i} \rhd \lambda] f_{y,1} \lambda^{-1})^{-1} \big[c_{z,i} \rhd ([c_{y,i} \rhd \lambda] f_{z,1} \lambda^{-1})^{-1}\big] d_{x,i} \big[c_{y,i} \rhd ([c_{z,i} \rhd \lambda] f_{y,1} \lambda^{-1})\big]\\
		& \hspace{3cm} ([c_{y,i} \rhd \lambda] f_{z,1} \lambda^{-1})=1_E.
	\end{align*}
	
	Finally, we can substitute in the labels $\partial(\lambda)s_{x,1}=v_x$, $c_{y,i}=v_y$, $c_{z,i}=v_z$, $d_{x,i}=w_x$, $[c_{z,i} \rhd \lambda] f_{y,1} \lambda^{-1}=w_y$ and $[c_{y,i} \rhd \lambda] f_{z,1} \lambda^{-1} =w_z$ to obtain
	\begin{align*}	
		[v_x \rhd w_x^{-1}] w_y^{-1} [v_z \rhd w_z^{-1}] w_x [v_y \rhd w_y] w_z=1_E,
	\end{align*}
	which gives us Condition \ref{Y4}, once we use the fact that the group $E$ is Abelian to rearrange the product. Therefore, we have shown that Conditions \ref{Y1}-\ref{Y4} are satisfied by our projector.

	Next we need to check that the conditions on the coefficients (Conditions \ref{C5}-\ref{C8} in Section \ref{Section_3D_Topological_Charge_Torus_Tri_nontrivial}) are satisfied. Once we convert Conditions \ref{C5}-\ref{C8} from conditions on the $T$ operators to conditions on the associated $Y$ operators using Equation \ref{Equation_define_Y}, the conditions are described by the equivalence relations
	\begin{align*}
		(v_x,w_y,w_z) &\sim(\partial(e)v_x,w_y [v_z \rhd e] e^{-1}, w_z [v_y \rhd e] e^{-1}) \label{Y5} \tag{Y5}\\
		(v_y,w_x,w_z) & \sim (\partial(e) v_y, w_x e^{-1} [v_z \rhd e], w_z [v_x \rhd e^{-1}] e) \label{Y6} \tag{Y6}\\
		(v_z,w_x,w_y) &\sim (\partial(r^{-1})v_z, w_x r^{-1} [v_y \rhd r], w_y [v_x \rhd r] r^{-1}) \label{Y7} \tag{Y7}\\
		(v_x,v_y,v_z,w_x,w_y,w_z) & \sim (av_xa^{-1},av_ya^{-1},av_za^{-1}, a \rhd w_x, a \rhd w_y, a \rhd w_z), \label{Y8} \tag{Y8}
	\end{align*}
	where coefficients of the $Y$ operators whose labels are related by these relations must have equal coefficients in the measurement operator. We first examine Condition \ref{Y5}, which looks similar to the relationship given in Equation \ref{Equation_bulk_class_equivalence_relation} between the representative element $(s_{x,1},f_{y,1},f_{z,1})$ of the class $\mathcal{E}$ of bulk $\mathcal{G}$-colourings and the other elements of that class. These other elements of the class appear in the labels of $Y$ due to the sum over $\lambda \in E$ in Equation \ref{Equation_torus_projector_definition_appendix_normalized}. This suggests that this sum over $\lambda$ will ensure that the projector satisfies Condition \ref{Y5}. In order to see that this is indeed the case, we change the dummy index $\lambda$ in the sum over $E$ for $\lambda' = \mu^{-1}\lambda$ (for some arbitrary $\mu \in E$) in Equation \ref{Equation_torus_projector_definition_appendix_normalized}, which gives us
	\begin{align*}
		\sum_{\lambda \in E}& Y^{(\partial(\lambda)s_{x,1}, \: c_{y,i}, \: c_{z,i},d_{x,i}, \: [c_{z,i} \rhd \lambda] f_{y,1} \lambda^{-1}, \: [c_{y,i} \rhd \lambda] f{z,1} \lambda^{-1})}(m)\\
		&=\sum_{\lambda'= \mu^{-1} \lambda \in E} Y^{(\partial(\mu) \partial(\lambda')s_{x,1}, \: c_{y,i}, \: c_{z,i},\: d_{x,i}, \: [c_{z,i} \rhd (\mu \lambda')] f_{y,1} {\lambda'}^{-1} \mu^{-1}, \: [c_{y,i} \rhd ( \mu \lambda')] f{z,1} {\lambda'}^{-1} \mu^{-1})}(m).
	\end{align*}	
	Then, because $\lambda'$ is just a dummy index, we can freely rename it to $\lambda$, so
	\begin{align*}
		\sum_{\lambda \in E}& Y^{(\partial(\lambda)s_{x,1}, \: c_{y,i}, \:c_{z,i},d_{x,i}, \: [c_{z,i} \rhd \lambda] f_{y,1} \lambda^{-1}, \: [c_{y,i} \rhd \lambda] f{z,1} \lambda^{-1})}(m)\\
		&= \sum_{\lambda \in E} Y^{(\partial(\mu) \partial(\lambda)s_{x,1}, \:c_{y,i}, \: c_{z,i},d_{x,i}, \: [c_{z,i} \rhd \mu] ([c_{z,i} \rhd \lambda] f_{y,1} \lambda^{-1}) \mu^{-1}, \: [c_{y,i} \rhd \mu] ([c_{y,i} \rhd \lambda] f{z,1} \lambda^{-1}) \mu^{-1})}(m).
	\end{align*}
	This shows that the $Y$ operators
	$$Y^{(\partial(\lambda)s_{x,1}, \: c_{y,i}, \: c_{z,i}, \: d_{x,i}, \: [c_{z,i} \rhd \lambda] f_{y,1} \lambda^{-1}, \: [c_{y,i} \rhd \lambda] f{z,1} \lambda^{-1})}(m)$$
	and
	$$Y^{(\partial(\mu) \partial(\lambda)s_{x,1}, \: c_{y,i}, \: c_{z,i},d_{x,i}, \: [c_{z,i} \rhd \mu] ([c_{z,i} \rhd \lambda] f_{y,1} \lambda^{-1}) \mu^{-1}, \: [c_{y,i} \rhd \mu] ([c_{y,i} \rhd \lambda] f{z,1} \lambda^{-1}) \mu^{-1})}(m)$$
	have the same coefficient in the projection operator, because they appear with equal weight in the sum. Writing this in terms of the labels $v_x$, $v_y$, $v_z$, $w_x$, $w_y$ and $w_z$ of the measurement operator, we see that this implies that the equivalence relation
	$$(v_x,w_y,w_z) \sim (\partial(\mu)v_x, [v_{z} \rhd \mu] w_y \mu^{-1}, [v_y \rhd \mu] w_z \mu^{-1})$$
	is satisfied, meaning that the equivalent labels appear with equal coefficient in the measurement operator. Therefore, Condition \ref{Y5} is satisfied.

	Next we will consider Conditions \ref{Y6}-\ref{Y8}. These can be consolidated into a single equivalence relation. Starting with Condition \ref{Y8}, we then apply Condition \ref{Y6} to obtain
	\begin{align*}
		(v_x,v_y,v_z,w_x,w_y,w_z) & \sim (av_xa^{-1},av_ya^{-1},av_za^{-1}, a \rhd w_x, a \rhd w_y, a \rhd w_z)\\
		& \sim (av_xa^{-1}, \partial(e)av_ya^{-1},av_za^{-1}, [(av_za^{-1}) \rhd e] [a \rhd w_x] e^{-1}, a \rhd w_y, e [a \rhd w_z] [av_xa^{-1} \rhd e^{-1}]).
	\end{align*}
	
	Finally we apply Condition \ref{Y7}, resulting in the relation
	\begin{align*}
		(v_x,v_y,v_z,w_x,w_y,w_z)& \sim (av_xa^{-1}, \: \partial(e)av_ya^{-1},av_za^{-1}, \: [(av_za^{-1}) \rhd e] [a \rhd w_x] e^{-1}, \: a \rhd w_y, e [a \rhd w_z] [av_xa^{-1} \rhd e^{-1}]) \\
		& \sim ( av_xa^{-1}, \: \partial(e)av_ya^{-1}, \: \partial(r^{-1})av_za^{-1},r^{-1} [(av_za^{-1}) \rhd e] [a \rhd w_x] e^{-1} (\partial(e)av_ya^{-1}) \rhd r,\\
		& \hspace{1cm} r^{-1} [a \rhd w_y] [(av_xa^{-1}) \rhd r],\: e [a \rhd w_z] [av_xa^{-1} \rhd e^{-1}] ).
	\end{align*}
	
	Looking at the expression $(\partial(e)av_ya^{-1}) \rhd r$, we note that when $E$ is Abelian the factor of $\partial(e)$ has no effect, because the second Peiffer condition (Equation \ref{Equation_Peiffer_2} in the main text) implies that $\partial(e) \rhd x =exe^{-1}=x$ for all $e, \:x \in E$. Dropping the factor of $\partial(e)$ and replacing the dummy variable $r$ with $t=r^{-1}$, we obtain the equivalence relation
	\begin{align}
		\big(v_x,v_y,v_z,w_x,w_y,w_z\big)& \sim \big(av_xa^{-1}, \: \partial(e)av_ya^{-1}, \: \partial(t)a v_za^{-1}, \: [(av_za^{-1}) \rhd e] t [a \rhd w_x] [(av_ya^{-1}) \rhd t^{-1}] e^{-1}, \notag \\
		& \hspace{1cm}t [a \rhd w_y] [(av_xa^{-1}) \rhd t^{-1}], \: e [a \rhd w_z] [(av_xa^{-1}) \rhd e^{-1}] \big), \label{Equation_Conditions_Y6-Y8}
	\end{align}
	where we have one such condition for each $a \in G$, $e \in E$ and $t \in E$ (note that we can obtain the original relations \ref{Y6}-\ref{Y8} by setting two of the variables to the identity of the associated groups). We then define the action of a triple $(a,t,e) \in (G,E,E)$ on a sextuple $(v_x,v_y,v_z,w_x,w_y,w_z)$ by
	\begin{align}
		(a,t,e): \big(v_x&,v_y,v_z,w_x,w_y,w_z\big) \notag \\
		&= \big(av_xa^{-1}, \: \partial(e)av_ya^{-1}, \: \partial(t)a v_za^{-1}, \: [(av_za^{-1}) \rhd e] t [a \rhd w_x] [(av_ya^{-1}) \rhd t^{-1}]e^{-1}, \notag\\
		& \hspace{1cm} \: t [a \rhd w_y] [(av_xa^{-1}) \rhd t^{-1}], \: e [a \rhd w_z] [(av_xa^{-1}) \rhd e^{-1}] \big), \label{Equation_torus_grou_action_def}
	\end{align}
	so that the equivalence relation in Equation \ref{Equation_Conditions_Y6-Y8} can be written as 
	\begin{equation}
		(v_x,v_y,v_z,w_x,w_y,w_z) \sim (a,t,e): (v_x,v_y,v_z,w_x,w_y,w_z)
	\end{equation}
	for all $(a,t,e) \in (G,E,E)$. This action is in fact a group action, as we will show shortly. Consider acting on the sextuple first with $(a,t,e)$ and then with a new triple $(a',t',e')$. We have
	\begin{align*}
		&(a',t',e'): \big((a,t,e): (v_x,v_y,v_z,w_x,w_y,w_z)\big)\\
		&= (a',t',e'):\big(av_xa^{-1}, \: \partial(e)av_ya^{-1}, \: \partial(t)a v_za^{-1}, \: [(av_za^{-1}) \rhd e] t [a \rhd w_x] [(av_ya^{-1}) \rhd t^{-1}]e^{-1},\\
		& \hspace{1cm} t [a \rhd w_y] [(av_xa^{-1}) \rhd t^{-1}], \: e [a \rhd w_z] [(av_xa^{-1}) \rhd e^{-1}] \big) \\
		&= \big(a'av_x a^{-1} {a'}^{-1}, \: \partial(e')a' \partial(e)av_ya^{-1}{a'}^{-1}, \:\partial(t')a' \partial(t)av_za^{-1} {a'}^{-1}, \\
		& \hspace{1cm}[(a'\partial(t)a v_za^{-1}{a'}^{-1}) \rhd e']t' \big[a'\rhd\big( [(av_za^{-1}) \rhd e] t [a \rhd w_x] [(av_ya^{-1}) \rhd t^{-1}]e^{-1}\big)\big] [(a'\partial(e)av_ya^{-1}{a'}^{-1}) \rhd {t'}^{-1}]{e'}^{-1},
		\\& \hspace{1cm} t' [a' \rhd (t [a \rhd w_y] [(av_xa^{-1}) \rhd t^{-1}])] [(a'av_xa^{-1}{a'}^{-1}) \rhd {t'}^{-1}],\\
		& \hspace{1cm}
		e' [a' \rhd ( e [a \rhd w_z] [(av_xa^{-1}) \rhd e^{-1}])] [(a'av_xa^{-1}{a'}^{-1}) \rhd {e'}^{-1}] \big).
	\end{align*}
	
	As we explained earlier, the factors of $\partial(t)$ and $\partial(e)$ which appear as part of an $\rhd$ action, such as in $(a'\partial(t)a v_za^{-1}{a'}^{-1}) \rhd e'$, have no effect when $E$ is Abelian (due to the second Peiffer condition, Equation \ref{Equation_Peiffer_2} in the main text). We can therefore drop these factors to obtain
	\begin{align}
		(a',t',&e'): \big((a,t,e): (v_x,v_y,v_z,w_x,w_y,w_z)\big) \notag\\
		&= \big(a'av_x a^{-1} {a'}^{-1}, \: \partial(e')a' \partial(e)av_ya^{-1}{a'}^{-1}, \:\partial(t')a' \partial(t)av_za^{-1} {a'}^{-1}, \notag \\ & \hspace{1cm}[(a'a v_za^{-1}{a'}^{-1}) \rhd e']t' \big[a'\rhd \big( [(av_za^{-1}) \rhd e] t [a \rhd w_x] [(av_ya^{-1}) \rhd t^{-1}]e^{-1}\big)\big] [(a'av_ya^{-1}{a'}^{-1}) \rhd {t'}^{-1}]{e'}^{-1}, \notag \\ & \hspace{1cm}
		t' [a' \rhd (t [a \rhd w_y] [(av_xa^{-1}) \rhd t^{-1}])] [(a'av_xa^{-1}{a'}^{-1}) \rhd {t'}^{-1}], \notag \\ & \hspace{1cm}
		e' [a' \rhd ( e [a \rhd w_z] [(av_xa^{-1}) \rhd e^{-1}])] [(a'av_xa^{-1}{a'}^{-1}) \rhd {e'}^{-1}] \big). \label{Equation_torus_group_action_product_1}
	\end{align}
	
	It seems that we can combine each term from the action of $(a',t',e')$ and the action of $(a,t,e)$ into single terms of the same form. For example it seems that we can combine $a'$ and $a$ into $a'a$. A few more algebraic manipulations will make this apparent. We start by grouping the factors of $a$ and $a'$. In the second element of the sextuplet, we have the expression $\partial(e')a' \partial(e)a$. Because $\partial$ maps to the centre of $G$, we could just commute $a'$ past $\partial(e)$ so that it is adjacent to $a$. However, it will actually be more useful to use the Peiffer condition $ a' \partial(e){a'}^{-1} = \partial(a' \rhd e)$ to write 
	$$\partial(e')a' \partial(e)a = \partial(e') \partial(a' \rhd e)a'a.$$
	Similarly, we have
	$$\partial(t')a'\partial(t)a = \partial(t') \partial(a' \rhd t)a'a$$
	for the analogous expression in the third element of the sextuplet. Inserting these expressions into Equation \ref{Equation_torus_group_action_product_1}, we have
	\begin{align}
		(a',& \: t',e'): \big((a,t,e): (v_x,v_y,v_z,w_x,w_y,w_z)\big) \notag\\
		&= \big((a'a)v_x (a'a)^{-1}, \: \partial(e')\partial(a' \rhd e)(a'a)v_y(a'a)^{-1}, \: \partial(t')\partial(a' \rhd t) (a'a)v_z(a'a)^{-1}, \notag \\ & \hspace{1cm} [((a'a) v_z(a'a)^{-1}) \rhd e']t' \big[a'\rhd\big( [(av_za^{-1}) \rhd e] t [a \rhd w_x] [(av_ya^{-1}) \rhd t^{-1}]e^{-1}\big)\big] [((a'a)v_y(a'a)^{-1}) \rhd {t'}^{-1}]{e'}^{-1}, \notag \\ & \hspace{1cm}
		t' [a' \rhd (t [a \rhd w_y] [(av_xa^{-1}) \rhd t^{-1}])] ((a'a)v_x(a'a)^{-1}) \rhd {t'}^{-1}, \notag \\ & \hspace{1cm}
		e' [a' \rhd ( e [a \rhd w_z] [(av_xa^{-1}) \rhd e^{-1}])] ((a'a)v_x(a'a)^{-1}) \rhd {e'}^{-1} \big). \label{Equation_torus_group_action_product_2}
	\end{align}
	
	Next we wish to unpack the expressions inside the $a' \rhd$ action, such as 
	$$a'\rhd( [(av_za^{-1}) \rhd e] t [a \rhd w_x] [(av_ya^{-1}) \rhd t^{-1}]e^{-1})$$
	in the fourth element of the sextuplet. We have
	\begin{align*}
		a'\rhd( [(av_za^{-1}) \rhd e]& t [a \rhd w_x] [(av_ya^{-1}) \rhd t^{-1}]e^{-1})\\
		&=[(a'av_za^{-1}) \rhd e] [a' \rhd t] [(a'a) \rhd w_x] [(a'av_ya^{-1}) \rhd t^{-1}] [a' \rhd e^{-1}],
	\end{align*}
	which we can rewrite as
	\begin{align*}
		a'\rhd( [(av_za^{-1}) \rhd e]& t [a \rhd w_x] [(av_ya^{-1}) \rhd t^{-1}]e^{-1})\\
		&=[(a'av_za^{-1}{a'}^{-1}) \rhd (a' \rhd e)] [a' \rhd t] [(a'a) \rhd w_x] [(a'av_ya^{-1}{a'}^{-1}) \rhd (a' \rhd t)^{-1}] [a' \rhd e^{-1}].
	\end{align*}
	
	Performing similar manipulations for the fifth and sixth elements of the sextuplets and substituting into Equation \ref{Equation_torus_group_action_product_2} gives us
	\begin{align*}
		(a',t',e'):& \big((a,t,e): (v_x,v_y,v_z,w_x,w_y,w_z)\big)\\
		&= \big((a'a)v_x (a'a)^{-1}, \: \partial(e')\partial(a' \rhd e)(a'a)v_y(a'a)^{-1}, \: \partial(t')\partial(a' \rhd t) (a'a)v_z(a'a)^{-1}, \\ & \hspace{1cm}[((a'a) v_z(a'a)^{-1}) \rhd e']t' [(a'av_za^{-1}{a'}^{-1}) \rhd (a' \rhd e)] [a' \rhd t] [(a'a) \rhd w_x] [(a'av_ya^{-1}{a'}^{-1}) \rhd (a' \rhd t)^{-1}]\\
		& \hspace{1.5cm} \times [a' \rhd e^{-1}] [((a'a)v_y(a'a)^{-1}) \rhd {t'}^{-1}]{e'}^{-1}, \\ & \hspace{1cm}
		t' [a' \rhd t] [(a'a) \rhd w_y] [(a'av_xa^{-1}{a'}^{-1}) \rhd (a' \rhd t)^{-1}] ((a'a)v_x(a'a)^{-1}) \rhd {t'}^{-1}, \\ & \hspace{1cm}
		e' [a' \rhd e] [(a'a) \rhd w_z] [(a'av_xa^{-1}{a'}^{-1}) \rhd (a' \rhd e)^{-1}] ((a'a)v_x(a'a)^{-1}) \rhd {e'}^{-1} \big).
	\end{align*}
	
	We then make use of the Abelian nature of $E$ to group the expressions corresponding to $e'$ and $e$. Similarly, we collect the expressions corresponding to $t'$ and $t$. This gives us
	\begin{align*}
		(a',t',e'):& \big((a,t,e): (v_x,v_y,v_z,w_x,w_y,w_z)\big)\\
		&= \big((a'a)v_x (a'a)^{-1}, \: \partial(e') \partial(a' \rhd e)(a'a)v_y(a'a)^{-1}, \: \partial(t') \partial(a' \rhd t) (a'a)v_z(a'a)^{-1}, \\ & \hspace{1cm} [((a'a) v_z(a'a)^{-1}) \rhd e'] [((a'a) v_z(a'a)^{-1}) \rhd (a' \rhd e)] t'[a' \rhd t] [(a'a) \rhd w_x] \\ & \hspace{1.5cm} \times[((a'a)v_y(a'a)^{-1}) \rhd (a' \rhd t^{-1})] [((a'a)v_y(a'a)^{-1}) \rhd {t'}^{-1}] [a' \rhd e]^{-1} {e'}^{-1},\\ & \hspace{1cm}
		t' [a' \rhd t] [(a'a) \rhd w_y] ((a'a)v_x(a'a)^{-1}) \rhd(a' \rhd t^{-1}) [((a'a)v_x(a'a)^{-1}) \rhd {t'}^{-1} ],\\ & \hspace{1cm}
		e' [a' \rhd e] [(a'a) \rhd w_z] [((a'a)v_x(a'a)^{-1}) \rhd (a' \rhd e)^{-1}] [((a'a)v_x(a'a)^{-1}) \rhd {e'}^{-1}] \big).
	\end{align*}
	
	We then use the relations $\partial(e')\partial(a \rhd e)= \partial( e' [a \rhd e])$ and $[g \rhd e'] [g \rhd (a\rhd e )]= g \rhd (e' [a \rhd e])$ for any $g \in G$, together with the equivalent relations for $t$ and $t'$, to write
	\begin{align*}
		(a',t',e'):& \big((a,t,e): (v_x,v_y,v_z,w_x,w_y,w_z)\big)\\
		&= \big((a'a)v_x (a'a)^{-1}, \: \partial(e'[a' \rhd e])(a'a)v_y(a'a)^{-1}, \: \partial(t'[a' \rhd t]) (a'a)v_z(a'a)^{-1}, \\ & \hspace{1cm} [((a'a) v_z(a'a)^{-1}) \rhd (e' [a' \rhd e])]t'[a' \rhd t] [(a'a) \rhd w_x] [((a'a)v_y(a'a)^{-1}) \rhd (t' [a' \rhd t])^{-1}] ( e' [a' \rhd e])^{-1},\\ & \hspace{1cm}
		t' [a' \rhd t] [(a'a) \rhd w_y] ((a'a)v_x(a'a)^{-1}) \rhd ({t'[a' \rhd t]})^{-1},\\ & \hspace{1cm}
		e' [a' \rhd e] [(a'a) \rhd w_z] ((a'a)v_x(a'a)^{-1}) \rhd {(e' [a' \rhd e])}^{-1} \big).
	\end{align*}
	From Equation \ref{Equation_torus_grou_action_def}, we can recognise this as the action of a single triple $(a'a, t' [a' \rhd t], e' [a' \rhd e])$ on our sextuple:
	\begin{align*}
		(a',t',e'):\big((a,t,e): (v_x,v_y,v_z,w_x,w_y,w_z)\big)&= (a'a, t' [a' \rhd t], e' [a' \rhd e]) : (v_x,v_y,v_z,w_x,w_y,w_z).
	\end{align*}
	
	This means that the combined action of the two triples $(a,t,e)$ and $(a',t',e')$ is the same as the action of a single triple $(a'a, t' [a' \rhd t], e' [a' \rhd e])$. This latter triple is the product of the two triples under the group product $\cdot$, as defined in Equation \ref{Equation_triple_product_definition}. That is $(a'a, t' [a' \rhd t], e' [a' \rhd e]) = (a',t',e') \cdot (a,t,e)$. We therefore have
	\begin{align*}
		(a',t',e'): \big((a,t,e): (v_x,v_y,v_z,w_x,w_y,w_z)\big)&= \big( (a',t',e') \cdot (a,t,e)\big) :(v_x,v_y,v_z,w_x,w_y,w_z),
	\end{align*}
	so the action of the triple on the sextuple respects the group product of the triples. We have therefore written the equivalence relation in Equation \ref{Equation_Conditions_Y6-Y8}, which represents a restriction that our projectors must satisfy, in terms of an action with a group structure (note that the symmetry, reflexivity and transitivity of the equivalence relation is then reflected by the inverse, identity and closure properties of the group). We aim to use this group structure to show that conditions \ref{Y6}-\ref{Y8} are satisfied by the projector. Consider the labels $(v_x,v_y,v_z,w_x,w_y,w_z)$ which appear in the $Y$ operators in our projector. As described previously, for a projector labelled by class $C$ of boundary $\mathcal{G}$-colourings and irrep $R$ of the associated stabiliser group, the labels of $Y$ take the form $(g, c_{y,i},c_{z,i},d_{x,i},e_1,e_2)$, where $(c_{y,i},c_{z,i},d_{x,i})$ is the $i$th element of class $C$. A necessary condition for the condition described by our equivalence relation Equation \ref{Equation_Conditions_Y6-Y8} to hold, is that when we act on this label with an arbitrary triple, we must obtain a new label of the same form. This means that the $(c_{y,i},c_{z,i},d_{x,i})$ part of the sextuple must remain in the class $C$ under the action of an arbitrary triple $(a,b_1,b_2)$, because only elements of the class can appear in the label of $Y$. Our action of the triple on the label is given by
	\begin{align}
		(a,b_1,b_2): \big(g&,c_{y,i},c_{z,i},d_{x,i},e_1,e_2\big) \notag \\
		&= \big(aga^{-1}, \: \partial(b_2)ac_{y,i}a^{-1}, \: \partial(b_1)a c_{z,i}a^{-1}, [(ac_{z,i}a^{-1}) \rhd b_2] b_1 [a \rhd d_{x,i}] [(ac_{y,i}a^{-1}) \rhd b_1^{-1}]b_2^{-1}, \notag \\ & \hspace{1cm} b_1 [a \rhd e_1] [(aga^{-1}) \rhd b_1^{-1}], b_2 [a \rhd e_2] [(aga^{-1}) \rhd b_2^{-1}] \big). \label{Equation_torus_group_action_1}
	\end{align}
	
	Under this action $(c_{y,i},c_{z,i},d_{x,i})$ becomes
	$$(\partial(b_2)ac_{y,i}a^{-1}, \partial(b_1)a c_{z,i}a^{-1}, [(ac_{z,i}a^{-1}) \rhd b_2] b_1 [a \rhd d_{x,i}] [(ac_{y,i}a^{-1}) \rhd b_1^{-1}]b_2^{-1}).$$
	We wish to know if this new element is also in the class $C$. Consider an element $(c_{y,j},c_{z,j},d_{x,j})$ of class $C$. The other elements of the class are then generated by the equivalence relation
	\begin{align*}
		(c_{y,j},c_{z,j},d_{x,j}) \overset{\mathrm{I}}{\sim} (a^{-1}&\partial(b_2^{-1})c_{y,j}a, \: a^{-1}\partial(b_1^{-1})c_{z,j}a,\\
		& a^{-1} \rhd (b_1^{-1}(c_{z,j} \rhd b_2^{-1})d_{x,j}(c_{y,j} \rhd b_1) b_2)) 
	\end{align*}
	for each $(a,b_1,b_2) \in (G,E,E)$. This means that there is some element $(c_{y,i},c_{z,i},d_{x,i})$ in $C$ such that
	$$(a^{-1}\partial(b_2^{-1})c_{y,j}a, \: a^{-1}\partial(b_1^{-1})c_{z,j}a, a^{-1} \rhd (b_1^{-1}(c_{z,j} \rhd b_2^{-1})d_{x,j}(c_{y,j} \rhd b_1) b_2))=(c_{y,i}, c_{z,i},d_{x,i}). $$
	Inverting this relation gives us
	\begin{align*}
		(c_{y,j},c_{z,j},d_{x,j}) &= (\partial(b_2)ac_{y,i}a^{-1}, \partial(b_1)ac_{z,i}a^{-1}, [c_{z,j} \rhd b_2] b_1 [a \rhd d_{x,i}] b_2^{-1} [c_{y,j} \rhd b_1^{-1}] )\\
		&= (\partial(b_2)ac_{y,i}a^{-1}, \partial(b_1)ac_{z,i}a^{-1}, [(\partial(b_1)ac_{z,i}a^{-1}) \rhd b_2] b_1 [a \rhd d_{x,i}] b_2^{-1} [(\partial(b_2)ac_{y,i}a^{-1}) \rhd b_1^{-1}] ).
	\end{align*}
	
	Then the factors of $\partial(b_1)$ and $\partial(b_2)$ in the $\rhd$ (for example in $(\partial(b_1)ac_{z,i}a^{-1}) \rhd b_2$ ) have no effect when $E$ is Abelian. This means that
	\begin{align*}
		(c_{y,j},c_{z,j},d_{x,j}) &= (\partial(b_2)ac_{y,i}a^{-1}, \partial(b_1)ac_{z,i}a^{-1}, [c_{z,j} \rhd b_2] b_1 a \rhd d_{x,i} b_2^{-1} [c_{y,j} \rhd b_1^{-1}] )\\
		&= (\partial(b_2)ac_{y,i}a^{-1}, \partial(b_1)ac_{z,i}a^{-1}, [(ac_{z,i}a^{-1}) \rhd b_2] b_1 a \rhd d_{x,i} b_2^{-1} [(ac_{y,i}a^{-1}) \rhd b_1^{-1}]).
	\end{align*}
	Therefore, we can also write the equivalence relation that defines the class $C$ as
	\begin{align*}
		(c_{y,i},c_{z,i},d_{x,i}) \overset{\mathrm{I}}{\sim} (\partial(b_2)ac_{y,i}a^{-1}, \partial(b_1)ac_{z,i}a^{-1}, [(ac_{z,i}a^{-1}) \rhd b_2] b_1 [a \rhd d_{x,i}] b_2^{-1} [(ac_{y,i}a^{-1}) \rhd b_1^{-1}]).
	\end{align*}
	We see that the right-hand side of this equivalence relation is just the same as the result of the action of the triple $(a,b_1,b_2)$ on the $(c_{y,i},c_{z,i},d_{x,i})$ part of the sextuple in Equation \ref{Equation_torus_group_action_1}, so this element stays in the class when acted on by $(a,b_1,b_2)$, as we require.

	Having seen that the equivalence relation defining the class $C$ arises naturally from our group action of the triple on the sextuple, it will be useful to see how other quantities related to $C$ appear in this formalism. In particular, the labels of the operators in our projector depend on certain stabiliser groups, so we wish to show how these stabiliser groups relate to our group action in order to show that the labels satisfy the condition from Equation \ref{Equation_Conditions_Y6-Y8}. Recall that for a boundary $\mathcal{G}$-colouring $(c_{y,i},c_{z,i},d_{x,i})$ in class $C$, we have a stabiliser group $Z_{c_{y,i},c_{z,i}}=Z_{C,i}$, consisting of classes of bulk $G$-colourings whose representative elements $(s_{x,1},f_{y,1},f_{z,1})$ (and indeed other elements) satisfy 
	\begin{equation}
		(c_{y,i},c_{z,i},d_{x,i})=(s_{x,1}^{-1} \partial(f_{z,1})^{-1} c_{y,i}s_{x,1}, \: s_{x,1}^{-1} \partial(f_{y,1})^{-1} c_{z,i}s_{x,1}, \: s_{x,1}^{-1} \rhd (f_{y,1}^{-1} (c_{z,i} \rhd f_{z,1})^{-1} d_{x,i} [c_{y,i} \rhd f_{y,1}] f_{z,1} )).
	\end{equation}
	
	If we invert this equation, we obtain
	\begin{equation}
		(\partial(f_{z,1})s_{x,1}c_{y,i}s_{x,1}^{-1},\partial(f_{y,1})s_{x,1}c_{z,i}s_{x,1}^{-1},[c_{z,i} \rhd f_{z,1}] f_{y,1} [s_{x,1} \rhd d_{x,i}]f_{z,1}^{-1} [c_{y,i} \rhd f_{y,1}^{-1}] )= (c_{y,i},c_{z,i},d_{x,i}). \label{Equation_stabiliser_condition_inverted}
	\end{equation}
	We can then use the relation $c_{z,i} = \partial(f_{y,1})s_{x,1}c_{z,i}s_{x,1}^{-1}$ to write
	\begin{align*}
		c_{z,i} \rhd f_{z,1} &= (\partial(f_{y,1})s_{x,1}c_{z,i}s_{x,1}^{-1}) \rhd f_{z,1}\\
		&= (s_{x,1}c_{z,i}s_{x,1}^{-1}) \rhd f_{z,1},
	\end{align*}
	where in the last step we used the Abelian nature of $E$ to discount the effect of the factor of $\partial(f_{y,1})$ in the $\rhd$ action. Similarly we have
	$$c_{y,i} \rhd f_{y,1} = (s_{x,1}c_{y,i}s_{x,1}^{-1}) \rhd f_{y,1}.$$
	Inserting these two relations into Equation \ref{Equation_stabiliser_condition_inverted} gives us
	\begin{equation}
		(\partial(f_{z,1})s_{x,1}c_{y,i}s_{x,1}^{-1},\partial(f_{y,1})s_{x,1}c_{z,i}s_{x,1}^{-1},[(s_{x,1}c_{z,i}s_{x,1}^{-1}) \rhd f_{z,1}] f_{y,1} [s_{x,1} \rhd d_{x,i}]f_{z,1}^{-1} [(s_{x,1}c_{y,i}s_{x,1}^{-1}) \rhd f_{y,1}^{-1}] )= (c_{y,i},c_{z,i},d_{x,i}). \label{Equation_stabiliser_condition_inverted_2}
	\end{equation}
	
	Let us examine how the element $(s_{x,1},f_{y,1},f_{z,1})$ from a class in the stabiliser group $Z_{C,i}$ acts via our group action on a sextuple that includes the corresponding element $(c_{y,i},c_{z,i},d_{x,i})$. From Equation \ref{Equation_torus_group_action_1}, we have 
	\begin{align}
		(s_{x,1},f_{y,1},f_{z,1}): \big(g,c_{y,i}&,c_{z,i},d_{x,i},e_1,e_2\big) \notag \\
		&= \big(s_{x,1}gs_{x,1}^{-1}, \: \partial(f_{z,1})s_{x,1}c_{y,i}s_{x,1}^{-1}, \: \partial(f_{y,1})s_{x,1} c_{z,i}s_{x,1}^{-1}, \notag \\
		& \hspace{1cm} [(s_{x,1}c_{z,i}s_{x,1}^{-1}) \rhd f_{z,1}] f_{y,1} [s_{x,1} \rhd d_{x,i}] [(s_{x,1}c_{y,i}s_{x,1}^{-1}) \rhd f_{y,1}^{-1}]f_{z,1}^{-1}, \notag \\ & \hspace{1cm}f_{y,1} [s_{x,1} \rhd e_1] [(s_{x,1}gs_{x,1}^{-1}) \rhd f_{y,1}^{-1}], \: f_{z,1} [s_{x,1} \rhd e_2] [(s_{x,1}gs_{x,1}^{-1}) \rhd f_{z,1}^{-1}] \big). \label{Equation_torus_group_action_stabiliser_element_1}
	\end{align}
	The part of this corresponding to class $C$ is 
	$$(\partial(f_{z,1})s_{x,1}c_{y,i}s_{x,1}^{-1}, \partial(f_{y,1})s_{x,1} c_{z,i}s_{x,1}^{-1}, [(s_{x,1}c_{z,i}s_{x,1}^{-1}) \rhd f_{z,1}] f_{y,1} [s_{x,1} \rhd d_{x,i}] [(s_{x,1}c_{y,i}s_{x,1}^{-1}) \rhd f_{y,1}^{-1}]f_{z,1}^{-1} ),$$
	which from Equation \ref{Equation_stabiliser_condition_inverted_2} we know to be equal to $(c_{y,i},c_{z,i},d_{x,i})$ (indeed this is equivalent to the definition of the stabiliser group). Therefore, the action of $(s_{x,1},f_{y,1},f_{z,1})$ on the sextuple is
	\begin{align}
		(s_{x,1},f_{y,1},f_{z,1}): \big(g&,c_{y,i},c_{z,i},d_{x,i},e_1,e_2\big) \notag\\
		&=(g', c_{y,i},c_{z,i},d_{x,i},e_1',e_2' ), \label{Equation_torus_group_action_stabiliser_element_2}
	\end{align}
	where
	$$(g',e_1',e_2') = (s_{x,1}gs_{x,1}^{-1},f_{y,1} [s_{x,1} \rhd e_1] [(s_{x,1}gs_{x,1}^{-1}) \rhd f_{y,1}^{-1}], f_{z,1} [s_{x,1} \rhd e_2] [(s_{x,1}gs_{x,1}^{-1}) \rhd f_{z,1}^{-1}]).$$
	
	We see that the part of the sextuple corresponding to the class $C$ is left invariant, but we also want to consider $(g',e_1',e_2')$. We will show that this is equal to
	$$(s_{x,1},f_{y,1},f_{z,1}) \cdot (g,e_1,e_2) \cdot (s_{x,1},f_{y,1},f_{z,1})^{-1},$$
	i.e., that it is the same as conjugation of $(g,e_1,e_2)$ by the element $(s_{x,1},f_{y,1},f_{z,1}) $. From Equation \ref{Equation_triple_product_definition}, we have
	$$(s_{x,1},f_{y,1},f_{z,1})^{-1} = (s_{x,1}^{-1}, s_{x,1}^{-1} \rhd f_{y,1}^{-1},s_{x,1}^{-1} \rhd f_{z,1}^{-1}).$$
	
	Therefore,
	\begin{align}
		(s_{x,1},f_{y,1},f_{z,1})& \cdot (g,e_1,e_2) \cdot (s_{x,1},f_{y,1},f_{z,1})^{-1}= (s_{x,1},f_{y,1},f_{z,1}) \cdot (g,e_1,e_2) \cdot (s_{x,1}^{-1}, s_{x,1}^{-1} \rhd f_{y,1}^{-1},s_{x,1}^{-1} \rhd f_{z,1}^{-1})\notag \\
		&=(s_{x,1}g, \: f_{y,1} [s_{x,1}\rhd e_1], \: f_{z,1} [s_{x,1} \rhd e_2]) \cdot (s_{x,1}^{-1}, \: s_{x,1}^{-1} \rhd f_{y,1}^{-1},s_{x,1}^{-1} \rhd f_{z,1}^{-1})\notag \\
		&= (s_{x,1}gs_{x,1}^{-1}, \: f_{y,1}[s_{x,1}\rhd e_1] (s_{x,1}g) \rhd (s_{x,1}^{-1} \rhd f_{y,1}^{-1}), \: (s_{x,1}g) \rhd (s_{x,1}^{-1} \rhd f_{z,1}^{-1}) )\notag\\
		&= (s_{x,1}gs_{x,1}^{-1}, \: f_{y,1}[s_{x,1}\rhd e_1] (s_{x,1}gs_{x,1}^{-1} \rhd f_{y,1}^{-1}), \: f_{z,1} [s_{x,1} \rhd e_2] (s_{x,1}gs_{x,1}^{-1} \rhd f_{z,1}^{-1}) )\notag\\
		&=(g',e_1',e_2'), \label{Equation_torus_group_action_stabiliser_element_3}
	\end{align} 
	so that the action of the triple $(s_{x,1},f_{y,1},f_{z,1})$ on $(g,e_1,e_2)$ is indeed the same as conjugation by $(s_{x,1},f_{y,1},f_{z,1})$. This means that if $(s_{x,1},f_{y,1},f_{z,1})$ lies in a class $X \in Z_{C,i}$ and $(g,e_1,e_2)$ belongs to a class $\mathcal{E} \in Z_{C,i}$, then $(g',e_1',e_2')$ belongs to a class $X\mathcal{E}X^{-1}$. We have therefore seen that triples belonging to the stabiliser group of the element $(c_{y,i},c_{z,i},d_{x,i})$ leave $(c_{y,i},c_{z,i},d_{x,i})$ (the part of the sextuple corresponding to class $C$) invariant and change the rest of the label (corresponding to an element of the stabiliser group $Z_{C,i}$) only by conjugation.

	If the stabiliser group of class $C$ is made of elements that preserve the class element $(c_{y,1},c_{z,1},d_{x,1})$, then the opposite of this would be the elements $(p_{x,i},q_{y,i},q_{z,i}) \in (G,E,E)$ which were defined in Equation \ref{Equation_define_p_and_q}. These are representatives that satisfy
	$$(c_{y,1},c_{z,1},d_{x,1})=(p_{x,i}^{-1}\partial(q_{z,i})^{-1}c_{y,1}p_{x,i}, p_{x,i}^{-1}\partial(q_{y,i}^{-1})c_{z,1}p_{x,i}, p_{x,i}^{-1} \rhd [q_{y,i}^{-1} (c_{z,i} \rhd q_{z,i}^{-1})d_{x,i} [c_{y,i} \rhd q_{y,i}] q_{z,i}]),$$
	and so can move us between the elements of the class $C$. Inverting this equation gives us
	$$(c_{y,i},c_{z,i},d_{x,i})= ( \partial(q_{z,i})p_{x,i}c_{y,1}p_{x,i}^{-1}, \partial(q_{y,i}) p_{x,i}c_{z,1} p_{x,i}^{-1}, [c_{z,i} \rhd q_{z,i}] q_{y,i} [p_{x,i} \rhd d_{x,1}] q_{z,i}^{-1} [c_{y,i} \rhd q_{y,i}]^{-1} ).$$
	
	Then 
	\begin{align*}
		c_{z,i} \rhd q_{z,i}&= (\partial(q_{y,i})p_{x,i}c_{z,1}p_{x,i}^{-1}) \rhd q_{z,i}\\
		&= (p_{x,i}c_{z,1}p_{x,i}^{-1}) \rhd q_{z,i}
	\end{align*}
	and similarly
	$$c_{y,i} \rhd q_{y,i} = (p_{x,i}c_{y,1}p_{x,i}^{-1}) \rhd q_{y,i},$$
	so that
	\begin{align}
		(c_{y,i},c_{z,i},d_{x,i})= ( \partial(q_{z,i})p_{x,i}c_{y,1}p_{x,i}^{-1}, \partial(q_{y,i}) p_{x,i}c_{z,1} p_{x,i}^{-1}, [(p_{x,i}c_{z,1}p_{x,i}^{-1}) \rhd q_{z,i}] q_{y,i} [p_{x,i} \rhd d_{x,1}] q_{z,i}^{-1} [(p_{x,i}c_{y,1}p_{x,i}^{-1}) \rhd q_{y,i}]^{-1} ). \label{Equation_p_and_qs_inverted}
	\end{align}
	
	We will shortly compare this to the action of the triple $(p_{x,i},q_{y,i},q_{z,i})$ via our group action, but we first wish to consider another aspect of $(p_{x,i},q_{y,i},q_{z,i})$. In addition to moving us between elements of $C$, $(p_{x,i},q_{y,i},q_{z,i})$ appears in the isomorphism between the stabiliser groups $Z_{C,i}$ and $Z_C$ that maps the class $\mathcal{E}$ of $Z_{C,i}$ to the class $[\mathcal{E}^{\text{stab.}}_C]_{i,i}$ of $Z_C$. It is natural that $(p_{x,i},q_{y,i},q_{z,i})$ appears in this isomorphism, because the isomorphism describes how the stabiliser group should change as we change the representative of $C$. As described by Equation \ref{Equation_definition_stabiliser_group_isomorphism}, if the class $\mathcal{E}$ has representative element $(s_{x,1},f_{y,1},f_{z,1})$, then the class $[\mathcal{E}^{\text{stab.}}_C]_{i,i}$ has representative element
	\begin{equation*}
		(s_{x,1}',f_{y,1}',f_{z,1}')= \big(p_{x,i}^{-1}s_{x,1}p_{x,i}, \: \: p_{x,i}^{-1} \rhd (q_{y,i}^{-1} f_{y,1} s_{x,1} \rhd q_{y,i}), \: \: p_{x,i}^{-1} \rhd (q_{z,i}^{-1} f_{z,1}(s_{x,1} \rhd q_{z,i})) \big).
	\end{equation*}
	Inverting this equation gives us
	\begin{equation*}
		\big(p_{x,i}s_{x,1}'p_{x,1}^{-1}, q_{y,i} [p_{x,i} \rhd f_{y,1}'] [s_{x,1} \rhd q_{y,i}^{-1}],q_{z,i} [p_{x,i} \rhd f_{z,1}'] [s_{x,1} \rhd q_{z,i}^{-1}]\big) = (s_{x,1},f_{y,1},f_{z,1}),
	\end{equation*}
	which implies that
	\begin{align}
		(s_{x,1},f_{y,1},f_{z,1})= \big(p_{x,i}s_{x,1}'p_{x,1}^{-1}, q_{y,i} [p_{x,i} \rhd f_{y,1}'] [(p_{x,i}s_{x,1}'p_{x,i}^{-1}) \rhd q_{y,i}^{-1}],q_{z,i} [p_{x,i} \rhd f_{z,1}'] [(p_{x,i}s_{x,1}'p_{x,i}^{-1}) \rhd q_{z,i}^{-1}]\big). \label{Equation_stabiliser_isomorphism inverted}
	\end{align}

	Now we consider the action of the triple $(p_{x,i},q_{y,i},q_{z,i})$ on the sextuple $(s_{x,1}',c_{y,1},c_{z,1},d_{x,1},f_{y,1}',f_{z,1}')$. Using Equation \ref{Equation_torus_group_action_1}, we have
	\begin{align}
		(p_{x,i},q_{y,i},q_{z,i}): \big(s_{x,1}'&,c_{y,1},c_{z,1},d_{x,1},f_{y,1}',f_{z,1}'\big) \notag \\
		&= \big(p_{x,i}s_{x,1}'p_{x,i}^{-1}, \: \partial(q_{z,i})p_{x,i}c_{y,1}p_{x,i}^{-1}, \: \partial(q_{y,i})p_{x,i} c_{z,1}p_{x,i}^{-1}, \notag \\ & \hspace{1cm} [(p_{x,i}c_{z,1}p_{x,i}^{-1}) \rhd q_{z,i}] q_{y,i} [p_{x,i} \rhd d_{x,1}] [(p_{x,i}c_{y,1}p_{x,i}^{-1}) \rhd q_{y,i}^{-1}]q_{z,i}^{-1}, \notag \\ & \hspace{1cm} q_{y,i} [p_{x,i} \rhd f_{y,1}'] [(p_{x,i}s_{x,1}'p_{x,i}^{-1}) \rhd q_{y,i}^{-1}], \: q_{z,i} [p_{x,i} \rhd f_{z,1}'] [(p_{x,i}s_{x,1}'p_{x,i}^{-1}) \rhd q_{z,i}^{-1}] \big).
	\end{align}
	
	Comparing this to Equations \ref{Equation_p_and_qs_inverted} and \ref{Equation_stabiliser_isomorphism inverted}, we see that we can write this as
	\begin{align}
		(p_{x,i},q_{y,i},q_{z,i}): \big(s_{x,1}'&,c_{y,1},c_{z,1},d_{x,1},f_{y,1}',f_{z,1}'\big) \notag \\
		&= (s_{x,1}, c_{y,i},c_{z,i},d_{x,i},f_{y,1},f_{z,1}), \label{Equation_torus_group_action_p_and_qs}
	\end{align}
	where $(s_{x,1},f_{y,1},f_{z,1})$ is the representative element of the class $\mathcal{E}$ whose associated class $[\mathcal{E}^{\text{stab.}}_C]_{i,i}$ is represented by $(s_{x,1}',f_{y,1}',f_{z,1}')$. Therefore, acting on the sextuple (which represents the label of a $Y$ operator in our projector) with $(p_{x,i},q_{y,i},q_{z,i})$ moves us between elements of the class $C$ and also simultaneously performs the transformation between the associated stabiliser groups.

	So far, we have written our Conditions \ref{Y6}-\ref{Y8} in terms of a group action of triples in $(G,E,E)$ on the labels of our $Y$ operators. We have then considered the action of particular classes of triples via this group action. Now we wish to show that any triple can be decomposed into a pair of triples, where one is a representative $(p_{x,i},q_{y,i},q_{z,i})$ and one is from the stabiliser group. That is, we wish to show that we can write an arbitrary triple $(a,b_1,b_2)$ as the product
	$$(a,b_1,b_2)= (s,f_1,f_2) \cdot (p_{x,i},q_{y,i},q_{z,i})$$
	for some $i$, where $(s,f_1,f_2)$ belongs to a class in the stabiliser group $Z_{C,i}$. To see this, consider the action of $(a,b_1,b_2)$ on a sextuple $(g,c_{y,1},c_{z,1},d_{x,1},e_1,e_2)$, where $(c_{y,1},c_{z,1},d_{x,1})$ is the representative element of class $C$. We showed earlier that the action can only change $(c_{y,1},c_{z,1},d_{x,1})$ to another element of the class $C$ (see the discussion near Equation \ref{Equation_torus_group_action_1}). Therefore, we can write the action of $(a,b_1,b_2)$ on this sextuple as
	$$(a,b_1,b_2): (g,c_{y,1},c_{z,1},d_{x,1},e_1,e_2) = (g',c_{y,i},c_{z,i},d_{x,i}, e_1', e_2'),$$
	for some index $i$ and elements $(g',e_1',e_2')$. The effect of $(a,b_1,b_2)$ on the part of the sextuple corresponding to $C$ is the same as the action of $(p_{x,i},q_{y,i},q_{z,i})$ given in Equation \ref{Equation_torus_group_action_p_and_qs}. We therefore consider writing
	$$(a,b_1,b_2) = (s,f_1,f_2) \cdot (p_{x,i},q_{y,i},q_{z,i}),$$
	where $$(s,f_1,f_2)= (a,b_1,b_2) \cdot (p_{x,i},q_{y,i},q_{z,i})^{-1}$$
	and we have not yet made any claims about the properties of $(s,f_1,f_2)$. Then 
	$$(p_{x,i},q_{y,i},q_{z,i}):(g,c_{y,1},c_{z,1},d_{x,1},e_1,e_2) = (g'',c_{y,i},c_{z,i},d_{x,i}, e_1'', e_2'')$$
	for some other elements $(g'',e_1'',e_2'')$. We can therefore alternatively write 
	\begin{align*}
		(a,b_1,b_2) : (g,c_{y,1},c_{z,1},d_{x,1},e_1,e_2) &= ((s,f_1,f_2) \cdot (p_{x,i},q_{y,i},q_{z,i})) : (g,c_{y,1},c_{z,1},d_{x,1},e_1,e_2)\\
		&= (s,f_1,f_2) : ((p_{x,i},q_{y,i},q_{z,i}): (g,c_{y,1},c_{z,1},d_{x,1},e_1,e_2))\\
		&=(s,f_1,f_2): (g'',c_{y,i},c_{z,i},d_{x,i}, e_1'', e_2'')\\
		&\mbeq (g',c_{y,i},c_{z,i},d_{x,i}, e_1', e_2').
	\end{align*}
	
	Therefore, the remnant piece $(s,f_1,f_2)$ leaves $(c_{y,i},c_{z,i},d_{x,i})$ invariant and so belongs to the stabiliser group $Z_{C,i}$. This means that we can indeed decompose an arbitrary element $(a,b_1,b_2)$ into a product of an element $(p_{x,i},q_{y,i},q_{z,i})$ and an element in the stabiliser group $Z_{C,i}$. This means that, if we can show that the coefficient of the $Y$ operator in the projector is invariant when we act on the label with these two types of elements, then the coefficient is also invariant when we act on the label with a general element and so Conditions \ref{Y6}-\ref{Y8} are satisfied. The projector is given by
	\begin{align*}
		P^{R,C}(m)=& \frac{|R|}{|E||Z_C|} \sum_{i=1}^{|C|} \sum_{\mathcal{E}\in Z_{C,i}} \sum_{m=1}^{|R|} \sum_{\lambda \in E} D^R_{m,m}([\mathcal{E}_C^{\text{stab.}}]_{i,i}) Y^{(\partial(\lambda) s_{x,1},\: c_{y,i}, \: c_{z,i}, \: d_{x,i}, \: [c_{z,i} \rhd \lambda] f_{y,1} \lambda^{-1}, \: [c_{y,i} \rhd \lambda] f_{z,1} \lambda^{-1} )}(m).
	\end{align*}
	
	We see that the coefficient for a given $Y$ operator $Y^{(\partial(\lambda) s_{x,1},\: c_{y,i}, \: c_{z,i}, \: d_{x,i}, \: [c_{z,i} \rhd \lambda] f_{y,1} \lambda^{-1}, \: [c_{y,i} \rhd \lambda] f_{z,1} \lambda^{-1} )}(m)$ is
	$$\frac{|R|}{|E||Z_C|} \sum_{m=1}^{|R|} D^R_{m,m}([\mathcal{E}_C^{\text{stab.}}]_{i,i}).$$
	The factor of $\frac{|R|}{|E||Z_C|}$ is a constant for a given projector, so we are interested in the $\sum_{m=1}^{|R|} D^R_{m,m}([\mathcal{E}_C^{\text{stab.}}]_{i,i})$ part of this expression. We wish to see which $Y$ operators have the same coefficient. We already showed that the factors involving $\lambda$ in the label of the $Y$ operator take us to different elements of the class $\mathcal{E}$ and that these different elements have the same coefficient. We will therefore ignore these factors in the following discussion. The label of the $Y$ operator with $\lambda=1_E$ is $(s_{x,1}, c_{y,i},c_{z,i},d_{x,i},f_{y,1},f_{z,1})$, where $(c_{y,i},c_{z,i},d_{x,i})$ is the $i$th element of class $C$ and $(s_{x,1},f_{y,1},f_{z,1})$ is the representative element of the class $\mathcal{E}$. We will start by considering the $Y$ operator with label $(s'_{x,1},c_{y,1},c_{z,1},d_{x,1},f'_{y,1},f'_{z,1})$, where $(s'_{x,1},f'_{y,1},f'_{z,1})$ is the representative element of a class $X \in Z_C$. The coefficient of this $Y$ operator is
	$$\sum_{m=1}^{|R|} D^R_{m,m}([X_C^{\text{stab.}}]_{1,1})=\sum_{m=1}^{|R|} D^R_{m,m}(X).$$
	
	We now wish to show that this coefficient is the same as the coefficient of the $Y$ operator with label 
	$$(a,b_1,b_2) : (s'_{x,1},c_{y,1},c_{z,1},d_{x,1},f'_{y,1},f'_{z,1}),$$
	for general $(a,b_1,b_2) \in (G,E,E)$. We decompose $(a,b_1,b_2)$ as
	$$(a,b_1,b_2)= (s,f_1,f_2) \cdot (p_{x,i},q_{y,i},q_{z,i})$$
	where $(s,f_1,f_2)$ is in a class $A$ of the stabiliser group $Z_{C,i}$. Then we consider first the action of $(p_{x,i},q_{y,i},q_{z,i})$. As described in Equation \ref{Equation_torus_group_action_p_and_qs}, we have
	$$(p_{x,i},q_{y,i},q_{z,i}):(s'_{x,1},c_{y,1},c_{z,1},d_{x,1},f'_{y,1},f'_{z,1})=(s_{x,1},c_{y,i},c_{z,i},d_{x,i},f_{y,1},f_{z,1}),$$
	where $(s_{x,1},f_{y,1},f_{z,1})$ is the representative element of the class $\mathcal{E}$ in $Z_{C,i}$ that satisfies $[\mathcal{E}_C^{\text{stab.}}]_{i,i}=X$ (recall that $X$ is the class containing $(s'_{x,1},f'_{y,1},f'_{z,1})$). What is the coefficient of this new label set? From our expression for the projector, the coefficient is given (up to constant factors) by
	$$\sum_{m=1}^{|R|} D^R_{m,m}([\mathcal{E}_C^{\text{stab.}}]_{i,i}).$$
	However, using the fact that $X=[\mathcal{E}_C^{\text{stab.}}]_{i,i}$, this is just
	$$\sum_{m=1}^{|R|} D^R_{m,m}(X),$$
	which is the same coefficient as the original $Y$ operator. We write this as
	\begin{equation}
		(s'_{x,1},c_{y,1},c_{z,1},d_{x,1},f'_{y,1},f'_{z,1}) \sim (p_{x,i},q_{y,i},q_{z,i}):(s'_{x,1},c_{y,1},c_{z,1},d_{x,1},f'_{y,1},f'_{z,1}), \label{Equation_projector_equivalence_p_and_qs}
	\end{equation}
	indicating that the coefficient is preserved under the action of $(p_{x,i},q_{y,i},q_{z,i})$. Next we need to show that the coefficient is unchanged by the further action of $(s,f_1,f_2)$. We have
	\begin{align*}
		\big((s,f_1,f_2) \cdot &(p_{x,i},q_{y,i},q_{z,i})\big):(s'_{x,1},c_{y,1},c_{z,1},d_{x,1},f'_{y,1},f'_{z,1})\\
		&= (s,f_1,f_2) : (s_{x,1},c_{y,i},c_{z,i},d_{x,i},f_{y,1},f_{z,1}).
	\end{align*}
	
	Because $(s,f_1,f_2)$ belongs to the stabiliser group, we can use Equation \ref{Equation_torus_group_action_stabiliser_element_2} to write
	\begin{align*}
		(s,f_1,f_2) : &(s_{x,1},c_{y,i},c_{z,i},d_{x,i},f_{y,1},f_{z,1})\\&= \big(ss_{x,1}s^{-1}, c_{y,i}, c_{z,i}, d_{x,i}, f_1 [s \rhd f_{y,1}] [(ss_{x,1}s^{-1}) \rhd f_1^{-1}], f_2 [s \rhd f_{z,1}] [(ss_{x,1}s^{-1}) \rhd f_2^{-1}] \big). 
	\end{align*}
	Then from Equation \ref{Equation_torus_group_action_stabiliser_element_3}, we know that $$(ss_{x,1}s^{-1}, f_1 [s \rhd f_{y,1}] [(ss_{x,1}s^{-1}) \rhd f_1^{-1}], f_2 [s \rhd f_{z,1}] [(ss_{x,1}s^{-1}) \rhd f_2^{-1}])$$
	is the same as $(s,f_1,f_2)\cdot (s_{x,1},f_{y,1},f_{z,1}) \cdot (s,f_1,f_2)^{-1}$. This means that this element belongs to the class $A\mathcal{E}A^{-1}$, where $(s,f_1,f_2)$ belongs to the class $A$ in $Z_{C,i}$ and $(s_{x,1},f_{y,1},f_{z,1})$ belongs to the class $\mathcal{E}$. Therefore, the coefficient for the $Y$ operator of label $(ss_{x,1}s^{-1}, f_1 [s \rhd f_{y,1}] [(ss_{x,1}s^{-1}) \rhd f_1^{-1}], f_2 [s \rhd f_{z,1}] [(ss_{x,1}s^{-1}) \rhd f_2^{-1}])$ is
	$$\sum_{m=1}^{|R|} D^R_{m,m}([(A\mathcal{E}A^{-1})_C^{\text{stab.}}]_{i,i}),$$
	from our expression for the projector. Using the isomorphism property of the map from $Z_{C,i}$ to $Z_C$, we can write this as
	$$\sum_{m=1}^{|R|} D^R_{m,m}([A^{\text{stab.}}_C]_{i,i} [\mathcal{E}_C^{\text{stab.}}]_{i,i} [A^{\text{stab.}}_C]_{i,i}^{-1}).$$
	However, $\sum_{m=1}^{|R|} D^R_{m,m}([\mathcal{E}_C^{\text{stab.}}]_{i,i})$ is the trace of the matrix, which is invariant under conjugation. This indicates that
	$$\sum_{m=1}^{|R|} D^R_{m,m}([A^{\text{stab.}}_C]_{i,i} [\mathcal{E}_C^{\text{stab.}}]_{i,i} [A^{\text{stab.}}_C]_{i,i}^{-1}) = \sum_{m=1}^{|R|} D^R_{m,m}([\mathcal{E}_C^{\text{stab.}}]_{i,i}),$$
	so the coefficient of the $Y$ operator with label 
	$$\big((s,f_1,f_2) \cdot (p_{x,i},q_{y,i},q_{z,i})\big):(s'_{x,1},c_{y,1},c_{z,1},d_{x,1},f'_{y,1},f'_{z,1})$$
	is the same as the one with label
	$$(p_{x,i},q_{y,i},q_{z,i}):(s'_{x,1},c_{y,1},c_{z,1},d_{x,1},f'_{y,1},f'_{z,1})$$
	and therefore (from Equation \ref{Equation_projector_equivalence_p_and_qs}) also the same as the coefficient of the $Y$ operator with label 
	$$(s'_{x,1},c_{y,1},c_{z,1},d_{x,1},f'_{y,1},f'_{z,1}).$$
	
	We have therefore shown that the $Y$ operators in the projector satisfy
	\begin{equation}
		(s'_{x,1},c_{y,1},c_{z,1},d_{x,1},f'_{y,1},f'_{z,1}) \sim (a,b_1,b_2):(s'_{x,1},c_{y,1},c_{z,1},d_{x,1},f'_{y,1},f'_{z,1}), \label{Equation_projector_equivalence_1}
	\end{equation}
	where $(a,b_1,b_2)$ is a general element of $(G,E,E)$ and the equivalence relation indicates that the coefficients of the $Y$ operators whose labels lie in the same equivalence class are the same. At the moment, we have only obtained this relation for $(c_{y,1},c_{z,1},d_{x,1})$, which is a specific element of the class $C$. We may therefore worry that this equivalence relation does not hold for arbitrary elements. However, we can obtain any element of class $C$ by applying an appropriate $(p_{x,i},q_{y,i},q_{z,i})$. Then given an arbitrary element $(s_{x,1},c_{y,i},c_{z,i},d_{x,i},f_{y,1},f_{z,1})$, to show that this the projector satisfies the equivalence relation 
	$$(s_{x,1},c_{y,i},c_{z,i},d_{x,i},f_{y,1},f_{z,1})\sim (a,b_1,b_2):(s_{x,1},c_{y,i},c_{z,i},d_{x,i},f_{y,1},f_{z,1}),$$
	we can write $(a,b_1,b_2) = (a,b_1,b_2) \cdot (p_{x,i},q_{y,i},q_{z,i}) \cdot (p_{x,i},q_{y,i},q_{z,i})^{-1}$. Then we know that
	$$(s_{x,1},c_{y,i},c_{z,i},d_{x,i},f_{y,1},f_{z,1})= (p_{x,i},q_{y,i},q_{z,i}): (s'_{x,1},c_{y,1},c_{z,1},d_{x,1},f'_{y,1},f'_{z,1})$$
	for some $(s'_{x,1},f'_{y,1},f'_{z,1})$, which implies that 
	\begin{align*}
		(s_{x,1},c_{y,i},c_{z,i},d_{x,i},f_{y,1},f_{z,1}) &\sim (s'_{x,1},c_{y,1},c_{z,1},d_{x,1},f'_{y,1},f'_{z,1})\\
		& = (p_{x,i},q_{y,i},q_{z,i})^{-1}:(s_{x,1},c_{y,i},c_{z,i},d_{x,i},f_{y,1},f_{z,1}) 
	\end{align*}
	from Equation \ref{Equation_projector_equivalence_1}. Then we can use Equation \ref{Equation_projector_equivalence_1} again to write 
	\begin{align*}
		(s'_{x,1},c_{y,1},c_{z,1},d_{x,1},f'_{y,1},f'_{z,1}) \sim& \big((a,b_1,b_2) \cdot (p_{x,i},q_{y,i},q_{z,i})\big):(s'_{x,1},c_{y,1},c_{z,1},d_{x,1},f'_{y,1},f'_{z,1})\\
		&= \big((a,b_1,b_2) \cdot (p_{x,i},q_{y,i},q_{z,i})\big) : \big((p_{x,i},q_{y,i},q_{z,i})^{-1} :(s_{x,1},c_{y,i},c_{z,i},d_{x,i},f_{y,1},f_{z,1})\big)\\
		&= \big((a,b_1,b_2) \cdot (p_{x,i},q_{y,i},q_{z,i}) \cdot (p_{x,i},q_{y,i},q_{z,i})^{-1}\big) :(s_{x,1},c_{y,i},c_{z,i},d_{x,i},f_{y,1},f_{z,1})\\
		&= (a,b_1,b_2):(s_{x,1},c_{y,i},c_{z,i},d_{x,i},f_{y,1},f_{z,1}).
	\end{align*}
	Therefore, 
	$$(s_{x,1},c_{y,i},c_{z,i},d_{x,i},f_{y,1},f_{z,1}) \sim (s'_{x,1},c_{y,1},c_{z,1},d_{x,1},f'_{y,1},f'_{z,1})$$
	and $$(s'_{x,1},c_{y,1},c_{z,1},d_{x,1},f'_{y,1},f'_{z,1}) \sim (a,b_1,b_2):(s_{x,1},c_{y,i},c_{z,i},d_{x,i},f_{y,1},f_{z,1}),$$
	which by the transitive property means that
	$$(s_{x,1},c_{y,i},c_{z,i},d_{x,i},f_{y,1},f_{z,1}) \sim(a,b_1,b_2):(s_{x,1},c_{y,i},c_{z,i},d_{x,i},f_{y,1},f_{z,1})$$
	for arbitrary $(a,b_1,b_2) \in (G,E,E)$. This establishes that the coefficient of the $Y$ operator is invariant under the group action on the label of the $Y$ operator and so Conditions \ref{Y6}-\ref{Y8} are satisfied.

	We have now established that all of our conditions (Conditions \ref{Y1}-\ref{Y8}) are satisfied and so the projectors do indeed belong to our space of measurement operators. Therefore, we have an orthogonal set of projectors for our space. Finally, we wish to show that they form a complete set of projectors. We want to show that $\sum_{C,R} P^{R,C}(m)$ is the identity, at least when acting in the space of unexcited membranes (the space that we project to before applying any $Y$ operators). We have
	\begin{align}
		&\sum_{\text{classes }C} \sum_{\substack{\text{irreps } R \\\text{ of } Z_C}} P^{R,C}(m) \notag\\
		& = \sum_{\text{classes }C} \sum_{\substack{\text{irreps } R \\\text{ of } Z_C}} \frac{|R|}{|E||Z_C|}\sum_{i=1}^{|C|} \sum_{\mathcal{E}\in Z_{C,i}} \sum_{m=1}^{|R|} \sum_{\lambda \in E} D^{R}_{m,m}([\mathcal{E}_C^{\text{stab.}}]_{i,i}) Y^{(\partial(\lambda)s_{x,1},c_{y,i},c_{z,i} d_{x,i}, \: [c_{z,i} \rhd \lambda] f_{y,1} \lambda^{-1}, \: [c_{y,i} \rhd \lambda] f_{z,1} \lambda^{-1})}(m). \label{Equation_torus_sum_projectors_1}
	\end{align}
	
	The sum over irreps $R$ of $Z_C$ allows us to make use of another orthogonality relation:
	\begin{align*}
		\sum_{\substack{\text{irreps } R \\\text{ of } Z_C}} \sum_m D^{R}_{m,m}([\mathcal{E}_C^{\text{stab.}}]_{i,i}) |R|&= \sum_{\substack{\text{irreps } R \\\text{ of } Z_C}} \sum_m D^{R}_{m,m}([\mathcal{E}_C^{\text{stab.}}]_{i,i}) \chi_R(1_{Z_C})\\
		&= \sum_{\substack{\text{irreps } R \\\text{ of } Z_C}} \chi_R([\mathcal{E}_C^{\text{stab.}}]_{i,i}) \chi_R (1_{Z_C}) \text{ (where $\chi_R$ is the character of the representation $R$)}\\
		&= |Z_C| \delta([\mathcal{E}_C^{\text{stab.}}]_{i,i},1_{Z_C}).
	\end{align*}
	Substituting this result into Equation \ref{Equation_torus_sum_projectors_1}, we have
	\begin{align}
		\sum_{\text{classes }C}& \sum_{\substack{\text{irreps } R \\\text{ of } Z_C}} P^{R,C}(m) \notag \\
		&= \sum_{\text{classes }C} \frac{1}{|E|}\sum_{i=1}^{|C|} \sum_{\mathcal{E}\in Z_{C,i}} \sum_{\lambda \in E} \delta([\mathcal{E}_C^{\text{stab.}}]_{i,i},1_{Z_C}) Y^{(\partial(\lambda)s_{x,1},c_{y,i},c_{z,i} d_{x,i}, \: [c_{z,i} \rhd \lambda] f_{y,1} \lambda^{-1}, \: [c_{y,i} \rhd \lambda] f_{z,1} \lambda^{-1})}(m).
		\label{Equation_torus_sum_projectors_2}
	\end{align}
	
	Therefore, we need to look at the case where $[\mathcal{E}_C^{\text{stab.}}]_{i,i}=1_{Z_C}$. This condition on $[\mathcal{E}_C^{\text{stab.}}]_{i,i}$ induces one on the index that we sum over, $\mathcal{E}_{C,i}$. Given that the relation between the two classes is an isomorphism, if $[\mathcal{E}_C^{\text{stab.}}]_{i,i}$ is the identity of $Z_C$ then $\mathcal{E}_{C,i}$ must be the identity of the group $Z_{C,i}$. We can verify this explicitly using the relationship between the representatives of the two classes (see Equation \ref{Equation_definition_stabiliser_group_isomorphism}):
	\begin{align*}
		p_{x,i}^{-1} s_{x,1} p_{x,i} &=1_G \iff s_{x,1}=1_G\\
		p_{x,i}^{-1} \rhd (q_{z,i}^{-1} f_{z,1} [s_{x,1} \rhd q_{z,i}]) & = 1_E \iff f_{z,1} =1_E\\
		p_{x,i}^{-1} \rhd (q_{y,i}^{-1} f_{y,1} [s_{x,1} \rhd q_{y,i}]) &= 1_E \iff f_{y,1}=1_E
	\end{align*}
	so that
	$$[\mathcal{E}_C^{\text{stab.}}]_{i,i}=1_{Z_C} \iff \mathcal{E}_{C,i}=1_{Z_{C,i}}.$$
	Applying this to Equation \ref{Equation_torus_sum_projectors_2}, we see that the sum of projectors becomes
	\begin{equation}
		\sum_{\text{classes }C} \sum_{\substack{\text{irreps } R \\\text{ of } Z_C}} P^{R,C}(m) = \sum_C \sum_{i=1}^{|C|} \sum_{\lambda \in E} \frac{|Z_C|}{|E||Z_C|} Y^{(\partial(\lambda),c_{y,i},c_{z,i},d_{x,i}, [c_{z,i} \rhd \lambda] \lambda^{-1}, [c_{y,i} \rhd \lambda] \lambda^{-1})}(m). \label{Equation_torus_sum_projectors_3}
	\end{equation}
	
	Now we take a closer look at the particular $Y$ operators that appear in this expression. Splitting the $Y$ operator into the constituent membrane and ribbon operators (using Equations \ref{Equation_T_operator_definition_1} and \ref{Equation_define_Y}), we have
	\begin{align}
		Y^{(\partial(\lambda),c_{y,i},c_{z,i},d_{x,i}, [c_{z,i} \rhd \lambda] \lambda^{-1}, [c_{y,i} \rhd \lambda] \lambda^{-1})}(m) &=T^{\big[ [c_{z,i} \rhd \lambda] \lambda^{-1}, \: [c_{y,i} \rhd \lambda^{-1}] \lambda, \: d_{x,i} \:,c_{y,i}^{-1}, \:c_{z,i}^{-1}, \:\partial(\lambda)\big]}(m) \notag\\
		& \hspace{-1cm} =B^{[c_{z,i} \rhd \lambda] \lambda^{-1}}(c_1)B^{[c_{y,i} \rhd \lambda^{-1}] \lambda}(c_2)C^{\partial(\lambda)}_T(m) \delta(\hat{e}(m),d_{x,i}) \delta (\hat{g}(c_1),c_{y,i}^{-1}) \delta(\hat{g}(c_2), c_{z,i}^{-1}). \label{Equation_Y_operator_identity}
	\end{align}
	
	Recall from Sections \ref{Section_3D_Topological_Charge_Torus_Tri_trivial} and \ref{Section_3D_Topological_Charge_Torus_Tri_nontrivial} that when we set up the restrictions on our membrane operators, one of them was not required for commutation with the energy terms, but instead generated equivalent operators (when acting on an unexcited membrane), so we summed over the equivalent operators to avoid over-counting. That condition came from the edge transforms on the outwards edges (those cut by the dual membrane) and generated terms of the form
	$$B^{[g_{c_2}^{-1} \rhd e] e^{-1}}(c_1) B^{[g_{c_1}^{-1} \rhd e^{-1}] e} C^{\partial(e)}_T(m)$$
	which has the same form as the term
	$$B^{[c_{z,i} \rhd \lambda] \lambda^{-1}}(c_1)B^{[c_{y,i} \rhd \lambda^{-1}] \lambda}(c_2)C^{\partial(\lambda)}_T(m)$$
	at the front of the last line of Equation \ref{Equation_Y_operator_identity} (with $e= \lambda$). This indicates that this term acts trivially on the space where the toroidal region is unexcited (it is equivalent to a series of edge transforms). We can therefore replace the term with the equivalent operator
	$$B^{1_E}(c_1)B^{1_E}(c_2)C^{1_G}(m)=I,$$
	which is just the identity operator. Therefore, our sum over projectors from Equation \ref{Equation_torus_sum_projectors_3} becomes
	\begin{align}
		\sum_{\text{classes }C} \sum_{\substack{\text{irreps } R \\\text{ of } Z_C}} P^{R,C}(m) &= \sum_{\text{classes }C} \sum_{i=1}^{|C|} (\sum_{\lambda \in E} \frac{1}{|E|}) \delta(\hat{e}(m),d_{x,i}) \delta(\hat{g}(c_1),c_{y,i}^{-1}) \delta(\hat{g}(c_2),c_{z,i}^{-1}) \notag \\
		&= \sum_{\text{classes }C} \sum_{i=1}^{|C|} \delta(\hat{e}(m),d_{x,i}) \delta(\hat{g}(c_1),c_{y,i}^{-1}) \delta(\hat{g}(c_2),c_{z,i}^{-1}). \notag
	\end{align}
	Then because $i$ indexes the elements of $C$, the sum over all classes $C$ of boundary $\mathcal{G}$-colourings and over the index $i$ for that class reaches all boundary $\mathcal{G}$-colourings. We therefore have
	\begin{align}
		\sum_{\text{classes }C} \sum_{\substack{\text{irreps } R \\\text{ of } Z_C}} P^{R,C}(m)	&= \sum_{\text{all boundary $\mathcal{G}$-colourings $(y,z,x)$}} \delta(\hat{e}(m),x) \delta(\hat{g}(c_1),y^{-1}) \delta(\hat{g}(c_2),z^{-1}). \label{Equation_torus_sum_projectors_4}
	\end{align}
	
	The boundary $\mathcal{G}$-colourings give all allowed values of $\hat{e}(m), \hat{g}(c_1)^{-1}$ and $\hat{g}(c_2)^{-1}$ given that the membrane is unexcited (i.e., all values that satisfy fake-flatness), so summing the Kronecker delta over all of the possible colourings gives the identity on the unexcited membrane space. To put that another way, we could extend the sum to add in all of the rest of the triples $(y,z,x)$ that are not included here, without affecting the action of our operator on our space (though of course this would change the action on a generic state). This would change our sum of projectors to
	$$\sum_{y,z \in G} \sum_{x \in E} \delta(\hat{e}(m),x) \delta(\hat{g}(c_1)^{-1},y) \delta(\hat{g}(c_2)^{-1},z)=I.$$
	Therefore, we have
	$$\sum_{\text{classes }C} \sum_{\substack{\text{irreps } R \\\text{ of } Z_C}} P^{R,C}(m) =I_{\text{restr.}},$$
	where $I_{\text{restr.}}$ is the identity on the space of unexcited membranes. This indicates that our projectors are complete in this space. We already showed that the projectors are orthogonal and the number of them match the dimension of our space, which indicates that we have correctly constructed the projectors for topological charge measured by a torus.

	\subsection{Topological charge within a sphere}
	\label{Section_sphere_topological_charge_appendix_full}

	In Section \ref{Section_Sphere_Charge_Reduced} of the main text we described how to measure the topological charge held within a sphere and the conditions that must be satisfied by the measurement operators, in the case where $E$ is Abelian and $\partial$ maps to the centre of $G$ (Case 2 in Table \ref{Table_Cases} of the main text). Then we defined projection operators that were a linear combination of the measurement operators. In this section we will prove that the conditions we gave for the measurement operators are correct. We will also prove that the projection operators given are indeed projectors and that they form a complete orthogonal set.

	To measure the charge within a sphere we take the same approach as we did with the torus in Sections \ref{Section_3D_Topological_Charge_Torus_Tri_trivial} and \ref{Section_3D_Topological_Charge_Torus_Tri_nontrivial}. Firstly we project to the part of our Hilbert space where there are no excitations on the measurement surface. Then we consider which operators we can apply in this region. As described in Section \ref{Section_Sphere_Charge_Reduced} of the main text, any closed ribbons that lie on the sphere are contractible on the sphere and hence any non-confined ribbon operators can be contracted to nothing. On the other hand, the confined ribbon operators will inevitably produce excitations along their path. In the case of the torus, we saw that applying such confined ribbons along the seams of the torus could cancel the boundary effects of the membrane operators applied on the torus, and therefore prevent excitations near the boundary. However, in the case of the sphere, where there are no non-contractible loops, the membrane operators do not have boundary effects. For example, recall from Section \ref{Section_3D_Topological_Charge_Torus_Tri_nontrivial} that a magnetic membrane operator applied on a torus produces plaquette excitations along its boundary due to the discontinuity in the paths defined across the membrane. This discontinuity means that certain paths to the same location cannot be deformed into one-another, which leads to the additional plaquette excitations. In the case of a sphere, all such paths can be deformed into one-another and so this discontinuity does not occur.

	In a similar way, on the torus the $E$-valued membrane and magnetic membrane operator can produce edge excitations, which can cancel the edge excitations of confined electric ribbon operators applied on the seams of the torus. However, when there are no non-contractible cycles this is not the case and so we cannot include the confined electric ribbon operators. As we described in Section \ref{Section_Sphere_Charge_Reduced} of the main text, this leaves us only with the magnetic membrane operator $C^{h}_T(m)$ and the $E$-valued membrane operator $L^e(m)=\delta(\hat{e}(m),e)$ applied over the whole surface of the sphere. As in the case of the torus, we take the start-points of these two operators to be the same. This is because having separate start-points for the different operators would require each operator to commute separately with the vertex transforms at their start-point. However, if the membrane operators individually commute with the vertex operator at the start-point, then we can move the start-point of one of these membrane operators without affecting the action of the operator (as we showed in Section \ref{Section_magnetic_membrane_central_change_sp} for the magnetic membrane operator and Section S-I C in Ref. \cite{HuxfordPaper2} for the $E$-valued membrane operator). In particular, we could then put the start-points of each membrane operator in the same position anyway. Therefore, without loss of generality, we can consider all of the start-points of the operators to be in the same location.

	Taking all of this into account, our measurement operators (omitting the initial projection to the space where the sphere is unexcited) take the form
	\begin{equation}
		\hat{O}=\sum_{e \in E} \sum_{h \in G} \alpha_{h,e}C^{h}_T(m)L^{e}(m).
	\end{equation} 
	Next we have to ensure that the measurement operator commutes with the Hamiltonian, which will restrict which sets of coefficients $\alpha_{h,e}$ are allowed. We first consider the blob energy terms. The $E$-valued membrane operator $L^e(m)$ commutes with all blob energy terms, but the magnetic part $C^h_T(m)$ might not commute with one particular blob term. Recall from Section \ref{Section_3D_MO_Central} of the main text that the operator $C^{h}_T(m)$ affects the value of the privileged blob, blob 0, that we define when we specify the magnetic membrane operator $C^h_T(m)$. In the case where the start-point lies on the direct membrane (as opposed to the case considered in Section \ref{Section_3D_Braiding_Central} of the main text, where we displace the start-point away from the membrane), the operator $C^{h}_T(m)$ takes the surface label of the blob from $1_E$ (if it is initially unexcited) to $[h \rhd \hat{e}(m)^{-1}] \hat{e}(m)$, as we proved in Section \ref{Section_Magnetic_Tri_Non_Trivial}. Combining this with the $E$-valued membrane operator $L^e = \delta(\hat{e}(m),e)$ fixes $\hat{e}(m)$ so that the surface label of blob 0 is given by $[h \rhd e^{-1}] e$. Requiring the blob to remain unexcited after the action of our operator therefore gives the restriction 
	\begin{equation}
		h \rhd e =e. \tag{R1}
	\end{equation}
	That is, the coefficient $\alpha_{h,e}$ must be zero for pairs $(h,e)$ not satisfying this restriction. Using sets of coefficients that do satisfy this condition ensures that the operator will commute with every blob energy term, because as already stated the blob term at blob 0 is the only one that could have been left excited by $C^{h}_T(m)L^e(m)$.

	The next energy terms that we consider are the edge transforms. The edge transforms on the surface (i.e, for edges lying in the direct membrane, rather than pointing upwards from the membrane) have no effect. This is clear for the transforms in the bulk of the membrane (i.e., away from any boundaries), since it is true for our open magnetic and $E$-valued membrane operators (as we showed in Section \ref{Section_ribbon_membrane_energy_commutation}). However, when we produce our closed membrane operator we have to ``close up" an open membrane operator by bringing the boundary of the open membrane together. This could leave a ``seam", where excitations may be present. This is a point that was relevant for the torus, where we saw that the edge transforms on the cycles of the torus were non-trivial (see Section \ref{Section_Torus_Charge} of the main text). However, for the sphere, which has no non-trivial cycles, this seam does not lead to any excitations and the membrane operator commutes with the edge transforms here (we can see that this must be the case by considering what would happen on the torus if we removed the non-trivial cycles or took their labels to be trivial in Equations \ref{Equation_torus_charge_edge_transform_cycle_1_surface_label} and \ref{Equation_torus_charge_edge_transform_cycle_2_surface_label}).

	In addition to the edge transforms for edges in the direct membrane, we must also consider edge transforms on the ``cut" edges, the edges that are affected by the magnetic membrane (see Figure \ref{fluxmembrane2} in the main text for a reminder). We will call these the outwards edges, because they point away from the direct membrane. We consider applying outward edge transforms on each outwards edge $i$, with the edge transform on edge $i$ labelled by $\hat{g}(s.p-v_i)^{-1} \rhd f$. Here $s.p-v_i$ is the path from the start-point of the operator to the end of the edge $i$ that lies on the membrane (the same path that we considered when looking at the action of the magnetic membrane operator on each edge). These edge transforms replicate the action of a magnetic membrane operator of label $\partial(f)$ (one of our condensed magnetic membranes), as we proved in Section \ref{Section_condensation_magnetic_centre_case} (in that case we considered open membranes, for which the membrane operator was replicated up to a ribbon operator around the boundary, which would be trivial for a spherical membrane). Therefore, combining this with our magnetic membrane operator labelled by $h$ gives us a total magnetic membrane operator labelled by $\partial(f)h$. However, because we project to the case where these edges are unexcited, we ensure that edge transforms such as these act trivially. This is because acting on a state where an edge $i$ is unexcited with $\mathcal{A}_i^f$ is equivalent to acting with $\mathcal{A}_i^f \mathcal{A}_i$, because such a state $\ket{\psi}$ satisfies $\mathcal{A}_i \ket{\psi} = \ket{\psi}$ (is an eigenstate of the edge term with eigenvalue 1). Then, as described in Section \ref{Section_Recap_3d} of the main text, $\mathcal{A}_i^f \mathcal{A}_i=\mathcal{A}_i$, so
	$$\mathcal{A}_i^f \mathcal{A}_i \ket{\psi}=\mathcal{A}_i \ket{\psi}= \ket{\psi}.$$
	This means that the action of the outwards edge transforms on the state must be trivial. Because the action of the edge transforms is equivalent to the action of $C^{\partial(f)}_T(m)$, this means that the action of $C^{\partial(f)}_T(m)$ must also be trivial on our state (the state where we projected to the energy terms on the sphere being satisfied). This means that the two membrane operators $C^{h}(m)$ and $C^{\partial(f)h}_T(m)$ actually have the same action on our state and should not be considered as giving us distinct topological charge measurement operators. We write this in terms of an equivalence relation as 
	\begin{equation}
		h \sim \partial(f)h, \tag{R2}
	\end{equation}
	meaning that we always include an equal superposition of $C^h_T(m)$ and $C^{\partial(f)h}_T(m)$ in our operator to reflect the fact that these operators are not distinct (instead of including a superposition, we could simply remember not to include the equivalent operators separately, but using a sum is a slightly neater way of dealing with this problem).

	Now consider the plaquette terms. Because there are no non-contractible cycles, and so the sphere has no potential boundaries, we know that the surface label of the sphere is always in the kernel of $\partial$ when fake-flatness is satisfied (because fake-flatness ensures that the $\partial$ of a surface label matches the inverse of the boundary path element). This gives us the condition that
	\begin{equation}
		\partial(e)=1_G. \tag{R3}
	\end{equation}
	Apart from this, all of the plaquette terms are automatically satisfied, including the plaquettes cut by the dual membrane of the magnetic membrane operator. As with the edge transforms, this is not obvious for the plaquettes near the location where we ``close up" our membrane. In Section \ref{Section_3D_Topological_Charge_Torus_Tri_nontrivial} we saw that this could lead to excitations for a toroidal surface, but taking the non-contractible cycles of the torus to have labels in $\partial(E)$ in Equations \ref{Equation_plaquette_holonomy_boundary_torus_1} and \ref{Equation_plaquette_holonomy_boundary_torus_2} to reflect the lack of non-contractible cycles on the sphere shows that the plaquette terms are satisfied on the sphere (because elements in $\partial(E)$ are in the centre of $G$ and therefore commute with $h$ in those equations).

	The final type of energy term to check is the vertex terms. Of these, only the vertex term at the mutual start-point of the operators that we apply may be excited by the $E$-valued and magnetic membrane operators, as we showed in Section \ref{Section_Magnetic_Tri_Nontrivial_Commutation} for the magnetic membrane operator and Ref. \cite{HuxfordPaper2} for the $E$-valued membrane operator. When we apply the vertex transform $A_{s.p}^{g^{-1}}$ at the start-point, we take the label of the magnetic membrane operator $h$ to $ghg^{-1}$ and the label of the $E$-valued membrane operator $e$ to $g \rhd e$, as shown in Section \ref{Section_Magnetic_Tri_Nontrivial_Commutation} (see Equation \ref{Equation_magnetic_membrane_tri_nontrivial_start_point_transform}) and Ref. \cite{HuxfordPaper2} (see Equation S11 in Section S-I C of the Supplemental Material). Therefore, in order to commute with the vertex term $A_{s.p}= \frac{1}{|G|} \sum_{g \in G} A_{s.p}^g$ we must always include an equal superposition of operators labelled by $(ghg^{-1}, g \rhd e)$ for each $g \in G$. We represent this condition with the equivalence relation 
	\begin{equation}
		(h,e) \sim (ghg^{-1}, g \rhd e). \tag{R4}
	\end{equation}

	With these restrictions, we reduce the task of counting the number of allowed topological charges to counting particular classes of pairs of elements $(h,e)$, with $h \in G$ and $e \in E$. In particular, we find classes of pairs that satisfy $h \rhd e =e$ (condition R1, from applying the blob condition) and $\partial(e) =1_G$ (condition R3, from fake-flatness), with the classes being defined by the equivalence relations $(h,e) \sim (\partial(f)h,e) $ (condition R2, from commutation with edge transforms) and $(h,e) \sim (ghg^{-1},g \rhd e)$ (condition R4, from commutation with the vertex transform at the start-point). We can combine the two types of equivalence relations into a single form of equivalence relation:
	\begin{equation}
		(h,e) \overset{\mathrm{S}}{\sim} (g\partial(f)hg^{-1},g \rhd e) \: \: \forall f \in E, \: \: g \in G. \label{Equation_sphere_equivalence_relation}
	\end{equation}
	Then the number of topological charges is the number of classes generated using the equivalence relation Equation \ref{Equation_sphere_equivalence_relation}, starting with pairs $h$ and $e$ that satisfy the restrictions R1 and R3 ($e$ is in the kernel of $\partial$ and $h \rhd e= e$). While we can use the restrictions and equivalence relations above to count the number of allowed charges, we also want to explicitly construct the projectors that measure each charge. To do this, we will first construct a simple basis for the allowed operators using the rules described above, then we perform a change of basis to give a set of projectors. By showing that the change of basis is reversible, we will establish a one-to-one correspondence between the projectors and the original basis, thus showing that the number of projectors matches the dimension of the space of allowed operators and that the projectors give a complete set of the charges that we can measure.

	First, we construct a simple basis for our measurement operators. From the restrictions we found above, the space of allowed operators is spanned by
	\begin{align*}
		\{T^{e,h}(m)= \sum_{g \in G} \sum_{f \in E} &C^{ghg^{-1}\partial(f)}_T(m)L^{g \rhd e}(m)\\
		& | \partial(e)=1_E, h \rhd e =e\}.
	\end{align*}
	We denote the set $\set{ h \in G | h \rhd e = e}$ by $Z_{e, \rhd}$ and make it a group with the multiplication inherited from $G$ (i.e., $Z_{e, \rhd}$ is a subgroup of $G$). Then the set of operators given above can be written as
	\begin{align*}
		\{T^{e,h}(m)= \sum_{g \in G} \sum_{f \in E} &C^{ghg^{-1}\partial(f)}_T(m)L^{g \rhd e}(m)\\
		& | \partial(e)=1_E, h \in Z_{e, \rhd} \}.
	\end{align*}
	
	While the space of measurement operators is spanned by the set of operators given above, this set contains redundancy, in that different values of $e$ and $h$ can give the same operator. For instance, given $e$ and $h$ we have that
	$$T^{e,h}(m)=T^{x \rhd e, xhx^{-1}}(m)$$
	for any $x \in G$, as a natural consequence of our equivalence relations. We therefore split the kernel of $\partial$ into ``$\rhd$-classes" defined by the relation that $e_1 \sim e_2$ if and only if there exists a $g \in G$ such that $g \rhd e_1 =e_2$. Then for each such class $C$ we pick a representative element of that class, $r_C$. We can define one measurement operator for each such class, with these operators then spanning the space of the allowed operators:
	\begin{equation}
		T^{C,h}(m) := T^{r_C,h}(m) = \sum_{g \in G} \sum_{f \in E} C^{ghg^{-1}\partial(f)}_T(m) L^{g \rhd r_C}(m). \label{Equation_sphere_measurement_operator_1_full}
	\end{equation}
	
	This is still not enough to completely eliminate the redundancy. Given two $G$ labels $h$ and $xhx^{-1}$ where $x$ is in $Z_{\rhd,r_C}$ (so that $x \rhd r_C=r_C$), the corresponding operators $T^{C,h}(m)$ and $T^{C,xhx^{-1}}(m)$ are the same:
	\begin{align*}
		T^{C,xhx^{-1}}(m)&=\sum_{g \in G} \sum_{f \in E}C^{gxhx^{-1}g^{-1}\partial(f)}_T(m)L^{g \rhd r_C}(m)\\
		&= \sum_{g \in G} \sum_{f \in E} C^{gxhx^{-1}g^{-1}\partial(f)}_T(m) L^{(gx) \rhd r_C}(m)\\
		& \hspace{1cm} \text{(using $x \rhd r_C =r_C$)}\\
		&= \sum_{g'=gx \in G} \sum_{f \in E} C^{g'h{g'}^{-1}\partial(f)}_T(m) L^{g' \rhd r_C}(m)\\
		&=T^{C,h}(m).
	\end{align*}
	
	This suggests that we should replace $h$ by a conjugacy class of $Z_{\rhd,r_C}$. However, we also have redundancy from the $\sum_{f \in E}$, which means that in fact $h$ and $xhx^{-1}\partial(w)$ (for any $w \in E$) will also give the same operator. We define classes in $Z_{\rhd,r_C}$ by the equivalence relation
	\begin{equation}
		h \sim xhx^{-1} \partial(w) 
		\label{union_coset_relation_full}
	\end{equation}
	for all $h, x \in Z_{\rhd, r_C}$ and $w \in E$. Then, rather than having a separate measurement operator for each element $h \in Z_{\rhd,r_C}$, we only have a distinct measurement operator for each class $D$ of $Z_{\rhd, r_C}$ defined by Equation \ref{union_coset_relation_full}. These classes are unions of cosets and in particular they are unions of the cosets related by conjugation. Therefore, these classes are like conjugacy classes of cosets, but where we take the union of the conjugacy-related cosets rather than having the cosets as the elements of the class. There is a one-to-one relation between these classes in $Z_{\rhd,r_C}$ and the conjugacy classes of cosets, that is the conjugacy classes of the quotient group $Z_{\rhd,r_C}/\partial(E)$ (note that $\partial(E)$ is a normal subgroup of $Z_{\rhd,r_C}$ due to the first Peiffer condition, Equation \ref{Equation_Peiffer_1} of the main text). Rather than implement our restriction to one measurement operator per class by choosing a representative of each class, we will extract an expression that manifestly depends only on the class rather than a particular choice of representative. From Equation \ref{Equation_sphere_measurement_operator_1_full}, we have
	$$T^{C,h}(m)=\sum_{g \in G} \sum_{f \in E} C^{ghg^{-1}\partial(f)}_T(m) L^{g \rhd r_C}(m),$$
	and we want to make this explicitly depend only on the class $D$ containing $h$. Firstly, this expression only depends on $f$ through $\partial(f)$, so we replace $\sum_{f \in E}$ with $\sum_{k \in \partial(E)}$ and drop the multiplicative factor $\frac{|E|}{|\partial(E)|}$ that this would result in (from the fact that multiple $f \in E$ give the same $\partial(f)$), to get
	$$\sum_{g \in G} \sum_{k \in \partial(E)} C^{ghg^{-1}k}_T(m) L^{g \rhd r_C}(m).$$

	Now for each element $e$ in the $\rhd$ class represented by $r_C$, we pick an element $q \in G$ such that $q \rhd r_C=e$. If we label the elements of the $\rhd$-class by $e_i$, where $i$ is an index that runs from 1 to the size of the class, then we denote the corresponding $q$ by $q_i$. We let $e_1=r_C$ and $q_1=1_G$. Then we can decompose any element $g \in G$ into a product $g=q_ix$, where $x$ is an element of $Z_{\rhd, r_C}$, in a unique way. To determine $q_i$, we find which element $e_i$ in $C$ is equal to $g \rhd r_C$ and the corresponding $q_i$ is the representative satisfying $e_i = q_i \rhd r_C$. Then we can uniquely determine $x =q_i^{-1}g$. This must be an element of $Z_{\rhd, r_C}$ because
	$$x \rhd r_C = q_i^{-1} \rhd (g \rhd r_C)= q_i^{-1} \rhd e_i = q_i^{-1} \rhd q_i \rhd r_C =r_C.$$
	
	We have seen that for each element $g \in G$ there is a unique decomposition into a pair $(q_i,x)$. Similarly, each pair $(q_i, x)$ obviously has a unique product $g=q_ix$. Therefore, the correspondence between pairs $(q_i,x)$ and elements $g \in G$ is one-to-one, and we can replace the sum over elements $g \in G$ with a sum over such pairs. To facilitate this, we denote the set of representatives $q_i$ by $Q_C$.

	Applying this to our measurement operator, we have
	\begin{align*}
		\sum_{g =qx \in G} \sum_{k \in \partial(E)} C^{ghg^{-1}k}_T(m)L^{g \rhd r_C}(m)
		&= \sum_{q \in Q_C} \sum_{x \in Z_{\rhd, r_C}} \sum_{k \in \partial(E)} C^{qxhx^{-1}q^{-1} k}_T(m) L^{(qx) \rhd r_C}(m).
	\end{align*}
	We can then use the fact that $x$ is in $Z_{\rhd, r_C}$ (by the definition of $x$) to write $(qx) \rhd r_C= q \rhd r_C$. We can also use the fact that $k \in \partial(E)$ is in the centre of $G$ to write $qxhx^{-1}q^{-1} k= qxhx^{-1}k q^{-1}$. Putting this together, we can write the measurement operator as
	\begin{align*}
		\sum_{g =qx \in G} \sum_{k \in \partial(E)} C^{ghg^{-1}k}_T(m)L^{g \rhd r_C}(m)
		&=\sum_{q \in Q_C} \sum_{x \in Z_{\rhd, r_C}} \sum_{k \in \partial(E)} C^{qxhx^{-1}kq^{-1}}_T(m) L^{q \rhd r_C}(m).
	\end{align*}
	
	Finally, we note that the element $xhx^{-1}k$ is always in the same equivalence class $D$ as $h$, where the class is defined by the equivalence relation Equation \ref{union_coset_relation_full} and all elements of the class $D$ can be written in this form. In fact, summing this expression over all $x$ and $k$ is equivalent to summing over all of the elements of the class (possibly summing over the elements multiple times, but the same number of times for each element of the class). To see this, consider an arbitrary element $h_1 \in D$. Then to count the number of times that $h_1$ appears in the sum over $x$ and $k$, we must calculate the quantity
	$$\sum_{x \in Z_{\rhd, r_C}} \sum_{k \in \partial(E)} \delta(xhx^{-1}k, h_1).$$
	
	By definition there exists some $x_1 \in Z_{\rhd, r_C}$ and $k_1 \in \partial(E)$ such that $h_1 = x_1h x_1^{-1} k_1$. Therefore, we can rewrite the sum as
	\begin{align*}
		\sum_{x \in Z_{\rhd, r_C}}& \sum_{k \in \partial(E)} \delta(xhx^{-1}k, h_1)\\
		&= \sum_{x \in Z_{\rhd, r_C}} \sum_{k \in \partial(E)} \delta(xhx^{-1}k, x_1hx_1^{-1}k_1)\\
		&=\sum_{x \in Z_{\rhd, r_C}} \sum_{k \in \partial(E)} \delta(x_1^{-1}xhx^{-1}kk_1^{-1}x_1,h).
	\end{align*}
	Using the fact that $k$ and $k_1$ are in the centre of $G$, we can commute $x_1$ next to $x^{-1}$ to obtain
	\begin{align*}
		\sum_{x \in Z_{\rhd, r_C}}& \sum_{k \in \partial(E)} \delta(xhx^{-1}k, h_1)\\
		&= \sum_{x \in Z_{\rhd, r_C}} \sum_{k \in \partial(E)} \delta(x_1^{-1}xhx^{-1}x_1kk_1^{-1},h).
	\end{align*}
	We can then define $x' =x_1^{-1}x$ and $k' =kk_1^{-1}$ and replace the sum over $x$ and $k$ with a sum over these primed variables (using the fact that we sum over groups in each case). This gives us
	\begin{align*}
		\sum_{x \in Z_{\rhd, r_C}}& \sum_{k \in \partial(E)} \delta(xhx^{-1}k, h_1)\\
		&= \sum_{x'=x_1^{-1}x \in Z_{\rhd, r_C}} \sum_{k'=kk_1^{-1} \in \partial(E)} \delta(x'hx^{\prime -1}k',h).
	\end{align*}
	However, this is identical to the expression that would count how many times $h$ appears in the sum. Therefore, $h_1$ appears the same number of times as $h$ in the sum, and because $h_1$ is an arbitrary element of the class $D$, the same result holds for any element in the class. This means that the sum over $x$ and $k$ is proportional to a sum over the elements $d \in D$. This tells us that our measurement operator satisfies
	\begin{align*}
		\sum_{q \in Q_C}& \sum_{x \in Z_{\rhd, r_C}} \sum_{k \in \partial(E)} C^{qxhx^{-1}kq^{-1}}_T(m) L^{q \rhd r_C}(m) \propto \sum_{q \in Q_C} \sum_{d \in D} C^{qdq^{-1}}_T(m) L^{q \rhd r_C}(m).
	\end{align*}

	Now we have rewritten the measurement operator in a way that explicitly only depends on the class $D$, without favouring specific elements within that class. We can then define a set of measurement operators, one for each class $C$ of the kernel and class $D$ of $Z_{\rhd,r_C}$,
	\begin{equation}
		T^{D,C}(m)= \sum_{q \in Q_C} \sum_{d \in D} C^{qdq^{-1}}_T(m) L^{q \rhd r_C}(m), \label{Equation_sphere_measurement_class_operator}
	\end{equation}
	which form a basis for the space of measurement operators. This tells us how many ``sphere" topological charges we have (one for each pair $C,D$) but we also want to create a set of projectors to definite topological charge, just as we did in Section \ref{Section_3D_Topological_Charge_Torus_Projectors} for the toroidal measurement surface. We claim that the projectors are given by
	\begin{equation}
		T^{R,C}(m)=\frac{|R|}{|Z_{\rhd,r_C}|} \sum_{D \in (Z_{\rhd,r_C})_{cl}} \chi_R(D) T^{D,C}(m),
		\label{Sphere_Projectors_appendix}
	\end{equation}
	where $R$ is an irrep of $Z_{\rhd,r_C}/\partial(E)$ of dimension $|R|$ and $(Z_{\rhd,r_C})_{cl}$ is the set of classes of $Z_{\rhd,r_C}$ defined by the equivalence relation in Equation \ref{union_coset_relation_full}. As discussed when we defined the classes, the equivalence class $D$ can be interpreted as a union of cosets related by conjugation, so each class is equivalent to a conjugacy class of the quotient group $Z_{\rhd,r_C}/\partial(E)$. There are the same number of irreps of a group as conjugacy classes, so the change of basis from classes to irreps preserves the number of operators (and indeed the change of basis is reversible). There is a slight subtlety here, in that the elements of the quotient group are the cosets and our classes are unions of cosets rather than sets of cosets. That is, we have defined our classes $D$ as subsets of the group $Z_{\rhd,r_C}$ rather than the quotient group. When we find the character of the class in irrep $R$, which is a representation of the quotient group, we need to find the conjugacy class of the quotient group that corresponds to the class. We simply see which cosets $D$ contains and take the character of one of those in $R$ (by conjugacy invariance of characters, every choice would give the same answer). For simplicity, we also define the character of an arbitrary element $g$ of $Z_{\rhd,r_C}$ in irrep $R$ of $Z_{\rhd,r_C}/\partial(E)$ to be the character of the coset that the element $g$ belongs to (effectively defining for each such irrep $R$ an irrep of $Z_{\rhd,r_C}$).

	In the rest of this section, we will check that the projectors defined in Equation \ref{Sphere_Projectors_appendix} are indeed a complete set of orthogonal projectors. First, we will demonstrate that they are orthogonal projectors. In order to do this, we must take a product of two of the projectors. This will be a linear combination of products of membrane operators, such as
	$$C^{h}_T(m)L^e(m) C_T^{g}(m)L^f(m) =C^h_T(m) \delta(\hat{e}(m),e) C^g_T(m) \delta(\hat{e}(m),f).$$
	The surface label $\hat{e}(m)$ commutes with the magnetic membrane operators, provided that we do not put blob 0 of $m$ inside the sphere (which would cause the blob ribbon operators that belong to $C^h_T(m)$ to pierce $m$) or move the start-point of $m$ outwards away from the membrane (which would cause the paths from the start-point to the plaquettes in $m$ to pass through the dual membrane). This commutation allows us to put the two magnetic membrane operators together. We know from Section \ref{Section_3D_Topological_Charge_Torus_Projectors} that
	$$C^h_T(m)C^g_T(m)=C^{hg}_T(m).$$
	In addition, the $E$-valued membrane operators satisfy $\delta(\hat{e}(m),e)\delta(\hat{e}(m),f) = \delta(e,f)\delta(\hat{e}(m),e)$. Therefore
	\begin{align*}
		C^{h}_T(m)L^e(m) C_T^{g}(m)L^f(m)& =C^h_T(m) \delta(\hat{e}(m),e) C^g_T(m) \delta(\hat{e}(m),f)\\
		&=C^h_T(m) C^g_T(m) \delta(\hat{e}(m),e) \delta(\hat{e}(m),f)\\
		&=C^{hg}_T(m) \delta(\hat{e}(m),e) \delta(e,f)\\ 
		&=C_T^{hg}(m)L^e(m) \delta(e,f).
	\end{align*}
	Therefore, the product of two of our projectors defined by Equation \ref{Sphere_Projectors_appendix}, with labels $(R,C)$ and $(R',C')$ is given by 
	\begin{align}
		T^{R,C}(m)T^{R',C'}(m) &= \frac{|R|}{|Z_{\rhd,r_C}|}\frac{|R'|}{|Z_{\rhd,r_{C'}}|} \sum_{D \in (Z_{\rhd,r_C})_{cl} } \sum_{D' \in (Z_{\rhd,r_{C'}})_{cl} } \sum_{d \in D} \sum_{d' \in D'} \sum_{q \in Q_C} \sum_{q' \in Q_{C'}} C^{qdq^{-1}}_T(m) L^{q \rhd r_C}(m) \notag \\
		& \hspace{0.5cm}C^{q'd'{q'}^{-1}}_T(m) L^{q' \rhd r_{C'}}(m)\chi_R(D) \chi_{R'}(D')\notag \\
		&= \frac{|R|}{|Z_{\rhd,r_C}|}\frac{|R'|}{|Z_{\rhd,r_{C'}}|} \sum_{D \in (Z_{\rhd,r_C})_{cl} } \sum_{D' \in (Z_{\rhd,r_{C'}})_{cl} } \sum_{d \in D} \sum_{d' \in D'} \sum_{q \in Q_C} \sum_{q' \in Q_{C'}} C^{qdq^{-1}q'd'{q'}^{-1}}_T(m) \notag \\ 
		& \hspace{1cm}L^{ q\rhd r_C}(m) \delta(q \rhd r_C,q' \rhd r_{C'}) \chi_R(D) \chi_{R'}(D'). \label{Equation_sphere_product_projectors_1_full}
	\end{align}

	Next, we note that if $q \rhd r_C=q' \rhd r_{C'}$, then $r_C$ and $r_{C'}$ belong to the same $\rhd$-class of the kernel of $\partial$. Therefore, $C=C'$ and so, because the $r_C$ are unique representatives of the $\rhd$-classes, $r_C=r_{C'}$. Furthermore, this means that $q \rhd r_C = q' \rhd r_C$. The $q$ are representatives so that each $e \in C$ is assigned a unique $q$ such that $e = q \rhd r_C$. If $q \rhd r_C$ is the same element as $q' \rhd r_C$ then they must have the same $q \in Q_C$, which means that $q' =q$. We can therefore rewrite the Kronecker delta $\delta(q \rhd r_C,q' \rhd r_{C'})$ as $\delta(C,C') \delta(q,q')$. Substituting this into Equation \ref{Equation_sphere_product_projectors_1_full} gives us
	\begin{align}
		T^{R,C}&(m)T^{R',C'}(m) \notag \\
		&= \frac{|R|}{|Z_{\rhd,r_C}|}\frac{|R'|}{|Z_{\rhd,r_{C'}}|} \sum_{D \in (Z_{\rhd,r_C})_{cl} } \sum_{D' \in (Z_{\rhd,r_{C'}})_{cl} } \sum_{d \in D} \sum_{d' \in D'} \sum_{q \in Q_C} \sum_{q' \in Q_{C'}} C^{qdq^{-1}q'd'{q'}^{-1}}_T(m) L^{q\rhd r_C}(m) \notag \\
		& \hspace{1cm}\delta(C,C') \delta(q,q') \chi_R(D) \chi_{R'}(D') \notag \\
		&= \delta(C,C') \frac{|R|}{|Z_{\rhd,r_C}|} \frac{|R'|}{|Z_{\rhd,r_C}|} \sum_{D \in (Z_{\rhd,r_C})_{cl} } \sum_{D' \in (Z_{\rhd,r_{C}})_{cl} } \sum_{d \in D} \sum_{d' \in D'} \sum_{q \in Q_C} C^{qdd'{q}^{-1}}_T(m) L^{q\rhd r_C}(m)\chi_R(D) \chi_{R'}(D'). \label{Equation_sphere_product_projectors_2_full}
	\end{align}
	
	Next we can replace the double sum $\sum_{D \in (Z_{\rhd,r_C})_{cl} } \sum_{d \in D}$ with a sum over all elements of $Z_{\rhd,r_C}$ (and do the same for $d'$ and $D'$), so we have
	\begin{align*}
		T^{R,C}&(m)T^{R',C'}(m) \\
		&= \delta(C,C') \frac{|R|}{|Z_{\rhd,r_C}|} \frac{|R'|}{|Z_{\rhd,r_C}|} \sum_{d, d' \in Z_{\rhd,r_C}} \sum_{q \in Q_C} C^{qdd'{q}^{-1}}_T(m) L^{q\rhd r_C}(m) \chi_R(d) \chi_{R'}(d'). 
	\end{align*}
	
	We then wish to use orthogonality relations to remove one of the irreps. In order to do so, we first introduce the dummy variable $\tilde{d}=dd' \in Z_{\rhd, r_C}$ and replace the sum over $d'$ with a sum over this new variable.
	\begin{align*}
		T^{R,C}&(m)T^{R',C'}(m) \\	
		&= \delta(C,C') \frac{|R|}{|Z_{\rhd,r_C}|} \frac{|R'|}{|Z_{\rhd,r_C}|} \sum_{d \in Z_{\rhd,r_C}} \sum_{\tilde{d}=dd' \in Z_{\rhd,r_C}} \sum_{q \in Q_C} C^{q\tilde{d}{q}^{-1}}_T(m) L^{q\rhd r_C}(m) \chi_R(d) \chi_{R'}(d^{-1} \tilde{d}).
	\end{align*}
	
	Next we use the fact that the character is the trace of the matrix representation to split $\chi_{R'}(d^{-1} \tilde{d})$ into contributions from $d^{-1}$ and $\tilde{d}$:
	\begin{align}
		T^{R,C}&(m)T^{R',C'}(m) \notag \\		
		&= \delta(C,C') \frac{|R|}{|Z_{\rhd,r_C}|} \frac{|R'|}{|Z_{\rhd,r_C}|} \sum_{d \in Z_{\rhd,r_C}} \sum_{\tilde{d}\in Z_{\rhd,r_C}} \sum_{q \in Q_C} C^{q\tilde{d}{q}^{-1}}_T(m) L^{q\rhd r_C}_T(m) \sum_{a=1}^{|R|} \sum_{b,c=1}^{|R'|} [D^{R}(d)]_{aa}[D^{R'}(d^{-1})]_{bc}[D^{R'}(\tilde{d})]_{cb}. \label{Equation_sphere_product_projectors_3_full}
	\end{align}
	
	Then we wish to use the Grand Orthogonality to evaluate $\sum_{d \in Z_{\rhd,r_C}} [D^{R}(d)]_{aa}[D^{R'}(d^{-1})]_{bc}$, because $d$ does not appear anywhere else in our expression. Because $R$ and $R'$ are defined as irreps of the quotient group $Z_{\rhd,r_C}/\partial(E)$ rather than of the group $Z_{\rhd,r_C}$, we must convert from elements of $Z_{\rhd,r_C}$ to cosets of the same group, using our definition that the matrix element for an element $d$ in $Z_{\rhd, r_C}$ in irrep $R$ is the same as the matrix element associated to the coset containing $d$. Then we have
	\begin{align*}
		\sum_{d \in Z_{\rhd, r_C}} [D^R(d)]_{aa} [D^{R'}(d^{-1})]_{bc}&= |\partial(E)| \sum_{\text{cosets $s$ of $Z_{\rhd,r_C}$ in $G$}} [D^R(s)]_{aa} [D^{R'}(s^{-1})]_{bc}\\
		&= |\partial(E)| \delta_{R,R'} \frac{|Z_{\rhd,r_C}/\partial(E)|}{|R|} \delta_{ab}\delta_{ac}\\
		&= \frac{|Z_{\rhd,r_C}|}{|R|} \delta_{R,R'} \delta_{ab}\delta_{ac},
	\end{align*}
	using the Grand Orthogonality Theorem. Inserting this into our earlier expression for the product of two projectors (Equation \ref{Equation_sphere_product_projectors_3_full}), we find that
	\begin{align*}
		T^{R,C}(m)T^{R',C'}(m)&= \delta(C,C') \frac{|R|}{|Z_{\rhd,r_C}|} \frac{|R'|}{|Z_{\rhd,r_C}|} \sum_{\tilde{d} \in Z_{\rhd,r_C}} \sum_{q \in Q_C} C^{q\tilde{d}{q}^{-1}}_T(m)L^{q\rhd r_C}(m) \sum_{a=1}^{|R|} \sum_{b,c=1}^{|R'|} \frac{|Z_{\rhd,r_C}|}{|R|} \delta_{R,R'} \delta_{ab}\delta_{ac} [D^{R'}(\tilde{d})]_{cb}\\
		&= \delta(C,C') \frac{|R|}{|Z_{\rhd,r_C}|} \sum_{\tilde{d} \in Z_{\rhd,r_C}} \sum_{q \in Q_C} C^{q\tilde{d}{q}^{-1}}_T(m)L^{q\rhd r_C}(m) \sum_{a=1}^{|R|} \delta_{R,R'} [D^{R}(\tilde{d})]_{aa}\\	
		&= \delta(C,C') \delta(R,R') \frac{|R|}{|Z_{\rhd,r_C}|}\sum_{D \in (Z_{\rhd,r_{C}})_{cl}} \chi^R(D) \sum_{q \in Q_C} \sum_{\tilde{d} \in D} C^{q\tilde{d}q^{-1}}_T(m) L^{q \rhd r_C}(m),
	\end{align*}
	where in the last step we used the fact that the characters are functions of the classes $D$. We can recognise 	
	$$\frac{|R|}{|Z_{\rhd,r_C}|}\sum_{D \in (Z_{\rhd,r_{C}})_{cl}} \chi^R(D) \sum_{q \in Q_C} \sum_{\tilde{d} \in D} C^{q\tilde{d}q^{-1}}_T(m) L^{q \rhd r_C}(m)$$
	as the projector $T^{R,C}(m)$ from Equation \ref{Sphere_Projectors_appendix}, and so we have
	\begin{equation}	
		T^{R,C}(m)T^{R',C'}(m)	= \delta(C,C') \delta(R,R') T^{R,C}(m).
	\end{equation}
	That is, the projectors defined by Equation \ref{Sphere_Projectors_appendix} are orthogonal.

	In additional to being orthogonal, these projectors are complete in our space (in that they sum to give identity when acting on the subspace where the membrane is unexcited), as we will now show. We have
	\begin{align*}
		\sum_{\text{classes }C} \sum_{\substack{\text{irreps }R\\ \text{ of }Z_{\rhd,r_C}/\partial(E)}} T^{R,C}(m) &= \sum_{\text{classes }C} \sum_{\text{irreps }R} \frac{|R|}{|Z_{\rhd,r_C}|} \sum_{D \in (Z_{\rhd,r_{C}})_{cl}} \chi_R(D) \sum_{d \in D} \sum_{q \in Q_C} C_T^{qdq^{-1}}(m) \delta(\hat{e}(m),q \rhd r_C)\\
		&= \sum_{\text{classes }C} \sum_{D \in (Z_{\rhd,r_{C}})_{cl}} \big(\sum_{\text{irreps }R} \chi_R(D) |R|\big) \frac{1}{|Z_{\rhd,r_C}|} \sum_{d \in D} \sum_{q \in Q_C} C_T^{qdq^{-1}}(m) \delta(\hat{e}(m),q \rhd r_C).
	\end{align*}
	
	We can then apply the column orthogonality relations for characters by writing
	$$\sum_{\text{irreps }R} \chi_R(D) |R|= \sum_{\text{irreps }R} \chi_R(D) \chi_R(1_N)$$
	where $1_N$ is the identity element of $|Z_{\rhd,r_C}/\partial(E)|$. Using the orthogonality relation then gives us
	\begin{align*}
		\sum_{\text{classes }C} \sum_{\substack{\text{irreps }R\\ \text{ of }Z_{\rhd,r_C}/\partial(E)}} T^{R,C}(m)&=\sum_{\text{classes }C} \sum_D \big(\sum_{\text{irreps }R} \chi_R(D) \overline{\chi}_R(1_N)\big) \frac{1}{|Z_{\rhd,r_C}|} \sum_{d \in D} \sum_{q \in Q_C} C_T^{qdq^{-1}}(m) \delta(\hat{e}(m),q \rhd r_C)\\
		&= \sum_{\text{classes }C} \sum_D \big(|Z_{\rhd,r_C}/\partial(E)| \delta(D,1_N)\big) \frac{1}{|Z_{\rhd,r_C}|} \sum_{d \in D} \sum_{q \in Q_C} C_T^{qdq^{-1}}(m) \delta(\hat{e}(m),q \rhd r_C).
	\end{align*}
	
	Now the only contribution is from the identity class $D=1_N$, which is just the coset $\partial(E)$. Therefore, we can replace the sum over $d \in D$ with a sum over elements in $\partial(E)$. Furthermore, the elements $d \in \partial(E)$ are in the centre of $G$, so $qdq^{-1}=d$. This means that we can write the sum over projectors as
	\begin{align*}
		\sum_{\text{classes }C} \sum_{\substack{\text{irreps }R\\ \text{ of }Z_{\rhd,r_C}/\partial(E)}} T^{R,C}(m)&= \sum_{\text{classes }C} \frac{1}{|\partial(E)|} \sum_{d \in \partial(E)} \sum_{q \in Q_C} C_T^{qdq^{-1}}(m) \delta(\hat{e}(m),q \rhd r_C)\\
		&=\sum_{\text{classes }C} \frac{1}{|\partial(E)|} \sum_{d \in \partial(E)} \sum_{q \in Q_C} C_T^{d}(m) \delta(\hat{e}(m),q \rhd r_C)\\
		&= \sum_{d \in \partial(E)} \frac{1}{|\partial(E)|} C_T^d(m) (\sum_{\text{classes }C} \sum_{q \in Q_C} \delta(\hat{e}(m), q \rhd r_C)).
	\end{align*}
	Then recalling that the classes $C$ partition the kernel of $\partial$, and the representatives $q \in Q_C$ relate the elements of $C$ to the representative $r_C$ (so that summing over $Q_C$ is equivalent to summing over elements of the class), the sum $\sum_{\text{classes }C} \sum_{q \in Q_C}$ is equivalent to a sum over all elements of the kernel. Therefore,
	\begin{align*}
		\sum_{\text{classes }C} \sum_{\substack{\text{irreps }R\\ \text{ of }Z_{\rhd,r_C}/\partial(E)}} T^{R,C}(m)&= \sum_{d \in \partial(E)} \frac{1}{|\partial(E)|} C_T^d(m) \big( \sum_{e \in \text{ker}(\partial)} \delta(\hat{e}(m),e)\big).
	\end{align*}
	
	The term $\sum_{e \in \text{ker}(\partial)} \delta(\hat{e}(m),e)$ is the identity when acting on the space where the membrane is unexcited, because in that subspace the surface label must be in the kernel of $\partial$ in order to satisfy fake-flatness. In addition $C_T^d(m)$ has label in the image of $\partial$ and is therefore a closed condensed magnetic operator. As we saw in Section \ref{Section_condensation_magnetic_centre_case}, condensed magnetic membrane operators are equivalent to a series of edge transforms in the region of the membrane, combined with a blob ribbon operator around the boundary of the membrane. In the case of a spherical membrane, this ribbon collapses and becomes trivial. This means that $C_T^d(m)$ is just equivalent to a set of edge transforms in the region of the membrane. Because these edge transforms can be absorbed into a state for which these edges are unexcited, $C_T^d(m)$ acts trivially in the subspace where the membrane is unexcited. This means that $\sum_{\text{classes }C} \sum_{\text{irreps }R} T^{R,C}(m)$ acts as the identity in the unexcited subspace. That is, we have shown that the sum over all the projectors is equal to the identity in the subspace where the sphere is unexcited, and so the set of projectors is complete in the subspace for which we can measure the topological charge.

	\subsubsection{Point-like topological charge of a higher-flux excitation}
	\label{Section_point_like_charge_higher_flux}
\begin{figure}[h]
		\begin{center}
			\begin{overpic}[width=0.9\linewidth]{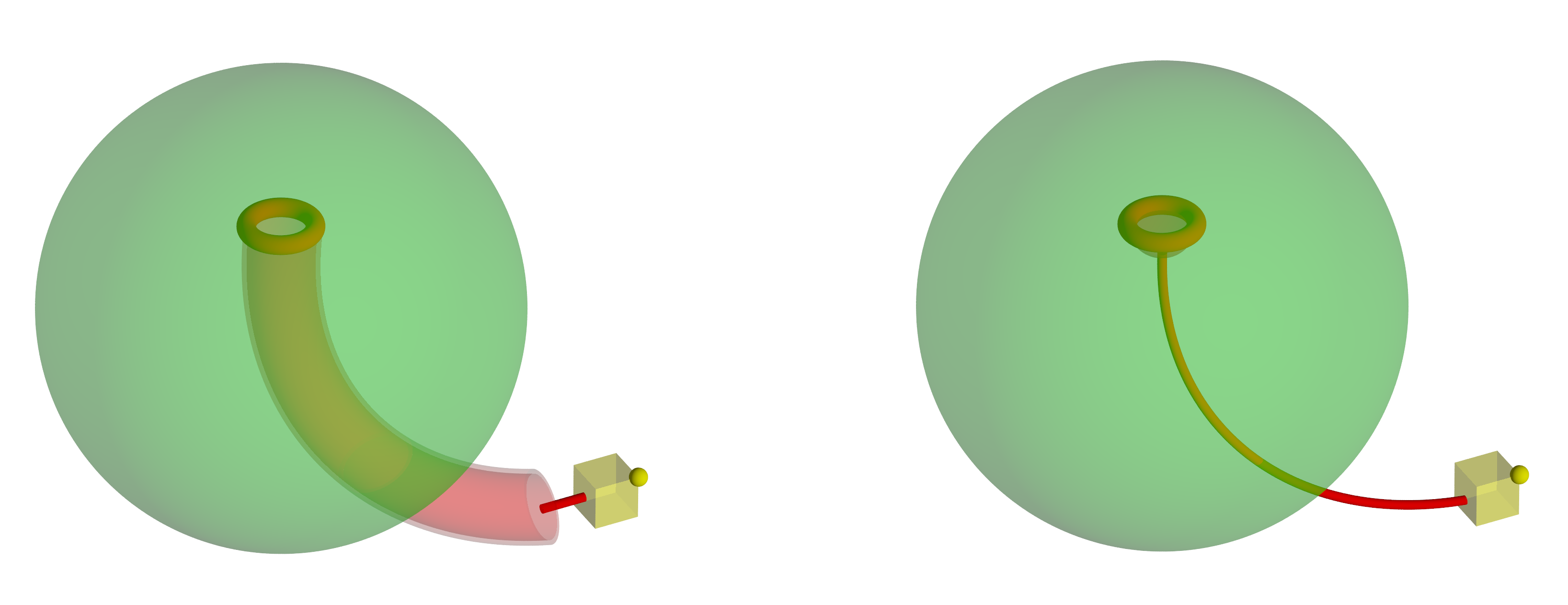}
				\put(25,2){$C^{h,e_m}(m)$}
				\put(42,5){blob 0$(m)$}
				\put(41.5,8){$s.p(m)$}
				\put(0,30){$T^{R,C}(c)$}
				
				\put(45,15){\Huge $\rightarrow$}
				\put(43,18){deform $m$}
			\end{overpic}
			
			\caption{We consider applying a topological charge measurement operator $T^{R,C}(c)$ on a sphere that encloses the loop-like excitation produced by a higher-flux membrane operator $C^{h,e_m}(m)$ (red). This will measure the point-like charge of the loop excitation, giving a non-zero result only if the excitation carries some charge labelled by $R$ and $C$. In order to calculate the result, it is convenient to deform $m$ so that the membrane lies entirely within the surface $c$ of the measurement operator. However, the start-point and blob 0 of $m$ remain outside $c$, so the operators still intersect. In this figure, the dual membrane of $m$ and $c$ is outside the direct membrane, so the edges cut by the dual membranes would protrude outwards from the surfaces shown. In the right-hand figure, the red curved line indicates the path from the start-point of $m$ to the rest of $m$, as well as the dual paths from blob 0 to $m$ on which the blob ribbon operators in $C^h_T(m)$ are applied. }
			\label{3D_sphere_charge_higher_flux}
			
		\end{center}
	\end{figure}

	In Section \ref{Section_3D_sphere_charge_examples} of the main text, we showed that the $E$-valued loop excitations possess a point-like topological charge that can be measured by a measurement operator on a spherical surface. In this Section, we will show that the same is true for the higher-flux excitations that generalize the magnetic and $E$-valued loop excitations, and we will demonstrate how this charge may be calculated. We consider the situation shown in Figure \ref{3D_sphere_charge_higher_flux}, where we apply a higher-flux membrane operator $C^{h,e_m}_T(m)=C^h_T(m) \delta(\hat{e}(m),e_m)$ and then apply a measurement operator $T^{R,C}(c)$ that encloses the loop-like excitation produced by the higher-flux membrane operator, but not the start-point or blob 0 of the membrane operator. That is, we consider the state
	$$T^{R,C}(c) C^{h,e_m}_T(m)\ket{GS},$$
	where $\ket{GS}$ is a ground state of the model.

	We wish to commute the measurement operator $T^{R,C}(c)$ to the right, so that it acts directly on the ground state. The measurement operator is a sum of pairs of operators $C^{g}_T(c)L^e(c)$ with different labels, so we first wish to consider the state
	$$C^{g}_T(c)L^e(c)C^{h,e_m}_T(m)\ket{GS},$$
	before we introduce the sum over appropriate labels $g$ and $e$. We start by using the topological property of the membrane operators to deform $C^{h,e_m}_T(m)$, just as we did for the $E$-valued membrane operator in Section \ref{Section_3D_sphere_charge_examples} of the main text, so that the dual membrane of $m$ is entirely enclosed within $c$, as shown in Figure \ref{3D_sphere_charge_higher_flux}. This means that the operator $C^{h,e_m}_T(m)$ will not affect the edges or plaquettes in the membrane $c$. However, the paths involved in $C^{h,e_m}_T(m)$ will pass through $c$, which causes the commutation relation with $C^{g}_T(c)L^e(c)$ to be non-trivial, as we will see shortly. For simplicity, we will continue to refer to the deformed membrane as $m$. We note that this commutation relation is analogous to the one considered in Section \ref{Section_braiding_higher_flux_higher_flux}, when we considered the braiding relations of two higher-flux membrane operators, except that we are considering the reversed direction of commutation and $c$ is closed.

	The next step is to separate $C^{h,e_m}_T(m)$ into its constituent parts. As defined in Section \ref{Section_3D_Braiding_Central} of the main text, we have
	$$C^{h,e_m}_T(m)= C^h_T(m) \delta(\hat{e}(m),e_m).$$
	Furthermore, the membrane operator $C^h_T(m)$ can be split into two parts as described in Section \ref{Section_3D_MO_Central} of the main text. The first part is $C^h_{\rhd}(m)$, which acts on the edges cut by the dual membrane of $m$ and on plaquettes that are cut by the dual membrane and have base-point on the direct membrane. The second part is a set of blob ribbon operators, one for each plaquette $p$ on the direct membrane of $m$, which pass from blob 0 of $m$ to the blob attached to the plaquette $p$. Then
	$$C^h_T(m) = C^h_{\rhd}(m) \prod_{p \in m} B^{f_h(p)}(\text{blob 0}(m) \rightarrow \text{blob }p),$$
	where $f_h(p)=[g(s.p(m)-v_0(p)) \rhd \hat{e}_p] [(h^{-1}g(s.p(m)-v_0(p))) \rhd \hat{e}_p^{-1}] $ and we added the subscript $h$ to $f$ to distinguish between the labels for the different magnetic membrane operators. In the expression for $f_h(p)$ above, $\hat{e}_p$ is the label of the plaquette $p$, assuming that it is oriented away from the dual membrane (so it matches the orientation of $m$ used in $\hat{e}(m)$). If the plaquette had the opposite orientation, we would instead replace the plaquette label with its inverse. Using this decomposition of the higher-flux membrane operators, we have
	\begin{align}
		C^{g}_T(c) L^e(c)C^{h,e_m}_T(m)\ket{GS}&=C^g_T(c) \delta( \hat{e}(c),e) C^h_{\rhd}(m) \prod_{p \in m} B^{f_h(p)}(\text{blob 0}(m) \rightarrow \text{blob }p) \delta(\hat{e}(m),e_m) \ket{GS}. \label{Equation_sphere_charge_higher_flux_commutation_1}
	\end{align}
	
	First, we note that due to the deformation of $m$, $C^h_{\rhd}(m)$ does not affect the labels of any of the edges or plaquettes in $c$, and so $C^h_{\rhd}(m)$ commutes with $\delta( \hat{e}(c),e)$, giving us
	\begin{align}
		C^{g}_T(c) L^e(c) C^{h,e_m}_T(m)\ket{GS}&= C^g_T(c) C^h_{\rhd}(m) \delta( \hat{e}(c),e) \prod_{p \in m} B^{f_h(p)}(\text{blob 0}(m) \rightarrow \text{blob }p) \delta(\hat{e}(m),e_m) \ket{GS}.\label{Equation_sphere_charge_higher_flux_commutation_2}
	\end{align}
	
	On the other hand, the blob ribbon operators $ B^{f_h(p)}(\text{blob 0}(m) \rightarrow \text{blob }p)$ do not commute with $\delta( \hat{e}(c),e)$ because they pass through $c$ (blob 0$(m)$ is outside $c$ whereas blob $p$ is inside $c$ for each plaquette $p$). Noting that the orientations of the blob ribbon operators match the orientation of $c$ (because the surface $c$ is taken to point inwards from the sphere and the blob ribbon operators similarly pass inwards through the sphere), we can use Equation \ref{Equation_blob_ribbon_on_surface_1} which states that the action of the blob ribbon operator on the surface label is given by
	$$B^{f_h(p)}(\text{blob 0}(m) \rightarrow \text{blob }p):e(c)= [g(s.p(c)-s.p(m)) \rhd f_h(p)^{-1}] e(c),$$
	and so the commutation relation with the surface label operator $\hat{e}(c)$ is
	$$\hat{e}(c) B^{f_h(p)}(\text{blob 0}(m) \rightarrow \text{blob }p) = B^{f_h(p)}(\text{blob 0}(m) \rightarrow \text{blob }p) [g(s.p(c)-s.p(m)) \rhd f_h(p)^{-1}] \hat{e}(c). $$

	This means that
	\begin{align}
		\hat{e}(c) \prod_{p \in m} B^{f_h(p)}(\text{blob 0}(m) \rightarrow \text{blob }p) &= \big[ \prod_{p \in m} B^{f_h(p)}(\text{blob 0}(m) \rightarrow \text{blob }p) \big] \big[ \prod_{p \in m} [g(s.p(c)-s.p(m)) \rhd f_h(p)^{-1}] \big] \hat{e}(c) \notag \\
		&= \big[ \prod_{p \in m} B^{f_h(p)}(\text{blob 0}(m) \rightarrow \text{blob }p) \big] \big[ g(s.p(c)-s.p(m)) \rhd \big( \prod_{p \in m}f_h(p)^{-1} \big) \big] \hat{e}(c). \label{Equation_sphere_charge_higher_flux_surface_c_blob_commutation_1}
	\end{align}
	We can simplify this by considering $\prod_{p \in m}f_h(p)^{-1}$. As we showed in Section \ref{Section_braiding_higher_flux_higher_flux} (see Equation \ref{Equation_product_f_1}), we have
	\begin{equation}
		\prod_{p \in m}f_h(p)^{-1} = [h^{-1} \rhd \hat{e}(m)] \hat{e}(m)^{-1}.
		\label{Equation_product_f_2}
	\end{equation}
	Then Equation \ref{Equation_sphere_charge_higher_flux_surface_c_blob_commutation_1} becomes
	\begin{align*}
		\hat{e}(c) \prod_{p \in m} &B^{f_h(p)}(\text{blob 0}(m) \rightarrow \text{blob }p)\\
		&= \big[ \prod_{p \in m} B^{f_h(p)}(\text{blob 0}(m) \rightarrow \text{blob }p) \big] \: \big[ g(s.p(c)-s.p(m)) \rhd \big( [h^{-1} \rhd \hat{e}(m)] \hat{e}(m)^{-1} \big) \big] \hat{e}(c),
	\end{align*}
	which means that
	\begin{align*}
		\delta( &\hat{e}(c),e) \prod_{p \in m} B^{f_h(p)}(\text{blob 0}(m) \rightarrow \text{blob }p)\\
		&= \big[ \prod_{p \in m} B^{f_h(p)}(\text{blob 0}(m) \rightarrow \text{blob }p) \big] \delta\bigg(\big[ g(s.p(c)-s.p(m)) \rhd \big( [h^{-1} \rhd \hat{e}(m)] \hat{e}(m)^{-1} \big) \big] \hat{e}(c), e\bigg).
	\end{align*}
	Substituting this into our overall commutation relation, Equation \ref{Equation_sphere_charge_higher_flux_commutation_2}, gives us
	\begin{align}
		C^{g}_T(c) L^e(c)& C^{h,e_m}_T(m)\ket{GS} \notag \\
		&= C^g_T(c) C^h_{\rhd}(m) \big[ \prod_{p \in m} B^{f_h(p)}(\text{blob 0}(m) \rightarrow \text{blob }p) \big] \delta\bigg(\big[ g(s.p(c)-s.p(m)) \rhd \big( [h^{-1} \rhd \hat{e}(m)] \hat{e}(m)^{-1} \big) \big] \hat{e}(c), e\bigg) \notag \\
		& \hspace{3cm} \delta(\hat{e}(m),e_m) \ket{GS}. \label{Equation_sphere_charge_higher_flux_commutation_3}
	\end{align}
	
	We can then use the operator $\delta(\hat{e}(m),e_m)$ to replace the operator $\hat{e}(m)$ with the group label $e_m$, to obtain
	\begin{align}
		&C^{g}_T(c) L^e(c)C^{h,e_m}_T(m)\ket{GS} \notag \\
		&= C^g_T(c) C^h_{\rhd}(m) \big[ \prod_{p \in m} B^{f_h(p)}(\text{blob 0}(m) \rightarrow \text{blob }p) \big] \delta\bigg(\big[ g(s.p(c)-s.p(m)) \rhd \big( [h^{-1} \rhd e_m] e_m^{-1} \big) \big] \hat{e}(c), e\bigg) \delta(\hat{e}(m),e_m) \ket{GS}. \label{Equation_sphere_charge_higher_flux_commutation_4}
	\end{align}
	Commuting $\delta(\hat{e}(m),e_m)$ to the left and rearranging the other Kronecker delta operator gives us
	\begin{align}
		C^{g}_T(c) L^e(c) C^{h,e_m}_T(m)\ket{GS}&= C^g_T(c) C^h_{\rhd}(m) \big[ \prod_{p \in m} B^{f_h(p)}(\text{blob 0}(m) \rightarrow \text{blob }p) \big] \delta(\hat{e}(m),e_m) \notag \\
		& \hspace{1.5cm} \delta\bigg( \hat{e}(c), e \big[ g(s.p(c)-s.p(m)) \rhd \big( [h^{-1} \rhd e_m]^{-1} e_m \big) \big]\bigg) \ket{GS}. \label{Equation_sphere_charge_higher_flux_commutation_5}
	\end{align}
	
	Next we wish to consider $ C^g_T(c) C^h_{\rhd}(m) $. We now split $ C^g_T(c)$ into its component parts, just as we did for $C^h_T(m)$. That is, we write
	$$C^g_T(c)= C^g_{\rhd}(c) \prod_{ p' \in c} B^{f_g(p')}(\text{blob }0(c) \rightarrow \text{blob }p').$$
	Because $ C^h_{\rhd}(m)$ does not affect any of the edges or plaquettes on $c$, and the blob ribbon operators
	$$B^{f_g(p')}(\text{blob }0(c) \rightarrow \text{blob }p')$$
	do not intersect with $m$ (or affect the paths from the start-point of $m$ to any edges cut by $m$), $ C^h_{\rhd}(m)$ commutes with the blob ribbon operators. On the other hand, $ C^g_{\rhd}(c)$ and $ C^h_{\rhd}(m)$ do not commute. We saw this in Section \ref{Section_braiding_higher_flux_higher_flux}, when we considered the braiding of two higher-flux membrane operators (the calculation in that case is just the reverse of the calculation we are considering in this section). The non-commutativity occurs because all of the paths from the start-point of $m$ to edges or plaquettes in the region of $m$ that appear in the operator $C^h_{\rhd}(m)$ pass through $c$. For example, the action of $C^h_{\rhd}(m)$ on an edge $i$ cut by the dual membrane of $m$ is
	$$C^h_{\rhd}(m):g_i = g(s.p(m)-v_i)^{-1}hg(s.p(m)-v_i)g_i$$
	in the case where $i$ points away from the direct membrane of $m$. This depends on the path element $g(s.p(m)-v_i)$. However, because the path passes through $c$, the path element is affected by $C^g_{\rhd}(c)$ according to
	$$C^g_{\rhd}(c):g(s.p(m)-v_i) = g(s.p(c)-s.p(m))^{-1}g^{-1}g(s.p(c)-s.p(m)) g(s.p(m)-v_i),$$
	as we showed in Section \ref{Section_electric_magnetic_braiding_3D_tri_trivial}. Defining $g_{[m-c]}= g(s.p(c)-s.p(m))^{-1}gg(s.p(c)-s.p(m))$, we can write this action on the path element as
	$$C^g_{\rhd}(c):g(s.p(m)-v_i) = g_{[m-c]}^{-1} g(s.p(m)-v_i).$$
	
	Similar paths occur in the action of $C^h_{\rhd}(m)$ on the plaquettes near $m$. In Section \ref{Section_braiding_higher_flux_higher_flux}, we saw that the action of $C^g_T(c)$ on these path elements leads to the commutation relation (from Equation \ref{Equation_higher_flux_braiding_magnetic_commutation})
	$$C^h_{\rhd}(m) C^g_T(c) = C^g_T(c) C^{g^{\phantom{-1}}_{[m-c]}hg_{[m-c]}^{-1}}_{\rhd}(m)$$
	and so
	\begin{equation}
		C^g_T(c) C^h_{\rhd}(m) = C^{g^{-1}_{[m-c]}hg_{[m-c]}^{\phantom{-1}}}_{\rhd}(m) C^g_T(c). \label{Equation_higher_flux_point_charge_magnetic_commutation}
	\end{equation}
	
	When inverting the commutation relation in the last step, we may worry that $g_{[m-c]}$ is an operator (because it depends on the path element $g(s.p(c)-s.p(m))$) and so depends on the order of multiplication of the operators. However, while the path element $g(s.p(c)-s.p(m))$ is affected by $C^g_T(c)$, $g_{[m-c]}$ is not. This is because the path element is just pre-multiplied by $g$ under the action of $C^g_T(c)$, which leads to the $g$ in $g_{[m-c]}$ being conjugated by itself, which is trivial.

	Using the commutation relation from Equation \ref{Equation_higher_flux_point_charge_magnetic_commutation} in Equation \ref{Equation_sphere_charge_higher_flux_commutation_5} gives us
	\begin{align}
		&C^{g}_T(c) L^e(c) C^{h,e_m}_T(m)\ket{GS} \notag\\
		&=C^{g^{-1}_{[m-c]}hg_{[m-c]}^{\phantom{-1}}}_{\rhd}(m) C^g_T(c) \big[ \prod_{p \in m} B^{f_h(p)}(\text{blob 0}(m) \rightarrow \text{blob }p) \big] \delta(\hat{e}(m),e_m) \notag\\
		& \hspace{0.5cm} \delta( \hat{e}(c), e \big[ g(s.p(c)-s.p(m)) \rhd \big( [h^{-1} \rhd e_m]^{-1} e_m \big) \big]) \ket{GS} \notag\\
		&= C^{g^{-1}_{[m-c]}hg_{[m-c]}^{\phantom{-1}}}_{\rhd}(m) C^g_{\rhd}(c) \big[\prod_{ p' \in c} B^{f_g(p')}(\text{blob }0(c) \rightarrow \text{blob }p') \big] \big[ \prod_{p \in m} B^{f_h(p)}(\text{blob 0}(m) \rightarrow \text{blob }p) \big] \delta(\hat{e}(m),e_m) \notag\\
		& \hspace{0.5cm}\delta( \hat{e}(c), e \big[ g(s.p(c)-s.p(m)) \rhd \big( [h^{-1} \rhd e_m]^{-1} e_m \big) \big]) \ket{GS}. \label{Equation_sphere_charge_higher_flux_commutation_6}
	\end{align}
	
	Next we must commute the blob ribbon operators from the two membrane operators past each-other. The blob ribbon operators from the different membrane operators do not commute. This is because the labels $f_g(p')$ of the blob ribbon operators from $C^g_T(c)$ depend on the labels $\hat{e}_{p'}$ of the plaquettes in $c$, but the blob ribbon operators from $C^h_T(m)$ pass through $c$ and so affect these plaquette labels. For simplicity, we will choose all of the blob ribbon operators from $C^h_T(m)$ to intersect $c$ at the same plaquette $q$ (we can do this because we can deform these blob ribbon operators due to there being no excitations in the region of the measurement operator). Then only the ribbon operator $B^{f_g(q)}(\text{blob }0(c) \rightarrow \text{blob }q)$ fails to commute with $ \prod_{p \in m} B^{f_h(p)}(\text{blob 0}(m) \rightarrow \text{blob }p)$. To find the commutation relation between these operators, we must first understand how the $B^{f_h(p)}(\text{blob 0}(m) \rightarrow \text{blob }p)$ operators affect the label of the plaquette $q$. The plaquette $q$ is taken to be aligned with $c$ and so also aligned with the blob ribbon operators from $C^h_T(m)$. Therefore, from the standard expression for the action of the blob ribbon operators, we have
	$$B^{f_h(p)}(\text{blob 0}(m) \rightarrow \text{blob }p): e_q = e_q [g(s.p(m)-v_0(q))^{-1} \rhd f_h(p)^{-1}]$$
	and so
	$$\hat{e}(q)B^{f_h(p)}(\text{blob 0}(m) \rightarrow \text{blob }p) = B^{f_h(p)}(\text{blob 0}(m) \rightarrow \text{blob }p) \hat{e}(q) [g(s.p(m)-v_0(q))^{-1} \rhd f_h(p)^{-1}]. $$

	Using the fact that $f_g(q) = [g(s.p(c)-v_0(q)) \rhd \hat{e}(q)] [(g^{-1}g(s.p(c)-v_0(q))) \rhd \hat{e}(q)^{-1}]$ we therefore have
	\begin{align*}
		&f_g(q) B^{f_h(p)}(\text{blob 0}(m) \rightarrow \text{blob }p)\\
		&= B^{f_h(p)}(\text{blob 0}(m) \rightarrow \text{blob }p)\big[g(s.p(c)-v_0(q)) \rhd (\hat{e}(q) [g(s.p(m)-v_0(q))^{-1} \rhd f_h(p)^{-1}]) \big]\\
		& \hspace{1cm} \big[(g^{-1}g(s.p(c)-v_0(q))) \rhd (\hat{e}(q) [g(s.p(m)-v_0(q))^{-1} \rhd f_h(p)^{-1}] )^{-1}\big] \\
		&= B^{f_h(p)}(\text{blob 0}(m) \rightarrow \text{blob }p) [g(s.p(c)-v_0(q)) \rhd \hat{e}(q)] [(g^{-1}g(s.p(c)-v_0(q))) \rhd \hat{e}(q)^{-1}]\\
		& \hspace{1cm} [ g(s.p(c)-v_0(q)) \rhd (g(s.p(m)-v_0(q))^{-1} \rhd f_h(p)^{-1}) ] [(g^{-1}g(s.p(c)-v_0(q))) \rhd ( g(s.p(m)-v_0(q))^{-1} \rhd f_h(p))] \\
		&= B^{f_h(p)}(\text{blob 0}(m) \rightarrow \text{blob }p) [g(s.p(c)-v_0(q)) \rhd \hat{e}(q)] [(g^{-1}g(s.p(c)-v_0(q))) \rhd \hat{e}(q)^{-1}] \\
		& \hspace{1cm}[ (g(s.p(c)-v_0(q))g(s.p(m)-v_0(q))^{-1}) \rhd f_h(p)^{-1} ] [(g^{-1}g(s.p(c)-v_0(q)) g(s.p(m)-v_0(q))^{-1}) \rhd f_h(p)]\\
		&= B^{f_h(p)}(\text{blob 0}(m) \rightarrow \text{blob }p) [g(s.p(c)-v_0(q)) \rhd \hat{e}(q)] [(g^{-1}g(s.p(c)-v_0(q))) \rhd \hat{e}(q)^{-1}] \\
		& \hspace{1cm}[ g(s.p(c)-s.p(m)) \rhd f_h(p)^{-1} ] [(g^{-1}g(s.p(c)-s.p(m))) \rhd f_h(p)].
	\end{align*}
	
	We can recognise $[g(s.p(c)-v_0(q)) \rhd \hat{e}(q)] [(g^{-1}g(s.p(c)-v_0(q))) \rhd \hat{e}(q)^{-1}]$ as $f_g(q)$, the original label blob ribbon operator from $C^g_T(c)$. This means that the effect of the blob ribbon operator associated to $p$ piercing the plaquette $q$ on the label of the blob ribbon operator associated to plaquette $q$ is
	\begin{align*}
		&f_g(q) B^{f_h(p)}(\text{blob 0}(m) \rightarrow \text{blob }p)\\
		&= B^{f_h(p)}(\text{blob 0}(m) \rightarrow \text{blob }p) f_g(q)[ g(s.p(c)-s.p(m)) \rhd f_h(p)^{-1} ] [(g^{-1}g(s.p(c)-s.p(m))) \rhd f_h(p)].
	\end{align*}
	Therefore, when we consider all of the blob ribbon operators $B^{f_h(p)}(\text{blob 0}(m) \rightarrow \text{blob }p)$ from $C^h_T(m)$, we have
	\begin{align*}
		f_g(q) &\prod_{p \in m} B^{f_h(p)}(\text{blob 0}(m) \rightarrow \text{blob }p) \\
		&= \big[ \prod_{p \in m} B^{f_h(p)}(\text{blob 0}(m) \rightarrow \text{blob }p) \big] f_g(q) \big(\prod_{p \in m} [ g(s.p(c)-s.p(m)) \rhd f_h(p)^{-1} ] [(g^{-1}g(s.p(c)-s.p(m))) \rhd f_h(p)] \big).
	\end{align*}
	
	Using our expression for $\prod_{p \in m} f_h(p)$ from Equation \ref{Equation_product_f_1}, we obtain 
	\begin{align*}
		f_g(q) &\prod_{p \in m} B^{f_h(p)}(\text{blob 0}(m) \rightarrow \text{blob }p) \\
		&= \big( \prod_{p \in m} B^{f_h(p)}(\text{blob 0}(m) \rightarrow \text{blob }p) \big) f_g(q) \\
		& \hspace{0.5cm} \big[ g(s.p(c)-s.p(m)) \rhd \big([h^{-1} \rhd \hat{e}(m)] \hat{e}(m)^{-1}\big) \big] \big[(g^{-1}g(s.p(c)-s.p(m))) \rhd \big( [h^{-1} \rhd \hat{e}(m)^{-1}] \hat{e}(m)\big)\big].
	\end{align*}
	Therefore, the commutation relation between the ribbon operators is given by
	\begin{align*}
		&B^{f_g(q)}(\text{blob }0(c) \rightarrow \text{blob }q) \prod_{p \in m} B^{f_h(p)}(\text{blob 0}(m) \rightarrow \text{blob }p) \\
		&= \big( \prod_{p \in m} B^{f_h(p)}(\text{blob 0}(m) \rightarrow \text{blob }p) \big) \\
		& \hspace{0.5cm} B^{ f_g(q) [ g(s.p(c)-s.p(m)) \rhd ([h^{-1} \rhd \hat{e}(m)] \hat{e}(m)^{-1}) ] [(g^{-1}g(s.p(c)-s.p(m))) \rhd ( [h^{-1} \rhd \hat{e}(m)^{-1}] \hat{e}(m))]} (\text{blob }0(c) \rightarrow \text{blob }q)\\
		&= \big[ \prod_{p \in m} B^{f_h(p)}(\text{blob 0}(m) \rightarrow \text{blob }p) \big] B^{ f_g(q)}(\text{blob }0(c) \rightarrow \text{blob }q) \\
		& \hspace{0.5cm} B^{[ g(s.p(c)-s.p(m)) \rhd ([h^{-1} \rhd \hat{e}(m)] \hat{e}(m)^{-1}) ] [(g^{-1}g(s.p(c)-s.p(m))) \rhd ( [h^{-1} \rhd \hat{e}(m)^{-1}] \hat{e}(m))]} (\text{blob }0(c) \rightarrow \text{blob }q),
	\end{align*}
	where in the last step we separated the blob ribbon operator corresponding to plaquette $q$ into a part with the ordinary label and an extra part with a label depending on $\hat{e}(m)$. Substituting this into our commutation relation Equation \ref{Equation_sphere_charge_higher_flux_commutation_6}, we obtain
	\begin{align}
		C^{g}_T(c) L^e(c)& C^{h,e_m}_T(m)\ket{GS} \notag\\
		&= C^{g^{-1}_{[m-c]}hg_{[m-c]}^{\phantom{-1}}}_{\rhd}(m) C^g_{\rhd}(c) \big[ \prod_{p \in m} B^{f_h(p)}(\text{blob 0}(m) \rightarrow \text{blob }p) \big] \big[\prod_{ p' \in c} B^{f_g(p')}(\text{blob }0(c) \rightarrow \text{blob }p') \big] \notag \\
		&\hspace{0.5cm} B^{[ g(s.p(c)-s.p(m)) \rhd ([h^{-1} \rhd \hat{e}(m)] \hat{e}(m)^{-1}) ] [(g^{-1}g(s.p(c)-s.p(m))) \rhd ( [h^{-1} \rhd \hat{e}(m)^{-1}] \hat{e}(m))]} (\text{blob }0(c) \rightarrow \text{blob }q) \notag \\
		& \hspace{0.5cm} \delta(\hat{e}(m),e_m) \delta( \hat{e}(c), e \big[ g(s.p(c)-s.p(m)) \rhd \big( [h^{-1} \rhd e_m]^{-1} e_m \big) \big]) \ket{GS}. \label{Equation_sphere_charge_higher_flux_commutation_7}
	\end{align}
	
	We can then use $\delta(\hat{e}(m),e_m)$ to replace $\hat{e}(m)$ with $e_m$ in the label of the extra blob ribbon operator, to obtain
	\begin{align}
		C^{g}_T(c) L^e(c)& C^{h,e_m}_T(m)\ket{GS}\\
		&= C^{g^{-1}_{[m-c]}hg_{[m-c]}^{\phantom{-1}}}_{\rhd}(m) C^g_{\rhd}(c) \big[ \prod_{p \in m} B^{f_h(p)}(\text{blob 0}(m) \rightarrow \text{blob }p) \big] \big[\prod_{ p' \in c} B^{f_g(p')}(\text{blob }0(c) \rightarrow \text{blob }p') \big] \notag\\
		& \hspace{0.5cm} B^{[ g(s.p(c)-s.p(m)) \rhd ([h^{-1} \rhd e_m] e_m^{-1}) ] [(g^{-1}g(s.p(c)-s.p(m))) \rhd ( [h^{-1} \rhd e_m^{-1}] e_m)]} (\text{blob }0(c) \rightarrow \text{blob }q) \notag \\
		& \hspace{0.5cm} \delta(\hat{e}(m),e_m) \delta( \hat{e}(c), e \big[ g(s.p(c)-s.p(m)) \rhd \big( [h^{-1} \rhd e_m]^{-1} e_m \big) \big]) \ket{GS}. \label{Equation_sphere_charge_higher_flux_commutation_8}
	\end{align}
	
	We then define
	\begin{align*}
		x&= [ g(s.p(c)-s.p(m)) \rhd ([h^{-1} \rhd e_m] e_m^{-1}) ] [(g^{-1}g(s.p(c)-s.p(m))) \rhd ( [h^{-1} \rhd e_m^{-1}] e_m)]\\
		&= g(s.p(c)-s.p(m)) \rhd \big(([h^{-1} \rhd e_m] e_m^{-1}) g^{-1}_{[m-c]} \rhd ([h^{-1} \rhd e_m^{-1}] e_m)\big)
	\end{align*}
	to write Equation \ref{Equation_sphere_charge_higher_flux_commutation_8} more concisely, as
	\begin{align}
		C^{g}_T(c) L^e(c)& C^{h,e_m}_T(m)\ket{GS} \notag\\
		&= C^{g^{-1}_{[m-c]}hg_{[m-c]}^{\phantom{-1}}}_{\rhd}(m) C^g_{\rhd}(c) \big[ \prod_{p \in m} B^{f_h(p)}(\text{blob 0}(m) \rightarrow \text{blob }p) \big] \big[\prod_{ p' \in c} B^{f_g(p')}(\text{blob }0(c) \rightarrow \text{blob }p') \big] \notag \\
		& \hspace{0.5cm} B^{x} (\text{blob }0(c) \rightarrow \text{blob }q) \delta(\hat{e}(m),e_m) \delta( \hat{e}(c), e \big[ g(s.p(c)-s.p(m)) \rhd \big( [h^{-1} \rhd e_m]^{-1} e_m \big) \big]) \ket{GS}. \label{Equation_sphere_charge_higher_flux_commutation_9}
	\end{align}
	
	The next step is to commute the blob ribbon operators $B^{f_h(p)}(\text{blob 0}(m) \rightarrow \text{blob }p)$ past $C^g_{\rhd}(c)$. These operators may fail to commute for two reasons. Firstly, the label $f_h(p)$ includes a path element $g(s.p(m)-v_0(p))$, which will be altered by $C^g_{\rhd}(c)$ because the path intersects with the membrane $c$. Secondly, even if the label were a constant the action of blob ribbon operators on a plaquette $a$ depends on a path $g(s.p(m)-v_0(a))$ which may pass through $c$ and so be affected by $C^g_{\rhd}(c)$, as we explained when considering braiding between blob ribbon operators and the higher-flux membrane operators in Section \ref{Section_braiding_higher_flux_blob}. In order to account for the first effect, we write 
	$$B^{f_h(p)}(\text{blob 0}(m) \rightarrow \text{blob }p) = \sum_{y \in E} B^y (\text{blob 0}(m) \rightarrow \text{blob }p) \delta(y, f_h(p)),$$
	so that we can separately consider how the label and the blob ribbon operator transform under commutation with $C^g_{\rhd}(c)$. To deal with the second effect, we split the dual path of the blob ribbon operator into two parts, the first part being the section up to the intersection with $c$ at blob $q$ and the second part being the rest of the dual path of the ribbon (note that the start-point of each section is $s.p(m)$, and only the dual path is split in this way). That is, we write
	$$B^{f_h(p)}(\text{blob 0}(m) \rightarrow \text{blob }p) = \sum_{y \in E} B^y (\text{blob 0}(m) \rightarrow \text{blob }q) B^y (\text{blob }q \rightarrow \text{blob }p) \delta(y, f_h(p)).$$
	
	As we discussed when considering the braiding between (constant-labelled) blob ribbon operators and higher-flux membrane operators in Section \ref{Section_braiding_higher_flux_blob}, the part of the ribbon operator ending at blob $q$ commutes with $C^g_{\rhd}(c)$. On the other hand, from Equation \ref{Equation_blob_ribbon_magnetic_commutation_appendix_1_reversed}, $B^y (\text{blob }q \rightarrow \text{blob }p)$ obeys the commutation relation
	$$B^{y}(\text{blob }q \rightarrow \text{blob }p) C^g_{\rhd}(c)= C^g_{\rhd}(c) B^{(g(s.p(m)-s.p(c)) g g(s.p(m)-s.p(c))^{-1}) \rhd y}(\text{blob }q \rightarrow \text{blob }p)$$
	and so
	\begin{align*}
		C^g_{\rhd}(c) B^{y}(\text{blob }q \rightarrow \text{blob }p)&= B^{(g(s.p(m)-s.p(c)) g^{-1} g(s.p(m)-s.p(c))^{-1}) \rhd y}(\text{blob }q \rightarrow \text{blob }p) C^g_{\rhd}(c)\\
		&= B^{g_{[m-c]}^{-1} \rhd y}(\text{blob }q \rightarrow \text{blob }p) C^g_{\rhd}(c).
	\end{align*}
	
	Note that, while $g_{[m-c]}$ is an operator, it commutes with $C^g_{\rhd}(c)$, as we discussed previously. Now we must consider how $\delta(y,f_h(p))$ is affected by commutation with $C^g_{\rhd}(c)$. We have
	$$f_h(p) = g(s.p(m)-v_0(p)) \rhd \hat{e}_p [(h^{-1}g(s.p(m)-v_0(p))) \rhd \hat{e}_p^{-1}].$$
	$C^g_{\rhd}(c)$ acts on the path element $g(s.p(m)-v_0(p))$ according to
	$$C^g_{\rhd}(c): g(s.p(m)-v_0(p)) = g_{[m-c]}^{-1} g(s.p(m)-v_0(p)),$$
	so that
	$$[g(s.p(m)-v_0(p)) \rhd \hat{e}_p] C^g_{\rhd}(c) = C^g_{\rhd}(c) [(g_{[m-c]}^{-1} g(s.p(m)-v_0(p))) \rhd \hat{e}_p]. $$
	
	Inverting this gives us
	$$C^g_{\rhd}(c) [g(s.p(m)-v_0(p)) \rhd \hat{e}_p] = [(g_{[m-c]} g(s.p(m)-v_0(p))) \rhd \hat{e}_p] C^g_{\rhd}(c)$$
	and so
	$$C^g_{\rhd}(c) f_h(p) = [(g_{[m-c]} g(s.p(m)-v_0(p))) \rhd \hat{e}_p ] [(h^{-1}g_{[m-c]}g(s.p(m)-v_0(p))) \rhd \hat{e}_p^{-1}] C^g_{\rhd}(c). $$
	We can write this as
	$$C^g_{\rhd}(c) f_h(p) = g_{[m-c]} \rhd( [g(s.p(m)-v_0(p)) \rhd \hat{e}_p] [(g_{[m-c]}^{-1}h^{-1}g_{[m-c]}g(s.p(m)-v_0(p))) \rhd \hat{e}_p^{-1}])C^g_{\rhd}(c).$$
	
	We see that the expression 
	$$[g(s.p(m)-v_0(p)) \rhd \hat{e}_p] [(g_{[m-c]}^{-1}h^{-1}g_{[m-c]}g(s.p(m)-v_0(p))) \rhd \hat{e}_p^{-1}]$$
	has the same form as $f_h(p)$, but with $h$ replaced by $g_{[m-c]}^{-1}hg_{[m-c]}$. We will therefore denote this by $f_{h'}(p)$, where $h' = g_{[m-c]}^{-1}hg_{[m-c]}$ is also the label we found for $C^h_{\rhd}(m)$ after the commutation with $C^g_T(c)$. Then
	$$C^g_{\rhd}(c) f_h(p) = g_{[m-c]} \rhd f_{h'}(p).$$
	
	Putting these relations together, we have 
	\begin{align*}
		C^g_{\rhd}(c) &B^{f_h(p)}(\text{blob 0}(m) \rightarrow \text{blob }p)\\
		&= C^g_{\rhd}(c) \sum_{y \in E} B^y (\text{blob 0}(m) \rightarrow \text{blob }q) B^y (\text{blob }q \rightarrow \text{blob }p) \delta(y, f_h(p))\\
		&= \sum_{y \in E} B^y (\text{blob 0}(m) \rightarrow \text{blob }q) B^{g_{[m-c]}^{-1} \rhd y}(\text{blob }q \rightarrow \text{blob }p) \delta(y, g_{[m-c]} \rhd f_{h'}(p) ) C^g_{\rhd}(c)\\
		&= B^{g_{[m-c]} \rhd f_{h'}(p)} (\text{blob 0}(m) \rightarrow \text{blob }q) B^{f_{h'}(p)}(\text{blob }q \rightarrow \text{blob }p) C^g_{\rhd}(c).
	\end{align*}
	Substituting this into our commutation relation, Equation \ref{Equation_sphere_charge_higher_flux_commutation_9}, gives us
	\begin{align}
		C^{g}_T(c) L^e(c) C^{h,e_m}_T(m)\ket{GS}&= C^{g^{-1}_{[m-c]}hg_{[m-c]}^{\phantom{-1}}}_{\rhd}(m) \big[\prod_{p \in m} B^{g_{[m-c]} \rhd f_{h'}(p)} (\text{blob 0}(m) \rightarrow \text{blob }q) B^{f_{h'}(p)}(\text{blob }q \rightarrow \text{blob }p) \big] \notag \\
		& \hspace{0.5cm} C^g_{\rhd}(c) \big[\prod_{ p' \in c} B^{f_g(p')}(\text{blob }0(c) \rightarrow \text{blob }p') \big] B^{x} (\text{blob }0(c) \rightarrow \text{blob }q) \notag \\
		& \hspace{0.5cm} \delta(\hat{e}(m),e_m) \delta( \hat{e}(c), e \big[ g(s.p(c)-s.p(m)) \rhd \big( [h^{-1} \rhd e_m]^{-1} e_m \big) \big]) \ket{GS}. \label{Equation_sphere_charge_higher_flux_commutation_10}
	\end{align}
	
	Next, we wish to commute $B^x (\text{blob }0(c) \rightarrow \text{blob }q)$ past $C^g_{\rhd}(c)$ to join the blob ribbon operators from $m$. Because this blob ribbon operator came from $C^g_T(c)$, its start-point is $s.p(c)$ and not $s.p(m)$. To avoid confusion when we put this blob ribbon operator next to the ones from $C^h_T(m)$, we will add $|s.p(c)$ to its argument and write the ribbon operator as $B^{x} (\text{blob }0(c) \rightarrow \text{blob }q|s.p(c))$. Now consider how this blob ribbon operator commutes with $C^g_{\rhd}(c)$. There are two issues to consider. Firstly, $x$ is an operator label and will not commute with $C^g_{\rhd}(c)$. Secondly, when the start-point of $c$ is on the direct membrane of $c$, the blob ribbon operators corresponding to $c$ do not commute with $C^g_c(m)$. For a constant-labelled blob ribbon operator $B^e (\text{blob }0(c) \rightarrow \text{blob }q|s.p(c))$, from Equation \ref{blob_ribbon_magnetic_commutation} in Section \ref{Section_Magnetic_Tri_Nontrivial_Commutation} we have
	$$C^g_{\rhd}(c) B^e (\text{blob }0(c) \rightarrow \text{blob }q|s.p(c))= B^{g \rhd e} (\text{blob }0(c) \rightarrow \text{blob }q|s.p(c)) C^g_{\rhd}(c).$$
	Then writing $B^{x} (\text{blob }0(c) \rightarrow \text{blob }q|s.p(c))$ as
	$$B^{x} (\text{blob }0(c) \rightarrow \text{blob }q|s.p(c))= \sum_{e \in E} B^{e} (\text{blob }0(c) \rightarrow \text{blob }q|s.p(c)) \delta(x,e),$$
	where $$x= g(s.p(c)-s.p(m)) \rhd \big(([h^{-1} \rhd e_m] e_m^{-1}) g^{-1}_{[m-c]} \rhd ([h^{-1} \rhd e_m^{-1}] e_m)\big),$$
	we have 
	$$C^g_{\rhd}(c) B^x (\text{blob }0(c) \rightarrow \text{blob }q|s.p(c))= \sum_{e \in E} B^{g \rhd e} (\text{blob }0(c) \rightarrow \text{blob }q|s.p(c)) C^g_{\rhd}(c) \delta(x,e).$$
	
	Next we must consider how $x$ commutes with $C^g_{\rhd}(c)$. $C^g_{\rhd}(c)$ affects the path element $g(s.p(c)-s.p(m))$ according to
	$$g(s.p(c)-s.p(m)) C^g_{\rhd}(c) = C^g_{\rhd}(c) g g(s.p(c)-s.p(m))$$
	and so
	$$C^g_{\rhd}(c) g(s.p(c)-s.p(m)) = g^{-1}g(s.p(c)-s.p(m))C^g_{\rhd}(c).$$
	Therefore,
	\begin{align*}
		C^g_{\rhd}(c) x &= C^g_{\rhd}(c)g(s.p(c)-s.p(m)) \rhd \big(([h^{-1} \rhd e_m] e_m^{-1}) g^{-1}_{[m-c]} \rhd ([h^{-1} \rhd e_m^{-1}] e_m)\big)\\
		&= (g^{-1}g(s.p(c)-s.p(m))) \rhd \big(([h^{-1} \rhd e_m] e_m^{-1}) g^{-1}_{[m-c]} \rhd ([h^{-1} \rhd e_m^{-1}] e_m)\big) C^g_{\rhd}(x)\\
		&= [g^{-1} \rhd x] C^g_{\rhd}(x).
	\end{align*}
	This means that
	\begin{align*}
		C^g_{\rhd}(c) B^x (\text{blob }0(c) \rightarrow \text{blob }q|s.p(c)) &= \sum_{e \in E} B^{g \rhd e} (\text{blob }0(c) \rightarrow \text{blob }q|s.p(c)) C^g_{\rhd}(c) \delta(x,e)\\
		&= \sum_{e \in E} B^{g \rhd e} (\text{blob }0(c) \rightarrow \text{blob }q|s.p(c)) \delta(g^{-1} \rhd x,e) C^g_{\rhd}(c)\\
		&= B^e (\text{blob }0(c) \rightarrow \text{blob }q|s.p(c)) C^g_{\rhd}(c),
	\end{align*}
	so that the blob ribbon operator commutes with $C^g_{\rhd}(c)$. We note that if we had displaced $s.p(c)$ away from the direct membrane, as we did for $m$, this relationship would still hold, because in that case the constant-labelled blob ribbon operator $B^e$ would commute with $C^g_{\rhd}(c)$ and so would the path element $g(s.p(c)-s.p(m))$ (and so $x$ would commute). Substituting this into Equation \ref{Equation_sphere_charge_higher_flux_commutation_10} gives us
	\begin{align}
		C^{g}_T(c) L^e(c) C^{h,e_m}_T(m)\ket{GS}&= C^{g^{-1}_{[m-c]}hg_{[m-c]}^{\phantom{-1}}}_{\rhd}(m) \bigg( \prod_{p \in m} B^{g_{[m-c]} \rhd f_{h'}(p)} (\text{blob 0}(m) \rightarrow \text{blob }q) B^{f_{h'}(p)}(\text{blob }q \rightarrow \text{blob }p) \bigg) \notag\\
		& \hspace{0.5cm}B^{x} (\text{blob }0(c) \rightarrow \text{blob }q|s.p(c))C^g_{\rhd}(c) \big[\prod_{ p' \in c} B^{f_g(p')}(\text{blob }0(c) \rightarrow \text{blob }p') \big] \notag \\
		& \hspace{0.5cm} \delta(\hat{e}(m),e_m) \delta\bigg( \hat{e}(c), e \big[ g(s.p(c)-s.p(m)) \rhd \big( [h^{-1} \rhd e_m]^{-1} e_m \big) \big]\bigg) \ket{GS} \notag \\
		&=C^{g^{-1}_{[m-c]}hg_{[m-c]}^{\phantom{-1}}}_{\rhd}(m) \bigg(\prod_{p \in m} B^{g_{[m-c]} \rhd f_{h'}(p)} (\text{blob 0}(m) \rightarrow \text{blob }q) B^{f_{h'}(p)}(\text{blob }q \rightarrow \text{blob }p) \bigg) \notag \\
		& \hspace{0.5cm}B^{x} (\text{blob }0(c) \rightarrow \text{blob }q|s.p(c))C^g_{T}(c) \delta(\hat{e}(m),e_m) \notag \\
		& \hspace{0.5cm} \delta( \hat{e}(c), e \big[ g(s.p(c)-s.p(m)) \rhd \big( [h^{-1} \rhd e_m]^{-1} e_m \big) \big]) \ket{GS}. \label{Equation_sphere_charge_higher_flux_commutation_11}
	\end{align}
	
	We will later consider combining this blob ribbon operator with the ones from $C^h_T(m)$. Before we do this however, we will commute $\delta(\hat{e}(m),e_m)$ past $C^g_{\rhd}(c)\big(\prod_{ p' \in c} B^{f_g(p')}(\text{blob }0(c) \rightarrow \text{blob }p') \big)$, so that all of the operators pertaining to $m$ are together. Because the membrane $m$ itself is fully enclosed by $c$, the blob ribbon operators on $c$ do not penetrate $m$ and so do not affect $\hat{e}(m)$. On the other hand, the paths from the start-point of $m$ to all of the plaquettes in $m$ pass through $c$ and so are affected by $C^h_{\rhd}(c)$. This means that the surface label of $m$ is affected by $C^h_{\rhd}(c)$. In Section \ref{Section_braiding_higher_flux_higher_flux}, when considering the braiding of two higher-flux membrane operators, we saw that this leads to the commutation relation
	\begin{align*}
		\delta(e_m,\hat{e}(m))C^{g}_T(c)&=C^{g}_T(c) \delta(e_m, \hat{g}(s.p(c)-s.p(m))^{-1}g^{-1}\hat{g}(s.p(c)-(m))\rhd \hat{e}(m))\\
		&=C^{g}_T(c)\delta (g_{[m-c]}^{-1} \rhd \hat{e}(m), e_m).
	\end{align*}
	
	Reversing the relation above gives us
	$$C^{g}_T(c) \delta (\hat{e}(m), e_m)= \delta (g_{[m-c]} \rhd \hat{e}(m), e_m) C^g_T(c).$$
	Substituting this into our overall commutation relation Equation \ref{Equation_sphere_charge_higher_flux_commutation_11}, we see that
	\begin{align}
		C^{g}_T(c) L^e(c) C^{h,e_m}_T(m)\ket{GS} &=C^{g^{-1}_{[m-c]}hg_{[m-c]}^{\phantom{-1}}}_{\rhd}(m) \bigg(\prod_{p \in m} B^{g_{[m-c]} \rhd f_{h'}(p)} (\text{blob 0}(m) \rightarrow \text{blob }q) B^{f_{h'}(p)}(\text{blob }q \rightarrow \text{blob }p) \bigg) \notag\\
		& \hspace{0.5cm} B^{x} (\text{blob }0(c) \rightarrow \text{blob }q|s.p(c)) \delta(g_{[m-c]} \rhd \hat{e}(m),e_m) C^g_{T}(c) \notag\\
		& \hspace{0.5cm} \delta( \hat{e}(c), e \big[ g(s.p(c)-s.p(m)) \rhd \big( [h^{-1} \rhd e_m]^{-1} e_m \big) \big]) \ket{GS}. \label{Equation_sphere_charge_higher_flux_commutation_12}
	\end{align}
	
	We can now tidy this expression up by combining $B^{x} (\text{blob }0(c) \rightarrow \text{blob }q|s.p(c))$ with the other blob ribbon operators. We start by moving its start-point to the start-point of $m$. As explained in Section \ref{Section_blob_ribbon_move_sp}, when we move the start-point in this way we must simultaneously change the label to $g(s.p(c)-s.p(m))^{-1} \rhd x$:
	$$B^{x} (\text{blob }0(c) \rightarrow \text{blob }q|s.p(c)) = B^{g(s.p(c)-s.p(m))^{-1} \rhd x} (\text{blob }0(c) \rightarrow \text{blob }q|s.p(m)).$$
	The label $x$ is 
	$$g(s.p(c)-s.p(m)) \rhd \big([h^{-1} \rhd e_m] e_m^{-1} \big[g^{-1}_{[m-c]} \rhd ([h^{-1} \rhd e_m^{-1}] e_m)\big]\big),$$
	so 
	$$g(s.p(c)-s.p(m))^{-1} \rhd x = [h^{-1} \rhd e_m] e_m^{-1} \big[g^{-1}_{[m-c]} \rhd ([h^{-1} \rhd e_m^{-1}] e_m)\big].$$
	
	We can use $\delta(g_{[m-c]} \rhd \hat{e}(m),e_m)$ to replace the constant label $e_m$ with the operator $\hat{e}(m)$, in order to match the form in which we express the labels of the other blob ribbon operators. The label $g(s.p(c)-s.p(m))^{-1} \rhd x$ is then given by
	\begin{align*}
		\big([h^{-1} \rhd (g_{[m-c]} \rhd \hat{e}(m))] &[g_{[m-c]} \rhd \hat{e}(m)]^{-1}\big) \big[g^{-1}_{[m-c]} \rhd \big([h^{-1} \rhd(g_{[m-c]} \rhd \hat{e}(m))^{-1}] [g_{[m-c]} \rhd \hat{e}(m)]\big)\big]\\
		&= \big[g_{[m-c]} \rhd \big([ (g_{[m-c]}^{-1}h^{-1}g_{[m-c]})\rhd \hat{e}(m)] \hat{e}(m)^{-1}\big)\big] \big[[(g^{-1}_{[m-c]} h^{-1}g_{[m-c]}) \rhd \hat{e}(m)^{-1}] \hat{e}(m)\big].
	\end{align*}
	This means that we can write the associated blob ribbon operator as
	\begin{align*}
		B^{x} &(\text{blob }0(c) \rightarrow \text{blob }q|s.p(c))\\
		&=B^{\big[g_{[m-c]} \rhd \big([ (g_{[m-c]}^{-1}h^{-1}g_{[m-c]})\rhd \hat{e}(m)] \hat{e}(m)^{-1}\big)\big] \big[[(g^{-1}_{[m-c]} h^{-1}g_{[m-c]}) \rhd \hat{e}(m)^{-1}] \hat{e}(m)\big]} (\text{blob }0(c) \rightarrow \text{blob }q|s.p(m)).
	\end{align*}
	
	From Equation \ref{Equation_product_f_2}, we can recognise the expression 
	$$[ (g_{[m-c]}^{-1}h^{-1}g_{[m-c]})\rhd \hat{e}(m)^{-1}] \hat{e}(m)$$
	as 
	$$\prod_{p \in m} f_{h'}(p),$$
	where $h'= g_{[m-c]}^{-1}hg_{[m-c]}$. Therefore, the blob ribbon operator is
	\begin{align*}
		B^{x} (\text{blob }0(c) \rightarrow \text{blob }q|s.p(c))&=B^{ \big[ \prod_{p \in m} g_{[m-c]} \rhd f_{h'}(p)^{-1} \big] \big[ \prod_{p \in m} f_{h'}(p) \big] } (\text{blob }0(c) \rightarrow \text{blob }q|s.p(m)).
	\end{align*}
	We split this blob ribbon operator into two parts (on the same ribbon), to give
	\begin{align*}
		B&^{ \big[ \prod_{p \in m} g_{[m-c]} \rhd f_{h'}(p)^{-1} \big] \big[ \prod_{p \in m} f_{h'}(p) \big] } (\text{blob }0(c) \rightarrow \text{blob }q|s.p(m))\\
		&= B^{ \big[ \prod_{p \in m} g_{[m-c]} \rhd f_{h'}(p)^{-1}\big] } (\text{blob }0(c) \rightarrow \text{blob }q|s.p(m)) B^{\big[\prod_{p \in m} f_{h'}(p)\big] } (\text{blob }0(c) \rightarrow \text{blob }q|s.p(m)).
	\end{align*}
	We then reverse the orientation of the dual path of the first part, which also inverts the label, to obtain
	\begin{align*}
		B&^{ \big[ \prod_{p \in m} g_{[m-c]} \rhd f_{h'}(p)^{-1} \big] \big[ \prod_{p \in m} f_{h'}(p) \big] } (\text{blob }0(c) \rightarrow \text{blob }q|s.p(m))\\
		&= B^{ \big[\prod_{p \in m} g_{[m-c]} \rhd f_{h'}(p)\big] } (\text{blob }q \rightarrow \text{blob }0(c)|s.p(m)) B^{ \big[\prod_{p \in m} f_{h'}(p)\big] } (\text{blob }0(c) \rightarrow \text{blob }q|s.p(m)).
	\end{align*}
	Then we split the ribbon operators into one part for each plaquette $p \in m$, to obtain
	\begin{align*}
		\bigg(\prod_{p \in m} B^{ g_{[m-c]} \rhd f_{h'}(p)} (\text{blob }q \rightarrow \text{blob }0(c)|s.p(m)) B^{ f_{h'}(p) } (\text{blob }0(c) \rightarrow \text{blob }q|s.p(m))\bigg).
	\end{align*}
	That is, we have found that
	\begin{align}
		B^{x}(\text{blob }0(c) &\rightarrow \text{blob }q|s.p(c)) \notag \\
		&= \bigg(\prod_{p \in m} B^{ g_{[m-c]} \rhd f_{h'}(p)} (\text{blob }q \rightarrow \text{blob }0(c)|s.p(m)) B^{ f_{h'}(p) } (\text{blob }0(c) \rightarrow \text{blob }q|s.p(m)) \bigg).
	\end{align}
	Substituting this into our commutation relation Equation \ref{Equation_sphere_charge_higher_flux_commutation_12}, we obtain
	\begin{align}
		C^{g}_T(c) L^e(c) C^{h,e_m}_T(m)\ket{GS} &=C^{g^{-1}_{[m-c]}hg_{[m-c]}^{\phantom{-1}}}_{\rhd}(m) \bigg(\prod_{p \in m} B^{g_{[m-c]} \rhd f_{h'}(p)} (\text{blob 0}(m) \rightarrow \text{blob }q) B^{f_{h'}(p)}(\text{blob }q \rightarrow \text{blob }p) \bigg) \notag \\
		& \hspace{0.5cm} \bigg(\prod_{p \in m} B^{ g_{[m-c]} \rhd f_{h'}(p)} (\text{blob }q \rightarrow \text{blob }0(c)|s.p(m)) B^{ f_{h'}(p) } (\text{blob }0(c) \rightarrow \text{blob }q|s.p(m)) \bigg) \notag \\
		& \hspace{0.5cm} \delta(g_{[m-c]} \rhd \hat{e}(m),e_m) C^g_{T}(c) \delta\bigg( \hat{e}(c), e \big[ g(s.p(c)-s.p(m)) \rhd \big( [h^{-1} \rhd e_m]^{-1} e_m \big) \big]\bigg) \ket{GS}. \label{Equation_sphere_charge_higher_flux_commutation_13}
	\end{align}
	
	Now we can see why we performed all these manipulations on $B^{x}(\text{blob }0(c) \rightarrow \text{blob }q|s.p(c))$. We now have two sets of blob ribbon operators, with each set including two blob ribbon operators for each plaquette $p$. Furthermore the labels of the blob ribbon operators from the two sets are the same. That is, for each blob ribbon operator 
	$$B^{f_{h'}(p)}(\text{blob }q \rightarrow \text{blob }p)$$
	we have another blob ribbon operator
	$$B^{ f_{h'}(p) } (\text{blob }0(c) \rightarrow \text{blob }q|s.p(m)).$$
	These blob ribbon operators have the same start-point ($s.p(m)$) and label, and their dual paths connect together, so we can concatenate these ribbon operators (as explained in Section \ref{Section_blob_ribbon_concatenate}), to obtain a single blob ribbon operator 
	$$B^{ f_{h'}(p) } (\text{blob }0(c) \rightarrow \text{blob }p).$$ 
	
	As we discussed in Section \ref{Section_Magnetic_Tri_Non_Trivial_Definition}, the label $f_{h'}(p)$ is in the kernel of $\partial$, because it has the form $e [z \rhd e^{-1}]$ for some $e \in E$ and $z \in G$ and $\partial(e)= \partial(z \rhd e)$ when $\partial$ maps to the centre of $G$. This means that the blob ribbon operator is topological (as explained in Section \ref{Section_Topological_Blob_Ribbons}) and so we do not need to worry about the precise position of the dual path. This means that (as we expect) the precise position of the plaquette $q$ is irrelevant.

	Similarly, for each blob ribbon operator
	$$B^{g_{[m-c]} \rhd f_{h'}(p)} (\text{blob 0}(m) \rightarrow \text{blob }q)$$
	we have a blob ribbon operator
	$$B^{ g_{[m-c]} \rhd f_{h'}(p)} (\text{blob }q \rightarrow \text{blob }0(c)|s.p(m)).$$
	Again, these blob ribbon operators can be concatenated into a single ribbon operator
	$$B^{ g_{[m-c]} \rhd f_{h'}(p)} (\text{blob }0(m) \rightarrow \text{blob }0(c)|s.p(m)).$$
	This means that our commutation relation Equation \ref{Equation_sphere_charge_higher_flux_commutation_13} becomes
	\begin{align}
		C^{g}_T(c) L^e(c)& C^{h,e_m}_T(m)\ket{GS} \notag \\
		&=C^{g^{-1}_{[m-c]}hg_{[m-c]}^{\phantom{-1}}}_{\rhd}(m) \bigg( \prod_{p \in m} B^{g_{[m-c]} \rhd f_{h'}(p)} (\text{blob 0}(m) \rightarrow \text{blob }0(c)) B^{f_{h'}(p)}(\text{blob }0(c) \rightarrow \text{blob }p) \bigg) \notag\\
		& \hspace{0.5cm} \delta(g_{[m-c]} \rhd \hat{e}(m),e_m) C^g_{T}(c) \delta( \hat{e}(c), e \big[ g(s.p(c)-s.p(m)) \rhd \big( [h^{-1} \rhd e_m]^{-1} e_m \big) \big]) \ket{GS}, \label{Equation_sphere_charge_higher_flux_commutation_14}
	\end{align}
	which no longer has any explicit reference to the arbitrary intersection plaquette $q$. We see that, as we discussed in the context of braiding between two higher-flux membrane operators in Section \ref{Section_braiding_higher_flux_higher_flux}, the blob ribbon operators of $m$ are diverted to blob 0 of $c$. The blob ribbon operators enter blob 0 of $c$ with one label and exit with another label. Apart from this splitting of the blob ribbon operators, the commutation relation between the two higher-flux membrane operators just changes the labels of the higher-flux membrane operators.

	Having seen the commutation relation, we now wish to replace $C^{g,e}_T(c)$ with a spherical measurement operator by taking an appropriate sum over $g$ and $e$. As explained in Section \ref{Section_sphere_topological_charge_appendix_full}, a spherical measurement operator $T^{R,C}(c)$ is defined by
	$$T^{R,C}(c) = \frac{|R|}{|Z_{\rhd,r_C}|} \sum_{d \in Z_{ \rhd, r_C}} \sum_{q \in Q_C} \chi_R(d) C_T^{qdq^{-1}, q \rhd r_C}(c).$$
	Here $C$ is a $\rhd$-class of the kernel of $\partial$ with representative $r_C$ (so that the elements $e$ of $C$ satisfy $e= g \rhd r_C$ for some $g \in G$). Then $Z_{\rhd,r_C}$ is the set of elements $x$ of $G$ satisfying $x \rhd r_C =r_C$, while $Q_C$ is a set of representatives from $G$ so that each element $e_i \in C$ has a unique $q_i \in Q_C$ such that $e_i =q_i \rhd r_C$. Then from Equation \ref{Equation_sphere_charge_higher_flux_commutation_14}, we have
	\begin{align}
		T^{R,C}(c)& C^{h,e_m}_T(m)\ket{GS} \notag\\
		&= \frac{|R|}{|Z_{\rhd,r_C}|} \sum_{d \in Z_{ \rhd, r_C}} \sum_{q \in Q_C} \chi_R(d) C^{(qdq^{-1})^{-1}_{[m-c]}h(qdq^{-1})_{[m-c]}^{\phantom{-1}}}_{\rhd}(m) \notag \\
		&\hspace{0.5cm} \bigg(\prod_{p \in m} B^{(qdq^{-1})_{[m-c]} \rhd f_{h'}(p)} (\text{blob 0}(m) \rightarrow \text{blob }0(c)) B^{f_{h'}(p)}(\text{blob }0(c) \rightarrow \text{blob }p) \bigg) \notag \\
		& \hspace{0.5cm} \delta((qdq^{-1})_{[m-c]} \rhd \hat{e}(m),e_m) C^{qdq^{-1}}_{T}(c) \delta( \hat{e}(c), [q \rhd r_C] \big[ g(s.p(c)-s.p(m)) \rhd \big( [h^{-1} \rhd e_m]^{-1} e_m \big) \big]) \ket{GS}. \label{Equation_sphere_charge_higher_flux_measurement_1}
	\end{align} 
	
	We can now start to evaluate this by applying the properties of closed and contractible membrane operators acting on the ground state. We know that the surface label of a contractible closed membrane in the ground state must be the identity. Therefore,
	$$\delta( \hat{e}(c), [q \rhd r_C] \big[ g(s.p(c)-s.p(m)) \rhd \big( [h^{-1} \rhd e_m]^{-1} e_m \big) \big]) \ket{GS} = \delta([q \rhd r_C], g(s.p(c)-s.p(m)) \rhd \big( [h^{-1} \rhd e_m] e_m^{-1} \big) )\ket{GS}.$$
	This is only non-zero when 
	$$[h^{-1} \rhd e_m] e_m^{-1} = (g(s.p(c)-s.p(m))^{-1}q) \rhd r_C,$$
	which means that $[h^{-1} \rhd e_m] e_m^{-1} $ is in the class $C$. In other words, the result of applying the measurement operator is only non-zero when $C$ is the class containing $[h^{-1} \rhd e_m] e_m^{-1} $. This determines one of the labels of the point-like topological charge of the higher-flux loop excitation. We then have
	\begin{align}
		T^{R,C}(c)& C^{h,e_m}_T(m)\ket{GS} \notag\\
		&= \frac{|R|}{|Z_{\rhd,r_C}|} \sum_{d \in Z_{ \rhd, r_C}} \sum_{q \in Q_C} \chi_R(d) \delta( [h^{-1} \rhd e_m]^{-1} e_m \in C) C^{(qdq^{-1})^{-1}_{[m-c]}h(qdq^{-1})_{[m-c]}^{\phantom{-1}}}_{\rhd}(m) \notag\\
		& \hspace{0.5cm} \bigg(\prod_{p \in m} B^{(qdq^{-1})_{[m-c]} \rhd f_{h'}(p)} (\text{blob 0}(m) \rightarrow \text{blob }0(c)) B^{f_{h'}(p)}(\text{blob }0(c) \rightarrow \text{blob }p) \bigg) \notag \\
		& \hspace{0.5cm} \delta((qdq^{-1})_{[m-c]} \rhd \hat{e}(m),e_m) C^{qdq^{-1}}_{T}(c) \delta( q \rhd r_C, g(s.p(c)-s.p(m)) \rhd \big( [h^{-1} \rhd e_m] e_m^{-1} \big) ) \ket{GS}. \label{Equation_sphere_charge_higher_flux_measurement_2}
	\end{align}
	
	Next we want to move $C^{qdq^{-1}}_T(c)$ so that it is applied directly on the ground state, by commuting it past
	$$\delta( [q \rhd r_C], \big[ g(s.p(c)-s.p(m)) \rhd \big( [h^{-1} \rhd e_m] e_m^{-1} \big) \big]).$$
	To do so, we note that
	$$g(s.p(c)-s.p(m)) C^{qdq^{-1}}_T(c) = C^{qdq^{-1}}_T(c) qdq^{-1} g(s.p(c)-s.p(m))$$
	and so
	\begin{align*}
		C^{qdq^{-1}}_{T}(c) & \delta( [q \rhd r_C], g(s.p(c)-s.p(m)) \rhd \big( [h^{-1} \rhd e_m] e_m^{-1} \big) )\\
		&= \delta( [q \rhd r_C], (qd^{-1}q^{-1}g(s.p(c)-s.p(m))) \rhd \big( [h^{-1} \rhd e_m] e_m^{-1} \big))C^{qdq^{-1}}_T(c).
	\end{align*}
	
	Then the magnetic membrane operator $C^{qdq^{-1}}_T(c)$ acting on a closed contractible membrane in the ground state is trivial, as we showed in Section \ref{Section_Topological_Magnetic_Tri_Nontrivial}. Therefore, Equation \ref{Equation_sphere_charge_higher_flux_measurement_2} becomes
	\begin{align}
		T^{R,C}(c) &C^{h,e_m}_T(m)\ket{GS} \notag\\
		&= \delta( [h^{-1} \rhd e_m]^{-1} e_m \in C) \frac{|R|}{|Z_{\rhd,r_C}|} \sum_{d \in Z_{ \rhd, r_C}} \sum_{q \in Q_C} \chi_R(d) C^{(qdq^{-1})^{-1}_{[m-c]}h(qdq^{-1})_{[m-c]}^{\phantom{-1}}}_{\rhd}(m) \notag \\
		& \hspace{0.5cm} \bigg(\prod_{p \in m} B^{(qdq^{-1})_{[m-c]} \rhd f_{h'}(p)} (\text{blob 0}(m) \rightarrow \text{blob }0(c)) B^{f_{h'}(p)}(\text{blob }0(c) \rightarrow \text{blob }p) \bigg) \notag \\
		& \hspace{0.5cm} \delta((qdq^{-1})_{[m-c]} \rhd \hat{e}(m),e_m) \delta( q \rhd r_C,  (qd^{-1}q^{-1}g(s.p(c)-s.p(m)))\rhd \big( [h^{-1} \rhd e_m] e_m^{-1} \big) ) \ket{GS}. \label{Equation_sphere_charge_higher_flux_measurement_3}
	\end{align}
	
	We can simplify the rightmost Kronecker delta slightly. We can act on both parts of the delta with $qdq^{-1} \rhd$ (using the fact that this is an isomorphism), to obtain
	\begin{align*}
		\delta( [q \rhd r_C], (qd^{-1}q^{-1}&g(s.p(c)-s.p(m)))\rhd \big( [h^{-1} \rhd e_m] e_m^{-1} \big) )\\
		&= \delta( [(qdq^{-1}) \rhd (q \rhd r_C)], g(s.p(c)-s.p(m))\rhd \big( [h^{-1} \rhd e_m] e_m^{-1} \big) )\\
		&=\delta( (qd) \rhd r_C, (g(s.p(c)-s.p(m)))\rhd \big( [h^{-1} \rhd e_m] e_m^{-1} \big) ).
	\end{align*}
	
	Then we note that $d$ is in $Z_{\rhd,C}$ and so $d \rhd r_C=r_C$ by the definition of $Z_{\rhd, r_C}$. Therefore, $(qd)\rhd r_c= q \rhd r_C$ and our original Kronecker delta from Equation \ref{Equation_sphere_charge_higher_flux_measurement_3} becomes
	\begin{align*}
		\delta( q \rhd r_C, (qd^{-1}q^{-1}g(s.p(c)-s.p(m)))\rhd \big( [h^{-1} \rhd e_m] e_m^{-1} \big) )&= \delta( q \rhd r_C, (g(s.p(c)-s.p(m)))\rhd \big( [h^{-1} \rhd e_m] e_m^{-1} \big) ).
	\end{align*}
	This means that
	\begin{align}
		T^{R,C}(c) C^{h,e_m}_T(m)\ket{GS} &= \delta( [h^{-1} \rhd e_m]^{-1} e_m \in C)\frac{|R|}{|Z_{\rhd,r_C}|} \sum_{d \in Z_{ \rhd, r_C}} \sum_{q \in Q_C} \chi_R(d) C^{(qdq^{-1})^{-1}_{[m-c]}h(qdq^{-1})_{[m-c]}^{\phantom{-1}}}_{\rhd}(m) \notag \\
		& \hspace{0.5cm} \prod_{p \in m} \big(B^{(qdq^{-1})_{[m-c]} \rhd f_{h'}(p)} (\text{blob 0}(m) \rightarrow \text{blob }0(c)) B^{f_{h'}(p)}(\text{blob }0(c) \rightarrow \text{blob }p) \big) \notag\\
		& \hspace{0.5cm}\delta((qdq^{-1})_{[m-c]} \rhd \hat{e}(m),e_m) \delta( q \rhd r_C,  g(s.p(c)-s.p(m))\rhd \big( [h^{-1} \rhd e_m] e_m^{-1} \big) ) \ket{GS}. \label{Equation_sphere_charge_higher_flux_measurement_4}
	\end{align}
	
	Having removed the operators $\hat{e}(c)$ and $C_t^{qdq^{-1}}(c)$ corresponding to the closed membrane $c$, we now wish to recombine the parts of the membrane operator on $m$ into a sum of higher-flux membrane operators. However, there is an obstruction to this in that the blob ribbon operators corresponding to $m$ are split into two parts with two different labels. First consider the blob ribbon operators with label $(qdq^{-1})_{[m-c]} \rhd f_{h'}(p)$. These blob ribbon operators all act on the same ribbon $(\text{blob 0}(m) \rightarrow \text{blob }0(c))$. We can therefore combine them into a single ribbon operator:
	\begin{equation}
		\prod_{p \in m} B^{(qdq^{-1})_{[m-c]} \rhd f_{h'}(p)}(\text{blob 0}(m) \rightarrow \text{blob }0(c)) = B^{(qdq^{-1})_{[m-c]} \rhd (\prod_{p \in m} f_{h'}(p))}(\text{blob 0}(m) \rightarrow \text{blob }0(c)). \label{Equation_magnetic_sphere_charge_combine_blob_ribbons}
	\end{equation}
	
	Now consider the label of this blob ribbon operator $(qdq^{-1})_{[m-c]} \rhd (\prod_{p \in m} f_{h'}(p))$. From Equation \ref{Equation_product_f_2}, we know that
	\begin{align}
		\prod_{p \in m} f_{h'}(p) &= [ {h'}^{-1}\rhd \hat{e}(m)^{-1}] \hat{e}(m)\notag \\
		&=[ (g_{[m-c]}^{-1}h^{-1}g_{[m-c]})\rhd \hat{e}(m)^{-1}] \hat{e}(m) \notag \\
		&=[((qdq^{-1})_{[m-c]}^{-1}h^{-1}(qdq^{-1})_{[m-c]})\rhd \hat{e}(m)^{-1} ] \hat{e}(m),
	\end{align}
	where in the last step we substituted the value of $g$ from our projector. Therefore, the label of the combined blob ribbon operator is
	\begin{align*}
		(qdq^{-1})_{[m-c]} \rhd (\prod_{p \in m} f_{h'}(p))&=(qdq^{-1})_{[m-c]} \rhd ( [ ((qdq^{-1})_{[m-c]}^{-1}h^{-1}(qdq^{-1})_{[m-c]})\rhd \hat{e}(m)^{-1}] [\hat{e}(m)])\\
		&= [(h^{-1}(qdq^{-1})_{[m-c]})\rhd \hat{e}(m)^{-1}] [(qdq^{-1})_{[m-c]} \rhd \hat{e}(m)].
	\end{align*}
	Then we can use the $\delta((qdq^{-1})_{[m-c]} \rhd \hat{e}(m),e_m)$ to replace the operator $\hat{e}(m)$ to give
	\begin{align}
		(qdq^{-1})_{[m-c]} \rhd \big(\prod_{p \in m} f_{h'}(p)\big)&=
		[(h^{-1}(qdq^{-1})_{[m-c]})\rhd \hat{e}(m)^{-1}] [(qdq^{-1})_{[m-c]} \rhd \hat{e}(m)^{-1}] \notag \\
		&= [h^{-1} \rhd e_m^{-1}] e_m. \label{Equation_magnetic_sphere_charge_ribbon_label}
	\end{align}
	
	Now consider the expression 
	$$\prod_{p \in m} f_{h'}(p),$$
	which we can write as 
	$$(qdq^{-1})_{[m-c]}^{-1} \rhd \bigg((qdq^{-1})_{[m-c]} \rhd \big(\prod_{p \in m} f_{h'}(p)\big)\bigg)$$
	by inserting the identity in the form of $(qdq^{-1})_{[m-c]}^{-1} \rhd (qdq^{-1})_{[m-c]} \rhd$. From the argument above, this is given by
	\begin{align*}
		\prod_{p \in m} f_{h'}(p)&= (qdq^{-1})_{[m-c]}^{-1} \rhd( [h^{-1} \rhd e_m^{-1}] e_m).
	\end{align*}	
	Recalling that
	$$(qdq^{-1})_{[m-c]}= (g(s.p(c)-s.p(m))^{-1}qdq^{-1} g(s.p(c)-s.p(m))),$$
	we therefore have
	\begin{align*}
		\prod_{p \in m} f_{h'}(p)&= (g(s.p(c)-s.p(m))^{-1}qd^{-1}q^{-1} g(s.p(c)-s.p(m))) \rhd ( [h^{-1} \rhd e_m^{-1}] e_m)\\
		&= (g(s.p(c)-s.p(m))^{-1}qd^{-1}q^{-1}) \rhd \big( g(s.p(c)-s.p(m)) \rhd ( [h^{-1} \rhd e_m^{-1}] e_m) \big).
	\end{align*}
	
	We can then use the Kronecker delta $\delta( [q \rhd r_C], g(s.p(c)-s.p(m))\rhd \big( [h^{-1} \rhd e_m] e_m^{-1} \big) )$ from Equation \ref{Equation_sphere_charge_higher_flux_measurement_4} to replace $g(s.p(c)-s.p(m)) \rhd ( [h^{-1} \rhd e_m^{-1}] e_m)$ with $(q \rhd r_C)^{-1}$, obtaining
	\begin{align*}
		\prod_{p \in m} f_{h'}(p)&= (g(s.p(c)-s.p(m))^{-1}qd^{-1}q^{-1}) \rhd (q \rhd r_C)^{-1}\\
		&= (g(s.p(c)-s.p(m))^{-1}qd^{-1}) \rhd r_C^{-1}.
	\end{align*}
	Then because $d$ is in $Z_{\rhd,C}$, $d \rhd r_C =r_C$ and so $r_C=d^{-1} \rhd r_C$ (and the same must be true of $r_C^{-1}$ because $d^{-1} \rhd$ is a group homomorphism on $E$). Therefore, 
	\begin{align*}
		\prod_{p \in m} f_{h'}(p)&= (g(s.p(c)-s.p(m))^{-1}q) \rhd r_C^{-1}\\
		&= g(s.p(c)-s.p(m))^{-1} \rhd (q \rhd r_C)^{-1}.
	\end{align*}
	
	We can then once again use the Kronecker delta $\delta( [q \rhd r_C], g(s.p(c)-s.p(m))\rhd \big( [h^{-1} \rhd e_m] e_m^{-1} \big) )$ from Equation \ref{Equation_sphere_charge_higher_flux_measurement_4}, this time to replace $q \rhd r_C$. We then have
	\begin{align*}
		\prod_{p \in m} f_{h'}(p)&=g(s.p(c)-s.p(m))^{-1} \rhd( g(s.p(c)-s.p(m))\rhd \big( [h^{-1} \rhd e_m^{-1}] e_m \big) )\\
		&=[h^{-1} \rhd e_m^{-1}] e_m.
	\end{align*}
	Then from Equation \ref{Equation_magnetic_sphere_charge_ribbon_label}, we see that
	\begin{equation}
		\prod_{p \in m} f_{h'}(p)= \prod_{p \in m} (qdq^{-1})_{[m-c]} \rhd f_{h'}(p).
	\end{equation}
	This means that we can replace the label $ \prod_{p \in m} (qdq^{-1})_{[m-c]} \rhd f_{h'}(p)$ of the combined ribbon operator with $ \prod_{p \in m} f_{h'}(p)$ in Equation \ref{Equation_magnetic_sphere_charge_combine_blob_ribbons} to obtain
	\begin{align*}
		B^{\big(\prod_{p \in m} (qdq^{-1})_{[m-c]} \rhd ( f_{h'}(p))\big)}(\text{blob 0}(m) \rightarrow \text{blob }0(c))&= B^{\prod_{p \in m} f_{h'}(p)}(\text{blob 0}(m) \rightarrow \text{blob }0(c))\\
		&= \prod_{p \in m} B^{f_{h'}(p)}(\text{blob 0}(m) \rightarrow \text{blob }0(c)),
	\end{align*}
	where in the last step we split the combined blob ribbon operator into ribbon operators for each plaquette $p$. These blob ribbon operators now have the same label as the other ribbon operator,
	$$B^{f_{h'}(p)} (\text{blob 0}(c) \rightarrow \text{blob }p),$$
	and have dual paths that can be concatenated. Therefore, we connect these blob ribbon operators into one blob ribbon operator for each plaquette $p$:
	\begin{align*}
		\bigg(\prod_{p \in m} B^{(qdq^{-1})_{[m-c]} \rhd f_{h'}(p)} (\text{blob 0}(m) \rightarrow \text{blob }0(c)) B^{f_{h'}(p)}(\text{blob }0(c) \rightarrow \text{blob }p) \bigg)& = \prod_{p \in m} B^{f_{h'}(p)}(\text{blob }0(m) \rightarrow \text{blob }p).
	\end{align*}
	
	Substituting this into our total state from Equation \ref{Equation_sphere_charge_higher_flux_measurement_4}, we have
	\begin{align}
		T^{R,C}(c) C^{h,e_m}_T(m)\ket{GS} &= \delta( [h^{-1} \rhd e_m]^{-1} e_m \in C) \frac{|R|}{|Z_{\rhd,r_C}|} \sum_{d \in Z_{ \rhd, r_C}} \sum_{q \in Q_C} \chi_R(d) C^{(qdq^{-1})^{-1}_{[m-c]}h(qdq^{-1})_{[m-c]}^{\phantom{-1}}}_{\rhd}(m) \notag\\
		& \hspace{0.5cm}\big[ \prod_{p \in m} B^{f_{h'}(p)}(\text{blob }0(m) \rightarrow \text{blob }p) \big] \delta((qdq^{-1})_{[m-c]} \rhd \hat{e}(m),e_m)\notag\\
		& \hspace{0.5cm} \delta( [q \rhd r_C], g(s.p(c)-s.p(m))\rhd \big( [h^{-1} \rhd e_m] e_m^{-1} \big) ) \ket{GS}. \label{Equation_sphere_charge_higher_flux_measurement_5}
	\end{align}
	
	If we rewrite $\delta((qdq^{-1})_{[m-c]} \rhd \hat{e}(m),e_m)$ as $\delta(\hat{e}(m),(qdq^{-1})^{-1}_{[m-c]} \rhd e_m)$, and note that $h' = (qdq^{-1})^{-1}_{[m-c]}h(qdq^{-1})_{[m-c]}^{\phantom{-1}}$, then we can recognise
	$$C^{(qdq^{-1})^{-1}_{[m-c]}h(qdq^{-1})_{[m-c]}^{\phantom{-1}}}_{\rhd}(m)\big[ \prod_{p \in m} B^{f_{h'}(p)}(\text{blob }0(m) \rightarrow \text{blob }p) \big] \delta( \hat{e}(m), (qd^{-1}q^{-1})_{[m-c]} \rhd e_m)$$
	as the higher-flux membrane operator 
	$$C^{(qdq^{-1})^{-1}_{[m-c]}h(qdq^{-1})_{[m-c]}^{\phantom{-1}}, (qd^{-1}q^{-1})_{[m-c]} \rhd e_m}_{T}(m).$$
	
	Therefore, we have
	\begin{align} 
		T^{R,C}(c)& C^{h,e_m}_T(m)\ket{GS}\notag \\
		&= \delta( [h^{-1} \rhd e_m]^{-1} e_m \in C) \frac{|R|}{|Z_{\rhd,r_C}|} \sum_{d \in Z_{ \rhd, r_C}} \sum_{q \in Q_C} \chi_R(d) C^{(qdq^{-1})^{-1}_{[m-c]}h(qdq^{-1})_{[m-c]}^{\phantom{-1}}, (qd^{-1}q^{-1})_{[m-c]} \rhd e_m}_{T}(m) \notag\\
		& \hspace{0.5cm}\delta( [q \rhd r_C], g(s.p(c)-s.p(m))\rhd \big( [h^{-1} \rhd e_m] e_m^{-1} \big) ) \ket{GS}. \label{Equation_sphere_charge_higher_flux_measurement_6}
	\end{align}
	
	We can simplify this expression further by considering $\delta( [q \rhd r_C], g(s.p(c)-s.p(m))\rhd \big( [h^{-1} \rhd e_m] e_m^{-1} \big) )$. From this, we already deduced that the class $C$ must contain $[h^{-1} \rhd e_m] e_m^{-1}$, otherwise we get a result of zero. We therefore extracted a factor of $\delta( [h^{-1} \rhd e_m] e_m^{-1} \in C )$ to write the Kronecker delta as
	$$\delta( [q \rhd r_C], g(s.p(c)-s.p(m))\rhd \big( [h^{-1} \rhd e_m] e_m^{-1} \big) ) = \delta( [h^{-1} \rhd e_m] e_m^{-1} \in C ) \delta( [q \rhd r_C], g(s.p(c)-s.p(m))\rhd \big( [h^{-1} \rhd e_m] e_m^{-1} \big) ).$$
	We can also obtain further information from the rightmost part of this Kronecker delta. To do so, we write
	$$\delta( [q \rhd r_C], g(s.p(c)-s.p(m))\rhd \big( [h^{-1} \rhd e_m] e_m^{-1} \big) )$$
	as
	$$\delta( (g(s.p(c)-s.p(m))^{-1}q) \rhd r_C, [h^{-1} \rhd e_m] e_m^{-1}).$$
	This will fix the value of $q$ in terms of $g(s.p(c)-s.p(m))$, $r_C$ and $ [h^{-1} \rhd e_m] e_m^{-1}$ and we can use this fact to remove the sum over $q$. However, the labels of our operators depend on the combination
	\begin{align*}
		(qdq^{-1})_{[m-c]} &= g(s.p(c)-s.p(m))^{-1}q dq^{-1}g(s.p(c)-s.p(m))\\
		&= (g(s.p(c)-s.p(m))^{-1}q) d(g(s.p(c)-s.p(m))^{-1}q)^{-1}.
	\end{align*}
	Therefore, it is convenient to consider the combination $g(s.p(c)-s.p(m))^{-1}q$. We can decompose this (just as we can decompose an arbitrary element of $G$) in terms of some $q' \in Q_C$ and some $x \in Z_{\rhd,C}$: 
	$$g(s.p(c)-s.p(m))^{-1}q =q' x,$$
	where $q'$ and $x$ are generally operators (they depend on $g(s.p(c)-s.p(m))$). Then the Kronecker delta becomes
	$$\delta( (q'x) \rhd r_C, [h^{-1} \rhd e_m] e_m^{-1}) = \delta(q' \rhd r_C, [h^{-1} \rhd e_m] e_m^{-1}),$$
	where we used the fact that $x \rhd r_C =r_C$ because $x \in Z_{\rhd,r_C}$ by definition. Then we have
	\begin{align} 
		T^{R,C}(c)& C^{h,e_m}_T(m)\ket{GS}\notag \\
		&= \delta( [h^{-1} \rhd e_m]^{-1} e_m \in C) \frac{|R|}{|Z_{\rhd,r_C}|} \sum_{d \in Z_{ \rhd, r_C}} \sum_{q' \in Q_C} \chi_R(d) C^{(q'xd(q'x)^{-1})^{-1}h(q'xd (q'x)^{-1}), \: (q'xd^{-1}(q'x)^{-1}) \rhd e_m}_{T}(m) \notag\\
		& \hspace{0.5cm}\delta( q' \rhd r_C, [h^{-1} \rhd e_m] e_m^{-1}) \ket{GS}. 
	\end{align}

	Because the $q \in Q_C$ are representatives such that there is a unique $q \in Q_C$ satisfying $ q \rhd r_C =e$ for any $e \in C$, we see that the value of $q'$ which contributes to the sum is completely fixed by the Kronecker delta, and the delta does not depend on any operators so the contributing value of $q'$ is also not an operator (on the other hand, $x$ is still an operator). We will denote this contributing value of $q'$ by $q(h,e_m)$. Then we can remove the sum over $q'$ by replacing $q'$ with the contributing value $q(h,e_m)$. This means that the result of our measurement is
	\begin{align}
		T^{R,C}(c) C^{h,e_m}_T(m)\ket{GS} &= \delta( [h^{-1} \rhd e_m] e_m^{-1} \in C ) \frac{|R|}{|Z_{\rhd,r_C}|} \sum_{d \in Z_{ \rhd, r_C}} \chi_R(d) \notag\\
		& \hspace{0.5cm} C^{(q(h,e_m)xdx^{-1}q(h,e_m)^{-1})^{-1}h(q(h,e_m)xdx^{-1}q(h,e_m)^{-1}), \: (q(h,e_m)xdx^{-1}q(h,e_m)^{-1})\rhd e_m}_{T}(m) \ket{GS}. \label{Equation_sphere_charge_higher_flux_measurement_7}
	\end{align}
	
	Now $x$ is in $Z_{\rhd,r_C}$ and $\chi_R$ is the character for an irrep of $Z_{\rhd,r_C}$, so $\chi_R(d)=\chi_R(xdx^{-1})$ because the character is a function of conjugacy class. We can then change the dummy variable $d \in Z_{ \rhd, r_C}$ with $xdx^{-1}=d'$ to obtain
	\begin{align}
		T^{R,C}(c) C^{h,e_m}_T(m)\ket{GS}&= \delta( [h^{-1} \rhd e_m] e_m^{-1} \in C ) \frac{|R|}{|Z_{\rhd,r_C}|} \sum_{d' \in Z_{ \rhd, r_C}} \chi_R(d') \notag\\
		& \hspace{0.5cm} C^{(q(h,e_m)d'q(h,e_m)^{-1})^{-1}h(q(h,e_m)d^{\prime -1}q(h,e_m)^{-1}), \: (q(h,e_m)d^{\prime -1}q(h,e_m)^{-1})\rhd e_m}_{T}(m) \ket{GS}. \label{Equation_sphere_charge_higher_flux_measurement_8}
	\end{align}
	
	This eliminates the operator dependence in the labels of the resulting higher-flux membrane operator. Determining which irreps $R$ give a non-zero result (i.e., which charges are carried by the excitation produced by $C^{h,e_m}_T(m)$, where the excitation could carry a mixture of charges because the excitation is not usually in an eigenstate of topological charge) is then a purely group-theoretic task. For example, consider the case where $[h^{-1} \rhd e_m] e_m^{-1}=1_E$, so that the excitation carries trivial 2-flux. In this case, the only class $C$ which gives a non-zero result is the trivial class $C= \set{1_E}$, so we can identify this as one of the charge labels for the excitation. When $C$ is trivial $Z_{\rhd,r_C}$ is the entire group $G$, while $Q_C$ is the trivial group containing only the identity. Then the result of the measurement is
	$$T^{R,C}(c) C^{h,e_m}_T(m)\ket{GS} = \delta( 1_E \in C ) \frac{|R|}{|G|} \sum_{d \in G} \chi_R(d') C^{(d')^{-1}h(d'), \: (d')\rhd e_m}_{T}(m) \ket{GS}.$$
	We see that this mixes higher-flux membrane operators with labels in a particular orbit defined by of pairs $(h,e_m)$ under the group action $d: (h,e_m) = (d^{-1}hd,d \rhd e_m)$. If we consider a higher-flux membrane operator made of a sum of different higher-flux membrane operators with labels in this orbit, then the coefficients can be decomposed in terms of certain irreps of $G$ (irreps which must be trivial for the subgroup of $G$ which acts trivially via the group action). Then using the Grand Orthogonality Theorem with this decomposition will give the contributing charges.

\end{document}